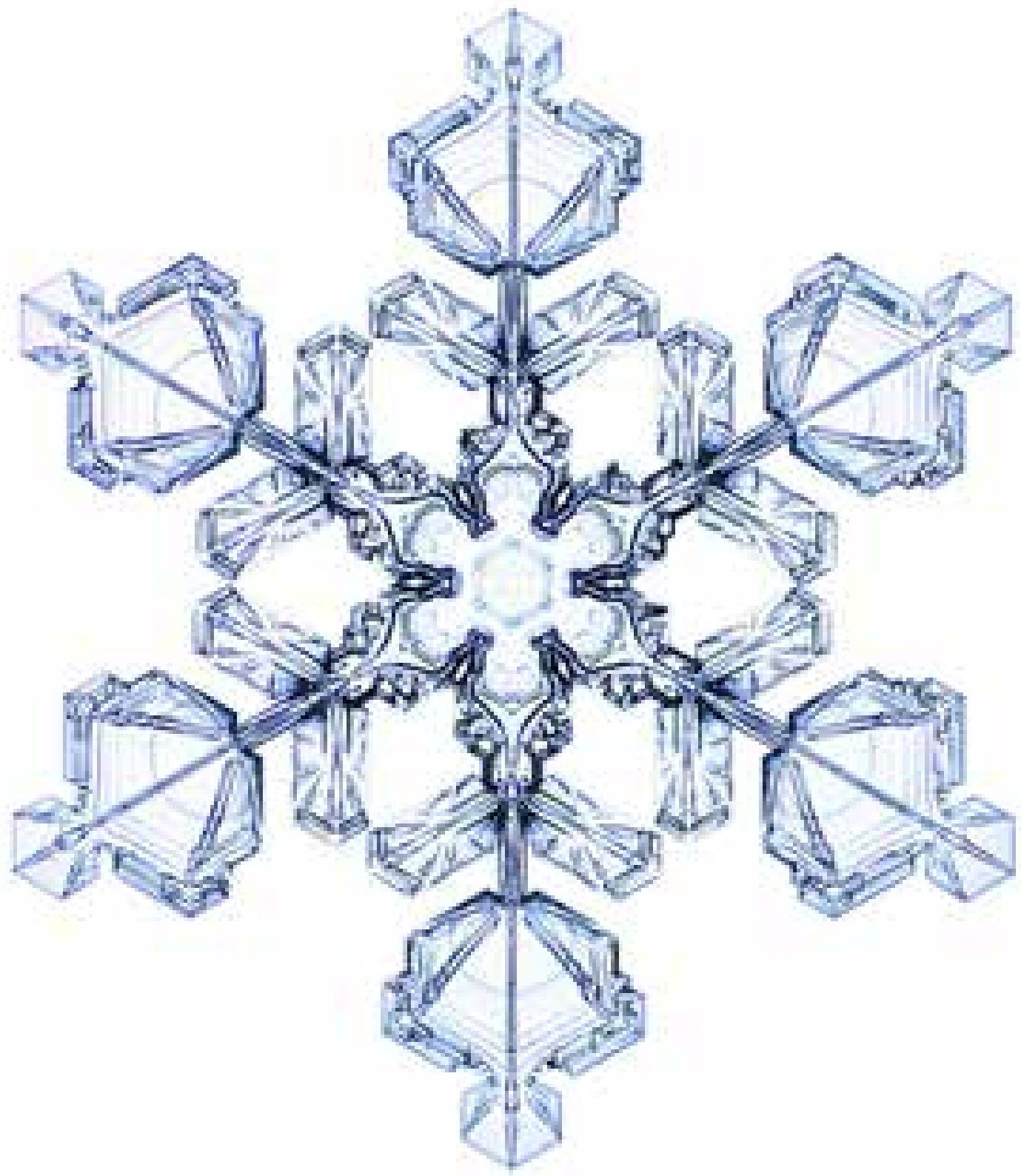

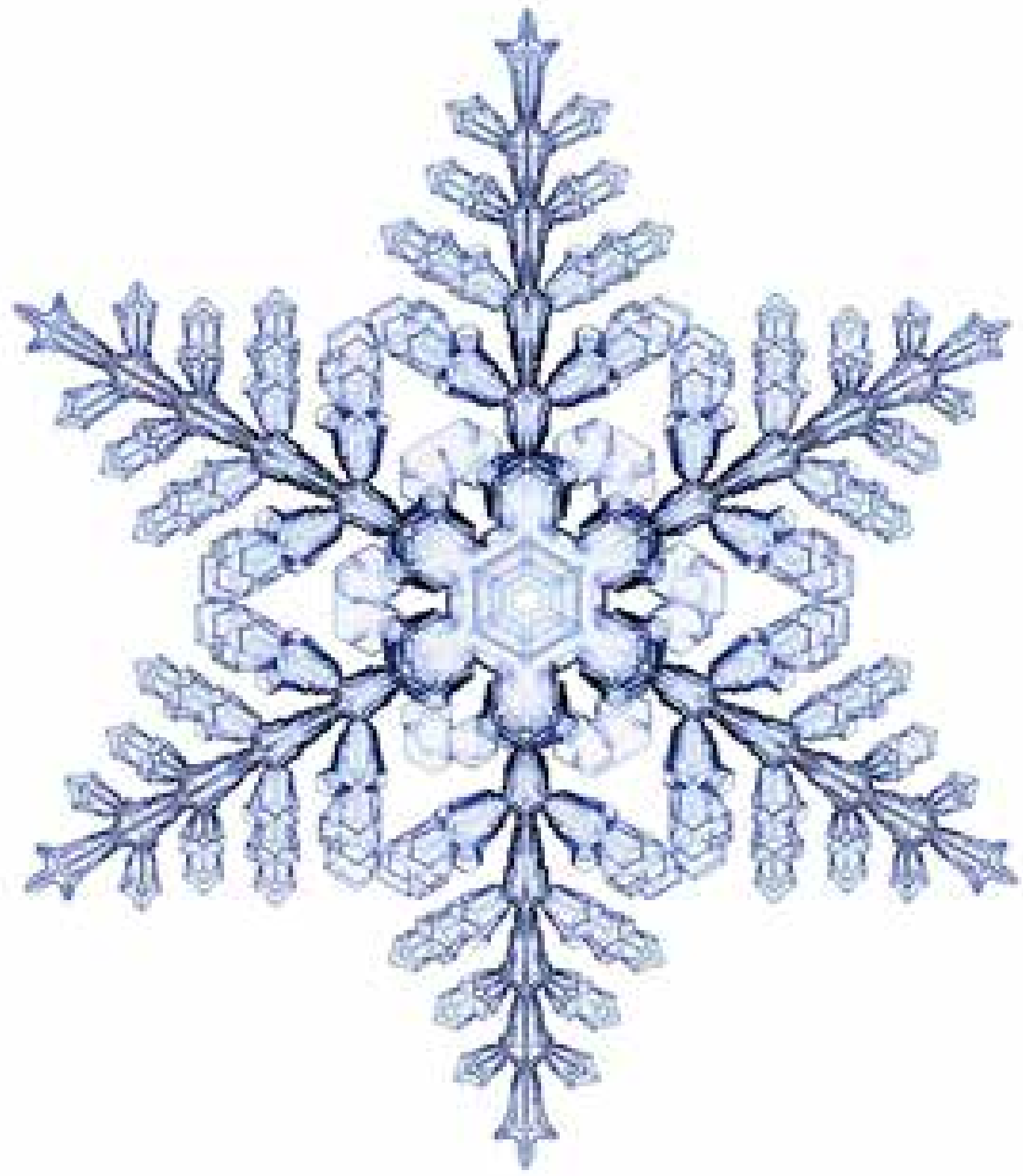

# Snow Crystals

- Kenneth G. Libbrecht -

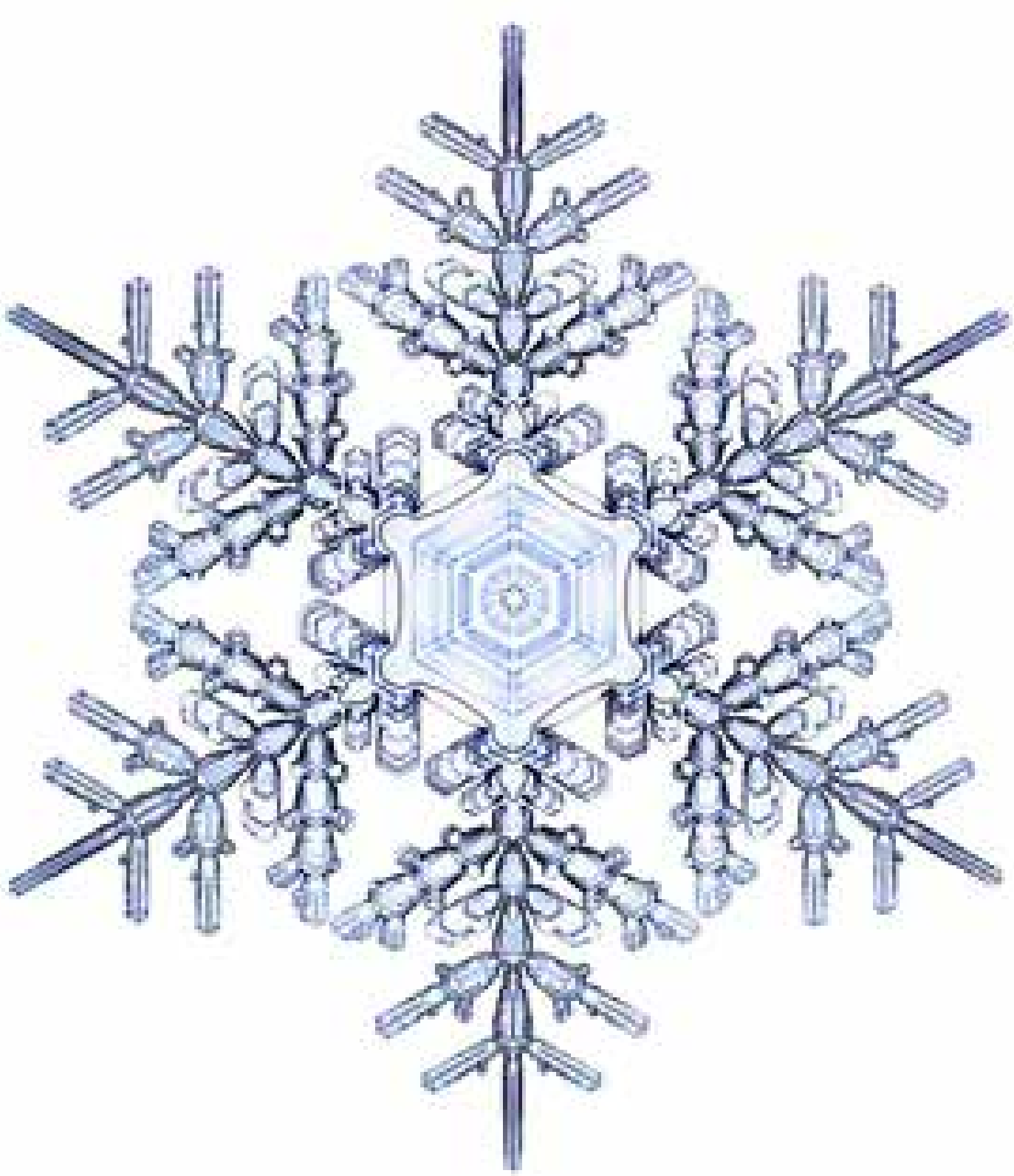

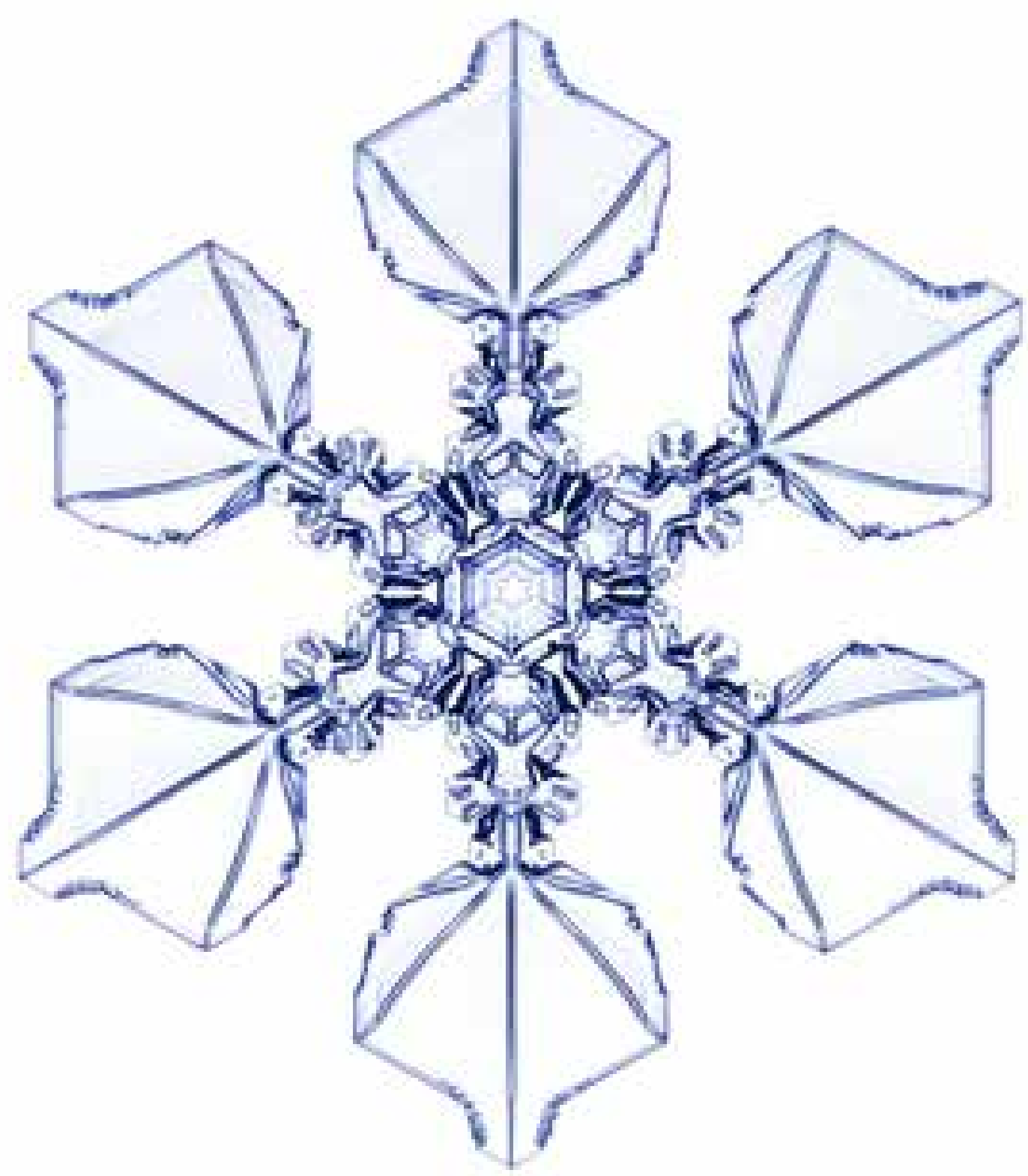

# SNOW CRYSTALS

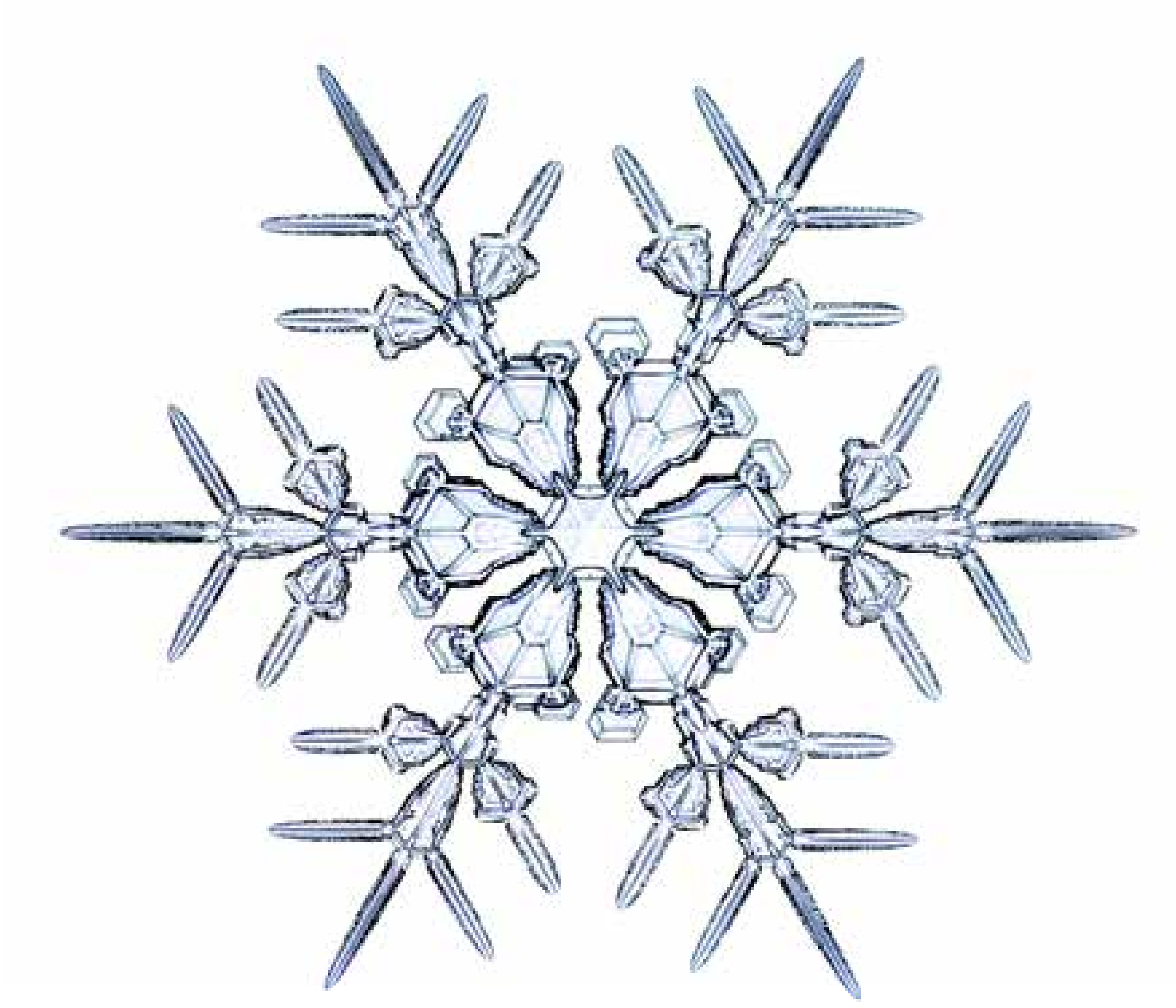

# Forward

Studying snow crystals is a somewhat unusual endeavor, so people often ask me what got me started on this path, and why I have kept at it for the past two decades. The short answer is simply that I find the science both fascinating and entirely worthy of attention. Snow crystal growth involves the coordinated molecular motions of water molecules undergoing a phase transition from vapor to ice, creating elaborate structures in the process. There is still a great deal about this process that we cannot fully explain.

One might imagine that the formation of ice crystals from water vapor would be a solved problem by now; it's not neurobiology, after all, just plain, ordinary ice. Nevertheless, it turns out that the physics underlying crystal growth in general is quite a tough nut to crack, and ice is a particularly intriguing example. Even now, well into the 21st century, our fundamental understanding of why snow crystals grow into the rich variety of structures we see falling from the clouds is remarkably primitive.

Part of me feels that the lowly snowflake has become something of an embarrassment to the scientific community. We can split the atom and sequence the human genome; but explaining the growth of a snowflake remains beyond our abilities. Every winter we see these icy works of art simply appearing, spontaneously, quite literally out of thin air. And yet we have no ready explanation as to why snowflakes look the way they do.

Examining the falling snow up close, one soon witnesses a remarkable menagerie of different crystal types, including thin plates, slender columns, and blocky prisms, all branched, hollowed, faceted, and patterned to varying degrees, often exhibiting a baffling degree of complexity, symmetry, and morphological diversity. How does all this work exactly? What forces result in such complex structures? Why do the crystals change so dramatically from one snowfall to the next? No one yet knows how to answer these questions. When you drill down into the details, the fundamental physical dynamics of snow crystal growth is both captivating and mysterious.

Another part of me feels that the physics of crystal growth is something we ought to know better. The manufacture of semiconductor crystals underlies the entire electronics industry, yet growing crystals is a bit like growing carrots – knowing how to do it is not the same as knowing how it works. I often think of snowflakes as a convenient case-study in the science of crystal growth; if we can figure out the molecular dynamics governing snow crystal formation, maybe that knowledge will have application in other areas.

Although crystal growth is an important area in materials science and engineering, my studies are not motivated by practical applications. My focus is instead on fundamental questions regarding the molecular physics of crystal growth. Applied research can certainly be highly rewarding; but contemplating the overarching scientific questions can be worthwhile also. History clearly teaches us that the knowledge gained from basic scientific pursuits often ends up being quite beneficial, even if one cannot always imagine right now how, when, or where those future benefits might arise.

On a related note, I always make a point of telling people that I am not spending any tax dollars on this research. I have always considered my snow-crystal studies to be something of a scientific hobby – interesting to me, but with no obvious financial payoff now or down the road. I figure with over seven billion people on the planet, and vast resources being spent on sports, entertainment, and all manner of generally unnecessary activities, maybe a few of us can be spared to contemplate the inner workings of a snowflake.

Although studying snowflakes is an unusual activity, I am certainly not the first to engage in it, and hopefully I will not be the last. Beginning with Johannes Kepler over four centuries ago, numerous scientists have put serious effort into understanding the details of how ice forms from water vapor, and how structures arise during the process. At any given time during that long history, you could usually find a handful of people pushing the field forward, bit by bit. It has never been a popular area of research, and it has attracted little support from the usual funding sources. But it seems there are always a few scientists willing to ponder the topic. My efforts build upon what my predecessors have accomplished over the years, and my sincere hope is that this book will provide a starting point for those wanting to continue studying the science of snow crystals into the future.

My foray into snow crystals began in 1995 during a conversation with Stephen Ross, who I had recently hired as a post-doctoral researcher in my lab at Caltech. Stephen had been working with electrodynamic ion trapping in his previous job, so one evening we were chatting about what new opportunities might lie in that direction. We both thought it might be worthwhile to study the growth of isolated single crystals that were levitated in an ion trap, as the physics of structure formation during crystal growth was not well understood. But it was mostly idle conversation that evening, and our attention was soon pulled back to projects we had underway in atomic physics.

Nevertheless, over the next few days I began musing about exactly what crystals one might examine in an ion trap, and my attention quickly turned to ice. If nothing else, it was certainly an inexpensive material to work with, with no unpleasant safety issues, and its freezing temperature was easily accessible as well. As an experimental physicist thinking about the general subject of crystal growth, it seemed reasonable to start with a relatively low-cost test case.

My interest now being piqued, I started doing a bit of online research to see what was known about the science of snowflakes. The internet was quite a new thing back then, and I soon found that it was the perfect tool for learning about an unusual subject like snow crystals. While most current scientific fields involve well-defined communities of researchers who have regular meetings and publish in established journals, studies about snowflakes have been relatively sporadic and isolated, with articles appearing in widely varied forums. Scientific interest has waxed and waned over several centuries, usually led by a few curious souls from here and there around the globe. Locating many of the relevant published scientific references was a nontrivial challenge.

The internet soon steered me toward an extraordinary book entitled *Snow Crystals, Natural and Artificial* published in 1954 by Japanese physicist Ukichiro Nakaya [1954Nak]. It quickly became apparent that I had to have this book, and I then discovered that rare-book dealers were early adopters of internet marketing, as they normally sold their wares to a widely dispersed clientele. It all sounds ordinary now; but locating a copy of a difficult-to-find, long-out-of-print book and purchasing it from a small shop halfway around the world with a few clicks was a marvelous experience at the time.

When my new purchase arrived, I was soon to board a flight with my wife and two young children to North Dakota, visiting some family there for Christmas, so I packed *Snow Crystals* along for some atypical holiday reading. Thumbing through the book on the plane, I saw my first photos of several capped columns, as Nakaya had photographic examples of many uncommon snow crystal types. Although I grew up in North Dakota, and I had plenty of first-hand experience with snow, I had never witnessed anything like the exotic capped column.

As luck would have it, it began snowing a few days later, so I braved the cold and went outside, magnifier in hand, to see what I could find. And, lo and behold, there I found my first capped column! As I delved more deeply into Nakaya's book, I soon decided that I had to set my sights on writing a popular-science book about snowflakes. People who live in snowy climates, as I had done for 18 years, ought to know more about these marvelous works of art floating down from the clouds.

By the fall of 1998, I had created a website devoted to snowflakes, which eventually morphed into what is now *SnowCrystals.com*. I posted examples of snowflake photographs I had found here and there, accompanied by brief descriptions of the science and history of snow-crystal research. As the internet was rapidly picking up in popularity during those days, and educational content was still scarce, my snowflake website received a fair bit of attention from all corners.

As my research continued, I soon found that little progress had been made in snowflake photography over the preceding 50 years. This was problematic for me, as I could hardly contemplate writing a popular-science book about snowflakes without including a representative collection of photographs. Wilson Bentley's photos were something of a standard, but they were over 100 years old, and their quality was rather poor by modern standards. Nakaya's photos were better, but they too were black-and-white and somewhat grainy. A few other photographers had taken additional snowflake photos, but overall the quality I was looking for was not to be found.

As a laboratory physicist, I was already experienced with optics and electronics, so I decided that I had to build a better snowflake photomicroscope. I experimented with different optical hardware and lighting methods, using small alum crystals as surrogate snowflakes to work out the photographic details. This soon led to a collaboration with Patricia Rasmussen in Wisconsin, who put the microscope to good use during the 2001-2 winter season, substantially raising the bar for high-resolution snowflake photography. Voyageur Press then worked with us to publish *The Snowflake: Winter's Secret Beauty* in the fall of 2003, just in time for Christmas.

Being the first-ever popular-science book about snowflakes, adorned with colorful photographs that were considerably higher in quality than past efforts, *The Snowflake* became an immediate hit, selling over 100,000 copies. Once winter arrived and snow started falling around the country, the book received a great deal of publicity from dozens of newspapers, often in full-page Sunday articles featuring photos of different types of snowflakes.

Building on this initial success, I soon made numerous improvements to my microscope, including fitting it into a rugged suitcase for traveling, calling it the *SnowMaster9000*. Although Southern California was my home, I decided it was time to take the plunge and become a serious snowflake photographer. This led to several expeditions to northern Ontario and central Alaska, including countless hours out in the cold photographing minute ice crystals. These new photos formed the basis for *The Little Book of Snowflakes*, which came out during the 2004 holiday season. This was a smaller, inexpensive gift book, and again it did quite well and sold over 100,000 copies.

During the following several years, I continued photographing snowflakes around the globe, and the subject remained quite popular in the media. Voyageur Press and I produced a new book every year, including *The Art of the Snowflake, The Secret Life of a*

*Snowflake, Ken Libbrecht's Field Guide to Snowflakes, The Magic of Snowflakes, Snowflakes, and The Snowflake: Winter's Secret Artistry.* I made numerous appearances to talk about the books, including on the Martha Stewart show, and I supplied snowflake images for numerous newspapers and magazines. I even had four snowflake photos on a set of U.S. first-class postage stamps (over 3 billion sold!), followed by Austrian stamps, Swedish stamps, and then again on U.S. bulk-mail stamps. It was all quite a thrilling experience.

After about a decade of snowflake everything for me, the phenomenon slowly quieted down and life got back to normal once again. Happily, with an influx of revenue from book royalties I was able to gear up my snowflake lab, to the point that I could start doing serious experimental research investigating the science of snow-crystal growth. This led to better measurements of the molecular attachment kinetics, studies using electric needle crystals, and making designer Plate-on-Pedestal snow crystals, topics that are discussed at some length in the chapters that follow.

My students and I made considerable progress on several scientific fronts, but not yet as much as I had hoped. I dreamed that it might be possible to "solve" the problem of snowflake growth, or at least make a big step forward. Alas, the lowly snowflake presents a rather rich and deep challenge. Like an onion, as you peel away layers, you find more layers. Of course, I and others are still pushing forward, so perhaps our big *Eureka!* moment is just around the corner. Science is generally more about steady, layer-by-layer progress than astonishing breakthroughs, but one never knows in this business.

During much of this time, my work on snow crystals was mostly a side project. My scientific interests have drifted over several decades from solar astrophysics to atomic/laser physics to gravitational physics and the LIGO (Laser Interferometer Gravitational-wave Observatory) project, and I dabbled with snowflakes when time permitted. Lately I have begun to realize that snow crystals are my new calling, so, starting around 2014, I have been focusing nearly all my efforts in this area. It remains, at least to me, a continually fascinating scientific endeavor.

I am fortunate to have worked with many talented undergraduate students from Caltech and other universities on my snow-crystal research, including Victoria Tanusheva, Mark Rickerby, Nina Budaeva, Robert Bell, Hannah Arnold, Timothy Crosby, Molly Swanson, Han Yu, Johanna Bible, Ryan Potter, Christopher Miller, Kevin Lui, Cameron Lemon, Sarah Thomas, Helen Morrison, and Benjamin Faber. Their determined efforts are much appreciated.

In the same vein, I have enjoyed countless enjoyable interactions with fellow snow/ice enthusiasts, colleagues and collaborators, including Walter Wick, David Griffeath, Janko Gravner, Don Komarechka, Alexey Kljatov, Patricia Rasmussen, Mary Ann White, Carol Norberg, Matthew Sturm, Ted Kinsman, James Kelly, Joseph Shaw. My editors at Voyageur Press, Michael Dregni and Todd Berger, were terrific to work with.

I am much indebted to Caltech for hiring me as a young professor and providing me gainful employment for what has been nearly my entire adult life. Caltech has provided ample lab space while allowing me full rein to explore this atypical line of research. Without Caltech's constant support, none of this work would have been possible.

Finally, my wife, Rachel Wing, and our two children, Max and Alanna, have been enthusiastic participants throughout this snowflake adventure, especially on our numerous snowflake-related vacations to such far-flung venues as northern Japan, Vermont, northern Ontario, northern Sweden, Alaska, and the mountains of California, all during the cold of winter. Thanks for the memories!

<div align="right">Kenneth Libbrecht<br>Pasadena, California<br>August 30, 2019</div>

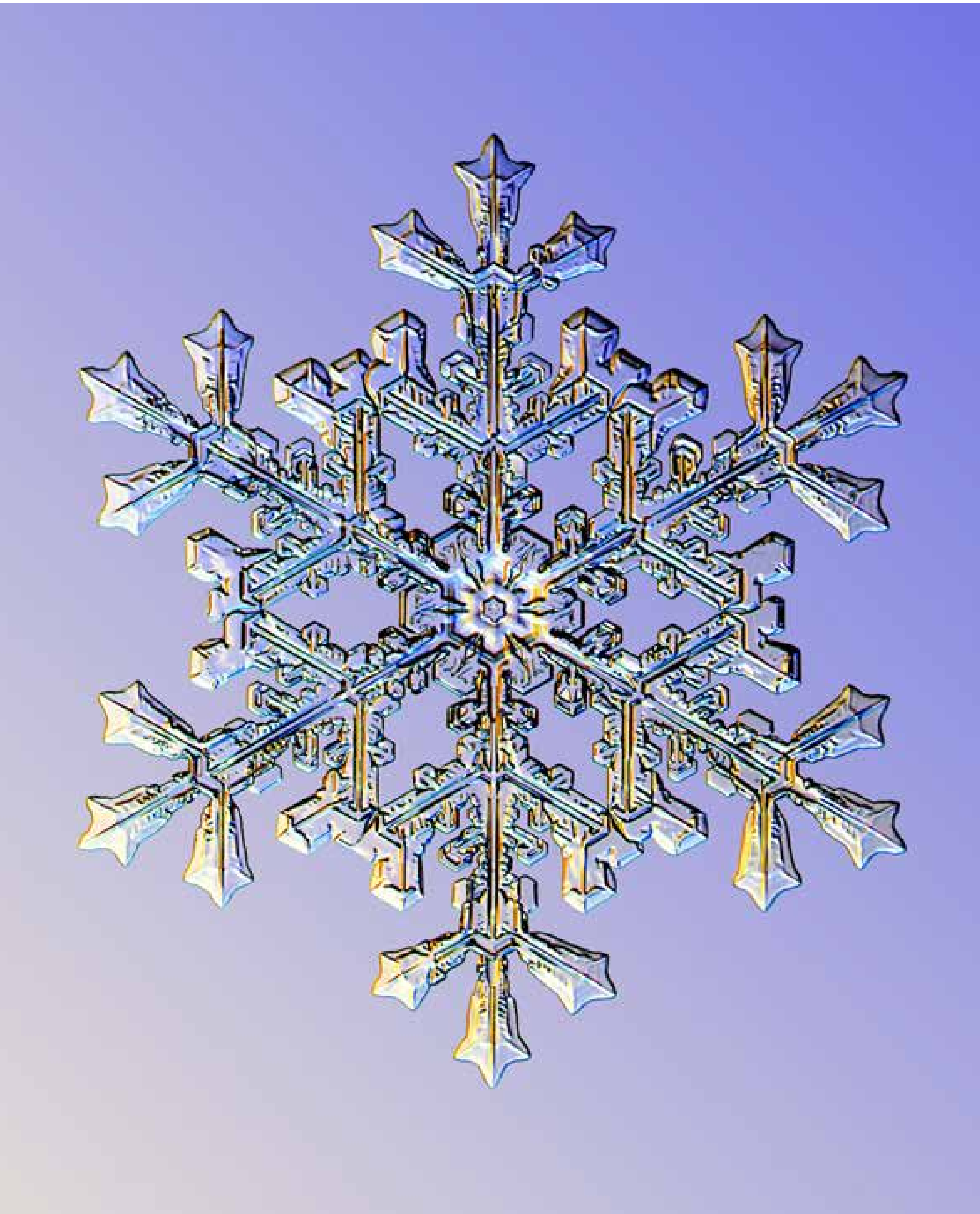

# Chapter 1

# Snow Crystal Science

*How full of the creative genius is the air*
*in which these are generated!*
*I should hardly admire more if real stars*
*fell and lodged on my coat.*
– Henry David Thoreau
*Journal, 1856*

This book is about the science of snowflakes. Its overarching objective is to explain why snowflakes grow into those remarkable crystalline structures that can be found floating down from the winter clouds. In these pages, I will attempt to answer some of the basic scientific questions one might ask while scrutinizing a newly fallen snowflake: Where do snowflakes come from? How does formless water vapor manage to arrange itself, spontaneously, into such a variety of amazingly ornate shapes? What physical processes guide the development of such elaborate, yet symmetrical, patterns? Why does all this happen the way it does?

**Facing Page: An exceptional stellar snow crystal, measuring about four millimeters from tip to tip, photographed by the author in Kiruna, Sweden.**

Comprehending the lowly snowflake is a surprisingly challenging task. The seemingly simple phenomenon of water vapor freezing into ice involves a veritable symphony of subtle molecular processes, from the diffusion of water molecules through the air to the complex attachment kinetics that govern how molecules assimilate into a rigid crystalline lattice. Explaining this intricate act of meteorological morphogenesis requires a rather deep dive into areas of mathematical physics, statistical mechanics, computational algorithms, and the many-body molecular dynamics of crystal growth. Even now, well into the 21st century, snowflake science is very much a work in progress, as several rather basic aspects of the surface structure and dynamics of ice at the molecular level remain quite mysterious.

When I first begin reading about this subject in the 1990s, I was immediately struck by just how little was really understood about snowflake formation. While many different types of snowflakes had been observed and cataloged over the years, there was no prevailing explanation for why different shapes appeared under different growth conditions.

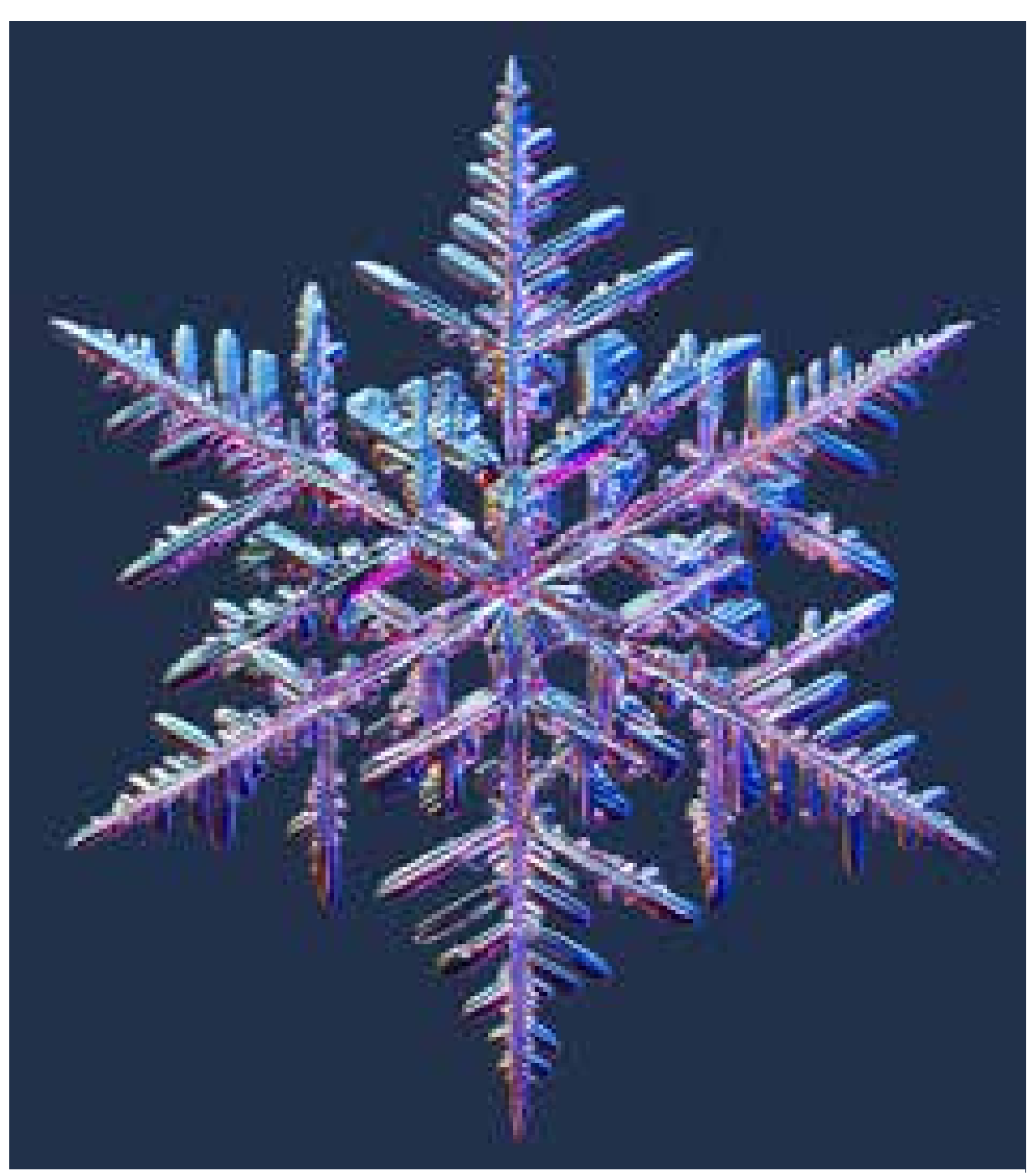

**Figure 1.1: Large stellar dendrites like this one are perennial holiday favorites, but these are only one type of snow crystal. Stellar dendrites are thin and flat in overall shape, and they only appear when the temperature in the clouds is near -15 C. The full menagerie of natural snow crystals is discussed in Chapter 10.**

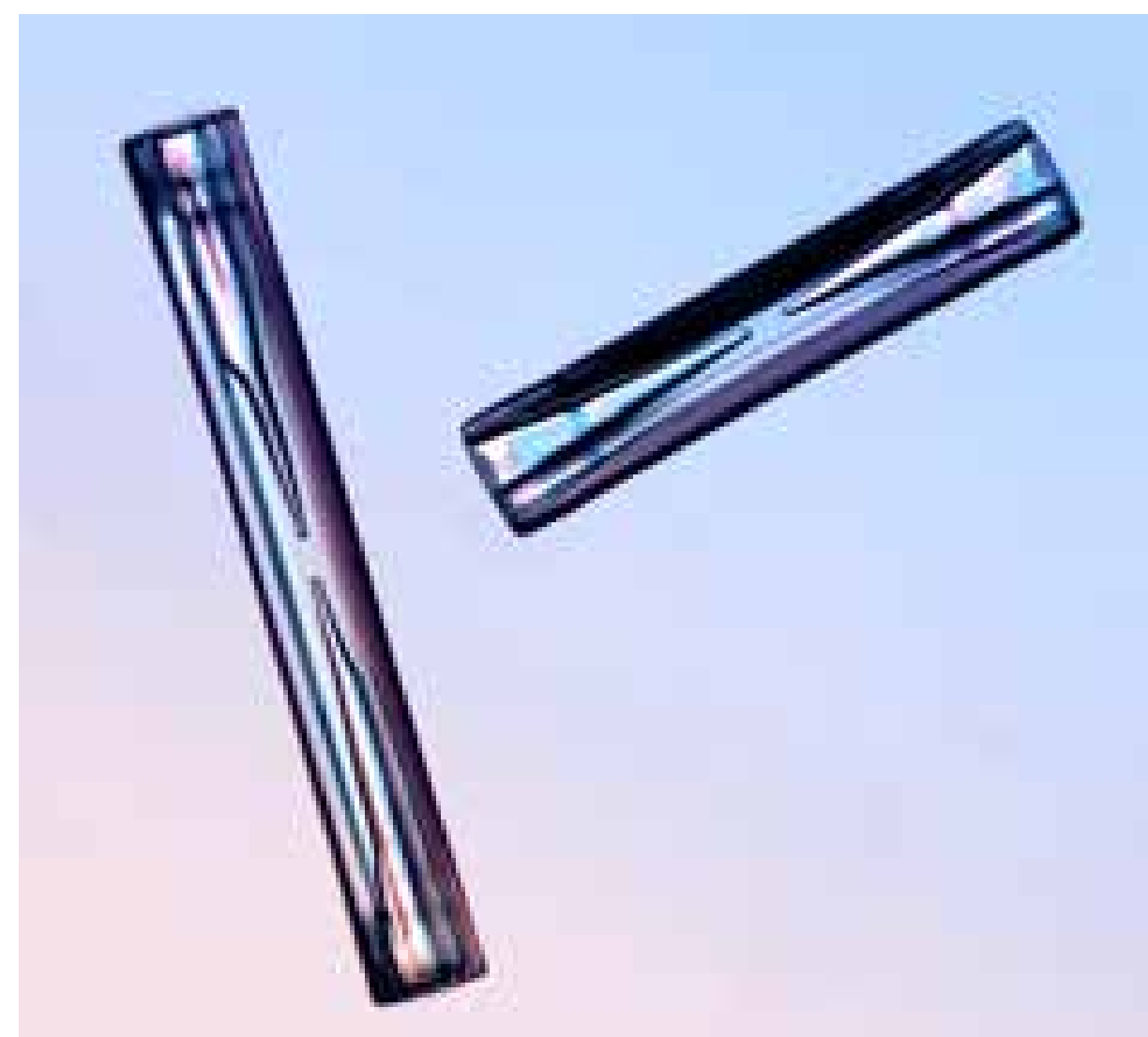

**Figure 1.2: At temperatures near -5 C, columnar snow crystals like these often appear. Their basic shape is a hexagonal column, like that of a standard wooden pencil. These particular examples are "hollow" columns that exhibit roughly conical hollow regions in each end.**

For example, thin plates and ornate stellar crystals mainly appear when the temperature is in a narrow range around -15 C (Figure 1.1), while slender columnar crystals form when the temperature is near -5 C (Figure 1.2). Why is this? I will expound at some length on this question in Chapter 4, as it is a long-standing puzzle, and I have developed a few new ideas aimed at answering it. But a complete understanding of even this straightforward observation remains elusive.

At first glance, the snowflake appears to be a somewhat basic natural phenomenon. It is made of little more than pure ice, and it assembles itself, quite literally, out of thin air. Yet trying to understand snowflake formation in detail will take us to the very cutting edge of contemporary science. The journey will be neither short nor easy, so we begin with the basics.

## 1.1 Complex Symmetry

I often use the term *snowflake* synonymously with *snow crystal*. The latter is a single crystal of ice, in which the water molecules are all lined up in a precise hexagonal array. Whenever you see that characteristic six-fold symmetry often associated with snowflakes, you are actually looking at a snow crystal.

A snowflake, on the other hand, is a more general meteorological term that can mean an individual snow crystal, a cluster of snow crystals that form together, or even a large aggregate of snow crystals that collide and stick together in mid-flight. Those large puff-balls you see floating down in warmer snowfalls are called snowflakes, and each is made from hundreds or even thousands of individual snow crystals. Snow crystals are commonly called snowflakes, and this is fine, like calling a tulip a flower.

A snow crystal is not a frozen raindrop; that type of precipitation is called sleet. Rather, a snow crystal forms out of water vapor in the atmosphere, as water molecules transition directly from the gaseous to the solid state. Complex structures emerge as the crystal grows, driven mainly by how water vapor molecules are transported to the developing crystal via diffusion, together with how readily impinging molecules stick to different ice surfaces.

## From Clouds to Crystals

To begin our study of snow-crystal formation, consider the life of a large, well-formed snowflake that falls from the winter clouds, like the one shown in Figure 1.1. The story begins as weather patterns transport and cool a parcel of moist air until its temperature drops below the *dew point*, meaning the relative humidity rises above 100 percent and the air becomes *supersaturated* with water vapor. When this happens, the gaseous water vapor in the air tends to condense out as liquid water.

Near the ground, the water vapor might condense as dew on the grass (which is why this temperature is called the dew point). At higher altitudes, however, the water vapor condenses into countless cloud droplets. The liquid droplets nucleate around a microscopic particles of dust, which are typically abundant in the atmosphere. Cloud droplets are so small – about 10-20 microns in diameter – that they can remain suspended in the air almost indefinitely.

If the cloud continues cooling and its temperature drops significantly below 0 C, then the liquid water droplets will start freezing into ice. Not all the droplets freeze at once, and none will freeze right at 0 C. Instead, the droplets become *supercooled* as their temperature drops, often remaining in a metastable liquid state for long periods of time. Some droplets will freeze when the temperature drops below -5 C, and most will freeze somewhere around -10 C. A hearty few may survive unfrozen at -20 C or below, but all will become solid ice before the temperature reaches -40 C.

The freezing temperature of a particular cloud droplet is determined in large part by the speck of dust it contains. Pure water can be cooled to nearly -40 C before freezing, while some materials (silver iodide in particular) will nucleate freezing as high as -4 C. Certain bacterial proteins can even promote freezing at temperatures as high as -2 C. These exotic materials are not much present in the atmosphere, however, so your average speck of dust will nucleate freezing around -10 C. Note that the character of an included dust particle usually has little effect on the final snow crystal shape, because it is microscopic in size and soon becomes buried within the ice.

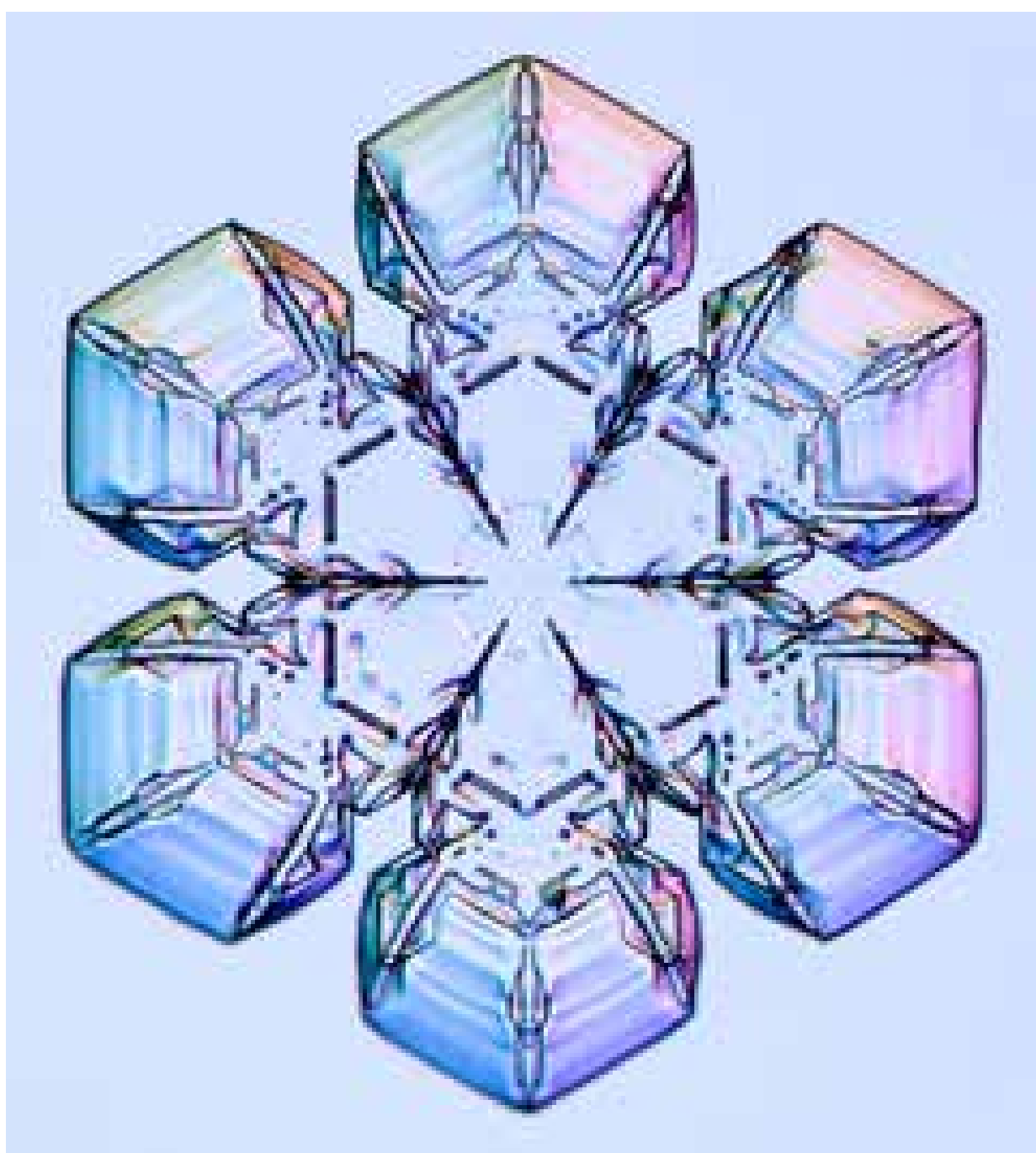

**Figure 1.3: The overall thin and flat shape of this crystal is what puts the "flake" in "snowflake." This unusual example exhibits an exceptionally striking six-fold symmetry in its complex surface patterns.**

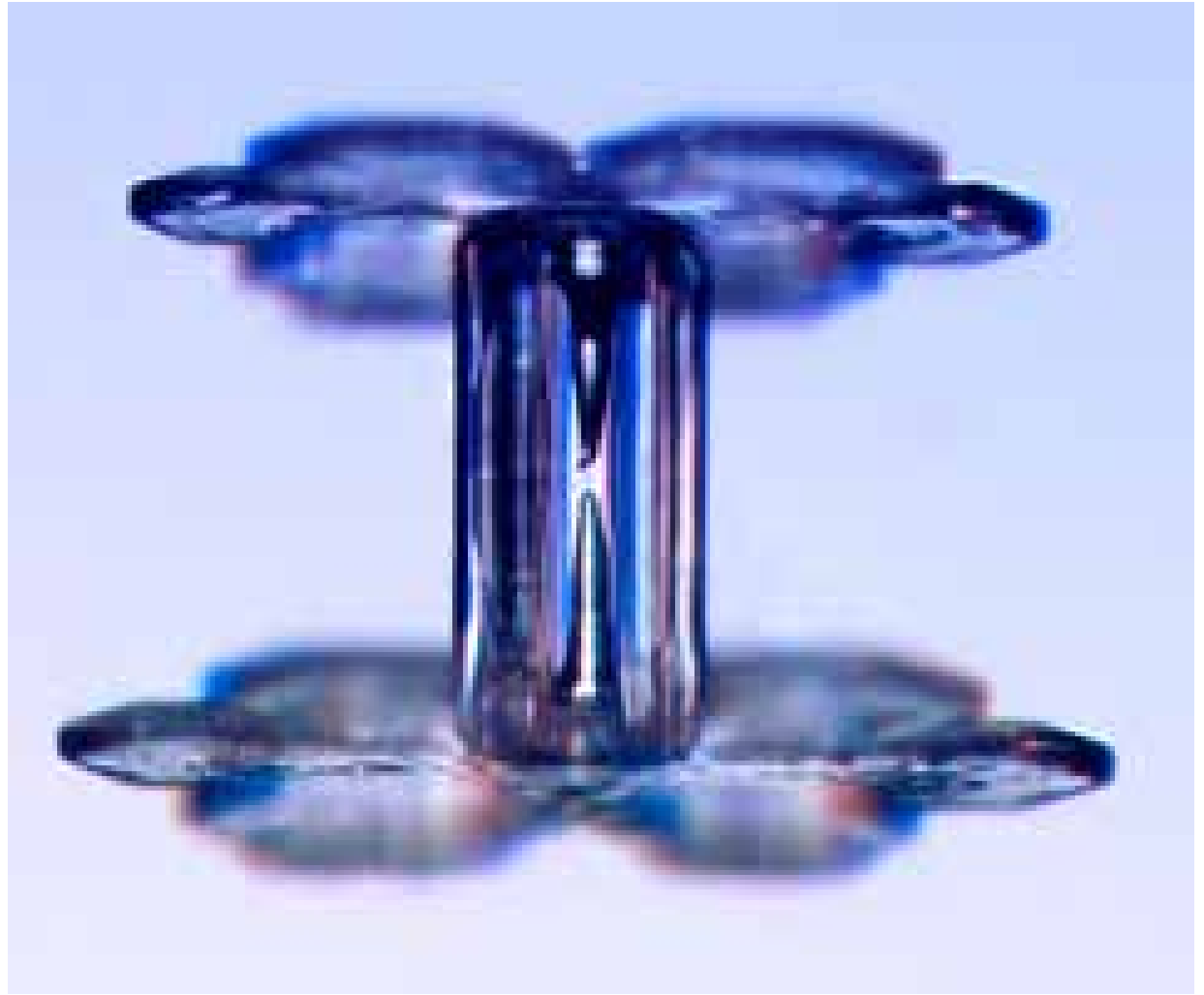

**Figure 1.4: This snow crystal is called a "capped column", consisting of a stout columnar crystal with a pair of plate-like crystals growing out from the two ends of the column. The initial column grew when the temperature was near -5 C, and later the plates formed when the column moved to a colder location. Complex snow crystals often develop in changing growth conditions.**

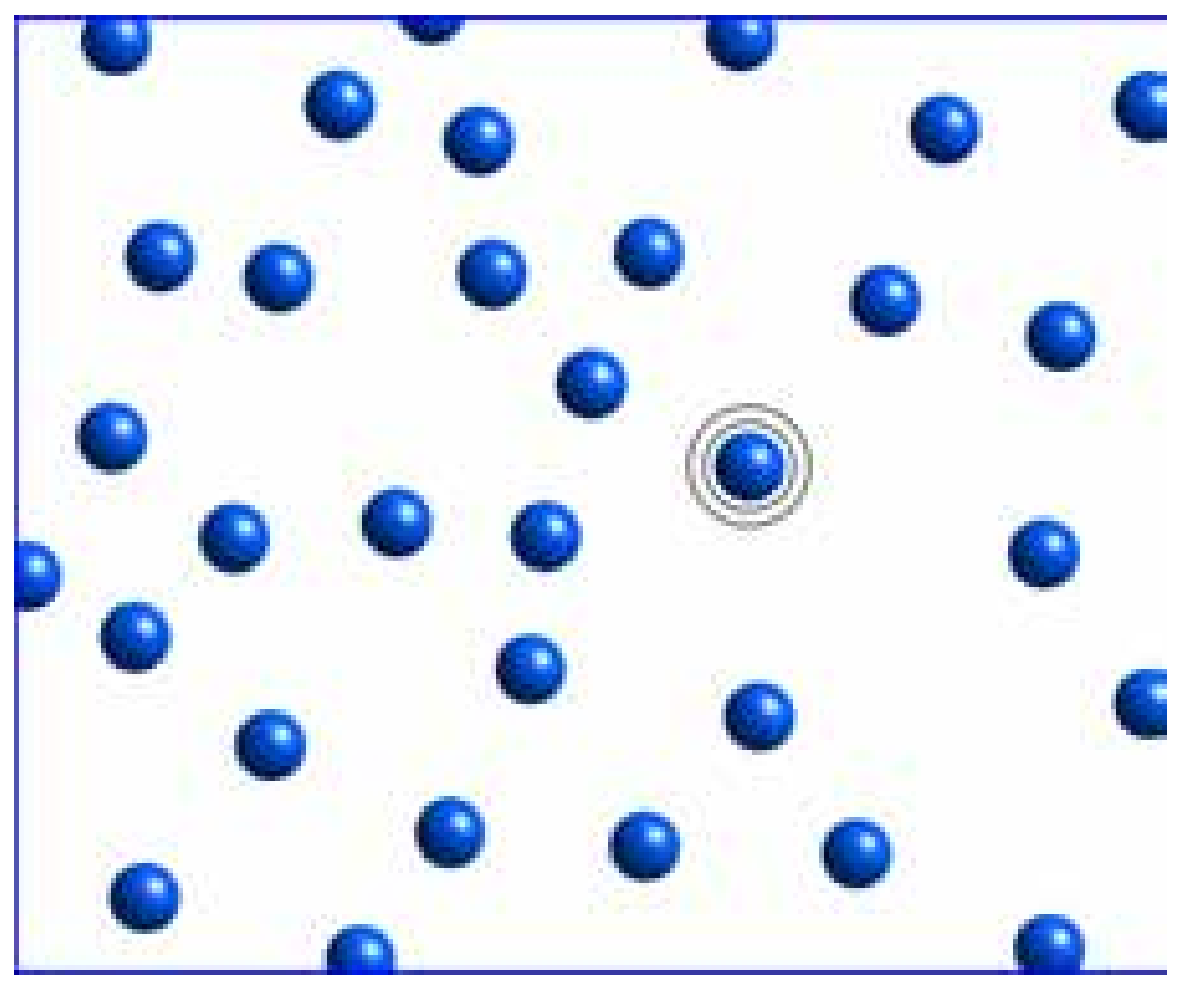

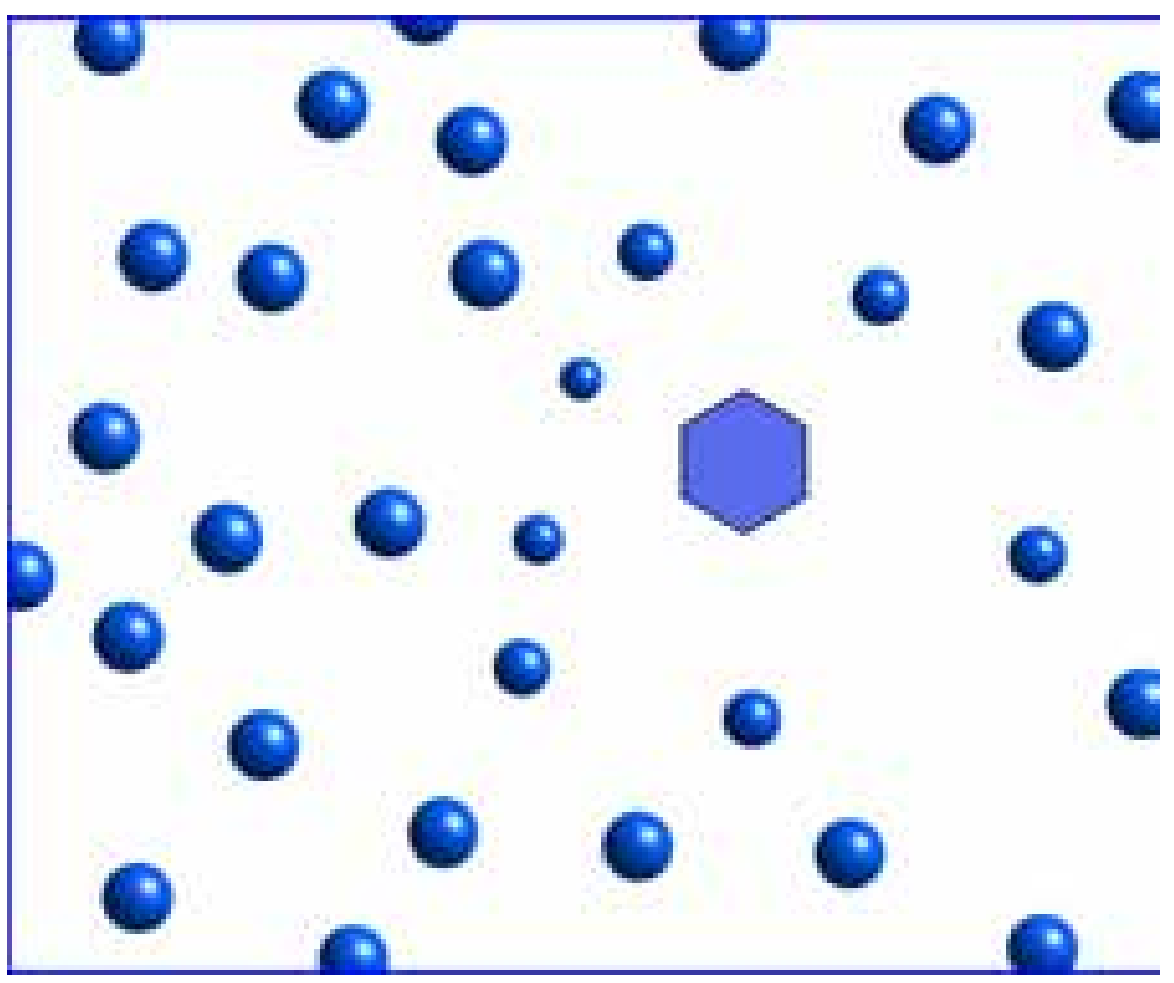

**Figure 1.5: A snowflake is born when a liquid cloud droplet freezes into ice (first sketch). The ice particle initially grows into a faceted prism (second sketch), as the growth is limited by anisotropic attachment kinetics on the ice surface. After the crystal grows larger, the diffusion of water molecules through the air causes branches to sprout from the six corners of the prism (third sketch). The growing crystal removes water vapor from the air, which is replenished by the evaporation of nearby water droplets (fourth sketch). About 100,000 cloud droplets evaporate to provide enough material to make one large stellar snow crystal. The flake continues growing inside the cloud until it becomes so heavy that it falls to earth.**

Once a cloud droplet freezes, it becomes an embryonic snow crystal that commences growing by absorbing water vapor from the air around it. Because the vapor pressure of liquid water is higher than that of solid ice (see

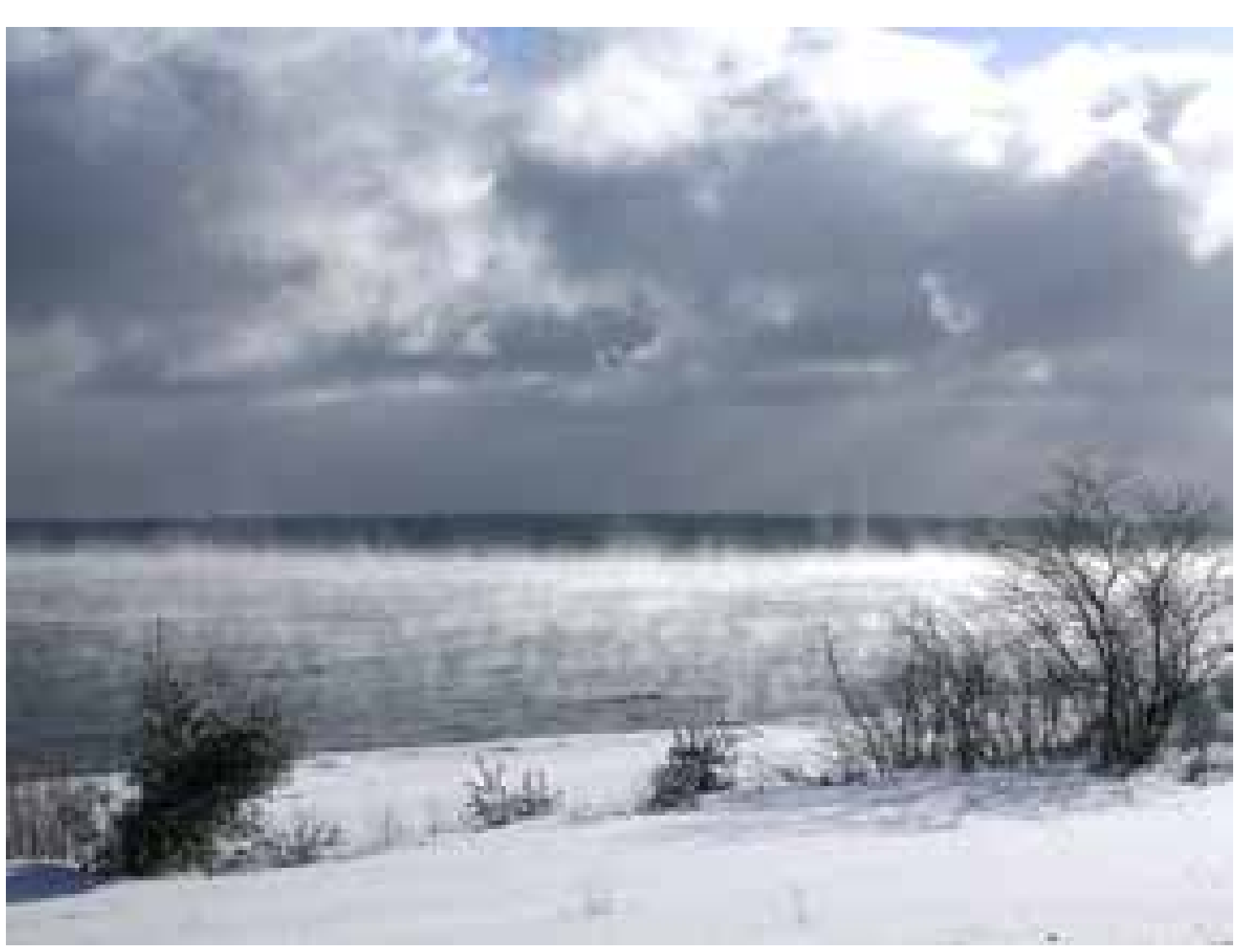

**Figure 1.6: (Left) A winter scene looking out over Lake Superior near Houghton, Michigan. Water vapor evaporating from the warm lake quickly condenses into mist droplets, because the air temperature is substantially colder than the water temperature. But the mist soon evaporates back to water vapor as it rises up from the lake. The vapor condenses once again into droplets at higher altitudes, forming thick clouds. Should the clouds cool down sufficiently, most of the liquid droplets will evaporate to feed the formation of snowflakes that fall back into the lake, completing the water cycle.**

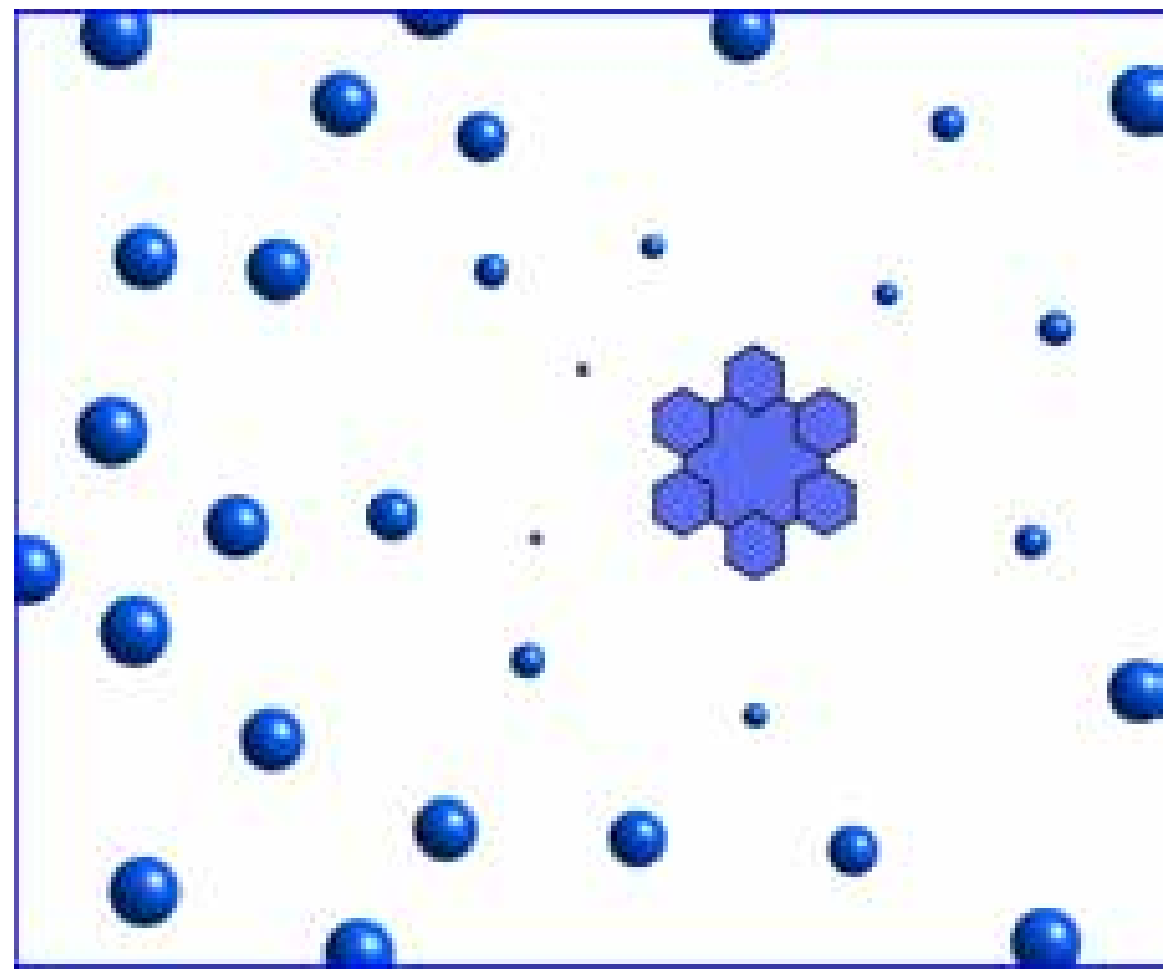

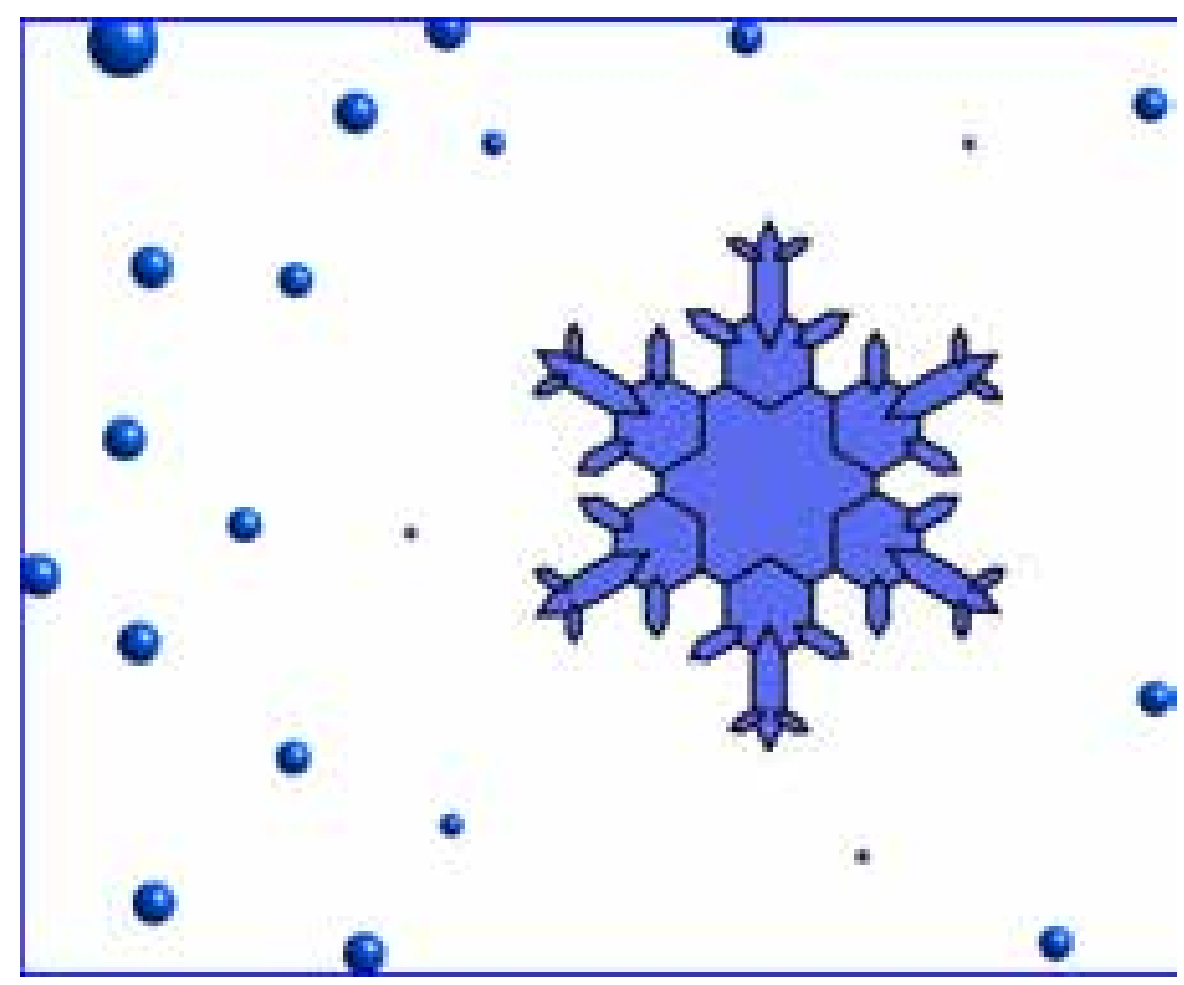

Chapter 2), the cloud droplets surrounding the nascent snowflake begin to evaporate away, as shown in Figure 1.5. During this process, there is a net transfer of water molecules from liquid water droplets to water vapor, and then from vapor to ice. About 100,000 cloud droplets will evaporate away to provide enough material to make one good-sized snowflake. This roundabout route is how most of the liquid water in a cloud freezes into solid ice.

As the temperature inside a cooling cloud falls substantially below -5 C, cloud droplets will freeze in large numbers, thus initiating a full-fledged snowfall. By the time the cloud has cooled to around -20 C, most of the liquid droplets will be gone, as some will have frozen and many will have evaporated away. At temperatures below -20 C, it is often said to be "too cold to snow," because nearly all the liquid cloud droplets will have already disappeared before the cloud cools to that temperature. And when no liquid water remains to feed growing snowflakes, there can be no snowfall.

## Faceting and Branching

Going back to our single, just-frozen droplet, it quickly absorbs water vapor from the air around it and grows into the shape of a small, faceted hexagonal prism, illustrated in Figure 1.7. The prism shape is defined by two *basal facets* and six *prism facets* that arise from the underlying six-fold symmetry of the ice crystal lattice, which is described in Chapter 2.

The molecular mechanism that creates this faceted prism shape is illustrated in Figure 1.8. Water vapor molecules strike the ice crystal everywhere on its surface, but they are more likely to stick when the surface is molecularly "rough", meaning it has a lot of dangling chemical bonds. The facet surfaces are special because they are aligned with the lattice structure of the crystal, so these surfaces exhibit fewer open molecular bonds. Thus, the facet surfaces accumulate water vapor at a lower rate than the rough surfaces, and this process soon yields a faceted ice prism. The rate at which impinging water molecules stick to various surfaces is called the *attachment kinetics*, and I discuss this subject in detail in Chapter 4.

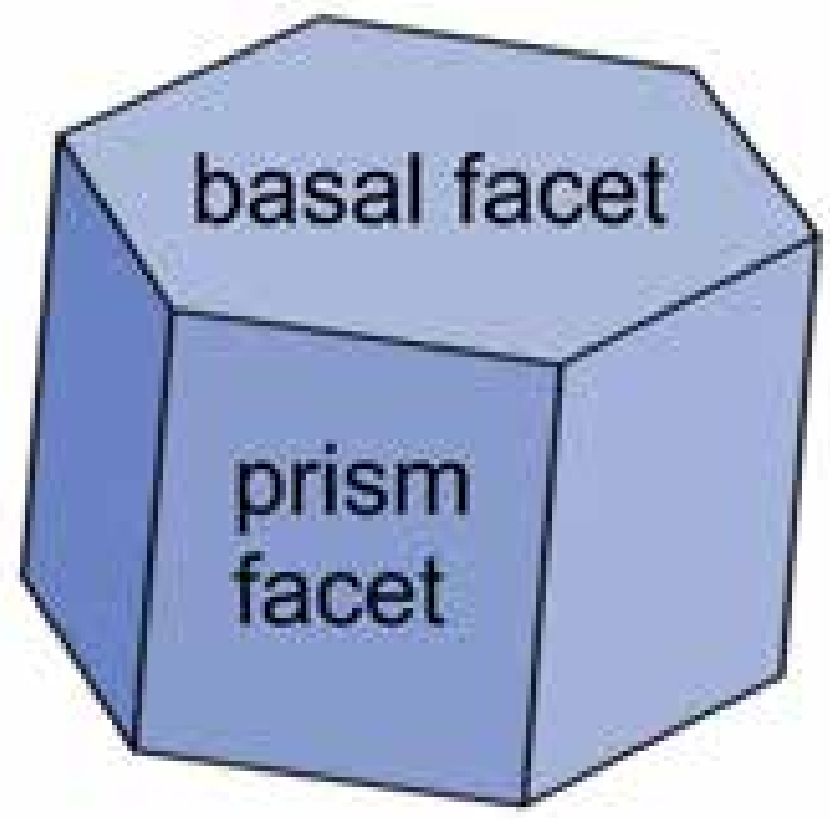

**Figure 1.7: The most basic shape of a snow crystal is a *hexagonal prism* with two basal facets and six prism facets. This shape arises because of the underlying hexagonal structure of the ice crystal lattice, as described in Chapter 2.**

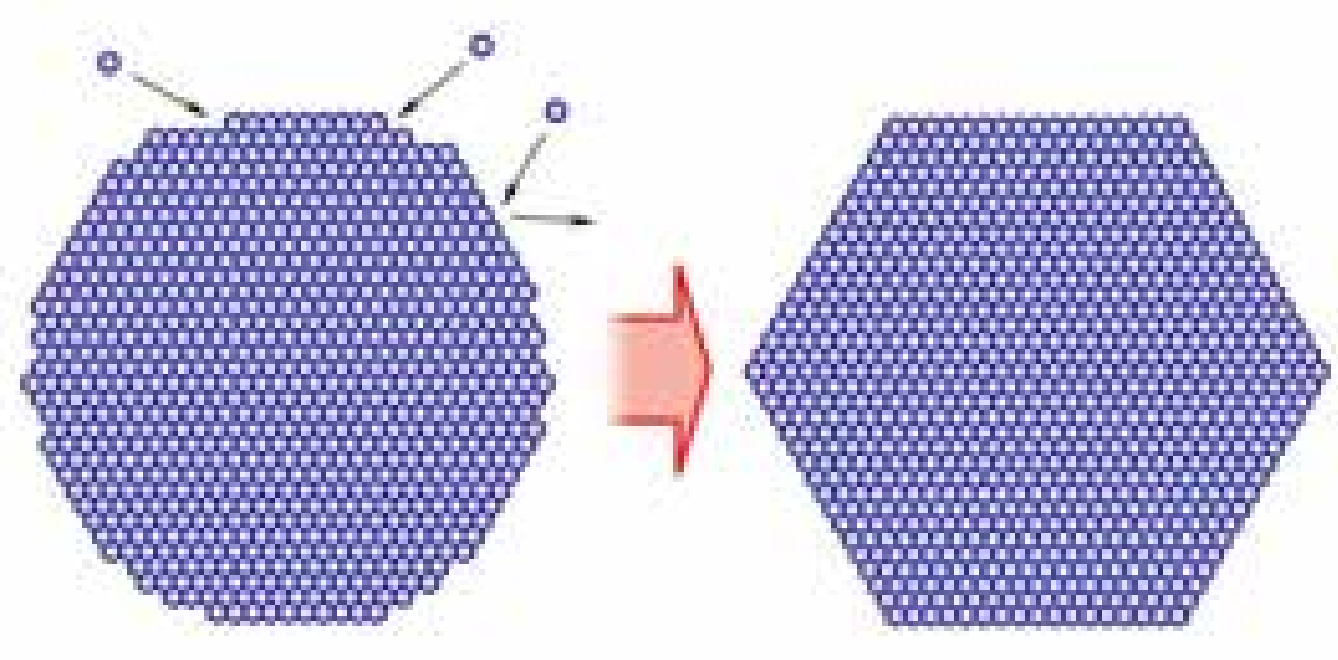

**Figure 1.8: When water vapor molecules strike a molecularly rough ice surface, they tend to stick. But when they strike a molecularly smooth facet surface, they are less likely to stick. As the crystal grows, soon the rough areas fill in, leaving a fully faceted ice prism.**

If the cloud temperature is near -15 C, which is often the case when clouds are producing snow, then the basal surfaces will accumulate material especially slowly, while water vapor will condense on the prism facets much more readily (see Chapter 4). Thus the nascent frozen droplet develops into a thin, flat, hexagonal plate, which is an early stage of what will eventually become a large stellar snow crystal.

As the small hexagonal plate grows larger, its six corners stick out slightly into the surrounding humid air, causing the tips of the hexagon to absorb water vapor a bit more quickly than other parts of the crystal. The faster growth makes the corners stick out farther still, causing them to grow even faster. This positive-feedback effect causes a set of six branches to sprout from the hexagonal plate, as illustrated in Figure 1.9. This *branching instability*, described in Chapter 3, is responsible for most of the complex structure seen in snow crystals.

Once the six branches begin to develop, most of the subsequent growth occurs near the branch tips, where the supply of water vapor is greatest. Moreover, the growth behavior of each branch is quite sensitive to the temperature and humidity in the air surrounding it. As the crystal travels through the inhomogeneous clouds, it experiences ever-changing conditions that modify how the crystal grows. Sometimes the branch tips become faceted, while at other times they may sprout additional sidebranches. It all depends on the growth conditions at any given time. The final shape of the branch, therefore, reflects on the entire history of its growth, which was determined by the meandering path it took through the atmosphere.

The six branches of a snow crystal develop in near synchrony simply because they all travel together through the cloud. Thus the six branches all experience essentially the same growth conditions at the same times, so all six develop into the same elaborate shape, as illustrated in Figure 1.10. Note that the growth of the six branches is not synchronized by any communication between them, but rather by their common history. And because no two snowflakes follow exactly the same path through the turbulent atmosphere, no two look exactly alike. (Although the full story of snowflake uniqueness is a bit more involved, as I describe later in this chapter.)

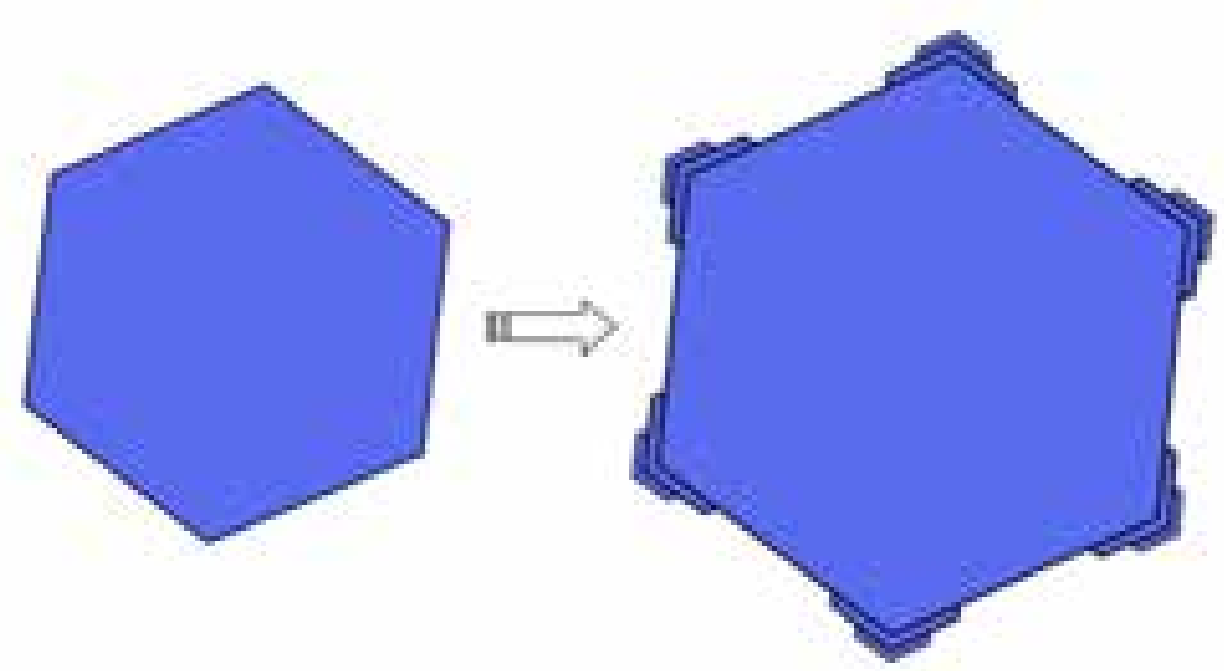

**Figure 1.9: The six corners of a thin hexagonal plate stick out into the humid air around it. Water vapor condenses preferentially on the corners as a result, making them stick out even farther. This leads to a *branching instability* (described in Chapter 3) that causes six branches to sprout from the corners of the hexagon.**

**Figure 1.10: (Right) The final shape of a complex stellar snow crystal depends on the path it traveled through the clouds. Sudden changes in the temperature and humidity around a crystal can cause abrupt changes in its growth behavior. However, because the six arms see the same changes at the same times, they grow in near synchrony. The final snow crystal thus exhibits a complex structure with an overall six-fold symmetry.**

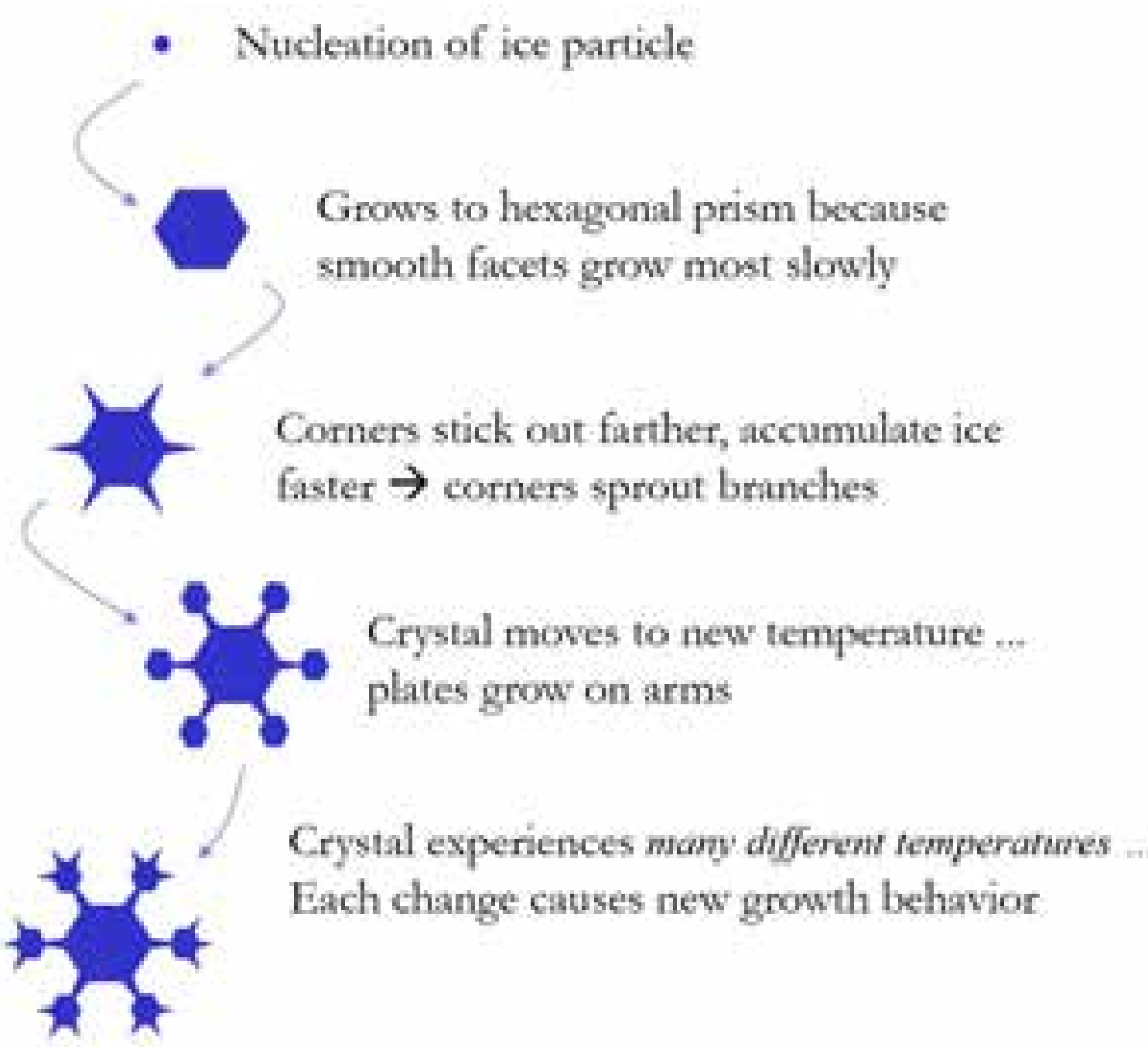

**Figure 1.11: (Below) A laboratory snow crystal grown using the Plate-on-Pedestal technique (see Chapter 9). When creating this snowflake, I imposed a series of abrupt changes in temperature and supersaturation in order to produce a complex, yet symmetrical, morphology.**

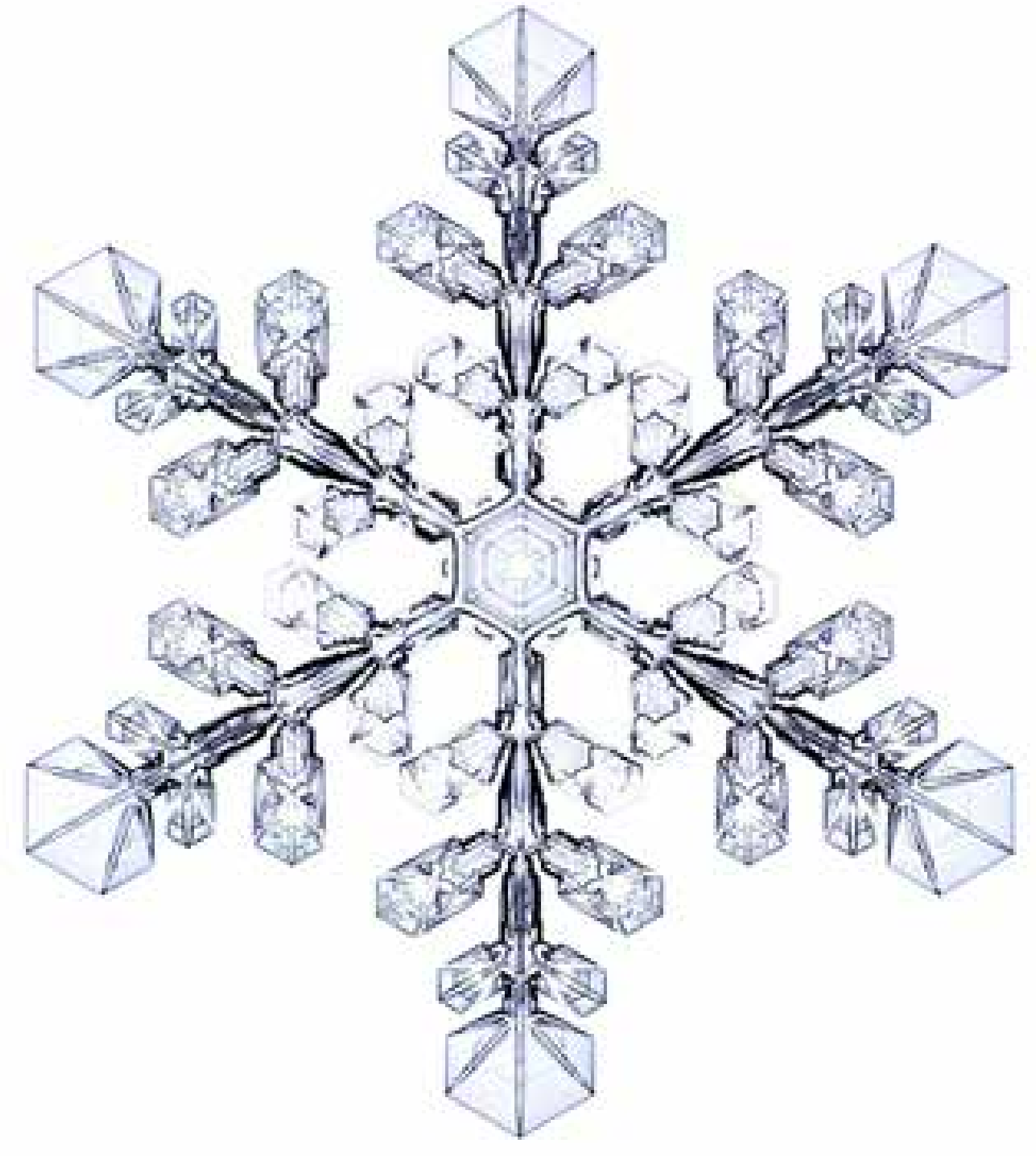

A complex, symmetrical snow-crystal design, as that shown in Figure 1.11, usually means that the crystal experienced many abrupt changes in environmental conditions as it developed. This particular example is a snow crystal I grew suspended in my lab, so I was able to observe directly how the growth responded as I changed the temperature and humidity. It took about 45 minutes to create this specimen, and I controlled the formation of its sidebranches and faceted features in real time by following the rules of snow-crystal growth. I describe the hardware for engineering these kinds of "designer" snowflakes in Chapter 9, along with some of the design rules.

Although this narrative describes the origin of complex symmetry in stellar snow crystals, it provides little insight into the formation of other morphological types, such as the columnar crystals shown in Figure 1.2. The full menagerie of natural snow crystals is presented in Chapter 10, and laboratory studies have found that these can be organized according to the *Nakaya diagram* shown in Figure 1.12. One common theme throughout this discussion is that the detailed shape of any snow crystal is determined mainly by the environmental conditions it experienced as it grew.

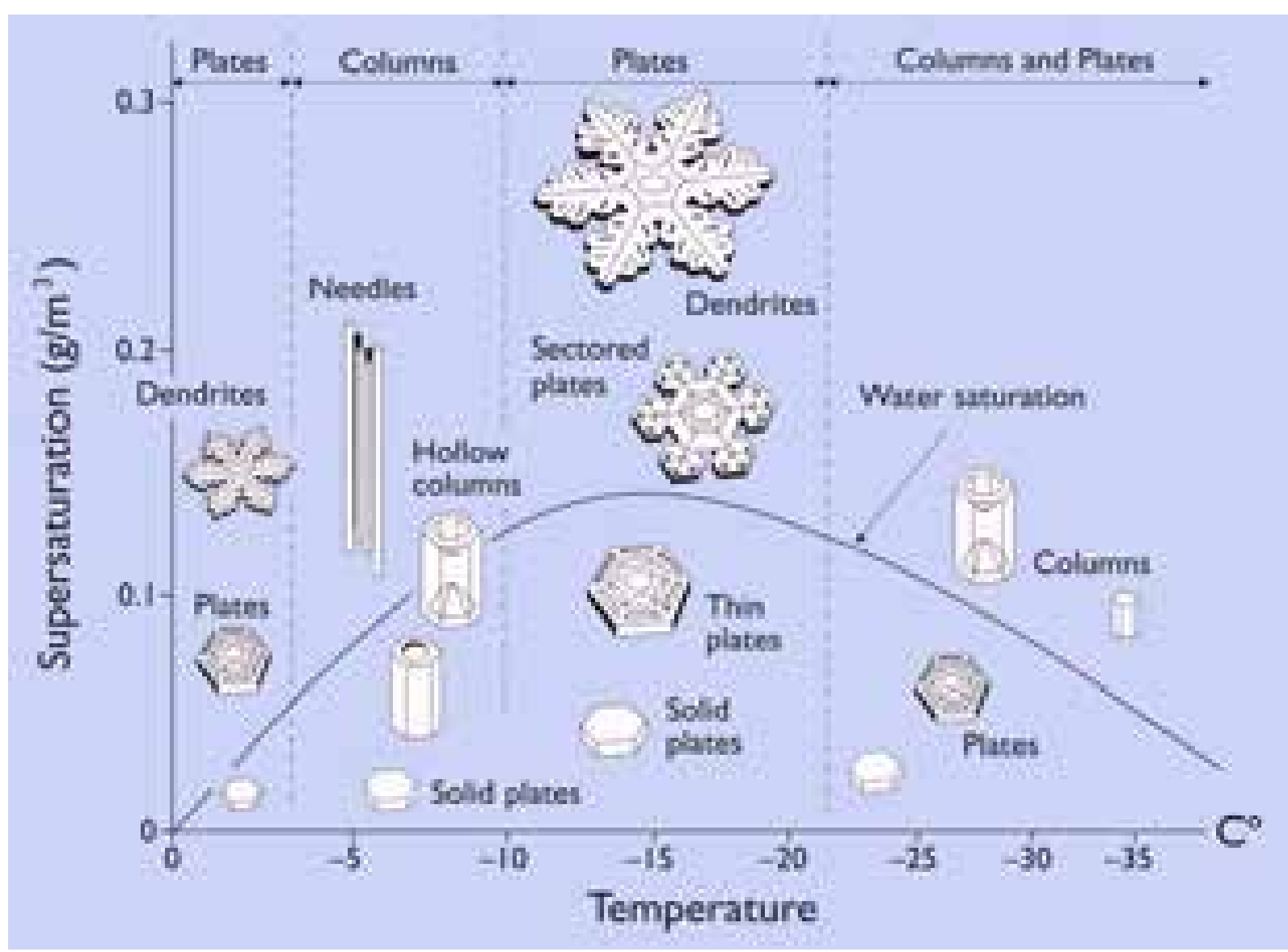

**Figure 1.12: The *Nakaya diagram* illustrates the morphologies of snow crystals that develop at different temperatures and supersaturations. The *water saturation* line shows the supersaturation that is typically found in a dense winter cloud made of liquid water droplets.**

The overarching goal of snow crystal science is to explain how all this works in detail. Developing empirical rules and recipes for creating different growth behaviors is fine for engineering designer snowflakes, but explaining why the recipes work in the first place presents quite a challenge. Researchers have been pondering the mechanisms underlying snow crystal formation for over 400 years, and the quest for true scientific understanding continues to this day.

## 1.2 A Brief History of Snow-Crystal Science

I like to think about the snow crystal as a case study of the scientific endeavor. Science is fundamentally about understanding the natural world, so snowflakes, being part of that world, deserve an explanation. Richard Feynman commented that *"Nature uses only the longest threads to weave her patterns, so each small piece of her fabric reveals the organization of the entire tapestry"* [1964Fey]. There is hardly a more fitting example of this truism than the intricate patterns of common snowflakes, as the full panoply of modern scientific knowledge is still not quite enough to explain their origin.

The study of snowflake science began when the distinctive six-fold symmetry of individual snow crystals was first recognized as something that could be investigated and possibly understood. Over time, this led to a greater scrutiny of what fell from the clouds, yielding early sketches that began to document the remarkable variety of different morphological types. With advances in technology, snow crystals were examined in greater detail using optical microscopy and further documented in extensive photographic studies. As more sophisticated scientific tools became available, researchers progressed from observations of natural snowfalls to scrutinizing laboratory-grown snowflakes, eventually leading to precision measurements of snow-crystal properties and growth rates, molecular-dynamics simulations, and investigations using computer-generated snowflakes.

In many ways, the snowflake story mirrors the historical development of science itself.

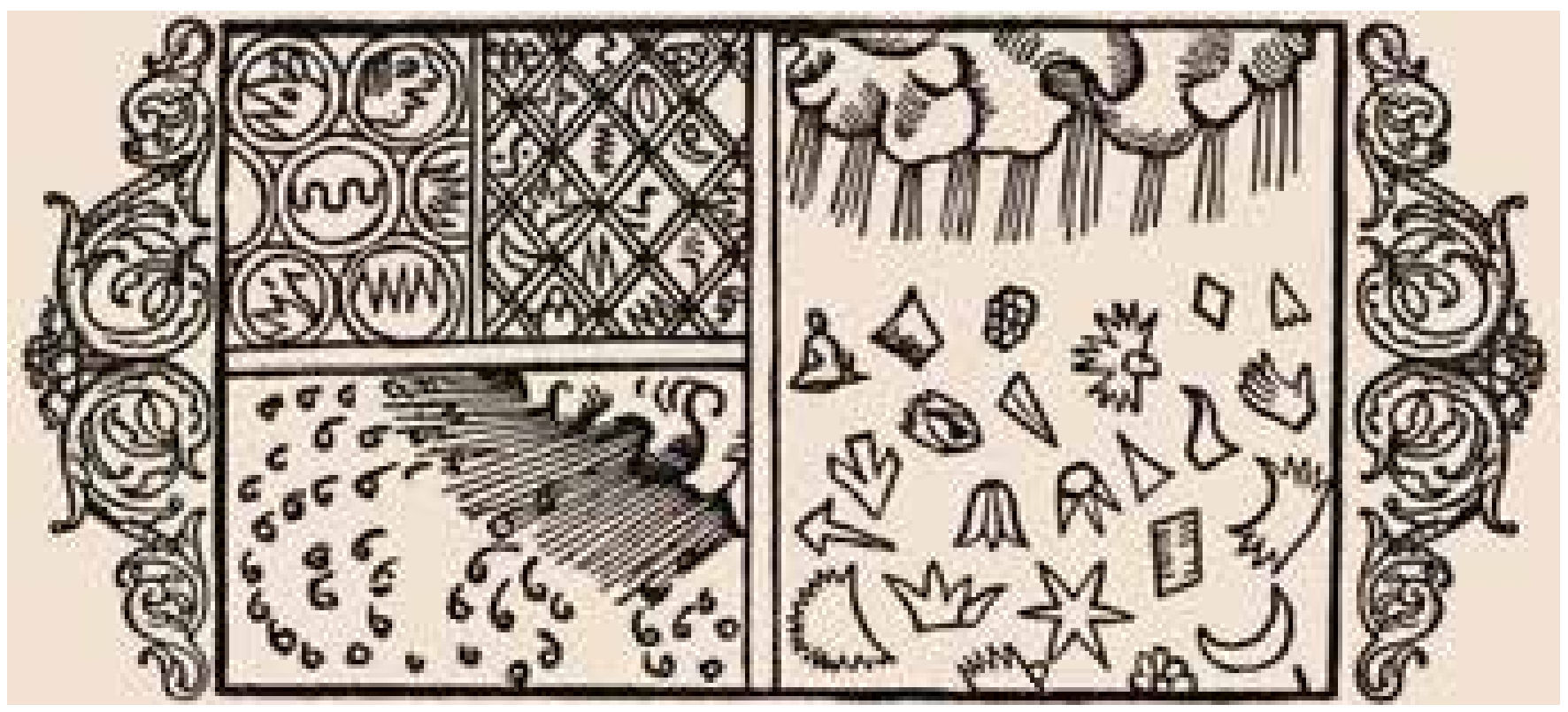

**Figure 1.13: This 1555 woodcut by Olaus Magnus was perhaps the first European illustration depicting a six-fold symmetrical snow crystal.**

European authors began documenting snowflakes many centuries after the first Asian accounts, and one oft-quoted reference is the woodcut shown in Figure 1.13, created by Olaus Magnus in 1555 [1982Fra]. It can be seen however, that the clergyman depicted snowflakes as having a curious assortment of odd shapes, including crescents, arrows, and even one that looked like a human hand, so perhaps this account does not quite warrant being called a historical first. It appears that English astronomer Thomas Harriot was the first in Europe to clearly identify and document the snowflake's six-fold symmetry in 1591 [1982Fra].

French philosopher and mathematician René Descartes recorded the first detailed account of snow crystal structures in his famous *Les Météores* in 1637, including the sketches shown in Figure 1.14. In his essay, Descartes described some remarkably thorough naked-eye observations of snow crystals, which included several uncommon forms [1982Fra]:

*After this storm cloud, there came another, which produced only little roses or wheels with six rounded semicircular teeth ...which were quite transparent and quite flat ...and formed as perfectly and symmetrically as one could possibly imagine. There followed, after this, a further quantity of such wheels joined two by two by an axle, or rather, since at the beginning these axles were quite thick, one could as well have described them as little crystal columns, decorated at each end with a six-petaled rose a little larger than their base. But after that there fell more delicate ones, and often the roses or stars at their ends were unequal. But then there fell shorter and progressively shorter ones until finally these stars completely joined, and fell as double stars with twelve points or rays, rather long and perfectly symmetrical, in some all equal, in others alternately unequal.*

Early observations of snow-crystal symmetry played a role in the creation of modern science by inspiring interest in the mathematical basis of Earth-bound natural phenomena. As laboratory-based science emerged, synthetic snowflakes revealed an intrinsic order in the observed diversity of snow crystal morphologies. And as the nanoscale structure of crystalline materials has become better characterized in the modern era, our understanding of the attachment kinetics governing snow crystal growth has improved as well. We can only guess as to what future scientific tools will be brought to bear in our quest to comprehend the inner workings of the lowly snowflake.

## Early Observations

The earliest account (of which I am aware) describing the six-fold symmetry of individual snow crystals was written in 135 BC by Chinese philosopher Han Yin [2002Wan], who commented: *"Flowers of plants and trees are generally five-pointed, but those of snow, which are call ying, are always six-pointed."* Subsequent Chinese authors mentioned snow-crystal symmetry as well, an example being the sixth-century poet Hsiao Tung, who penned, *"The ruddy clouds float in the four quarters of the cerulean sky. And the white snowflakes show forth their six-petaled flowers."*

In this passage, we can see snowflakes influencing—in their own small way—the early development of modern science. Descartes was clearly impressed with the geometrical perfection he saw in snow crystal forms, with their flat facets and hexagonal symmetry. Pondering this and other observations, he went on to reason how the principles of geometry and mathematics play a central role in describing the natural world. Although we take this for granted now, using mathematics to explain ordinary phenomena was still something of a novel idea at the time, and a major step forward in science.

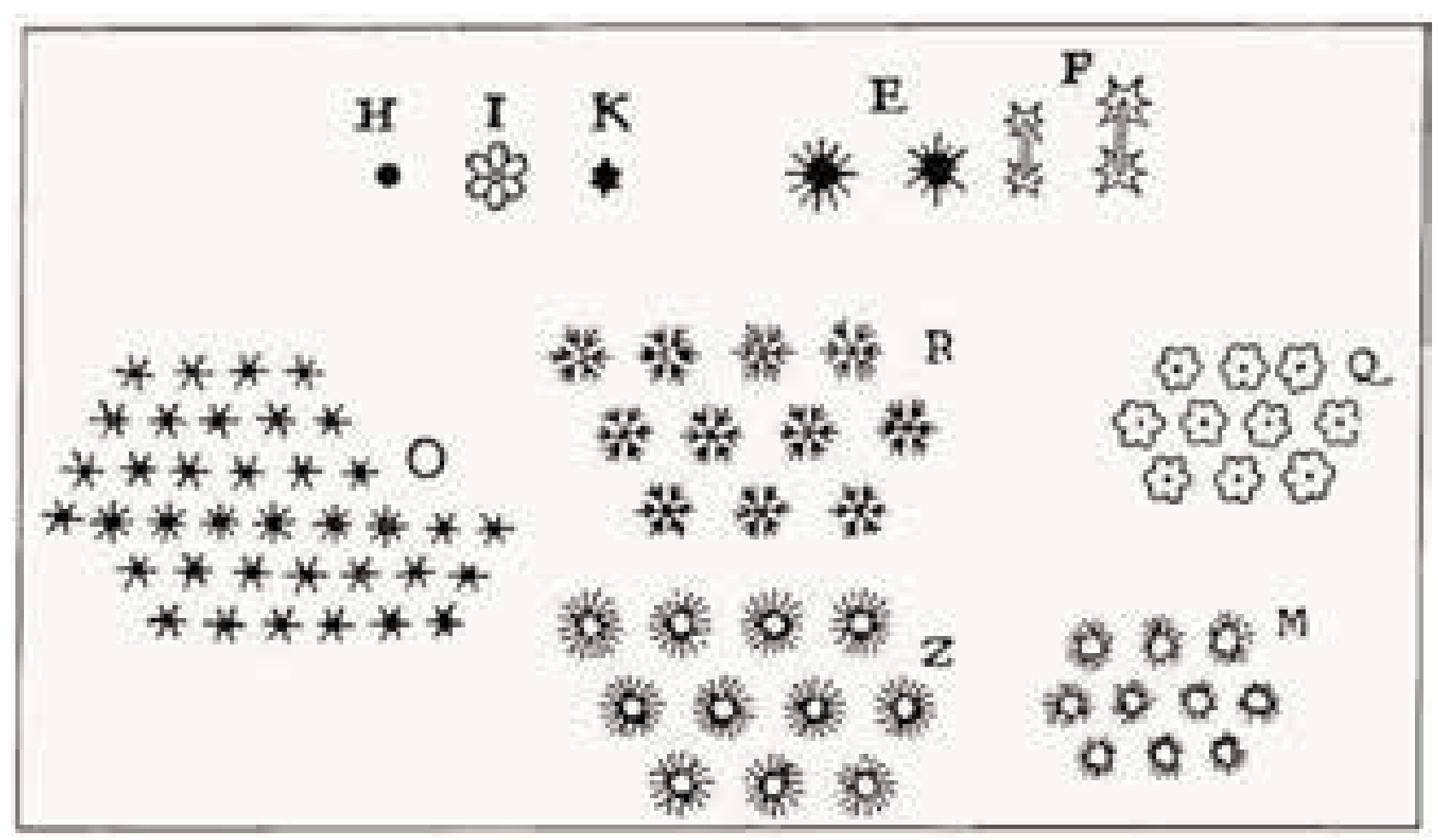

**Figure 1.14: Rene Descartes made some of the first accurate sketches of different snow crystal morphologies 1637, including observations of capped columns (group F in this sketch) [1637Des].**

## Emerging Science

The first scientist to speculate about an actual theory to explain the six-fold symmetry of snow crystals was German astronomer and mathematician Johannes Kepler. In 1611, Kepler presented a small treatise entitled *The Six-Cornered Snowflake* to his patron, Holy Roman Emperor Rudolf II, as a New-Year's Day gift. In his treatise, Kepler contrasted the six-fold symmetry of snowflakes with similar symmetries found in flowers. He deduced that the similarities must be in appearance only, because flowers are alive and snowflakes clearly are not:

*Each single plant has a single animating principle of its own, since each instance of a plant exists separately, and there is no cause to wonder that each should be equipped with its own peculiar shape. But to imagine an individual soul for each and any starlet of snow is utterly absurd, and therefore the shapes of snowflakes are by no means to be deduced from the operation of soul in the same way as with plants.*

Kepler saw that a snowflake is a relatively simple thing, made only from ice, compared to the baffling complexity of living things. He offered, therefore, that there might be some relatively simple organizing principle that was responsible for snow-crystal symmetry. Drawing upon correspondence with Thomas Harriot, Kepler noted that stacking cannonballs also yielded geometric structures with six-fold symmetry, and he further surmised that there might be a mathematical connection between these two phenomena. There was certainly a germ of truth in this reasoning, as the geometry of stacking water molecules lies at the heart of snow-crystal symmetry. But this was long before the atomistic view of matter had been developed, so Kepler could not carry the cannonball analogy very far.

Kepler realized that the genesis of crystalline symmetry was a worthy scientific question, and he also recognized the similarity between snow crystals and mineral crystals, as they both exhibited symmetrical faceted structures. At the end of his treatise, however, Kepler accepted that the science of his day was not advanced enough to explain any of it. He was certainly correct in this conclusion, for three centuries would pass before scientists knew enough about atoms, molecules, and their arrangement in solid materials to finally answer Kepler's 1611 query.

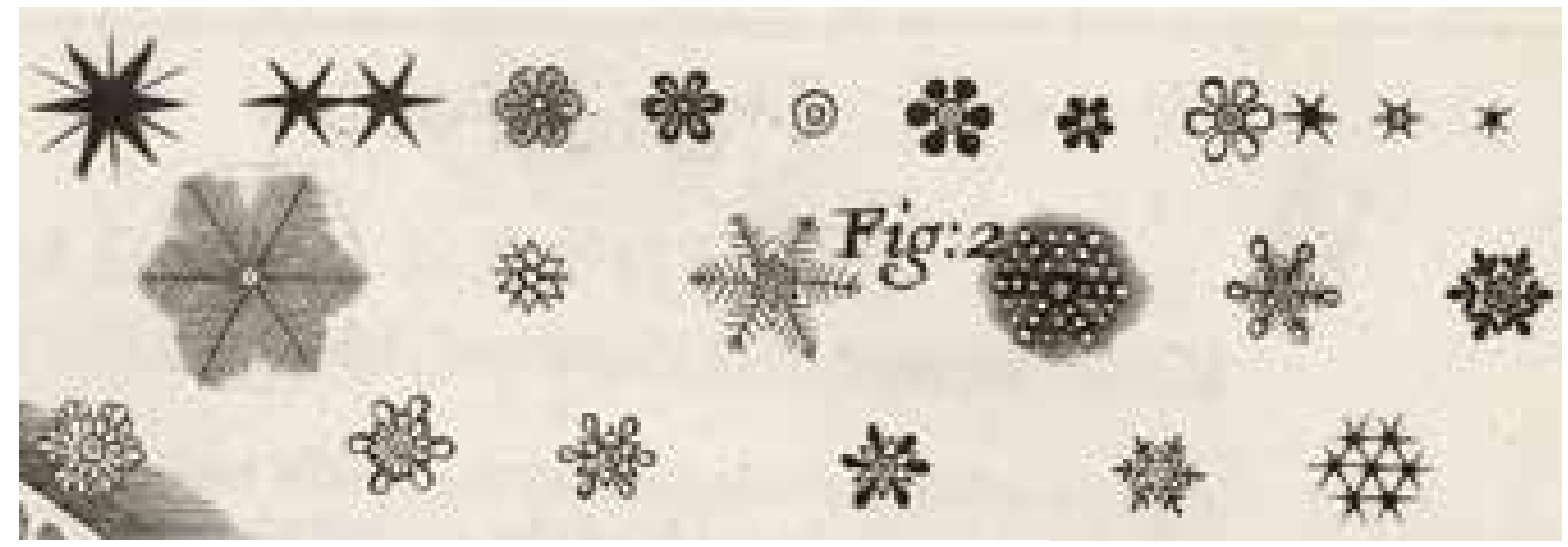

**Figure 1.15: Robert Hooke sketched these observations of snowflakes 1665, enabled by his newly invented microscope [1665Hoo].**

## Microscopic Observations

The invention of the microscope in the mid-seventeenth century quickly led to more and better snowflake observations. English scientist and early microscopist Robert Hooke sketched snowflakes (see Figure 1.14) and practically everything else he could find for his book *Micrographia*, published in 1665 [1665Hoo]. Although his microscope was crude by modern standards, Hooke's drawings nevertheless began to reveal the complexity and intricate symmetry of snow-crystal structure, details that could not be detected with the unaided eye.

As the quality and availability of optical magnifiers improved, so did the accuracy of snow crystal drawings. By the mid-nineteenth century, a number of observers around the globe had recorded the diverse character of snow crystal forms, and one notable example is shown in Figure 1.16. Given the ephemeral nature of a snowflake, however, observers inevitably relied on memory to complete their sketches. As a result, even the best snow crystal drawings lacked detail and were not completely faithful to their original subjects.

**Figure 1.16: (Below) English explorer William Scoresby made these sketches during a winter voyage through the Arctic, which he recounted in 1820 [1820Sco]. These are the first drawings that accurately depicted many features of snow-crystal structure, as well as several rare forms, including triangular crystals and capped columns. Scoresby also noted that the cold arctic climate produced more highly symmetrical crystals than were typically seen in Britain.**

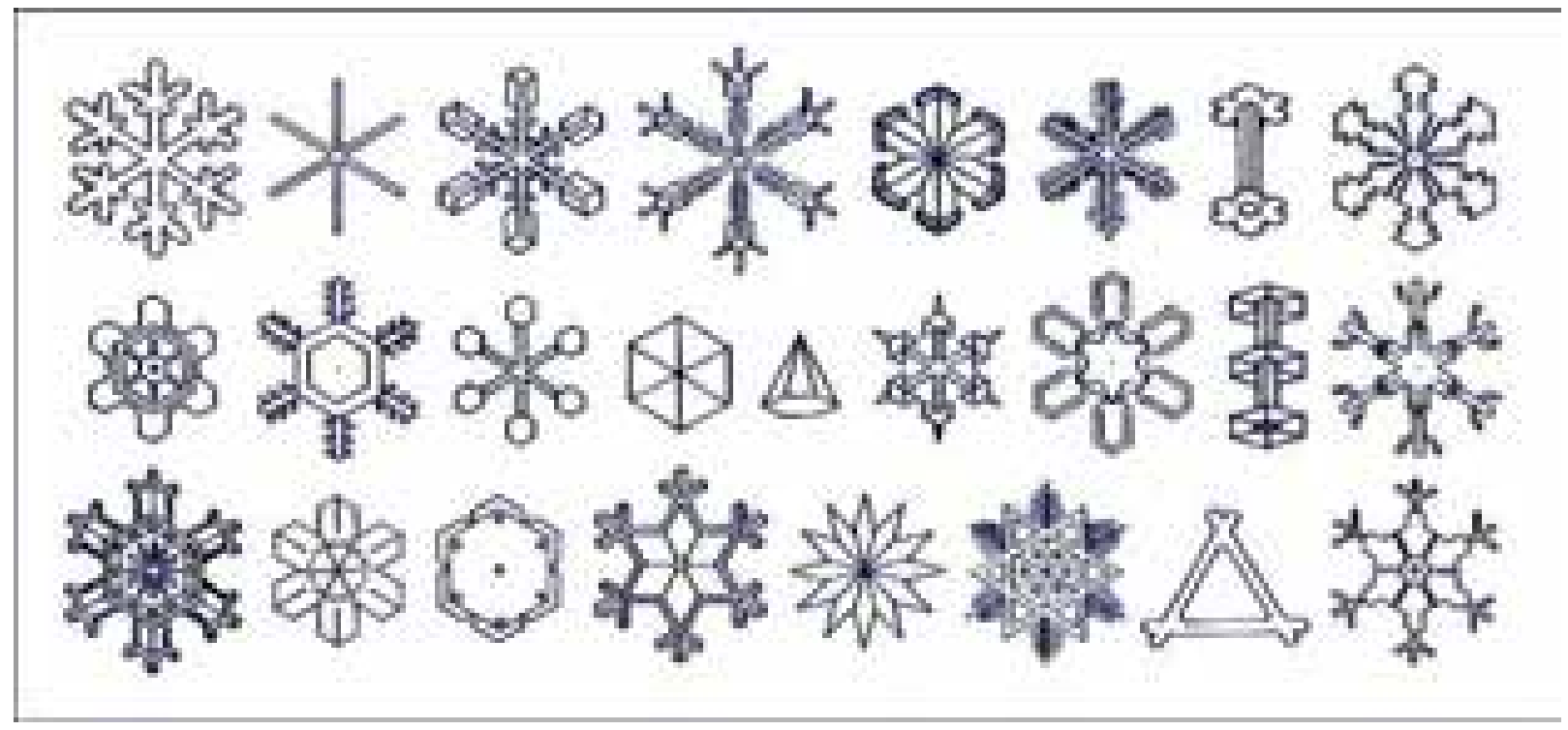

## Snowflake Photography

It took Wilson Bentley, a farmer from the small town of Jericho, Vermont, to create the first photographic album of falling snow, thus awakening the world to the hidden wonders of snowflakes. Bentley became interested in the microscopic structure of snow crystals as a teenager in the 1880s, and he soon began experimenting with the new medium of photography as a means of recording what he observed. He constructed an ingenious mechanism for attaching a camera to his microscope for this purpose, and he succeeded in photographing his first snow crystal in 1885, when he was 19 years old.

To say Bentley was dedicated to the task is an understatement. Snowflake photography became his lifelong passion, and over the course of forty-six years he captured more than 5,000 snow crystal images, each on a four-inch glass photographic plate. He resided his entire life in the same Jericho farmhouse, photographing snowflakes each winter using the same equipment he constructed as a teenager. Figure 1.17 shows

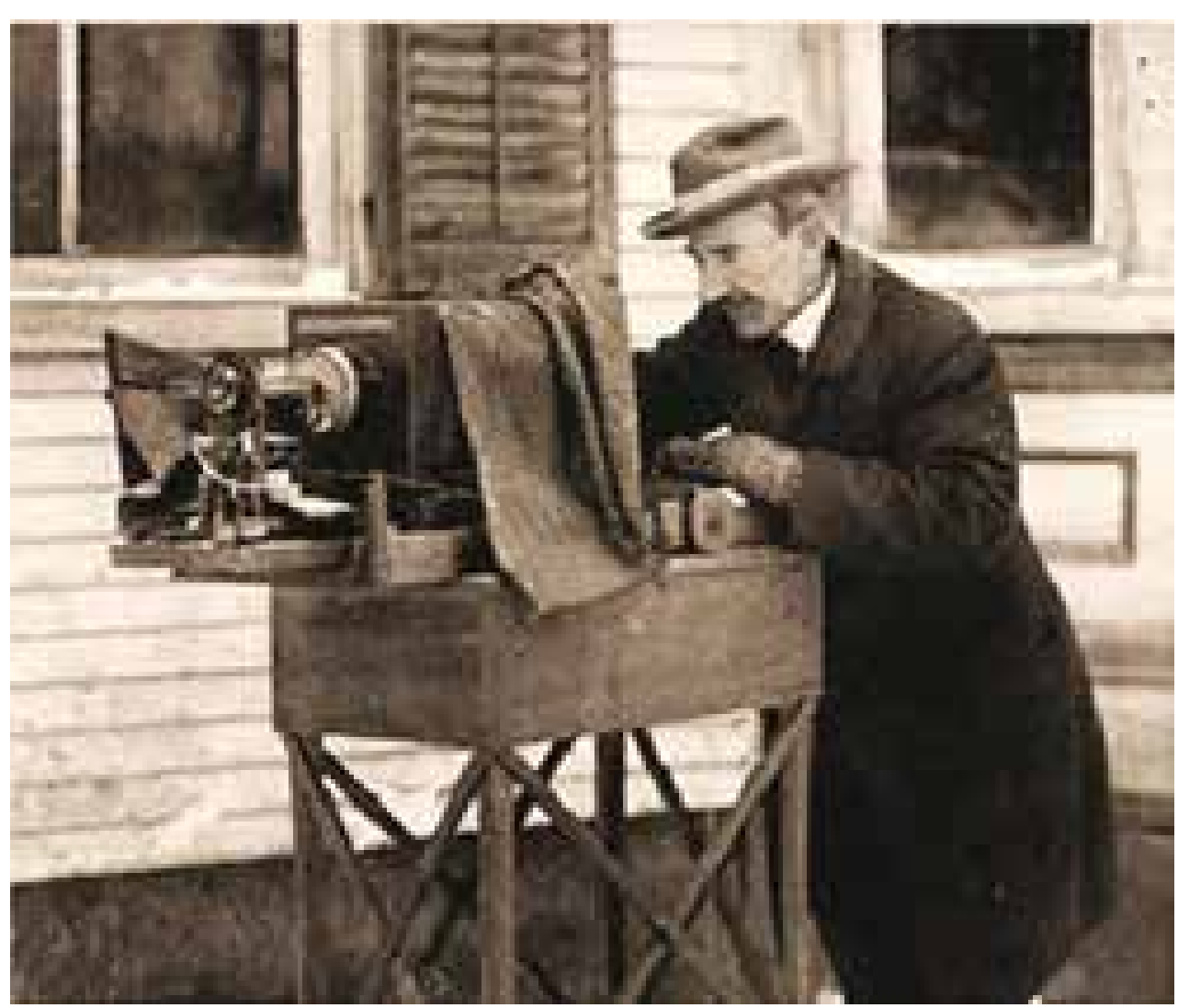

**Figure 1.17: Vermont farmer Wilson Bentley first developed the art of snowflake photography in the 1880s, eventually producing a large album of images. He is shown here with his specially built snow-crystal photo-microscope [1931Ben].**

Wilson Bentley demonstrating his apparatus, although the grass at his feet suggests there were no snowflakes to be found that day.

While Bentley usually presented his photographs as white snowflakes on a black background, as shown in Figure 1.18, the original photos had a bright background. A snow crystal is made of pure ice, which is clear, not white. When illuminated from behind, as Bentley did, a snow-crystal photo exhibits a somewhat low-contrast "bright on bright" appearance (see Chapter 11). To increase the contrast, Bentley made a copy of each photographic negative and painstakingly scraped away the emulsion from the background areas. A print made from the modified negative then yielded a white snowflake on a black background, as shown in Figure 1.18. Bentley preferred this high-contrast look, so he modified most of his photos using this technique. Some have accused Bentley of altering his photos to augment what nature had provided, but he did not hide the fact that he processed his photos this way. And he was always quick to point out that he never changed the snow-crystal images themselves during this process.

One aspect of his work that Bentley rarely emphasized is that large, symmetrical stellar snow crystals are not the norm (see Chapter 10). Over the course of an entire winter season, he only photographed about a hundred specimens on average, reserving his expensive emulsions for only the most photogenic snow crystals he could find. Modern automated cameras that photograph falling snow without any selection bias confirm that well-formed stellar crystals are exceedingly rare [2012Gar].

Bentley's photographs appeared in numerous publications over several decades, providing for many their first look at the inner structure and symmetry of snow crystals. And with thousands of snowflakes, all unique, the world was exposed to their incredible variety as well. The now-familiar old chestnut that no two snowflakes are exactly alike appears to have had its origin in Bentley's photographs.

**Figure 1.18: (Below) These are just a few of the thousands of snowflake photographs taken by Wilson Bentley between 1865 and 1931. The original photos showed bright crystals against a bright background, as the clear snowflakes were illuminated from behind. The photos were modified by essentially cutting each crystal out and placing it on a black background [1931Ben].**

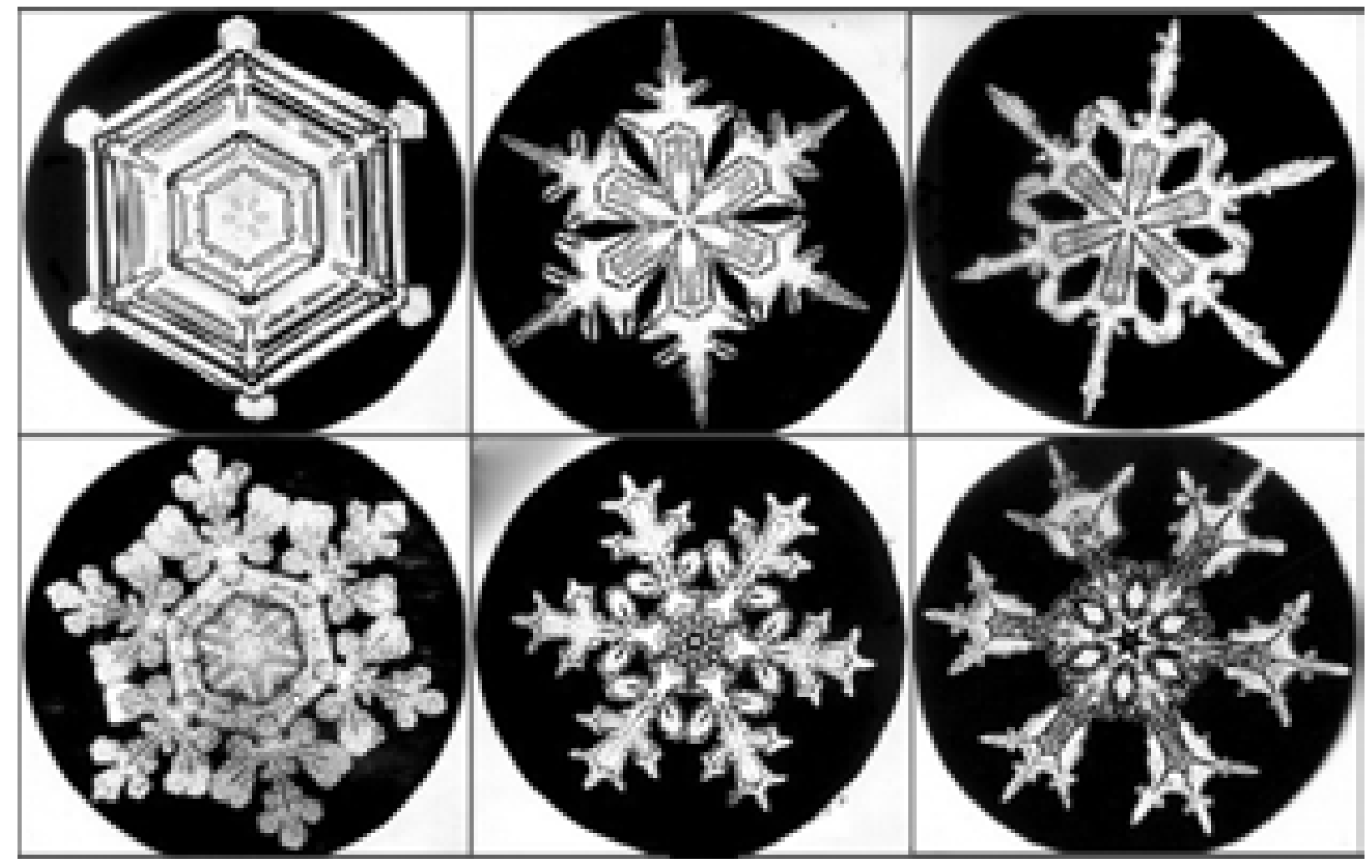

In the late 1920s, Bentley teamed with W. J. Humphreys, chief physicist for the United States Weather Bureau, to publish his magnum opus containing more than 2,000 snow crystal photographs [1931Ben]. The book appeared in November 1931, and the 66-year-old Vermont farmer died of pneumonia just a few weeks later. In the decades that followed this seminal work, many others have taken up the challenge of capturing the structure and beauty of snow crystals using photography, and I describe some techniques and results in Chapter 11.

## Crystallography

The word crystal derives from the Ancient Greek *krystallos*, meaning "ice" or "rock ice." Contrary to what the definition implies, *krystallos* was not originally used to describe ice, but rather the mineral quartz. The early Roman naturalist Pliny the Elder described clear quartz *krystallos* as a form of ice, frozen so hard that it could not melt. Pliny was quite mistaken on this point, as quartz is not a form of ice, nor is it even made of water. Nevertheless, after nearly 2,000 years, Pliny's misunderstanding is still seen in the language of the present day. If you look in your dictionary, you may find that one of the definitions for crystal is simply "quartz."

While mineral collectors have admired beautiful crystalline specimens for millennia, understanding the origin of their faceted structures required a bona fide scientific breakthrough. In 1912, German physicist Max von Laue and co-workers discovered that when X-rays were shone through a crystal of copper sulphate, the crystal acted like a grating and produced a diffraction pattern that could be measured on photographic film. Australian-born British physicists William Henry Bragg and William Lawrence Bragg (father and son) soon developed a mathematical theory showing how the atomic structures of crystalline materials could be ascertained from these diffraction patterns, thus creating the field of crystallography.

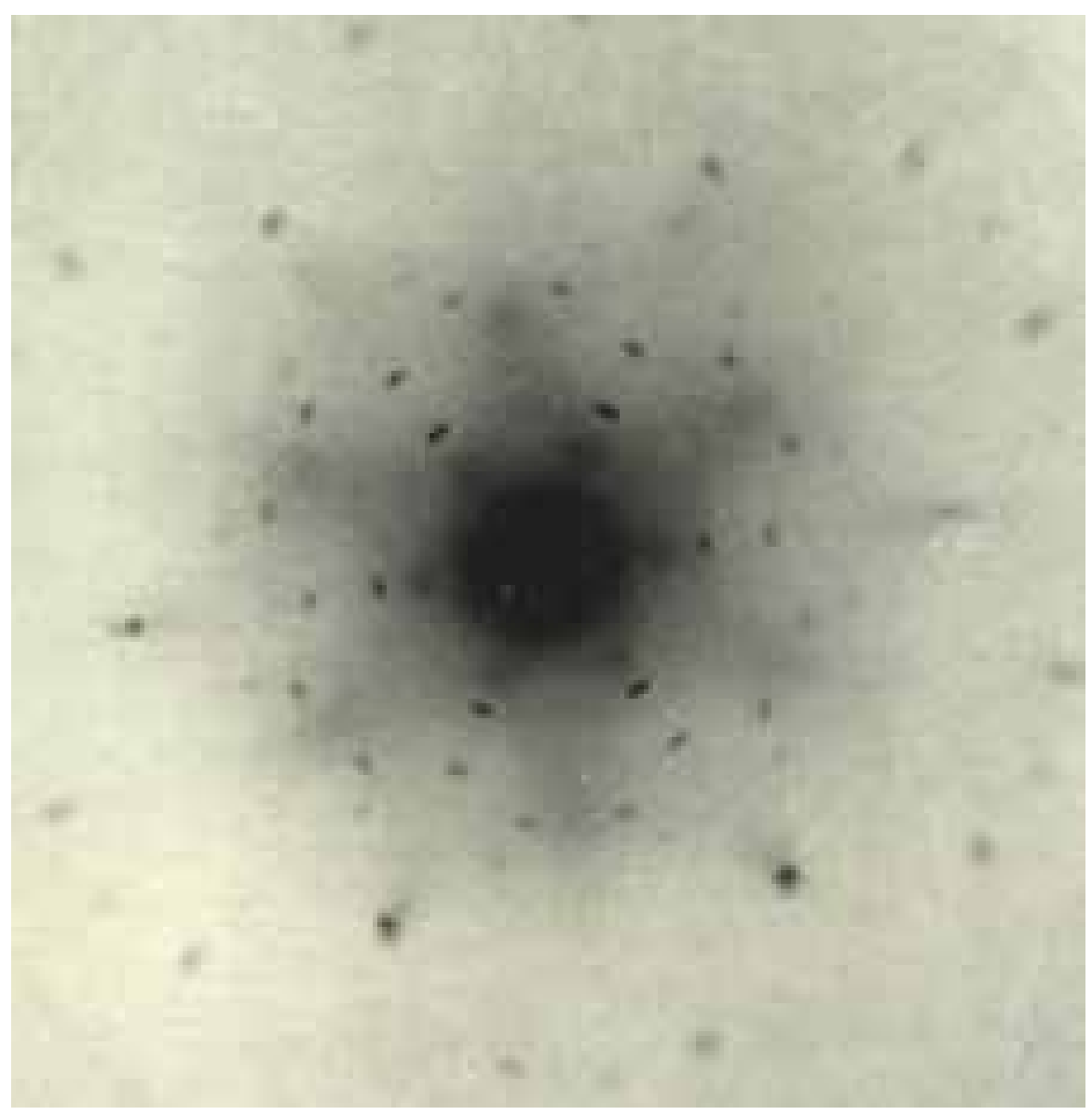

**Figure 1.19: This X-ray diffraction pattern was made by a block of ice, used by William Barnes to derive the hexagonal lattice structure of the ice crystal [1929Bar].**

Working with the Bragg's in their Cambridge laboratory, William Barnes used X-ray crystallography to determine the structure of ice for the first in 1929 [1929Bar], discovering the now-familiar hexagonal lattice of ice Ih, which is the normal form of environmental ice found in snow crystals (see Chapter 2). Figure 1.19 shows Barnes's discovery photograph. In subsequent studies over many decades, numerous additional solid phases of water have been discovered and characterized, mostly at very high pressures. Three hundred years after Kepler's initial musings, scientists had finally proven that the geometry of stacking was indeed the underlying source of the snowflake's six-fold symmetry.

In the decades that followed these early crystallographic discoveries, the development of quantum mechanics and quantum chemistry have allowed precise *ab initio* calculations of the water molecule electronic and atomic structure, including two-body and higher order interactions between water molecules. And from these fundamental quantum-mechanical calculations, researchers have been able to reproduce the known structures of water in many of its solid phases. As a result, the lattice structure of ice Ih, and thus snow crystals, is now well understood at the most fundamental physical level.

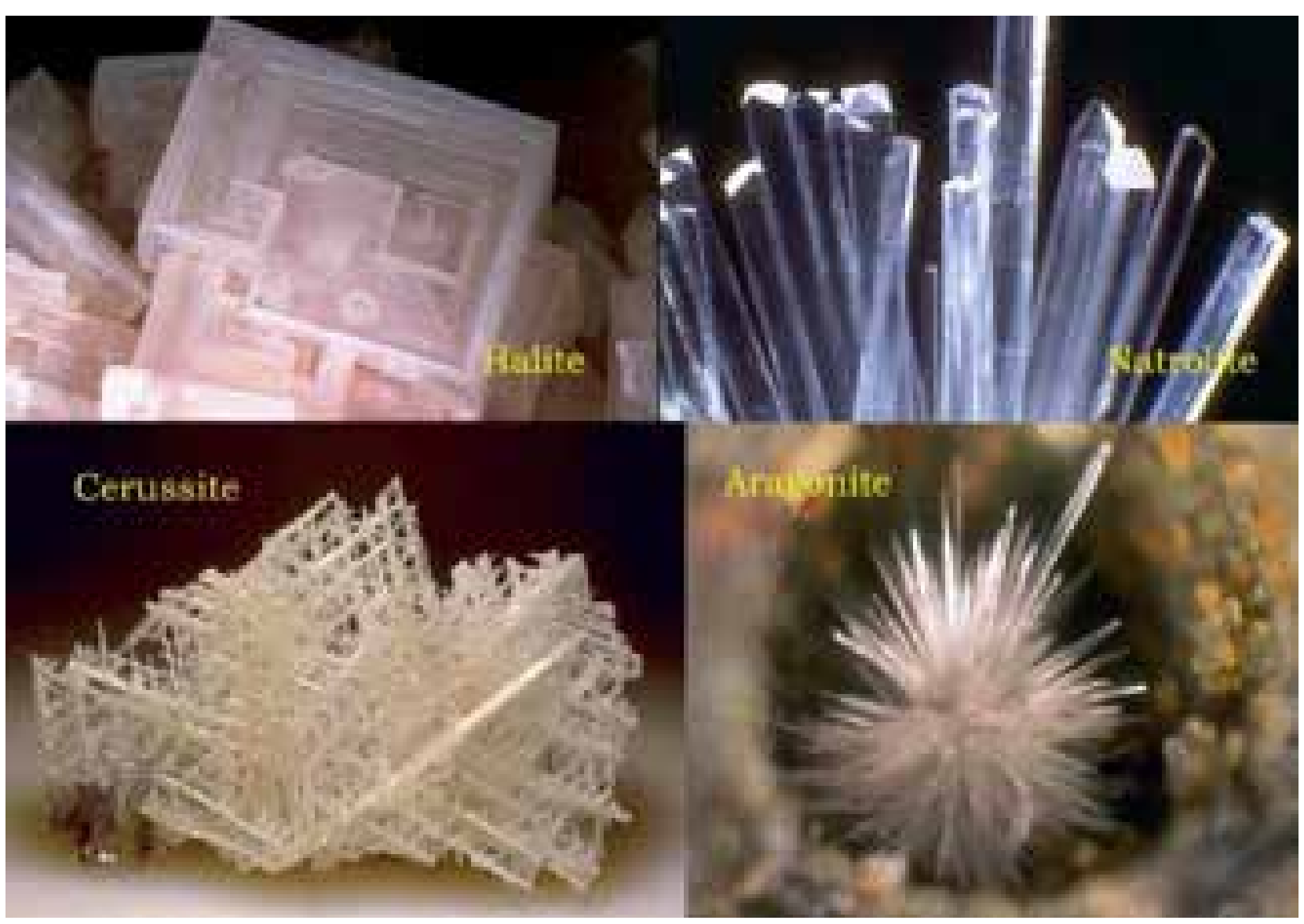

**Figure 1.20: Many mineral crystals grow into faceted morphologies under the right conditions, as seen in these examples. The lattice structure determines the overall symmetry of these crystalline forms, but the attachment kinetics is largely responsible for the appearance of faceted surfaces.**

## Attachment Kinetics

While the six-fold symmetry of a snowflake ultimately derives from the symmetry of the ice crystal lattice, how the nanoscale structure of the molecular matrix translates into the large-scale morphology of a growing crystal is a separate matter. For example, quartz and copper are both crystalline minerals, but quartz often exhibits striking faceted features that reveal its lattice structure, while copper rarely does. Why? The answer lies in the physical processes that govern the formation of faceted surfaces, collectively called the *surface attachment kinetics*.

Around the beginning of the 20th century, scientists began examining the physics of solidification using the newly developed laws of statistical mechanics, which were being developed by James Clerk Maxwell, Ludwig Boltzmann, J. Willard Gibbs, Amedeo Avogadro, Lord Kelvin, and other scientific luminaries throughout the 19th century. An early result came from German physicist Heinrich Hertz [1882Her] and independently from Danish physicist Martin Knudsen [1915Knu], who calculated the growth rate of a solid from its vapor phase (like ice from water vapor) as a function of the net flux of vapor molecules striking the solid surface. The resulting Hertz-Knudsen law provides the starting point for the surface attachment kinetics I describe in Chapter 4.

Some decades later, however, it was becoming clear that the Hertz-Knudsen law did not provide a good description of the growth of faceted crystalline surfaces. The net flux of molecules striking the surface was only one factor determining the growth rate; another was the probability that an impinging molecule would permanently attach to the surface and become part of the bulk crystal lattice. This probability, ranging from zero to one, is now called the *attachment coefficient*, also discussed at length in Chapter 4. Figure 1.8 shows how an anisotropic attachment coefficient produces faceted crystal growth, and this mechanism is also responsible for the appearance of faceted minerals, such as those shown in Figure 1.20.

Beginning around the 1930s, physicists I. N. Stranski [1928Str], R. Kaischew [1934Str], R. Becker and W. Döring [1935Bec], M. Volmer [1939Vol], and others pushed the field forward by developing a detailed a statistical-mechanical theory describing the nucleation and subsequent growth of one-molecule-high *terraces* on flat faceted surfaces. Many additional scientists fortified this theory in the following decades, notably W. K. Burton, N. Cabrera, and F. C. Frank [1951Bur], building it into the modern theory of crystal growth and surface attachment kinetics that is described in modern textbooks [1996Sai, 1998Pim, 2002Mut, 2004Mar].

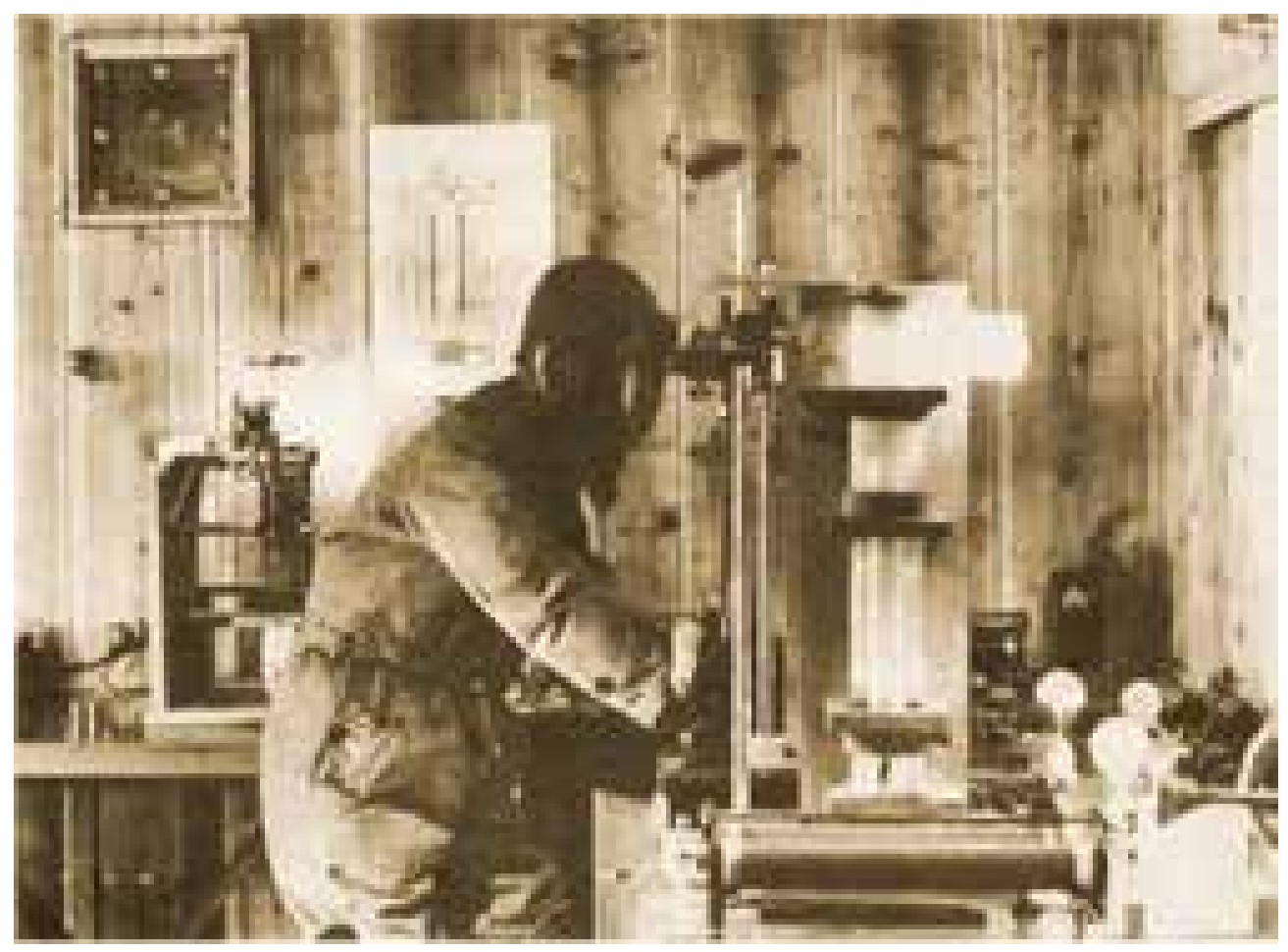

**Figure 1.21: Japanese physicist Ukichiro Nakaya working in his refrigerated snow-crystal laboratory in Hokkaido University [1954Nak].**

In snow crystal growth, the attachment kinetics are a major factor in determining the growth rates and resulting morphologies in different environmental conditions. For example, the main difference between a thin-plate snow crystal (Figures 1.1 and 1.3) and a columnar snow crystal (Figure 1.2) lies mainly in the attachment kinetics. More broadly, the large-scale structure of nearly every snowflake is shaped to a large degree by how the surface attachment kinetics changes with temperature and other factors.

It is a common misconception to think that crystallography explains crystal growth, but this is far from the truth. Crystallography refers to the lattice structure of crystalline materials, and this is entirely a statics problem describing the lowest-energy molecular configuration in equilibrium. Crystal growth, on the other hand, is a dynamical problem involving many-particle interactions in systems far from equilibrium. Modern science is good at statics problems, but less so with many-body dynamics problems.

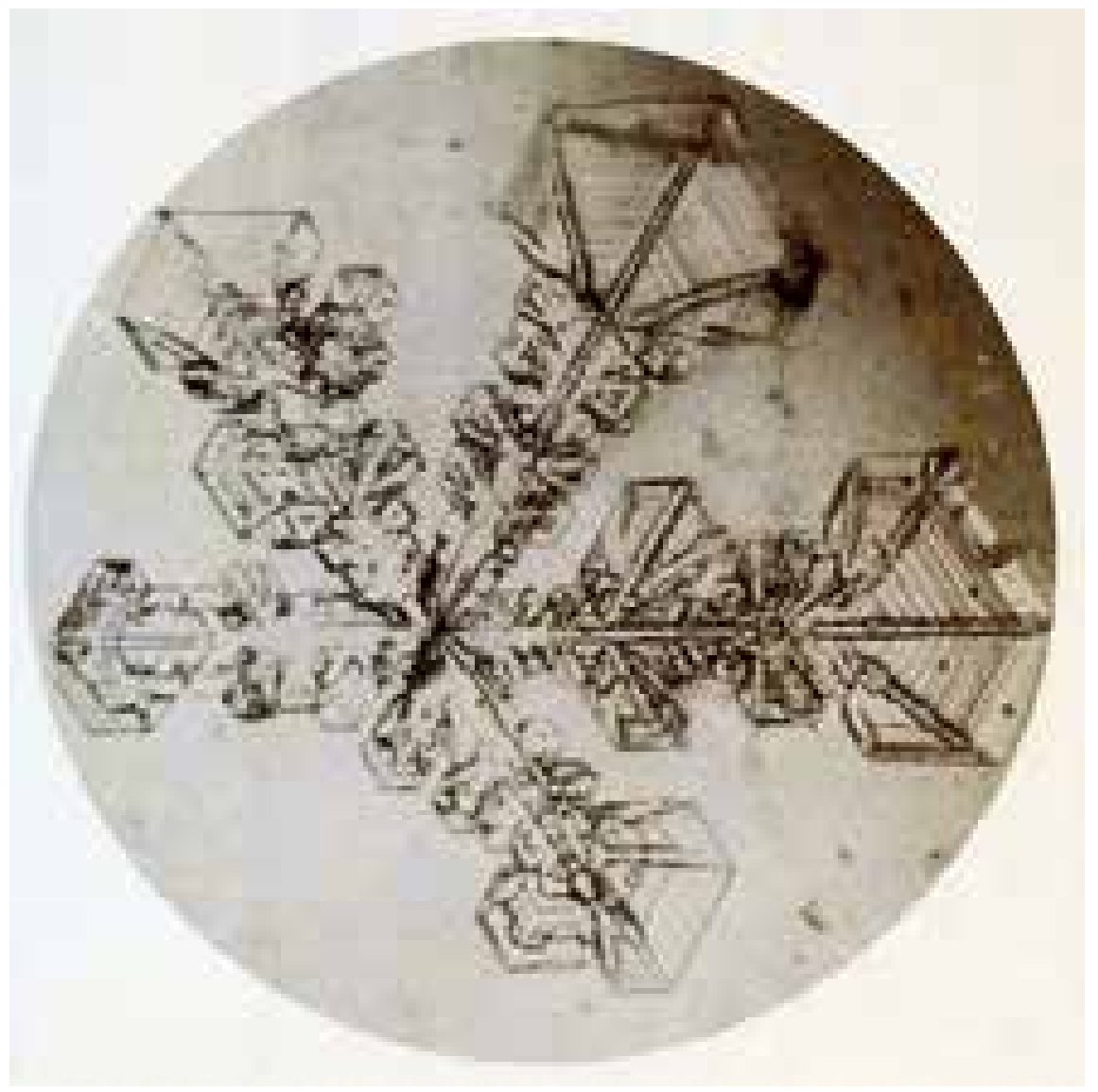

**Figure 1.22: The world's first synthetic snow crystal, grown by Ukichiro Nakaya on March 12, 1936 [1954Nak].**

For this reason, the crystallography of ice has been essentially solved for decades, while many important aspects the ice attachment kinetics remain quite puzzling. Terrace nucleation theory from the 1930s nicely explains many aspects of snow-crystal attachment kinetics, but certainly not all. As I describe in Chapter 4, creating a comprehensive model of snow-crystal attachment kinetics is very much a work in progress, with many unsolved problems still outstanding.

## The Nakaya Diagram

Japanese physicist Ukichiro Nakaya conducted the first true scientific investigation of snow crystals at Hokkaido University in the 1930s. Motivated by the abundant snowfalls in Hokkaido, and inspired by Wilson Bentley's photographs, Nakaya began his investigations by cataloging the different types of falling snow. Unlike Bentley, Nakaya looked beyond stellar crystals and focused his attention on describing the full range of different snowflake types, including columns, needles, capped columns, and other less-common forms.

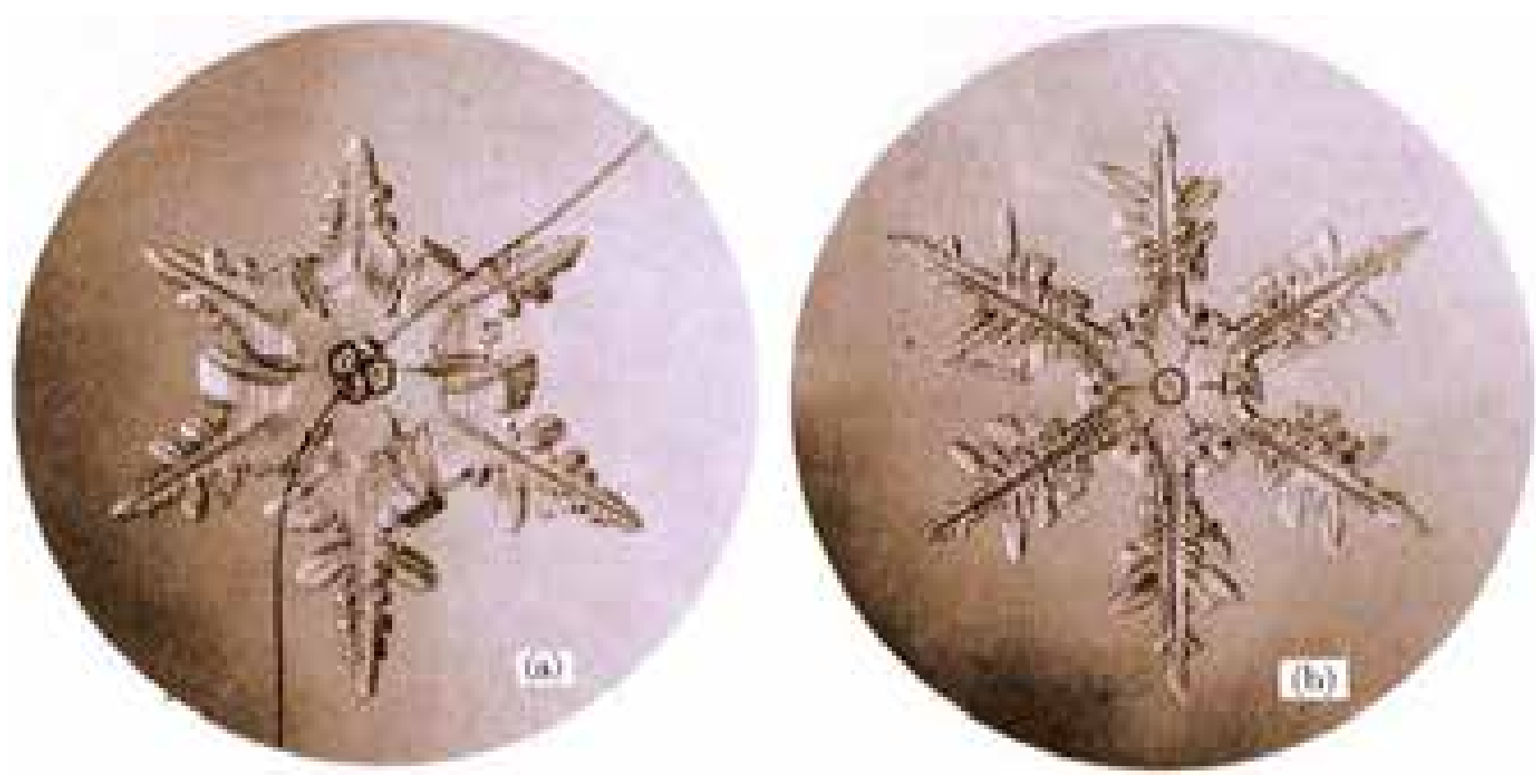

**Figure 1.23: A comparison between a synthetic snow crystal grown on a thin fiber near -15 C (left) with a natural snow crystal (right) [1954Nak].**

Nakaya thus produced the first photographic documentation of the broader menagerie of falling snow.

While learning a great deal from direct observations, Nakaya quickly realized that laboratory experiments would be essential for better understanding the origin of what he saw falling from the clouds. To this end, he constructed a walk-in freezer laboratory at Hokkaido, which he used for a variety of experimental investigations. Prominent among them, Nakaya created the world's first laboratory-grown snowflakes in his lab, shown in Figures 1.21 through 1.23.

Nakaya spent years examining how his synthetic snow crystals grew and developed as he varied the temperature and supersaturation within his growth chamber, soon combining all his observations into what is now called the *snow crystal morphology diagram*, or the *Nakaya diagram*, shown in Figure 1.24. Subsequent researchers have further refined and clarified the morphology diagram, yielding the progression of improved versions shown in Figure 1.25. More recently, Bailey and Hallett extended these results with additional observations exploring temperatures down to -70 C [2009Bai, 2012Bai].

Nakaya's morphology diagram was immediately recognized as being like a Rosetta Stone for snowflakes. With it, one can translate the shape of a falling snow crystal into a description of its growth history. Upon seeing a slender needle crystal, for example, one can deduce that it must have grown in high humidity at a temperature near -5 C. A large stellar crystal suggests growth near -15 C, and the amount of sidebranching provides an indication of the level of supersaturation it experienced.

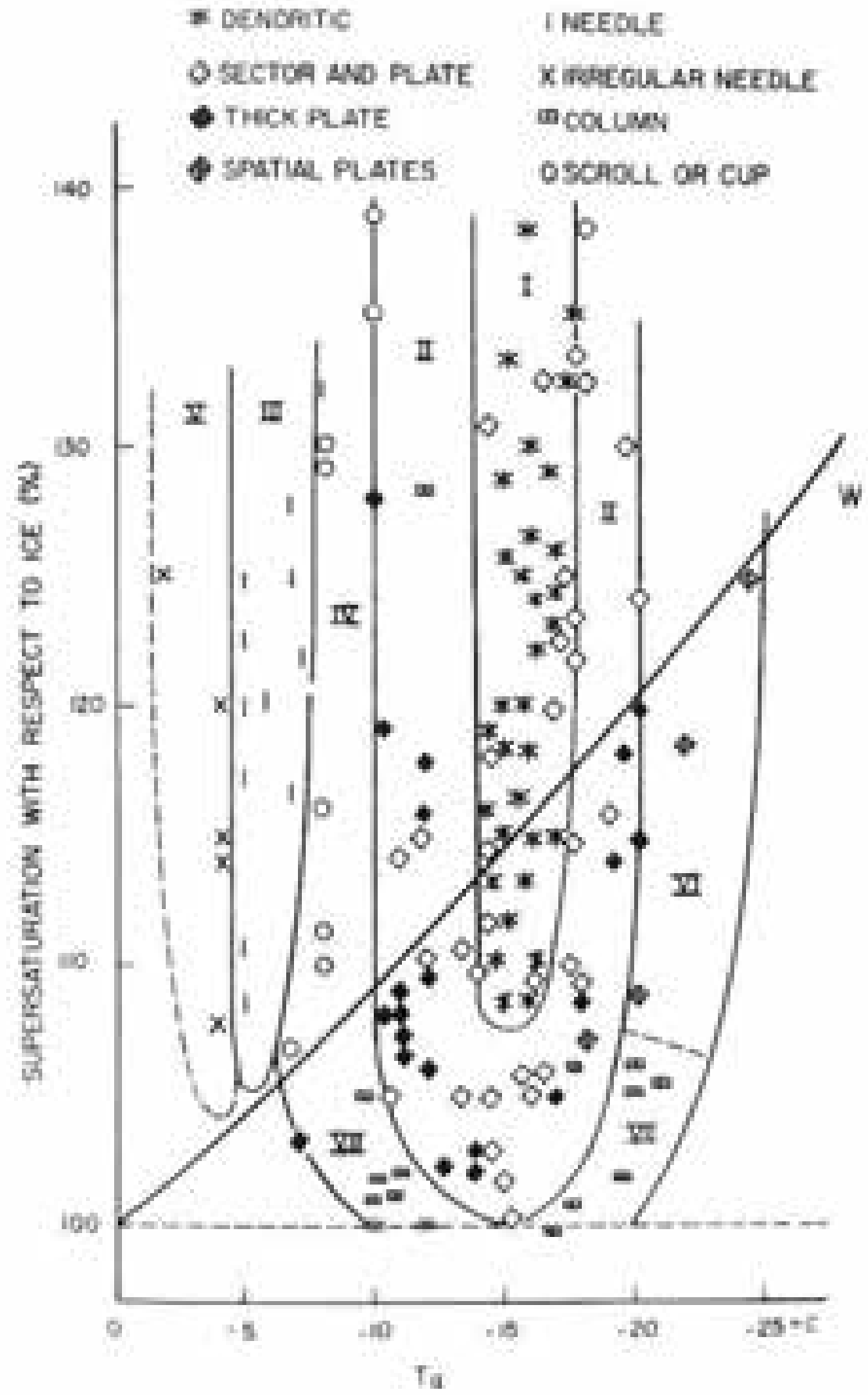

**Figure 1.24: Nakaya's *Snow Crystal Morphology Diagram* [1954Nak, 1958Nak] illustrates the different types of snow crystals that grow under different environmental conditions. For example, large stellar plates only form in a narrow temperature range around -15 C, while slender needle crystals only appear near -5 C.**

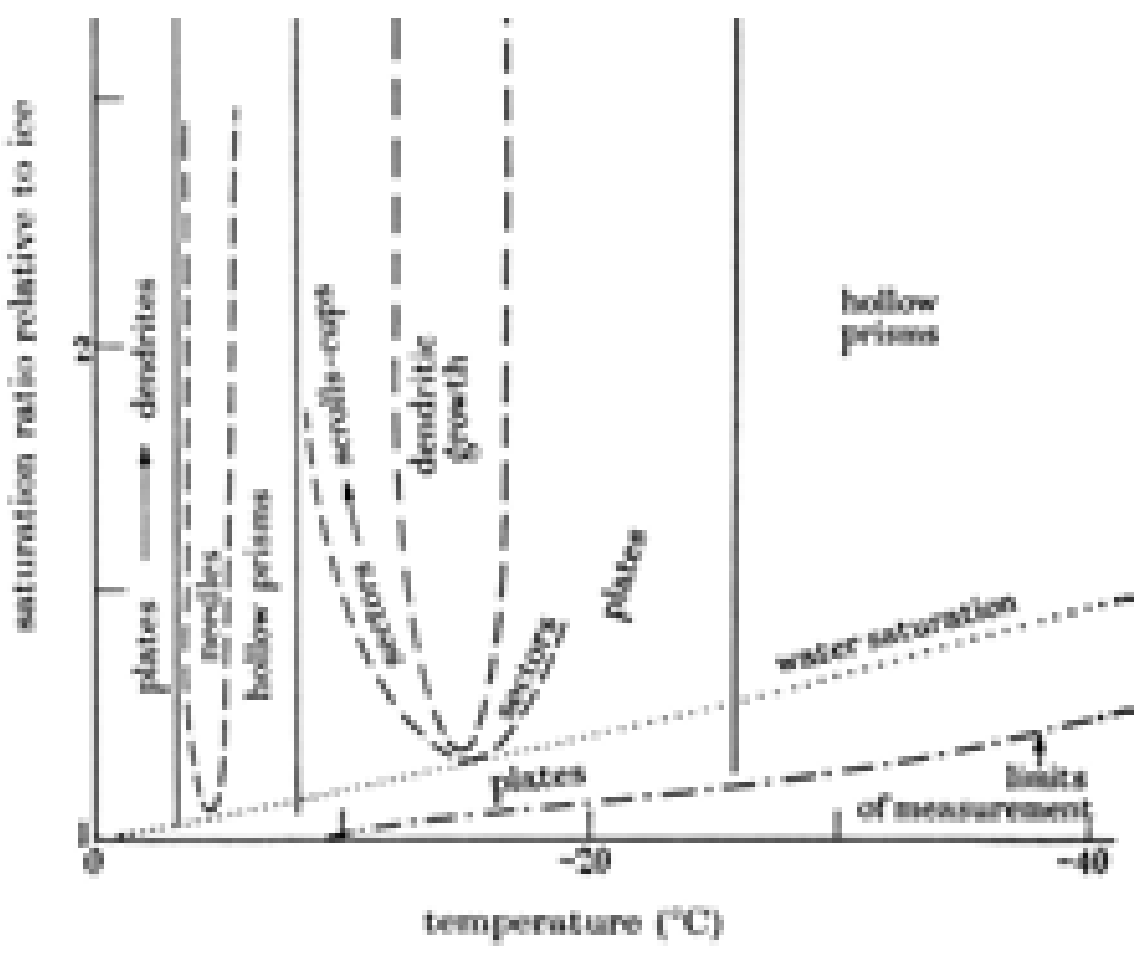

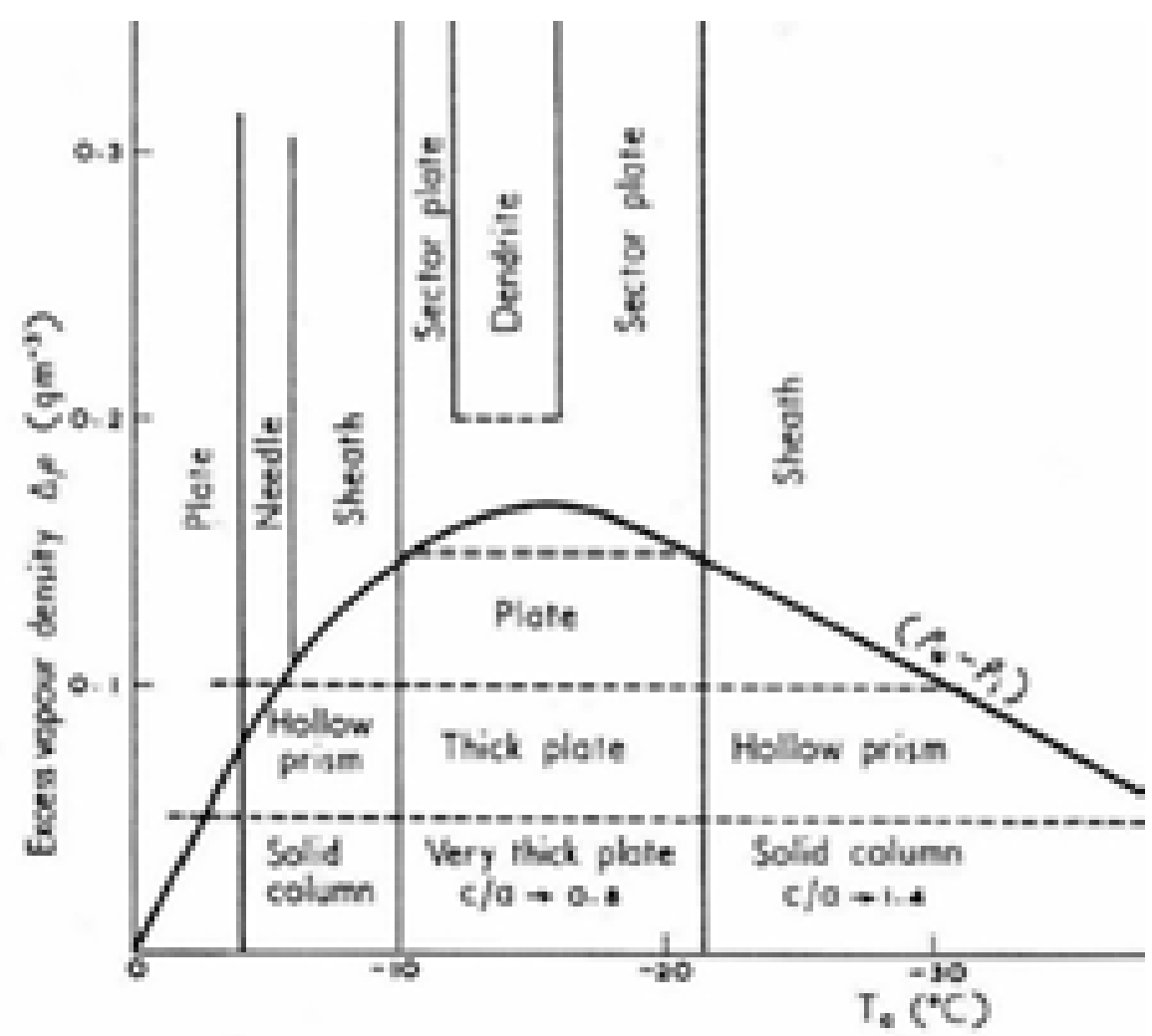

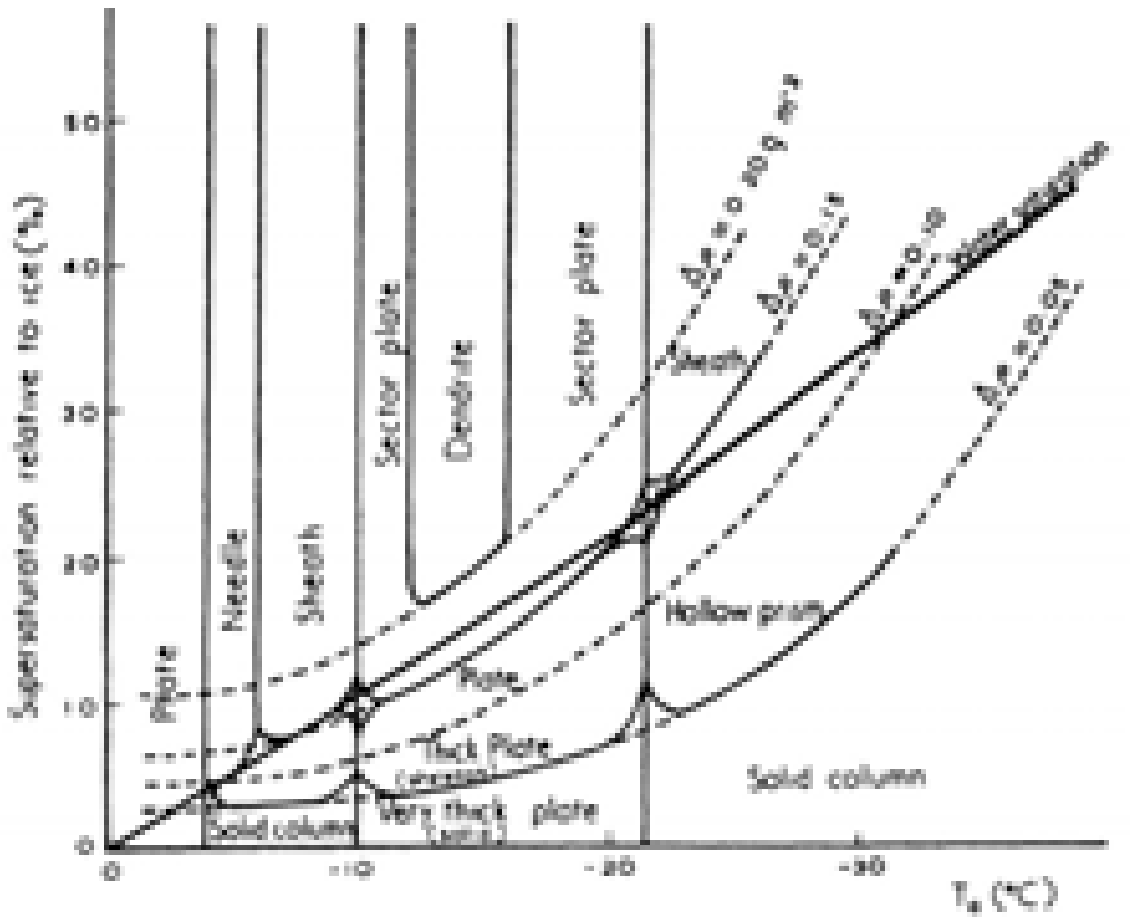

**Figure 1.25: Several published versions of the Nakaya diagram observed by different investigators: [1958Hal] (top), [1961Kob] (middle), and [1990Yok] (bottom).**

The formation of a capped column can likewise be understood by a change in growth conditions with time. First a crystal begins growing in a region of the clouds where the temperature is around -5 F, so it develops into a columnar crystal. Then the wind carries it a different region where the temperature is closer to -15 C, and at that point plates begin growing on the ends of the column.

Nakaya liked to remark that snowflakes are like "hieroglyphs from the sky." With the morphology diagram, one can connect the shape of a snow crystal to the atmospheric conditions it experienced as it formed. An spectator on the ground can thus decipher the observed crystal morphology to reveal the conditions of the clouds in which it formed, like a kind of meteorological hieroglyphics.

The morphology diagram also tells us that snow crystal growth is remarkably sensitive to temperature. Even a change of a few degrees can dramatically alter its growth behavior, and this helps explain why snowflakes have such a remarkable diversity of shapes. We will come back to the morphology diagram many times in this book, as it has become an essential tool for understanding the variable nature of snow crystal formation.

## Crystal Dendrites

In 1917, Scottish zoologist D'Arcy Wentworth Thompson published *On Growth and Form*, in which he pondered the physical, biological, and mathematical origins of complex structures in Nature [1917Tho, 1961Tho]. While confessing that crystal growth was somewhat outside the province of his book, Thompson commented:

*… yet snow-crystals … have much to teach us about the variety, the beauty and the very nature of form. To begin with, the snow-crystal is a regular hexagonal plate or thin prism; that is to say, it shows hexagonal faces above and below, with edges set at co-equal angles of 120º. Ringing her changes on this fundamental form, Nature superadds to the primary hexagon endless combinations of similar plates or prisms, all with identical angles but varying lengths of side; and*

*she repeats, with an exquisite symmetry, about all three axes of the hexagon, whatsoever she may have done for the adornment and elaboration of one.*

In his celebrated treatise, Thompson used extensive examples to focus scientific attention on the central question of how complex structures arise spontaneously in natural systems. Humans tend to create intricate objects via a subtractive process, beginning with bulk material and carving it into a final desired form, following a preconceived design. Thus a human-crafted snow crystal might take shape as illustrated in Figure 1.26, which is clearly not how it works in nature. At the opposite end of the fabrication spectrum, living things develop into amazingly sophisticated organisms quite spontaneously, using the additive process of growth. Thompson strove to comprehend the underlying physical and chemical principles that guide the development of living organisms, thus pioneering what has become the field of developmental biology.

Like Kepler 300 years before him, however, Thompson found that the whole of biological structure formation presented a challenging problem, to say the least. An easier approach, therefore, might be to consider something like the snowflake, which exhibits an interesting degree of spontaneous structure formation, but in a far simpler physical system. Over time, physicists too began to appreciate that the patterns arising during solidification offered a worthy phenomenon to investigate. Just as the hydrogen atom was a first step toward understanding the complex chemistry of large biomolecules, perhaps the physical origin of structure formation during solidification would provide insights into systems with greater complexity.

A significant step forward in this direction was made in 1964 when American physicists William W. Mullins and Robert F. Sekerka realized that *growth instabilities* are often associated with pattern-forming systems, and with solidification in particular. In their seminal paper [1964Mul], the authors showed that many of the simplest solutions to the equations describing diffusion-limited growth were mathematically unstable to perturbations. And, importantly, these growth instabilities would drive the spontaneous formation of complex structures, as I discuss in Chapter 3.

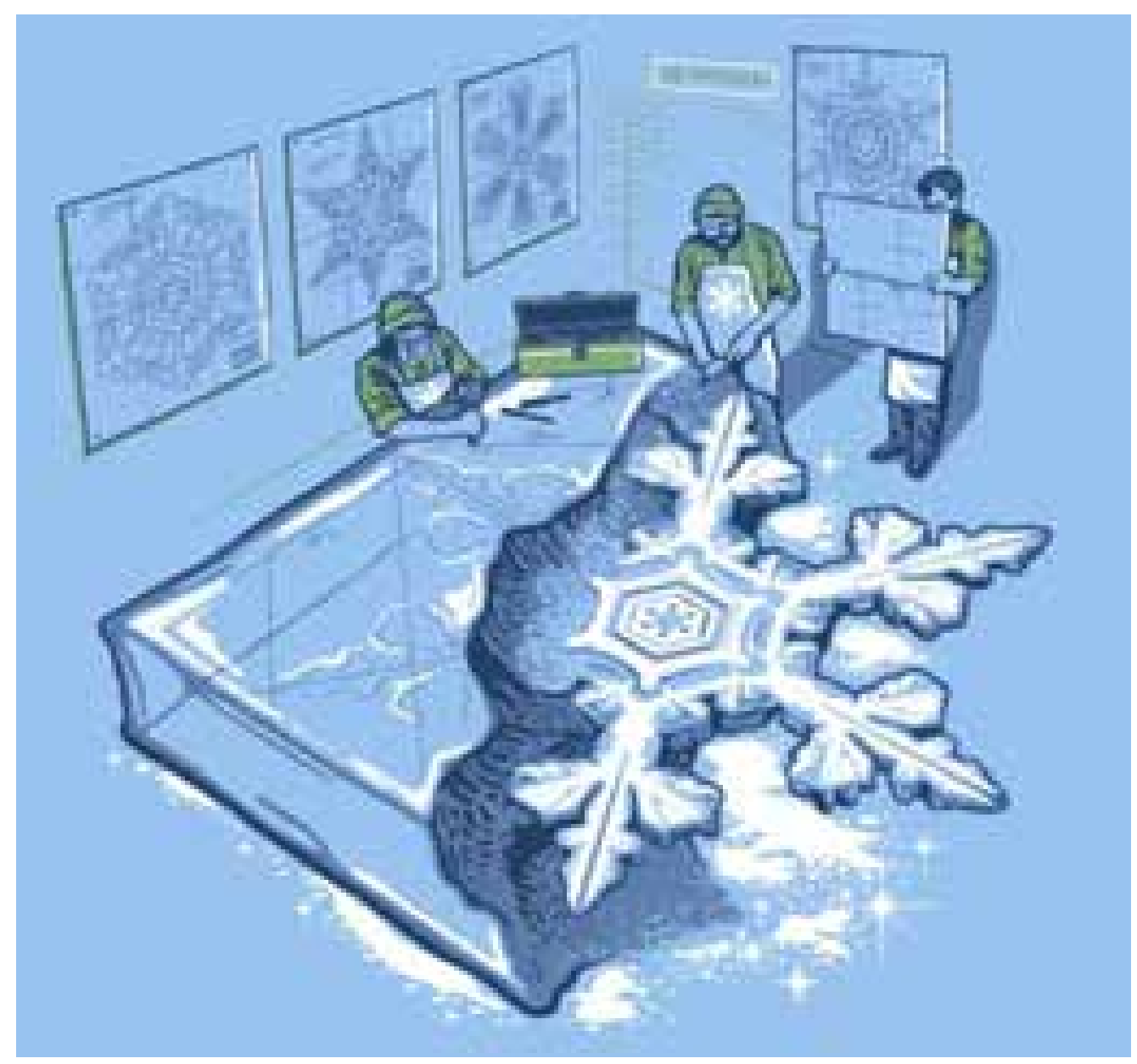

**Figure 1.26: The wrong way to make a snowflake. While human artisans sculpt by removing material, nature creates complex structures through a process of self-assembly during growth. The design of a snowflake emerges as it grows and develops in the clouds. Its morphology does not follow from any predetermined blueprint, but rather results from the changing external conditions it experiences while it forms. (Adapted from an image at threadless.com/product/688/no_repeats, created by Christopher Buchholz.)**

For example, in the case of snow crystal growth, a minimal solution to the diffusion equation is that of a sphere that grows radially outward. Water vapor diffuses to the crystal, deposits on the ice surface, and the radius of the sphere increases with time. Although this is a mathematically sound solution to the growth equations, it is not a stable solution. Small perturbations in the shape of the sphere soon lead to the formation of small bumps that grow and develop into complex, branched structures. For an initially faceted snow-crystal plate, the same effect yields six branches sprouting from its hexagonal corners, as illustrated in Figure 1.9.

This spontaneous branching process that arises during solidification – now known as the *Mullins-Sekerka instability* (see Chapter 3) – plays a central role whenever diffusion limits the solidification of materials. And it is responsible for essentially all the complex morphological features seen in snow crystals.

Although dendritic structures had been described in a broad range of physical and biological systems by D'Arcy Thompson and others for many decades, the underlying causes of these forms was beyond the reach of early scientific knowledge. Counting the petals on a flower was one thing; explaining their existence was another matter entirely. Indeed, comprehending even quite simple biological structures remains a largely intractable problem to this day. Mullins and Sekerka showed, however, at least for simple physical systems, that it was possible to make some progress toward understanding how complex structures arise spontaneously in non-equilibrium systems.

A systematic study of growth instabilities in laboratory solidification was undertaken in the 1970s by American materials scientist Martin Glicksman and others, who examined the growth of dendritic structures when liquids cooled and solidified [1976Gli, 1981Hua]. In an extensive series of beautiful experiments, Glicksman et al. made detailed measurements of structure formation during the freezing of liquid succinonitrile, choosing this material for its convenient properties that are generally comparable to common metals. In particular, succinonitrile has a simple crystalline structure, it is transparent, and it freezes near room temperature.

When unconstrained by any container walls, Glicksman found that freezing often yielded branched structures like that shown in Figure 1.27, with growth characteristics that depended mainly on crystal symmetry and the degree of supercooling of the liquid. Similar branching is seen in some stellar snow crystals, like the one shown in Figure 1.1, and this same kind of dendritic growth behavior was observed to be quite ubiquitous during solidification from both liquids and vapors over a broad range of materials.

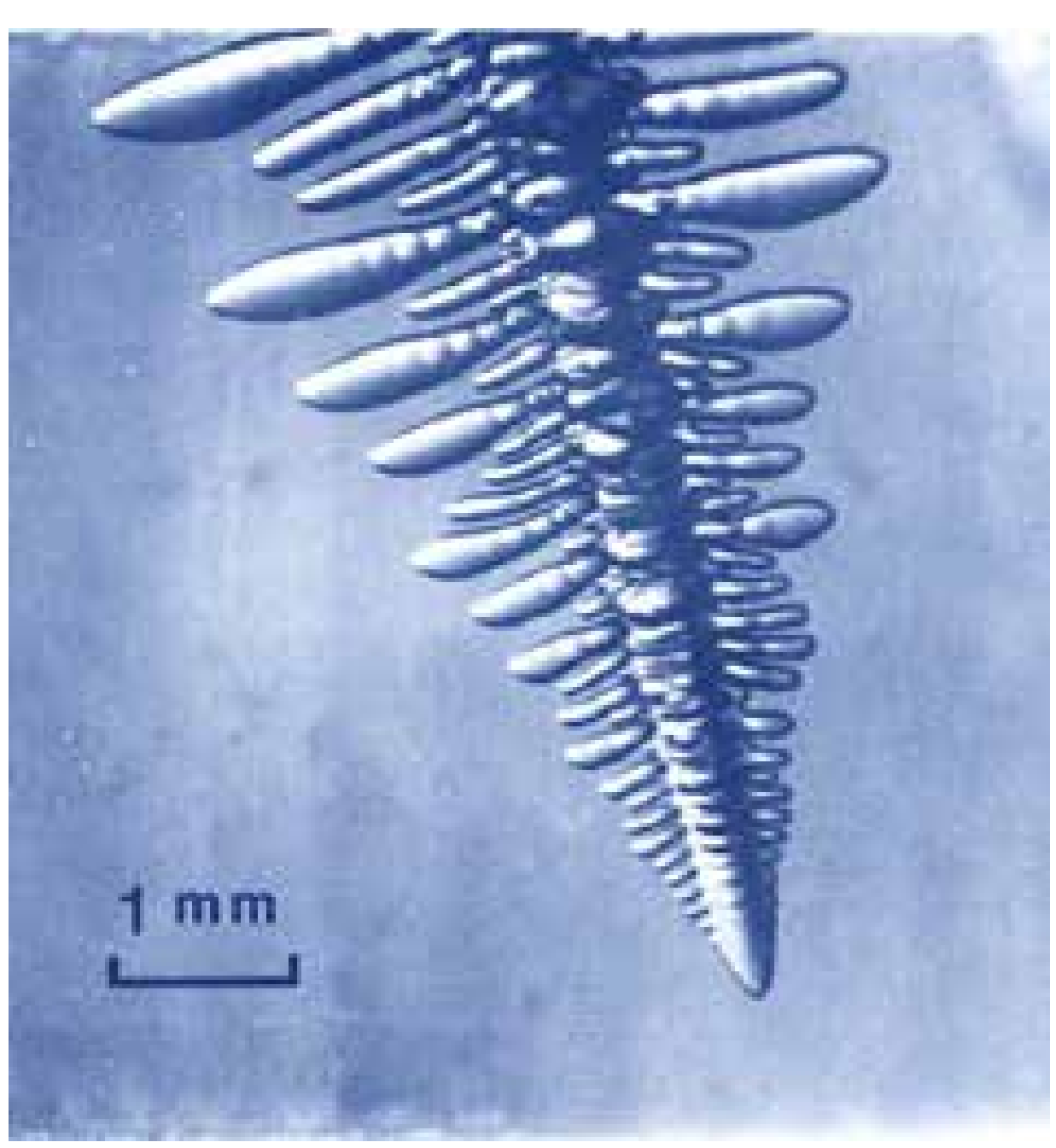

**Figure 1.27: This photo shows a dendritic crystal of succinonitrile growing into a supercooled melt of the same material. Under constant conditions, the tip advances at a constant growth velocity while the radius of curvature of the tip remains constant [1981Hua].**

The work of Glicksman and others soon called attention to the Mullins-Sekerka instability and its consequences for structure formation during crystal growth. There followed a concerted push by physicists, material scientists, and applied mathematicians to form a self-consistent theory describing the characteristics of the diffusion-limited growth of dendritic structures. Efforts in the 1980s led by James Langer [1978Lan, 1980Lan, 1989Lan], Hans Müller-Krumbhaar, Efim Brener, Herbert Levine, and others eventually yielded what has become known as *solvability theory* (see Chapter 3), which explains many of the overarching characteristics of dendritic crystal growth.

### Quantitative Snow Crystals

The push to understand dendritic growth in the 1980s laid the foundation for understanding the development of complex structures during solidification, including snow crystal formation. First, diffusion was clearly identified as one of the key processes affecting growth behaviors: primarily thermal diffusion for the case of growth from the melt, while primarily particle diffusion for the case of snow crystals in air. Second, several material properties were identified as playing major roles as well: primarily anisotropic surface energies for the case of unfaceted growth from the melt, while primarily anisotropic attachment kinetics for the case of faceted snow crystal growth.

The development of solvability theory also created a coherent theoretical framework for further scientific investigation of structure formation during solidification, and of snow crystal growth in particular. While pure mathematical treatments like solvability theory could describe some overall characteristics in dendritic growth, it was soon realized that numerical modeling would be the only way to fully describe the formation of complex structures. It is now clear that computational snow crystals will play a large role in future studies of snow crystal growth, combining diffusion physics with molecular models of the attachment kinetics to produce numerical simulations of snow crystal growth. I explore this computational frontier in Chapter 5.

More broadly, Glicksman et al. also demonstrated that careful quantitative investigations are essential for pushing forward our understanding of the physics underlying structure formation during solidification. This was the case for succinonitrile, and it is proving to be abundantly true for snow crystal growth as well. Morphological studies are a fine first step, but precision measurements of growth rates under controlled conditions are the future, particularly when compared with sophisticated physical modeling.

Quantitative studies of snow crystal growth have a long history, but early measurements exhibited broad inconsistencies between different measurements, and were found to be significantly affected by systematic errors [2004Lib]. I believe we have made some progress toward managing these measurement errors [2012Lib] and obtaining reliable growth measurements [2013Lib, 2017Lib], but numerous experimental challenges remain. I describe many of these issues in detail in Chapters 4 and 7. Closing the loop to make quantitative comparisons between measurements and computational models has only just begun, and I discuss some recent progress on this front in Chapter 8.

## 1.3 Twenty-First-Century Snowflakes

My primary goal with this book is to help carry the torch forward as snow crystal science advances through the 21st century. Review papers are useful in this regard [2017Lib, 2005Lib, 2001Nel, 1987Kob], but their inevitable page limitations make it difficult to give the subject a proper treatment. I found much inspiration in Nakaya's book [1954Nak] when I first began studying snowflakes, and it is still a fascinating read. But there has been no comparable volume on the subject since that time, so, after nearly 70 years, it is definitely time for an update.

When written by a single author, books often give a somewhat biased view of a subject, and this book is no exception in that regard. I find certain sub-topics especially interesting, and so I tend to dwell perhaps too long in these areas. On the other hand, I give too little attention to other subjects when feel less qualified to expound upon them. My intentions are good, and I have made some attempt to provide a broad overview of snow crystal science. But there are only so many hours in a day for in-depth research, so one tends to write about the topics one knows best.

## Reductionism to Holism

My overarching goal in snow-crystal science is a combination of reductionism and computational holism. The reductionism side aims to break down the physics of snow crystal growth into its constituent parts and processes, including things like crystal structure, attachment kinetics, and diffusion-limited growth. Each of these areas can be isolated and examined separately, perhaps right down to the molecular level, with the hope of developing precise mathematical models of all the relevant physical processes.

Some of the reductionist pieces are already well understood, while others remain quite puzzling. For example, the statistical mechanics of diffusion is well known for ideal gases, and the ideal-gas approximation is more than adequate for describing snow crystal growth (see Chapter 3). On the other hand, the ice surface structure is rather poorly known, and the attachment kinetics is a challenging problem even today (see Chapter 4). In principle, however, it is possible to isolate, investigate, and ultimately comprehend all the relevant physical process involved in snow crystal growth.

Reductionism, however, is not sufficient to describe all of snow-crystal science. Characterizing all the pieces of a puzzle and assembling the puzzle are two different endeavors. Learning the fundamental laws of quantum physics does not immediately explain everything in the field of chemistry, because understanding how atoms assemble into molecules is a separate problem from understanding individual atoms alone. Similarly, comprehending the formation of a complete snow crystal is not the same as characterizing the separate physical processes involved in its growth. Holism in this case is not so much that the whole is greater than the sum of the parts. Rather, seeing the whole requires that you are able to assemble the parts.

Holism thus compels us to create a numerical simulation that incorporates all the known physical processes involved in ice crystal growth, with an accuracy sufficient to yield a realistic computational snow crystal. In principle, using a large enough computer, it would be straightforward to create the necessary algorithms. But the devil is in the details, and computational models involve a lot of details. Numerical inaccuracies and instabilities can be problematic, and even the fastest supercomputers cannot come even close to realizing molecular resolution in large-scale phenomena. Numerical techniques for simulating solidification are evolving rapidly, and I examine the current state-of-the-art for creating computational snow crystals in Chapter 5.

Finally, quantitative comparisons between computational snow crystals and synthetic laboratory crystals grown under carefully controlled conditions are also essential for making progress in snow-crystal science. Creating a numerical model that generates structures that look like snowflakes is necessary, but certainly not sufficient. Only when theory and experiment agree with one another, with suitable precision over a broad range of circumstances, can we begin to believe that our physical description of snow crystal formation is correct.

Some might argue that creating an accurate computer simulation of a growing snow crystal would not constitute a true understanding of the underlying phenomenon. Debating this point would require a precise definition of the word "understanding," which is itself a nontrivial undertaking. Snow crystal formation involves a multitude of complex physical processes acting over a broad range of length and time scales. It may indeed be the case that our small brains cannot fully absorb all aspects of what is happening. If that is true, then we will have little choice but to let our machines do the heavy lifting for us. I would argue that a detailed computer model that reproduces laboratory snow crystals with high fidelity is probably as close to a true understanding as we are likely to achieve. We can start with that anyway.

As of this writing, many aspects of this overarching scientific strategy only now taking shape. Our understanding of the attachment kinetics is improving (Chapter 4), computational models are just now becoming viable (Chapter 5), and detailed comparisons between experiments and theoretical models are almost nonexistent (Chapter 8). But there appears to be no serious roadblocks impeding the path forward, and progress is being made rapidly on all fronts.

## The Road Ahead

Like most technical subjects, it is impossible to present snow crystal science in a completely linear fashion. Different topics are invariably interconnected to some extent, so one cannot fully appreciate any individual chapter in this book without having at least some understanding of the material presented in all the other chapters. Science is like that. In an effort to minimize the confusion that may result, therefore, I begin here with a brief synopsis of some key areas that are most important in the science of snow crystals.

**Ice Crystal Structure – Chapter 2**

While the lattice structure of crystalline ice is well characterized in the bulk, the molecular structure of the ice surface remains largely undetermined. Surface premelting has been observed down to temperatures of about -15 C (perhaps lower), and this phenomenon almost certainly plays a large role in snow crystal growth. However, there is little theoretical understanding of surface premelting, and its effect on crystal growth is largely unstudied.

The terrace step energy factors directly into terrace nucleation, which plays a major role in the growth of faceted ice surfaces. Moreover, like the bulk energy and surface energy, the terrace step energy is a fundamental material property of crystalline ice in equilibrium. As such, it may be possible to use molecular dynamics simulations to calculate step energies directly from known molecular interactions. This would be a major advance in our understanding of snow crystal science, forging a direct chain from growth measurements → nucleation dynamics → terrace step energies → water molecular interactions → fundamental chemical physics.

**Diffusion-Limited Growth – Chapter 3**

While anisotropic attachment kinetics bring about the formation faceted snow-crystal surfaces, the slow diffusion of water vapor molecules through air is responsible for the growth of elaborately branched structures. Faceting and branching (i.e., attachment kinetics and diffusion-limited growth) are the two primary factors controlling snow crystal formation, and their complex interplay is what yields the full menagerie of snow crystal morphologies.

The physics of particle diffusion is well described by the statistical mechanics of ideal gases, so this aspect of snow crystal science is essentially a solved problem. However, applying this theory to the growth of complex structures continues to be a nontrivial challenge. Analytic solutions are suitable for especially simple examples like growing spheres and parabolic needle-like forms, and these are extremely useful for examining scaling relations and revealing the relative importance of competing factors in overall growth behaviors. But numerical modeling is needed to reproduce the complexity seen in all but the simplest snow crystals.

**Attachment Kinetics – Chapter 4**

The greatest obstacle currently preventing the creation of accurate numerical simulations of growing snow crystals is our poor understanding of the attachment kinetics. The ubiquitous appearance of faceted surfaces on growing snow crystals results directly from how rapidly water vapor molecules attach to different surfaces (see Figure 1.8), so the role of anisotropic attachment kinetics is certainly one of the most important aspects of the snowflake story. But our overall picture of the physical processes involved remains a work in progress.

The underlying origin of this challenge is obvious: the many-body molecular processes governing attachment at the ice surface are extremely complicated. As I describe in Chapter 4, current experimental evidence suggests that the attachment kinetics depend on a multitude of factors, including temperature, water vapor supersaturation, surface orientation relative to the crystal axes, background gas pressure, and even the size of a faceted surface. Making sense of all this is a fascinating undertaking that lies at the forefront of modern snow-crystal science.

**Computational Snow Crystals – Chapter 5**

Numerical models of solidification have been around since the 1980s, but only around 2005 did researchers begin demonstrating model structures that exhibited both branching and faceting that resembled natural snow crystals. Several numerical difficulties arise when the attachment kinetics are highly anisotropic, exhibiting deep cusps at the facet angles, and dealing with these requires specialized techniques. At the time of this writing, the existing models have reproduced reasonable-looking structures, but only when some non-physical assumptions are adopted. Resolving these issues is the subject of current research.

A number of different numerical strategies have been investigated, including front-tracking methods, phase-field techniques, and cellular automata systems, as described in Chapter 5. The cellular automaton models seem to be especially adept at dealing with strongly anisotropic attachment kinetics, and these appear to be winning the race to develop physically realistic computational snow crystals. This area of snowflake science is evolving rapidly, however, and it is difficult to predict how future advances in numerical algorithms will impact the different modeling strategies.

**Laboratory Snow Crystals – Chapter 6**

Quantitative experimental verification is essential to confirm that our computational algorithms are creating physically realistic models of snow crystal growth. Simply observing morphological similarity between models and some types of natural snow crystals is not sufficient. A comprehensive model should reproduce the full growth behavior of complex snow crystals as a function of time, including growth rates as well as morphological development.

Achieving this level of scientific veracity requires a range of experimental tools that allow *in situ* observations of growing snow crystals over a wide range of environmental conditions. In Chapter 6, I describe a variety of experimental tricks and techniques that have been developed over the years, and I examine their various merits and drawbacks. As with the ongoing development of computational techniques, there exists considerable opportunity in this area for developing precision instruments for innovative snow crystal investigations.

**Simple Ice Prisms – Chapter 7**

Creating a comprehensive physical model of the attachment kinetics requires precision measurements of the growth of simple ice prisms as a function of temperature, supersaturation, background gas pressure, crystal size, and other parameters. Experience to date indicates that the attachment kinetics are too complex to be deduced by simple physical reasoning or a few easy experiments. Instead, careful measurements of growth rates will be needed, and considerable attention must be paid to acknowledging, understanding, and controlling systematic errors in these measurements.

Minimizing systematic effects arising from particle diffusion further necessitates working with the smallest possible crystal samples, preferably no larger than a few tens of microns in overall size. Producing, handling, observing, and measuring these samples requires special experimental considerations that are presented in Chapter 7.

**Electric Ice Needles – Chapter 8**
While tiny ice prisms are best suited for studies of the attachment kinetics, larger laboratory snow crystals are desired for examining the development of complex morphologies in comparison with 3D numerical models. One especially promising technique for observing such crystals is by growing them on the ends of slender "electric" ice needles in a dual diffusion chamber, as this allows stable support, in situ observation, and a well-defined initial seed-crystal geometry. The construction and operation of apparatus is described in detail in Chapter 8.

**Plate-on-Pedestal Snow Crystals – Ch. 9**
This specialized laboratory technique exploits the Edge-Sharpening Instability (see Chapter 4) to create thin plate-like snow crystals atop small, blocky ice-prism "pedestals." While not as flexible or scientifically valuable as the electric-needle method described in Chapter 8, PoP snow crystals are nearly ideal for recording high-resolution images of growing stellar-plate crystals. This apparatus has yielded the first photographs of stellar snow crystals that exhibit qualities that are overall superior to the best natural specimens, including better symmetry and sharper faceted features. The PoP technique has also yielded the highest-quality videos of growing stellar snowflakes thus far produced, along with the first observations of "identical-twin" snow crystals.

**A Field Guide to Snowflakes – Chapter 10**
This chapter examines the full menagerie of natural snow crystal types with illustrative sketches and abundant photographic examples. Several classification schemes are presented, along with descriptions of a variety of common snow-crystal structural features. If printed separately, this naturalist's guide to falling snow allows for convenient outdoor examination and identification of falling snow crystals.

**Snowflake Photography – Chapter 11**
Capturing quality imagery of natural snow crystals in cold conditions presents some unusual challenges for aspiring snowflake photographers. Finding especially photogenic specimens is nontrivial, handling them can be challenging, attaining suitable magnification requires special lenses, and lighting is problematic because single-crystal ice is quite transparent. This chapter looks at each of these issues in detail and presents examples of a variety of innovative techniques that have been pioneered by the community of snowflake photographers.

**Ice from Liquid Water – Chapter 12**
Although ice growing from liquid water is a ubiquitous phenomenon that plays numerous important roles in our environment, is has received remarkably little scientific attention, substantially less than ice growing from water vapor. In particular, the attachment kinetics at the ice/water interface has not been especially well measured or modeled. This topic is somewhat outside of the scope of this book, but I present a brief overview of the science of freezing water in this final chapter.

## Brief but not Forgotten

As described in the paragraphs immediately preceding, much of this book focuses on areas that are most important for understanding and advancing snow-crystal science – attachment kinetics, particle diffusion, mathematical methods, computational algorithms, and experimental techniques. The topics listed next are somewhat less important and are therefore mentioned relatively briefly in comparison.

**Heat Diffusion.** Solidification from the melt, such as liquid water freezing to ice, is typically greatly affected by latent heating and the diffusion of heat generated at the solidification front. In snow crystal growth, however, latent heating is a relatively small perturbation, as the effects of heat diffusion are dwarfed by those from particle diffusion. In fact, reducing latent heating to zero in snow-crystal models has only a relatively minor effect on snow crystal growth at temperatures quite close to 0 C. Most of the snow crystal morphology diagram would be unchanged if latent heating could be turned off entirely. Analytical modeling strongly supports these statements, as does the experimental evidence (see Chapter 3). Someday we will have to address the dual-diffusion problem (heat diffusion and particle diffusion), but the addition of this complication is unwarranted at present, given our poor understanding of the attachment kinetics.

**Surface Energy.** Here again, solidification from the melt is often strongly influenced by surface-energy (a.k.a. surface-tension) effects. In particular, the anisotropic surface energy at the solid/liquid interface plays a large role in stabilizing dendritic growth (see Chapter 3). In snow crystal growth, however, surface energy effects are dwarfed by effects from anisotropic attachment kinetics. In computational snow-crystal models (see Chapter 5), setting the surface energy to zero has some effect when the supersaturation is extremely low, but under typical conditions this is negligible. Because so much solidification work has been done with liquid/solid systems (for example, Figure 1.27), researchers sometimes erroneously try to apply the same physical principles (heat diffusion + surface energy) to snow-crystal growth, even though the latter is mainly defined by (particle diffusion + attachment kinetics).

As a general rule-of-thumb, all occurrences of faceted crystal growth arise from anisotropic attachment kinetics. Surface-energy effects are simply too weak to produce faceted growth in all but the most extreme circumstances (that are rarely seen during crystal growth). Some crystals exhibit faceted shapes in equilibrium, but growing crystals are usually far from equilibrium, to the point that growth forms are completely different from equilibrium forms.

Likewise, it is a common misconception that faceted morphologies appear because faceted surfaces have the lowest surface energies. This is a misleading statement because often both the surface energy and the attachments kinetics are low on faceted surfaces. But anisotropic attachment kinetics are by far the dominant effect causing faceting.

**Chemical Effects.** Surface chemistry can strongly affect attachment kinetics, and it is well known that chemical impurities can strongly influence snow-crystal growth (see Chapter 4). Unfortunately, there is essentially no theory to guide this topic, and the experimental data are all over the map. This promises to be a fascinating area of study someday, but it is difficult to make progress without a solid theoretical foundation upon which to build.

**Molecular-Dynamics Simulations.** In the future, it may be possible to calculate the ice attachment kinetics directly from MD simulations, but this appears to be a rather distant goal. A major step in that direction may be calculating terrace step energies using these methods, and I discuss that possibility briefly in Chapter 4. Although MD simulations have already contributed substantially to our understanding of ice surface structures, and it appears likely that they will also be important for modeling attachment kinetics, so far their direct influence on snow crystal science has not been great. Plus this is a highly technical field that I am not qualified to review, so I do not delve into this area in much detail in this book.

## 1.4 Why Study Snowflakes?

I include this section because I hear this question surprisingly often, almost every time I give a lecture about the science of snowflakes. It can take different forms, such as "What motivates your research?", or "What applications might come from it?", but the underlying sentiment is mostly "What is any of this good for?" At first, I was somewhat taken aback by these questions, as I had previously spent much of my professional career in the field of astrophysics, and no one had ever asked me what that was good for.

The difference, as far as I can tell, is mainly one of expectations. Astronomy is about exploring the Universe, so there is no expectation of any Earthly applications. But materials science is a branch of engineering, and engineering is all about technology. Physics is often split into "basic" physics and "applied" physics, but there is no "basic" engineering. Therefore, because I study the materials-science aspects of ice growth, people naturally assume that I must have some applications in mind.

I learned early on to always inform my audience that I have never spent any of their tax dollars on snowflakes. Some folks are surprisingly upset that such a possibility would exist. People seem to imagine that I have a large team of crack researchers in my lab, performing this somewhat valueless research at their expense. But the team is mostly just me, and I have a day job. I occasionally take on an undergraduate student or two if I can get them for free, but, to a large extent, this work is little more than my scientific hobby.

I have written several popular-science books about snowflakes, and these have generated some funds for equipment, but otherwise the enterprise is mainly just me and my credit card. Of course, I am wholly indebted to Caltech for paying my salary and allowing this foolishness to continue on site, and I truly appreciate the freedom that comes with being a professor at a research-oriented university.

I like to call what I do "forbidden" research, because there is no way any of the usual funding agencies will touch it. And that goes for private donors as well, who generally opt for grander projects. Fortunately, the work I do is quite inexpensive, and hardly anyone else studying snowflakes has direct funding either. One of the main reasons humanity still does not better understand how snowflakes work is that very few people have studied it over the years. Typically, there are a handful of interested souls around the globe at any given time thinking about this problem, but that is about it.

Of course, people often feel I should put my position and my creativity into more worthy pursuits, say curing cancer. Alas, that particular track is not an option for me, as I have no idea how to cure cancer, plus the field is limited almost entirely by funding, as there are plenty of brilliant researchers with novel ideas ready to pursue. Also, as a society, we spend large sums on sports, music, movies, and similar pursuits. So perhaps some of those funds could also be better spent looking for a cancer cure. Perhaps each of us could divert some of our personal resources to such lofty goals. Personally, I think hobbies are fine; mine is just a tad off the norm.

Of course, I could point out that there are countless examples where seemingly worthless scientific endeavors led to wildly successful applications that no one foresaw. And it is not even difficult to imagine potential benefits arising from studying snowflakes. For example, it should be possible to grow large diamond crystals in your bathtub, or at least such a feat may be ultimately possible. It certainly does not violate any laws of physics. The problem is just that no one yet knows how to do it. Meanwhile, bacteria sometimes grow crystals using somewhat mysterious biochemical techniques in a process called biomineralization. Perhaps we could manufacture large diamond crystals cheaply if we only knew more about the fundamental physics and chemistry of surface attachment kinetics. And this is what I mostly study with snow crystal growth. It is an interesting thought, but my research is definitely not motivated by the possibility of growing large diamonds.

No, my real motivation is simply that I find the phenomenon of snow crystal growth to be fascinating in its own right. The basic science behind all this is what I find most appealing. These ornate structures simply fall from the sky, yet we cannot yet say exactly why they look the way they do. Science is about understanding the natural world, and this, in my opinion anyway, includes snowflake formation. So why not? With over seven billion people on the planet, I figure maybe a few of us can be spared to look into this matter.

My hope with this book is to provoke a bit of interest in the topic, and perhaps stimulate a few others to pick up the mantle of snow-crystal science. If you desire large research grants and a broad community of like-minded scholars with which to interact, well, this is not the field for you. Better you should join the mega-project nearest you. But if you possess some hermit-like tendencies and have a bit of time on your hands, then you might make your mark in this unusual field. Small-scale science has become old-fashioned in modern times, but it remains quite enjoyable. And there is still a little room in the world for trying to understand a puzzling phenomenon just because it falls from the sky.

## 1.5 No Two Alike?

Is it really true that no two snowflakes are alike? This is another a question I hear a lot. It's a funny question, almost like a Zen *koan* – if two identical snowflakes fell, my inquisitive friend, would anyone know? And can you ever be sure that no two are alike, as you cannot possibly check them all to find out?

Although there is indeed a certain level of unknowability to the question of snowflake alikeness, as a physicist I find that I can shed some light on this issue. And, as I will demonstrate, the answer depends to a large extent on what you mean by the question. (Physics does occasionally have its Zen-like qualities.)

The short answer to the question is yes – it is indeed extremely unlikely that any two complex snowflakes will look exactly alike. It is so unlikely, in fact, that even if you looked at every one ever made, over all of Earth's history, you would almost certainly not find any exact duplicates.

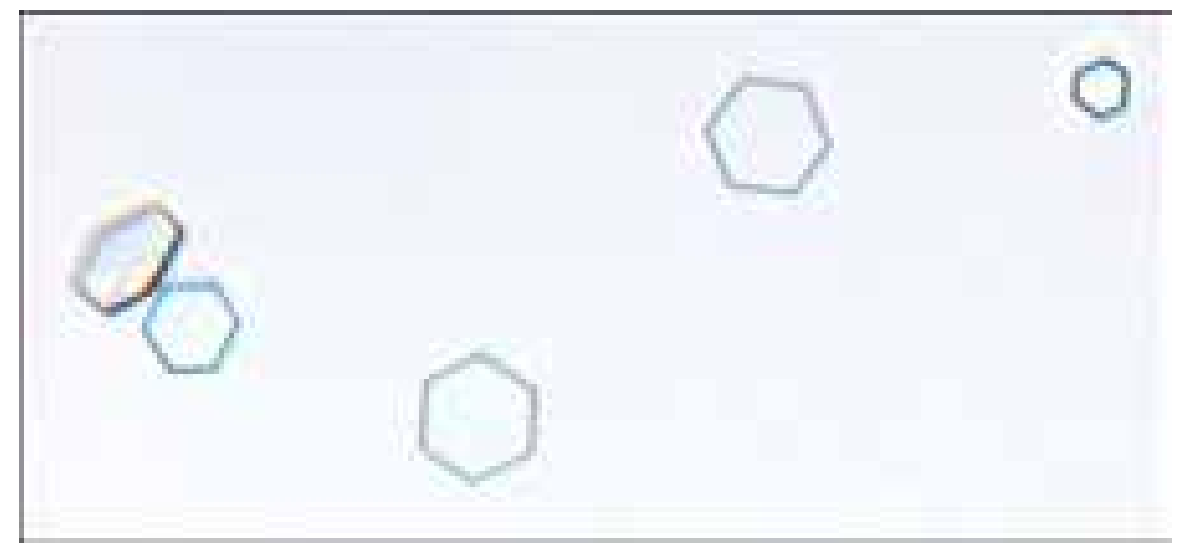

**Figure 1.28: This laboratory photo shows several small, thin-plate snow crystals that grew while falling freely in air and then landed on a transparent substrate (see Chapter 6). Because the crystals have a simple hexagonal shape, one can easily find a pair of nearly identical specimens next to one another, like the two centered here.**

The long answer is a bit more involved, however, as it depends on just what you mean by "alike," and it even depends on just how you define a "snowflake." To begin diving into this, I will claim that it is possible that two nano-snowflakes could be exactly alike. When developing the theory of quantum mechanics, physicists discovered that some things in Nature are precisely, perfectly alike – indistinguishable is the proper technical term. For example, our understanding of elementary particles indicates that all electrons are exactly the same, absolutely indistinguishable from one another. This is one of the cornerstones of quantum physics, and alikeness in this arena is a profound concept. Indistinguishability is part of what defines a truly elementary particle.

A water molecule is considerably more complex than an electron, and not all water molecules are exactly alike. If we restrict ourselves to water molecules that contain two ordinary hydrogen atoms and one ordinary $^{16}$O atom, then again physics tells us that all such water molecules are exactly alike. However, about one molecule out of every 5000 naturally occurring water molecules will contain an atom of deuterium in place of one of the hydrogens, and about one in 500 will contain an atom of $^{18}$O instead of the more common $^{16}$O. And these rogue atoms can be distinguished from their common cousins.

Because a typical small snow crystal might contain $10^{18}$ water molecules, we see that about $10^{15}$ of these will be isotopically different from the rest. These unusual molecules will be randomly scattered throughout the snow crystal, giving it a unique design. The probability that two snow crystals would have exactly the same placement of these isotopic anomalies is very, very, very small. Even with $10^{24}$ snow crystals being made per year, the probability that two will be exactly identical within the lifetime of the Universe is so low that it is essentially zero.

Thus, at some very pure level, no two snow crystals are exactly alike because of these isotopic differences. However, an exception (there are few absolute statements in science) would be a snow crystal with only a handful of molecules. If we assemble an ice crystal of only six molecules, for example, then it could easily happen that each of the six will contain two ordinary hydrogen atoms and one ordinary $^{16}O$ atom. Furthermore, a cluster of only six molecules will only have a few stable configurations. Therefore, there is a reasonable probability that six ten-molecule snow crystals would be exactly alike, quantum-mechanically indistinguishable. But perhaps six molecules does not a snowflake make.

I might add that even if we restrict ourselves to isotopically pure water molecules, it is still extremely improbable that two macroscopic snow crystals would be exactly alike. When a crystal grows, its molecules do not always stack together with perfect regularity, so a typical snow crystal contains a huge number of crystal dislocations, which again are scattered throughout the crystal in a random fashion. One can then argue, like with the isotopes, that the probability of two crystals growing with exactly the same pattern of dislocations is vanishingly small. And again, one has the exception of few-molecule crystals, which can easily be free of dislocations.

Another part of the alikeness story is that small snow crystals can at least look alike, even if they are not exactly identical down to the last molecule. So let us relax our definition of alikeness and say that two snow crystals are alike if they just look alike in an optical microscope (the smallest features one can see in an optical microscope are about one micrometer in size, which is about 10000 times larger than an atom). With this relaxed definition, everything changes.

It is quite easy, for example, to find simple hexagonal prisms falling from the sky, and it is especially easy to make such simple crystals in the laboratory (see Chapter 6). Crystals with such simple shapes often look quite similar to one another, and Figure 1.28 show two nearly identical-looking snowflakes that happened to fall next to one another in my lab. It is not hard to imagine that if you sifted through a reasonable number of Antarctic snow crystals (which tend to have simpler shapes, because the climate is so cold and dry) you would find two that were optically identical under a microscope.

As the morphology of a snow crystal becomes more complex, however, the number of possible ways to make it soon becomes staggeringly large. To see just how rapidly the possibilities increase, consider a simpler question – how many ways can you arrange books on your bookshelf? With three books, there are six possible arrangements, and you can easily sketch all of them for yourself. Increasing to 15 books, there are 15 choices when you place the first book on the shelf, then 14 for the second, 13 for the third, and so on. Multiply it out and there are over a trillion ways to arrange just 15 books. With a hundred books, the number of possible arrangements goes up to just under $10^{158}$, which is about $10^{70}$ times larger than the total number of atoms in the entire known universe!

If you gaze at a complex snow crystal under a microscope, you can often pick out a hundred separate features if you look closely. Because all those features could have developed differently, or could have appeared in slightly different places, the math ends up being similar to that with the books. The exact calculation would depend on the details, along with how you define a feature and other details. But clearly the total number of possible ways to make a complex snow crystal is almost unfathomably large. Thus, it is virtually impossible that any two complex snow crystals, out of all those made over the entire history of the planet, have ever looked exactly alike.

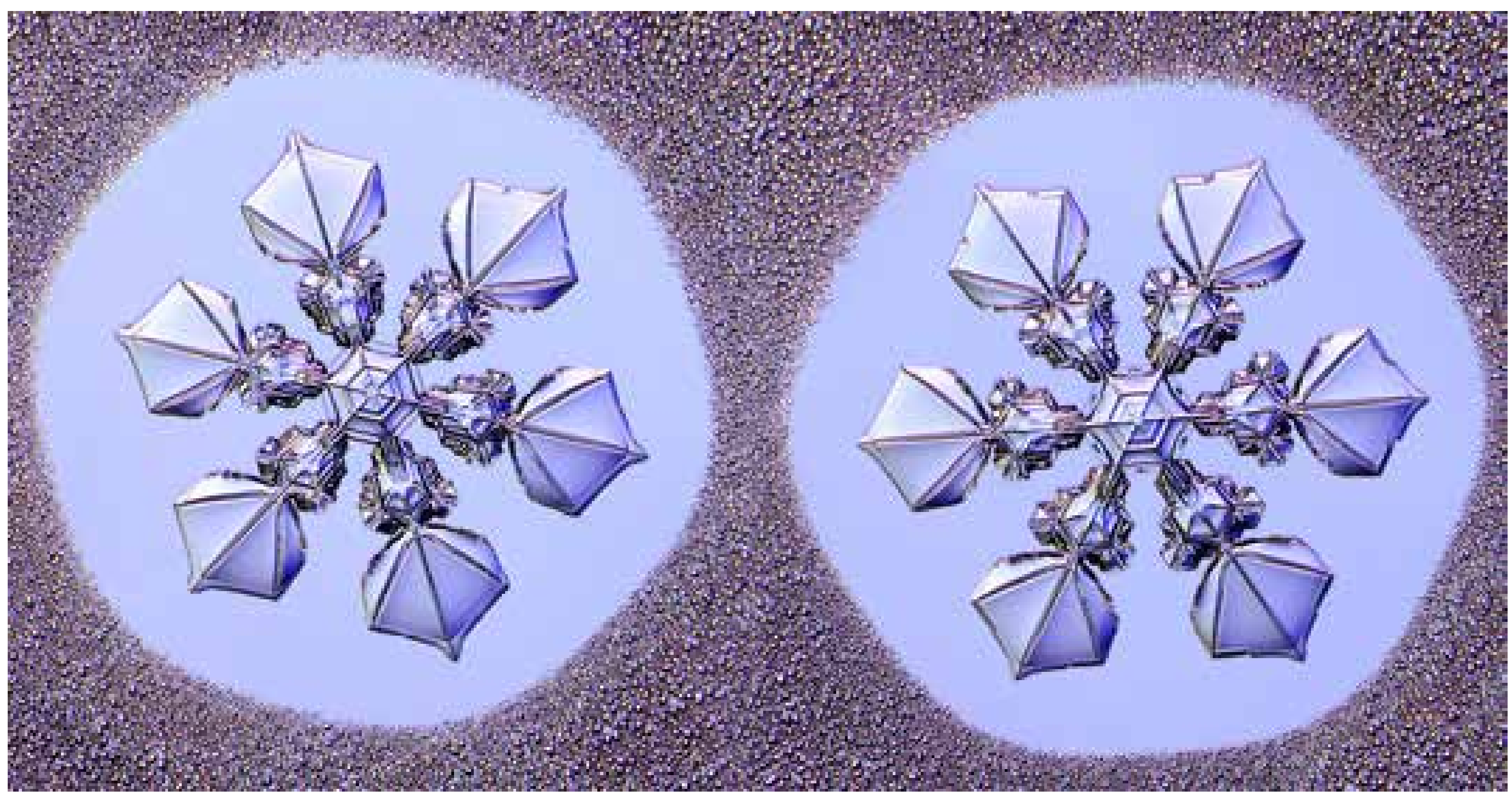

**Figure 1.29: A pair of laboratory-grown "identical-twin" snow crystals, surrounded by a field of water droplets. These grew side-by-side on a fixed transparent substrate using the Plate-on-Pedestal technique described in Chapter 10. Because both crystals experienced the same growth conditions as a function of time, they grew into nearly identical shapes. Analogous to identical-twin people, these similar snowflakes are clearly related somehow, but they are not precisely equal in every detail.**

The story of snowflake alikeness takes an amusing turn when you start looking as laboratory-grown crystals. As I described early in this chapter, the final shape of a complex snow crystal is largely determined by the path it traveled through the clouds as it formed. Because the air is usually turbulent to some extent, even under calm conditions, the paths of different snow crystals are typically quite meandering. Trajectories that bring two crystals close to one another at a particular time will likely soon diverge, separating them again by large distances.

In the laboratory, on the other hand, it is possible to place two seed crystals near one another on a fixed substrate, and then subject them to the same growth conditions as a function of time. Doing this with some care yields results like that shown in Figure 1.29. As these crystals were developing, I occasionally subjected them to abrupt changes in temperature and/or supersaturation. Because both crystals saw the same changes at the same times, they responded with synchronized growth behaviors. I like to call these "identical-twin" snowflakes in analogy to identical-twin people. They are clearly so alike that there must have been some underlying connection between them, yet they are not identical in an absolute sense.

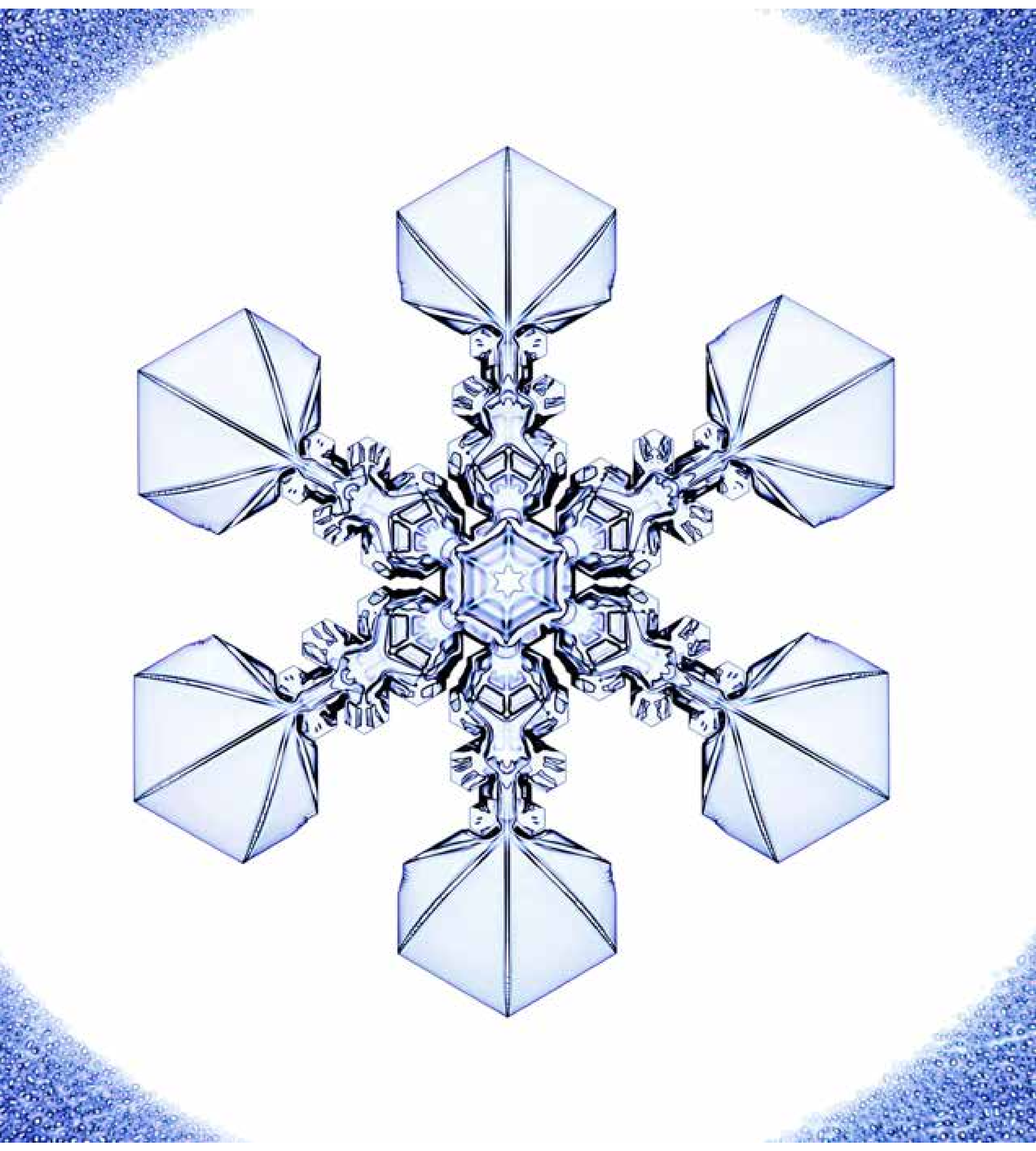

# Chapter 2

# Ice Crystal Structure

*The chief forms of beauty*
*are order*
*and symmetry*
*and definiteness.*
– Aristotle
*Book XIII (350 B.C.)*

The lattice structure and physical properties of ice are foundational elements needed to understand the formation of snow crystals. Several books about the material science of ice have appeared over the years (e.g. [1970Fle], [1974Hob], [1999Pet]), so we will not delve into all of the multifarious mechanical, thermal, chemical, electrical, and optical properties of this fascinating material. Mostly our discussion will be restricted to those ice properties that factor directly or indirectly into snow crystal growth.

**Facing page: A synthetic snow crystal, 2.3 mm from tip to tip, grows while supported above a transparent substrate, surrounded by a field of small, supercooled water droplets. Chapter 10 describes how snow crystals like this one can be created in the laboratory.**

One principal focus will be on the molecular structure of the ice crystal lattice, as this underlies the well-known six-fold symmetry seen in snow crystals. The lattice construction also elucidates several varieties of crystal twinning that have been observed in natural snow crystals. Throughout this discussion, we will find that two-dimensional projections of the molecular lattice are remarkably useful for visualizing different aspects of the ice crystal structure.

The molecular organization of the ice/vapor interface is another important theme in this chapter, because the surface is where growth takes place. The basal and prism facets are identified with specific lattice planes for which the ice/vapor interface is particularly "smooth" on a molecular scale. This characteristic not only gives each facet a lower-than-average surface energy, it also hinders the attachment kinetics of impinging molecules to the surface. The latter property, more than the former, controls the formation of snow crystal facets.

Thermal equilibrium will be a shared feature of most of the topics discussed in this chapter, distinguishing it substantially from those that follow. Crystal growth is a non-equilibrium, dynamically driven process, which is largely why it is so difficult to understand and quantify. Solving the many-body problem for a system in equilibrium involve statics, energy minimization, and equilibrium statistical mechanics, and these topics are generally tractable, at least in principle. Crystal growth, on the other hand, involves molecular dynamics, energy flow and non-equilibrium statistical mechanics, and in many instances even the basic physical theories are not well understood. For this reason, we begin our scientific discussion of snow-crystal formation by examining the most relevant properties of ice in equilibrium.

## 2.1 The Phase Diagram

The full phase diagram of water is a complicated beast, as shown in Figure 2.1, including some 15 known forms of ice, with perhaps more to come as ever-higher pressures are being explored. Although there is considerable scientific interest in the crystallography and stability of the various exotic states of ice, these lie beyond the scope of this book. Our concern will be the basic, no-frills, ice/vapor transition, circled in red in Figure 2.1. Although just a small slice of the water phase diagram, its complex dynamical behavior is already more than enough to fill a book.

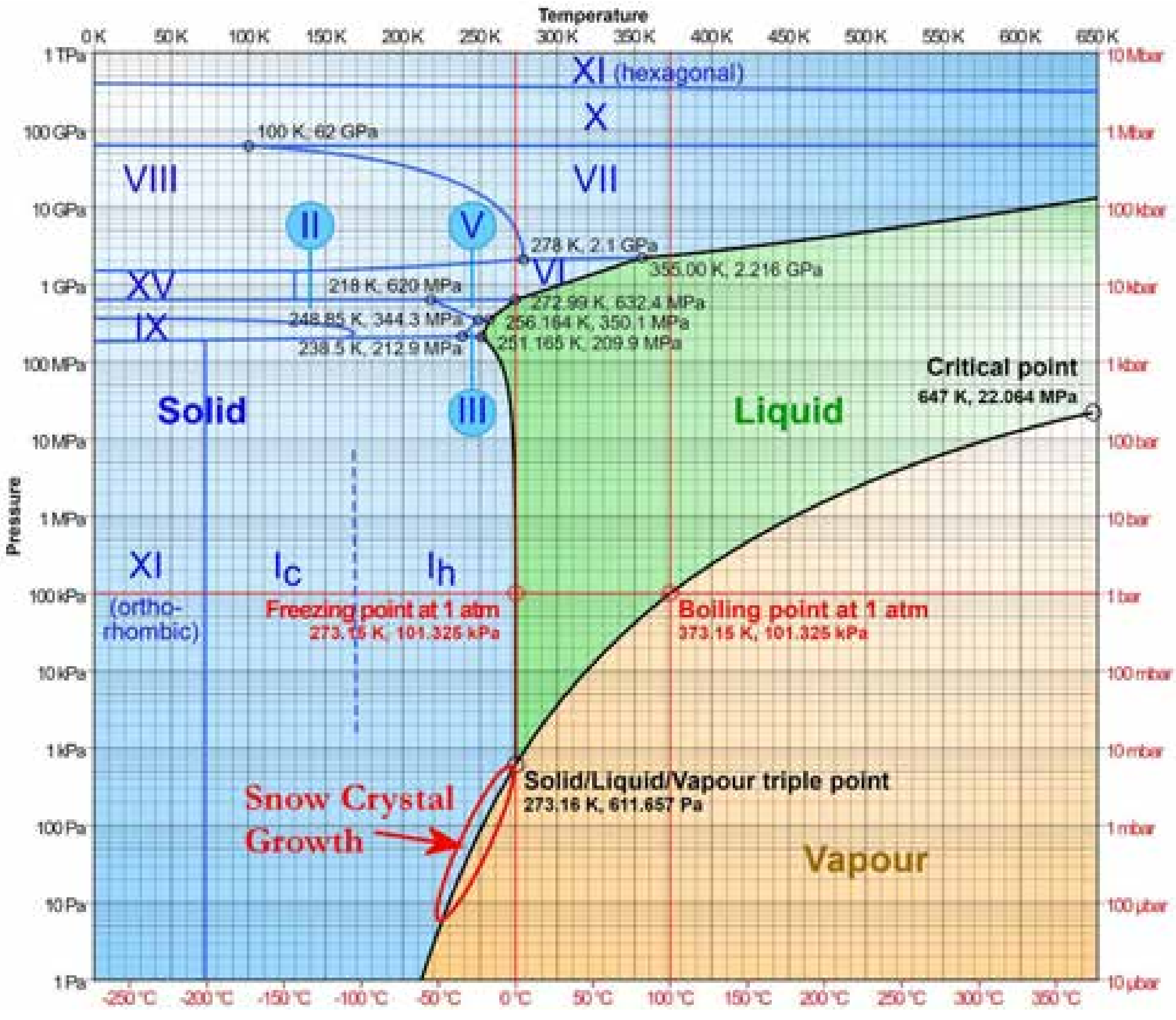

**Figure 2.1: The phase diagram of water as a function of temperature and pressure. Snow crystal growth occurs on the ice/vapor boundary below the triple point, mostly at temperatures between 0 C and -40 C. Note that most of the known phases of ice exist only at extremely high pressures. Image from https://en.wikipedia.org/wiki/Ice.**

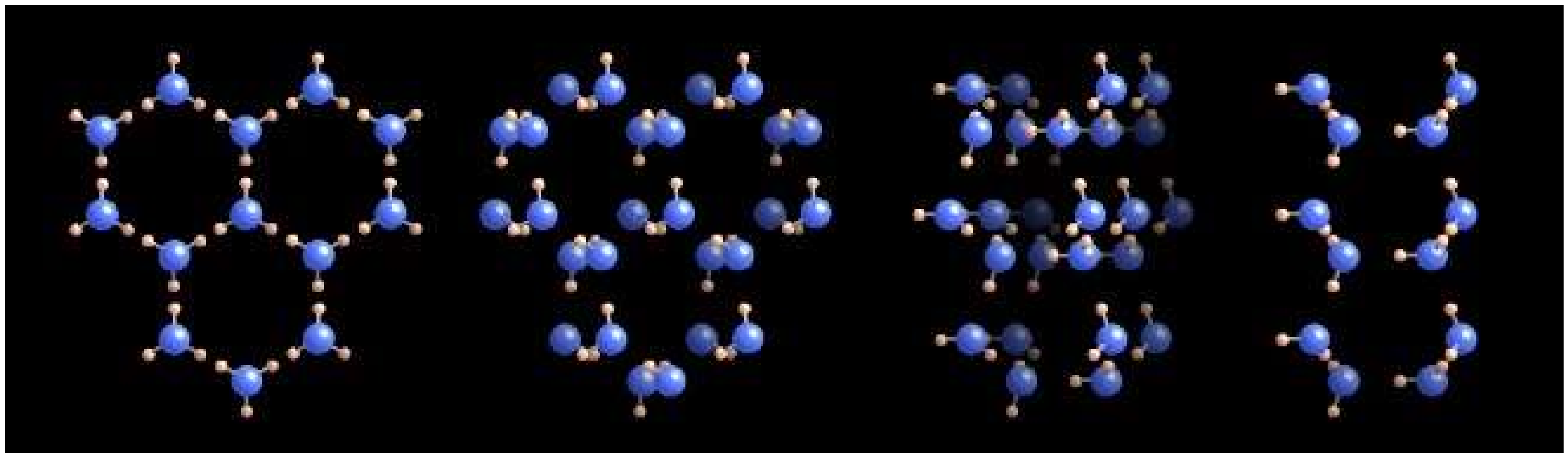

**Figure 2.2: A molecular model of the ice Ih crystal rotated by 0, 18, 72, and 90 degrees about the [$01\bar{1}0$] axis. The large blue spheres represent oxygen atoms, and the smaller spheres represent hydrogen atoms. The first image is looking down the c-axis, revealing the hexagonal lattice structure.**

The phase diagram in Figure 2.1 only delineates the boundaries between the different equilibrium phases of ice, water, and water vapor, depicting the lowest-energy state at each temperature and pressure. As such, the chart conceals the phenomenological richness inherent in transitions between the different states. Each line in Figure 2.1 tells us under what conditions water will change phase, but it says nothing about the nonequilibrium physical processes that define the character of that phase transition. It is likely that there are many fascinating stories to tell about the growth dynamics that must occur along every line segment in Figure 2.1. We restrict our attention to one particular phase transition because it has proven itself worthy of study, plus the others remain almost completely unexplored.

## 2.2 Ice Crystallography

The word "crystal" derives from the Greek *krystallos*, which means "ice" or "clear ice." Despite its meaning, the word was not originally used to describe ice, but rather the mineral quartz. Pliny the Elder, an early Roman naturalist, described clear quartz *krystallos* as a form of ice, frozen so hard that it could not melt. He was of course mistaken on this point; quartz is not a form of ice, nor is it even made of water. Nevertheless, Pliny's misinterpretation is still felt in the language of the present day, after nearly 2000 years. If you look in your dictionary, you may find that one of several definitions of the word crystal is simply "the mineral quartz."

The usual scientific definition of crystal is any material in which the atoms and molecules are arranged in an ordered lattice (although liquid crystals, quasi-crystals, and other uses of the word can be found in the scientific literature). Figure 2.2 shows a molecular model of ice Ih, the only form of ice commonly found on the Earth's surface, including in snow crystals. Note how two hydrogen atoms closely flank each oxygen atom, so the trio forms an essentially intact $H_2O$ molecule. The ice crystal, therefore, can be considered as a collection of whole water molecules arranged in a lattice structure.

Strong molecular bonds bind the individual $H_2O$ molecules, and these in turn are connected together by weaker O-H bonds to form the crystal. In the terminology of chemical bonding, covalent bonds tightly bind the two hydrogens in each water molecule, while weaker hydrogen bonds connect adjacent water molecules. This arrangement is described by the Bernal–Fowler rules [1933Ber], placing exactly one hydrogen atom between each adjacent pair of oxygen atoms, as shown in Figure 2.2. Of course, this is all better visualized using a full 3D bonding model, and one can purchase model kits specifically for water ice.

While the arrangement of oxygen atoms is fully described by the hexagonal lattice structure in ice Ih, the placement of hydrogen atoms using the Bernal–Fowler rules allows a certain degree of ambiguity, as is demonstrated by example in Figure 2.3. It is possible to twist the individual $H_2O$ molecules around into a large number of possible arrangements while maintaining two strong and two weak O-H bonds on each oxygen atom. This ambiguity appears to play no important role in snow crystal growth, but it is a basic feature of the ice Ih lattice.

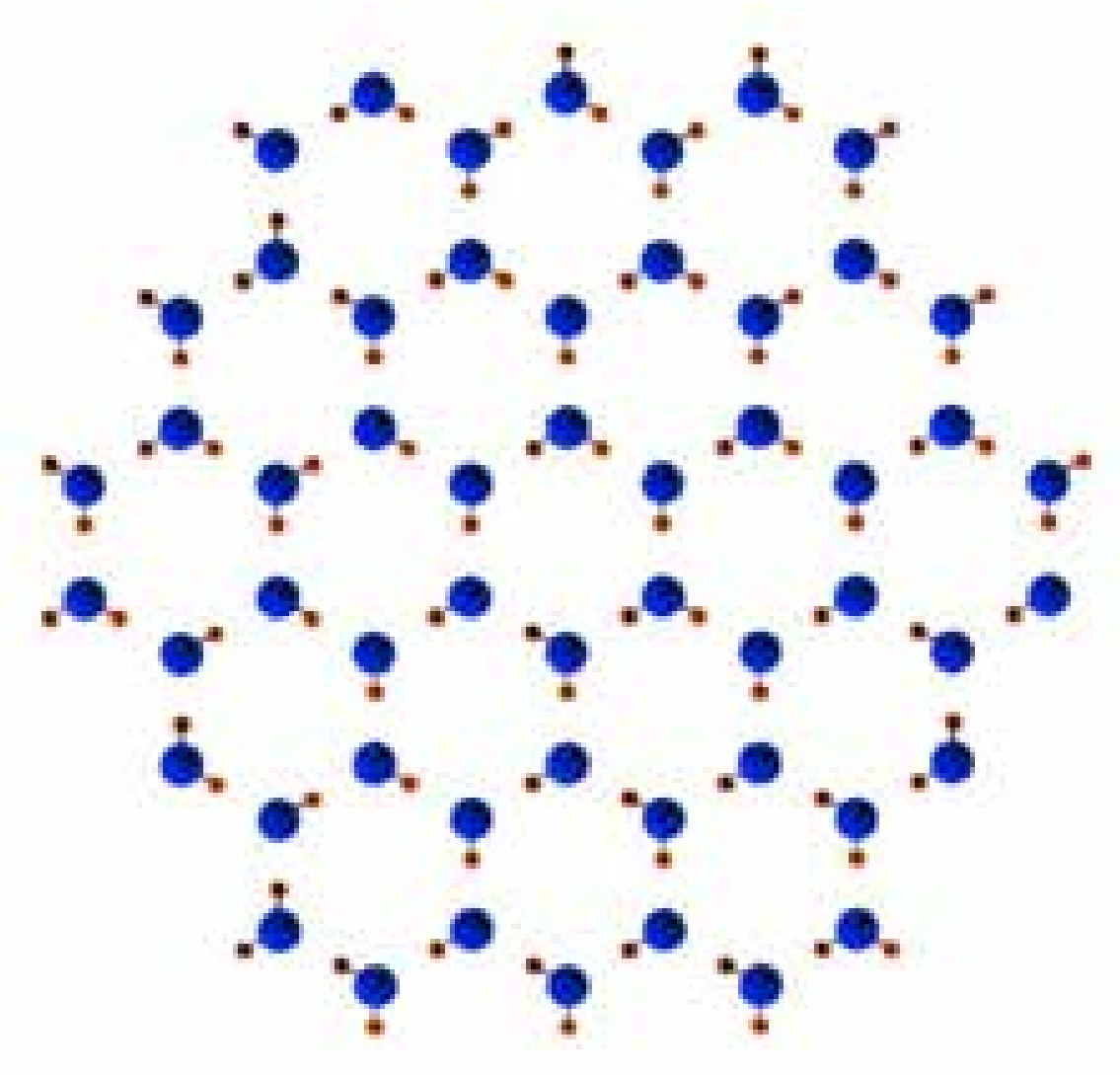

**Figure 2.3: A 2D toy-model example of the Bernal–Fowler rules in ice Ih (although the full 3D rules do not visualize especially well in 2D). The basic idea sketched here is that there are many possible ways to orient the individual $H_2O$ molecules in the crystal while keeping exactly one hydrogen atom between each adjacent pair of oxygen atoms, with each hydrogen forming one strong and one weak bond.**

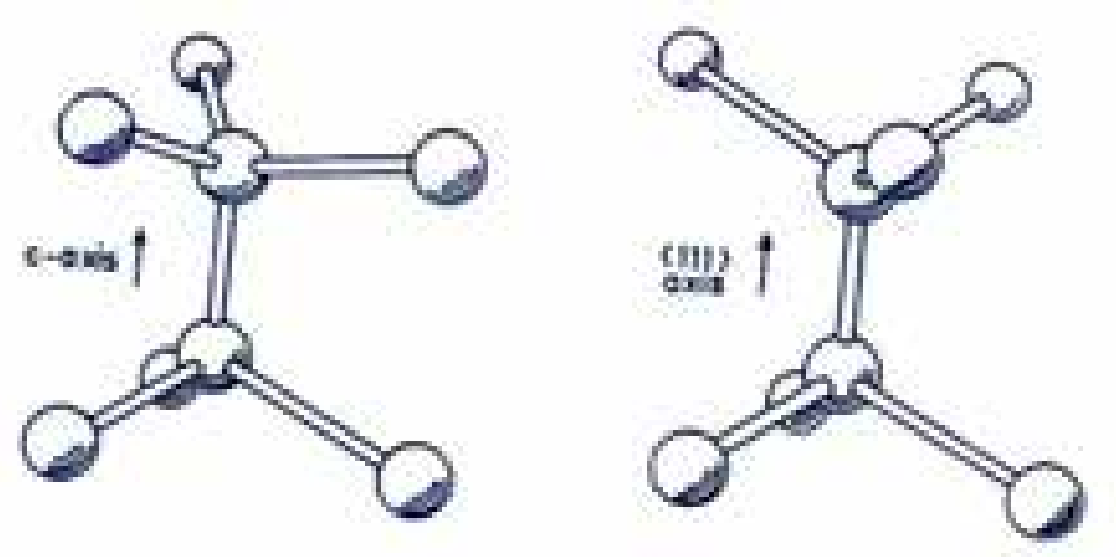

**Figure 2.4: The essential lattice difference between hexagonal ice Ih (left) and cubic ice Ic (right) is a twist in some of the bonds, specifically those between the basal layers in ice Ih. Here the balls and sticks represent oxygen and hydrogen atoms, respectively.**

## Hexagonal and Cubic Ice

The two O-H bonds in a free $H_2O$ molecule meet at an angle of 104.5 degrees, which is close to the tetrahedral angle of 109.5 degrees. Because of this near match, the four O-H bonds emanating from each oxygen atom in the ice crystal (two tight bonds, two weaker bonds) are essentially in a tetrahedral arrangement. Given this bond structure, there are two ways to form a crystal lattice: the hexagonal (ice-like) structure and a cubic (diamond-like) structure, as shown in Figure 2.4. The difference between these two structures comes down to a twist in the bonding of adjacent tetrahedra separating the basal planes.

Adopting the ice Ih bonding shown in Figure 2.4 gives the normal ice lattice shown in Figure 2.2, which has hexagonal symmetry. Choosing the ice Ic bonding throughout the lattice yields a structure with a cubic symmetry called ice Ic. Both crystal structures can be found in the phase diagram in Figure 2.1, but only ice Ih is stable under ordinary environmental conditions. Nevertheless, as we will see below, cubic ice bonding appears to play a small role in normal snow crystal structures.

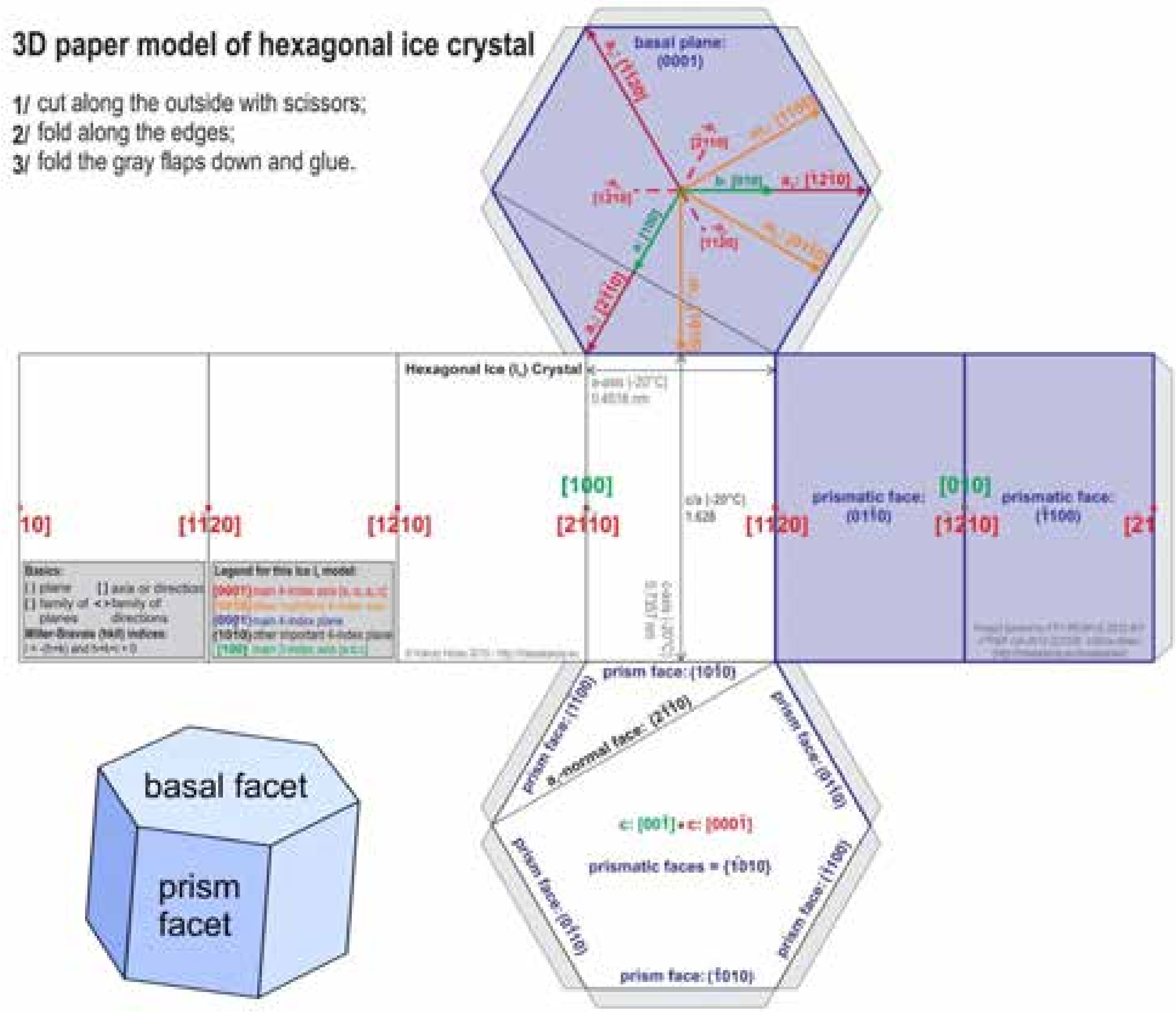

**Figure 2.5: The Miller and Miller-Bravais indices displayed on an unfolded hexagonal prism of ice Ih. Image by Maurine Montagnat and Thomas Chauve, www.hidaskaroly.eu/insidestrain/ice/ice.html**

Stacking spheres is another way to see the difference between the Ih and Ic lattice structures. Beginning with a flat surface, the first layer of spheres is optimally stacked in the usual hexagonal close-packed arrangement. The next layer goes on top of the first with no ambiguity; there is only one way to place a second layer of spheres on top of the first. With the third layer, however, there are two choices in its placement. If the first sphere is placed directly over a sphere in the first layer, the Ih lattice results (assuming this same choice is made for all subsequent layers). Shifting the third layer slightly, so the 3$^{rd}$-layer spheres are not directly over spheres in the first layer results in the Ic structure (again, assuming this choice is made consistently). If some layers use the Ih placement and others use the Ic placement, this is called a "stacking disordered" structure.

Although cubic ice Ic has been created in the laboratory at low temperatures, no examples of faceted "cubic" snow crystals have been made, although this would certainly be an interesting sight. Moreover, to my understanding, pure cubic ice Ic has not even been definitively observed in the laboratory or in nature; laboratory samples are generally stacking-disordered to a substantial degree and are thus essentially mixtures of ice Ic and Ih.

It appears that ice Ic has a lower surface energy than ice Ih, causing it to nucleate more readily from liquid water under certain conditions [2005Joh]. There is even evidence that ice Ic can be found as minute ice grains in high-altitude clouds, although this is not known with certainty. The line between ice Ih and ice Ic in Figure 2.1 is dotted, not solid, representing that this "pseudo" phase boundary (not a real phase boundary in the usual thermodynamic sense of a first-order phase transition) is not so well understood.

## Lattice Projections

When referring to the various planes and axes in crystal lattices, it is customary to use either the 3-axis Miller indices or (more usually) the 4-axis Miller-Bravais indices, and both are shown in Figure 2.5 for ice Ih. Curly brackets refer to families of planes, including the $\{0001\}$ basal facets and the $\{\bar{1}010\}$ family of prism facets. Parentheses refer to specific planes, including the basal facet $(0001)$ and the six prism facets $(1\bar{1}00)$, $(10\bar{1}0)$, $(01\bar{1}0)$, $(\bar{1}100)$, $(\bar{1}010)$, and $(0\bar{1}10)$. Square brackets denote directions, for example the c-axis $[0001]$ that is perpendicular to the basal face, or the a-axis $[11\bar{2}0]$ that is perpendicular to the $(11\bar{2}0)$ face. All six a-axes point to corners of the hexagonal prism, as shown in Figure 2.5. The $\{11\bar{2}0\}$ planes are sometimes called the secondary prism faces, although growing ice crystals do not form facets on these planes.

Although the $\{0001\}$ basal facets and the $\{\bar{1}010\}$ prism facets are by far the most common faceted surfaces seen in snow crystals, the $\{10\bar{1}1\}$ pyramidal facets have also been observed, as shown in Figure 2.6. Little is known about the growth of pyramidal facets, but the evidence suggests that they form only rarely, at quite low temperatures (below -20 C) and perhaps only in air (or, more to the point, not in a near-vacuum environment).

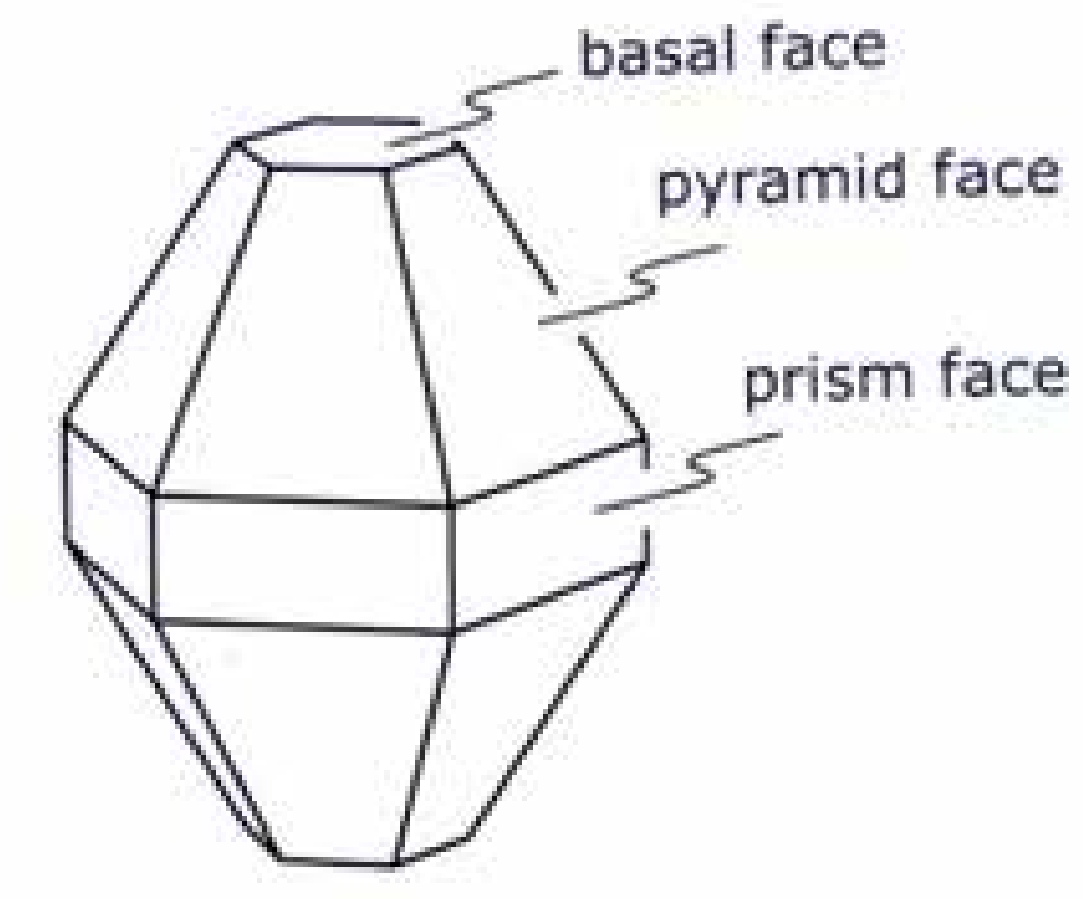

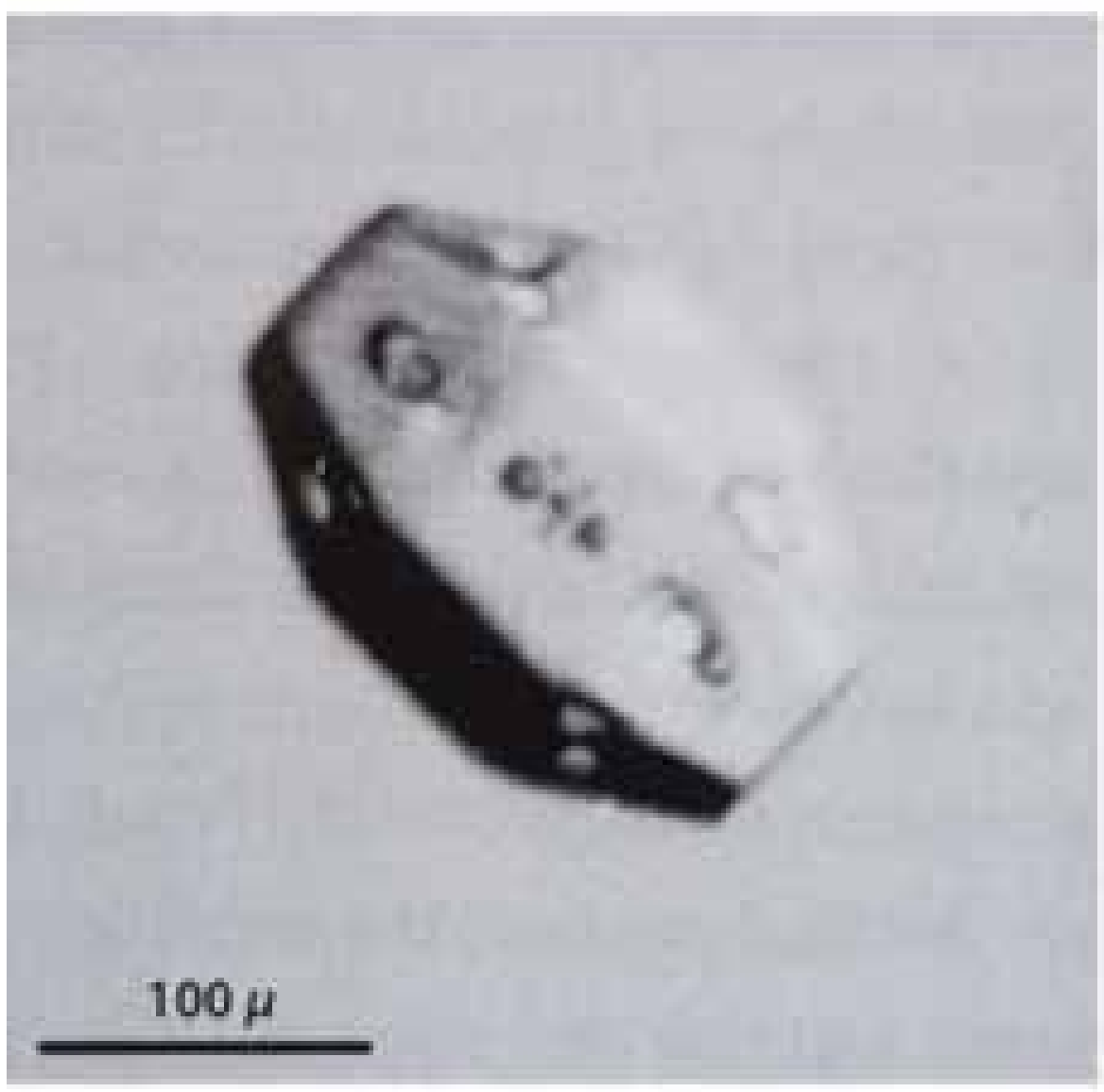

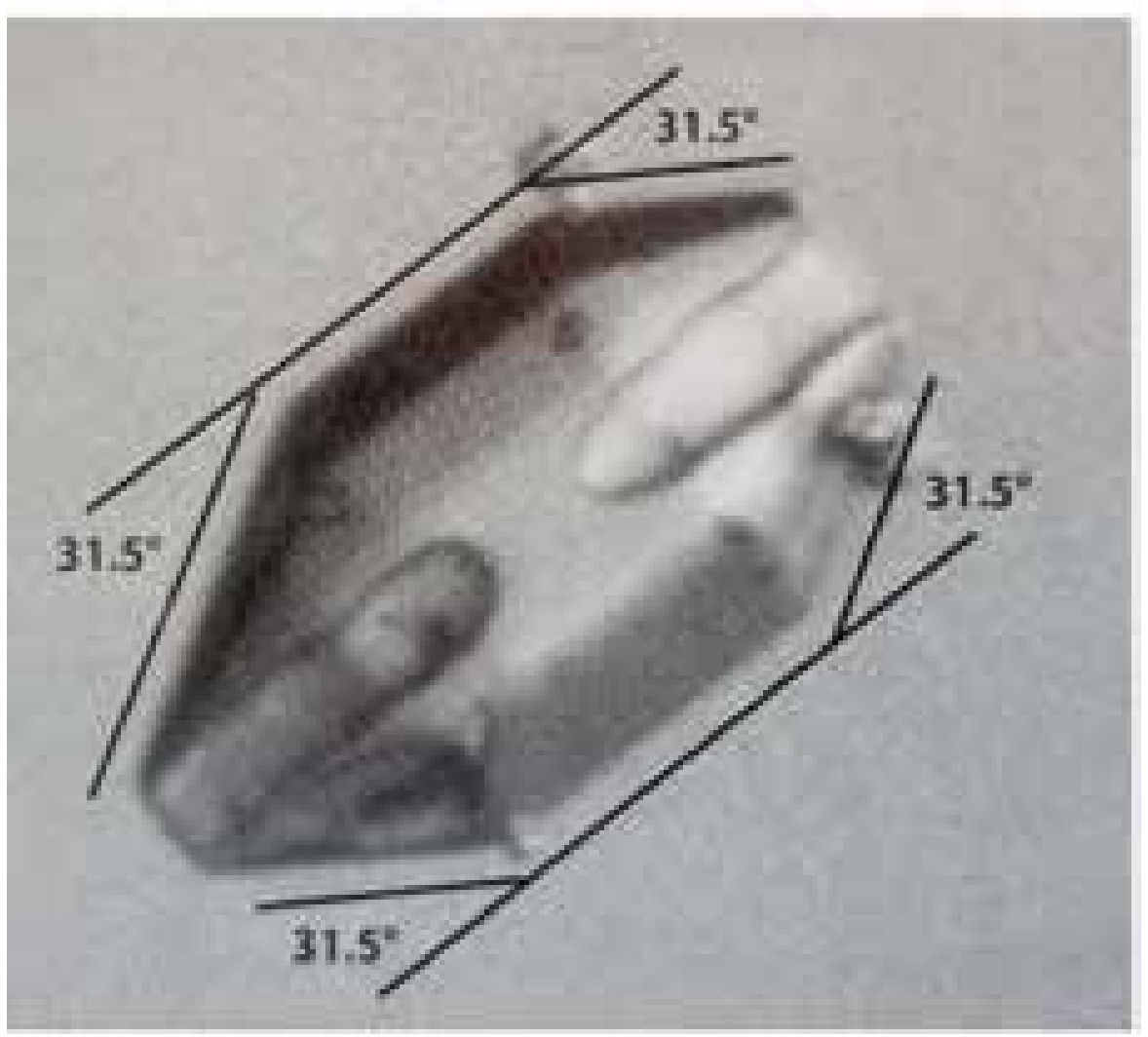

**Figure 2.6: South-pole snow crystals displaying pyramidal facets. Image from [2006Tap].**

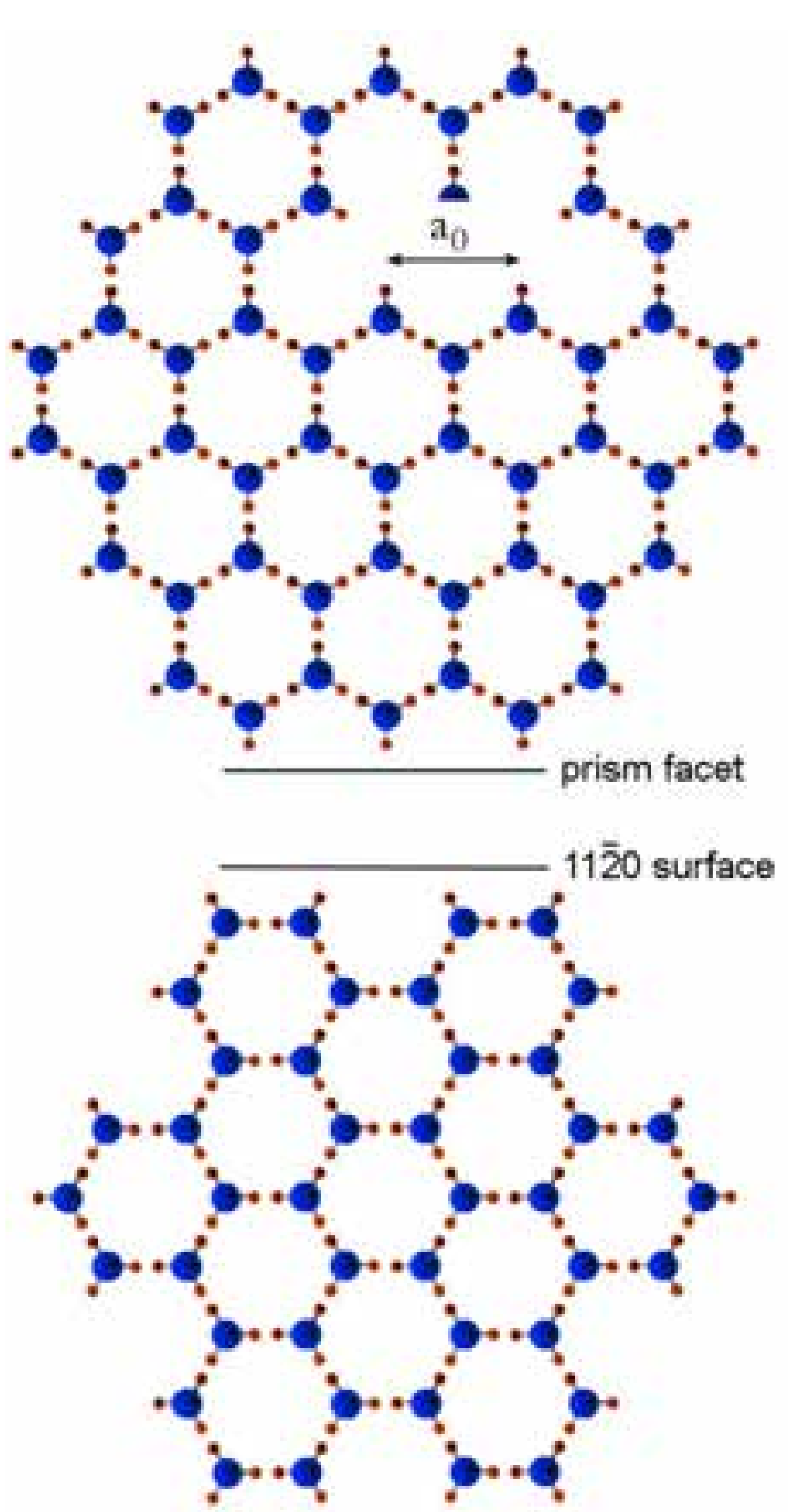

**Figure 2.7: Two projections of the ice Ih lattice looking down along the [0001] direction (perpendicular to a basal facet). A prism $\{\bar{1}010\}$ facet and a $\{11\bar{2}0\}$ surface are labeled. (The latter is sometimes called a secondary prism surface, although it does not exhibit any known faceting.)**

To relate the Miller-Bravais indices to the ice crystal facets, it is useful to examine several 2D projections of the 3D lattice structure of ice Ih. For example, Figure 2.7 shows the lattice structure of the $\{\bar{1}010\}$ prism facets and the $\{11\bar{2}0\}$ surfaces, both looking down along the c-axis. Since the early days of X-ray crystallography, it had been thought that the $\{\bar{1}010\}$ surfaces coincided with the well-known prism facets seen in snow crystals, as the $\{\bar{1}010\}$ surfaces have a somewhat simpler lattice structure compared to the $\{11\bar{2}0\}$ surfaces. Although this early assessment was indeed correct, it was only definitively confirmed by direct observation rather recently [2017Bru]. Figure 2.8 shows another lattice projection that includes the basal, prism, and pyramidal facets. The ice lattice parameters $a_0$ and $c_0$ are defined in Figures 2.7 and 2.8, and measurements near 0 C give

$$a_0 = 0.452 \text{ nm} \tag{2.1}$$

$$c_0 = 0.736 \text{ nm}$$

and the respective spacings between basal and prism layers are then

$$x_{basal} = \frac{c_0}{2} = 0.37 \text{ nm} \qquad (2.2)$$

$$x_{prism} = \left(\frac{\sqrt{3}}{2}\right) a_0 = 0.39 \text{ nm}$$

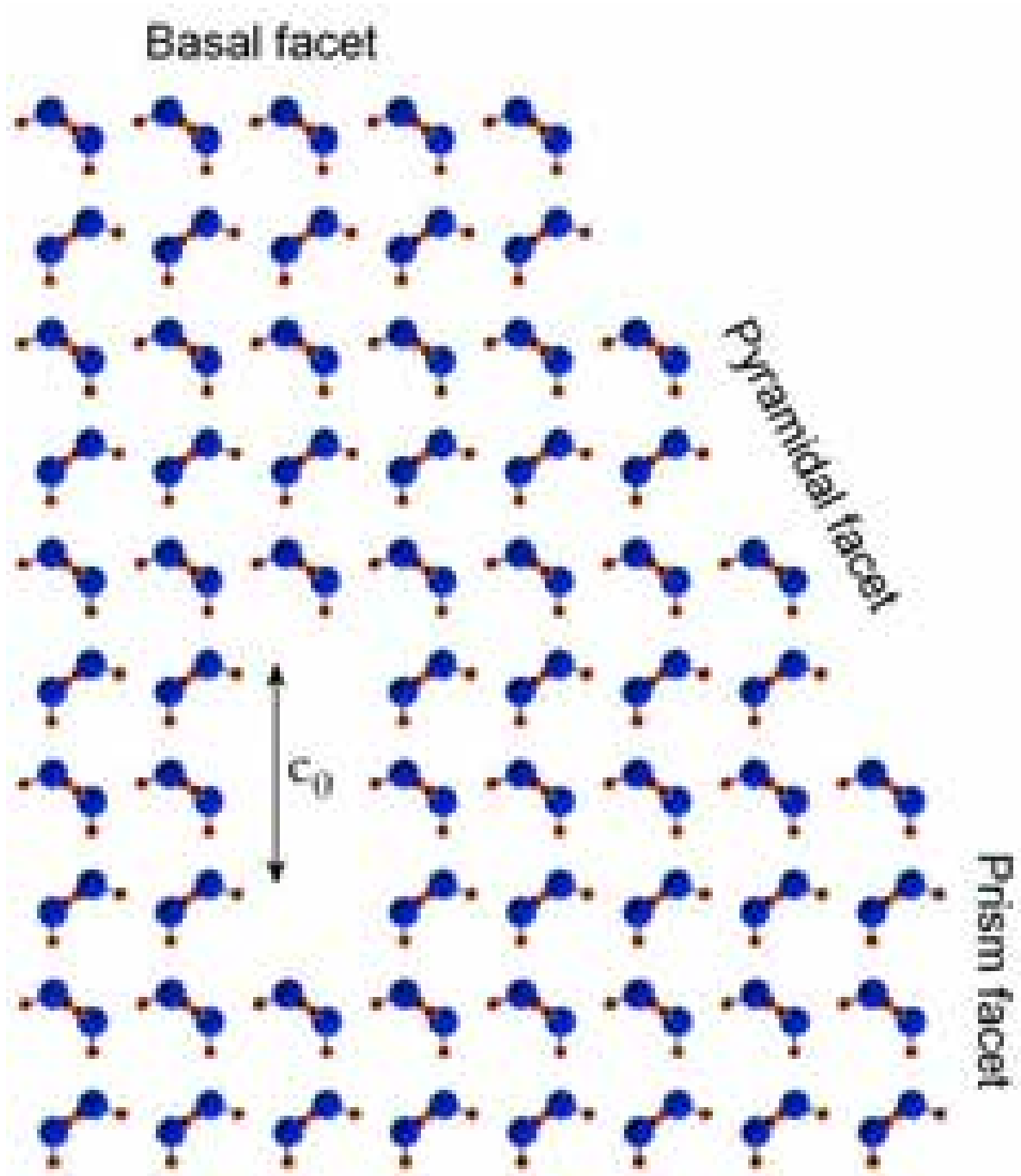

**Figure 2.8: A projection of the ice Ih lattice looking along the $[11\bar{2}0]$ direction, showing the basal, prism, and pyramidal facets. The angle between the prism and pyramidal facets is $\tan^{-1}(\sqrt{3}a_0/2c_0) = 28.0$ degrees. (The 31.5-degree angle seen in Figure 2.6 is from looking along the $[\bar{1}010]$ direction, which is not parallel to a pyramidal facet.)**

## Terrace Steps

As we will see in Chapter 4, the nucleation of new molecular terraces is a key factor in the formation of snow-crystal facets, and terrace nucleation is governed by terrace step energies. It is useful, therefore, to examine the molecular structure of terrace steps, and several lattice projections that do so are shown in Figures 2.9, 2.10, and 2.11. While the surface energies of the facet planes are generally lower than non-faceted surfaces, this anisotropy in the surface energy seems to play only a minor role in snow-crystal faceting. The anisotropy in the attachment kinetics (see Chapter 4) is much stronger, and this is the more important factor governing snow-crystal growth rates and faceting.

The fact that the terrace steps are relatively shallow on the $\{11\bar{2}0\}$ surfaces, as shown in

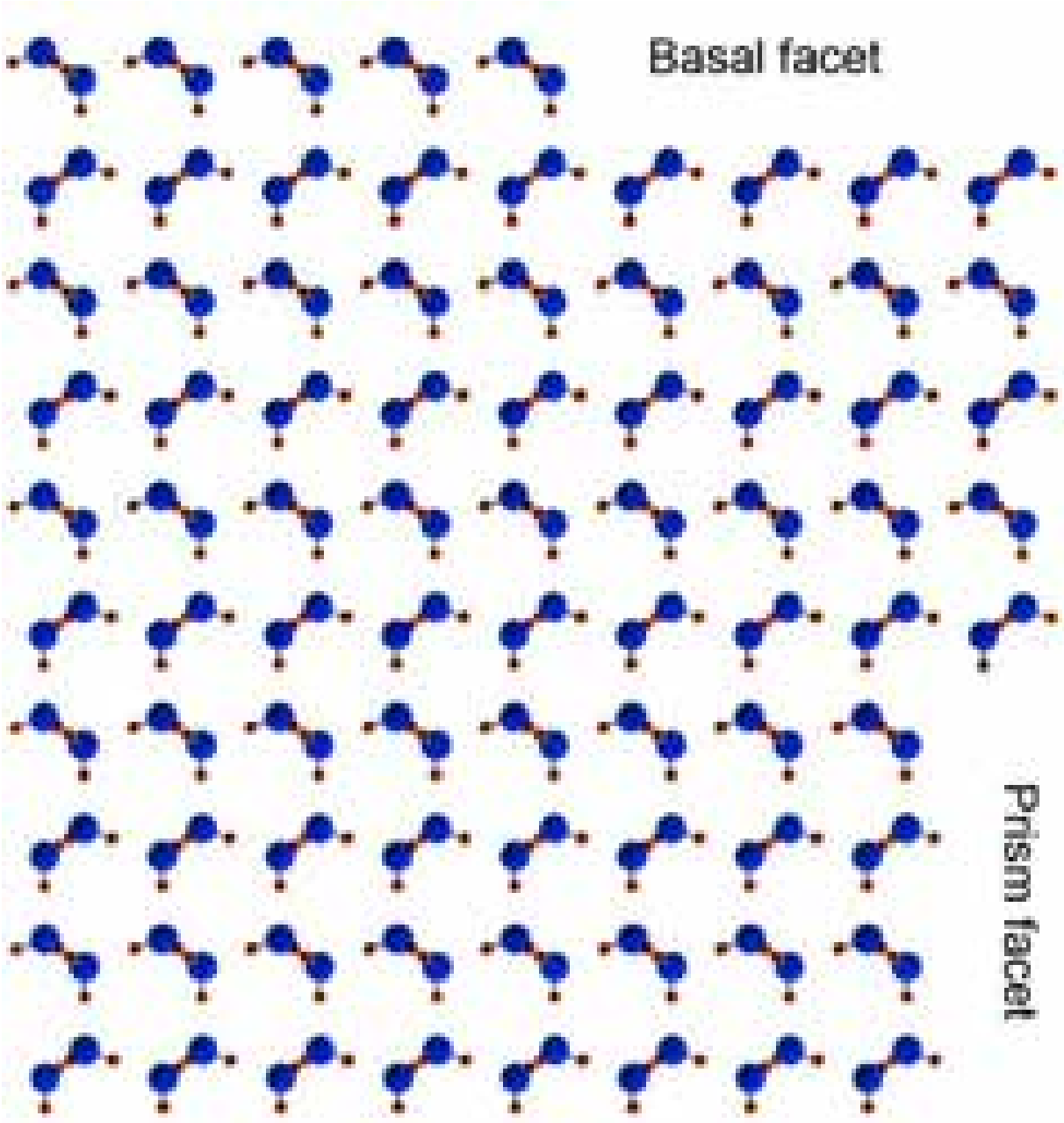

**Figure 2.9: A projection of the ice Ih lattice looking along the $[11\bar{2}0]$ direction, showing basal and prism facets and terrace steps. The spacing between basal layers is $c_0/2 = 0.37$ nm, while the spacing between prism layers is $(\sqrt{3}/2)a_0 = 0.39$ nm.**

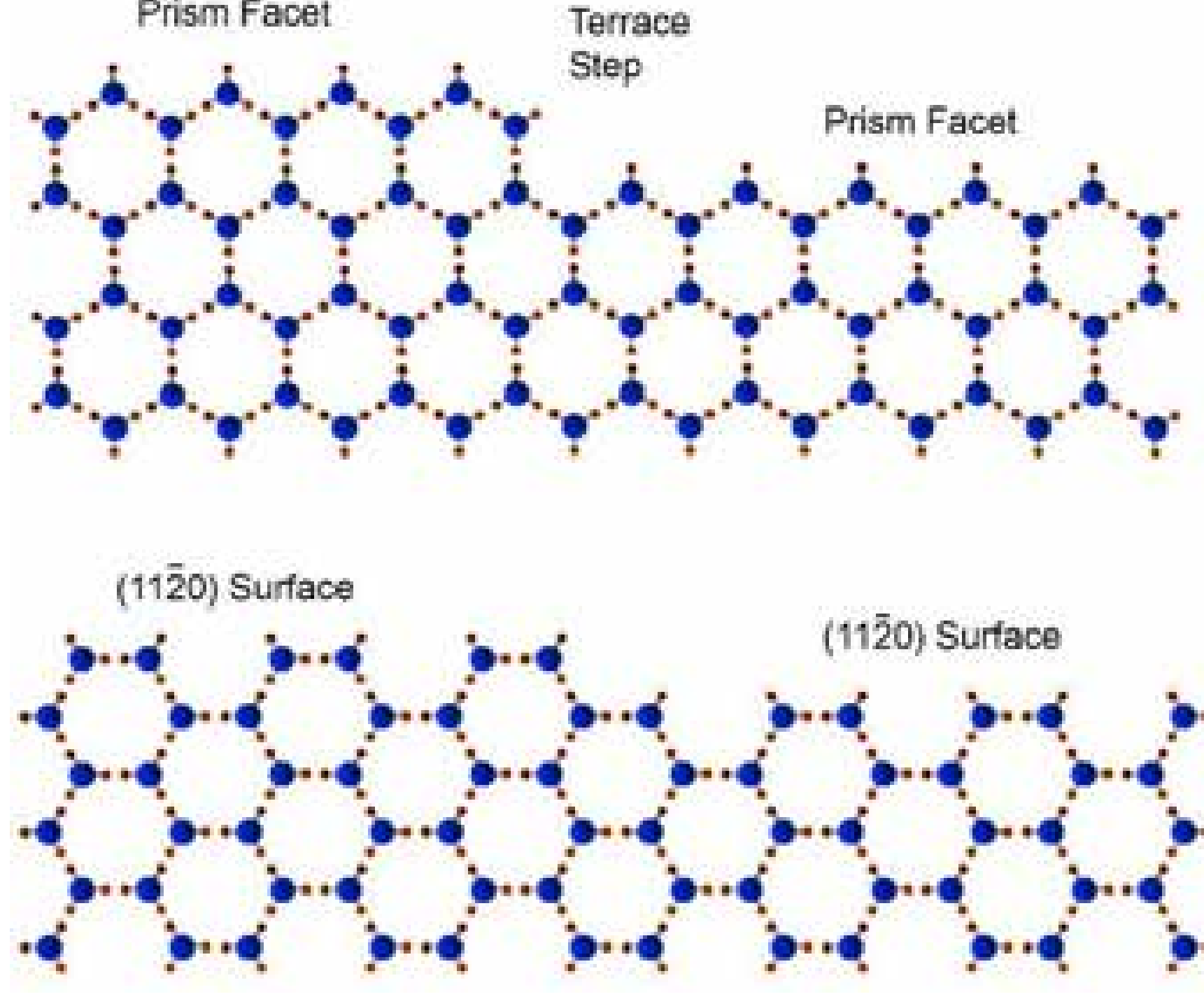

**Figure 2.10 (left): Terrace steps on a prism $\{\bar{1}010\}$ facet and a $\{11\bar{2}0\}$ surface, as seen looking along the $[0001]$ direction. Note that the prism step is more distinct, and admolecules would likely bind much more strongly at a terrace step than on the prism facet. In contrast, the $\{11\bar{2}0\}$ step is less distinct, and binding at the step edge is likely not much stronger than elsewhere on the surface. The larger terrace step in the top sketch suggests stronger faceting on the prism surface, as is observed.**

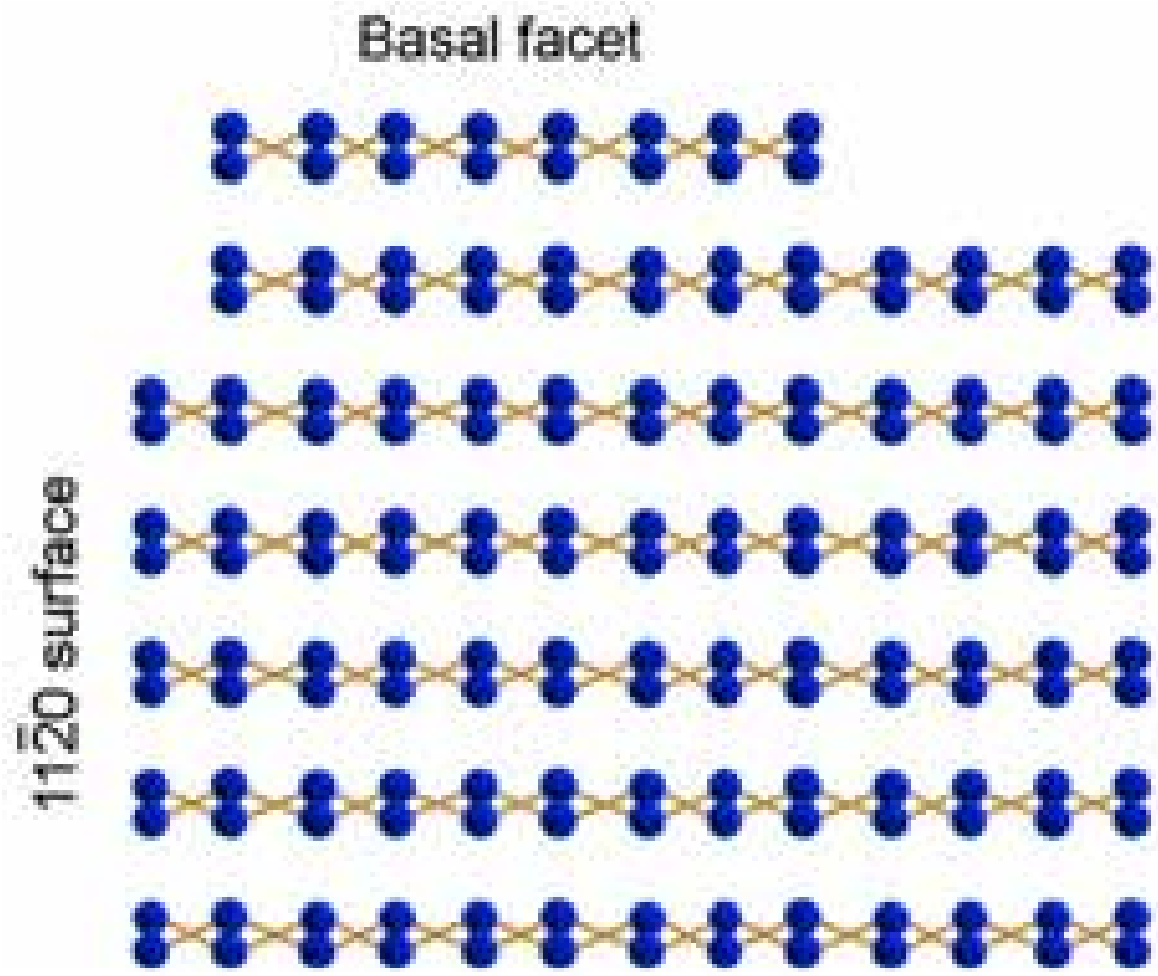

**Figure 2.11: A projection of the ice Ih lattice looking along the [$\bar{1}$010] direction, showing basal and (11$\bar{2}$0) surfaces and terrace steps. From this perspective, it is easy to see why the basal terraces are sometimes called molecular *bilayers*.**

Figure 2.10, likely explains the absence of faceting on these surfaces. In contrast, the larger terrace steps on the basal and prism surfaces creates a substantial nucleation barrier that promotes strong faceting on these surfaces. This line of reasoning further suggests that the pyramidal step energy likely becomes substantial at low temperatures, as pyramidal facets apparently only form in quite cold conditions. These are both speculative statements, however, as only the basal and prism step energies have been measured to date. Even then, as we will see in Chapter 4, there is some mystery surrounding their values in air versus in near-vacuum.

As a word of caution, it should be remembered that these sketches of lattice projections are oversimplified representations of the molecular surface structures of ice, valid only at extremely low temperatures. At typical temperatures associated with snow crystal growth, thermal fluctuations can easily distort, or even completely rearrange, the lattice structure at the surface, and the high vapor pressure of ice means that molecules are continuously leaving and reattaching to the lattice at a prodigious rate. Even a background gas like air may affect the ice surface characteristics, as we will see in Chapter 7. Although these lattice sketches can be quite useful for visualization purposes, real ice surfaces are neither rigid nor static.

## 2.3 Surface Premelting

Figure 2.12 shows a dramatic example of how real ice surfaces can deviate from rigid lattice structures, in this case via a phenomenon known as surface premelting [1999Pet, 2006Das, 2007Li, 2014Lim]. Because water molecules near an ice surface are less tightly bound compared to those in the bulk, a disordered *quasi-liquid layer* (QLL) appears on the surface near the melting point. First predicted by Michael Faraday in 1859 [1859Far], much recent research has been aimed at better understanding this enigmatic phenomenon.

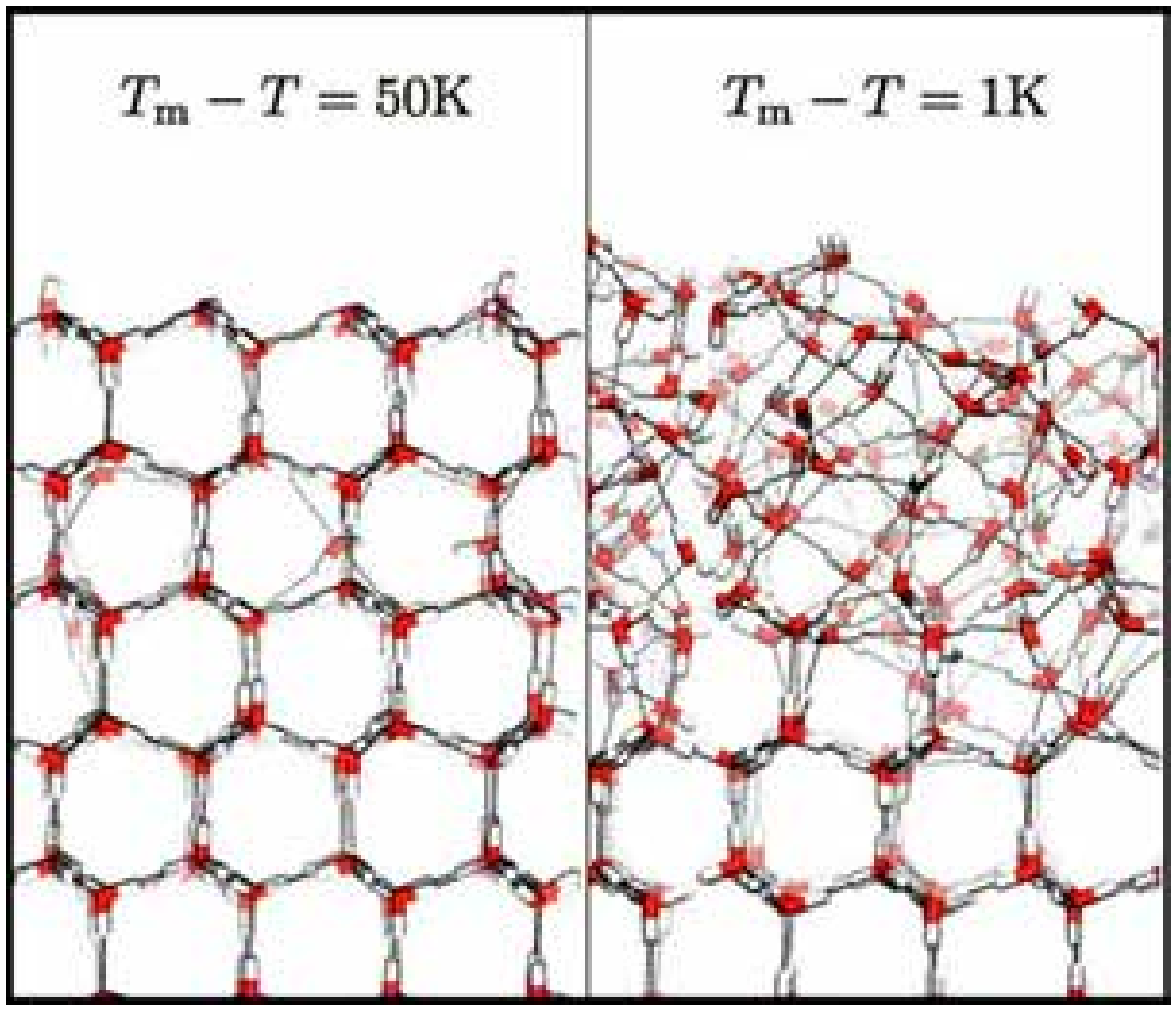

**Figure 2.12: A molecular-dynamics simulation from [2014Lim] demonstrating surface premelting. At temperatures far below the melting point (left), the bulk lattice structure persists all the way up to the surface (here showing an edge-on view of a basal facet). Near the melting point, however, the top molecular layers lose their ordered structure, forming an amorphous premelted layer, also called a quasi-liquid layer (QLL).**

One way of thinking about surface premelting is through the *Lindemann criterion* [1910Lin], which states that a solid will melt if thermal fluctuations of the intermolecular distance become larger than approximately 10-15 percent of the average distance. This empirical rule applies (roughly) to a broad range of materials, and one expects that thermal fluctuations will be larger near the surface, owing to the reduced binding there. One can turn this into a useful toy model [2005Lib], but more-sophisticated theoretical treatments of the phenomenon have been developed [2006Das, 2014Lim]. At present, the structure and dynamics of surface premelting is quite poorly understood.

Surface premelting is an equilibrium phenomenon that demonstrates that the phase boundary between the liquid and solid phases of a material is only precisely defined when the sample has infinite extent (called "bulk" material). For nanometer-scale clusters, when the QLL thickness becomes comparable to the size of the system, the cluster melting point can be substantially lower than the bulk melting temperature. There is considerable current theoretical and experimental interest in the topic of cluster premelting, and finite-sample thermodynamics more generally, and this work is related to surface premelting in ice.

Based on the experimental and theoretical evidence to date, the thickness of the ice QLL (defined by some appropriate parameterization of the molecular disorder relative to a rigid lattice) becomes roughly equal to the monolayer thickness at around -10 C, diverging logarithmically as the temperature approaches the melting point. There is, unfortunately, a great deal of uncertainty regarding the detailed structure and general behavior of premelting in ice. Many experimental surface probes have found convincing evidence for premelting, but different experiments measure different surface properties, and it is not always clear how to compare different results. Molecular dynamics simulations provide many insights into the detailed structure of the QLL (as seen in Figure 2.13), but quantitative comparison with experiments remains a nontrivial challenge.

It has long been speculated that surface premelting in ice plays a major role in snow crystal growth, perhaps explaining much of the growth behavior seen in the snow-crystal morphology diagram [1987Kob, 1984Kur1]. While it seems quite likely that surface premelting is important, defining what its various effects actually are exactly has been problematic. For the present, there seems to be no concrete, well-established physical connection between surface melting and snow crystal growth, although this situation may change at any time.

Surface premelting becomes especially pronounced at temperatures above -1 C, when the QLL becomes quite thick and perhaps subject to instabilities that result in a nonuniform QLL thickness [2015Asa]. This topic relates to a long-debated question of whether water completely "wets" ice at 0 C. Put another way, does a water drop on ice remain confined with a non-zero contact angle, or does it spread until it covers the ice surface in a layer of uniform thickness? This question is difficult to answer because experimental observations are always potentially affected by surface impurities that could affect the wetting behavior. The evidence suggests that the ice/water contact angle at 0 C is small but nonzero, although the subject has not been fully settled yet. Snow crystal growth at temperatures between -1 C and 0 C could be phenomenologically quite interesting, and few observations have been made to date in this regime.

A particularly promising direction for understanding the ice surface is the use of molecular-dynamics simulations to measure important properties directly from the known chemical interactions between constituent molecules. The same MD simulations that impressively demonstrate the structural behavior of surface premelting can also be used to calculate terrace step energies [2012Fro], at least in principle. This allows for a ready comparison with experiments, as step energies have recently been measured over a broad range of temperatures [2013Lib]. Moreover, because the step energy is a static, equilibrium quantity, the prospects for making accurate numerical estimates look excellent. Comparing the structure and energetics of terrace steps with the accompanying surface premelting may lead to many valuable insights about the molecular origins of the large-scale physical properties of ice surfaces.

## 2.4 Ice Energetics

Many ice properties factor into the physics of snow crystal growth, notably the bulk, surface, and step energies of the ice crystal. The terrace step energies are particularly important for determining growth rates, so these will receive special attention throughout this book. For completeness, however, we document here a list of particularly relevant ice energetics.

### Bulk Energies

The bulk properties refer to sample sizes that are large enough that surface effects are all negligible. The bulk energies include the specific heats of the water/vapor transition (in this case often called evaporation, condensation, or vaporization, depending on conventions), the ice/vapor transition (typically called sublimation or deposition) and the ice/water transition (melting, freezing, or fusion). Each of these is the amount of energy needed to cross a line in the water phase diagram (Figure 2.1), as all of these are simple first-order phase transitions. Because water is an extremely well-studied material, all these quantities have been accurately measured over a broad range of temperatures and pressures. Near the triple point of water, we have (using the notation s/l/v = solid/liquid/vapor)

$$L_{sv} \approx 2.8 \times 10^6 \text{ J/kg} \tag{2.3}$$

$$L_{lv} \approx 2.5 \times 10^6 \text{ J/kg}$$

$$L_{sl} \approx 0.33 \times 10^6 \text{ J/kg}$$

and we see $L_{sv} \approx L_{sl} + L_{lv}$, as we would expect at the triple point. The specific heats vary somewhat with position on the respective phase boundaries, but these differences are not very important for our focus on snow crystal growth.

Related useful quantities include heat capacities (here at constant pressure) of water vapor, liquid water, and ice:

$$c_{p,wv} \approx 2.0 \text{ kJ/kg} \cdot \text{K} \tag{2.4}$$

$$c_{p,water} \approx 4.2 \text{ kJ/kg} \cdot \text{K}$$

$$c_{p,ice} \approx 2.1 \text{ kJ/kg} \cdot \text{K}$$

thermal conductivities

$$\kappa_{wv} \approx 0.02 \text{ W/m} \cdot \text{K} \tag{2.5}$$

$$\kappa_{water} \approx 0.6 \text{ W/m} \cdot \text{K}$$

$$\kappa_{ice} \approx 2.3 \text{ W/m} \cdot \text{K}$$

and material densities

$$\rho_{ice} \approx 917 \text{ kg/m}^3 \tag{2.6}$$

$$\rho_{water} \approx 1000 \text{ kg/m}^3$$

Both the ice and water densities can be assumed constant for our purposes (although both change somewhat with temperature). The equilibrium vapor density, on the other hand, depends strongly on temperature, so requires special treatment.

| $Temp$ | $P_{water}$ | $P_{ice}$ | $c_{sat}$ | $\sigma_{water}$ | $v_{kin}$ | $\eta$ | $\kappa$ | $X_0$ | $C_{diff}$ |
|---|---|---|---|---|---|---|---|---|---|
| $C$ | $mbar$ | $mbar$ | $\#/m^3$ | | $\mu m/sec$ | | | $\mu m$ | $K^{-2}$ |
| -40 | 0.19 | 0.13 | 3.99E+21 | 0.474 | 17.0 | 0.1091 | 0.029 | 0.153 | 0.0059 |
| -39 | 0.21 | 0.14 | 4.44E+21 | 0.460 | 19.0 | 0.1082 | 0.032 | 0.152 | 0.0058 |
| -38 | 0.23 | 0.16 | 4.95E+21 | 0.446 | 21.2 | 0.1072 | 0.036 | 0.152 | 0.0057 |
| -37 | 0.26 | 0.18 | 5.50E+21 | 0.433 | 23.6 | 0.1061 | 0.039 | 0.152 | 0.0056 |
| -36 | 0.28 | 0.20 | 6.12E+21 | 0.419 | 26.3 | 0.1052 | 0.043 | 0.151 | 0.0055 |
| -35 | 0.31 | 0.22 | 6.79E+21 | 0.406 | 29.3 | 0.1044 | 0.048 | 0.151 | 0.0054 |
| -34 | 0.35 | 0.25 | 7.54E+21 | 0.392 | 32.5 | 0.1036 | 0.052 | 0.151 | 0.0054 |
| -33 | 0.38 | 0.28 | 8.35E+21 | 0.379 | 36.1 | 0.1026 | 0.057 | 0.15 | 0.0053 |
| -32 | 0.42 | 0.31 | 9.25E+21 | 0.366 | 40.1 | 0.1017 | 0.063 | 0.15 | 0.0052 |
| -31 | 0.46 | 0.34 | 1.02E+22 | 0.353 | 44.5 | 0.1010 | 0.069 | 0.15 | 0.0051 |
| -30 | 0.51 | 0.38 | 1.13E+22 | 0.340 | 49.3 | 0.1003 | 0.076 | 0.149 | 0.0050 |
| -29 | 0.56 | 0.42 | 1.25E+22 | 0.326 | 54.5 | 0.0993 | 0.083 | 0.149 | 0.0049 |
| -28 | 0.61 | 0.47 | 1.38E+22 | 0.314 | 60.3 | 0.0983 | 0.091 | 0.149 | 0.0048 |
| -27 | 0.67 | 0.52 | 1.52E+22 | 0.301 | 66.6 | 0.0976 | 0.1 | 0.148 | 0.0048 |
| -26 | 0.74 | 0.57 | 1.68E+22 | 0.289 | 73.6 | 0.0967 | 0.109 | 0.148 | 0.0047 |
| -25 | 0.81 | 0.63 | 1.85E+22 | 0.276 | 81.2 | 0.0957 | 0.118 | 0.148 | 0.0046 |
| -24 | 0.88 | 0.70 | 2.03E+22 | 0.264 | 89.5 | 0.0953 | 0.13 | 0.148 | 0.0045 |
| -23 | 0.96 | 0.77 | 2.23E+22 | 0.252 | 98.6 | 0.0947 | 0.142 | 0.147 | 0.0045 |
| -22 | 1.05 | 0.85 | 2.45E+22 | 0.239 | 108.5 | 0.0935 | 0.154 | 0.147 | 0.0044 |
| -21 | 1.15 | 0.94 | 2.69E+22 | 0.228 | 119.3 | 0.0931 | 0.168 | 0.147 | 0.0043 |
| -20 | 1.25 | 1.03 | 2.95E+22 | 0.215 | 131.2 | 0.0922 | 0.182 | 0.146 | 0.0042 |
| -19 | 1.37 | 1.14 | 3.24E+22 | 0.204 | 144.0 | 0.0912 | 0.198 | 0.146 | 0.0042 |
| -18 | 1.49 | 1.25 | 3.54E+22 | 0.192 | 158.0 | 0.0906 | 0.215 | 0.146 | 0.0041 |
| -17 | 1.62 | 1.37 | 3.88E+22 | 0.180 | 173.3 | 0.0898 | 0.233 | 0.146 | 0.0040 |
| -16 | 1.76 | 1.51 | 4.24E+22 | 0.169 | 189.9 | 0.0891 | 0.253 | 0.145 | 0.0040 |
| -15 | 1.91 | 1.65 | 4.63E+22 | 0.157 | 207.9 | 0.0885 | 0.275 | 0.145 | 0.0039 |
| -14 | 2.08 | 1.81 | 5.06E+22 | 0.146 | 227.4 | 0.0878 | 0.298 | 0.145 | 0.0039 |
| -13 | 2.25 | 1.98 | 5.52E+22 | 0.135 | 248.7 | 0.0872 | 0.323 | 0.144 | 0.0038 |
| -12 | 2.44 | 2.17 | 6.02E+22 | 0.124 | 271.8 | 0.0864 | 0.349 | 0.144 | 0.0037 |
| -11 | 2.64 | 2.38 | 6.56E+22 | 0.113 | 296.7 | 0.0859 | 0.378 | 0.144 | 0.0037 |
| -10 | 2.86 | 2.60 | 7.15E+22 | 0.102 | 323.8 | 0.0851 | 0.408 | 0.144 | 0.0036 |
| -9 | 3.10 | 2.84 | 7.78E+22 | 0.091 | 353.1 | 0.0843 | 0.44 | 0.143 | 0.0036 |
| -8 | 3.35 | 3.10 | 8.46E+22 | 0.081 | 384.7 | 0.0837 | 0.475 | 0.143 | 0.0035 |
| -7 | 3.62 | 3.38 | 9.20E+22 | 0.070 | 419.0 | 0.0831 | 0.512 | 0.143 | 0.0035 |
| -6 | 3.91 | 3.68 | 9.99E+22 | 0.060 | 455.9 | 0.0824 | 0.552 | 0.143 | 0.0034 |
| -5 | 4.21 | 4.01 | 1.08E+23 | 0.050 | 495.9 | 0.0818 | 0.595 | 0.142 | 0.0034 |
| -4 | 4.54 | 4.37 | 1.18E+23 | 0.040 | 539.0 | 0.0812 | 0.64 | 0.142 | 0.0033 |
| -3 | 4.90 | 4.76 | 1.28E+23 | 0.030 | 585.3 | 0.0806 | 0.689 | 0.142 | 0.0033 |
| -2 | 5.27 | 5.17 | 1.38E+23 | 0.020 | 635.5 | 0.0800 | 0.741 | 0.141 | 0.0032 |
| -1 | 5.68 | 5.62 | 1.50E+23 | 0.010 | 689.4 | 0.0793 | 0.795 | 0.141 | 0.0032 |
| 0 | 6.11 | 6.11 | 1.62E+23 | 0.000 | 747.4 | -- | 0.824 | 0.141 | 0.0031 |

**Table 2.1: Several handy physical quantities that commonly appear in the study of snow crystal growth, listed as a function of temperature.**

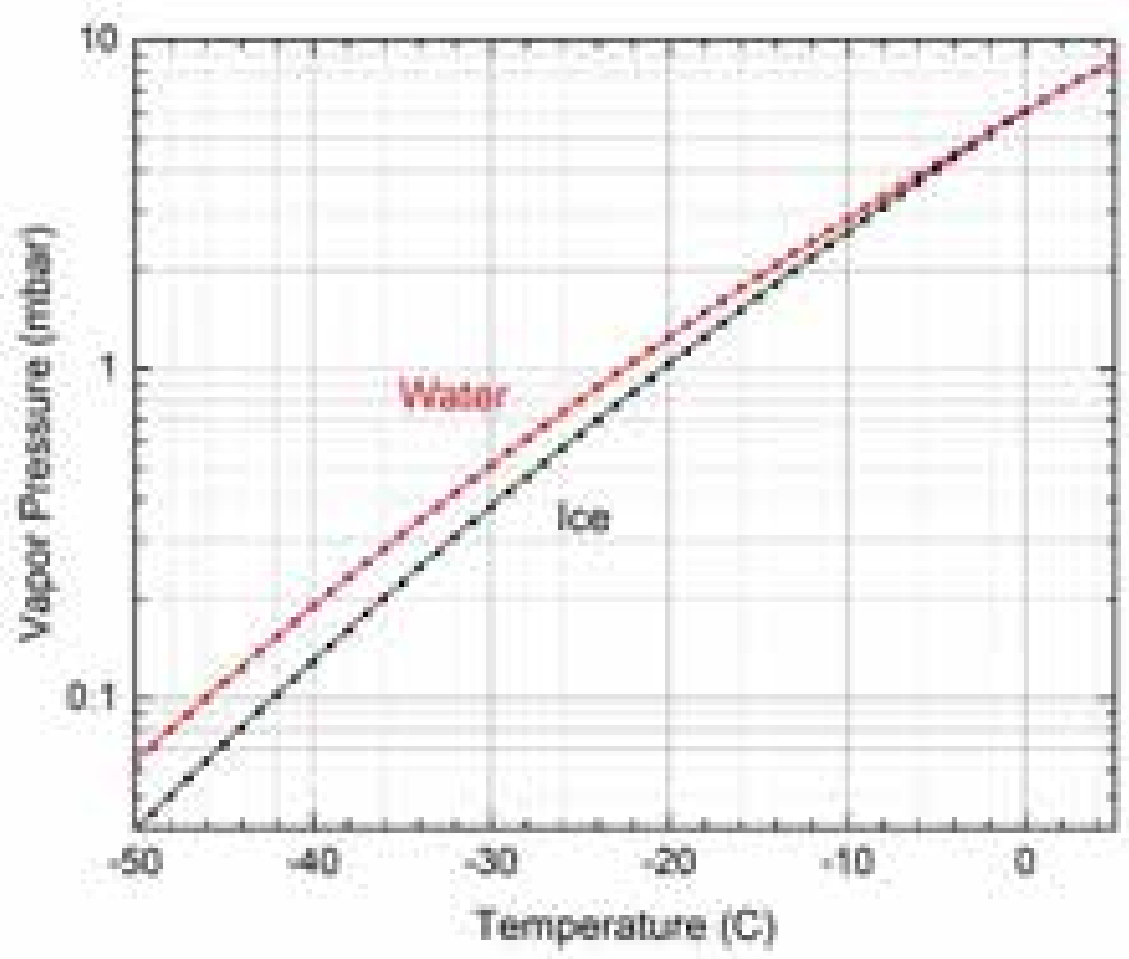

**Figure 2.13: The vapor pressure of ice and supercooled water as a function of temperature. The points are measured values, and lines show the approximate Arrhenius models described in the text.**

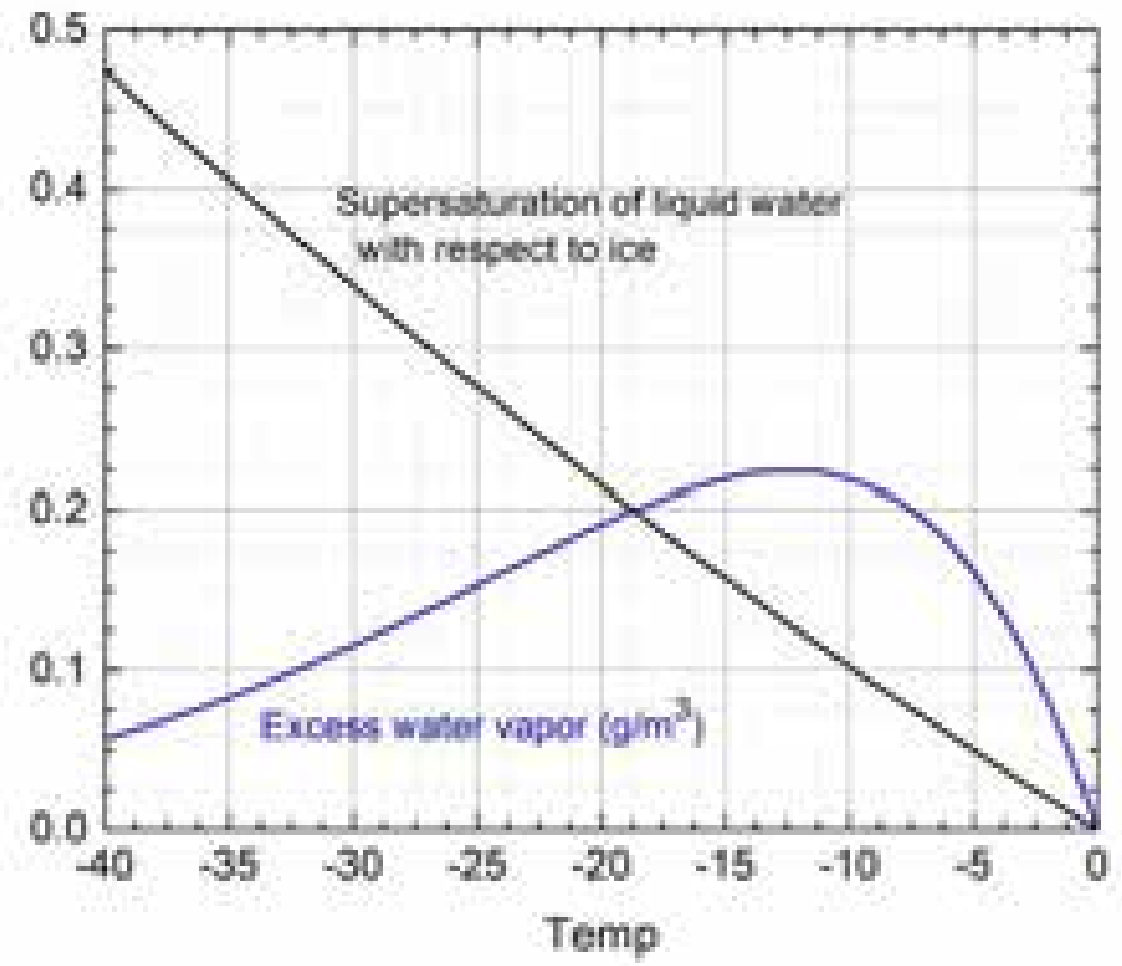

**Figure 2.14: The supersaturation of supercooled water with respect to ice, given by $\boldsymbol{\sigma_{water}(T) = [c_{water}(T) - c_{sat}(T)]/c_{sat}(T)}$. This is also plotted as the "excess" water vapor, equal to $\boldsymbol{[c_{water}(T) - c_{sat}(T)]m_{water}}$.**

## Vapor Pressure and Related Quantities

The equilibrium (saturated) vapor pressure of ice and water can be written in the Arrhenius form

$$c_{sat} \approx C(T)\exp\left(-\frac{\ell}{kT}\right) \qquad (2.7)$$

where $\ell \approx Lm_{mol}$ is the latent heat per molecule and $C(T)$ is a weak function of temperature. Table 2.1 gives measurements of the vapor pressure of water and ice [1971Mas] along with other useful quantities, and Figure 2.13 shows the vapor pressure data along with the following Arrhenius curves (slightly modified to better fit the data):

$$p_{ice} \approx 3.7e10 \cdot \exp(-6150/T_K) \qquad (2.8)$$

$$p_{water} \approx (2.8e9 + 1700T_C^3) \cdot \exp(-5450/T_K)$$

where $T_K$ is the temperature in Kelvin and $T_C$ is the temperature in Celcius. Figure 2.14 shows the supersaturation of supercooled liquid water relative to ice

$$\sigma_{water} = \frac{c_{sat,water} - c_{sat,ice}}{c_{sat,ice}} \qquad (2.9)$$

along with the "excess" water vapor mass density plotted as a function of temperature

$$[c_{sat,water}(T) - c_{sat,ice}(T)]m_{water} \qquad (2.10)$$

The data in the table and these two plots are often useful for understanding the physics underlying snow crystal growth, and for estimating experimental quantities.

## Surface Energies

The surface energy, simply put, is the amount of energy needed to create an interface between two material phases. For example, breaking a piece of ice in two requires breaking the chemical bonds holding the ice together. The amount of energy needed to do this is proportional to the new surface area created during the break, and this defines the surface energy. If the material is intrinsically anisotropic, like a crystal lattice, then the surface energy could be anisotropic as well, depending on the angle of the surface relative to the crystal axes.

Because liquid water is an amorphous material, the water/vapor surface energy (also known as the surface tension of water) is isotropic, equal to

$$\gamma_{lv} \approx 76 \text{ mJ/m}^2 \tag{2.11}$$

near the triple point. This decreases with increasing temperature, dropping to $\gamma_{lv} \approx 72$ mJ/m$^2$ at 25 C and $\gamma_{lv} \approx 59$ mJ/m$^2$ at 100 C. The water/vapor surface energy is known to quite high accuracy, being determined from observations of the oscillation frequencies of liquid droplets.

The ice/water surface energy is best measured from the homogeneous nucleation of ice from supercooled water droplets as a function of temperature, yielding

$$\gamma_{sl} \approx 30 \pm 5 \text{ mJ/m}^2 \tag{2.12}$$

near the triple point. This number is somewhat model dependent, as it assumes a good understanding of nucleation theory together with extrapolations of measured properties of liquid water down to temperatures as low as -40 C. The uncertainty in $\gamma_{sl}$ given above is a rough estimate based on the data presented in [2015Ick]. The value of $\gamma_{sl}$ decreases with decreasing temperature, down to roughly $\gamma_{sl} \approx 20$ mJ/m$^2$ at temperatures near -40 C, again with considerable uncertainty [2015Ick].

The ice/water surface energy $\gamma_{sl}$ is likely nearly isotropic, although, to my knowledge, the dependence of $\gamma_{sl}$ on orientation angle relative to the crystal axes has not been directly measured. The basal surface is expected to have the lowest surface energy, and there we know that the terrace step energy (see below) is $\beta_{sl,basal} \approx 5.6 \times 10^{-1}$ J/m. Even if we pack steps as tightly as possible on the surface, with a spacing equal to the molecular size $a \approx 0.3$, this gives an additional surface energy

$$\begin{aligned} \Delta\gamma_{sl} &\approx \gamma_{rough} - \gamma_{sl,basal} \\ &\approx \frac{\beta_{sl,basal}}{a} \approx 2 \text{ mJ/m}^2 \end{aligned} \tag{2.13}$$

and this is likely an overestimate because it neglects step-step interactions that can lower the surface energy through surface relaxation. It appears likely, therefore, that the anisotropy in $\gamma_{sl}$ is no more than a few percent, which is comparable to the surface-energy-anisotropy of many metals and other simple crystalline materials.

The ice/vapor surface energy $\gamma_{sv}$ is more difficult to measure than either $\gamma_{lv}$ or $\gamma_{sl}$, plus it has not received as much experimental or theoretical attention. Surface wetting measurements have produced the best measurements of $\gamma_{sv}$ [1974Hob, 1999Pet], but these can be quite susceptible to surface contamination, so the measurements probably deserve a conservatively high uncertainty. Nevertheless, the measurements appear to be consistent with Antonow's relation, which states that

$$\begin{aligned} \gamma_{sv} &\approx \gamma_{sl} + \gamma_{lv} \\ &\approx 106 \pm 15 \text{ mJ/m}^2 \end{aligned} \tag{2.14}$$

near the triple point. This estimate for the measurement uncertainty is somewhat subjective, based on an examination of existing data from various sources. The ice/vapor surface energy $\gamma_{sv}$ is also likely nearly isotropic, as we discuss below. Surface-energy anisotropies are best determined from measurements of the equilibrium crystal shape, which have not yet been observed for ice in water or in vapor (discussed in Section 2.6 below).

### Terrace Step Energies

Terrace step energies factor into the rate of nucleation of new molecular terraces on faceted ice surfaces, which is one of the most important processes in snow crystal growth. In analogy to the surface energy, the step energy is the amount of energy needed to create the edge, or step, of a molecular terrace on a crystal facet. For example, separating a single island terrace into two smaller islands requires energy to break the molecular bonds holding the terrace edges together. The amount of separation energy needed is proportional to the length of new terrace step created, and this defines the terrace step energy. As with the bulk energies and surface energies, the step energies are equilibrium properties of the material.

For ice surfaces in water, only the basal step energy has been accurately measured from ice-growth measurements [1958Hil, 1966Mic, 2014Lib], yielding $\beta_{sl,basal} \approx 5.6 \times 10^{-13}$ J/m near the triple point. This step energy provides a substantial nucleation barrier that results in basal faceting and the formation of thin, plate-like ice crystals when liquid water freezes in an unconstrained fashion at low supercooling (see the pond crystals described in Chapter 10). The Hillig measurement [1958Hil] was especially well crafted, with a careful examination of possible systematic errors, and it has not been improved upon (in my opinion) by any subsequent measurement.

Prism faceting is not observed in ice grown from liquid water near the triple point, and $\beta_{sl,prism}$ has not yet been measured, being much smaller than $\beta_{sl,basal}$. However, prism faceting has been observed at very high pressures [2005Mar], when the ice/water phase transition occurs near -20 C (see Figure 2.1), suggesting that $\beta_{sl,prism}$ becomes non-negligible in that region of the phase diagram.

The ice/vapor step energies on both the basal and prism facets have been measured with good accuracy at temperatures ranging from -2 C to -40 C, yielding the results shown in Figure 2.15. Whether these data survive the test of time as well as the Hillig number remains to be seen, as the potential for systematic errors is significant [2013Lib]. Here again, these measurements were made using nucleation-limited ice-growth measurements, inferring the step energies using classical nucleation theory, as described in Chapter 4. The central importance of the step energies (and, more generally, attachment kinetics) in snow crystal growth cannot be overstated, and this topic is examined in much detail throughout this book.

In snow crystal growth, prism faceting is greatly diminished above -2 C, and this has been interpreted as a roughening transition on the prism surface [1991Elb]. Given the step energy measurements shown in Figure 2.15, we see that this phenomenon is better interpreted as a gradual reduction, not an abrupt transition, in $\beta_{prism}$ with increasing temperature. At what point the faceting disappears completely will likely depend on $\beta_{prism}$ and other growth factors, such as the supersaturation at the growing surface.

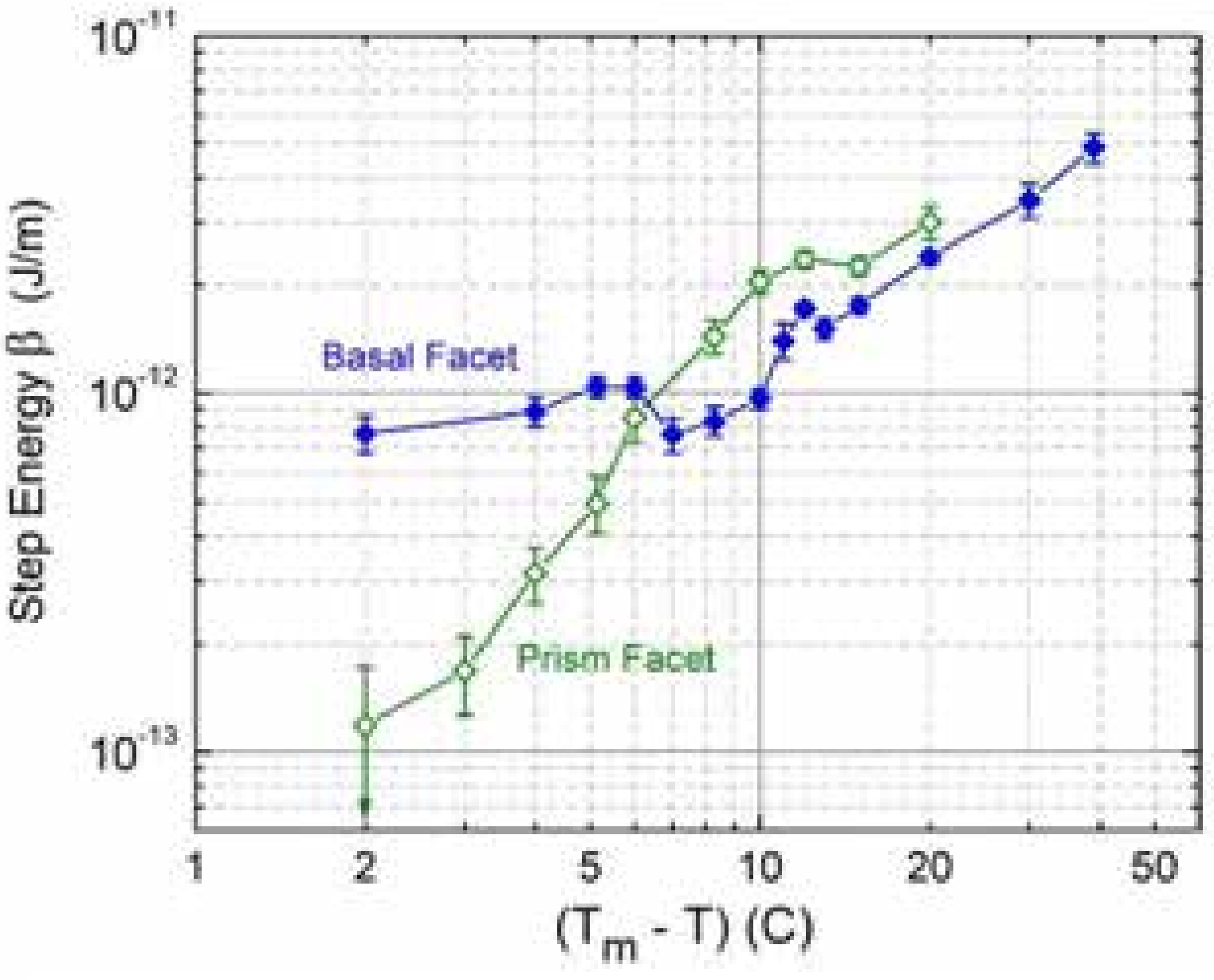

**Figure 2.15: Measured ice/vapor terrace step energies (in near-vacuum conditions) for the basal and prism facets as a function of temperature [2013Lib].**

Although we have confined our discussion in this section to the basal and prism step energies, the appearance of pyramidal facets at low temperatures suggests that the step energy of a pyramidal terrace becomes significant in this regime. If this step energy increases with decreasing temperature like the basal and prism facets, as is shown in Figure 2.15, then this might explain why the pyramidal facets are not more prevalent at higher temperatures.

The data shown in Figure 2.15 refer to the step energies in near-vacuum conditions, when the ice/vapor interface is unfettered by possible interactions with any additional background gases. As we will see in Chapters 3 and 7, the experimental evidence suggests that the attachment kinetics on faceted prism surfaces are significantly altered in the presence of air at normal atmospheric pressures. This may mean that the step energy on this facet is also changed in the presence of air, although there is no obvious mechanism that would bring about such a change. At present, our understanding of the ice surface structure is not sufficient to explain these observations, so the mystery remains for another day.

## A Correspondence near 0 C

It is no coincidence that $\beta_{sl,basal}$ on the ice/water interface (as shown in Figure 2.15) is quite similar to $\beta_{sv,basal}$ on the ice/vapor interface as the temperature approaches 0 C [2014Lib]. In the presence of extensive surface premelting, the ice/vapor surface can be approximated as two nearly separate interfaces: an ice/QLL interface and a QLL/vapor interface, as sketched in Figure 2.16. The QLL/vapor interface is expected to behave like a water/vapor interface, with no nucleation barrier and no crystalline step energy. As a result, the location of the ice/vapor step energy must be at the ice/QLL interface. Because the properties of the QLL are expected to approach that of bulk liquid water as the QLL thickness diverges, we expect that the ice/vapor step energy should approach the ice/water step energy as the temperature approaches 0 C. This correspondence between the ice/water and ice/vapor step energies near the triple point also applies to the prism surface, where both values become very small (perhaps zero) as the temperature approaches the melting point.

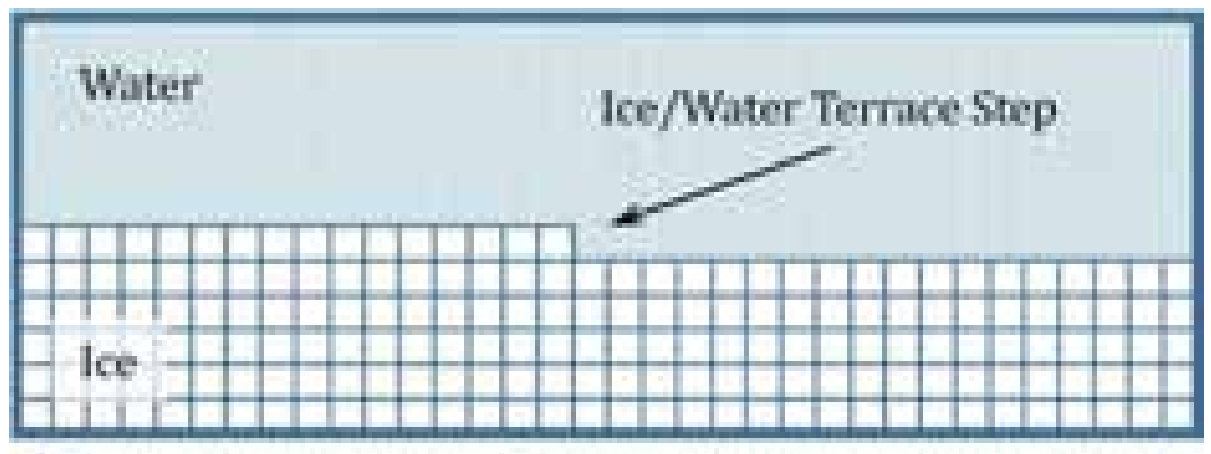

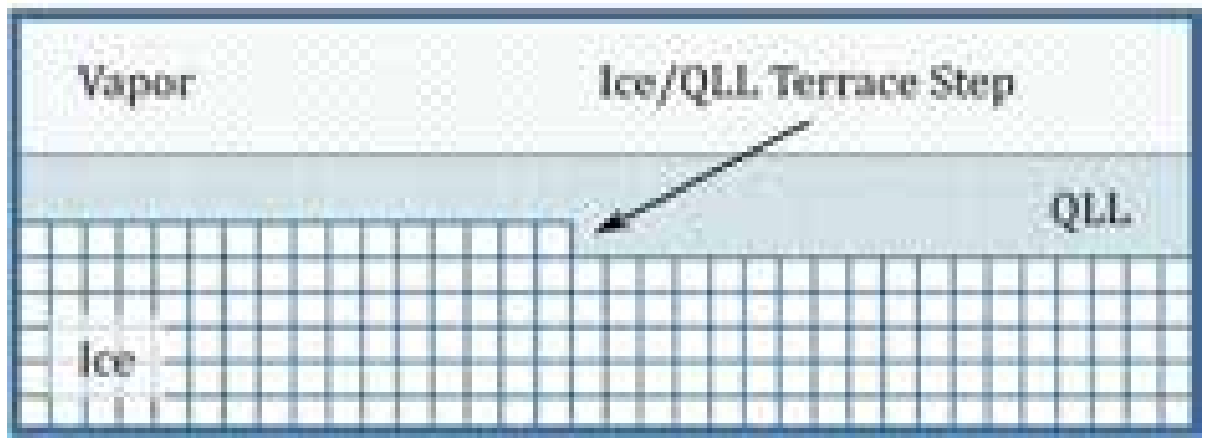

**Figure 2.16: At temperatures just below the melting point, surface premelting turns the ice/vapor interface into an ice/QLL/vapor interface, where crystal terrace steps are localized essentially at the ice/QLL interface (top). Inasmuch as the QLL resembles bulk water in its properties, the ice/QLL step energy should nearly equal the ice/water step energy (bottom). As discussed in the text, step-energy measurements at the ice/water and ice vapor interfaces do seem to show a good correspondence as the temperature approaches 0 C.**

## Surface Relaxation

It is useful to examine the relationships that exist between bulk, surface, and step energies, and what the numbers suggest about surface relaxation processes. For example, in a naïve chemical bond picture, one would calculate an ice/vapor surface energy of approximately

$$\gamma_{0,sv} \approx \frac{1}{6} a \rho_{ice} L_{sv} \approx 130 \ \mathrm{mJ/m^2} \qquad (2.15)$$

which follows because (naïvely) vaporizing bulk ice creates a total surface area of $6a^2$ per molecule. Although the reasoning is crude, it does yield a remarkably accurate estimate.

Moreover, one expects that the actual value of $\gamma_{sv}$ would be lower than $\gamma_{0,sv}$ because of surface relaxation. Slicing a crystal in two without making any adjustments to the crystal lattice at the surface yields $\gamma_{0,sv}$. However, once such a cut has been made, the surface molecules can rearrange themselves, pointing their exposed bonds inward to some degree, relaxing the system to a lower total energy. The result is that $\gamma_{0,sv}$ provides an approximate upper bound on $\gamma_{sv}$.

Looking at this another way, creating a surface gives the surface molecules more freedom to move about, as they are no longer constrained as tightly as in the bulk lattice. With this additional freedom, the molecules can "spread out" into the space above the cut, resulting in a density profile that is not a step function, but rather exhibits a more gradual decline, and this behavior is seen in molecule-dynamics simulations [2018Moh].

Turning to step energies, we see that creating an abrupt terrace edge of length $\ell$ with no surface relaxation (again in the naïve rigid-lattice picture) creates an additional surface area equal to $\ell a$, yielding an "unrelaxed" step energy

$$\beta_0 \approx a\gamma_{sv} \approx 3 \times 10^{-11} \text{ J/m} \tag{2.16}$$

Once again, surface relaxation reduces $\beta_0$ to the observed $\beta_{sv}$ shown in Figure 2.15. Unlike the surface energy $\gamma_{0,sv}$, however, here we see that surface relaxation results in a sizable reduction in the step energy, as $\beta_{sv}$ is much smaller than $\beta_0$, especially at higher temperatures.

One way to think about surface relaxation of step energies is with the simple geometrical model shown in Figure 2.17. Replacing the abrupt step by a gradual step of width $w$, we see that the additional surface area depends on $w$, reducing to zero as $w \to \infty$. In this picture, the surface relaxation that reduces the step-energy from $\beta_0$ to $\beta_{sv}$ is equated with a "softening" (increased width) of the terrace step. Although this geometrical picture is likely of little use in a quantitative analysis of the surface structure, it might be helpful for guiding one's thinking about terrace step energies. For example, this geometrical picture suggests that accurately calculating the step energy using a numerical crystal-lattice-relaxation technique will likely require a physical model size of order $(\beta_0/\beta_{sv})a$, which is quite large for the smaller step energies seen in Figure 2.15.

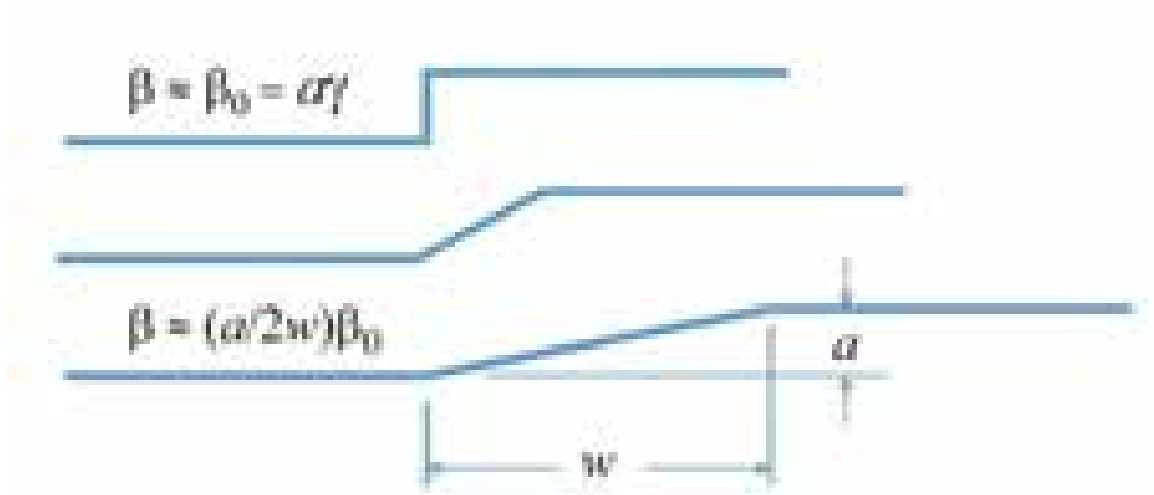

**Figure 2.17: A simple geometrical model describing how surface relaxation reduces the terrace step energy. In this picture, the additional surface area added by a terrace step depends on the step width, and this additional surface area is taken to be the step energy.**

It is likely that the observed change in $\beta_{sv}$ with temperature is related to structural changes associated with surface premelting. At the lowest temperatures in Figure 2.15, surface premelting is essentially absent, and the ice lattice becomes so rigid that there is little surface relaxation of the step energy. This would explain why $\beta_{sv}$ tends toward $\beta_0$ with decreasing temperature. At the other extreme, near the melting point, the ice/vapor interface becomes an ice/QLL/vapor interface, and the ice/QLL step energy tends toward the ice/water step energy as the temperature approaches 0 C.

From this reasoning, we see that the high- and low-temperature limits seen in Figure 2.15 form a reasonably self-consistent physical picture, at least qualitatively. Of course, understanding the full form of $\beta_{sv}(T)$ on the basal and prism surfaces remains on ongoing challenge, and there is certainly room for greater theoretical understanding of terrace step energies on faceted ice surfaces.

## 2.5 Gibbs-Thomson Effect

There are two ways to see that the vapor pressure of a convex ice surface is slightly higher than that of a flat surface. In terms of molecular attachments, a molecule on a convex surface is not adjacent to as many other molecules as one on a flat surface, simply from geometrical considerations. The convex-surface molecule is thus less tightly bound, resulting in a higher vapor pressure compared to a flat surface. This picture helps in visualizing the Gibbs-Thomson phenomenon, but it is difficult to quantify without precisely adding up all the binding energies.

The second approach involves including the surface energy in a calculation of the vapor pressure. For an ice sphere of radius $R$, pulling one molecule off the sphere reduces its surface area by an amount $\delta A = 2/c_{ice}R$, as the size of the sphere is reduced slightly with the loss of one molecule. This, in turn, results in a reduction in the surface energy by an amount $\delta E = 2\gamma_{sv}/c_{ice}R$, where $\gamma_{sv}$ is the ice/vapor surface energy.

Including this additional energy term in the Arrhenius equation (Equation 2.1, see also Appendix B) gives the modified equilibrium vapor pressure

$$\begin{aligned} c_{eq}(R) &\approx C(T)\exp\left(-\frac{\ell - \delta E}{kT}\right) \\ &\approx c_{sat}(1 + d_{sv}\kappa) \end{aligned} \qquad (2.17)$$

where $d_{sv} = \gamma_{sv}/c_{ice}kT \approx 1\ nm$ and $\kappa = 2/R$ is the curvature of the spherical surface. For a smooth but non-spherical surface, the curvature is defined as $\kappa = 1/R_1 + 1/R_2$, where $R_1$ and $R_2$ are the two principal radii of curvature of the surface. Note that this expression reduces to the normal flat-surface vapor pressure $c_{sat}$ when $R \to \infty$, as it must. The analysis is more complicated when $\gamma_{sv}$ is anisotropic, but the functional for is generally similar to the above expression as long as the surface-energy anisotropy is small, as it is for ice. Replacing $\kappa = 2/R$ with $\kappa = 2G/L$, where $L$ is the size of the crystal and $G$ is a geometrical factor of order unity, gives a reasonable approximation for most simple shapes.

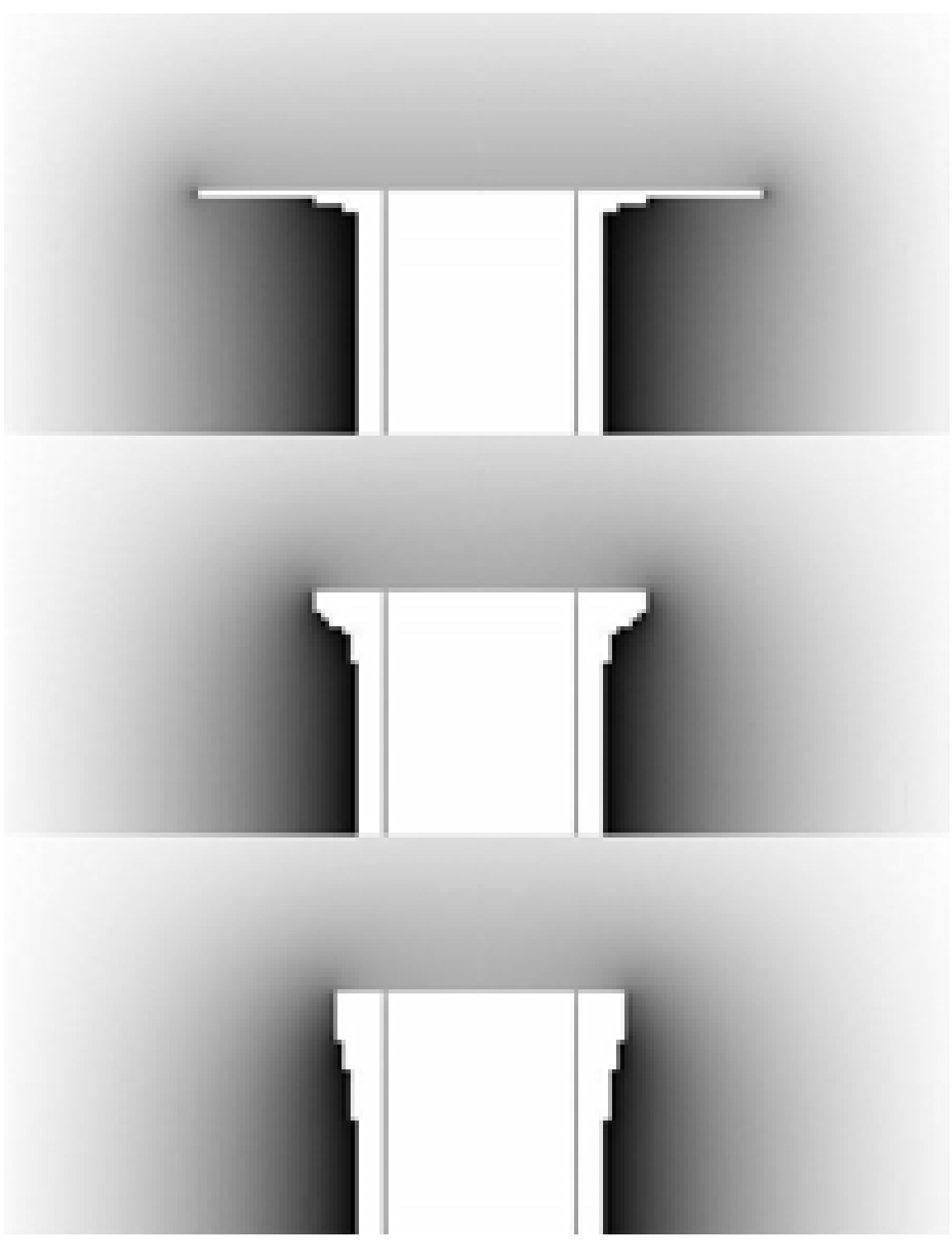

**Figure 2.18: Three examples modeling the growth of a thin snow-crystal plate on the end of a columnar crystal (here shown in side view) [2013Lib1]. Setting $d_{sv} = 0$ (top model) yields an unrealistic 0.1-micron-thick plate growing out from the column when the applied supersaturation is low. Using a realistic value of $d_{sv}$ (bottom model), the plate-growth is suppressed. Without including the Gibbs-Thomson effect, snow-crystal models would not give physically realistic results at exceedingly low supersaturations. In many environmental circumstances, however, the Gibbs-Thomson effect is so small that it is essentially negligible.**

This additional vapor pressure can have a significant effect on snow crystal growth, especially when the supersaturation is low and the surface curvature is high. For example, taking $R = 1\ \mu m$ gives a change in the effective supersaturation of $\Delta\sigma = 2d_{sv}/R \approx 0.1$ %. Under many normal circumstances in snow crystal growth, this small perturbation of the supersaturation is negligible. However, modeling reveals that the Gibbs-Thomson effect plays a large role in preventing the growth of thin plates at especially low supersaturations, as demonstrated in Figure 2.18. More about how the Gibbs-Thomson effect factors into snow-crystal modeling will be presented in Chapter 5.

## 2.6 Equilibrium Shape

The equilibrium crystal shape (ECS) of an isolated snow crystal is the shape that minimizes the total surface energy at constant volume. For a perfectly isotropic surface energy, the ECS is a perfect sphere, as this shape minimizes the surface area and thus minimizes the surface energy. If $\gamma_{facet} < \gamma_{unfaceted}$, however, then it becomes energetically favorable to increase the facet surface areas relative to the unfaceted areas. If $\gamma_{facet}$ is only slightly smaller than $\gamma_{unfaceted}$, then the ECS becomes essentially a spherical shape with small faceted "dimples". The dimples become larger with increasing anisotropy, and, when the surface-energy anisotropy is sufficiently high, the ECS becomes a fully faceted prism.

To see how much surface-energy anisotropy is needed to produce a fully faceted ECS, it is useful to consider the simple 2D examples shown in Figure 2.19, assuming only two possible surface energies: $\gamma_{facet}$ on the square facets and $\gamma_{unfaceted}$ on all other surfaces. (These are actually perimeter energies in 2D, but for clarity of notation we will continue to call them surface energies.) It is straightforward to show that the square has a lower total surface energy than its inscribed circle if we have $\gamma_{facet} < (\pi/4)\gamma_{unfaceted} = 0.79\gamma_{unfaceted}$. Similarly, the hexagonal shape in Figure 2.19 has a lower surface energy than its inscribed circle if $\gamma_{facet} < (\pi\sqrt{3}/6)\gamma_{unfaceted} = 0.91\gamma_{unfaceted}$. Extending this calculation (see Appendix B), one can show that the corners of square or hexagonal equilibrium shapes will not exhibit any rounding if the same inequalities hold.

Although this 2D exercise is far from adequate for describing 3D equilibrium shapes, it does tell us that we can expect an ECS that is fully faceted, or nearly so, if the facet surface energy is of order 10-20% lower than the (maximal) surface energy of unfaceted surfaces. If the surface-energy anisotropy is less than 10-20%, the ECS will be a faceted prism with rounded corners, and the degree of rounding will depend on the amount of the anisotropy. More precise ECS calculations can be done using the Wolff construction, and there is much discussion of this in the scientific literature and in textbooks on crystal growth [1996Sai, 2004Mar]. Figure 2.20 shows on example of a 2D Wolff construction, comparing the ECS with the corresponding surface-energy anisotropy.

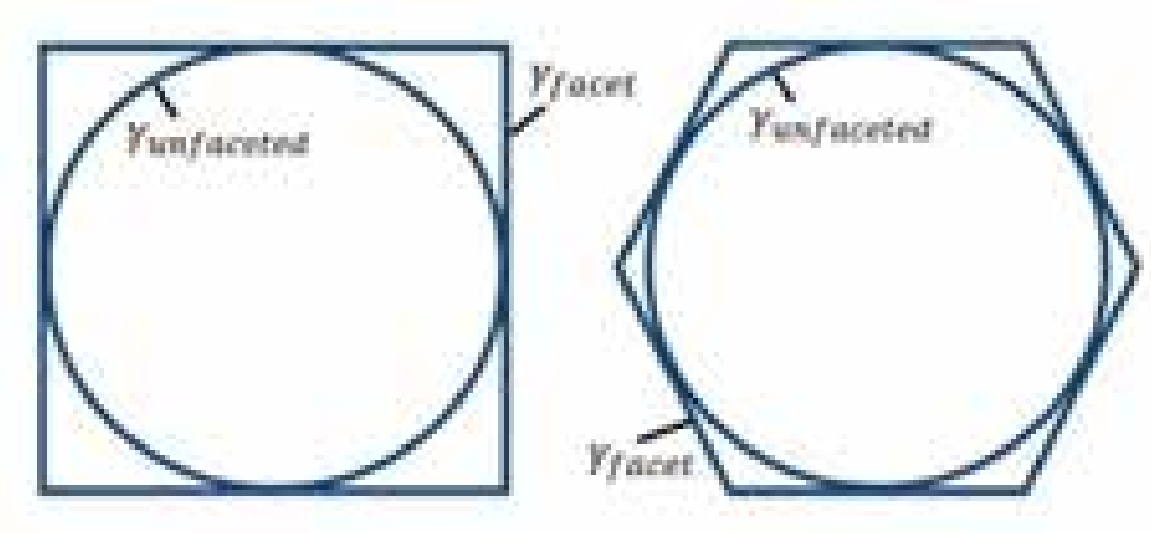

**Figure 2.19: Basic limits on equilibrium shapes in 2D. The square shape has a lower surface energy than its inscribed circle if $\gamma_{facet} < (\pi/4)\gamma_{unfaceted} = 0.79\gamma_{unfaceted}$, and the hexagon has a lower surface energy if $\gamma_{facet} < (\pi\sqrt{3}/6)\gamma_{unfaceted} = 0.91\gamma_{unfaceted}$.**

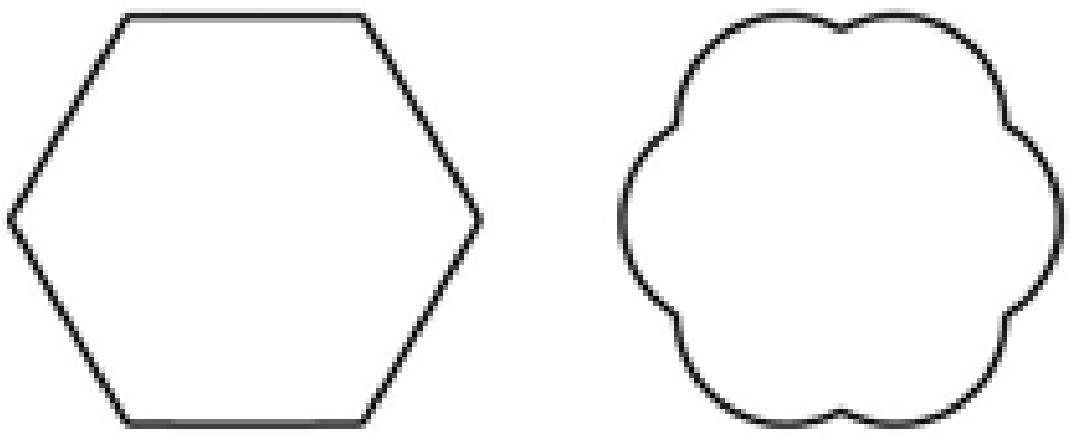

**Figure 2.20: An example plot showing the ECS (left) together with a polar plot of the surface energy (right). These diagrams may not be representative of ice, as the ECS has not been definitively measured. The evidence suggests that the ice ECS is nearly spherical above -10 C, but may become somewhat or even fully faceted at lower temperatures.**

The ice/vapor ECS is sometimes reported to be fully faceted at temperature below -10 C, while exhibiting basal facets but no prism facets at higher temperatures [1997Pru], as this was the conclusion of the sole experimental investigation of the subject [1985Col]. However, the time needed to reach full equilibrium is so long (see below) that the true ice/vapor ECS has probably not been definitely observed. Snow-crystal growth forms are often strongly faceted, as are "negative" ice crystals (ice bubbles formed by evacuation [1965Kni, 1993Fur]; see Chapter 6), but these are both far from equilibrium.

One substantial, albeit circumstantial, piece of experimental evidence in favor of a non-faceted ECS is the apparent absence of any observations of prismatic, or even slightly faceted, bubbles in ice [2010Dad, 2016Feg]. The ECS of a equilibrated bubble in ice would be the same as the snow-crystal ECS, and researchers have been examining ice bubbles for decades, especially in glaciers and ancient ice cores. If the bubble ECS was faceted, one expects that there would be at least some photographic evidence to support this. Unfortunately, bubbles in natural ice are usually not particularly spherical either, so the conclusions one can draw from bubble observations regarding the ice ECS are limited.

A more quantitative piece of evidence can be gleaned from the measured step energies in Figure 2.15. Neglecting step-step interactions, the half-angle of a faceted ECS dimple is approximately [2001Bon]

$$\theta \approx \frac{\beta}{a\gamma_{facet}} \qquad (2.18)$$

in the limit of small $\theta$. Taking $\beta \approx 10^{-12}$ J/m from Figure 2.15 and $\gamma_{facet} \approx 0.1$ J/m$^2$ gives $\theta \approx 0.03$ radians, which is quite a tiny faceted dimple [2012Lib2]. The data in Figure 2.15 suggest, however, that the ice/vapor ECS might become faceted at lower temperatures, at least to some degree. The question of the ice/vapor ECS will likely only be definitively answered with additional direct observations.

## Approach to Equilibrium

It takes a substantial amount of time for an ice crystal in an initially arbitrary shape to reach its equilibrium shape, as this involves the transport of molecules between different regions on the crystal surface. Observations of bubble migration in ice [2010Dad] indicate that vapor transport (and not other mechanisms, such as surface migration) is the dominant path to equilibration, so the equilibration time can be estimated using the Gibbs-Thomson effect. Beginning with a slightly non-spherical shape of overall radius $R$, the equilibration time in air is approximately (see Appendix B)

$$\tau_{eq} \approx \frac{c_{ice}R^3}{2c_{sat}d_{sv}D} \qquad (2.19)$$

which is equal to about eight days for $R = 50\ \mu m$.

Lowering the pressure $P$ increases the diffusion constant as $D \sim P^{-1}$, until the vapor transport becomes limited by attachment kinetics instead of diffusion (see Chapter 3). In the kinetics-limited case, the equilibration time becomes

$$\tau_{eq} \approx \frac{R^2}{2\alpha v_{kin} d_{sv}} \qquad (2.20)$$

which becomes a more-favorable 2 hours for $R = 50\ \mu m$ and using representative parameters at -15 C. The result changes somewhat for faceted or other complex shapes, but replacing $R$ with the overall crystal size gives a good approximation to the equilibration time for most circumstances. In the kinetics-limited case, $\alpha$ can become extremely small on faceted surfaces, which may be problematic in some experimental circumstances. In nearly all snow-crystal growth scenarios, the equilibration times above are much longer than typical growth times $\tau_{growt} \approx R/v$. This fact suggests that surface-energy effects are less important than effects from attachment kinetics in most snow crystal growth circumstances (except at extremely low supersaturations, as was mentioned above).

## 2.7 Twinning

When two single crystals grow together with a specific orientation between their respective lattices, this is called crystal *twinning*. Many mineral crystals exhibit twinning in various forms, and ice is no exception. Two questions immediately arise with crystal twinning: 1) what defines the orientation between the twin crystals? and 2) what circumstances bring about the different twinned states?

The first of these questions is usually the easier to answer, as it involves statics and energetics. A twinned state is typically a metastable state: a local energy minimum that the crystal structure fell into during the early phases of its early growth, from which it cannot reach the lowest-energy configuration. The second question is one of dynamics, and therefore more difficult to answer, involving how the crystal's nucleation and growth history happened to produce a twinned state. Estimating the probability that a particular twinned state will form under different circumstances is an extremely challenging task. Our goal here will be relatively modest: simply to report on some examples of snow-crystal twins and try to explain their structures as best we can. Although certainly not a well-studied topic, snow-crystal twinning has been discussed in the scientific literature over many decades [1971Iwa, 1978Fur, 1987Kob, 2011Kik, 2013Kik].

### Columnar Twins

Figure 2.21 shows a typical example of a columnar twin crystal. This is the most common form of twinning found in natural snow crystals, and the most easily explained from ice crystallography. If you look carefully when columnar crystals are falling near -5 C, you are likely to find some twin columns in the mix. As shown in Figure 2.22, the two columns fit perfectly together at the twin plane by simply replacing the usual hexagonal bonds with a plane of cubic bonds. This can be nicely demonstrated using a 3D molecular model as well.

A columnar twin plane likely originates when the initial crystal nucleation (typically from a liquid water droplet) produces some stacking disorder and at least one plane of cubic bonds. Nucleation can be a somewhat violent event, as the supercooling just prior to nucleation is often quite high. Thus, it is perhaps not surprising that some lattice disorder can arise during this phase. The initial ice growth soon warms the liquid, reducing its supercooling, so the subsequent growth quickly settles into the energetically favorable ice Ih configuration. By the time the initial droplet is all frozen and growth from water vapor commences, two normal columns are growing out from the nucleation site.

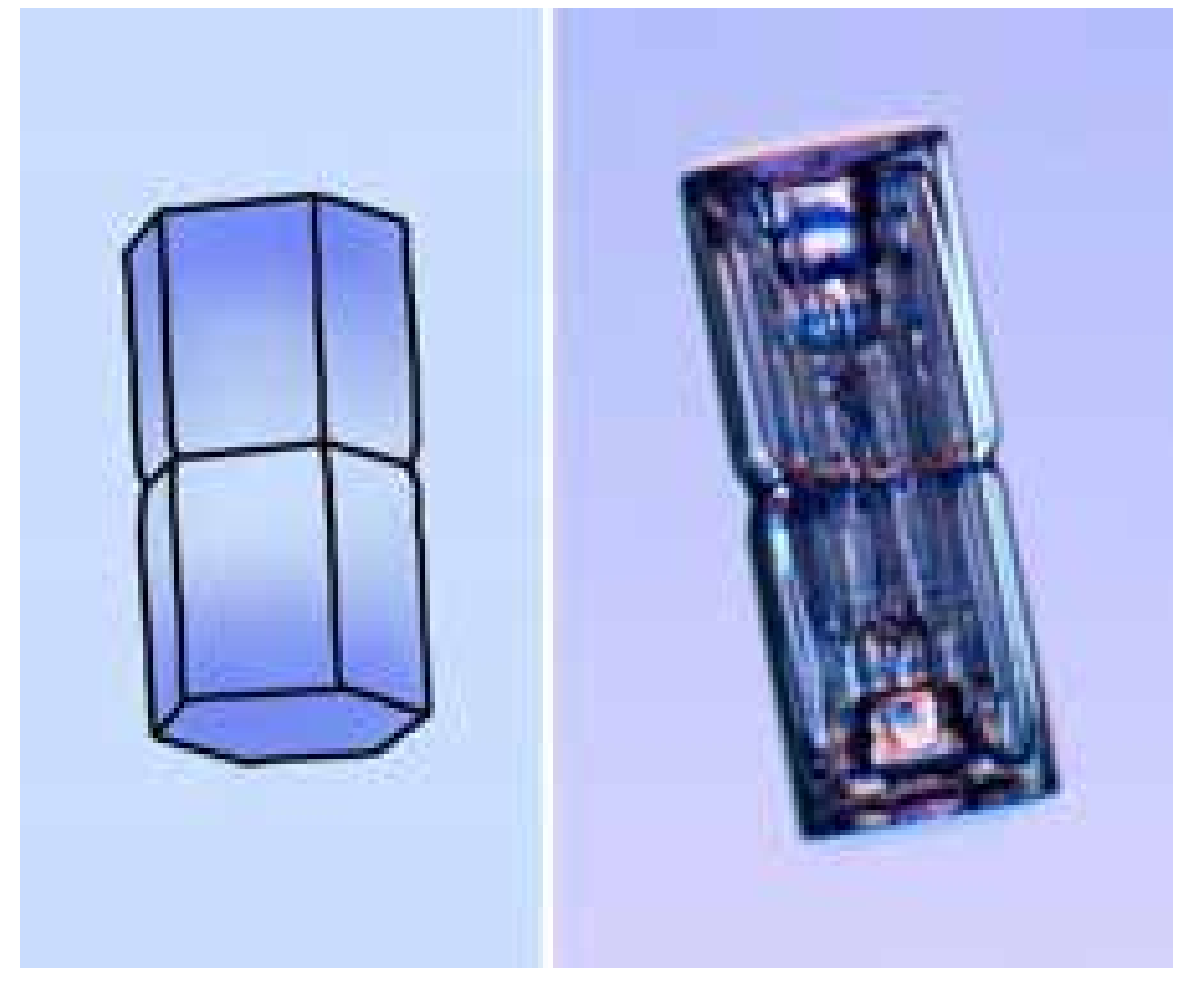

**Figure 2.21: Columnar Twins.**
**A sketch (left) and photograph (right) of a columnar twin snow crystal. These are essentially two ordinary columnar crystals connected by a *twin plane* between them. The crystal structure is relatively weaker in the twin plane, so sublimation often produces an *evaporation groove* between the two columns. The presence of the evaporation groove identifies this as a twinned column, which otherwise looks just like a normal single-crystal column.**

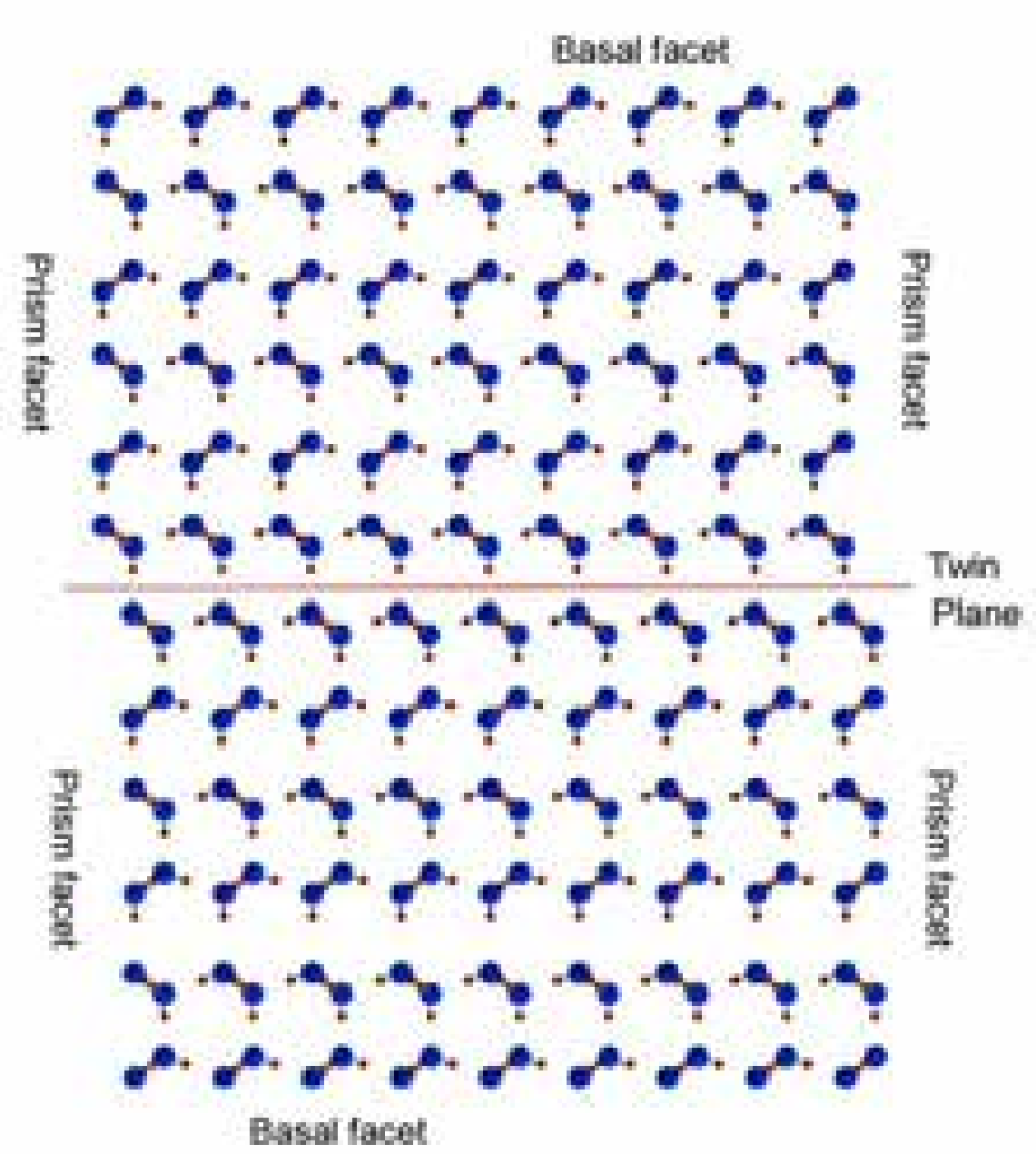

**Figure 2.22: Columnar Twins.**
**The crystal-lattice structure of a columnar twin crystal. Comparing this with Figure 2.4, it becomes apparent that the twin plane consists of a layer "cubic" bonds between water molecules, whereas the rest of the structure is made from the typical "hexagonal" bonds. Note also the small offset between the columns, necessary to accommodate the cubic bond structure. The twin plane could consist of multiple, randomly arranged layers of cubic/hexagonal bonds (i.e., a stacking-disordered region), rather than the single cubic layer shown here.**

Because ice Ih is energetically favorable over ice Ic, the cubic bond plane comes with an energy cost, and this slightly increases the vapor pressure at the twin plane. As a result, when the crystal begins slowly sublimating away (as usually happens after it falls from the clouds, or when it is being photographed), the twin plane sublimates faster than the surrounding prism facets, yielding an *evaporation groove* that appears like a belt around the columns at the twin plane, as shown in Figure 2.23. An evaporation groove is a characteristic marking that identifies a columnar twin; otherwise, the twinned column is essentially indistinguishable from a normal columnar snow crystal.

During the nucleation and growth of a columnar crystal, the most likely scenario is that no twin planes form. Thus, most columns are un-twinned single crystals. The next most likely scenario is that there is some stacking disorder during nucleation, resulting in a twin plane. Perhaps this plane containing a single, clean, cubic layer, as shown in Figure 2.22, but it may contain several randomly stacked layers. In any case, the situation eventually sorts itself out, yielding two normal columns on either side of the twin plane.

Because both columns grow outward at about the same rate, the evaporation groove is usually near the midpoint of the twinned column. However, one column may grow a bit faster than the other, so the twin symmetry need not be perfect in every case. Creating more than one twin plane in a single structure would be quite unlikely, and I know of no photographic examples of a simple hexagonal column exhibiting multiple evaporation grooves.

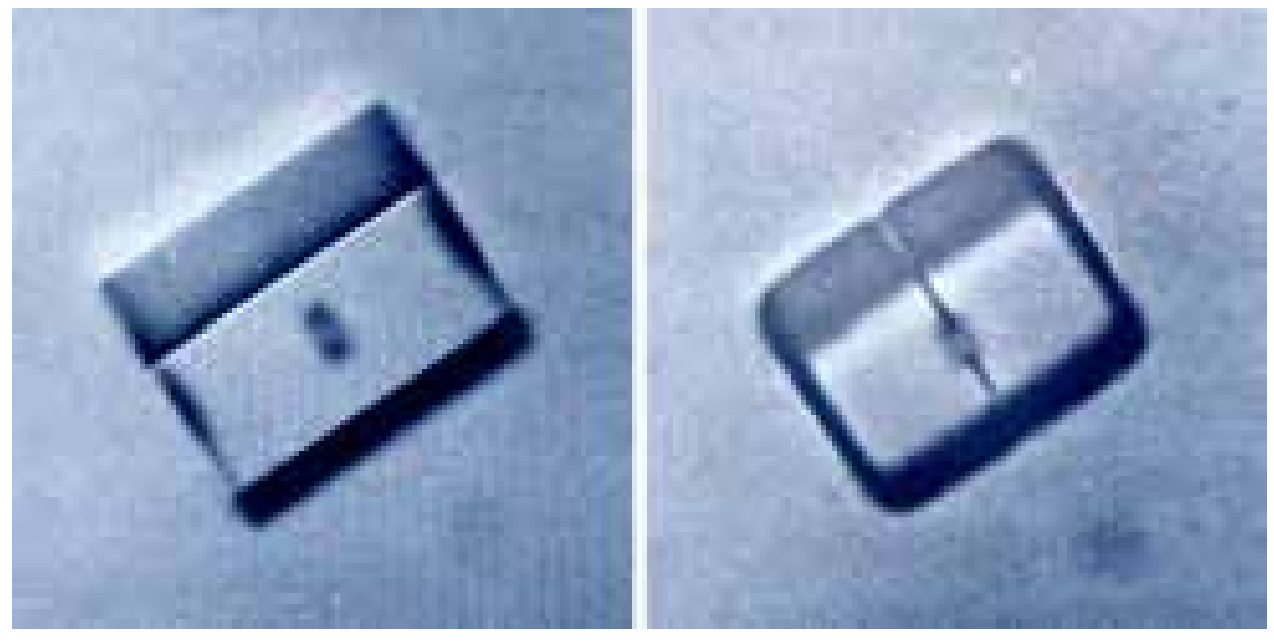

**Figure 2.23: Columnar Twins.**
**When a columnar twin is growing (left), the facet corners are sharp and the twin plane is not readily apparent. When it begins to sublimate (right), the corners become rounded and the evaporation groove deepens. Image from [1987Kob].**

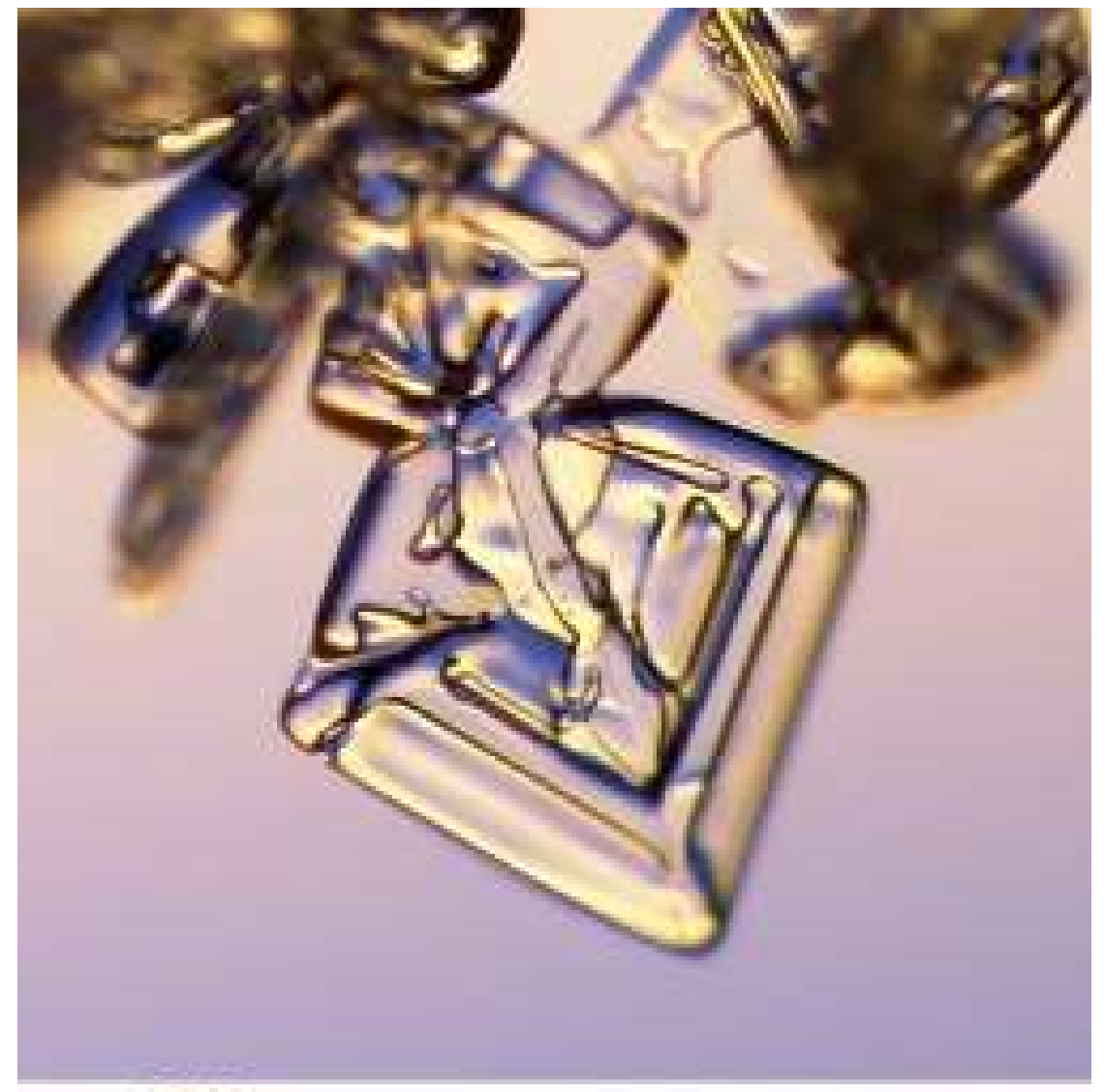

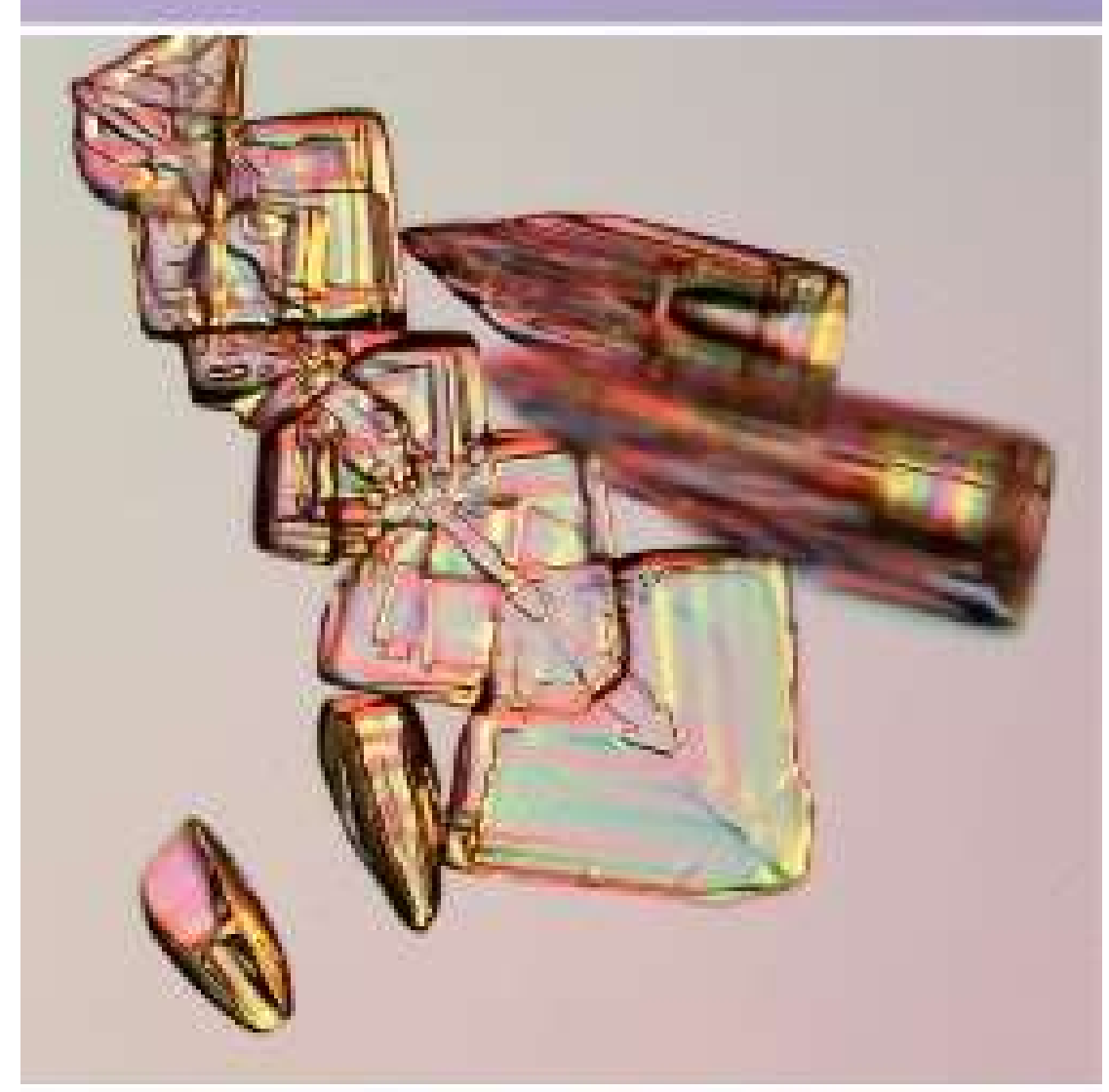

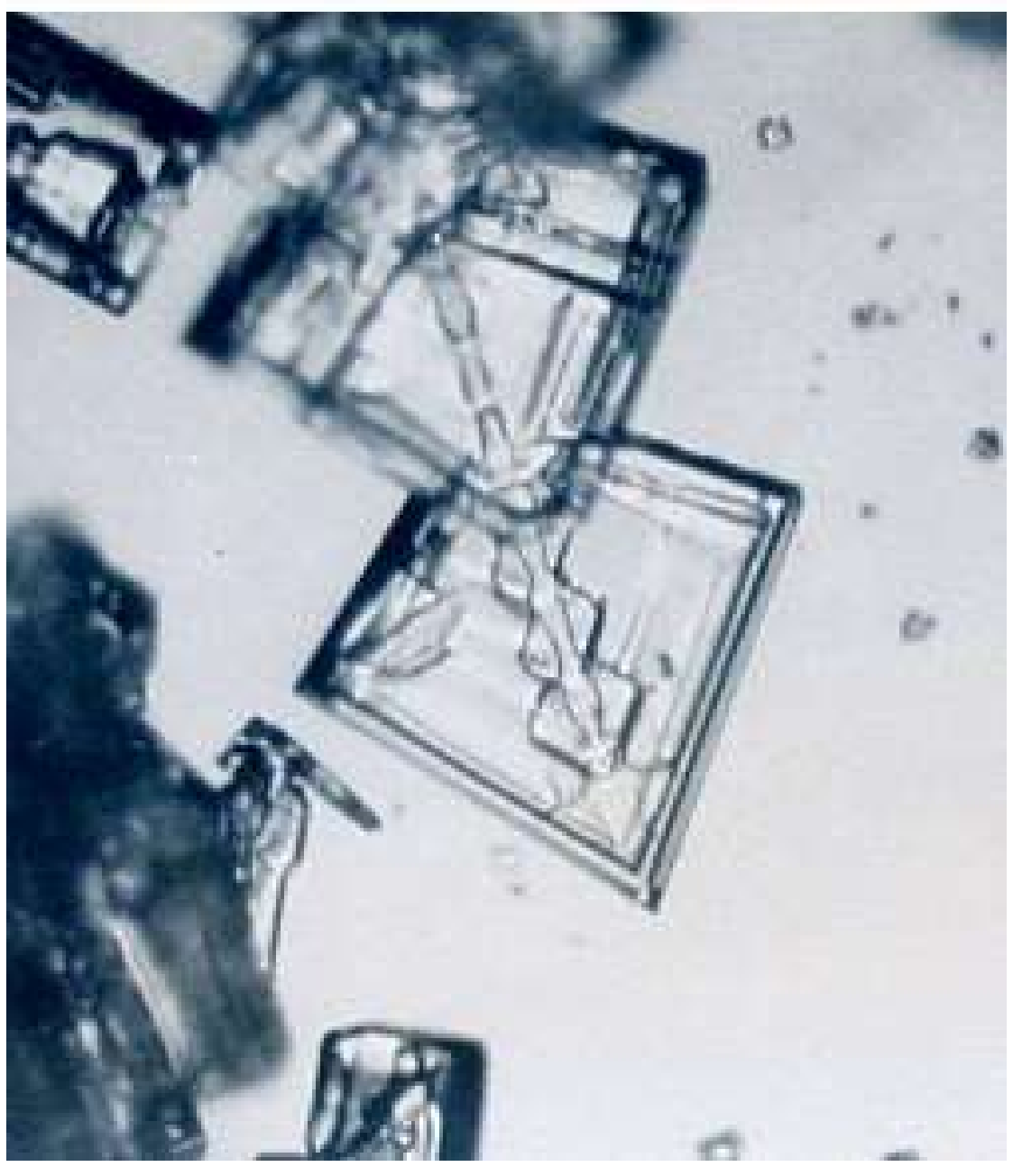

**Figure 2.24: Arrowhead Twins Variant I.**
**Left: These three photos show examples of arrowhead twin snow crystals with 78-degree apex angles, from [2003Lib2] (top two photos) and [2006Tap] (bottom photo). These examples were all found falling together will hollow columns, indicating that these twinned crystals grew at temperatures near -5 C.**

## Arrowhead Twins

Figure 2.24 shows several photographs of one variant of *arrowhead* snow-crystal twins, and the sketch in Figure 2.25 describes their faceted structure. The crystal surface consists mainly of fast growing faceted basal edges, slow growing prism edges, and slow growing prism faces. With this geometry, the basal edges flanking the apex experience the fastest growth, extending the apex forward. By comparison, the prism surfaces accumulate water vapor very slowly.

Figure 2.26 shows a crystal lattice model that explains the structure of this arrowhead twin variant, giving a theoretical apex angle of $2\tan^{-1}(c_0/2a_0) = 78.3$ degrees, in good agreement with observations. Arrowhead twins tend to be one-sided, in that a single apex emerges from some messy polycrystalline nucleus. To my knowledge, there are no known photographic examples of a "clean" arrowhead twin arising from a simple initial state, in contrast to columnar twins.

Arrowhead twins are generally quite small (seldom more than a millimeter in size), appearing in warm snowfalls alongside hollow columns. A good way to spot them is to let crystals fall onto some glass slides for a few minutes, and then scan the slides using a low-power microscope. Arrowhead twins are uncommon, but they have an easily recognizable shape, so they can be found if you go searching for them. Of course, large, well-formed specimens are exceedingly rare. To my knowledge, no one has ever published examples of any arrowhead twins that were grown in the laboratory.

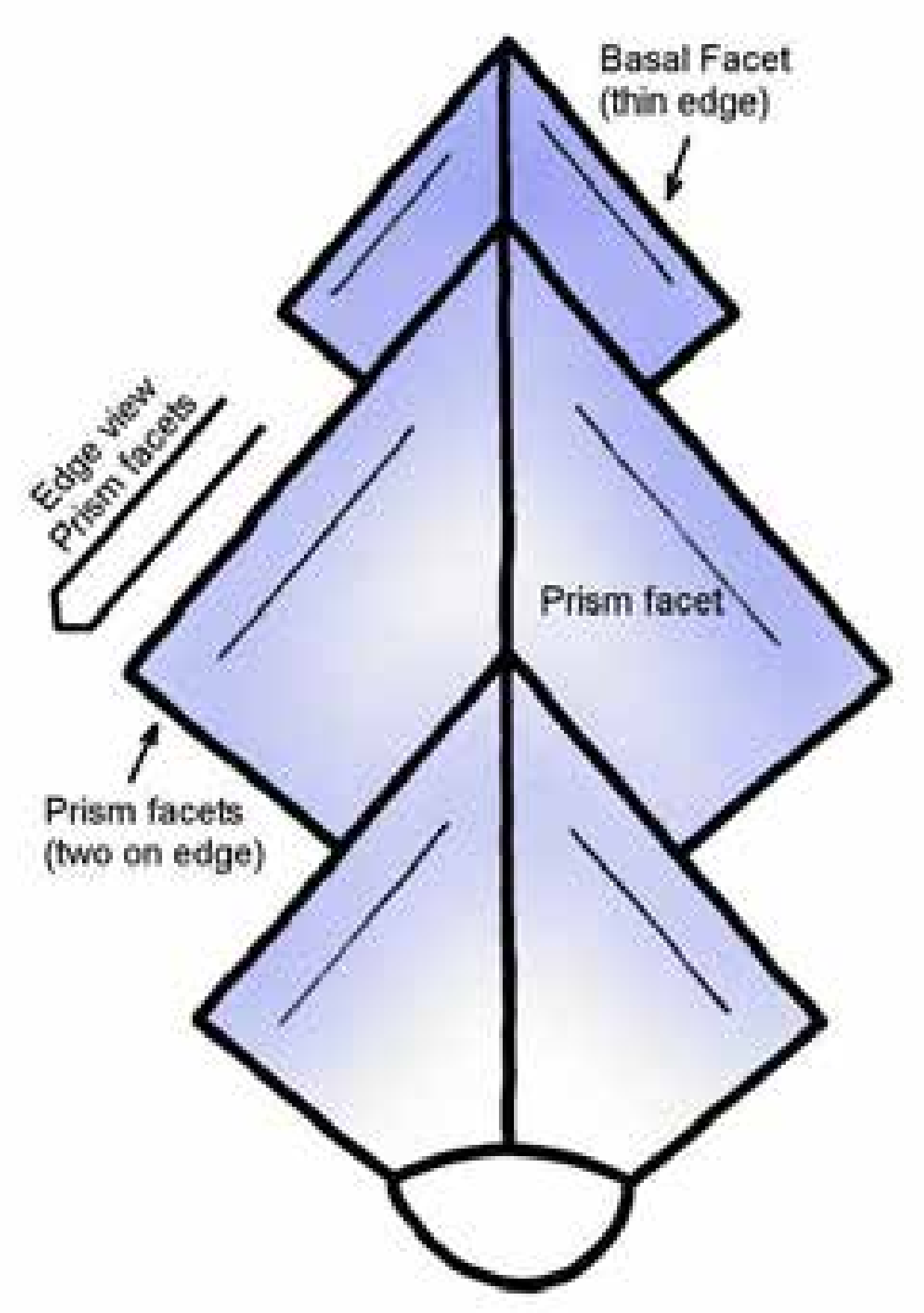

**Figure 2.25: Arrowhead Twins Variant I.**
**Above: This sketch illustrates the peculiar faceted structure of a 78-degree arrowhead twins. It is a thin-plate crystal (seen face-on in the photos, and in this sketch), but the two faces of the plate are prism facets, not the usual basal facets seen in thin hexagonal plates. A pair of narrow prism facets make up each of the lower-facing edges, while the top-facing edges are narrow basal facets. Note that all the prism surfaces grow slowly near -5 C, while the basal edge grows rapidly. Once begun, the basal growth yields the observed plate-like structure. An arrowhead twin typically emerges from some unknown initial structure, here shown as an ill-formed blob as the bottom of the sketch.**

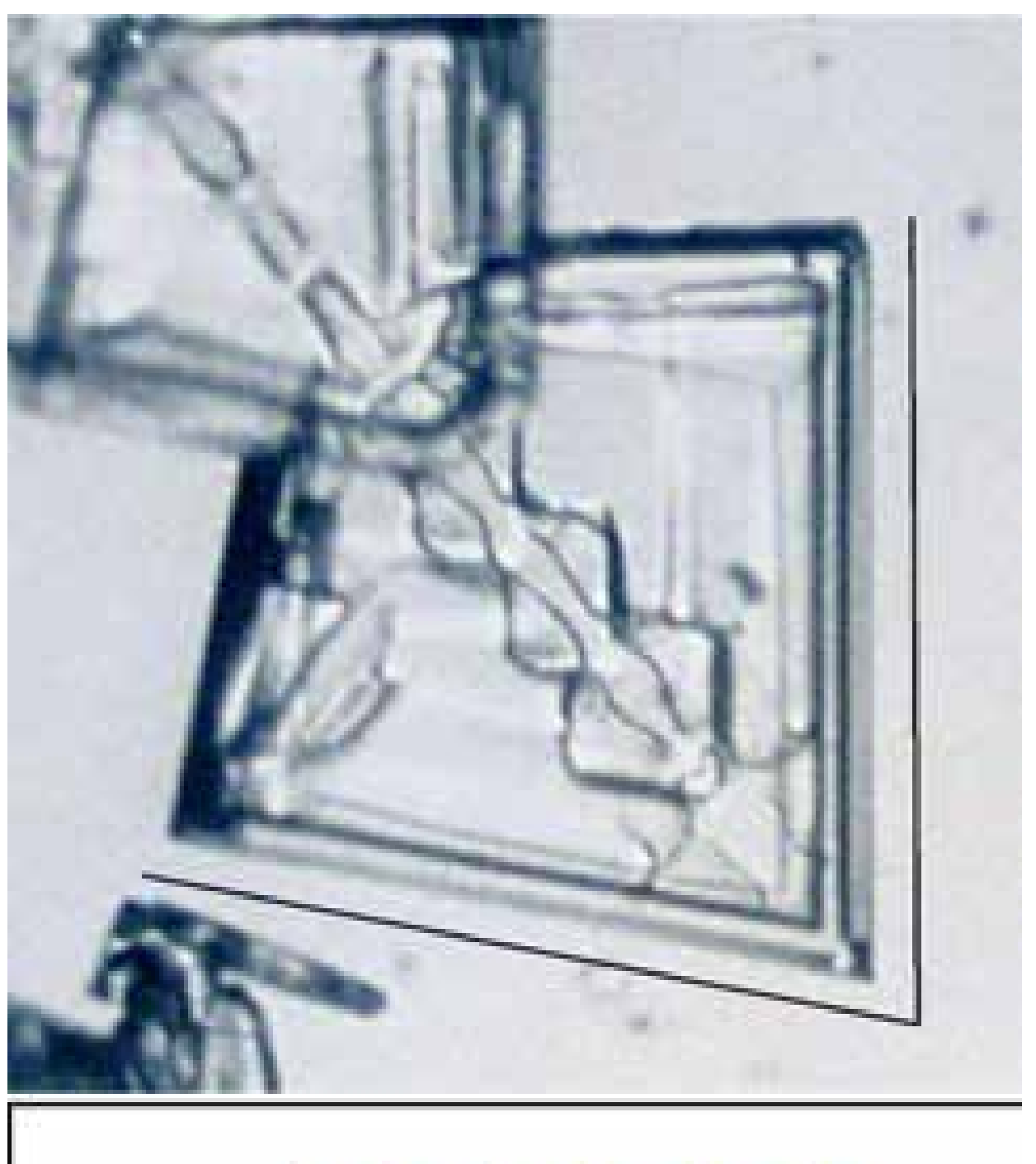

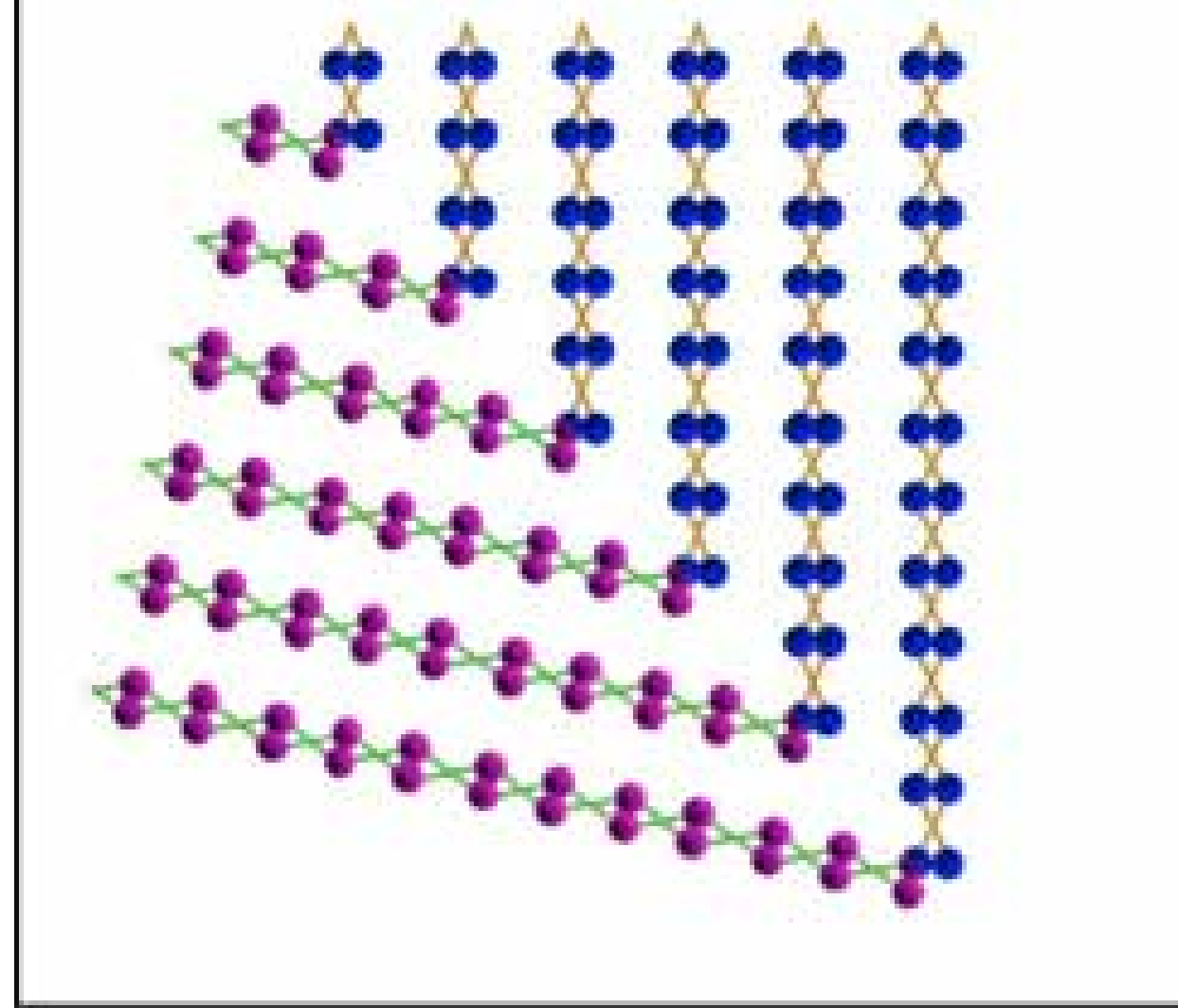

**Figure 2.26: Arrowhead Twins Variant I.**
**Above: The crystal lattice projection in the lower sketch is looking down along the $[\bar{1}010]$ axis, so looking down perpendicular to a prism facet. This is called a *contact twin*, with the two crystals meeting at a reflection plane. This model gives a theoretical apex angle of $2\tan^{-1}(c_0/2a_0) = 78.3$ degrees. A pair of black lines separated by 78-degrees has been superimposed on the above photograph [2006Tap], showing good agreement with theory.**

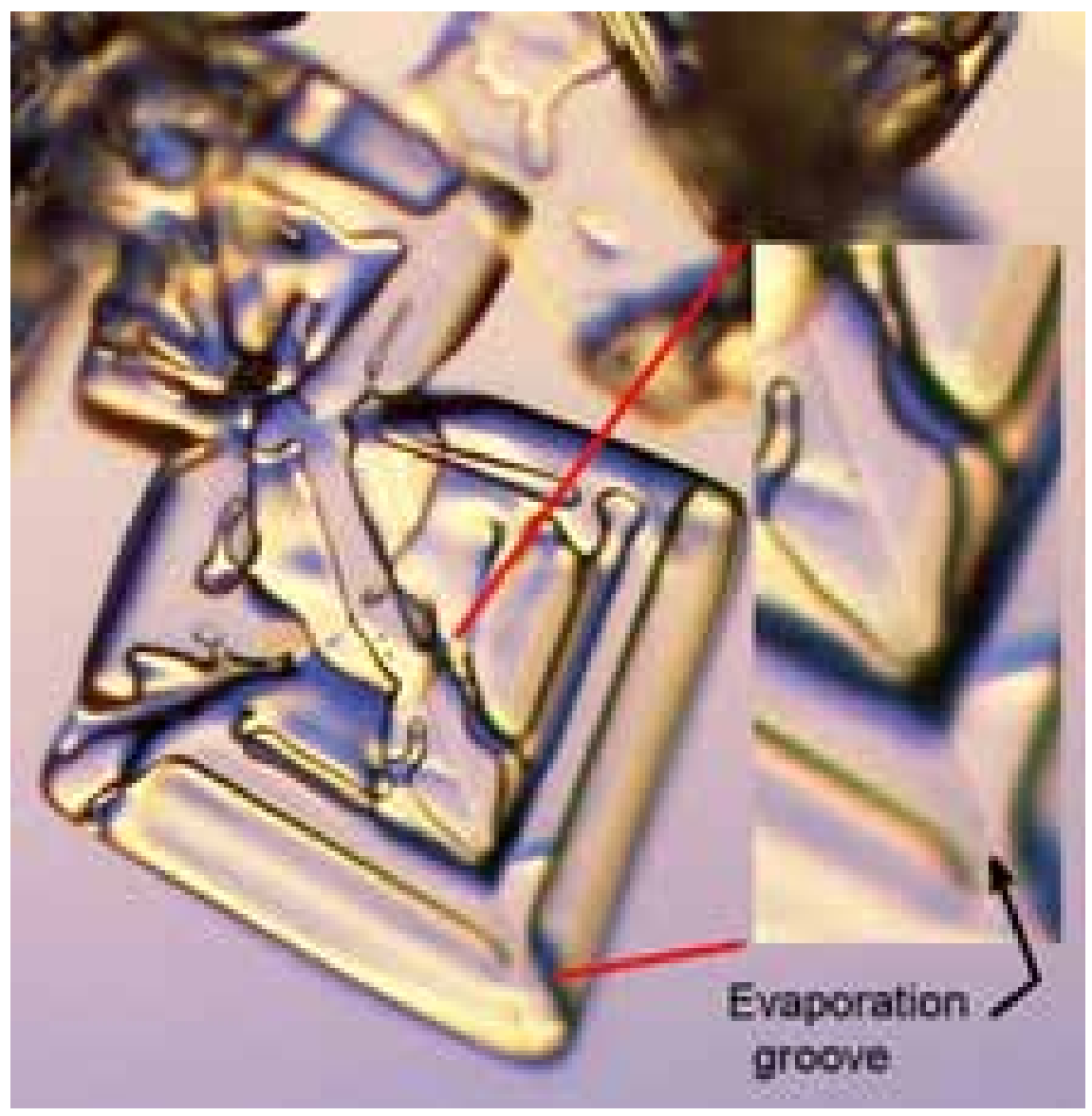

**Figure 2.27: Arrowhead Twins Variant I.**
**This photo shows a weak evaporation groove along the twin plane in an arrowhead crystal. The straightness and sharpness of this feature suggests its crystalline origin, in contrast to the other growth features in the crystal. Image by Patricia Rasmussen [2003Lib2].**

Figure 2.28 shows a second variant of arrowhead twinning, similar to Variant I but with a different apex angle. Comparing Figures 2.26 and 2.28, it becomes apparent that the twinning mechanisms are quite similar, and the geometry of Variant II gives it an apex angle of $2\tan^{-1}(c_0/3a_0) = 57.0$ degrees.

Figure 2.29 shows a third arrowhead variant, this time with an apex angle just slightly below 90 degrees. The model does not give as good a lattice match as the previous two arrowhead variants, but the theoretical apex angle of $2\tan^{-1}(3c_0/5a_0) = 88.7$ degrees agrees well with one of the best photographic specimens, as shown in Figure 2.30.

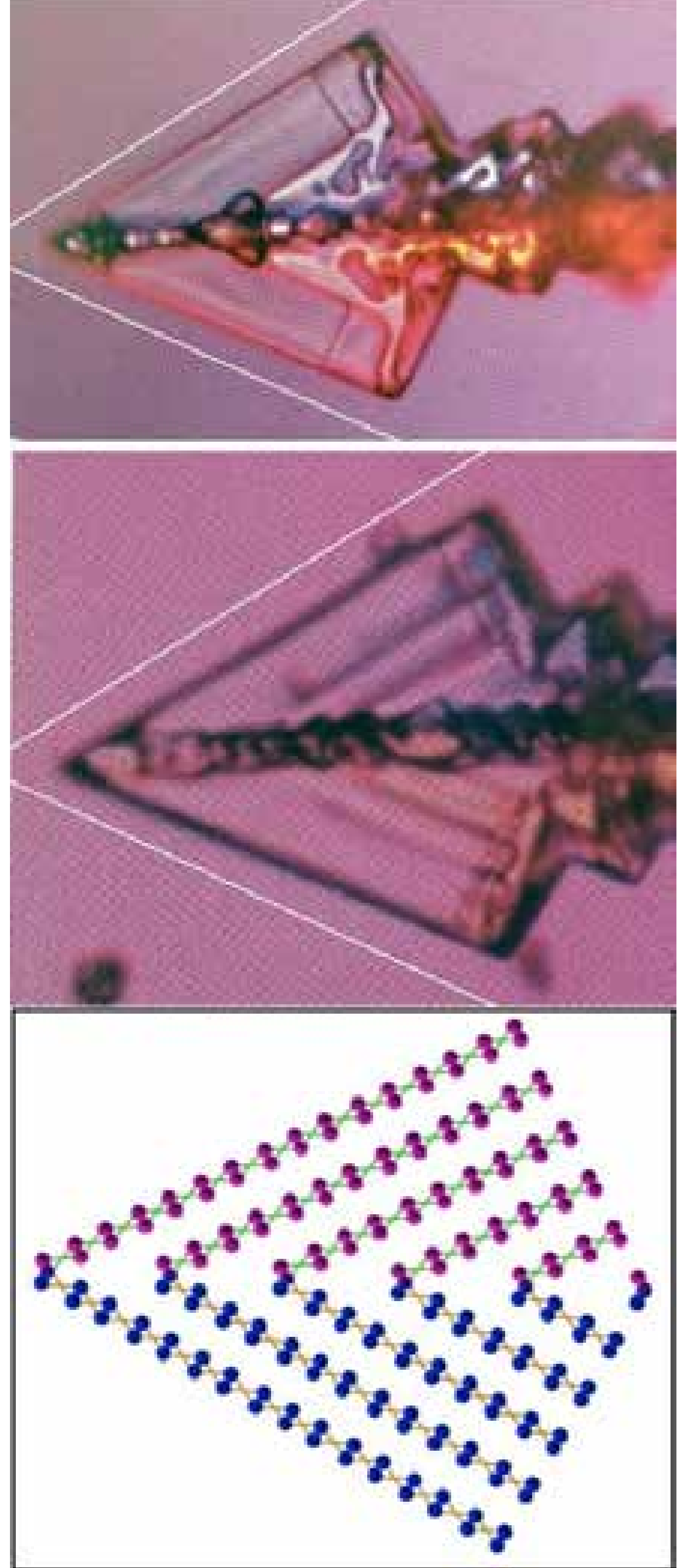

**Figure 2.28 Arrowhead Twins Variant II.**
**The top and middle photos above show two examples of this arrowhead-twin variant [2011Kik]. The lattice projection in the lower sketch is similar to that in Figure 2.26, giving a theoretical apex angle of $2\tan^{-1}(c_0/3a_0) = 57.0$ degrees. The pairs of white lines in the photos subtend this angle, showing good agreement with theory.**

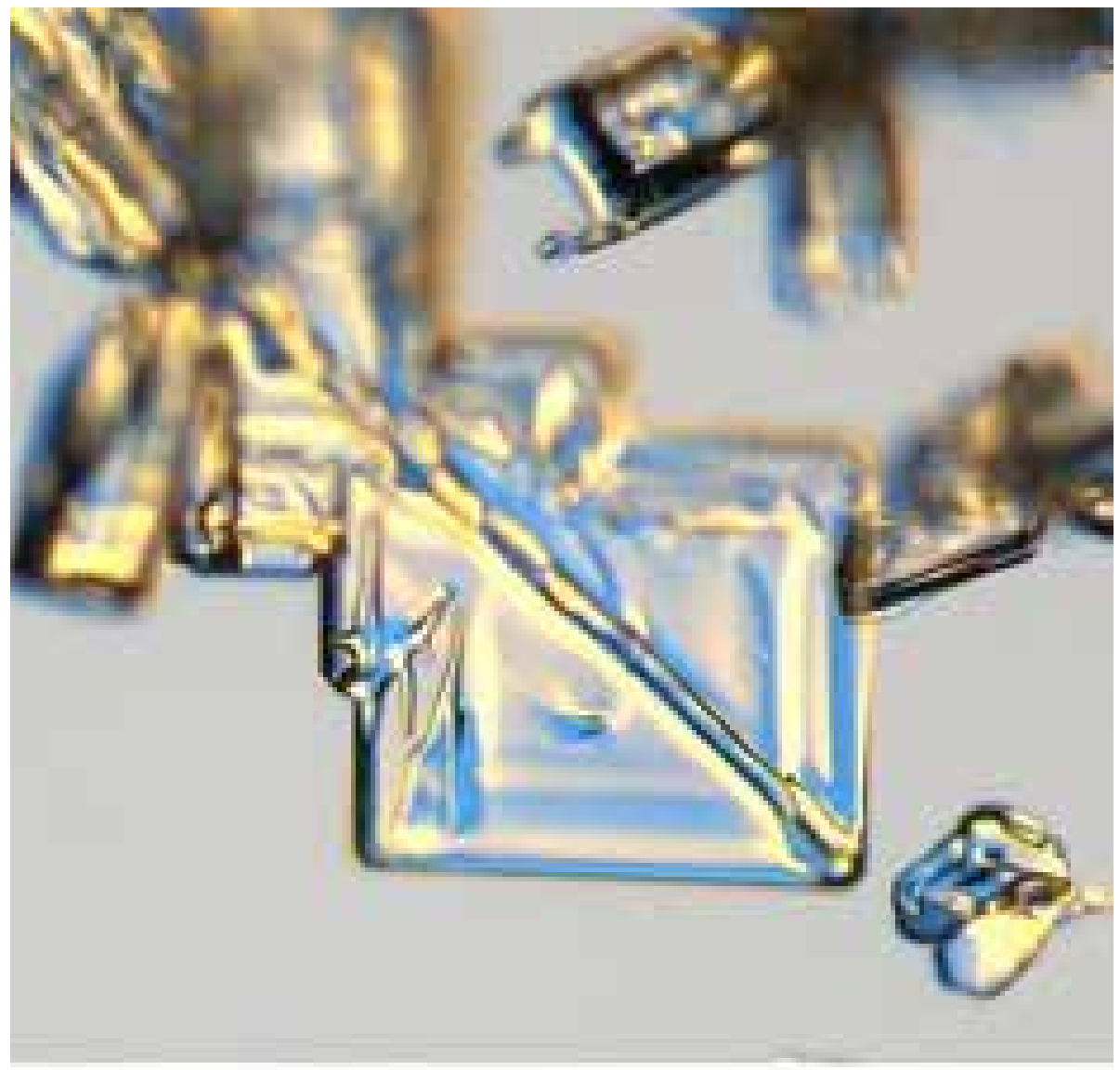

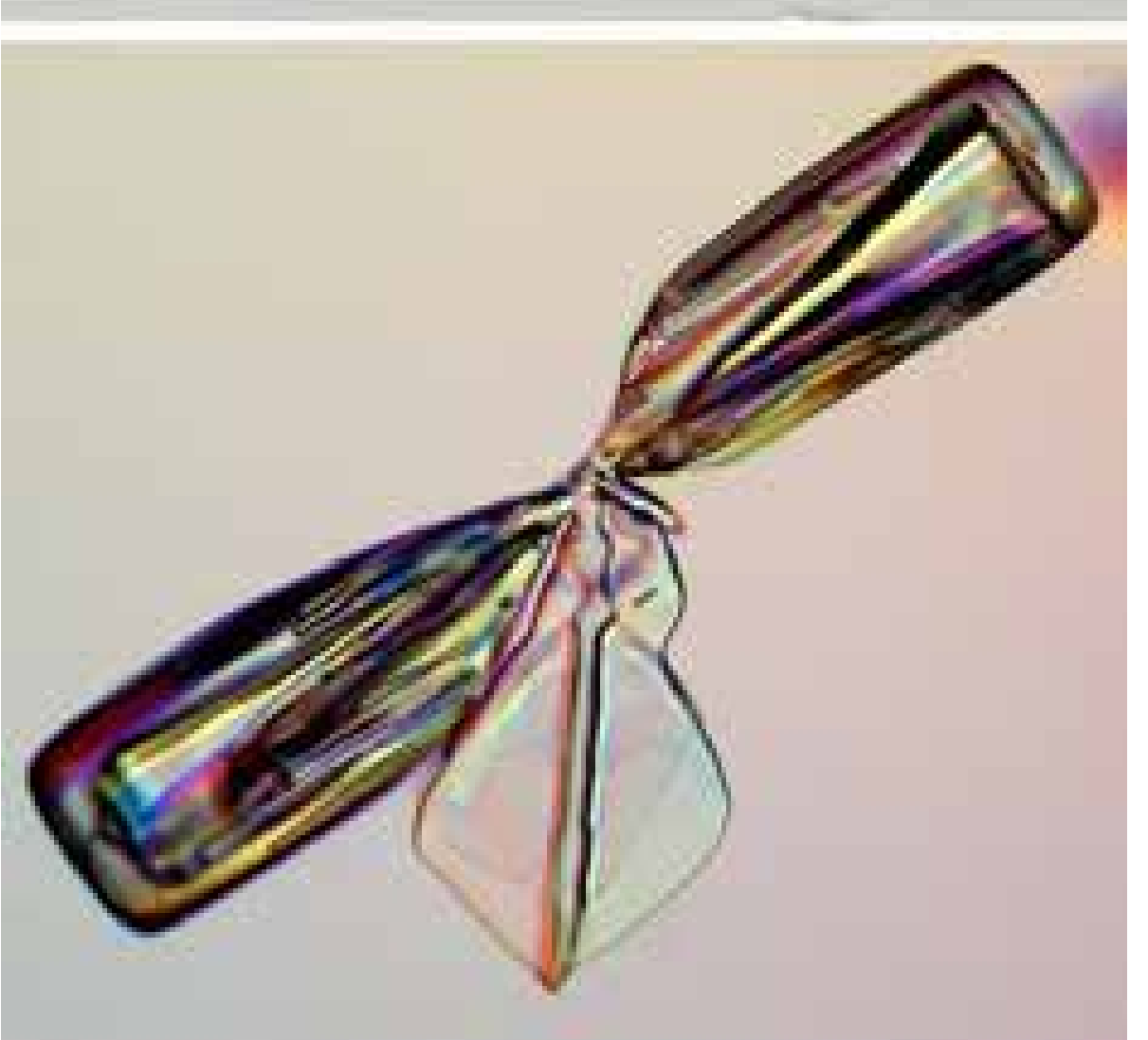

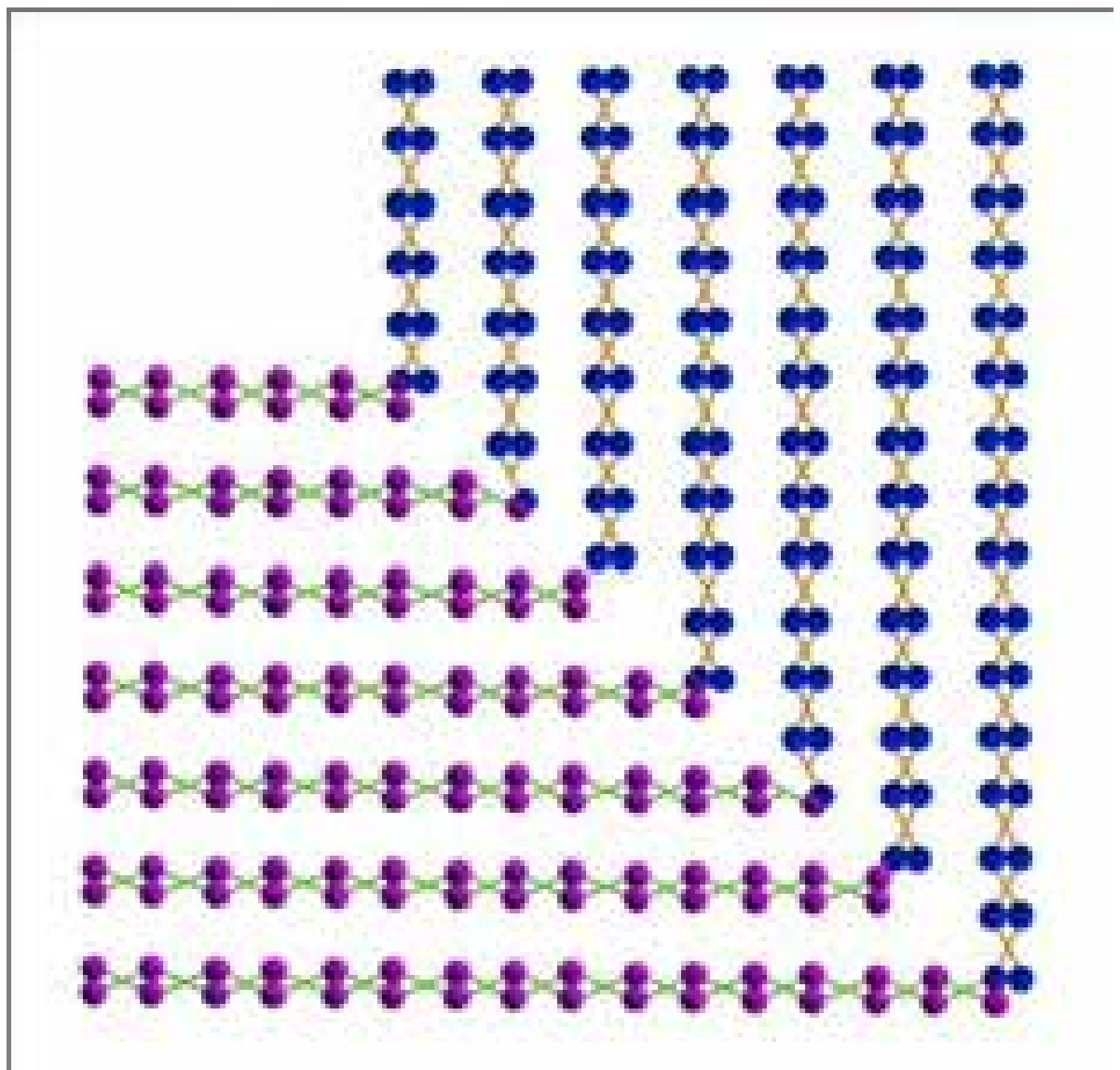

**Figure 2.29: (Left) Arrowhead Twins Variant III. This variant of an arrowhead twin displays an apex angle that is close to 90 degrees, and the model structure shown in the sketch has a theoretical value of $2\tan^{-1}(3c_0/5a_0) = 88.7$ degrees. There are few quality photographic examples of this arrowhead variant, and no laboratory-grown specimens. Nevertheless, to the limits of what has been measured, this model is in good agreement with observations. Images by Patricia Rasmussen [2003Lib2].**

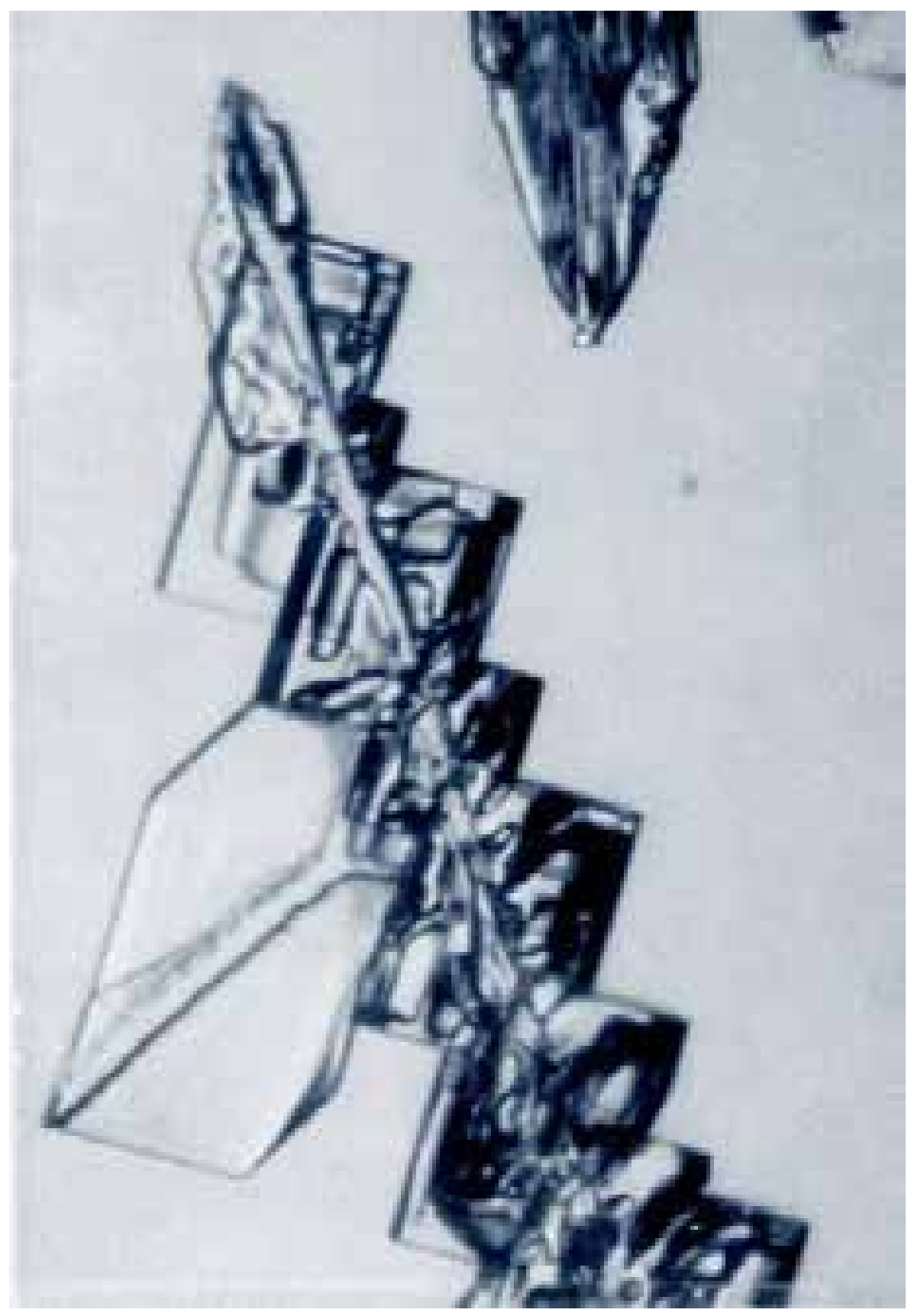

**Figure 2.30: Arrowhead Twins Variant III. Walter Tape photographed this sharply faceted example of an arrowhead twin near the South Pole [2006Tap]. A careful measurement of the crystal (by KGL) yielded an apex angle of 88.5±0.5 degrees, in good agreement with the model shown in Figure 2.29.**

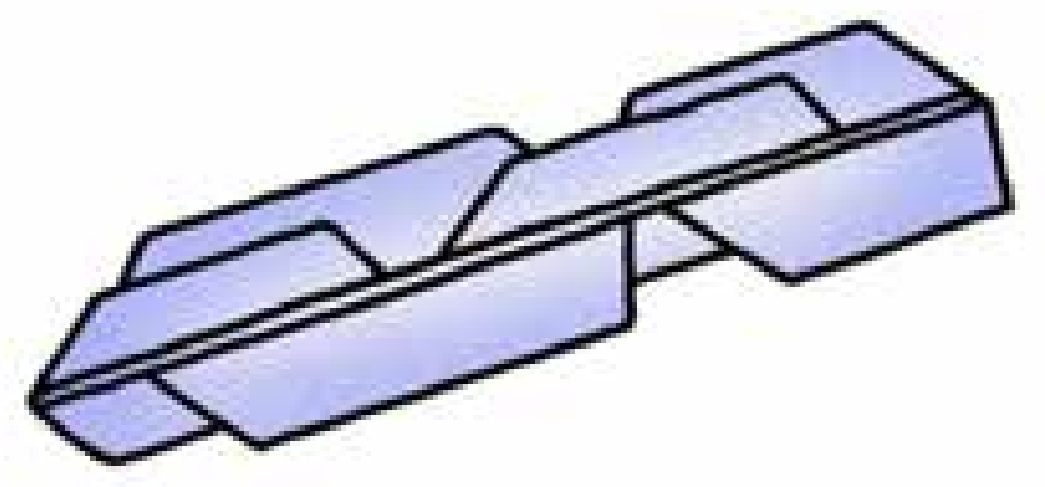

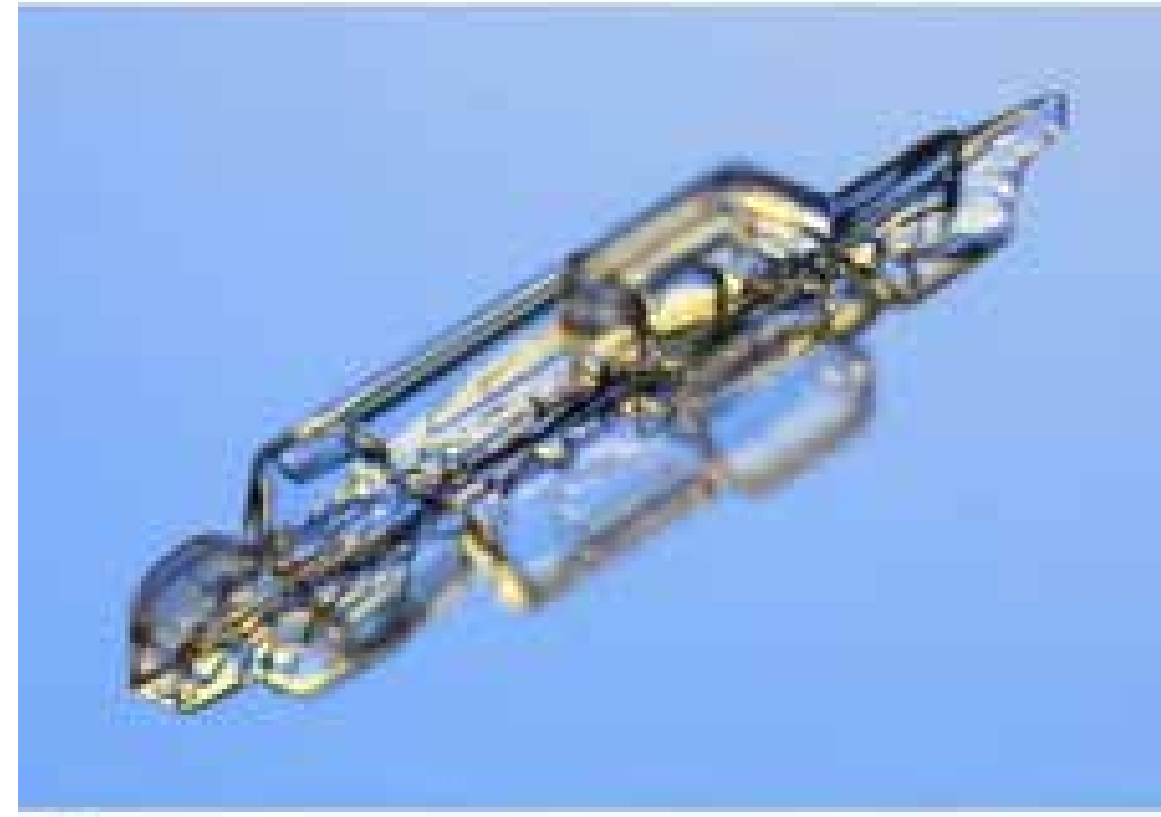

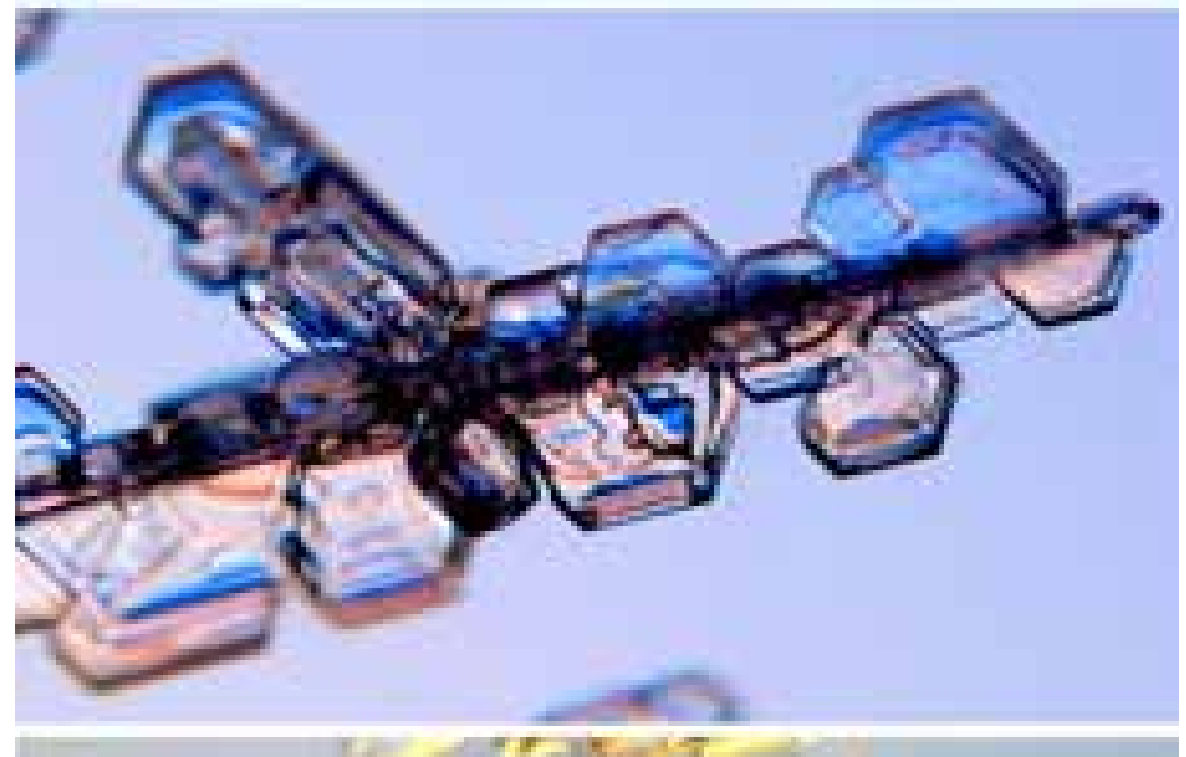

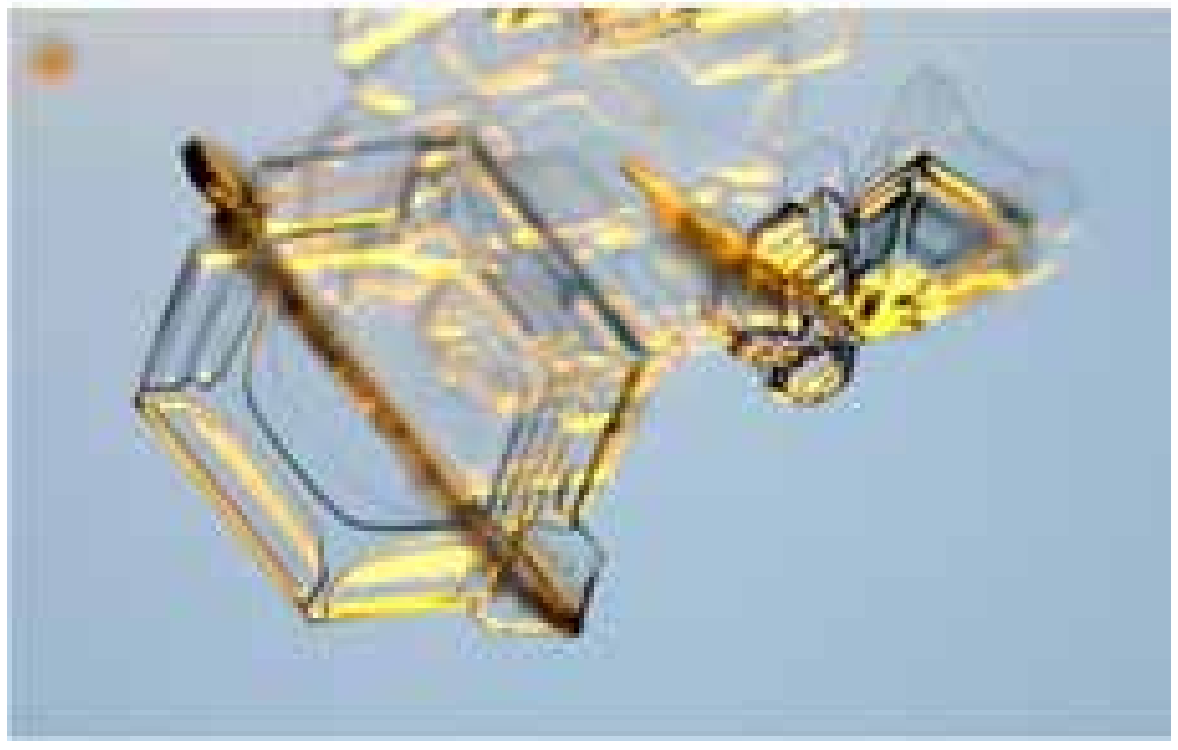

**Figure 2.31: Crossed Plates Variant I.**
**These photos show examples of natural twin-plate crystals, along with a sketch of their overall structure. Note that one prism facet edge is parallel to the intersection axis of the twinned crystals. Top, bottom images by Patricia Rasmussen [2003Lib2].**

## Crossed Plates

Figure 2.31 shows several examples of another class of snow crystal twinning, this time in the form of crossed plate-like crystals. These can be found in nature, although specimens are relatively uncommon, quite small, and most are rather poorly formed, as shown in the figure. They seem to form mostly at temperatures near -2 C. Crossed plates have been seen in the laboratory as well, but there are few good photographs to show here

Figure 2.32 shows a proposed crystal structure for this first crossed-plate variant. The twin angle is $2\tan^{-1}(c_0/\sqrt{3}a_0) = 86.5$ degrees, which is consistent with the observations showing plates crossing at roughly 90 degrees. Unfortunately, there do not appear to be any photographic examples that have allowed a precise measurement of the angle between the crossed plates.

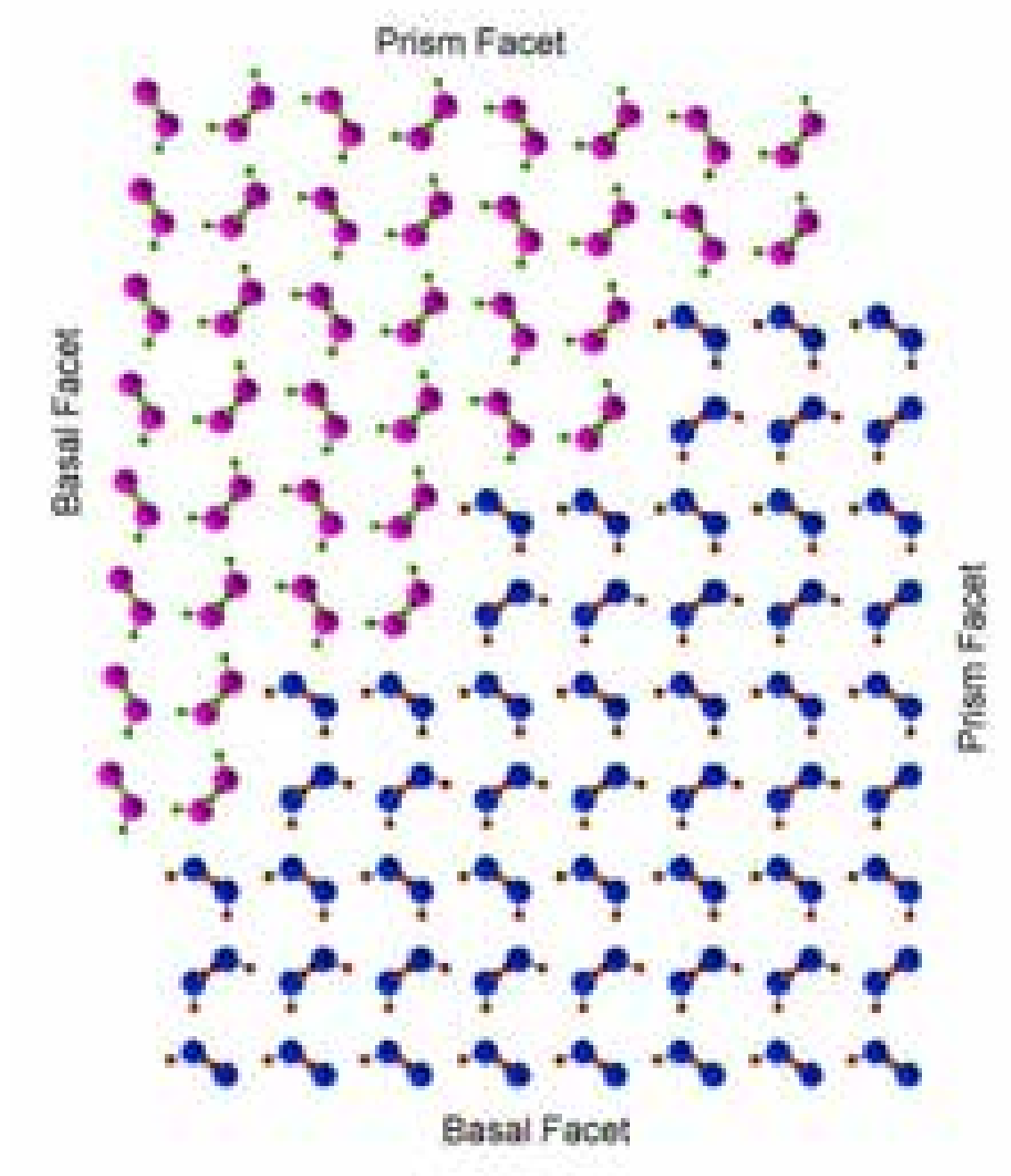

**Figure 2.32: Crossed Plates Variant I.**
**A possible lattice model for Variant-I crossed-plate twinning. This is a contact twin where the theoretical angle between basal facets is $2\tan^{-1}(c_0/\sqrt{3}a_0) = 86.5$ degrees.**

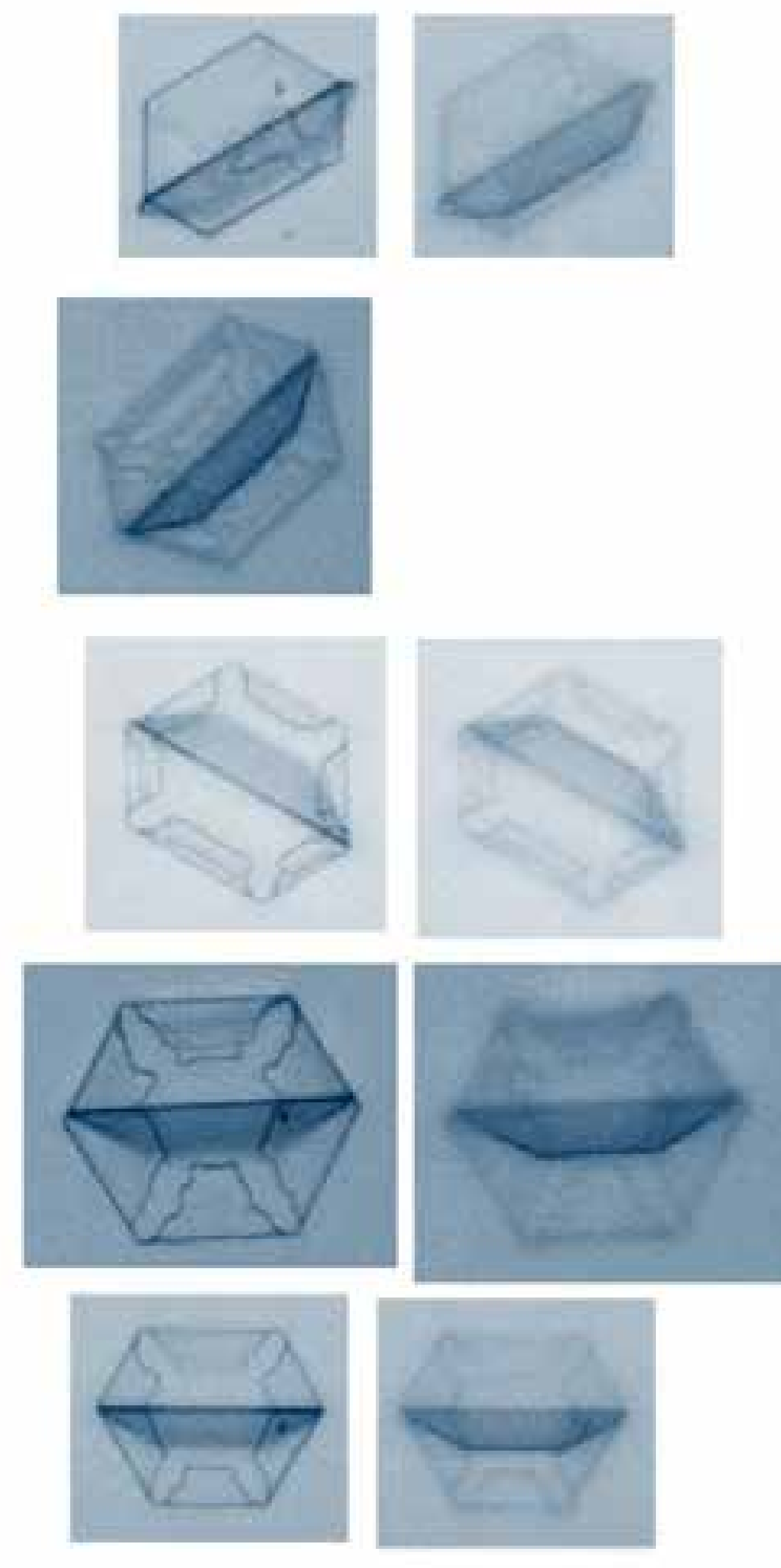

**Figure 2.33: Crossed Plates Variant II. These images show several snow crystal twins grown in air in a free-fall chamber near -10 C. Side-by-side pairs of images show the same crystal with a different microscope focus, one image focusing on the flat plate resting on the substrate and the other image focusing on the top edge of the twin plate. Photos by the author.**

Figure 2.33 shows another crossed-plate variant that is observed quite readily in free-falling laboratory crystals grown in air near -10 C, this time exhibiting a twin angle of about 70 degrees. This form was first documented by Kunimoto Iwai [1971Iwa], who proposed the lattice model shown in Figure 2.34 with a theoretical angle between plates of $\tan^{-1}(\sqrt{3}c_0/a_0) = 70.5$ degrees. This is not a contact-twin model, however, so one crystal is not a mirror reflection of the other about the twin plane. As a result, the connection points (circled in Figure 2.34) have theoretical spacings of $\sqrt{3a_0^2 + 9c_0^2}$ in the top crystal and $3\sqrt{3}a_0$ in the bottom crystal. Although these values only differ by about 0.2 percent, this geometry does lead to an inevitable lattice mismatch as the twin plane propagates outward.

Kobayashi and Kuroda [1987Kob] pointed out that this crossed-plate variant could originate from an ice Ic seed crystal, as shown in Figure 2.35. Presumably this seed appeared during the initial nucleation process, after which it stimulated the twin crystals shown in the figure. As the subsequent vapor growth produced two crossed plates, the minute seed was soon buried by the faster-growing ice Ih. This cubic-nucleation model nicely explains why such a high-order twinning should occur so readily, and it supports the hypothesis that stacking disordered crystals, containing a mix of Ih and Ic bonding, can play a significant role in snow-crystal nucleation.

There are several additional observations of snow crystals with rather odd geometries that have been reported in the literature, and these might be explained as variants of twinning beyond those described here. To date, however, these observations are rather poor, so it is perhaps premature to extend our discussion of twinning much further. As with many other aspects of snow-crystal science, better observations may yield additional surprises in the future.

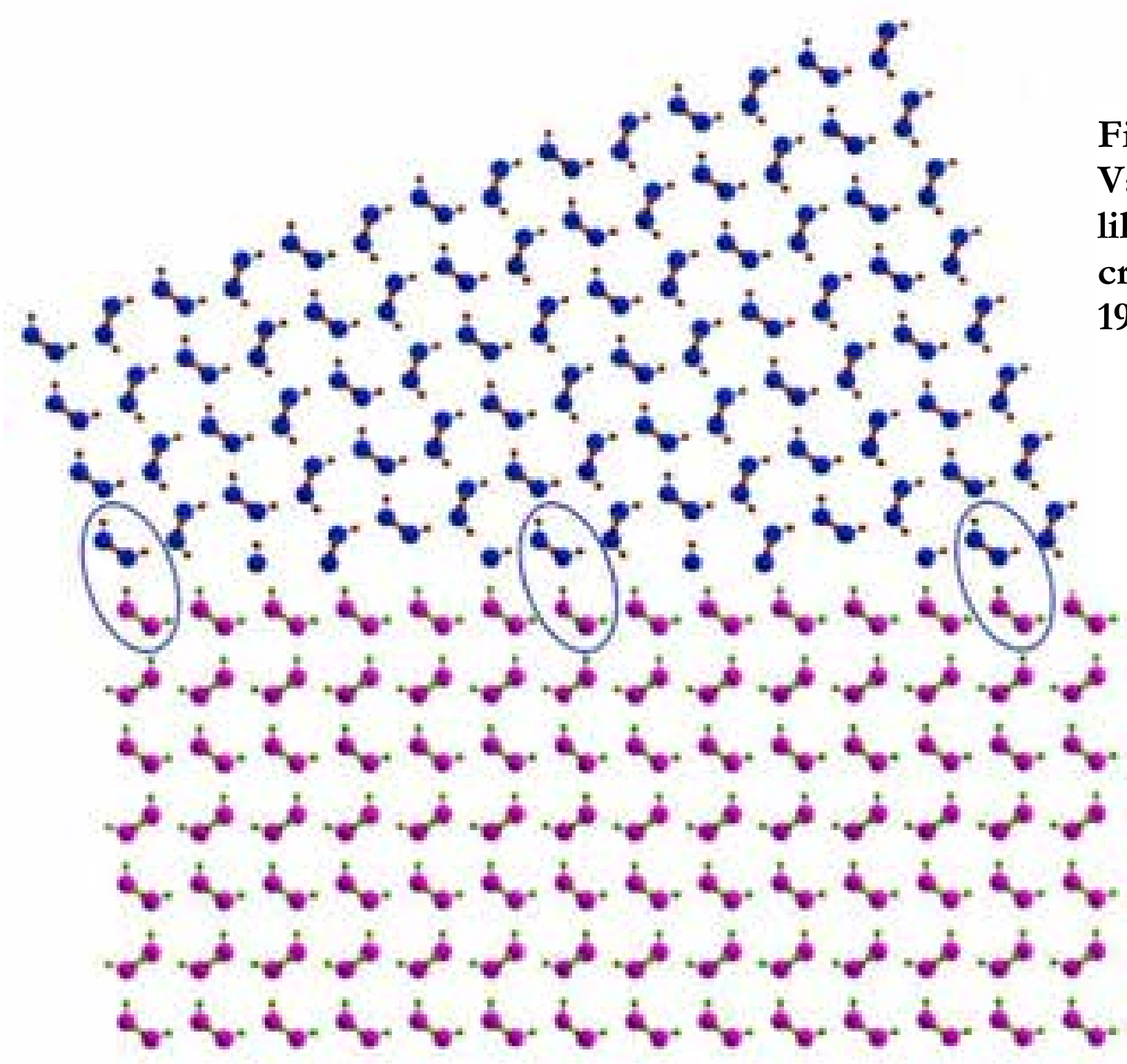

**Figure 2.34: (Left) Crossed Plates Variant II. This diagram shows the likely crystal structure for variant-II crossed-plate twinning [1971Iwa, 1978Fur].**

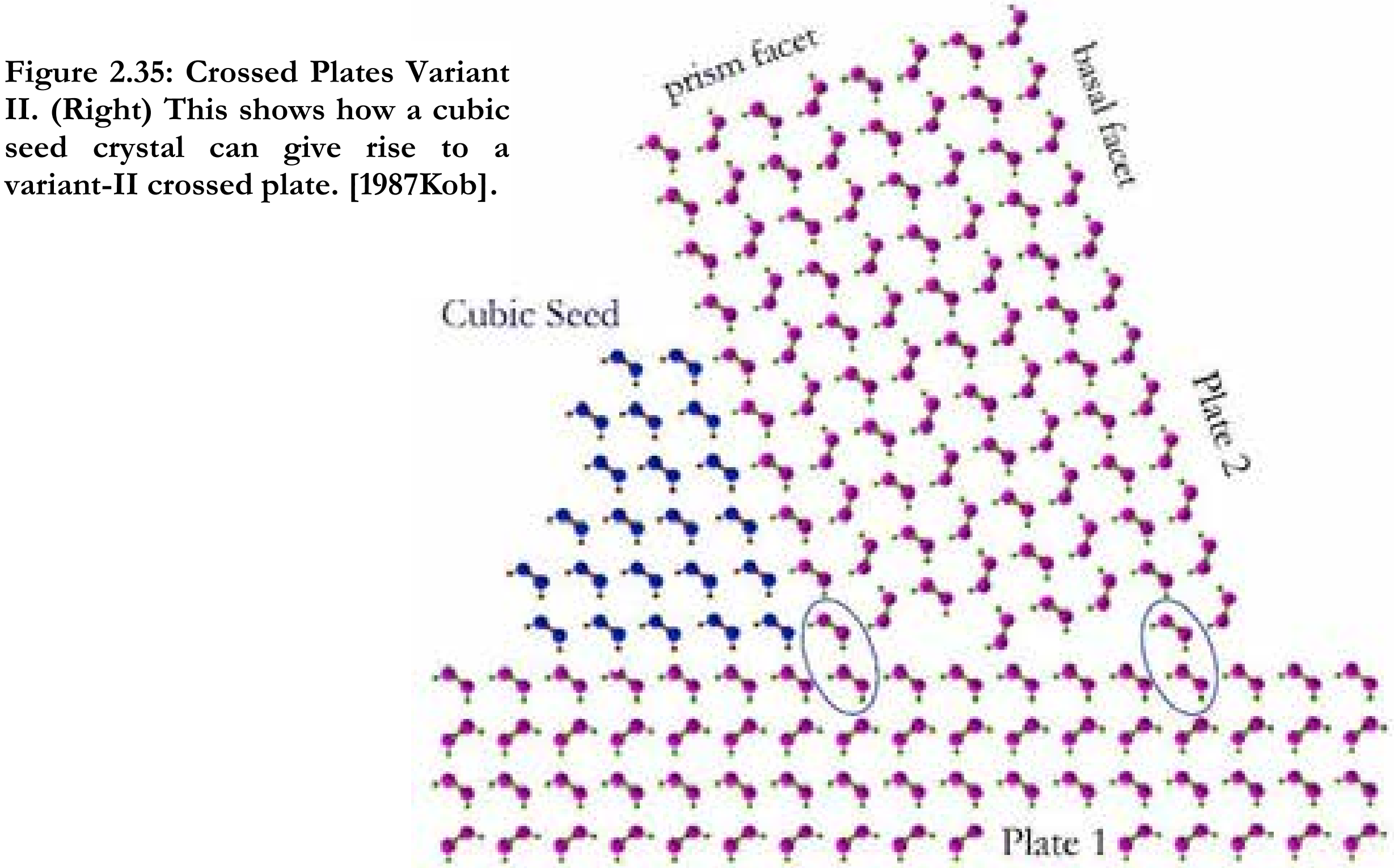

**Figure 2.35: Crossed Plates Variant II. (Right) This shows how a cubic seed crystal can give rise to a variant-II crossed plate. [1987Kob].**

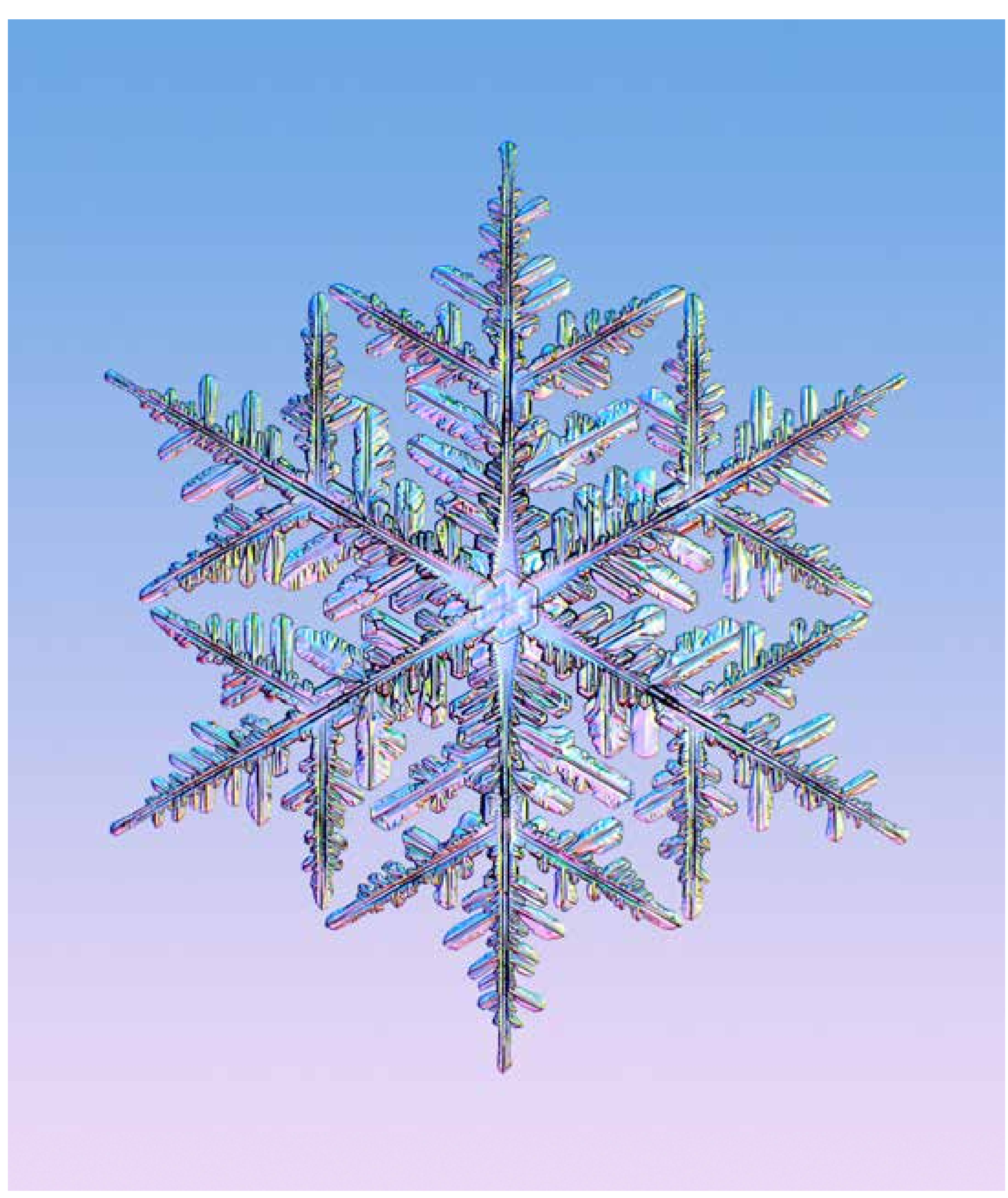

# Chapter 3

# Diffusion-Limited Growth

*If the Lord Almighty had consulted me before embarking on Creation, I should have recommended something simpler.*
– Alphonso the Wise
*attributed, ~1250*

When a snow crystal grows in the atmosphere, it does so by removing water vapor molecules from the air in its vicinity. To continue growing, more water molecules must diffuse through the surrounding air, making their way into the depleted region near the crystal. Because diffusion is a slow process, it tends to limit the development of the crystal, so we say its growth is *diffusion-limited*. As we will discuss in this chapter, diffusion-limited growth is responsible for the creation of branches and other structures, and it is of great importance in the formation of complex patterns in snow crystals.

**Facing Page: This stellar snow crystal displays complex sidebranching brought about by diffusion-limited growth. It also experienced a major stimulated-sidebranching event when the primary branches were about half their final length. (photo taken by the author in Kiruna, Sweden)**

The word diffusion derives from the Latin *diffundere*, meaning to spread out over time. The diffusion of water molecules in air results from the normal thermal jostling of air and water molecules, which tends to mix the two species together. If the water-vapor density is not spatially uniform, then the random molecular motions will, on average, transport water molecules from higher-density regions to lower-density regions. Therefore, as a growing snow crystal consumes water vapor molecules in its vicinity, more will flow inward toward the crystal from afar, providing additional material for continued growth.

Diffusion is a common phenomenon in everyday life, although we may not readily notice diffusion in action, especially when it involves invisible gases like air and water vapor. Diffusion in colored liquids can be more easily visualized, as shown in Figure 3.1. Most people are familiar with material dispersing out from a central source, but perhaps less so with diffusion toward a central sink, as shown in the illustration. Particle diffusion always tends to mix materials together, so the net diffusive transport is always from high to low densities. More quantitatively, the net flow of material arising from diffusion is along a density gradient and is proportional to the size of the gradient.

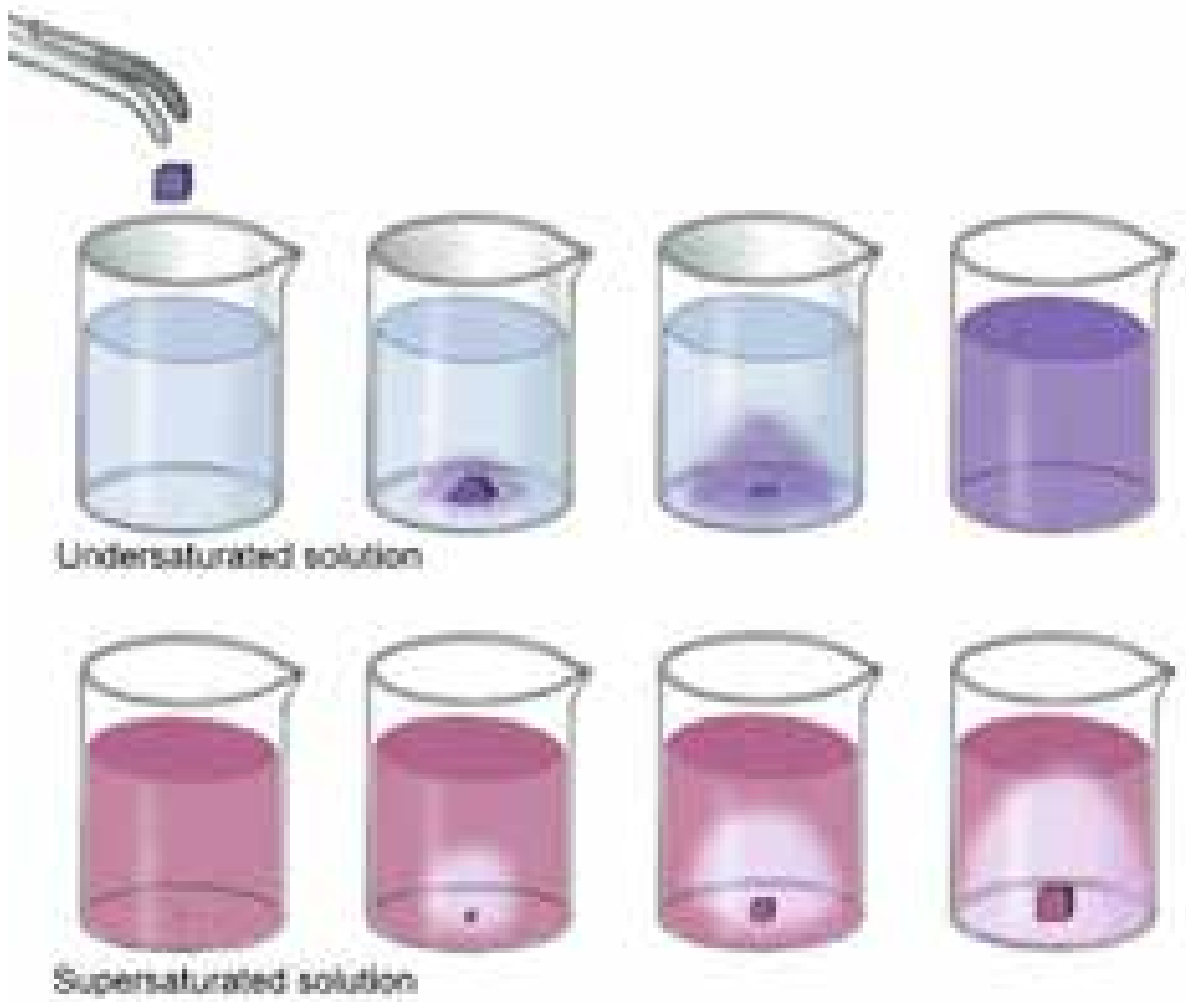

**Figure 3.1: The process of diffusion, shown operating in two directions. A crystal dropped into an undersaturated solution will dissolve (top row). Diffusion will then slowly spread the dissolved material throughout the solution. In contrast, a seed crystal placed into a supersaturated solution (bottom row) will grow as diffusion carries material to the crystal and depletes the solution nearby. The first case is analogous to a snow crystal sublimating in undersaturated air, and the second case is analogous to a snow crystal growing in supersaturated air. Adapted from an image found at https://byjus.com/jee/diffusion-of-gases/.**

There are actually two types of diffusion involved in snow crystal growth: particle diffusion and heat diffusion. The latter arises when latent heat is generated by deposition and then diffuses away through the air. As I will show below, the effects of heating and heat diffusion are relatively small compared to particle diffusion. Except in a few special circumstances, one can describe the growth of snow crystals to a good approximation by neglecting heat diffusion entirely, and that will be my default assumption unless otherwise indicated.

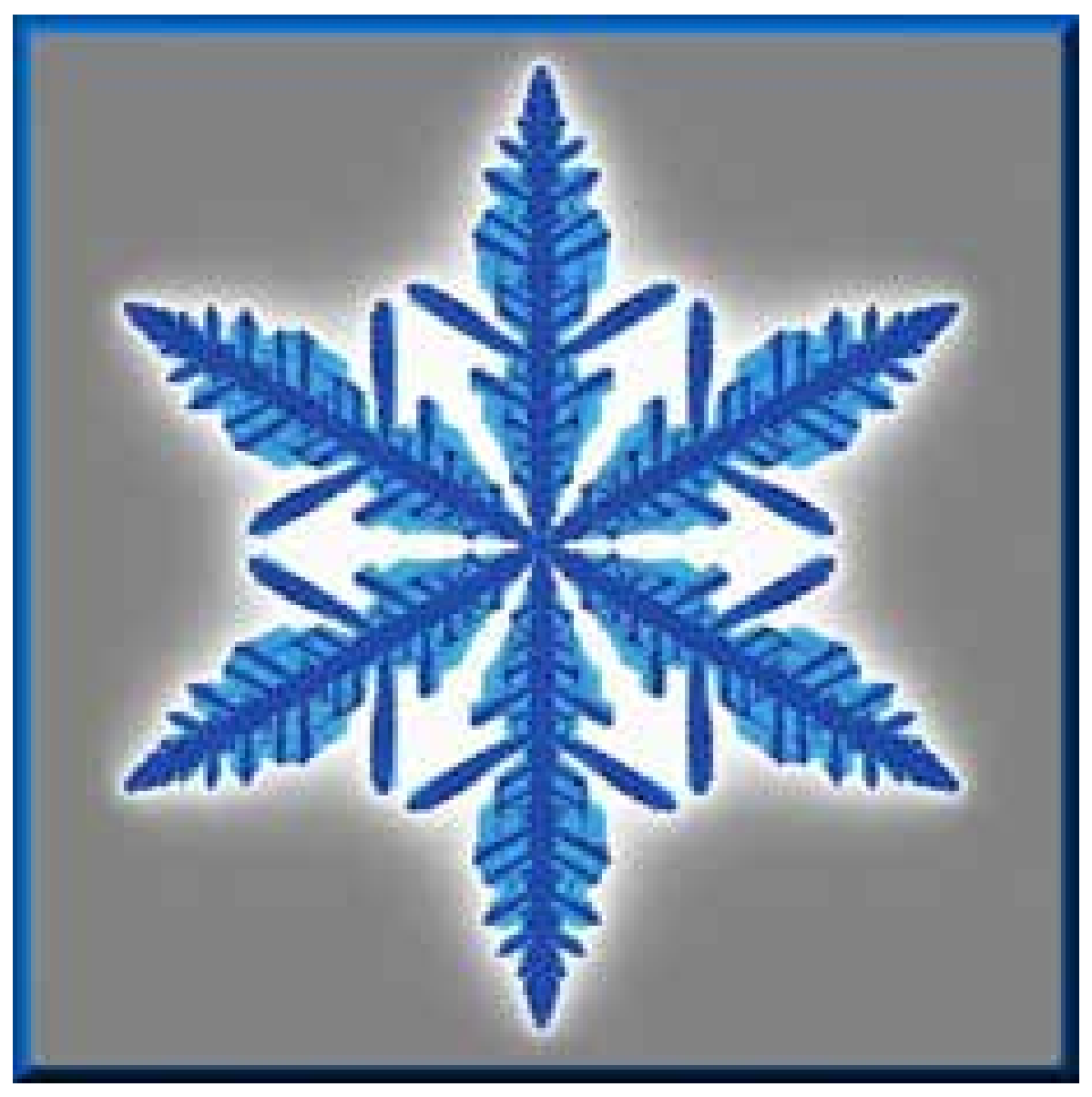

**Figure 3.2: This 2D numerical simulation [2008Gra] demonstrates the depletion of water vapor around a snow crystal. The supersaturation is constant (grey) far away from the crystal, but it drops to near zero (white) at the growing crystal surface. The supersaturation gradient produces a diffusion-driven inward flow of water vapor that continuously supplies material for the growing crystal.**

Figure 3.2 illustrates how a growing ice crystal depletes the water-vapor density around it, creating a supersaturation gradient. As seen in this computer simulation, the gradient is highest near the tips of a growing crystal, resulting in a high flow of water vapor at these points, and thus fast growth at the tips. In the interior parts of the crystal, the supersaturation gradients are lower, providing lower water-vapor flow and slower growth. The result can be seen in movies of growing snow crystals (both in computer models and in laboratory observations), as the outer regions grow outward quickly while the interior structures evolve more slowly.

Figure 3.3 shows another illustration of the depletion of water vapor around a growing snow crystal, this time in a laboratory setting. Water droplets condense on non-ice surfaces whenever the supersaturation is above $\sigma_{water}$, because this means that the humidity is above the dew point (see Chapter 2). When humid air is blown down onto the growing snow crystal in Figure 3.3, a fog of water droplets condenses on the substrate around the crystal because $\sigma > \sigma_{water}$ in that outer region. But the supersaturation is depleted near the growing crystal, giving $\sigma < \sigma_{water}$ and no condensed droplets in the inner region. The boundary between the inner and outer regions reveals where $\sigma \approx \sigma_{water}$ near the substrate surface. While this laboratory image nicely illustrates water vapor depletion around a growing snow crystal, it would require a rather sophisticated 3D numerical simulation to reproduce the $\sigma \approx \sigma_{water}$ contour seen in Figure 3.3.

**Figure 3.3: This laboratory-grown POP snowflake (see Chapter 9) indirectly shows the depletion of water vapor around a growing crystal. As moist air blows down on the substrate supporting the crystal, a fog of tiny water droplets condenses onto its surface. No droplets condense near the ice crystal, however, because the water vapor density is lower in that region. The boundary between these two regions shows where the humidity passes through the dew point (or, equivalently, the supersaturation passes through $\sigma_{water}$), which is when water droplets begin to condense.**

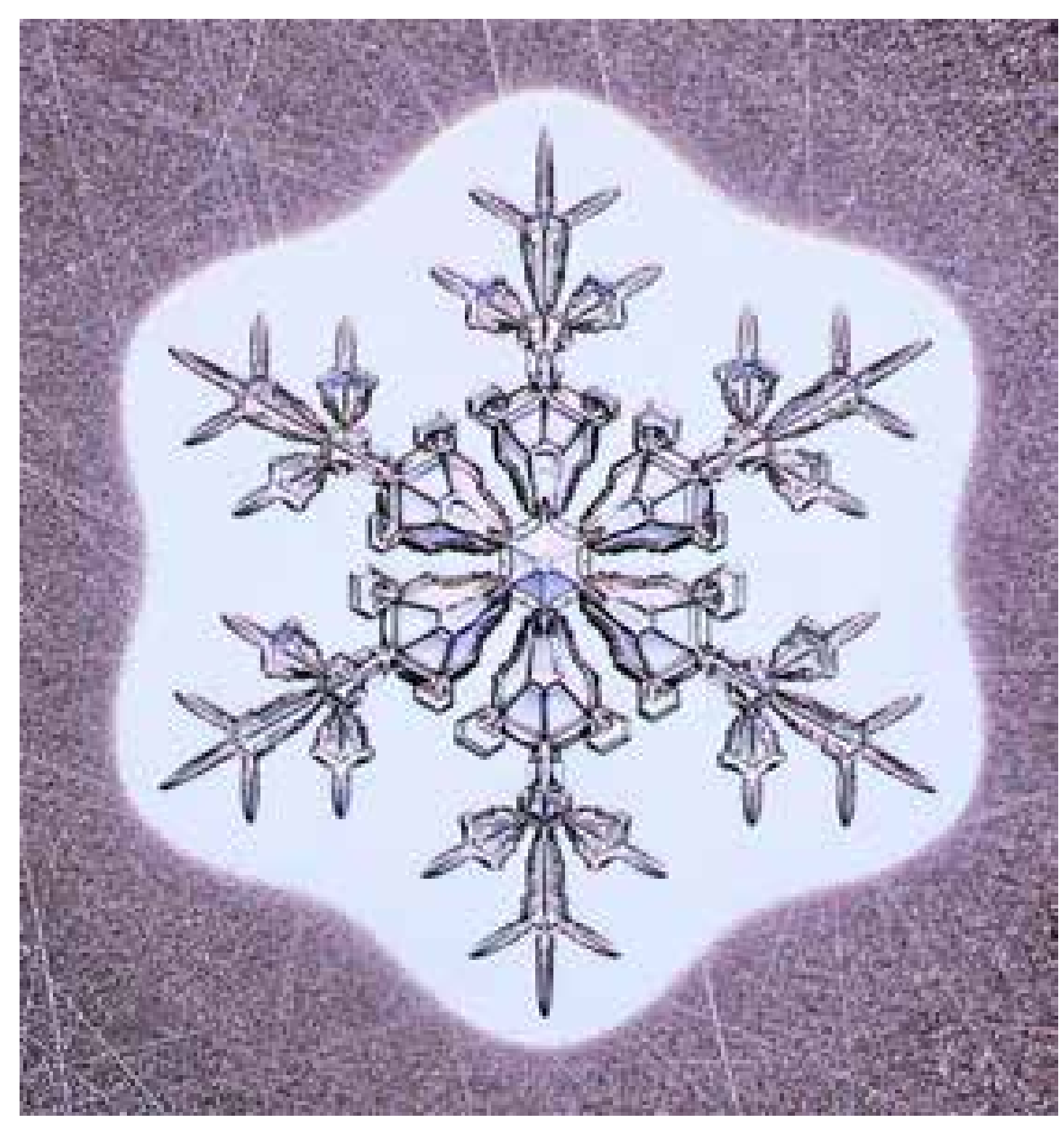

Large-scale air flow can also transport and mix water vapor in air, and these flows operate in addition to diffusion. Wind and turbulence thus affect snow-crystal growth, sometimes substantially, and I discuss this topic later in the chapter. Plain particle diffusion is by far the most important transport process, however, and it operates even in still air. To a first approximation, therefore, we can ignore large-scale air flow and focus our attention on understanding particle diffusion and how it affects snow crystal growth.

The process of diffusion is defined mathematically by the classical diffusion equation. Unlike attachment kinetics, where our grasp of the underlying molecular dynamics is depressingly poor, the essential physics of diffusion is extremely well understood, and has been for over a century. Calculating the effects of diffusion in complex geometries (for example, surrounding a branched snow crystal) remains a challenging computation problem, but at least we have a firm grasp of the underlying physics.

One goal in this chapter is somewhat qualitative in nature: to describe the phenomenology of diffusion-limited growth as it pertains to the specific case of snow-crystal formation. Phenomenological descriptions are not always the best way to understand the underlying science, as they can involve rough approximations and empirical descriptions. Our brains, however, are very much tuned to visual inputs, such as graphs, sketches, and photographs, and less so to mathematical formulae. So a well-crafted phenomenological description is not without pedagogical value. Moreover, I have always found it useful to develop a basic mental picture of a physical phenomenon under study, unfettered by words or equations, as this often provides a helpful intuitive grasp of the subject.

A second goal in this chapter is to develop the quantitative side of diffusion-limited growth, writing down the relevant equations and then outlining techniques used to solve them. This mathematical background will be needed in the next chapter, when we address the problem of making numerical models of growing snow crystals. Computational simulations have become essential in this field, as analytical calculations are simply inadequate for dealing with the morphological richness inherent in the diffusion-limited growth of snow-crystal structures that are both faceted and branched.

Our overarching game plan is a straightforward example of modern reductionist science: 1) break down a complex phenomenon (snow crystal formation) into its simplest component pieces, 2) study and understand those pieces as best one can by using realistic physical models informed by precise measurements, and 3) reassemble the pieces into a computer simulation that recreates the original phenomenon and agrees with quantitative observations. As I have been striving to execute this plan over the years, I have found that both quantitative and qualitative perspectives are necessary to fully understand and appreciate the science of snow crystal formation, so I try to address both in this chapter.

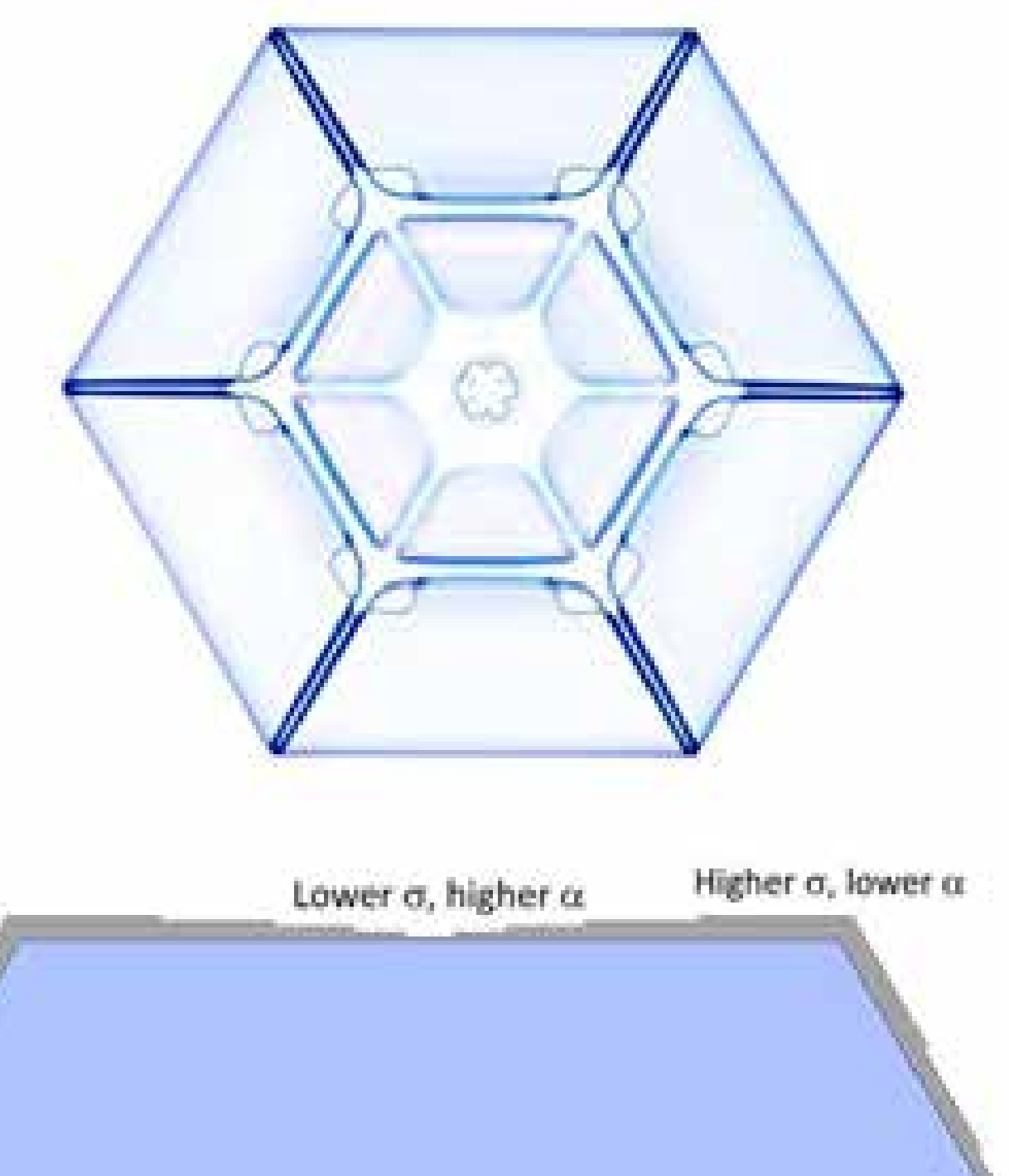

**Figure 3.4: The prism facets on this POP snow crystal (top photo) look straight, but they must be slightly concave at the molecular scale (bottom sketch). The hexagonal corners stick out farther into the humid air, so $\sigma$ is slightly higher at the corners than at the facet centers. At the same time, the density of molecular terrace steps is higher nearer the facet centers, making $\alpha$ slightly higher there (the surface lines in the sketch represent molecular layers). Because the growth rate is proportional to $\alpha\sigma$, the entire facet surface grows upward at a constant rate, maintaining its flat appearance.**

## 3.1 Faceting and Branching

A good starting point is to examine how particle diffusion affects the transition from faceted to branched snow-crystal growth. Faceting is driven primarily by attachment kinetics, as I described in Chapter 3, and faceted snow crystals grow readily even when diffusion effects are negligible, as in near-vacuum environments. But the interplay of faceting and branching is a central theme in snow crystal growth, so we begin our discussion with the growth of the simple faceted crystal shown in Figure 3.4.

As we discussed above, a growing crystal absorbs water vapor around it, creating a depleted region in the supersaturated air. In other words, the supersaturation asymptotes to some constant value $\sigma_\infty$ far from the crystal and decreases as one approaches the crystal. Because the six corners of the hexagonal crystal in Figure 3.4 stick out farther into the humid air, the corners typically experience a slightly higher supersaturation than that found near the facet centers. This phenomenon is sometimes called the *Berg effect* [1938Ber].

Beginning with a perfectly flat, faceted surface, a higher supersaturation at the hexagonal corners makes the corners grow more rapidly than the facet centers. Soon the faceted surface is no longer perfectly flat, but becomes slightly concave at the molecular level, as illustrated in the sketch in Figure 3.4. New molecular terraces nucleate preferentially near the corners, where the supersaturation is highest, and the steps subsequently grow inward toward the facet centers. Again, because the supersaturation is highest near the corners, the terrace steps grow fastest there, slowing down as they approach the facet centers.

This change in step velocity causes the terrace steps to bunch up near the facet centers, as shown in Figure 3.4. Because the terrace edges readily absorb water-vapor molecules (faster than faceted surfaces with no steps), the increased step density near the facet centers means that the attachment coefficient $\alpha$ is higher at the facet centers than near the corners. Thus, the corners experience a slightly higher $\sigma_{surf}$ and lower $\alpha$, while the facets centers experience a slightly lower $\sigma_{surf}$ and higher $\alpha$, as shown in the figure.

This process of nucleation and motion of molecular steps results in stable, self-regulating facet growth. There is a negative feedback that maintains the precise concave shape needed to keep the growth velocity, equal to $v_n = \alpha v_{kin} \sigma_{surf}$, constant across the entire facet surface. If the surface became too flat, the corners would grow a bit faster and increase the surface curvature. If the surface became too concave, the facet-center region would grow a bit faster and again restore the correct concave shape. There is a stable point in the curvature that is set by the supersaturation, crystal size, attachment kinetics, and perhaps other parameters. This process is one of the simplest examples of spontaneous structure formation, as the rules of crystal growth bring about the sustained, stable growth of a faceted crystal.

Molecular steps are too small to be observed even with high-power optical microscopy, so the concave facet shape is normally imperceptible. In a similar vein, one cannot easily observe the supersaturation variations around the crystal, as water vapor is an invisible gas. When you look at faceted snow crystals by eye, or with a magnifier or microscope, all you can see is that the faceted surfaces appear as smooth and flat as a pane of glass.

## Transition to Branching

This picture of facet formation takes on a new twist as the growth rate increases, or as the crystal grows larger. In either of these cases, the faceting mechanism continues working only until $\alpha \approx 1$ at the facet centers. Once this happens, $\alpha$ can no longer increase, and this causes the self-regulating process described above to break down. At some point, the facet centers will no longer be able to keep pace with the corners, and the facet will no longer be able to maintain its flat appearance. When this happens, branches sprout from the hexagonal corners, as shown in Figure 3.5.

This transition from faceted to branched growth tends to be rather abrupt. Once the hexagonal tips sprout branches, they quickly grow outward and leave the regions between the branches far behind. Exactly when the transition occurs depends on several factors. A general rule of thumb is that faceting dominates when: 1) crystals are small, 2) the degree of anisotropy in the attachment kinetics is large, and 3) the growth is slow. As these three qualitative criteria become lessened, branched growth becomes more likely.

For example, if $\alpha_{facet}$ is just below unity, then the facet stability is weak, as there can be little difference between the value of $\alpha$ at the corners and at the facet centers. In this case, branches form readily and will sprout from quite small crystals even when they are growing slowly. If $\alpha_{facet} \ll 1$, on the other hand, then faceted growth with be highly stable and a crystal may grow quite large before branches sprout. Of course, the value of $\alpha_{facet}$ depends on the surface supersaturation $\sigma_{surf}$, which in turn depends on the size of the crystal, the functional form of $\alpha_{facet}(\sigma_{surf})$, and the far-away supersaturation $\sigma_{\infty}$. So, determining the exact point at which branches appear becomes a nontrivial problem.

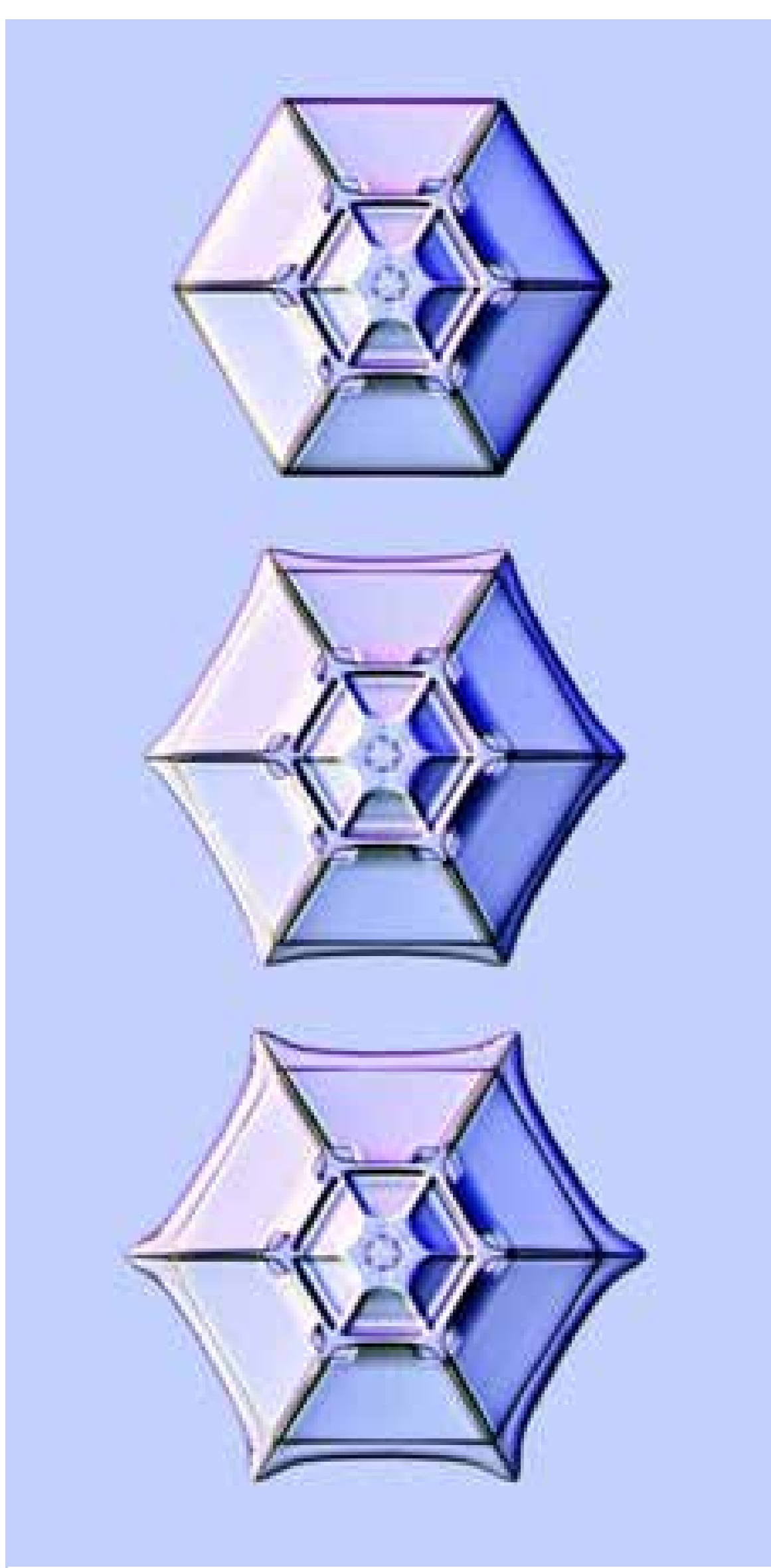

**Figure 3.5: This series of photographs shows branches sprouting from the six corners of a hexagonal snow crystal when the applied supersaturation was increased. Branching like this becomes more likely when a hexagonal crystal is large and/or its growth rate is fast.**

The formation of symmetrical branched snow crystals often results from the fact that faceting is more stable on smaller crystals. When a nascent snow crystal begins its existence, it is small and tends to grow into a faceted hexagonal prism. As it grows larger, at some point six branches will sprout simultaneously from the six corners of the hexagon, as seen in Figure 3.5. Whenever you see a branched snow crystal with six-fold symmetry, it is almost certain that the primary branches first sprouted in unison from a faceted ice crystal.

## The Branching Instability

The transition from stable faceted growth to branching is just one example of a more general phenomenon in diffusion-limited growth called the *Mullins-Sekerka Instability*, named after its discoverers William Mullins and Robert Sekerka [1963Mul, 1964Mul]. I also like to call it the *Branching Instability* because this is a more descriptive moniker, and using it avoids unnecessary jargon. The basic idea is shown in Figure 3.6 for the case of ice growth when $\alpha \approx 1$, but the Mullins-Sekerka instability applies to many other systems as well, whenever growth is limited by diffusion.

The hallmark of any instability is positive feedback, and Figure 3.6 illustrates this for the special case of an initially flat ice surface with $\alpha \approx 1$. If a small bump randomly appears on the surface, then the top of the bump sticks out slightly into the humid air above it. As a result, slightly more water vapor in the air diffuses to the top of the bump than diffuses to the flat surface around it. With this slight enhancement in material transported to it, the top of the bump grows slightly faster than its surroundings, and so it grows taller. Soon the bump sticks out even farther than it did before, causing it to grow even faster, which makes it stick out still farther, and so on. In this way, positive feedback yields a branching instability.

Note that the uniform surface growth shown in the top panel in Figure 3.6 is a perfectly valid solution to the diffusion equation; it is just not a stable solution. Any deviation from perfect flatness, no matter how small, will grow larger. Thus, the Mullins-Sekerka instability will eventually turn a simple growing structure into a complex, branched structure with an ever-changing morphology. When you get to the heart of the matter, this growth instability is why the simple process of water vapor freezing into ice creates the beautiful, complex snow crystal patterns we observe falling from the winter clouds.

Going back to Figure 3.5, we see that faceting provides a stabilizing influence that initially inhibits the Mullins-Sekerka instability. Anisotropic attachment kinetics provides a negative feedback that reduces perturbations and maintains the ever-so-slightly concave shape of the surface, as I described above. When there is no anisotropy in the attachment kinetics (for example when $\alpha \approx 1$), then the surface is always susceptible to the Mullins-Sekerka instability, and this is the case shown in Figure 3.6. In the opposite extreme, when there is a large anisotropy in the attachment kinetics ($\alpha_{facet} \ll 1$), then faceted growth will continue for quite some time.

The growth behavior of snow crystals is most often determined by a combination of branching and faceting, with the details depending on the crystal size, growth rate, attachment kinetics, and other factors. Faceting dominates in some regions of parameter space, while branching dominates in other regions. The complex interplay of the processes of faceting and branching is what gives snow crystal growth its especially rich phenomenology.

Another aspect of this topic to note is that abrupt transitions are a common hallmark of instabilities. For a simple analogy, consider a cylindrical rod balancing upright on one of its flat ends on a table. Add a bit of damping to the problem by placing the upright rod in a small bowl of honey. If you perturb the rod, perhaps by giving it a gentle tap, it will wobble a bit and then go back to its upright, balanced state. Thus, the motion of the rod is stable to minor perturbations. As you increase the length of the rod, however, it becomes less stable. For a sufficiently tall, slender rod, even a small tap will topple it. Moreover, when it topples, it will topple over completely; the transition from upright to not upright will be abrupt and dramatic. The transition from faceted to branched growth is similar: once branches sprout from the corners of a faceted prism, they grow rapidly outward and leave the facets far behind.

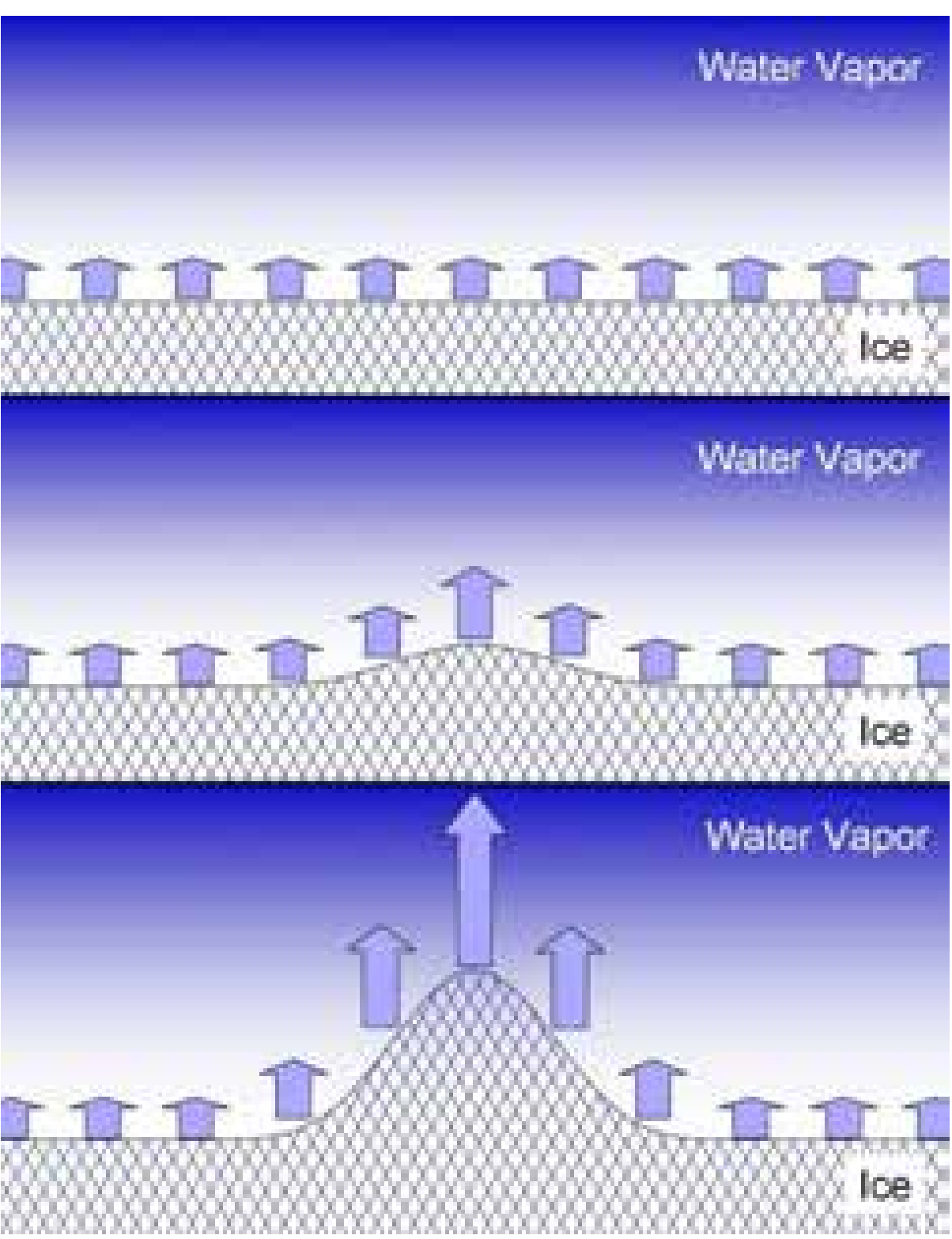

**Figure 3.6: The diffusion-limited growth of an initially flat surface with $\alpha \approx 1$ (top sketch) is susceptible to the Mullins-Sekerka instability, also known as the branching instability. If a small bump appears on the surface (center), it will stick out farther into the supersaturated medium, so the top of the bump will grow slightly faster than the surrounding flat surface. This initiates a positive feedback effect, causing the bump to become larger, increasing its relative growth rate even more (bottom). This illustration assumes zero anisotropy in the attachment kinetics ($\alpha \approx 1$) to eliminate the possibility of faceting and thus emphasize the Mullins-Sekerka instability.**

The branching instability is also well known for its repeated application in the formation of complex dendritic structures. Once a branch sprouts and grows outward, perturbations on its surfaces may again become amplified, thereby sprouting additional sidebranches, as illustrated in Figure 3.7. In principle, this could lead to sidebranches on the sidebranches, and so on, yielding intricate structures.

Dendritic snow crystals forming near -15 C are often characterized by a set of six primary branches that are decorated with copious sidebranches, as shown in Figure 3.8. Side-sidebranches are sometimes seen, although they are somewhat rare. More common is a mixture of both faceting and branching behaviors on a single crystal, producing the endless morphological variations we associate with snowflakes.

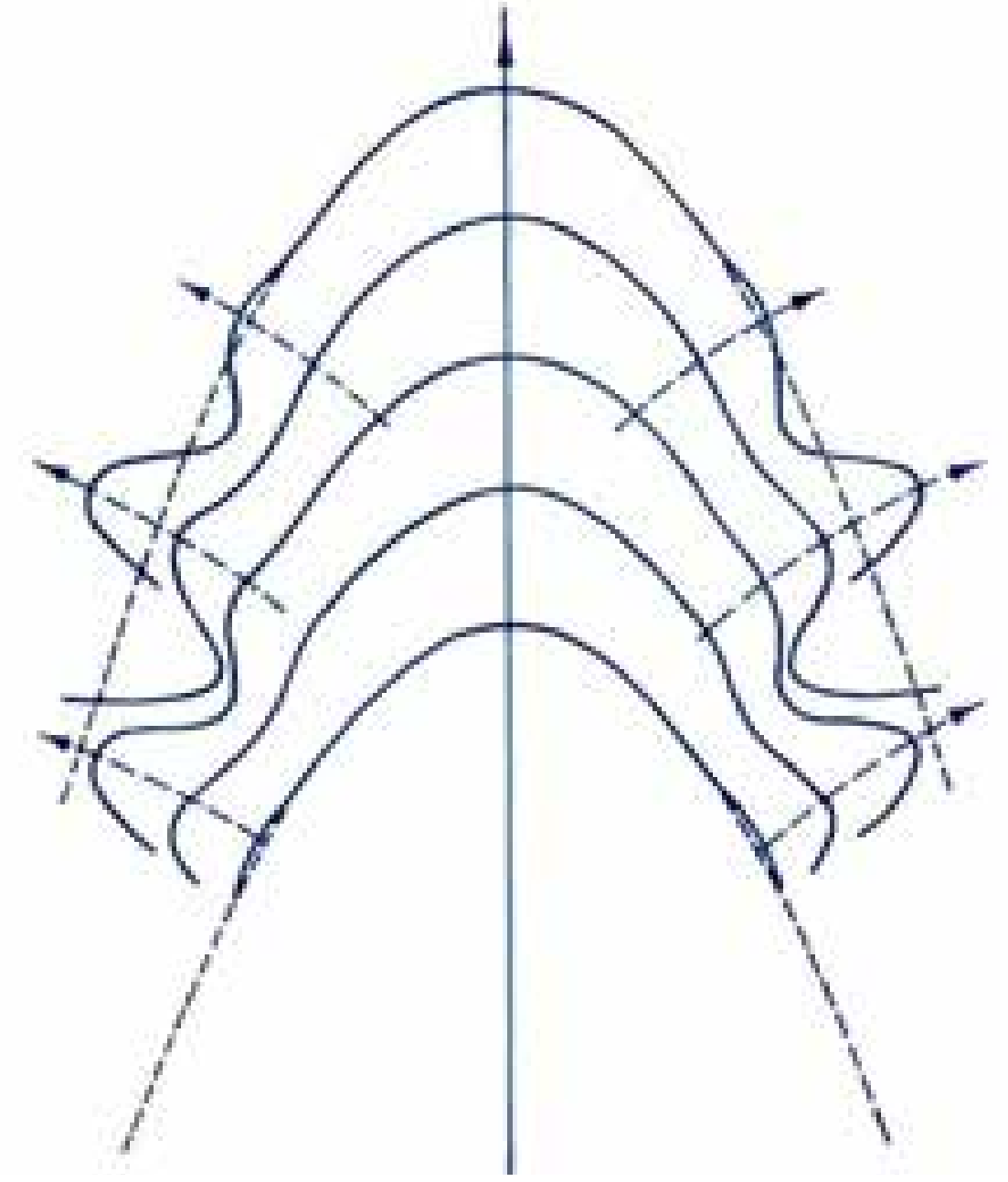

**Figure 3.7: A schematic diagram showing the creation of dendritic sidebranches resulting from repeated application of the Mullins-Sekerka instability [1980Lan]. Random perturbations on the sides of the growing tip typically yield somewhat erratically placed sidebranches. However, there is often a characteristic length scale involved in the formation of these perturbations, resulting in a somewhat regular spacing between adjacent sidebranches.**

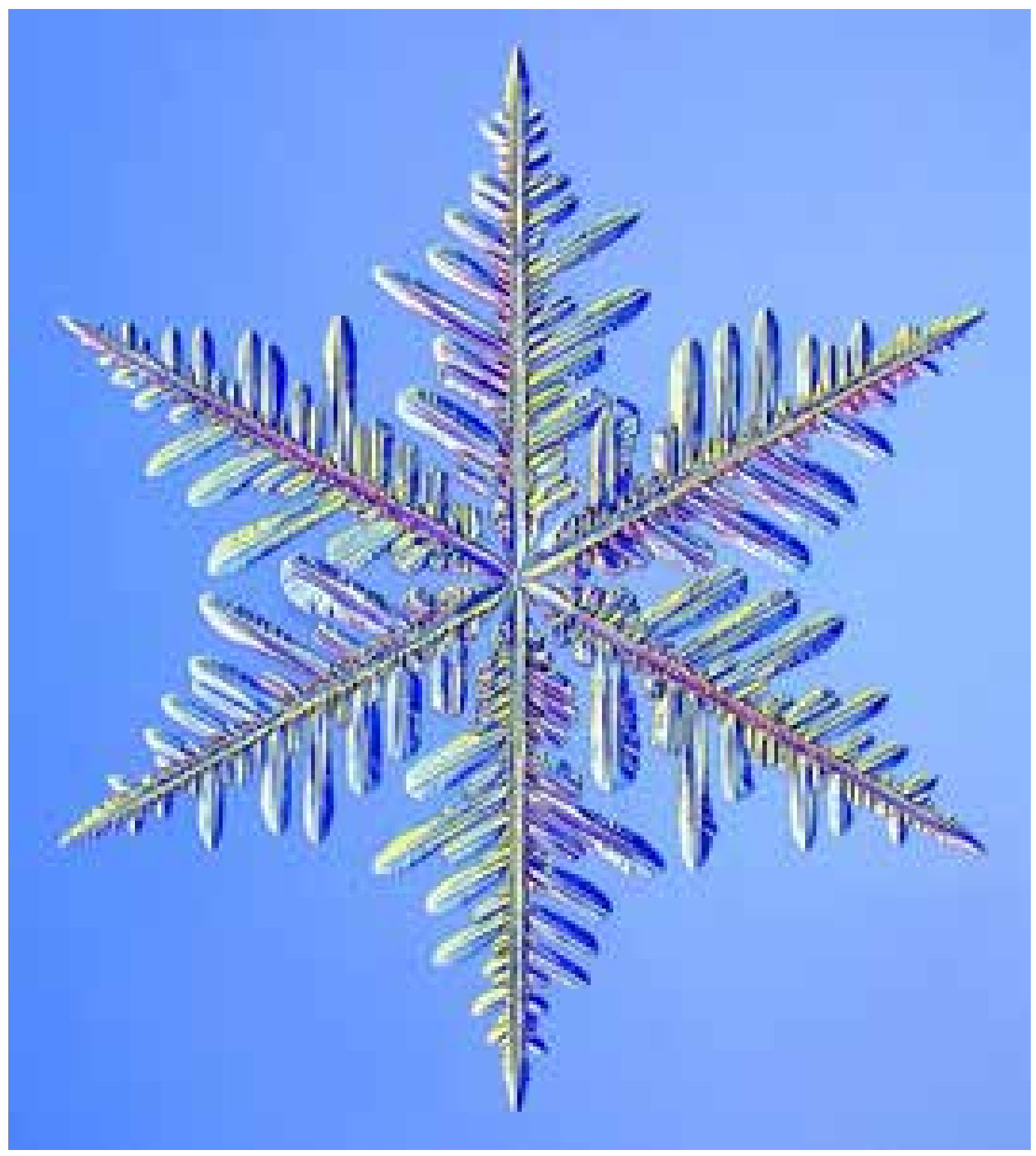

**Figure 3.8: Both branching and faceting play large roles in fernlike stellar dendrites like this one. The branching instability is clearly responsible for the copious sidebranching, and for the largely random sidebranch placement. However, this crystal began its life as a faceted hexagonal prism, because the six primary branches must have spouted from the small prism's six corners. Moreover, the crystal is thin and flat, indicating strong basal faceting even in the presence of highly developed dendritic branching.**

The chaotic nature of the branching instability becomes apparent in crystals like that shown in Figure 3.8. The six primary branches must have simultaneously sprouted from the corners of a tiny hexagonal prism, as the symmetry of these branches reflects the initial faceted order of the crystal. The sidebranches, on the other hand, arose from perturbations that occurred near the primary branch tips. Because these perturbations were somewhat random in nature, the sidebranches are largely uncorrelated among the primary branches, and even the two sides of a single primary branch exhibit quite different sidebranches. Thus, the detailed six-fold symmetry of the crystal in Figure 3.8 is quite poor. It is possible, however, to stimulate symmetrical sidebranches with not-so-random perturbations, and I discuss this phenomenon later in the chapter.

### Additional Snow Crystal Growth Instabilities

Nature abounds with dynamical instabilities in nonequilibrium systems, even though they are not much discussed in early science teaching. For example, when sunlight heats the ground, the adjacent air becomes warmer than the air above it, and this situation is unstable to convection. The convective instability drives the wind, the clouds, and much of our weather. When the resulting wind blows over a still lake, the surface of the lake becomes unstable to the formation of ripples and waves. When waves reach the shore, they become unstable and break. Whenever you see any kind of complex structure in nature, it is a good bet that some dynamical instabilities were involved in its formation. We do not talk about them much in science teaching because their behavior can be extremely complicated and difficult to understand. Nevertheless, understandable or not, instabilities are everywhere in the natural world.

In snow crystal formation I count three distinct growth instabilities that have been found so far:

**The Mullins-Sekerka instability** is clearly the big player, producing branched structures and a host of competitive growth behaviors in complex snow crystals. Because the underlying physics in this phenomenon is well understood, the branching instability is very-much amenable to computer modeling.

**The edge-sharpening instability** (see Chapter 3) is also present in Figure 3.8, putting the "flake" in snowflake by keeping this crystal thin and flat. Being an especially complicated phenomenon involving bulk diffusion, surface diffusion, and attachment kinetics, the ESI remains something of a scientific puzzle.

**The electric-needle instability** (see Chapter 8) provides an excellent tool for studying snow crystal formation under well controlled initial conditions. This phenomenon is reasonably well understood, being essentially an extension of normal dendritic growth, but several important aspects remain unstudied.

As we continue to probe more deeply into the physics of snow crystal formation, there are likely additional dynamical instabilities lurking in the varied molecular processes occurring on the ice surface, still waiting to be discovered.

## 3.2 Free Dendrites

When the branching instability is operating in high gear, dendritic structures with copious sidebranches are the norm. Snow crystal dendrites appear whenever the supersaturation is sufficiently high, and their morphology can be quite different at different temperatures. The most well-known example is the fernlike stellar dendrite, for example that shown in Figure 3.8. These are the largest and fastest growing natural snow crystals, and they readily form at temperatures near -15 C.

Individual fernlike dendritic branches like those shown in Figure 3.9 are easy to grow in the lab as well, and they exhibit several characteristic traits: 1) The branching structure is mainly confined to a flat plane because of strong basal faceting; 2) The tip of the dendritic branch grows outward at a constant tip velocity $v_{tip}$ that increases approximately linearly with the far-away supersaturation, so $v_{tip} \sim \sigma_\infty$; 3) The radius of curvature of the tip is equal to about $R_{tip} \approx 1$ micron, a value that is roughly independent of supersaturation; 4) Each distinct sidebranch grows out at an angle of 60 degrees relative to the primary branch; 5) New sidebranches typically spout at a characteristic distance from the tip that is roughly several times $R_{tip}$; 6) The sidebranch spacing is generally erratic with little correlation on either side of a primary branch.

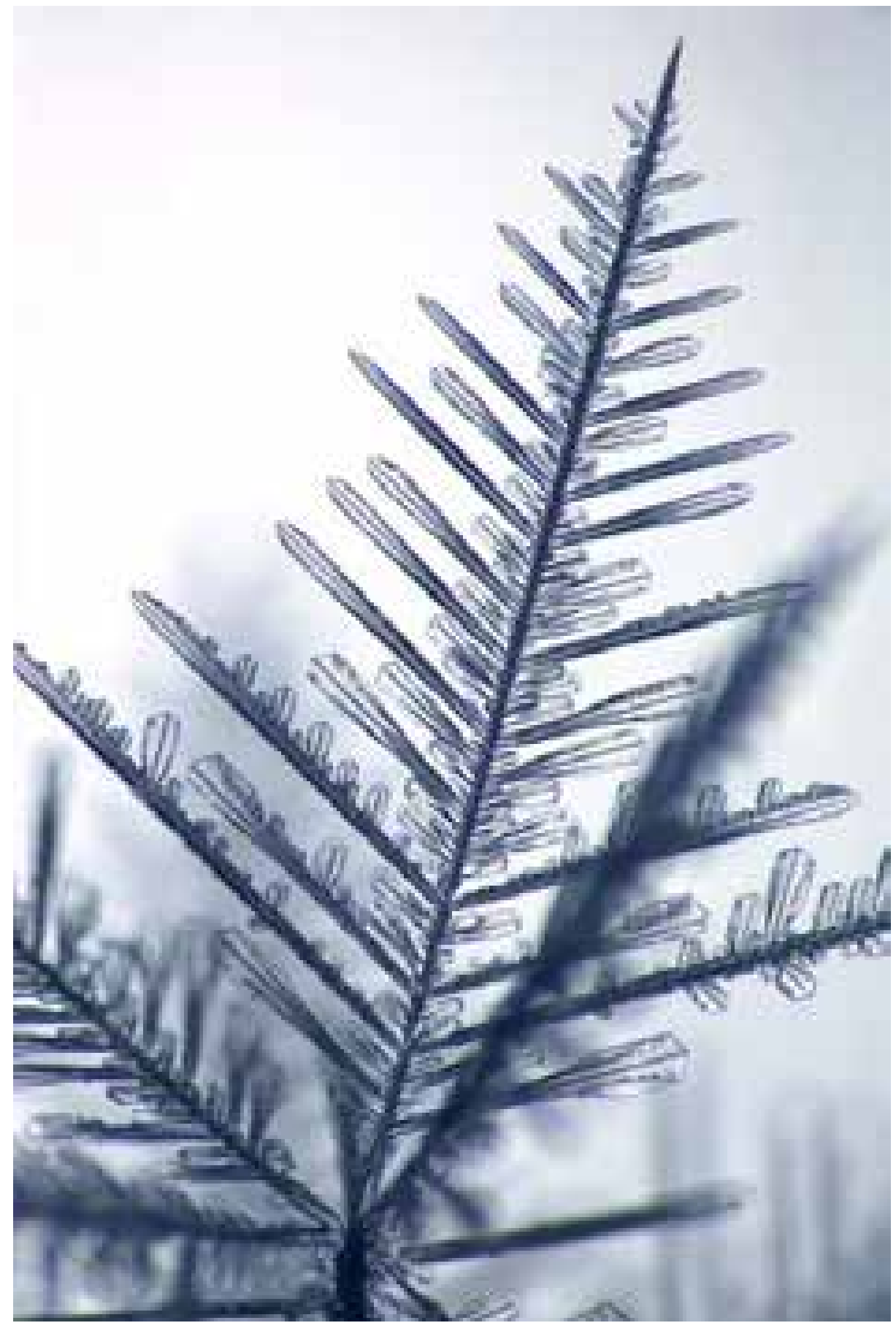

**Figure 3.9: Several fernlike dendrites growing out from the tip of a wire substrate at a temperature of -15 C (only a single dendrite is in sharp focus in this photo). Although they are oriented randomly with respect to the wire, each complex dendritic structure is mostly confined to a thin plane by slow basal growth.**

At lower air pressures, $R_{tip}$ increases with decreasing pressure, and snow-crystal dendritic structures are generally more complex at higher air pressures [1976Gon]. I delve a bit more into the mathematical aspects of dendrite formation in the section on solvability theory below.

An individual branch like the one in Figure 3.9 is called a "free" dendrite because it is a self-assembling structure that can be considered free from the constraints of container walls or competing crystals. The branch grows out into open space with a uniform far-away supersaturation $\sigma_\infty$. Moreover, the overall tip morphology of a free dendrite is essentially independent of time. If you photographed the near-tip structure at different times, you would find that the photos all looked about the same. The detailed placement of the sidebranches with respect to the tip is different at different times, but the overall morphology remains constant. Also, the initial origin of the dendritic branch largely unimportant; once it becomes fully developed, the branch automatically assumes its characteristic shape and properties.

## Fishbone Dendrites near -5 C

While fernlike dendrites near -15 C are something of a canonical snow crystal form, other dendritic structures appearing at different temperatures are also worthy of attention. The "fishbone" dendrites shown in Figures 3.10 through 3.12 are especially pronounced, as they grow rapidly near -5 C and make up the "fishbone peak" often seen in snow crystal diffusion chambers (see Chapters 6 and 8).

While they look quite different, fishbones are also free dendrites with many of the same characteristic traits as the fernlike dendrites just described. However, the sidebranches are not conveniently confined to a nearly flat plane, and thus their structure is not so easy to convey using a single photograph. Moreover, the formation of fishbone dendrites requires high supersaturation levels that do not occur in nature, so they are entirely a laboratory creation [2009Lib1].

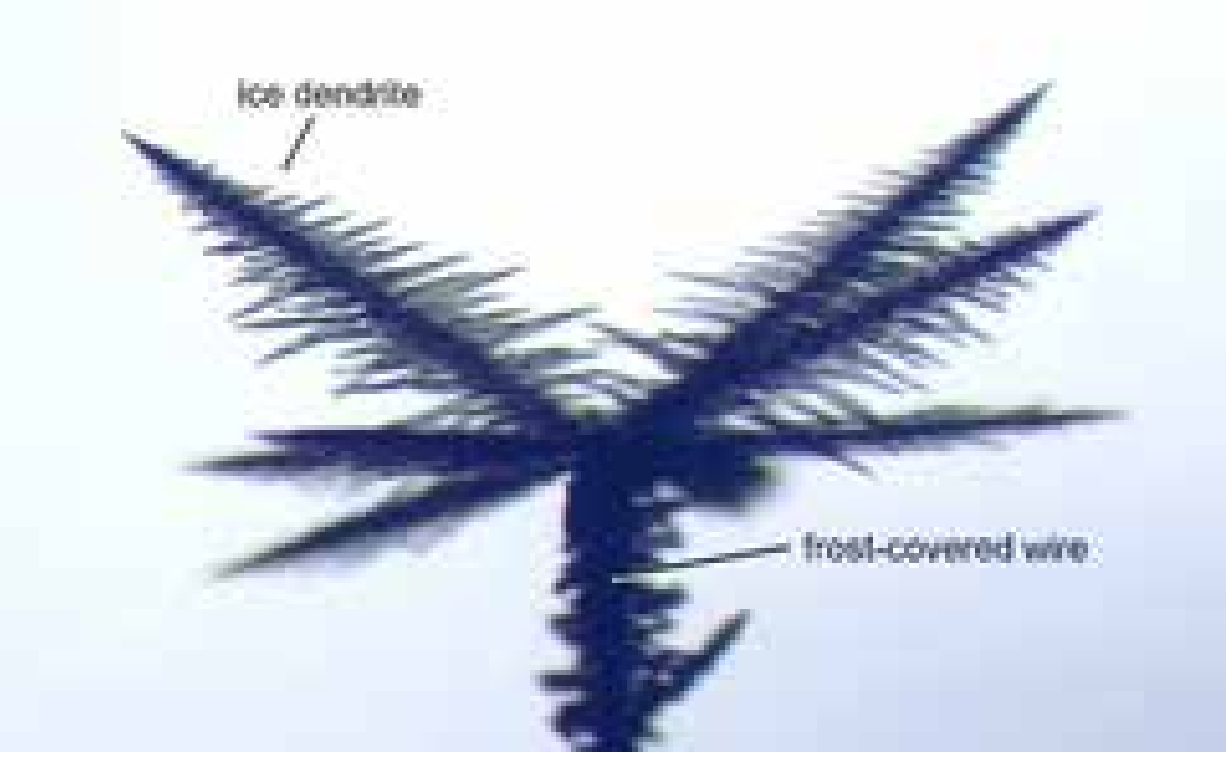

**Figure 3.10: Several "fishbone" dendrites growing out from a frost-covered wire at a temperature of -5 C. Note the overall similarities between the different dendritic branches.**

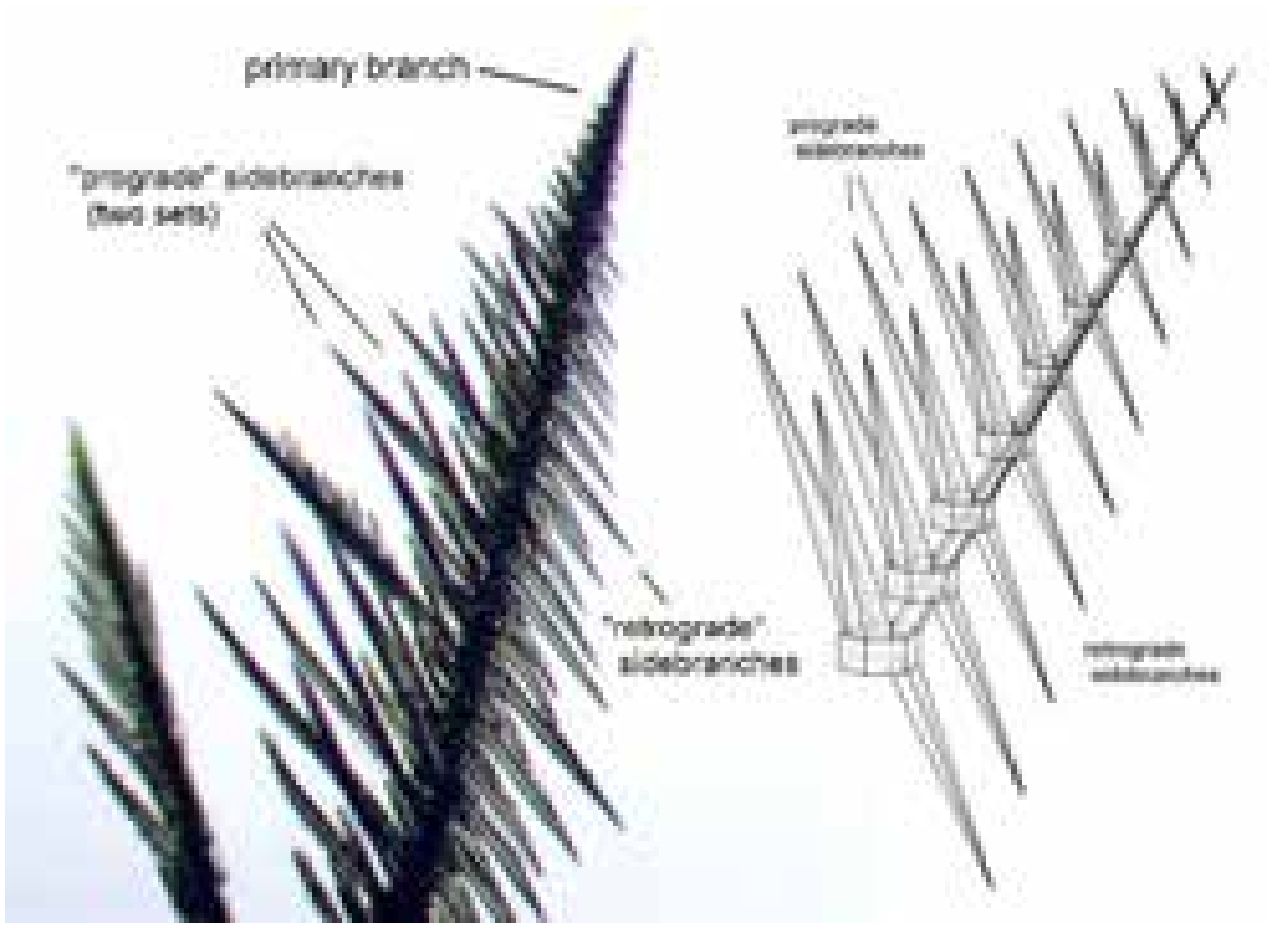

**Figure 3.11: The sidebranch structure of a fishbone dendrite is more three-dimensional than a fernlike dendrite. This figure compares a photograph with a sketch that shows the orientations of the different sidebranches with respect to the ice crystal axes (defined by the hexagonal prisms in the sketch). Unlike fernlike dendrites, fishbone dendrites are clearly not confined to a flat growth plane.**

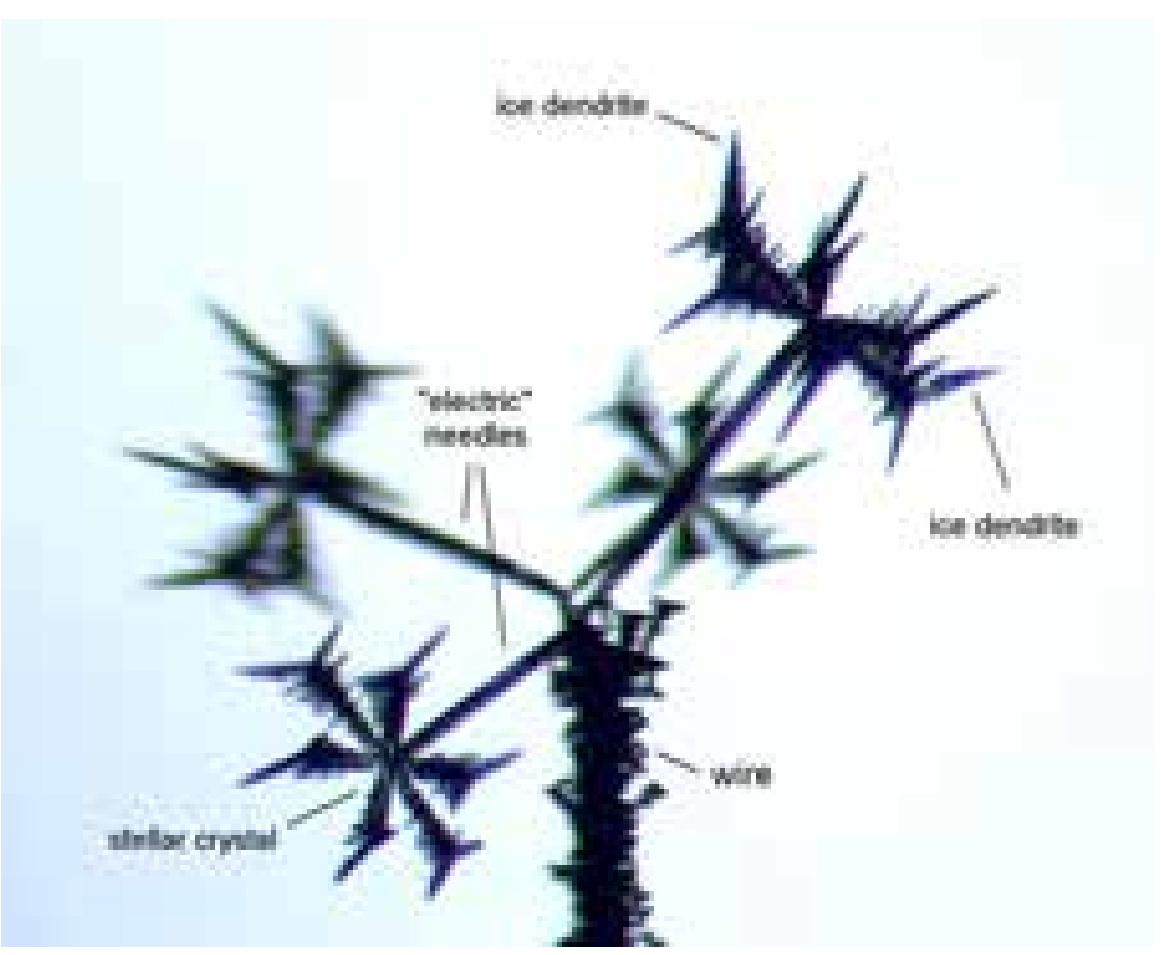

**Figure 3.12: The crystal structure of fishbone dendrites can be difficult to discern from optical photographs like those in the previous two figures. Observing their growth on c-axis e-needles (see Chapter 8) allows an unambiguous determination of the sidebranch orientations with respect to the crystal axes, as described in [2009Lib1].**

These two special cases – fernlike dendrites near -15 C and fishbone dendrites near -5 C – are the fastest growing and most distinctive examples from the family of snow crystal dendritic structures. I will not discuss here the full spectrum of free dendrites that appear in the snow crystal morphology diagram, but photographs from a broad range of temperatures are presented in Chapter 8.

Note that the direction of dendrite tip growth in snow crystals depends on both temperature and supersaturation. Fernlike dendrites have $v_{tip}$ aligned with the crystal a-axis, but typically the growth direction is not aligned with any particular axis but is determined by details of the attachment kinetics. In particular, the growth direction depends on the ratio of $\alpha_{basal}$ to $\alpha_{prism}$, which depends on both temperature and supersaturation at the growing tip. Also notable is that no snow crystal free dendrites grow along the c-axis, although e-needles (see Chapter 8) can be coaxed to grow in that direction using chemical vapor additives.

## Ice Dendrites Growing from Liquid Water

It is useful, at this point, to expand our horizons beyond the snow crystal realm by examining ice free dendrites growing from liquid water. These appear readily in nature in the form of pond crystals, like that shown in Figure 3.13, and several different dendrite tip morphologies are illustrated in Figure 3.14.

While snow crystals grow in the two-dimensional parameter space of temperature and supersaturation, ice growing from liquid water is mainly described by a single variable, the *undercooling*, which is the far-away temperature of the water bath in which the ice dendrites grow. Pressure is the missing second parametric variable in this case, missing because few experiments have examined ice

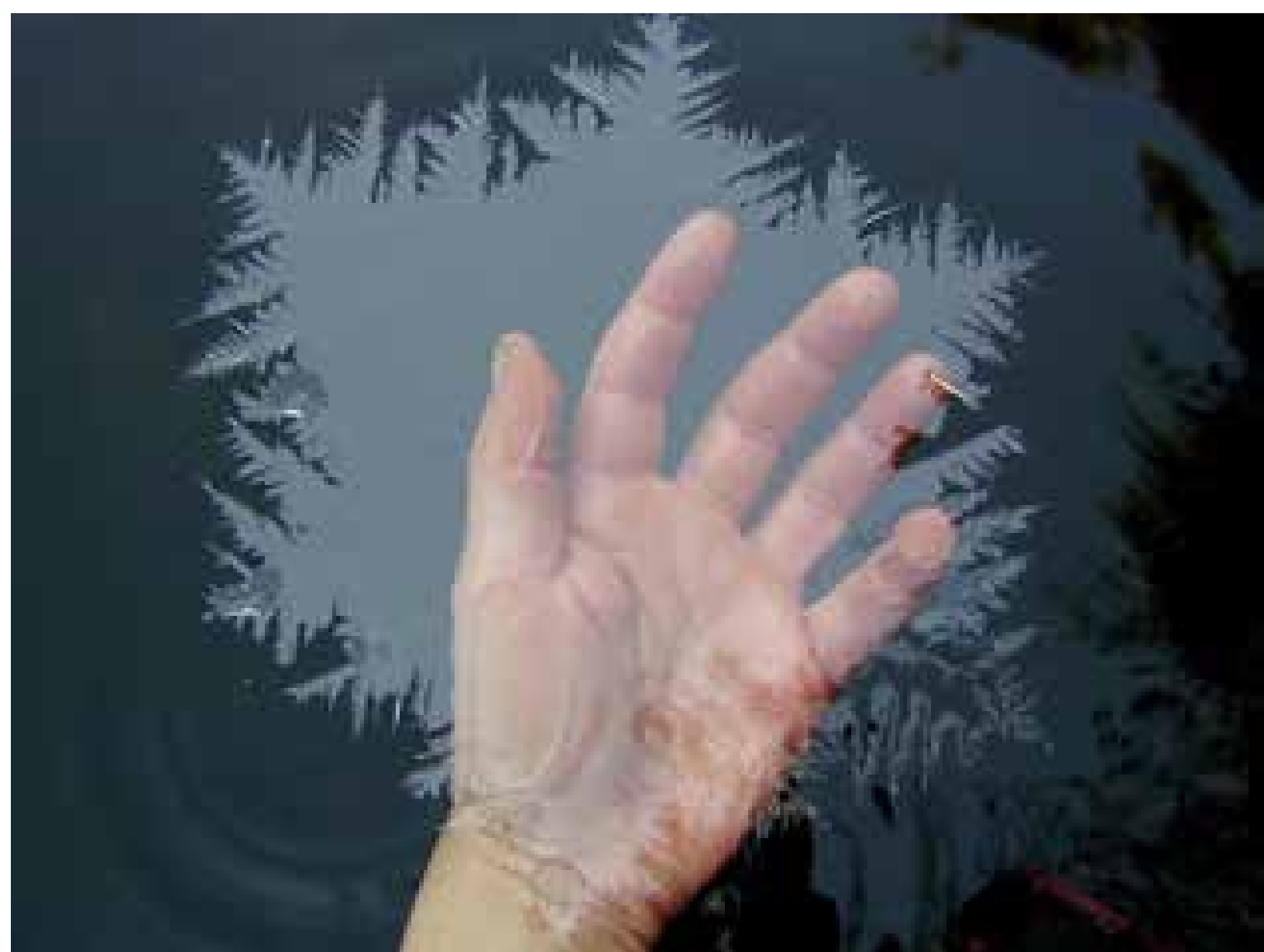

**Figure 3.13: This photo shows a plate-like dendritic ice crystal that formed on the surface of a still pond. It shows the same six-fold symmetry of snow crystals, arising from the underlying symmetry of the ice lattice. It also exhibits dendritic sidebranching brought about by diffusion-limited growth, in this case the diffusion of latent heat created during solidification. Strong basal faceting is responsible for the overall plate-like structure, indicating the importance of attachment kinetics in the formation of this crystal. [Photo by Bathsheba Grossman.]**

growth at the high pressures needed to see interesting effects [1997Mar, 2005Mar].

Ice growth in liquid water bears a good resemblance to the solidification of metals from their melts, and the latter subject has received much attention from metallurgists in the scientific literature. In both cases, the growth is strongly limited by the diffusion of latent heat generated during solidification, and in both cases dendritic structures appears as a result. Figure 3.15 shows an example of a dendrite forming in succinonitrile, which is transparent and often used as a metal analog. Unlike ice from liquid water, however, there is negligible faceting in succinonitrile and other metal analogs, meaning that attachment kinetics is not an important factor in these systems.

## A Brief Classification of Solidification Systems

To avoid confusion regarding which physical effects are important and which can be neglected, it is useful to list the different types of systems in which dendritic structures arise from diffusion-limited crystal growth. One soon finds that while there is a large scientific literature on dendrite formation, and solidification more generally, a great deal of it cannot be directly applied to snow crystal growth. The underlying physics is so dissimilar in the various materials that each system must be approached differently.

**Unfaceted Solidification from the Melt.** A great deal of scientific literature on solidification can be found in this category. Succinonitrile and pivalic acid are two oft-studied materials, popular because they are easy to work with and are considered good proxies for simple metallurgical systems. Listing the dominant physical effects that need to be considered, in order of important, we obtain:

1) Thermal diffusion. Removing latent heat is a major consideration in solidification from the melt, so the thermal diffusion equation is of paramount importance. Dendritic structures in this system arise mainly from thermal-diffusion-limited growth.

2) Surface energy. Although less important than thermal diffusion, this physical effect sets the scale for $R_{tip}$ and therefore the overall structure of a free dendrite. Together, thermal diffusion and surface energy effects define the primary features seen in metallurgical solidification.

3) Anisotropic surface energy. As we discuss later in this chapter, stable dendrites require

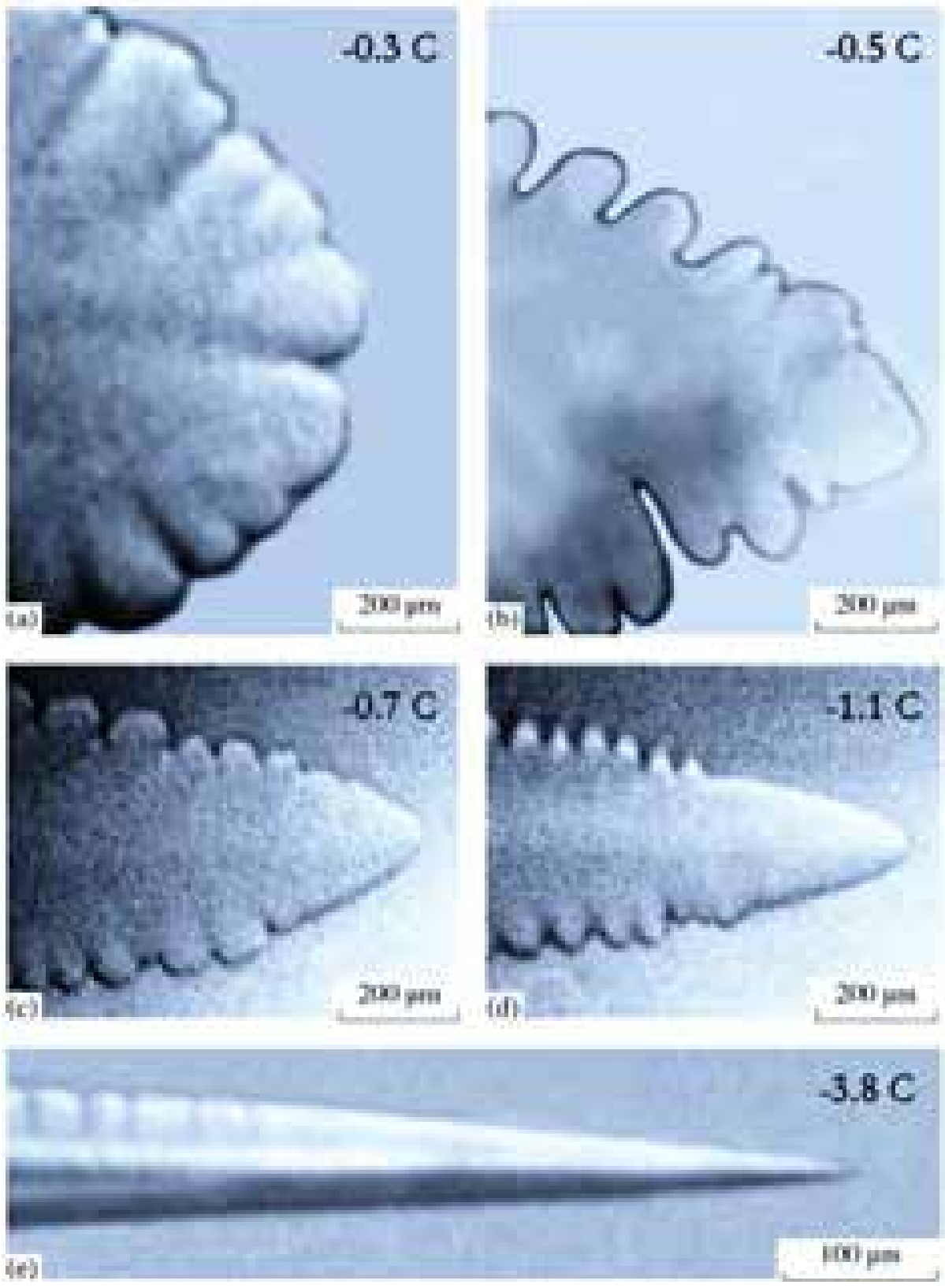

**Figure 3.14: Several optical photographs showing ice free-dendrite growth from liquid water as a function of the bath temperature [2004Shi]. Near the freezing point (-0.3 C), the growth is relatively slow, the tip radius is large, and tip splitting can occur. As the bath temperature decreases, the growth rate increases and the tip radius decreases, yielding somewhat different morphological features.**

some anisotropy in the boundary conditions, and here that is provided by an anisotropic surface energy.

4) Attachment kinetics. Usually neglected entirely, as the attachment kinetics is often so fast that it does not limit growth significantly. This applies only to systems that exhibit no faceting, which is true for succinonitrile (Figure 3.15) and most metals and metal analogs.

5) Particle diffusion. Also usually neglected, because particle diffusion is not needed to bring material to the solidification surface, as liquid is always present. Thus, particle diffusion does not limit the growth significantly.

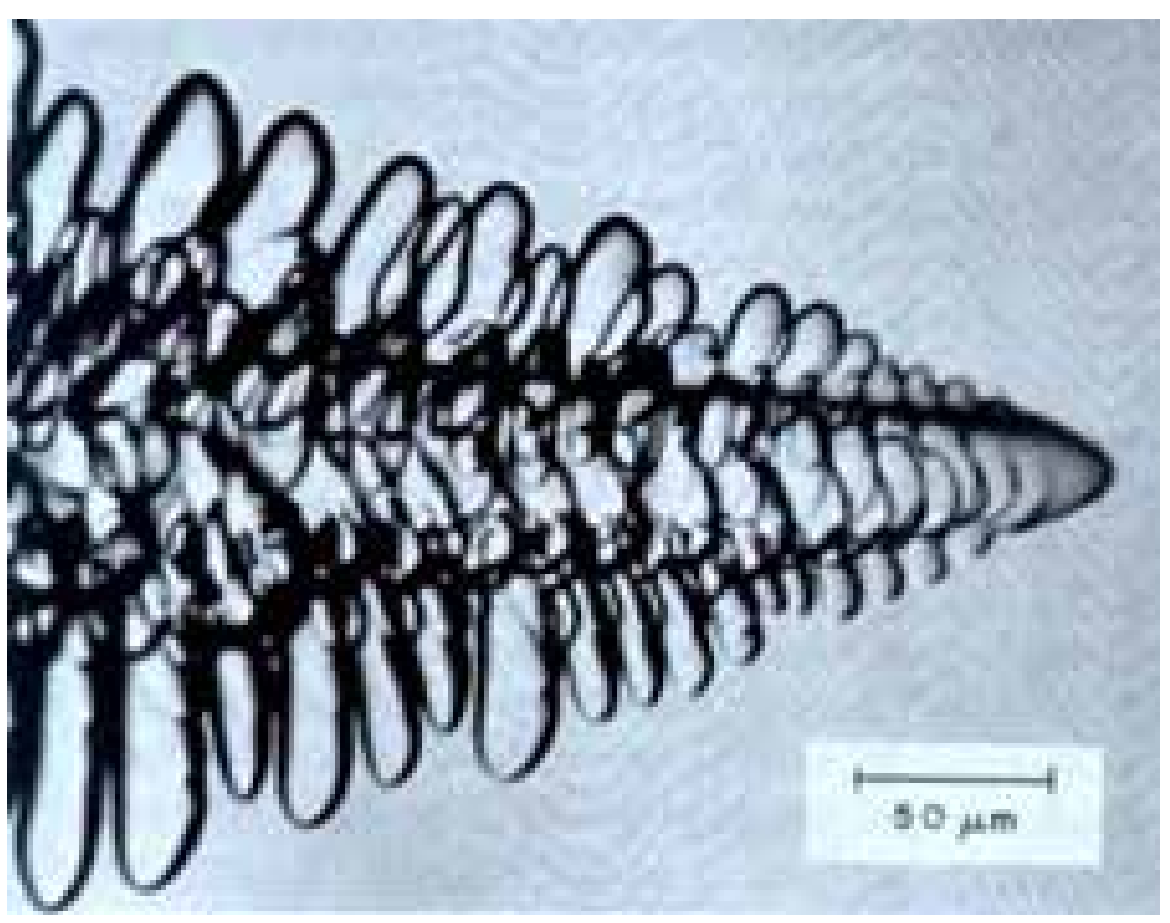

**Figure 3.15: A free dendrite forming in succinonitrile from its melt [1976Gli]. This transparent material is often used as a metal analog in studies of solidification, as it forms dendritic structures that are typical for solidification from the melt when attachment kinetics is not an important factor.**

**Faceted Solidification from Vapor.** Snow crystal formation stands out as perhaps the most studied example of solidification of a high-vapor-pressure system. Material science, usually classified as a branch of engineering, has little interest in materials that readily evaporate away, so high-vapor-pressure materials have received remarkably little attention over the years. Again, listing the dominant physical effects, in order of importance, we obtain:

1) Particle diffusion. In air, particle diffusion transports water vapor molecules to the ice surface, and this slow process greatly limits growth. Particle diffusion is responsible for branching and essentially all the complex structure seen in snow crystals.

2) Attachment kinetics. In the formation of snow crystal free dendrites, attachment kinetics sets the scale for $R_{tip}$ and the overall branched structure, as I describe in detail below. Together, particle diffusion and attachment kinetics define the primary features seen in snow crystal growth.

3) Anisotropic attachment kinetics. Snow crystal attachment kinetics are generally highly

anisotropic, and this tends to yield strongly faceted structures.

4) Thermal diffusion. A minor effect compared to particle diffusion, thermal diffusion is often neglected. It plays a role close to 0 C, but this can often by approximated by a simple rescaling of $\sigma_\infty$ (as is discussed later in this chapter).

5) Surface energy. Almost negligible because surface-energy effects are dwarfed by similar effects arising from attachment kinetics. However, the surface energy is necessary in modeling to avoid unphysical results at very low supersaturations (see Chapter 5).

6) Anisotropic surface energy. Negligible. Surface energy effects are small to begin with, plus the surface energy anisotropy in ice is quite small. Any residual effects are dwarfed by similar effects from anisotropic attachment kinetics.

7) Surface diffusion. This is nominally part of the attachment kinetics, but surface diffusion introduces nonlocal effects that are not included with a simple attachment coefficient. It is not yet known how important surface-diffusion effects are in snow crystal growth.

**Chemical Vapor Deposition.** So much work has been done with CVD that it deserves a separate listing all its own, although in principle is could be included in previous categories. The primary focus in CVD systems has been on technological applications, so these materials almost always exhibit low vapor pressures. Theoretical descriptions of CVD often make an implicit assumption of zero vapor pressure from the outset, which greatly simplifies the theory but also immediately changes the underlying physics compared to high-vapor-pressure systems. Thus, the vast literature on CVD systems often has little relation to snow crystal growth, so I do not delve into the topic at all in this book. I encourage the reader to sample some of the excellent books devoted to CVD systems, but keep in mind that a nonzero vapor pressure is not much considered in these studies.

**Unfaceted Solidification from Vapor.** To my knowledge (quite limited in this case), the formation of unfaceted free dendrites from vapor has received little scientific attention. I performed a few experiments using carbon tetrabromide, as this seemed to be a convenient test system, but little came out of those observations. Important physical effects could include all the items mentioned above – particle diffusion, heat diffusion, surface energy, and attachment kinetics, all to varying degrees depending on the specific material under consideration. With few practical applications, substantial experimental challenges, and especially complex input physics, it is perhaps little surprise that vapor solidification of free dendrites has not been a popular research topic.

**Faceted Solidification from the Melt.** Ice growth from the melt falls into this category, as is clearly indicated by the presence of strong basal faceting like that shown in Figure 3.13 (see also [2017Lib] and references therein). The dominant physical effects are the same as with unfaceted solidification from the melt, except now one must include effects from both anisotropic attachment kinetics and anisotropic surface energy; neither is negligible compared with the other. Unfortunately, this complicates matters substantially and creates a pretty full plate on the theory side.

Again, there has been very little scientific research focusing on faceted dendritic growth from the melt. Most research into solidification from the melt tends to ignore attachment kinetics entirely, as this simplifies the mathematics, and this is a reasonable assumption with unfaceted metallic systems. The ice/water system is a good example of the current state of affairs in this category; basal faceting clearly plays an important role, but attachment kinetic at the ice/water interface has received little attention (see Chapter 12). Several ice/water solidification studies ignore attachment kinetics entirely. Given the ease of creating and studying ice structures from liquid water, and the many recent advances in numerical modeling, this topic seems ripe for additional experimental research.

## Sidebranch Competition

Another manifestation of the Mullins-Sekerka instability can be seen in the development of sidebranches after they sprout near a dendrite tip. For example, consider the fernlike stellar dendrite shown in Figure 3.8. As the individual sidebranches grow longer, each competes with its neighbors for available water vapor. If one branch becomes slightly longer than others nearby, then it sticks out farther into the humid air and shields its neighbors. Diffusion brings the longer branch a greater supply of water vapor at the expense of the neighboring branches. Soon the long branch shoots ahead while its immediate neighbors are greatly stunted. The underlying physical effect is essentially the same as with the Mullins-Sekerka instability described above.

Over time, this competition plays out on many length scales, so the spacing between the fastest-growing sidebranches becomes ever larger, as a few players become dominant by appropriating available resources at the expense of their neighbors. The result can be seen in the sidebranching characteristics of most fernlike stellar dendrites, including the example shown in Figure 3.8. Diffusion-driven competition between neighboring structures is a common feature in snow crystal growth dynamics, and many examples can be found scattered throughout this book.

In socio-economic circles, a similar phenomenon is sometimes called the *Matthew Effect*, from the biblical quote: "For unto every one that hath shall be given, and he shall have abundance: but from him that hath not shall be taken away even that which he hath." Matthew 25:29.

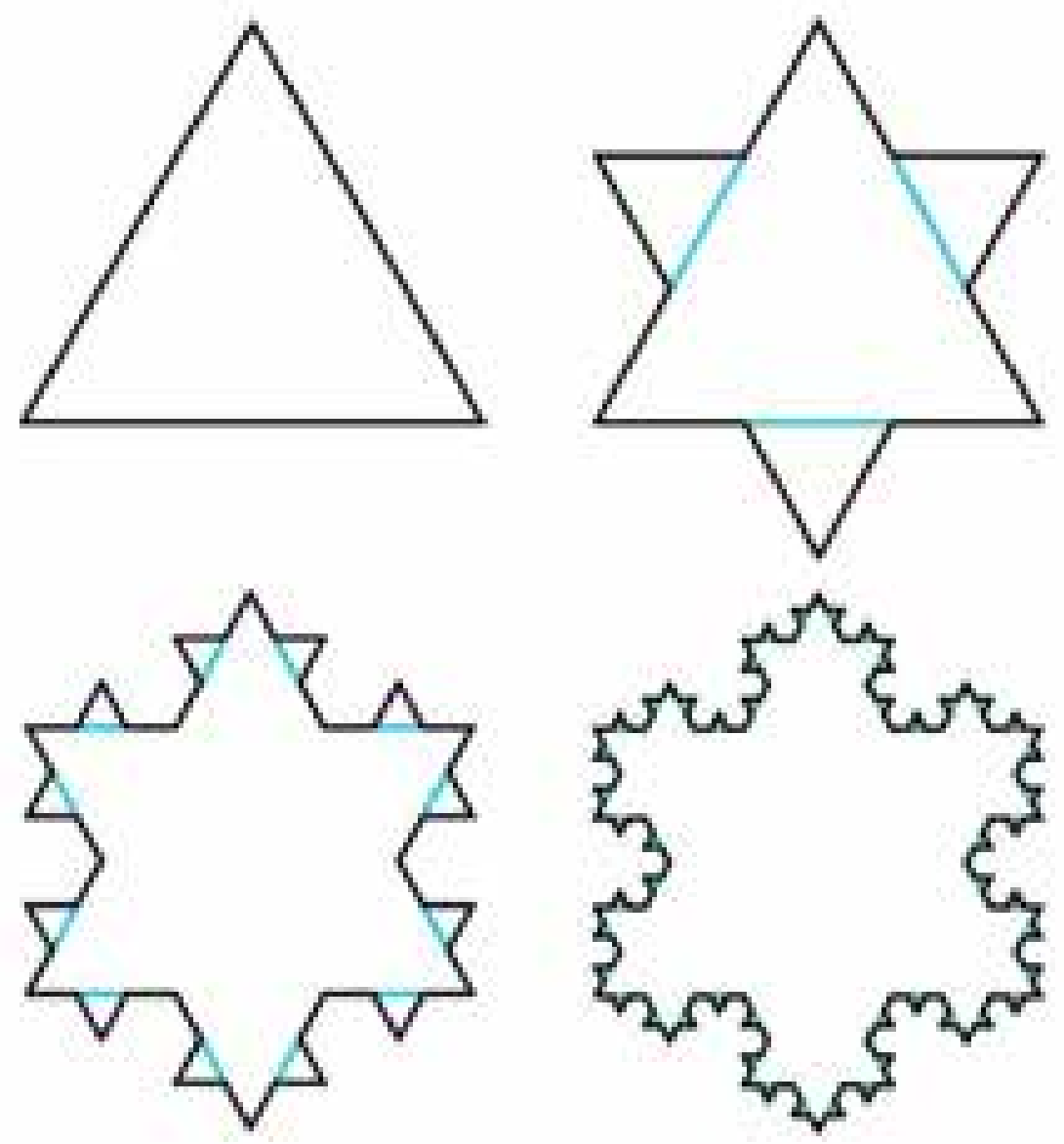

**Figure 3.16: This sketch shows the construction of geometrical curve known as the *Koch snowflake*, first described by Swedish mathematician Helge von Koch in 1904. It is one of the earliest known examples of a *fractal structure*. As ever-smaller triangular sidebranches are attached *ad infinitum*, the area of the Koch snowflake converges to 8/5 times the area of the original triangle, while its perimeter length diverges to infinity. Consequently, the Koch snowflake has a finite area bounded by an infinitely long perimeter. Although this fractal structure bears some resemblance to a stellar dendrite snow crystal, there is little real connection between fractal mathematics and the physics of snow crystal formation.**

## Fractal Structure

I have not found that the concepts of fractal mathematics add much to our understanding of snow-crystal formation and structure. Perhaps this view results from my perspective as an experimental physicist focusing on the material-science and crystal-growth aspects of this problem. But my bias is reinforced by the fact that fractal mathematics provides essentially no predictive power when it comes to understanding snow crystal formation.

Nevertheless, snow crystals do exhibit some fractal characteristics. The most apparent of these is a degree of self-similarity in the formation of dendritic structures, as illustrated in Figures 3.16 and 3.17. Primary branches yield sidebranches, and these can yield side-sidebranches, etc. If a dendrite sidebranch develops sufficiently, its overall structure will be indistinguishable from the central branch from which it arose. And the same would be true of side-sidebranches, if they matured to the same extent.

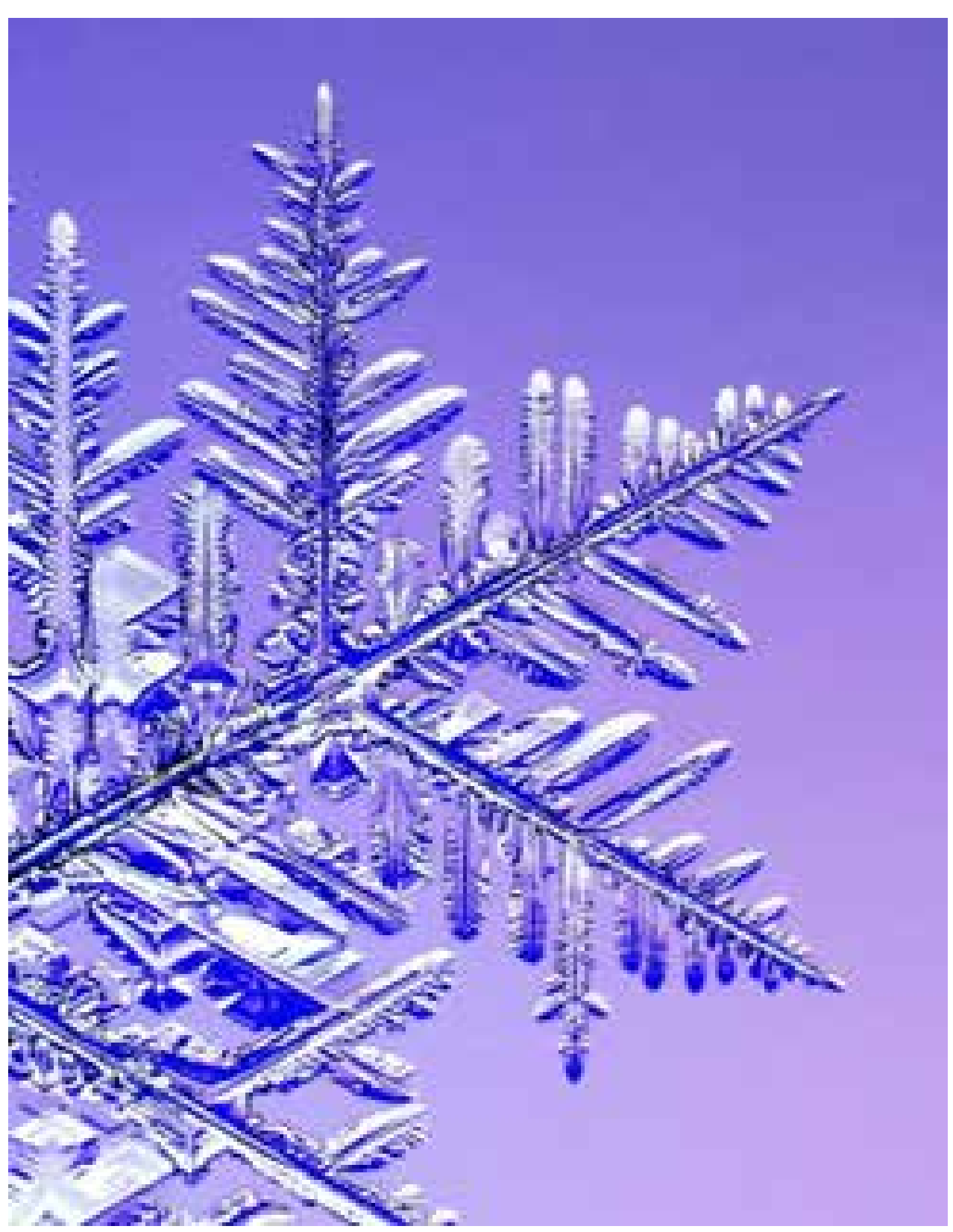

**Figure 3.17: This photo shows a close-up of one section of an exceptionally large fernlike stellar dendrite. The crystal exhibits a somewhat self-similar fractal structure with sidebranches begetting side-sidebranches, and even a few side-side-sidebranches.**

Observations generally reveal that the degree of self-similarity seen in snow crystal structure is relatively minor. Moreover, the concept of self-similarity does not provide a physics-based explanation of the formation of the dendritic structure in the first place – that requires the Mullins-Sekerka instability. Explaining snow crystal structure requires a broad understanding of diffusion-limited growth, attachment kinetics, and ultimately the molecular dynamics of the ice-crystal surface. Saying that a snowflake has some self-similar fractal characteristics is an accurate description, but, by itself, this description does not provide many useful insights that allow one to comprehend the underlying physical phenomenon.

## 3.3 Diffusion Physics in Snow Crystal Growth

Now that we have examined a few of the more prominent morphological effects of diffusion on snow crystal growth, it is time to delve into the underlying mathematics. Most textbooks on mathematical physics derive the diffusion equation and examine its solution, and I will assume that the reader already has a basic understanding of diffusion physics. My focus in the following discussion, therefore, will be on describing how to apply the diffusion equation to the specific problem at hand. In addition, I have relegated some of the more tedious mathematical details to Appendix A, focusing here mainly on physical concepts and specific model systems. A useful list of variables and physical constants can be found near the front of this book.

### Particle Diffusion

We begin with the particle diffusion equation that describes the transport of water molecules through the air

$$\frac{\partial c}{\partial t} = D_{air} \nabla^2 c \qquad (3.1)$$

where $c(\vec{x})$ is the number density of water molecules, $D_{air}$ is the diffusion constant for water molecules in air, and $\vec{x}$ is the position vector. For typical atmospheric conditions, $D_{air} \approx 10^{-5}$ m$^2$/sec.

If the temperature is equal to a fixed value everywhere (the isothermal approximation), then Figure 3.1 can be rewritten in terms of the supersaturation as

$$\frac{\partial \sigma}{\partial t} = D_{air} \nabla^2 \sigma \qquad (3.2)$$

where $\sigma(\vec{x})$ is defined by

$$\sigma(\vec{x}) = \frac{c(\vec{x}) - c_{sat}}{c_{sat}} \qquad (3.3)$$

and $c_{sat}$ is the saturated water vapor density, equal to $c$ above a flat ice surface in equilibrium with the vapor phase. Here we assumed that $c_{sat}$ is a constant independent of $\vec{x}$, which is true in the isothermal approximation. Because the values of both $c$ and $\sigma$ vary with position around a growing crystal, I often refer to the *fields* $c(\vec{x})$ and $\sigma(\vec{x})$.

In addition to the isothermal approximation, we can also employ a quasi-static approximation that reduces Figure 3.2 to Laplace's equation

$$\nabla^2\sigma = 0 \qquad (3.4)$$

This is a good approximation because the relaxation timescale for water vapor diffusion in air is typically very short compared to the growth time of a snow crystal.

To see this, consider suddenly placing a snow crystal into a uniform body of preexisting supersaturated air. Immediately the crystal will begin growing and thereby create around it a region somewhat depleted of water vapor. The size of this depleted "hole" in the water vapor density will be a few times larger than $R$, the size of the crystal, and its creation will take a time roughly equal to $\tau_{diffusion} \approx R^2/D$. (This is a well-known result from diffusion physics. To make the notation more compact I often use $D$ in place of $D_{air}$.) If we put in some typical numbers, taking $D = D_{air} \approx 2 \times 10^{-5}\ \mathrm{m^2/sec}$ and $R \approx 1\ \mathrm{mm}$, we obtain $\tau \approx 50\ \mathrm{msec}$.

Meanwhile it takes a time $\tau_{growt} \approx 2R/v_n$ for a snow crystal to grow appreciably, where $v_n$ is the growth velocity. The ratio of the relaxation time to the growth time is called the *Peclet number*, defined as $p_{Peclet} = Rv_n/2D$, and its value is typically less than $10^{-4}$ for a snow crystal growing in air. (Note that the Peclet number is usually *not* small for solidification from the melt, so the quasi-static approximation is not valid in those systems. Solidification from the vapor generally yields much smaller Peclet numbers than solidification from the melt.)

The take-away message from all this is that the depleted region around a snow crystal adjusts itself almost instantaneously to changes in the crystal shape, and this is equivalent to the quasi-static approximation. Adopting Figure 3.4 affords a substantial simplification in the mathematics, allowing us to assume that the water-vapor field surrounding a growing snow crystal is always in its completely relaxed state. This state changes as the crystal grows, but we need not worry about the relaxation process itself.

## Boundary Conditions

To solve the diffusion equation, we also need to supply appropriate boundary conditions. These are nontrivial for snow crystal growth, so we need to consider them with some care. We present a summary here, with more details presented in Appendix A.

**Far-away Boundary.** One often-used boundary condition is to assume that the supersaturation is equal to some fixed value $\sigma_\infty$ far from the growing crystal. The term "far-away" in this context usually means at a distance much larger than the size of the growing crystal in question.

This boundary condition works well in 3D if the growing crystal is small in all three dimensions. It is possible, however, to apply this boundary condition incorrectly, and people sometimes do. Assuming a simple far-away boundary may not work with infinitely long cylinders, infinitely large walls, large dendritic structures, or other system geometries. We will encounter examples of such cases later in this chapter.

**Ice-covered Walls.** In many experimental situations, a boundary might consist of an ice-covered surface at some temperature $T$. Assuming the ice is neither growing or sublimating appreciably, the vapor pressure will equal the equilibrium value, $c \approx c_{sat}(T)$, at the ice-covered surface. For an isothermal environment, this means $\sigma \approx 0$ at the ice surface.

Even here, we must be a bit careful with this boundary condition. For the isothermal case, the surface boundary condition is given more accurately as

$$\sigma_{surf} \approx \frac{v}{\alpha v_{kin}} \tag{3.5}$$

where $v$ is the growth velocity of the ice on the surface. This can be very close to zero for a large ice-covered reservoir wall, but it is often *not* a good assumption to take $\sigma \approx 0$ at the surface of a small, isolated ice crystal. In any case, the only time $\sigma_{surf}$ is exactly zero is in equilibrium, when the growth velocity is also zero, as shown in Figure 3.5.

If the temperature varies in an experimental system, then we must be careful about the definition of $\sigma$ itself, as $c_{sat}$ is temperature dependent. In this case, the boundary condition is best left as $c \approx c_{sat}(T)$ at the surface of a large, ice-covered wall, as written above.

**Mass Flux.** To look at additional boundary conditions, we need to understand the flow of material in a diffusing system. In particle diffusion, there is always a net particle flux associated with a density gradient, in our case given by $F = D(\hat{n} \cdot \vec{\nabla} c) = D c_{sat}(\hat{n} \cdot \vec{\nabla} \sigma)$. (This is also a standard result from basic diffusion physics found in textbooks.) Note that the equals sign goes both ways: if there is a gradient in the water-vapor density in air, it necessarily results in a flow of water-vapor molecules given by $F$. Likewise, if there is a net diffusive flow of water-vapor molecules through air, it must be accompanied by a density gradient $\vec{\nabla} c$.

**Ice-free Walls.** If particles cannot flow into or out of an ice-free wall in an experimental chamber, then zero net particle flux implies zero density gradient at the wall. This means that the boundary condition at an ice-free wall is given by

$$\left(\frac{\partial \sigma}{\partial n}\right)_{wall} = 0 \tag{3.6}$$

where $(\partial \sigma / \partial n)_{wall}$ is the gradient of the supersaturation in the direction of the surface normal.

**Mass Conservation.** If a snow crystal is growing in air, then there must be a particle flux into the surface of the ice, as the flow of particles is what supplies the growth. Doing the math (see Appendix A), this yields a surface boundary condition

$$v_n = \frac{c_{sat} D}{c_{ice}} \left(\frac{\partial \sigma}{\partial n}\right)_{surf} \tag{3.7}$$

where $v_n$ is the growth velocity of the crystal normal to the surface and $(\partial \sigma / \partial n)_{surf}$ is the normal gradient of the supersaturation just above the ice surface. Combining this with the Hertz-Knudsen relation (see Chapter 3)

$$v_n = \alpha v_{kin} \sigma_{surf} \tag{3.8}$$

then gives the surface boundary condition as

$$X_0 \left(\frac{\partial \sigma}{\partial n}\right)_{surf} = \alpha \sigma_{surf} \tag{3.9}$$

where

$$X_0 = \frac{c_{sat}}{c_{ice}} \frac{D}{v_{kin}} \tag{3.10}$$

This is called a *mixed* boundary condition because it involves both the value and gradient of $\sigma$ at the surface.

In some circumstances, it is reasonable to just assume $\sigma_{surf} \approx 0$ at the surface of a growing snow crystal, and one occasionally sees this assumption in the literature. But it is an oversimplification that often obscures interesting aspects of snow-crystal growth. Assuming $\sigma_{surf} = 0$ on a growing crystal can never be absolutely accurate, as Figure 3.5 would then imply zero growth. Quantifying this discussion is important, and I defer that topic to the section below on spherical solutions. Figure 3.9 is the usual boundary condition needed at the surface of a growing snow crystal.

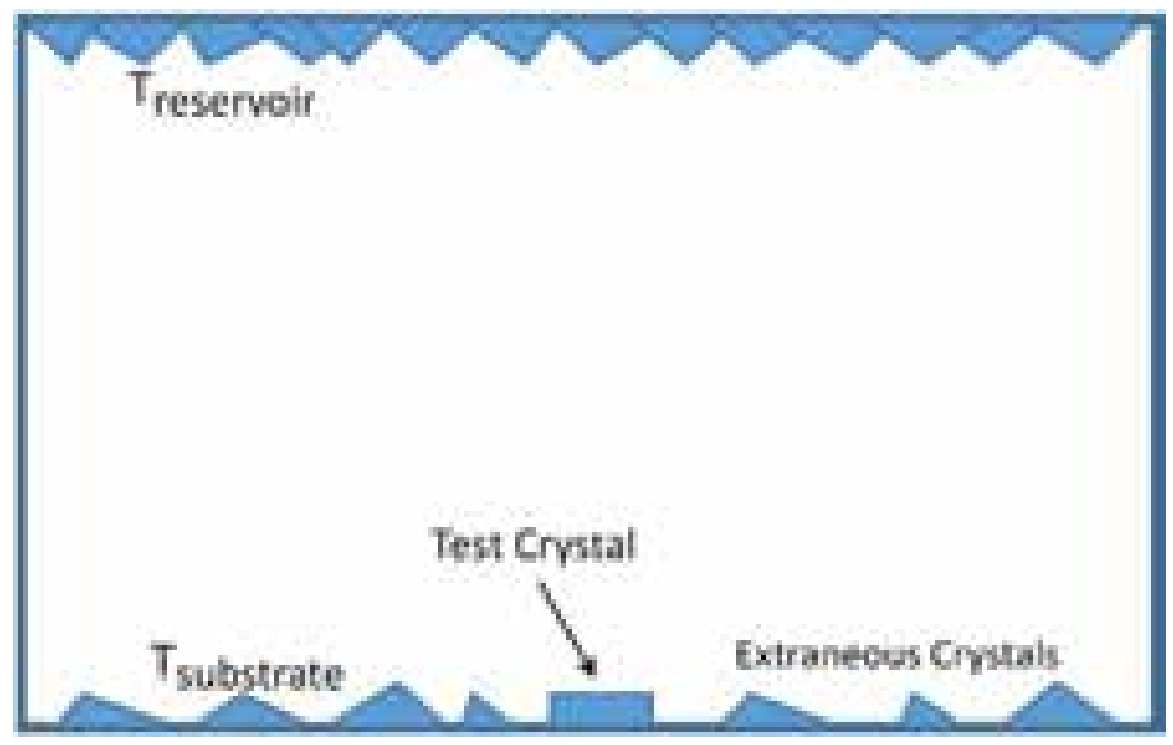

**Figure 3.18: An example of a poorly designed experimental set-up. As described in the text, the boundary conditions are so ill-defined that it is essentially impossible to do a quantitative analysis of the growth of the test crystal. Removing the extraneous crystals largely remedies this situation, but only if all of them are removed.**

## Example: A Basic Ice-Growth Experiment

Looking through the older scientific literature, researchers sometimes grew ice crystals in air and tacitly assume $\sigma_{surf} \approx \sigma_\infty$, which can be a shockingly poor assumption (for example, see [2004Lib, 2015Lib]). Making this assumption without a careful consideration of the diffusion equation and its solution can lead to wholly incorrect conclusions regarding the ice attachment kinetics and other matters.

To see this in detail, consider the ice-growth experimental chamber diagrammed in Figure 3.18. The basic idea in this thought experiment is to measure the growth of a small test crystal under well-controlled environmental conditions, specifically with well-known supersaturation boundary conditions around the test crystal. If the boundary-conditions are well defined, then one can model the $\sigma(\vec{x})$ field around the growing test crystal and learn something useful about the underlying growth physics. If the boundary conditions are not carefully constrained using proper experimental design, however, the results obtained may be of little scientific value.

To analyze the experiment shown in Figure 3.18, first consider the case where we remove all the ice crystals from the bottom substrate while leaving a good coating of ice on the reservoir surface. Being an ice-covered wall, the reservoir surface has the boundary condition $c = c_{sat}(T_{reservoir})$. The other walls are free of ice, giving $(\partial c/\partial n)_{wall}$ on those surfaces. Solving the diffusion equation for this situation yields the simple solution $c = c_{sat}(T_{reservoir})$ throughout the chamber, as this satisfies the diffusion equation and all the boundary conditions. Moreover, this solution for $c(\vec{x})$ is independent of $T_{substrate}$ or the temperatures of any of the walls.

Now note that the air just above the substrate surface has a temperature $T = T_{substrate}$. Thus, the supersaturation just above the substrate will be

$$\sigma_1 = \frac{c_{sat}(T_{reservoir}) - c_{sat}(T_{substrate})}{c_{sat}(T_{substrate})} \quad (3.11)$$

which will be greater than zero if $T_{reservoir} > T_{substrate}$.

Next, place a single, small test crystal on the substrate (with none of the "extraneous" crystals shown in Figure 3.18), and assume that the test-crystal temperature is equal to the substrate temperature (typically a good assumption, as $\kappa_{ice}$ is quite large). As long as the test crystal is much smaller than the substrate/reservoir spacing, we see that taking $\sigma \approx \sigma_\infty = \sigma_1$ at distances far from the growing crystal is an excellent approximation. The test crystal depletes the supersaturation in its vicinity, but not at distances many times larger than its size.

Assuming $\sigma_\infty = \sigma_1$ at distances far from the test crystal (but still near the substrate), we are then ready to solve the diffusion equation to model the test-crystal growth, presumably for comparison with measurements of the same. This can be done by adding a mirror reflection of the entire experiment about the substrate plane. Combining the real problem with its mirror reflection yields a single test crystal (twice as tall as the real crystal) surrounded entirely by $\sigma_\infty$, and this reflected geometry automatically satisfies the boundary condition on the substrate surface, namely $(\partial\sigma/\partial n)_{wall} = 0$. (Creating such a mirror reflection is a common trick used for obtaining analytic solutions of the diffusion equation.)

From this discussion we see that the set-up in Figure 3.18 (with no "extraneous" crystals) gives us a fighting chance of making a solid, quantitative, scientifically interesting measurement of the crystal growth rate in a known $\sigma_\infty$. We still must model the test crystal, with the mixed boundary condition at its surface, but at least the rest of the system is well characterized. Thus, as long as we have just a single, isolated test crystal on the substrate, this is a good experimental set-up.

All this changes if we add the "extraneous crystals" shown in Figure 3.18. Now the boundary condition on the substrate surface is a mess, as each extraneous crystal provides an additional, unknown water-vapor sink. Given that some of these unseen crystals have rough surfaces with $\alpha \approx 1$, and we do not know where the extraneous crystals are placed on the substrate, the boundary conditions are overall ill-defined, and we are left with an intractable diffusion problem. It is likely that $\sigma \approx 0$ around the test crystal, but we cannot know for sure.

Even a few extraneous crystals in this situation can make a big difference, even if the filling fraction is very small (I will quantify that statement using a model system later in this chapter). One thing is certain, however, and that is that assuming $\sigma \approx \sigma_\infty = \sigma_1$ may not be a good way to proceed. A surprising number of ice-growth researchers have not paid enough attention to a careful vapor-diffusion analysis, and this helps explain why there are such large discrepancies in experimental results presented in the literature [2004Lib]. This issue was most acute in the pre-computer era, but even some recent experiments have suffered from similar problems [2015Lib].

## Heat Diffusion

The solidification of water molecules at a growing snow crystal surface releases latent heat that increases the surface temperature and thus slows growth. The temperature rise is countered by the diffusion of heat away from the surface through the surrounding air, producing another type of diffusion-limited growth. Heat diffusion is less important than particle diffusion in snow crystal growth, so it is rightfully ignored in most numerical models, at least for the time being. Nevertheless, the separate contributions of heat and particle diffusion have been observed at least once [2016Lib], so one will have to face the full dual-diffusion problem (particle plus heat diffusion) at some point in the future.

Heat diffusion is described by the thermal diffusion equation

$$\frac{\partial T}{\partial t} = D_{therm}\nabla^2 T \qquad (3.12)$$

where $T(\vec{x})$ is the temperature field surrounding the crystal and

$$D_{therm} = \frac{\kappa_{air}}{\rho_{air} c_{p,air}} \qquad (3.13)$$

(see Appendix A and the list of constants near the front of this book). For typical atmospheric conditions, $D_{therm} \approx 2 \times 10^{-5}$ m$^2$/sec, and the fact that $D_{therm} \approx D_{air}$ reflects the universal nature of diffusion through ideal gases.

The quasi-static approximation applies for heat diffusion as it does for particle diffusion, giving

$$\nabla^2 T = 0 \qquad (3.14)$$

to quite high accuracy. Equating the heat flux away from the crystal surface to the heat generated gives the surface boundary condition

$$\kappa_{air}\left(\frac{\partial T}{\partial n}\right)_{surf} = v_n \rho_{air} L_{sv} \quad (3.15)$$

With all the relevant diffusion equations and boundary conditions now in hand, we can proceed to examine their simplest analytic solutions.

## 3.4 The Spherical Solution

A spherical snowflake is like the hydrogen atom in atomic physics – simple enough to be solvable analytically, yet remarkably useful for gaining intuition that can be applied to more challenging scenarios. If you really want to understand the growth of snow crystals, with all their branching, faceting, and other complex structures and growth behaviors, I recommend starting your quest with the simplest possible example – the growth of a spherical ball of ice. I examine this problem in detail in Appendix A and present the results here with a focus on the physical concepts and most useful results.

### Kinetics plus Diffusion

The spherical problem can be solved analytically and exactly, and I like to start with the minimum physics needed to describe the basic problem. Thus, let us begin by including particle-diffusion-limited growth and attachment kinetics with a constant $\alpha$ on the surface of the sphere. This addresses the heart of the problem without a lot of unnecessary complications. The solution of the diffusion equation gives (see Appendix A)

$$\sigma(r) = \sigma_\infty - \frac{R}{r}\left(\sigma_\infty - \sigma_{surf}\right) \quad (3.16)$$

where

$$\sigma_{surf} = \frac{\alpha_{diff}}{\alpha + \alpha_{diff}}\sigma_\infty \quad (3.17)$$

with

$$\alpha_{diff} = \frac{X_0}{R} \quad (3.18)$$

The crystal growth velocity is then

$$v_n = \left(\frac{\alpha\alpha_{diff}}{\alpha + \alpha_{diff}}\right) v_{kin}\sigma_\infty \quad (3.19)$$

There are two limiting cases that deserve special attention:

**I) Kinetics-limited growth**
This limit applies when

$$\alpha \ll \alpha_{diff} \quad (3.20)$$

and gives

$$v_n \approx \alpha v_{kin} \sigma_\infty \quad (3.21)$$

$$\sigma_{surf} \approx \sigma_\infty \quad (3.22)$$

As the name implies, kinetics-limited growth depends on $\alpha$ but is independent of $D$ or $X_0$.

Because $X_0$ is typically about 0.15 microns in normal air, $\alpha_{diff}$ is quite small for even a small natural snow crystal. For this reason, kinetics limited growth usually applies in air only when $\alpha$ is extremely small. At low pressure, however, $X_0$ can be substantially larger, so kinetics-limited growth is more likely to apply in near-vacuum conditions.

**II) Diffusion-limited growth**
This limit applies when

$$\alpha_{diff} \ll \alpha \quad (3.23)$$

and gives

$$v_n \approx \frac{c_{sat}D}{c_{ice}R}\sigma_\infty \quad (3.24)$$

$$\approx \frac{X_0}{R}\, v_{kin}\sigma_\infty$$

$$\sigma_{surf} \approx \frac{\alpha_{diff}}{\alpha}\ \sigma_\infty \qquad (3.25)$$

$$\approx \frac{X_0}{\alpha R}\ \sigma_\infty$$

Figure 3.19 shows some example solutions for $\sigma(r)$ using a constant $R$ and several different values of $\alpha$.

As a general rule, looking beyond the spherical solution, faceting becomes a dominant growth characteristic in the kinetics-limited regime, while branching tends to dominate in the diffusion-limited regime. Thus, tiny snow crystals (small $R$) tend to grow into simple faceted prisms, as do slow-growing crystals (small $\alpha$). Crystals grown at low pressures (large $D$) often grow as simple prisms for the same reason. Conversely, branching tends to dominate over faceting in the diffusion-limited regime. Although the spherical solution is of little use for describing the detailed formation of complex snow crystals, it is invaluable for understanding different limiting behaviors.

The spherical solution also tells us that that $v_n$ is independent of $\alpha$ in the diffusion-limited regime, but $\sigma_{surf}$ is not. We also see that $\sigma_{surf}$ generally becomes smaller as $R$ becomes larger. However, $\sigma_{surf}$ never reduces fully to zero for a growing crystal, as zero supersaturation would be equivalent to a zero-growth equilibrium state.

It is also worth noting that while the spherical solution is a perfectly correct and accurate solution to the diffusion equation, in real life it is not a stable solution. Diffusion-limited spherical growth is subject to the Mullins-Sekerka instability, eventually producing dendritic structures.

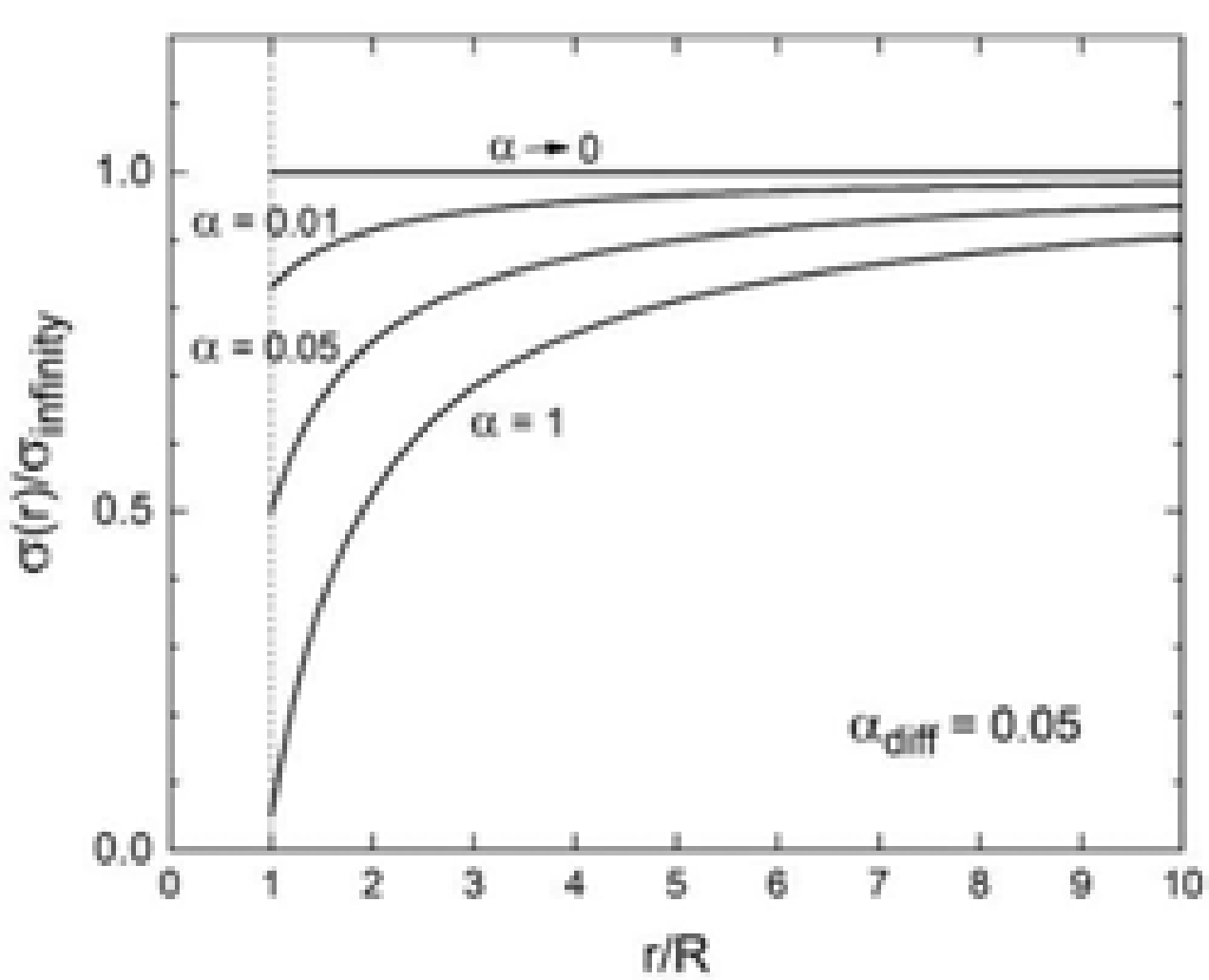

**Figure 3.19: The solution to the diffusion equation for the growth of a spherical snow crystal with $\alpha_{diff}$=0.05. When $\alpha \to 0$, the growth is kinetics limited and $\sigma_{surf} \approx \sigma_\infty$. Then as $\alpha$ increases, the growth becomes more diffusion limited and $\sigma_{surf}$ decreases.**

## Kinetics, Diffusion, and Heating

When latent heating is included in the spherical problem, we must then simultaneously solve both the heat and particle diffusion equations, which is a substantially more difficult problem. Notably, the isothermal approximation clearly no longer holds, so $c_{sat}$ is not a simple constant and one must be quite careful with the definition of the supersaturation field $\sigma(\vec{x})$. The details are provided in Appendix A, and here I jump straight to the solution for the growth velocity, which can be written in the same basic form as Figure 3.19:

$$v_n = \left(\frac{\alpha\alpha_{diff,heat}}{\alpha + \alpha_{diff,heat}}\right) v_{kin}\sigma_\infty \qquad (3.26)$$

where

$$\alpha_{diff,heat} = \frac{X_0}{R}\frac{1}{1+\chi_0} \qquad (3.27)$$

and $\chi_0$ is a dimensionless parameter

$$\chi_0 = \frac{\eta D L_{sv}\rho_{ice}}{\kappa_{air}}\frac{c_{sat}}{c_{ice}} \qquad (3.28)$$

as described in Appendix A, with all the variables in Figure 3.28 evaluated at $T_\infty$. In addition, the surface temperature of the growing spherical crystal is given by

$$\Delta T = \frac{1}{\eta} \frac{\alpha}{\alpha + \alpha_{diff,heat}} \frac{\chi_0}{1 + \chi_0} \sigma_\infty \qquad (3.29)$$

where $\Delta T = T_{surf} - T_\infty$. If $\alpha_{diff} \ll \alpha$, so the growth is purely diffusion-limited (which gives the maximum $\Delta T$), then Figure 3.29 reduces to

$$\Delta T \approx \frac{1}{\eta} \frac{\chi_0}{1 + \chi_0} \sigma_\infty \qquad (3.30)$$

which, interestingly, is independent of the crystal radius $R$.

A first take-away message from this analysis is that heat diffusion plays a somewhat minor role in snow crystal growth compared to particle diffusion. The relevant variable $\chi_0$ equals about 0.8 at -1 C, drops to about 0.4 at -10 C, and it continues falling with colder temperatures. If the growth is mainly kinetics-limited, then neither particle or heat diffusion matters much (in air at normal pressures). If the growth is mainly diffusion-limited, then Figure 3.26 becomes

$$v_n \approx \alpha_{diff} v_{kin} \frac{\sigma_\infty}{1 + \chi_0} \qquad (3.31)$$

which means that the main effect of heating can be incorporated into a simple rescaling of $\sigma_\infty$.

This is a significant result; once we can create realistic computer models of snow crystal growth incorporating only particle diffusion and attachment kinetics, then adding heat diffusion can be done to a reasonably good approximation simply taking $\sigma_\infty \to \sigma_\infty/(1 + \chi_0)$ in the same models. The take-away message is that we should probably ignore heating (in atmospheric snow crystal growth) until we first solve the problem including just particle diffusion and attachment kinetics. One step at a time.

Qualitatively, we can understand the heating effects from the underlying physics. Deposition generates latent heat, and this warms the growing snow crystal until a balance is reached, when the heat carried away by diffusion equals that generated. The increased crystal temperature then lowers the effective supersaturation by changing $c_{sat}$ at the surface. Moreover, the heat conductivity of ice is much higher than that of air, so the whole snow crystal heats nearly uniformly. When all this is considered in the spherical solution, we see that thermal diffusion effects, although not always negligible, are not nearly as important as particle diffusion and attachment kinetics for snow crystal growth in air.

This situation changes at low pressures, however. To first order, $D$ is inversely proportional to background gas pressure $P$, so particle diffusion speeds up considerably at low pressures. But $\kappa_{air}$ is roughly independent of $P$ down to quite low pressures (until the mean free path of air molecules becomes larger than other scales in the problem). For this reason, $\chi_0 \sim P^{-1}$ and heating effects can become quite pronounced at lower pressures. In [1972Lam], for example, the growth behavior of somewhat large ice crystals was substantially influenced by heating effects. Smaller crystals resting on a conducting substrate are less effected by heating, however, even in near vacuum conditions [2012Lib, 2013Lib]. We examine this point in detail in Chapter 7.

## Experimental Verification

Although diffusion theory is well understood, it is nevertheless good to see an experimental verification, for no other reason than to obtain a "reality check" to make sure one is on the correct theoretical track. Producing a suitable experiment is nontrivial, however, as spherical growth is generally unstable to the Mullins-Sekerka instability, plus just getting to an interesting place in parameter space is not a simple task.

I was able to validate the particle+heat diffusion model using measurements of the growth of long ice needles [2016Lib], and the results are shown in Figure 3.20. Although needles are certainly not spheres, the mathematics of cylindrical growth is nearly identical to that of spherical growth, as I describe below. Moreover, slightly tapered cylinders have the desirable property that $\alpha$ is large enough that $\alpha_{diff} \ll \alpha$ is valid and the growth is mainly diffusion limited.

With (almost) no adjustable parameters [2016Lib], experiment and theory were found to agree nicely. These data confirm the relative roles of particle and heat diffusion, demonstrating that the net effect of heating is greater near the melting point, reflecting the dependence of $\chi_0$ on temperature. To my knowledge, this is the first and only quantitative result demonstrating that snow-crystal growth is indeed limited by a combination of particle and heat diffusion.

While this experiment nicely demonstrates that latent heating plays a role in snow crystal growth in air, I reiterate that it a relatively modest perturbation when compared with the dominant effects of particle diffusion and attachment kinetics. Someday, we will need to solve the combined particle+heat double-diffusion problem in full 3-D to explain all the subtleties of snow-crystal growth. But that day is not yet upon us.

Throughout most of this book, therefore, I have largely ignored latent heating and heat diffusion (except, of course, when dealing with systematic errors in precision growth experiments). A scientist's life, after all, is one of successive approximations, especially in the world of condensed-matter systems. Until we have a better understanding of the attachment kinetics over the full range of growth conditions, it is reasonable to (mostly) ignore heating effects, at least for the immediate future.

## Kinetics, Diffusion, Heating, and Surface Energy

Rounding out our analysis of spherical ice growth, we add surface energy by including the Gibbs-Thomson effect, yielding the expression

$$v_n = \left( \frac{\alpha \alpha_{diff,heat}}{\alpha + \alpha_{diff,heat}} \right) v_{kin} (\sigma_\infty - \frac{2 d_{sv}}{R}) \tag{3.32}$$

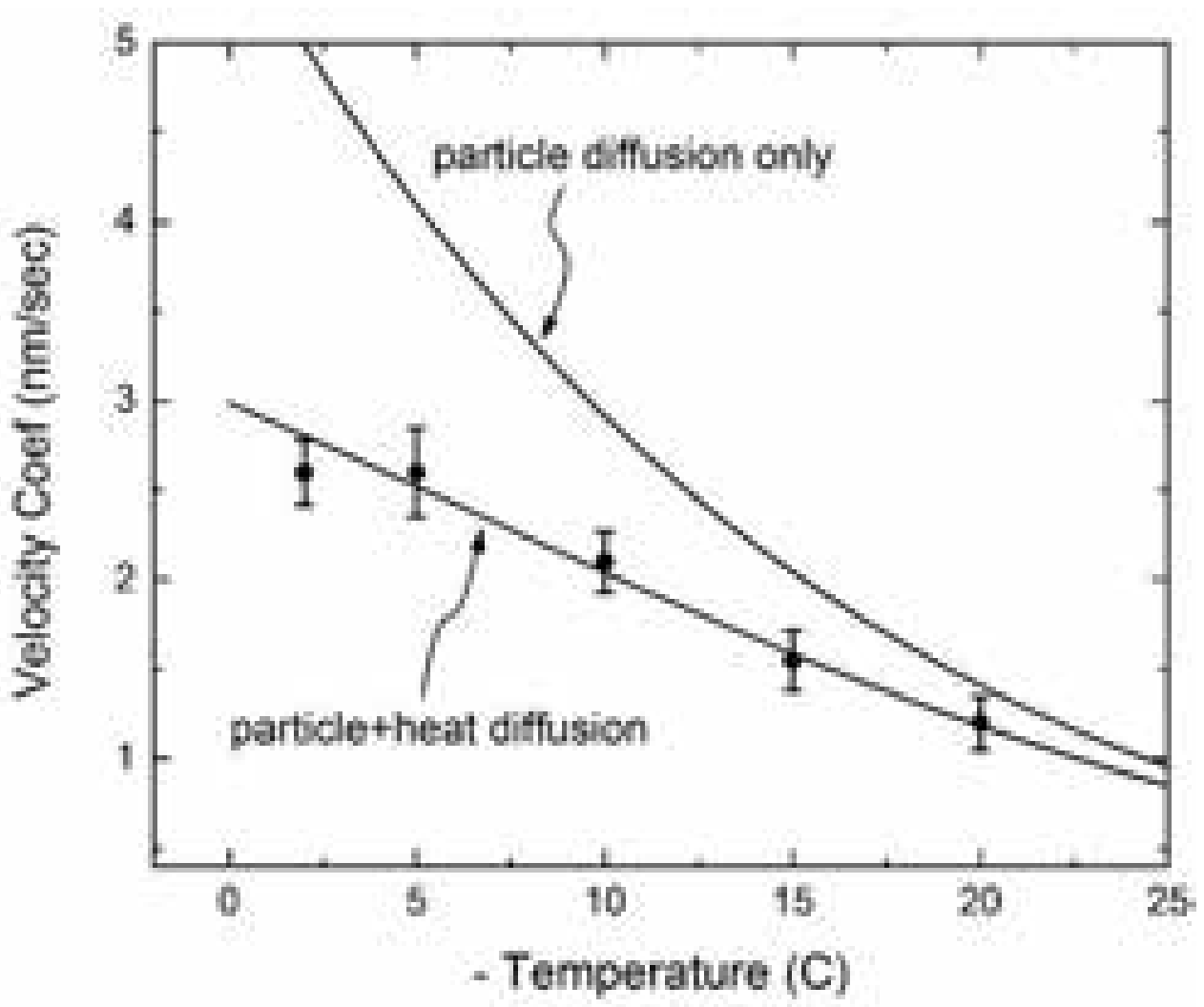

**Figure 3.20: Measurements of the radial growth of thin ice needles, together with an analytical model that included only particle diffusion (top line), plus a similar model that included both particle and heat diffusion (lower line) [2016Lib]. The plotted velocity coefficient is equal to the cylinder growth rate at a fixed far-away supersaturation and a fixed cylinder radius of five microns, as described in [2016Lib]. The measurements show good agreement with the particle+heat diffusion model, confirming the temperature-dependent reduction in growth rate caused by latent heating.**

where the origin of the $d_{sv}$ term is described in Chapter 2 (see also Appendix A). If one wishes to ignore heating effects, $\alpha_{diff,heat}$ can be replaces with $\alpha_{diff}$.

Under most snow-crystal growth conditions, the added Gibbs-Thomson term is a minor effect, especially with large crystals or fast growth rates. With a fernlike stellar dendrite, for example, the tip radius is $R \approx 1\ \mu$m, so $2d_{sv}/R \approx 0.002$, while the supersaturation is typically $\sigma_\infty > 0.1$. Thus, surface energy likely has only a small effect, and the same can be said for most snow crystals growth scenarios.

There are times when surface energy is an important factor, however. During the growth of exceptionally thin plates at low supersaturations, it appears that the Gibbs-Thomson effect does limit the growth and prevent the formation of even thinner plates than those observed. I discuss this further in Chapter 5.

Once again, the spherical solution is quite useful for examining the relative sizes of different physical effects, and for making suitable simplifying approximations in other analyses. Moreover, analytical solutions also play an important role in testing quantitative computation models of snow-crystal growth, verifying that the simulations reproduce known analytical results with acceptable accuracy (see Chapter 5).

## Finite Outer Boundary

Bringing the outer boundary in from infinity complicates the analysis, but the finite-boundary case is useful for validating numerical models to make sure they obtain correct quantitative results. Including both particle diffusion and attachment kinetics, the solution becomes [2013Lib1]

$$\sigma(r) = \sigma_{out} - \left(\frac{R'}{r} - \frac{R'}{R_{far}}\right)\sigma_{out} \quad (3.33)$$

where

$$R' = \left[\frac{\gamma}{R} - \frac{1}{R_{far}}\right]^{-1} \quad (3.34)$$

and

$$\gamma = \frac{\alpha + \alpha_{diff}}{\alpha_{diff}} \quad (3.35)$$

This then gives the crystal growth velocity

$$v_n = \left(\frac{\alpha\alpha_{diff}}{\alpha + \alpha_{diff}}\right) v_{kin}\sigma_\infty \left[1 - \frac{R}{\gamma R_{far}}\right]^{-1} \quad (3.36)$$

and we see that this reduces to Figure 3.19 when $R_{far} \rightarrow \infty$, as it must.

## Example: Growth-Chamber Supersaturation Estimation

With the spherical solution in hand, we are now able to better examine the experimental chamber shown in Figure 3.18, and to make a quantitative estimate of the supersaturation near the test crystal in that chamber. To this end, we re-draw that chamber and model the extraneous crystals with an array of ice hemispheres of radius $R$, as shown in Figure 3.21.

We assume these crystals are all identical in size and shape, and they are arranged on a square grid with a separation $L$ between the crystals, as shown. As before, the top surface is covered with ice crystals at a temperature of $T_{reservoir}$, and these crystals serve as the source of water vapor in the experiment.

Defining $\sigma_{bottom}$ to be the effective supersaturation at the bottom of the chamber, as shown in Figure 3.21, we see that Figure 3.11 gives $\sigma_{bottom} = \sigma_1$ in the absence of any bottom crystals, as I described above. But we will have $\sigma_{bottom} < \sigma_1$ when ice crystals are present on the substrate. Assuming $R \ll L \ll H$, the spherical solution gives the growth velocity $v \approx (c_{sat}/c_{ice})(D/R)\sigma_{bottom}$ for each of the hemispherical crystals. Adding up all the water vapor consumed by these crystals, there must be a downward flux given by $F \approx (2\pi R D c_{sat}/L^2)\sigma_{bottom}$ flowing from the top of the chamber to the bottom. But this flux must also equal $F = Dc_{sat}(\sigma_1 - \sigma_{bottom})/H$, and equating these two expressions yields

$$\sigma_{bottom} \approx \frac{1}{1 + \left(\frac{2\beta H}{R}\right)} \sigma_1 \qquad (3.37)$$

where we have defined the filling fraction $\beta = \pi R^2/L^2$ as the fraction of the total area of the substrate covered with ice crystals.

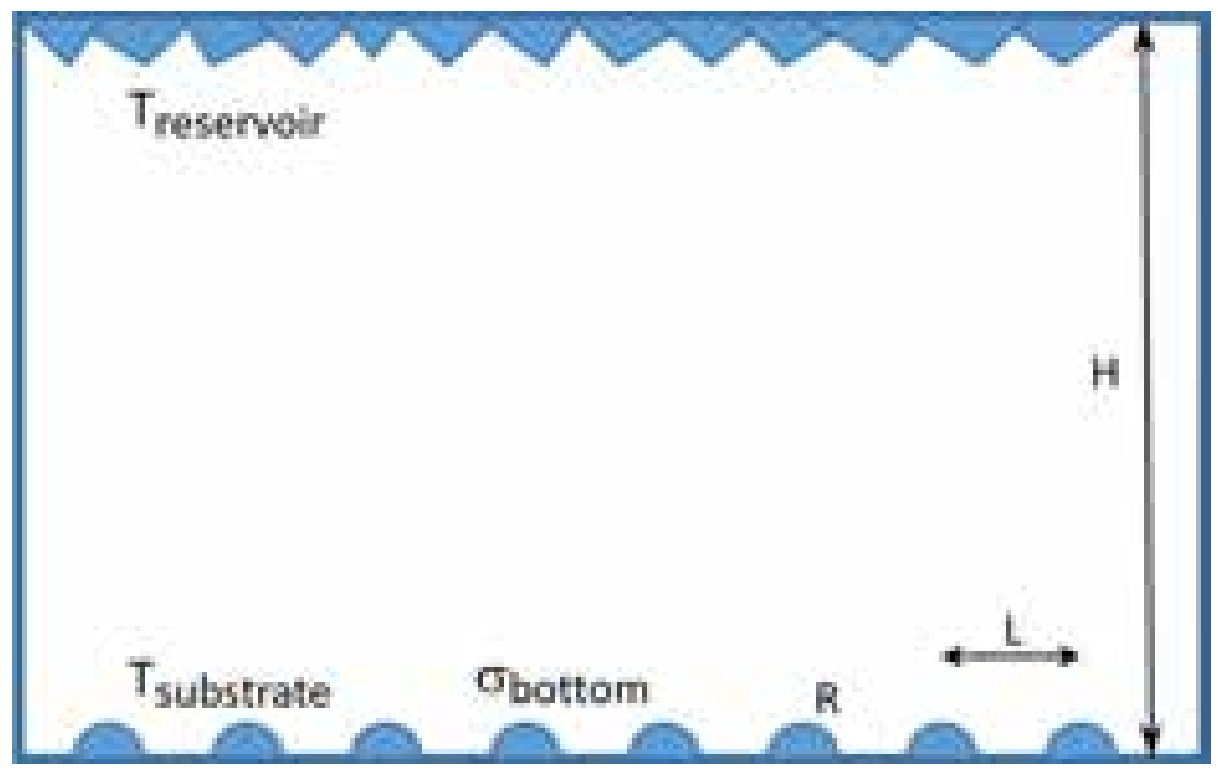

**Figure 3.21: Analytic solutions of the diffusion equation can be quite useful for estimating the supersaturation in an experimental ice growth chamber, such as in this example. Even if there is only a low density of ice crystals on the bottom of the chamber, the supersaturation there can be reduced to surprisingly low levels.**

At this point the difficulty in building a good ice growth experiment begins to become apparent. For example, assume $H = 2.5$ cm and $R = 50$ microns for typical chamber and crystal sizes. Then if we want to keep $2\beta H/R < 0.2$, so $\sigma_{bottom}$ is close to the ideal expectation, this requires a filling fraction of $\beta < 0.0002$. In other words, the filling factor must be *very* small before the experiment approaches the ideal case with $\sigma_{bottom} \approx \sigma_1$.

Even a seemingly small filling fraction of $\beta = 0.01$ yields $\sigma_{bott} = 0.09\sigma_1$ with these values of $H$ and $R$. As a result, even small uncertainties in the placement of crystals on the substrate may produce large uncertainties in the estimate of $\sigma_{bottom}$. And once $\sigma_{bottom}$ is not well determined, it becomes nearly impossible to produce meaningful, quantitative experimental results.

Interestingly, this exercise shows that $\sigma_{bottom}$ goes to zero as $H$ increases to infinity, regardless of the crystal size or filling fraction. So $\sigma_{bottom}$ is likely going to be especially small if the reservoir is far from the growing crystals. It may seem like a somewhat counterintuitive result, but only because one's intuition can easily be wrong regarding solutions of the diffusion equation.

While studying the ice-growth literature over the years, I found a surprising number of papers in which the authors did not adequately account for diffusion effects in their experiments, rendering their results somewhat unreliable [2004Lib]. In a more recent example, I found that a reanalysis of the diffusion problem could change the reported measurements by over two orders of magnitude [2015Lib].

The main take-away message from this example is that it remarkably easy to make mistakes involving diffusion analyses, so one must proceed with considerable caution when estimating supersaturations in ice growth experiments. Full 3D numerical simulations of the diffusion field within a growth chamber are best, but one can learn a great deal by a careful application of analytic solutions to the diffusion equation.

## 3.5 Additional Analytic Solutions

While the spherical solution is the best starting point for any quantitative discussion of diffusion limited growth, several other analytic solutions can be found. In this section I examine some of these additional solutions and their application.

### Cylindrical Growth

I have found that the analytic solution for an infinitely long growing cylinder is surprisingly useful when examining measurements of the growth of electric needle crystals (see Chapter 8), so I present this case in Appendix A and mention the results here. The diffusion analysis is analogous to the spherical case, the main change being to work in a cylindrical coordinate system. Once again, the solution can be written in a form like Figure 3.19, giving the radial growth velocity

$$v_n = \left(\frac{\alpha \alpha_{diffcyl}}{\alpha + \alpha_{diffcyl}}\right) v_{kin} \sigma_\infty \qquad (3.38)$$

where

$$\alpha_{diffcyl} = \frac{1}{B} \frac{X_0}{R_{in}} \qquad (3.39)$$

and $B = \log(R_{out}/R_{in})$, where $R_{in}$ is the radius of the cylinder and $R_{out}$ is the radius of the far-away boundary (see Appendix A). Note that one cannot assume $R_{out} \to \infty$ in this solution without encountering a logarithmic divergence, a feature that is well known from cylindrical electrostatics problems. It is straightforward to extend the analysis to include latent heating and heat diffusion, and the resulting model is the one compared with experimental data in Figure 3.20. The cylindrical solution is also useful for validating numerical models, as described in [2013Lib1].

## The Ivantsov Solution

Continuing the progression from spherical and cylindrical coordinates, one can also find a solution to the diffusion equation in parabolic coordinates in either 2D or 3D. While parabolic coordinates seldom come up in physics problems, in this case the solution yields a parabolic crystal morphology that exhibits a constant tip radius $R_{tip}$ and constant tip growth velocity $v_{tip}$ that looks a lot like the observed behavior in free dendrite growth. The parabolic solution was discovered in 1947 by Russian physicist G. P. Ivantsov [1947Iva], and it provides many valuable insights into the formation and structure of free-dendrite snow crystals.

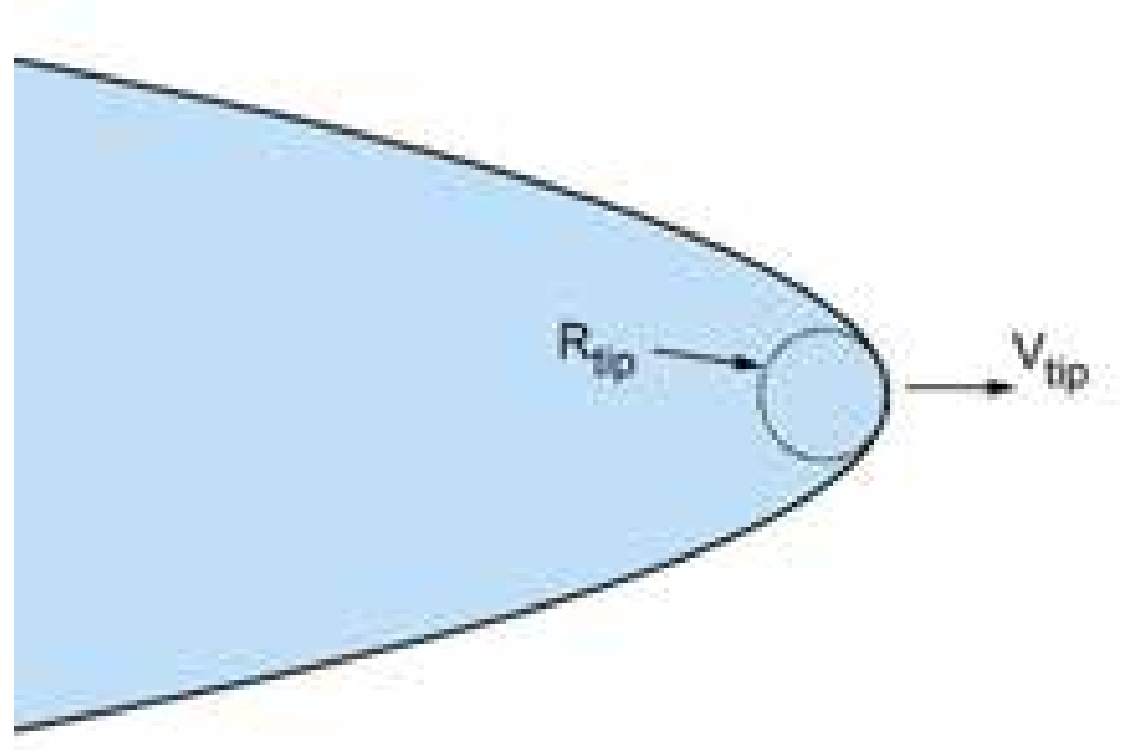

**Figure 3.22: The Ivantsov solution to the diffusion equation describes a crystalline paraboloid of revolution with a constant parabolic shape and tip radius $R_{tip}$ that grows forward with a constant velocity $v_{tip}$. If viewed from a frame of reference that moves in the growth direction with velocity $v_{tip}$, the system would appear completely static.**

The full 3D Ivantsov solution takes the form of a needle-like paraboloid-of-revolution that is parameterized by its tip radius $R_{tip}$, as shown in Figure 3.22. For purely diffusion-limited growth, solving the diffusion equation (for either particle or heat diffusion separately) yields that this entire paraboloid grows at a constant velocity $v_{tip}$ in the direction of the needle axis, while $R_{tip}$ and the full parabolic shape of the crystal remain unchanged in time.

We will not describe the derivation of the Ivantsov solution here but present the primary result only. For the case of ice growing from water vapor in air, neglecting surface energy and attachment kinetics ($\alpha_{diff} \ll \alpha \approx 1$), the growth velocity is given by [1996Sai, 2002Lib]

$$v_{tip} = \frac{2D}{BR_{tip}} \frac{c_{sat}}{c_{ice}} \sigma_{far} \tag{3.40}$$

$$= \frac{2}{B} \frac{X_0}{R_{tip}} v_{kin} \sigma_{far}$$

where $B = \log(\eta_{far}/R_{tip})$, $\eta_{far}$ is the distance to the far-away boundary (using a parabolic coordinate system with standard variables $(\xi, \eta, \varphi)$) and $\sigma_{far}$ is the supersaturation at $\eta_{far}$. This equation is analogous to Figure 3.24, but with one noteworthy distinction: in the spherical case, $R$ increases as the crystal grows, while $R_{tip}$ remains constant in the Ivantsov solution. And again, with the cylindrical solution above, we cannot assume an outer boundary at infinity, as this produces a logarithmic divergence in $B$.

As with the spherical solution, the Ivantsov solution is unstable with respect to the Mullins-Sekerka instability. However, the Ivantsov parabola provides an approximate description of free dendrite growth, where we see that the shape and growth near the dendrite tip are roughly like that shown in Figure 3.22. The Mullins-Sekerka instability creates sidebranching away from the tip, but the tip itself is reasonably well described by the Ivantsov solution.

Although sidebranches clearly complicate the picture, the Ivantsov form is a remarkably robust solution to the diffusion equation. Thus, the large-scale outline of a typical free dendrite is roughly parabolic, and usually the structure near the tip is smoother with a nearly parabolic form. Put another way, the Ivantsov solution creates a self-assembling free-dendrite morphology with constant $v_{tip}$ that is generally insensitive to perturbations from other growth effects. This property helps to explain why free-dendrite growth in diffusion-limited solidification is such a commonly observed phenomenon.

Looking at the snow crystal free-dendrite morphologies in Figures 3.9 and 3.11, clearly the Ivantsov solution only goes so far when describing real crystal growth. The same can be said of the spherical and cylindrical morphologies. In all cases, the analytic solutions are useful mainly as limiting cases and for examining overall trends regarding different parameters and growth behaviors. Most of all, analytic solutions are a good way to build one's intuition and understanding about which physical processes can be safely neglected in what circumstances. Reproducing actual snow crystal structures and growth measurements with any real fidelity, however, will require numerical modeling.

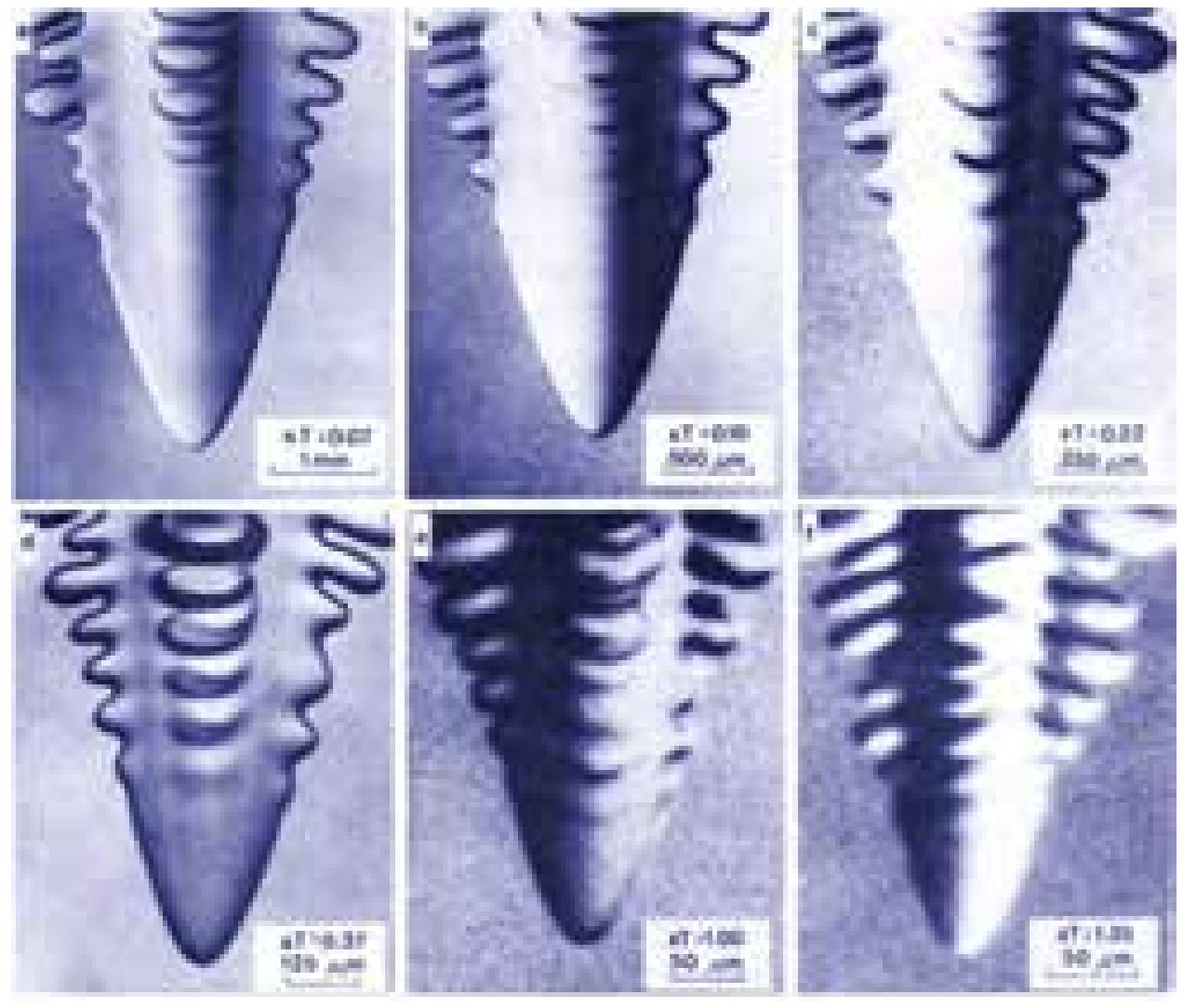

**Figure 3.23: A series of photographs showing the tips of free dendrites growing during the solidification of liquid succinonitrile (a clear, waxy material that melts at 57 C). As the supercooling $\Delta T$ of the liquid increases, $R_{tip}$ decreases while $v_{tip}$ increases, their product $R_{tip}v_{tip}$ satisfying the Ivantsov relation for thermal diffusion. Meanwhile the overall growth behavior and dendrite tip morphology remains nearly independent of $\Delta T$. (Image from [1981Hua].)**

## 3.6 Solvability Theory

For roughly a decade around the 1980s, there was a concentrated effort to create an analytical model of free-dendrite growth, and the result became known as *solvability theory* [1988Kes, 1988Sai, 1989Lan, 1991Bre]. The primary goal of this endeavor was to derive $v_{tip}$ and $R_{tip}$ directly from basic physical principles and intrinsic material properties, reproducing measurements from a broad range of materials. This research effort was stimulated in part by a series of beautiful, quantitative observations of dendritic solidification from the melt by Martin Glicksman and others, with one experimental example shown in Figure 3.23.

It was realized early on that the Ivantsov relation provides only a relation between $v_{tip}$ and $R_{tip}$, without specifying either. In fact, the Ivantsov solution is a family of solutions, specifying $v_{tip}$ for a given $R_{tip}$, as can be seen in Figure 3.40 for the snow crystal case. This physical indeterminacy became known as the *selection problem*. If diffusion alone does not define $v_{tip}$ and $R_{tip}$ uniquely, what does?

### The Selection Problem

Resolving the selection problem required some additional physics beyond diffusion alone, and the two leading candidates were surface energy and attachment kinetics. In solidification from the melt, surface energy turns out to be the dominant effect, and many theoretical treatments in the literature ignore attachment kinetics for that reason. Solidification from the vapor phase has been much less studied, but here it appears that attachment kinetics dominates over surface energy, as we will see shortly with the snow crystal case.

My goal in this chapter is not to provide an in-depth review of all aspects of solvability theory, but rather to outline its basic results as applied to snow crystal formation. To this end, I will ignore all heating effects, as particle diffusion is more important than heat diffusion, and the former by itself is enough to develop a crude version of vapor-growth solvability theory. Using this theory, I will then show that attachment kinetics is likely more important than surface energy in the theory, allowing us to additionally ignore surface energy while keeping attachment kinetics, which is opposite to the melt-growth case.

I begin with a perturbation expansion of the spherical solution, Figure 3.32, neglecting heating ($\alpha_{diff,heat} = \alpha_{diff}$) and assuming that the growth is mainly diffusion limited ($\alpha_{diff} \ll \alpha$), which gives

$$v_n \approx \frac{X_0}{R} v_{kin} \left( \sigma_\infty - \frac{2d_{sv}}{R} - \frac{\sigma_\infty}{\alpha} \frac{X_0}{R} \right) \quad (3.41)$$

For a typical fernlike dendrite tip (taking $R = R_{tip} \approx 1\ \mu$m, $\sigma_\infty \approx 1$, and $\alpha \approx 1$), we find that the second and third terms in this equation are indeed small compared to $\sigma_\infty$, justifying the perturbation expansion.

For the next step, I assume that the near-hemispherical tip of a parabolic ice dendrite behaves much like spherical growth, so I can do an analogous perturbation expansion of the Ivantsov solution, Figure 3.40, giving (see Appendix A)

$$v_{tip} \approx \frac{2X_0}{BR_{tip}} v_{kin} \left( \sigma_{far} - \frac{R_{GT}}{R_{tip}} - \frac{v_{tip}}{\alpha v_{kin}} \right) \quad (3.42)$$

$$\approx \frac{2X_0}{BR_{tip}} v_{kin} \left( \sigma_{far} - \frac{R_{GT}}{R_{tip}} - \frac{\sigma_{far}}{\alpha} \frac{R_{kin}}{R_{tip}} \right)$$

where we have defined $R_{GT} = 2d_{sv} \approx 2$ nm and $R_{kin} = 2X_0/B \approx 35$ nm.

From this expansion, we can begin to see the essential physics underlying the dendrite selection problem. Referring to Figure 3.6, we see that the Mullins-Sekerka instability generally promotes the growth of bumps on top of broad, flat surfaces. Zooming in on the end of a dendrite tip, it stands to reason that the Mullins-Sekerka instability would also promote the growth of a smaller bump on top of a broad dendrite tip. Taking this reasoning to its logical conclusion, we see that the Mullins-Sekerka instability would, if no other forces intervened, sharpen a dendrite tip indefinitely, driving $R_{tip} \to 0$.

The available intervening forces are those found in Figure 3.42, specifically in the second and third terms of this expression. As $R_{tip} \to 0$, these terms both become so large that they are no longer small compared to $\sigma_{far}$. At some value of $R_{tip}$, therefore, one or both of these forces halts any further tip sharpening.

The nature of these two stabilizing effects can also be reasonably well understood from the underlying physics. The Gibbs-Thomson effect states that the equilibrium vapor pressure of ice increases as $R^{-1}$ on a spherical surface, and this effectively reduces the driving supersaturation at the tip. Following the math through, this gives the $R_{GT}$ term in Figure 3.42. The negative sign means that this is a stabilizing force that prevents runaway tip sharpening.

The kinetics term $R_{kin}$ arises from the fact that a finite surface supersaturation $\sigma_{surf} > 0$ is needed to drive crystal growth, and this $\sigma_{surf}$ increases with $v_{tip}$, which is proportional to ${R_{tip}}^{-1}$ to first order in this perturbation analysis. In effect, there is a supersaturation "penalty" for fast growth, and this also serves to prevent $R_{tip} \to 0$. The fact that $R_{GT} \ll R_{kin}$ suggests that the kinetics term in Figure 3.42 is more important than the surface-energy term for selecting the final dendrite tip radius in typical snow-crystal scenarios.

## Snow Crystal Dendrites

Extending this qualitative discussion of Figure 3.42 into a rigorous theory is not a simple task, which is why it took a concerted effort to develop solvability theory. Although my comprehension of the theory is not thorough, it appears that the final result can be expressed in a fairly simple form [1988Kes, 1988Sai, 1989Lan, 1991Bre, 2002Lib]. The answer, however, depends on the relative importance of the surface-energy and attachment-kinetics terms in Figure 3.42.

For the snow-crystal case, we neglect surface energy (on the grounds that $R_{GT} \ll R_{kin}$), and solvability theory then yields the relationship

$$v_{tip} R_{tip}^2 \approx \frac{4\sigma_{far} v_{kin} X_0^2}{s_0 B \alpha} \qquad (3.43)$$

where $s_0$ is a dimensionless constant called the *solvability parameter*. This second mathematical relationship, in addition to the Ivantsov solution, allows one to uniquely determine both $R_{tip}$ and $v_{tip}$ as a function of intrinsic material properties and external growth conditions.

Combining Equations 3.40 and 3.43 yields

$$R_{tip} \approx \frac{2X_0}{s_0 \alpha} \qquad (3.44)$$

$$v_{tip} \approx \frac{s_0}{B} \alpha v_{kin} \sigma_{far}$$

and at this point it is beneficial to compare the theory with experimental observations.

Figure 3.24 show measurements of $v_{tip}$ as a function of $\sigma_{far}$ for fernlike free dendrites growing near -15 C. The data support a linear dependence $v_{tip} \sim \sigma_{far}$, and the low-resolution tip images are at least consistent with $R_{tip}$ being independent of $\sigma_{far}$, so both observed trends agree with Figure 3.44. A fit to the data assuming $B \approx 10$ yields $R_{tip} \approx 1\ \mu$m and $\alpha s_0 \approx 0.25$ [2002Lib]. Similar data for fishbone free dendrites yields $R_{tip} \approx 1.5\ \mu$m and $\alpha s_0 \approx 0.2$ [2002Lib].

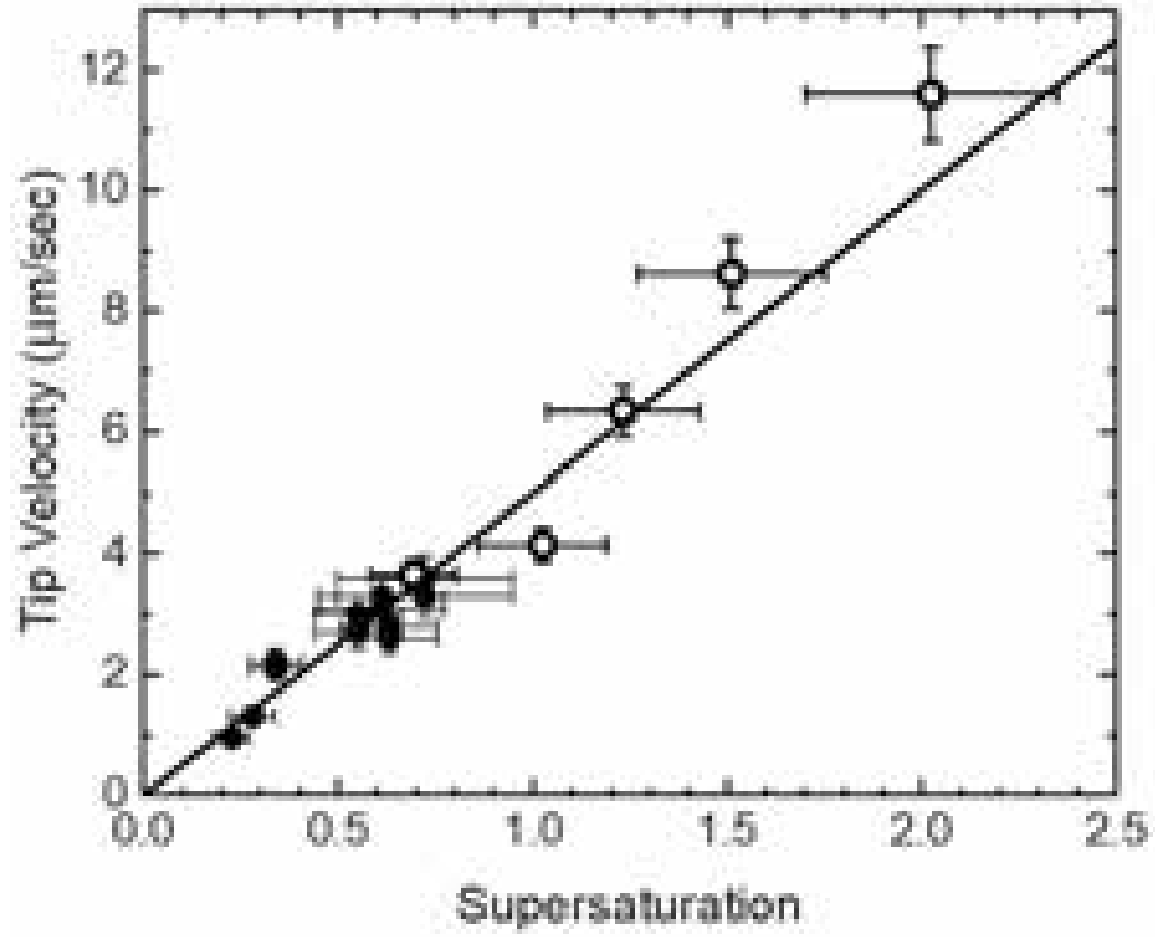

**Figure 3.24: Measurements of the tip velocity of fernlike free dendrites growing near -15 C as a function of the far-away supersaturation. The data indicate a linear relationship between these variables, and the line shows $v_{tip} = 5\sigma_{far}$ μm/sec. This supports the theoretical expectation that attachment kinetics provides the dominant selection perturbation in solvability theory for the case of snow-crystal free dendrite growth. (The data are from [2002Lib] with additional later measurements added to the plot.)**

Note that these values of $R_{tip}$ were measured in air, and we see from Figure 3.44 that theory indicates $R_{tip} \sim X_0$. This implies that $R_{tip}$ should be roughly inversely proportional to the background air pressure. It would be straightforward to confirm this prediction, but to my knowledge it has not yet been done. However, experiments have revealed finer structural details in snow crystals grown at higher pressures, supporting this solvability theory prediction [1976Gon].

This analysis of snow-crystal free dendrite growth comes with some caveats, however. Solvability theory indicates that the value of $s_0$ depends on the detailed properties of the most important stabilization term, namely the attachment kinetics in this case. This is problematic because the attachment kinetics are not well known from independent measurements, and they may depend on the growth conditions at the tip surface, specifically the near-surface supersaturation. This means that the theory is somewhat under-constrained due to a poor knowledge of material properties, so we should not read too much into the linear trend seen in Figure 3.24. This issue is a manifestation of a more general problem with solvability theory: an analytic theory including just a few basic parameters may not be sufficient to describe a complex phenomenon like free dendrite growth.

Note that had we ignored the attachment-kinetic perturbation and instead kept the surface-energy perturbation in solvability theory, the result would have included the scaling $R_{tip} \sim \sigma_{far}$ and $v_{tip} \sim \sigma_{far}^2$. The above caveats notwithstanding, Figure 3.24 does not agree with such a quadratic dependence, and this supports the notion that that attachment kinetics is the more important stabilizing mechanism, in agreement with the sizes of the relevant terms in Figure 3.42.

## Ice Dendrites Growing from Liquid Water

The formation of ice dendrites from freezing liquid water is a substantially different problem than the snow crystal case. In melt solidification, growth is driven by an undercooling $\Delta T$ (a.k.a. supercooling) and is limited primarily by latent heat diffusion. The Peclet number is generally large, so the Laplace approximation to the diffusion equation is not justified. In solvability theory describing solidification from the melt, the dominant stabilization term is typically surface tension, while attachment kinetics can often be neglected.

For the succinonitrile system shown in Figure 3.23, for example, the dendrite tip radius scales as $R_{tip} \sim \Delta T^{-1}$ and the tip velocity scales as roughly $v_{tip} \sim \Delta T^{2.7}$ [1981Hua]. (If the Peclet number were small, the velocity scaling would likely change to $v_{tip} \sim \Delta T^2$, in accordance with the previous discussion.) This behavior is nicely explained by the relevant version of solvability theory that includes a large Peclet number and neglects attachment kinetics. The growth of ice dendrites from liquid water also fits this theory quite well, as seen in Figure 3.25 (see also [1978Lan, 1980Lan]).

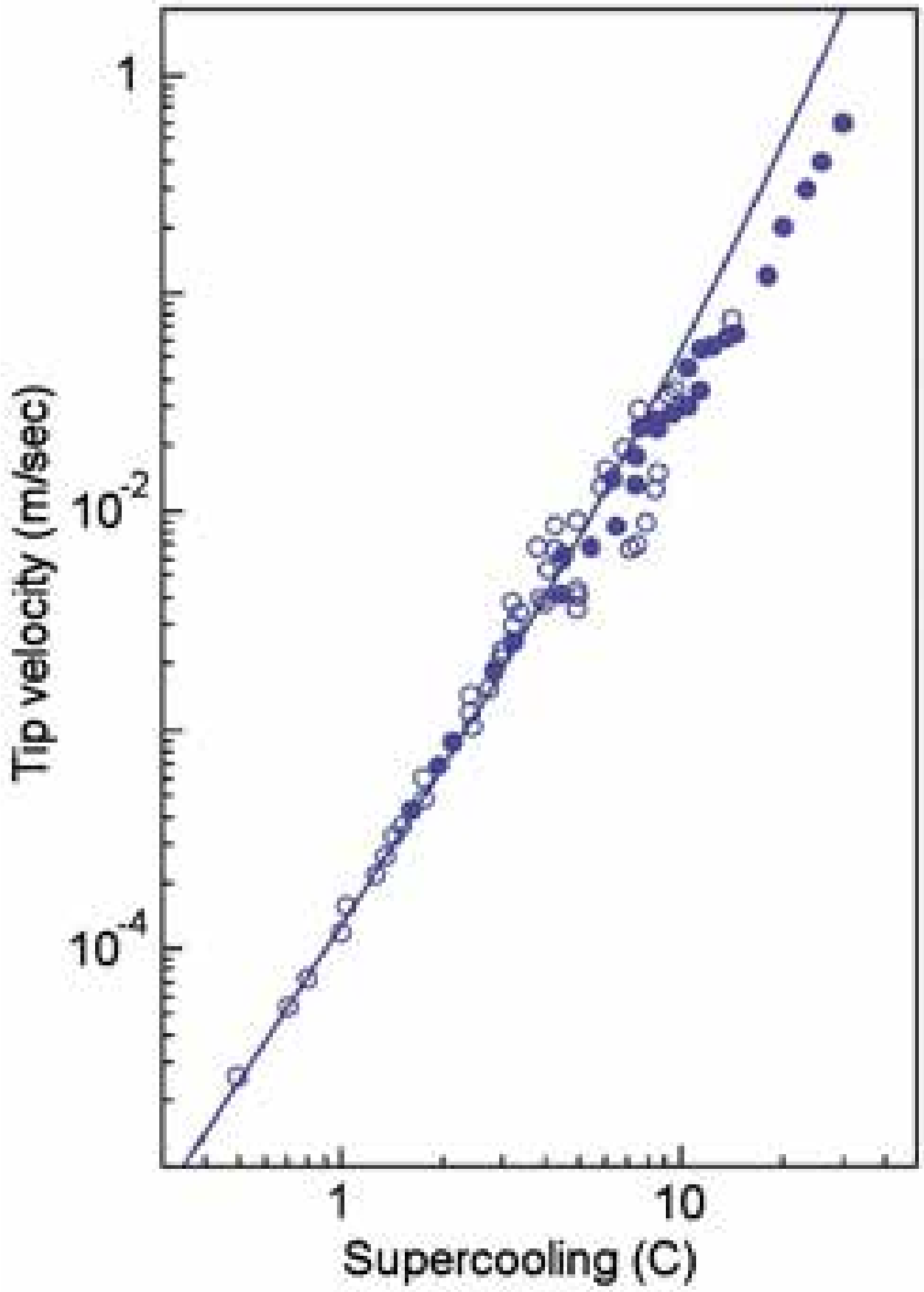

**Figure 3.25: Measurements of the free dendrite tip velocity for ice growing from liquid water as a function of the far-away supercooling. The line shows a calculation from solvability theory. (Figure adapted from [2005Shi], with additional context given in [2017Lib].)**

Here again, the theory comes with some caveats. In the case of ice dendrites growing from liquid water, we know that attachment kinetics play an important role, because the dendrite morphology is that of a flat plate over a broad range of undercoolings, indicating strong basal faceting. Neglecting attachment kinetics in solvability theory, therefore, is at least a somewhat questionable assumption in the water/ice case. And, once again, the underlying material properties are not known well enough to constrain the theory adequately, and numerical modeling methods are not yet sophisticated enough to include the combined effects of diffusion, surface energy, and attachment kinetics. As a result, even though theory and experiment seem to agree nicely in both Figures 3.24 and 3.25, the full story has perhaps not been told in either case.

## Anisotropy and Tip Splitting

Another important discovery from solvability theory is that $s_0$ depends on the *anisotropy* of the surface physics that stabilizes the dendrite tip radius. For perfectly isotropic systems, even the initial premise of a stable, Ivantsov-like parabolic tip structure turns out to be incorrect. With perfect isotropy, the Mullins-Sekerka instability brings about not only sidebranches, but also tip splitting.

Tip splitting behavior is best seen in computer simulations of dendritic growth, which have improved greatly since the founding days of solvability theory. Figure 3.26 shows a growing dendritic system for which the anisotropy was varied in different runs. With no anisotropy, the dendritic branches exhibited frequent tip splitting that resulted in a complex "seaweed-like" structure. Above some threshold anisotropy, dendrites with stable tip structures appeared.

This dependence on anisotropy appears to be a general property of diffusion-limited dendrite formation, present over a broad range of different physical systems. When the surface stabilization forces are sufficiently anisotropic, Ivantsov-like dendrites appear with stable tip structures, as seen in snow-crystal dendrites. As the anisotropy is turned down, first tip-splitting begins to occur only occasionally, and it becomes more frequent until the growth transitions to completely random seaweed-like structures as the anisotropy decreases to zero.

Tip splitting is largely absent in snow-crystal dendrites, owing to the exceptionally large underlying anisotropy in the attachment kinetics. Nevertheless, Figure 3.26 shows an example of tip splitting in a rapidly growing fernlike stellar dendrite, indicating $\alpha_{prism} \approx 1$ when the surface supersaturation is sufficiently high. This general behavior fits the model for nucleation-limited attachment kinetics presented in Chapter 3. I have also witnessed some dendritic tip splitting at temperatures near 0 C when the supersaturation is high and chemical contaminants are present, again indicating $\alpha_{prism} \approx 1$ under those conditions.

Figure 3.14 shows some evidence of tip splitting for ice/water solidification, but only at quite low supercooling. In this case we need to consider the anisotropy in both the attachment kinetics and the surface energy, and neither of these is known well from other measurements.

Note that the basal anisotropy is relatively high under all growth conditions, for both ice/water and ice/vapor, owing to a finite basal step energy at 0 C (see Chapter 3). Thus, one expects a complete absence of basal tip

**Figure 3.26: In crystal growth, some kind of surface anisotropy, usually in the attachment kinetics or surface energy, is necessary to prevent tip splitting and create stable free-dendrite growth (far left). For perfectly isotropic systems, seaweed-like structures emerge (far right). This numerical simulation shows a morphological transition between these two states, in this case when the underlying anisotropy possesses six-fold symmetry. (Image from [2006Gra1].)**

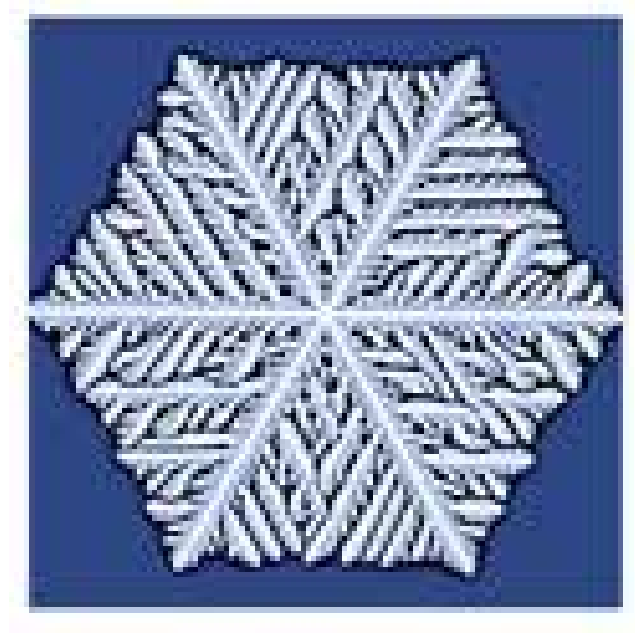
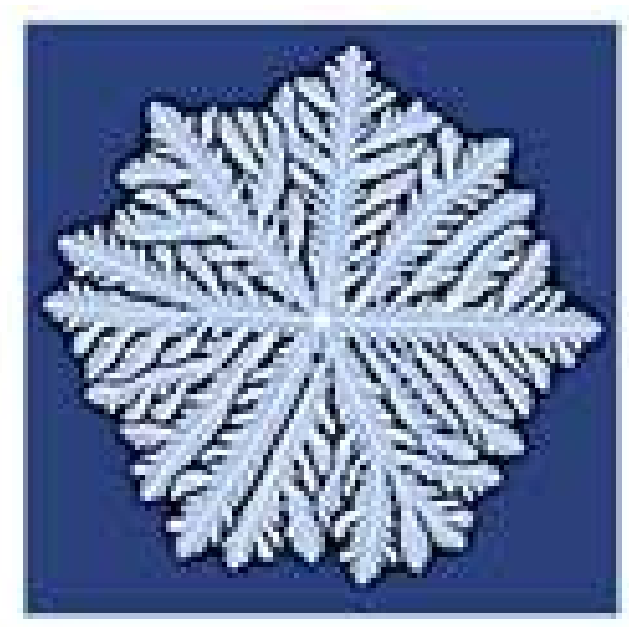
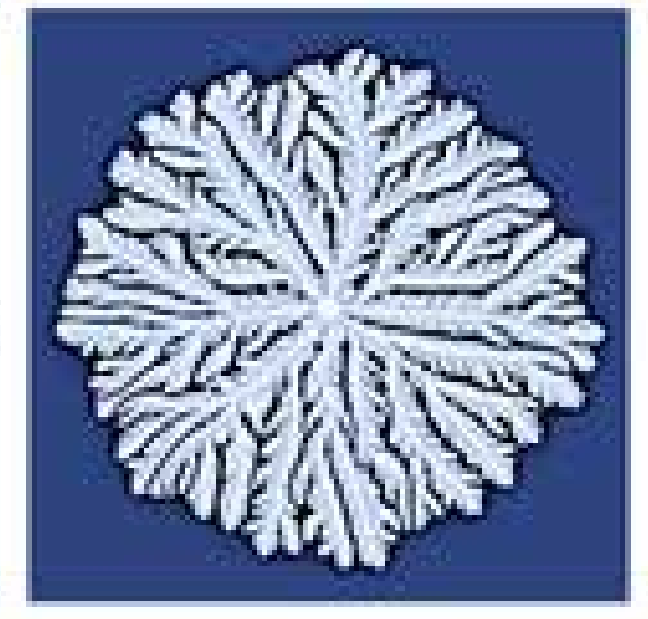
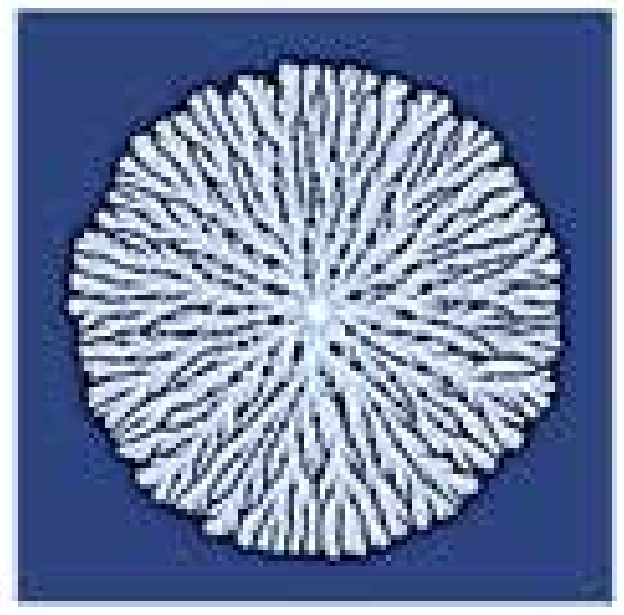

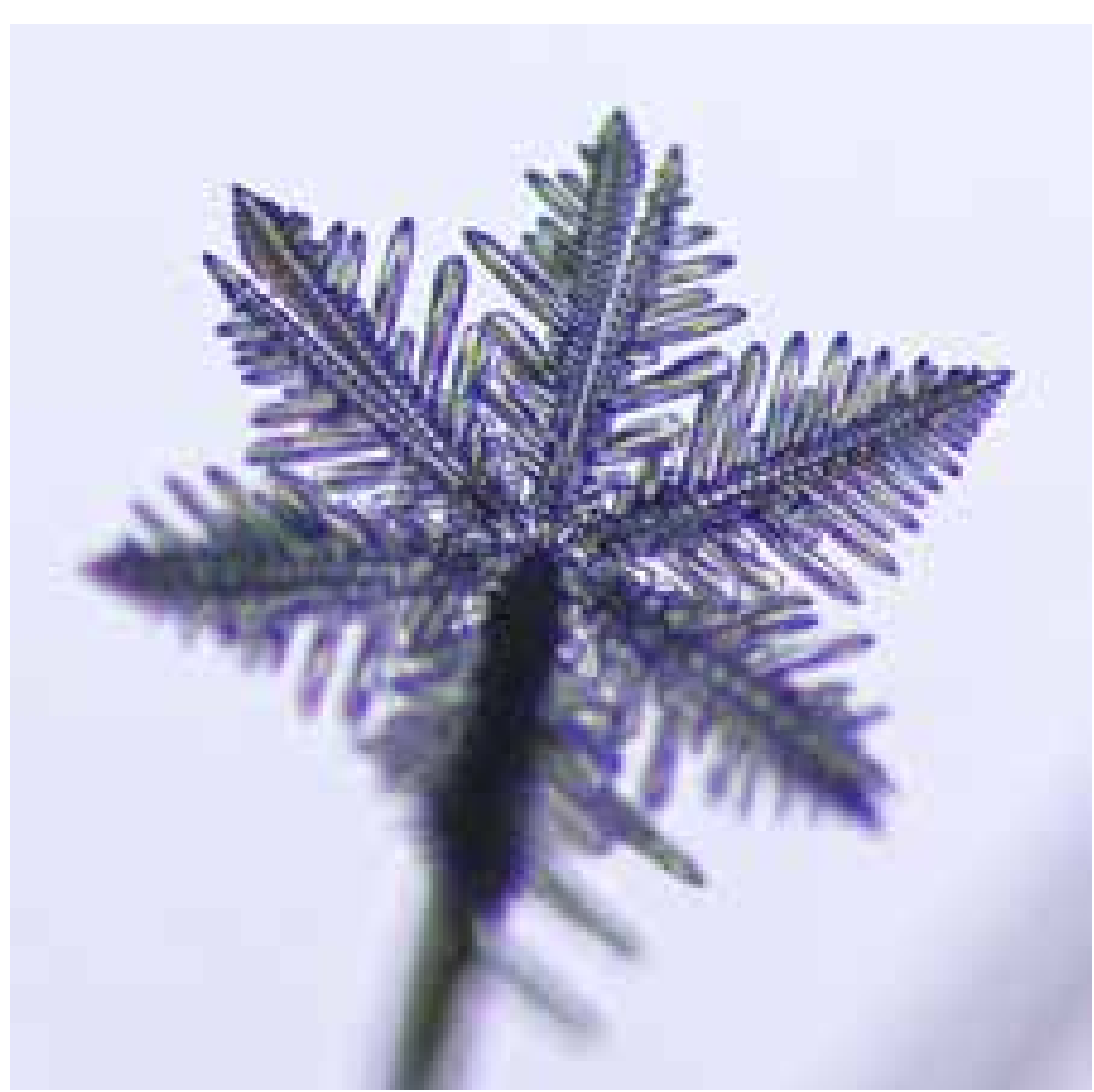

**Figure 3.27: An example of tip splitting in rapidly growing fernlike dendrites at $\sigma_{far} \approx 1.3$ near -15 C. In these conditions, $\alpha_{prism}$ is close to unity and the anisotropy in the kinetic coefficient becomes quite low. Notably, the tip splitting in this example occurred when the crystal was small, and the surface supersaturation was therefore higher than at later times (as indicated in Equation 3.25).**

splitting, and this expectation is consistent with observations.

## 3.7 Snow Crystal Aerodynamics

We next turn our attention to how aerodynamics can affect snow crystal growth and morphologies [2009Lib3, 1982Kel, 2002Wan, 1999Fuk, 1997Pru, 2009Lib4]. In normal air, the motion of falling crystals can align their orientation relative to the horizon, change their growth rates, and even alter their growth morphologies, although typically all these effects are rather small perturbations compared to normal growth behaviors. Our main goal in this section is to outline the basic physical processes using analytic models, estimating the importance of the various effects over a range of growth conditions.

Throughout this discussion, it is important to remember that wind speed relative to the ground is not the relevant parameter in the problem, but rather wind speed relative to the crystal in question. A small snow crystal may be carried by the wind for long distances, but, like a speck of dust, it mostly travels along with the moving air around it. Thus, while wind blowing over a stationary snow crystal in the lab may strongly perturb its growth [1982Kel], we cannot apply these results until we understand the velocity of air flow around a freely falling snow crystal.

### Drag and Terminal Velocity

Gravity creates a net velocity between crystals and air, and a falling snow crystal quickly reaches its terminal velocity in still air. The viscous drag on a snow crystal is well described by Stokes drag at low velocities, given by

$$F_{Stokes} = 6\pi\mu R_H u \qquad (3.45)$$

where $F_{Stokes}$ is the drag force, $R_H$ is the hydrodynamic radius of the object, $\mu$ is the dynamical viscosity of air, and $u$ is the flow velocity. For a spherical particle, $R_H$ equals the radius $R$ of the sphere.

At the velocity increases, the flow becomes turbulent, adding a component to the drag that is proportional to $u^2$. Assuming a thin disk morphology with radius $R$ and thickness $T$ (a satisfactory model for a plate-like snow crystal), the drag force becomes

$$F_{drag} \approx 6\pi\mu R u + \frac{\pi}{2}\rho_{air} R^2 u^2 \qquad (3.46)$$

to a reasonable approximation, where $\rho_{air}$ is the density of air [2009Lib3]. The two terms in this expression are equal when the Reynolds number $R_e$ is about 24, where I take

$$R_e = \frac{2\rho_{air} u R}{\mu} = \frac{2uR}{\nu_{kinematic}} \qquad (3.47)$$

and $\nu_{kinematic} = \mu/\rho_{air}$ is the kinematic viscosity.

The falling thin-disk crystal reaches its terminal velocity $u_{term}$ when $F_{drag} = mg$,

where $m = \pi R^2 T \rho_{ice}$ is the mass of the crystal, giving

$$u_{term} \approx \frac{1}{6}\frac{\rho_{ice} g}{\mu} RT \quad (low\ R_e) \tag{3.48}$$
$$\approx 8\left(\frac{R}{100\ \mu m}\right)\left(\frac{T}{10\ \mu m}\right)\ \text{cm/sec}$$

for the case of small crystals falling at low Reynolds number and

$$u_{term} \approx \left(\frac{2T\rho_{ice} g}{\rho_{air}}\right)^{1/2} \quad (high\ R_e) \tag{3.49}$$
$$\approx 40\left(\frac{T}{10\ \mu m}\right)^{1/2}\ \text{cm/sec}$$

for larger crystals moving at high Reynolds number [2009Lib3]. For these thin-disk crystals, the transition from low to high Reynolds number terminal velocity occurs when the crystal radius exceeds

$$R_{transition} \approx 450\left(\frac{10\ \mu m}{T}\right)^{1/2}\ \mu m \tag{3.50}$$

Figure 3.28 shows an example of the terminal velocity of a 2-$\mu m$ thick disk as a function of its radius.

Comparing terminal velocity calculations with observations is not especially fruitful, unfortunately. The theory is well understood for small crystals with simple shapes, while most measurements have been obtained using larger crystals with complex, rather poorly characterized morphologies and sizes. Nevertheless, the extensive measurements of fall velocities in a vertical flow chamber made by Fukuta and Takahashi [1999Fuk] seem to be consistent with the above theory, given the substantial uncertainties involved.

## Horizontal Alignment

Over a range of snow crystal sizes and morphologies, drag forces align falling crystals relative to the horizon. The resulting alignment is well known in natural snow crystals, as it is essential for explaining many distinctive features in atmospheric halos [2006Tap, 1990Tap, 1980Gre]. Thin disks, for example, often align with the basal surfaces in a horizontal orientation (vertical c-axis), while slender columns align with a horizontal c-axis. In some instances, columns may align further with two prism facets in a horizontal orientation, known as the Parry orientation [2006Tap]. In some rare halo observations, models suggest widespread crystal alignments as precise as a few degrees relative to the horizon.

Focusing on plates, theory suggests that the smallest crystals will not align unless their terminal velocities are larger than surrounding turbulent air flows that perturb their fall and orientation. Meanwhile large plates are unstable to various fluttering and tumbling instabilities when the Reynolds number exceeds $R_e \approx 100$. The latter regime applies to crystals with sizes of about 1 mm or more, while Figure 3.28 shown two models for turbulent air velocities [2009Lib3]. Unless the air is exceptionally still, snow crystals are likely to exhibit good alignment only in roughly the 0.1-1 mm size range.

**Figure 3.28: The terminal velocity curve on this plot shows the fall velocity of a 2-$\mu m$-thick snow crystal disk as a function of disk radius, while approximate scaling with disk thickness is given by Equations 3.48 and 3.49. The turbulence curves show two models of root-mean-squared air velocities when the average air speed in 1 m/sec. When $R$ is large enough that the terminal velocity curve is above the turbulence curves, then gravity can align the crystal horizontally [2009Lib3].**

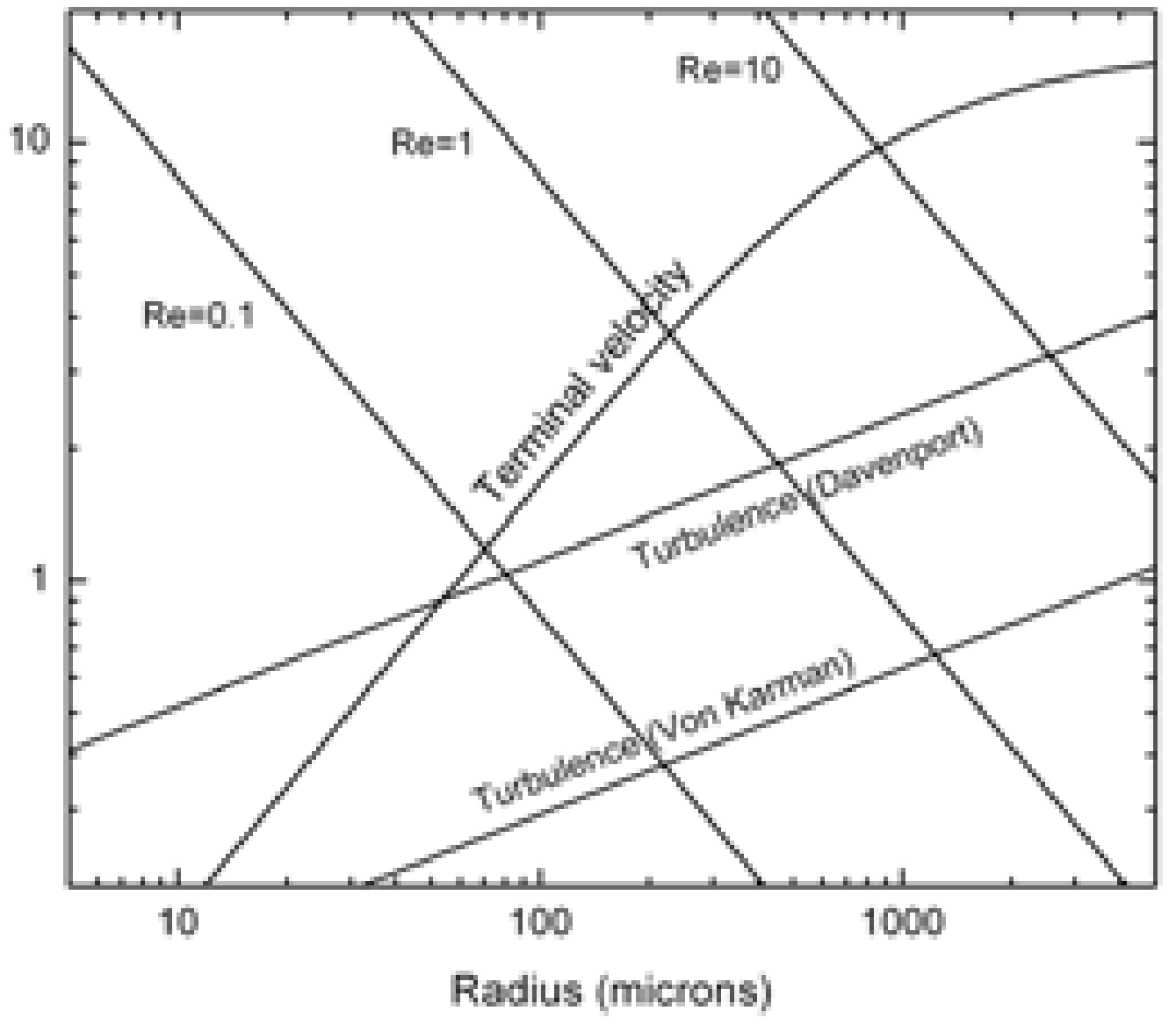

## The Ventilation Effect

When supersaturated air flows around a snow crystal, its growth rate increases as the flow essentially enhances the diffusion of water vapor molecules to its surface, and this phenomenon is called the *ventilation effect* [1982Kel, 1997Pru]. The magnitude of the growth change can be estimated by comparing the diffusion time and the flow time. The diffusion timescale for water molecules diffusing a distance $L$ through the air is

$$\tau_{diffusion} \approx \frac{L^2}{D} \qquad (3.51)$$

A growing crystal significantly reduces the supersaturation in its vicinity only out to a distance comparable to its size, so we take $L$ to be the approximate size of the crystal. Meanwhile the time it takes for air to flow the same distance $L$ is

$$\tau_{flow} \approx \frac{L}{u} \qquad (3.52)$$

If the flow velocity is low and $\tau_{diffusion} \ll \tau_{flow}$, then diffusion creates a depleted region around the crystal before the air flows by. In this case, the ventilation effect would be negligible, as we would expect when $u \rightarrow 0$. We expect, therefore, that air flow would significantly affect the crystal growth only when $\tau_{flow} < \tau_{diffusion}$, which is equivalent to the regime $R_e > 1$.

In a somewhat more in-depth analysis incorporating studies of liquid droplet growth in the literature, I found that that the growth rate of a spherical snow crystal is enhanced by a factor

$$f_v \approx 1 + 0.1 R_e \quad (R_e < 1) \qquad (3.53)$$
$$f_v \approx 0.8 + 0.3 R_e^{1/2} \quad (R_e > 1)$$

to a reasonable approximation [2009Lib3].

Applying this to a specific example, consider the fernlike stellar dendrite shown in Figure 3.8. This is a common snow crystal morphology, and an examination of the calibrated photo reveals that the initial branching instability occurred when the crystal radius was no larger than $R \approx 30\ \mu m$. From Figure 3.28, the Reynolds number of the air flow around this nascent crystal was likely about $R_e \approx 0.1$, giving an enhancement factor of $f_v \approx 1.01$, meaning that the ventilation effect was likely negligible when the first branching event occurred. Strong turbulence might have increased this enhancement, but well-formed crystals like the one in Figure 3.8 rarely survive long in windy, turbulent conditions.

As this crystal grew larger, the Reynolds number of the flow around it increased, and the crystal morphology became dominated by the six fernlike dendritic branches. Then the air flow likely aligned the crystal so its basal faces were nearly horizontal, and the flow past each tip was roughly perpendicular to the growth direction. The ventilation effect is more difficult to analyze in this case, but the sharp-tipped geometry leads to a substantially higher ventilation effect compared to the spherical case. At terminal velocity for this crystal, the ventilation effect would produce roughly a 25 percent increase in tip growth velocity [2009Lib3].

Combining alignment and ventilation effects, it is possible that aerodynamics plays a role in promoting the high symmetry of some snow crystal structures. This is likely a small effect, a supposition that is supported by the fact that the vast majority of snow crystals do not exhibit a high degree of six-fold symmetry. Nevertheless, aerodynamic alignment can lead to tumbling instabilities that would tend to enhance symmetrical growth of several crystal morphologies. Some possibilities along these lines are discussed in [2009Lib3, 1999Fuk].

It has also been suggested that aerodynamic effects may promote the growth of triangular snow crystals, through a combination of alignment and ventilation effects [2009Lib4]. However, as discussed in the next section, the origin of triangular plate snow crystals is still a bit of a mystery, and it is not yet clear if aerodynamics plays a major role in their development.

The bottom line in this discussion is that aerodynamics can play a role in snow crystal growth dynamics, but it is usually a rather minor one. Small crystals are the least susceptible to aerodynamic effects, although remarkably precise crystal alignments are possible in rare and especially calm conditions.

# 3.8 Growth Behaviors

Snow crystal morphologies are determined mainly by the interplay of two physical processes: attachment kinetics and particle diffusion. Attachment kinetics brings about ordered, faceted surfaces with sharp edges and corners, defined precisely by the crystal lattice structure. Diffusion brings about instability, yielding complex structures and the chaotic sidebranching seen in dendritic growth. These are the competing forces of order and chaos that drive the formation of snow crystals. In some circumstances, additional physical effects from heat diffusion, surface energy, air flow, etc., may also be important; but typically, these effects are small and often negligible.

In this section I examine a selection of snow crystal morphological features in some detail and attempt to describe how each originates from the combined effects of attachment kinetics and particle diffusion. This undertaking would be best accomplished with the help of corresponding numerical simulations, but here the state-of-the-art is somewhat unreliable. Computational snow crystals do not yet reproduce real snow crystal structures with good fidelity (see Chapter 5).

One motivation in this section, therefore, is to qualitatively describe various growth behaviors that might be investigated more quantitatively in future numerical simulations. Another motivation is to develop an overarching physical intuition regarding the underlying causes of snow crystal formation, as this is helpful for making additional progress in the field. And last, but not least, it is simply pleasing to have an essential understanding of some of the puzzling characteristics often seen in snow crystal structures.

## Aspect Ratios and Anisotropy

As a general rule, I have found that the large-scale aspect ratio of a snow crystal – here defined as the ratio of the overall size of a crystal along the c-axis to that along an a-axis – reflects the anisotropy in the underlying attachment kinetics. For example, the formation of thin plates invariably requires $\alpha_{basal} \ll \alpha_{prism}$, while the formation of slender columns requires $\alpha_{basal} \gg \alpha_{prism}$. Although qualitative in nature, this aspect-ratio rule applies throughout the menagerie of different snow-crystal types.

An important aspect of this rule is that diffusion effects alone cannot yield structures with extreme aspect ratios, like thin plates or slender columns. Numerical studies have revealed that while the Mullins-Sekerka instability may yield complex dendritic branching, the overall aspect ratios of the resulting structures are determined by anisotropies in the underlying attachment kinetics or surface energies. A well-known example is illustrated in Figure 3.26, where isotropic surface physics yielded the seaweed-like structure shown, exhibiting intricate branching but an overall round structure.

Another important corollary of this rule is that extreme aspect ratios seen in the solidification of real materials do not arise from surface energy anisotropy. Highly anisotropic surface energies are mostly a theoretical fiction (except perhaps in exotic materials). As discussed in Chapter 2, simple solids (such as ice or metals) generally exhibit modest anisotropies in the surface energy that are too small to produce large aspect ratios during solidification. In particular, the extreme aspect ratios seen in snow crystals and plate-like pond crystals are the result of highly anisotropic attachment kinetics, and not from highly anisotropic surface energies.

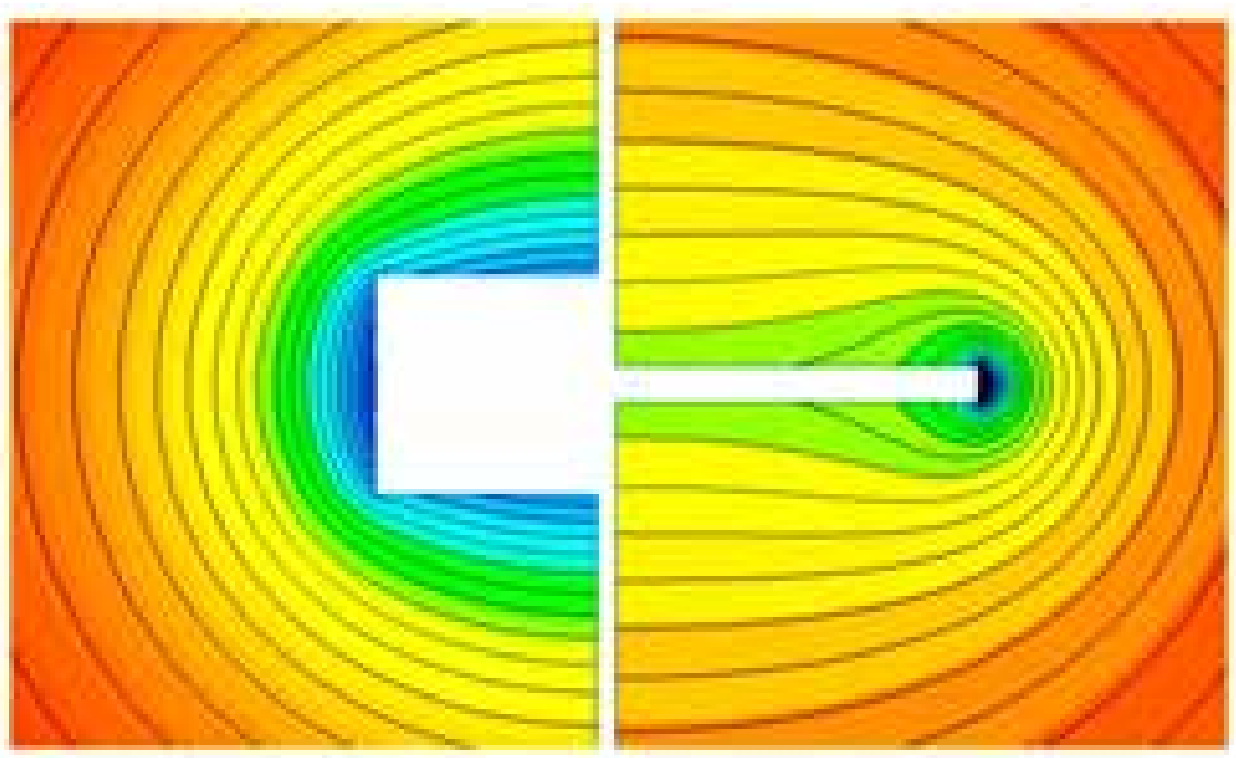

**Figure 3.29: Calculated contour plots of supersaturation levels around two growing ice crystals, shown here in (r,z) coordinates. Around a blocky crystal (left), the supersaturation is highest near the corners of the faceted prism, a phenomenon called the Berg effect. Around a thin-plate crystal, the contour lines are tightly bunched at the plate edge, and the supersaturation is highest at the center of the basal facets. The model on the left assumed $\alpha_{basal} \approx \alpha_{prism} < 1$, while the model on the right assumed $\alpha_{basal} \ll \alpha_{prism}$.**

Figure 3.29 shows a nice illustration of what low and high anisotropy in the attachment kinetics looks like around a faceted snow crystal. In the case of a nearly isometric faceted prism (left side of Figure 3.29), $\alpha_{basal}$ and $\alpha_{prism}$ are roughly equal in magnitude, so the basal and prism facets grow at about the same rates. Thus, the overall aspect ratio is close to unity, indicating $\alpha_{basal} \approx \alpha_{prism}$.

Of course, the presence of strong basal and prism faceting indicates anisotropy in the sense that both $\alpha_{basal} < 1$ and $\alpha_{prism} < 1$. The overall aspect ratio of the crystal, however, is determined mainly by $\alpha_{basal}/\alpha_{prism}$. In this case, we see that the supersaturation near the surface is highest at the corners of the prism, essentially because the corners stick out farther into the supersaturated air, a phenomenon called the Berg effect (see also Figure 3.4).

Looking at the other side of Figure 3.29, the growth of the thin plate crystal resulted because $\alpha_{bas} \ll \alpha_{prism}$. In keeping with the general rule, the extreme aspect ratio for this crystal resulted from highly anisotropic attachment kinetics. Now wee see that the surface supersaturation is substantially higher on the basal facet than on the prism facet, contrary to the usual expectation from the Berg effect. Even though the prism edge sticks out farther into the supersaturated air, $\sigma_{surf}$ is lowest there. Thus, the normal Berg effect does not apply in cases when the anisotropy in the attachment kinetics is sufficiently high.

The contour lines around the thin plate in Figure 3.29 also show that the supersaturation gradient is highest near the prism facet. This makes sense because the prism surface is growing rapidly, which requires a high influx of water vapor molecules. And this implies a steep supersaturation gradient, as the diffusion equation tells us that particle flux is proportional to $\hat{n} \cdot \vec{\nabla} \sigma$. In contrast, the particle flux and supersaturation gradient is low near the center of the basal facets.

Additional models like these reveal that the strong correlation between aspect ratio and anisotropy in the attachment kinetics applies over a broad range of growth conditions. If, for example, one begins with a thin plate crystal and then changes parameters so that $\alpha_{basal} \approx \alpha_{prism}$, then the subsequent growth will not maintain the thin-plate structure. Instead the edges of the plate would thicken over time and the overall aspect ratio would tend toward unity. Diffusion-limited growth generally pushes morphologies toward small overall aspect ratios, and this trend is usually countered only by a strong anisotropy in the attachment kinetics.

## Stellar Dendrites near -15 C

Although often overlooked, the primary morphological feature of a stellar dendrite snow crystal is the fact that it is thin and flat. The aspect ratio can be as low as 0.01 for a large thin plate, and this extreme aspect ratio is what puts the "flake" in "snowflake." Following the discussion above, this structure immediately demands that $\alpha_{basal} \ll \alpha_{prism}$. Moreover, rounding on the branch tips indicates that $\alpha_{prism} \approx 1$ there, because rounding indicates that rough and faceted surfaces are growing at about the same rate. If $\alpha_{prism}$ were substantially below unity, prism faceting would be more prevalent. Thus, just looking at a large stellar dendrite with rounded branch tips reveals that $\alpha_{prism} \approx 1$ at the tips and $\alpha_{basal} < 0.01$.

Stellar dendrite snow crystals are a good illustration of the complex interplay between branching and faceting. Because $\alpha_{basal} \ll 1$, basal faceting dominates the c-axis dimension of the crystal structure. At the opposite extreme, the fact that $\alpha_{prism} \approx 1$ means that prism faceting is quite weak and susceptible to branching and sidebranching. Thus, both the aspect ratio and the degree of sidebranching are determined by the attachment coefficients on the two primary facet surfaces.

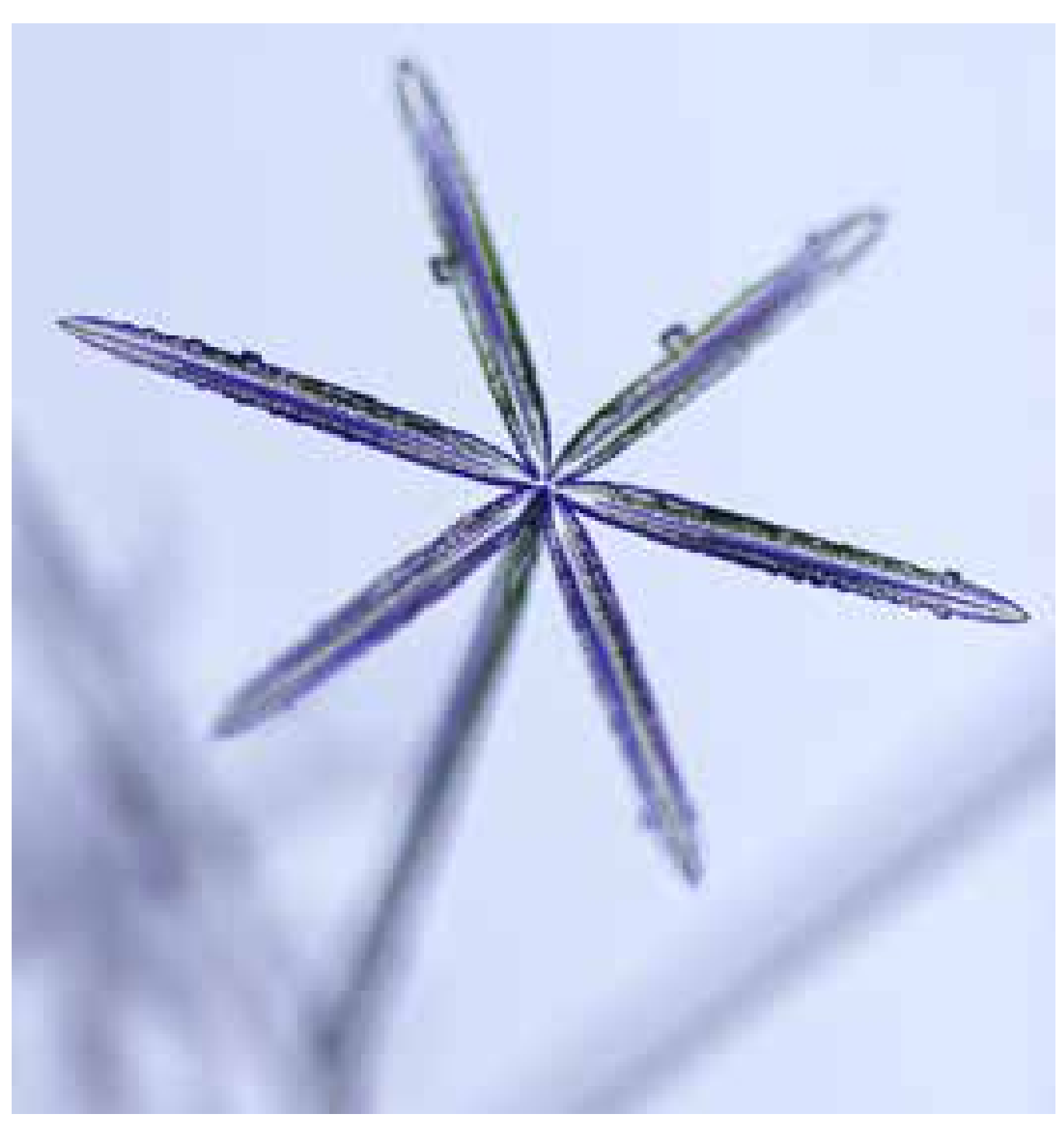

**Figure 3.30: Simple stars present a case where large-scale morphology is directed by an unusual characteristic of anisotropic attachment kinetics. Rounding near the branch tips indicates $\alpha_{prism} \approx 1$ on the outermost prism surfaces, but faceting on the branch sides indicates $\alpha_{prism} < 1$ on those prism surfaces. This difference explains why sidebranching is suppressed on this crystal. The Edge-Sharpening Instability (see Chapter 3) may be present at the branch tips, although high-fidelity 3D numerical simulations will likely be needed to fully understand this seemingly simple morphology.**

If $\sigma_\infty$ around the growing crystal is high, then $\sigma_{surf}$ becomes relatively high as well, sending $\alpha_{prism} \rightarrow 1$ as described in Chapter 3, stimulating copious sidebranching. But if $\sigma_\infty$ and $\sigma_{surf}$ are lower, then $\alpha_{prism}$ is lower and the branches exhibit greater prism faceting. This contributes to why higher $\sigma_\infty$ yields more complex branched structures.

Similarly, $\sigma_{surf}$ is typically highest near the branch tips, so these are often rounded, while $\sigma_{surf}$ is lower near the crystal center, yielding more prism faceting in the central region. Indeed, photographs of stellar dendrite crystals often reveal greater prism faceting in the inner parts of the crystals.

We see that many morphological characteristics of stellar dendrites can be explained from the properties of $\alpha_{basal}$ and $\alpha_{prism}$ as functions of $\sigma_{surf}$. This can be turned around as well; the values of $\alpha_{basal}$ and $\alpha_{prism}$ can be inferred, at least to a rough approximation, by an informed examination of the crystal morphology. In the words of Yogi Berra, you can observe a lot just by watching.

## Faceting and Anisotropy

The above discussion illustrates another general rule in snow crystal growth – faceting requires a high anisotropy in the attachment kinetics, specifically $\alpha_{basal} \ll 1$ for basal faceting or $\alpha_{prism} \ll 1$ for prism faceting. Put another way, the higher $\alpha_{facet}$ becomes, the more likely it will be that a faceted surface will be susceptible to some form of the branching instability. When $\alpha_{facet} \approx 1$, faceting is no longer possible, yielding rounded (unfaceted) surfaces and branched structures.

This fundamental feature of the Mullins-Sekerka instability helps explain the increasing morphological complexity with increasing $\sigma_\infty$, which is one of the principal characteristics of the snow crystal morphology diagram. To see this, start with the fact that $\alpha_{facet}$ typically increases monotonically with $\sigma_{surf}$ (see Chapter 3) and add to this the fact that $\sigma_{surf}$ generally increases with $\sigma_\infty$. The result is that branching becomes more prevalent at higher $\sigma_\infty$, thus leading to more complex snow crystal morphologies at higher supersaturations.

## Hollow Columns and Needles near -5 C

Once again, the primary morphological feature of snow crystal columns and needles is their large aspect ratio, which can be 20 or more for an especially slender needle. The same anisotropy rule applies, so an overall columnar shape indicates $\alpha_{prism} \ll 1$ for moderate supersaturations near -5 C. One can create a simple diffusion model for a faceted column, analogous to the models in Figure 3.29, and the results are similar to what was discussed above. If the anisotropic in the attachment kinetics is sufficiently high, then the supersaturation around a slender column is lowest near the basal surfaces, accompanied by steep supersaturation gradients that drive the fast basal growth.

Fully faceted prisms are the norm when $\sigma_\infty$ is sufficiently low, but hollow columns form near -5 C as the supersaturation increases to intermediate values. The basic hollowing mechanism is a form of the Mullins-Sekerka instability, as illustrated in Figure 3.31. Diffusion-limited growth causes $\sigma_{surf}$ to be higher at the edges of a basal facet compared to the basal facet center, and soon the facet edges grow upward and leave the facet centers behind, resulting in conical hollow regions on both ends of the column. This behavior is analogous to the formation of branches shown in Figure 3.5, except now the edges of the basal surface remain faceted, or nearly so, as the hollows develop. Figure 3.32 shows an example of a natural hollow columnar snow crystal. In this crystal, the hollow regions changed their growth behavior as the external conditions changed, yielding a wavy structure in the shape of the hollows. Because both ends of the column experienced the same growth conditions as a function of time, the shape variations on the two ends of the column are nearly symmetrical.

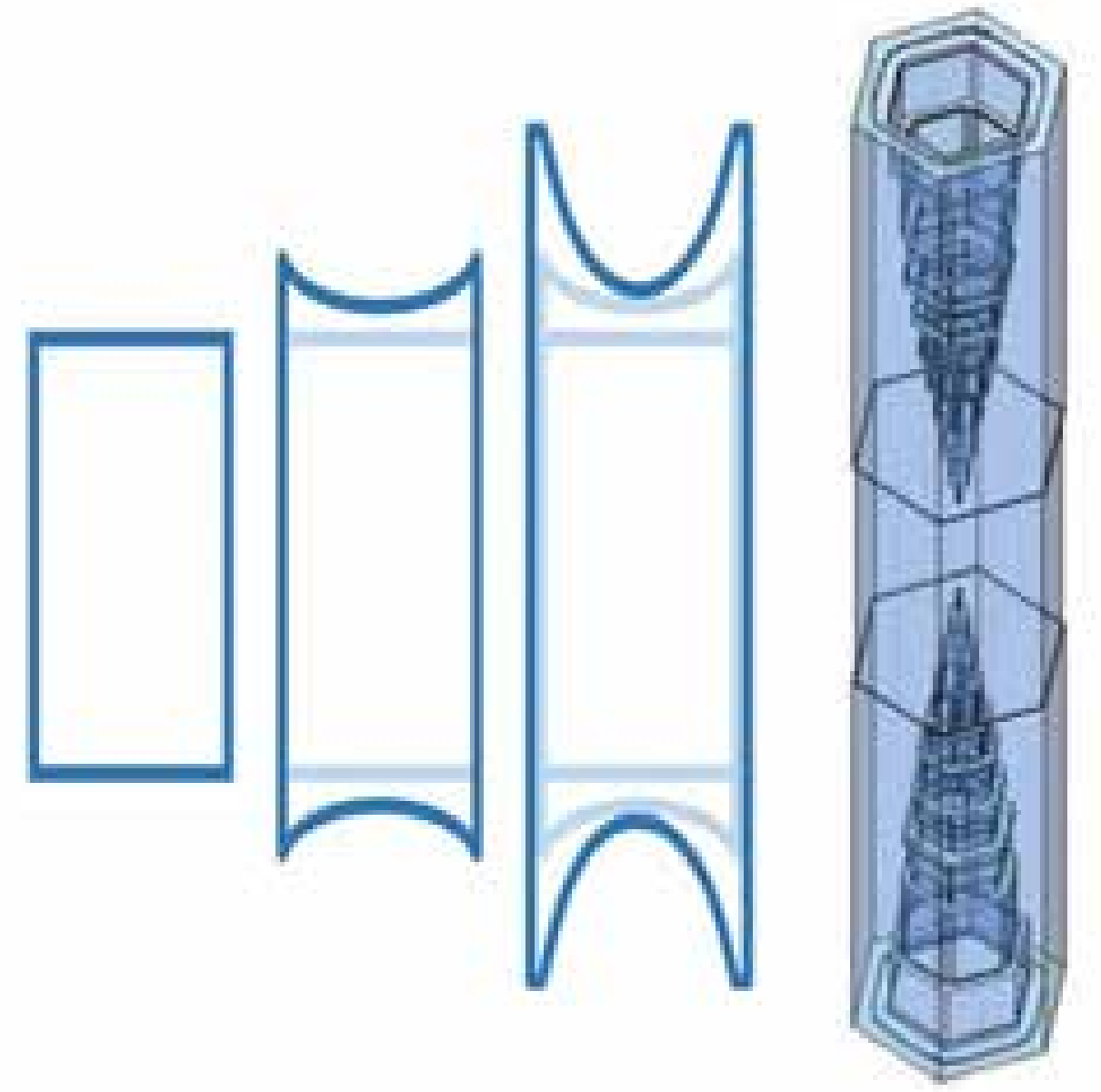

**Figure 3.31: A schematic diagram illustrating the transition from solid columnar growth (first sketch, showing a side view of a solid column) to the formation of a hollow columnar snow crystal (third sketch) via the Mullins-Sekerka instability. The image on the right shows a 3D numerical simulation of hollow-column growth, from [2009Gra].**

As the growth of a hollow column continues, eventually the basal edges will no longer be able to maintain their faceted shape as they too succumb to the branching instability. When this happens, the basal edges can split into slender needles, as shown in Figure 3.33. Note how the initial conical voids are still present near the center of this crystal, exemplifying the transition from a solid column at the earliest stage of growth to a hollow column for a time, and finally to a set of needle-like branches sprouting from the basal corners. Note also that the very center of a hollow column can never itself be hollow, as there would be no mechanism that would yield such a structure from a small seed crystal.

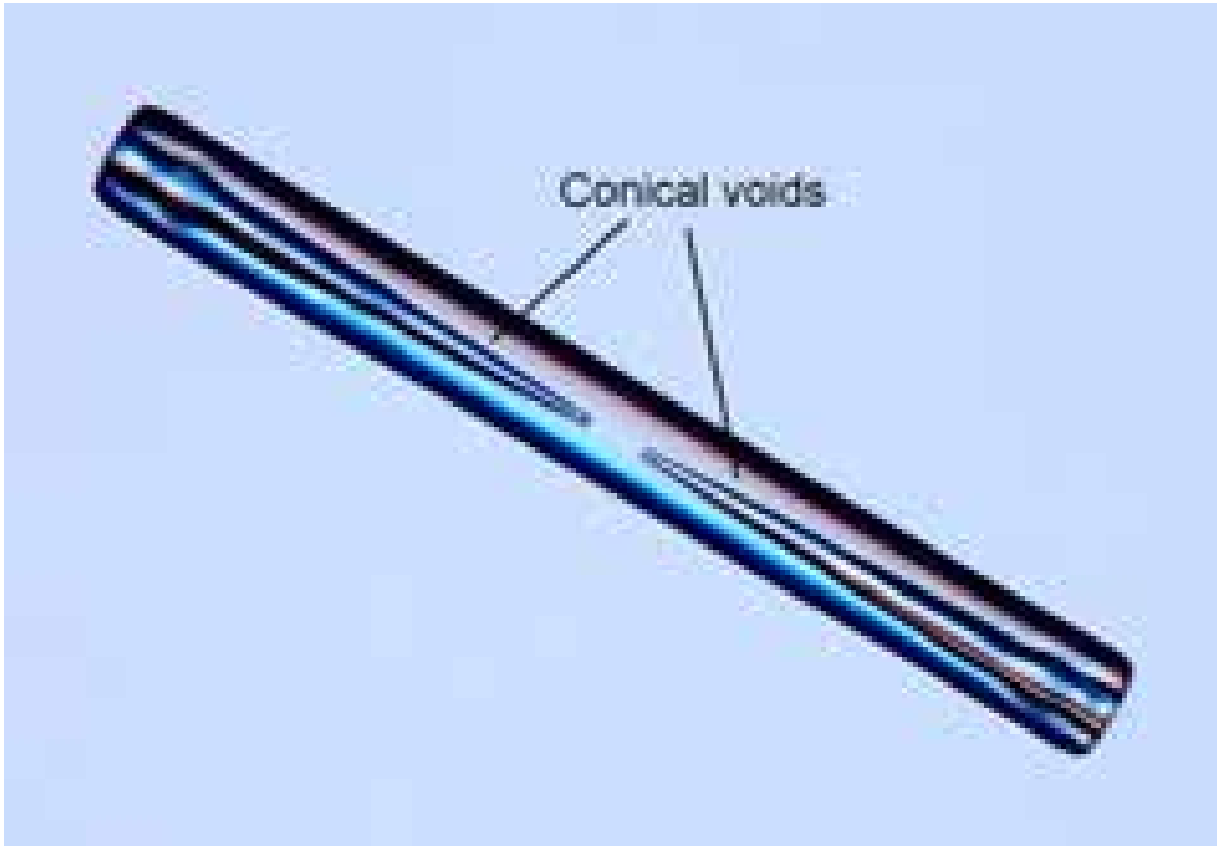

**Figure 3.32: This hollow column snow crystal shows a characteristic matched pair of conical hollow voids in the ice.**

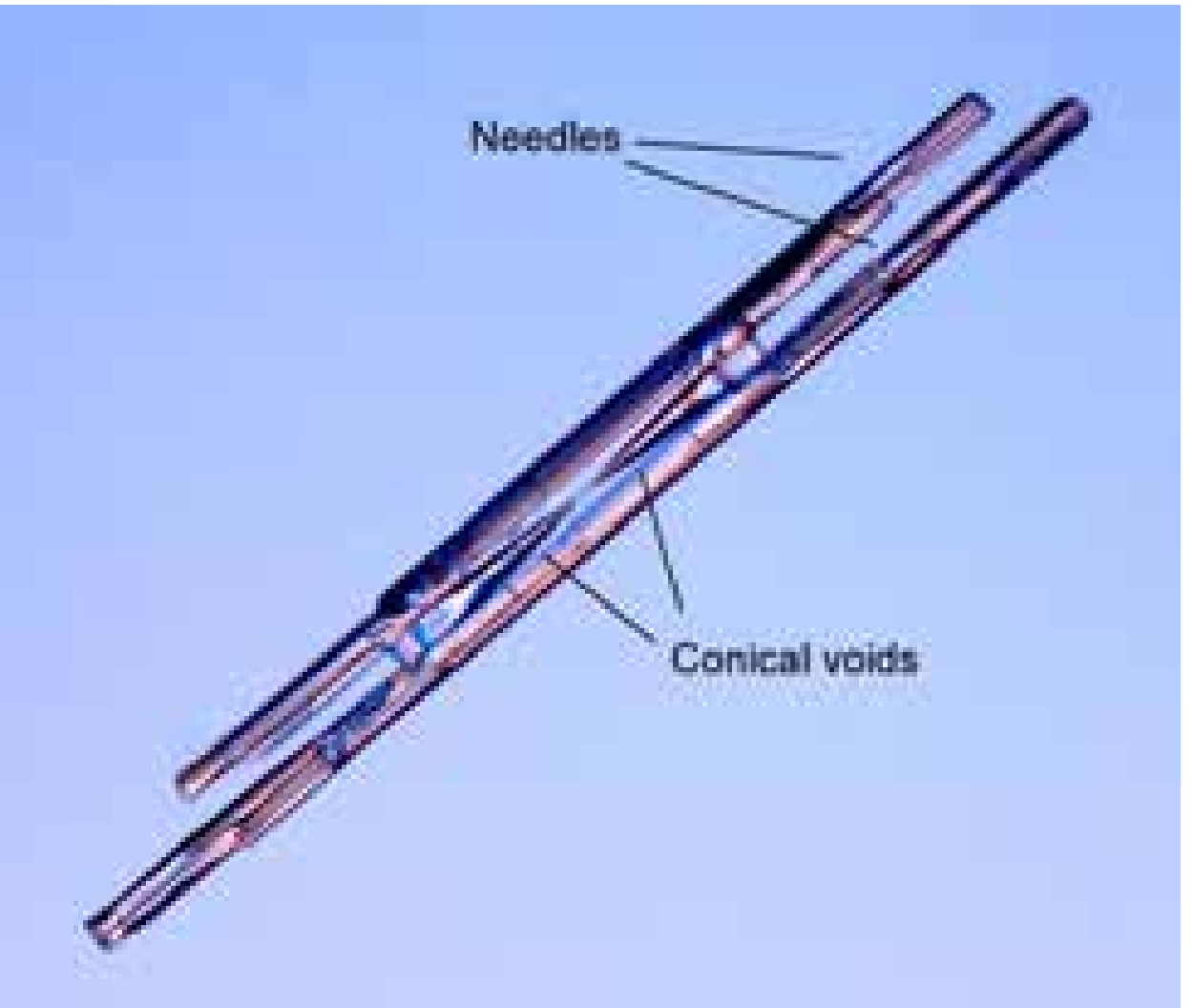

**Figure 3.33: This crystal began as a solid column when it was small, but soon transformed into a hollow column, leaving behind central conical voids in the ice as it grew. Later, the corners of the basal edges sprouted branches, here taking the form of slender ice needles.**

The successful numerical simulation of faceted hollow columns was an excellent early achievement for the 3-D cellular-automata method [2009Gra], which I describe in Chapter 5. Moreover, hollow columnar behavior can be reproduced and studied in the lab using electric needle crystals, as described in Chapter 8. Making quantitative comparisons between laboratory observations and numerical models is therefore now feasible, but not much work along those lines has been done to date.

It is also amusing to speculate what columnar crystals growing near -5 C might develop into if the supersaturation were increased to very high levels. Fishbone dendrites are the natural morphology in these conditions, just like fernlike dendrites are the natural form at -15 C in high $\sigma_\infty$. Just as a fernlike stellar dendrite is essentially six fernlike dendrite branches connected at a common center, Figure 3.34 shows twelve fishbone dendrites growing from the twelve corners of a columnar prism.

The existing observations of fishbone dendrites in the lab suggests $\alpha_{basal} \approx \alpha_{prism}$ for these crystals, yielding the near-unity aspect ratio seen in Figure 3.34. The conditions needed to create such a crystal are not wholly out of reach, so consider this a prediction for future laboratory observations. Alas, the high supersaturation levels necessary to produce fishbone dendritic crystals are not be found in nature.

## Bubbles in Columns

Under the right circumstances, the conical hollows in a hollow-column snow crystal can develop into enclosed bubbles, as illustrated in the two examples in Figure 3.35. I have also created enclosed columnar bubbles in the lab using electric needles, and an example is shown near the end of Chapter 8. In all these cases, the first step is to create a hollow column, as described previously, followed by a period of growth at lower supersaturation that seals off the hollow ends.

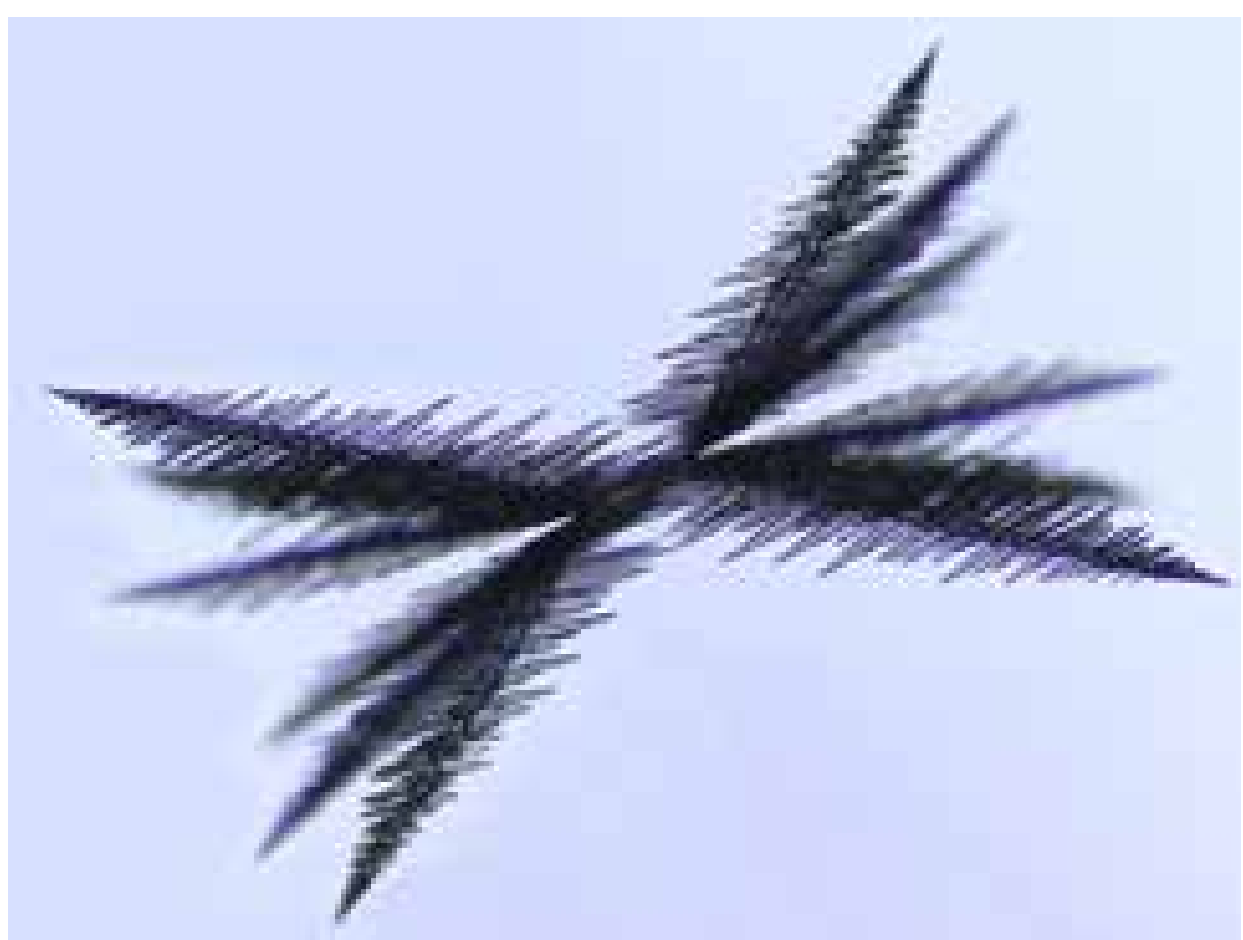

**Figure 3.34: The author's conception showing what a snow crystal growing at -5 C with very high supersaturation probably looks like. Here an initial simple prism rapidly branched into twelve fishbone dendrites. Because $\alpha_{prism}$ is only slightly lower than $\alpha_{basal}$ in these extreme conditions, the overall aspect ratio of this crystal is near unity. This image was created by modifying a photograph of electric needle growth at -5 C and $\sigma_{\infty} = 128\%$, as described in Chapter 8. Actual freely growing crystals at such high supersaturations have not yet been observed, either in the lab or in nature.**

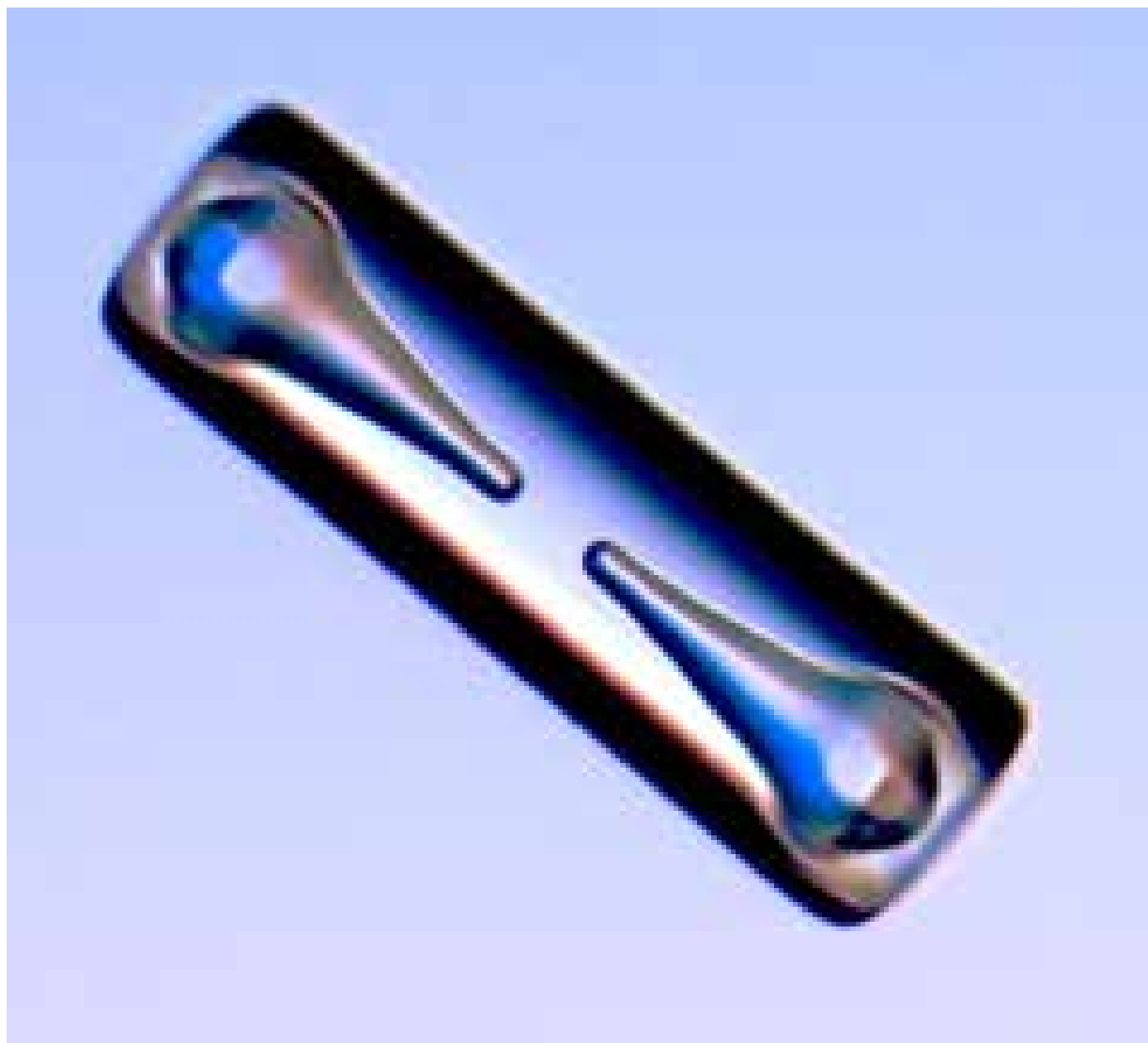

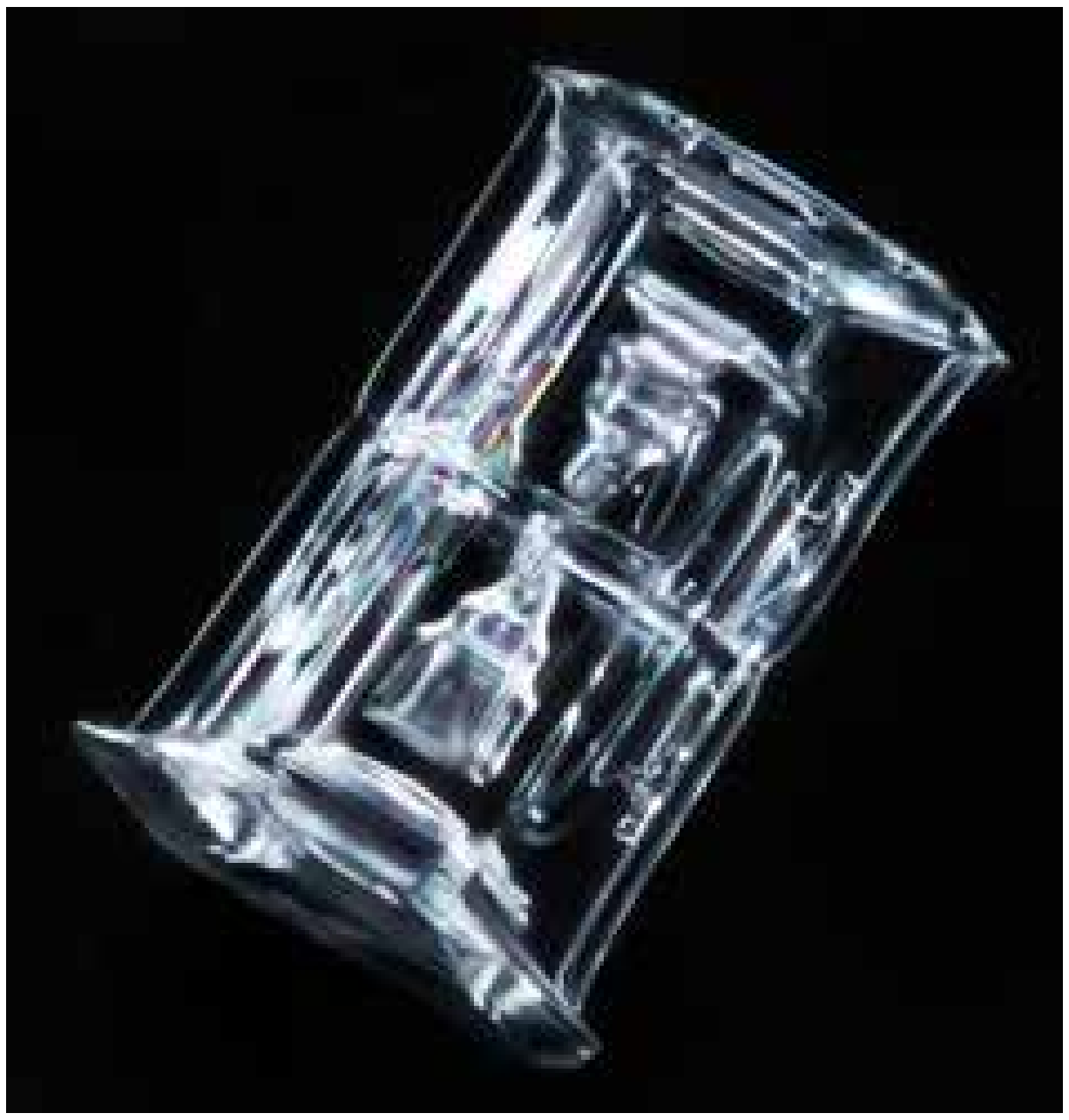

**Figure 3.35: Two photographs of natural snow crystals showing enclosed bubbles in ice columns. The top image was captured by the author, the bottom by Don Komarechka [2013Kom].**

Figure 3.36 illustrates how different nucleation dynamics on convex and concave surfaces facilitates the sealing-off process. The outer surfaces of a growing column soon become faceted because of the usual nucleation barrier that makes $\alpha_{prism} \ll \alpha_{rough} \approx 1$, yielding a hexagonal column. Inside the hollow region, however, there are always interior corners at which there is no nucleation barrier. In Figure 3.36, for example, the red hexagon (representing an idealized molecular cell) can readily attach at the corner shown, as this position is essentially the same as the edge of a terrace step on a faceted surface. More generally, the growth of concave surfaces is never limited (in a global sense) by a nucleation barrier.

When a hollow column is exposed to a relatively low supersaturation, therefore, the strong nucleation barrier on the outer faceted surfaces slows additional growth. But the inner surfaces lack this nucleation barrier, so they grow readily under the same conditions. At the same time, diffusion brings more water vapor molecules to the columnar ends than to regions deep inside the hollows, so growth at the ends is preferred. Thus, the inner surfaces near the columnar ends grow fastest, soon sealing off the conical hollow regions to form enclosed bubbles.

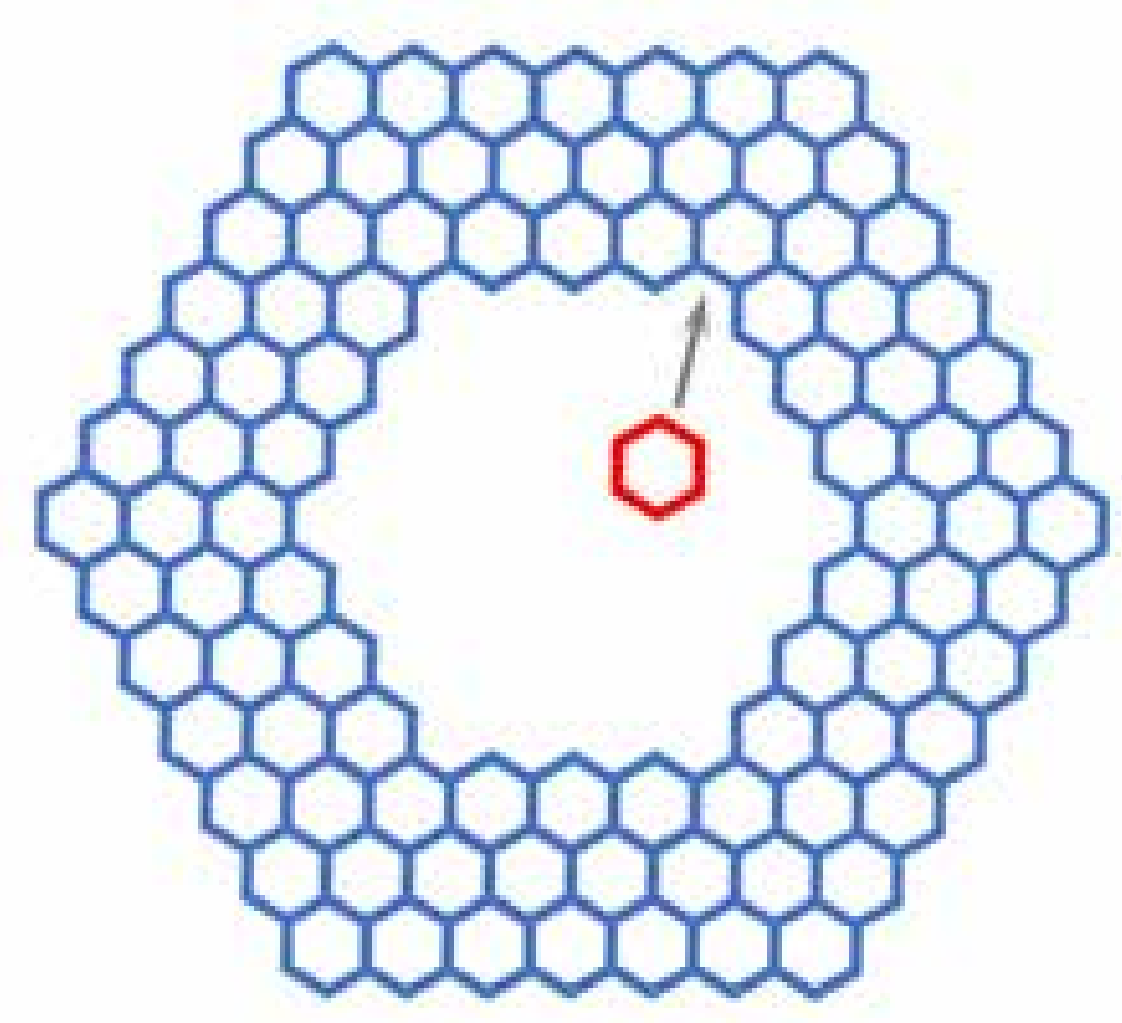

**Figure 3.36: A diagram of the end of a fictitious nanoscopic hollow columnar snow crystal, with hexagons representing molecular cells in the ice lattice. While a nucleation barrier prevents growth of the outer faceted surfaces, there is no nucleation barrier on the inner surfaces because molecules can always attach at interior corners (red hexagon).**

## Hollows & Bubbles in Plates

Hollow columns appear when three conditions are met: 1) $\alpha_{prism} \ll \alpha_{basal}$, 2) $\alpha_{basal} \approx 1$, and 3) the supersaturation is not too low (which would yield solid columns) and not to high (which would yield needle-like crystals). These conditions are typically restricted to temperatures around -5 C, so this is why hollow columns are most prevalent at this region of the snow crystal morphology diagram.

Hollow plates appear when these same conditions are met, but with $\alpha_{prism}$ and $\alpha_{basal}$ exchanged. The underlying physics is essentially the same as with hollow columns, and Figures 3.37 and 3.38 show two examples of hollow plates, one with enclosed bubbles.

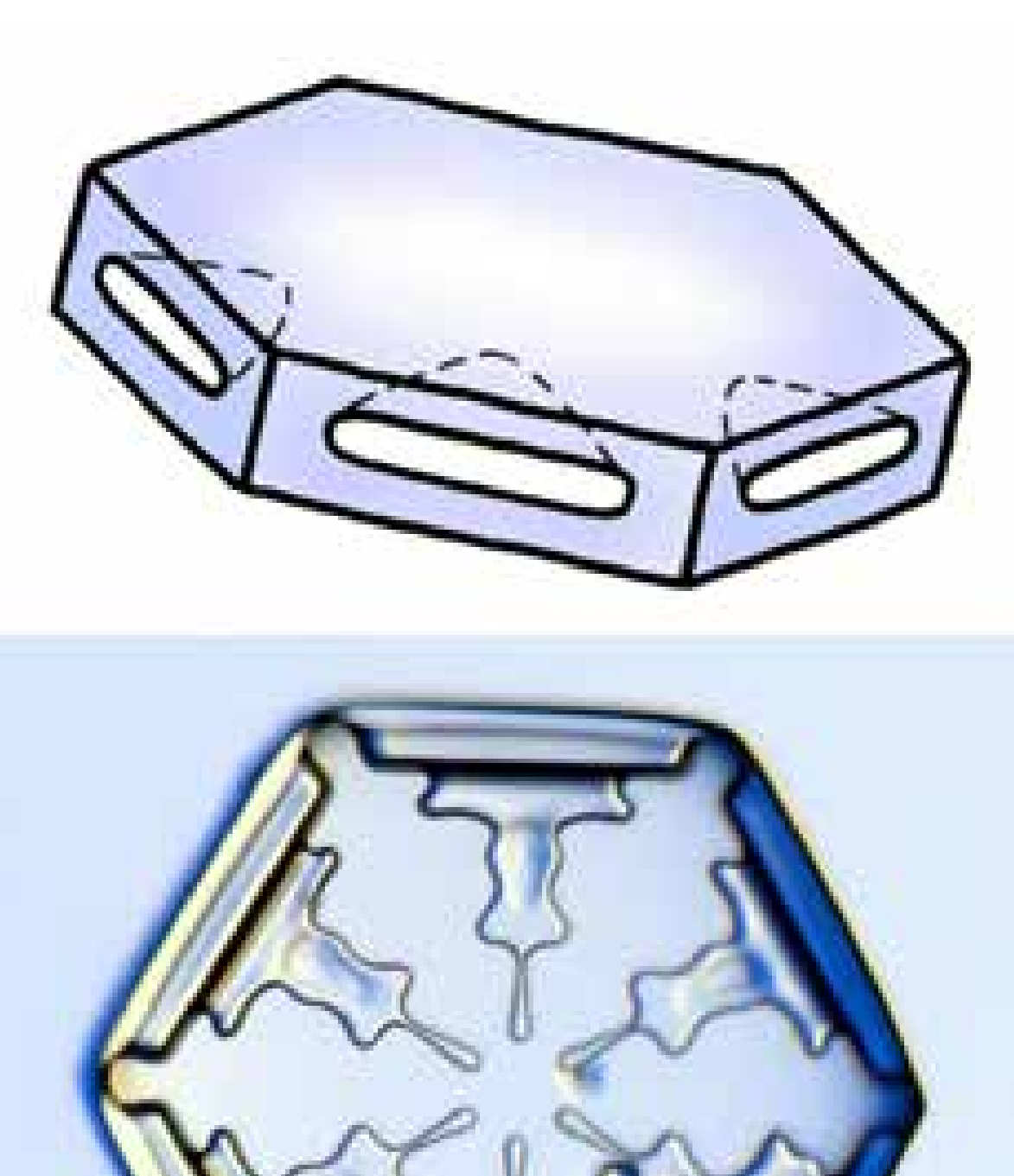

**Figure 3.37: The essential geometry of a hollow-plate snow crystal (top) and a photograph of a natural snow crystal with deep hollows. Note that the six hollow regions are separated by solid ice at the hexagonal corners. The photo exhibits oddly shaped hollow regions reflecting the changing conditions that the crystal experienced during its growth.**

Figure 3.39 shows a hollow plate during as it forms at $(T,\sigma) = (-16\ C, 16\%)$.

Growing in constant environmental conditions, the geometry of the crystal in Figure 3.39 tells a story about its formation. On the prism facets, new molecular layers first nucleate at the corners, as the supersaturation is highest there. The terraces then grow inward, initially forming full faceted prism surfaces near the hexagonal tips. A terrace growth instability then causes the terraces to split, each new layer slightly enlarging the prism hollow region. The resulting hollow-plate morphology seen in Figure 3.39 is similar to the basic form shown in the sketch in Figure 3.37. I believe that reproducing this robust growth morphology would be another interesting challenge for 3D numerical modeling.

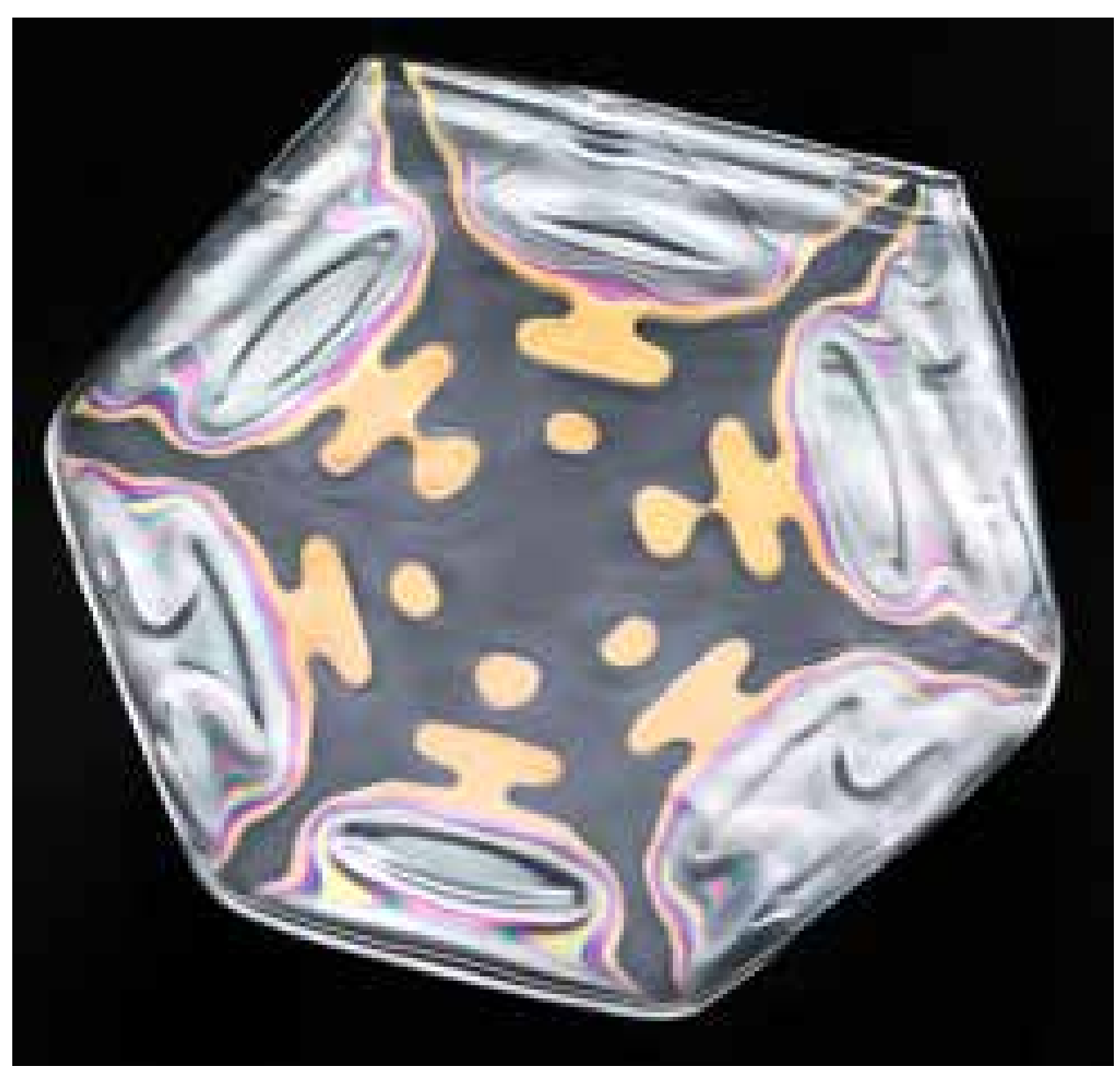

**Figure 3.38: Another natural snow crystal exhibiting deep hollow regions in each of the prism surfaces. Near the center of the crystal, some of the hollows have closed off to form thin bubbles in the ice. The colors arise from optical interference between reflections off the top and bottom surfaces of the hollows/bubbles, which are separated by about one wavelength of light. Photo by Don Komarechka [2013Kom].**

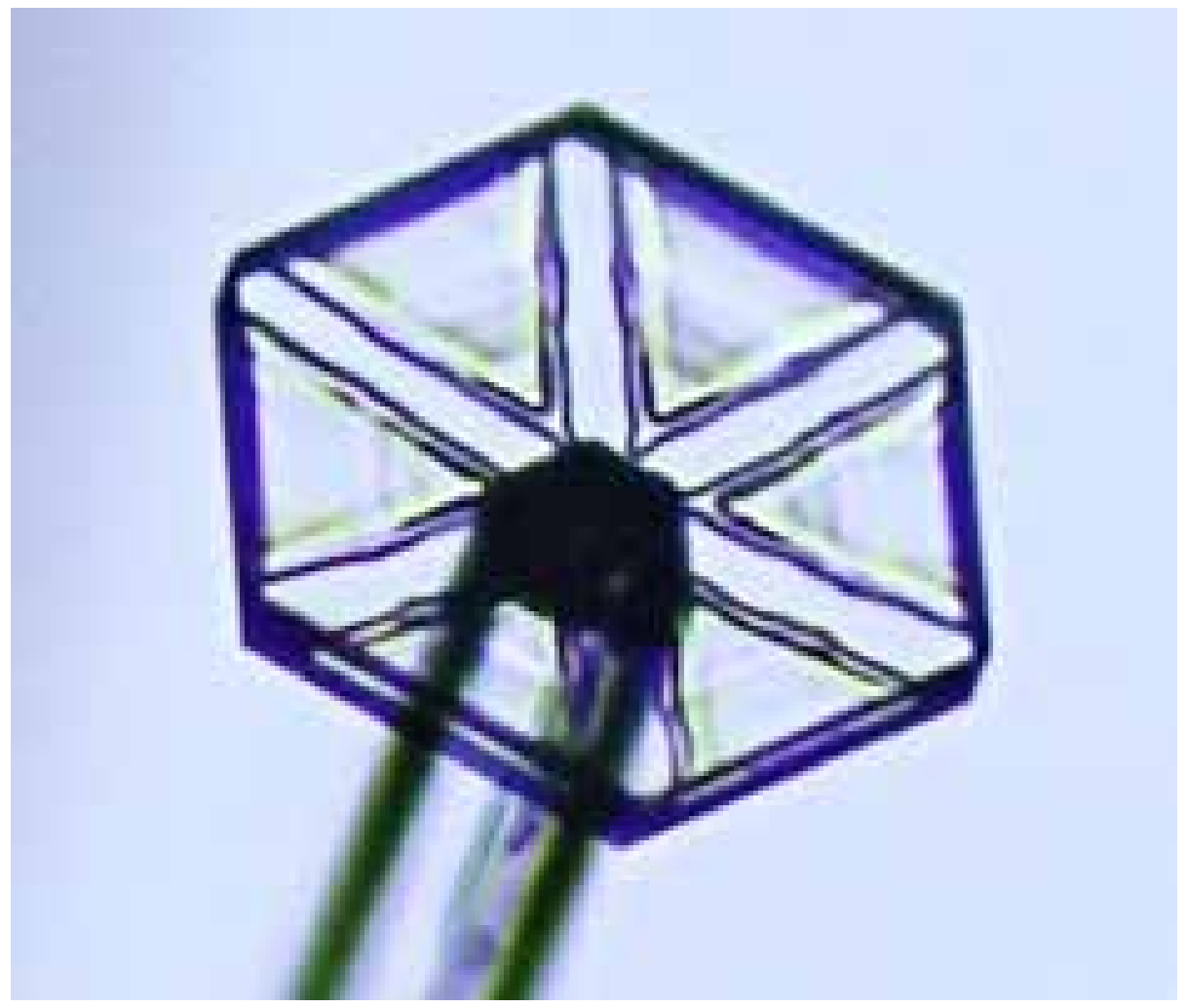

**Figure 3.39: A laboratory snow crystal growing on an e-needle with $(T, \sigma)$ = (-16 C, 16%). Under these constant conditions, the hollow regions grew as roughly triangular segments separated by solid ice spokes.**

If a hollow region evolves into an enclosed bubble, the void becomes essentially a closed system unaffected by the supersaturation field surrounding the crystal. In this isolated state (neglecting any temperature gradients in the crystal), the bubble would naturally evolve toward its equilibrium shape, which is likely nearly spherical (see Chapter 2). However, relaxation toward equilibrium is significantly hindered by a nucleation barrier on the interior faceted surfaces, as shown in Figure 3.40. Here we see that while the growth of interior concave surfaces is never limited by a nucleation barrier, evaporation is limited by a nucleation barrier. For this reason, even an exceedingly thin bubble in a plate-like snow crystal may retain its nonequilibrium shape for long periods of time.

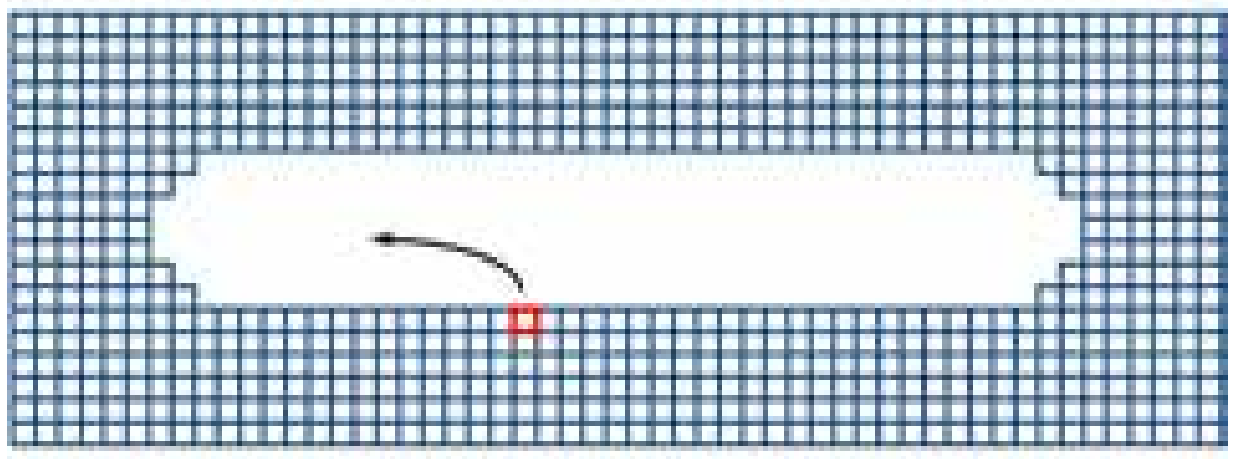

**Figure 3.40: Even if the equilibrium shape of an enclosed bubble is spherical, a faceted bubble may evolve exceedingly slowly toward that final form. It is difficult to remove molecules from a fully faceted surface (red), and this presents a strong hole-nucleation barrier that can greatly slow equilibration. For this reason, even very thin bubbles in plate-like crystals can retain their shape for long periods of time.**

## Ridges and Sectored Plates

Ridge structures are commonly found in both natural and synthetic snow crystals, and a particularly simple example is shown in Figure 3.41. Here the six ridges are thick linear features in an otherwise thin plate. The ridges divide the hexagonal plate into six equal sectors, like slices of a hexagonal pie, so these are called sectored plates. Sectored-plate snow crystals typically grow near -15 C at intermediate supersaturations, when the growth conditions are nearly constant in time. Laboratory observations reveal that ridges are typically associated with slightly convex basal surfaces, as shown in the sketch in Figure 3.41.

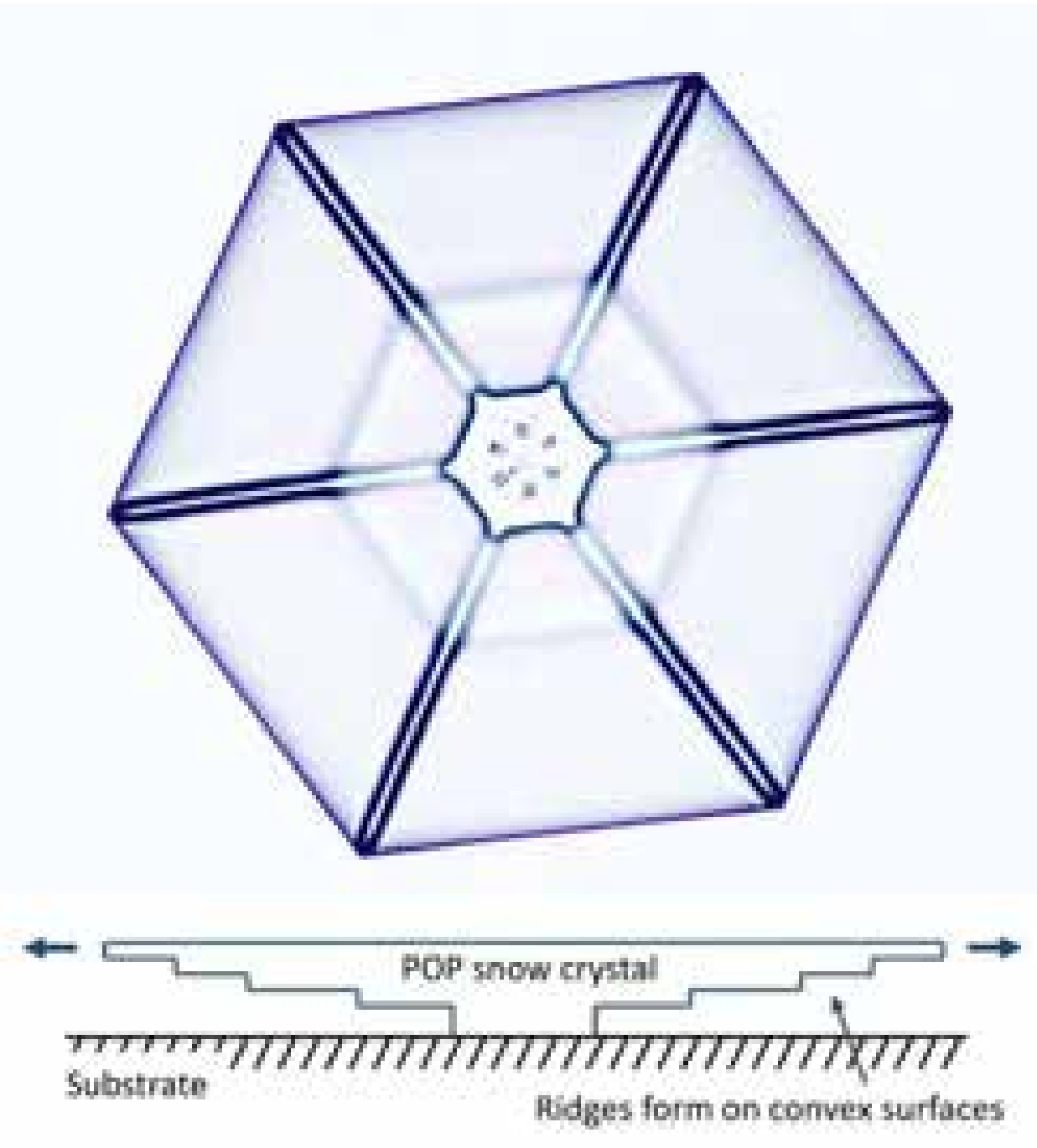

**Figure 3.41: A small Plate-on-Pedestal (POP) snow crystal exhibiting simple ridges that originate at the six faceted corners as the plate grows outward. As shown in the accompanying sketch, the top basal surface of this crystal is essentially flat and unfeatured, while the ridges and other visible structural features exist on the convex lower surface of the crystal.**

Figure 3.42 shows a diagram of the growth process that leads to ridge formation. The convex basal surface includes a series of regularly spaced molecular terrace steps, and the step spacing defines the slope of the surface, like a set of contour lines on a topographic map. As the top basal surface grows upward (see Figure 3.41), the overall crystal thickness increases, and more steps appear on the lower surface. The lower surface grows much more slowly, as it is shielded by the substrate below.

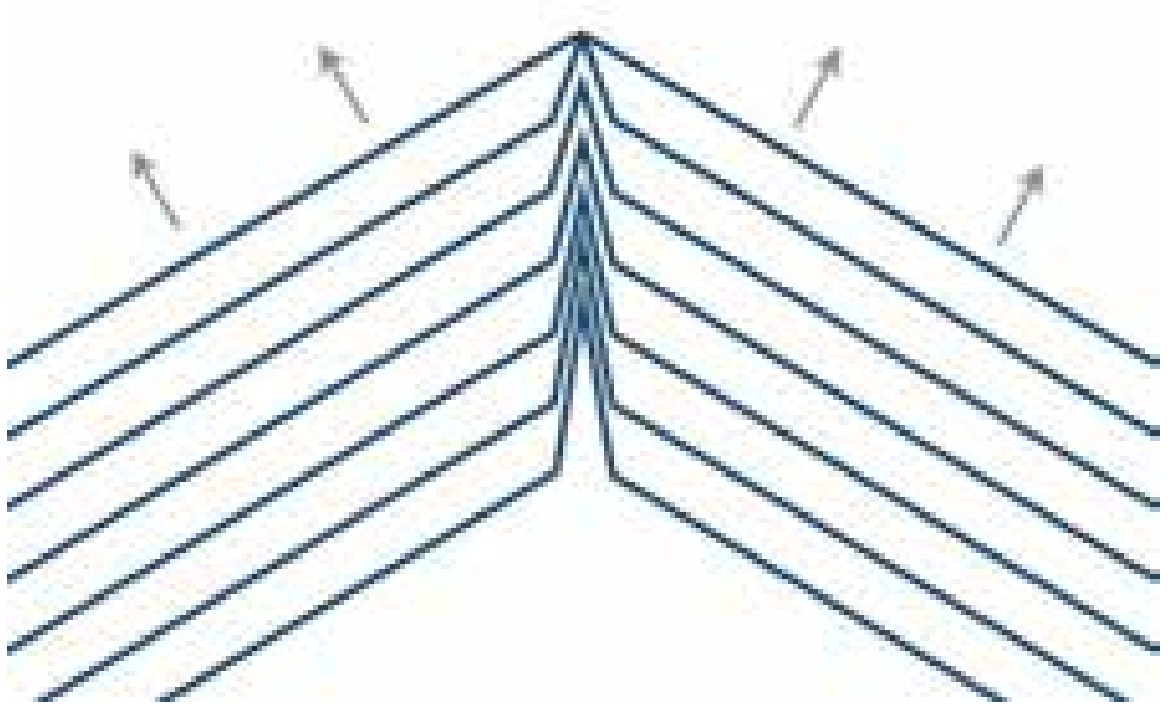

**Figure 3.42: A sketch showing the development of a snow-crystal ridge on a convex basal surface. Lines represent molecular steps defining individual terraces. As the prism facet edges grow outward (arrows), diffusion enhances the growth of the terrace corners, leading to ridge formation. This terrace-branching model naturally explains why ridges are generally associated with convex basal surfaces.**

As the faceted prism edges grow outward (arrows in Figure 3.42) the lower terrace edges grow outward also, although the step velocity need not be the same as the edge velocity. A two-dimensional manifestation of the Mullins-Sekerka instability comes into play on the molecular steps, enhancing the corner growth, as diffusion brings a greater supply of water vapor to the step corners. Each terrace corner thus sprouts a one-molecule-high "branch", as shown in Figure 3.42, and these linear branches combine to form a macroscopic ridge. Note that the closely spaced contours around the ridge indicate its steep vertical sides, as in a topographic map.

One can speak of the one-dimensional attachment kinetics associated with a molecular step, and how anisotropic step attachment kinetics could lead to step "faceting." It appears, however, that any existing anisotropies are negligibly small, so there is little or no inhibition to the formation of ridges arising from attachment kinetics. If this is indeed the case, then step growth rates will be largely driven by the local surface

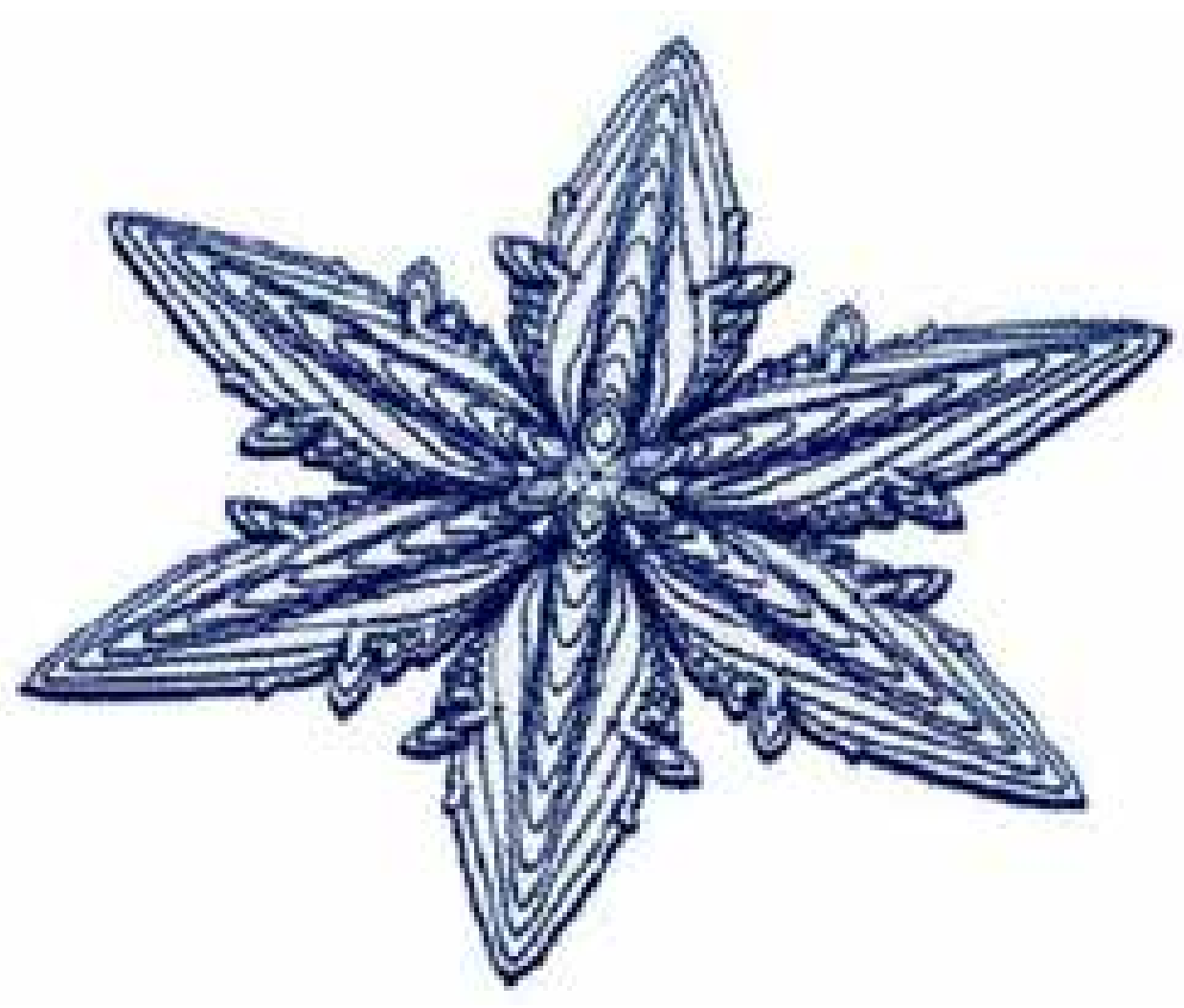

**Figure 3.43: This numerical model of a growing stellar crystal exhibits clear ridging on the primary branches. The micron-scale steps in the model are orders of magnitude larger than molecular steps, but the underlying diffusion-driven ridge growth is essentially the same. It appears that ridge formation is readily seen in both natural snow crystals and numerical modeling, at least using the cellular automata method. The robustness of ridging seems to reflect the basic diffusion physics underlying the phenomenon, which is insensitive to other material parameters. Image from [2014Kel].**

supersaturation, which is limited by water-vapor diffusion.

The appearance of ridges on plate-like snow crystals appears to be always, or nearly always, associated with slightly convex basal surfaces. Because a POP crystal is intrinsically asymmetrical, as water vapor is supplied from above the substrate, the upper surface of a POP crystal stays flat or slightly concave, containing no ridges. But the bottom surface is convex and often displays strong ridge structures. For this reason, POP crystals are well suited for observing ridge growth.

In contrast, the two basal surfaces of a fernlike stellar dendrite (like the one shown in Figure 3.8) are possibly both slightly convex in shape, with each exhibiting ridges. I have not seen any direct evidence of this kind of double-sided ridging, but it is likely quite common. With many natural snow crystals, the basal surface curvature will depend on its growth history, aerodynamic effects, and other factors, so ridges may appear on one, both, or neither of the basal surfaces.

The relatively simple nature of ridge formation means that ridges are readily found in 3D numerical simulations of snow-crystal growth using cellular automata, as shown in Figures 3.43 and 3.44. It is not necessary to include accurate attachment kinetics in these models to produce ridge-like structures, as ridging requires only that diffusion limits the growth of basal terraces. Ridges are less apparent in phase-field models [2012Bar, 2017Dem], but the underlying physics in these models (using highly anisotropic surface energies) is generally inappropriate for modeling snow-crystal growth, as we discuss in Chapter 5. Once again, many insights await when we move beyond the demonstration phase and are finally able to make direct quantitative comparisons between snow crystal models and experiments.

**Figure 3.44: (Below) Another 3D numerical model showing clear ridge structures on slightly convex basal surfaces. Image from [2009Gra].**

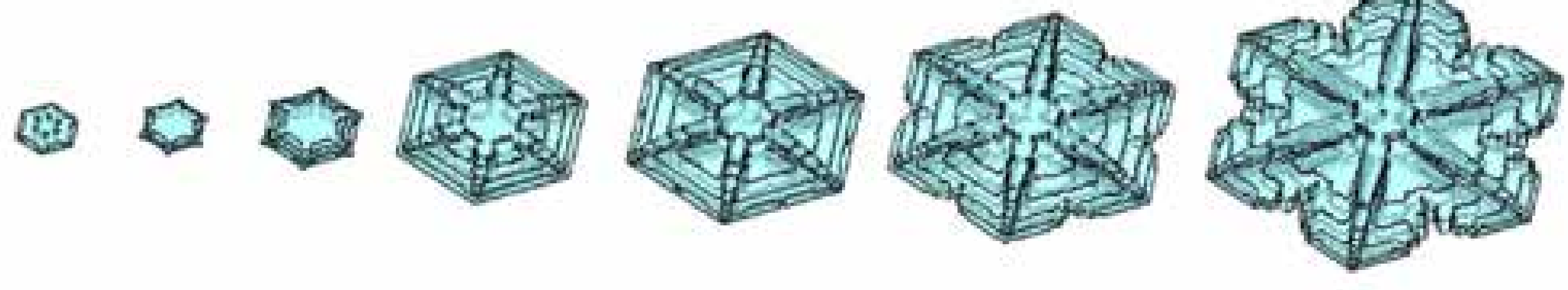

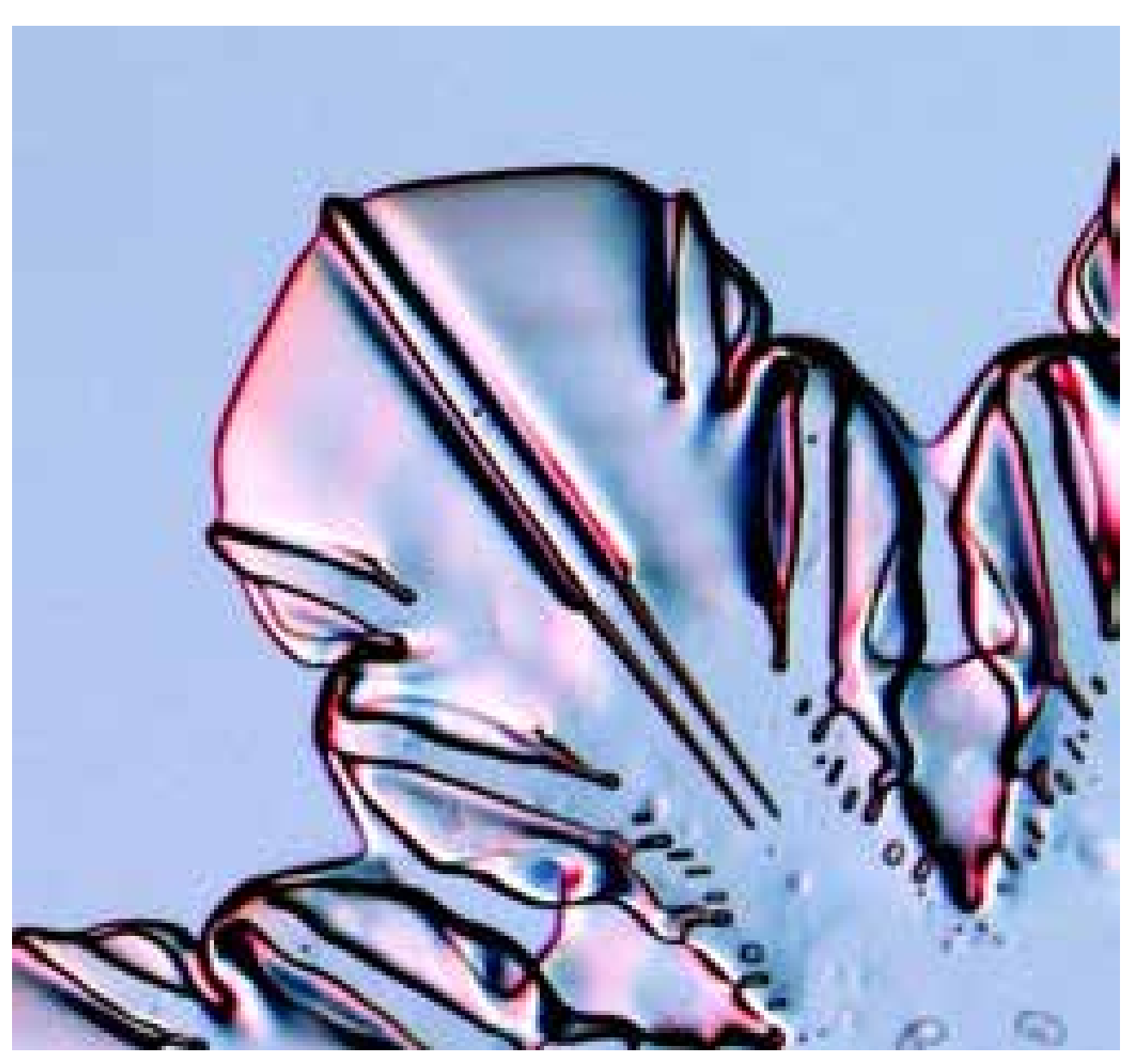

**Figure 3.45: This photo shows a partially sublimated sectored-plate snow crystal that exhibits a pair of "grooves" flanking the central ridge. Similar features can be found in many snow crystals, although usually they are not as distinctive as in this example.**

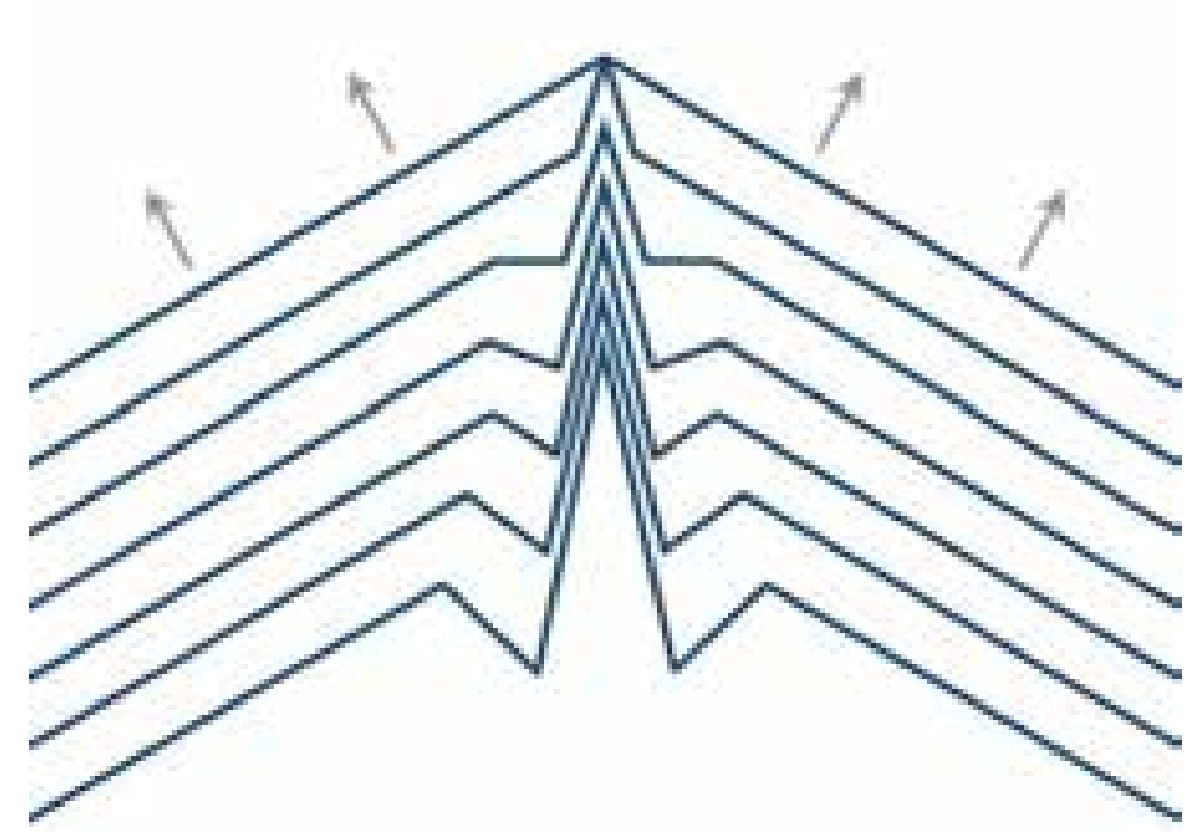

**Figure 3.46: A sketch showing the development of a snow-crystal ridge flanked by two grooves on a convex basal surface. As the ridge structure develops, it depletes the water-vapor supply in its proximity. This suppresses the advancements of steps near the ridge, thereby creating grooves.**

## Ridges with Grooves

In many instances, snow-crystal ridges are flanked by linear "grooves" that are long, shallow depressions in the ice on either side of a ridge, as shown in Figure 3.45. The formation of these grooves appears to be yet another example of the Mullins-Sekerka instability relating to step growth, an additional feature on top of basic ridge formation described above.

Once a ridge begins to form, as shown in Figure 3.46, it sticks up above the basal surface surrounding it, and $\alpha$ on the sides of the ridges is close to unity. The ridge growth thus attracts a great deal of water vapor, depleting the air near it. As diagrammed in the figure, the presence of the high-$\alpha$ ridge means that the growth velocity of a step far from the ridge is larger than the velocity of the same step adjacent to the ridge. This rather subtle dance of step advancements, choreographed by particle diffusion around the growing crystal, results in ridges flanked by parallel grooves.

I suspect that chemical impurities in the air may further aid in the formation of these grooves, along with perhaps some of the small pits seen in Figure 3.45 and other snow crystals. Impurities are not readily incorporated into the ice lattice, and an advancing molecular step will tend to push impurity molecules ahead of it as it grows. Thus, although the average density of impurity molecules on the ice surface may not be high, step motion could redistribute and concentrate those impurities that are present.

Looking at ridge and groove formation, as thousands of steps march along during the process, their collective motion will tend to push impurities into the grooves, where they will remain, stuck on the ice surface. The concentrated layer of chemical crud could then substantially impede further ice growth, and the grooves would remain etched into the ice. Additional laboratory experiments would be needed to investigate whether chemical impurities really have such effects on snow crystal surface features.

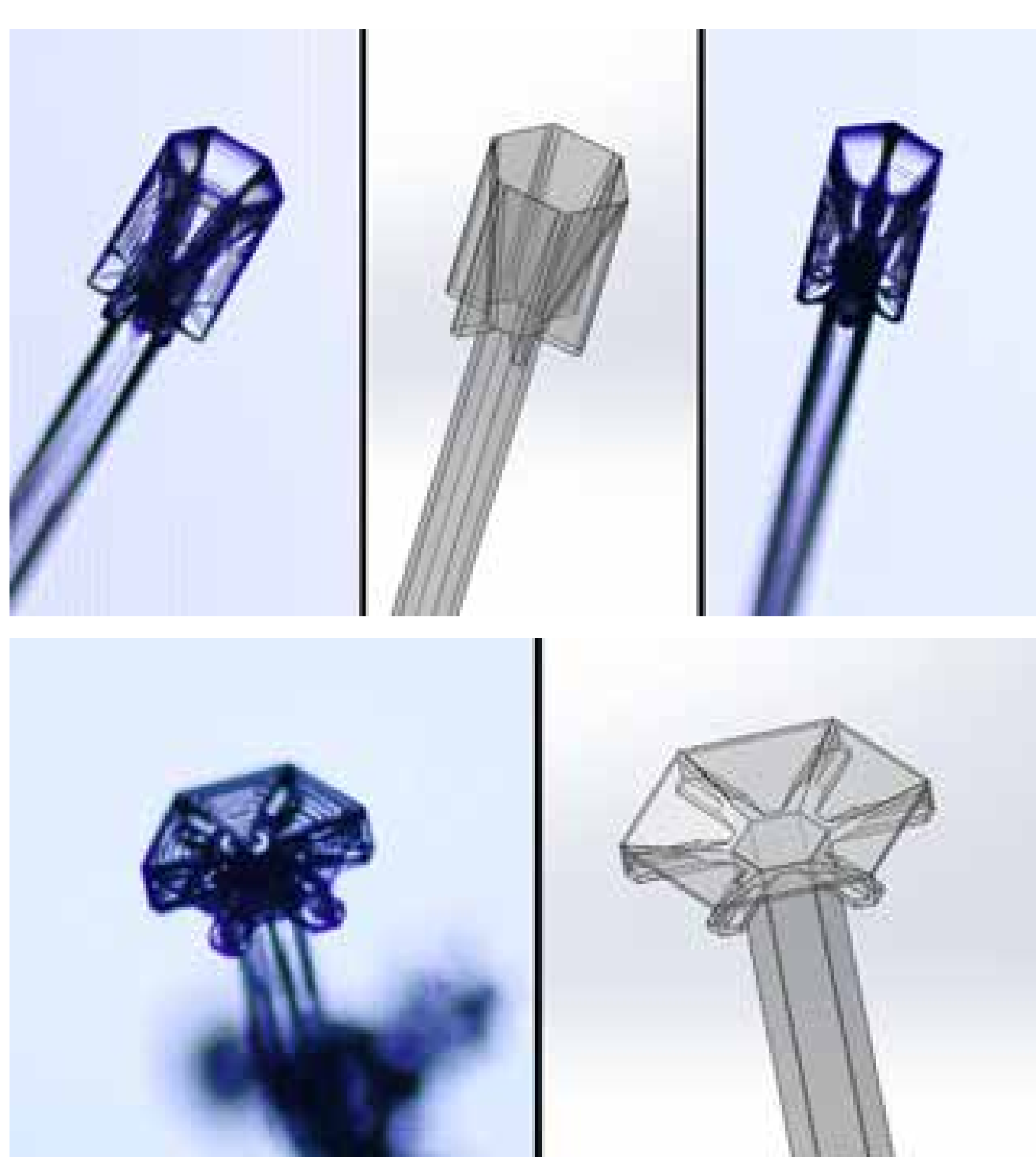

**Figure 3.47: Cups with Fins. The photos on the left and right show two views of a cup-shaped snow crystal growing on the end of a slender ice column. The 3D drawing (center) illustrates the main structural features, including six plate-like "fins" that are related to snow crystal ridges. The laboratory crystals were grown on an e-needle (see Chapter 8) near -7 C. [SolidWorks drawings by Ryan Potter.]**

**Figure 3.48: I-Beams. The photo on the left shows a snow crystal growing on an e-needle near $(T, \sigma)$ = (-9 C, 16%), and its overall structure is illustrated in the drawing on the right. Here the plate has an intermediate cone angle, yielding short fin-ridges that subsequently developed plate-like extensions, giving an overall "I-beam" ridge structure. This morphology is remarkably robust on e-needles, occurring over a broad range of growth conditions (see Chapter 8). [SolidWorks drawings by Ryan Potter.]**

## Ridges on Cones and Cups

The detailed structure of ridges in snow crystals depends a great deal on the "cone angle" of the plate on which they grow. The previous discussion assumed a small cone angle, by which I mean a nearly flat basal surface that is slightly convex in overall shape. (Of course, this is not a true cone in the strict geometrical sense, but a roughly cone-like shape made from six slightly tapered flat surfaces.) This morphology includes the trains of propagating steps shown in Figures 3.42 and 3.46 that are necessary to produce ridge structures.

One can extend the discussion further to include steeper cone angles, progressing from nearly flat plates to cup-like structures such as those shown in Figure 3.47. The outer surfaces of a cup also include trains of molecular steps, similar to those described above except with much higher step densities. The same ridge-formation instability applies, but now the ridges develop into the pronounced "fins" shown in Figure 3.47. Note that because the supersaturation is relatively low below the fast-growing cup edge, the fins develop nearly faceted prism surfaces.

Figure 3.48 shows another ridge morphology that readily occurs on e-needles over a broad range of conditions when the cone angle is intermediate between plates and cups. Here the ridges grow out to form what are essentially stubby fins, but then plate-like extensions grow out from the base of the fins, yielding what I call an "I-beam" structure. This feature can be found in natural snow crystals as well, but it is especially clear on e-needles because the growth conditions can be kept constant for long periods of time. The lower plates on the I-beams are another example of how readily thin plates emerge in snow crystal growth.

All the ridge structures described above are easily created in the laboratory under constant growing conditions, especially on e-needles. Moreover, the ridge morphology is quite robust, with different variations appearing over a broad range of temperatures and supersaturations. It is difficult to explain their structure simply, owing to the complex balance of faceting, branching, and step motions that must be happening. For this reason, however, ridges should prove to be a good test of future 3D numerical modeling techniques. Once models are able to reproduce these kinds of complex structures, especially with growth rates that match observations, we will finally be able to say that we have made serious progress toward providing realistic simulations of snow crystal structure formation.

## Step Bunching and Inwardly Propagating Rings

Just as ridge-like structures are common on slightly convex basal plates growing under constant environmental conditions, inward-propagating rings readily form on slightly concave basal plates, and one example is shown in Figure 3.49. Similar rings appear quite frequently on POP snow crystals (see Chapter 9), owing to their unique geometry of thin plates that are slightly conical in overall shape, as illustrated in Figure 3.50.

The asymmetrical POP construction often produces outward-propagating terrace steps on the lower basal surface and inward-propagating steps on the upper basal surface. Ridges then develop on the lower convex surface, as described above, while rings appear on the upper concave surface. There is essentially no interaction between the ridges and rings, and indeed these structures appear to be independent of one another in Figure 3.49.

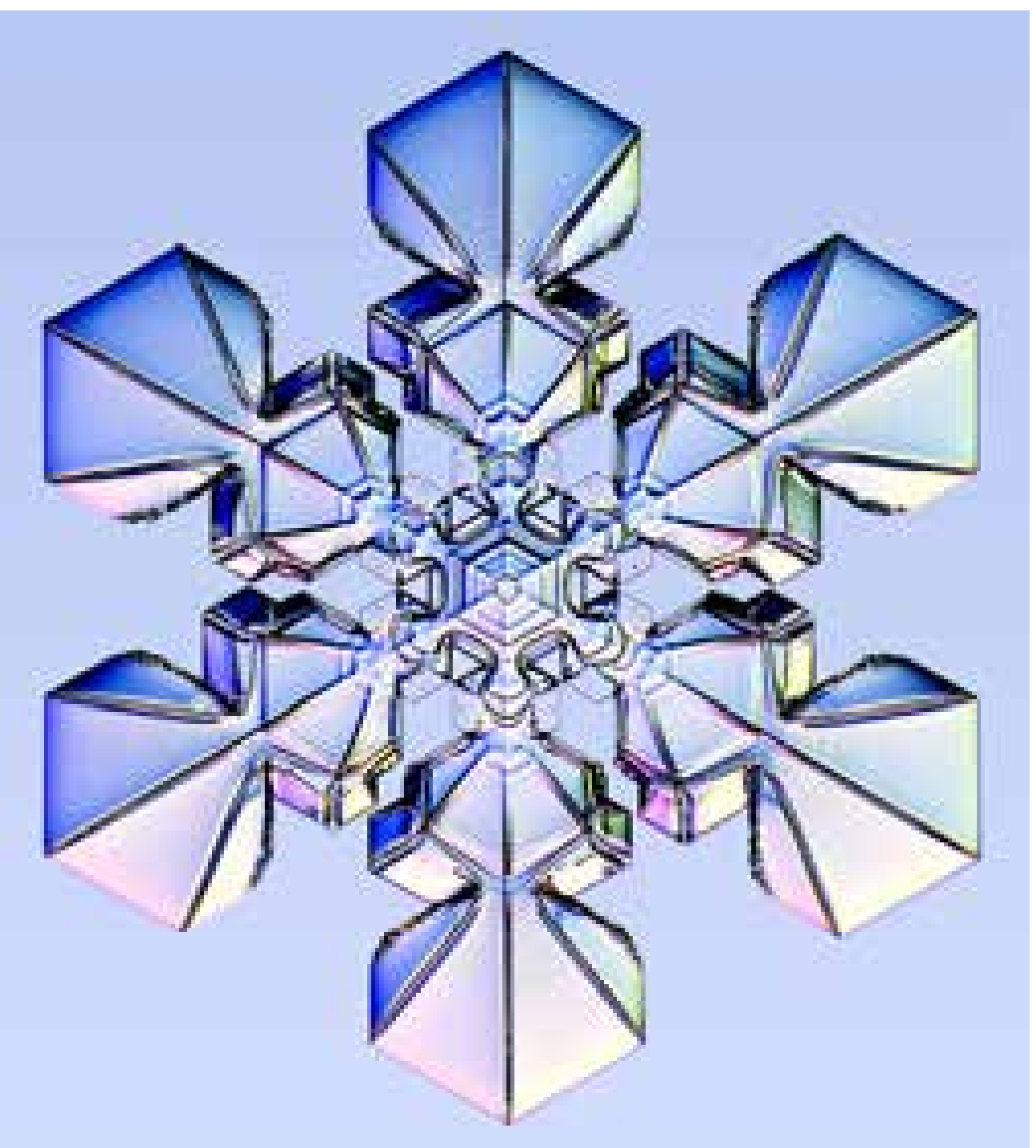

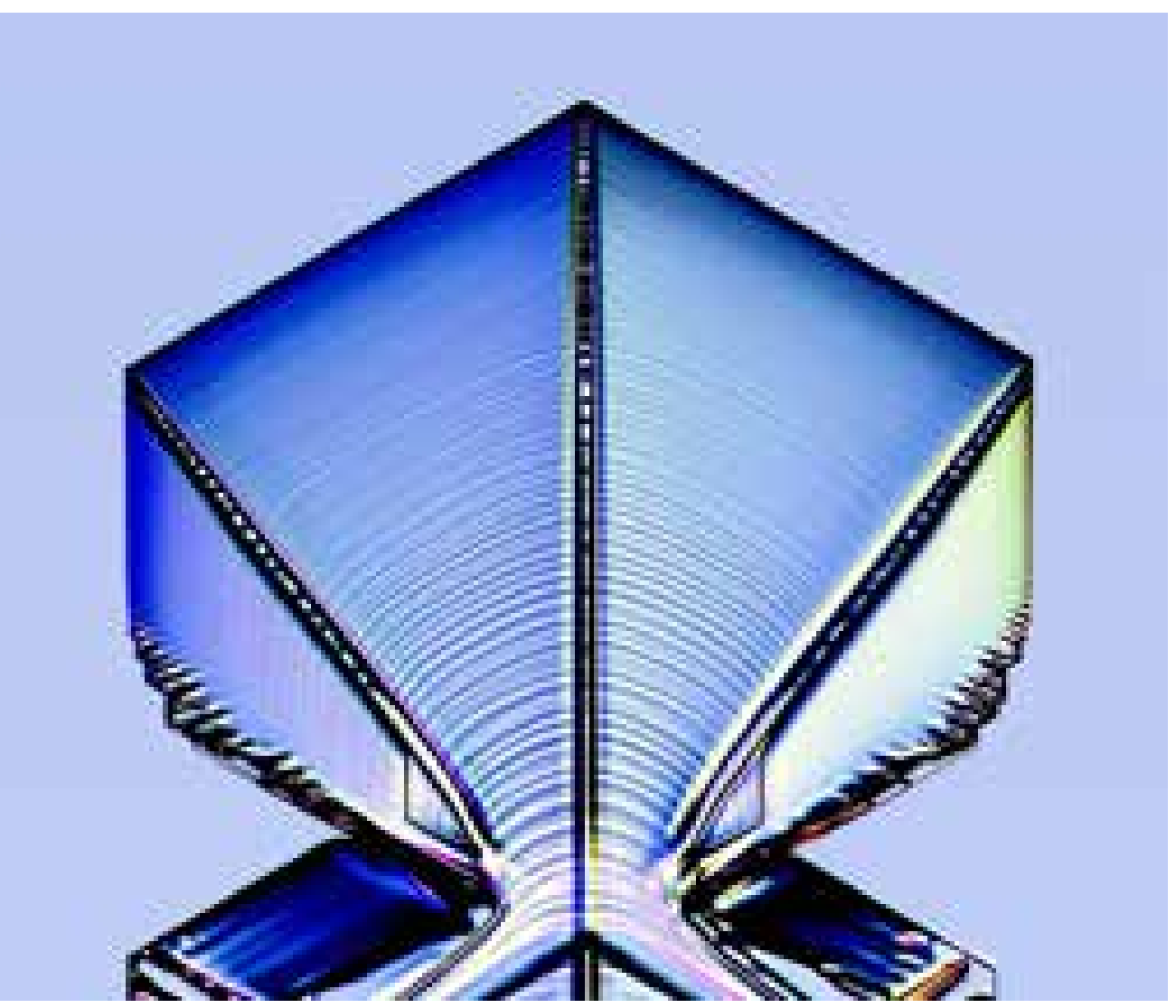

**Figure 3.49: The upper photo shows a large laboratory-grown POP snow crystal, while the lower photo shows a close-up of a single sectored-plate branch. Molecular steps on the top basal surface nucleate near the outer edge of the crystal and subsequently grow toward the center. The molecular steps bunch together to produce a series of inward-propagating ring-like features. Near-constant growth conditions are required to produce such a uniform set of growth rings.**

Under uniform growth conditions, one might naively expect that a steady creation of new terraces on the top basal surface might yield a simple vicinal surface with roughly uniform spacing between steps. In fact, while such a surface is a valid solution to the diffusion equation, it is not a stable solution. The ubiquitous Mullins-Sekerka instability, along with possible additional effects from molecular surface diffusion, results in a phenomenon called *step bunching*. As the name implies, isolated steps soon bunch together to form *macrosteps* that are large enough that they can be seen using optical microscopy, as illustrated in Figure 3.49.

To my limited understanding, step bunching can be the result of several different physical effects, and disentangling these for the case of ice growth is not a trivial task. Bulk diffusion almost certainly plays a role via the Mullins-Sekerka instability, but surface diffusion effects might be important also. Because there is no clear model of step bunching in ice, it is not yet possible to calculate the average macrostep height or (equivalently) the spacing between macrosteps for a given vicinal angle. As with so many features in snow crystal growth, macrostep phenomena are easily observable, but not so easily understood in detail.

Note that diffusion effects on inward- and outward-propagating steps are markedly different. In Figure 3.50(b), for example, diffusion makes inward-propagating steps (on a slightly concave basal surface) evolve toward a circular shape. Because water vapor diffuses in from the supersaturated air surrounding the crystal, the step growth is faster for steps nearer the outer edges. Thus, any deviation from a circular shape is corrected by the growth dynamics. For inward-propagating steps, therefore, diffusion-limited growth brings about a stabilizing effect that maintains a simple circular shape.

In contrast, Figure 3.50(c) shows how diffusion-limited growth on outward-propagating steps (on a slightly convex basal surface) yields a form of the branching instability. The water-vapor supersaturation is highest near the outer corners of the crystal, so terrace step branches soon form at each corner, and repeated branching on multiple steps leads to the formation of macroscopic ridges, as described previously. A small change in the basal surface geometry, from slightly concave to slightly convex, thus yields a large change in overall growth behavior.

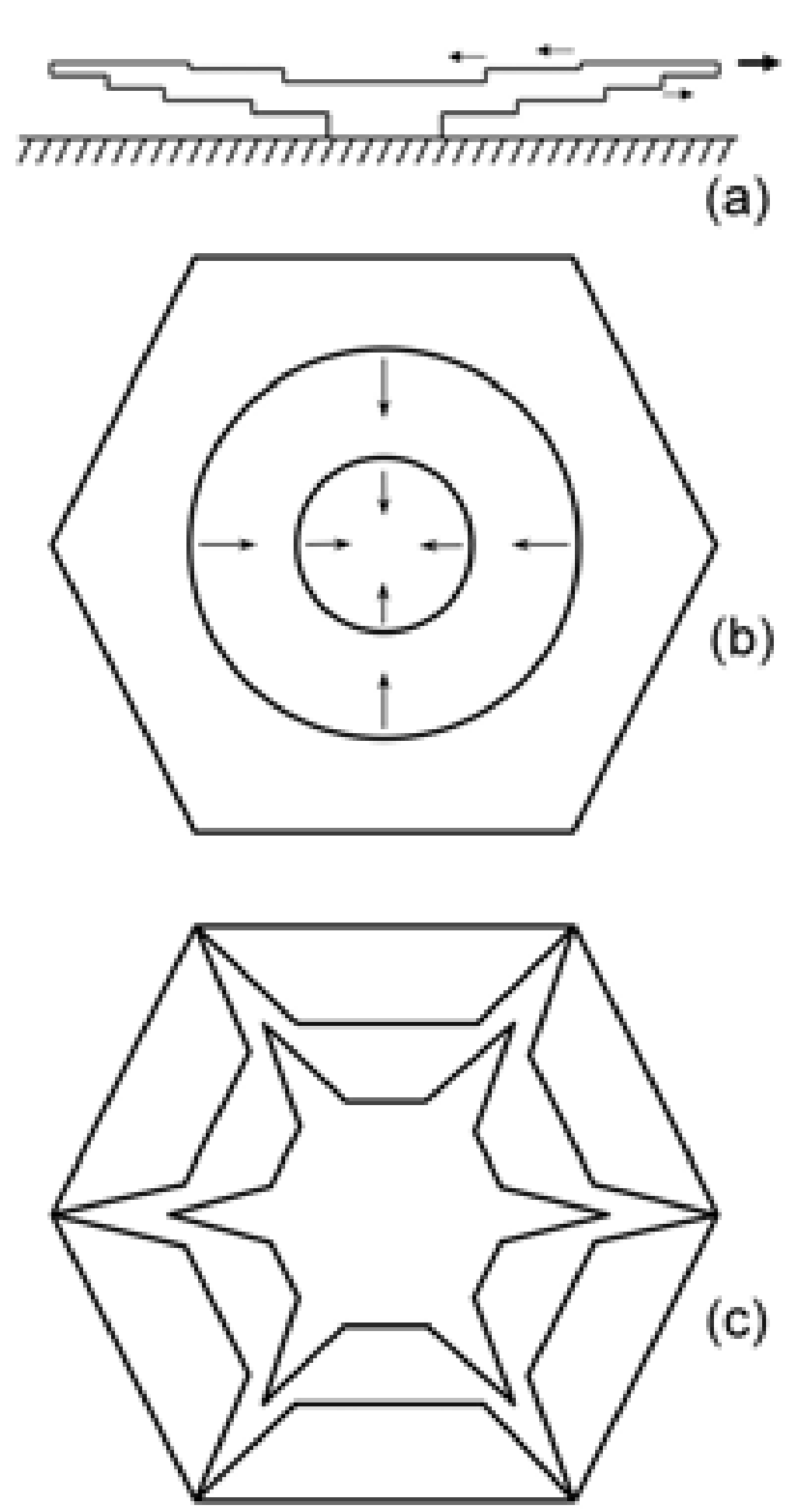

**Figure 3.50: (a) The overall geometry of a POP crystal, as seen from the side. Steps represent molecular terraces on the basal surfaces. (b) Growth on the concave top surface yields nearly circular inward-propagating steps. Step bunching turns a series of one-molecule-high steps into a coarser series of macrosteps that can be seen using optical microscopy. (c) Faster step advancement near the hexagonal corners yields ridges on the lower convex surface.**

## Ribs on Plates

While ridges and inward-propagating rings readily appear under constant environmental conditions, other common snow crystal structural features require changing conditions for their formation. One prominent example is the creation of hexagonal "ribs" like those shown in Figures 3.51 and 3.52. In Figure 3.51, the ribs form a set of hexagonal rings where the ice is a bit thicker than elsewhere in the plate. In Figure 3.52, the ribs are restricted to the crystal's outer plate-like extensions, where they exhibit the same overall hexagonal structure.

In both cases, the ribs are accompanied by ridges that divide the plates into sectors. Both ribs and ridges are frequently found in natural snow crystals as well, as described in Chapter 10. As with other growth phenomena in this section, ribs are especially nicely demonstrated using laboratory-grown POP snow crystals, where the growth conditions can be well controlled and quickly modified at will.

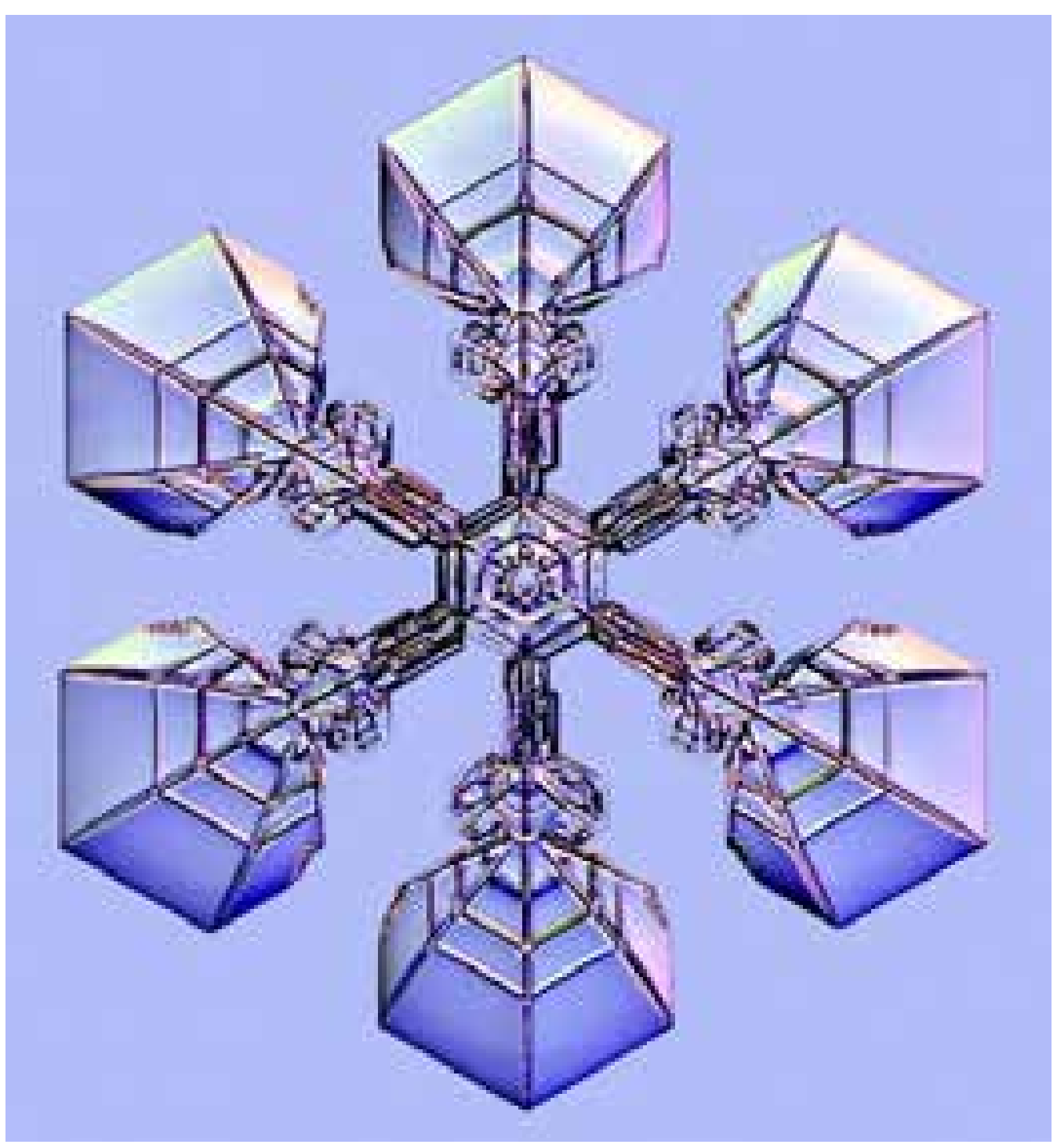

**Figure 3.52: This POP snow crystal exhibits broad, plate-like extensions growing on the ends of narrow branches. Here again, the two prominent sets of ribs in the plates did not appear spontaneously but were induced by twice lowering the supersaturation to form a rib, then increasing it back again to resume the thin-plate growth.**

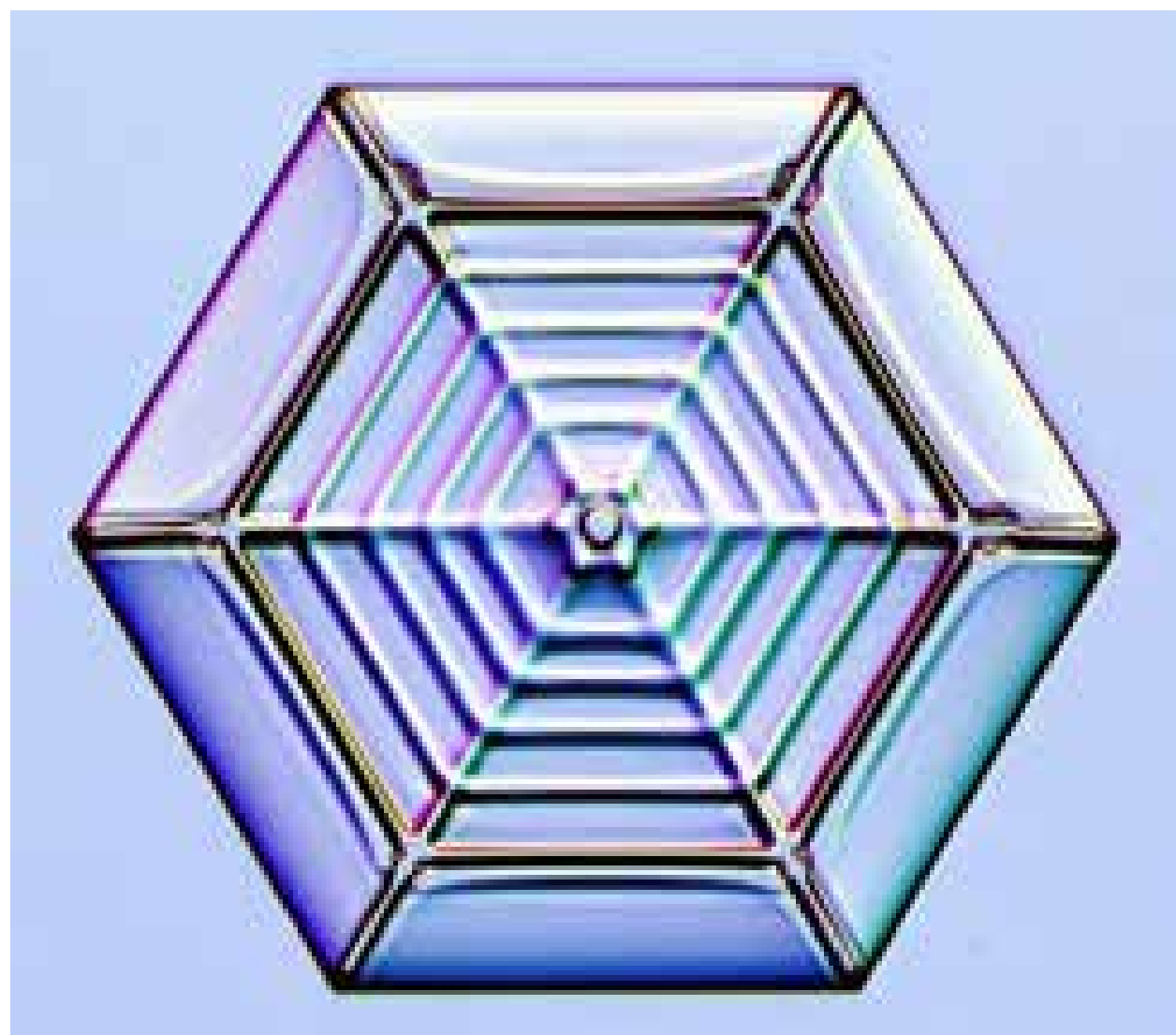

**Figure 3.51: While growing this POP snow crystal, I periodically reduced and then increased the supersaturation, yielding the inner "spider-web" structure of ribs and ridges. After that, I added an extra-thick rib and then let the thin plate grow out unencumbered. Hexagonal rib patterns like these are typically associated with changes in external growth conditions.**

Figure 3.53 illustrates how a temporary drop in supersaturation surrounding a snow crystal can lead to the formation of a rib on a growing plate. The top sketch shows the outer edge of a POP crystal growing at a relatively high supersaturation near -15 C. These conditions result in the formation of thin plates, and the sketch shows a plate that is flat on the upper surface and slightly convex on the lower surface, which is typical of POP crystals (see Chapter 9). This thin-plate morphology continues as long as the supersaturation remains high.

Upon lowering the supersaturation (second sketch in Figure 3.53), the edge-sharpening instability (ESI) is diminished, yielding subsequent growth as a thicker plate.

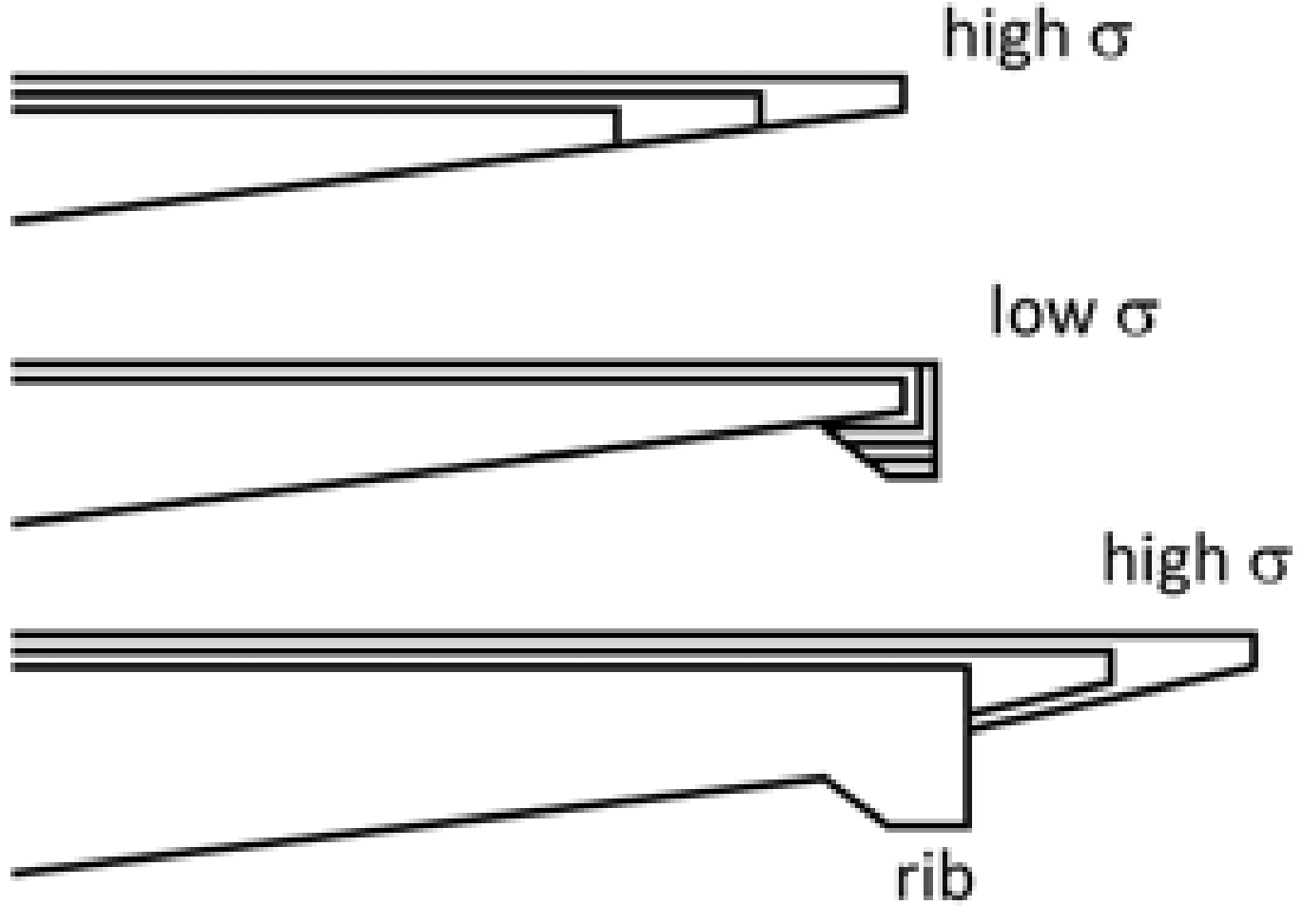

**Figure 3.53: This series of sketches chronicles the formation of a snow-crystal rib on the underside of a POP snow crystal. When the supersaturation is high (top), thin-plate growth results. Upon lowering the supersaturation (middle), the crystal grows more slowly, and its edge develops a thick rim. Restoring high supersaturation (bottom), thin-plate growth commences from the upper edge of the rim, leaving a rib structure behind on the underside of the plate.**

Note that the faceted basal and prism surfaces grow slowly at low $\sigma$ because of the usual nucleation barriers on those surfaces. The underside of the plate begins as a vicinal surface, however, on which there is no nucleation barrier. Thus, the underside grows relatively quickly, especially near the edge of the plate, as illustrated in the figure. Soon a thick "rim" of ice emerges on the edge of the plate.

Increasing the supersaturation to its previous high level (third sketch in Figure 3.53), the ESI again kicks in and a thin plate grows out from the upper edge of the thicker rim. As this thin plate grows outward, it shields further growth below it, leaving a thick rib structure behind.

The qualitative explanation in Figure 3.53 was easily confirmed by observing the growth of POP crystals in real time while adjusting the supersaturation in the process. Both ribs and ridges were easily created, and it was straightforward to confirm that both these features were confined to the lower convex basal surfaces. The multitude of features seen in natural snow crystals are much more difficult to interpret, as illustrated in Figure 3.54. This crystal exhibits a cacophony of rib-like features that reflect the ever-changing and unknown conditions in which it grew.

**Figure 3.54: (Below) This natural snow crystal apparently experienced many variations in supersaturation that resulted in a complex set of rib-like surface features. Because the crystal tumbled through the atmosphere as it grew, the ribs likely formed on both basal surfaces, in contrast to the simpler ribs seen on POP crystals.**

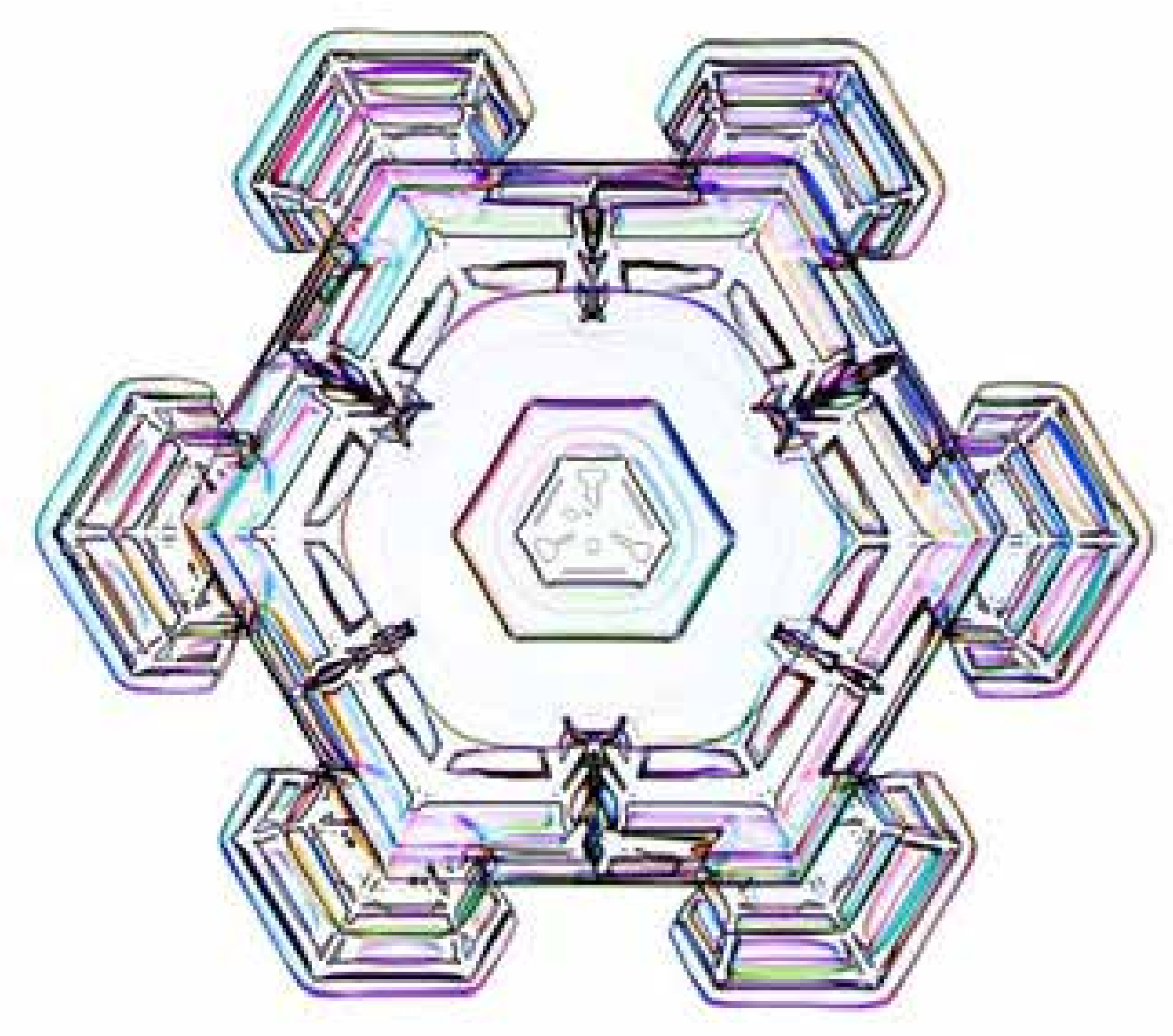

## Induced Sidebranching And Complex Symmetry

Changes in environmental conditions can also profoundly affect the growth of dendritic snow crystals, and one particularly important phenomenon is *induced sidebranching*, illustrated in Figure 3.55 for the case of a dendrite tip growing near -15 C. A key point in this discussion is that branches tend to sprout from the corners of faceted prism, as was discussed previously in conjunction with Figure 3.5. This results in orderly branching in that the primary branches sprout simultaneously on each corner of the prism.

But if $\alpha_{prism}$ is high and the growth is sufficiently rapid, then there is no prism faceting, so sidebranches sprout somewhat chaotically, yielding fernlike stellar dendrites like that in Figure 3.8. There is little order in this process, with essentially no correlation between branches, or even between different sides of an individual branch.

In induced sidebranching, a period of slow growth allows faceting at the branch tip, yielding three faceted corners on each branch. Increasing the supersaturation then causes sidebranches to sprout simultaneously from each of these three corners, and likewise from the three corners on each of the six branches. This results in a large-scale coordination of sidebranching over the entire snow crystal, as shown in the growth of the POP crystal in Figure 3.56.

This example shows in detail that no communication between the different branches was needed to induce simultaneous sidebranching; instead the event was stimulated by an abrupt change in the externally applied environmental conditions. It is straightforward to introduce many abrupt changes as a POP snow crystal is growing, and one result is shown in Figure 3.57. Essentially all the large-scale symmetry seen in this example was created by induced-sidebranching events or other longer-time changes in the growth conditions. This kind of large-scale complex symmetry generally does not arise if the growth conditions are held constant in time.

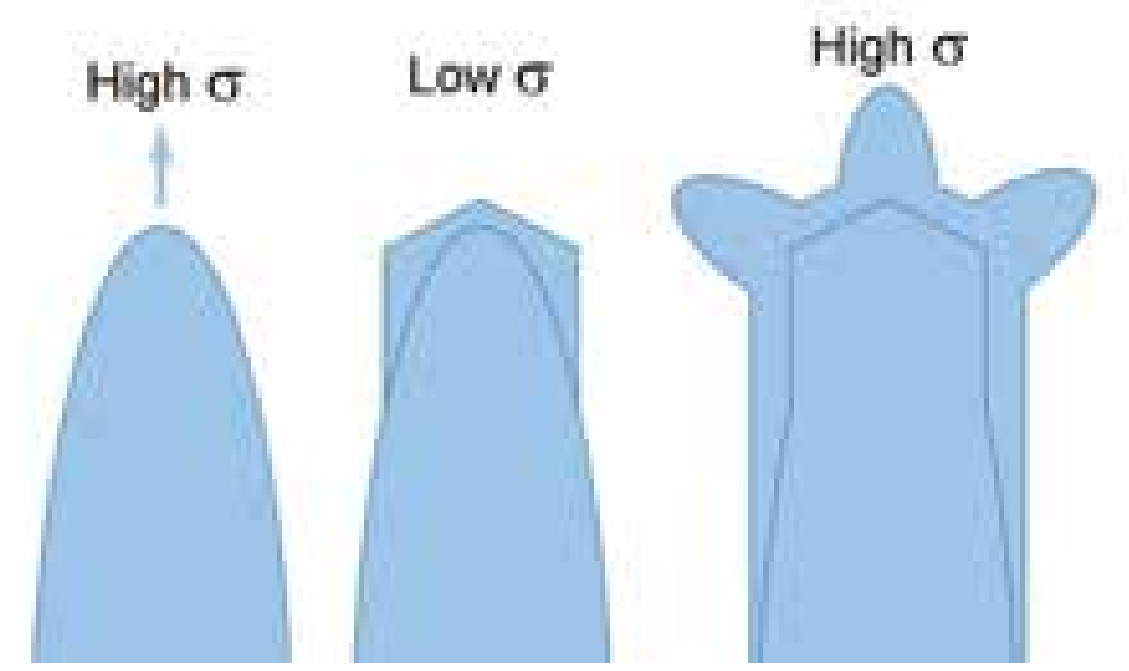

**Figure 3.55: Induced Sidebranching. When the supersaturation is sufficiently high near -15 C (left), $\alpha_{prism}$ will be near unity at the tip of a growing dendrite, yielding a rounded shape with little prism faceting. Upon lowering the supersaturation (middle), $\alpha_{prism}$ is reduced and the tip becomes faceted. Increasing the supersaturation once again (right), branches sprout from the three exposed corners of the faceted tip. One branch continues in the primary direction while the other two become sidebranches. Looking at the whole crystal, this mechanism creates a coordinated set of sidebranches on all six primary branches. Induced sidebranching is thus responsible for much of the complex symmetry seen in stellar dendrite snow crystals.**

Having made numerous movies of growing POP snow crystals, they have an almost magical appearance because the viewer cannot discern the temperature or humidity from the images alone. Watching the video, sidebranches appear simultaneously on all the primary branches from no apparent cause. Making the movie is a different experience, however, as I consciously change the growth conditions to produce different effects at different times, with predictable outcomes. This experience makes it abundantly clear that the choreography and symmetry of a complex snow crystal is almost entirely determined by time-varying externally applied growth conditions.

The photograph at the beginning of this chapter exhibits a great deal of chaotic dendrite growth that produced the helter-skelter sidebranching typical of fernlike stellar dendrites. But the largest set of sidebranches, occurring at the same location on all six primary branches, were almost certainly the product of an induced-sidebranching event.

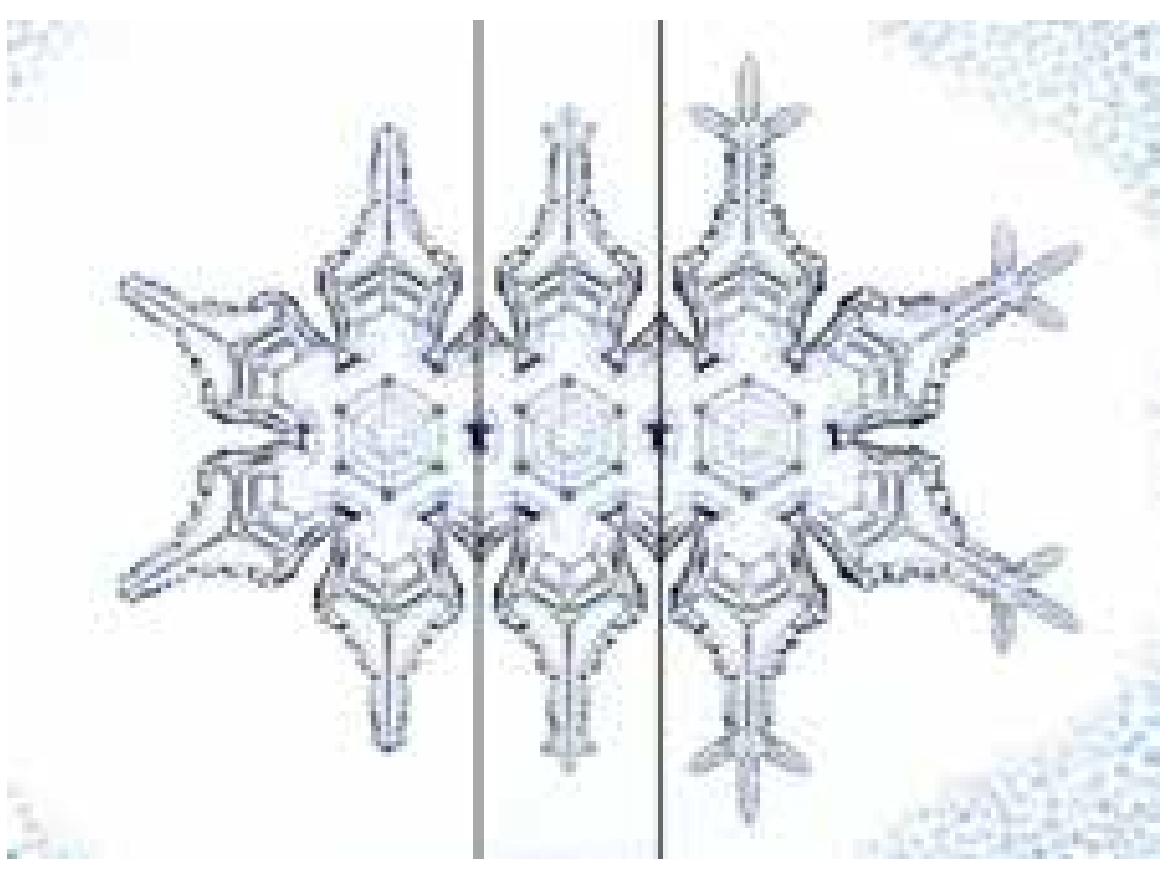

**Figure 3.56: This composite photograph shows the phenomenon of induced sidebranching on a laboratory-grown POP snow crystal. The left image shows the crystal after long branch tips were first grown out at high supersaturation, and then the branch tips became faceted after a short period of low supersaturation. Increasing the supersaturation again then caused central branches and sidebranches to sprout simultaneously on the tips of all the primary branches.**

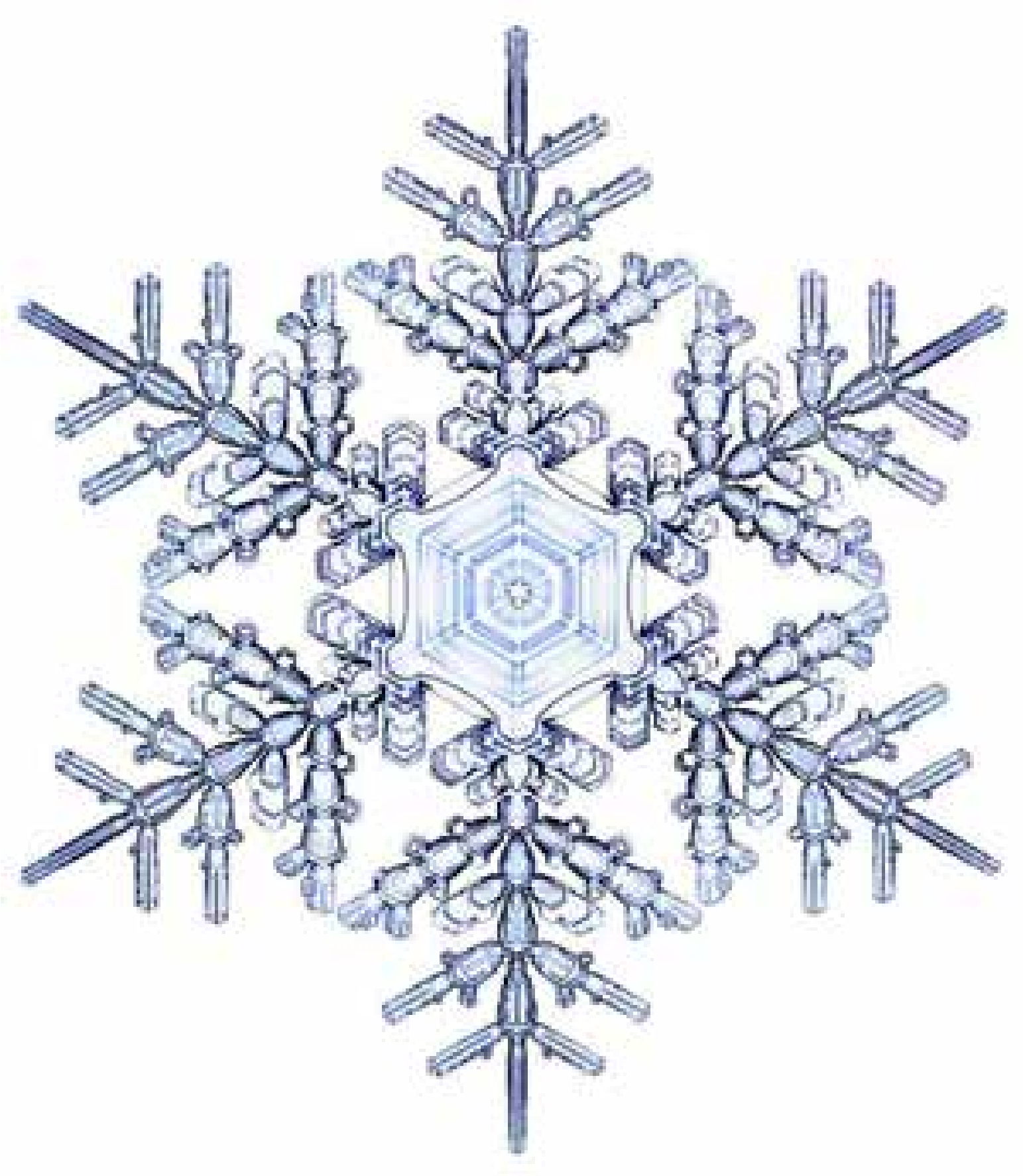

## Tridents and Triangular Snow Crystals

While six-fold symmetry is a snow-crystal hallmark, small plates often exhibit a three-fold symmetry like that shown in Figure 3.58. Note that the angles of the prism facets are the same as with a normal hexagonal prism, but now there are alternating long and short facets, giving the overall appearance of a truncated equilateral triangle. Triangular plates like these can often be found together with hexagonal plates in natural snowfalls, although the latter are always much more common.

We did a brief study looking at the statistics of triangular plates by growing small plate-like snow crystals in a free-fall growth chamber in air near -10 C with $\sigma_\infty \approx 1.4$ percent [2009Lib4]. Small hexagonal plates are the normal morphology under these conditions [2008Lib1, 2009Lib], but about five percent of the crystals exhibited a truncated triangular morphology. Figure 3.59 shows some examples of these and other non-hexagonal morphologies observed. Examples with nearly perfect equilateral-triangle morphologies were readily found in this sample.

**Figure 3.57: The high degree of complex symmetry seen in this POP snow crystal did not emerge spontaneously; I imposed it using a series of induced-sidebranching events. Induced sidebranching is the primary mechanism that coordinates the growth of sidebranching on stellar snow crystals, both in the lab and in nature.**

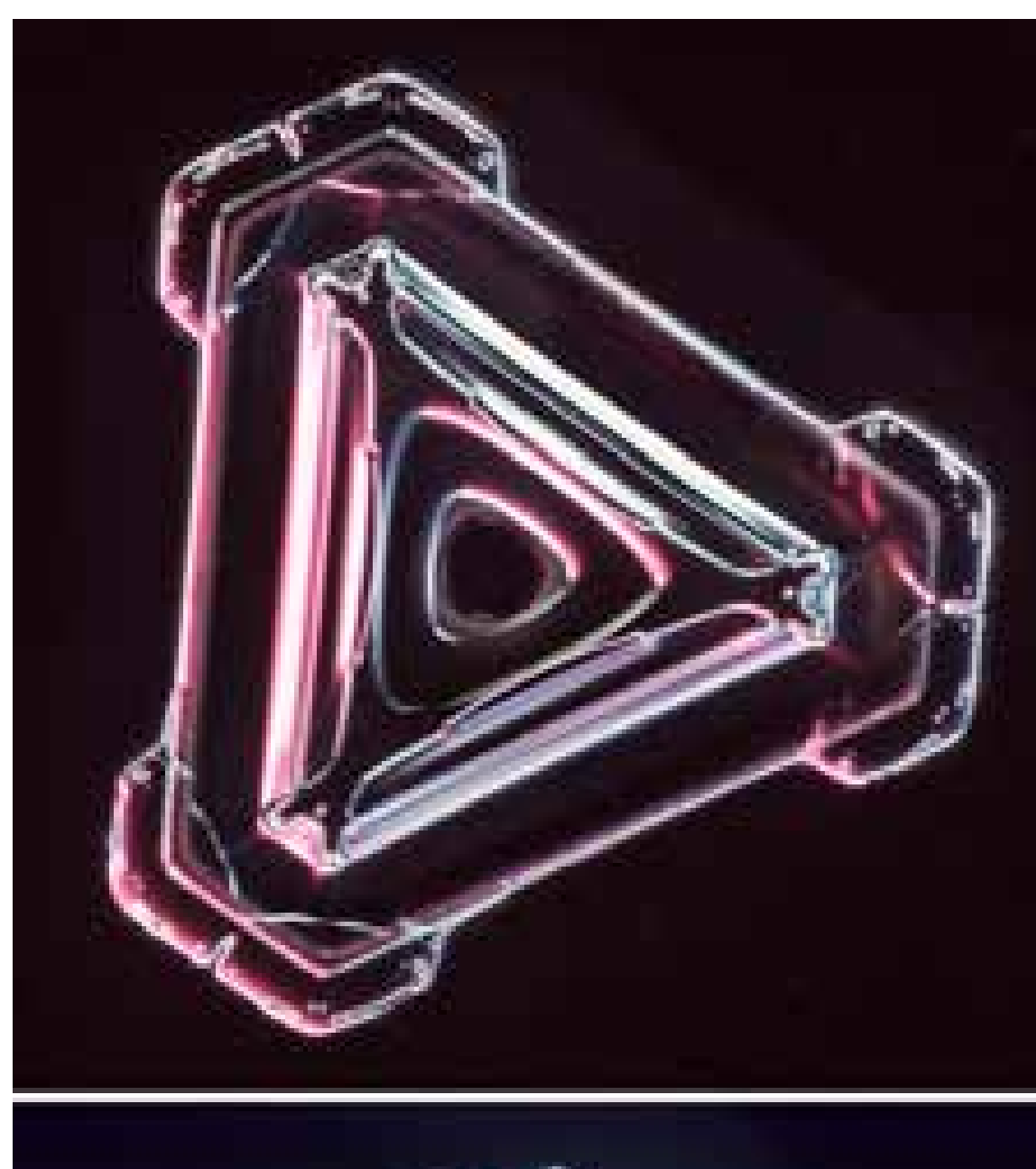

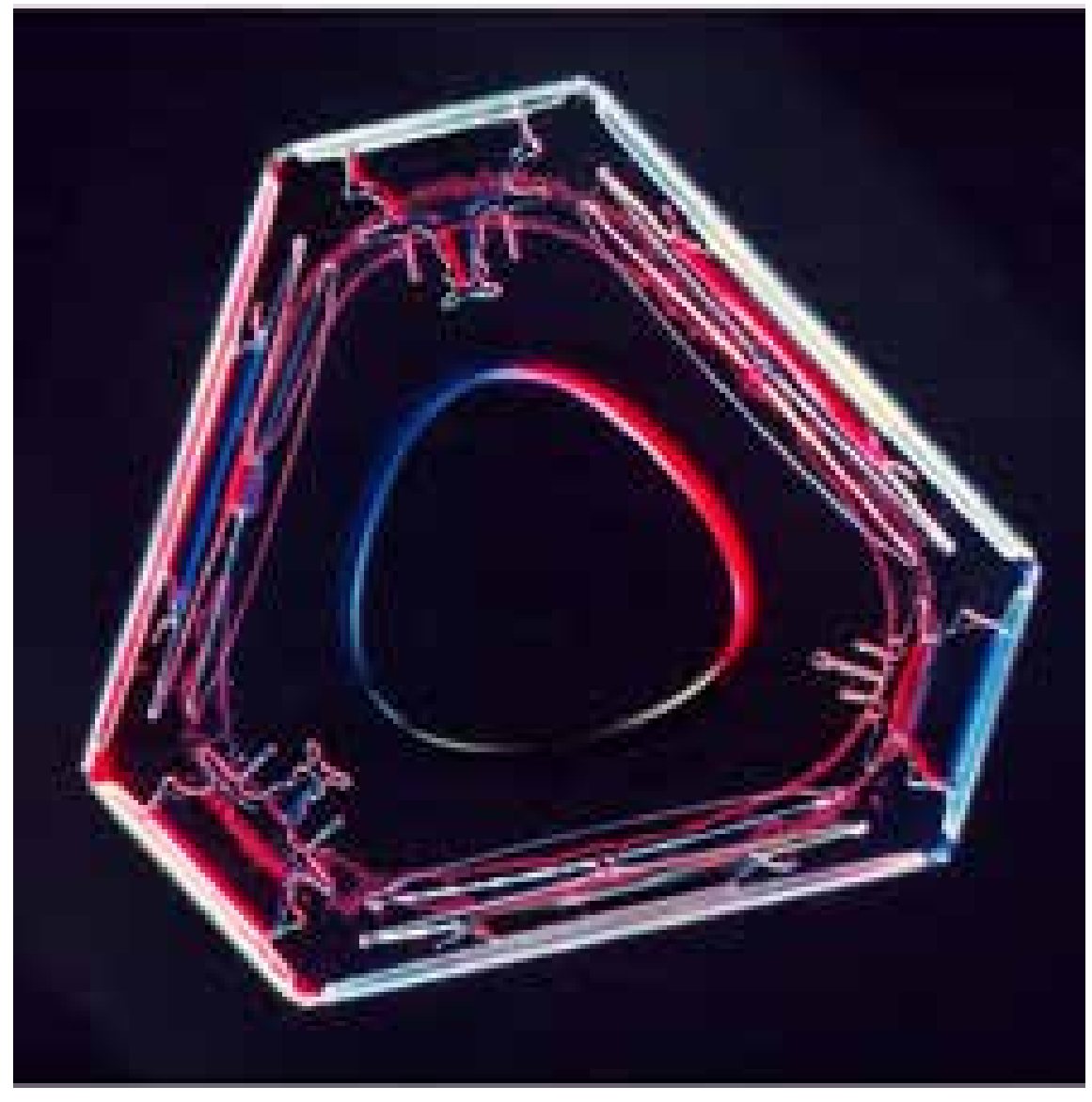

**Figure 3.58: These natural snow crystals are shaped like small truncated triangular plates. They typically appear together with much larger numbers of normal hexagonal plates.**

We first measured an unbiased sample of all simple plate-like crystals and defined a "hexagonality" parameter $H = L_1/L_6$ as the ratio of the length of the shortest side to that of the longest side. While $H = 1$ for a perfect hexagonal prism, we found that any crystal with $H > 0.75$ had a generally hexagonal appearance by eye, and Figure 3.60 shows our measured $H$ distribution. While many crystals in our sample exhibited a roughly hexagonal appearance, these data indicate that near-perfect hexagons $(H \approx 1)$ were somewhat rare.

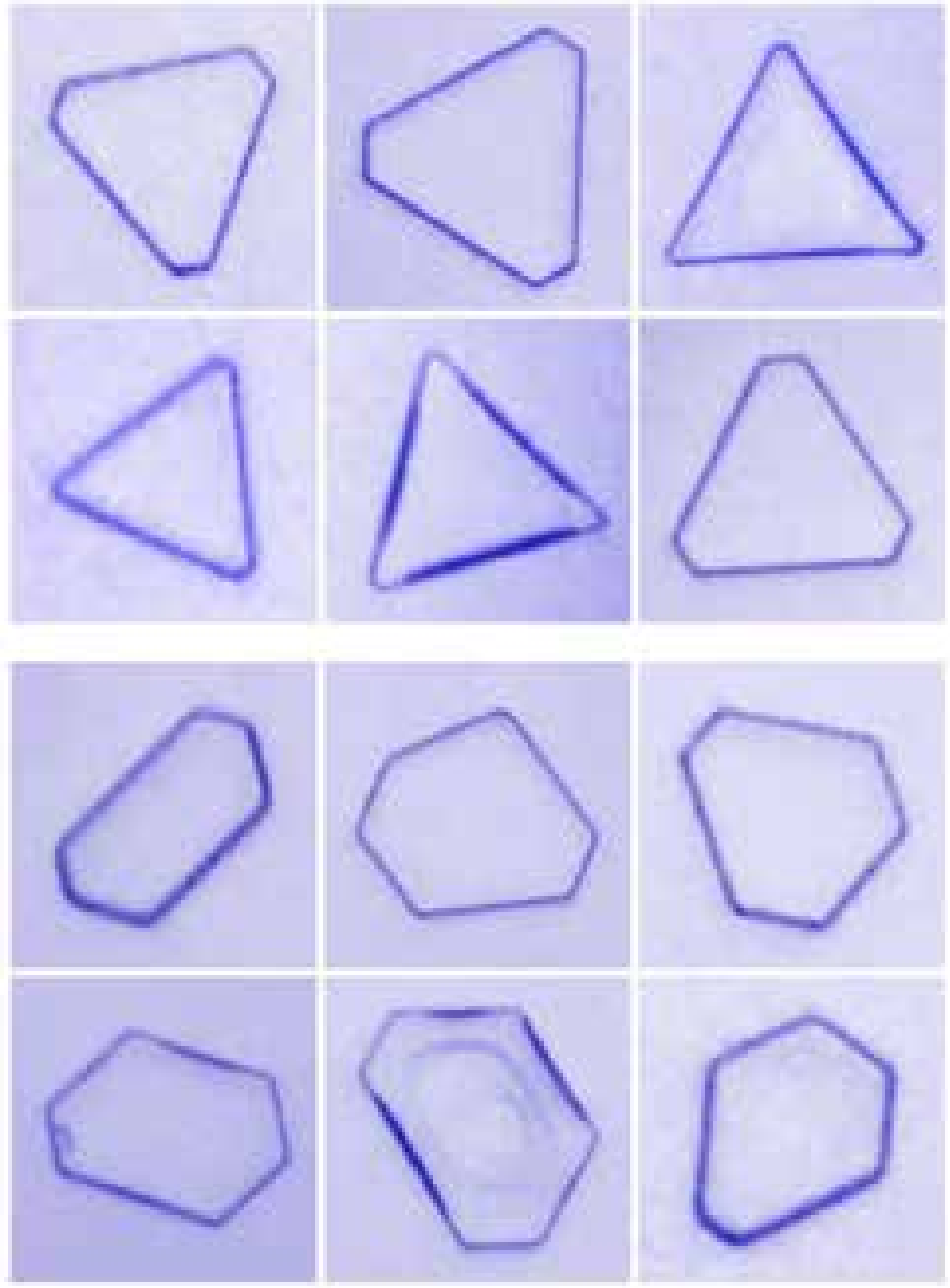

**Figure 3.59: A selection of non-hexagonal plates observed in a free-fall growth chamber at -10 C with $\sigma_\infty \approx 1.4$ percent. The top six images show crystals with an overall triangular symmetry, while the lower six images show crystals that are neither hexagonal nor triangular in appearance.**

We then examined a larger sample from which we rejected crystals with $H > 0.33$, and in this non-hexagonal sample we defined a "triangularity" parameter $T = L_3/L_4$ as the ratio of the lengths of the third and fourth longest sides. A truncated triangular morphology would have a small value of $T$, while $T \to 0$ for a near-perfect equilateral triangle.

Figure 3.61 shows the $T$ distribution we measured from our data, showing a sizable peak at low values, quantifying our visual impression that truncated triangular morphologies were the most common among non-hexagonal shapes. We then devised a Monte Carlo model in which we generated crystals where the perpendicular growth velocity of each facet was chosen from the same random distribution. From these crystals we selected ones with $H < 0.33$ and calculated the $T$ parameter for each. Comparing the model and data in Figure 3.61, we conclude that crystals with a triangular morphology (small $T$) are much more common than one would expect from random growth perturbations of normal hexagonal crystals.

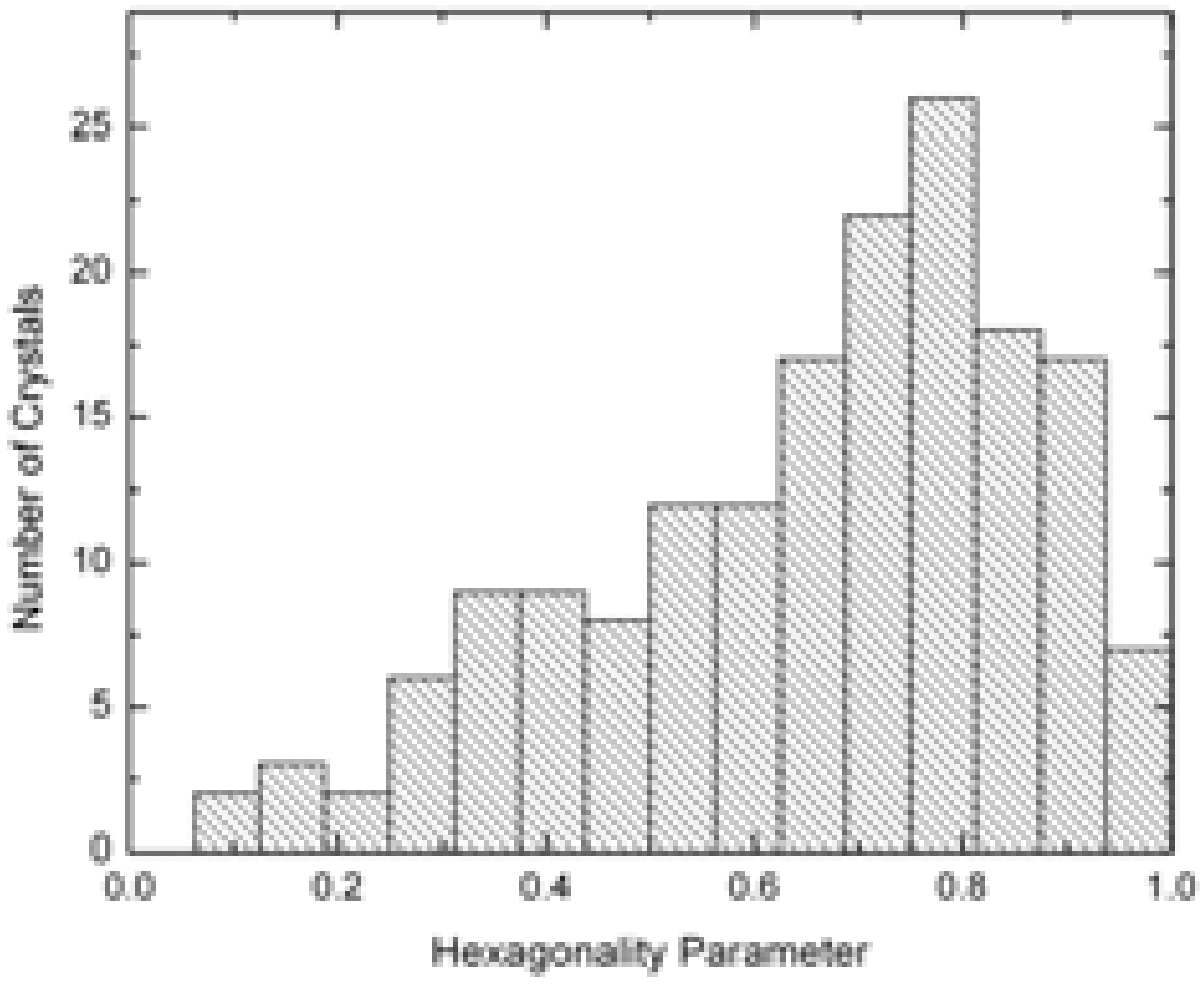

**Figure 3.60: The distribution of measured values of the hexagonality parameter $H$ for an unbiased sample of small plate-like crystals grown at -10 C. These data show that most plates are visually hexagonal in shape, but near-perfect hexagons are rare.**

The point of this exercise is to show, quantitatively, that there really is something special about the triangular morphology. Of all the other possible non-hexagonal shapes (some of which are shown in Figure 3.59), those with overall three-fold symmetry are by far the most numerous. Given that the underlying ice crystal symmetry is unchanged, so all six facets are essentially identical at the molecular level, the question then becomes what forces guide the development of triangular plates?

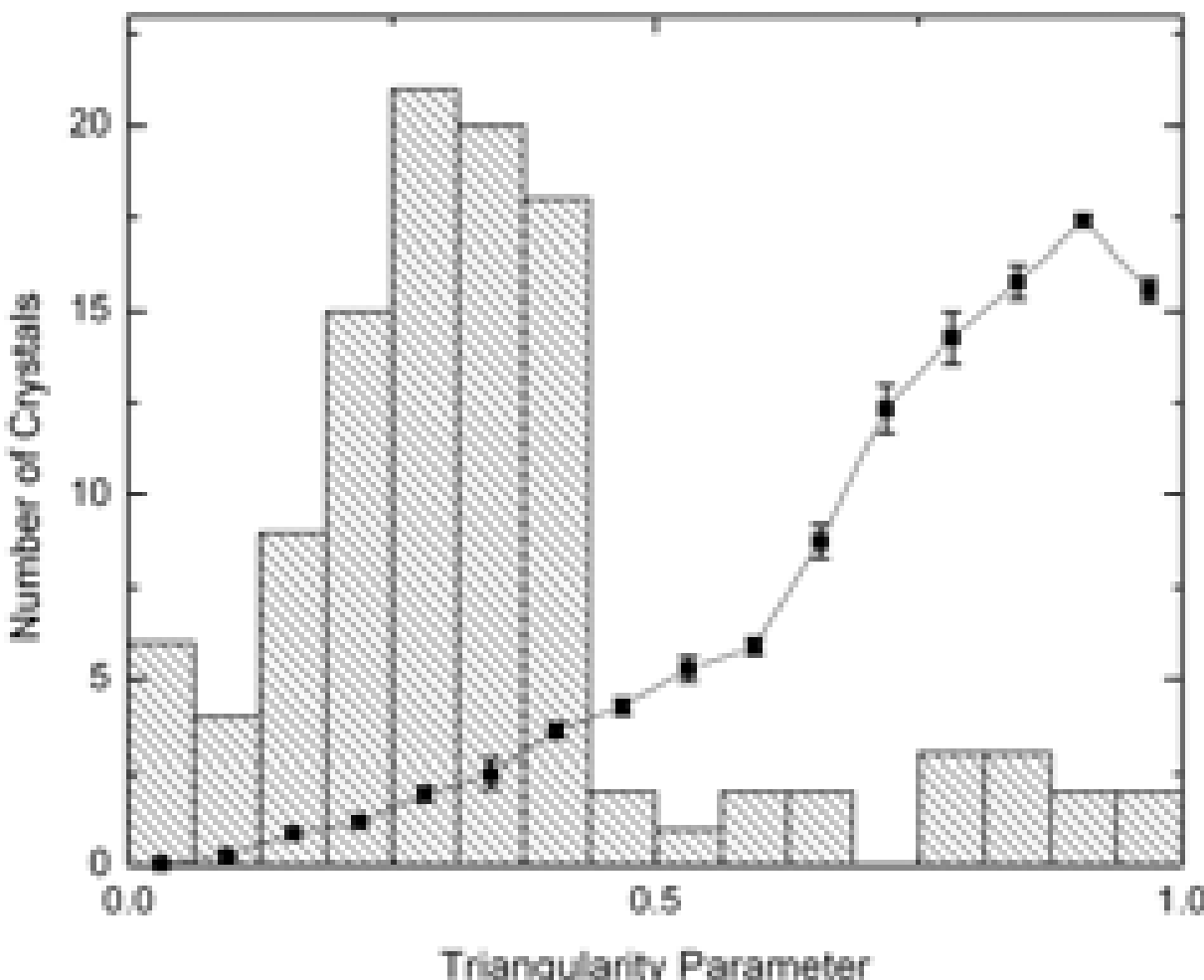

**Figure 3.61: The distribution of measured values of the triangularity parameter $T$ for a sample of plate-like crystals with $H < 0.33$ grown at -10 C. The line shows a Monte-Carlo model that assumes random growth perturbations of a hexagonal plate, again for crystals with $H < 0.33$ [2009Lib4]. These data indicate that truncated triangular plates are far more common than other non-hexagonal simple plates.**

To address this question, it is useful to also look at the formation of dendritic crystals forming from hexagonal columns near -5 C, and two examples are illustrated in Figure 3.62. The high-$\sigma$ crystal developed into a six-pronged "witch's broom" shape as fishbone dendrites sprouted from each of the six corners of the initial hexagonal column. In contrast, only three branches developed in the low-$\sigma$ crystal, giving it a three-pronged "trident" shape. Here again the crystal exhibits a three-fold symmetry, and more than half of all crystals grown under these conditions exhibited the same trident morphology.

The formation of tridents can be nicely explained from the diffusion-driven competition between the different branches, as illustrated in Figure 3.63. Beginning with six identical branches, assume that one grows out a bit faster than the others, just by random chance. This branch then sticks out farther into the supersaturated air and shields the growth of its nearest neighbors slightly. The larger branch thus grows faster while its nearest neighbors are soon left behind, this process being yet another manifestation of the Mullins-Sekerka instability.

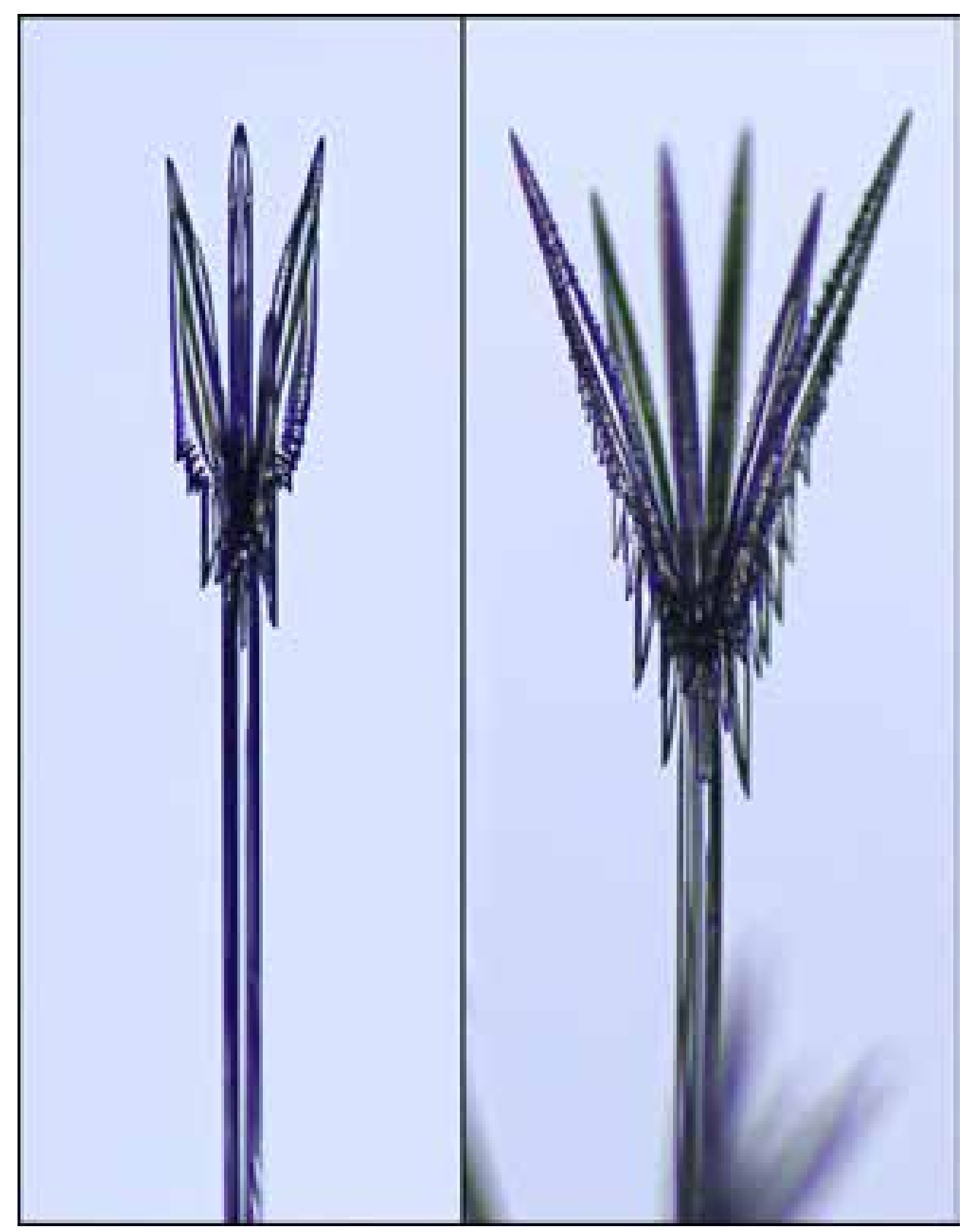

**Figure 3.62: (Left) A "trident" snow crystal forming on the end of an ice e-needle (see Chapter 8) in air with $(T, \sigma)$ = (-5 C, 32%). Starting from a hexagonal column, only three branches grew to a discernable length. (Right) A similar crystal growing with $(T, \sigma)$ = (-5 C, 64%). In this case the opening angle between the branches is larger and all six grew out from the initial columnar crystal.**

Of the remaining three branches, the outer two receive slightly more water vapor as a result of their two stunted neighbors, so they too grow out faster, leaving their middle neighbor behind. Assuming this diffusion dance plays out quickly, then a trident crystal emerges. It is left as an exercise for the reader to show that a trident also results if initially one branch grew slightly slower than the others.

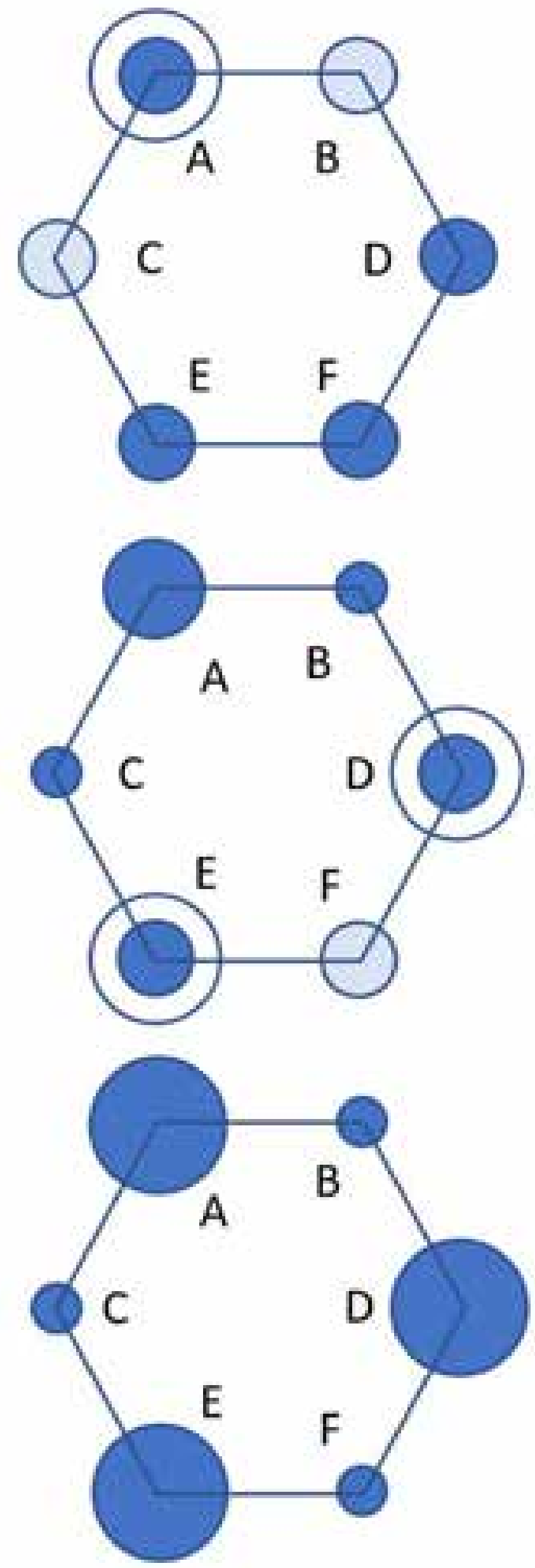

**Figure 3.63: Trident Formation. (Top) If one of six primary branches (branch A) extending from a columnar crystal becomes a bit taller than the others, then the Mullins-Sekerka instability will enhance its growth, while its immediate neighbors (branches B and C) will be shielded. (Middle) As branch A grows taller and branches B and C are left behind, D and E will be more exposed to the supersaturated air and will thus grow faster, shielding branch F. (Bottom) In time, branches A, D, and E will dominate while B, C and F are strongly shielded, yielding a trident-shaped crystal.**

This mechanism also explains why the low-$\sigma$ crystal in Figure 3.62 developed into a trident while the high-$\sigma$ crystal retained all six branches. In the low-$\sigma$ case, the opening angle of the branches was small, so the competition between branches was strong and persistent. In the high-$\sigma$ case, the opening angle was greater, and the growth rate was faster, so the branches quickly grew apart and the competition between them was weaker.

A key feature in this discussion is that three-fold symmetry is generally more stable than six-fold symmetry, at least regarding diffusion-limited growth. If a hexagonal plate is perturbed slightly toward a trigonal symmetry, then the Mullins-Sekerka instability will reinforce this perturbation, growing it to larger scales. But this process does not work in reverse; perturbing a triangular crystal slightly does not produce a hexagonal crystal via this mechanism. In the same vein, tridents are more stable than six-branched witch's brooms.

Given this one-way stability feature, all that is needed to turn a hexagonal crystal into a trigonal one is the initial perturbation. In the case of tridents, the likely perturbation mechanism is illustrated in Figure 3.63. The case for triangular plates is not so clear, but we described a possible aerodynamic mechanism in [2009Lib4]. This may not be the correct mechanism, however, and I suspect that there may be additional physics to consider in this matter. Nevertheless, it appears that three-fold symmetry in snow-crystal formation generally arises from diffusion-limited growth.

## Negative Snow Crystals

Figure 3.64 illustrates the growth of a "negative" snow crystal created by attaching a vacuum pump to a thin capillary tube inserted into a block of single-crystal ice. Water vapor is extracted through the capillary, leaving behind a void in the ice. Facets appear because it is especially difficult to remove water molecules from a perfectly faceted surface, as each is tightly bound by neighbors on all sides. It is comparatively much easier to remove molecules from a terrace step, as step-edge molecules have fewer nearest neighbors and are thus less tightly bound. Beginning with an arbitrarily shaped void, molecules are preferentially removed from terrace edges, eventually leaving behind a perfectly faceted void in the shape of a hexagonal prism.

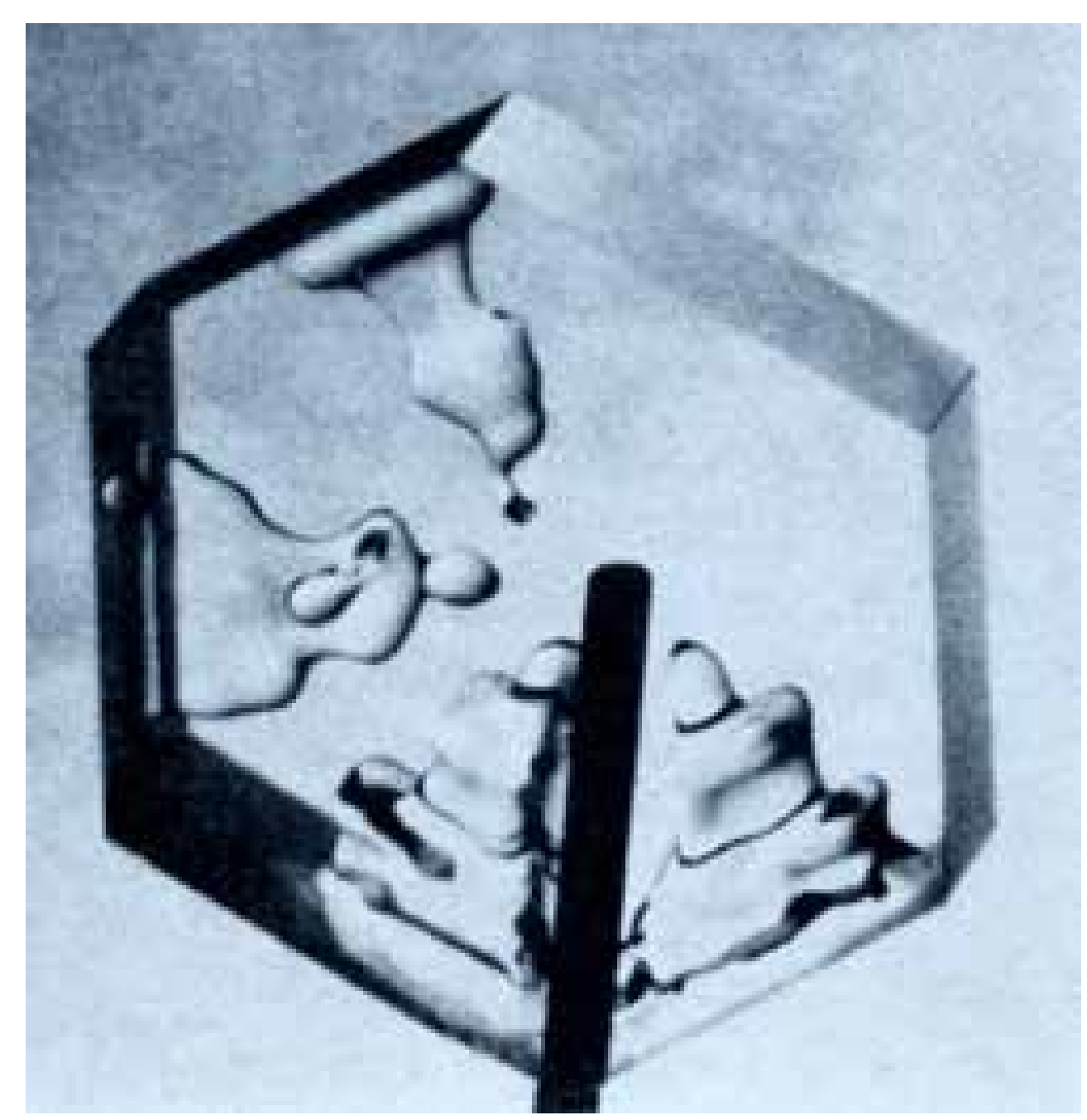

**Figure 3.64: A photograph of a "negative" snow crystal growing at -14 C [1965Kni]. The capillary tube (black) has a diameter of 0.45 mm. The overall faceted shape arises from a strong nucleation barrier in the sublimation kinetics, while the protrusions result from thermal-diffusion-limited growth. As with normal snow crystal growth, faceting is a nonequilibrium process dominated by molecular kinetics, while surface energy effects (which define the equilibrium crystal shape) are likely negligible.**

Note that the faceted shape does not arise from surface energy effects, and the minimum energy shape is likely nearly spherical (see Chapter 2). As with snow crystal growth, the formation of a prismatic void is a dynamical effect resulting from sublimation kinetics (which are related to attachment kinetics). In this case, the sublimation kinetics includes a strong nucleation barrier preventing the sublimation of molecules from internal faceted surfaces.

Note also that the time needed for this prismatic void to relax to its equilibrium shape might be exceedingly long. If the capillary were somehow extracted to leave behind a clean faceted void, then relaxation to a spherical shape would mean removing molecules from the facet surfaces and depositing them in the corners. This process is strongly suppressed by the nucleation barrier, greatly increasing the relaxation time. This slow equilibration also applies to bubbles in ice, as illustrated in Figure 3.40.

The formation of negative snow crystals is also affected by thermal diffusion, resulting in the peculiar protruding shapes seen within the void in Figure 3.64. Beginning with a perfectly faceted void, removing material via sublimation cools the ice, and the extracted heat must be replaced by thermal diffusion from the surrounding medium. The prism corners, sticking out farther into the ice, are more efficiently heated by diffusion, so the corners sublimate more quickly than the facet centers. As this process continues and the void grows larger, ice protrusions extend from the facet centers into the void.

Negative snow crystals have received relatively little study [1965Kni, 1993Fur], in part because they are somewhat difficult to grow under well-defined environmental conditions. Determining the undersaturation with useful accuracy is challenging, and negative crystals tend to be substantially larger than normal snow crystals. Quantitative growth measurements of negative crystals are subject the same kinds of systematic errors discussed in Chapter 7, and these are generally smaller with normal snow crystals. Nevertheless, a careful investigation of the growth and equilibration dynamics of negative ice crystals as a function of temperature could yield many useful insights.

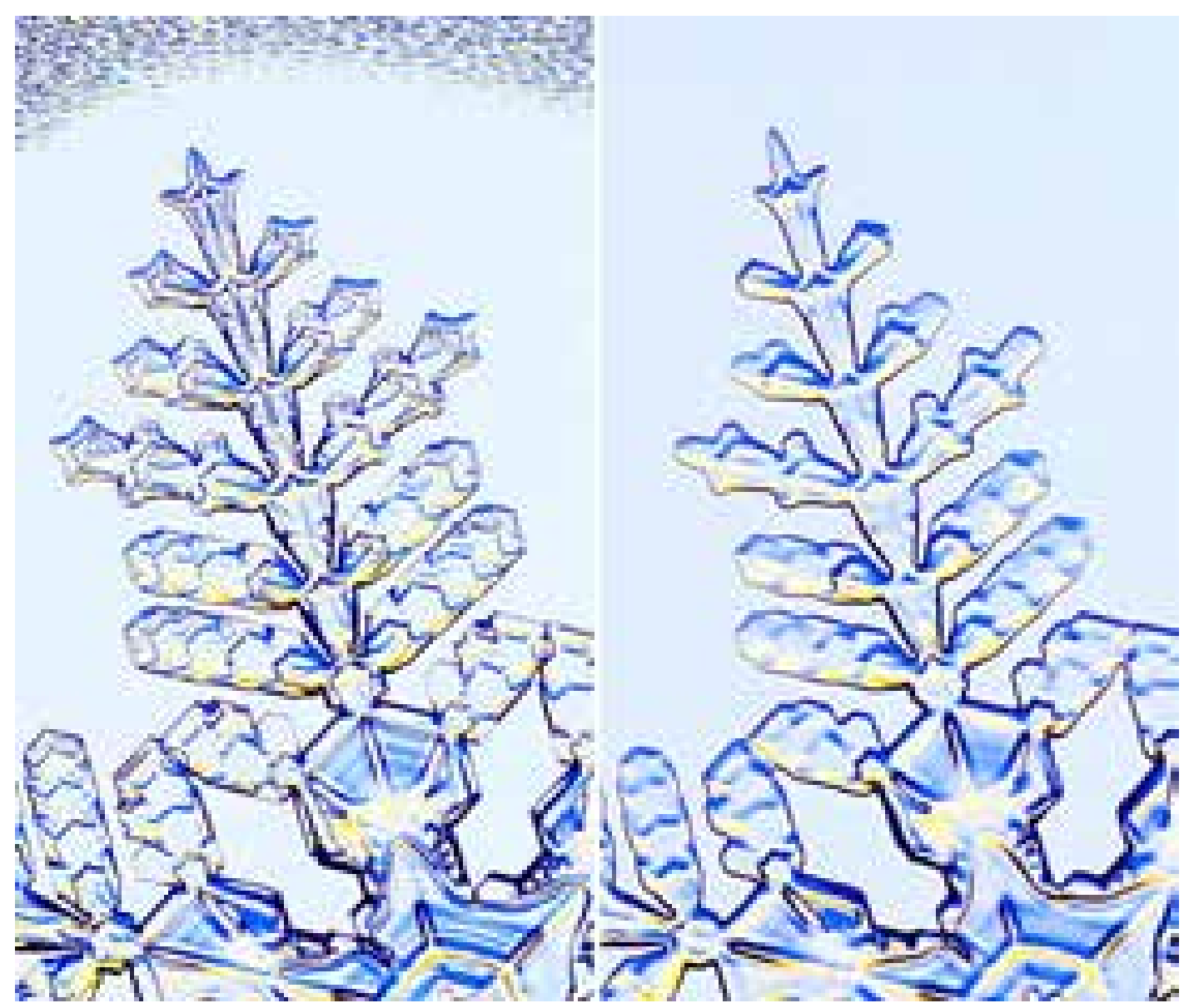

**Figure 3.65: (Left) A photograph of a single branch of a growing POP snow crystal. (Right) The same POP crystal a few minutes later, after reducing the humidity to sub-saturation levels, at which point the ice stopped growing and began sublimating away.**

## Sublimation

If the water vapor pressure in air surrounding a snow crystal is lower than the equilibrium vapor pressure of ice, then $\sigma_\infty < 0$ and sublimation will begin removing molecules from the ice surface, as illustrated in Figure 3.65. The term sublimation refers to the phase transition taking the solid directly into vapor, in this case when the temperature remains below 0 C (see Chapter 2). In sub-saturation conditions, the excess vapor near the surface must be carried away by diffusion, so the process of diffusion-limited sublimation is governed by the same physics described above, including both particle and heat diffusion.

One substantial difference between sublimation and deposition is that there are no nucleation barriers for the sublimation of convex surfaces, which means most of the surfaces seen in Figure 3.65. Thus, while faceting is a major player in snow crystal growth, it is largely absent in snow crystal sublimation as most surfaces exhibit $\alpha \approx 1$. For this reason, the sublimating crystal in

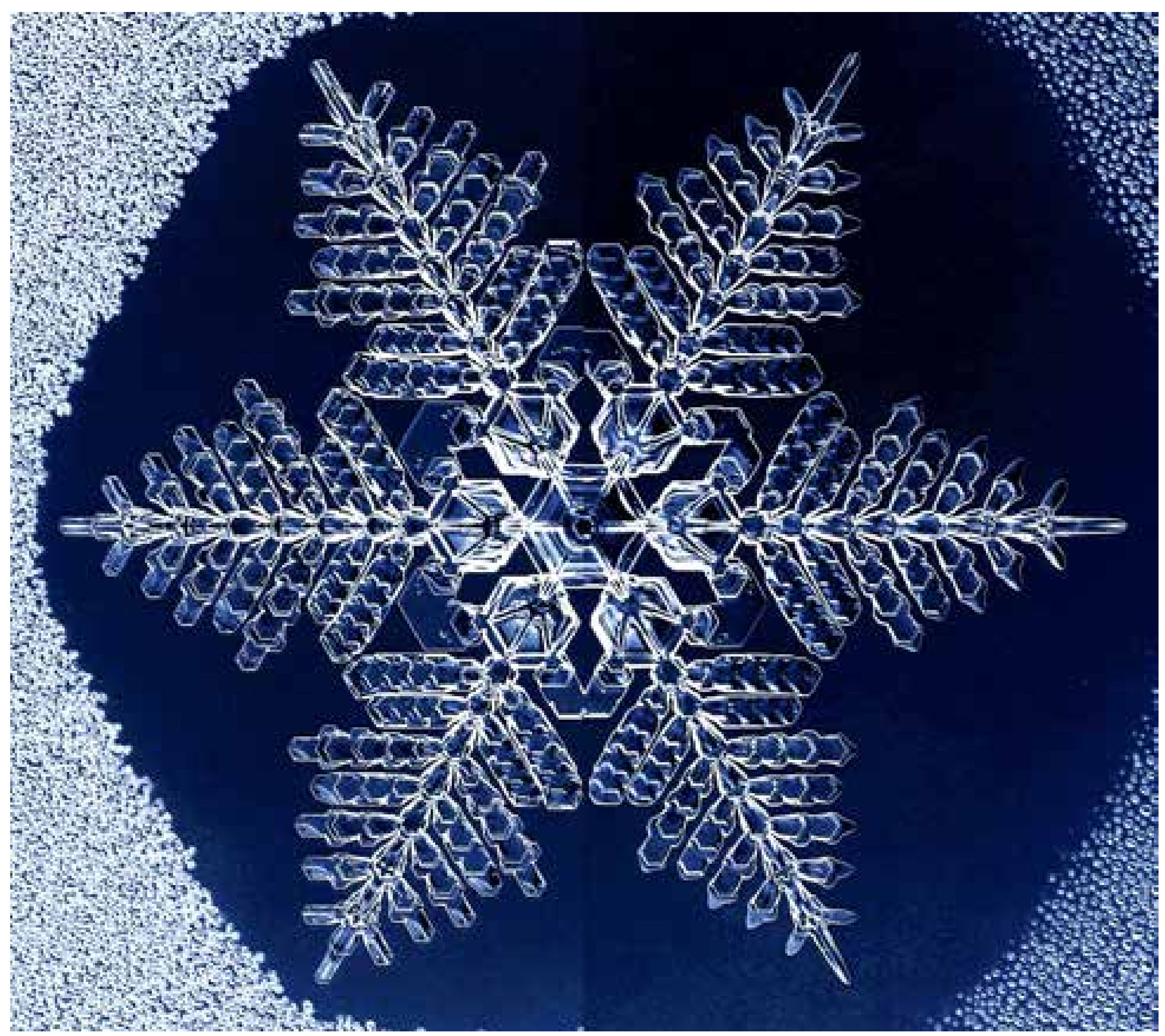

**Figure 3.66: A composite image showing a POP crystal as it grows (left side) and sublimates (right side).**

Figure 3.65 exhibits mostly rounded surfaces, while the growing crystal has a generally sharper appearance. In diffusion-limited sublimation, crystal features that stick out farthest into the sub-saturated air sublimate fastest, so sharp corners and edges quickly become rounded.

Photographs of natural snow crystals often show rounded edges because they begin sublimating once they leave the supersaturated clouds for their final descent through sub-saturated air. When snow clouds are quite high in the sky, falling crystals often have a "travel-worn" appearance for this reason. Snow crystal photography can be especially rewarding when the clouds are close to ground level, revealing sharply faceted features. Laboratory-grown POP crystals exhibit generally sharper, more vibrant structural features because they are photographed as they are growing.

Occasionally people will capture multiple photographs of a natural snow crystal as it sublimates away under the camera lens, and then claim the time series running in reverse shows a growing snowflake. Of course, taking a set of photographs is much easier than actually growing a snow crystal in the lab, but a trained eye can quickly identify the telltale signs of sublimation.

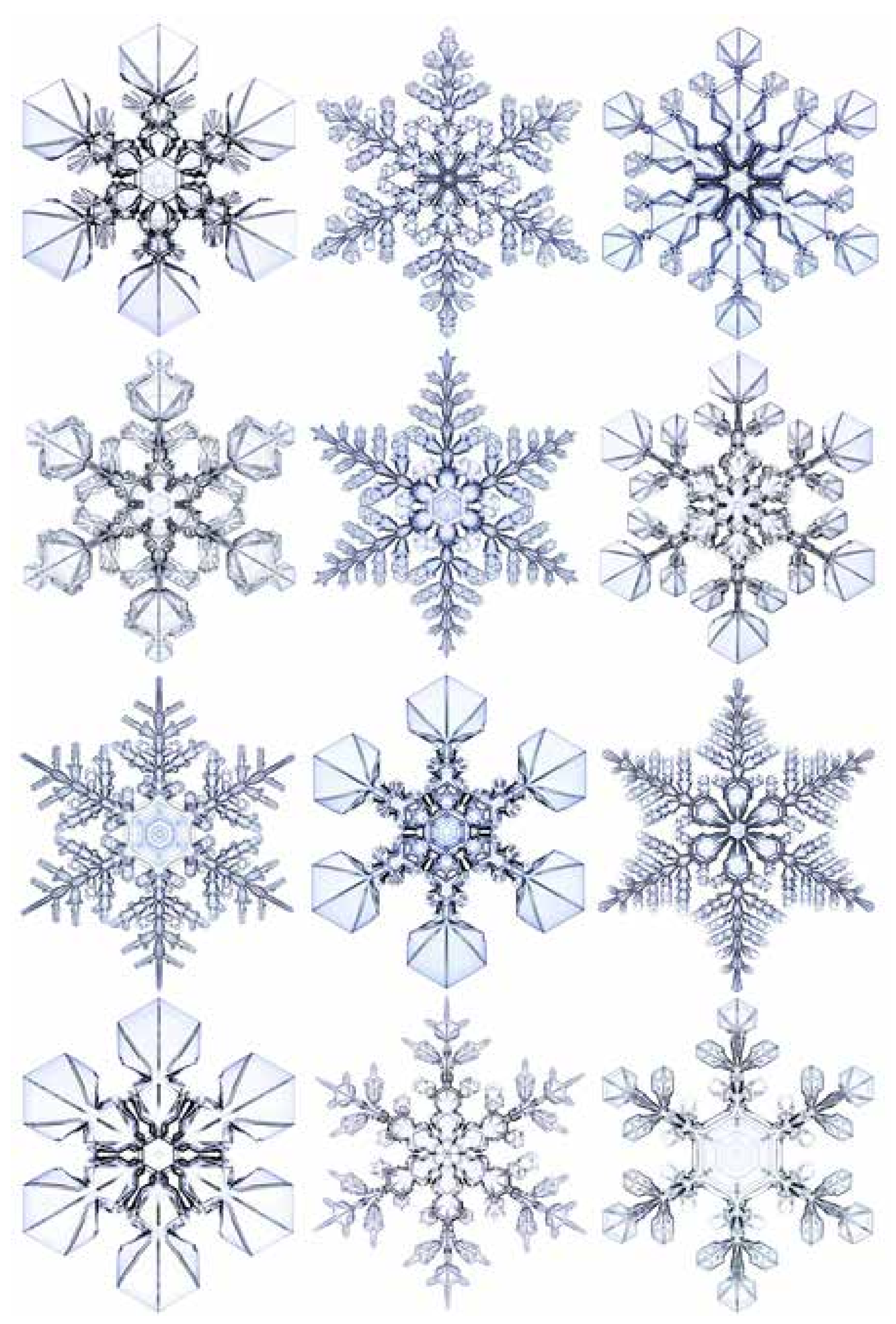

# Chapter 4

# Attachment Kinetics

*Like a great poet, Nature knows how to produce the greatest effects with the most limited means.*

– Heinrich Heine
*Pictures of Travel, 1871*

In snow crystal growth, the *attachment kinetics* describe how water vapor molecules striking an ice surface become incorporated into the crystal lattice. The attachment process is ultimately determined by the complex molecular interactions that jostle incident water vapor molecules into position so they can bind to the existing lattice structure. Because this many-body molecular dance is both intricate and unseen, there is much about it that we do not understand, even at a basic qualitative level. Nevertheless, the attachment kinetics represent one of the most important parts of the snow crystal story, driving the formation of faceted ice surfaces and other large-scale structural features. It has been well known for many decades that particle diffusion and surface attachment kinetics are the two primary physical processes governing snow crystal formation [1982Kur, 1984Kur1, 1990Yok, 2017Lib].

**Facing page: Several photographs of laboratory-grown Plate-on-Pedestal snow crystals (see Chapter 9). Attachment kinetics are responsible for the overall thin-plate structure of these crystals along with their sharply faceted features.**

Whether a snow crystal develops into a thin stellar plate or a slender columnar form reflects how rapidly water vapor molecules bind to different ice surfaces. With a platelike crystal, water molecules attach readily to the edges of the plate but attach only slowly to the basal faces. The opposite is true for long columns. The appearance of plates and columns at different temperatures in the Nakaya diagram (Chapter 1) is thus an especially intriguing aspect of snow crystal formation that derives mainly from the *anisotropy* in the molecular attachment kinetics.

My principal goal in this chapter is to define and quantify a comprehensive physical model of the molecular attachment kinetics at the ice/vapor interface. I will outline a suitable theoretical framework, examine what has been learned from experiments, and I will attempt to develop at least a qualitative picture of the detailed molecular processes involved. Many-body molecular dynamics is a complex subject, however, and previous attempts to understand the attachment kinetics have not fared particularly well (as discussed below). But a good model provides a springboard for future discussion and makes quantitative predictions that can be tested in future experiments. In the science of complex systems, progress often begins with a model, even an imperfect one. For the lack of a better name, I will refer to the specific model advanced in this chapter as the *Comprehensive Attachment Kinetics* (CAK) model.

Because this book is all about the growth and formation of snow crystals, our model of the attachment kinetics should provide a foundation for explaining the Nakaya diagram as well as quantitative ice-growth experiments (Chapters 7 and 8), and it should provide a suitable parameterization for creating computational models of growing snow crystals (Chapter 5). As we will quickly discover, however, achieving these goals is a tall order. The underlying molecular physics is often unclear, and a complex model will be needed to encompass a variety of different growth regimes. Like it or not, understanding snow crystal formation is not a trivial undertaking.

I believe that the model presented below represents significant progress toward understanding the many nuances of the Nakaya diagram and other aspects of snow crystal growth. However, some of its core ideas are still relatively new and not thoroughly tested. The model has some important merits (in my opinion), but it is far from being widely accepted canon. Indeed, some aspects of the model are speculative and may not withstand additional experimental and theoretical scrutiny. Nevertheless, this is the best solution I have found after considerable study of this thorny problem, so I present it here in its current form. Developing a comprehensive model of the ice/vapor attachment kinetics remains very much a work in progress, as one can say about many scientific endeavors.

## 4.1 Ice Kinetics

We begin our discussion with the Hertz-Knudsen relation [1882Her, 1915Knu, 1996Sai, 1990Yok], which we write as

$$v_n = \alpha v_{kin} \sigma_{surf} \tag{4.1}$$

where $v_n$ is the crystal growth velocity perpendicular to the growing surface, $\alpha$ is a dimensionless *attachment coefficient*, $\sigma_{surf} = (c_{surf} - c_{sat})/c_{sat}$ is the water vapor supersaturation at the surface, $c_{surf}$ is the water-vapor number density just above the surface, $c_{sat} = c_{sat}(T)$ is the saturated number density of an ice/vapor surface in equilibrium at temperature $T$, and

$$v_{kin} = \frac{c_{sat}}{c_{ice}} \sqrt{\frac{kT}{2\pi m_{mol}}} \tag{4.2}$$

is the *kinetic velocity*, in which $m_{mol}$ is the mass of a water molecule, $c_{ice} = \rho_{ice}/m_{mol}$ is the number density of ice, and $\rho_{ice}$ is the mass density of ice. Values of several of these quantities as a function of temperature are given in Chapter 2 and in Appendix A. Throughout the discussion, I will assume that any background gases surrounding a growing snow crystal, such as air and water vapor, are well described by the ideal gas laws in statistical mechanics. Given this assumption, which is highly accurate in most situations, one can work in terms of the water vapor molecular number density or the water vapor partial pressure, as the two are proportional (at constant temperature). I prefer the former, so $c_{sat}$, $c_{surf}$, and $c_{ice}$ appear throughout this book.

Equations 4.1 and 4.2 derive from the basic tenets of statistical mechanics [1996Sai, 1965Rei], and I will assume that the reader is generally familiar with this area of fundamental physics. In a nutshell, these equations come from considering two water vapor fluxes: the flux of molecules incident on a surface, equal to $c_{surf}\langle v_{mol}\rangle$, where $\langle v_{mol}\rangle$ is an average molecular velocity, and the flux leaving the surface from sublimation, equal to $c_{sat}\langle v_{mol}\rangle$. The difference between these two fluxes defines the growth velocity, and the statistical physics of ideal gases gives the appropriately weighted average velocity $\langle v_{mol}\rangle$ that appears in $v_{kin}$. Note that Equation 4.1 includes the trivial case of a vapor/solid interface in equilibrium; if the supersaturation $\sigma_{surf}$ is zero, then the growth rate must also be zero.

Much of the interesting molecular physics involved in snow crystal growth is wrapped up in the attachment coefficient $\alpha$, whose value lies in the range $0 \leq \alpha \leq 1$. One can think of $\alpha$ as a sticking probability, equal to the probability that a water vapor molecule striking the ice surface becomes assimilated into the crystal lattice. The value of $\alpha$ may depend on $\sigma_{surf}$, $T$, surface orientation relative to the crystal axes, and perhaps other factors. Ice surfaces that are "rough" at the molecular level typically exhibit $\alpha_{rough} \approx 1$, as water vapor molecules striking a rough surface usually become immediately indistinguishable from those in the existing ice lattice. Meanwhile, it is common to find $\alpha_{facet} \ll 1$ on "smooth" faceted surfaces, as these expose fewer open molecular binding sites, reducing the average sticking probability. With this definition of the attachment coefficient, it is generally true that thin platelike crystals (including stellar plates and stellar dendrites) form only when $\alpha_{basal} \ll \alpha_{prism}$ while slender columnar forms require $\alpha_{prism} \ll \alpha_{basal}$.

## Molecular Processes

Theoretical models of the attachment kinetics typically begin with an atomistic (or, in our case, molecular) picture of the crystal surface structure and dynamics, as illustrated in Figure 4.2. This sketch depicts several molecular processes that can occur on a growing ice surface:

1) Deposition (a.k.a. adsorption) – when a water vapor molecule strikes the surface and sticks. Molecules that are loosely attached to faceted surfaces are called *admolecules*, while tightly bound molecules are simply considered part of the underlying solid.
2) Sublimation (a.k.a. desorption) – when thermal fluctuations cause molecules to leave the surface and join the vapor phase. Isolated admolecules are especially likely to sublimate, often doing so before ever becoming tightly bound to the crystal lattice.
3) Surface diffusion – random motions of admolecules along a crystal surface. Diffusion along faceted surfaces can be especially substantial, as lateral motion is only weakly inhibited by molecular binding.
4) Attachment – when a diffusing admolecule encounters a terrace step (typically from the lower terrace) and becomes incorporated into the ice lattice. Detachment from a terrace step (typically onto the lower terrace) yields an isolated admolecule, and this process is often a precursor to sublimation.
5) Ehrlich-Schwoebel barrier – a potential barrier that inhibits admolecule motion between terraces. For an admolecule to leave an upper terrace and attach to a lower terrace step (see sketch), it would first have to detach from the upper terrace. The attachment to the lower edge is energetically favorable, but the initial detachment from the upper terrace is not. The resulting *Ehrlich-Schwoebel barrier* tends to suppress these over-step transitions. Thus, surface diffusion between separate terraces is suppressed compared to diffusion confined to the surface of a single terrace.
6) Terrace nucleation – when several admolecules on a faceted surface come together to form a new molecular layer, or terrace. This process is required to form new terraces on the topmost terrace of a facet surface.

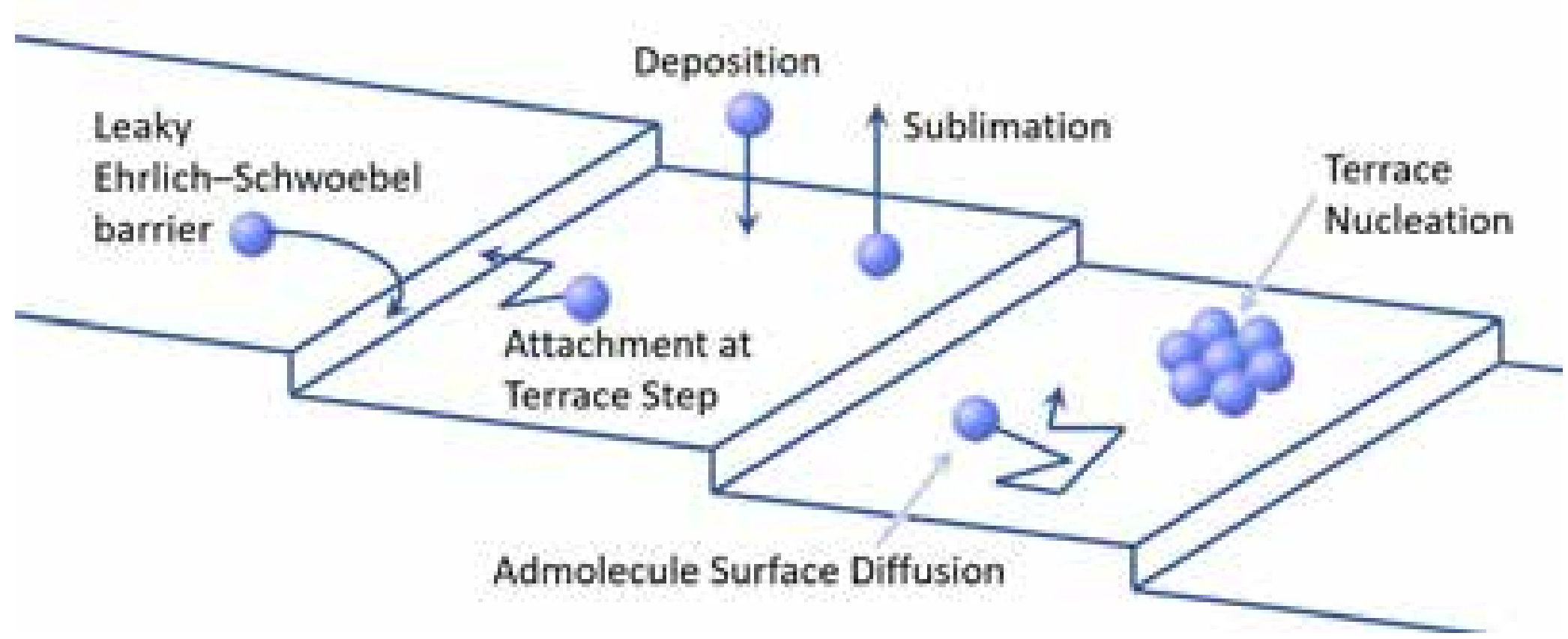

Figure 4.2: The attachment kinetics are governed by a variety of molecular processes occurring on the surface of a growing crystal, including those illustrated here.

This cartoon molecular picture of a crystal surface is too simplistic to provide a full kinetic description of ice, even at a qualitative level. Ice has a high vapor pressure, is not made from simple spherical molecules, and the ice surface experiences surface premelting at high temperatures (Chapter 2). As a result, the above list of molecular processes likely leaves out much important many-body physics. Molecular-dynamics simulations can help develop our understanding of real ice surfaces, but we are still far from fully understanding the structure and dynamics of the ice/vapor interface. Nevertheless, experiments suggest that this cartoon picture can aid our intuition when describing many important aspects of snow crystal growth, so we adopt it as a reasonable starting point.

## Surface Characteristics

Three types of surfaces play especially important roles in the discussion of snow crystal attachment kinetics: *faceted*, *rough*, and *vicinal* surfaces:

**Faceted Surfaces.** A faceted crystalline surface is defined by its low Miller indices, as described in Chapter 2. Low-index facets tend to have well-defined molecular terraces, and these surfaces best resemble the sketch shown in Figure 4.2. A perfect faceted surface can be thought of as being molecularly "flat" in that it contains no terrace steps or dislocations. The detailed molecular structures of the principal basal and prism facets are described in Chapter 2, as are the bilayer terrace steps on those surfaces.

An important feature of any faceted surface is that its molecular structure includes fewer dangling molecular bonds than a non-faceted surface. One result of this tighter molecular structure is a lower surface energy, but this fact plays a relatively minor role in snow crystal growth dynamics. The overall surface-energy anisotropy is just too weak to significantly affect snow-crystal growth or facet development. On the other hand, the anisotropy in the attachment kinetics can be enormous, as faceted surfaces often exhibit $\alpha_{facet}$ values that are orders of magnitude lower than on non-faceted surfaces. In contrast to the surface energy, the highly anisotropic attachment kinetics typically plays a major role in determining the development of snow crystal structures.

**Rough Surfaces.** A rough surface contains a high density of terrace steps, giving it a high density of dangling molecular bonds and a high attachment coefficient. In the case of ice surfaces, the experimental evidence suggests that water vapor molecules striking a rough surface are immediately indistinguishable from molecules in the ice lattice, which is another way of saying that incident molecules are immediately incorporated into the ice lattice. And this, by definition, means $\alpha_{rough} \approx 1$. This commonly encountered rough-surface limit is sometimes called *fast kinetics*. The attachment kinetics at a water/vapor interface is also near unity.

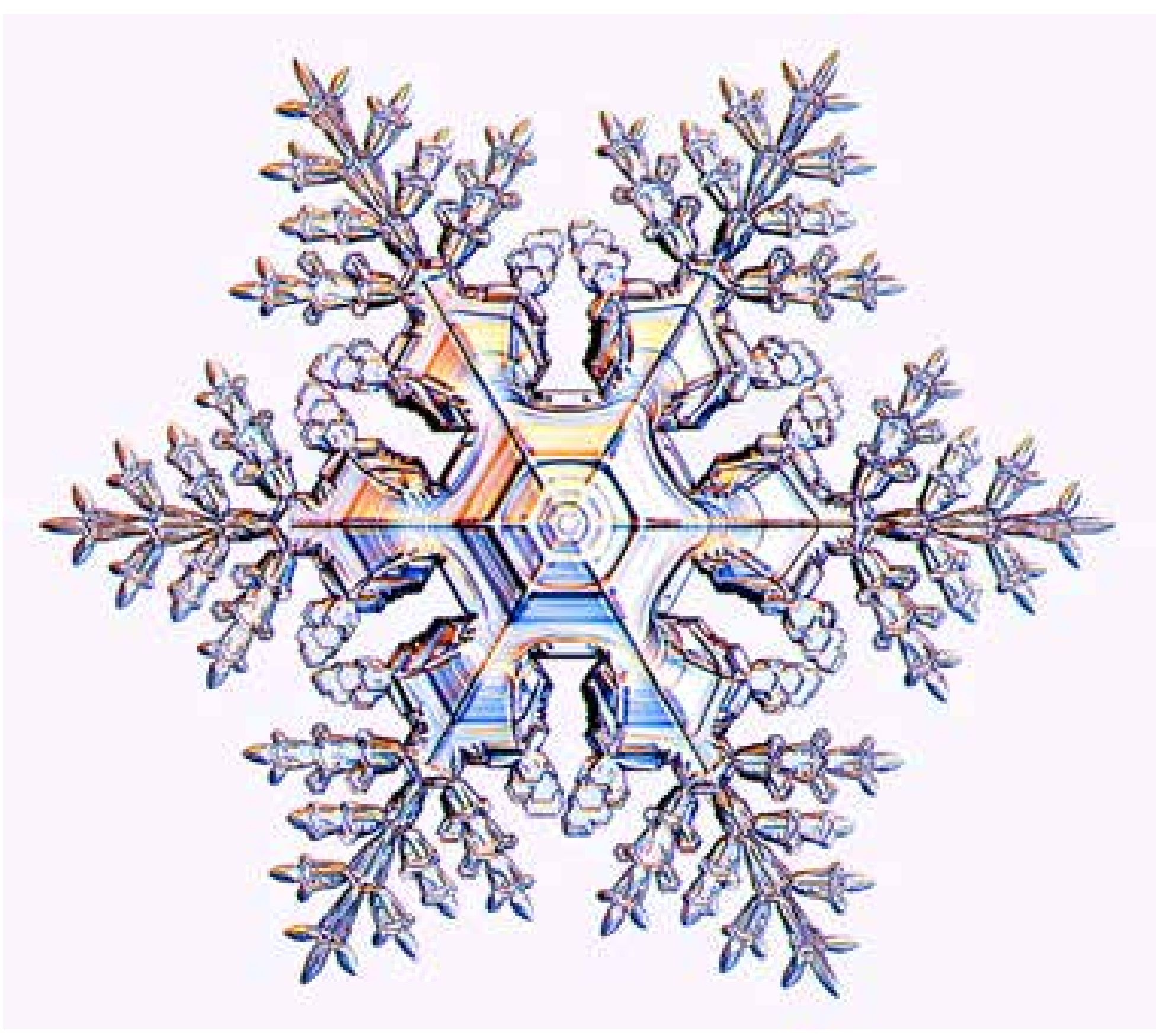

Figure 4.3. Facet-Dominated Growth. The attachment kinetics on any snow crystal can be divided into three classes of surfaces: top basil terraces described by $\alpha_{basal}$, top prism terraces described by $\alpha_{prism}$, and all other surfaces described by $\alpha \approx \alpha_{rough} \approx 1$. The overall platelike structure of this crystal, as well as much of its dendritic branching, is determined by the first two surface classes. Thus, the attachment kinetics on the topmost facet terraces strongly influences the morphological development of the entire structure.

**Vicinal Surfaces.** A vicinal surface is essentially a flat surface cut at a slight angle relative to a faceted surface. Figure 4.2 illustrates a vicinal surface, which includes a series of terrace steps with an average spacing $\ell_{vicinal}$, where $\ell_{vicinal}$ depends on the vicinal angle. If $\ell_{vicinal}$ is less than the mean diffusion length $x_{diff}$ (the typical distance admolecules on a faceted surface will diffuse before sublimating), then most admolecules will encounter a terrace step and attach. Therefore, $\alpha_{vicinal} \approx 1$ when $\ell_{vicinal} < x_{diff}$, and $\alpha_{vicinal} \rightarrow \alpha_{facet}$ when $\ell_{vicinal} \rightarrow \infty$. Note that the process of surface diffusion is usually implicitly incorporated into the attachment coefficient. It is also generally assumed, as part of the local form of the attachment kinetics defined by Equation 4.1, that $x_{diff}$ can be considered small compared to most macroscopic snow crystal structures.

## Facet-Dominated Growth

In ice growth from water vapor, the attachment kinetics are often highly anisotropic with deep cusps in $\alpha(\theta,\phi)$ at the principal facet angles, where $(\theta,\phi)$ is the angular orientation of the surface normal relative to the crystal lattice axes. It is not uncommon to have $\alpha_{facet} < 0.01$ while $\alpha_{vicinal} \approx 1$ at a vicinal angle of just a degree or two. As we will see in Chapter 5, such a deep, cusp-like anisotropy can lead to challenging numerical stability problems in computational snow-crystal modeling. But this strong anisotropy also helps bring about the rich variety of observed snow crystal patterns.

As we saw in Chapter 1, this basic model of anisotropic molecular attachment kinetics immediately explains the formation of snow crystal facets. Because $\alpha_{facet} \ll \alpha_{rough}$, the rough surfaces quickly accumulate material and fill in, while the faceted surfaces accumulate material at a much slower rate. The appearance of macroscopic facets in most natural crystalline materials, including mineral crystals, typically results from highly anisotropic attachment kinetics via this mechanism. (Although on most commercial gemstones, facets are fabricated with a grinder.)

Beyond facet formation, the attachment kinetics on the basal and prism surfaces also play a large role in guiding the formation of even complex, large-scale snow crystal structures (Figure 4.3). Essentially all non-faceted ice/vapor surfaces can be described with $\alpha \approx \alpha_{rou} \approx 1$ to a reasonable degree of accuracy. Thus, the only thing left to define the overall morphology of a snow crystal is $\alpha_{basal}$ and $\alpha_{prism}$. On dendritic structures exhibiting complex curved surfaces, it is often the attachment kinetics on the topmost, sometimes tiny, basal and prism terraces that guide the overall growth behavior. If not for the high anisotropy in the attachment kinetics, expressed mainly in $\alpha_{basal}$ and $\alpha_{prism}$, the menagerie of snow crystal forms would be absent its marvelous diversity. I call this situation *facet-dominated growth* because the slowest-growing, faceted surfaces tend to define the overall growth morphology, even when the growth is also strongly diffusion limited. In this circumstance, understanding snow crystal morphologies largely comes down to creating a suitable comprehensive model of the attachment kinetics on the basal and prism facets.

## 4.2 Large-Facet Attachment Kinetics

Because the basal and prism facets play such an important role in defining the growth behavior and structural development of many snow crystals, our first task is to examine the attachment kinetics on these surfaces. We begin with the ideal case of a large facet surface of effectively infinite lateral extent, so we can ignore any nonlocal, edge-related effects. We further assume a perfect crystalline structure, free from dislocations and other lattice defects. This is a good place to begin the discussion, as large faceted surfaces are common on many snow crystals. Moreover, both natural and laboratory-grown snow crystals frequently (although not always) appear to be essentially free from significant lattice imperfections.

As illustrated in Figure 4.2, isolated admolecules on a faceted ice surface are not yet fully incorporated into the crystalline lattice, owing to their relatively weak binding. In the absence of nearby terrace edges to bind to, a typical admolecule will reside on the surface for only a short time before thermal fluctuations send it back into the vapor phase. On large faceted surfaces, sustained crystal growth requires the *nucleation* of new molecular terraces. In this circumstance, $\alpha_{facet}$ is mainly determined by *nucleation-limited attachment kinetics*.

The facet surface is typically a maelstrom of molecular activity, as admolecules are continually coming and going, diffusing along the surface, and interacting with one another. Small terrace islands are constantly forming and disintegrating, growing as admolecules attach to their edges and shrinking as molecules thermally detach and diffuse away. Small terrace islands are the least stable and frequently break up via thermal fluctuations, but these are also the most likely to form via chance encounters. Larger islands are less likely to form but generally survive longer before breaking up. The nucleation of a new, permanent terrace occurs when an island appears with a radius larger than some critical size $R_{crit}$ that depends on the local supersaturation. Once such a stable terrace forms, it will usually continue to grow indefinitely as more admolecules diffuse to its edges and attach. The facet attachment coefficient depends on the value of $R_{crit}$, the rate at which stable terraces nucleate, and how rapidly stable terraces accumulate additional admolecules.

If the terrace nucleation rate is exceptionally low, then a single terrace may nucleate and grow until it covers the entire facet surface before the next new terrace appears. This is called *layer-by-layer* growth, and it is generally not so important when considering large facets (and certainly not for our ideal facets of infinite extent). For most growing snow crystals, a large faceted surface will contain many stable terraces of various sizes at all times, and this situation is called a *multinucleation model*.

## Terrace Nucleation Theory

On a faceted ice surface, the equilibrium vapor pressure of a small island of admolecules is higher than the normal saturated vapor pressure. Using an argument like that used to derive the Gibbs-Thomson effect in Chapter 2, the equilibrium vapor pressure of a circular island terrace of radius $R$ is

$$c_{eq} \approx c_{sat}\left(1+\frac{a^2\beta}{RkT}\right) \qquad (4.3)$$

where $a$ is the molecular size, $k$ is the Boltzmann factor, $T$ is the surface temperature, and $\beta$ is the step energy of the terrace edge. Thus, for a terrace island to be stable against sublimation, the supersaturation near the surface must be at least $\sigma_{surf} = a^2\beta/RkT$. Turning this around, an island terrace will achieve long-term stability only if its radius is greater than $R_{crit} = a^2\beta/\sigma_{surf}kT$. Putting in some typical numbers, $\beta \approx 10^{-12}$ J/m and $\sigma_{surf} \approx 1$ percent, a barely-stable island terrace might have a radius of about $R_{crit} \approx 10a$ and contain roughly 300 water molecules.

The growth rate $v_n$ of a faceted surface, and thus the attachment coefficient $\alpha_{facet}$, is tied directly to the rate at which new terraces appear and the rate at which existing terraces grow via admolecule attachment. The statistical mechanics describing these processes has been much studied over many decades, yielding a well-established *classical nucleation theory* that is described in detail in essentially all textbooks on crystal growth [e.g., 1994Ven, 1996Sai, 1998Pim, 2002Mut]. In three dimensions, classical nucleation theory describes the homogeneous nucleation of liquid droplets, while in two dimensions the same theory applies to the nucleation of island terraces on faceted crystal surfaces. The derivation of nucleation theory is quite involved, and I cannot improve upon the existing textbook treatments. In this book, therefore, I simply quote several salient features in the theory and apply it to the case of snow crystal growth.

Jumping straight to the main result in a multinucleation model, the creation and growth of new terraces yields an attachment coefficient that can be written, to a reasonable approximation, as [1996Sai]

$$\alpha(\sigma_{surf}) = Ae^{-\sigma_0/\sigma_{surf}} \qquad (4.4)$$

for the growth of a faceted surface, where $A$ and $\sigma_0$ are dimensionless parameters, with

$$\sigma_0(T) = \frac{S\beta^2a^2}{k^2T^2} \qquad (4.5)$$

Here I have included a dimensionless geometrical factor $S \approx 1$ to absorb several small theoretical factors (for example, the difference between $a$ and the actual terrace thickness). Given the substantial uncertainties in our current knowledge of $\beta$, the exact value of $S$ is not a great concern at this time, so I assume $S = 1$.

Nucleation theory generally indicates that $A$ will depend weakly on $\sigma_{surf}$, but I will mostly neglect any such dependence, as it is dwarfed by the strongly varying $\exp(-\sigma_0/\sigma_{surf})$ factor. Therefore, I take the parameters $A(T)$, $\sigma_0(T)$, and $\beta(T)$ to all be independent of $\sigma_{surf}$, and the resulting function in Equation 4.4 seems to provide a good fit to existing measurements of faceted ice growth (Chapter 7).

As a note of caution, I point out that theoretical models of terrace nucleation invariably include a variety of implicit, simplifying assumptions regarding molecular surface structure and dynamics, and some of these assumptions may not be justified for ice. For example, the theory usually begins with the basic surface molecular picture illustrated in Figure 4.2, which does not include surface premelting. While it is well known that premelting is an important structural characteristic of ice crystal surfaces near 0 C, we do not know how this phenomenon modifies the dynamics of terrace nucleation. Nucleation theory was developed mainly for low-vapor-pressure solids like metals and semiconductors, and it is, I believe, not thoroughly tested experimentally outside of this realm.

These caveats notwithstanding, some aspects of nucleation theory appear to be quite robust in the sense that they are largely insensitive to many surface characteristics. The exponential factor $\exp(-\sigma_0/\sigma_{surf})$ is an especially robust feature in nucleation theory, along with the relationship between $\sigma_0$ and $\beta$ given in Equation 4.5. These aspects of the theory are essentially independent of details pertaining to how admolecules diffuse along a faceted surface, the admolecule residence time, how terraces grow, and the number of stable terraces that are present on the surface at any given time. Importantly, over a broad range of surface characteristics, the terrace step energy $\beta$ is the only parameter that has a substantial effect on $\sigma_0$.

Because of this robust feature in nucleation theory, observing the functional form $\alpha \sim \exp(-\sigma_0/\sigma_{surf})$ in ice growth experiments is a strong indication that the growth rate is limited primarily by the nucleation of new terraces. In this case, one can use measurements of $\alpha(\sigma_{surf}, T)$ to extract the step energy function $\beta(T)$, which is otherwise difficult to measure. Moreover, $\beta$ is a fundamental material property of any faceted surface, like the surface energy. In principle, the step energy could be determined from purely equilibrium measurements (i.e., independent of crystal growth dynamics) or even by detailed calculations that determine crystal structure and energetics from known molecular interactions. To date, however, growth-rate experiments have yielded the best measurements of $\beta$ on the ice facets (Chapter 7).

One of the marvelous aspects of terrace nucleation theory is that it reduces a complex surface-dynamical process to essentially a single equilibrium quantity $\beta$. All the specific molecular-dynamics details regarding admolecule deposition, sublimation, surface diffusion, and bonding at terrace steps become largely irrelevant. The distinctive functional form $\alpha \sim \exp(-\sigma_0/\sigma_{surf})$ depends only on the terrace step energy. Terrace nucleation theory thus provides a simple, accurate, and quantitative physical model that plays an important role in the ice/vapor attachment kinetics.

## Model Parameters from Measurements

How well the terrace nucleation model applies to real ice growth can only be determined by detailed comparisons with experimental observations. Early measurements clearly exhibited some of the characteristic traits of nucleation-limited growth [1972Lam, 1983Bec, 1989Sei, 1998Nel], and the evidence became ever stronger as improved experimental techniques yielded greater measurement precision [2013Lib, 2019Lib3]. For example, Figure 4.4 illustrates the exponential rise in growth velocity with $\sigma_{surf}$ that is the hallmark of a nucleation-limited model. Here the data points show measurements of the perpendicular growth velocity of a prism facet surface as a function of the near-surface supersaturation $\sigma_{surf}$. These data were taken at -15 C with a background air pressure of 20 mbar using the VIG apparatus (Chapter 7). The line shows $v_n = \alpha v_{kin} \sigma_{surf}$ with $\alpha(\sigma_{surf}) = \exp(-\sigma_0/\sigma_{surf})$ and $\sigma_0 = 3$ percent. The inset image shows the test crystal immediately after the growth measurements were completed [2013Lib]. Many additional experimental results like this suggest that the terrace-nucleation model accurately describes the growth of large, defect-free basal and prism faceted surfaces over essentially all conditions relevant to snow crystal formation. This is a strong statement, and it represents my experienced, albeit perhaps not entirely unbiased, interpretation of the available published data.

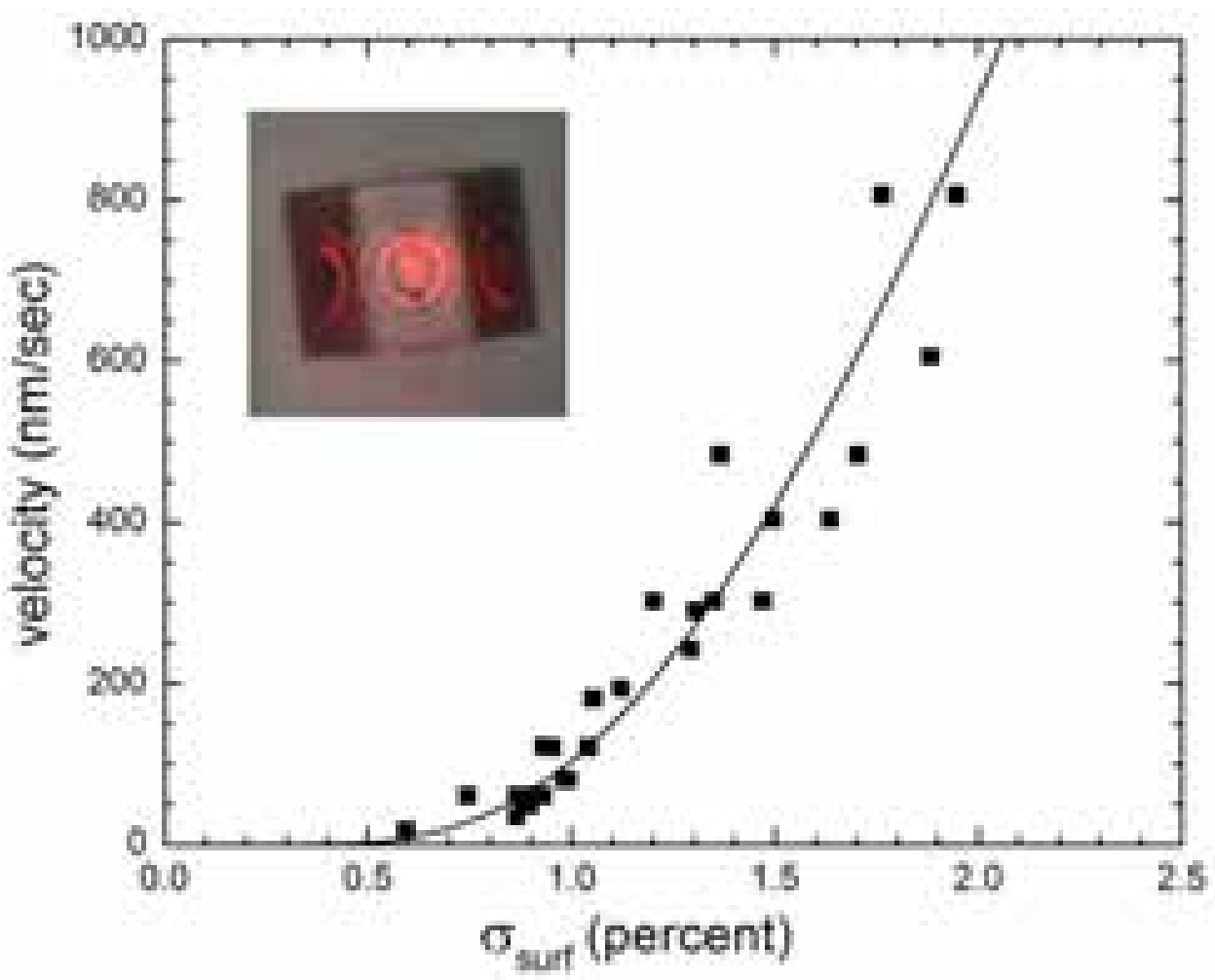

Figure 4.4: Ice-growth measurements (data points) are well described by a terrace nucleation model (curve) [2013Lib]. Chapter 7 describes these data in detail.

As a first step in developing the Comprehensive Attachment Kinetics (CAK) model in this chapter, I combine results from the best available ice-growth experiments to produce the best-estimate extracted model parameters $A(T)$, $\sigma_0(T)$, and $\beta(T)$ shown in Figures 4.5 and 4.6. For the sake of pedagogy, I display these largely empirical results in the form of smooth curves, as this approach allows a less-cluttered discussion of the underlying physical processes, thus facilitating our continued development of the CAK model below. This somewhat oversimplified presentation glosses over experimental uncertainties and the inevitable inconsistencies between published data sets, and I examine some of those issues in Chapter 7. For now, suffice it to say that these curves are probably reasonably accurate in an absolute sense, but may require some modification in the light of future improved experiments.

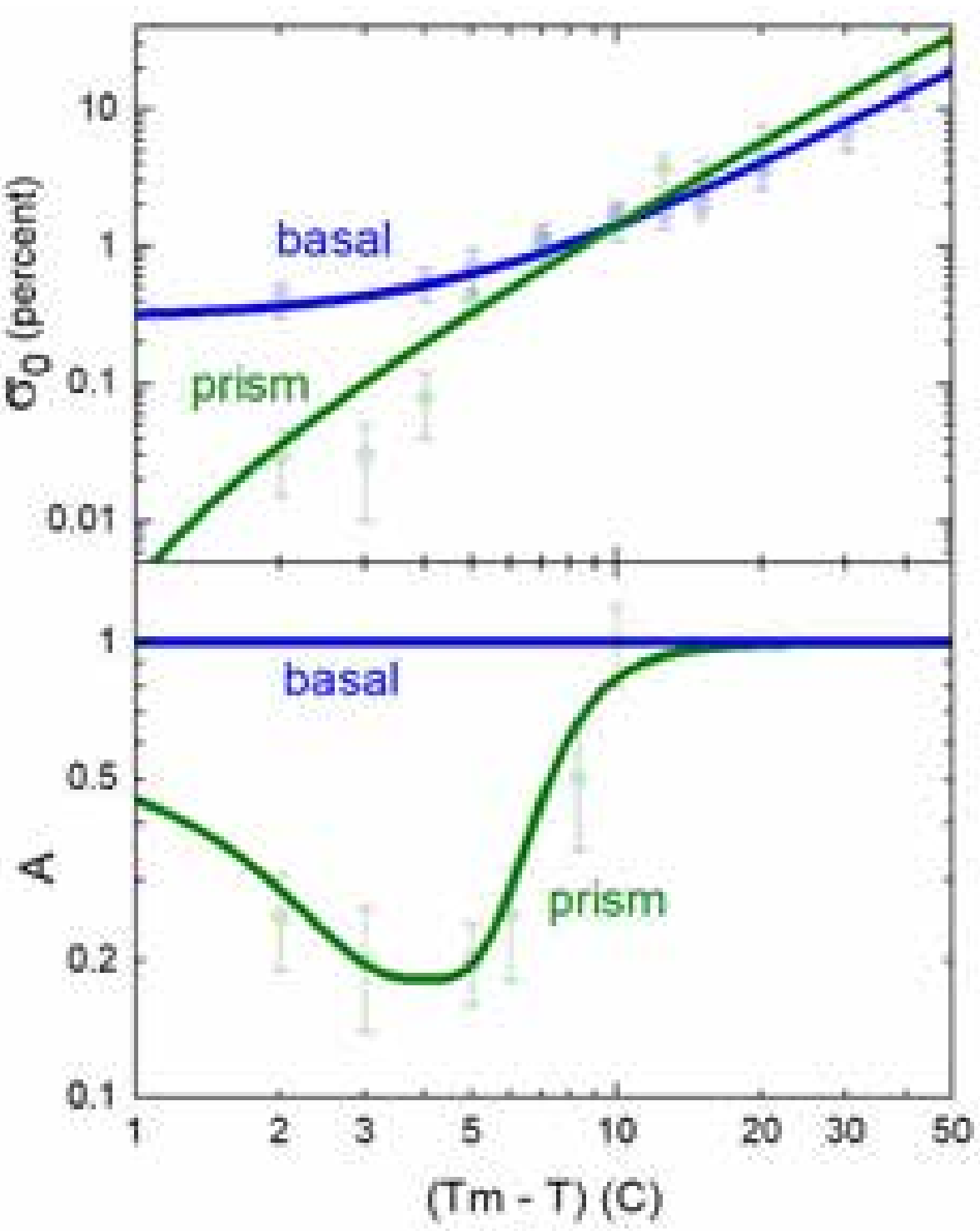

Figure 4.5: The smooth curves in these graphs show best-estimate CAK model functions $\sigma_0(T)$ and $A(T)$ on large basal and prism facets as a function of temperature, where $T_m = 0$ C is the ice melting point. The respective attachment coefficients are then given by $\alpha(\sigma_{surf}) = A\exp(-\sigma_0/\sigma_{surf})$. The underlying experimental data and measurement techniques used to obtain these curves are described in Chapter 7. Because these curves parameterize the large-facet attachment kinetics over a broad range of growth conditions, they provide the foundation for the CAK model of the snow crystal attachment kinetics.

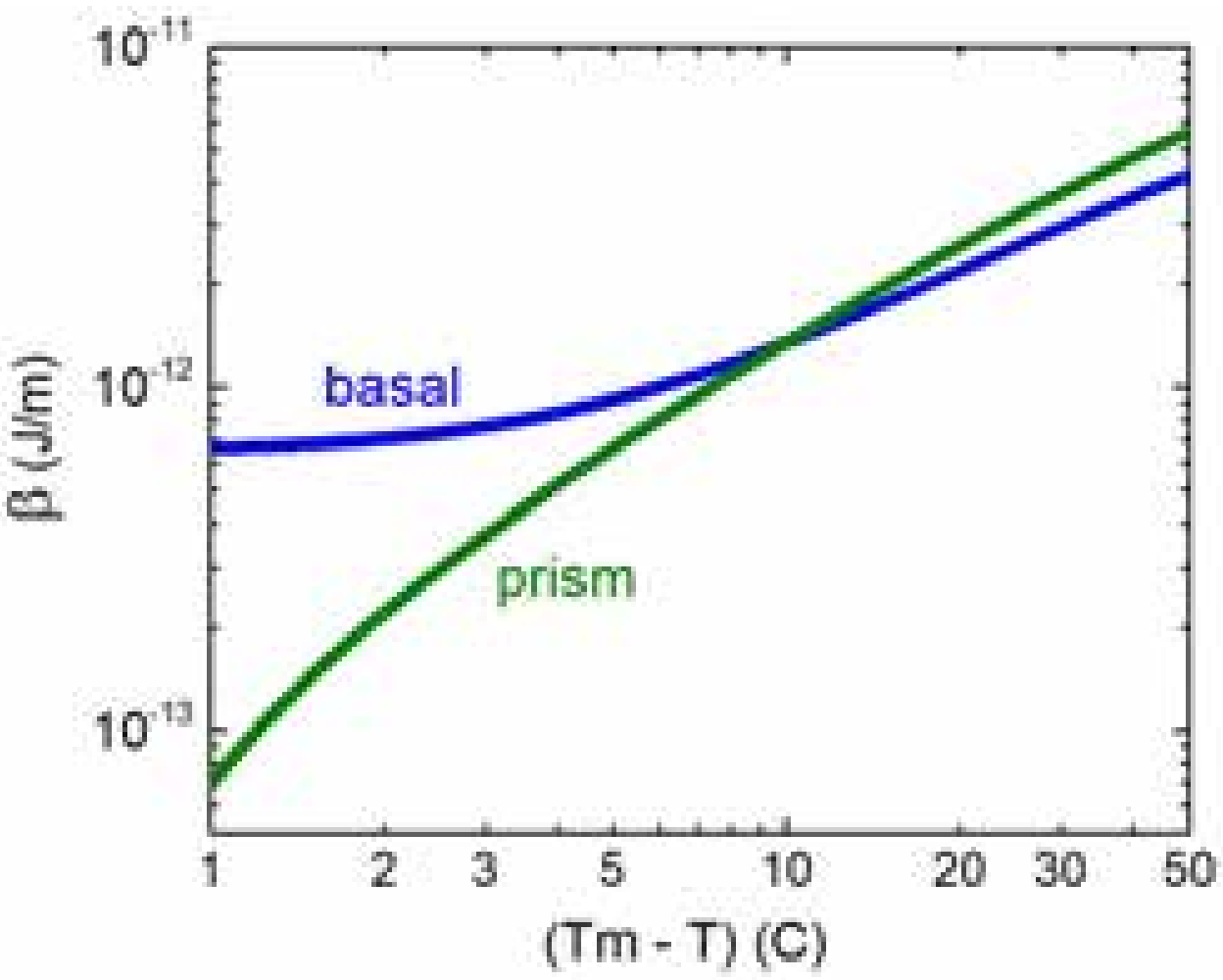

Figure 4.6: These curves show CAK model estimates for the step energies $\beta(T)$ for large basal and prism facets as a function of temperature, extracted from $\sigma_0(T)$ using Equation 4.5. While derived from dynamical growth measurements, these step energies represent fundamental equilibrium properties of the ice crystal surface.

The set of curves in Figures 4.5 and 4.6 parameterize the attachment kinetics describing large, defect-free faceted surfaces in the CAK model, and they are little more than parameterized fits to experimental data. Measurements of $v_n(\sigma_{surf}, T)$ like those shown in Figure 4.4 are first converted to $\alpha(\sigma_{surf}, T)$ using Equation 4.1, and then fits to Equation 4.4 are used to extract the parameters $A(T)$ and $\sigma_0(T)$. The step energies $\beta(T)$ are further obtained from $\sigma_0(T)$ using Equation 4.5. The assumed functional forms came from terrace-nucleation theory, and they seem to provide quite a good fit to ice-growth data over a broad range of $(\sigma_{surf}, T)$. The CAK model includes an implicit assumption that the attachment kinetics does not depend on background gas pressure or other chemical effects. In the spirit of an ideal large-facet kinetics model, I assume that water is the only chemical species in the problem. How well this pure model describes real snow crystal growth is an important experimental question, and I discuss this issue later in this chapter and in Chapter 7.

The measured terrace step energies are a foundational element in the CAK model, and these represent equilibrium material properties of the ice surface, as I discussed in Chapter 2. At low temperatures, surface premelting is essentially absent on both the basal and prism facets, so the step energies should tend toward the rigid-lattice value of $\beta_0 = a\gamma_{sv} \approx 3 \times 10^{-11}$ J/m, and this statement is reasonably consistent with the data. At high temperatures, the basal step energy should tend toward the ice/water step energy $\beta_{sl,basal} \approx 5.6 \pm 0.7 \times 10^{-13}$ J/m (Chapter 2), while the prism step energy should tend toward $\beta_{sl,prism} \ll \beta_{sl,basal}$. And again, these statements appear to fit the data. Put another way, although we do not have a precise model of the terrace step energies for either the ice/vapor or ice/water interfaces, the measurements seem to paint a reasonably consistent picture of the underlying physics. Thus, while the step energies in Figure 4.6 reflect empirical measurements, the results seem sensible overall.

## Frustrated QLL Kinetics

The measured $A(T)$ behavior is not as important as the terrace step energies, but it is a significant feature in the measurements, and it does not suggest a ready physical explanation. Going back to the ice-growth data that yielded Figure 4.5, the values of both $A_{bas}$ and $A_{prism}$ generally reflect the growth behaviors at high $\sigma_{surf}$, specifically when $\sigma_{surf} \gg \sigma_0$ and the $\exp(-\sigma_0/\sigma_{surf})$ factor is near unity. In this fast-growth regime, terrace nucleation is rapid, so the surface contains a relatively high density of terrace steps, even though it remains faceted on large scales. The abundance of terrace steps means that the surface begins to resemble a molecularly rough surface. In this situation, surface diffusion will quickly transport admolecules to nearby terrace steps where they will become incorporated into the ice lattice. Thus, on both the prism and basal facets, it is natural to expect $\alpha \to \alpha_{rough} \approx 1$ when $\sigma_{surf} \gg \sigma_0$, and this expectation implies $A \approx 1$. And, indeed, we see $A_{basal} \approx 1$ at all temperatures in Figure 4.5 (Chapter 7), along with $A_{prism} \approx 1$ at temperatures below -10 C. Consequently, it appears that the high-temperature region with $A_{prism} < 1$ is something of an anomaly.

To explain this anomalous $A_{prism}$ behavior, the short answer is that this is an empirical observation with no obvious physical explanation. Nevertheless, the temperature dependence suggests that surface premelting may be involved, because the transition to $A_{prism} < 1$ happens right about where surface premelting becomes important. Therefore, undeterred by mere ignorance, let us follow this thread a bit further and look at the attachment kinetics in the presence of surface premelting when the nucleation barrier is low. In the extreme case of a thick QLL, assume that the surface has an overall ice/QLL/vapor structure (as inferred from MD simulations), and further assume that attachment is fast at the QLL/vapor interface. If these reasonable assumptions are true, then the relevant physics giving us $A_{pris} < 1$ must be happening at the ice/QLL interface. And this suggests that we take a quick foray into the ice/water attachment kinetics.

While the fundamental theory of solid/liquid attachment kinetics is not well developed, the Wilson-Frenkel model of melt growth provides an overarching view of the underlying physics [1900Wil, 1932Fre, 1996Sai]. In a nutshell, the solid/liquid kinetics is limited by the mobility that allows a near-surface liquid molecule to move into position so it can become incorporated into the solid lattice. In the Wilson-Frenkel model, this mobility translates into an attachment coefficient that is inversely proportional to the liquid viscosity, which is proportional the liquid diffusion constant [1996Sai].

In ice growth from liquid water, prism growth rates are extremely fast, so this suggests that the ice/QLL kinetics would be fast as well, again implying $A_{prism} \approx 1$. Because this is not observed to be the case above -10 C, it suggests that the molecular diffusion constant in the QLL is substantially slower than in bulk water. This is not a shocking statement to make, as the QLL is a nanoscale thin layer, plus its structure is somewhat constrained by its proximity to the ordered ice lattice [2018Lou]. However, neither theory nor experiments tell us much for certain about QLL diffusion, so we cannot say how this might give us the observed $A_{prism} < 1$ behavior in Figure 4.5, or why there is no similar behavior with $A_{basal}$. Nevertheless, a crude "frustrated QLL kinetics" model (the word model being a bit generous here) may provide at least a qualitative physical picture of the underlying molecular processes.

Figure 4.7 shows a graphical representation of this frustrated-kinetics model, comparing the ice/water case with that for an ice/QLL/vapor interface. Although this sketchy line of reasoning provides us no means to calculate $A_{prism}$, it does agree with at least one recent MD simulation [2018Lou] showing that the near-surface QLL viscosity is up to 1000 times higher than bulk water, and is especially high on the prism facet. Although simulating ice surfaces is not an exact science at present, this result suggests that it may be possible to investigate the frustrated-QLL model further using MD simulations. Moreover, the model immediately makes a significant experimental prediction. Because the mobility must increase with QLL thickness in this model, it predicts that $A_{prism} \to 1$ as $T \to 0$ C. The upward trend in $A_{prism}$ at the highest temperatures in Figure 4.5 largely reflects this CAK model bias, although there is a weak indication that this trend is present in the data. If the upward trend is confirmed in targeted experimental investigations at higher temperatures, this would support the model, crude as it is. Moreover, the frustrated-kinetics model becomes useful once again for interpreting the growth of ice dendrites at high

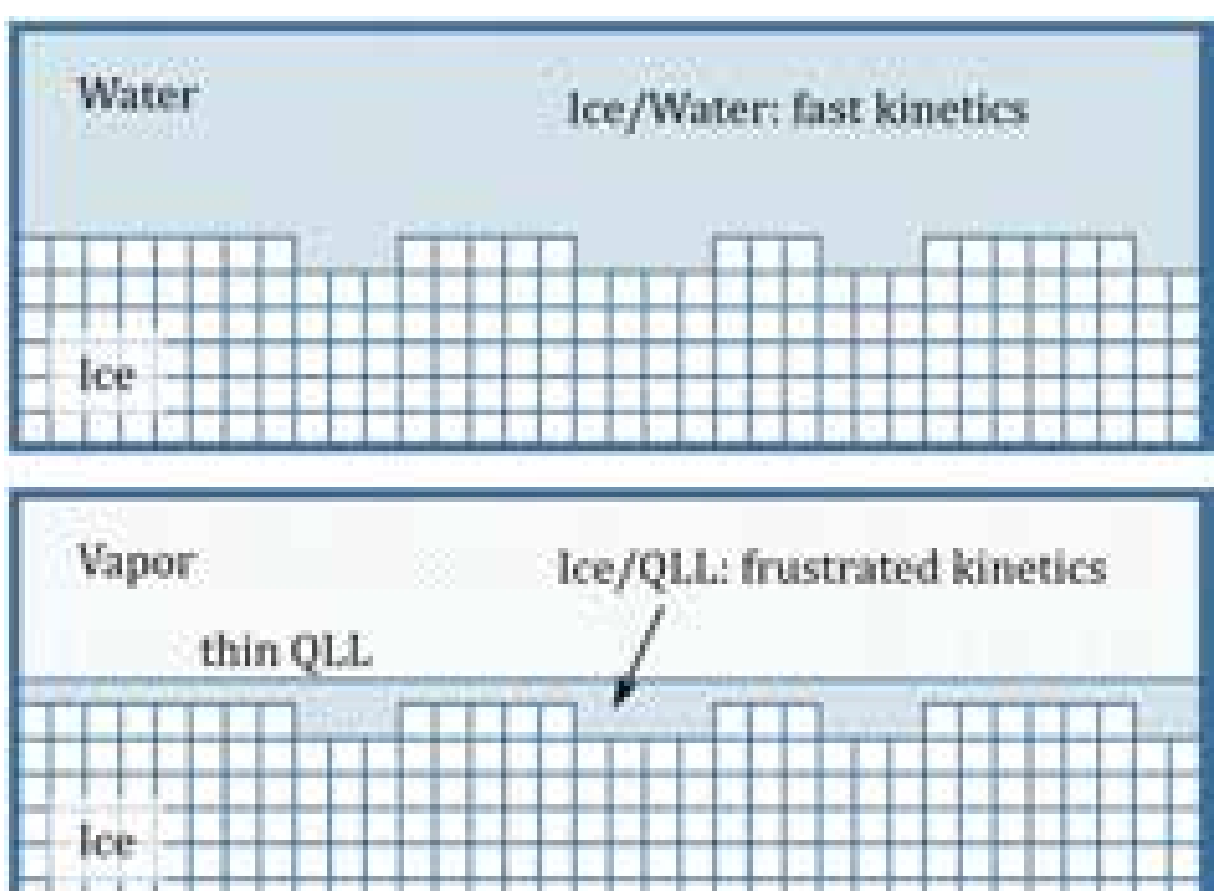

Figure 4.7: A cartoon picture illustrating how the inhibited mobility of QLL molecules might explain the observed $A_{prism} < 1$ behavior seen in Figure 4.5. In the absence of a nucleation barrier (when $\sigma_{surf} \gg \sigma_0$), ice growth is limited by how quickly liquid (or quasi-liquid) molecules can diffusion into position to be incorporated into the ice lattice. The data suggest that perhaps there is a high mobility in the ice/water system (top) but a substantially lower mobility near the ice/QLL interface (bottom).

temperatures, which I discuss further in the SDAK-2 section below.

An alternative explanation for $A_{prism} < 1$ might come from classical terrace-nucleation theory [1996Sai], which indicates that $A_{prism}$ is proportional to the surface diffusion length $x_{diff}$ on prism facets. If $x_{diff}$ is substantially lowered by the onset of surface premelting, this could also explain the observed $A_{prism} < 1$ behavior. However, a reduction in $x_{diff}$ is not altogether dissimilar to a reduction in QLL mobility, so perhaps the overall picture is roughly the same in both explanations. And again, it remains a mystery why $A_{prism}$ is substantially different from $A_{basal}$ at high temperatures. But this behavior can certainly be investigated further with additional targeted experiments and MD simulations [2018Lou, 2020Llo].

We can add an additional telling observation to this story, as prism faceting is readily observed at -0.5 C in air, as illustrated in Figure 4.8. This faceting could not result from a high nucleation barrier, because the growth data indicate that $\sigma_0$ is too low to cause faceting at the estimated $\sigma_{surf}$. Instead, this high-temperature prism faceting appears to arise because $A_{prism} < 1$ at high temperatures, providing a direct, independent confirmation of this feature in the data [1991Elb].

Because Figures 4.5 and 4.6 describe the attachment kinetics on perfect faceted surfaces having infinite lateral extent, the CAK model curves are relatively clean and well-defined. There must exist, for example, definite values for $\beta_{basal}(T)$ and $\beta_{prism}(T)$, just as there are definite values for the facet surface energies $\gamma_{basal}(T)$ and $\gamma_{prism}(T)$. These are all fundamental properties of the ice crystal, determined by the lattice structure and water molecular interactions. The actual values of $\beta_{basal}(T)$ and $\beta_{prism}(T)$ shown in Figure 4.6 are imperfect to some degree, owing to experimental uncertainties, and the step energies may vary slightly with the orientation angles of the steps, step curvature, and other

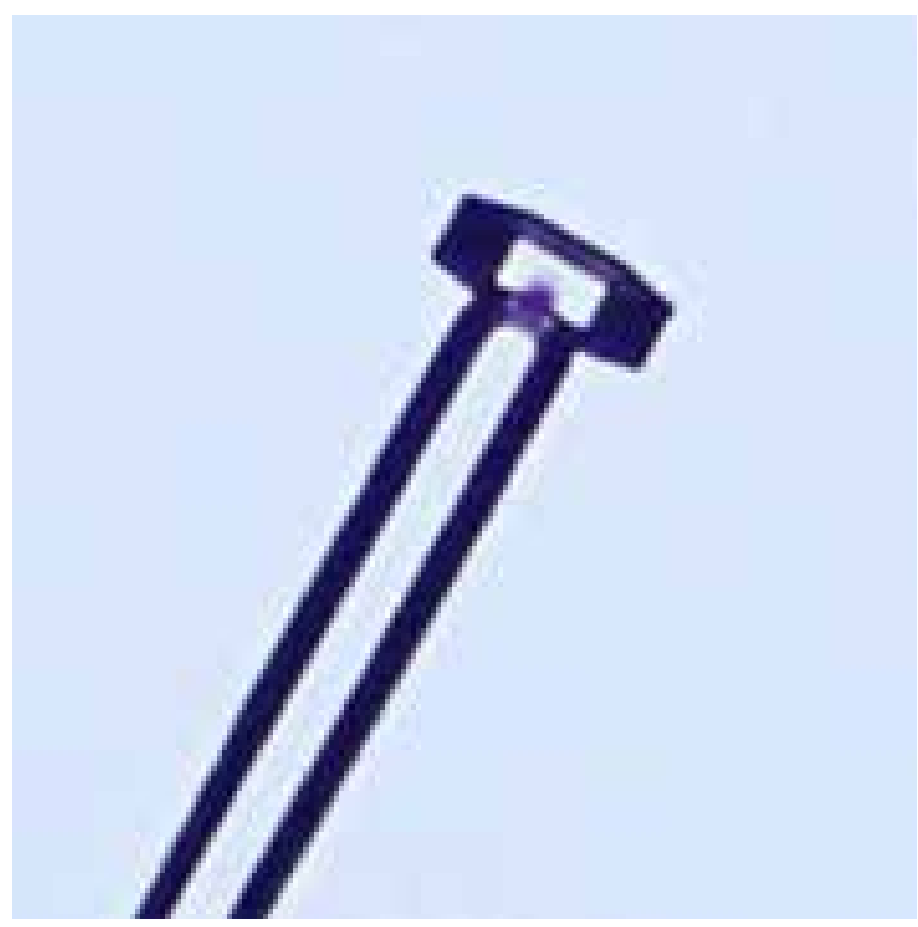

Figure 4.8: This image shows a blocky plate crystal growing on the end of a c-axis electric ice needle (Chapter 8) in air at -0.5 C, with $\sigma_{surf} \approx 0.1$ percent. Because $\sigma_{surf} \gg \sigma_0$ under these conditions, the strong prism faceting did not result from a nucleation barrier but arose because $\alpha_{prism} \approx A_{prism} < 1$ at this temperature, as shown in Figure 4.5. The CAK model proposes a "frustrated QLL kinetics" phenomenon to explain this low $A_{prism}$ behavior, although the underlying molecular physics here is not known with certainty.

minor factors. But well-defined step energies must exist in the defect-free, large-facet limit.

The parameters $A(T)$ and $\sigma_0(T)$ are somewhat model dependent, as they assume that a terrace nucleation model is generally correct for the growth of faceted ice surfaces. But this assumption is quite strongly supported by a fair amount of experimental evidence at this point (Chapter 7), and defect-free crystals seem to be the norm over this temperature range. The somewhat odd behavior of $A_{prism}(T)$ is a complication, but this is a relatively minor issue that mainly affects fast-growing prism surfaces at high temperatures. Even including that regime, however, ice-growth measurements generally tell us that the terrace nucleation model parameterized in Figures 4.5 and 4.6 provides quite a good approximation for the growth of the principal basal and prism faceted surfaces in the large-facet limit.

The large-facet limit, however, is only one aspect of the attachment kinetics. A great variety of snow crystal morphologies, including thin plates, hollow columns, and essentially all dendritic structures, exhibit sharply curved features that contain tiny basal and prism facets with last-terrace dimensions that are often as small as 50 nanometers. Even these minute facets can be important drivers of the overall growth morphology, however, as essentially all non-faceted surfaces exhibit $\alpha \approx \alpha_{roug} \approx 1$. One might imagine that the addition of edge effects or other factors on such small facets could yield attachment kinetics that are quite different compared to large facets. In our quest to understand the full range of snow crystal morphologies, therefore, we must address the attachment kinetics on these diminutive faceted surfaces.

## 4.3 Structure-Dependent Attachment Kinetics

To begin our examination of the growth of small faceted surfaces, consider the edge of a thin hexagonal plate, as illustrated in Figure 4.9. The edge terminates on a narrow prism facet, and, from basic geometry, the width of the last prism terrace on a curved edge is $w \approx \sqrt{8Ra}$, where $R$ is the radius of curvature of the edge and $a$ is the size of a water molecule. Experimental measurements, together with input from solvability theory (Chapter 3) suggest edge and tip radii that are often in the 1-2 μm range for a broad range of snow crystal morphologies growing in normal air. Slower growth generally yields larger $R$, as does growth at lower air pressures. But fast-growing structures in air do not generally develop structures with $R$ values much lower than 1-2 μm. Thus $w$ can be as low as about 50 nanometers on many snow crystal structures, which is less than 200 molecules across. (Although electric ice needles have substantially sharper tips, as described in Chapter 8.)

This lower limit puts us squarely in the mesoscopic regime, as a 200-molecule-wide facet would likely have roughly the same molecular surface structure as a large facet, even in the presence of significant surface premelting. On the other hand, lateral molecular transport from the facet edges might affect the overall growth behavior in comparison to that of a large faceted surface. Theoretical considerations alone cannot tell us whether edge effects will significantly alter the attachment kinetics, but rather basic observations of snow crystal morphologies indicate that the large-facet attachment kinetics described above cannot be the whole story.

Thin-plate crystals like that shown in Figure 4.9 are common near -15 C, and their very existence requires $\alpha_{prism} \gg \alpha_{basal}$. Basic numerical modeling further indicates that the large aspect ratio of a thin-plate crystal cannot be a simple diffusion effect but must be caused by highly anisotropic attachment kinetics (Chapter 3). The large-facet curves in Figure 4.5, however, clearly indicate that $\alpha_{prism} \approx \alpha_{basal}$ at -15 C, thereby precluding the formation of thin plates. This discrepancy is immediately problematic because our large-facet attachment kinetics model is largely empirical, supported by quite solid experimental measurements. Either something is very wrong with the large-facet measurements, or small facets have a different story to tell. Delving into this discrepancy over some years, I reached the conclusion that the facet attachment kinetics must depend strongly on the local mesoscopic crystal structure, specifically the facet width $w$, a phenomenon I call *Structure-Dependent Attachment Kinetics* (SDAK) [2003Lib1].

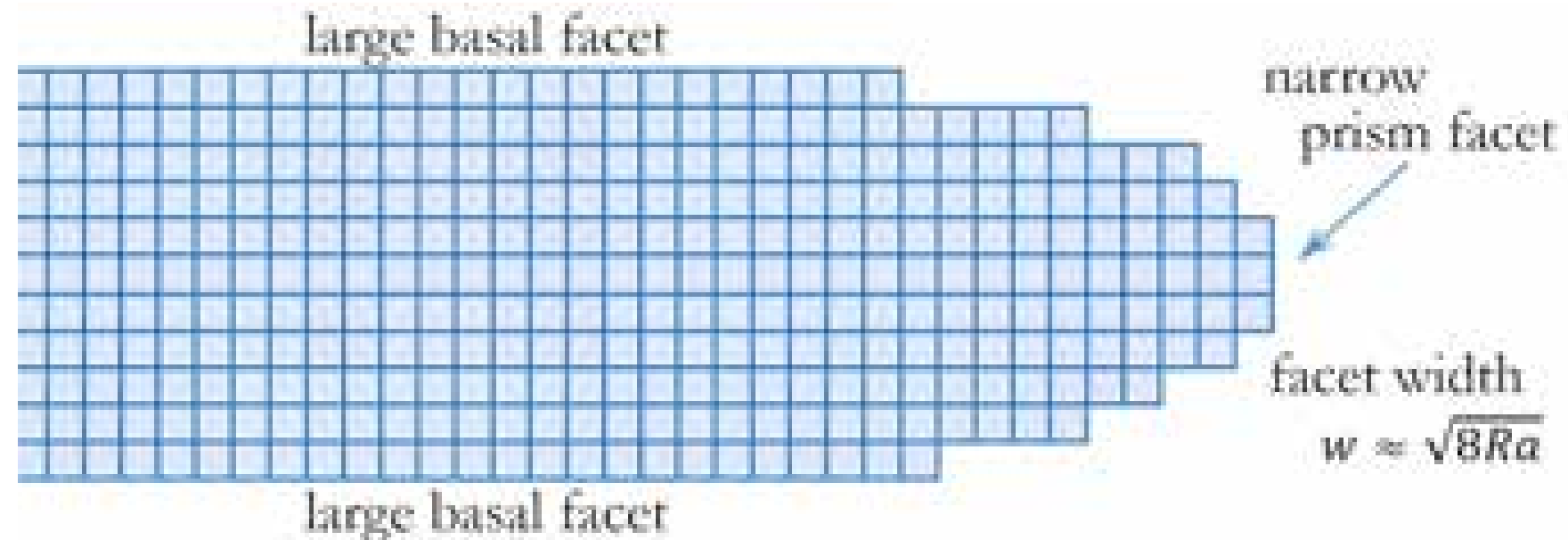

Figure 4.9: The shape of a thin hexagonal platelike crystal is largely defined by the growth of the basal and prism facets. While the basal facets are quite large, the six prism facets have widths $w \approx \sqrt{8Ra}$, where $R$ is the radius of curvature of the edge and $a$ is the molecular size. In many common circumstances, the attachment kinetics on these extremely narrow facets is markedly different from the attachment kinetics on large faceted surfaces.

Supposing that this SDAK hypothesis is correct, our next challenge is to understand the physical origin of a facet-width-dependent attachment kinetics. Ice surface physics is already challenging in the large-facet limit, and it does not become simpler when any number of edge-related physical effects may come into play. While this might be a good point to throw up our hands and declare that no clear solution is possible, such a declaration provides little insight into possible paths forward. Instead I will plunge ahead and describe two possible mechanisms I have developed for SDAK effects. Both have merits in different growth regimes, but neither is solidly rooted in well-understood molecular physics. These models should therefore be considered as hypotheses to be tested with additional experimental and theoretical investigations. I have been going down this path of hypothesis testing recently using numerous precision ice-growth measurements, and, so far anyway, the results are quite promising. But a long road lies ahead to fully develop and verify these SDAK models.

## SDAK-1: Enhanced Terrace Nucleation

As a first venture into developing an SDAK model, let us consider more carefully the structure and growth of a prism edge on a thin platelike crystal. If we establish the thin-plate morphology and then turn down the supersaturation, so the crystal sits in near-equilibrium conditions, then Figure 4.10a illustrates the resulting rounded edge structure. The value of $R_{edge}$ is set by the initial conditions, and the discussion in the previous several paragraphs suggests that 1-2 μm is a typical value for an especially thin edge. If we set $\sigma_\infty \approx \sigma_{surf} \approx 2d_{sv}/R_{edge}$ around this crystal, then the Gibbs-Thomson effect provides that the edge of the plate will neither grow nor shrink with time. If we further assume a large nucleation barrier having $\sigma_{0,basal} \gg \sigma_{surf}$, then the basal facets will remain static as well. The resulting lack of any significant growth is consistent with our assumption of $\sigma_\infty \approx \sigma_{surf}$ in this near-equilibrium situation.

The shape of the edge will be rounded, as shown in Figure 4.10a, as this minimizes the local surface energy. And, again from Chapter 2, the time needed for the edge to reach this rounded state is just a few seconds in air. Eventually, the overall platelike shape would evolve toward the (roughly spherical) equilibrium crystal shape, but the time required for this relaxation is extremely long. Thus, Figure 4.10a represents a quasi-static, near-equilibrium condition.

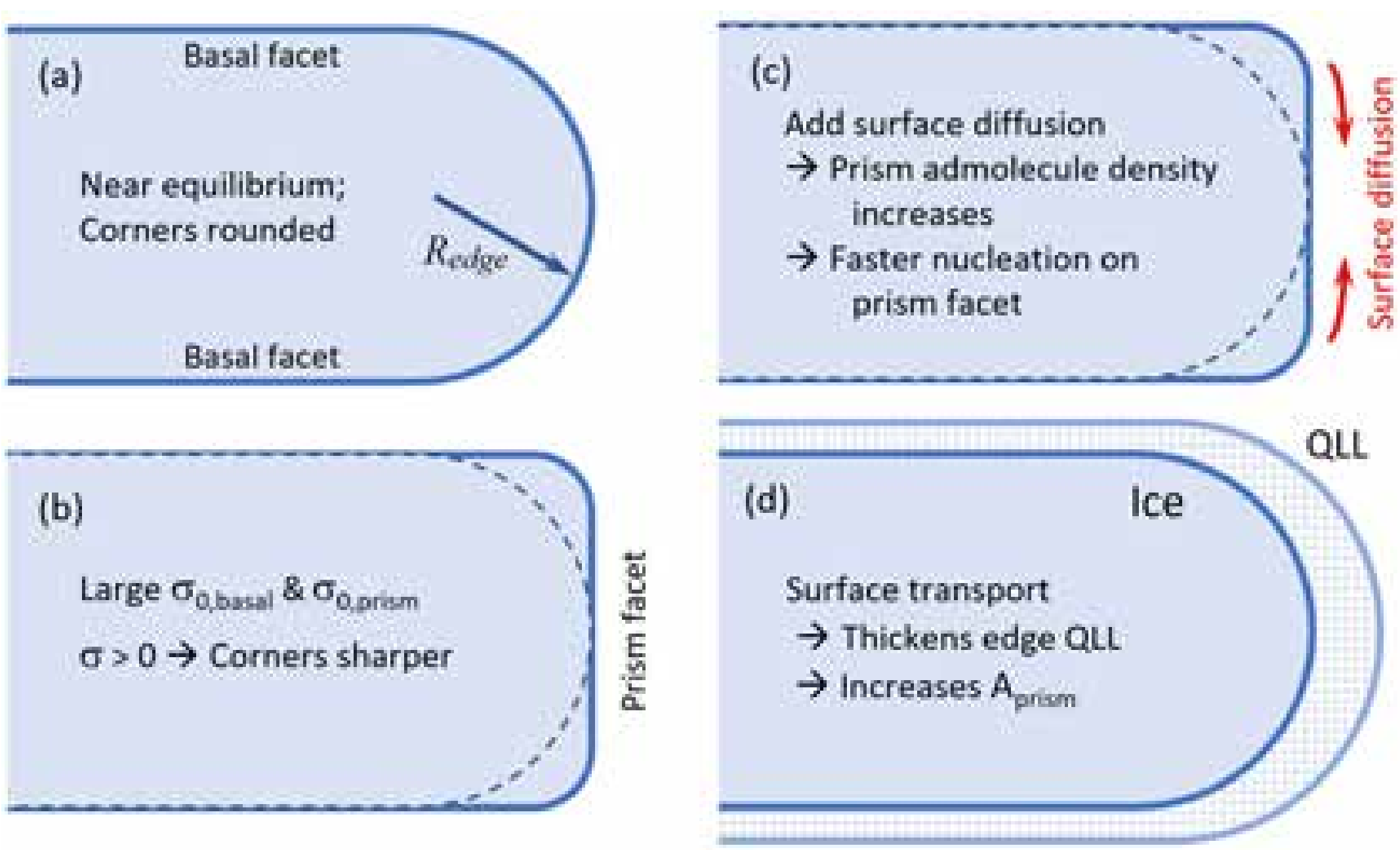

Figure 4.10: A possible physical model of an SDAK phenomenon on the thin prism edge of a platelike snow crystal [2019Lib1]. As described in detail in the text, a surface-energy-driven surface diffusion can increase the admolecule density on the narrow prism facet, enhancing the normal rate of terrace nucleation. As a result, the attachment kinetics on the narrow prism facet is different from that on a large prism facet, which is the essence of the SDAK phenomenon.

Next let us increase the supersaturation, as illustrated in Figure 4.10b. Assume that the nucleation barriers are still quite large on both facets, so they experience little growth. The only significant change is that the Gibb-Thomson effect now yields a new quasi-equilibrium state in which the corner radii are smaller than $R_{edge}$. This means that the increased supersaturation now prevents the plate edge from assuming its rounded, near-equilibrium shape. Thus, the high applied supersaturation constitutes a nonequilibrium condition that then supports a static, nonequilibrium edge shape, in this case one with sharpened corners.

For our final step, illustrated in Figure 4.10c, we now introduce the possibility of surface diffusion onto the prism facet. The nonequilibrium edge shape means that a transfer of molecules from the sharpened corners to the prism facet (arrows) will reduce the total surface energy of the system. Thus, if significant surface diffusion is allowed, there will be an energetically driven flow of water molecules onto the prism surface. These additional admolecules will not immediately attach to the prism surface, as there is still a nucleation barrier to overcome. But they will increase the overall admolecule surface density, and this disturbs the normal deposition/sublimation balance on the surface. With a greater admolecule surface density, terrace nucleation is statistically more likely, and this results in a lowering of the effective nucleation barrier, and thus a higher $\alpha_{prism}$. With some additional reasonable assumptions [2019Lib1], the growth increase behaves essentially as if $\beta_{prism}$ were lower. Note that the real $\beta_{prism}$ is unchanged, as the lattice structure of the facet is essentially equal to that of a large facet. But the nucleation rate goes up nevertheless, behaving much like a surface with a lower $\beta_{prism}$. As a result, because of this facet-edge effect, the effective $\sigma_{0,prism}$ goes down even when $\beta_{prism}$ remains unchanged. This phenomenon cannot occur on large facets, because large facets, by definition, are not influenced by edge effects.

This picture also suggests that the increase in admolecule surface density will be more pronounced as the facet width $w$ decreases, because the additional admolecules will quickly disperse over the available top-terrace surface. The narrower the facet surface, the higher the resulting admolecule density. Once again, this cartoon picture does not allow us to calculate the SDAK effect in any detail, as this is a complex surface-dynamical process. But it does suggest at least a plausible molecular mechanism for an SDAK phenomenon.

Notably, this SDAK mechanism cannot yield a value of $\alpha_{prism}$ above unity. As the nucleation barrier drops and $\alpha_{prism} \rightarrow 1$, the Gibbs-Thomson effect will no longer support the non-equilibrium edge shape with sharpened corners. This only happens when the prism nucleation barrier is significant enough to suppress prism growth. Once $\alpha_{prism} \approx 1$, the edge reverts to an overall rounded shape that no longer provides an energetically driven surface flow. In a nutshell, this SDAK mechanism effectively lowers $\sigma_{0,prism}$ until $\alpha_{prism} \rightarrow 1$, but it cannot yield $\alpha_{prism} > 1$.

## Surface Premelting and the Ehrlich-Schwoebel barrier

Continuing with this line of reasoning, our next question is when to expect a significant amount of corner-to-facet surface diffusion. With a traditional rigid-lattice surface structure, this type of edge-crossing surface diffusion is suppressed by the Ehrlich-Schwoebel (E-S) barrier (Figure 4.11), a phenomenon that is commonly discussed in crystal-growth textbooks [1996Sai, 1998Pim, 2002Mut]. From Chapter 2, the ice surface tends to resemble this rigid structure at especially low temperatures, when surface premelting is absent. We expect, therefore, that surface diffusion like that shown in Figure 4.10c would be effectively suppressed by the E-S barrier at low temperatures. In this case, terrace nucleation on narrow facets would be no different than on large facets.

As the temperature increases, however, surface premelting eventually begins to develop, increasing the amount of disorder in the top molecular layers of the crystal surface. When the temperature reaches some critical value $T_{onset}$, when the surface disorder is sufficiently large, it stands to reason that the Ehrlich-Schwoebel barrier will become quite leaky. At this temperature, I have proposed that the above SDAK mechanism becomes important, effectively lowering the nucleation barrier and thereby increasing $\alpha_{prism}$ [2019Lib1].

At still higher temperatures, surface diffusion becomes even more prevalent and a thick QLL develops atop the crystalline lattice,

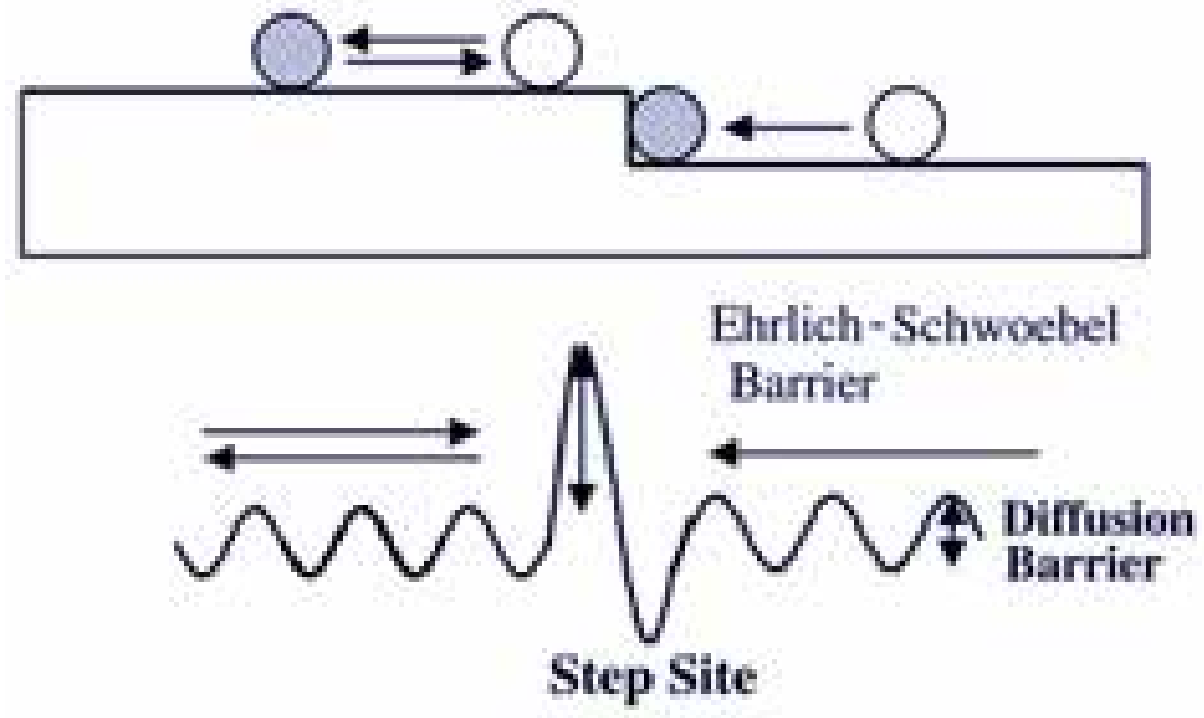

Figure 4.11: On a rigid lattice, admolecule diffusion over abrupt terrace steps is suppressed by the Ehrlich-Schwoebel barrier. An admolecule crossing over the step edge will (temporarily) be in a position nearly removed from the surface, which is energetically quite unfavorable. This creates a potential barrier that impedes such crossings. The step structure will become less rigid in the presence of surface premelting, however, which should substantially reduce the Ehrlich-Schwoebel barrier. Image from [2002He].

as illustrated in Figure 10d. In this regime, the ice/vapor terrace nucleation model is replaced with an ice/QLL model that starts to resemble terrace nucleation at an ice/water interface. The concept of an admolecule surface density begins to lose its meaning at the ice/QLL interface, as the molecular dynamics is more influenced by QLL bulk diffusion, as in the Wilson-Frenkel picture discussed above. Although there should be plenty of surface diffusion at temperatures well above $T_{onset}$, when the E-S barrier is essentially absent, this additional transport will have a relatively minor effect on the nucleation barrier. Thus, at temperatures substantially above $T_{onset}$, we again expect that terrace nucleation on small facets will be little changed from the large-facet model.

Putting all this together, this (admittedly inexact) model predicts the existence of an "SDAK dip" on each of the large-facet $\sigma_0$ curves, as illustrated in Figure 4.12. Here I have chosen different values of $T_{onset}$ for the basal and prism facets to impose agreement with the known ice-growth behaviors from the Nakaya diagram and other growth measurements. Moreover, the values of the SDAK curves must depend on $w$ and perhaps other factors, as the very nature of the SDAK hypothesis is that the attachment kinetics depends on the mesoscopic crystal structure. The specific curves shown in Figure 4.12 represent rough estimates for $R_{edge}$ in 1-2 μm range, as these are typical for snow-crystal growth in air, as discussed above. These portions of the curves are drawn as dotted lines to signify their substantial uncertainties.

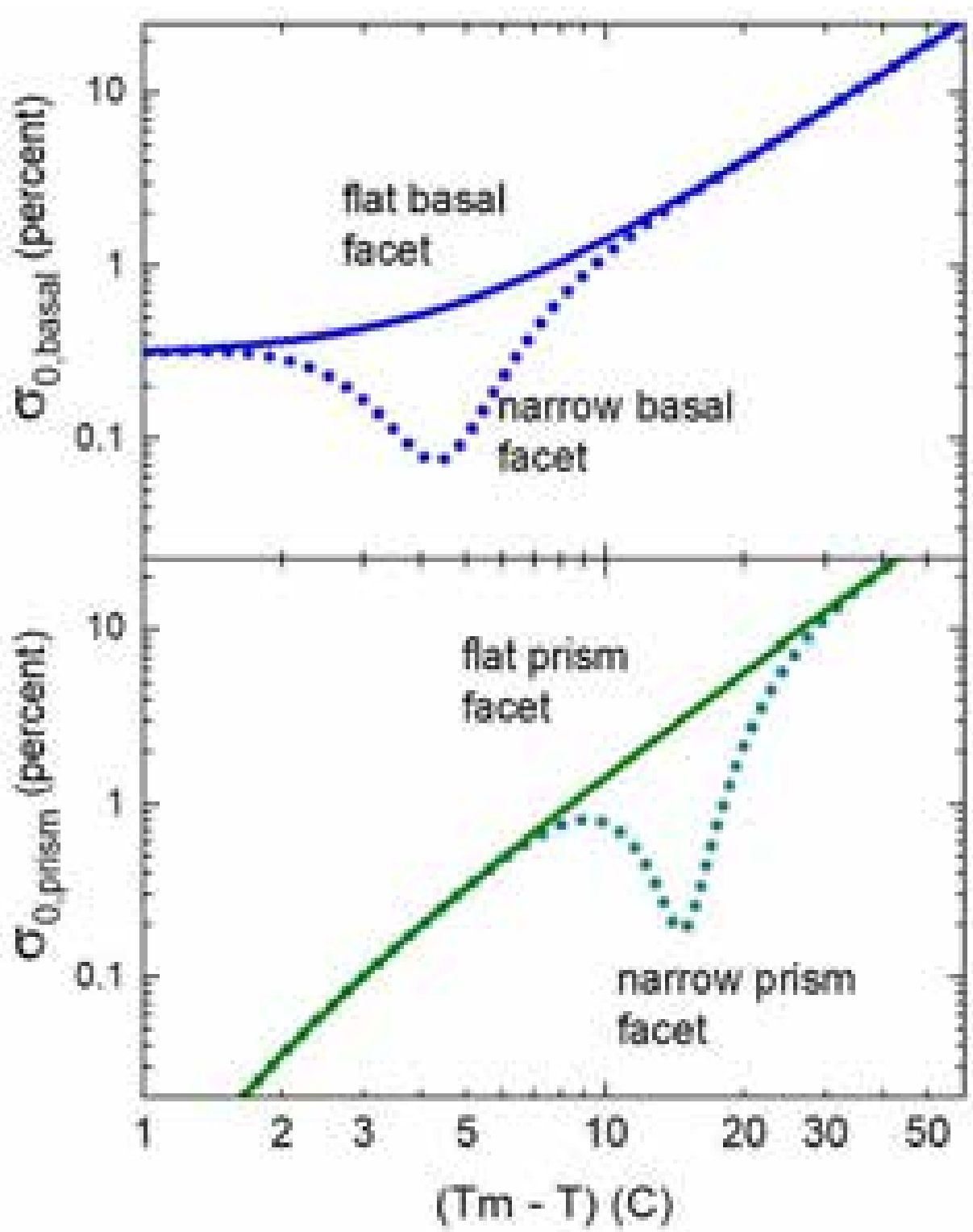

Figure 4.12: The solid lines in these two graphs are reproduced from Figure 4.5, again illustrating the CAK terrace-nucleation model for the attachment kinetics on large basal and prism facets. The dotted lines show the "SDAK dips" described in the text. Near -5 C, the narrow basal facets on hollow columns exhibit a much reduced effective $\sigma_{0,basal}$, explaining the rapid growth of these features. Near -15 C, narrow prism facets exhibit a similarly low effective $\sigma_{0,prism}$, explaining the growth of thin plates at that temperature.

Although the SDAK model is certainly wanting in that it does not make precise calculable predictions for $\alpha_{basal}$ and $\alpha_{prism}$, it nevertheless makes some surprisingly concrete predictions. For example, the CAK model with the SDAK phenomenon indicates that thick plates should be the primary habit at -5 C when the supersaturation is low, because $\alpha_{prism} > \alpha_{basal}$ on large facets at that temperature. But hollow columns can also grow at -5 C if the basal edges are thin, owing to the SDAK effect. Indeed, as described below, a close look at a broad range of ice-growth data indicates that thick plates do readily form at -5 C, as do hollow columns, and the two forms can even grow concurrently under certain conditions [2012Kni, 2019Lib2].

Plotting the two narrow-facet curves together yields the result in Figure 4.13, revealing the full impact of this SDAK modeling exercise. With suitable choices for $T_{onset}$ on the basal and prism facets, we have created a set of model curves that reproduces the overall temperature dependence seen in the Nakaya diagram. According to this CAK model, the various transitions between platelike and columnar growth are caused largely by changes in the effective nucleation barriers on *narrow* basal and prism facets. If correct, the SDAK phenomenon thus plays a crucial role in defining the long-mysterious temperature transitions observed in snow crystal morphology.

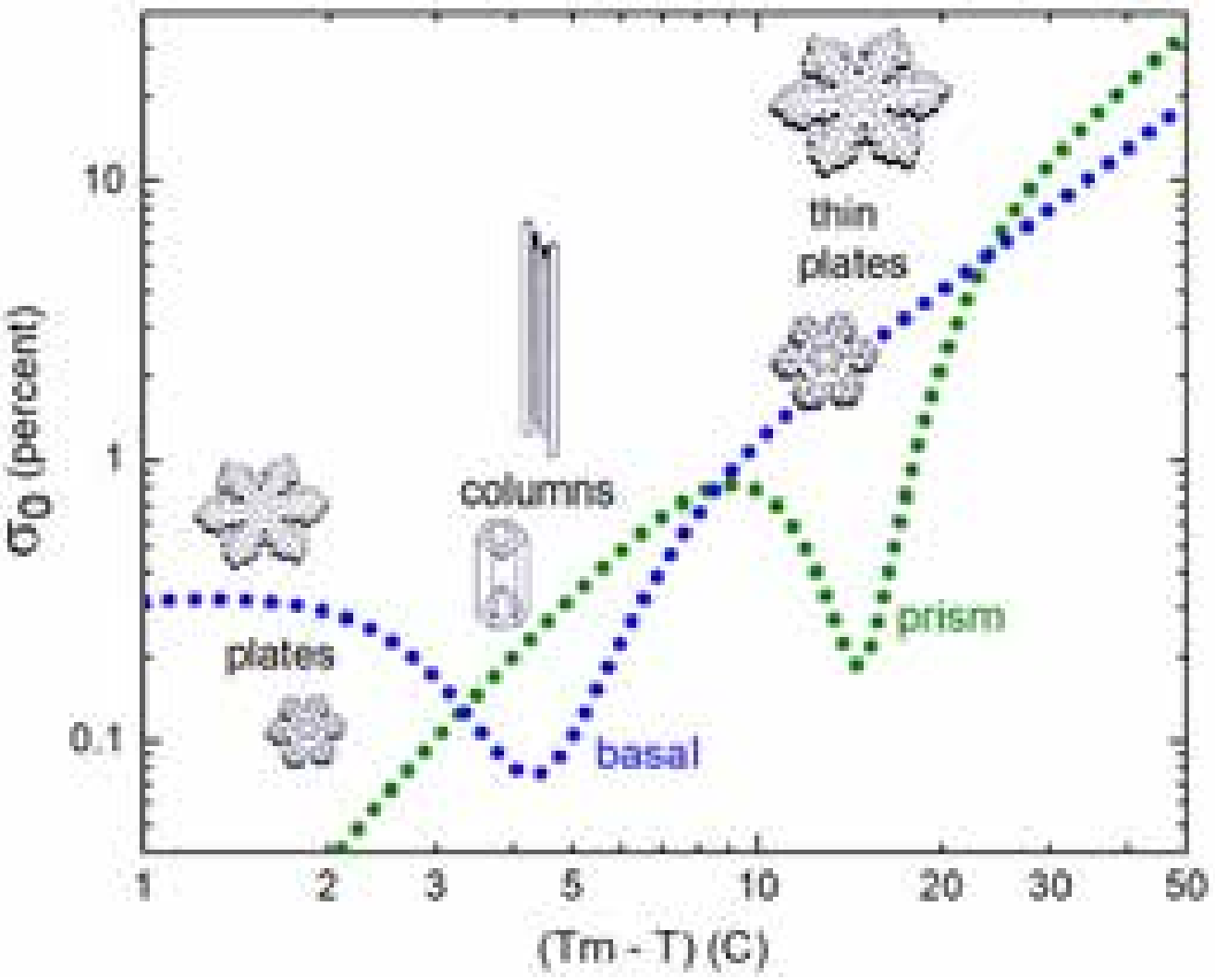

Figure 4.13: Plotting the SDAK curves from Figure 4.12 together, we can compare the basal and prism nucleation barriers as a function of temperature. A lower nucleation barrier (i.e., a lower $\sigma_0$) means faster attachment kinetics on the corresponding facet surface. This demonstrates that the CAK model exhibits the same transitions between platelike and columnar growth that are seen in the Nakaya diagram (Chapter 1). Beyond this qualitative agreement with morphologies, however, the CAK model also provides good quantitative agreement with many precision ice-growth measurements [2019Lib2, 2020Lib].

Of course, the critical reader may balk at the speculative nature of this model, along with its complexity and ill-defined curves. Fair enough. What the reader may not immediately see, however, is that the model is largely driven by empirical data from a variety of ice-growth measurements. The large-facet curves are well grounded in experiments, and the SDAK dips, or something similarly complex, are required to explain a substantial body of additional data. Creating a comprehensive model of the attachment kinetics that can explain the Nakaya diagram, along with a variety of additional ice-growth measurements, is a challenging task that requires a complex solution. As I will discuss below, there are essentially no other viable alternatives to the CAK model at present, at least not if one requires a reasonable consistency with recent experimental data. Although "something else" is always a viable alternative hypothesis, the CAK model, including the SDAK effect, provides at least a plausible working model of the attachment kinetics over a broad range of conditions. Moreover, the CAK model makes abundant quantifiable predictions regarding growth rates and morphological behaviors that can be tested with straightforward experimental investigations, as I describe below.

## SDAK-2: Enhanced QLL Kinetics

The SDAK mechanism illustrated in Figure 4.10 suggests an additional consequence of enhanced surface diffusion that might change the attachment kinetics at high temperatures. When the QLL is well developed, the transport of water molecules from the facet corners to narrow prism facets could become great enough to increase the nominal thickness of the QLL on the prism surfaces. If so, then the increased QLL thickness should tend to alleviate the frustrated QLL kinetics described in the large-facet model above. This additional mechanism – call it SDAK-2 – will thus increase $A_{prism}$ on small prism facets, as illustrated in Figure 4.14. Here again, the SDAK-2 curve is drawn as a dotted line, signifying that its exact location is not well defined, and it will change with $w$ and $\sigma_{surf}$. The model mainly makes the qualitative prediction that $A_{prism} \rightarrow 1$ on especially small, fast-growing prism facets on the ends of sharply tipped dendritic structures. The SDAK-2 phenomenon thus reconciles $A_{prism} < 1$ seen in large-facet growth with additional experimental evidence pointing to $A_{prism} \rightarrow 1$ on narrow facets [2019Lib2]. There are perhaps other physical mechanisms that could produce a similar growth behavior, but this model provides a ready explanation for $A_{prism}$ being especially large on the smallest prism facets.

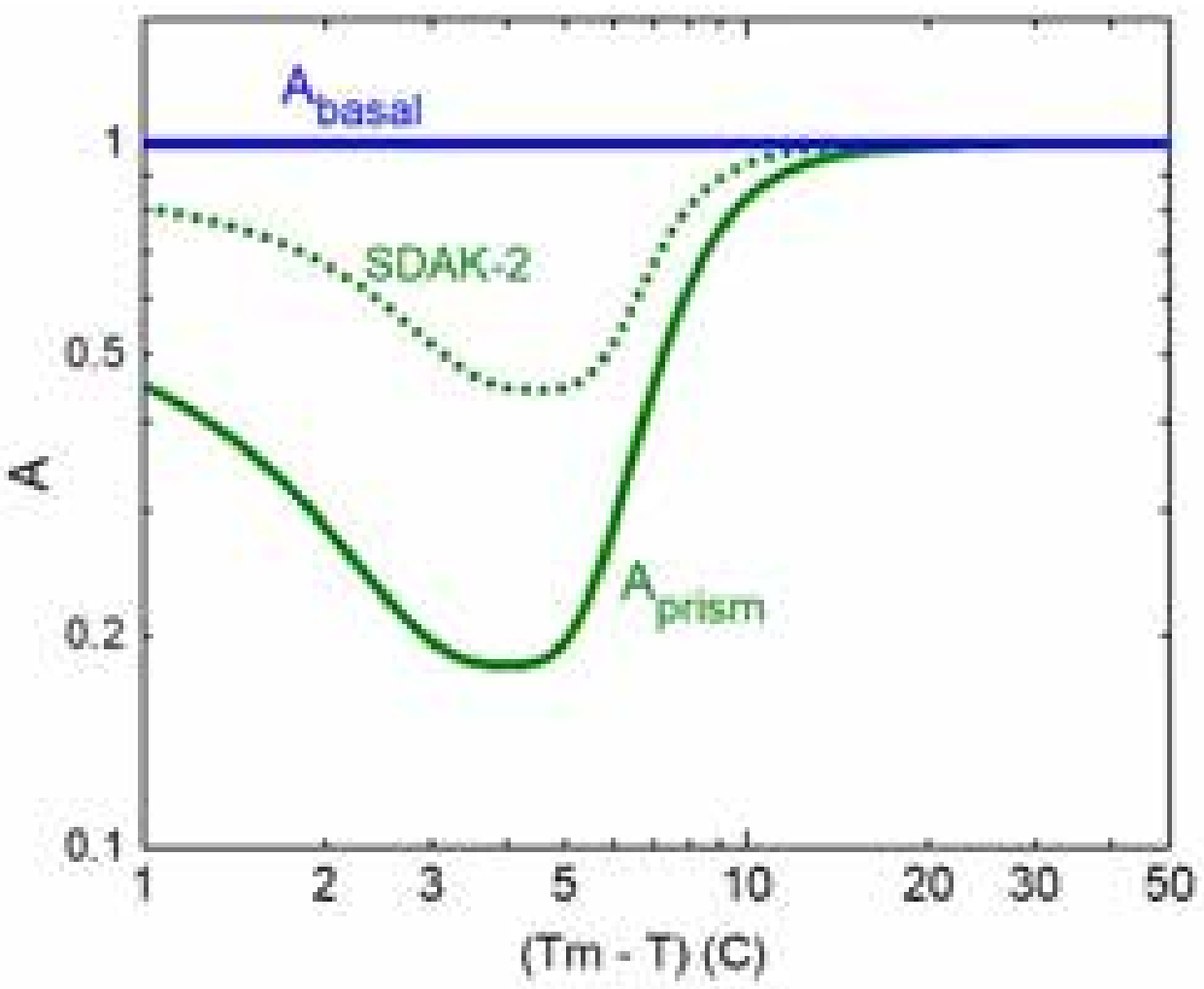

Figure 4.14: The SDAK-2 mechanism increases the value of $A_{prism}$ on small prism facets at temperatures above -10 C, as approximated here by the dotted curve. This curve is expected to depend on supersaturation, such that $A_{prism} \rightarrow 1$ at especially high $\sigma_{surf}$. The solid curves are reproduced from Figure 4.5. This change in $A_{prism}$ mainly affects the growth of sharp-tipped dendritic structures, which are often observed in high-supersaturation experiments.

Note that the SDAK-2 effect only becomes important on small prism facets at high growth rates, and it has little effect on the overall morphology diagram shown in Figure 4.13. As a result, SDAK-2 is overall less important than SDAK-1, as the former is mostly needed to reproduce the observed growth of sharply tipped dendritic structures at temperatures above -10 C. This is an easily studied region, however, as such crystals grow reliably in air at high supersaturations.

## The Edge-Sharpening Instability

A particularly fascinating consequence of the SDAK hypothesis is a phenomenon I call the *Edge-Sharpening Instability* (ESI), which is illustrated in Figure 4.15. As the name implies, the ESI tends to sharpen basal and prism edges via a positive feedback effect that takes place during diffusion-limited growth. I believe that this growth instability is largely responsible for the remarkably robust appearance of thin plates near -15 C and the similarly robust formation of hollow columns near -5 C. If this addition to the CAK model is correct, then the ESI is one of the most important physical processes shaping the growth of atmospheric snow crystals.

Looking at thin-plate growth near -15 C in Figure 4.15 (shown as a plate-on-pedestal morphology), the essential starting point for the ESI is the hypothesis that $\alpha_{prism}$ on an edge-like prism facet depends strongly on the width of the top prism terrace (the SDAK hypothesis). For a broad facet, $\alpha_{prism}$ is well described by the nucleation-limited model with the parameters given in Figure 4.5. On a narrow prism edge, $\alpha_{prism} \rightarrow 1$ as $w \rightarrow 0$, where $w$ is the edge width. The SDAK mechanism described above would bring about this behavior, but the precise physical mechanism is not important, as long as it operates over a narrow range of temperatures near -15 C.

If the SDAK hypothesis is correct, then diffusion-limited growth naturally brings about the positive-feedback effect illustrated in Figure 4.15. For example, when the applied supersaturation $\sigma_\infty$ is low and the facet growth rates are corresponding low, then stable facets form, as shown in the middle sketch in the figure. On both the basal and prism facets, new terraces mostly nucleate near the exposed corner, where the supersaturation is highest owing to diffusion effects. This yields trains of basal and prism terrace steps propagating from the corners to the facet centers. The facet surfaces both becomes slightly concave in the process, and this overall growth morphology describes stable, faceted crystal growth (Chapter 3).

As $\sigma_\infty$ is increased, the crystal grows faster as terraces nucleate more readily at the corners,

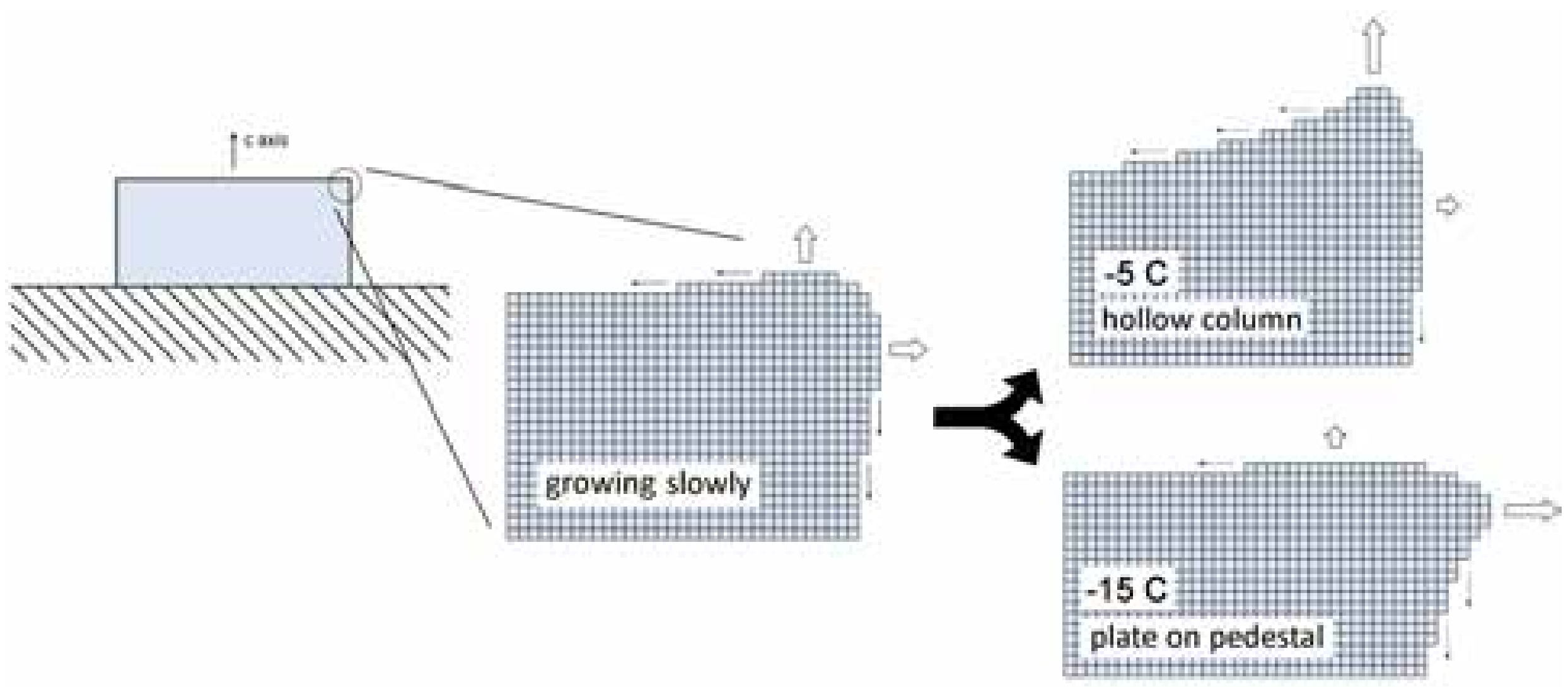

Figure 4.15: The Edge-Sharpening Instability (ESI). This sketch illustrates how the SDAK phenomenon can bring about an edge-sharpening growth instability on either the basal or prism facets. As described in detail in the text, as the uppermost terrace becomes narrower, the attachment coefficient on that terrace can increase via the SDAK mechanism, thus increasing the growth and further narrowing the upper terrace. This brings about a positive feedback effect, and thus a growth instability, that promotes the formation of sharp edges, either as hollow columns (upper right sketch) or thin plates (lower right sketch). In the comprehensive attachment kinetics (CAK) model presented in this chapter, the ESI mechanism in air is largely responsible for the formation of thin plates at -15 C and hollow columns at -5 C.

and the facets become more concave. The faster growth means that the terrace steps become more closely spaced and the width of the uppermost basal and prism terraces becomes smaller. Because $\sigma_{0,prism}$ decreases as the top prism terrace width decreases (our hypothesized SDAK behavior), the nucleation rate increases as well, so more prism terraces appear, and thus the width of the top prism terrace decreases further still. The result is a growth instability brought about by a positive feedback effect – the prism edge sharpens, $\sigma_0$ decreases, the edge sharpens more, $\sigma_0$ decreases more, etc. Above some threshold value of $\sigma_\infty$, the process runs away and a thin plate forms atop the initial blocky crystal. This is the ESI, which I exploit to create the Plate-on-Pedestal laboratory snow crystals described in Chapter 9.

Note that the ESI is *not* equivalent to the normal Mullins-Sekerka instability (MSI) that is a part of diffusion-limited growth (Chapter 3). Rather, the ESI can be thought of as an extension of the MSI that comes about when the SDAK phenomenon is incorporated into the overall growth process. The MSI by itself (without SDAK) will certainly cause branching that results in the formation of complex dendritic structures. But diffusion-limited growth and the MSI alone will *not* produce morphologies with extreme aspect ratios, like thin plates. The formation of thin plates requires highly anisotropic attachment kinetics, specifically $\alpha_{prism}/\alpha_{basal} \gg 1$, which results from the SDAK phenomenon. The underlying physics is both complicated and somewhat subtle because several physical processes are all happening simultaneously, but I believe that the ESI is the best explanation for the robust formation of thin plates and hollow columns in atmospheric snow crystals.

Being a growth instability, the ESI helps explain the somewhat extreme morphological features one often observes in natural snow crystals, as illustrated in Figure 4.16. As the first image illustrates, it is not uncommon to find capped columns with surprisingly thin endplates, suggesting an abrupt transition from columnar to platelike growth. These crystals likely experience gradually changing environmental conditions in the clouds, so one might expect more gradual morphological transitions. But the presence of a bone fide growth instability can handily explain these abrupt transitions, as the ESI naturally brings about rapid changes in $\alpha_{prism}/\alpha_{basal}$, which the MSI cannot accomplish by itself. Moreover, once the transition from columnar to platelike growth has occurred, the SDAK mechanism keeps $\alpha_{prism}$ high on the plate edges, reinforcing the development of thin plates on the capped columnar crystal.

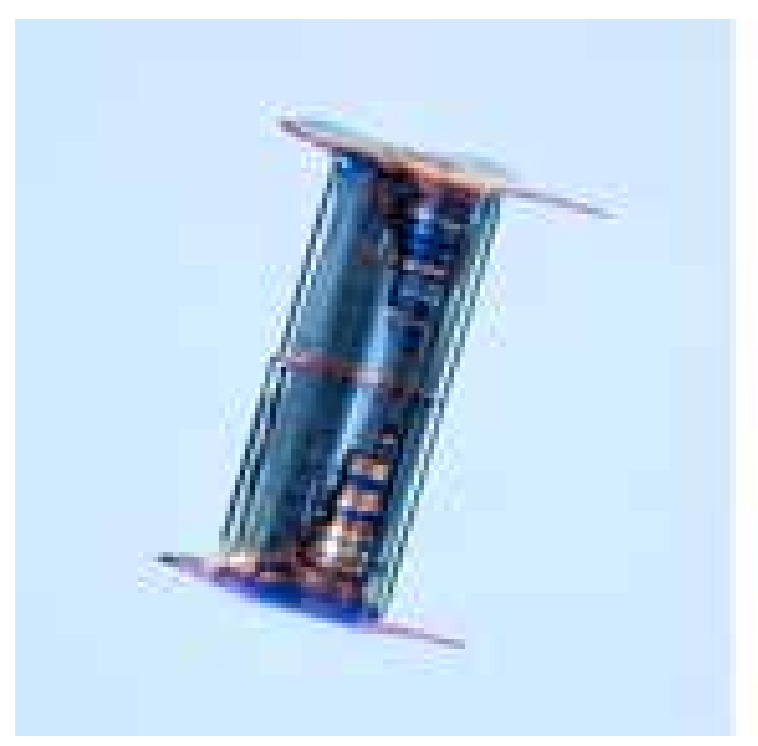

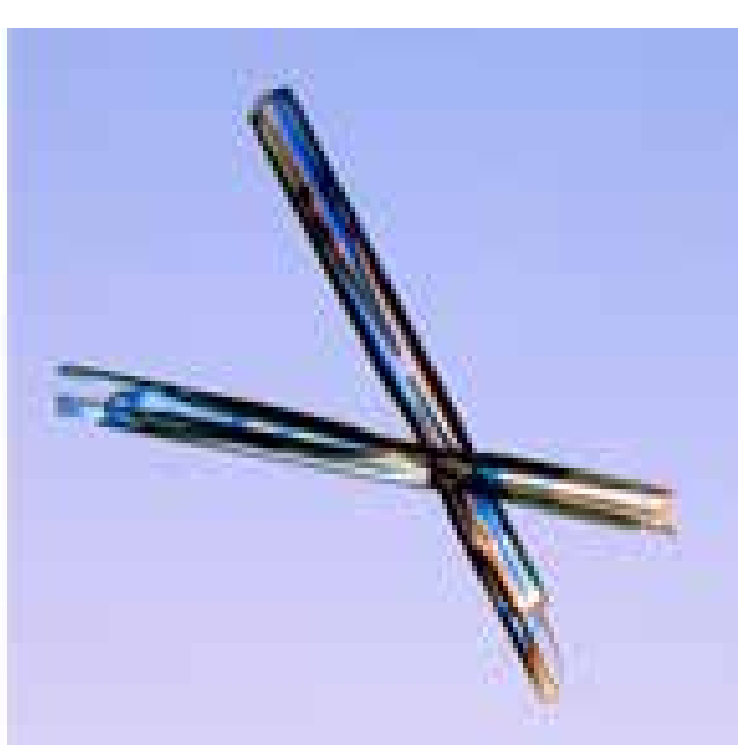

Figure 4.16: The Edge-Sharpening Instability (ESI) can explain many distinctive morphological features in natural snow crystals. For example, the oft-abrupt transition from columnar to platelike growth seen in many capped columns (left) suggests the ESI operating on the prism surfaces near -15 C. Similarly, the remarkably thin, sheath-like edges on many hollow columns (right) can arise when the ESI operates on basal surfaces near -5 C. It is difficult to explain these commonly observed snow-crystal features without invoking the ESI mechanism.

In the second image, the ESI also provides a ready explanation for the formation of thin-walled hollow columns near -5 C. These crystals appear over a narrow range of supersaturations, as $\sigma_{\infty}$ must be high enough to initiate the ESI mechanism, but not so high that the sheath-like edges split to yield a complex needle cluster. Note also that natural snow crystals can be a bit tricky to interpret, as the final morphology of a given crystal depends on its entire growth history. In the case of hollow columns, sheath-like crystal edges may develop a thicker character if the supersaturation later decreases, thus complicating the final interpretation. Laboratory observations, therefore, are generally better suited to examining the dynamics of the ESI under controlled conditions.

As a growth instability, the ESI naturally yields quite thin basal and prism edges in air, as are often observed. Once the instability kicks in, it often does not stop until the edge thickness reduces to a width of 1-2 microns. As a result, this edge thickness appears over a broad range of growth conditions, including thin-plate crystals, thin-walled hollow columns, and fast-growing dendritic structures. Solvability theory (Chapter 3) provides some insights as to why this particular size scale is so prevalent but computational modeling in tandem with experimental observations under controlled conditions will be needed to investigate this phenomenon in detail.

The nature of the ESI mechanism also suggests that interesting effects should be found when growing complex snow crystal structures as a function of background gas pressure. Because the ESI couples the attachment kinetics to diffusion-limited growth, this opens a new dimension in the exploration of structures that are both faceted plus branched. Takehiko Gonda found that both structural complexity and overall aspect ratios were pressure-dependent [1976Gon], but it appears that the supersaturation may have also varied with applied pressure in that work, making it difficult to obtain a quantitative interpretation of the observations. Controlling the temperature, supersaturation, and background gas pressure all simultaneously is challenging in the laboratory but exploring the ESI over a broad range of conditions could yield many interesting insights regarding our general understanding and modeling of snow crystal growth.

### Parameterized Kinetics

While the large-facet attachment kinetics can be examined by measuring the growth of simple ice prisms (Chapter 7), observations of complex snow crystal structures will be needed to investigate the SDAK and ESI phenomena. These can be readily grown in air on the tips of electric ice needles (Chapter 8) but analyzing these complex structures will be challenging. Simple platelike or hollow-column structures can be approximated by cylindrically symmetrical forms, but complex dendritic crystals will require full 3D computational modeling (Chapter 5). In either case, parameterized expressions describing the attachment kinetics will be needed as model inputs.

As described above, the large-facet attachment kinetics can be parameterized by Equation 4.4, as this functional form provides good agreement with experimental measurements. Including SDAK effects necessitates parameterizing the attachment coefficient as $\alpha(\sigma_{surf}, T, w)$ or $\alpha(\sigma_{surf}, T, \kappa)$, where $w$ is the width of the top terrace and $\kappa$ is the surface curvature. Note that $w$ and $\kappa$ are related, so the choice of independent variable is largely a matter of computational convenience. Unlike with terrace nucleation, however, theory does not provide any clear functional form for the SDAK effect.

Because the ESI tends to yield structures with tip radii in the 1-2 μm range in air, the precise functional form for the SDAK effect may not be essential for producing realistic computational models. As long as the chosen $\alpha(\sigma_{surf}, T, w)$ results in a suitable growth instability, the details may not matter as much as the attachment coefficients on the narrow 1-2 μm tips and edges. For example, in [2015Lib2] we used the functional form $\sigma_0 = \sigma_{0,\infty}[1 - \exp(-w/w_0)]$, where $\sigma_{0,\infty}$ is the broad-facet value and $w_0$ is an adjustable model parameter. This allowed us to reproduce the ESI transition to platelike growth at -15 C reasonably well, as the behavior that mattered most was having $\sigma_0 \to 0$ as $w \to 0$.

At present, 3D computational models are not sufficiently advanced to reproduce the full menagerie snow crystal structures with good fidelity, but that situation is improving rapidly (Chapter 5). On a parallel track, our ability to create complex snow crystals in controlled environments with reproducible initial conditions is also improving rapidly (Chapter 8). Soon it should be possible to combine these technologies and fully explore the SDAK and ESI phenomena, no doubt generating many new insights into how these peculiar aspects of the attachment kinetics affect snow crystal structure formation.

## 4.4 Explaining the Nakaya Diagram

At this point is beneficial to take a step back and examine just how difficult it has been to understand the enigmatic Nakaya diagram, even at a basic qualitative level. Ever since its discovery (Chapter 1), researchers have struggled to create a comprehensive physical model that describes its overall features. The increase in morphological complexity with greater supersaturation generally arises from diffusion-limited growth (Chapter 3), so this aspect is reasonably well explained. But the various transitions between platelike and columnar growth with temperature are brought about by changes in the attachment kinetics, and this aspect of the Nakaya diagram presented a challenge right from the outset. The overall empirical behaviors of the morphology transitions were quickly reproduced and extended by several researchers, so the observational side was quickly well established. But the quest for a comprehensive physical model continues to this day.

### Previous Attempts

Basil Mason and collaborators made an early attempt at developing a suitable model of the attachment kinetics in the 1960s by reporting measurements of admolecule diffusion lengths $x_{dif}$ $(T)$ on both the basal and prism facets as a function of temperature [1958Hal, 1963Mas]. These observations suggested a temperature dependence in the relative growth rates of the primary facets that roughly explained the several transitions between platelike and columnar growth seen in the Nakaya diagram. The Mason et al. model enjoyed some early popularity, but it has generally not withstood the test of time for several reasons. On the experimental side, the underlying measurements relied on observations of macrostep growth velocities in air to extract $x_{diff}$ values, and this technique is now known to be prone to systematic errors. Macrostep growth rates are strongly affected by bulk diffusion and other subtle effects, and even today it is exceeding difficult to measure admolecule surface diffusion lengths with good absolute accuracy [2014Asa, 2015Lib]. Moreover, recent measurements have achieved a reasonable degree of consensus that terrace nucleation, not surface diffusion, is the primary factor limiting facet growth in most circumstances [2019Har, 2017Lib, 2013Lib].

On the theory side, one can add that measurements of $x_{diff}(T)$ alone would not constitute a comprehensive model of the attachment kinetics unless the model included at least some sketch of the underlying molecular physics that explained the observations. In this same vein, while empirical measurements of $\alpha_{basal}$ and $\alpha_{prism}$ as a function of growth conditions are an important step forward, such measurements alone do not constitute a comprehensive physical model.

Toshio Kuroda and R. Lacmann (KL) created a new comprehensive model in the 1980s incorporating a molecular picture of how attachment kinetics might vary with temperature-dependent changes in surface premelting [1982Kur, 1984Kur]. The authors postulated several speculative, but physically plausible, transitions in the ice surface structure as a function of temperature, and further postulated that these transitions occurred at different temperatures on the basal and prism facets. By adjusting several model parameters, rough agreement with the observed plate/column transitions in the Nakaya diagram was obtained.

In many respects, the KL model was prescient in several of its features. Kuroda and Lacmann clearly recognized the separation between large-scale diffusion effects and the localized attachment kinetics, with the latter depending mainly on the near-surface supersaturation $\sigma_{surf}$. The authors further realized the importance of determining $\sigma_{surf}$ in ice-growth experiments via careful diffusion modeling. The KL model was also the first to incorporate surface premelting as a major physical component, and to postulate a facet-dependent premelting behavior as an integral part of the model. My selective positioning of the SDAK dips, postulated from facet-dependent premelting, was an idea adopted from the KL model.

Beyond these clear successes, the KL model does not provide an adequate comprehensive description of the attachment kinetics. With its focus on surface roughening phenomena at the ice/QLL/vapor interface, the KL model is generally inconsistent with subsequent observations identifying terrace nucleation as the primary mechanism limiting the growth of large facets. Moreover, it now appears that a more complex set of physical mechanisms, including substantial differences between large and small facets, is necessary to explain all the different snow-crystal growth behaviors that have been revealed by newer experiments.

The CAK model presented in this chapter builds upon these past efforts, responding to much additional input from recent precision ice-growth experiments. The model thus does a better job reproducing a broader spectrum of measurements, but at a cost of considerably increased physical complexity. Although the CAK model may not be correct in every detail, I believe that the following features will likely withstand the test of time:

1) Terrace nucleation on large facets. The experimental evidence has become quite strong that ice-growth data are well represented by a terrace-nucleation model over a broad range of growth conditions. The specific curves in Figures 4.5 and 4.6 may require some tweaking as experimental measurements improve, but the importance of terrace nucleation seems to be a solid conclusion.
2) $\alpha_{rough} \approx 1$ on most nonfaceted surfaces. Experiments generally support this blanket statement, and there are no definitive measurements (to my knowledge) indicating $\alpha_{rough}$ values significantly below unity. Here again, this may change with improved measurements, and the transition from rough to faceted (i.e., the attachment kinetics on vicinal surfaces) has not been well explored.
3) Structure Dependent Attachment Kinetics (SDAK). The realization that the attachment kinetics on small facets cannot be entirely the same as on large facets constituted something of a breakthrough in my thinking on this subject. Once I began examining large and small facets separately, I encountered many "ah, that makes sense now" moments. The specific SDAK models presented above may be overthrown or discarded by future research, but I have come to believe that the SDAK concept is essential for explaining snow crystal growth.
4) A rich SDAK phenomenology. If the preceding points are correct, then it soon becomes inevitable that the full SDAK picture will be complicated and difficult to understand in detail. The Edge-Sharpening Instability (ESI) is one example. Developing an accurate model of the attachment kinetics in this regime will be challenging, likely requiring a substantial research effort requiring a combination of carefully controlled laboratory observations with full 3D computational modeling of complex snow crystal structures.
5) Facet-dependent surface premelting. This seems to be the only reasonable way to explain the observed transitions between platelike and columnar growth as a function of temperature. The concept was postulated in the KL model, and I incorporated it into the CAK model as well. Surface premelting likely exhibits some differences on the basal and prism facets, so this is a plausible assumption. MD simulations generally see similar premelting behaviors on the two facets, but the calculations may not be accurate enough to see small facet-dependent effects. Moreover, no one has suggested a viable alternative that would yield a physically sensible, yet sufficiently complex, model of the attachment kinetics.
6) Data driven features. Beyond terrace nucleation, there is little solid theoretical guidance for modeling the attachment kinetics, and cartoon sketches can only take one so far. Progress must be driven largely by empirical reasoning, especially using precision measurements of ice growth rates under well-controlled conditions. I have begun a critical examination of the CAK model using a variety of ice-growth measurements [2019Lib2, 2020Lib], and the results so far are quite promising. But much additional work needs to be done along these lines.

My objective with the CAK model presented in this chapter is not to declare a full solution to the attachment kinetics problem, but rather to move the ball forward while stimulating additional progress in this area. A quantitative "working" model, even an imperfect one, is far better than no model at all, as it can serve to suggest specific, targeted experimental and theoretical investigations.

As described in Chapter 1, a primary motivator for this book is a desire to grow computational snow crystals that accurately reproduce what one finds in nature. Developing a physical model of the attachment kinetics is an important prerequisite in this task, as computation models require physical inputs. Because tracking every molecule is clearly impossible, even the best model can only provide approximate parameterizations of the functions $\alpha_{basal}$ and $\alpha_{prism}$ as a function of $\sigma_{surf}$, $T$, $w$, and perhaps other variables. While no such parameterizations can attain absolute accuracy, one can hope to capture the essential physics and create plausibly realistic computational snow crystals. At the very least, a good model of the attachment kinetics should agree with ice-growth experiments and provide a reasonable explanation of the Nakaya diagram. The CAK model presented in this chapter strives to attain these objectives.

## Beyond the Nakaya Diagram

One fact that becomes glaringly obvious when thinking about a comprehensive model of the attachment kinetics is that the traditional Nakaya diagram cannot capture the full richness of snow crystal formation. Describing typically observed growth morphologies in air on a temperature/supersaturation plane simply cannot do justice to such a complex physical phenomenon. Background pressure is another interesting environmental factor, and initial conditions play a role via the SDAK effect. Overall crystal size is important as well, as dendritic structures are generally less prevalent on smaller crystals. It is important, therefore, to move beyond morphologies and develop new ways to visualize the physics and phenomenology that describes snow crystal growth.

I have found that plotting $\alpha_{basal}$ and $\alpha_{prism}$ as a function of $\sigma_{surf}$ at constant temperature provides one useful complement to the Nakaya diagram, and some illustrative examples are shown in Figure 4.17. These curves were derived from the CAK model and include several separate "branches" that designate different growth regimes. Solid lines show the large-facet kinetics, while dotted lines indicate the two SDAK effects. Drawn this way, one begins to see some of the morphological complexity inherent in the CAK model.

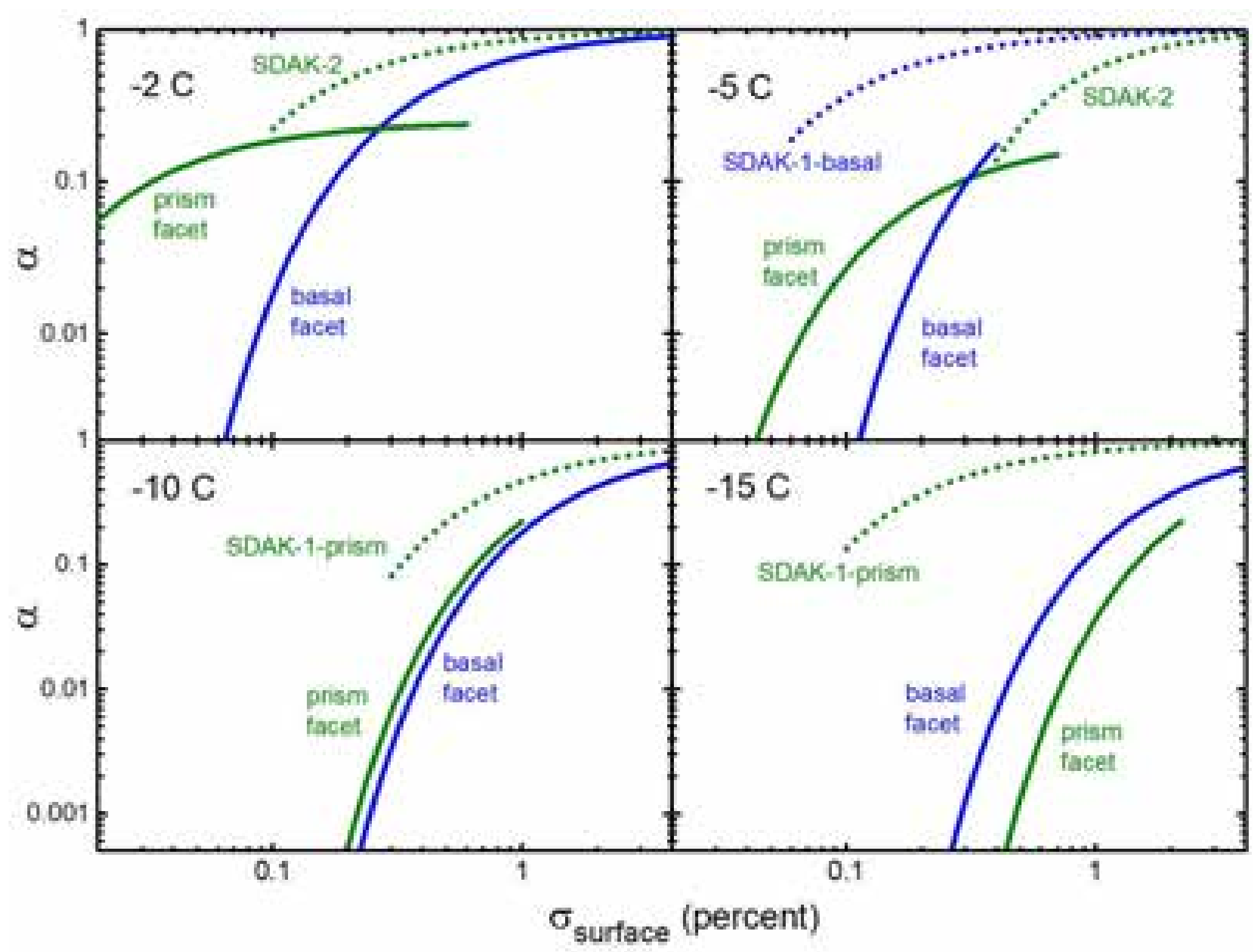

Figure 4.17: The inherent richness of the CAK model, and the attachment kinetics more generally, becomes apparent when displayed as constant-temperature $\alpha(\sigma_{surf})$ plots. The different "branches" represent different growth regimes identified by their corresponding physical effects. The graph shown for -5 C is particularly noteworthy, as it indicates that both platelike and columnar forms can appear at $\sigma_{surf}$=0.15%, a feature that is supported by observations. Plots like these are especially useful for comparing with all different types of ice-growth experiments.

At -5 C, for example, the appearance of four separate branches leads to several distinct morphological behaviors. Both platelike and columnar forms can develop at $\sigma_{surf} = 0.15$ percent, the former being simple plates with large facets, the latter being hollow columns or needles with narrow basal facets. And both forms are readily observed, offering persuasive support for the CAK model [2012Kni, 2019Lib2]. The ESI tends to stimulate hollow-column growth at relatively high $\sigma_\infty$ in normal air, while platelike forms are more prevalent at lower $\sigma_\infty$ and lower air pressures. At the highest $\sigma_{surf}$, the two SDAK curves at -5 C yield "fishbone" dendritic structures (Chapter 3) with a tip growth axis that varies with growth rate (Chapter 8). The full range of observed growth behaviors at -5 C is reasonably consistent with the CAK model but is difficult to explain otherwise.

At -15 C, Figure 4.17 indicates that simple prisms with large facets should be nearly isometric in form, and such crystals are readily observed in vacuum. But the ESI quickly yields rapidly growing prism edges in air, resulting in the formation of thin plates. Dendritic forms remain nearly platelike even at high growth rates, owing to the strong SDAK effect on prism facets together with the large nucleation barrier on basal facets. Morphologies at -2 C are generally platelike at all supersaturations.

While morphology studies are a good starting point, there is much more to be learned by exploring snow crystal growth as a function of background gas pressure and generally covering a broader range of environments and initial conditions. And, of course, quantitative experiments are a must for making detailed comparisons with models of the attachment kinetics. Graphical representations of growth phenomenology may soon become even more sophisticated once we begin making full 3D structural comparisons between computational models and advanced laboratory observations of complex snow crystal structures.

## 4.5 The Morphological Nexus at -5 C

Because snow crystal growth at -5 C involves a confluence of all the varied growth behaviors identified in the CAK model, this temperature provides an especially thorough test of many of the most important aspects of the attachment kinetics, including large-facet growth and both SDAK effects. In this section, I examine data from a variety of ice-growth experiments, including large and small facets, all focusing on the morphological nexus that appears at -5 C.

### Large-Facet Growth

Beginning with large facets, both the VIG and VPG experiments (Chapter 7) provide measurements of $\alpha_{basal}$ and $\alpha_{prism}$ as a function of $\sigma_{surf}$, and these results are shown together in Figure 4.18. Several immediate conclusions can be reached from these data:

1) The VIG and VPG experiments show excellent agreement at -5 C, even though they use significantly different measurement strategies and completely different hardware. Moreover, the basal and prism facets were measured separately in the VIG experiment, using different crystals in different runs, while both facets were measured simultaneously on smaller ice prisms in the VPG experiment. Obtaining clear agreement from such different experiments is a noteworthy achievement all its own, given the considerable measurement uncertainties

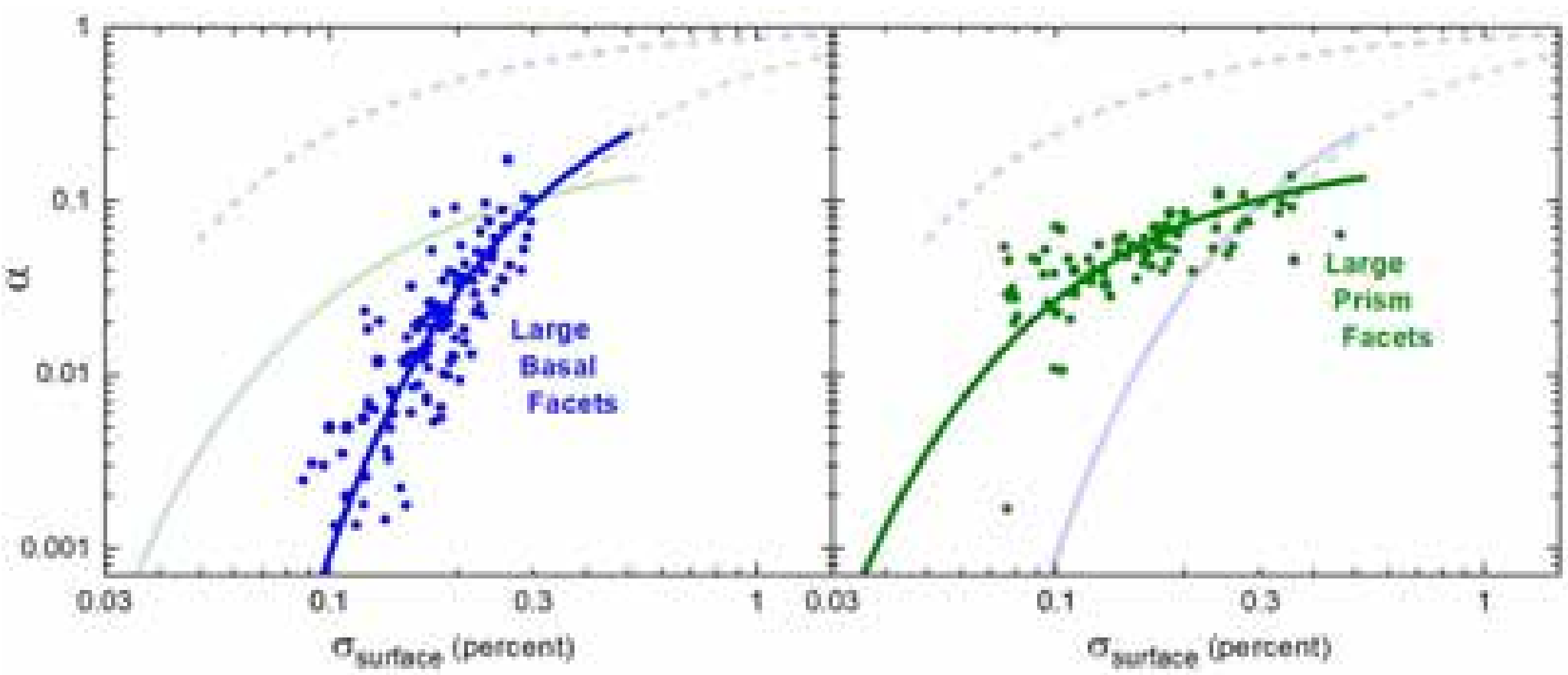

Figure 4.18: Data points in these two graphs show measurements of $\alpha_{basal}(\sigma_{surf})$ and $\alpha_{prism}(\sigma_{surf})$ at -5 C in a low-pressure environment. Lines in both graphs are from the CAK model for -5 C, while nonessential curves have been faded in the different panels. These observations reveal that platelike habits are the norm for simple ice prisms at -5 C, with thinner plates forming at lower supersaturations. These plots combine measurements from the VIG and VPG experiments described in Chapter 7.

and discrepancies found in earlier ice-growth experiments.

2) There is no obvious air-pressure dependence in either $\alpha_{basal}$ or $\alpha_{prism}$ as a function of $\sigma_{surf}$ in these data, at least over the limited range that was measured (see Chapter 7). This supports our assumption in the CAK model that air at a pressure of one bar has little effect on the attachment kinetics.
3) The data all strongly support the terrace-nucleation mechanism for the large-facet attachment kinetics in the CAK model. Note also that the data support $A_{basal} \approx 1$ and $A_{prism} < 1$ at this temperature, as described above. Constraining the model to have $A_{prism} = 1$ is not excluded completely by the measurements, but both data sets show quite similar trends that prefer $A_{prism} < 1$.

As shown in these examples at -5 C, precision measurements of simple prisms provide the foundation for the CAK model, in which large-facet growth is limited by terrace nucleation over a broad range of environmental conditions. The functional form of the model curves comes from terrace-nucleation theory, but the model parameters $A(T)$ and $\sigma_0(T)$ are obtained entirely from empirical data.

## Bimodal Behavior

The VPG experiment at -5 C allows another look at how the SDAK phenomenon can yield remarkably different growth morphologies at a single temperature. In a near-vacuum environment at low supersaturations, platelike growth is the norm, and Figure 4.19 illustrates that even quite extreme aspect ratios can be observed. According to the CAK model, even thinner plates should appear at lower supersaturations, but it appears that residual substrate interactions may be preventing these from forming in the VPG apparatus [2019Lib2]. Thin plates at low supersaturations require an exceedingly low basal nucleation rate (with growth rates below 1 nm/sec), and substrate-induced basal nucleation could set a practical limit on making thin plates on any substrate. A levitation experiment at -5 C might yield thinner plates than those shown in the figure, but small amounts of chemically induced basal nucleation may also be problematic at the lowest growth rates. In any real experimental environment, there may be some small rate of heterogeneous terrace nucleation that is difficult to eliminate.

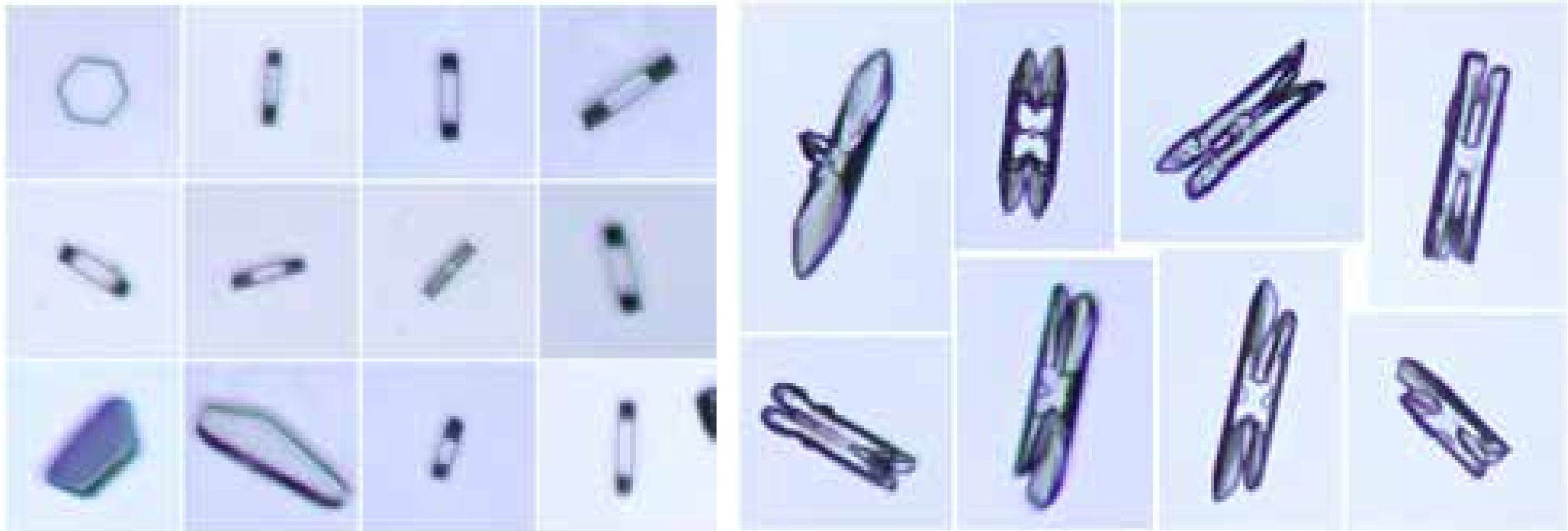

Figure 4.19: Left: A selection of thin platelike crystals growing at -5 C at low pressure in the VPG apparatus with $\sigma_{surf} \approx 0.1$ percent. These examples illustrate that quite thin plates can be grown at -5 C. Right: Several hollow columnar crystals grown in air at -5 C using the VPG apparatus. Although the overall morphologies are distorted by their intersection with the substrate, it appears that the Edge-Sharpening Instability is largely responsible for the formation of thin basal edges on these crystals.

Growing crystals in air at higher supersaturations yields the hollow columnar crystals shown in Figure 4.19, their overall shape being somewhat distorted by the fact that they are resting on the substrate. The sharp basal edges observed on these crystals suggests that the Edge-Sharpening Instability (ESI) played a major role in determining their growth morphology. As discussed above, the development of the CAK model was driven largely by the empirical necessity to explain the diversity of observations like these.

## Fishbones, Tridents, Needles, and Columns

The -5 C story becomes especially intriguing when one also considers the small prism and basal facets appearing on complex crystal structures growing in air. Figure 4.20 illustrates several examples including hollow columns (4% and 8%), needles (16%), and fishbone dendrites (32%, 64%, and 128%). Although we cannot yet model the growth of these elaborate morphologies in detail, it is still possible to glean some useful information from growth ratios. For the hollow columnar and needle crystals, for example, it is straightforward to measure $v_{basal}$ and $v_{prism}$ at the end of each column from a time series of images. Because the sides of the columns contain large faceted prism surfaces, the CAK model suggests that their growth should be determined by terrace nucleation, so the known $\alpha_{prism}(\sigma_{surf})$ in that case means we can ascertain $\sigma_{surf}$ from the measured $v_{prism}$. Of course, this is a model-dependent statement, and it would be better if a full diffusion model determined $\sigma_{surf}$ directly, as we did with the VIG and VPG experiments. But the diffusion corrections are too large for this to be practical at present, so the indirect, model-dependent determination will have to suffice. I refer to this as a "witness-surface" analysis, as the (presumably) known large-facet prism growth behavior provides a witness that allows a determination of $\sigma_{surf}$.

The next step in the witness-surface analysis is to assume that $\sigma_{surf}$ on the prism facet near the top corner of the columnar crystal is approximately equal to $\sigma_{surf}$ on the basal surface at the same corner. This is reasonable when the two surfaces are small and proximate, although the equality might be somewhat off if large supersaturation gradients are present. With this approximation for $\sigma_{surf}$, it then becomes possible to determine

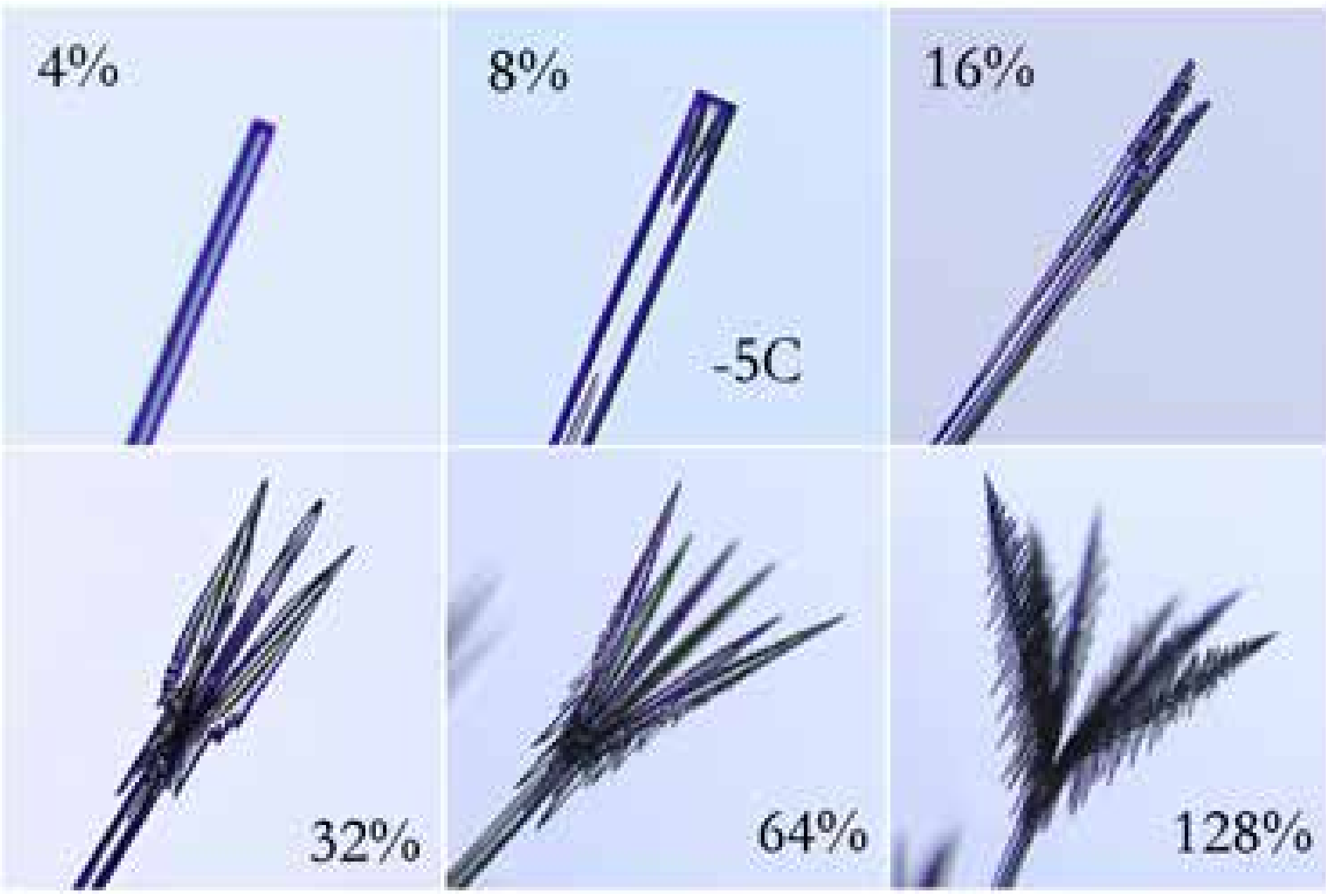

Figure 4.20: Several examples of crystals growing at -5 C in air on the ends of “electric” ice needles (Chapter 8). Each image is labeled with the supersaturation $\sigma_\infty$ far from the needle tip. At 4%, the initial needle thickens to form a solid column with slight basal hollowing, while hollow-column growth develops quickly at 8%. At 16%, a hollow column has split to form needle clusters. At 32%, trident structures are common (Chapter 3). At 64% and 128%, fishbone dendrites (Chapter 3) exhibit rapid growth.

$\alpha_{basal}(\sigma_{surf})$ from the measured $v_{basal}$, thus yielding the first three data points in Figure 4.21. Note that $v_{basal}/v_{prism} \gg 1$ for columnar and needle growth, fixing $\alpha_{basal}/\alpha_{prism}$ in this figure, and our main model-dependent statement was estimating $\sigma_{surf}$ by putting the $\alpha_{prism}$ points on the large-facet terrace nucleation model curve. Note also that we could not have applied the witness-surface analysis in the opposite direction, as the basal facet is small on a hollow column, so its growth is not well known from the CAK model.

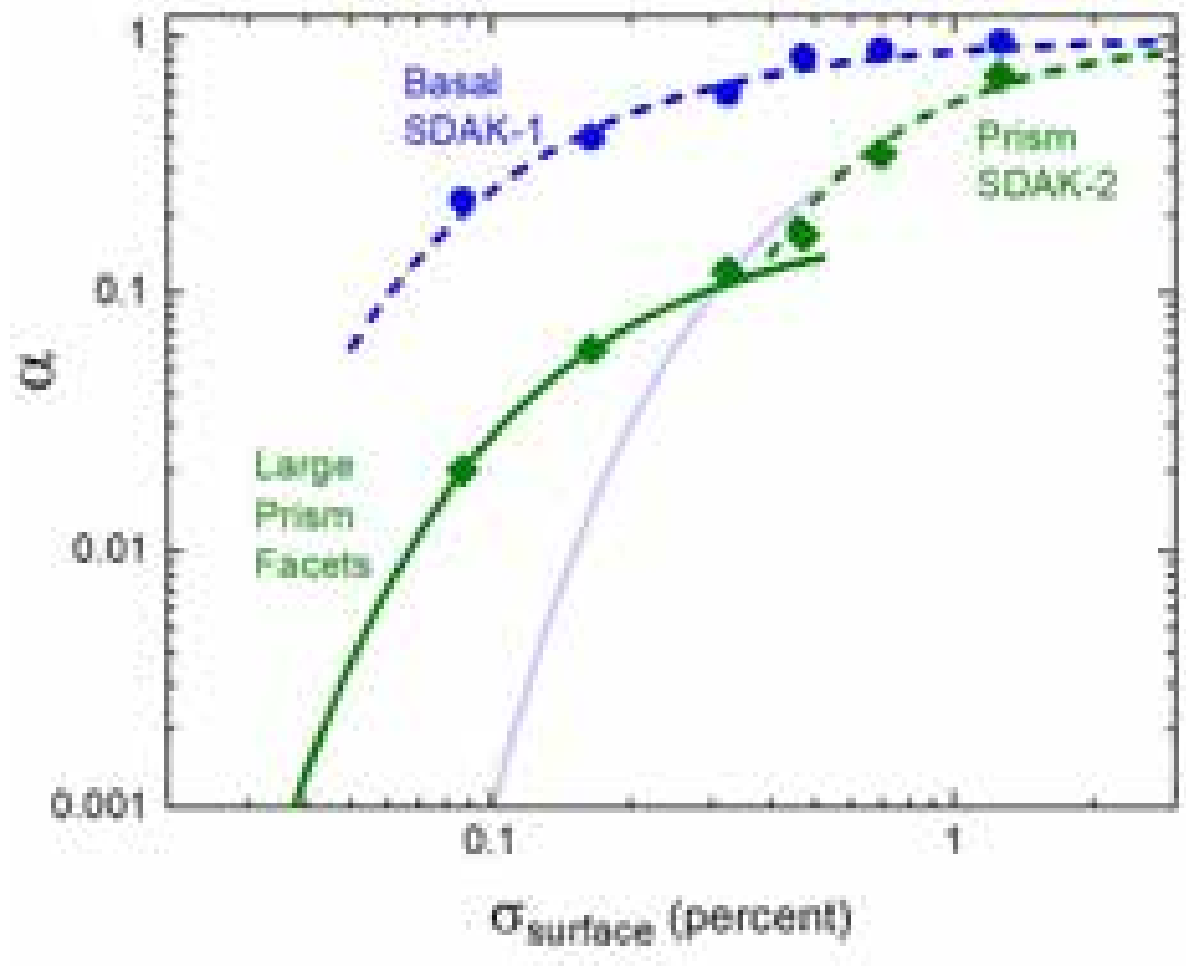

Figure 4.21: Analysis of the snow crystals growing at -5 C in air on the ends of electric ice needles shown in Figure 4.20. These data all show fast basal kinetics compared to large basal facets, resulting from the basal SDAK-1 mechanism. Prism growth also increases at high supersaturations from the SDAK-2 phenomenon. The lines are again from the CAK model at -5 C.

Perhaps the most interesting part of this analysis is that it quantifies the bimodal behavior at -5 C, showing that both platelike and columnar crystals can form under quite similar conditions. Platelike crystals grow readily in near-vacuum and in air, while columnar crystals only appear when the basal facets are small, and this only happens in air because of the ESI. With the CAK model including the ESI, it all makes a reasonable amount of sense. Without it, however, explaining the full range of columnar and platelike growth behaviors would be quite challenging.

The analysis is a bit trickier for the three fishbone dendrites, but again we can extract some useful information from growth ratios. Each of these crystals exhibits a measurable tip velocity and growth direction, where the latter is measured relative to the c-axis defined by the initial electric-needle crystal. Thus, from a time series of images, it is straightforward to measure $v_{basal}$ and $v_{prism}$ at the fishbone tips. The witness-surface analysis runs into problems, however, because now both the basal and prism surfaces are tiny on these sharply pointed crystals, so neither would be well described by large-facet growth. Nevertheless, starting with the 128% fishbone, the fast growth points to $\alpha_{basal}$ being near unity, and this is consistent with the SDAK-1 basal curve in Figure 4.21. Assuming $\alpha_{basal} \approx 1$, the witness surface analysis then yields $\alpha_{prism} \approx 1$ as well, simply reflecting the large opening angle of this set of fishbone dendrites (or, equivalently, $v_{basal} \approx v_{prism}$). This allows us to bootstrap our way along, as the $\alpha_{basal}$ points can be interpolated onto a reasonable SDAK-1 curve, in turn giving a set of $\alpha_{prism}$ points from an approximate witness-surface analysis. Note that the fishbone angle changes with $\sigma_{surf}$, and this means that the SDAK-2 points diverge from the SDAK-1 points at lower supersaturations. With the trident morphology at 32%, for example, we see growth that is midway between columnar and fishbone behaviors.

Also note that the $\sigma_{surf}$ values in Figure 4.21, as determined from the witness-surface analysis, drop by about a factor of two between data points, as one would expect from the corresponding drop in $\sigma_\infty$ between the different crystals. The Ivantsov solutions for parabolic growth (Chapter 3) predicts $\sigma_{surf}/\sigma_\infty \approx 2X_0/BR_{tip}$, and using $B \approx 10$ and $R_{tip} \approx 1.5$ µm gives $\sigma_{surf}/\sigma_\infty \approx 1/50$, which is not too far from what comes out of the witness-surface analysis. Thus, the numbers hold together reasonably well in an absolute sense, even if it is not yet possible to solve the diffusion equation with sufficient accuracy to make an absolute measurement. Although the witness-surface analysis is model dependent and only uses growth ratios, the results describe a self-consistent pattern of growth behaviors that support the CAK model overall. Figure 4.22 shows all the different CAK model growth behaviors present at -5 C.

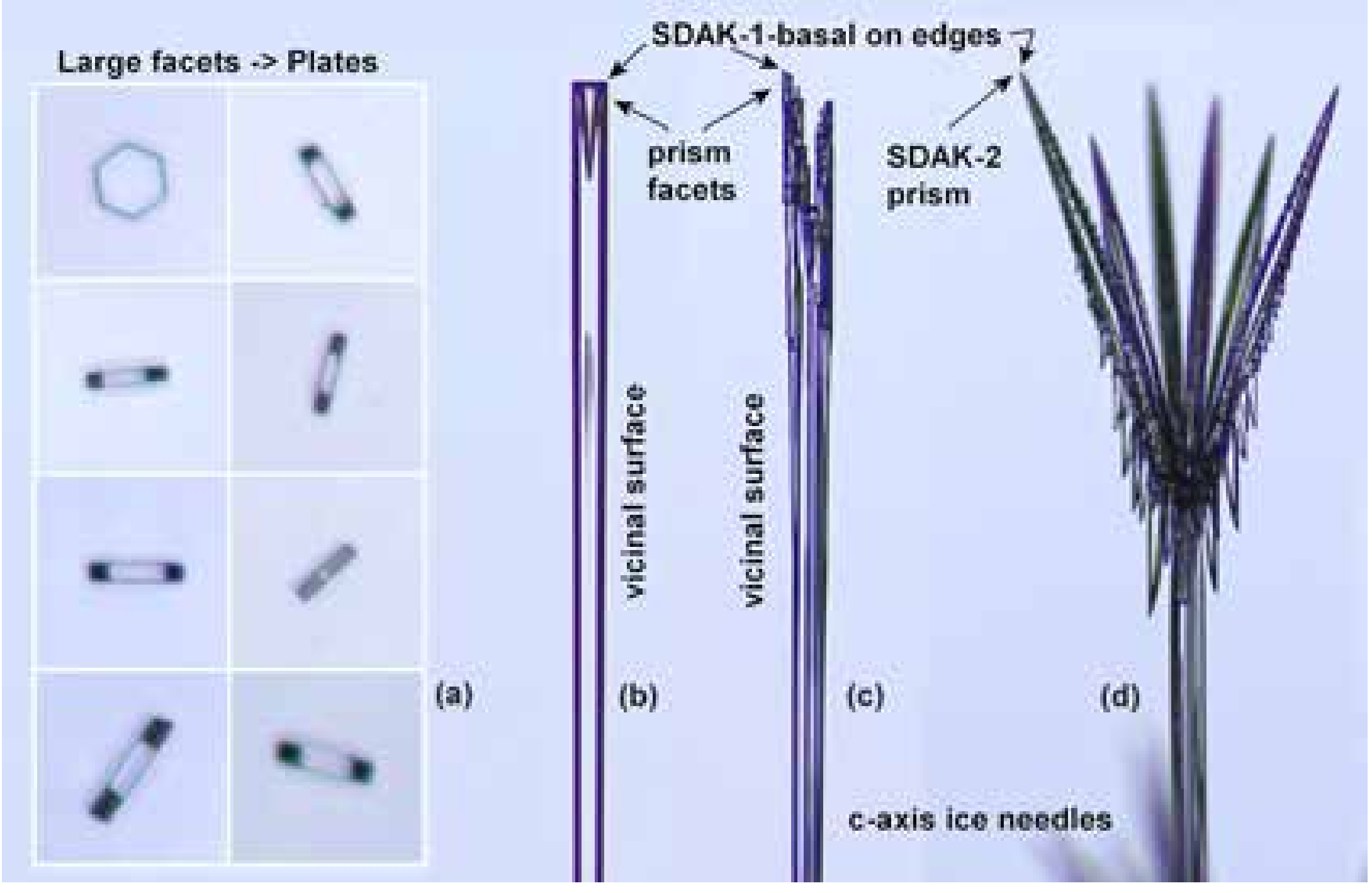

Figure 4.22. This photo collage illustrates several snow crystal growth behaviors at -5 C. (a) With simple prisms grown in vacuum, thick plates are the norm. (b) Starting with a c-axis needle in air (Chapter 8), the basal ESI from the SDAK-1 phenomenon brings about sharp basal edges and a hollow column. (c) At a higher supersaturation, the hollow column splits into a needle cluster. (d) At still higher growth rates, the SDAK-2 mechanism increases the prism kinetics, yielding dendritic structures where the opening angle increases with supersaturation. All four CAK branches in Figure 4.21 are needed to reproduce these different growth behaviors.

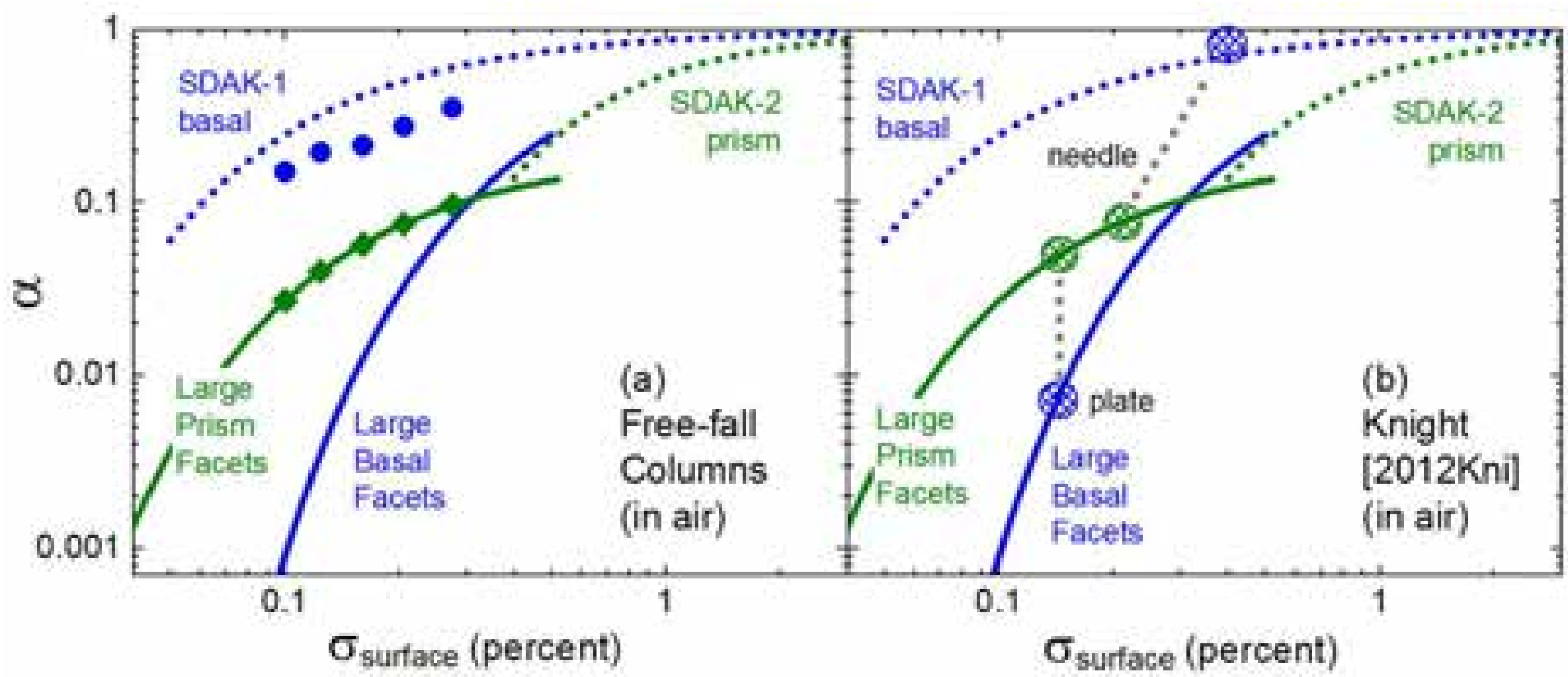

Figure 4.23: Additional snow-crystal growth data obtained at a fixed temperature of -5 C [2020Lib2]. (a) Data from a free-fall growth chamber, again analyzed using the witness-surface method. (b) A rough analysis of observations in [2012Kni] that demonstrated the simultaneous growth of platelike and needle structures at -5 C. The data can also be explained using the CAK model, albeit with substantial experimental uncertainties. Together with Figures 4.18 and 4.21, these data exhibit a remarkably diverse range of growth behaviors, and all can be reasonably well explained with the CAK model.

Figure 4.23(a) shows another data set, this time for columnar crystals growing in ordinary air in a free-fall growth chamber (Chapter 6). Although the experiment produced estimates of $\sigma_\infty$, the diffusion corrections are so high that it is not possible to determine $\sigma_{surf}$ with good accuracy. For this reason, the graph shows the result of another witness-surface analysis, once more assuming that the prism growth of the columnar crystals is given by the terrace-nucleation curve in the CAK model. And again, we see that the basal growth lies near the SDAK-1 curve, reflecting the fact that the hollow columnar crystals exhibited sharp basal edges. The free-fall measurements again support the CAK model to a reasonable degree, considering the various data and analysis uncertainties.

Figure 4.23(b) shows a crude analysis of some observations at -5 C reported by Knight [2012Kni], in which he describes the simultaneous growth of platelike and needle crystals in air. The nature of the observations makes it impossible to estimate $\sigma_{surf}$ with any real accuracy, so the data points shown in the figure should be taken as rough estimates only. Nevertheless, with some common-sense reckoning it is possible to place these observations on the CAK model, even if this cannot be done with great accuracy. The exercise mainly shows that these observations are at least consistent with the CAK model, and the simultaneous observation of platelike and needle forms certainly confirms the bimodal growth behavior at -5 C.

Summarizing all these results at -5 C, we find:

1) When large prism or basal facets are present, their growth fits the large-facet behavior provided by the CAK model, as illustrated in Figure 4.18.
2) On hollow columns, the narrow basal facets grow rapidly, in a manner that is consistent with the SDAK-1 mechanism, while the large prism facets are still described by large-facet growth (Figures 4.21 and 4.23).
3) On the tips of fishbone dendrites, the basal and prism growth behaviors are consistent with the SDAK-1 and SDAK-2 phenomena (Figure 4.21).
4) The CAK model explains the simultaneous formation of platelike and needle crystals under certain conditions (Figure 4.23).
5) By treating broad and narrow facets differently, the CAK model reasonably explains a convoluted tangle of morphological observations and quantitative ice growth data at -5 C.

## 4.6 Snow Crystal Cartography

The full spectrum of measurements at -5 C illustrates how several quite different growth behaviors can all be explained with the same comprehensive model of the attachment kinetics. These results at -5 C are especially intriguing, as so many physical effects come into play at this temperature, thus providing an excellent overall test the CAK model. I have begun to call this analysis "snow crystal cartography," as one is able to map out the different growth regimes in remarkable detail using a variety of different experimental and data-analysis strategies. Remarkably, even with just a simple witness-surface analysis for the fast-growth data, requiring no complex 3D diffusion modeling, we can explore the different SDAK branches in the CAK model. Although much remains to be done, a compelling picture of snow crystal growth phenomenology is beginning to emerge from these studies.

Figure 4.24 provides another example using data compiled at -2 C [2020Lib]. The CAK model is generally simpler at this temperature, as the SDAK-1 effect is no longer significant, so only platelike and dendritic morphologies appear under all measured growth conditions. In keeping with expectations from the Nakaya diagram, no columnar forms are present at -2 C. Once again, we have found that displaying the measurements on a log-log plot showing $\alpha$ as a function of $\sigma_{surf}$ provides a useful tool for exploring the different growth regimes, and again the data generally support the CAK model. At the time of this writing, we are examining additional temperatures following this same methodology, a process that will likely take some years to complete to satisfaction.

## 4.7 Dislocation-Mediated Growth

Up to this point, we have ignored lattice dislocations and other lattice imperfections in this chapter, focusing first on the ideal case of the attachment kinetics on a flawless underlying crystal. Real crystals inevitably have some flaws, however, so we now examine to what extent these can affect the attachment kinetics. If lattice dislocations are present on a faceted surface, then the terrace nucleation mechanism described above may be augmented by a faster mechanism involving screw dislocations. The first sketch in Figure 4.25 shows the lattice structure of a screw dislocation, which begins as a defect in the ideal lattice configuration. As admolecules attach to the exposed terrace edge, it develops into the characteristic spiral pattern shown in the second sketch in the figure. This spiral pattern can persist indefinitely, as the dislocation will not "heal" as the crystal grows. By providing a continuous source of molecular steps, a screw dislocation can thus yield substantial growth rates even when $\sigma_{surf}$ is far below that required for normal terrace nucleation. The theory behind this mechanism is described in most crystal growth textbooks [1996Sai, 1998Pim, 2002Mut], yielding $\alpha \sim \sigma_{surf}$ and a perpendicular growth velocity $v_n \sim \sigma_{surf}^2$.

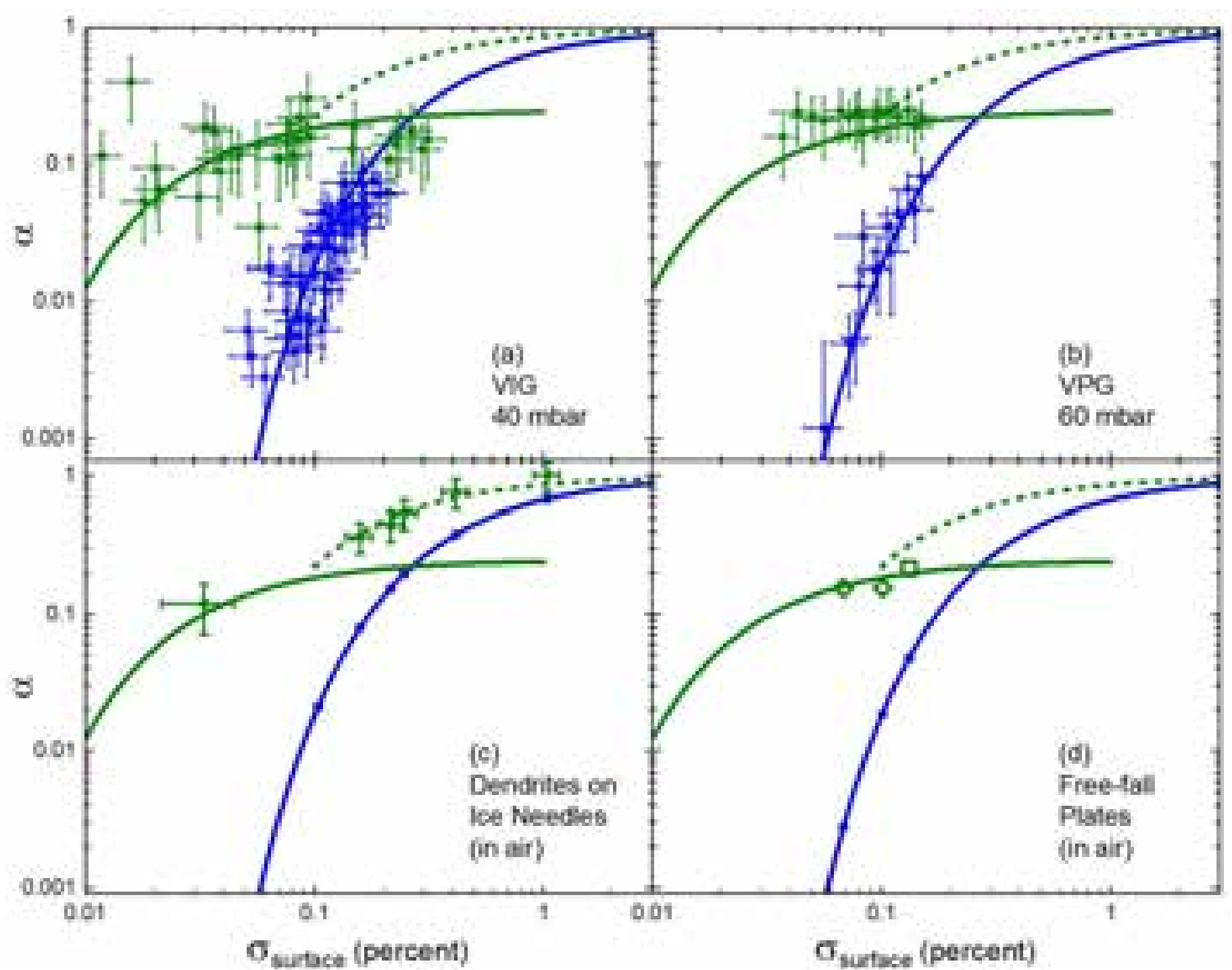

Figure 4.24: A collection of snow-crystal growth data obtained at a fixed temperature of -2 C [2020Lib]. (a) Data from the VIG experiment at low background pressure, showing good agreement with the CAK model. (b) Similar data from the VPG experiment, confirming the earlier VIG results and the CAK model. (c) Observations of ice crystals growing on the ends of electric ice needles, analyzed using the witness-surface method on the basal facets. These data show rapid prism growth at high supersaturations, in agreement with the SDAK-2 phenomenon in the CAK model. (d) Additional data from a free-fall growth chamber, again analyzed using the witness-surface method. Although the overall phenomenology at -2 C is not as rich as at -5 C, the data show quite good agreement with the CAK model on all fronts.

F. C. Frank suggested that the simple observation of symmetrical hexagonal prisms likely indicates the absence of dislocation-mediated growth on the six prism facets [1982Fra]. Because the facet surfaces all grow at equal rates, either there are no dislocations present on any of the facets or there must be at least one dislocation on each of them. If screw dislocations were present on some of the facet surfaces but not all, then the growth rates would vary, and the overall prism morphology would not show hexagonal symmetry. Although asymmetrical prisms can be found [1979Kik], most are nearly symmetrical. As a high dislocation density seems improbable on small crystals with simple morphologies, the logical conclusion is that most prism facet surfaces are free of dislocations, or at least those that greatly alter the facet growth rates.

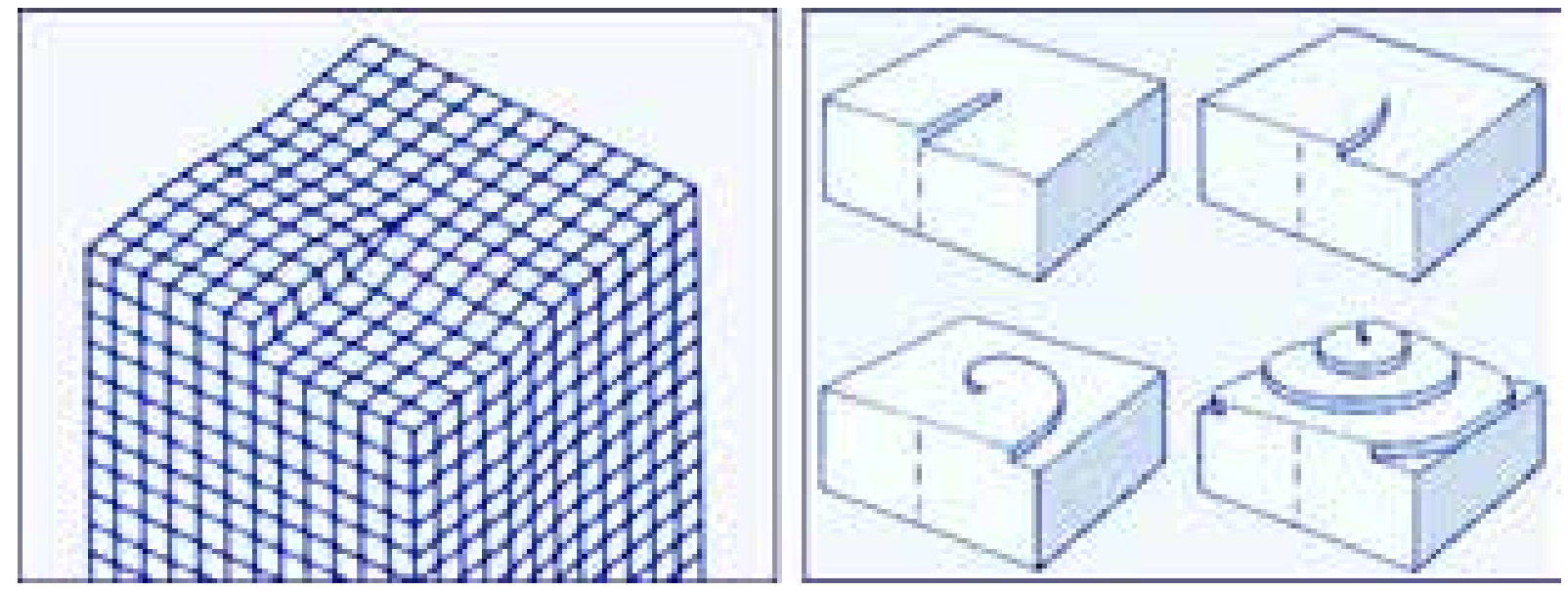

Figure 4.25: Left: The lattice structure of a screw dislocation. Right: As the dislocation edge grows from admolecule attachment, it creates a spiral pattern that can propagate indefinitely, yielding growth in the absence of additional terrace nucleation.

Beyond this simple observation, there is also considerable experimental evidence indicating that many laboratory-grown snow crystals, particularly simple prisms and other forms exhibiting especially clean morphologies, are generally free from dislocations that affect their growth. The example in Figure 4.4, which is representative of many similar crystal specimens, is quite well described by a nucleation-limited model, while a dislocation-mediated growth model gives a substantially poorer fit to the data. Some simple prisms do exhibit anomalous growth behaviors, but up to 90 percent of well-formed samples are consistent with nucleation-limited growth [2013Lib, 2019Lib2], at least at temperatures above -20 C. As with polycrystalline forms, dislocations appear to be more common as the temperature decreases [2004Bai, 2009Bai].

While many fascinating crystal-growth phenomena are mediated by dislocations [2013Men], this subject is somewhat beyond the scope of this book. Most of this chapter examined the attachment kinetics on essentially perfect single crystals, and the resulting CAK model is already quite complex. For this reason, we leave a fuller discussion of the effects of dislocations on snow crystal growth for another day. Instead, throughout most of this book, I focus on the growth of defect-free ice crystals.

## 4.8 Chemical Vapor Effects

In the preceding sections of this chapter, I made the implicit assumption that the ice/vapor attachment kinetics was completely unaffected by any physical or chemical effects associated with the surrounding medium. In our discussion of the ESI, for example, particle diffusion influenced the nucleation of new terraces by modifying $\sigma_{surf}$ around the crystal, but the presence of air did not change the attachment kinetics directly. In other words, I assumed that $\alpha_{basal}$ and $\alpha_{prism}$ had no intrinsic dependence on the background gas pressure or composition. Now I examine this assumption more closely.

### Air Effects

A first question is whether ordinary clean air affects the attachment kinetics in any significant way. This matters not only for natural snow crystals, but also for the many laboratory ice-growth experiments that have been performed in laboratory air, sometimes at reduced pressure. Beginning with nitrogen and oxygen, Henry's Law data show that both these gases have rather low solubilities in water, and adsorption data reveal that they do not stick well to ice surfaces either. Moreover, recent molecular-dynamics simulations indicate that nitrogen gas at 1 bar appears to have little direct effect on the structure and dynamics of surface premelting [2019Llo]. These considerations, although somewhat indirect, all suggest that normal air probably has a negligible influence on the attachment kinetics. Carbon dioxide has a higher solubility in water, but it has a low concentration in air, and snow crystals grown in $CO_2$ at one bar do not show any obvious abnormalities (see below).

Nevertheless, some experiments have indicated that ice-growth rates are influenced by the presence of air, to an extent beyond that caused by normal particle-diffusion-limited growth. For example, Beckmann et al. [1983Bec] found that the slower growth rates of ice crystals in air were due in part to a reduction in the attachment kinetics, and similar conclusions were reached by Kuroda and Gonda [1984Kur1]. Recently I suggested that air-dependent effects were especially strong on prism facets at -5 C, which could explain the growth of columnar crystals at that temperature [2016Lib1]. Looking more closely at these experiments, however, I now believe that all three suffered from faulty diffusion analyses. The diffusion corrections are quite large in air, and relatively small systematic errors can easily be mistaken for a pressure-dependent attachment kinetics. The best way around this problem is to examine especially small crystals, and then only in a low-$\alpha$ regime where diffusion effects are relatively minor. In that restricted regime, newer measurements show no significant dependence of the attachment kinetics on air pressure at -5 C [2019Lib2]. Given our current state of knowledge, several pieces of evidence appear to be reliable:

1) Basic chemical-physics reasoning suggests that nitrogen, oxygen, and other molecular species in air are generally inert with respect to ice adsorption. MD simulations appear to support this conclusion as well.
2) When diffusion effects are especially small (low $\alpha$, small crystals), newer experiments suggest negligible changes in $\alpha$ with gas pressure (Chapter 7). In this regime, the attachment kinetics is dominated by terrace nucleation on large facets.
3) When $\alpha_{diff} \ll \alpha$ in ice-growth experiments, it becomes exceedingly difficult to separate diffusion-limited growth from kinetics-limited growth, so results can be influenced by quite small systematic errors.
4) The CAK model can explain the morphology diagram and other ice-growth measurements without requiring that the attachment kinetics change with gas pressure. While this model may not be entirely correct, it suggests that gas-dependent kinetics are not obviously required to explain the full range of observed snow crystal growth behaviors.

There is much need for additional investigation in this area, as large swaths of parameter space have not yet been adequately explored by experiments. At present, however, it appears that the molecular dynamics determining the ice/vapor attachment kinetics is little affected by a background gas of ordinary air.

## Chemical-Dependent Attachment Kinetics

In contrast to inert gases, including clean air, the presence of chemically active vapor additives can dramatically change ice growth rates and morphologies. There is much experimental evidence supporting this general conclusion, and little understanding of any of it. This is a fascinating area for continued investigation, and a constant concern that unwanted chemical impurities can lead to erroneous experimental conclusions.

This area of research has a long history, as Vonnegut, Hallett and Mason [1948Von, 1971Mas, 1958Hal] found that the addition of just 10 ppm of butyl alcohol in air yielded columnar growth at -20 C instead of the usual platelike growth. Schaefer [1949Sch, 1971Mas] further observed that vapors of ketones, fatty acids, silicones, aldehydes, and alcohols could all change ice growth morphologies to varying degrees. Nakaya, Hauajima, and Mugurama [1958Nak] observed that even trace silicone vapor in air caused columnar crystals to grow at -15 C instead of thin plates. Hallett and Mason [1958Hal] found that the addition of camphor vapor in air could yield columnar ice crystals at all temperatures in the range $-40$ C < T < 0 C. These authors also observed that isobutyl alcohol in air changed ice growth near -15 C from plates to columns and then back to plates again as the concentration was increased. Anderson, Sutkoff, and Hallett [1969And] found that Methyl 2-Cyanoacrylate in air could change the morphology from platelike dendrites to needles at -15 C. Libbrecht, Crosby, and Swanson [2002Lib] found that acetic acid and other vapors promoted the c-axis growth of “electric” needle crystals in air near -5 C, thus yielding a useful tool for studying snow crystal growth (Chapter 8). Knepp, Renkens and Shepson [2009Kne] observed various morphological changes caused by acetic acid vapor in air, even in concentrations as low as 1 ppm. Libbrecht and Bell [2011Lib] examined snow-crystal morphologies as a function of temperature for a range of chemical additives as a function of concentration. From all these reports, we can summarize some of the principal findings:

1) In nitrogen gas at one bar, most chemical additives at concentrations below 10 ppm produce no clearly observable changes in ice crystal growth morphologies [2011Lib]. As a rough rule of thumb, if the air has no discernable odor, it likely has little effect on snow crystal formation at one bar.
2) Ice growth in air, nitrogen, helium, argon, hydrogen, carbon dioxide, and methane gases at a pressure of one bar yield roughly identical crystal morphologies as a function of temperature [1959Heu, 2008Lib1], suggesting that these gases are essentially chemically inert.
3) Growth in ultra-clean nitrogen gas was not significantly different from growth in ordinary laboratory air [2011Lib]. These first three points suggest that trace impurities in ordinary air do not play a large role in snow crystal growth.
4) Chemical additives generally tend to promote the growth of columnar crystals over platelike crystals.
5) Nitric acid vapor or nitrous oxide tends to promote the growth of triangular crystals near -15 C [1949Sch].
6) The most effective chemicals for producing growth modification are those having strong polar properties [1949Sch].
7) Chemical additives generally have an especially large, detrimental effect on thin plates and platelike dendritic growth at -15 C [2011Lib]. In general, thin plates at -15 C seem to grow best in inert gases at pressures of one bar or above.
8) Chemical effects are generally more pronounced at lower temperatures, and ice-growth experiments performed at $T < -20$ C are especially prone to unwanted influences from trace chemical contaminants.

Given our poor understanding of ice growth without the complicating effects of chemical additives, it should come as no surprise that there is essentially no theoretical understanding, even at a qualitative level, of how chemical additives alter growth rates and change growth morphologies. It appears likely that these impurities adsorb onto the ice surface where they modify the attachment kinetics in a variety of ways, but there is little theoretical guidance relating to how. While the adsorption of chemical vapors on ice has been much studied from the standpoint of atmospheric chemistry, relatively little attention has been given to chemical modification of the attachment kinetics. This remains a fascinating research direction, although somewhat hampered by a lack of theoretical guidance. Antifreeze proteins strongly affect the attachment kinetics at the ice/water interface, but that fascinating topic is only peripherally related to the ice/vapor interface.

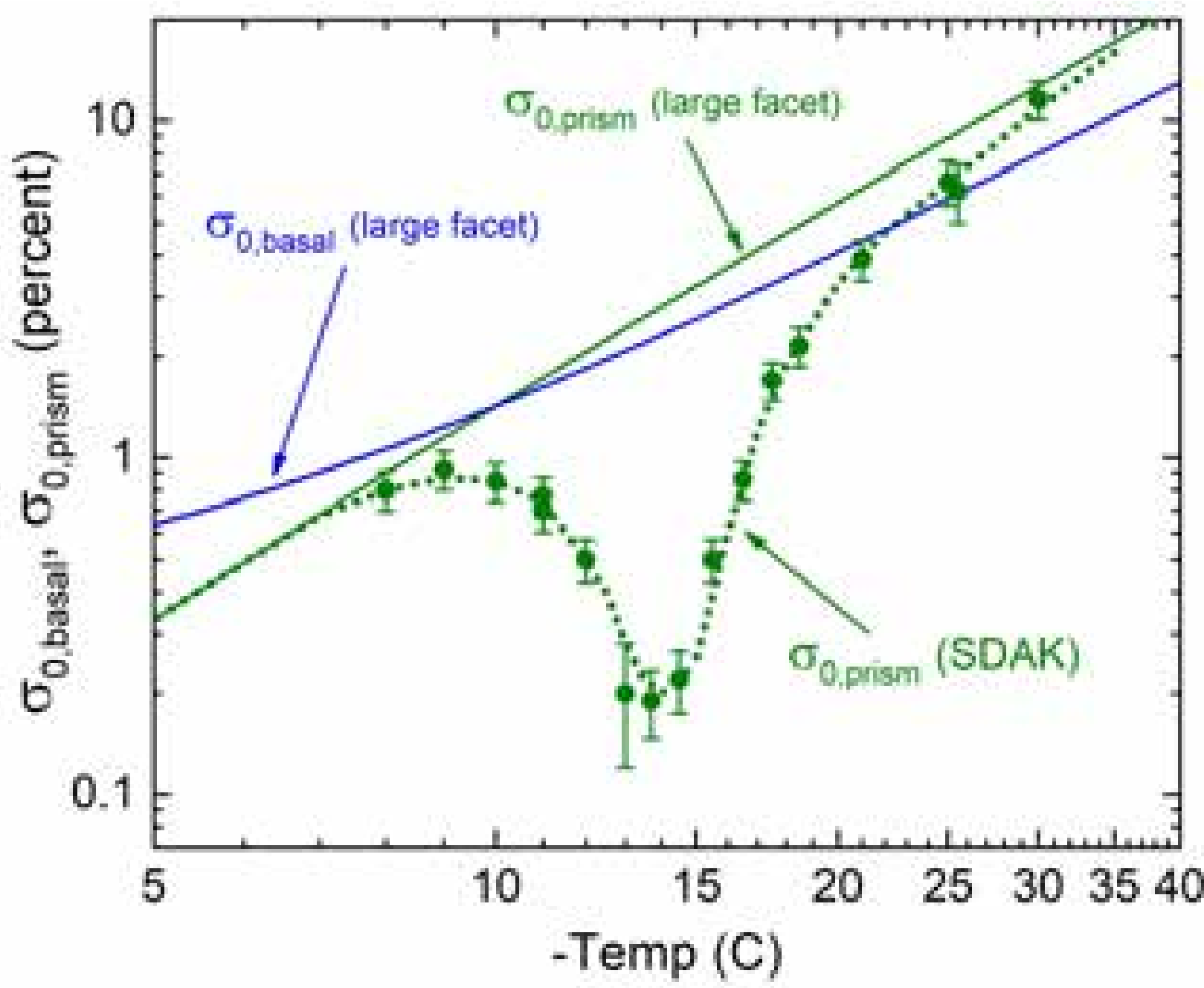

Figure 4.26: Measured values of $\sigma_{0,prism,SDAK}(T)$ as a function of temperature (data points), revealing a substantial "SDAK dip" centered near -14 C [2020Lib1]. For comparison, the solid lines show the CAK model parameters $\sigma_{0,basal}$ and $\sigma_{0,prism}$ for broad-facet growth.

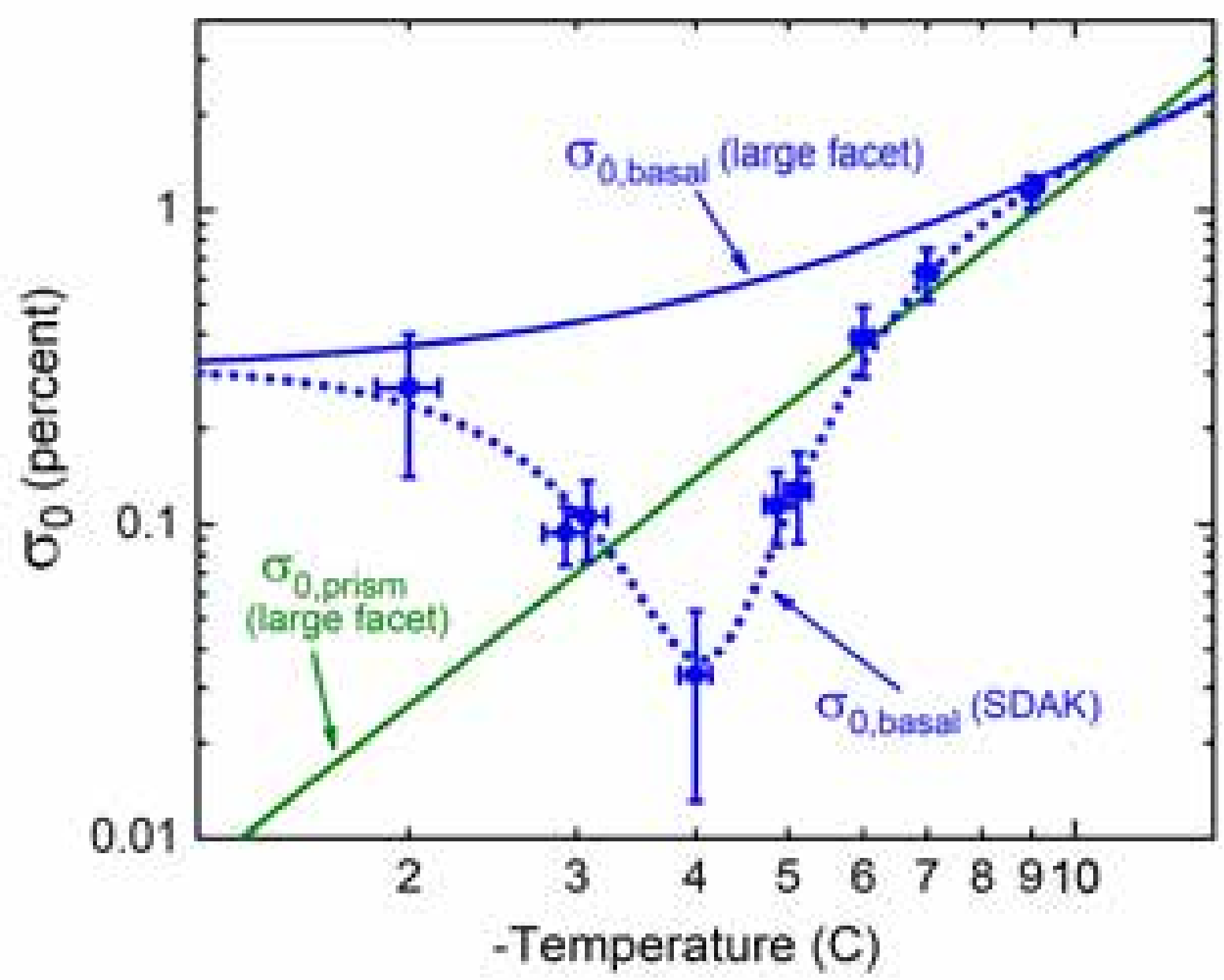

Figure 4.27: Measurements of the basal SDAK parameter $\sigma_{0,basal,SDAK}$ as a function of temperature, revealing a substantial "SDAK dip" centered near -4 C [2020Lib2]. For comparison, the solid lines show the CAK model parameters $\sigma_{0,basal}$ and $\sigma_{0,prism}$ for broad-facet growth.

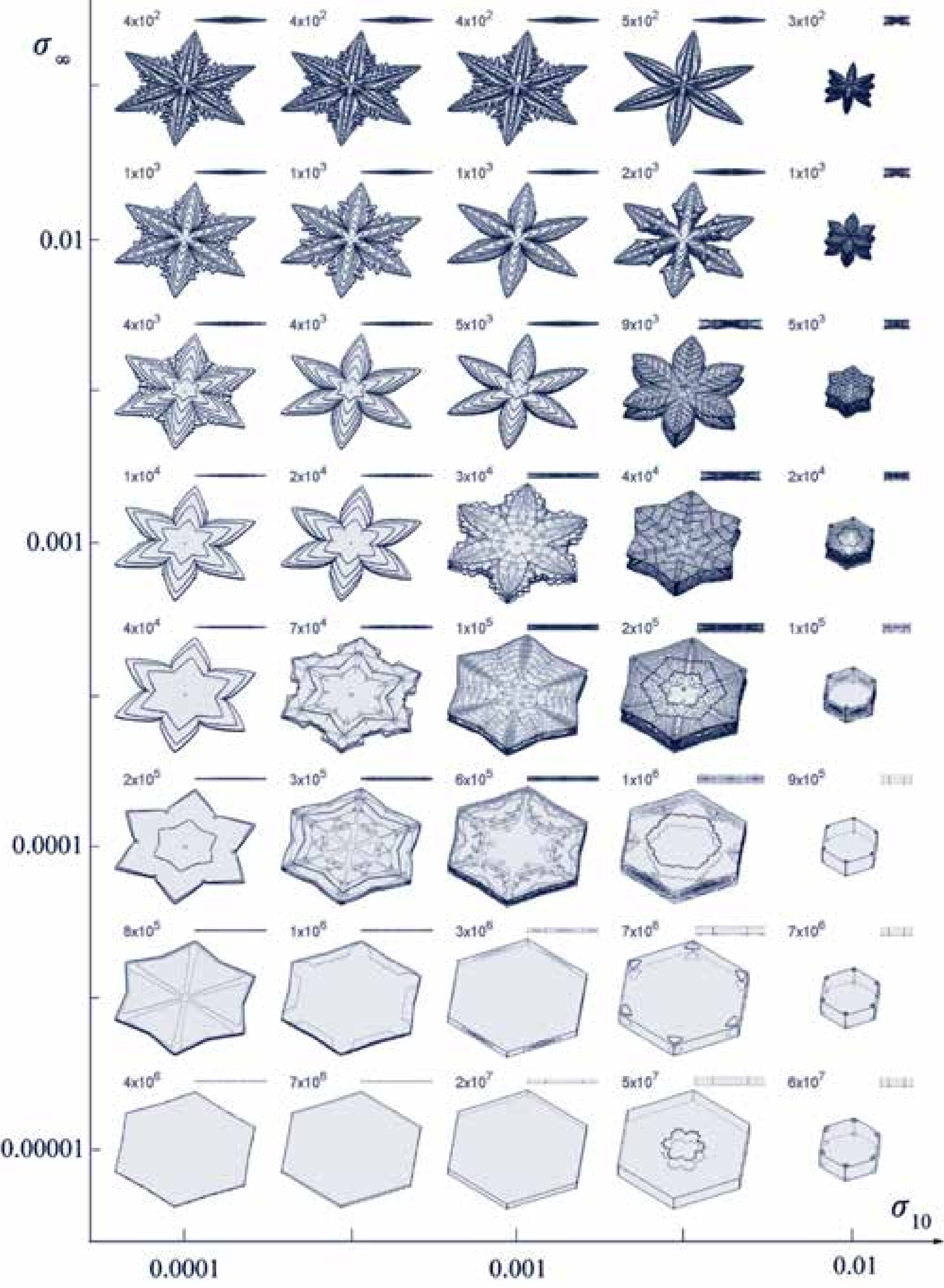

# CHAPTER 5

# COMPUTATIONAL SNOW CRYSTALS

*Nature is an endless combination and repetition of a very few laws. She hums the old well-known air through innumerable variations.*
– Ralph Waldo Emerson
*Essays, Lectures and Orations, 1851*

Computational modeling has become an important tool in contemporary science, and once again we find that the snow crystal presents a fascinating microcosm of modern scientific investigation. Being an intrinsically complex phenomenon, one does not simply develop a "theory" of snow crystal formation, at least not in the simplest sense of the word. Instead it is necessary to first break the problem down into its constituent parts, to better understand the variety of physical processes acting over different scales. Then one reassembles those parts to replicate the phenomenon as a whole, and that requires computational modeling.

**Facing Page: An array of snow-crystal models generated using the cellular-automaton method [2014Kel]. Different values of the supersaturation $\sigma_\infty$ (vertical axis) and $\sigma_{10}$ (horizontal axis) were used, the latter being a nucleation parameter in $\alpha_{prism}(\sigma_{surf})$. The basal attachment coefficient $\alpha_{basal}(\sigma_{surf})$ was the same for all models shown. Numbers give physical growth times in seconds.**

The scientific method remains intact in this view, but now the computational model becomes the hypothesis to be tested, as it predicts specific snow crystal structures for given environmental and physical inputs. If the hypothesis agrees with experimental measurements over a broad range of conditions, then we can rightly say that we have solved the problem to a large degree.

In the preceding chapters we focused on reductionist science in the classical sense, studying the different pieces of the snow crystal puzzle in isolation – the crystal structure and material properties of ice in Chapter 2, the molecular dynamics of attachment kinetics in Chapter 4, and the physics of diffusion-limited growth in Chapter 3. Using physical insights gained from these studies, we now examine numerical techniques that allow us to grow computational snow crystals.

In principle, building a computer model of a growing snow crystal is straightforward enough. Starting with a small digital ice crystal, first numerically solve the diffusion equation around it, assuming all the proper boundary conditions. From this solution, extract the growth rate at all points on the surface. Use this information to "grow" the crystal a small amount to yield a slightly larger crystal. Then repeat. After many iterations, the crystal develops into a complex morphology that hopefully resembles a laboratory snow crystal grown under the same physical conditions. Alas, although it all may sound straightforward in principle, developing appropriate numerical algorithms that can accomplish this task is far from trivial.

When setting out to create a snowflake simulator, one soon encounters error-propagation problems, numerical instabilities, uncertain geometrical factors, and numerous other issues that must be addressed. Moreover, a variety of shortcuts and approximations are required if one is to produce a realistic code with finite spatial resolution and a reasonable running time. Some quantities like local curvature may not even be precisely defined on complex polygonal surfaces, and the growth of faceted surfaces presents some unique challenges related to local geometry as well. As is often the case in science, the devil is in the details, and numerical modeling involves a lot of details.

Several different classes of computational strategies have been developed over the years for simulating a range of solidification problems, and each technique comes with its own strengths and weaknesses. Several of these methods have been applied to the specific problem of snow-crystal growth, but only with limited success to date. Developing a robust numerical method that reproduces crystal growth that is both branched and faceted remains very much a work in progress.

Because this book is about the science of snow crystal formation, the present chapter will focus on numerical modeling techniques that strive to produce physically realistic simulations, not just pretty pictures of snowflakes. Our objective is to create computational models that can be compared with experimental observations in a quantitative fashion, reproducing both growth rates and morphologies over a broad range of conditions. Moreover, the model underpinnings should derive from sound molecular and statistical physics to the greatest possible degree, rather than *ad hoc* parameterizations.

Importantly, a successful computational model should reproduce the full menagerie of snow crystal structures as a function of external growth conditions, including temperature, supersaturation, background gas pressure, and any other factors we care to include. Generating digital structures that resemble stellar snowflakes is a fine start, but this alone is not a bone fide scientific objective. The ultimate goal in this chapter is to develop a true physical model of snow crystal formation.

## 5.1 A Progression of Snow Crystal Models

The overarching topic of structure formation during solidification has received much attention in the scientific literature, and numerous reviews are available [2017Jaa, 2002Boe, 2016Kar, 2018Che]. The various algorithms and computational techniques can be mathematically quite sophisticated, and I am by no means an expert in this broad and technical field. To limit the scope of this chapter, therefore, I will mostly restrict the discussion to research efforts that have examined the specific problem of snow crystal growth.

### Packard Snowflakes

In 1986, Norman Packard described one of the first attempts to model structure formation during solidification using cellular automata (CA) methods [1986Pac]. Although Packard's CA rules were not physically derived, they revealed a rich variety of morphological structures that developed during growth, including the *Packard snowflake* shown in Figure 5.1. Packard's iterative cellular automata could be considered something of an extension of the ideas behind the Koch snowflake [1904Koc] (see Chapter 3), enabled by the easy availability of personal computers in the 1980s. These early models were intriguing for their ease in generating complex structures from simple governing rules, but they contained only a superficial relation to the actual physical process of solidification.

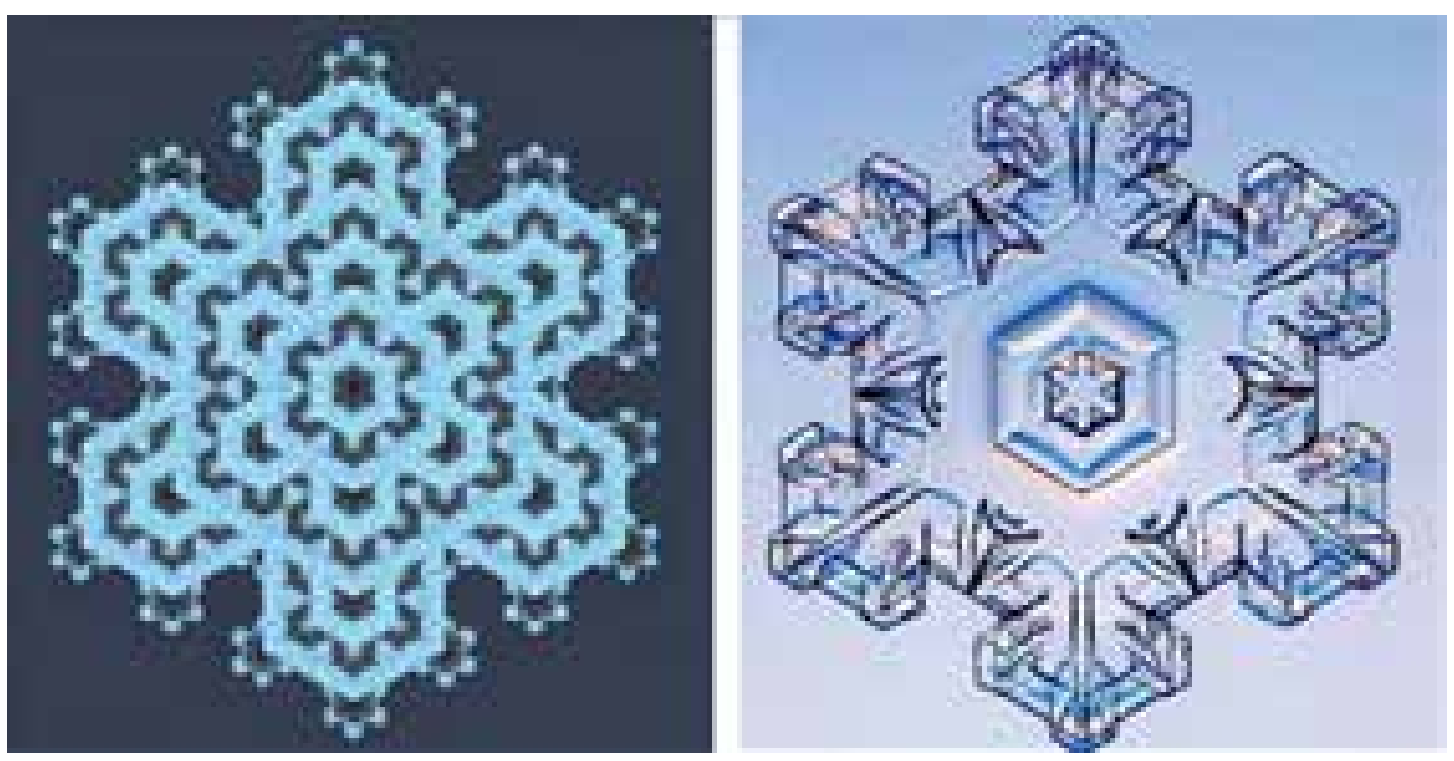

**Figure 5.1: A *Packard snowflake* (left), generated using simple nearest-neighbor rules in a cellular automaton, compared to a photograph of a natural snowflake (right). Although there are obvious structural similarities between the two images, the model has little basis in the physical processes underlying snow crystal growth. (Image adapted from [2008Gra].)**

## Diffusion-Limited Aggregation

In a landmark early paper modeling diffusive transport, Thomas Witten and Leonard Sander [1981Wit] examined the formation of metal-particle aggregates via a random-walk process that they called diffusion-limited aggregation (DLA). In their model, individual particles traversed a fixed grid in random small steps until they encountered a solid surface and stuck, thus simulating a crude form of diffusion-limited solidification. Being especially easy to implement on small computers, the DLA method was quickly adapted and applied across many fields to a wide range of physical phenomenon.

Rong-Fu Xiao, J. Iwan Alexander, and Franz Rosenberger carried the DLA method a step further by incorporating cellular-automata rules that attempted to simulate anisotropic attachment kinetics and molecular surface diffusion [1988Xia]. With a suitable adjustment of their model parameters, the authors demonstrated a clear transition from faceted to dendritic growth morphologies, as seen in Figure 5.2. Moreover, this transition resulted from the competing processes of particle diffusion and attachment kinetics, which is essentially the current paradigm of snow crystal formation. When applied to a fixed triangular grid, the Xiao et al. DLA model was the first to convincingly demonstrate this central snow-crystal morphological transition using rational (albeit not entirely accurate) physical foundations.

**Figure 5.2: A progression from faceted prism growth (a) to dendritic growth (d) in a two-dimensional DLA model [1988Xia]. This work was the first to demonstrate a morphological transition of this nature resulting from the competing processes of diffusion-limited growth and surface attachment kinetics.**

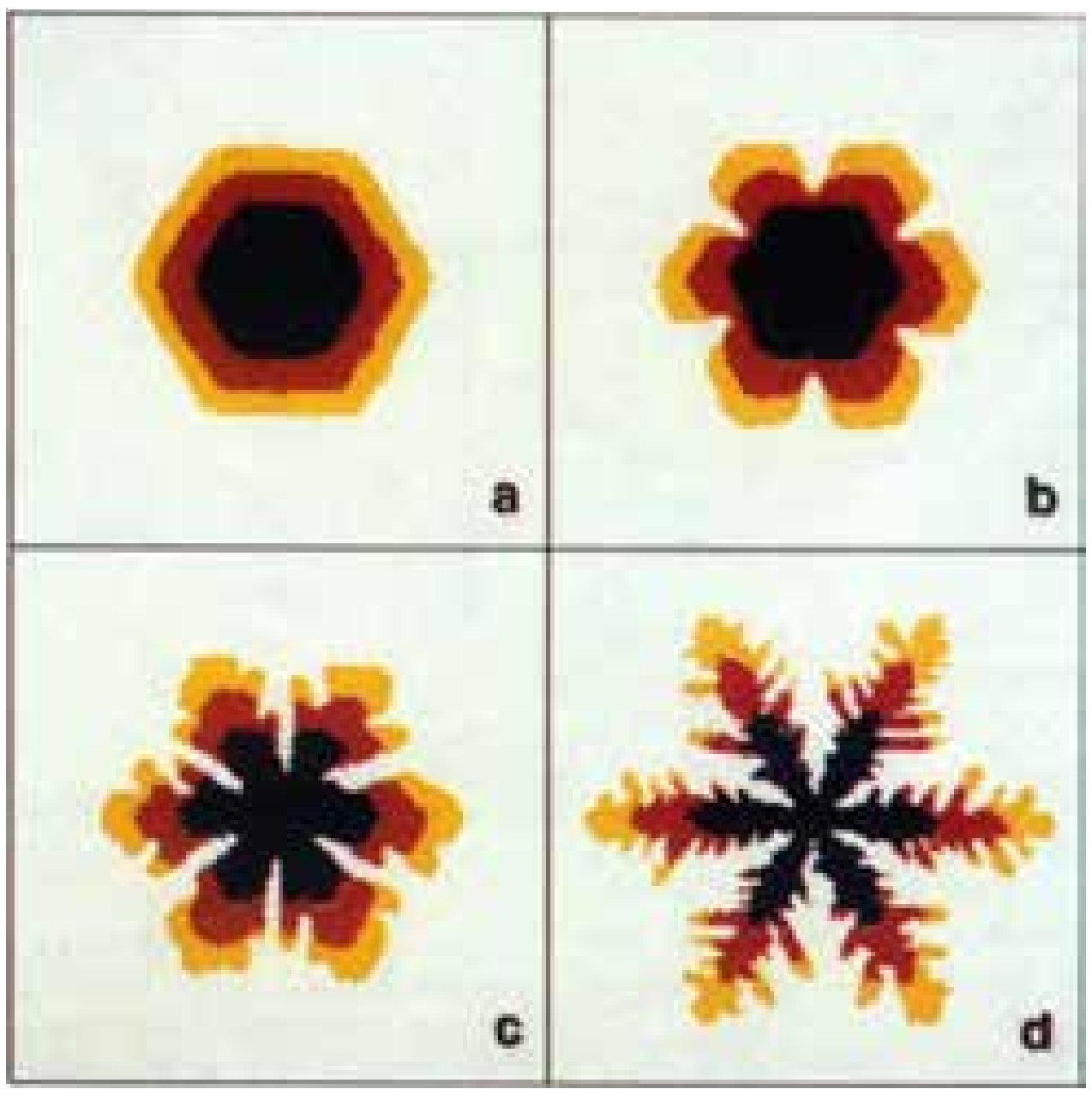

## Improving Physical Inputs

In 1990, Etsuro Yokoyama and Toshio Kuroda presented the first significant attempt to create a comprehensive physical model of snow-crystal growth dynamics [1990Yok]. By combined a novel molecular model of temperature-dependent attachment kinetics [1982Kur] with a numerical method for solving the diffusion equation, the authors sought to recreate the growth behavior of actual snow crystals under realistic environmental conditions.

While previous investigators had simulated general growth behaviors using *ad hoc* parameterizations, Yokoyama and Kuroda modeled the specific phenomenon of snow crystal growth from water vapor, including a careful examination of all the physical processes involved. Notably, the authors incorporated the known physical properties of ice and water vapor, allowing a direct quantitative comparison between simulated snow crystals and laboratory experiments. As the authors stated in their abstract [1990Yok]: "We propose a model of pattern formation in the growth of snow crystals that takes into account the actual elemental processes relevant to the growth of crystals, i.e., a surface kinetic process for incorporating molecules into a crystal lattice and a diffusion process."

In terms of numerical techniques, the authors began with the differential equations describing the diffusion of water-vapor molecules in air, along with a reasonable estimate for the boundary conditions at the crystal surface, including an attachment coefficient with deep cusps at the facet angles. The diffusion equation was solved using a Green's function method that generated the supersaturation field around the crystal along with the growth velocity at each point on the surface. The solidification front was then propagated in small steps to grow a two-dimensional snow crystal, as illustrated in Figure 5.3.

This paper, I believe, was the first to recognize the central importance of a detailed molecular model of the attachment kinetics for understanding snow crystal formation. Kuroda and his collaborators were also pioneering in their early realization that latent heating and surface-energy effects were relatively minor compared to the dominant processes of particle diffusion and surface attachment kinetics, as I discussed in detail in Chapter 3.

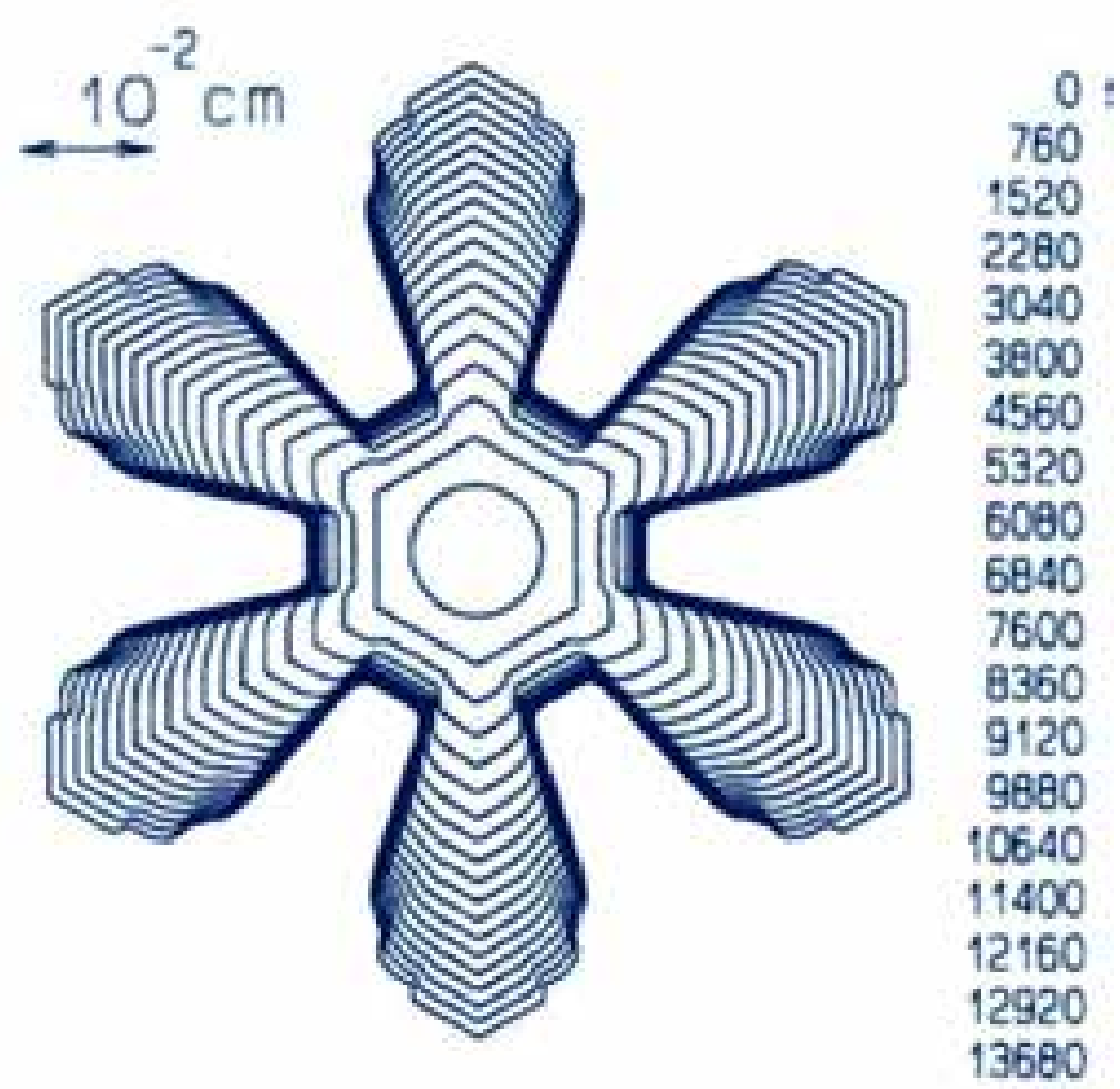

**Figure 5.3: This numerical model by Yokoyama and Kuroda [1990Yok] exhibits an initial transition from a round seed crystal to a faceted plate, followed by the formation of six primary branches. Unlike earlier investigations, these authors created the first detailed physical model of the specific phenomenon of snow crystal growth from water vapor.**

The computational model in Figure 5.3 exhibits the initial growth of a faceted prism followed by the development of six primary branches, both well-known phenomena in snow crystal formation. Note also that the model depicts actual physical sizes at real physical times, as is needed for comparison with experimental measurements. This was only a 2D simulation, using a largely incorrect model of the attachment kinetics (see Chapter 4), so it was not yet suitable for direct comparisons with experiments. Moreover, the Green's-function method used to solve the diffusion equation was inefficient compared to modern numerical techniques. Nevertheless, the authors' careful examination of the relevant physical processes was a substantial step toward developing a physically accurate model of snow crystal growth.

## Front-Tracking

Soon after these early modeling efforts, the field expanded rapidly as several innovative mathematical techniques were developed and applied to investigations of a variety of solidification phenomena. At first these studies focused mainly on solidification from the melt, which has metallurgical applications and involves relatively small anisotropies in surface physics (see Chapter 3). Just recently, however, the field has begun expanding into modeling crystal growth that exhibits both faceting and branching, including snow crystal growth.

**Figure 5.4: (below) A 3D front-tracking model of a plate-like snow crystal, showing a transition from faceted to branched growth [2012Bar].**

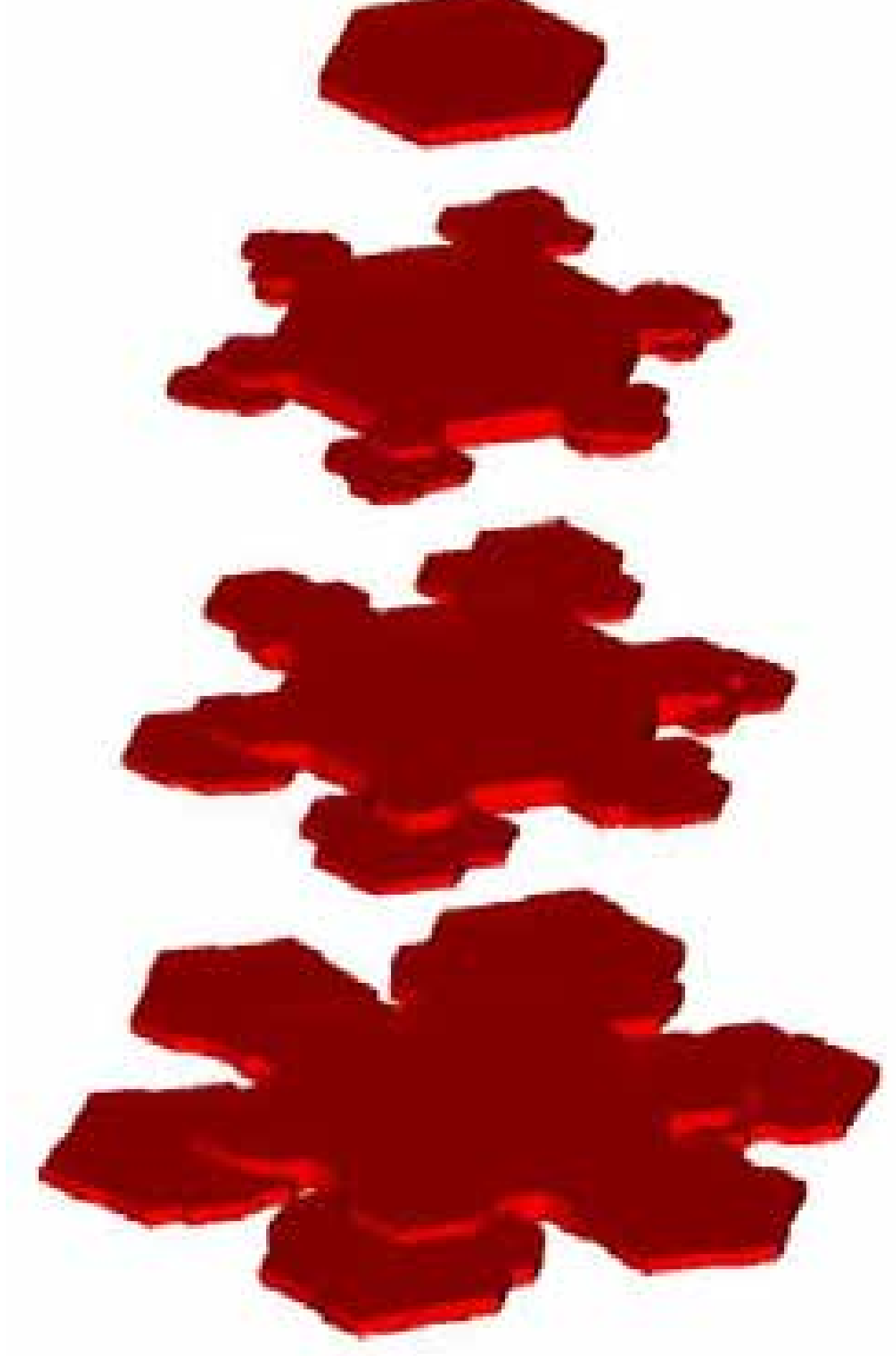

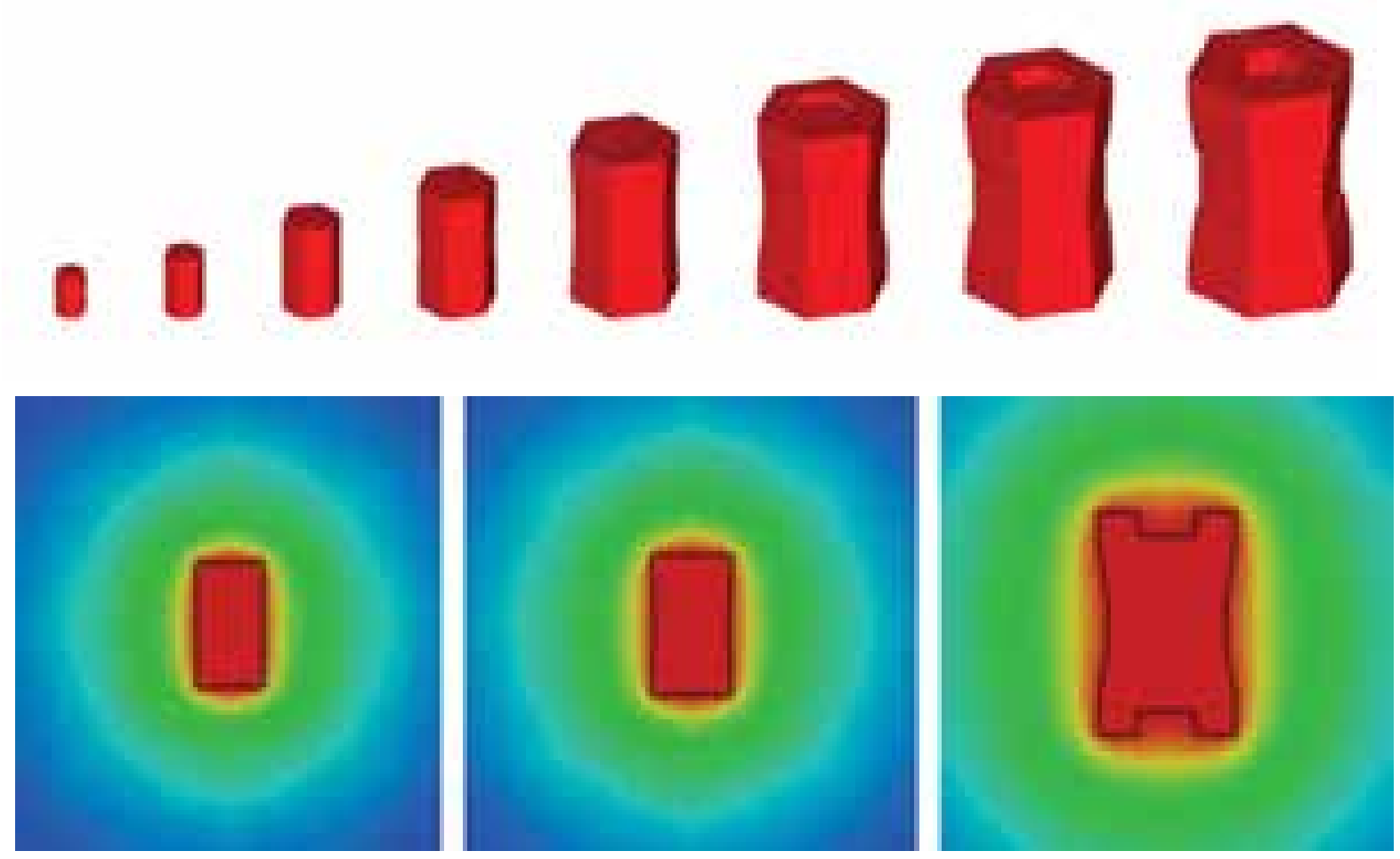

**Figure 5.5: A 3D front-tracking model of a hollow columnar snow crystal, showing a transition from faceted to hollow growth [2012Bar]. The top images show renders of the full 3D model as it developed, while the lower images show cross-sectional views that also depict the supersaturation field around the crystal.**

In 2012, Barrett, Garcke, and Nürnberg presented numerical simulations of growing snow crystals using a finite-element method in which the ice surface was approximated using an adaptive polygonal mesh [2012Bar]. This is one variant of a *front-tracking* strategy, as it defines a sharp solidification front between solid ice and the water-vapor field surrounding it [1996Sch, 2010Bar].

Figure 5.4 shows one example of a 3D simulation of a growing snow crystal from [2012Bar]. As with the Yokoyama and Kuroda model, Figure 5.4 again exhibits the initial formation of a faceted prism followed by a transition from faceted to branched growth. Figure 5.5 shows another example modeling the growth of a hollow column.

The authors concluded in this study that a substantial surface-energy anisotropy was necessary to produce faceted growth in their models, while anisotropy in the attachment coefficient was not enough to produce faceting. I believe that this conclusion is likely not correct, as the simulations in [2012Bar] examined only an extremely weak anisotropy in the attachment kinetics, far weaker than what is now expected (see Chapter 4). It is well known in materials science that highly anisotropic attachment kinetics promotes faceting, and this appears to be the dominant underlying cause of snow crystal faceting as well. Nevertheless, the work presented in [2012Bar] does demonstrate that their modern front-tracking numerical model can generate 3D structures that are both faceted and branched, which is a substantial step forward.

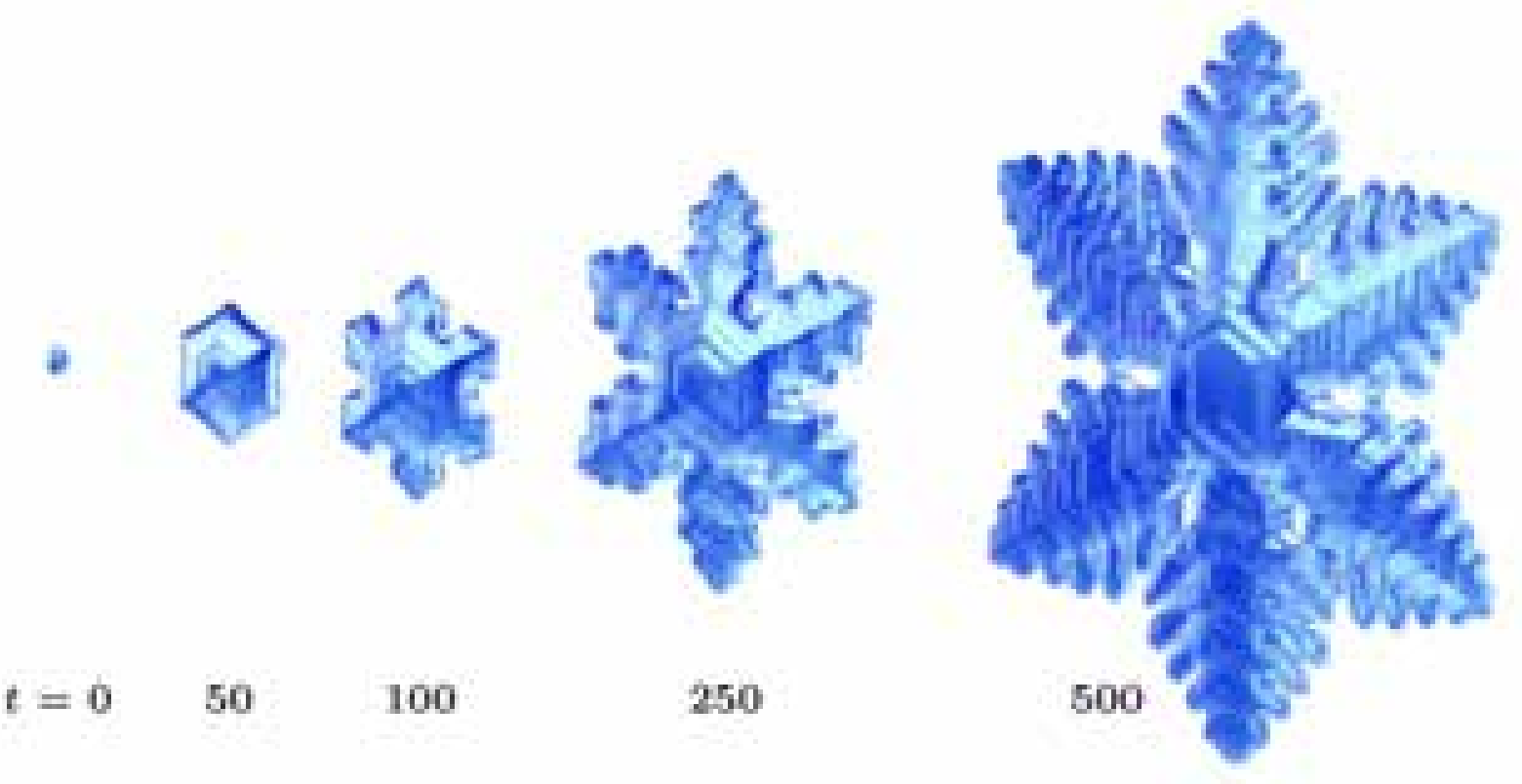

**Figure 5.6: A 3D phase-field model of a plate-like snow crystal, again showing a transition from faceted to branched growth, along with well-developed sidebranching [2017Dem1].**

## Phase-Field Snow Crystals

In 2017, Demange, Zapolsky, Patte and Brunel demonstrated a novel phase-field method for simulating snow crystal growth [2017Dem, 2017Dem1]. In contrast to front-tracking approaches, this method defines an artificial *phase field* parameter that equals -1 for the water-vapor phase and +1 for the ice phase, and this parameter varies smoothly between these values across a spatially diffuse interfacial region (spanned by at least several pixels in the model). By eliminating the sharp solidification boundary in this way, phase-field models can employ generally simpler numerical propagation algorithms [1996Kar, 1998Kar, 2002Boe, 2017Jaa].

In the phase field technique, the diffusion equation and its accompanying boundary conditions are replaced with a set of nonconservative phase-field equations. These equations represent a phenomenological description of the underlying microscopic interfacial physics that reduces to the correct physical description of the growth problem in the sharp-interface limit [1998Kar]. Once the proper phase-field equations have been determined, they are used to evolve the entire phase field in a uniform fashion, so no explicit front tracking is required.

Figure 5.6 shows an example of a 3D simulation of a growing snow crystal from [2017Dem] that again illustrates several features representative of stellar snow crystals, including the transition from faceted to branched growth, well-developed sidebranching, and rib-like surface markings. The authors were also able to reproduce several other commonly observed snow-crystal structures, as illustrated in Figure 5.7 [2017Dem1].

As with [2012Bar], however, the underlying physical parameters used in [2017Dem] were not realistic. For example, the attachment kinetics function was only weakly anisotropic and did not include the known basal nucleation barrier described in Chapter 4. The high degree of surface anisotropy needed to produce faceting instead came from the surface energy, which is likely not an accurate physical model for snow crystal dynamics. The Peclet number was also orders of magnitude higher in the model than in real snow crystal growth. These important technical points notwithstanding, the authors clearly demonstrated the potential of the phase-field method for modeling growth that is both faceted and branched, a necessary condition for creating accurate simulations of snow crystal growth.

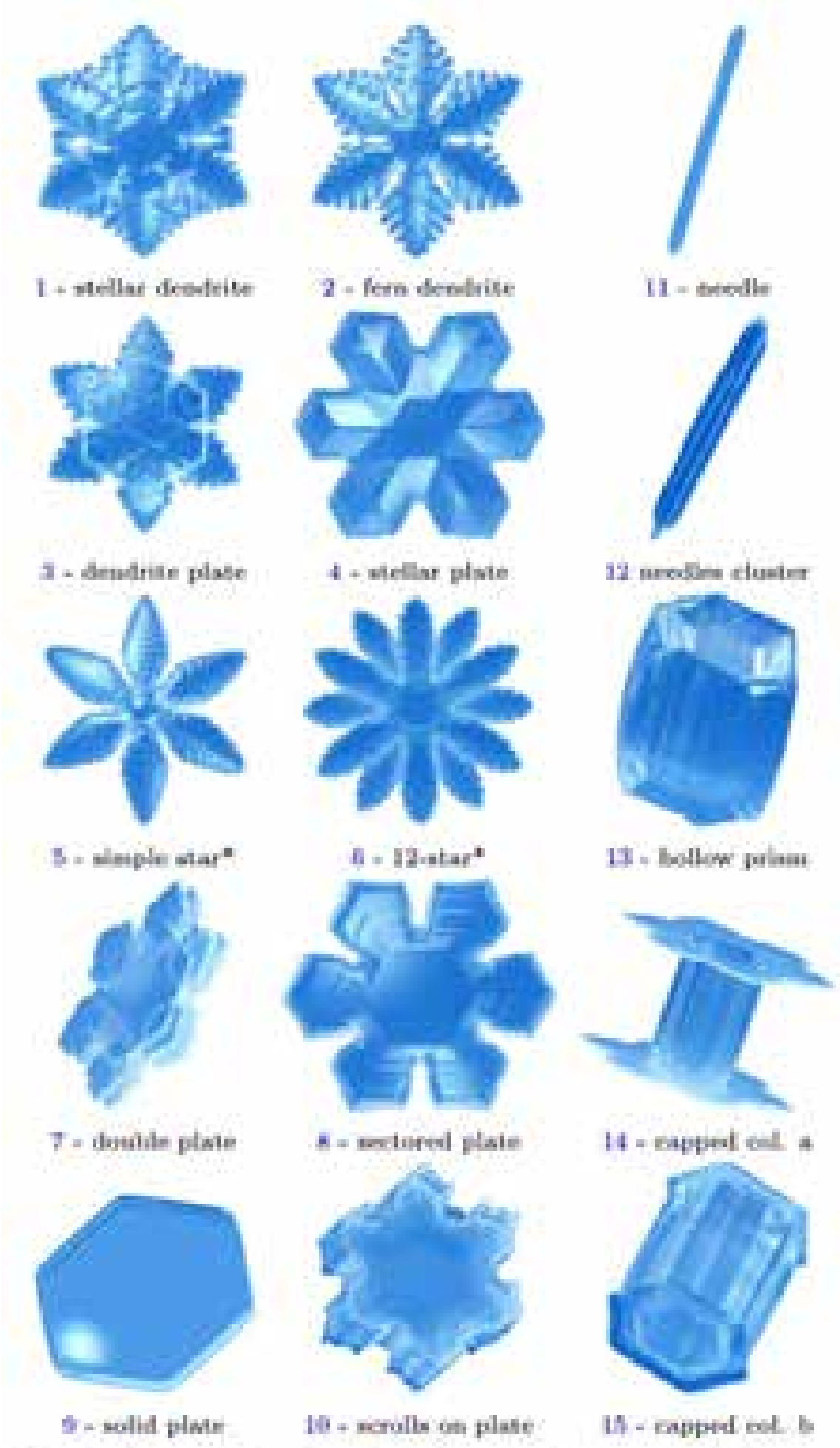

**Figure 5.7: Additional 3D phase-field models of several well-known snow crystal morphologies, from [2017Dem1].**

## Cellular Automata

Of the various computational strategies that have been applied to solidification problems so far, the cellular automata (CA) method has demonstrated the most promise (in my opinion) for providing a powerful research tool for investigating the physical dynamics of snow crystal growth. Much like the early Packard snowflakes [1986Pac], a CA model begins by defining a fixed grid having the same hexagonal symmetry as the ice-crystal lattice. Individual cells (a.k.a. pixels) on the grid are labeled as either ice or vapor, with vapor pixels having a value proportional to the water-vapor supersaturation. A set of CA "rules" evolves the supersaturation field with time and determines how pixels change their state from vapor to ice.

A CA model can describe physically realistic snow crystal growth if the rules are carefully chosen to simulate the actual physical processes involved. Both the mathematical structure and the numerical implementation of CA techniques are generally simpler than other simulation strategies, plus the results to date suggest that the CA method is better suited for handing the highly anisotropic attachment kinetics present in snow crystal growth.

Clifford Reiter first demonstrated the potential for creating realistic CA simulations of snow crystal growth when he presented a simple 2D model that yielded several snowflake-like structures, including those shown in Figure 5.8 [2005Rei]. Ning and Reiter described additional 3D models in [2007Nin].

Reiter's algorithms implemented nearest-neighbor rules that solved Laplace's equation in the region surrounding the snow crystal, thus accurately modeled the diffusion of water-vapor toward the growing crystal. The rules governing the conversion of vapor to ice had little basis in solidification physics, however, so the Reiter model did not describe the formation of actual snow crystals in a meaningful way. But it was a first step in what has turned out to be a very fruitful direction.

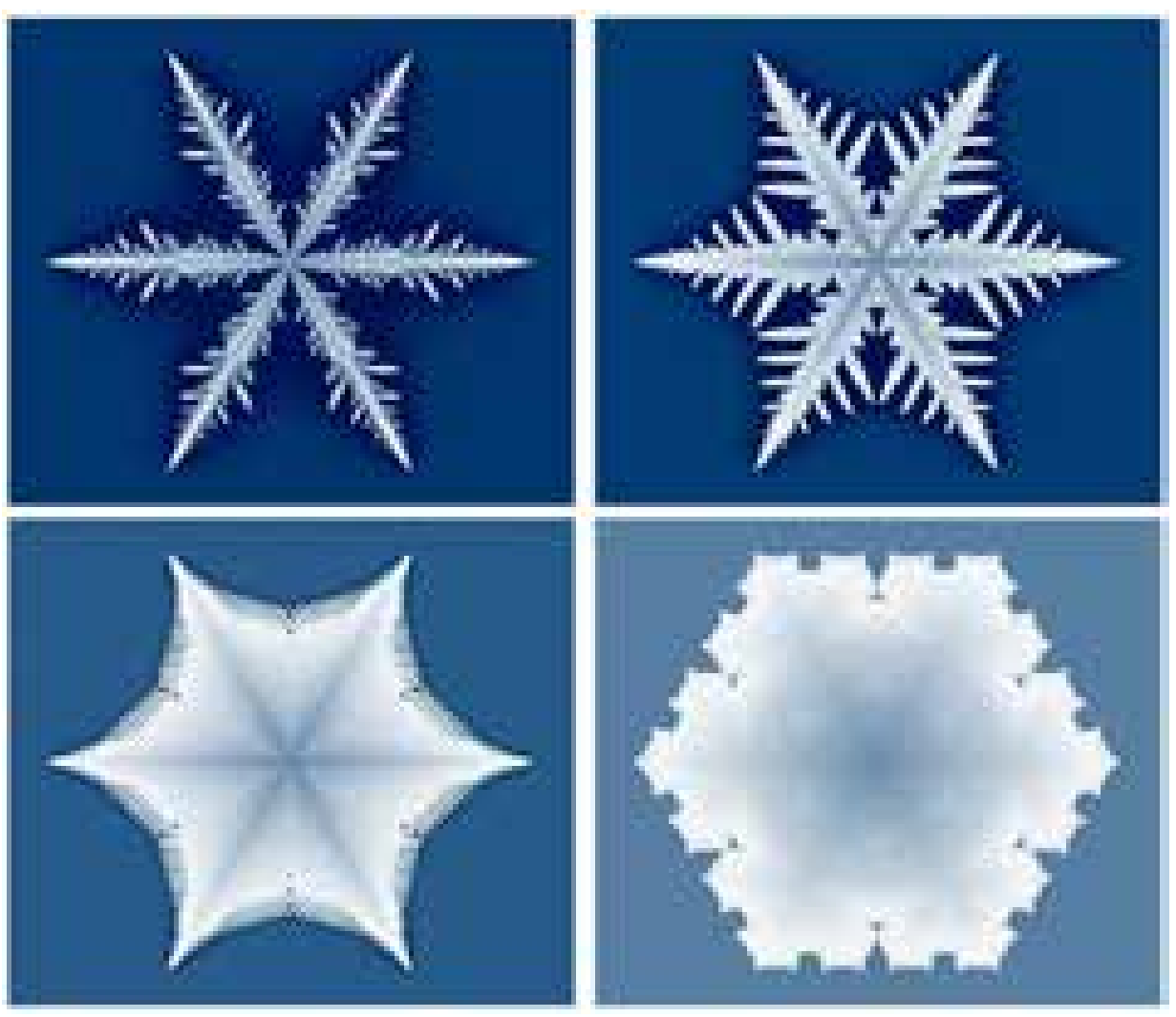

**Figure 5.8: Several 2D cellular automata models of diffusion-limited growth on a six-fold symmetrical lattice, exhibiting faceted and branched structures [2005Rei].**

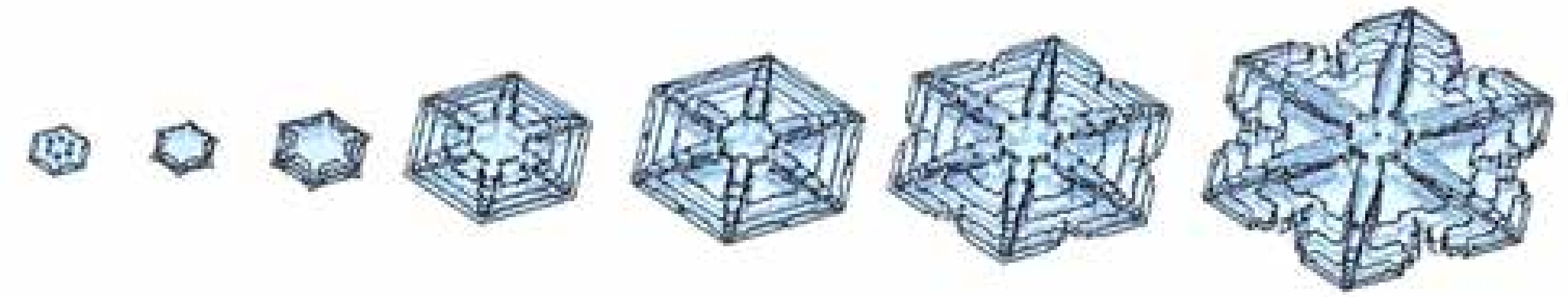

Janko Gravner and David Griffeath greatly expanded these ideas in a series of papers [2006Gra, 2008Gra, 2009Gra], the latest of which demonstrated a full 3D snow-crystal simulator that generated a remarkable diversity of snow-crystal-like morphologies, including details that had hitherto not been seen in any numerical simulations. The appearance of robust ridge-like structures on several stellar-plate morphologies is especially noteworthy, as these are also robust features in real snow crystals. A few representative examples from [2009Gra] are shown in Figures 5.9 and 5.10. Rendering a 3D model to produce a 2D image is also a nontrivial challenge, and Figures 5.11 and 5.12 show some particularly artistic renderings of Gravner-Griffeath snow crystals done by Antoine Clappier.

**Figure 5.10: A 3D cellular automaton model reproducing the development of a sectored plate snow crystal [2009Gra] The ridge formation on the convex basal surfaces is especially noteworthy in its resemblance to real snow crystal ridges.**

**Figure 5.9: (below) Several 3D cellular automata models by Gravner and Griffeath [2009Gra] demonstrating structures that are both faceted and branched, reproducing many morphological features found in real snow crystals.**

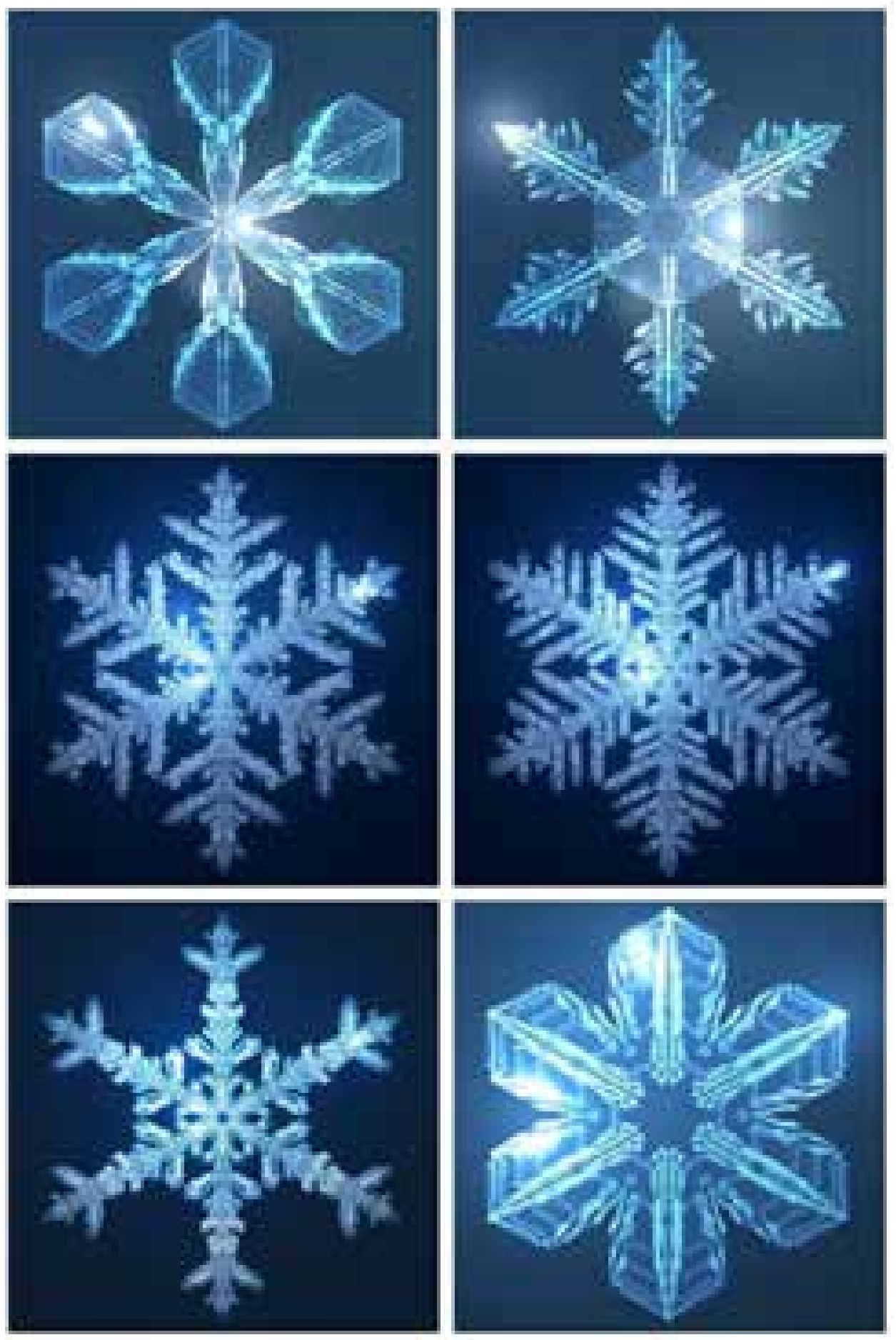

The Gravner-Griffeath work was (in my opinion) a significant breakthrough in modeling snow crystal growth, as it so clearly demonstrated the great potential of the cellular-automata method, especially for full 3D simulations. In additional to modeling several common snow crystal types, many surface structural details matched those seen on natural crystals to a remarkable degree. All previous numerical models of solidification had shown little or no adeptness for generating structures that are simultaneously faceted and branched, and this problem is still present to some degree in several of the computational strategies described above. The CA method, on the other hand, appears to be almost ideally suited for handling faceted+branched structures produced by diffusion-limited growth with highly anisotropic attachment kinetics.

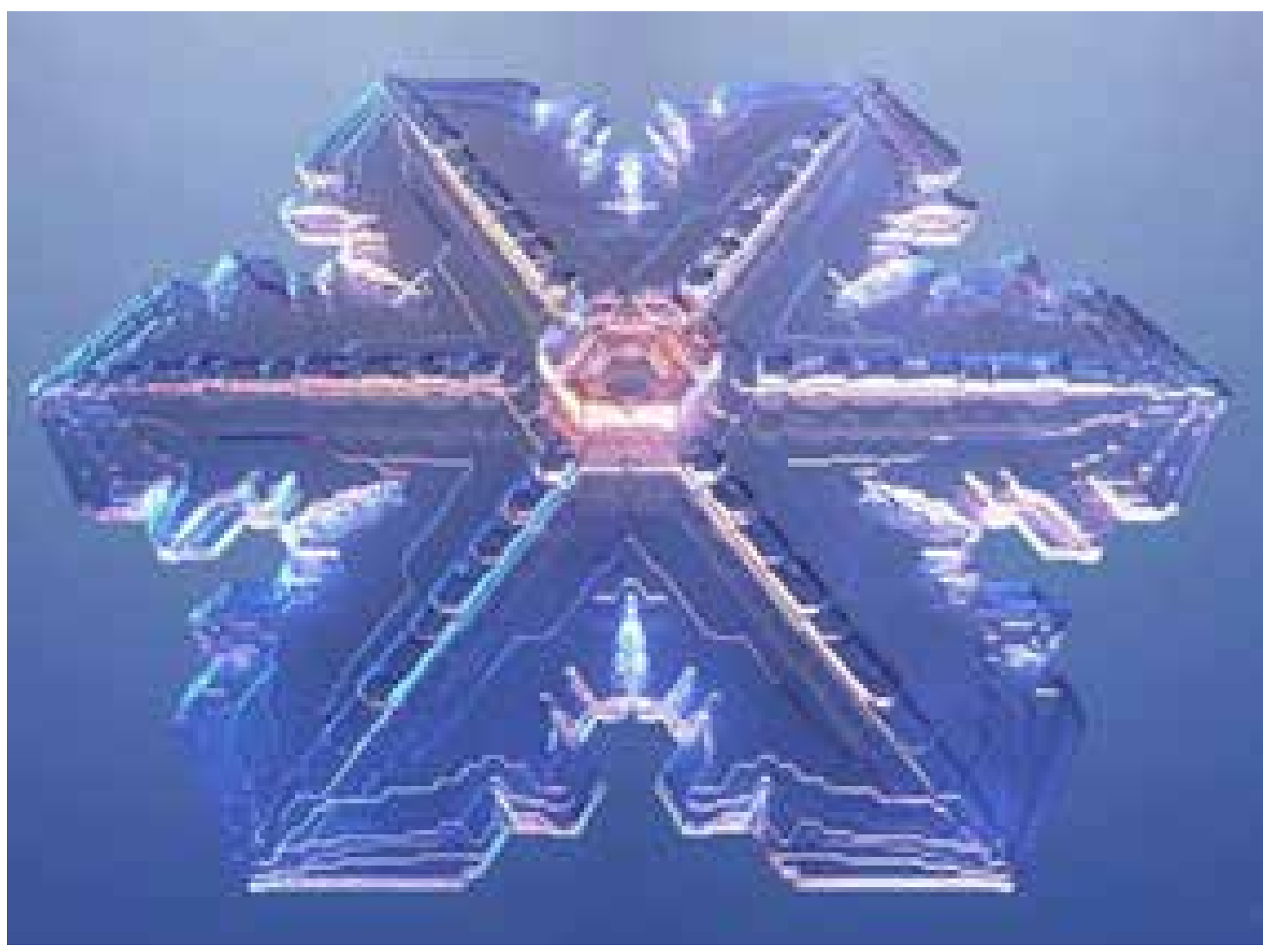

**Figure 5.11: A Gravner-Griffeath 3D snow crystal model beautifully rendered by Antione Clappier.**

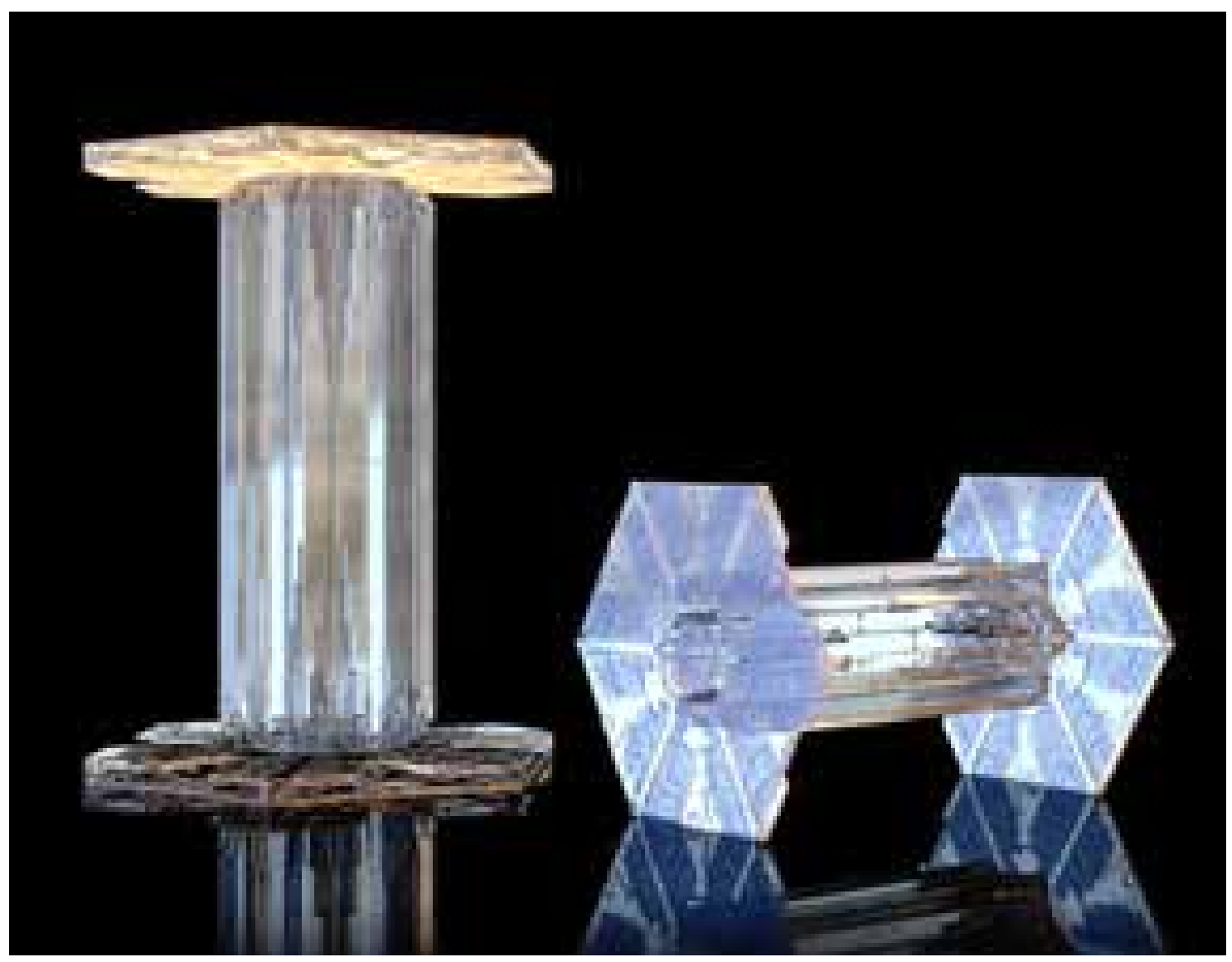

**Figure 5.12: An abrupt change in model parameters can yield the formation of capped columns with axial hollows, as illustrated in this Clappier-rendered Gravner-Griffeath 3D model.**

However, a substantial remaining problem with the Gravner/Griffeath model [2009Gra] was that it used a parameterized set of CA rules that were largely *ad hoc* and even nonphysical in nature. For example, the model imposed an artificial boundary condition setting $\sigma_{surf} = 0$ on all ice surfaces. This is technically true only in equilibrium, and thus is not correct for any growing snow crystal. Setting $\sigma_{surf} = 0$ may be a reasonable approximation for diffusion-limited growth in some circumstances (see Chapter 3), but it is not a suitable general assumption to make.

Kelly and Boyer [2014Kel] further pointed out that the Gravner/Griffeath model does not always obey mass continuity in the attachment step, in that crystal growth is not accompanied by a corresponding removal of water vapor from the air. Moreover, there was no clear relationship between the parameterized CA rules and the known physical properties of ice crystal attachment kinetics. Thus, while the parameters in the Gravner/Griffeath algorithm could be adjusted to yield remarkably realistic snow crystal structures, the surface boundary conditions were not appropriate for a physically accurate model.

## Physically Realistic Cellular Automata

The problem of creating a CA snow crystal model with physically derived rules was soon addressed by Libbrecht [2008Lib, 2013Lib1], who further investigated the incorporation of surface-energy effects [2013Lib1], surface diffusion [2015Lib1], and the edge-sharpening instability [2012Lib3, 2015Lib2] in a CA model with anisotropic attachment kinetics. Using a 2D model of cylindrically symmetrical 3D growth, this allowed some of the first direct, quantitative comparisons of simulated snow crystal growth with laboratory measurements [2015Lib2], as I describe below.

James Kelly and Everett Boyer made substantial additional progress by developing a full three-dimensional CA model with sound physical foundations, thus beginning a systematic study of 3D snow-crystal growth behavior as a function of a parameterized attachment kinetics [2013Kel, 2014Kel]. Some results from this work are shown at the front of this chapter and in Figure 5.13.

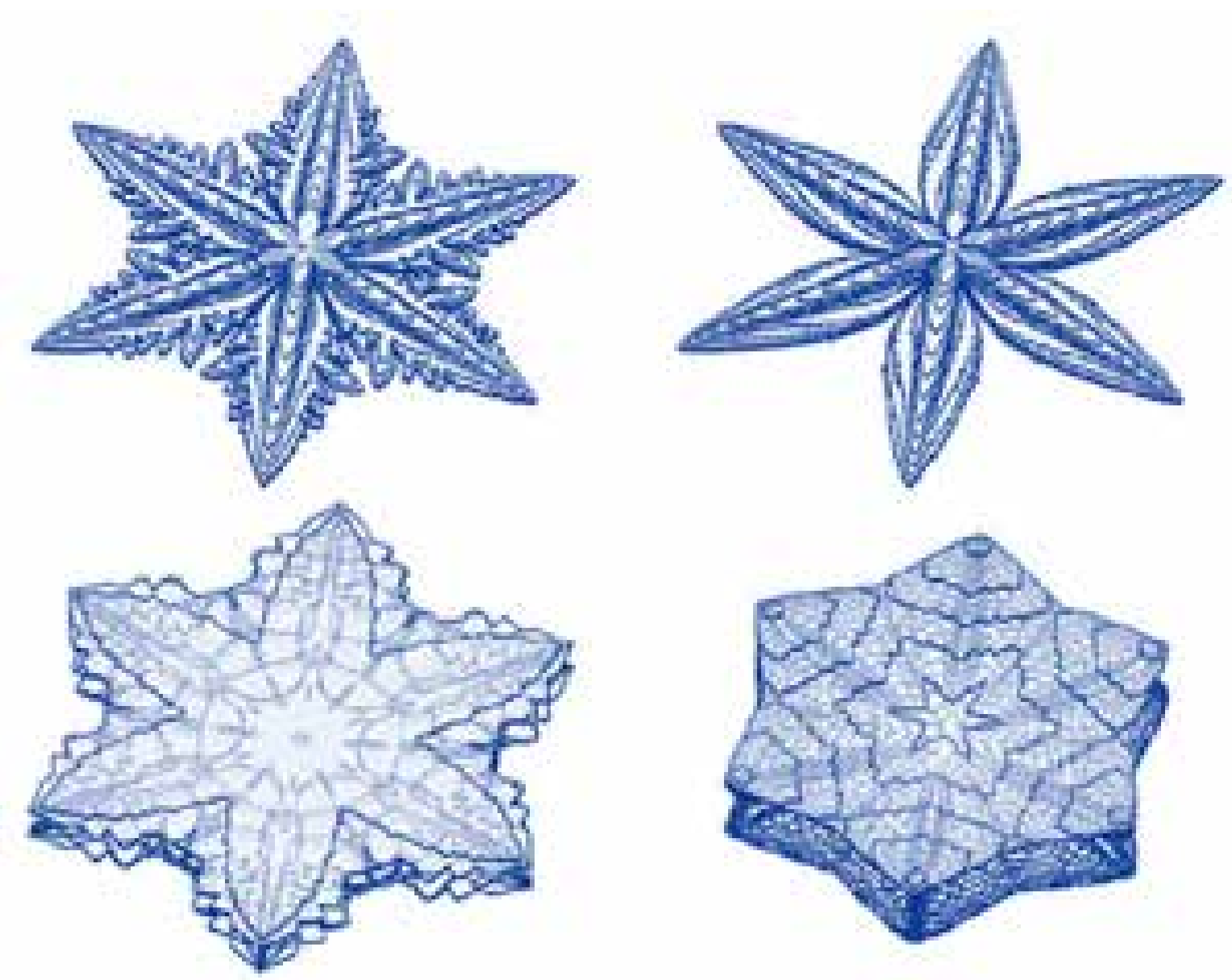

**Figure 5.13: A few representative 3D snow crystal models created by Kelly and Boyer [2014Kel], using CA rules derived from physically realistic calculations. More examples can be found at the front of this chapter. Again, these models exhibit several morphological features found in real snow crystals.**

It is now becoming clear, from these recent model studies, that perhaps the biggest impediment to creating accurate computational snow crystals is our poor understanding of the surface attachment kinetics over a broad range of growth conditions (see Chapter 4). Nevertheless, as further physical insights and model improvements are realized [2016Li], I expect that the CA technique will become the method of choice for modeling snow crystal growth, at least in the near term, and I discuss the specific algorithms and physical underpinnings in more detail below.

## Comparing Different Computational Methods

Although cellular automata models have produced the most impressive snow crystal results to date, other techniques show great promise as well. Numerical modeling of structure formation during solidification is a rapidly evolving field, so it makes sense at this point to briefly compare the different computational approaches.

The first thing to note is that all the existing computational techniques can solve the particle diffusion equation in free space with ease, especially as the Laplace approximation simplifies the problem considerably (see Chapter 3). The phase-field, front-tracking, and CA models all take different mathematical approaches to solving the diffusion equation, but the results are all basically the same, and all are highly accurate. The main differences between techniques lie not in solving the free-space diffusion equation, but rather how the surface boundary conditions are handled and how surface growth is propagated.

In many respects, polygonal front tracking methods seem the most natural when dealing with a continuum phenomenon like crystal growth. A distinct solidification front makes perfect sense for snow crystal growth, as there is an extremely sharp transition between the vapor and solid phases at the ice surface, just a few molecular layers in thickness (see Chapter 2). Because the molecular size is so small compared to even the smallest morphological features being modeled, a continuum model with a sharp interface is an excellent approximation.

Moreover, a polygonal surface is a reasonable computational model for almost any morphological situation, and the underlying surface physics is well defined on such a surface. As long as the grid is small enough, a front tracking algorithm should be capable of modeling all manner of solidification problems, including snow crystal growth.

One disadvantage with front-tracking, however, is the algorithmic complexity involved with deriving and continually adapting the polygonal solidification surface and the polygonal mesh that surrounds it. I have not worked in this area myself, but my impression is that it took many years to develop the computational tools needed to manage the diffusion equation with its nontrivial surface boundary conditions on an ever-adapting polygonal mesh. On the other hand, now that the required algorithms have been established, perhaps it is straightforward to apply them to a new physical system like snow crystal growth. But it does appear to be a somewhat daunting task.

Perhaps the biggest uncertainty in creating a suitable front-tracking code for snow crystal growth involves dealing with highly anisotropic attachment kinetics and faceting. The facet planes are unusual in that $\alpha_{basal}$ and $\alpha_{prism}$ can be *much* smaller than $\alpha_{vicinal}$, even when the vicinal angle is extremely low. Put another way, the attachment coefficient $\alpha(\theta_{surf})$ as a function of surface angle may have extremely sharp and deep cusps at the facet angles. This likely requires some special treatment of the facet surfaces, as Yokoyama and Kuroda noted even in their early examination of snow crystal modeling using front tracking [1990Yok]. It certainly does not seem unfeasible that one could incorporate highly anisotropic attachment kinetics into a front-tracking model. However, it has not been done to date, so we cannot say for sure how difficult such a task might be.

In contrast to front-tracking models, phase-field techniques are typically applied on a fixed coordinate grid, thus avoiding the use of complex polygonal meshes. This may also facilitate the preferred treatment of faceted surfaces, as the grid coordinates can easily be defined to be along facet planes. As with front-tracking, however, the issue of faceted growth remains a substantial uncertainty regarding our desire to model snow crystal growth, as highly anisotropy attachment kinetics have not yet been adequately explored in phase-field models.

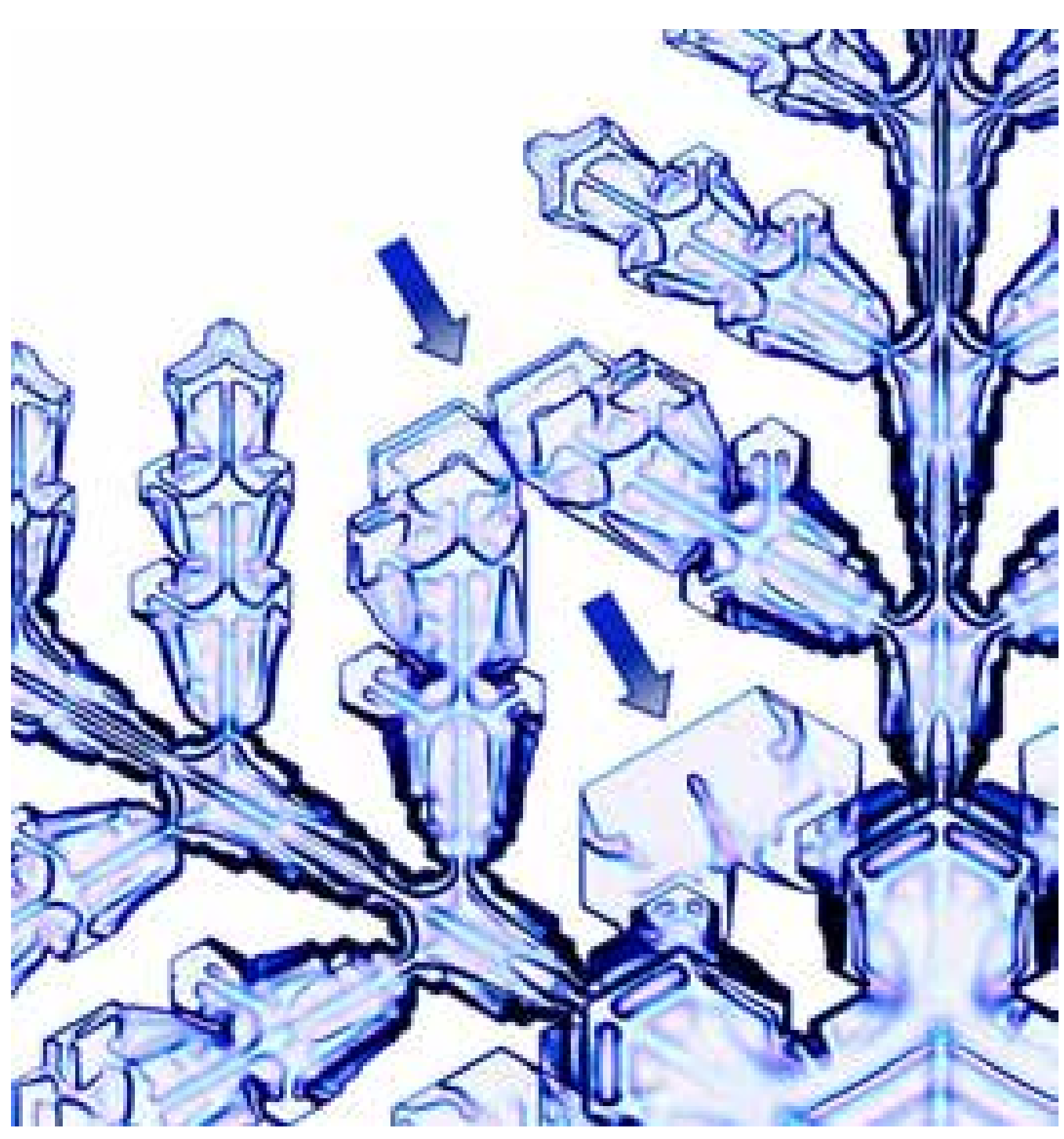

**Figure 5.14: A laboratory-grown POP snow crystal with two instances of sidebranch mergers (arrow). The upper merger occurred relatively recently before the photo was taken, so the separate branch edges are still clearly seen. The lower merger is older, and the individual sidebranches have grown together into a single flat plate.**

One disadvantage with phase-field models is the difficulty inherent in defining the phase-field equations. Once again, I have not worked in this area, but my impression is that deriving an appropriate set of phase-field equations from a specific parameterization of the surface boundary conditions is not a trivial task. It is unclear (to me) that finding phase-field equations that properly incorporate highly anisotropic attachment kinetics will be entirely straightforward. And when addition physical features are added, like surface energy effects and surface diffusion, the problem becomes that much more challenging. Once again, this is not necessarily a show-stopper for phase-field models, just an uncertainty in that the problem has not yet been adequately studied.

One excellent advantage of the phase-field method is that it nicely handles the merging of solidification fronts. In metallurgical applications, for example, one expects that a melt will eventually fully solidify into a solid block, and this process likely involves the merging of numerous independent solidifying structures into a final matrix of solid domains. Moreover, the material properties of the solidified metal may depend strongly on the size, structure, and arrangement of the domains. This kind of domain merging happens naturally in a phase-field model but is something of a computational nightmare in a front-tracking model.

Alas, this merging advantage is not so important in modeling snow crystal growth. Sidebranches do often merge, and one example is shown in Figure 5.14. But merging events like these are not a central feature in snow crystal growth compared to more basic morphological features. As the authors nicely described in [2014Bar1], "The main advantage of phase field methods over direct front tracking methods is that they intrinsically allow for topological changes. However, for the problem of solidification and dendritic growth as considered in this paper, topological changes are rare."

Cellular automata models are not especially popular in metallurgical solidification modeling, and yet they have demonstrated a remarkable aptitude for modeling snow crystal structures. A big reason for this success is that CA models can easily incorporate highly anisotropic attachment kinetics. By defining a grid with the same symmetry as the underlying ice crystal lattice, it is straightforward to give special treatment to faceted surfaces, as this is practically built into the model structure. One important downside of this rigid grid structure, however, is that it is nearly impossible to create a CA model that does *not* include some level of intrinsic numerical anisotropy in the surface boundary conditions, as I will describe in some detail below.

Another outstanding feature of cellular-automata models is that they are remarkably simple to define and build, plus the run times are relatively short. Moreover, the CA rules can be derived fairly easily from physical foundations, allowing straightforward parameterizations of the attachment kinetics and other physical effects. This will become apparent as I focus the majority of the remainder of this chapter on developing CA techniques specifically for modeling snow crystal growth.

## Facet-Dominated Growth

Much of the scientific literature on solidification modeling focuses on metallurgical systems, where the material anisotropies (mostly in the surface energy) are quite small, perhaps a few percent. However, as we learned from solvability theory (see Chapter 3), even these small anisotropies are critical in determining dendritic growth morphologies. For this reason, cellular automata techniques are a poor choice for modeling metallurgical solidification. In the opposite extreme, however, CA models appear to be quite well suited for highly anisotropic problems like snow crystal growth.

Snow crystal growth is somewhat unique in the field of solidification modeling because of the importance of highly anisotropic attachment kinetics. To my knowledge, snow crystal growth is the only highly anisotropic system that has received much attention, either theoretical or experimental, from the standpoint of understanding the basic physics of solidification and structure formation. Beginning with the careful studies of dendritic growth by Glicksman and others in the 1980s (see Chapter 3), nearly all substantial scientific efforts aimed at numerical solidification modeling were focused on weakly anisotropic metallurgical systems.

In a typical metallurgical system growing from the melt, the Peclet number is high, growth is largely limited by thermal diffusion, weakly anisotropic surface energy dominates the surface boundary conditions, and attachment kinetics is either weakly anisotropic or ignored altogether. In these systems, dendritic structures typically exhibit no faceting whatsoever, and it is imperative that computational models include very low intrinsic numerical anisotropies.

Snow crystal growth is, in many ways, a completely different problem. The Peclet number is very small, particle diffusion is more important than heat diffusion, surface energy effects are almost negligible, and anisotropic attachment kinetics is a central player in bringing about highly faceted dendritic structures. In snow crystal growth modeling, both $\alpha_{prism}$ and $\alpha_{basal}$ are often small and highly dependent on $\sigma_{surf}$, while one can reasonably assume $\alpha \approx 1$ on nearly all nonfaceted surfaces. I call this a "facet-dominated" growth regime, as the overall growth rates and morphologies are largely defined by the growth of the faceted surfaces.

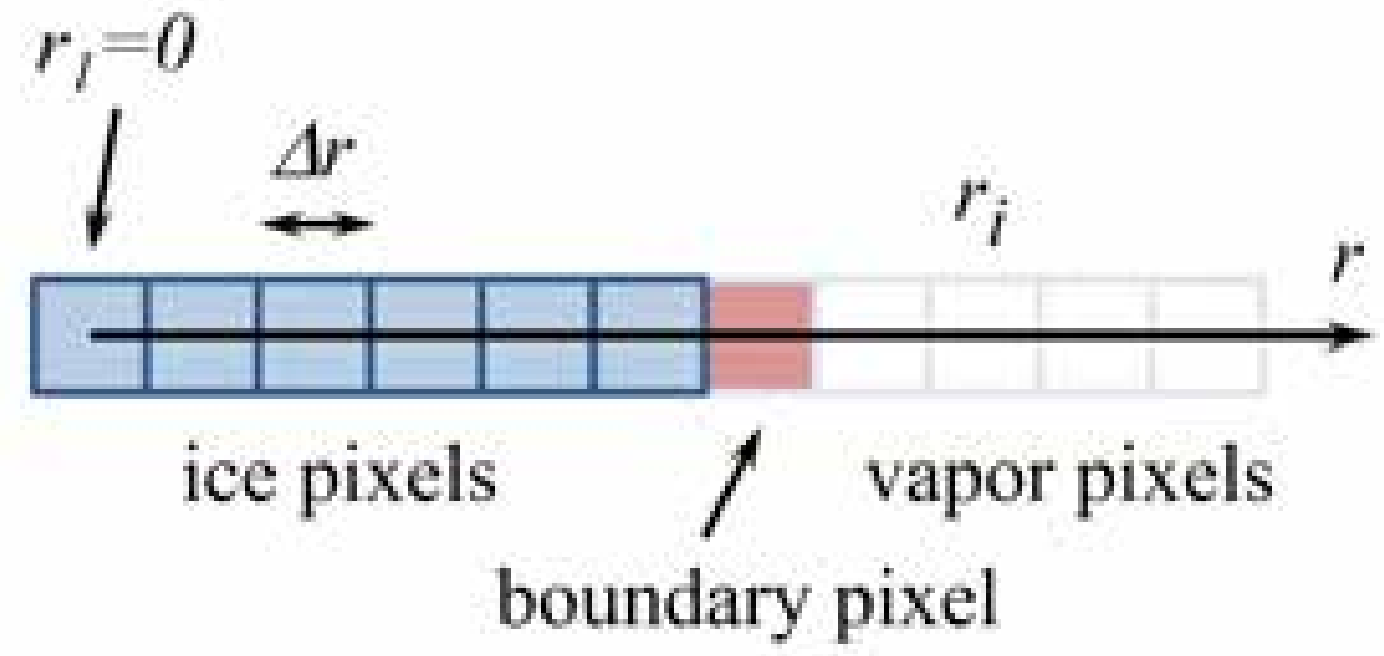

**Figure 5.15: The radial layout of cells (a.k.a. pixels) for a one-dimensional spherical cellular-automaton model. At any given time, the model consists mainly of ice pixels (blue) and vapor pixels (white). The red "boundary" pixel is a vapor pixel that neighbors an ice pixel.**

Modeling facet-dominated growth requires an especially accurate treatment of the facet dynamics, which means a careful handling of anisotropic attachment kinetics. On the other hand, a somewhat sloppy treatment of non-faceted surfaces may be tolerable. Thus, although it is not clear that one can build a perfect CA model even in principle, it may nevertheless be possible to build a CA model that reproduces most snow crystal morphologies with reasonable fidelity. The early results look quite promising, but the only real way to know for sure is to start building physically accurate models that allow quantitative comparison with careful experimental observations.

## 5.2 Spherical Cellular Automata

I focus the remainder of this chapter on cellular-automata models, as they are the leading contender for creating realistic computational snow crystals, at least in the short term. Opinions may differ on this, as other numerical strategies are promising as well, and we may uncover serious inherent limitations in the CA technique with additional study. Nevertheless, given how little modeling effort has focused on strongly faceted solidification, and how rapidly the field is evolving, I suspect the best strategy at this point is simply to dive in and see how far the cellular-automata method can take us.

The discussion below derives mainly from work I have done in developing CA models for snow crystal growth using physically derived CA rules [2008Lib, 2012Lib3, 2013Lib1, 2015Lib1, 2015Lib2], combined with excellent model improvements from Kelly and Boyer [2013Kel, 2014Kel], as well as a few additional (unpublished) tweaks I have been investigating recently. My main focus with these models has been not just on morphologies, but on developing quantitative CA rules that accurately reflect the underlying physical processes that govern snow crystal growth.

Little is finished in this field at present, especially because our understanding of the attachment kinetics is quite poor (Chapter 4). As targeted experiments address this problem (Chapter 7), my hope is that better computational models will soon allow detailed comparisons with complex morphologies grown under well-known conditions, especially on electric ice needles (Chapter 8). The ultimate goal is that all these efforts in parallel will eventually combine to yield a truly comprehensive model of snow crystal formation.

## The Diffusion Equation

For pedagogical reasons, I like to begin the discussion by creating a numerical model of the simplest possible physically interesting one-dimensional problem – the growth of a spherical crystal governed solely by particle diffusion and attachment kinetics. There is hardly any need for numerical modeling of this system, of course, as an exact analytic solution exists and was described in Chapter 3. But fully understanding spherical growth is always a good beginning before modeling more complex systems.

The first step in any cellular automata model is to define the cells, which I usually call pixels, as shown in Figure 5.15. For our spherical system we assume a set of radial pixels for which the pixel center is located at $r_i = (i-1)\Delta r$ for integer $i$ with $1 \leq i \leq N$. (Of course, other numbering conventions could be used to equal effect.) We further assume that pixels with $r_i < r_b$ are filled with ice, while pixels with $r_i \geq r_b$ are filled with vapor. We call the $i = b$ pixel a "boundary" pixel because it is filled with vapor but is adjacent to an ice pixel. The vapor pixels typically include a background gas of air, and each is labeled with the water-vapor supersaturation $\sigma_i = \sigma(r_i)$ at that location.

In the vapor surrounding the crystal we write the particle diffusion equation in spherical coordinates

$$\frac{\partial \sigma}{\partial t} = \mathrm{D}\nabla^2 \sigma \tag{5.1}$$

$$= D\left(\frac{2}{r}\frac{\partial \sigma}{\partial r} + \frac{\partial^2 \sigma}{\partial r^2}\right)$$

and on our radial grid this becomes

$$\sigma(r_i, \tau + \Delta\tau) = (1 - 2\Delta\tau)\sigma(r_i)$$
$$+\Delta\tau\left[\left(1 + \frac{\Delta r}{r_i}\right)\sigma(r_{i+1}) + \left(1 - \frac{\Delta r}{r_i}\right)\sigma(r_{i-1})\right] \tag{5.2}$$

where $\tau = Dt/(\Delta r)^2$. Note that the $(1 \pm \Delta r/r_i)$ terms arise from the spherical coordinate system, reflecting the fact that the volume in a $\Delta r$ shell increases with $r_i$. These terms introduce a potential problem in dividing by $r_i = 0$, but we will ignore this because the central pixel will always be part of the seed crystal in our model. A 1D cartesian model would avoid the $(1 \pm \Delta r/r_i)$ terms, but I prefer to work with a model that describes a real physical system, in this case the growth of a spherical ice crystal.

At this point we recognize that snow crystal growth is described by a very low Peclet number, as I described in Chapter 3. This means that the supersaturation field around a crystal relaxes very rapidly compared to the crystal growth time, so we can solve the particle diffusion equation while assuming a non-moving crystal surface. In our CA model, this means we can iterate Equation 5.2 with fixed boundaries until $\sigma(r_i)$ converges to a stationary solution of Laplace's equation. We do this without yet worrying about the actual growth of the crystal, because, as far as particle diffusion is concerned, the crystal is growing so slowly that it is essentially stationary.

For computational efficiency, we would like to relax $\sigma(r)$ using the smallest possible number of iterations of Equation 5.2, so we would like to choose $\Delta\tau$ to be as large as possible. Taking $\Delta\tau = D\Delta t/(\Delta r)^2 = 1/2$ seems to be about optimal, as larger values can lead to numerical instabilities. As a bonus, this choice sets one term in Equation 5.2 equal to zero, so the optimal propagation equation becomes

$$\sigma(r_i, k+1) =$$
$$\frac{1}{2}\left(1 + \frac{\Delta r}{r_i}\right)\sigma(r_{i+1}, k) + \frac{1}{2}\left(1 - \frac{\Delta r}{r_i}\right)\sigma(r_{i-1}, k) \tag{5.3}$$

where here we have replaced $\tau$ with a simple integer indexing variable $k$. At each instant in time, we simply iterate Equation 5.3 to determine the correct supersaturation field $\sigma(r)$ surrounding the crystal at that time.

From a computational perspective, note that Equation 5.3 can be performed using highly efficient vector processing. The vectors $(1 \pm \Delta r/r_i)$ are constant and need only be calculated once at the beginning of a modeling run. The vectors $\sigma(r_{i\pm1})$ are rapidly computed using a simple permutation of $\sigma(r_i)$, and vector operators can perform the arithmetic in Equation 5.3 using optimized parallel-processing algorithms built into the compiler. While optimizing efficiency is of little concern for a 1D spherical calculation, it becomes quite important in 3D codes with high spatial resolution.

## Boundary Conditions

For the outer boundary condition far from our growing spherical crystal, we assume a constant value $\sigma(r_N) = \sigma_{far}$, where $r_N = r_{far}$ is the position of the outer boundary. This is easily implemented in our cellular automaton algorithm by simply applying Equation 5.3 to all $\sigma(r_i)$ out to $\sigma(r_{N-1})$.

As described in Chapter 3, we have a mixed boundary condition at the crystal surface:

$$X_0 \left(\frac{\partial \sigma}{\partial n}\right)_{surf} = \alpha \sigma_{surf} \tag{5.4}$$

where

$$X_0 = \frac{c_{sat}}{c_{ice}} \frac{D}{v_{kin}} \tag{5.5}$$

On our CA model grid, Equation 5.4 becomes

$$\sigma_b = \sigma_{b+1} \left(1 + \alpha(\sigma_b) \frac{\Delta r}{X_0}\right)^{-1} \tag{5.6}$$

to first order in $\Delta r$, where $\sigma_b$ is the supersaturation in the boundary pixel.

Note that determining the value of $\alpha(\sigma_b)$ can be included in the convergence process by generalizing Equation 5.6 to give the propagation equation

$$\sigma(r_b, k+1) = \sigma(r_{b+1}, k) \left(1 + \alpha\big(\sigma(r_b, k)\big) \frac{\Delta r}{X_0}\right)^{-1} \tag{5.7}$$

Doing this allows one to assume any desired functional form for $\alpha(\sigma_{surf})$ without having to solve Equation 5.6 analytically.

In summary, calculating the supersaturation field means iterating Equation 5.3 for $b < i < N$ using $\sigma(r_N) = \sigma_{far}$, while simultaneously iterating Equation 5.7 for $i = b$. For any physically realistic scenario, this should converge to give the full solution $\sigma(r_i)$ that satisfies Laplace's equation with the proper boundary conditions.

## Convergence Criterion

A next question is how long to continue the iterative propagation of Equations 5.3 and 5.7. The supersaturation field $\sigma(r)$ should converge exponentially with time, which unfortunately means that it never actually reaches the exact solution. One common practice is to iterate until $\Delta\sigma/\sigma < \varepsilon$ at each step for all pixels, where $\varepsilon$ is a chosen constant in the model. If small values of $\sigma$ produce unnecessarily long convergence times, another reasonable criterion is $\Delta\sigma/(\sigma + \varepsilon_1 \sigma_{far}) < \varepsilon$, where $\varepsilon_1$ is another chosen constant. Little tweaks like this can yield significant reductions in run time with no great loss in accuracy, so it is often worth experimenting with different possibilities.

Because $\sigma(r)$ usually decreases as a crystal becomes larger (see the spherical solution in Chapter 3), the overall crystal growth rate tends to be larger than the exact analytic solution, with the difference being greater for higher $\varepsilon$ [2013Lib1]. Determining the overall accuracy of a growth model as a function of $(\varepsilon, \varepsilon_1)$ is easily accomplished using one of the exact analytic solutions presented in Chapter 3. Moreover, this is an important determination to make, as there is always a trade-off between accuracy and code running time.

## Growth Steps

Once we have calculated the supersaturation field $\sigma(r)$ at some instant in time, the next step is to use this solution to grow the crystal out a small amount. In the spherical CA model, this means turning a boundary pixel into an ice pixel, and we use the known surface growth rate (see Chapter 3)

$$v_n = \alpha v_{kin} \sigma_{surf} \\ = \alpha(\sigma(r_b)) v_{kin} \sigma(r_b) \qquad (5.8)$$

which indicates that ice growth would "fill" the boundary pixel in a time

$$\delta t = \frac{\Delta r}{v_n} \qquad (5.9)$$

and this time interval is easily calculated from the known supersaturation field $\sigma(r_i)$.

Putting everything together, growing a spherical snow crystal using this 1D CA model involves the following steps:

**1)** Set up the physical parameters and initial conditions, including the initial seed crystal.

**2)** Iterate Equations 5.3 and 5.7 until reaching convergence, yielding the supersaturation field $\sigma(r_i)$ around the crystal.

**3)** Promote the boundary pixel (there is only one in this 1D model) into an ice pixel while advancing the real time by $\delta t$ in Equation 5.9. Promote the next vapor pixel to a boundary pixel.

**4)** Either stop the model at this point or go back to step (2).

The final result of this process is a series of time steps giving $R(t)$, the size of the crystal as a function of time, along with $\sigma(r_i, t)$. Note that while the growth steps have a uniform size $\Delta r$, the time steps are not uniform in duration.

Note also that this CA model is completely deterministic, including no random processes of any kind. It also excludes evaporation, so once a vapor pixel becomes an ice pixel, it cannot go back again. Finally, the model only includes bulk diffusion in air together with surface attachment kinetics, ignoring all heating and surface-energy effects.

These model attributes are put in place mainly to simplify the calculations at this point, and other choices are possible. In particular, I discuss surface-energy, surface diffusion, and other physical effects in more detail below. I neglect thermal effects entirely in this discussion, however, as they play only a relatively minor role in snow crystal growth (as described in Chapter 3). One could make a dual-diffusion CA model that would incorporate both particle and thermal diffusion, but that requires a significant increase in complexity that is best left for another day.

## Adaptive Boundary Matching

One problem with any computational diffusion model is that the outer boundary is only a finite distance from the growing crystal, while often the outer boundary condition is specified at infinity, $\sigma(r \to \infty) = \sigma_\infty$. If we use the outer boundary condition $\sigma(r_{far}) = \sigma_{far}$ as described above, then the best results will be obtained by making $r_{far}$ as large as possible. However, for computational efficiency, one would like to keep $r_{far}$, small, as that decreases the total volume of modeled space.

One way to address this problem is with an adaptive grid, increasing the pixel size with distance from the crystal. This works because $\sigma(r)$ changes rapidly only near the crystal surface, so a coarser grid can be used far from the surface. With an adaptive grid, a small number of pixels can be used to model a large volume of space efficiently. But an adaptive grid introduces additional computational complexity and overhead, and it may interfere with one's ability to make full use of highly efficient parallel processing algorithms. Simply changing $\Delta r$ with $r_i$ is a fine approach with a 1D model, but extending this idea to higher dimensions becomes problematic.

Another relatively easy approach to the far-away-boundary problem is to keep $\Delta r$ constant while making $r_{far}$ reasonably small, but then adjust $\sigma_{far}$ as the crystal grows. To see how this works, start with the analytic solution for spherical growth presented in Chapter 3. Because the full supersaturation field $\sigma(r)$ is known in the spherical model, it is straightforward to show that

$$\sigma(r_{far}) = \sigma_\infty - \frac{dV/dt}{4\pi r_{far} X_0 v_{kin}} \quad (5.10)$$

where

$$\frac{dV}{dt} = 4\pi R^2 v_n \quad (5.11)$$

is the volume change per unit time for a spherical crystal with radius $R$. This expression is dictated by conservation of mass, which requires that the flux of water vapor diffusing toward the crystal must equal the rate at which vapor turns into ice.

From this knowledge of the exact spherical solution, we can write a propagation equation for $\sigma(r_{far})$

$$\sigma(r_{far}, k+1) = \sigma_\infty - \frac{dV/dt(k)}{4\pi r_{far} X_0 v_{kin}} \quad (5.12)$$

and this operation would be performed between steps (3) and (4) listed above. As the model crystal develops, $\sigma(r_{far})$ would adapt to the changing crystal size and growth behavior.

Note that this is an iterative procedure; once the supersaturation field $\sigma(r_i)$ around the crystal is known, this allows a calculation of $dV/dt$ at that point in time. Performing Equation 5.12 then sets up $\sigma(r_{far})$ for the next time step in the series. As long as the crystal grows slowly, the process will converge to yield a reasonable approximation of $\sigma(r_{far})$ at each time step. Once again, the overall accuracy of this adaptive outer boundary could be examined by comparing model results with the analytic solution of the diffusion equation.

For typical conditions, we expect (from the analytic solution for a spherical crystal) that $dV/dt{\sim}R$ for diffusion-limited growth and $dV/dt{\sim}R^2$ for kinetics-limited growth (see Chapter 3). In both cases, $dV/dt$ is small at early times, so Equation 5.12 yields $\sigma(r_{far}) \approx \sigma_\infty$. This makes sense, as presumably $r_{far}$ is much larger than the initial seed crystal. Then $\sigma(r_{far})$ decreases as the crystal grows larger, as one would expect.

This adaptive outer boundary method essentially "matches" the CA solution to the known analytic solution beyond $r_{far}$. While this procedure is somewhat trivial for the 1D spherical model (because the analytic solution is already known at all $r$), it becomes useful when working in higher dimensions, as we will see in the next section.

## 5.3 Two Dimensional Cylindrically Symmetric Cellular Automata

Having set the stage by exploring the simplest one-dimensional model, the next obvious step is to move up to a two-dimensional model. A 2D model introduces additional complexity and new physical effects compared to the 1D model, and it introduces additional model developments along several fronts.

While it is tempting to jump straight away to a full 3D model, we will soon find that the 2D case provides a valuable test system for addressing many nontrivial issues. Also, from the standpoint of practical pedagogics, a 2D model can be easily described using 2D sketches, which display well on a printed page and are relatively easy to comprehend. Communicating ideas relating to full 3D structures, on the other hand, can be something of a visual challenge.

Focusing, therefore, on 2D models, I have found that a cylindrically symmetrical system is the best choice for exploring the physics of snow crystal growth. A planar model is another 2D option that is often explored, but such a model is of little use for examining solidification physics.

For example, consider 2D models of stellar crystals like those shown in Figure 5.8. These models to not describe real snow crystals, but they are adequate models of infinitely long bars with snowflake-shaped cross sections. Solving the infinite-bar problem in 3D is identical to solving just the 2D cross-section. While this is a fine exercise, there are no real snow crystals that have anything like this kind of complex extruded morphology. So this kind of 2D flat-plate model is of little actual use when examining the physics of real snow crystals.

In contrast, 2D cylindrically symmetric models can include simple disks, simple columns, hollow columns, disks on columns, and other morphologies that serve as reasonable proxies for real snow crystals. While cylindrically symmetry definitely has its limitations, I have found that it actually works quite well for describing simple snow crystal morphologies. Thus, unlike the flat-plate 2D model, a cylindrically symmetrical 2D model connects much better to the real physics of snow crystal growth.

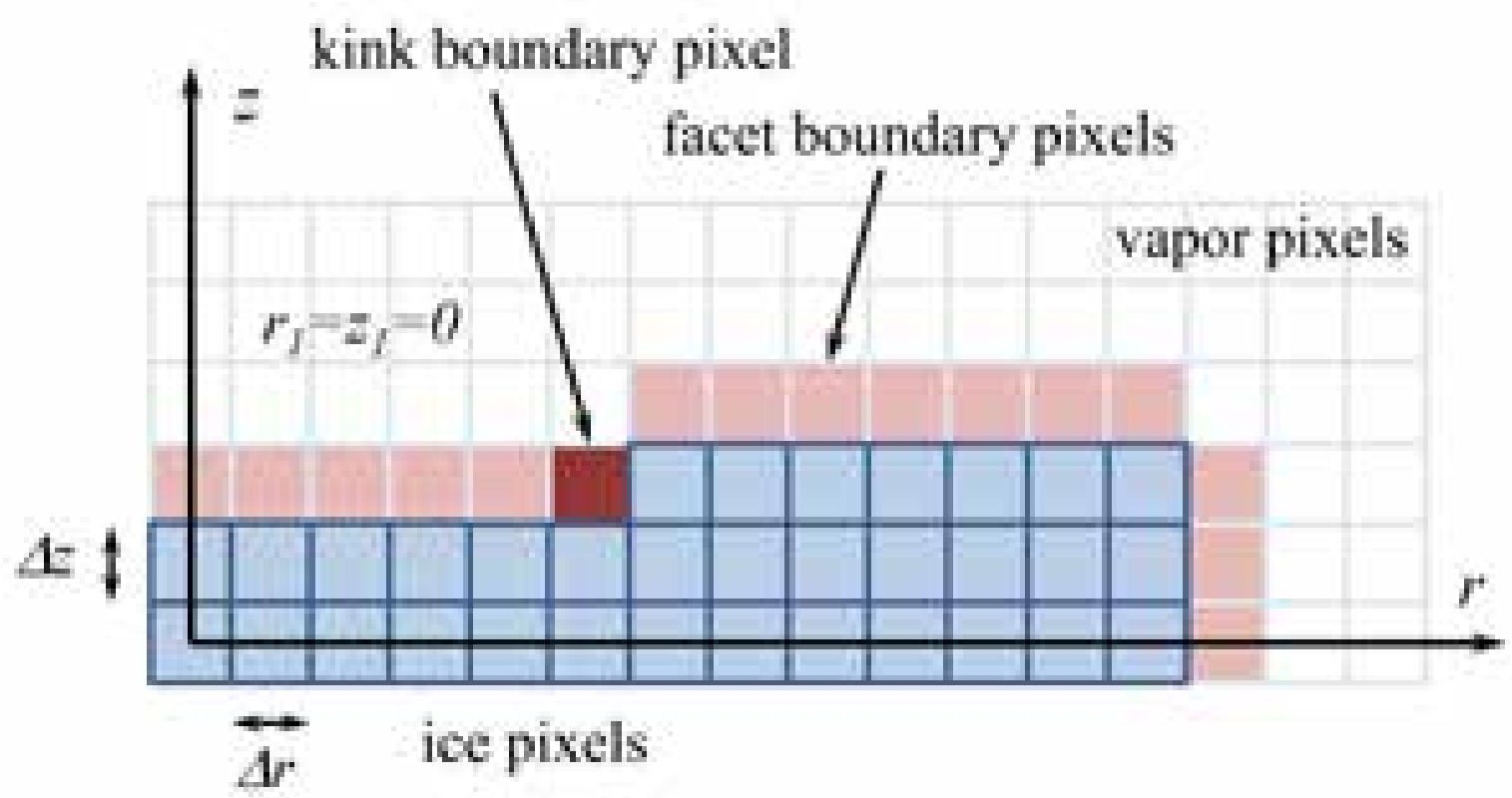

**Figure 5.16: A pixel geometry for a 2D cylindrically symmetric cellular automaton model. Ice pixels are shown as blue while vapor pixels are white. The "facet" boundary pixels (adjacent to faceted ice surfaces) are pink, while this particular model includes a single red "kink" boundary pixel that touches two ice pixels. For convenience I often take $\Delta r = \Delta z = \Delta x$, which defines the generic pixel size $\Delta x$.**

In a 2D cylindrically symmetrical model, a simple hexagonal plate is approximated by a thin disk. The six prism facets on the hexagonal plate are thus replaced by a single cylindrical "facet", while the basal facets are essentially unchanged. Particle diffusion around a thin disk is about the same as that around a hexagonal plate, and there is a good correspondence between the basal and prism attachment coefficients in the two cases [2015Lib2]. In particular, the same $\alpha_{pri}$ describing growth on the six prism surfaces of a hexagonal plate can be used for the single edge of the circular disk.

Transforming a hexagonal plate to a thin disk involves a small geometrical correction, but, other than that, the cylindrically symmetric disk is a tolerably good physical representation of a hexagonal plate. The same is true for snow crystal columns, hollow columns, and capped columns. For all these simple morphologies, cylindrically symmetrical models are quite well suited for investigating growth dynamics and attachment kinetics. Snow crystals grown on electric needles are also well suited for study using cylindrically symmetric models (see Chapter 8). Of course, dendritic structures and other complex morphologies will require full 3D modeling, but I have always found that difficult physics problems are best solved one step at a time.

Moving forward, Figure 5.16 shows a typical pixel geometry for a 2D cylindrically symmetrical cellular-automaton model. The position of the center of each pixel is $(r_i, z_j)$, where $r_i = (i-1)\Delta r$ and $z_j = (j-1)\Delta z$ for all $(i, j)$ ranging from $(1,1)$ to $(N_r, N_z)$. (Once again, different coordinate conventions are also possible.) I usually choose $\Delta r = \Delta z = \Delta x$, thus defining $\Delta x$, as this simplifies the mathematics and is also a reasonable choice for a realistic snow crystal model. The physical size of $\Delta x$ is somewhat arbitrary, but we will see below that $\Delta x$ should not be much greater than $X_0$ if one wishes to accurately reproduce small-scale snow crystal structures. It appears that there is little to be gained, however, in choosing $\Delta x < X_0$.

Although not apparent in Figure 5.16, most pixels in a CA model are vapor pixels, and each of these is assigned a supersaturation $\sigma_{i,j} = \sigma(r_i, z_j)$. As with the 1D model, the supersaturation field is determined by particle diffusion together with the appropriate boundary conditions, and the diffusion equation in cylindrical coordinates is

$$\frac{\partial \sigma}{\partial t} = \mathrm{D}\nabla^2\sigma \tag{5.13}$$
$$= D\left(\frac{1}{r}\frac{\partial \sigma}{\partial r} + \frac{\partial^2\sigma}{\partial r^2} + \frac{\partial^2\sigma}{\partial z^2}\right)$$

for $\sigma(r, z, t)$. As we did with the 1D case, we project this equation onto the 2D cellular-automaton grid to obtain

$$\sigma_{i,j}(\tau + \Delta\tau) = (1 - 4\Delta\tau)\sigma_{i,j}$$
$$+\Delta\tau\left[\left(1 + \frac{\Delta r}{2r_i}\right)\sigma_{i+1,j} + \left(1 - \frac{\Delta r}{2r_i}\right)\sigma_{i-1,j}\right]$$
$$+\Delta\tau\left[\sigma_{i,j+1} + \sigma_{i,j-1}\right] \tag{5.14}$$

Choosing $\Delta\tau = 1/4$ simplifies this expression by eliminating the first term, then yielding the propagation equation

$$\sigma_{i,j}(k+1) =$$
$$\frac{1}{4}\left[\left(1 + \frac{\Delta r}{2r_i}\right)\sigma_{i+1,j}(k) + \left(1 - \frac{\Delta r}{2r_i}\right)\sigma_{i-1,j}(k)\right]$$
$$+\frac{1}{4}\left[\sigma_{i,j+1}(k) + \sigma_{i,j-1}(k)\right] \tag{5.15}$$

In the limit of low Peclet number, iterating this equation to convergence will yield the static supersaturation field $\sigma(r_i, z_j)$ that satisfies Laplace's equation. Similar to the 1D case, the choice of $\Delta\tau = 1/4$ is a good one in that it produces rapid convergence without introducing numerical instabilities that can be problematic with higher values of $\Delta\tau$.

## A Facet-Kink Model

A central feature of any finite-element computational model is that one must define a mathematical system that operates at finite resolution while providing a good physical representation of what is essentially a continuum system. For crystal growth, this means that the mathematics must somehow deal with both nanometer physics at scales much smaller than $X_0$ (e.g., the molecular dynamics that governs attachment kinetics) and mesoscale physics at scales at and above $X_0$ (including particle diffusion around the crystal and other processes).

With a cellular automaton model, the specific CA rules need to derive from nanoscale physics but run accurately on a mesoscale grid. In particular, the attachment kinetics rules must be parameterized so that they can be applied at the much larger scale of the cellular automaton, and how one does this is not always immediately obvious. Dealing with this broad range of physical scales is one of the most difficult aspects of modeling snow crystal growth.

These issues mostly play out at the crystal boundary, and in 2D there is no obvious, simple choice for accurately specifying the boundary conditions or growth rules. One of the simpler ways to tackle this problem using cellular automata is with what I call a *facet-kink* model. In the 2D cylindrically symmetric case, this means defining the two classes of boundary pixels shown in Figure 5.16.

Each *facet boundary pixel* is a vapor pixel bordered by exactly one nearest-neighbor ice pixel (out of four nearest-neighbor positions, neglecting all farther-away positions). Meanwhile *kink boundary pixels* are vapor pixels bordered by exactly two ice pixels, as shown in Figure 5.16. Boundary pixels with three or even four neighboring ice pixels are also possible, but only the facet and kink boundary pixels play important roles in simple growth morphologies. How one treats 3- and 4-neighbor boundary pixels is not so important; even just turning them immediately into ice as soon as they appear does not greatly affect the overall model dynamics except perhaps in somewhat convoluted morphologies.

The characterization of the boundary pixels is an important part of any CA model when we address the diffusion boundary conditions, which is coming up next. The facet-kink CA model is especially simple in that the character of each boundary pixel is defined solely from its nearest neighbors. This local definition makes for easy bookkeeping, but we will soon find that this simplicity requires some compromises in physical accuracy.

## Boundary Conditions

The outer boundary is typically defined by a constant far-away supersaturation, so we set $\sigma_{i,j} = \sigma_{far}$ whenever $i = N_r$ or $j = N_z$, and this is easily implemented into the model by applying Equation 5.15 only out to $i = N_r - 1$ and $j = N_z - 1$. There are some numerical issues that must be dealt with along the $(0, z)$ and $(r, 0)$ axes, but these are fairly minor bookkeeping details that are discussed in [2008Lib, 2013Lib1]. It is typical to use reflection boundary conditions at $z = 0$, so the physically modeled space then includes $-z_{max} \leq z \leq z_{max}$ and $0 \leq r \leq r_{max}$.

The surface boundary conditions for faceted boundary pixels are similar to the 1D case described above, and Equation 5.6 becomes

$$\sigma_{i,j} = \sigma_{i+1,j}\left(1 + \alpha_{prism}(\sigma_{i,j})\frac{\Delta x}{X_0}\right)^{-1}$$
$$\sigma_{i,j} = \sigma_{i,j+1}\left(1 + \alpha_{basal}(\sigma_{i,j})\frac{\Delta x}{X_0}\right)^{-1} \quad (5.16)$$

where $\alpha_{prism}$ and $\alpha_{basal}$ are the attachment coefficients for the two principal facets. For simplicity, the particular index notation here is for facets that face in the $+r$ and $+z$ directions, and we have assumed $\Delta r = \Delta z = \Delta x$. These then become propagation equations that are similar in form to Equation 5.7.

For a kink boundary pixel, the optimum boundary condition can be estimated by examining the growth of the 45-degree surface orientation shown in Figure 5.17. Because only kink boundary pixels are present on this surface, the continuum boundary condition Equation 5.4 can be expressed in two essentially equivalent forms

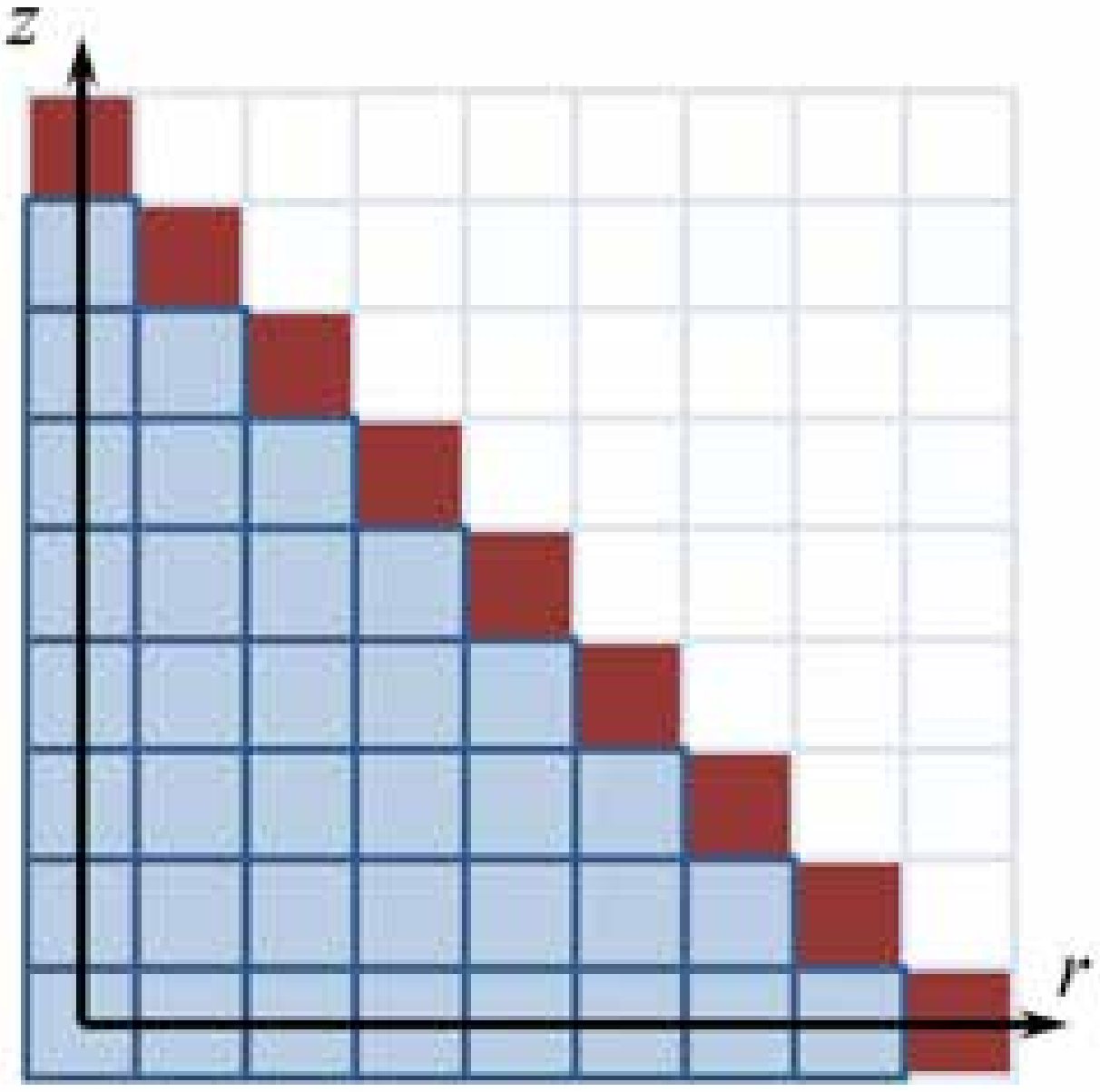

**Figure 5.17: The "45-degree" surface depicted here includes only kink boundary pixels. This can be taken as an essentially flat surface in the limit of a large crystal, giving a good approximation for determining the CA rules governing kink boundary pixels.**

$$\sigma_{i,j} = \sigma_{i+1,j+1}\left(1 + \alpha_{kink}(\sigma_{i,j})\frac{\sqrt{2}\Delta x}{X_0}\right)^{-1}$$
or
$$\sigma_{i,j} = \sigma_{opp}\left(1 + \alpha_{kink}(\sigma_{i,j})\frac{\Delta x}{\sqrt{2}X_0}\right)^{-1} \quad (5.17)$$

where $\sigma_{opp} = (\sigma_{i+1,j} + \sigma_{i,j+1})/2$ and $\alpha_{kink}$ is the appropriate attachment coefficient. Both expressions can be derived from Equation 5.4 using the geometry illustrated in Figure 5.17, and the two equations are equal to first order in $\Delta x$. If the 45-degree surface is essentially flat (the small $\Delta x$ limit), these expressions provide an accurate model of the surface boundary condition. Because a surface made from kink sites is molecularly rough, we expect $\alpha_{kink} \approx 1$. Note that the additional factor of $\sqrt{2}$ compared to Equation 5.16 is simply a geometrical factor coming from the tilted geometry of the 45-degree surface on the CA grid.

As with the 1D model, the 2D propagation equations can be iterated until some suitable convergence criterion is satisfied. This process solves Laplace's equation in the space surrounding the growing crystal, thus yielding the supersaturation field $\sigma(r_i, z_j)$ in all vapor and boundary pixels at a fixed time.

## Growth Steps

The next step in the model is to use the known supersaturation field to calculate the crystal growth rates at each point and define appropriate CA rules for turning boundary pixels into ice pixels. If all goes well, these the CA propagation equations and growth rules will generate physically accurate computational snow crystals.

As with the 1D model, we define a growth step as occurring when a single boundary pixel transforms into an ice pixel [2014Kel], and the newly defined surface then requires a new calculation of the supersaturation field. The main difference between the 1D and 2D models is now there are many boundary pixels to consider simultaneously.

To keep an ongoing account of the crystal growth at each point on the ice surface, we assign a numerical "filling factor" $f_b$ to each boundary pixel, where we assign an integer index $b$ to label the boundary pixels. Whenever a vapor pixel becomes a new boundary pixel, $f_b$ for that pixel is set to zero. As the model develops, each $f_b$ increases with time at a rate that derives from the crystal growth rate at its position. When a filling factor increases to unity, then that boundary pixel turns to ice.

After relaxing the supersaturation field to produce $\sigma(r_i, z_j)$ throughout the space above the crystal, we can again use Equation 5.8 to calculate the growth velocity along the surface normal. For a facet boundary pixel, the time required to "fill" the remainder of each boundary pixel becomes

$$\delta t_b = \frac{\Delta x}{v_n}(1 - f_b) \qquad (5.18)$$

while for kink boundary pixels we again examine the 45-degree surface in Figure 5.17 to obtain

$$\delta t_b = \frac{\Delta x}{\sqrt{2} v_n}(1 - f_b) \qquad (5.19)$$

Note that the additional $\sqrt{2}$ is again a geometrical factor associated with the 45-degree surface. From the entire set of time intervals $\delta t_b$, we choose the smallest one, $\delta t_{b,min}$, and then fill each boundary pixel for this amount of time, giving

$$f_b \rightarrow f_b + \frac{v_n}{\Delta x} \delta t_{b,min} \qquad (5.20)$$

for all facet boundary pixels and

$$f_b \rightarrow f_b + \frac{\sqrt{2} v_n}{\Delta x} \delta t_{b,min} \qquad (5.21)$$

for all kink boundary pixels. In doing this, one filling factor will reach $f_b = 1$ while all the others will increase but remain below unity. After updating the filling factors and turning one boundary pixel to ice, we then locate the new boundary pixels (assigning them a filling factor of zero) and proceed with calculating the next supersaturation field.

One pleasant feature of the CA method is that it is remarkably easy to write down physically realistic (albeit not entirely accurate) rules and transcribe them into relatively simple iterative algorithms. In general, other front-tracking and phase-field techniques require a substantially greater mathematical sophistication and a commensurate increase in programming effort. Unfortunately, the relative simplicity of a CA model brings with it some deficiencies in terms of accuracy, which we examine next. How serious all these problems are, and how well they can be addressed by developing more advanced CA rules, remains a topic for additional research.

## Numerical Anisotropy

If cylindrically symmetrical snow crystal growth could be modeled accurately using only faceted surfaces together with the 45-degree surface shown in Figure 5.17, then the facet-kink model described above would be adequate to solve this 2D problem. As $\Delta x \rightarrow 0$, the facet-kink CA rules satisfy the surface boundary conditions and growth rates to high accuracy on these surfaces. Problems arise, however, when one considers other surfaces.

Consider, for example, the 2:1 surface shown in Figure 5.18. In the small-$\Delta x$ limit, this is a simple vicinal surface, so a solution of the diffusion equation (assuming an infinite surface and ignoring the Mullins-Sekerka instability) would yield uniform planar growth with a growth velocity $v_n = \alpha v_{kin} \sigma_{surf}$. The facet-kink model, however, cannot reproduce this simple result, even if $\alpha$ is constant on all surfaces.

I have done some numerical modeling of surfaces like these [2013Lib1], and I find that the model growth rates are typically off by about 10 percent, depending on the model details. The growth rates are correct for the basal and prism facets, and for the 45-degree surface, but generally not for other vicinal surfaces. If one removes all the $\sqrt{2}$ factors in the above discussion, the maximum error may be as high as 40 percent.

In an absolute sense, a 10-percent growth-rate error is not so bad, as experiments are typically not able to determine $\sigma_{far}$ to even this level of accuracy. The problem arises because this is an anisotropic error. If snow crystal growth is sufficiently facet-dominated, as I described above, then perhaps a small intrinsic anisotropy in the model will have little importance in its overall morphological development. Even a $\sqrt{2}$ anisotropic error may not have much of a detrimental effect, as seen in [2014Kel].

Some features in snow crystal growth, however, may simply be impossible to reproduce with this level of intrinsic anisotropy. One example might be tip-splitting (see Chapter 3), as this phenomenon arises when the attachment kinetics is especially isotropic. Subtle features in ridge formation and other common snow crystal morphological features may also be adversely affected by built-in anisotropies in the facet-kink model. The only way to answer these questions will be to make detailed comparisons between computational and laboratory snow crystals over a broad range of growth conditions.

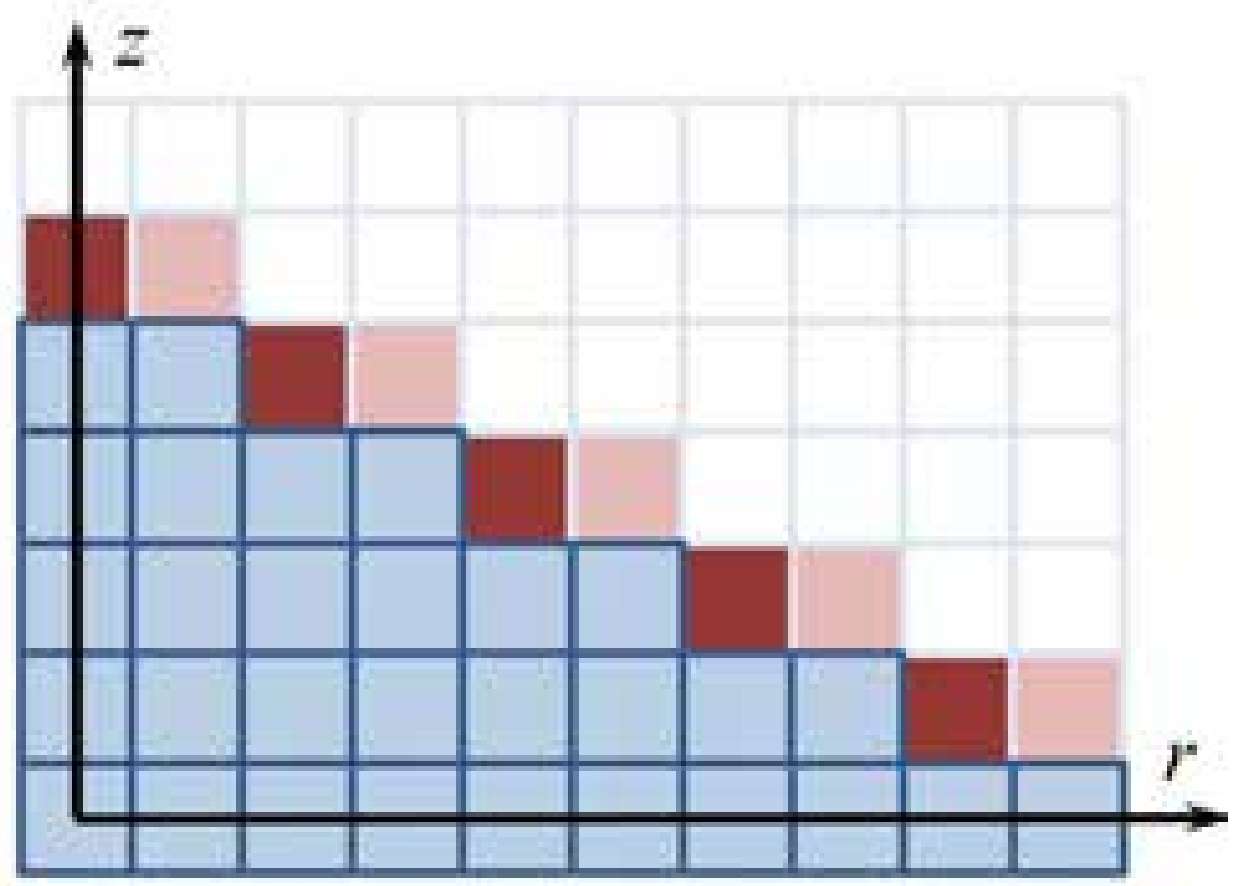

**Figure 5.18: This "2:1" vicinal surface includes equal numbers of facet and kink boundary pixels. A facet-kink CA model with a constant $\alpha$ yields growth-rate errors about ten percent, depending on model parameters [2013Lib1]. These errors present an intrinsic anisotropy in the facet-kink model that cannot be corrected by increasing the model resolution. How such a built-in anisotropy affects the overall morphological development of computational snow crystals is not yet known.**

## A Facet-Vicinal Model

One way to reduce the intrinsic anisotropies is to devise an improved set of cellular automata rules. The facet-kink model uses only nearest-neighbor interactions to determine the boundary conditions, and we can do better by incorporating non-local effects, at the expense of increased algorithmic complexity. One possibility in 2D is what I call a *facet-vicinal model*.

The basic idea in the facet-vicinal model is to define a new parameter $L$, equal to the width of the terrace "ledge" associated with each boundary pixel, in integer pixel units. Figure 5.19 shows one example of a vicinal surface with $L = 3$ for each boundary pixel. In this example, all the terrace ledges are basal surfaces facing the +z direction. On a more complex surface, one would simply count how many adjacent boundary pixels make up a single ledge, and that value of $L$ would be assigned to all the boundary pixels making up that ledge. Although more complicated than the facet-kink model, this degree of nonlocal bookkeeping is not an onerous computational task.

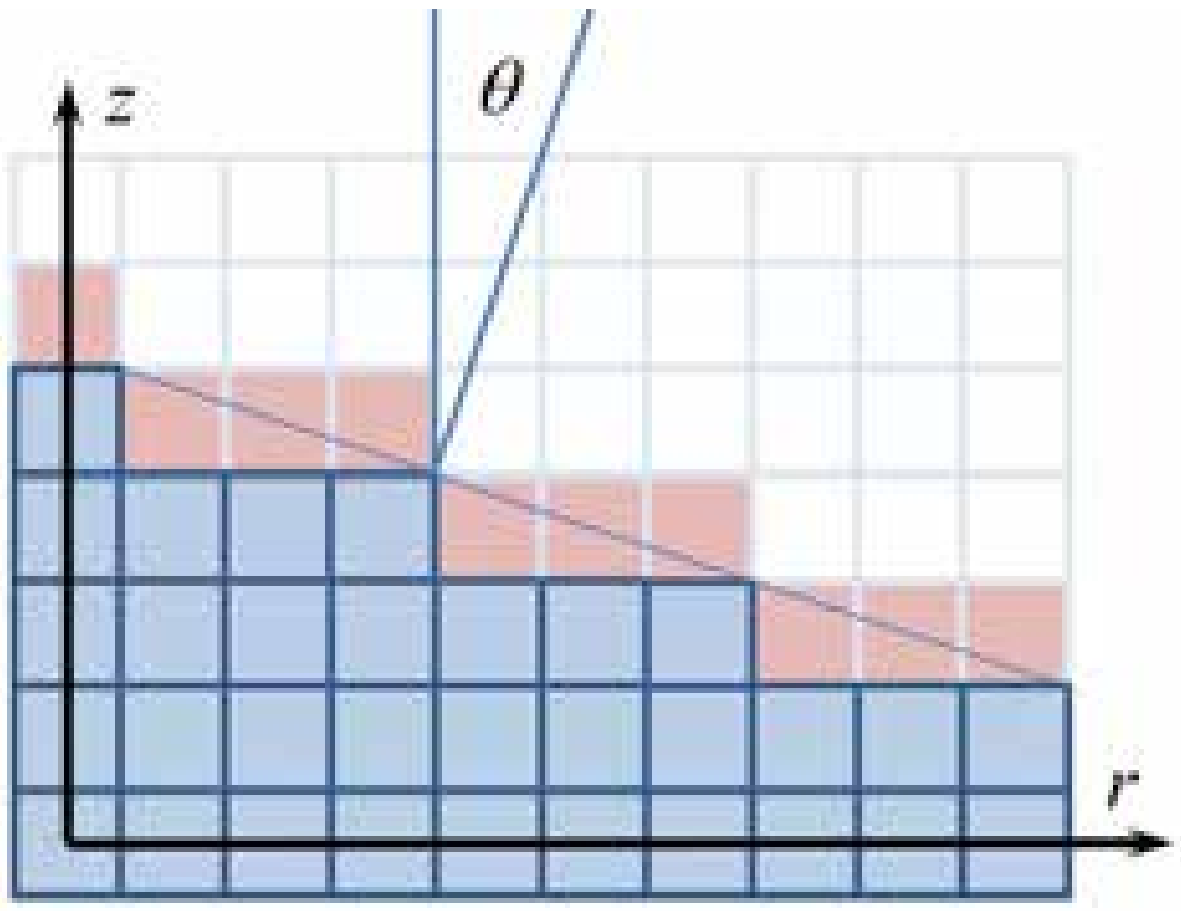

**Figure 5.19: A vicinal surface in which all the terrace ledges have a ledge width $L = 3$, and the vicinal angle $\theta$ is given by $\tan(\theta) = 1/L$. For any individual terrace facing the +z direction (as shown here), the ledge width $L$ is defined as the number of adjacent boundary pixels in its row, and that value of $L$ is assigned to all boundary pixels in that row.**

From the ledge width $L$, the vicinal angle $\theta$ shown in Figure 5.19 is given by $\tan(\theta) = 1/L$, and the attachment coefficient would be specified as $\alpha_{vicinal}(\sigma_{i,j}, \theta)$, being a function of both the surface supersaturation and the vicinal angle. For simplicity I now assume $0 < \theta < 45$ degrees, as generalization to angles outside this range is straightforward.

For any vicinal surface like that shown in Figure 5.19, the boundary conditions are again derived from the continuum boundary conditions

$$X_0 \left(\frac{\partial \sigma}{\partial n}\right)_{surf} = \alpha \sigma_{surf} \tag{5.22}$$

with

$$\left(\frac{\partial \sigma}{\partial n}\right)_{surf} = \hat{n} \cdot \nabla \sigma$$
$$= A_c \frac{\sigma_{i,j+1} - \sigma_{i,j}}{\Delta x} + A_s \frac{\sigma_{i+1,j} - \sigma_{i,j}}{\Delta x} \tag{5.23}$$

and again we are assuming $\Delta r = \Delta z = \Delta x$ and $\sigma_{i,j} = \sigma(r_i, z_j)$, and we define $A_c = \cos(\theta)$ and $A_s = \sin(\theta)$. With this, the propagation equation for the supersaturation in any boundary pixel becomes

$$\sigma_{i,j} = \frac{A_c \sigma_{i,j+1} + A_s \sigma_{i+1,j}}{A_c + A_s + \alpha_{vicinal}(\sigma_{i,j}, \theta) \dfrac{\Delta x}{X_0}} \tag{5.24}$$

This expression is a generalized form of the boundary conditions described above in Equations 5.16 and 5.17 (and agreement can be seen by noting that $\sigma_{i,j+1} + \sigma_{i+1,j} = \sigma_{i,j} + \sigma_{i+1,j+1}$ to first order in $\Delta x$). However, while Equations 5.16 and 5.17 applied with high accuracy to only three surfaces, Equation 5.24 is accurate for all flat vicinal surfaces with any vicinal angle. Thus, with some increase in bookkeeping, we have a new boundary condition with substantially reduced intrinsic anisotropy.

The growth algorithm is similar to that described above, but with Equations 5.18 and 5.19 replaced with

$$\delta t_b = A_c \frac{\Delta x}{v_n} (1 - f_b) \tag{5.25}$$

for the growth of +z boundary pixels with $0 < \theta < 45$ degrees, like those shown in Figure 5.19.

For the special case of a flat vicinal surface, such as that shown in Figure 5.19, we see that all the boundary pixels shown have identical properties. Because $\sigma_{surf}$ will be nearly constant along this flat surface, all the boundary pixels will turn to ice pixels at essentially the same time, and this will preserve the vicinal character of the surface. In the limit of small $\Delta x$, I expect that the facet-vicinal model will provide a much improved model behavior compared to a facet-kink model.

The facet-vicinal CA model has not yet been tried, as only facet-kink models have so far been demonstrated for snow-crystal growth. I suspect that facet-vicinal, or some improved version of this model, will soon displace facet-kink for modeling snow crystal growth. But, as the saying goes, it is hard to predict, especially the future. New developments in modeling faceted crystal growth are appearing rapidly on all fronts at present.

## Monopole Matching

Extending the outer boundary to infinity can again be accomplished, to a reasonable approximation, using the known analytical solution for spherical growth. The essential idea is the same as was described above, but in place of Equation 5.11 we use

$$\frac{dV}{dt} = \sum 2\pi r_b \frac{\Delta x^2}{\delta t_b} \qquad (5.26)$$

where the sum is over all boundary pixels. This yields the adaptive outer boundary

$$\sigma_{i,j}(\rho_{far}) \rightarrow \sigma_\infty - \frac{dV/dt}{4\pi\rho_{far}X_0 v_{kin}} \qquad (5.27)$$

where $\sigma_{i,j}$ refers to an outer boundary pixel and $\rho_{far} = \sqrt{r^2 + z^2}$ is the distance to the outer boundary point.

This iterative outer boundary assignment matches the outer boundary to an optimal spherical solution, so would be quite accurate for the case of spherical growth. For the general case, it can be considered a monopole approximation of the correct outer boundary. One can imagine extending this to higher-order multipole matching, but I will not elaborate further on that possibility here. For a sufficiently distant model boundary, the monopole approximation is probably good enough for most purposes, allowing a fairly realistic extension to an infinite outer boundary.

## Surface Diffusion and the FSD Approximation

In its most basic form, the facet-kink CA model described above does an especially poor job describing the growth of low-angle vicinal surfaces. As illustrated in Figure 5.20, surface diffusion on faceted surfaces transports admolecules to kink sites, and this process can greatly increase the attachment coefficient near terrace steps (assuming $\alpha_{kink} \gg \alpha_{facet}$). This bit of physics is absent in the facet-kink model, where all facet boundary pixels are described by $\alpha_{facet}$, even if they are right next to kink sites [2015Lib1].

When incorporating this physical effect into a CA model, the increase in the attachment coefficient extends over a distance of approximately $(x_{surf}/a)$ pixels, where $x_{surf}$ is the surface diffusion length and $a$ is the size of a water molecule [2015Lib1]. Although $x_{surf}$ is not well known on faceted ice surfaces, one expects $(x_{surf}/a) > 30$ (see Chapter 2), which can present a sizable increase in the attachment coefficient over large vicinal surfaces in a CA model.

**Figure 5.20: (below) On a low-angle vicinal surface, molecules can diffuse along a faceted surface to reach kink sites where they readily attach. Thus $\alpha \approx \alpha_{facet}$ far from a kink site, while $\alpha \approx 1$ within one surface-diffusion length from a kink site. Here we have assumed a high Ehrlich–Schwoebel barrier that prevents diffusion over the tops of terrace steps.**

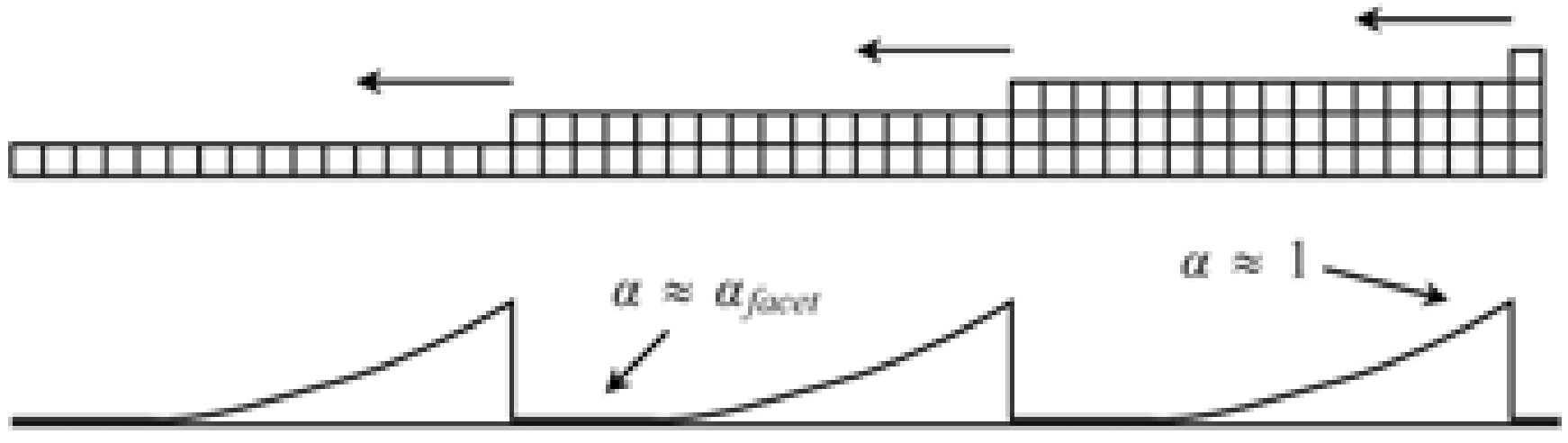

It is not outrageous to speak of a fast-surface-diffusion (FSD) approximation, where one simply assumes $\alpha \approx 1$ on all terraces that include a kink site. In this approximation, most of the ice surface would have $\alpha \approx 1$, while only small faceted "island" terraces, which I also call "upper terraces", would be described by $\alpha_{basal}$ or $\alpha_{prism}$. I suspect that the FSD approximation may be a fairly good representation of the actual attachment kinetics, and this would be an interesting regime to explore with models.

The bare facet-kink CA model, on the other hand, could be called a low-surface-diffusion (LSD) approximation, as it neglects surface diffusion entirely. This is almost certainly a poor approximation of the actual ice surface physics, although, once again, the actual surface diffusion lengths have not been well determined.

## Concave Growth

The growth of shallow concave plates is perhaps an interesting testing ground for exploring the accuracy of CA models. As shown in Figure 5.21, the issue of insufficient spatial resolution becomes especially acute with this geometry. Watching a CA model evolve in real time, one sees a peculiar time dependence in the model that does not happen in real life. With a perfectly faceted plate (a1 in the figure), $\alpha_{basal}$ is low across the entire plate, and the slow growth of this surface yields a high $\sigma_{surf}$ above it (see Chapter 3).

As soon as a kink site appears, however (a2 in the Figure 5.21), $\alpha$ increases substantially (especially with a FSD model), and thus $\sigma_{sur}$ drops. Moreover, this all happens essentially instantaneously, as this is how fast the supersaturation field responds in the Laplace approximation. Once the terrace fills in and is replaced by a fully faceted surface, $\alpha$ and $\sigma_{surf}$ again change instantaneously.

This is a curious sight to watch, as clearly it does not accurately model what must happen around an actual crystal (b in Figure 5.21). With full molecular resolution, even a shallow concave surface contains hundreds of terrace steps all marching inward. The attachment coefficient is thus $\alpha_{basal}$ near the plate edge and $\alpha \approx 1$ elsewhere, while $\sigma_{surf}$ over the surface would change little with time as the steps progress.

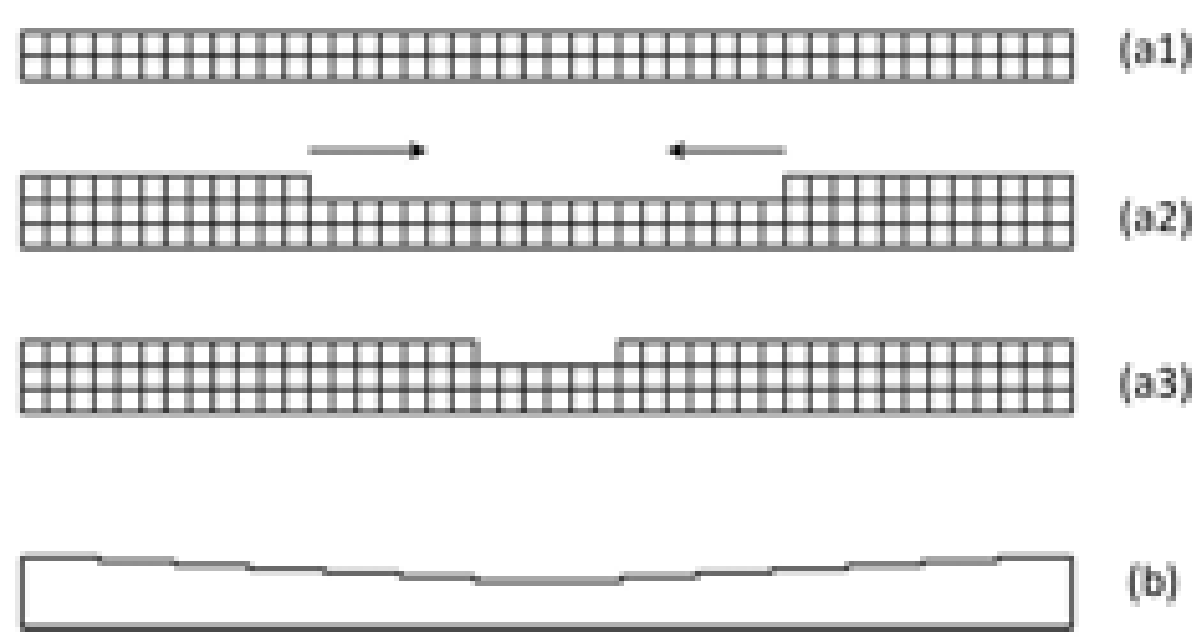

**Figure 5.21: The growth of shallow concave basal surface is difficult to reproduce with good accuracy in a cellular-automata model. In the model (a1,a2,a3), a new terrace nucleates at the edges of the plate and then grows inward. In real life (b), there is a continuous series of terraces that nucleate at the plate edge and propagate inward.**

This issue would go away with sufficient resolution in the CA model, and it would not be a problem if the concave depression is deep enough. But it is something to worry about. In contrast, a front-tracking model would likely handle this scenario much better. Even at low spatial resolution, the front-tracking model would look more like (b) in Figure 5.21, with a short faceted region at the outer edge and a segmented concave region within. The freedom to build a surface out of short line segments, rather than small blocks, gives a substantial advantage to the front-tracking model in this case.

## The Gibbs-Thomson Effect

So far in our CA modeling discussion we have ignored surface energy effects, focusing mainly on vapor diffusion and attachment kinetics, the latter depending on surface diffusion. The resulting models are likely reasonable approximations in many situations, but our examination of solvability theory in Chapter 3 suggests that surface-energy effects become important when $\alpha$ is large and $\sigma$ is low. This is borne out in model investigations like that shown in Figure 5.22 [2013Lib1].

In this modeling exercise, $\alpha_{basal} \ll \alpha_{prism} = 1$ and $\sigma_{\infty}$ was quite low, yielding the growth of a one-pixel-thick plate from the edge of a columnar crystal (top image in Figure 5.22). A single pixel measured 0.15 μm in this model, and the high curvature of the plate edge would create a large Gibbs-Thomson effect. Clearly the emergence of this ultra-thin plate would have been suppressed by surface energy effects (see Chapter 2), so this result is not physically plausible.

There are several ways to avoid this problem in a CA model. One is simply to avoid regions of parameter space where low $\sigma$ and high $\alpha$ can occur simultaneously. This is not especially difficult to arrange, and the model crystal in Figure 5.22 was something of an anomalous case. Another approach is to increase the CA pixel size $\Delta x$ to the point that the Gibbs-Thomson effect is negligible even with one-pixel-thick structures. And again, this is not especially difficult to arrange, but it is not a very satisfying approach to the problem. Of course, a better solution is to add the correct surface-energy physics into the CA model, thereby obviating the need to avoid certain areas of parameter space.

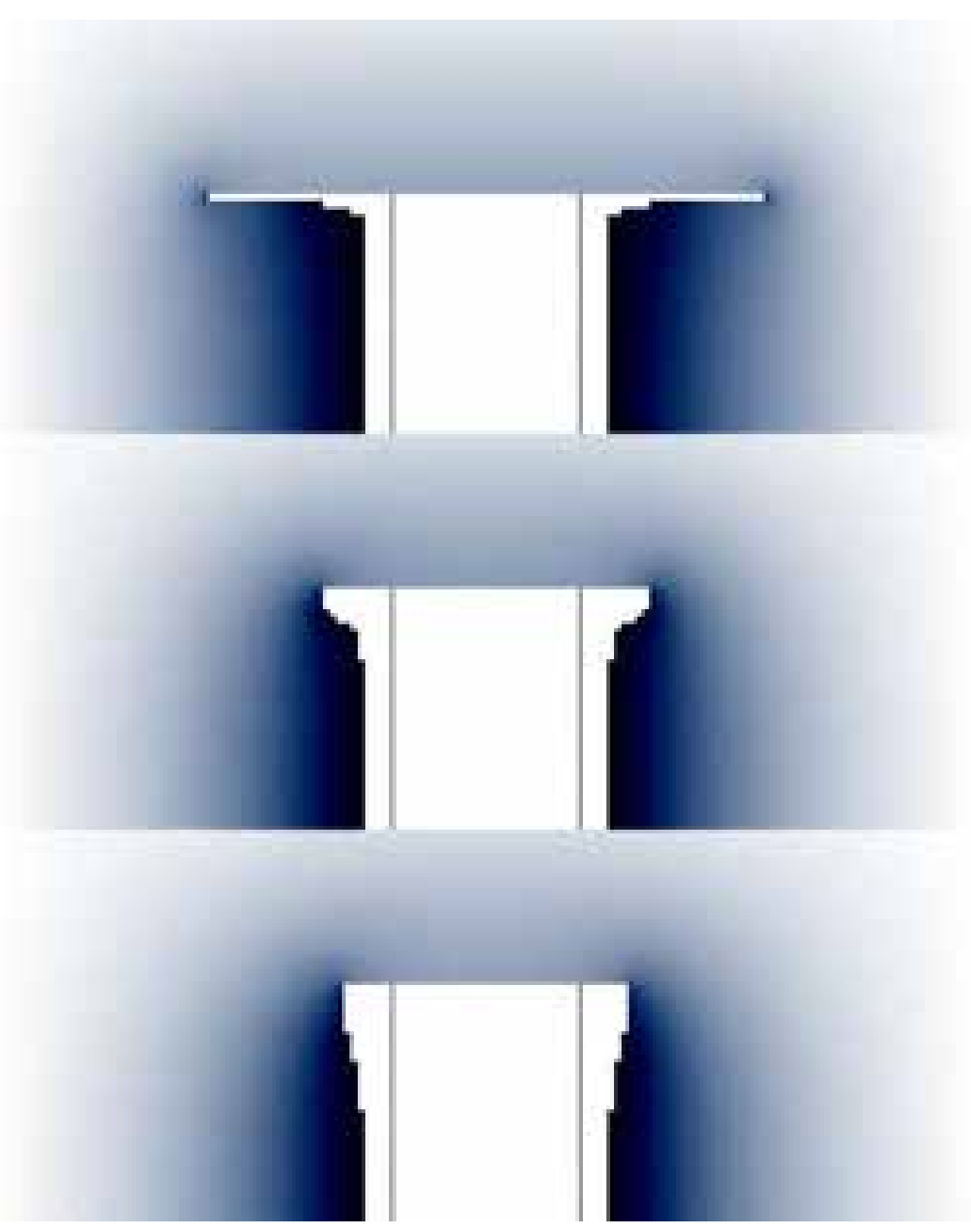

**Figure 5.22: An illustration of the Gibbs-Thomson effect in a cylindrically symmetric CA model of a plate growing from the edge of a column [2013Lib1]. With zero surface energy (top), a one-pixel-thick (0.15 μm) plate grows from the edge of the column, which is not a physically plausible solution. A Gibbs-Thomson length of $d_{sv} = 0.3\ nm$ (middle) or $d_{sv} = 1\ nm$ (bottom) suppresses the thin-plate growth. Vertical lines show the original seed crystal, while the supersaturation around the crystal is proportional to the image brightness.**

Because the Gibbs-Thomson effect is quite small in snow crystal growth, it is sufficient to approximate it rather crudely, as this is enough to eliminate one-pixel-wide plates and other non-physical model occurrences. One way to accomplish this quite easily in the CA model is to use the widths of the outermost terraces, which, owing to their extreme positions, do not include any kink pixels. The values of $L_{rmax}$ and $L_{zmax}$ (see Figure 5.23) can be used to estimate the edge curvatures, and the precise algorithm used is not very important.

The Gibbs-Thomson effect can be ignored in calculating the supersaturation field, as its effect is negligibly small. It need only be included in the calculation of the pixel growth, specifically replacing the usual $v_n = \alpha v_{kin} \sigma_b$ with $v_n = \alpha v_{kin}(\sigma_b - d_{sv}\kappa)$, using the known Gibbs-Thomson parameter $d_{sv}$ and a roughly estimated curvature $\kappa$. Although the outer terrace widths are not extremely accurate curvature indicators, Figure 5.22 demonstrates that this method is sufficient to suppress the formation of structures with especially high surface curvature.

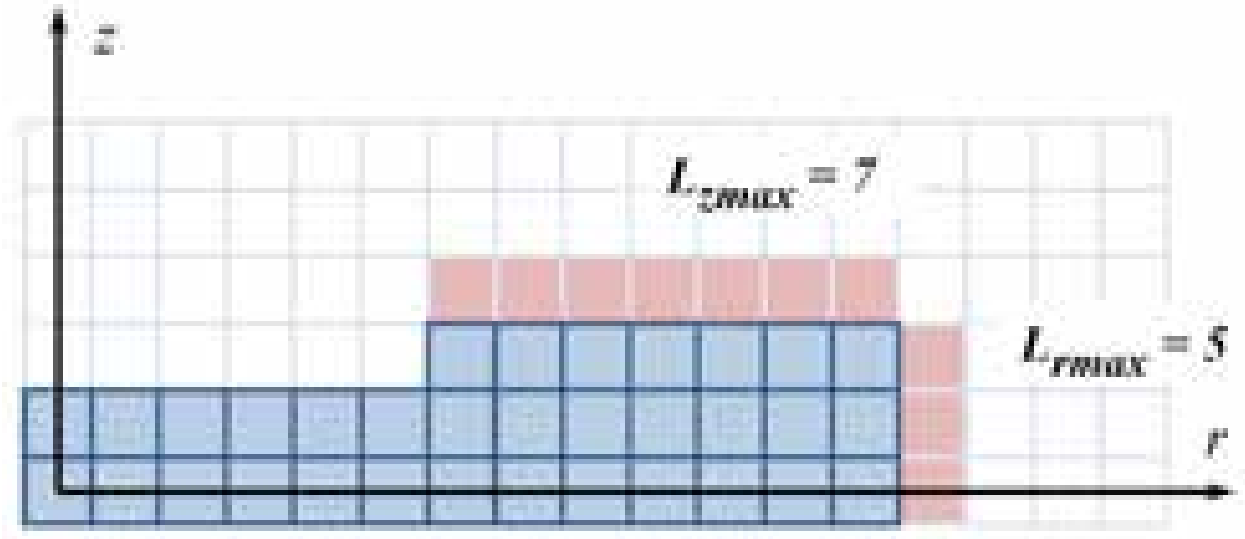

**Figure 5.23: A CA model in which the "outer" basal and prism boundary pixels are shown in pink. These pixels represent upper terraces with no adjoining kinks sites, and the terrace widths can be used to roughly estimate the edge curvature for including the Gibbs-Thomson effect.**

## Minimum Feature Sizes

While the CA grid size $\Delta x$ is somewhat arbitrary, it should not be made too large if one expects to reproduce realistic snow crystal structures. As we saw in our discussion of solvability theory (see Chapter 3), the characteristic radius of curvature of a growing dendrite tip is roughly

$$R_{tip} \approx \frac{2X_0}{s_0\alpha} \tag{5.28}$$

and measurements of ice dendrites in air have yielded $R_{tip} \approx 1\ \mu\text{m}$ and $\alpha s_0 \approx 0.25$ for fernlike dendrites near -15 C and $R_{tip} \approx 1.5\ \mu\text{m}$ and $\alpha s_0 \approx 0.2$ for fishbone dendrites growing near -5 C [2002Lib]. In both cases the tip structure was quite rounded, suggesting $\alpha \approx 1$.

In addition to dendrite tips, interferometric measurements of thin plates growing in a free-fall chamber showed thicknesses down to 1 μm at temperatures near -2 C and again near -12 C [2008Lib1]. Column diameters as low as a few microns were also observed near -5 C [2009Lib]. These observations all suggest that a grid size of a few times $X_0$ should be sufficient to reproduce essentially all snow crystal structures, where $X_0 \approx 0.15\ \mu m$ in normal air. Finer structures are likely suppressed by the Gibbs-Thomson effect and perhaps by additional surface diffusion effects.

## Edge Sharpening Instability

The outer facet widths defined in Figure 5.23, and the surface curvatures derived from them, can also be used to examine the Edge-Sharpening Instability (ESI, see Chapter 4). The basic idea here is to make the attachment coefficient depend on curvature $\kappa$, just as the effective supersaturation depends on $\kappa$ via the Gibbs-Thomson effect. These are two very different physical effects, but both can be included in the CA model using the outer facet widths.

Libbrecht et al. [2015Lib2] explored this idea somewhat by comparing CA models with experimental measurements of thin plates forming on electric needles near -15 C. Some additional details are presented in Chapter 8, with the results generally supporting the ESI and its CA models. While much additional work is needed in this area, this result certainly suggests that much could be learned from continued careful comparisons between CA models and experimental measurements.

## A Scaling Relation

If we switch variables from physical dimensions $(r, z)$ to scaled dimensions $(\xi_r, \xi_z) = X_0^{-1}(r, z)$, this converts the surface boundary condition, Equation 5.6, to dimensionless form. Laplace's equation is essentially unaltered by this variable change, so our CA models will proceed equally well in dimensionless coordinates, while physical time intervals change from $\delta t = \Delta x/\alpha v_{kin}\sigma$ to $\delta t = X_0\Delta\xi/\alpha v_{kin}\sigma$. What this all means is that we have a scaling relation for growth as a function of the diffusion constant $D$, or equivalently a scaling relation with air pressure $P$ [2013Lib1].

Because $X_0 \sim D \sim P^{-1}$, we see that increasing the air pressure by a factor of two will result in a crystal that grows half as large in a time that is twice as long compared to growth at the original pressure. This assumes all other aspects of the model, for example the attachment coefficients, are unchanged as a function of pressure and crystal size.

Immediately this scaling relation explains some prominent characteristics of snow crystal growth as a function of pressure. At low pressures, for example, crystals grow rapidly into faceted, prismatic shapes, even when the crystals are quite large. In normal air, on the other hand, initially faceted prisms quickly branch into dendritic morphologies, and dendritic shapes at higher pressures have been observed to show generally finer structural features [1976Gon].

At least at a qualitative level, the observed pressure dependencies are nicely explained by this simple scaling relation. Besides just morphologies, however, the scaling relation also makes clear predictions regarding growth rates as a function of pressure, although these have not yet been experimentally confirmed.

This scaling relation comes with numerous caveats, however, as it assumes that all other factors (other than particle diffusion) are independent of pressure and physical scale. For example, the scaling relation requires that $\alpha$ be independent of pressure, while the evidence suggests that $\alpha_{prism}$ depends rather strongly on pressure near -5 C (see Chapter 4). In addition, as particle diffusion becomes rapid at lower pressures, heat diffusion begins to dominate as a factor that limits growth, and this complicating factor would negate the simple scaling relation.

Other physical effects, including surface energy, surface diffusion, and the edge-sharpening instability, may also affect pressure scaling in various regions of parameter space. Thus, although expressing the problem in dimensionless coordinates brings some mathematical appeal, I find it tends to obscure the physics as well. And there is no getting around the fact that snow crystal growth is not just a mathematical problem, as it involves a variety of physical processes acting over many length scales.

## Comparison with Experiments

Although cellular-automata models have some inherent shortcomings, they can do a pretty good job producing realistic snow crystal morphologies and growth rates. This first became abundantly apparent when 3D models yielded morphological structures that resembled real snow crystals to a much higher degree than other models, including ridging and other features. In this section I want to describe how CA models have fared quite well in comparisons with experimental observations as well, at least to the limited degree to which the models have been tested.

To date, no 3D models of snow crystal growth, of any kind, have been subjected to detailed comparisons with experimental observations. That day is coming, but so far only 2D cylindrically symmetrical models have been examined as a means to reproduce snow crystal growth measurements.

Figure 5.24 shows one of my favorite early examples of a CA model matching the formation of a thin-plate snow crystal growing on the end of an electric ice needle (see Chapter 8). The cylindrically symmetric CA model cannot reproduce the hexagonal faceting or ridge features, as these would require full 3D modeling. But it does reproduce the slightly concave plate growth and the shielding of the columnar growth just below the plate. In terms of overall morphological features, the model seems to get the details right.

Moreover, both the morphology and growth measurements were both adequately reproduced using one set of model parameters. After some tweaking of the outer-boundary supersaturation and the attachment coefficients, the model could be made to fit the growth measurements quite well, as can be seen in the figure. A clear result from this exercise was that $\alpha_{prism} \approx 1$ was essential to fit the data and morphology, matching our expectations based on the observation that the plate sprouts branches if the supersaturation is raised only slightly higher than was used in this experiment.

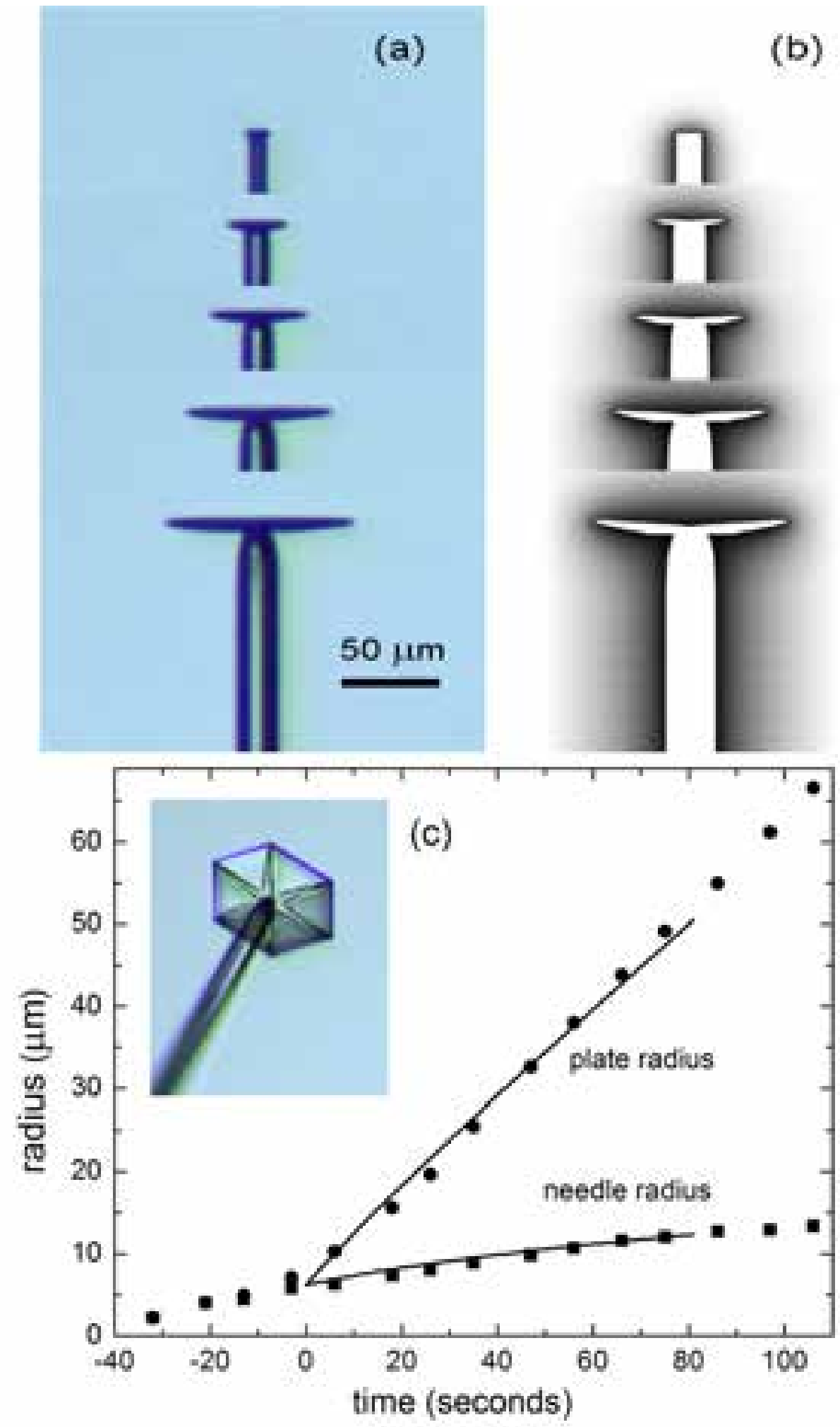

**Figure 5.24: These figures illustrate a quantitative comparison a 2D CA model with experimental data [2008Lib, 2013Lib]. (a) A composite image made from five photographs shows the growth of a plate-like snow crystal on the end of an electric ice needle, viewed from the side. (b) A cylindrically symmetrical 2D CA model reproduces the observations, also showing the water-vapor diffusion field around the crystal. (c) A quantitative comparison of experimental data (points) and the computational model (lines) shows good agreement. The inset photo shows the crystal in (a) from a different angle.**

Libbrecht et al. [2015Lib2] performed a series of measurements like the one shown in Figure 5.24, investigating the formation of thin plates on electric needles as a function of the far-away supersaturation level. This experiment is presented in some detail as a case-study in Chapter 8, as it illustrates the potential for using electric needles in quantitative studies of snow crystal growth.

This experiment is also a good example of how cylindrically symmetrical CA models can be used to analyze precise growth measurements to reach substantial, quantitative physical conclusions. As described in [2015Lib2], the data support the Edge-Sharpening Instability described in Chapter 4, indicating the need for some kind of morphology dependence in the attachment kinetics. The ESI model is still a hypothesis in need of additional testing, but it is abundantly clear that comparing CA models with experimental observations has much potential for yielding interesting scientific progress regarding the physics of snow crystal growth.

In another recent paper, I examined the role of surface diffusion in CA modeling by comparing several models with the formation of blocky prisms on electric needles near -10 C [2015Lib1]. As illustrated in Figure 5.25, the model morphology was substantially improved by including a high degree of surface diffusion, equivalent to the fast-surface-diffusion (FSD) approximation described above. Again, more data and more modeling are needed to draw firm conclusions, but we are making some progress. I have found that, even with 2D cylindrically symmetric models, the challenge of making models that agree with experiments requires some creative thinking regarding the behavior and physical origins of the surface attachment coefficients. The development of physically derived 3D models holds even greater promise for interesting results.

The growth of snow crystals on electric needles is especially suitable for quantitative modeling, as I describe in in Chapter 8. As another illustration of this statement here, note that Figure 5.24 shows a single thin plate emerging from the tip of a slender ice needle, and this morphology is nicely reproduced in the CA model calculation. Contrast this simple plate-on-needle morphology with the last two images in Figure 5.13, which illustrate what I call the “double-plate” problem.

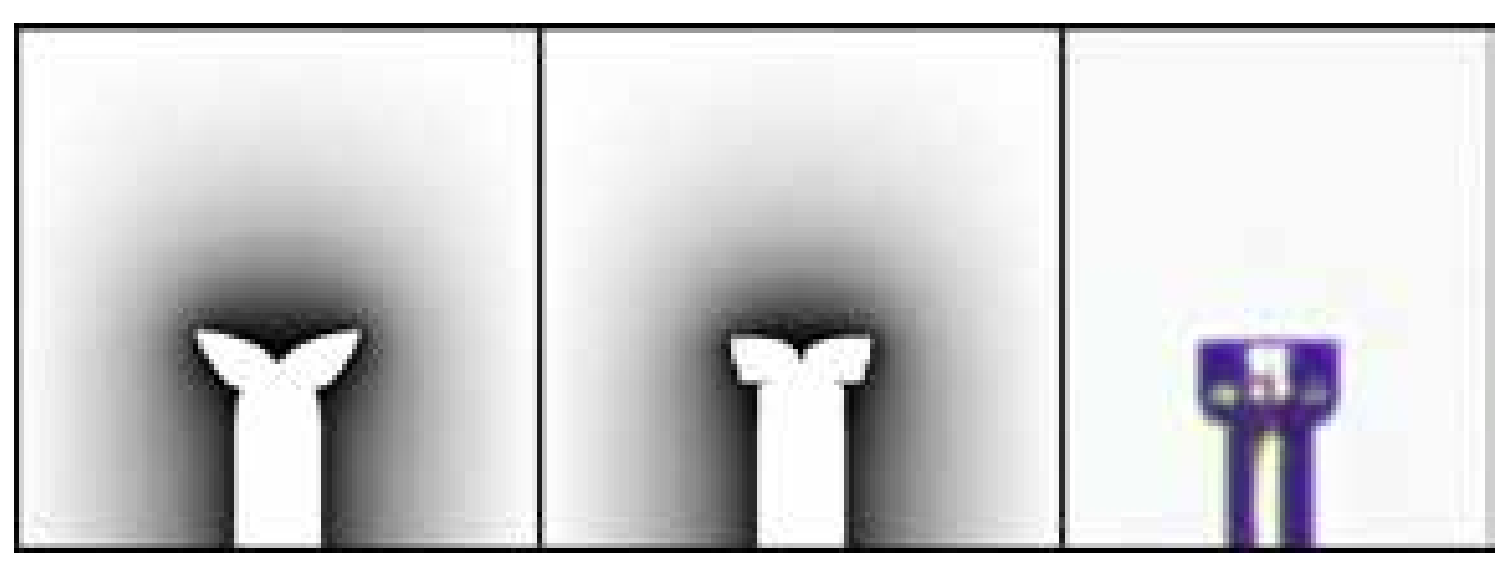

**Figure 5.25: The image on the right shows a blocky prism forming on the end of an electric ice needle near -10 C [2015Lib1]. Other than a few minor surface markings, the block has a relatively simple morphology with essentially no basal hollowing. The image on the left shows an attempt to model the blocky growth using a simple facet-kink CA model. With reasonable model adjustments, it was not possible to eliminate the substantial basal hollowing. The center image shows another attempt using a fast-surface-diffusion approximation, yielding an improved morphology with less basal hollowing. The CA models could not produce a sensible model with no basal hollowing, however, so perhaps some additional physics is needed in the model.**

Even beginning with small seed crystals, the models in Figure 5.13 both show the subsequent growth of double-plate crystals. In real life, these two plates would compete for water vapor (see Chapter 3), and usually one plate would soon overshadow the other. The result of this competition would be one large and one small plate, a phenomenon that is often observed in natural snow crystals (see Chapter 10).

The double-plate problem leads to something of a disconnect between model snow crystals and those grown in laboratory experiments. As seen in Figure 5.13, double plates readily form in models, but the unavoidable competition resulting from their close proximity prevents their realization in experiments. This problem is avoided using electric needles as seed crystals, because then only one plate emerges from the needle tip, both in experiments and models. I have found that the emergence of thin plates is an especially interesting phenomenon to explore, owing to unusual physical effects like the edge-sharpening instability, and the underlying physics is best studied if one avoids the double-plate problem by using electric ice needles.

## The 2D Future

Although clearly 3D modeling will be preferred in the long run, I believe that 2D cylindrically symmetrical models have substantial merits for investigating the physics underlying snow crystal growth dynamics, including:

1) With one fewer dimension, the run times for a 2D code are much faster than a 3D code. This allows one to run dozens or hundreds of models fairly quickly, which is highly beneficial when making detailed comparisons between models and experiments.

2) With a simpler geometry and fewer special cases to deal with, a 2D code is easier to write and modify than a 3D code. This makes it generally easier to incorporate additional physics like the Gibbs-Thomson effect, the edge-sharpening instability, and surface diffusion. Thus a 2D model is more practical for investigating the overall importance of these effects in conjunction with experimental observations.

3) A 2D cylindrically symmetric model can provide a reasonably accurate approximation for simple snow crystal morphologies, including simple plates and columns, hollow columns, capped columns, and simple forms growing on electric needles. Thus the 2D model is well suited for examining basic morphological changes in growth behavior with temperature, supersaturation, and background gas pressure. Low-pressure growth, exhibiting overall simpler structures than at higher pressures, is especially amenable to 2D modeling.

4) Most of what we have learned to date about snow crystal attachment kinetics has been from measurements of small crystals with relatively simple morphological structures. Here again, a 2D model should be sufficient for further investigations along these lines.
5) Working with a 2D CA model is a good prelude to building a full 3D CA model, as the 2D model already exhibits interesting behaviors and puzzling quirks, and learning these will benefit future efforts with full 3D modeling.

## 5.4 Three-Dimensional Cellular Automata

As described in the previous sections, 1D and 2D models give one an instructive perspective into many of the good and bad aspects of modeling snow crystal growth using cellular automata. CA models are generally simple to construct and fast to run, but it is difficult to remove the mathematical anisotropies that are essentially hard-wired into the fixed grid and CA rules.

Several of the foibles associated with the CA technique are relatively easy to see and understand in 2D cylindrically symmetric models, as I attempted to describe above. 3D models have not yet been abundantly explored, and the extra dimension will likely introduce even more hidden quirks that have not yet been discovered. Nevertheless, 3D modeling is the ultimate goal, so we examine that now. Much of this section is based on work done by Gravner and Griffeath [2009Gra] and by Kelly and Boyer [2014Kel], together with some additional embellishments derived from my own research.

### A 3D Hexagonal Grid

Figure 5.26 illustrates a hexagonal grid of cells appropriate for a 3D cellular-automaton snow-crystal model. In this grid, each pixel has eight nearest neighbors: two in the vertical direction and six in the horizontal direction, where here we use "horizontal" as a somewhat generic term referring to all directions perpendicular to the c-axis. As with the previous 1D and 2D models, the cells are labeled as ice, vapor, or boundary pixels.

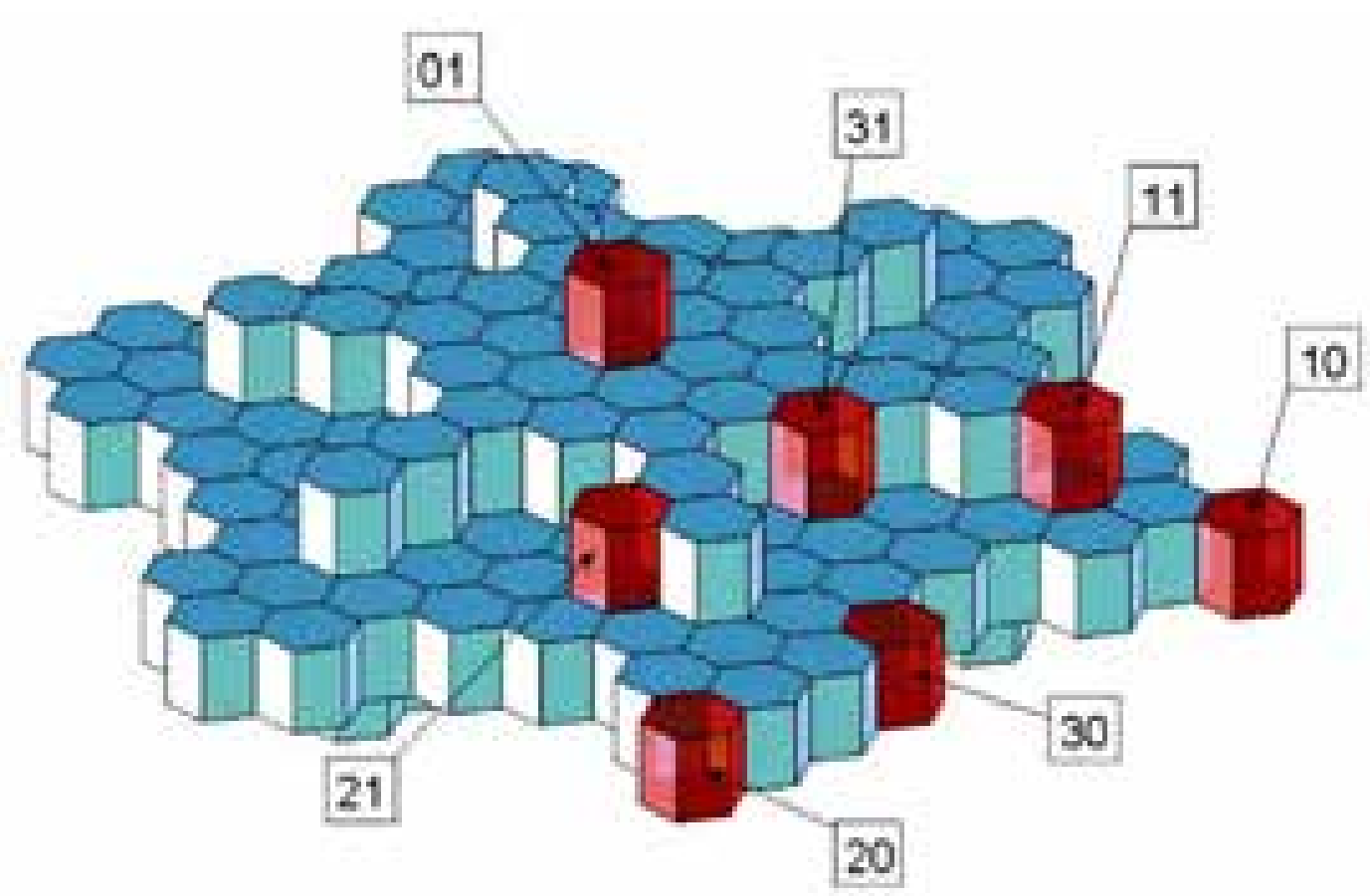

**Figure 5.26: A 3D hexagonal grid of cells for a cellular-automata snow-crystal model. Here the blue-green pixels represent ice and the red pixels show a few representative boundary pixels. Vapor pixels are not shown. The boundary pixels are labeled with [HV] nearest-neighbor data, where H is the number of adjacent horizontal ice pixels and V is the number of adjacent vertical ice pixels. Image from [2009Gra].**

Figure 5.27 shows a convenient mapping that takes a honeycomb structure in the horizontal plane to a simple rectilinear grid, which can be useful for bookkeeping purposes in the various CA algorithms. Note the definition of the spacing $\Delta x$ between prism-facet terraces shown in Figure 5.27. This is different from the definition in [2014Kel], for reasons that will become apparent when we discuss boundary conditions below. We also define $\Delta z$ to be the spacing between basal terraces, and we usually assume $\Delta z = \Delta x$.

Like the previous CA models in this chapter, our 3D model will be completely deterministic, including no random walks or random probabilities of any kinds, and evaporation will not be included. Running the model twice with the same initial conditions will produce the exact same result. Such a deterministic model must always exhibit perfect six-fold bilateral snow crystal symmetry, simply because the input physics and external boundary conditions contain this same symmetry by definition.

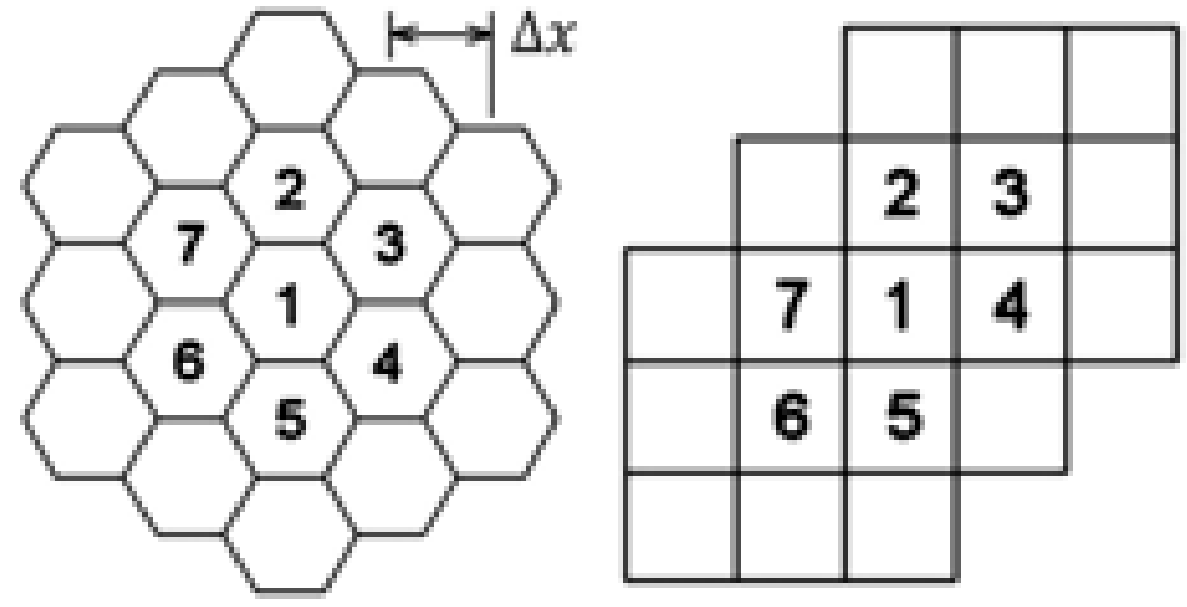

**Figure 5.27: A coordinate mapping that connects a 2D hexagonal grid to a 2D Cartesian grid, with numbers showing corresponding pixels. Note the definition of the horizontal coordinate scale $\Delta x$, equal to the spacing between prism facet terraces. We typically assume $\Delta x = \Delta z$, the latter being the distance between basal facet terraces.**

Because of this intrinsic symmetry, the CA model need only include 1/24th of the total physical space being modeled, with reflection boundary conditions recreating the full space from the 1/24th slice. A reflection boundary condition is applied about the $z = 0$ plane, as it was with the 2D model above, and similar boundary reflections occur at the edges of the 30-degree wedge shown in Figure 5.28. As one might expect, there is considerably more bookkeeping involved in a 3D model than a 2D model, which is simply the price one has to pay for the added complexity.

## Boundary Pixel Attributes

Another complicating issue with a 3D model is the plethora of different boundary pixel types, as illustrated in Figure 5.26. Moreover, to encompass all of the varied physical processes governing snow crystal growth, it is necessary to consider both non-local and nearest-neighbor interactions, as I discussed somewhat in the 2D model above. For this reason, it is necessary to describe the different types of boundary pixels and their various attributes with some care.

For example, a [01] boundary pixel indicates a position on a basal facet, and we also want to label this pixel with information relating to its surroundings beyond its nearest neighbors. One approach to accomplishing this is to count of the number of boundary pixels in each of the six directions out from the pixel in question, staying in the same basal plane, as illustrated in Figure 5.29. Doing so yields six integer pixel lengths $\pm L_i$, where the value is positive if the line of boundary pixels ends with a ledge (a terrace step approached from the top) and the value is negative if the line ends in a kink (a terrace step approached from below).

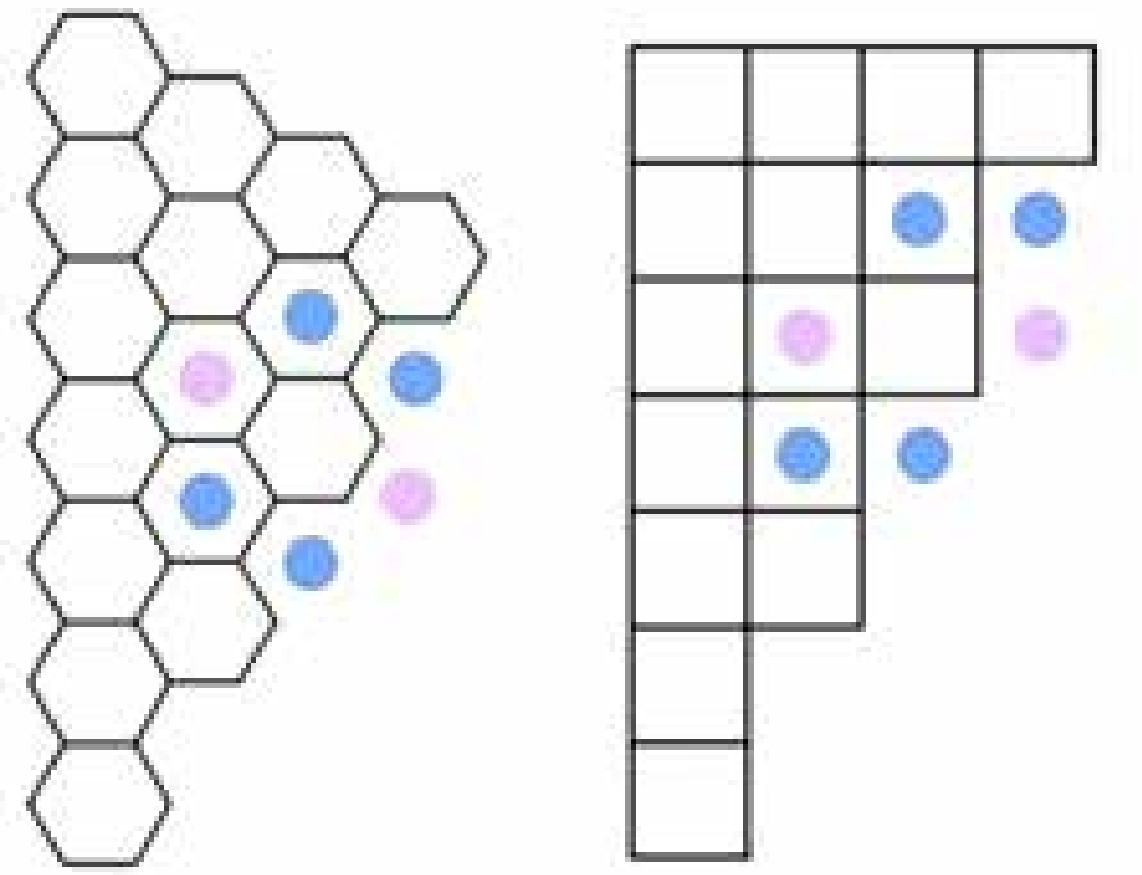

**Figure 5.28: This diagram shows the 1/12th slice of horizontal space needed to model a symmetrical snow crystal. Reflection boundary conditions apply on the two long edges, and colored dots illustrate some reflected pixels on the "ragged" edge. The short upper edge is the far-away boundary of the model.**

From these six lengths, one can extract quite a lot of useful information about the crystal structure near that boundary point, including:

1) If all the $L_i$ are positive, then the boundary pixel in question lies on an "upper" terrace, which is identified as having no ice terraces on top of it. On the faceted upper-terrace surface, $\alpha_{basal}$ is the appropriate attachment coefficient. As described in Chapter 4, the best functional form for this term is $\alpha_{basal} = A\exp(-\sigma_0/\sigma_{surf})$, where $A$ and $\sigma_0$ are physical parameters included in the model.

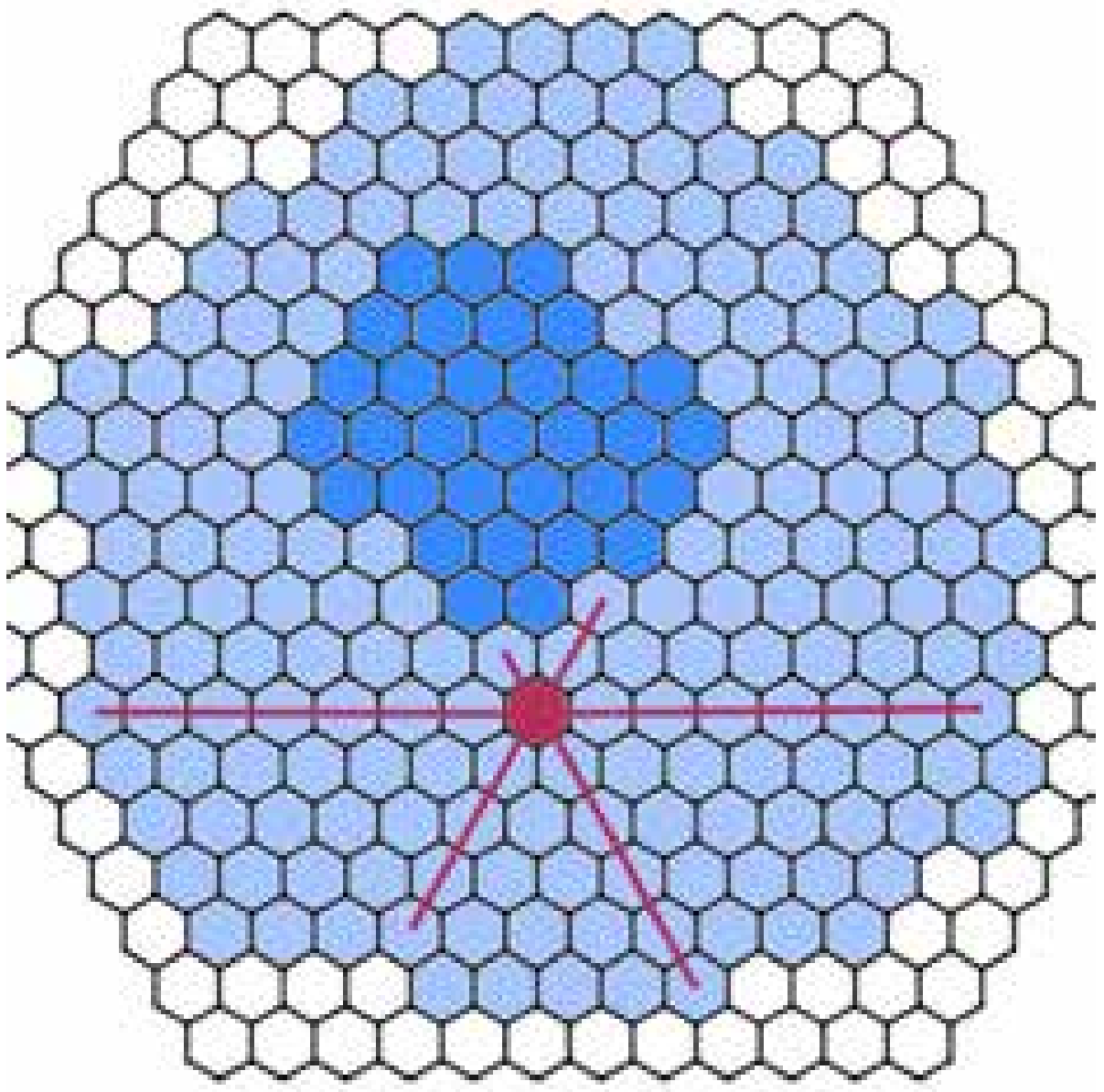

**Figure 5.29: This diagram of several basal terraces shows a topmost "upper" basal terrace in dark blue, the next lowest terrace in light blue, and the terrace below that in white. The red cell illustrates a representative boundary pixel that lies atop the light-blue terrace. Counting the number of same-terrace boundary pixels in the six directions shown yields the $L_i$ vector [-1,-2,+7,+5,+4,+7].**

2) If at least one of the $L_i$ is negative and small, then surface diffusion can carry admolecules to kink sites where they are readily adsorbed. In this case, the boundary pixel is best described by $\alpha \approx 1$.
3) If all the $L_i$ are positive and two opposing $L_i$ are both small, then the boundary pixel may lie on a thin basal edge (as one might find on the edge of a hollow column). This may change the value of $\alpha_{basal}$ through the ESI mechanism, if one wishes to incorporate such an effect in the model. Additionally, a thin edge suggests a high surface curvature, which lowers the effective supersaturation via the Gibbs-Thomson effect, and one may also wish to incorporate this effect into the model.
4) The six $\pm L_i$ can also be used to infer something about the vicinal angle of the surface near the boundary pixel in question. As described in the 2D model above, such knowledge is useful for turning a facet-kink model into a facet-vicinal model, and the latter has substantially reduced intrinsic model anisotropy.

All these possibilities exist for every [01] boundary pixel, which is just one type of boundary pixel, and perhaps the simplest type at that, as seen Figure 5.26. One must face an unfortunate reality in snow crystal that the underlying growth physics is complicated, so many different effects must be considered before a computational model will reproduce real life. It is not obvious at this point which physical effects must be included to high precision, which can be ignored altogether, and which are necessary but only to a rough approximation. Labeling each [01] boundary pixel with the six $\pm L_i$ is one way to incorporate a fair amount of flexibility into the model, which can then be used to explore different physical effects.

Moving on, [20] boundary pixels describe prism facets, so, like the [10] boundary pixels, it is important to characterize their surroundings carefully. Again, we define four lengths $\pm L_i$ by measuring the distances to the nearest ledges or kinks in each of the four vertical and horizontal directions. The discussion is then essentially identical to that for the [10] boundary pixels, except that we know that modeling of the prism-facet edges of thin plates is even more likely to involve some unusual physics like the ESI effect, the Gibbs-Thomson effect, etc. In particular, suppressing the growth of one-pixel-thick plates via the Gibbs-Thomson effect should be incorporated into the model.

Although the facet surfaces must be described carefully in any realistic snow crystal model, we can be a bit more cavalier regarding many of the remaining boundary pixels. For example, simply setting $\alpha = 0$ for all [10] pixels should be fine, as the molecular attachment at isolated [10] tips will be weak. At the same time, one can likely assume $\alpha = 1$ for all "kink-dominated" boundary pixels, such as [30], [40], [21], etc., as these are all tight-binding sites.

Snow crystal growth is largely facet-dominated, as I described above, so these model simplifications are likely acceptable over a broad range of growth conditions.

Note that the boundary pixel attributes must be recalculated after each growth step in the model. Every time a single boundary pixel turns to an ice pixel, the boundary geometry changes along with many of the $\pm L_i$ around it. This issue is usually handled by defining all the boundary pixels anew after each growth step and immediately recalculating all the attributes for the set. Some computational savings could be realized, however, by only recalculating boundary-pixel attributes near the position of the last growth step, as this is the only region where the boundary changes significantly during that step. It is not obvious that this savings would be substantial, but every little bit helps.

## Laplace Approximation

As discussed previously in this chapter, the low Peclet number associated with snow crystal growth means that the particle diffusion equation turns into Laplace's equation, and growth modeling can be divided into separate diffusion and growth steps. This latter point, first made by Kelly and Boyer [2014Kel], provides a substantial simplification in CA modeling.

The first step is to assume a static crystal surface and iterate to a solution of Laplace's equation in the space surrounding the crystal. For a 3D model, the optimal propagation equation (see Equation 5.15) becomes

$$\sigma_a(k+1) = \frac{1}{9}\sum_{i=1}^{6}\sigma_i(k) + \frac{1}{6}\sum_{i=7}^{8}\sigma_i(k) \qquad (5.29)$$

where $\sigma_i$ is a vapor pixel and the sum is over its eight nearest neighbors. As usual with CA models, this is the simple part, and the level of precision is mainly limited by the number of iterative steps computed.

## Outer Boundary and Monopole Matching

The next-easiest part of the model is the outer boundary, and again we can use monopole matching to extend the outer boundary to infinity to a satisfactory approximation. This avoids complications associated with an adaptive grid, allowing the use of a constant grid spacing with the outer boundary that is fairly close to the growing crystal. How close depends on the overall accuracy desired, as a close outer boundary will distort the supersaturation field to some extent.

For our 3D grid, the outer boundary is defined as (see Equations 5.10 and 5.27)

$$\sigma_B(\rho_{far}) \to \sigma_\infty - \frac{dV/dt}{4\pi\rho_{far}X_0 v_{kin}} \qquad (5.30)$$

where $\sigma_B$ refers to an outer boundary pixel and $\rho_{far}$ is the distance from the model's physical center to that outer boundary pixel location.

The rate of change of the total volume is given by (see Equation 5.26)

$$\frac{dV}{dt} = \sum \frac{G_1 \Delta x^3}{\delta t_b} \qquad (5.31)$$

where the sum is over all surface boundary pixels and $G_1 = 2/\sqrt{3}$ so the numerator is equal to the volume of a single pixel in our model, assuming $\Delta x$ as defined in Figure 5.27 along with $\Delta z = \Delta x$. The $\delta t_b$ are defined below.

## The Surface Boundary and Facet Dominated Growth

Most of the important physics in our model rests in the surface boundary conditions, so it is important that we define these to reflect the correct underlying physical processes as accurately as possible. This is most easily done for a simple basal surface, where the continuum surface boundary condition (see Chapter 3)

$$X_0\left(\frac{\partial\sigma}{\partial n}\right)_{surf} = \alpha\sigma_{surf} \qquad (5.32)$$

becomes (upper basal surfaces only)

$$\sigma_b = \sigma_{b+1}\left(1+\alpha(\sigma_b)\frac{\Delta x}{X_0}\right)^{-1} \qquad (5.33)$$

on our CA grid, where $\sigma_b$ is the supersaturation in the basal boundary pixel, $\sigma_{b+1}$ is the supersaturation in the vapor pixel just above the boundary pixel, and $\alpha(\sigma_b) = \alpha_{basal}$ is the attachment coefficient at the boundary pixel.

Note that this disagrees with the Kelly and Boyer boundary condition (Equation 11 in [2014Kel]), as the latter is incorrect for a flat basal surface. Note also that Equation 5.33 should be used as part of the iterative process of defining the supersaturation field, as was discussed in connection with Equation 5.7 above. When applied in this way, any functional form for $\alpha(\sigma_b)$ can be used, regardless of complexity. Both $\sigma_b$ and $\alpha(\sigma_b)$ should converge smoothly to the correct result during this iterative process.

Unfortunately, the 3D grid geometry is such that Equation 5.33 cannot be used for other boundary pixels, even on a prism facet. We therefore define a generalized boundary condition that is valid for all boundary pixels

$$\sigma_b = \sigma_{opp}\left(1+\alpha(\sigma_b)\frac{G_b\Delta x}{X_0}\right)^{-1} \qquad (5.34)$$

where $\sigma_b$ is any boundary pixel, $\sigma_{opp}$ is the average supersaturation in all vapor pixels that "oppose" ice pixels for this boundary pixel (see Figure 5.30), and $G_b$ is a dimensionless geometrical factor that must be defined for each boundary pixel. Conveniently, $G_b = 1$ for both [20] (prism facet) and [10] (basal facet) boundary pixels, which is why we defined $\Delta x$ as we did in Figure 5.27.

Admittedly, calculating $\sigma_{opp}$ is a chore for every boundary pixel on every iterative step of the relaxation process, but this is why we have computers. This could be done in a straightforward manner by using

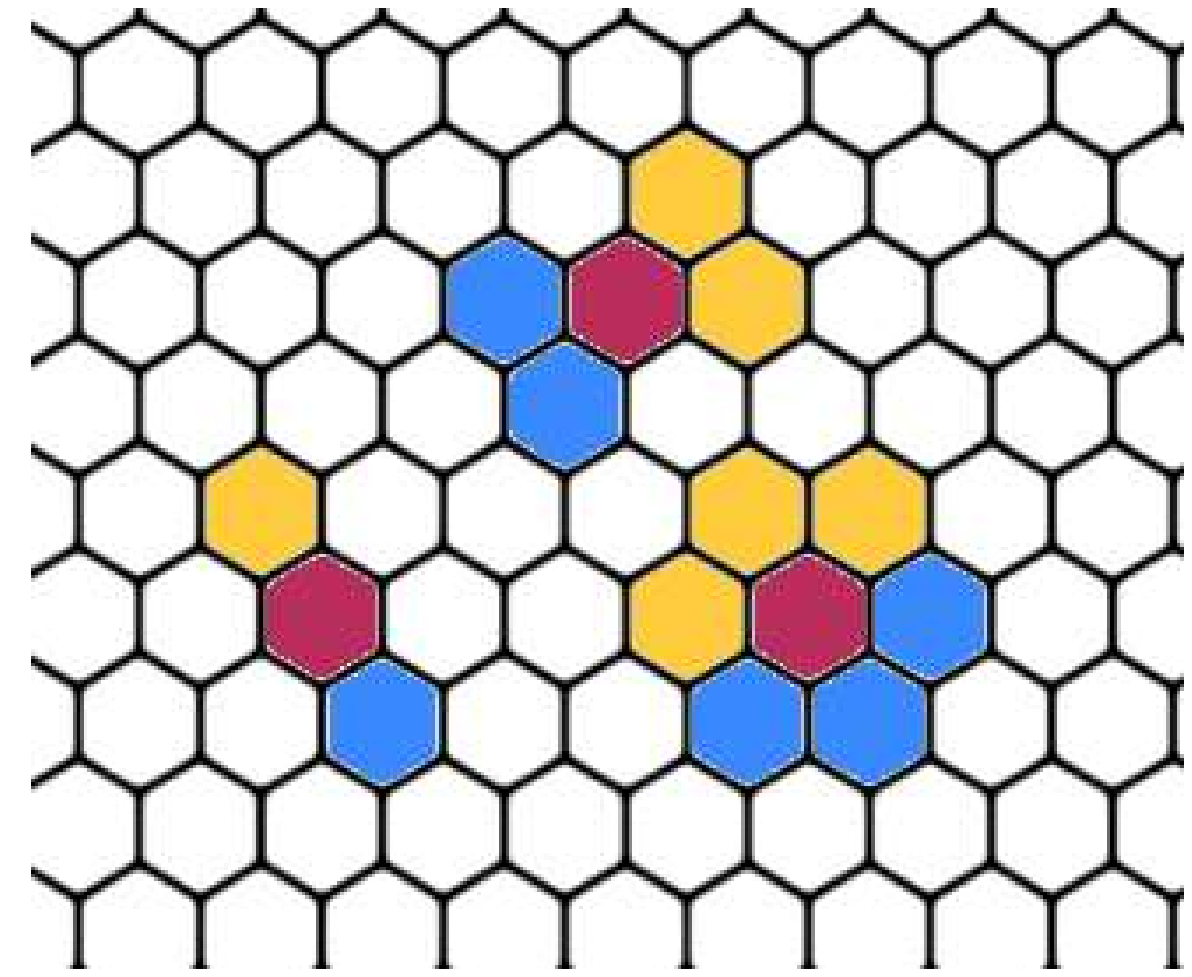

**Figure 5.30: This diagram illustrates the calculation of $\sigma_{opp}$, the supersaturation averaged over all pixels that "oppose" a given boundary pixel. In each of this examples, the red cell is a boundary pixel, blue cells are ice pixels, and the yellow cells "oppose" the ice pixels. The value of $\sigma_{opp}$ is calculated by averaging the supersaturations in the yellow cells. Note that this 2D diagram does not show additional cells in the vertical directions. In general, a boundary pixel labeled by [HV] nearest neighbors (see Figure 5.26) would have H+V "opposing" cells.**

$$\sigma_{opp} = \frac{1}{N_{weigh}}\sum_{i=1}^{8} M_i\sigma_i \qquad (5.35)$$

where $M_i$ is a weighting vector equal to 1 for an "oppose" pixel and 0 otherwise, and $N_{weight} = \sum M_i$. As with other boundary pixel attributes, the $M_i$ can be calculated once after each growth step. This vector then allows for rapid calculation of $\sigma_{opp}$ at each step in the Laplace iteration of the supersaturation field.

For facet-dominated growth, the values of the $G_b$ on non-facet boundary pixels may not greatly change the outcome of a model run. It might be beneficial to do the geometry correctly for a [21] boundary pixel, however, much like it was with the 45-degree surface in the 2D model above. A few added touches like this can substantially reduce the intrinsic anisotropy in the CA model, and this could improve how some familiar snow crystal features are reproduced in the model.

Another thought is to use $G_b = 1$ on [01] and [20] facet surfaces while leaving it as a constant, but adjustable, model parameter on all other surfaces to see what happens. It appears that some additional research is needed to determine the best course of action here. Greater algorithmic complexity can reduce intrinsic model anisotropies, but the additional effort may not greatly influence the outcomes for facet-dominated growth. Most important at this point is simply to make sure that the surface boundary condition in Equation 5.34 reproduces the correct attachment physics on the facet surfaces with the highest possible precision.

## Growth Steps

As with the 2D model described above, there are additional geometrical factors inherent in how we define a growth step. We begin with a generic functional form for the time required to "fill" the remainder of each boundary pixel, generalizing Equation 5.18 to

$$\delta t_b = \frac{H_b \Delta x}{v_n}(1 - f_b) \qquad (5.36)$$

where $f_b$ is the filling factor described above (equal to 0 when a boundary pixel first appears and 1 when a boundary pixel turns to ice),

$$\begin{aligned} v_n &= \alpha v_{kin} \sigma_{surf} \\ &= \alpha(\sigma_b) v_{kin} \sigma_b \end{aligned} \qquad (5.37)$$

is the growth velocity normal to the ice surface, and $H_b$ is a dimensionless geometrical factor near unity.

As with Equation 5.34, $H_b = 1$ for both [20] (prism facet) and [10] (basal facet) boundary pixels. This accurately describes facet growth partly because of our choice of $\Delta x$ in Figure 5.27, along with $\Delta z = \Delta x$, and partly because the surface normals lie along the grid axes for both primary facets.

As with the $G_b$ factors, determining $H_b$ for non-faceted boundary pixels is difficult to do with high accuracy, as the correct surface normal is not easily determined in a CA model. I described this same problem above with the 2D model, where the facet-vicinal model appears to achieve reasonably accurate geometrical factors for all flat vicinal surfaces, thus substantially reducing the built-in anisotropies in the CA model.

Extending the facet-vicinal ideas to a 3D model would be challenging, as one would have to deal with quite a bit of additional geometrical complexity. I can imagine a number of numerical strategies for calculating approximate $G_b$ and $H_b$ using the $\pm L_i$ data, but nothing that would be both simple and accurate.

At this stage in our understanding of 3D CA snow crystal models, I suspect that it is premature to worry too much about the intrinsic anisotropies and geometrical precision wrapped up in these geometrical factors. Perhaps simply using $G_b = H_b = 1$ for all boundary pixels is good enough to produce reasonable model results. It is valuable to recognize that these geometrical factors exist, and that errors in their calculation result in intrinsic anisotropies in the model. Understanding the extent and importance of this problem, however, I leave for another day.

## Attachment Coefficients on Faceted Surfaces

I believe that the most important challenge right now is to incorporate accurate attachment coefficients in snow crystal models, particularly on the primary facet surfaces. This will be something of an iterative scientific process, as detailed 3D modeling will help us better understand the attachment kinetics, which, in turn, will allow us to build better models, and so on.

As I described in Chapter 4, targeted experiments measuring $\alpha_{basal}$ and $\alpha_{prism}$ directly are necessary but probably not sufficient, as the attachment kinetics appear to depend on supersaturation, temperature, air pressure, terrace width, and perhaps other factors yet unknown. Growing and modeling full 3D structures in air and other gases at different pressures will likely be needed to fully comprehend what is going on with the attachment kinetics. And we are only just beginning to execute this scientific program in earnest.

Given that our understanding is quite limited at present, a good first step is to parameterize $\alpha_{basal}$ and $\alpha_{prism}$ with $\alpha_{facet} = A\exp(-\sigma_0/\sigma_{surf})$, where $A$ and $\sigma_0$ are model parameters that may depend on a variety of other factors. This functional form clearly applies at low background gas pressure (see Chapter 3), and it is likely a good approximation at higher pressures also. However, both $A$ and $\sigma_0$ may depend on pressure to some degree, especially near -5 C.

Whatever the form chosen, the faceted $\alpha_{basal}$ and $\alpha_{prism}$ apply only to upper-terrace boundary pixels or those that are far removed from any kink sites that can dramatically increase $\alpha$ via surface diffusion (see below). In terms of the $\pm L_i$ data, all these parameters must be greater than zero, or the smallest negative $L_i$ must be greater than some prescribed limit (which is a parameter in the model).

It is an unfortunate reality in snow crystal modeling that we do not fully understand attachment kinetics even on simple faceted ice surfaces, neither empirically nor at a basic physics level. Needless to say, this makes precision model-building something of a challenge at present, but we proceed regardless, one step at a time.

## Upper Terrace Effects

Sharp edges are somewhat common features in snow crystal growth, and it appears that these require some special attention in CA models. The edges of thin plates are especially prominent near -15 C, yielding narrow prism facets, but narrow basal facets can also be found on the edges of sheath-like hollow columns. In both cases, the radii of curvature of the edge surfaces appear to be few times $X_0$ in extreme cases, although the dimensions are not known with great precision.

We have regularly observed thin plates in normal air with overall thicknesses down to one micron using accurate interferometric measurements [2008Lib1, 2009Lib], and this thickness puts an upper limit on the edge curvature for these crystals. Sheath-like edges appear to be a few times thicker, but I am not aware of any accurate measurements for these crystals.

In a curvature class all their own are the electric ice needles (see Chapter 8), where the tips of c-axis needles have exhibited radii of curvature down to 100 nm in the most extreme cases [2002Lib]. Here the small radii were not measured directly, but inferred from the needle growth velocity using the Ivantsov relation (see Chapter 3), which probably yields accuracies of a factor of two or better. Incorporating electric-field effects in CA models is beyond the scope of this chapter, but suffice it to say that even 100-nm surface curvatures are not impossible to realize in snow crystal growth experiments.

There are two primary physical effects that one should consider when modeling these high-surface-curvature features: the Gibbs-Thomson effect (Chapter 2) and the edge-sharpening instability (ESI; Chapter 4). The former is a well-understood phenomenon, but appears to play a relatively small role in snow crystal growth. The latter is something of a hypothesis at present, barely tested and with uncertain physical origins, but it appears to play a substantial role in the formation of thin plates near -15 C. In both cases, however, it seems prudent at this point to include these phenomena in our modeling efforts, if for no other purpose than to test how much they influence snow crystal growth rates and morphologies.

I believe that both these physical phenomena can be incorporated into a CA model, to a reasonable approximation anyway, by considering only the upper terraces of faceted surfaces. The upper terraces are easily identified for a given boundary pixel by requiring that all the $L_i$ be positive, as was described above. Moreover, the curvature of the surface at a given point can be estimated from the $L_i$ data as well, for example by the sum of opposing $L_i$. The precise algorithm is probably not especially important, and one can imagine different approaches. What likely does matter is consistently identifying large arrays of high-curvature surfaces, for example the edges of thin plates that remain thin as the crystal grows.

In the ESI effect, a high-curvature prism surface (in the CA case, an upper prism terrace with at least one narrow width) would exhibit an attachment coefficient that is far larger than the usual $\alpha_{prism}$ associated with broad prism facets (see Chapter 3). This can be incorporated into the model by letting $\alpha_{pris}$ depend on the upper facet width, as extracted from the $L_i$ boundary pixel attributes. This factors into $\alpha(\sigma_b)$ in Equation 5.34, and we see that $\alpha(\sigma_b)$ changes automatically with every update of the boundary pixel attributes. The same goes for the growth step calculation in Equations 5.36 and 5.37.

For the Gibbs-Thomson effect, the surface curvature need only be used to adjust the growth step, so Equation 5.37 is replaced by the general form

$$v_n = \alpha(\sigma_b) v_{kin}(\sigma_b - d_{sv}\kappa) \qquad (5.37)$$

where $\kappa$ is again estimated from the $L_i$ boundary pixel attributes. Because the Gibbs-Thomson effect is not very strong in snow crystal growth, it is probably not necessary to provide an extremely accurate algorithm for estimating $\kappa$. Likewise, it is probably not necessary to include the Gibbs-Thomson effect when solving Laplace's equation to determine the supersaturation field around the growing crystal, as the perturbation arising from surface curvature is so small.

Throughout this discussion, we see that the key element for including the Gibbs-Thomson effect and investigating the ESI mechanism in our 3D CA model is the initial step of generating boundary pixel attributes that include non-local information via the $\pm L_i$ data. Only nearest-neighbor terms were used in previous 3D CA models [2009Gra, 2014Kel], but it is becoming clear (in my opinion) that non-local surface structure will be necessary before CA models will be able to match experimental observations of snow crystal growth rates and morphologies.

## Vicinal Surfaces and the FSD Approximation

While non-local information can be important on upper-terrace faceted surfaces, it is even more important on vicinal surfaces, which include essentially all non-upper-terrace surfaces in the CA model. The basic idea here is that admolecules on normal terraces (not upper terraces) can diffuse along the surface until they reach kinks sites where they are readily absorbed. Thus surface diffusion increases the attachment kinetics to $\alpha \approx 1$ for surface locations closer than $x_{diff}$, the surface diffusion length (see Chapter 2), to the nearest kink sites.

As I discussed above in connection with Figure 5.20, the important parameter for CA modeling is not $x_{surf}$, but $x_{surf}/a$, as surface diffusion effectively operates over $x_{surf}/a$ cells in the CA model. In a practical sense, this means that $\alpha \approx 1$ should be assumed on most non-upper-terrace surfaces in the model. Upper-terrace surfaces are described by $\alpha_{facet}$, as are surfaces where $|L_i| \gg x_{surf}/a$ for all negative $L_i$. But all other surfaces are best described by $\alpha \approx 1$. This is the essence of the Fast-Surface-Diffusion (FSD) approximation described above. Because essentially all non-faceted surfaces have $\alpha \approx 1$, the overall growth dynamics are largely determined by the behavior of the faceted surfaces.

Incorporating surface diffusion is relatively straightforward using the $\pm L_i$ boundary pixel attributes. If the distance to the nearest kink site is less than some prescribed value $L_{sd}$ in the model, then one simply sets $\alpha \approx 1$ for that boundary pixel and proceeds. It is likely, given our current knowledge of surface diffusion, that this will set $\alpha \approx 1$ over much of the surface for all but the most cleanly faceted crystals. Nevertheless, the existing measurements of $x_{surf}$ are quite poor, and the value could be quite different on the basal and prism facets. So $L_{sd}$ is perhaps best left as two adjustable parameters (one for each facet) at present.

Note that even a quite leaky Ehrlich–Schwoebel barrier does not mean that $\alpha \approx 1$ on upper-terrace surfaces. Moreover, the $x_{surf}/a$ logic described above does not apply to upper terrace surfaces like it does to other surfaces in CA models. If in the case of a leaky Ehrlich–Schwoebel barrier (see Chapter 2), admolecules near the edges of upper terraces could diffuse to an edge and be absorbed by a lower kink site. Physically, however, this only happens within a distance $x_{surf}$ from the edge, meaning it does not affect the entirety of a large facet area.

Assuming $x_{surf} \approx 10\, nm$ (see Chapter 2), we see that a leaky Ehrlich–Schwoebel barrier has hardly any effect on the growth of a large facet surface. It is reasonable to assume, therefore, that $\alpha_{facet}$ on an upper terrace in our CA model is also negligibly affected by a leaky Ehrlich–Schwoebel barrier. (An extremely narrow facet may be affected by a leaking E-S barrier, however; this possibility is discussed in Chapter 4 in conjunction with the ESI effect.)

## Quantitative Modeling

It is a telling statement that no 3D snow crystal model to date has ever been directly compared to quantitative experimental data. One interpretation of this state of affairs is that 3D modeling is hard, and that is certainly true. In particular, 3D modeling of faceted+branched crystal growth has only become possible in recent years.

Another interpretation of the above statement is that making precise measurements of growing snow crystals is difficult, and that is also true. I am particularly fond of the e-needle method for producing precise measurements of 3D structures in known growth conditions, and that too is a relatively new development.

The most likely reason that 3D models have not been compared with experiments, however, is likely just that snow crystal growth has not received much attention over the years. Other areas of materials science have commanded greater attention, commensurate with their potential for commercial applications, as is to be expected. Nevertheless, numerical models of crystal growth have become simpler with improved computational power, and progress in numerical techniques has been accelerating rapidly in recent years.

Scientists have pondered the formation of snow crystals for 400 years, ever since Kepler first puzzled over their six-fold symmetry [1611Kep]. The field has been pushed forward in fits and starts by a small number of interested researchers, making slow but steady progress over time. My own research suggests that all the pieces are just now beginning to come together, as we recently learned how to produce good experimental observations and how to build quite reasonable numerical models. The time is ripe to combine quantitative experimental observations of growth rates and morphological behaviors with corresponding computational modeling studies.

Our understanding of the underlying physics is incomplete at present, but that is what makes the scientific challenge especially interesting. The best way to proceed (in my opinion) is simply to plunge forward with both modeling and quantitative growth studies, comparing one with the other to see what works and what does not. The entire morphology diagram (see Chapter 8 in particular) lies waiting for a definitive explanation. There is much opportunity on this scientific track for greatly improving our understanding of snow crystal growth dynamics, and, hopefully, our understanding of crystal growth in general.

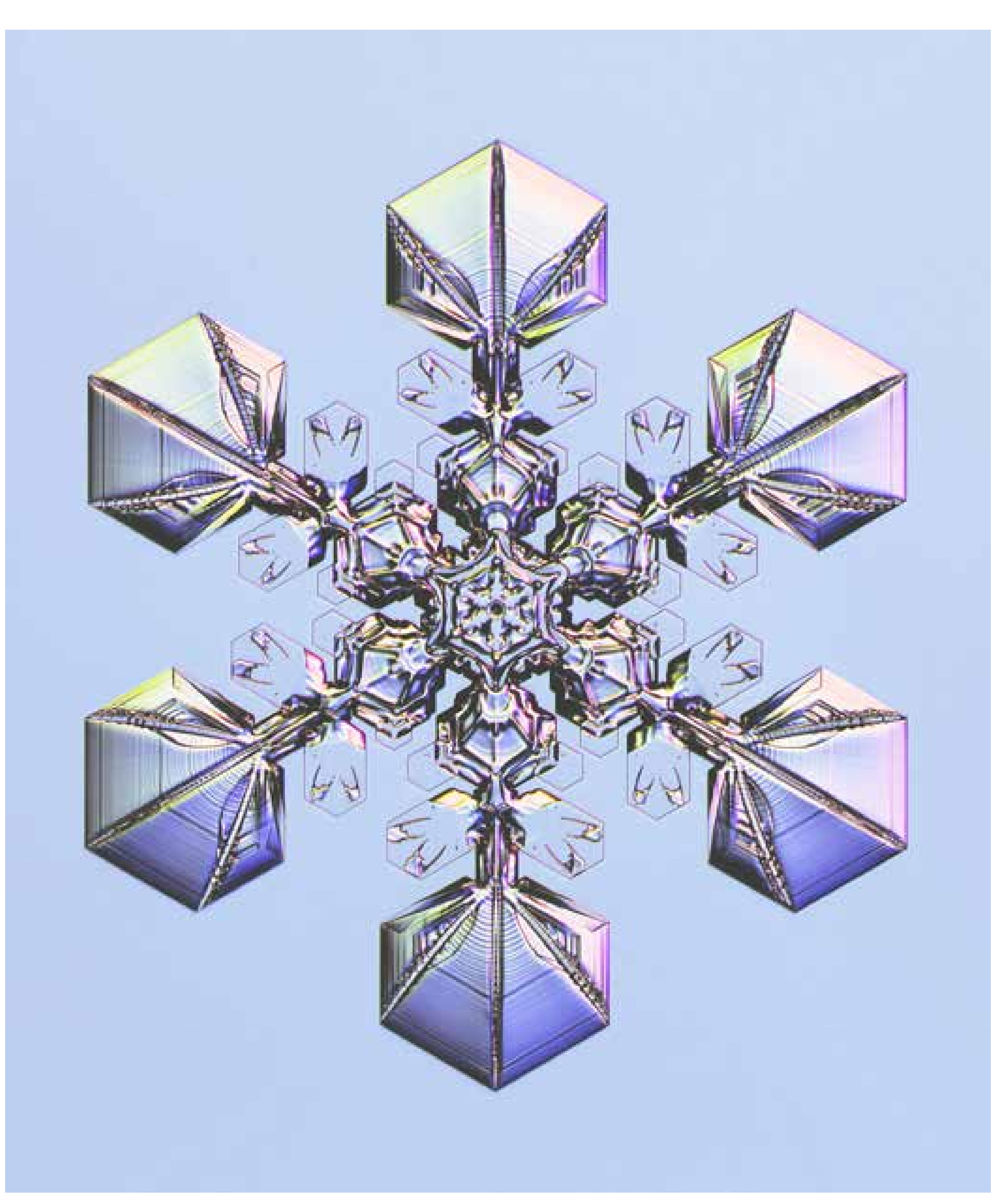

# Chapter 6

# Laboratory Snow Crystals

*A scientist does not study nature because it is useful; he studies it because he delights in it, and he delights in it because it is beautiful.*
– Henri Poincaré
*Science and Method, 1908*

In this chapter I examine the technology required to create synthetic snow crystals in the laboratory. The task involves a fairly broad skill set, encompassing mechanical and thermal engineering, optics, electronics, vacuum equipment, and photo-microscopy. Access to a machine shop is highly desired, and it certainly helps to have reasonably deep pockets and a general aptitude for building things. Although many key hardware components are commercially available, it often takes considerable design and engineering effort to assemble the pieces into a complete and functional snow-crystal growth chamber.

I was fortunate in that I began my work in snow-crystal research with a solid background in experimental physics and astrophysics, including considerable laboratory experience in the areas mentioned above. Nevertheless, as with any endeavor, there was a great deal I had to learn along the way. During several decades growing snow crystals for a variety of purposes, I have picked up quite a few tricks and techniques that have made a world of difference. In this chapter, therefore, I present an assortment of experimental methods and hardware recommendations that may be of use to the next generation of snow-crystal scientists and artists.

I include artists in the previous sentence because there are really two motivations for developing better synthetic snow crystals – science and art. Most of this book is devoted to the science side, where creating synthetic snow crystals allows researchers to examine how growth behaviors depend on temperature, supersaturation, air pressure, chemical additives, and a host of other parameters. Precise growth measurements can also be compared with analytic models and sophisticated computer simulations to better

**Facing Page: A synthetic snow crystal, measuring about three millimeters from tip to tip, grown in the laboratory by the author.**

understand how different physical processes guide snow crystal formation in detail.

On the artistic side, creating laboratory snow crystals provides a novel look at the exotic beauty of this fascinating phenomenon. Snowflakes falling from the sky are already quite striking when viewed under a microscope, but even the best specimens are inevitably a bit travel-worn, with rounded, partially sublimated features. Our view from the ground only hints at the unseen splendor of these icy works of art within the winter clouds.

By creating laboratory snow-crystal nurseries, one can observe razor-sharp facets, highly symmetrical branching, and a degree of morphological perfection that is rarely, if ever, found in nature. Figure 6.1 shows one example using the Plate-on-Pedestal technique described in Chapter 9. Beyond still photos, this method also allows *in situ* observation during the entire growth process, yielding some remarkable time-lapse videos of snow-crystal formation. I believe that there is considerable untapped artistic potential in this dynamic, self-assembled form of ice sculpture.

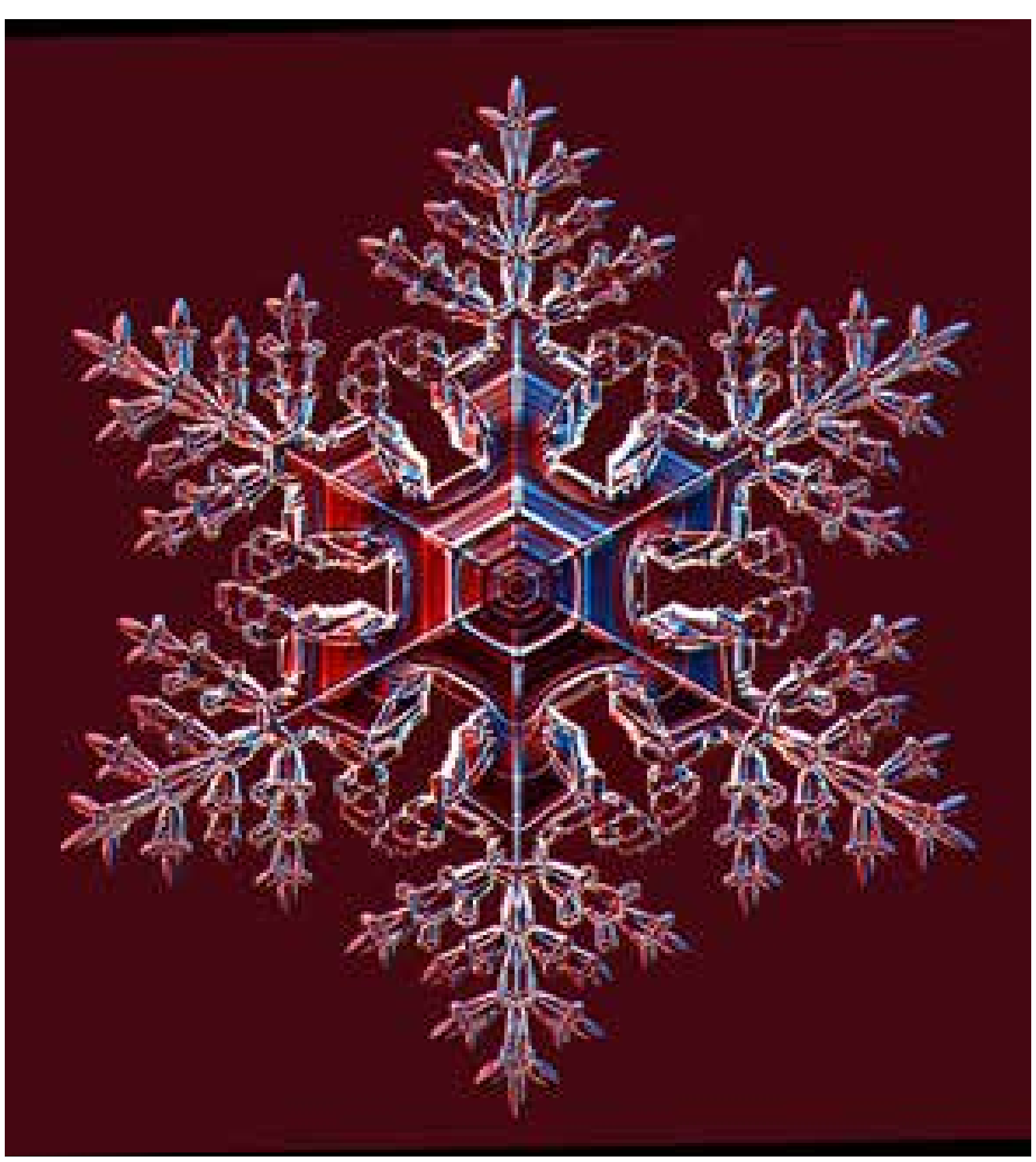

**Figure 6.1: In addition to its utility for scientific investigations, growing synthetic snow crystals is also desirable for its artistic value. This laboratory-grown snowflake shows a degree of symmetry and overall morphological perfection exceeding that found in natural snow crystals.**

When making synthetic snow crystals, artistic creativity mingles with the natural rules governing spontaneous structure formation. It is not possible to fabricate just any crystal design, as the underlying physical processes come with constraints. Instead of carving a solid block of material with full creative license, the snow-crystal artist adjusts temperature and humidity to direct the formation of desired features, letting Nature sculpt the ice as it grows. The art of creating synthetic snow crystals presents, in my opinion, a fascinating and largely unexplored fusion of art and science.

## Artificial versus Synthetic

Before proceeding further, I wish to clarify the distinction between *artificial snow* and *synthetic snow crystals*. The former is made at ski resorts by producing a spray of liquid water droplets and freezing them as quickly and cheaply as possible. Usually water is mixed with compressed air and shot out through atomizing nozzles in snow-making machines like the one shown in Figure 6.2. The compressed air cools as it expands, thus cooling the water droplets and causing them to freeze. There are many engineering tricks for making this process work effectively while minimizing the electric bill from the air compressors.

Artificial snow looks about like what you would expect – small globules of ice, with none of the ornate structure seen in snow crystals. Figure 6.3 shows an example of some artificial snow I photographed right after it fell from a snow-making machine. These ice particles do not resemble real snow crystals because they froze from liquid water, not from water vapor. They are essentially small sleet particles, which are basically frozen raindrops.

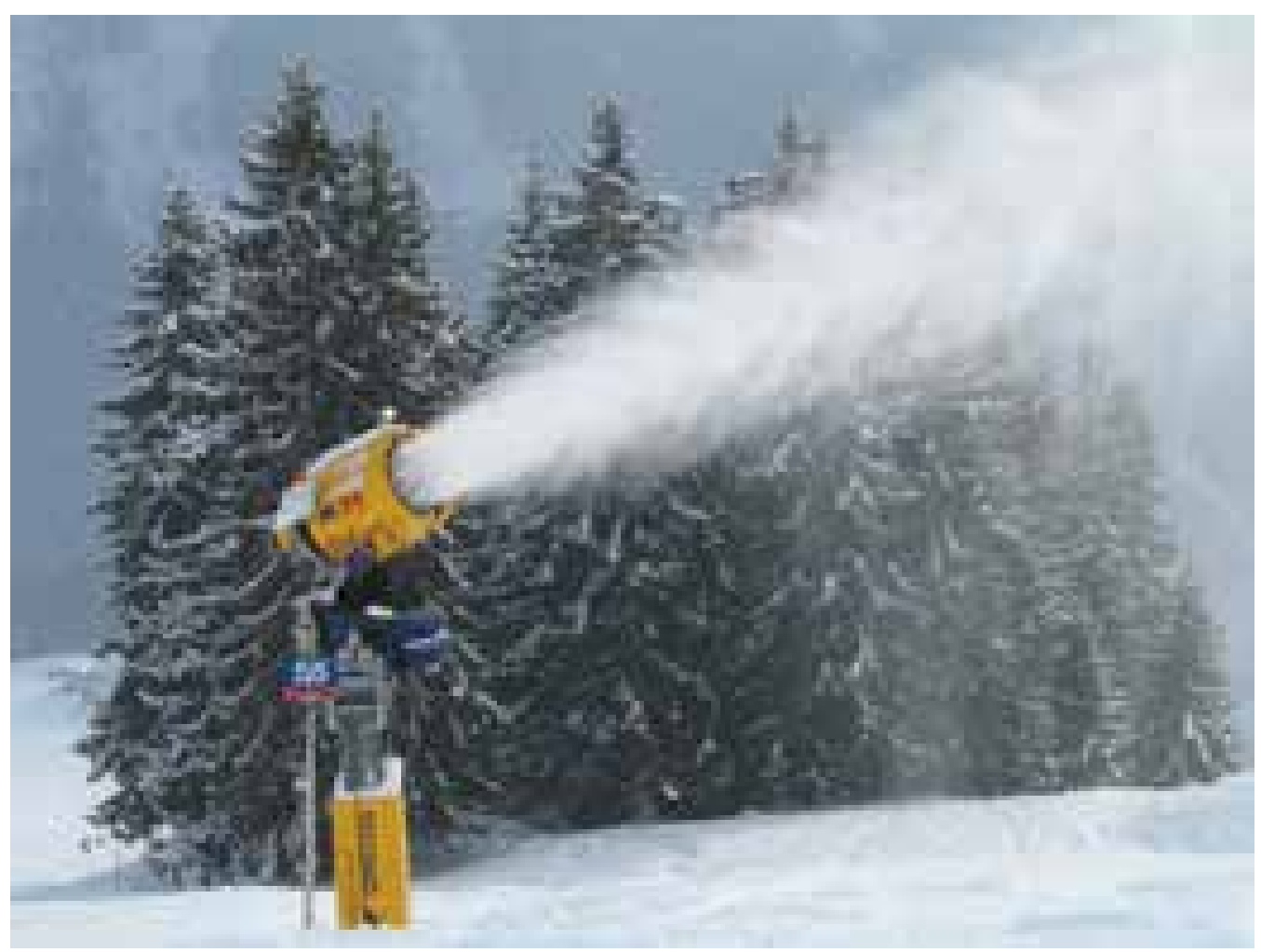

**Figure 6.2: Artificial snow is made by rapidly freezing a spray of tiny water droplets in a snow-making machine.**

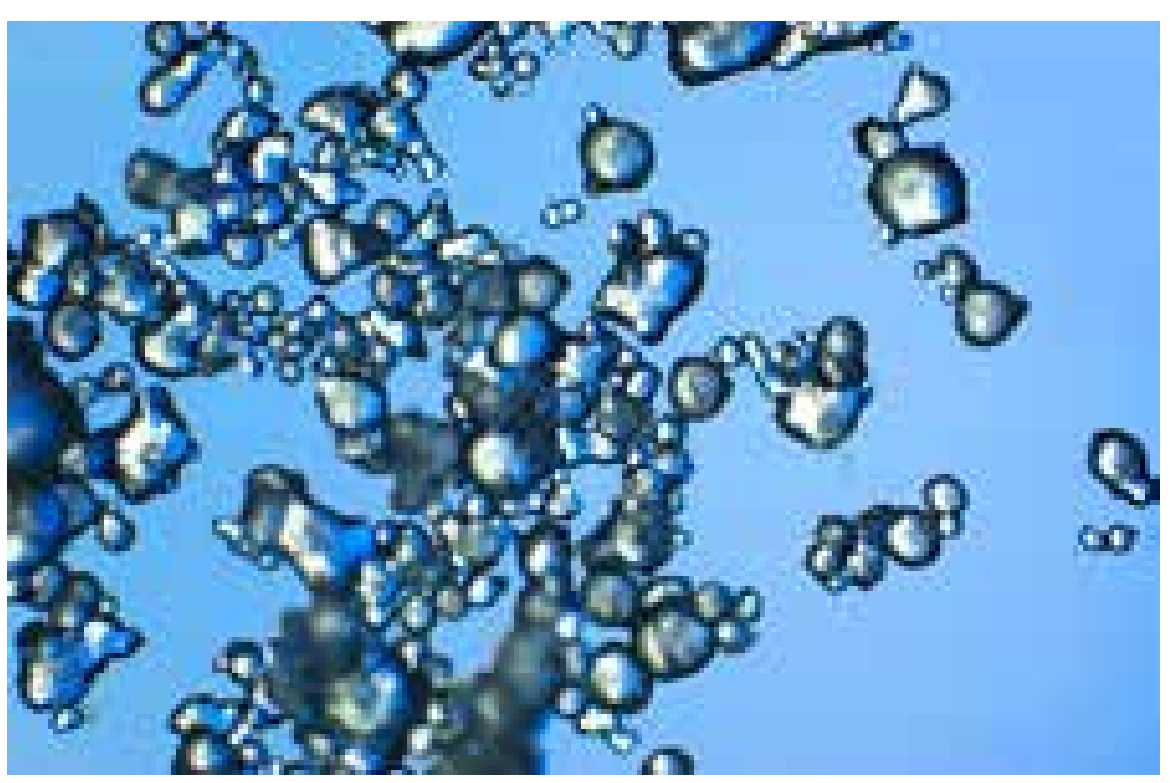

**Figure 6.3: Artificial snow, the kind made at ski resorts, is not real snow at all, but frozen water droplets. Because these do not grow from water vapor, they do not have the ornate structures seen in snow crystals.**

Synthetic snow crystals grow from water vapor, so they look essentially like real snow crystals. And they are real; we just call them synthetic because they grew in a laboratory environment. Labeling them synthetic is done merely to distinguish them from natural snow crystals that formed in the clouds.

By the way of an analogy, I would say that an animatronics dinosaur at Disneyland is artificial. It is made to look like a dinosaur, move like a dinosaur, and sound like a dinosaur, but clearly it is not real. A Jurassic-Park dinosaur, on the other hand, would be synthetic. It was created by unnatural means, but it is still a real skin-and-bones dinosaur.

## 6.1 Building Blocks

One reason scientists travel as much as they do is to keep up with the latest technological advances in their fields. Often this is best accomplished by visiting other people's labs, attending tech shows, or just chatting over dinner at conferences. Publishing these kinds of details in scientific journals is not as popular as it was in decades past, as every page of journal space is desired to convey new data and theoretical ideas. Thus, as the complexity of scientific instrumentation has increased with time, and the need to document and communicate the tricks of the trade has grown with it, actual dissemination of this information has become more difficult. I begin this chapter, therefore, with some discussion of the basic hardware needed to grow synthetic snow crystals.

### Refrigeration

The first ingredient needed for a proper cold chamber is a reliable source of cold. At the very least, a temperature range that extends down to -20 C is required, but most times it is desirable to reach as low as -40 C. Going below -40 C begins to require actual cryogenic hardware, which is another can of worms entirely, and I will not consider such low temperatures here. Largely for reasons of cost and difficulty, snow crystal growth at cryogenic temperatures and high pressures has not yet been well studied.

My workhorse refrigeration device is a recirculating chiller, specifically the RS33LT from SP Scientific shown in Figure 6.4. With few exceptions, I now have one of these hooked up to all my cold chambers. This chiller can remove up to 60 Watts of heat at -40 C, and up to 275 Watts at -20 C, by circulating coolant from a built-in pump and two-liter reservoir. I usually use methanol as the working fluid, connecting a snow-crystal chamber to the chiller using ordinary Tygon or polyethylene tubing. Once the plumbing is assembled, I can then simply set the desired temperature and the chiller's servo control does the rest automatically, maintaining a fixed coolant temperature to an accuracy of a few tenths of a degree. There are cheaper refrigeration alternatives available, but it is hard to beat the accuracy and convenience of a digitally controlled recirculating chiller.

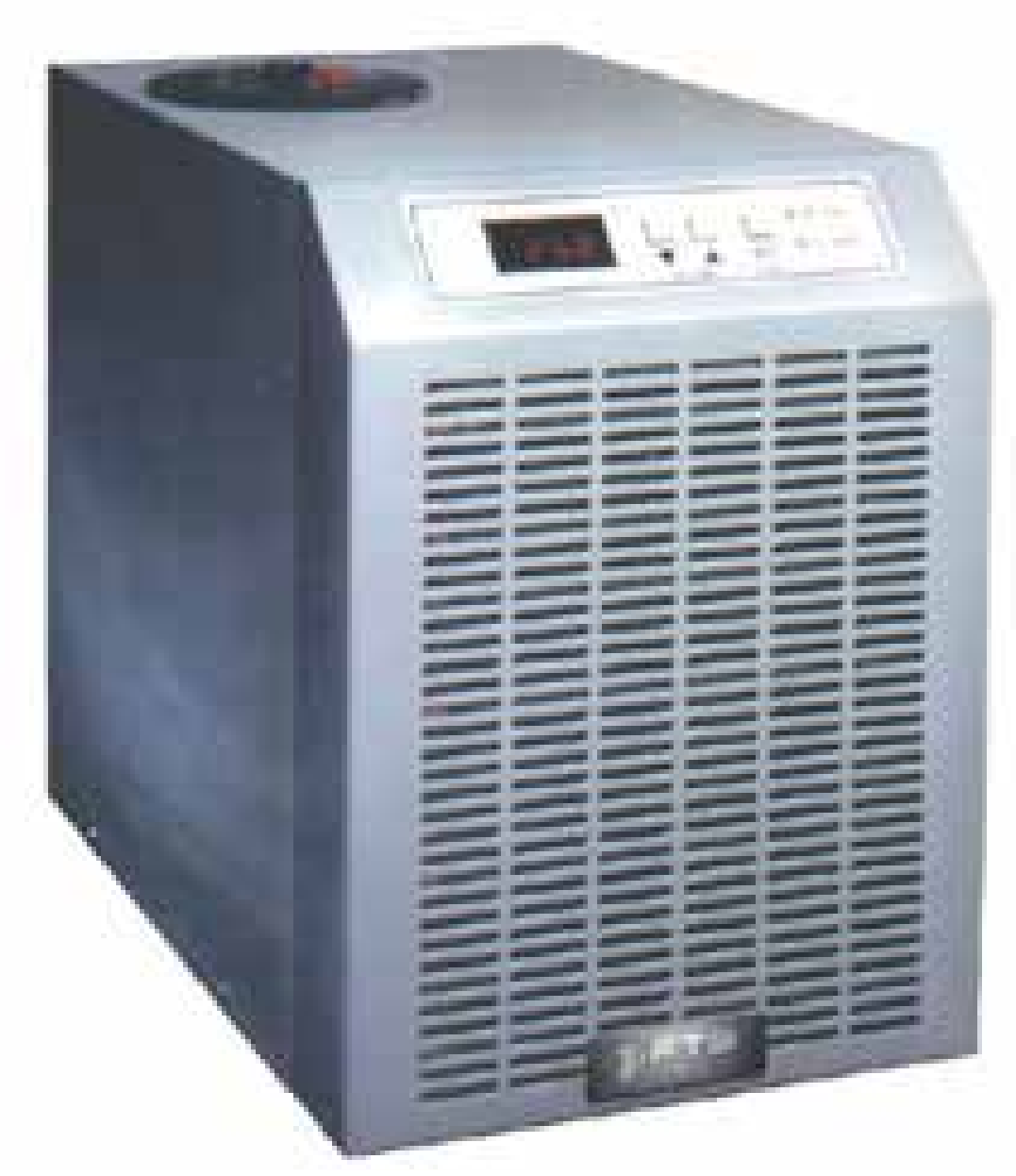

**Figure 6.4: A model RS33LT recirculating chiller from FTS Thermal Products at SP Scientific. This device is my main refrigerator for cooling snow-crystal growth chambers down to as cold a -40 C.**

Figure 6.5 shows the design of a general-purpose cold plate I often use as a starting point in my snow-crystal experiments. The large plate in this example is a 0.5x12x18-inch anodized aluminum breadboard from Thorlabs, supplied with an array of tapped holes on 1-inch centers. The plate shown in the figure is unmodified except for a large hole that allows optical access to the growth chamber.

A loop of 3/8-inch copper refrigerator tubing provides cooling from a recirculating chiller. The tubing is first soldered into a loop as shown using plumbing elbows, and the loop is then tested to give reliable, leak-free fluid flow. The pipe is then additionally soldered to two 1/8" thick copper plates, each rough-cut with plenty of attachment holes for good heat transfer from the breadboard.

I usually mount the assembled cold plate horizontally, with the cooling pipe on the bottom. This leaves the top surface of the breadboard open, with a large number of tapped holes for attaching the rest of the cold chamber. This overall cold-plate design is quite versatile, and similar aluminum breadboards are available in a broad range of sizes.

Another construction technique I have come to appreciate involves the use of aluminum "T-rail" like the example shown in Figure 6.6. These are inexpensive, extruded square bars with mounting T-slots on all sides, available in a great variety of shapes and sizes. Notably, some include central holes running along the lengths of the bars that are ideal for cooling from a recirculating chiller. I typically use an NPT (National Pipe Thread) tap in the central hole and screw in a Swagelok pipe fitting to plumb each bar.

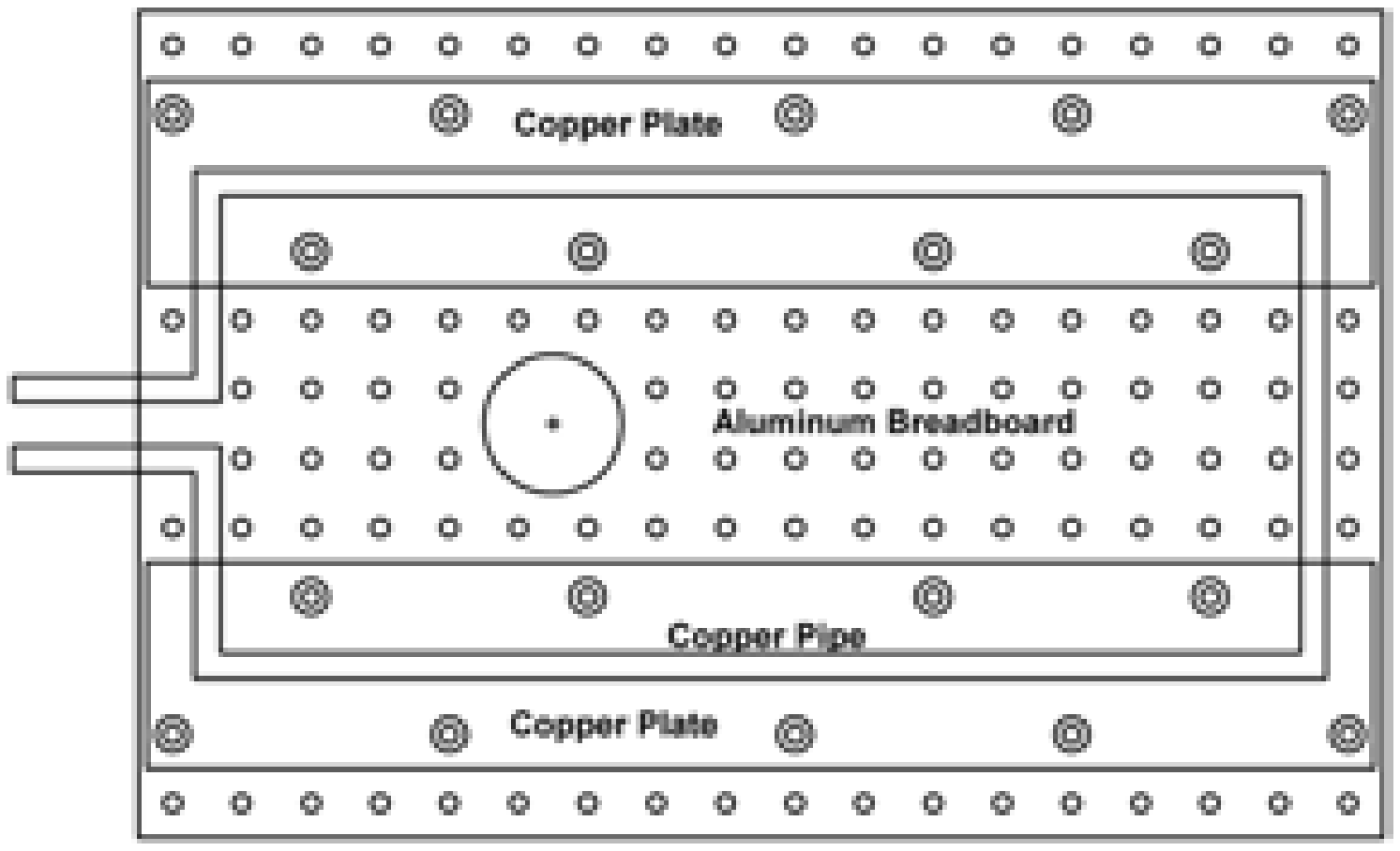

**Figure 6.5: A temperature-controlled aluminum breadboard, complete with an array of tapped holes, can serve as versatile starting point for many snow-crystal growth chambers. Cold fluid from a recirculating chiller flows through the copper pipe to cool the breadboard.**

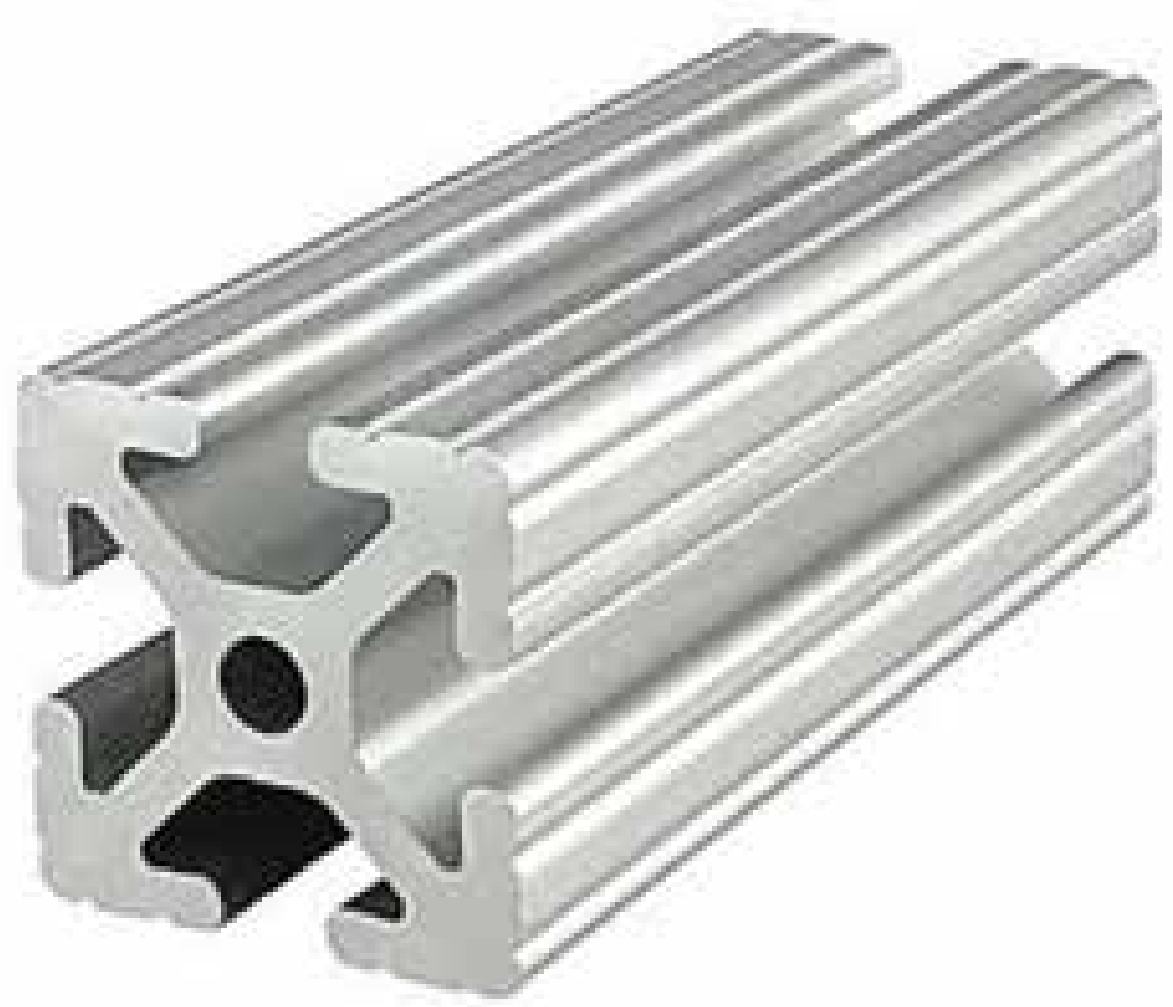

**Figure 6.6: An example of an extruded aluminum "T-rail" bar. Special nuts go into the four slots for general-purpose assembly, and fluid from a recirculating chiller can be passed through the central hole for convenient and extremely effective cooling.**

In one of my larger snow-crystal chambers, for example, I used T-rail to build a simple rectangular box frame. I piped coolant through the four longest edges of the box (using NPT fittings on the ends of the rails), and then used commercial T-rail angle brackets to connect eight shorter pieces of T-rail to form the box. To the frame, I then attached 1/8-inch thick aluminum panels to form the walls of the box. The result is an inexpensive, temperature-regulated chamber with easily removable side panels, suitable for any number of snow-crystal-growth projects.

## Insulation & Condensation

Condensation is a nuisance problem that must also be addressed when growing snow crystals. The dew point is usually around 10 C in the lab, so water vapor from the air will condense onto any surface with a temperature below that point, either as water or frost. My usual solution is to add an insulating layer of 1.5-inch-thick styrofoam sheet around a cold chamber. I sometimes use silicone caulk to hold this box together, but Mylar duct tape is handy also. (Mylar tape is the kind often recommended for actual ducts; cloth duct tape is a poor substitute.) In many scenarios, I like to use masking tape to hold the styrofoam panels together, as this makes for easy disassembly and reassembly as the apparatus is constantly being modified. A reasonably air-tight seal is usually necessary to keep air, and thus condensing water vapor, away from the cold chamber. Foam tubes used for household pipe insulation are also a common staple in the cold lab.

For optical access into my cold chambers, I often use viewport cells like the one shown in Figure 6.7. The overall construction consists of an acrylic or polycarbonate tube with a set of AR-coated optical windows held in place with silicone caulk or 5-minute epoxy. This version includes a third, inner window that is in good thermal contact with the cold aluminum wall of the chamber, providing an additional layer of thermal isolation. In less critical applications, the simpler two-window viewport cell alone is sufficient.

Although an evacuated or nitrogen-filled viewport cell would be superior, I have found that ordinary air inside the cell causes few problems. There is little interior water vapor in the cell to begin with, and this typically condenses onto the coldest interior side-wall surfaces rather than onto the windows. Nevertheless, the side plug seen in Figure 6.7 can be useful for blowing dry nitrogen gas into the cell to remove interior condensation that occasionally seeps in through cracks in the window seals.

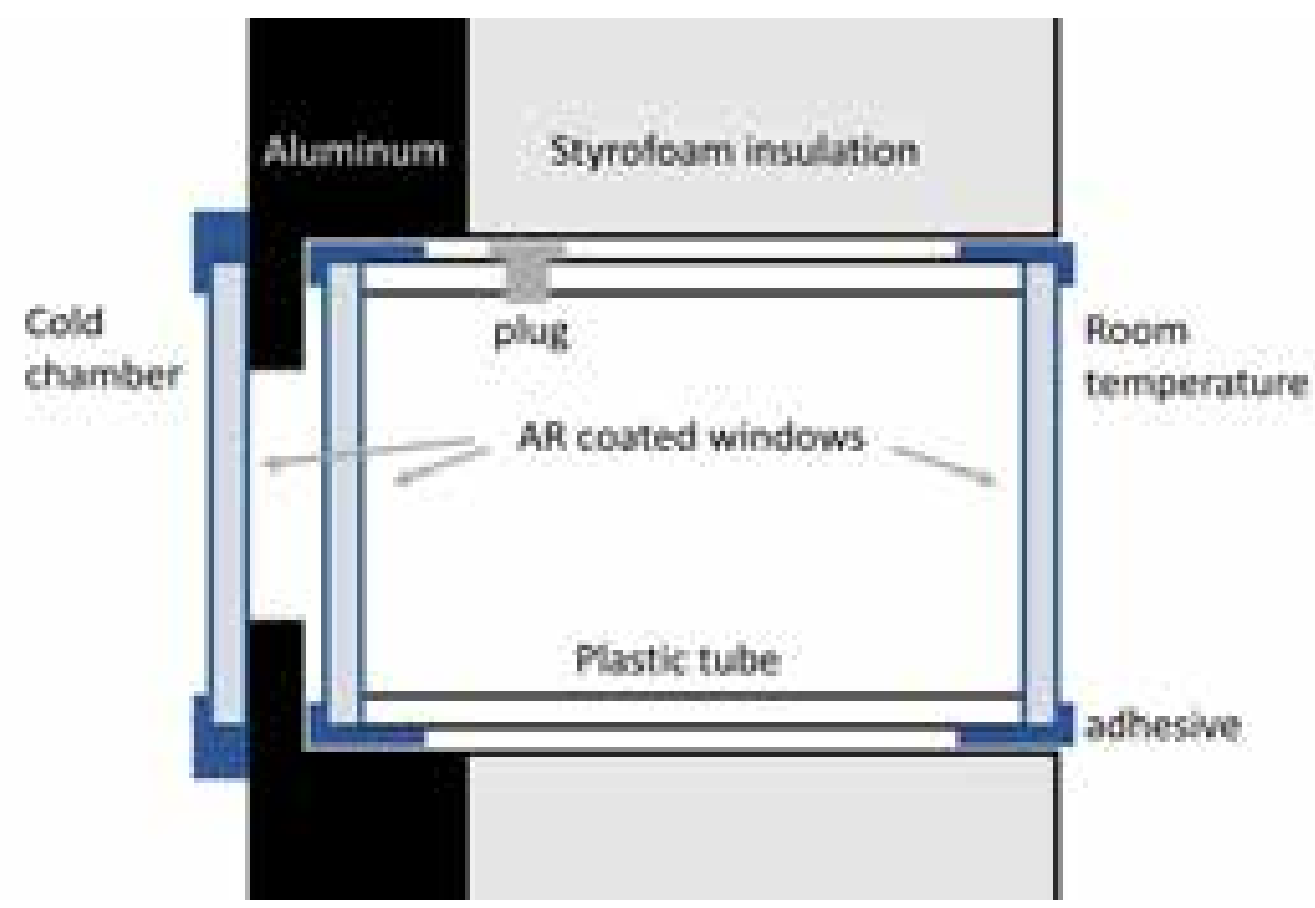

**Figure 6.7: A sketch showing a viewport cell installed into the wall of a cold chamber. This design allows optical access while minimizing heat flow into the chamber and preventing condensation of water vapor from the room.**

When high-resolution microscopy is needed, I might replace the inner window in Figure 6.7 with a mounted microscope objective. This brings a long-working-distance objective close to the growing snow crystals, while sending the resulting image out to a camera kept in the room-temperature environment. Although my microscope objectives were not designed specifically for sub-freezing temperatures, I have used them down to -40 C without problems.

Some researchers have avoided many of these experimental complexities by using a room-size cold laboratory. Building such a large freezer is an expensive option, but it does obviate the need for complicated viewports and careful insulation. On the other hand, it also requires that the experimenter work in bitter cold conditions, which has obvious practical disadvantages. Personally, I feel that careful experimental design of an insulated chamber is almost always preferable to spending long hours sitting in the cold.

## Vacuum Technology

In many scientific investigations, it is desirable to grow snow crystals at different background gas pressures, especially at lower pressures. For example, this is often necessary when investigating the attachment kinetics, as working at reduced pressures decreases the effects of particle diffusion on growth rates (see Chapter 3). Going down this road requires a great deal of specialized hardware, however, including vacuum chambers, pumps, and gauges.

A typical laboratory vacuum system begins with a stainless-steel chamber consisting cylinders, crosses, and tees connected by vacuum flanges. These components are robust and commercially available, but they are designed for far lower pressures than one normally requires for studying snow crystals. Cheaper, home-built alternatives can also work well, although standard vacuum components are certainly convenient.

In snow crystal studies, there is little need to achieve pressures below the water-vapor saturation pressure, which is typically around one millibar. This is considered a very high pressure in the vacuum world, so any kind of O-ring or other gaskets are fine for creating robust vacuum seals on flanges and other components, and epoxy seals are often adequate as well. Manufacturers produce a large range of vacuum flanges, fittings, and other parts needed to design a system for almost any need, if you have the budget. As there are already numerous books and internet resources discussing vacuum chambers and vacuum seals, I will not go into details here. It is good to remember, however, that working at millibar pressures may not require a lot of fancy and expensive hardware.

One issue I have with commercial vacuum fittings is that most viewport windows have remarkably poor optical quality. The glass is not flat, not AR coated, and wholly inadequate for inclusion in a high-resolution imaging system. To get around this problem, I often build my own vacuum windows as shown in Figure 6.8. A key feature in this simple design is an optical-grade coated window bonded to a flat flange with a large overlap area. Flat-against-flat is usually a good bonding strategy if one wants to avoid vacuum leaks, as it keeps leakage paths long. Most silicone caulks cycle well down to low temperatures, and I include spacers to ensure that the caulk layer is thick enough to have some flex during thermal cycling. None of this is exactly rocket science, but getting some of these small details right can save considerable time in the long run.

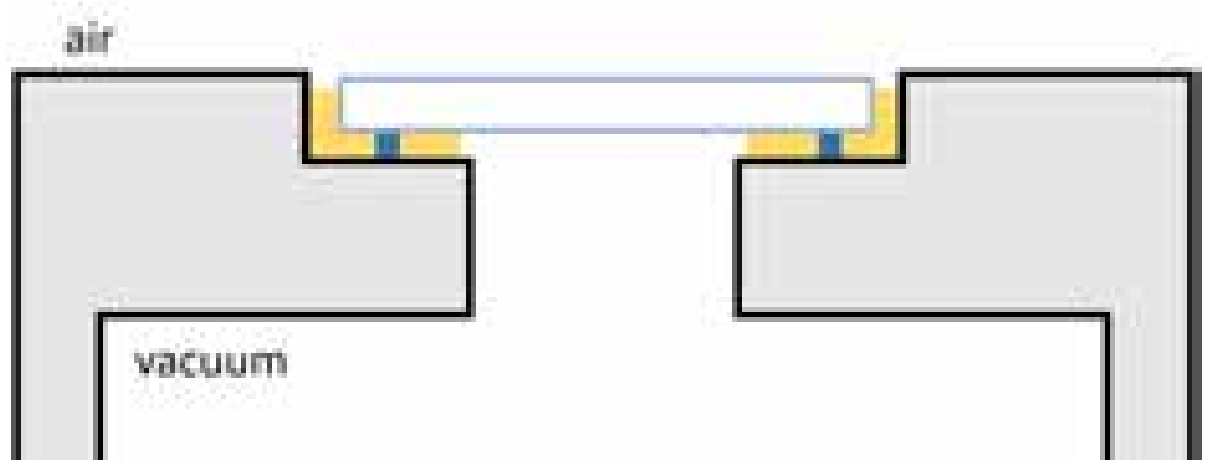

**Figure 6.8: A simple and inexpensive, yet quite effective, design for attaching an optical-quality vacuum window to a vacuum flange. The large-area flat-against-flat adhesive seal (yellow) works well for preventing vacuum leaks. Spacers (blue) keep the adhesive layer thick, so it can flex during thermal cycling without losing adhesion.**

Diaphragm pumps are a reasonable route for reaching down to one millibar, as they are relatively cheap (if that word can be applied to vacuum pumps) and they have a reasonable pumping speed. One disadvantage is that their base pressures are not especially low, so check the spec sheet before you buy. A positive feature is that diaphragm pumps are oil-free, so generally they will not introduce a lot of contaminants into the vacuum chamber.

Vacuum gauge technology has been advancing quite rapidly lately with the advent of micromachining techniques, so I would not hazard to guess what the best gauge choices will be even a few years from now. Thermocouple gauges are often used in the 1-1000 mbar range, but their sensitivity and signal level changes with gas composition, which can be inconvenient. Capacitance gauges are one of the best choices for overall sensitivity and utility in this pressure range, although they remain quite expensive.

## Chemical Contamination

Chemical vapor contamination is always a wild card when making quantitative measurements of snow crystal growth. As I describe in Chapter 4, even fairly low levels of vapor contaminants can dramatically alter the attachment kinetics, and the underlying chemical mechanisms are not at all understood at even a qualitatively level. This presents an excellent direction for additional research, but making progress is difficult without at least some reliable theoretical guidance. My approach to date has been to try to figure out the physics of snow crystal growth with no added chemistry first … one step at a time.

With this in mind, it is important to make sure that one's snow crystal growth chamber is free from all chemical vapor contaminants, at least to the degree that this is possible. One rule-of-thumb I like to use is the "smell test" – if you can smell anything in a growth chamber, then the contamination level might be too high. Unfortunately, if you take any sealed chamber that has been unopened for some hours or days, chances are high that you will smell something if you open it up and put your nose inside. Even after thorough cleaning and baking, using only stainless steel or glass components, there always seems to be some residual odors present.

The best solution I have found to this problem is to continuously trickle clean gas into a growth chamber during operation, typically replacing the air inside every half-hour or so, depending on the chamber size and geometry. I create this gas stream by taking air from a commercial oil-free compressor and passing it slowly through a column of activated charcoal, usually coconut-husk charcoal that is manufactured for the purpose of removing chemical contaminants from air. The set-up for this is simple and inexpensive to build, and it seems to work quite well. Clean-air generators can be purchased at substantial cost, and liquid-nitrogen boil-off is another option. But these are probably both overkill, as there is always some low level of outgassing from the air handling system and chamber walls that cannot be entirely eliminated.

## Temperature control

Careful temperature measurement and control is another essential skill when growing synthetic snow crystals. For measuring temperature, my preferred sensor is a precision two-kilo-ohm (2k) bead thermistor. Other temperature sensors are available, such as thermocouples and RTDs, but those are better suited for higher temperature applications. Precision 2k thermistors are available for as little as $10 apiece from a number of vendors, and their small size with ±0.1 C absolute accuracy make them ideally suited for installation in cold chambers.

The resistance of a thermistor increases with decreasing temperature at a rate of about 4.5% per degree, so a 2k room-temperature thermistor will perform well down to -40 C. An operating current of 10 μA is about optimal for temperature sensing. At higher currents, the voltage drop across the sensor at low temperatures might saturate the meter. Plus, the operating current can heat the thermistor from within, which is clearly undesirable in a temperature sensor. With many commercial thermistor controllers, one can select the operating current and then enter the Steinhart-Hart coefficients so the measured thermistor resistance is converted directly to degrees C.

For precisely controlling temperature, I typically use thermoelectric modules, as these are capable of both heating and cooling over a broad range of temperatures. I used to build my own temperature controllers [2009Lib2], but commercial units have improved so much recently that it is now almost pointless to do this yourself. I have been especially satisfied with 12V/10A temperature controllers from Arroyo Electronics, as they accept generic thermistor sensors and can provide high currents for thermoelectric modules.

## SUPERSATURATION

Determining the supersaturation surrounding a growing snow crystal is an especially challenging experimental problem. In contrast to temperature, supersaturation is an intrinsically nonequilibrium quantity, so even the presence of the measuring device can greatly affect the outcome. For example, if a hygrometer sensor is placed inside a cold chamber, then ice will eventually condense and grow on the sensor, removing water vapor from the nearby air, and thus influencing the measurement. Any growing ice, even the very crystal one is trying to measure, can substantially change the supersaturation in its vicinity.

One way to deal with the supersaturation problem is by creating a metastable, isothermal environment that has fixed, non-zero supersaturation. One example is simply a closed box cooled to some temperature below 0 C with an unfrozen puddle of water inside. If the temperature is not too low, the water can remain unfrozen long enough for the system to come into vapor equilibrium at a uniform temperature. Then the supersaturation with respect to liquid water would be identically zero, but the supersaturation with respect to ice would be greater than zero, specifically $\sigma_{water} = (c_{sat,water} - c_{sat})/c_{sat}$, where $c_{sat,water}$ is the equilibrium vapor pressure of supercooled water at this temperature (see Chapter 2).

If one uses a water solution instead of pure water, it is possible to obtain $\sigma$ values lower than $\sigma_{water}$, as described by [1996Nel] for a solution of lithium salt in water. The essential physics is that chemicals dissolved in water reduce the equilibrium water vapor pressure above the water, which can be calculated using Raoult's Law. Nelson and Knight went on to use this method when examining layer nucleation [1998Nel], and it has much additional potential for creating a pseudo-equilibrium ice-growth environment with a precisely known supersaturation.

A variation of the metastable-liquid method is to create a cloud of water droplets in the laboratory, thus directly simulating natural clouds. Pure water yields a supersaturation equal to $\sigma_{water}$, and Yamashita used this approach to produce some excellent measurements of snow crystal growth as a function of temperature at this supersaturation [1987Kob, Figure 11]. Clouds made from salt-water droplets could yield supersaturation levels below $\sigma_{water}$, but I believe this technique has not been used to obtain published growth data.

Air sampling from a cold chamber is another way to determine the supersaturation, for example using differential hygrometry to compare sampled air with a known reference [2008Lib4]. Converting a measurement of water vapor content to local supersaturation can still be tricky, however, because one needs to know quite accurately the local temperature as well as the water vapor content. Air sampling is not without its uses for measuring supersaturation, but it is not ideal for making precision measurements of ice growth dynamics.

My colleagues and I used differential hygrometry calibration to obtain a series of measurements of freely falling snow crystals in air as a function of time, temperature, and supersaturation over a fairly broad range of conditions [2008Lib4, 2009Lib]. Although the absolute accuracy of the supersaturation determination was somewhat uncertain, I believe that these remain some of the better measurements of absolute growth rates in air. There is general agreement between these measurements and the Yamashita results, but in both cases I would estimate that the overall supersaturation uncertainties are probably not better than 20-40 percent.

## Diffusion Modeling

After considerable work on this problem, I have found that careful modeling of a well-designed cold chamber is often the best way to determine the supersaturation with high accuracy. An especially simple example is shown in Figure 6.9. In the absence of the test crystal, the supersaturation just above the substrate will be (see Chapter 3)

$$\sigma_1 = \frac{c_{sat}(T_{reservoir}) - c_{sat}(T_{substrate})}{c_{sat}(T_{substrate})} \quad (6.1)$$

which will be greater than zero if $T_{reservoir} > T_{substrate}$.

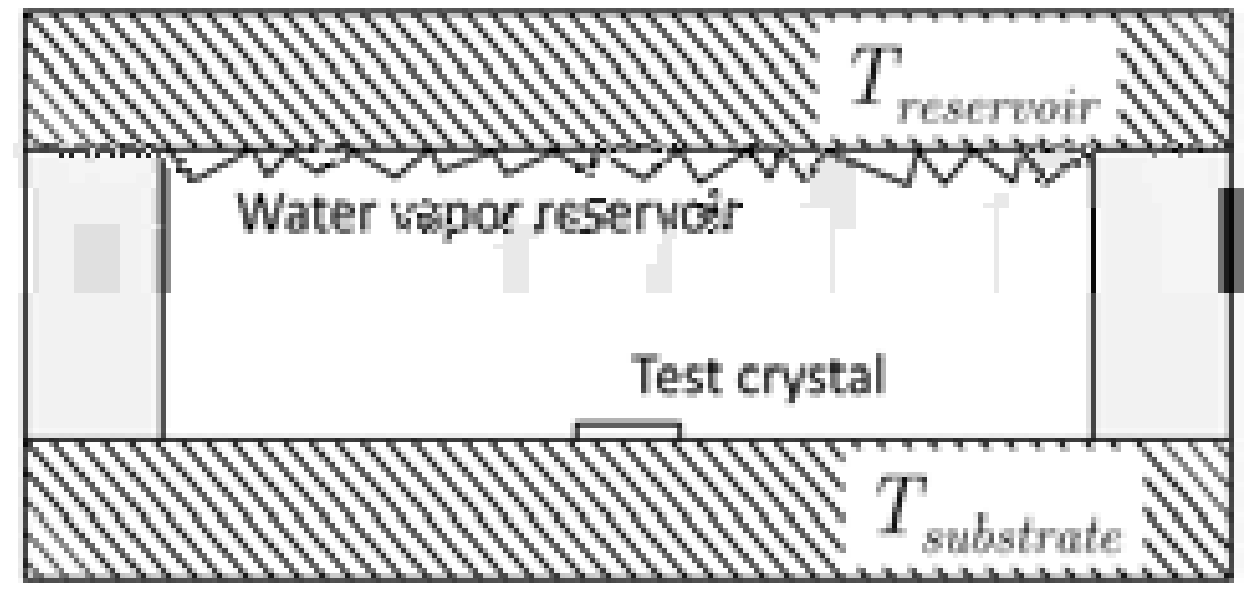

**Figure 6.9: This dual-temperature chamber contains a layer of ice that acts as a water-vapor reservoir along with an ice test crystal to be measured. Both the reservoir and test-crystal surfaces are carefully temperature regulated. The supersaturation around a small test crystal is easily calculated from the two temperatures, going to zero when there is no temperature difference between the top and bottom surfaces.**

If a small test crystal is placed on the substrate, then it will grow as if surrounded by the far-away boundary condition $\sigma_\infty = \sigma_1$. However, if the test crystal is too large, or if there are additional ice crystals on the substrate, then the actual supersaturation quickly deviates from Equation 6.1, as I discuss in some detail in Chapter 7. Libbrecht and Rickerby used this method to make some of the most accurate snow-crystal growth-rate measurements to date [2013Lib].

Diffusion modeling can be applied to a variety of different ice growth chambers to calculate supersaturation levels. In all cases, the first step is to solve the diffusion equation for heat transfer, thus determining the static temperature field $T(\vec{x})$ at all interior points within the chamber. With stable interior air and well-defined temperature boundary conditions on all the chamber walls, it is straightforward to create a numerical model of the temperature profile throughout.

The next step is to solve the particle diffusion equation within the chamber, thus determining the static water-vapor number density $c(\vec{x})$ at all interior points. If the chamber walls are all well coated with frost, then again one has well-defined boundary conditions, as the water vapor density near each frosted surface is equal to the saturated value $c_{sat}(T)$ at the known surface temperature. Thus, once again it is straightforward to create a static numerical model of the water vapor number density throughout the interior of the chamber.

Once the temperature and number-density fields have been determined, one can then calculate the supersaturation within the cold chamber from

$$\sigma(\vec{x}) = \frac{c(\vec{x}) - c_{sat}\big(T(\vec{x})\big)}{c_{sat}\big(T(\vec{x})\big)} \qquad (6.2)$$

at each position $\vec{x}$ within the chamber. Uncertainties in the calculated $\sigma(\vec{x})$ can be estimated from uncertainties in the modeled fields $c(\vec{x})$ and $T(\vec{x})$. With a simple chamber design and well-characterized wall temperatures, the supersaturation can be determined with quite high accuracy using diffusion modeling.

Two caveats are worth mentioning when modeling the supersaturation. First, it is essential that the air inside the cold chamber be stable with respect to convection. This requires a positive (or zero) vertical temperature gradient and no horizontal temperature gradients within the chamber. The lack of horizontal gradients can be a nontrivial requirement that must be built into the chamber design with some care.

The second caveat involves the water vapor density boundary conditions at the chamber walls. If a surface is covered in frost, then the boundary condition is $c(\vec{x}) = c_{sat}\big(T_{surf}\big)$ just above the surface. And, if a surface is completely clear of frost, then the boundary condition is a vanishing water vapor density gradient normal to the surface, so $dc/dn = 0$ just above the surface. If a surface is only partially covered with frost, however, then the boundary condition is ill-defined. In this case, neither the water vapor field nor the interior supersaturation can be precisely determined. The take-away message is simply that all surface boundary conditions need to be precisely defined, which is often difficult to accomplish in practice.

It is important to remember that all growing ice crystals within an experimental chamber will affect the water-vapor field $c(\vec{x})$, including the test crystal one is trying to measure. The only way to determine $c(\vec{x})$ accurately is by solving the heat and/or particle diffusion equations, taking into account all the relevant boundary conditions. This can be done analytically for a few simple cases, as described in Chapter 3 and Appendix B, and these analytic solutions are invaluable for making quick estimates during the initial design of a growth chamber. For more complex geometries, however, one must resort to numerical modeling.

## Nucleation

Another challenge in creating synthetic snow crystals is getting the process started, preferably when and where you want it. Exposing nearly any sufficiently cold surface to supersaturated air will nucleate the formation of ice, but usually in the form of frost, which is many small ice crystals crowded together. Other methods are needed to create a single, isolated seed crystal of ice that can then grow unhindered into a large snow crystal.

When Ukichiro Nakaya first sought to create laboratory snow crystals in the 1930s, he quickly ran into the frost problem. He tried suspending a variety of thin filaments in his cold chamber, including silk, cotton, fine wires, and even a strand of spider's web. In each case, he soon found the filaments coated with a thick layer of frost. He eventually discovered that a desiccated rabbit hair was fairly well suited for growing isolated crystals [1954Nak].

Nakaya surmised that sparsely spaced nodules on an individual rabbit hair served as ice nucleation sites, while oils on the hair otherwise reduced the probability of condensation. Over some range of $\sigma$ levels, it was therefore possible to nucleate a single, isolated snow crystal on a well-desiccated rabbit hair, although not every attempt yielded a suitable seed crystal. If an isolated crystal did form, it removed water vapor from the air in its vicinity, thus reducing $\sigma$ and preventing the nucleation of additional crystals nearby. This was a marvelously creative idea for supporting and growing large, isolated snow crystals, but it had a number of drawbacks, and I believe no one has since reproduced Nakaya's rabbit-hair method.

Researchers have used a variety of other nucleation methods over the years, and each has its benefits and drawbacks. For example, a short pulse of highly supersaturated air can nucleate small ice crystals on a substrate, as described in [1982Bec2]. Particles of silver-iodide smoke can also serve as nucleation sites [1982Gon, 1994Gon], and a fleck of dry ice dropped into supersaturated air will readily nucleate ice crystals as well [1981Sch]. Later in this chapter I describe how oriented ice crystals can form epitaxially on a covellite (CuS) and other crystalline substrates, and how isolated snow crystals can be grown on the ends of thin glass capillary tubes. Chapter 8 describes the use of long, slender electric ice needles as seed crystals. Aizenberg at al. [1999Aiz] have demonstrated an intriguing technique that uses patterned templates to nucleate regular arrays of small crystals on a substrate, but this technique has not yet been adapted to work with ice.

## Expansion Nucleator

My favorite general-purpose method for initiating ice crystal growth is the expansion nucleator shown in Figure 6.10. In this device, pressurized air feeds into the nucleator via an air hose and a flow restrictor, supplied by a commercial oil-free air compressor. The air pressure is typically 15-30 psi, and I pass the compressed air through a column of activated charcoal grains to absorb any remaining chemical impurities. With the solenoid valve closed, pressurized air fills the nucleator body in a few seconds.

Abruptly opening the solenoid valve allows the compressed air in the nucleator body to expand rapidly into the surrounding air. The expansion is impossible to calculate precisely, as the flow is turbulent, non-adiabatic, and overall quite ill-defined. Nevertheless, the rapid expansion cools the air, thus greatly increasing the supersaturation at localized positions within the outflowing air, and this is sufficient to nucleate the growth of numerous tiny ice crystals.

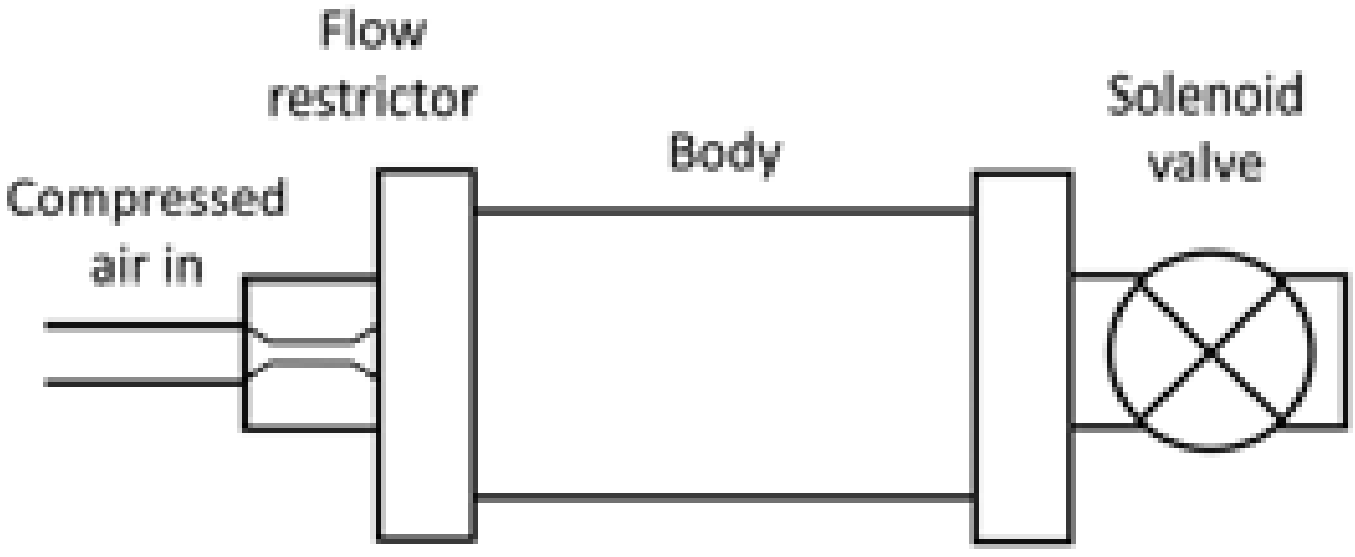

**Figure 6.10: An *expansion nucleator* for creating small ice crystals, consisting of a 5-cm length of pipe (body) flanked by a narrow constriction and a solenoid valve. When the valve is pulsed opened, the rapid expansion of air reduces its temperature sufficiently to nucleate a number of minute ice crystals. The length of the body portion of the nucleator is about five centimeters.**

The expansion nucleator relies on the fact that normal room air invariably contains a good amount of water vapor along with an ample supply of dust particles that act as ice nucleation sites. It is only necessary to increase the air pressure until ice crystals appear. A higher body temperature requires more cooling of the air, and thus higher air pressures, so I usually keep the nucleator at around -15C. At this temperature, no crystals form if the pressure is much below 10 psi, while 30 psi is usually high enough to form some thousands of minute crystals each time the valve is pulsed open.

In many ways, the expansion nucleator is like snow from a winter cloud, except much faster. The rapid expansion cools the air, initially forming water droplets on dust particles in the air, much like the formation of a cloud in the atmosphere. As in a cloud, the first nucleation step produces liquid water droplets, rather than ice crystals, in accordance with Ostwald's Step Rule. The nucleated droplets only last a fraction of a second, however, as continued expansion provides additional cooling, causing some of the droplets to freeze. The nascent ice particles quickly absorb water vapor around them, causing any remaining water droplets to evaporate away. The whole process is a bit like a miniature, high-speed snowfall. Each discharge of the nucleator results in a puff of tiny ice crystals.

## Oriented Ice Crystals

It is often desirable to have large, single-crystal specimens of bulk ice in the lab for a variety of functions, and how one obtains such samples is not immediately obvious. A century ago, researchers used to "mine" the Mendenhall glacier in Alaska for this purpose, as high-purity ice crystals could be found there with sufficient searching. Fortunately, a remarkably easy method for creating large single-crystal specimens in a top-loading household freezer was described by Knight [1996Kni], obviating the need for polar ice mining.

In Knight's method, one simply fills an open, insulated container with water, places it in the freezer, and waits. Water vapor first evaporates from the water's surface and deposits a bit of frost on the lid of the freezer above the water container. Over time the water cools to below freezing, and at some point (if all goes well) a single crystal of frost will break off the lid and fall into the water. If the water temperature is just below freezing, the ice seed will grow out as a thin disk crystal (see Chapter 12).

As it floats on the surface of the water, buoyancy forces automatically orient the growing disk so its c-axis points in the vertical direction as the edges of the disk grow outward. Soon the ice disk expands and covers the surface, preventing nucleation by subsequent falling frost crystals. Because the container is insulated, the rest of the water freezes slowly, its thickness increasing at a rate of about one centimeter per day.

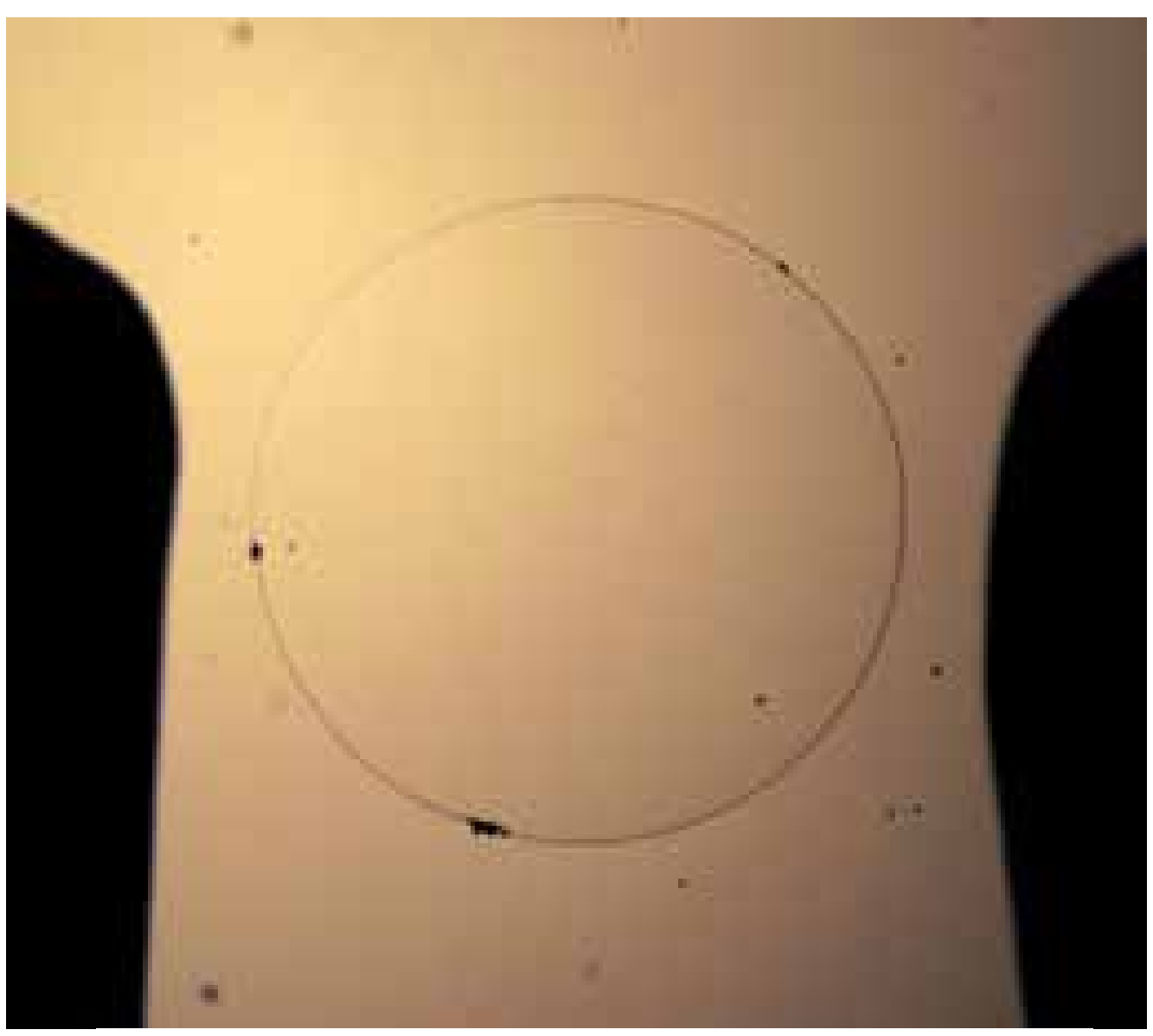

**Figure 6.11: A 2-mm-diameter disk of ice grows outward on the surface of a thin film of slightly supercooled water covering a glass plate. (The dark regions are copper support arms glued to the glass.) The c-axis of the oriented ice crystal is aligned perpendicular to the glass surface.**

This experimental procedure is not particularly well controlled, but usually there will be some sections of single-crystal ice on the surface, with the crystalline c-axis oriented vertically. Viewing the sheet between crossed polarizers verifies that no domain boundaries are present. This technique could be better developed, but I can verify that it works surprisingly well with little effort. Ice samples with higher purity, or with chemical dopants, can be grown using the Czochralski method [2017Bru, 1973Bil], but with a substantially greater investment of effort.

I have also created a miniature, on-demand version of Knight's method using an expansion nucleator. First a temperature-controlled glass window is treated with a hydrophilic coating (Rain-X) and wetted with a film of water. The window is then placed, water side up, in a supersaturated chamber, and its temperature is set to just below 0 C. A small puff of air containing seed crystals from the nucleator is passed over the disk, such that a single crystal will randomly drop onto the water's surface.

With a bit of luck, surface tension and buoyancy will align the crystal with respect to the water film, and a small disk of ice will slowly grow outward over the glass, as shown in Figure 6.11. This disk is a single crystal of ice oriented with its crystalline c-axis perpendicular to the glass surface. I have used this technique in an experiment growing snow crystals on glass capillaries (described below), but the method may have other useful applications as well.

## Substrate Interactions

I often grow snow crystals on sapphire substrates because this is one of the few transparent materials that has a high thermal conductivity. As an added bonus, sapphire is extremely hard and therefore resistant to scratching. Small sapphire windows are readily available and not too expensive, although it is often best to pay extra for windows that have been cut with the c-axis of the crystal perpendicular to the window surface, in order to avoid unwanted birefringent effects.

Whatever the choice of substrate material, it is important to remember that contact with a supporting surface may alter the growth behavior of snow crystals. One common mechanism is enhanced terrace nucleation at the ice/substrate contact line, as illustrated in Figure 6.12. A clean glass or sapphire substrate typically exhibits a contact angle near 90 degrees, so this mechanism can often influence the growth of perpendicular basal facets.

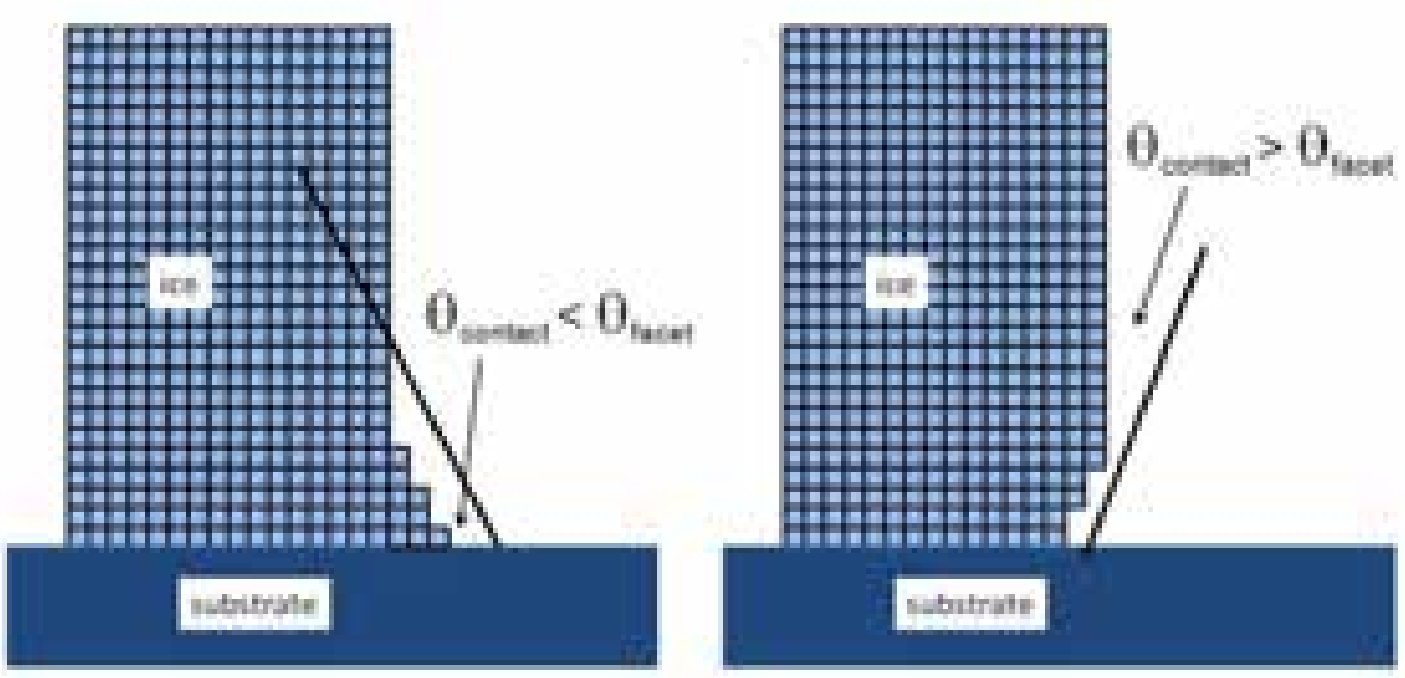

**Figure 6.12: If the ice/substrate contact angle is less than the facet/substrate angle (left sketch), then new terraces will be nucleated at the ice/substrate contact line. With a larger contact angle (right sketch), new terrace nucleation can only occur on the free faceted surface, away from the contact region. The enhanced terrace nucleation in the first case can greatly increase the growth rates of facets contacting the substrate.**

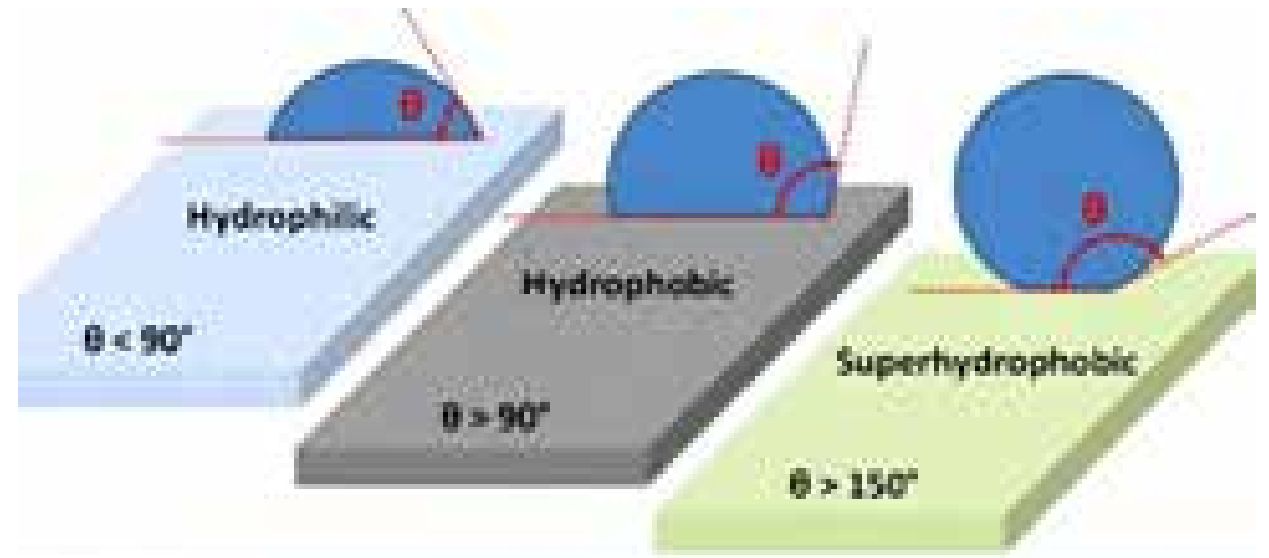

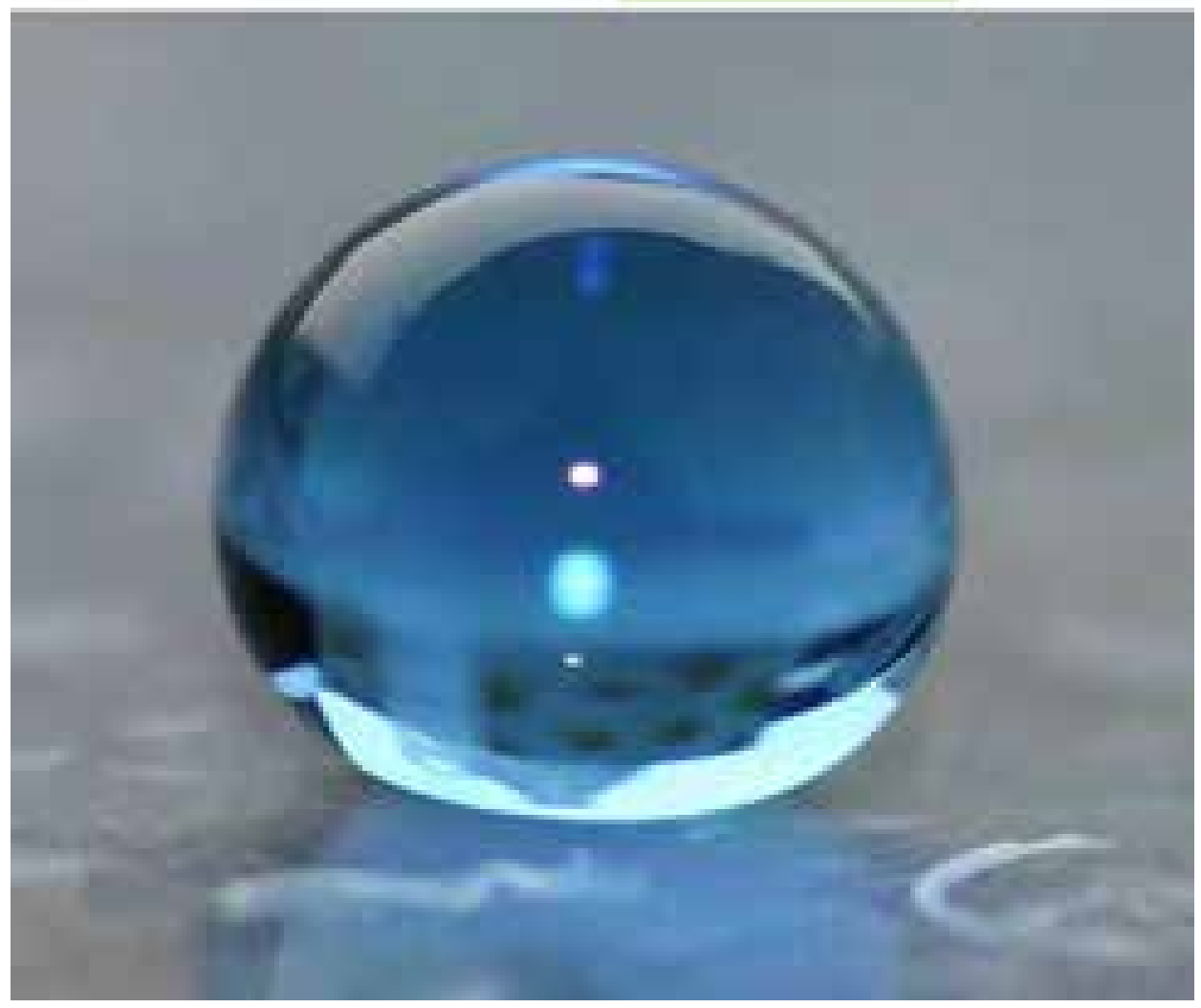

**Figure 6.13: A superhydrophobic substrate surface could greatly reduce unwanted substrate interactions. (Images from [2016Lui] and mandegar.info.)**

The use of superhydrophobic coatings could substantially reduce this particular form of substrate interaction. As illustrated in Figure 6.13, superhydrophobic surfaces with contact angles in excess of 150 degrees are now routinely demonstrated. To my knowledge, they have not yet made their way into studies of ice-crystal growth, but they soon will. This subfield of coating engineering is rapidly evolving, and I expect that easily fabricated transparent superhydrophobic substrates will soon be readily available.

Another type of substrate interaction arises from heat conduction through the ice to the cold substrate. For especially large and rapidly growing ice crystals, this can lead to a temperature gradient across the ice that distorts its growth behavior. This effect appears to have been an issue in [1972Lam], but it is usually negligible for small, slowly growing crystals.

## Optical Microscopy

The most straightforward approach to measuring ice crystal morphology and growth is via optical microscopy. The growth velocity derives from observations of the crystal size as a function of time, and velocity measurements can be combined with a knowledge of the supersaturation to examine the attachment kinetics. Optical images, moreover, are ideal for investigating morphologies and morphological transitions. High quality microscope objectives are readily available, and modern digital cameras produce excellent images even in low light levels.

Commercial bench microscopes typically use a two-step approach to imaging. First the microscope objective focuses an image onto an intermediate aperture, the purpose of which is to block all light except for that coming from around the desired subject. This first image is then reimaged onto the camera sensor by a second lens group. This technique greatly reduces problems arising from scattered light in the optical path, but it requires a second high-quality lens for reimaging.

I often take the direct approach of using the microscope objective on the end of a long extension tube. This places the first image directly on the camera sensor, without reimaging, which is a big gain in optical simplicity. However, it does tend to increase problems associated with scattered light, so I address these issues using any or all of these methods:

1) Use a field stop near the subject to block extraneous light. Then only the area around the test crystal is brightly illuminated.
2) Image the incident light onto the subject, again only illuminating a small area around the test crystal. This can generally be done using inexpensive optics that are not part of the imaging optical system.
3) Add baffles inside the extension tube.
4) Coat the inside of the extension tube with light absorbing material.

The first two items on this list reduce the amount of light that does not strike the test crystal but might still make its way into the microscope. This light contributes nothing to the desired image, but will add scattered light that reduces contrast on the final image. The last two items help decrease scattered light that strikes the imaging sensor.

In most of my ice growth experiments, I find it beneficial to build some of the optical system right into the growth chamber. In particular, I usually mount a microscope objective within the cold chamber, so I can place it at a desired distance away from the growing crystals. Most microscope objectives are not rated for use below 0 C, but I have never encountered any problems in temperatures down to -40 C. Condensation from room air is a constant concern, so I often use optical-grade windows in the imaging system to keep this problem in check.

Once the diffraction limit is reached, one always faces the question of optical resolution versus depth of focus. For example, I often use a Mitutoyo long-working-distance 10X objective for crystals with sizes in the range 5-50 microns, as it has a 0.28 numerical aperture, 1.0-micron resolving power, and a depth of focus of 3.5 microns. Although this objective is corrected for imaging at infinity, it works quite well without any additional lenses. The low depth-of-field can be problematic with this objective, however, as crystals thicker than 3.5 microns will not focus well, and this tends to get in the way of achieving the rated resolving power.

For somewhat larger crystals, or for imaging flat, plate-like crystals (where the depth of focus can be quite small), I am a big fan of the Mitutoyo long-working-distance 5X objective. This lens has a 0.14 numerical aperture, 2.0-micron resolving power, and a depth of focus of 14 microns, producing bitingly sharp images of stellar plate snow crystals in particular. Even for quite small crystals, the depth-of-field issue means that the 10X is not substantially better than the 5X in many circumstances.

In larger growth chambers, I often build in a 3X Mitutoyo Compact Objective. This has a working distance of 78 mm when imaging at infinity, becoming longer when imaging to shorter distances without a secondary lens. The resolving power is 2.5 microns, which is adequate for larger snow crystals, especially those with complex morphologies, and the depth of focus is a comfortable 23 microns.

In nearly all of my work, I often use focus stacking to achieve a higher effective depth of focus for a given resolution. Focusing by hand is adequate for stacking just a few images, and this can make a surprisingly large difference in overall image quality. For better consistency when acquiring a greater number of images, I use a StackShot automated focus-stacking rail that makes the image acquisition process quite simple. For post-processing, there are a variety of software tools available (for example Helicon Focus) for combining images. I discuss optical microscopy techniques further in Chapter 11.

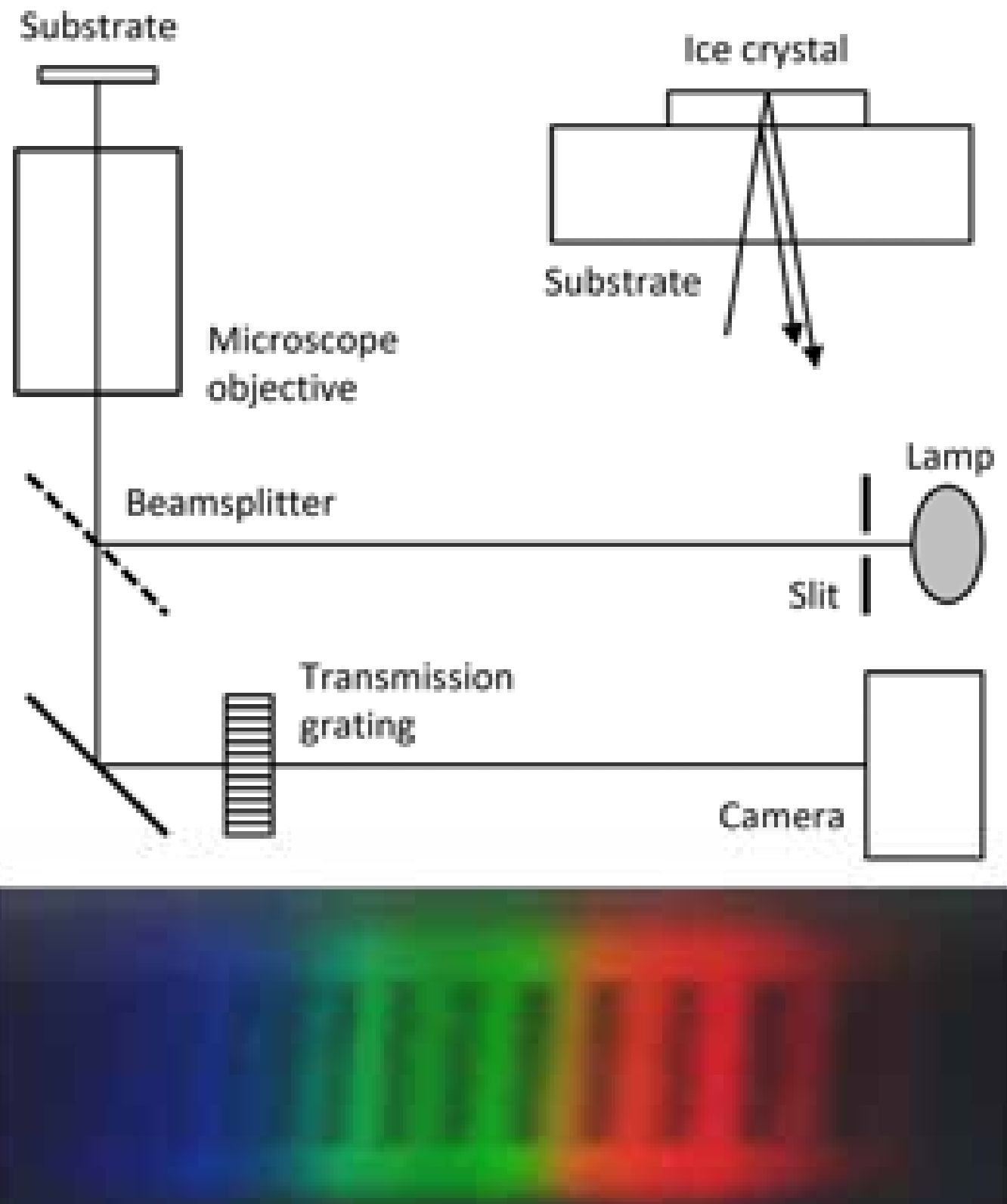

**Figure 6.14: (Top) An optical setup for measuring the thickness of thin ice crystal plates using white-light interferometry, described further in the text. The sketch on the upper right shows the interfering reflections from the substrate/ice and ice/air interfaces. (Bottom) An example white-light spectrum, in true color, showing a series of dark interference fringes. The absolute thickness of the ice crystal can be derived from the measured spacing between the optical fringes [2013Lib]. The growth velocity can be determined from the lateral motion of the fringes across the image.**

## Optical Interferometry

When the size of an ice crystal becomes comparable to the wavelength of light, optical interferometer becomes another useful measurement tool [1990Gon1, 1993Fur1, 1994Gon]. My favorite example is determining the thickness of thin, plate-like crystals using white-light interferometer, as illustrated in Figure 6.14. The essential idea is to interfere a reflection from the ice/substrate interface with a second reflection from the nearby ice/air interface, as illustrated in the upper right corner of the figure. Both reflections have roughly the same amplitude because the index jumps at the two interfaces are similar.

The interference of the two reflections will depend on the wavelength of the incident light, ranging from constructive to nearly fully destructive. A transmission grating disperses the reflected light to reveal a pattern of fringes. Note that the dark line segments in Figure 6.14 are images of the slit after destructive interference. The ends of the line segments indicate the edges of the thin ice prism.

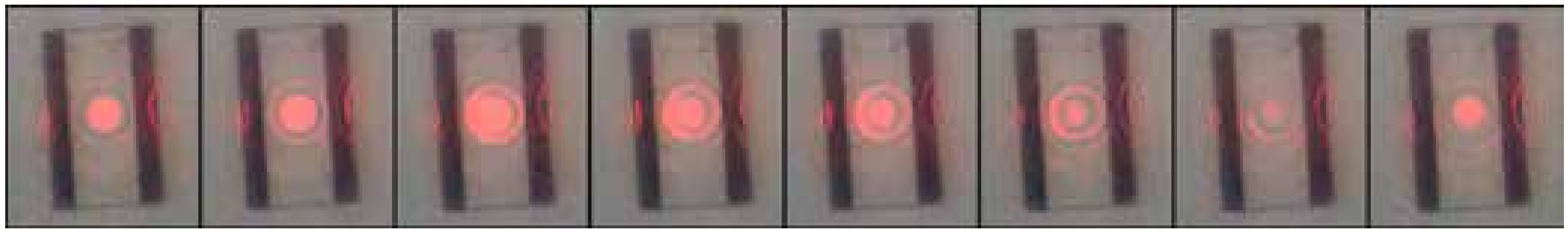

**Figure 6.15: A series of images of a columnar ice prism showing a reflected spot from a (extremely low power) Helium-Neon laser beam. As the crystal grows (left to right), the laser spot first increases and then decreases in brightness, owing to interference between reflections from the ice/substrate and ice/air interfaces.**

The spacing between the fringes in Figure 6.14 depends on the thickness of the ice crystal, with thicker crystals producing more closely spaced fringes. Working through the math, the ice-crystal thickness $h$ is given by

$$h = \frac{\lambda^2}{2n\Delta\lambda} \qquad (6.3)$$

where $\lambda$ is the wavelength of light, $\Delta\lambda$ is the fringe spacing, and $n$ is the index of refraction of ice. One convenient way to calibrate the image scale is to shine different color laser pointers through the slit and record the positions of the reflected light.

Note that the use of white-light interferometry allows for absolute thickness measurements, and one can achieve sub-micron precision with careful calibration. As a crystal grows thicker, the fringes move laterally across the image and the fringe spacing becomes smaller. The fringe motion can be determined with especially high accuracy, allowing velocity measurements down to 1 nm/sec in ideal cases.

The astute reader may note that the ice crystal at the focus of the microscope objective acts as a "cat's eye" reflector. A nearly collimated light beam enters the objective and reflects from the flat surfaces at the microscope's focus, then exiting through the objective as a nearly collimated beam. Reflections from other surfaces, such as the bottom surface of the substrate, do not take place in the focal plane and thus do not yield collimated exit beams. The cat's-eye effect explains why only the interference pattern from the ice crystal reflections appears on the camera sensor; interference patterns involving other reflections in the optical system are suppressed. However, this same effect means that especially thick crystals produce poor fringe patterns. In practice, it becomes difficult to discern fringe patterns when the crystal thickness is more than a few times the depth-of-focus of the microscope objective.

For thicker crystals, white-light interferometry becomes difficult as the fringe contrast becomes too low to observe. In this case, one can use direct laser interferometry by replacing the lamp and slit in Figure 6.14 with a collimated laser beam, and removing the transmission grating. Simple imaging then shows a bright spot superimposed on the crystal, as shown in Figure 6.15. As the crystal thickness increases, the laser intensity rises and falls, again from interference between reflections from the ice/substrate and ice/air interfaces. This technique does not yield absolute thickness measurements, and velocity measurements are far less precise than using white-light interferometry. Note that the laser intensity is far too low to significantly affect the crystal temperature or growth dynamics.

For crystals smaller than about 10 microns, the cyclical brightness oscillations from laser interferometry can be difficult to interpret. I have found that multiple reflections within the ice, especially for slender columnar crystals (substantially smaller than that shown in Figure 6.15), can produce a puzzling variety of brightness patterns. Once the reflecting surfaces are larger than the laser spot size, as illustrated in Figure 6.15, the oscillating brightness signal matches expectations from basic plane-wave interferometry theory.

When applied to simple ice prisms with clean faceted surfaces, white-light and laser interferometry are quite valuable in precision growth experiments, as I describe in Chapter 7. Laser interferometry can also be applied to larger snow crystals with complex structures [1993Fur1, 1994Gon], but interpreting the resulting measurements can be quite challenging compared to simple prisms.

## Electron Microscopy

Beyond optical imaging and measurement techniques, electron microscopy has also been adopted to examine ice crystal structure and growth. William Wergin and collaborators made an extensive study of natural snow crystals that were collected, transported in liquid nitrogen, sputter coated with a several-nanometer thickness of platinum (to provide a conductive surface), and imaged using a low-temperature scanning electron microscope (LTSEM) [1995Wer]. The authors found that this processing caused little damage to the snow crystals, and Figure 6.16 shows an example imaged using these techniques [2003Erb]. The side-by-side comparison in Figure 6.17 illustrates how the opaque character of the LTSEM image provides an exceptionally clear view of fine-scaled surface features. In both these figures, however, the snow crystals were quite large overall.

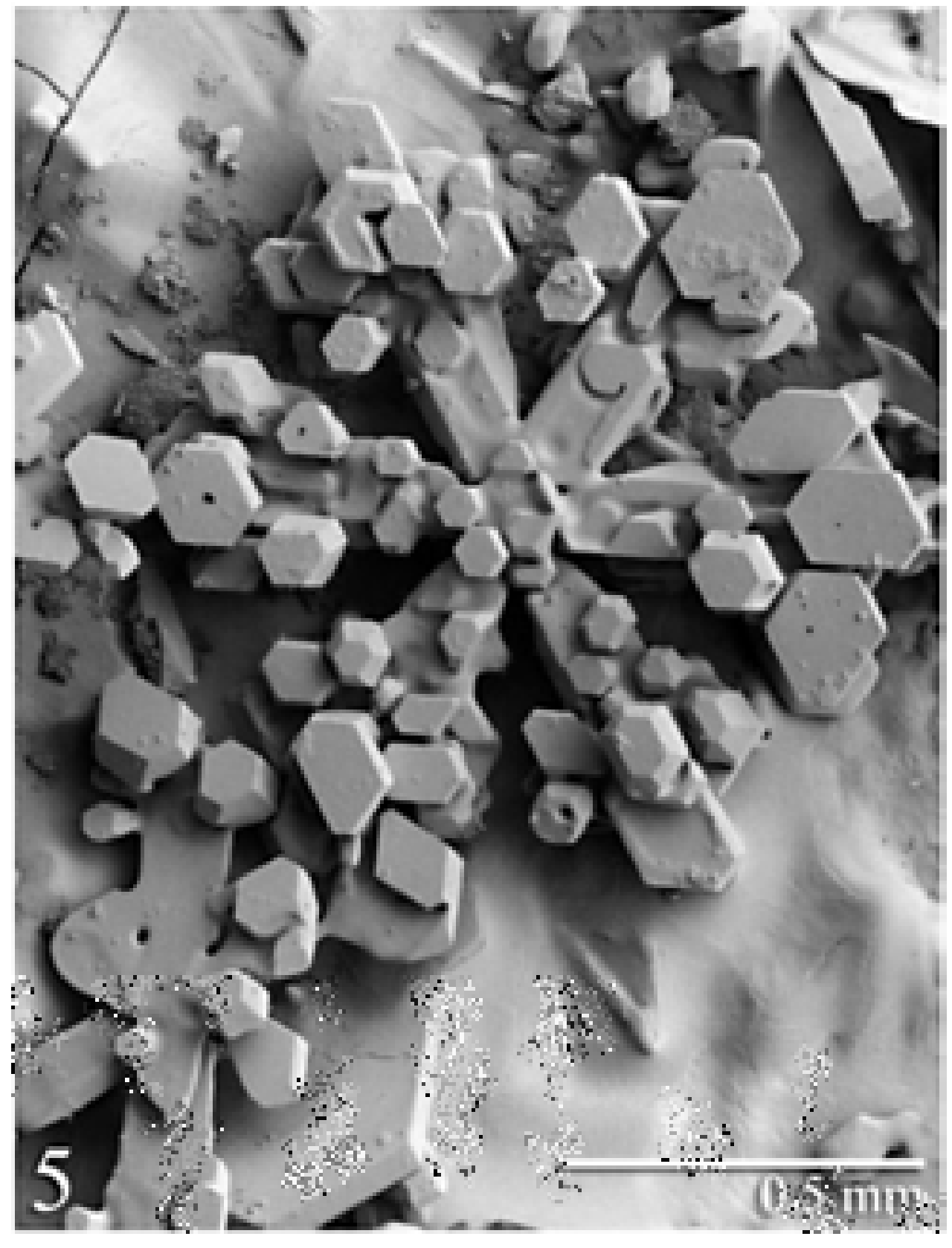

**Figure 6.16: An electron micrograph of a stellar snow crystal covered with numerous small hexagonal prisms. Note that the facets of the small prisms all align with the underlying stellar crystals, indicating epitaxial growth [2003Erb].**

**Figure 6.17: (Below) A side-by-side comparison of a platinum-coated snow crystal from an optical microscope (left) and an LTSEM [2003Erb].**

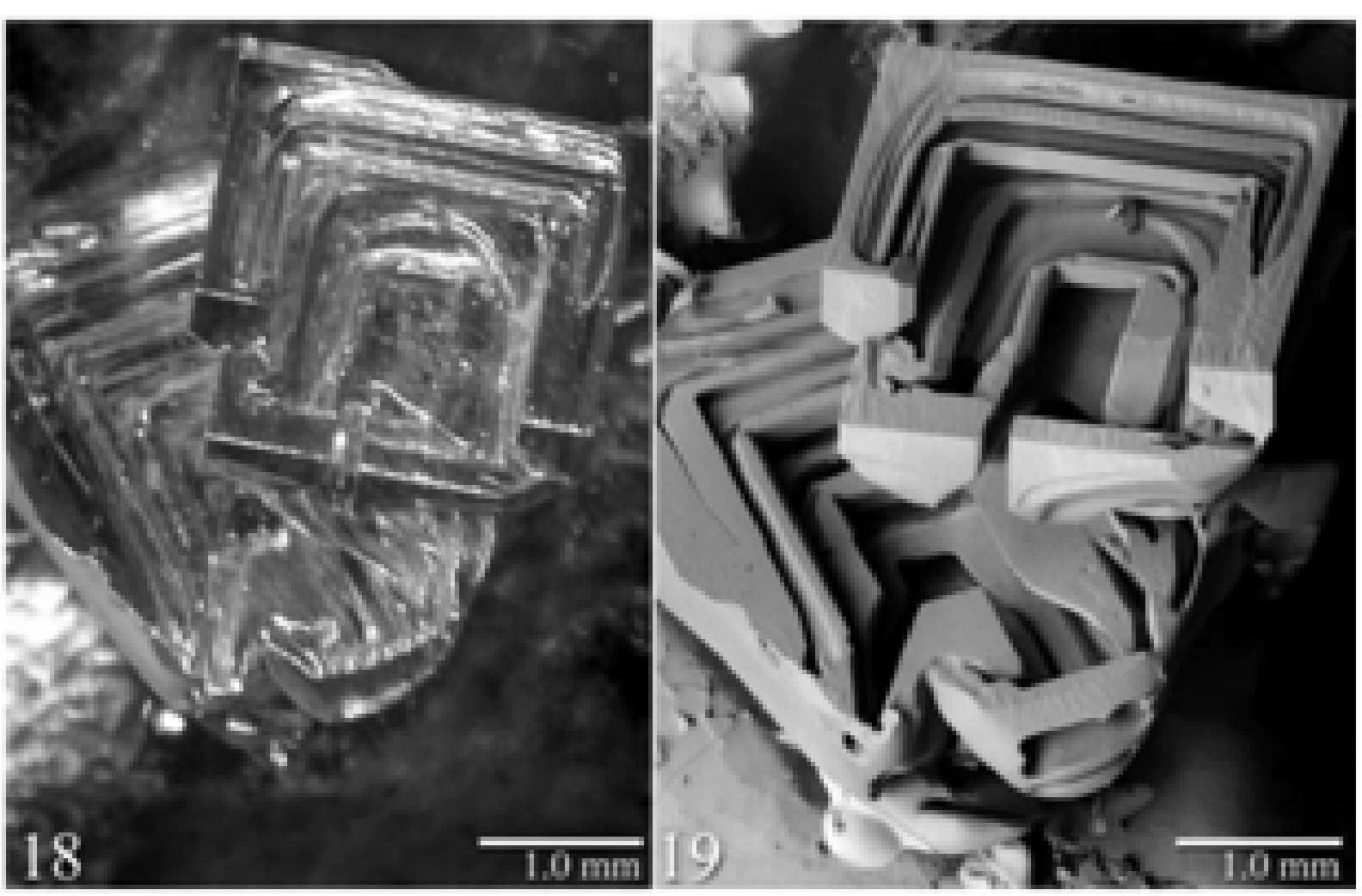

More recently, researchers have developed environmental scanning electron microscopy (ESEM) that allows direct imaging of uncoated snow crystals in humid, low-pressure environments. The example ESEM image Figure 6.18 reveals small-scale surface structures at a resolution clearly superior to what has been obtained with optical imaging. It is also possible to observe *in situ* growth and sublimation on the ESEM stage, as shown in Figure 6.19 [2010Pfa]. These authors also observed the development of peculiar "trans-prismatic strands" during sublimation. Clearly ESEM investigations have great potential for examining small-scale snow-crystal surface structures.

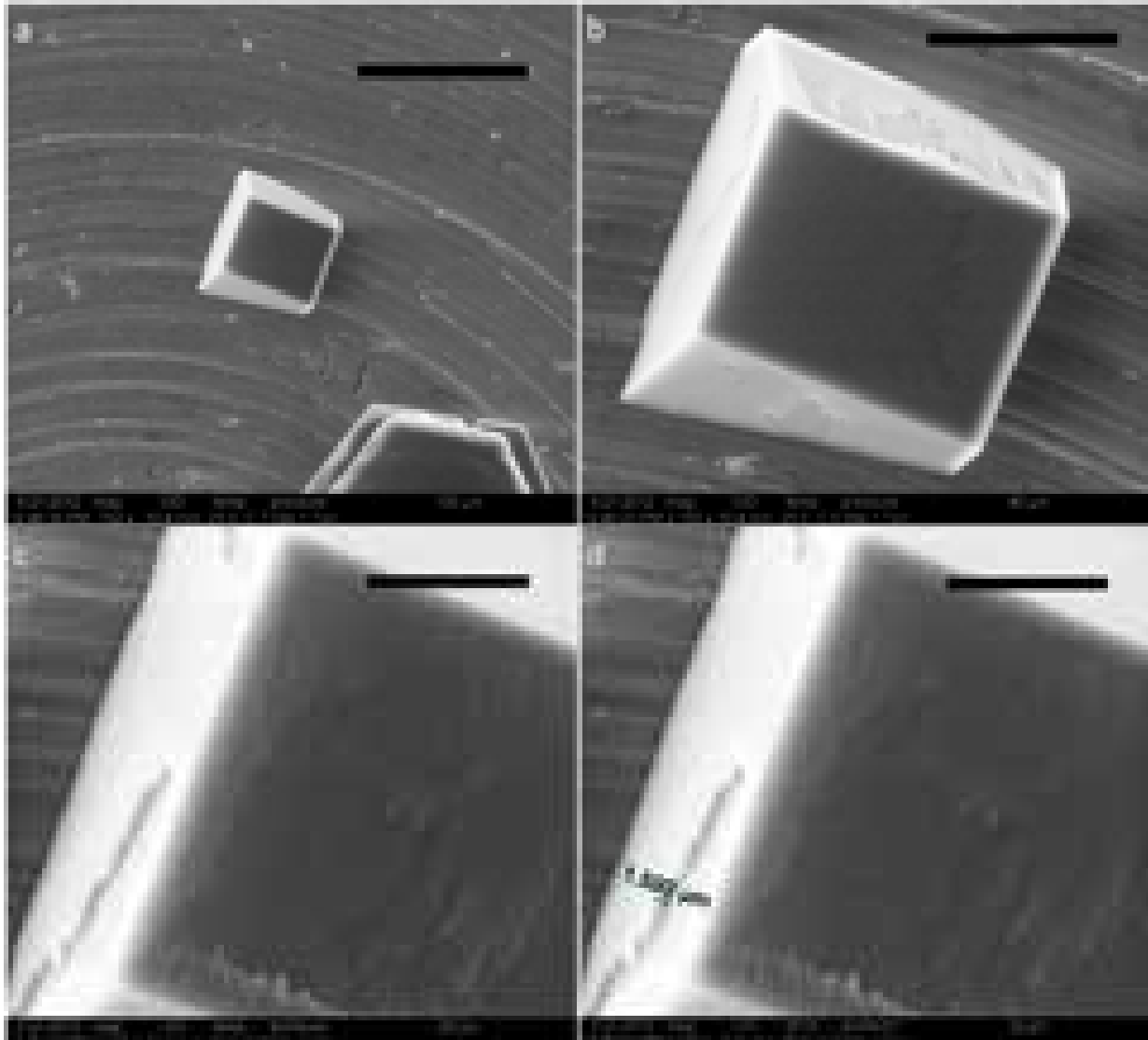

**Figure 6.18: Several ESEM images of a simple prismatic ice crystal that reveal fine surface structures, such as the 1.6 μm step seen in the lower right image. The black scale bars are 100, 40, 20, and 20 μm long (adapted from [2014Mag]).**

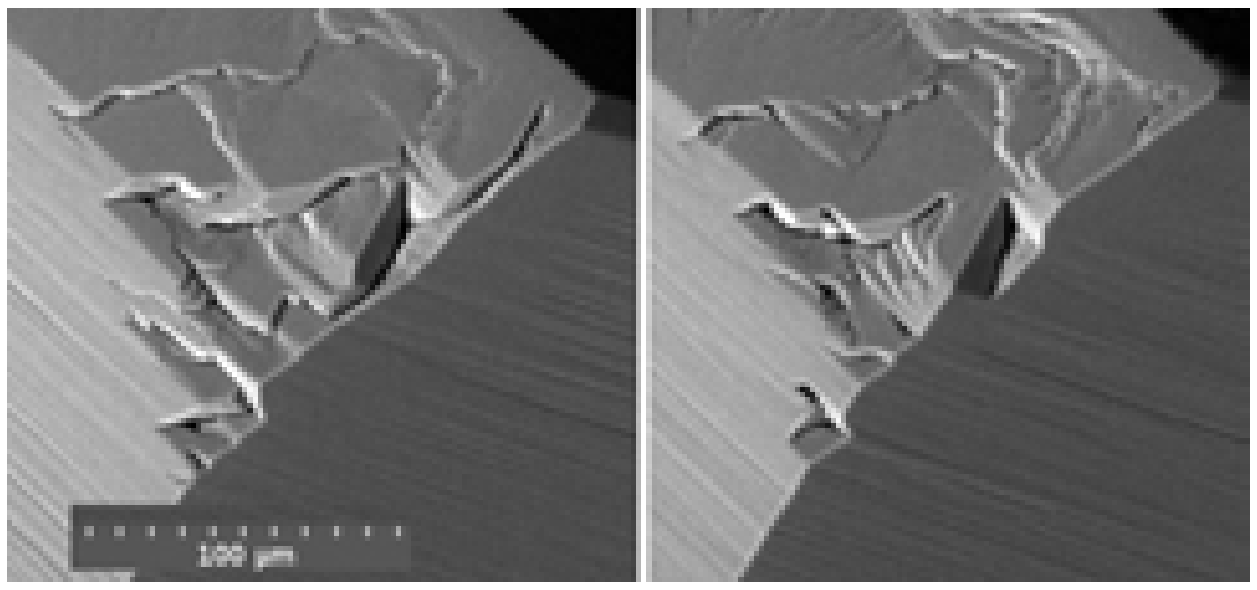

**Figure 6.19: ESEM images before (left) and after (right) a period of ice growth that filled in part of the unfaceted corner (adapted from [2010Pfa]).**

## Imaging Terrace Steps

Optical imaging of individual terrace steps on faceted mineral crystals was first demonstrated in the early 1950s using precision interferometry [1950Gri, 1951Ver], but related techniques were only applied to ice crystal surfaces recently [2010Saz]. The image shown in Figure 6.20 was obtained using laser confocal microscopy combined with differential interference contrast microscopy (LCM-DIM), and the authors clearly verified that the observed features were one-terrace-high molecular steps.

Subsequent studies using LCM-DIM have investigated the surface premelting structure near the triple point [2016Mur] and ice wetting by liquid water (including the water/ice wetting contact angle) [2015Asa, 2016Asa]. Measurements of step velocities have also been used to determine the surface diffusion length for admolecules on faceted surfaces [2014Asa, 2018Ino], although these measurements may have been affected by a lower-than-estimated supersaturation near the ice surface [2015Lib]. (I discuss modeling this difficult-to-determine parameter in Chapter 3, specifically in association with Equation 4.37 and Figure 4.21.)

Scanning probe microscopy has also been used to image terrace steps on many faceted crystalline surfaces, but so far not on ice [1997Pet, 1998Dop, 1998Pit, 2001Zep]. Surface premelting appears to interfere with molecular-scale resolution, perhaps by surface-tension forces, but the details are not yet well known.

## Snow Crystal Growth Chambers

Having examined some of the hardware components and measurement techniques used in snow-crystal studies, we now turn our attention to growth chambers. Specifically, I focus on experimental chambers used to investigate the physics underlying snow crystal growth and structure formation.

In broad brush strokes, growth chambers can be divided into three categories by their intended purposes: 1) making precise measurements of small ice prisms needed to better understand the attachment kinetics; 2) making measurements of larger snow crystals exhibiting complex structures for comparing with computational modeling; and 3) creating designer snow crystals mainly as an artistic pursuit. I discuss the first of these categories in greater detail in Chapter 7, one especially promising apparatus from the second category in Chapter 8, and a means of growing designer snow crystals in Chapter 9.

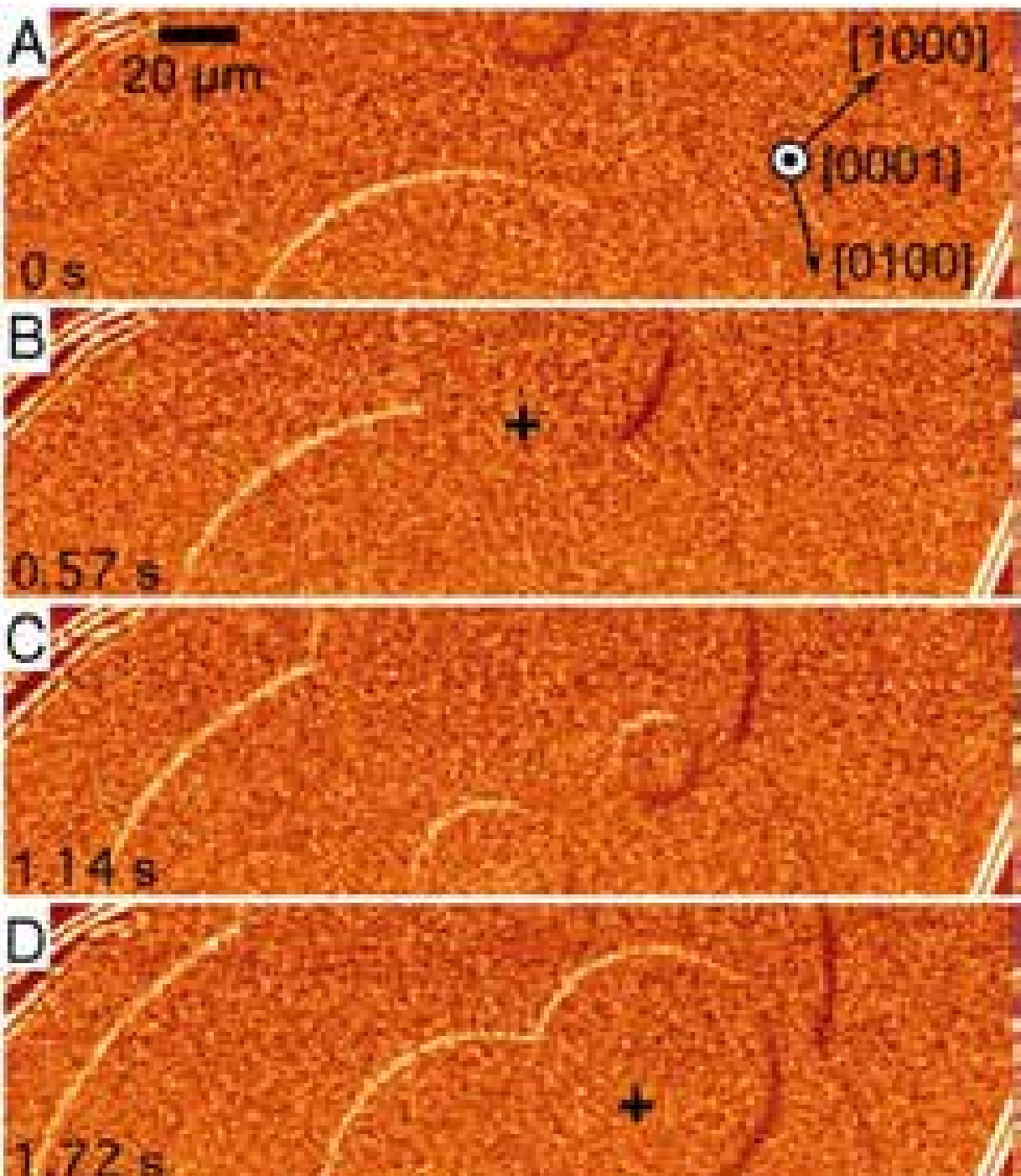

**Figure 6.20: Photomicrographs of terrace steps on the basal face of an ice crystal that display the nucleation and growth of new terraces [2010Saz].**

In the first two cases, I always stress the importance of quantitative measurements of growth rates in addition to observations of morphological features and development. This follows my stated objective throughout this book of creating a comprehensive physical model of snow crystal growth dynamics, including a detailed understanding of the attachment kinetics and a computational model that is capable of reproducing experimental observations over a broad range of environmental conditions.

For these scientific goals, it is not sufficient to develop techniques for high-resolution imaging, crystal handling, supersaturation control, and other considerations as separate achievements. To produce useful measurements, all these feats be accomplished together and integrated into a useable growth chamber. No one apparatus can serve all functions in this regard, so different strategies must be employed for different targeted investigations. Haphazardly growing snow crystals is relatively easy, but making precision measurements that further our overarching understanding of snow crystal formation is quite challenging. So we begin with an introduction to some of the different strategies used in snow crystal growth chambers.

## 6.2 Free-Fall Chambers

One of the technically simplest ways to grow synthetic snow crystals is by letting them fall through the air as they grow, essentially imitating the formation of atmospheric snowflakes. Of course, a laboratory growth chamber is vastly smaller than a winter cloud, so we can expect that freely falling synthetic snow crystals will be substantially smaller than the natural variety. But the size different is not as great as one would naively expect, which we can show from an understanding of snow-crystal growth and aerodynamics.

### Terminal Velocities

From Chapter 3 and [2009Lib3], the terminal velocity of a small spherical ice crystal of radius $R$ is

$$u_{term} \approx \frac{2}{9}\frac{\rho_{ice}g}{\mu}R^2 \qquad (6.4)$$

If we that assume the growth of the sphere is purely diffusion-limited at a pressure of one atmosphere, then the growth velocity is

$$v \approx \alpha_{diff} v_{kin} \sigma_\infty \qquad (6.5)$$
$$\approx \frac{X_0}{R} v_{kin} \sigma_\infty$$

and integrating this gives the crystal radius as a function of time

$$R \approx \sqrt{2 X_0 v_{kin} \sigma_\infty t} \qquad (6.6)$$

Using typical values $v_{kin} \approx 300$ μm/sec and $\sigma_\infty \approx 0.01$, integrating the terminal velocity over time gives a fall distance of $h \approx 1$ meter after a fall time of $T \approx 140$ seconds, at which point the crystal radius is $R \approx 11$ μm. Moreover, these results scale as $R \sim h^{1/4}$ and $T \sim h^{1/2}$. Comparing a 1-km and a 1-meter fall distance, we see that the resulting crystals will only be about six times smaller with the shorter fall distance. The details will depend on the attachment kinetics, crystal morphology, and other factors, but the overall conclusion is that, even with a modest fall distance in a laboratory free-fall chamber, crystal diameters of several tens of microns are easily achievable in a relatively short amount of time.

## Cloud Chambers

A basic top-loading household freezer is perhaps the least expensive means of creating freely falling synthetic snow crystals, albeit small ones. The set-up illustrated in Figure 6.21 was first demonstrated by Vincent Schaefer and colleagues in the 1940s as an easy demonstration of basic cloud physics. Opening the top of the freezer and simply breathing down into it produces a visible cloud of water droplets, as water vapor from your breath condenses on dust particles in the cold air.

This cloud-making process is the same as when you "see your breath" outside on a cold day, although the freezer cloud will be more stable. The cold air sinks stably into the freezer, and the cloud droplets will float inside for many minutes before slowly turning into frost on the freezer walls. The liquid water droplets supersaturate the air in this cold cloud, which becomes an excellent nursery for growing snowflakes.

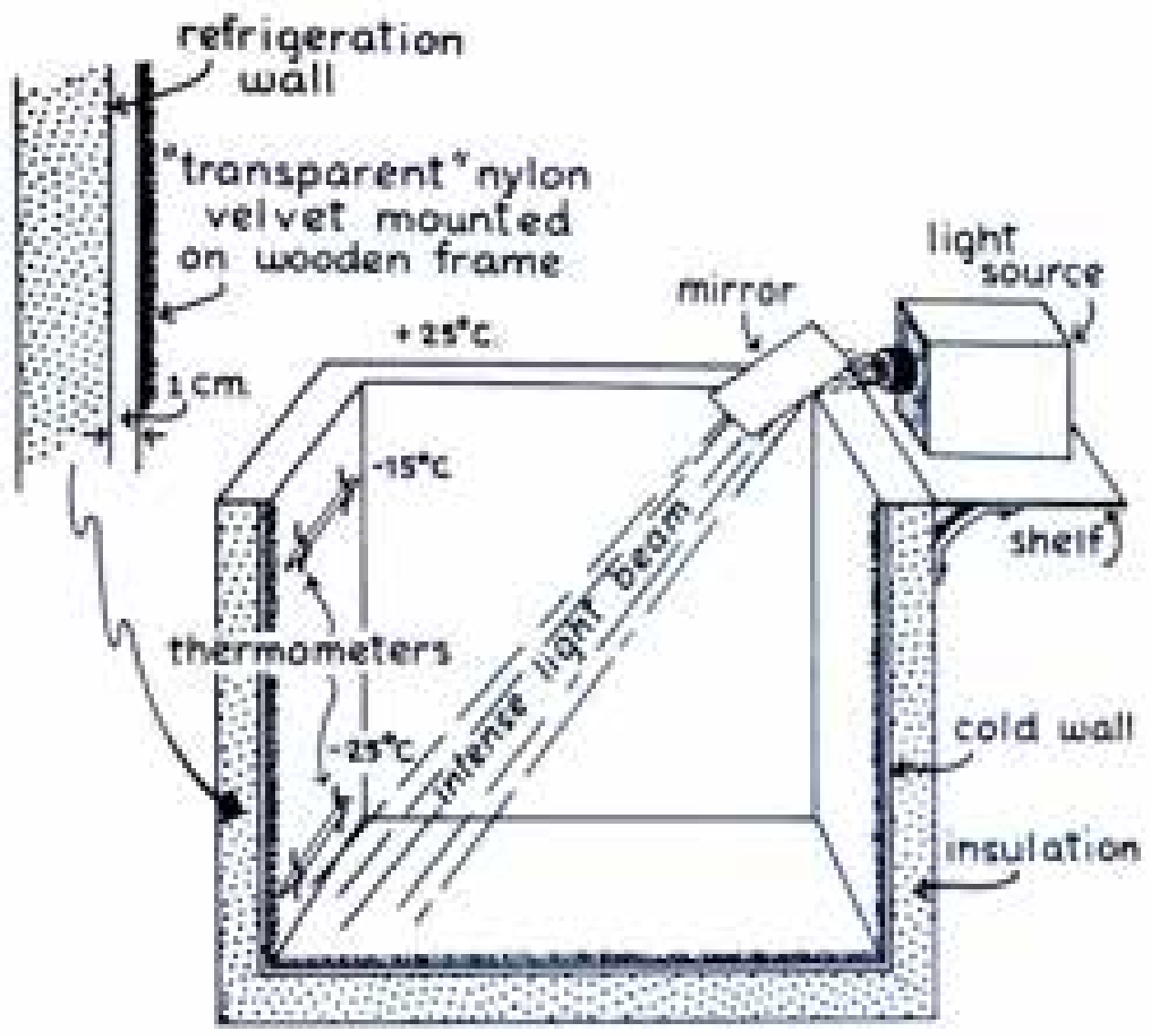

**Figure 6.21: A top-loading household freezer can be used to demonstrate several meteorological processes involved in the formation of snow crystals. Breathing down into the freezer produces a cloud of supercooled water droplets that is essentially a small-scale winter cloud. Dropping a grain of dry ice into the cloud nucleates a flurry of tiny, sparkling snow crystals. A green laser pointer is especially effective for observing the reflected bright glints from tiny faceted crystals. Image from [1981Sch].**

Nucleation is needed to produce floating snow crystals, and this is a nontrivial step. Popping cold bubble-pack using gloved hands can work as a budget expansion nucleator, but it does not usually produce copious numbers of crystals. A better, yet still inexpensive, method is to drop small flecks of dry ice into the floating freezer cloud. Dry ice sublimates at a temperature of -78 C, so ice crystals readily nucleate near its surface. By repeatedly breathing into the freezer and dropping in flecks of dry ice, one can create a veritable flurry of tiny snow crystals. Amusingly, Schaefer also described mixing iodine vapor with copious lead pollution from car exhaust in city air to form lead-iodide ice nucleators [1981Sch]. Times change.

Freezer temperatures are usually set to around -15 C, making an ideal environment for growing thin, hexagonal snow-crystal plates. By shining a bright flashlight or laser pointer into a cloud of freezer snowflakes, you can readily see sparkles from the flat basal facets. Covering the walls of the freezer with dark cloth greatly enhances the view. A cloud of water droplets looks like a rather dull gray fog, but a cloud of faceted plates above a black background makes a beautiful swirl of sparkling diamond dust.

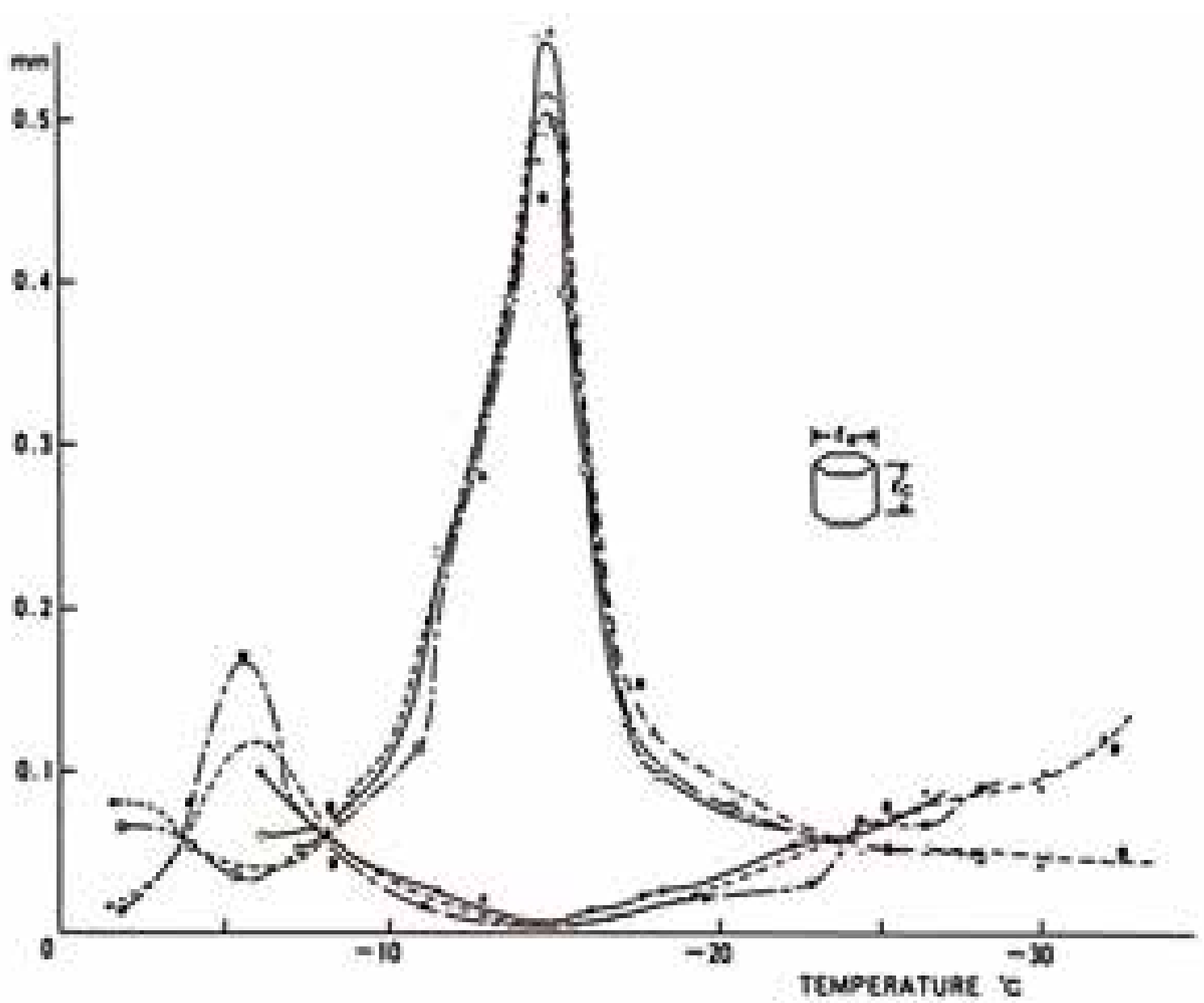

**Figure 6.22: Measurements of the diameters and thicknesses of snow crystals after growing in air for 200 seconds in a free-fall cloud chamber [1987Kob]. Note that the variation in crystal aspect ratio as a function of temperature matches expectations from the snow crystal morphology diagram.**

Viewing the tiny freezer snowflakes directly requires a microscope. As described in the calculation above, the size of a typical crystal in a free-fall chamber is typically some tens of microns, smaller than the diameter of a human hair, so quite high magnification is needed just to see the overall hexagonal form. Moreover, the microscope would have to be kept cold, and probably at the bottom of the freezer where the crystals fall.

Advancing beyond a household freezer, a research-grade cloud chamber can be constructed from a large, temperature-controlled tank, injecting water droplets from an ultrasonic humidifier to produce a cloud that soon equilibrates to the temperature of the tank. Nucleation is easily accomplished using the expansion nucleator described above, and the falling crystals can then be viewed using an upward-facing microscope built into the bottom of the chamber.

Figure 6.22 shows some cloud-chamber growth measurements in air from Yamashita [1987Kob], of snow crystals that resulted from a fall time of 200 seconds. To my knowledge, these remain some of the best data illustrating the growth of small snow crystals in air at a supersaturation equal to $\sigma_{water}$. As I discussed in Chapter 4, however, many additional measurements as a function of supersaturation and air pressure are needed to fully comprehend the attachment kinetics.

Cloud chambers offer numerous beneficial features when making precise measurements of snow crystal growth rates. In a static cloud of water droplets, the supersaturation quickly settles to a nearly constant value of $\sigma_{water}$ as the water vapor in the air reaches equilibrium with the numerous droplets, thus yielding a well-defined supersaturation at a well-defined temperature. It would be most beneficial to expand upon the Yamashita experiment by using salt-water droplets to produce different supersaturations, and observing over a range of air pressures and fall times. A serious limitation in such an experiment, however, is that one cannot readily create low supersaturation values in a cloud chamber.

## Convection Chambers

Another simple method for creating free-fall snow crystals is the convection chamber illustrated in Figure 6.23. Here a heated water reservoir provides a source of water vapor via evaporation and also stimulates convection that carries the water vapor upward. Turbulent convection mixes the air to yield a steady-state supersaturated environment. The expansion nucleation generates seed crystals on demand, which then float freely in the supersaturated air as they grow. After a few minutes, the crystals grow large and some fall onto the substrate for observation.

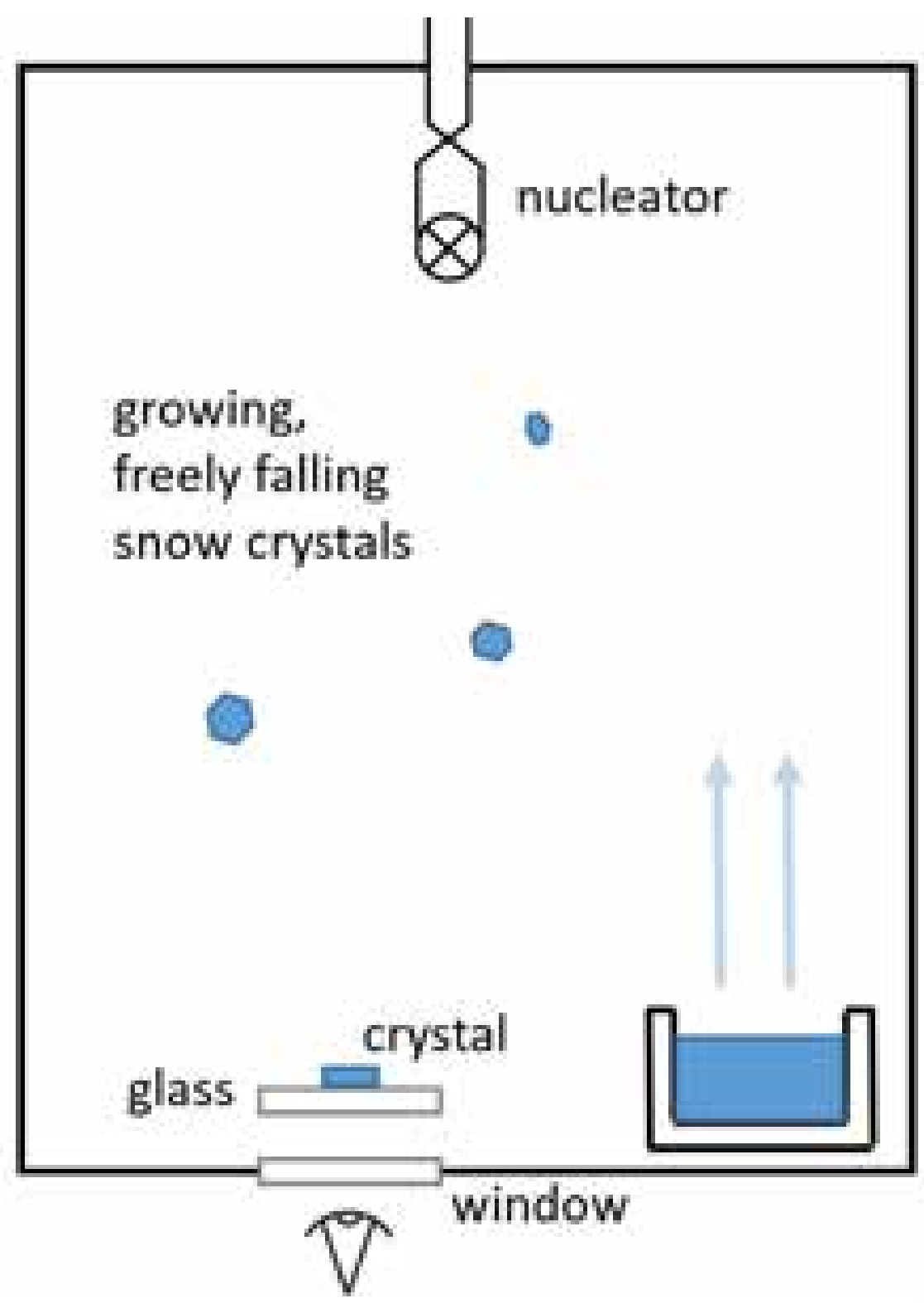

**Figure 6.23: A free-fall convection chamber. The expansion nucleator near the top of the chamber (described in Figure 6.8) produces microscopic seed crystals that subsequently grow and slowly fall in the supersaturated air. After a few minutes of growth, the crystals fall to the bottom of the chamber, where some land on a glass substrate for observation.**

Changing the temperature of the water reservoir changes the supersaturation within the chamber, giving the convection chamber added flexibility compared to a cloud chamber, and achieving a supersaturation range from zero to near $\sigma_{water}$ is straightforward. However, the variable supersaturation as a function of water temperature must be calibrated, and this is a challenging task. In [2008Lib4] we presented a calibration method using differential hygrometry measurements of air sampled from the chamber, giving the results shown in Figure 6.24. This calibration depends on the geometry of the chamber and water reservoir, but with care it was possible to obtain supersaturation calibration with an absolute accuracy of roughly 20 percent.

Although convection chambers are a good compliment to cloud chambers, their overall accuracy remains on the low side for investigating attachment kinetics. Differential hygrometry can be influenced by systematic errors, and neither the temperature nor supersaturation in a convecting body of air is as uniform as one would like. In spite of these issues, however, the convection-chamber results presented in [2008Lib1, 2009Lib] remains some of the best quantitative data covering a broad range of temperatures and supersaturations in air. Figure 6.25 shows some photographic examples of snow crystals grown in a convection chamber.

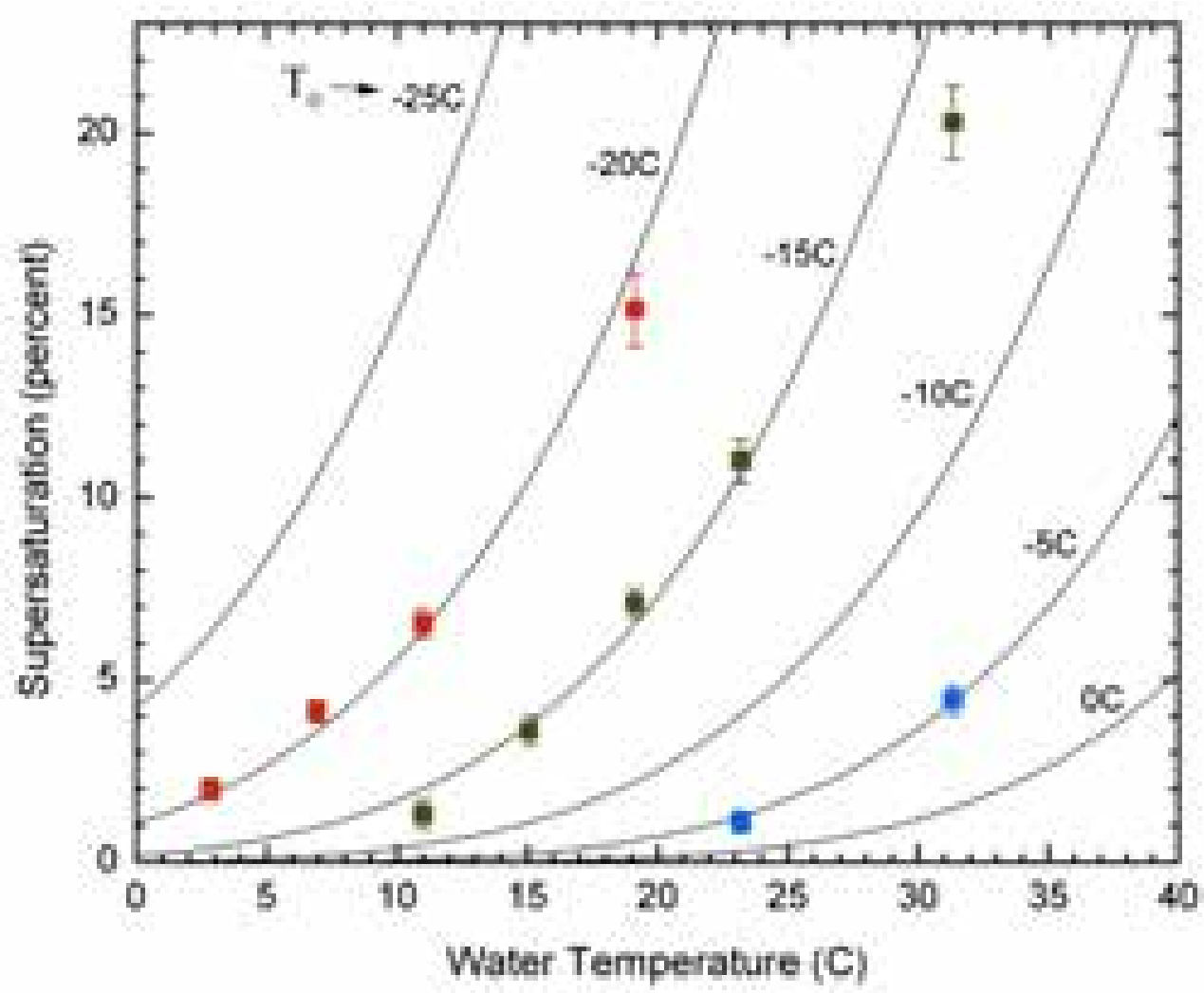

**Figure 6.24: The supersaturation in a 1-meter high free-fall convection chamber [2008Lib4] as a function of the water reservoir temperature $T_{water}$, at several values of the chamber temperature $T_0$. Data points show the supersaturation measured using differential hygrometry at -5 C, -15 C, and -20 C. Lines show an empirical model of $\sigma(T_0, T_{water})$ defined in [2008Lib4].**

**Figure 6.25: Examples of snow crystals grown in a free-fall convection chamber at temperatures of -2 C (top), -5 C (middle) and -15 C (bottom). The scale bar in the lower left corner of the image is 50 microns in length. The variation in crystal size and morphology at each temperature reflects inhomogeneities in temperature and supersaturation within the chamber. Overall, however, the morphologies agree with expectations from the snow-crystal morphology diagram.**

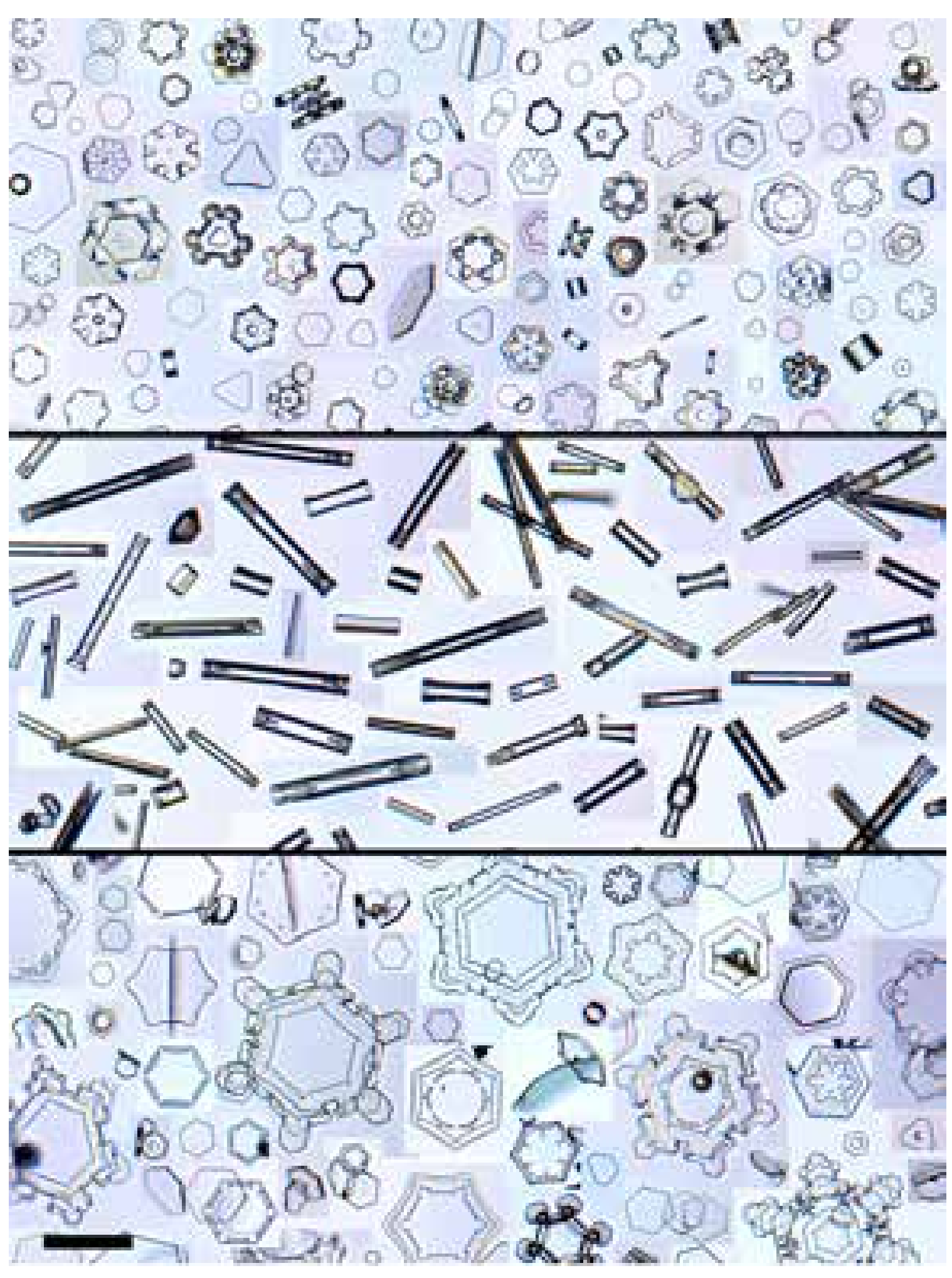

## A Seed-crystal Generator

One of my favorite applications of the free-fall convection chamber is to produce a copious and on-demand supply of ice seed crystals. The basic idea is a chamber like that shown in Figure 6.23, but without the observation hardware. By opening the nucleator valve about once every ten seconds during operation, a continuous cloud of fresh ice crystals can be found floating in the chamber. Small crystals grow quickly at first, roughly with $R{\sim}T^{1/2}$, while larger crystals fall to the floor of the chamber. These effects reduce the number of very small or very large crystals that can be found floating in the chamber, yielding a size distribution that typically peaks at around 20-50 microns. As seen in Figure 6.25, thin plates or slender columns can be produced, depending on the chamber temperature.

When I need a seed crystal for a nearby experiment, I push some air out of the free-fall chamber and waft it over a waiting cooled substrate, such that a few crystals randomly fall onto it. I then examine the substrate under a microscope and position a suitable seed crystal for subsequent study. Some examples of experiments using these free-fall seed crystals are described in Chapters 7 and 9.

When a free-fall seed-crystal chamber is set up with a temperature near -15 C and modest supersaturation levels, roughly half of the crystals are well-formed simple hexagonal plates, the remainder being malformed to some degree, perhaps from mid-air collisions, polycrystalline nucleation, or other factors.

## High Cleanliness and High Throughput

Two of my favorite characteristics of free-fall chambers are their intrinsic self-cleaning nature and their high throughput. On the cleanliness side, chemical contamination is a wild card in any ice growth experiment, as one is never sure how clean is clean enough. Seed crystals that sit around for long periods, require a lot of handing, or experience sublimation/regrowth cycling are apt to become contaminated with surface impurities. And this can cause unwanted changes in growth behavior that can be difficult to identify [2008Lib2].

Free-fall chambers are self-cleaning in that ice crystals are created in large numbers and quickly discarded. It typically takes just a few minutes for a given crystal to fall to the bottom the chamber where it remains for the duration of a run. The growing crystals, both in the chamber and covering the walls, absorb chemical contaminants from the chamber air and thus remove them from further influence. I believe that chemical contamination effects have been, and continue to be, underestimated in many snow-crystal experiments. Free-fall growth chambers are a good way to minimize these effects.

Another problem one encounters when studying snow crystals is that there is a large variation in growth behaviors under ostensibly constant growth conditions, and much of this remains poorly understood. Some crystals are likely influenced by dislocations, while seed-crystal variations may be a factor as well. Again, this is something of a wild-card in all experiments.

Free-fall growth chambers allow easy observation of hundreds of crystals in a single run, giving the experimentalist a fighting chance of making some sense of the distributions of observed growth behaviors. For example, some bimodal distributions have been reported [2008Lib1], but overall such effects have not been much studied. They are likely important in some areas, and free-fall chambers may be the best way to explore these issues.

## A Laminar-Flow Chamber

To address the problem of the finite fall distance in laboratory cloud chambers, Tsuneya Takahashi and Norihiko Fukuta developed an ingenious system for levitating a falling snow crystal in a vertical flow of air [1988Tak, 1991Tak, 1999Fuk]. A slightly tapered flow tube gently pushed a growing snow crystal toward the tube's central axis, and the laminar flow rate was continually adjusted to keep the crystal's vertical position fixed in an observation region as it grew. A fog of water droplets was added to the flowing air to keep the supersaturation at $\sigma_{water}$. (Unlike a static cloud chamber, however, the rapidly flowing air in the laminar-flow chamber may not always have time to fully equilibrate to $\sigma_{water}$, creating a possible systematic error in the known supersaturation.)

Using this laminar-flow chamber, the authors of the above papers reported extensive observations of snow crystals growing in air for up to 30 minutes, covering a range of temperatures from -2 C to -24 C, including the examples shown in Figure 6.26. This unique data set provides the best record to date of snow crystals growing in controlled environmental conditions quite close to what can be found in dense clouds. If an adjustable humidification method could be worked out, the laminar-flow chamber could perhaps be used over a range of supersaturations as well. I believe that the electric-needle technique presented in Chapter 8 is probably better suited for comparison with numerical modeling studies, but different options have different strengths and weaknesses, so all possibilities should be considered for future investigations.

**Figure 6.26: Example snow crystals grown in a laminar-flow chamber, from [1999Fuk].**

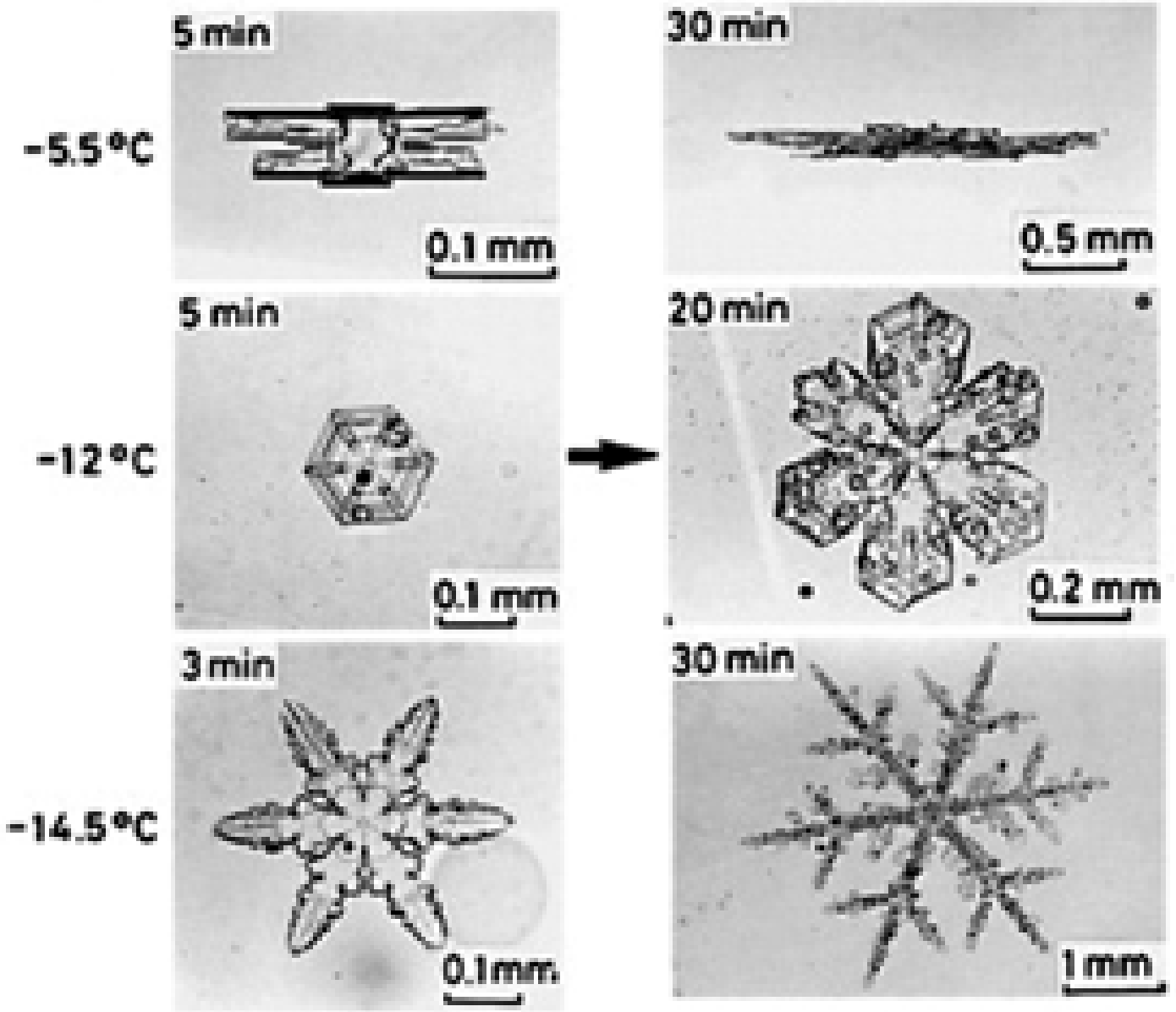

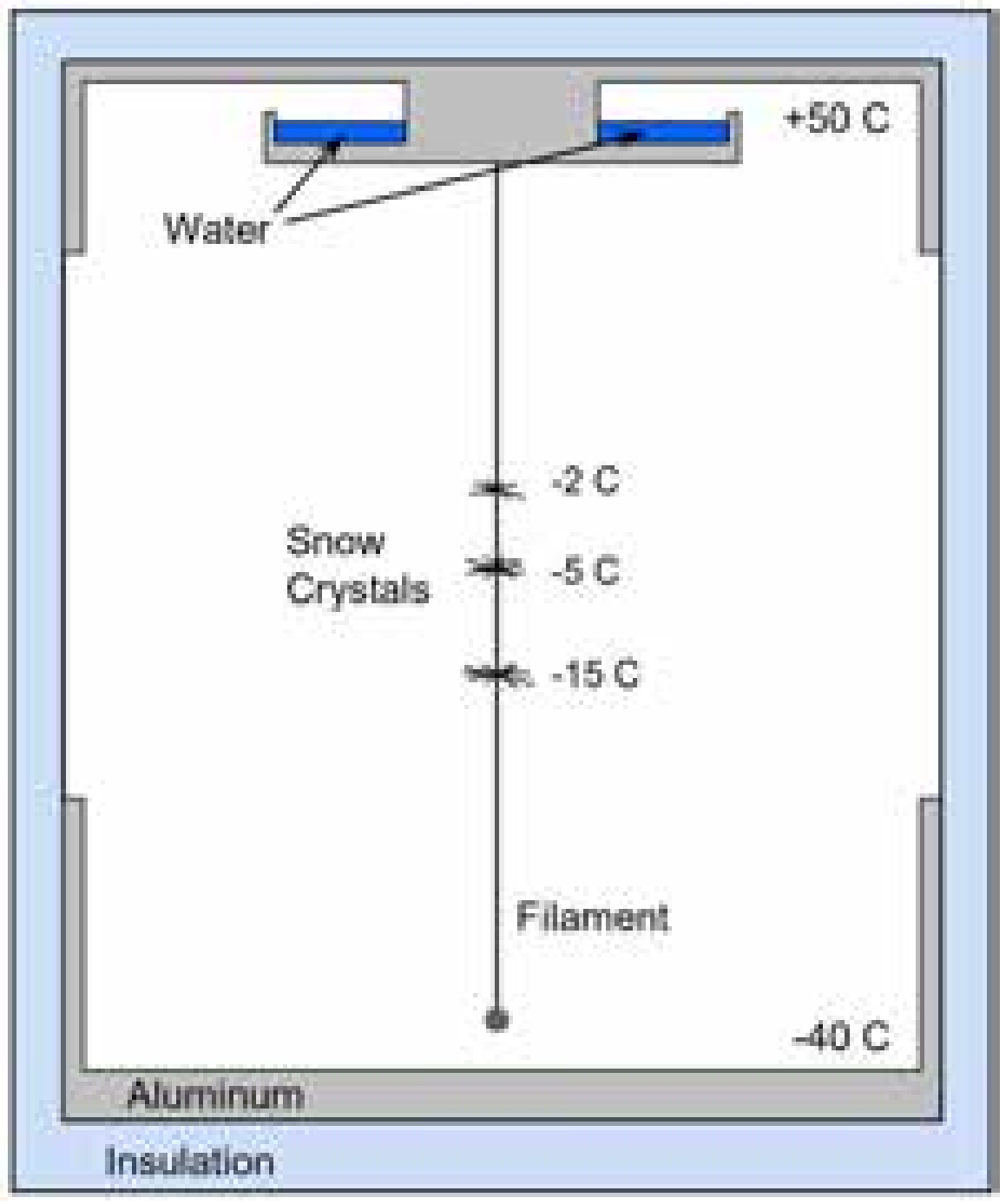

**Figure 6.27: This snow-crystal diffusion chamber creates a steep vertical temperature gradient with the top warmer than the bottom. Water vapor evaporates from the warm water reservoir and diffuses downward into the colder air below, yielding highly supersaturated air in the interior of the chamber. In this illustration, snow crystals grow at different temperatures on a vertical filament.**

## 6.5 Diffusion Chambers

Figure 6.27 shows an example of a snow-crystal diffusion chamber. The basic construction is an insulated box with a strong temperature gradient from top to bottom, with the top being hot and the bottom cold. Water vapor evaporates from the liquid water reservoir at the top of the chamber and diffuses downward into the colder air below, causing the air to become supersaturated. If any kind

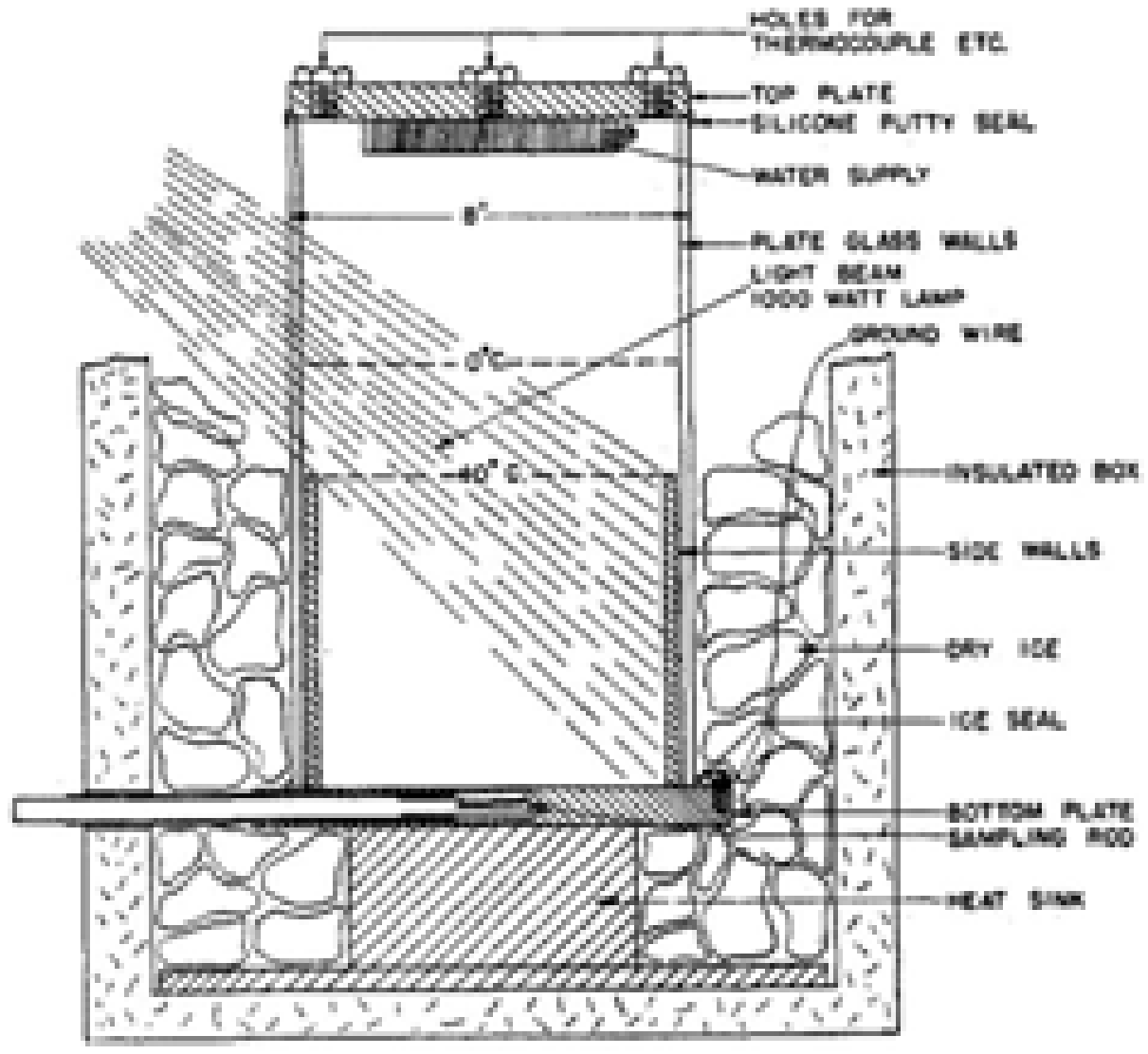

**Figure 6.28: Perhaps the first snow-crystal diffusion chamber, cooled with dry ice [1952Sch].**

of substrate is placed in this region, snow crystals will nucleate and grow.

The example shown in Figure 6.27 is also called a *continuous diffusion chamber*, as the supersaturation remains constant in time (unless perturbed by growing snow crystals). It reaches a peak near the center of the chamber, dropping to zero at the chamber walls, as these are soon covered with frost crystals.

The basic physical concepts in a diffusion chamber have been known since the cloud-chamber work of Wilson and others in the early 20$^{th}$ century. Early continuous snow-crystal diffusion chambers, like the one shown in Figure 6.28, were used by Schaefer [1952Sch] and by Hallett and Mason [1958Hal] in numerous studies of snow crystal growth dynamics and the morphology diagram.

One clever trick with diffusion chambers is to hang a vertical filament down the central axis of the chamber, as illustrated in Figure 6.27. Snow crystals will grow all along the filament, allowing one to view the different growth morphologies as a function of temperature, displaying many aspects of the snow-crystal morphology diagram at once [1958Hal]. When the supersaturation is high, fast-growing dendritic crystals (see Chapter 3) tend to grow out faster than blocky crystals, often yielding three distinct clusters along the filament at -2 C, -5 C, and -15 C, as shown in Figures 6.29 and 6.30.

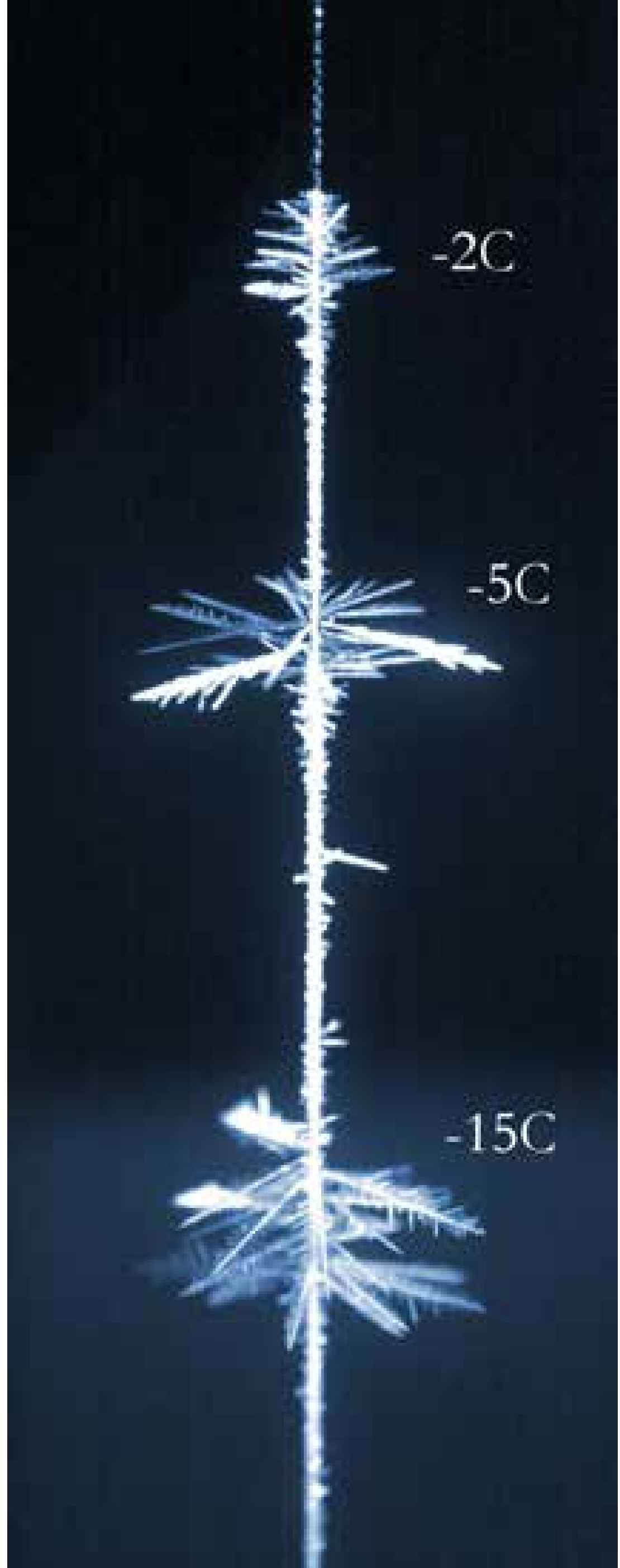

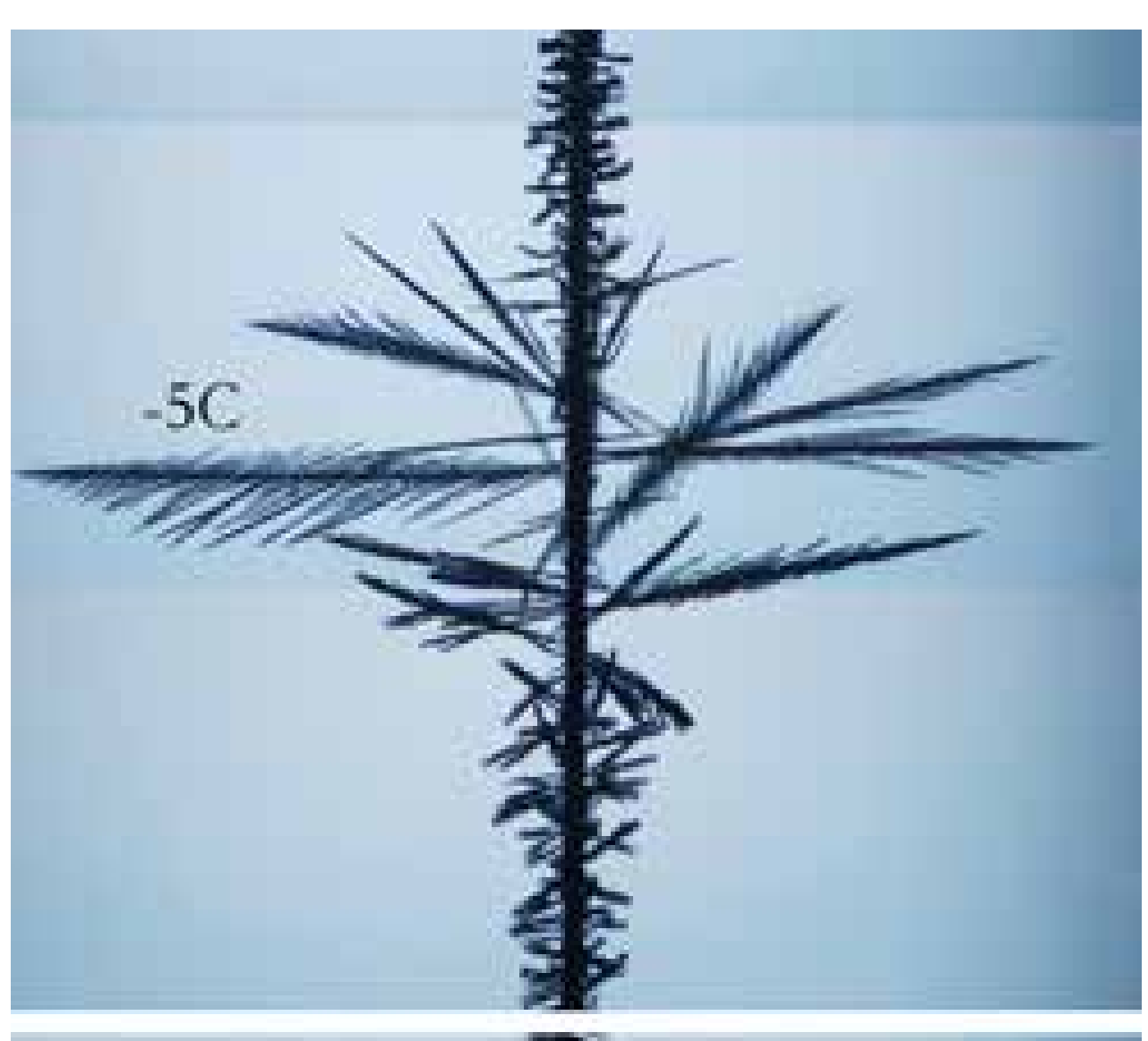

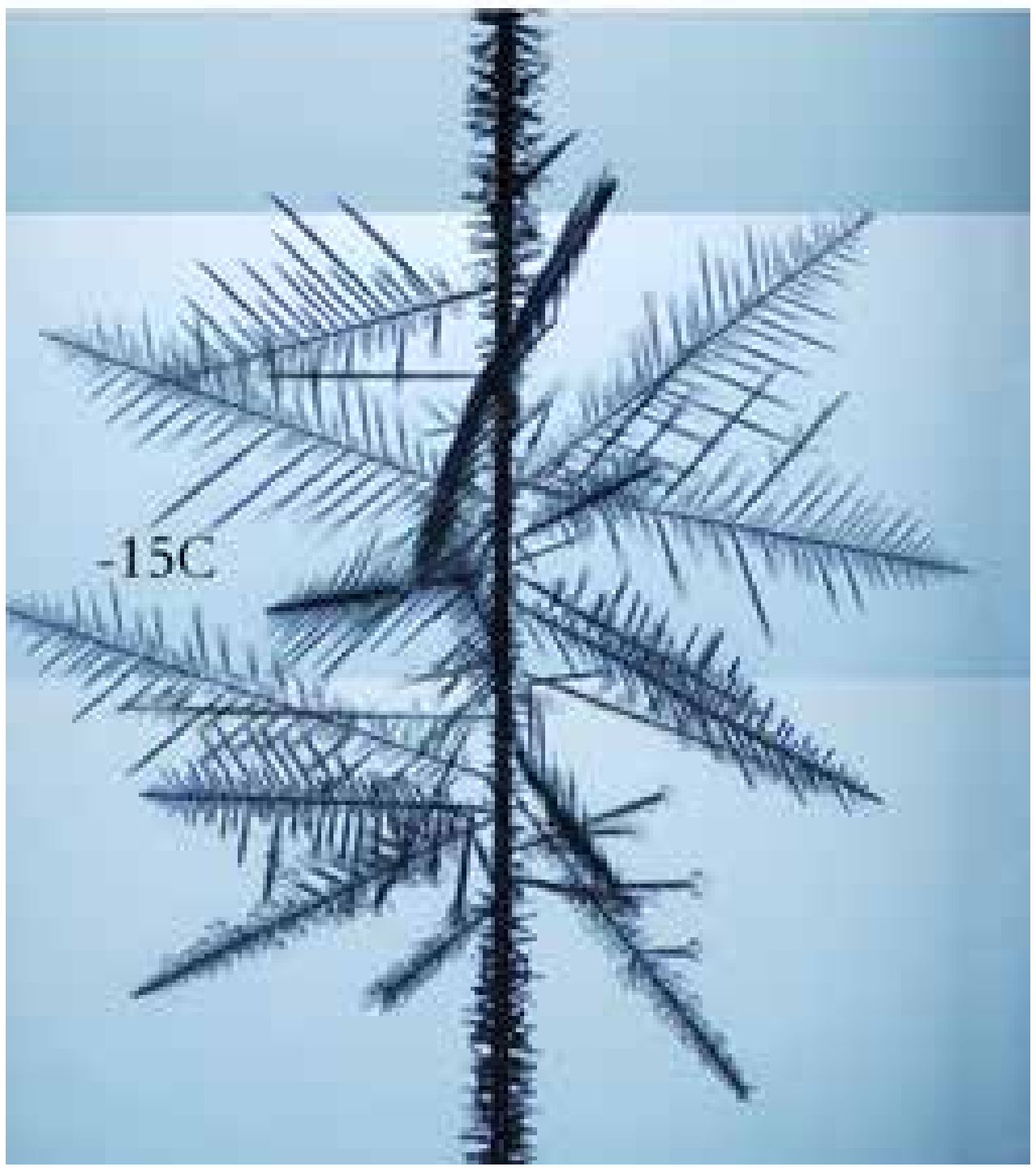

**Figure 6.30: Higher resolution images of a string of snow crystals similar to that shown in Figure 6.29. The cluster at -5 C is made of "fishbone" dendrites (see Figure 6.31), while the cluster at -15 C is made from individual branches of fernlike stellar dendrites. As soon as these fast-growing crystals extend out away from the filament, they tend to shield water vapor from reaching smaller crystals nearer the filament, stunting their growth.**

**Figure 6.29: A string of snow crystals growing at high supersaturation in a diffusion chamber like that shown in Figure 6.27. The fastest-growing dendritic crystals cluster at -2 C, -5 C, and -15 C. At other temperatures, the crystals tend to be blocky in form with slower growth rates.**

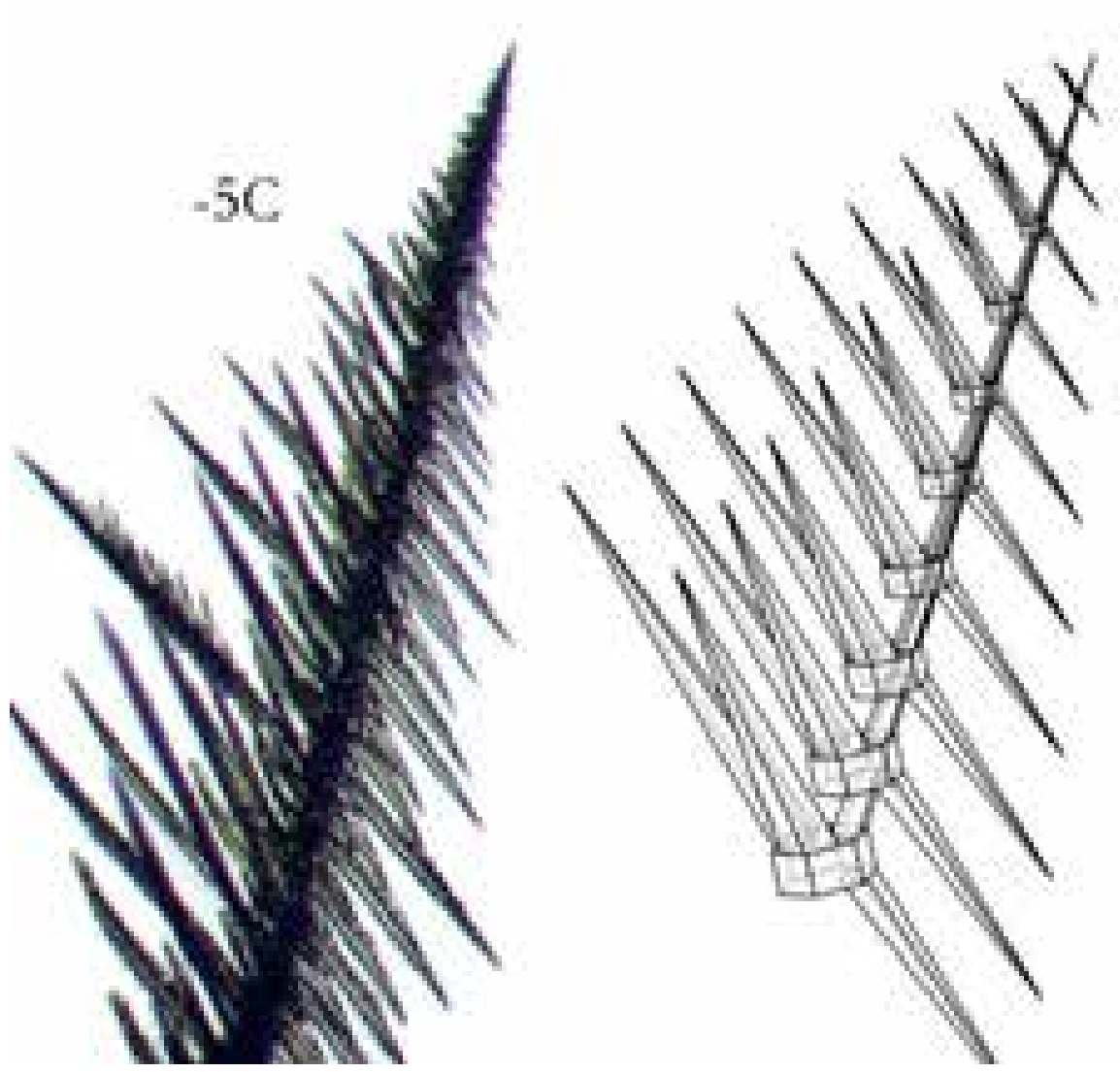

**Figure 6.31: "Fishbone" dendritic snow crystals form at temperatures near -5 C in highly supersaturated air, such as in Figure 6.30. They grow readily in snow-crystal diffusion chambers, but are not found in nature. The sketch on the right illustrates their overall structure, with hexagonal prisms showing the crystal axes [2009Lib1].**

## Clam Shell Design

The "clam-shell" aluminum pieces at the top and bottom of the apparatus shown in Figure 6.27 are there to conduct heat and tailor the temperature profile within the chamber. The supersaturation generally increases as the vertical temperature gradient $dT/dz$ increases, and clam-shell structures are often included to create high supersaturation levels.

Figure 6.32 shows an extreme example of a clam-shell diffusion chamber, designed to create an exceptionally high water-vapor supersaturation level at its central point. It is straightforward to solve the static heat-diffusion equation within this chamber, giving "naive" constant-temperature surfaces similar to those illustrated in Figure 6.32. I call these naive because this clam-shell design is unstable to convection, so the static diffusion equation does not apply in this situation. For this reason, it is not easily possible to accurately calculate either the temperature or the supersaturation profile in this chamber. This is a systemic problem with many snow-crystal diffusion chambers: while they are easy to build and deliver high supersaturations, it is often nearly impossible to model their characteristics with accuracy.

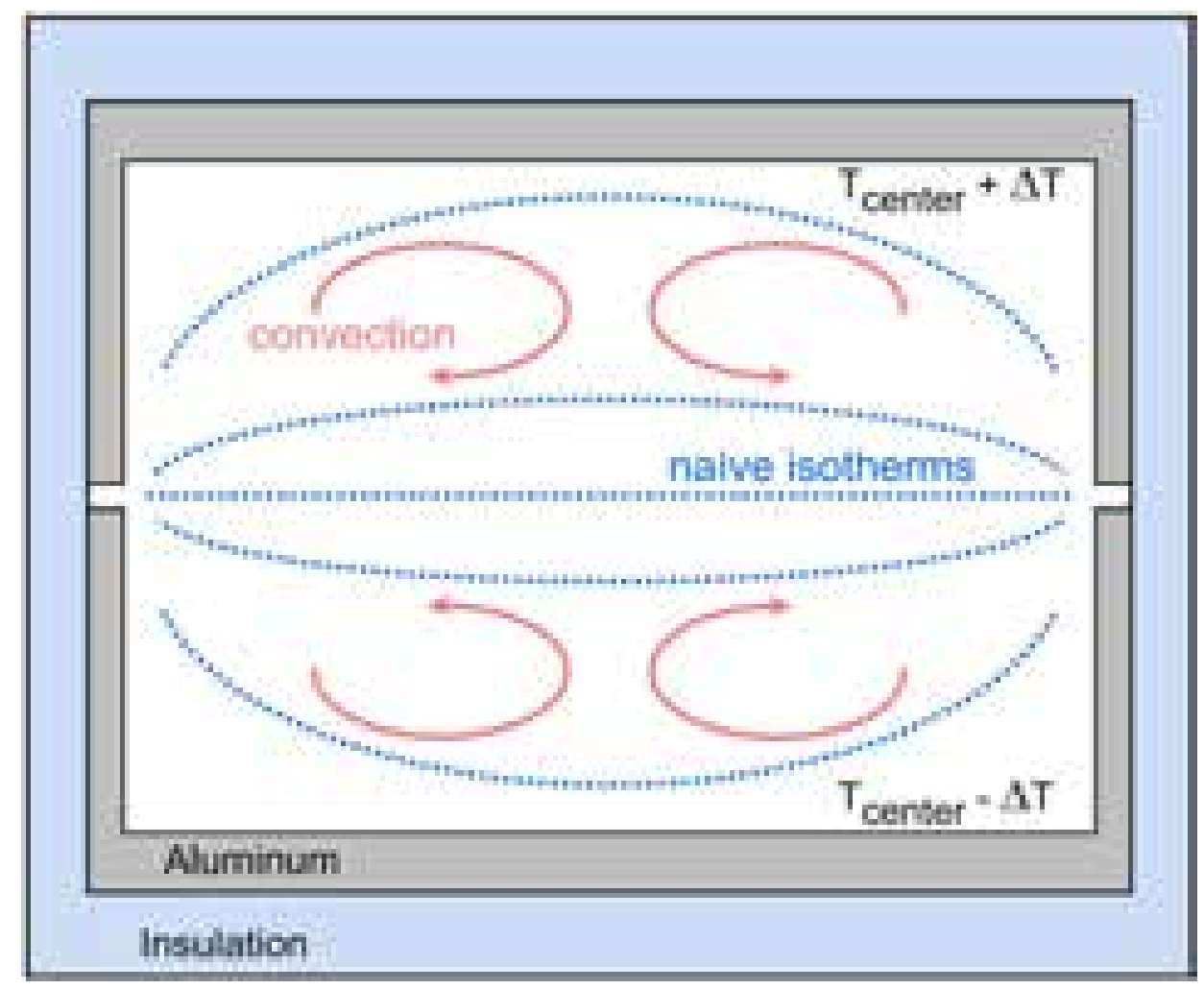

**Figure 6.32: A "clam-shell" diffusion chamber designed to create a high supersaturation level at the center of the chamber. Unfortunately, the system is unstable to convection, making it quite difficult to accurately characterize the temperature and supersaturation within the chamber.**

## Linear Gradient Diffusion Chamber

The Figure 6.33 shows a "linear-gradient" diffusion chamber that is amenable to accurate thermal and supersaturation modeling. In this design, the included stainless-steel plates have a modest thermal conductivity that results in a linear temperature gradient $dT/dz \approx const$ along the walls. Solving the heat diffusion equation then yields $dT/dz \approx const$ in the interior region as well, and this can be verified by direct measurements. With entirely horizontal isothermal surfaces, this thermal profile is stable against convection, so the static heat-diffusion equation will describe the correct 3D temperature field $T(\vec{x})$ within the chamber.

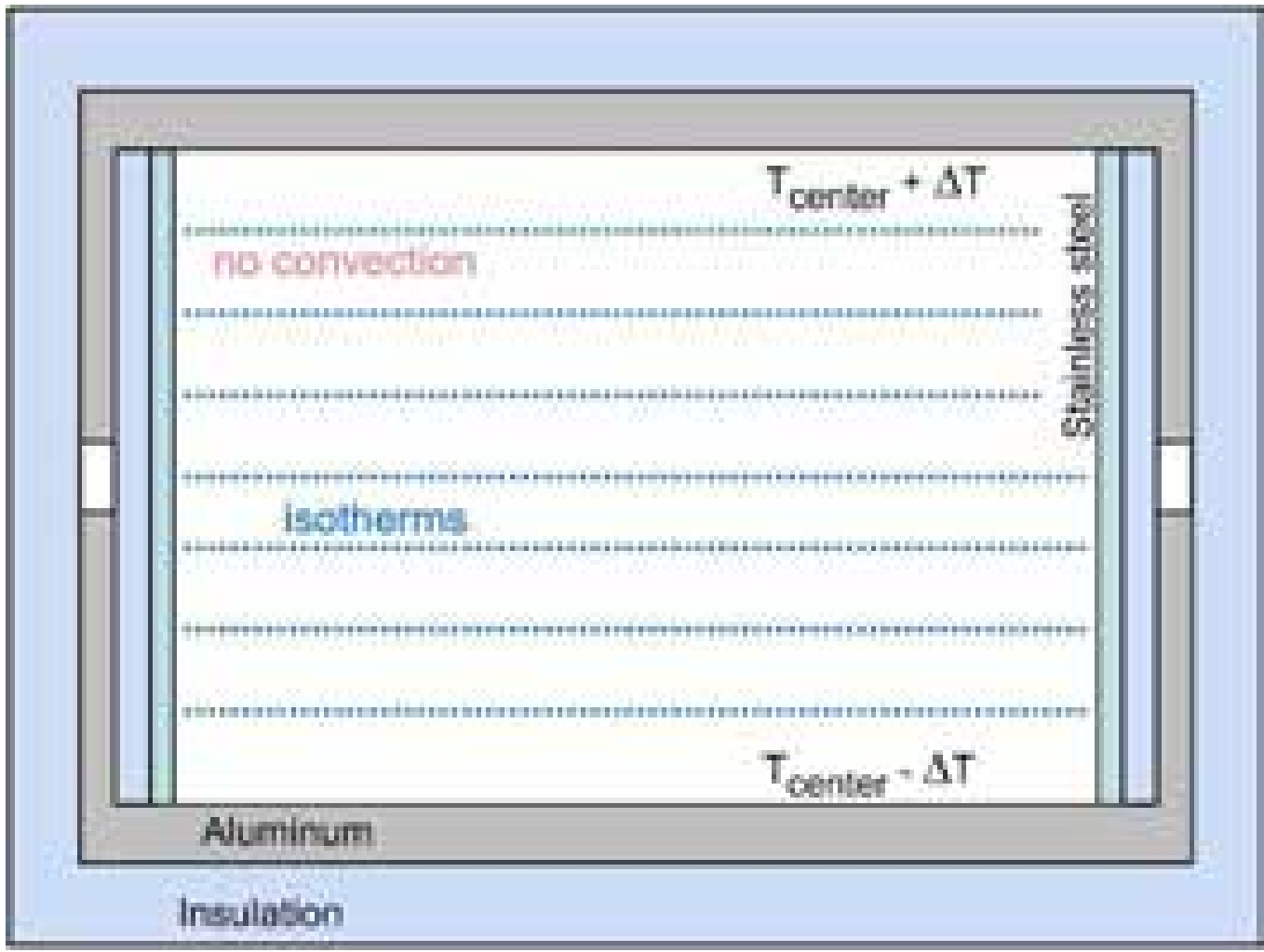

**Figure 6.33: A "linear-gradient" diffusion chamber, designed using (moderately conducting) stainless-steel wall to produce a linear temperature profile. With this simpler thermal structure, it becomes possible to model the supersaturation within the chamber with a quite high accuracy.**

With a stable, well-characterized temperature profile, it then becomes possible to solve the particle diffusion equation to obtain the supersaturation field $c(\vec{x})$ throughout the chamber. In a nutshell, the diffusion equation becomes Laplace's equation (because $\partial c/\partial t = 0$), with the boundary conditions $c = c_{sat}(T_{surf})$ on all surfaces. Solving this diffusion problem yields the supersaturation field $c(\vec{x})$ and thus the interior supersaturation

$$\sigma(\vec{x}) = \frac{c(\vec{x}) - c_{sat}\big(T(\vec{x})\big)}{c_{sat}\big(T(\vec{x})\big)} \qquad (6.7)$$

If we take the side walls in Figure 6.33 out to infinity, giving a simple parallel-plate diffusion chamber, then the diffusion equation yields a simple linear gradient for the particle density, giving

$$\sigma_{cent} \approx \frac{1}{2}\frac{1}{c_{sat}(T_{cent})}\frac{d^2 c_{sat}}{dT^2}(T_{cent}) \cdot (\Delta T)^2$$
$$\approx 0.0032(\Delta T)^2 \qquad (6.8)$$

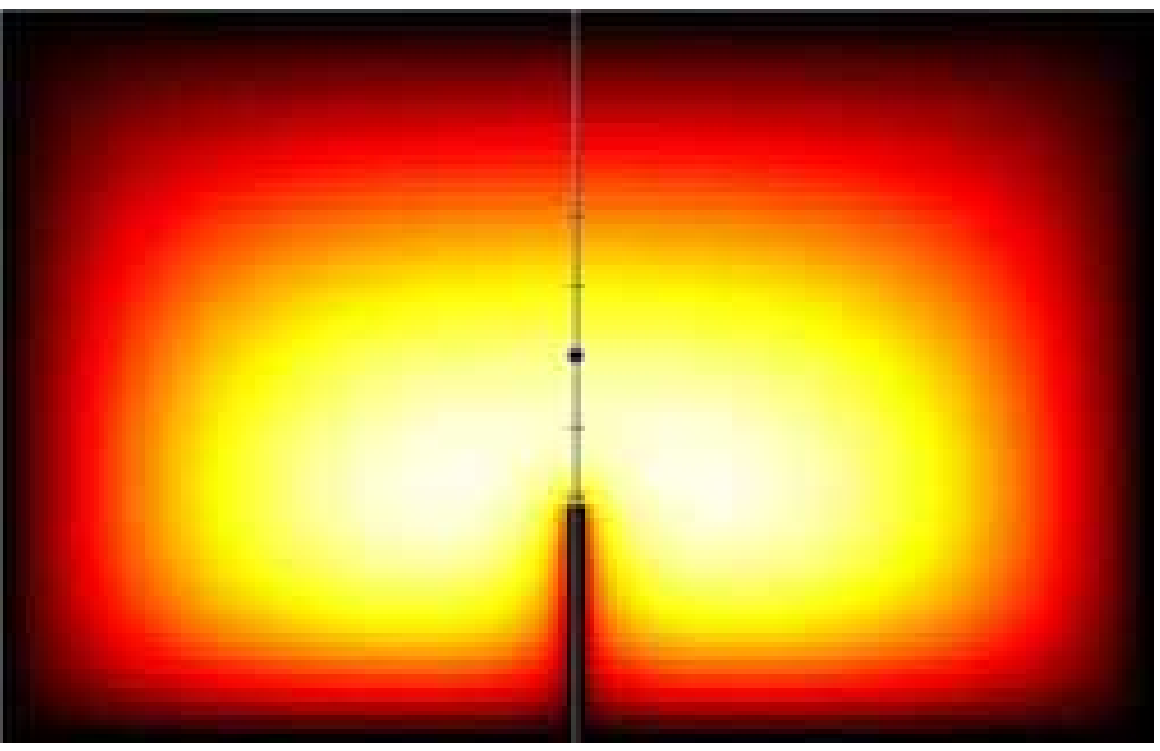

**Figure 6.34: A supersaturation model of a linear-gradient diffusion chamber like that illustrated in Figure 6.33. Note that $\sigma$ goes to zero (black) on all (frost-covered) surfaces, in this case including a central stem that supports the growing snow crystal. With such careful supersaturation modeling, it becomes possible to make quantitative comparisons between crystal growth measurements and theory [2016Lib]. (See also Chapter 8.)**

at the center point midway between the two plates, where $\Delta T = (T_{top} - T_{bottom})/2$ [2014Lib1]. Putting the side walls back as sketched in Figure 6.33 requires a numerical solution of Laplace's equation to determine $c(\vec{x})$. Figure 6.34 shows an example of a supersaturation model I created for one of my diffusion chambers [2016Lib]. The linear-gradient diffusion chamber is a good example of using careful modeling to determine supersaturations in a snow-crystal growth chamber.

## 6.5 Other Techniques

I end this chapter by briefly listing a few additional experimental techniques that have been developed for studying snow crystals or related subjects. Many of them have not yet been used in extensive laboratory investigations, but all have potential for opening up novel research directions.

### Microparticle Ion Trapping

Electrodynamic trapping of charged aerosol particles was first developed in the 1950s, and this levitation technique was applied to snow-crystal research by Brian Swanson and colleagues in 1999 [1999Swa]. Their apparatus is illustrated in Figure 6.35, using AC electric fields to provide trapping forces, together with DC electric fields that counter the downward force of gravity, yielding stably trapped ice crystals. Figure 6.36 illustrates the basic quadrupole geometry of the AC electric fields in a microparticle ion trap. Particle sizes were in the 10-100 μm range, carrying charges of typically 0.1-0.5 pC.

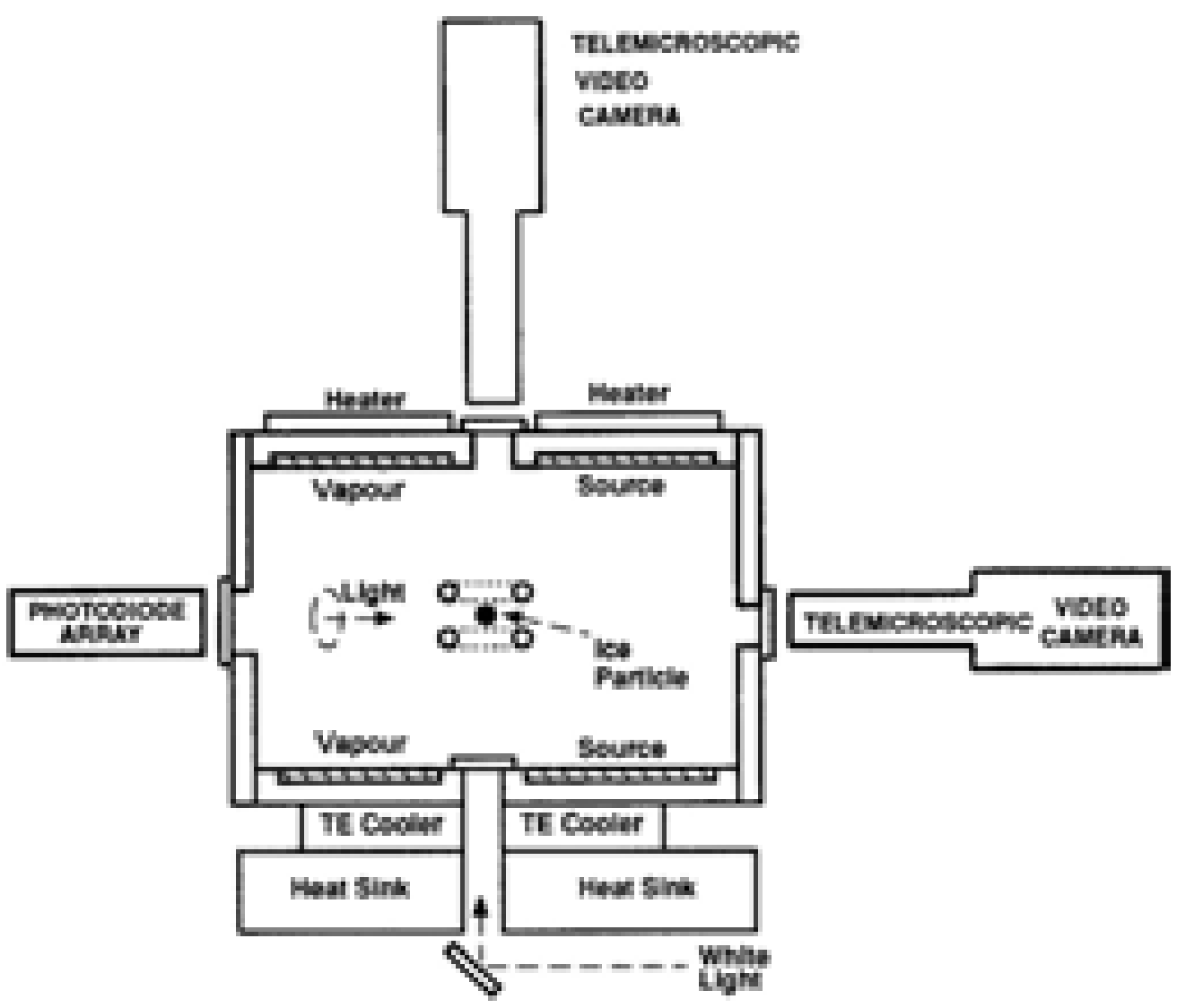

**Figure 6.35: An electrodynamic levitation chamber for examining the growth of ice crystals [1999Swa]. A pair of ring electrodes driven with a high-voltage 60-Hz sinusoidal signal traps a charged ice particle, while the surrounding diffusion chamber creates a supersaturated environment. The closely spaced pair of rings creates a quadrupole electric field around its central point, much like that from a single ring ion trap illustrated in Figure 6.36.**

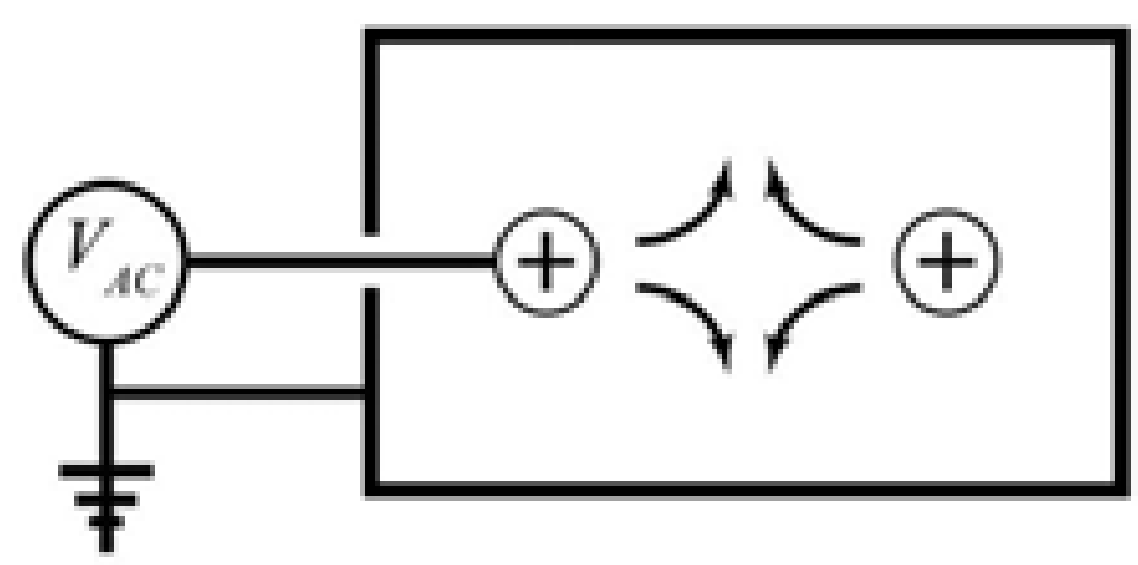

**Figure 6.36: A single ring electrode (seen here from the side, labeled ⊕) inside a conducting box creates a 3D quadrupolar electric field near its center, shown by the arrows. Applying a sinusoidal voltage to the ring creates electrodynamics forces that push charged particles toward the ring center. (Image from [2018Lib].)**

Subsequent development of this technique has led to different electrode geometries and other improvements [2003Bac, 2016Har]. For example, Figure 6.37 shows an electrode geometry that traps ice crystals at the center of a plane-parallel diffusion chamber, allowing precise control of temperature and supersaturation surrounding the particle. The physics underlying microparticle ion trapping is beyond the scope of this book, but a summary can be found in [2018Lib] together with techniques for building simple ion traps for laboratory demonstrations.

**Figure 6.37: (Right) In this electrodynamic trap [2016Har], a trapping quadrupole electric field oscillating at 60 Hz is delivered via a set of "button" electrodes, while a constant vertical electric field balances gravity. The parallel-plate diffusion-chamber geometry allows for precise modeling of the supersaturation at the position of a trapped ice crystal.**

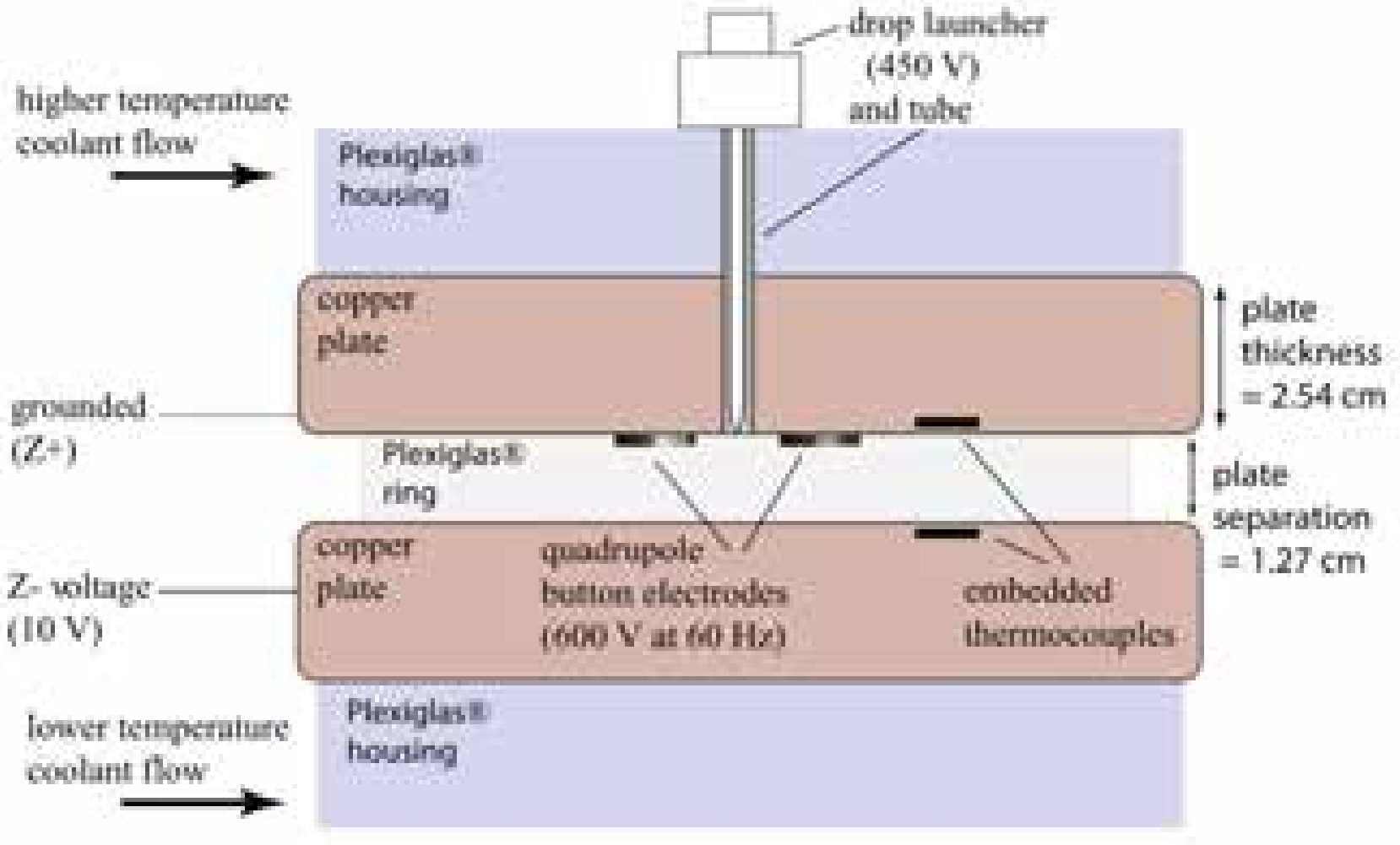

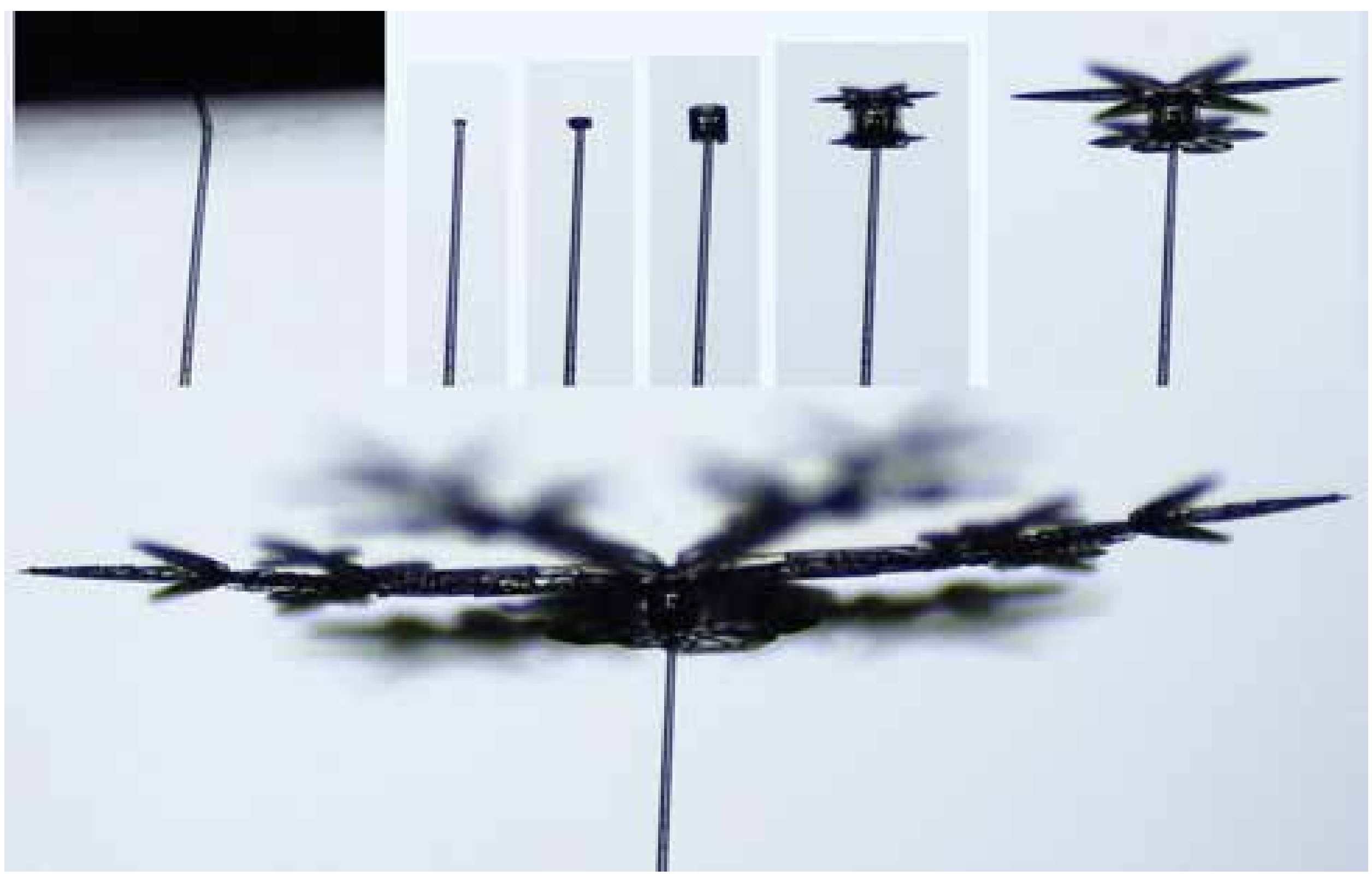

**Figure 6.38: Growing an oriented snow crystal on a 20-μm-diameter capillary tube. In the upper left image, the basal surface of an ice crystal touches the tip of a water-filled capillary, freezing the water with the ice c-axis (nearly) collinear with the capillary axis. With the ice plate removed, the other images show subsequent development of the crystal in a diffusion chamber. The stellar snow crystal in the lower image measures 3 mm from tip to tip, but this crystal eventually grew to 12 mm from tip to tip. (Photos by the author.)**

Electrodynamic trapping has a clear advantage that it allows observations of single, isolated, levitated ice prisms growing under well controlled environmental conditions, which is a somewhat ideal experimental set-up. Moreover, particle mass changes can be precisely determined by monitoring the levitating electric fields. Some disadvantages include difficulties with loading seed crystals, and the general slow throughput that comes with examining individual particles. This can be contrasted with the dual-temperature chamber discussed in Chapter 7, which must contend with substrate interactions, and free-fall chambers that present a generally less controlled environment. As usual, the apparatus choice in any experiment will reflect the specific scientific goal being targeted.

## Capillary Snow Crystals

Filamentary support of snow crystals has been popular ever since Nakaya's first experiments with rabbit hair (see Chapter 1), and the current state-of-the-art in this direction has been the development of thin capillary tubes for this purpose [1996Nel]. Heating and drawing capillaries with tip diameters down to 10 μm has been a staple of laboratory art for many decades, and suitable commercial micropipettes have recently become available also. Creating an isolated ice crystal at the tip can be accomplished by freezing water in the capillary from below [1996Nel], or by using ice transfer from the top as shown in Figure 6.38.

Thin capillaries offer simple and robust snow-crystal support that introduces minimal interference of subsequent growth, making it a useful technique for creating especially large specimens. The technique is rather slow and laborious however, and substantial substrate interactions may still be problematic, making the electric-needle method (see Chapter 8) generally better suited for quantitative studies.

## Gas Mixing & Expansion

Another way to create supersaturated air is to thoroughly mix two air flows containing saturated air at different temperatures. The resulting supersaturation is easy to calculate, but ensuring proper mixing on short timescales is a nontrivial experimental problem. I have tried to use two-flow mixing to create well-defined growth conditions, but never had much luck getting sufficiently stable results. Nevertheless, the technique may find use in future snow-crystal experiments.

Controlled adiabatic expansion is another route to create supersaturated air, and this method goes all the way back to the early cloud-chamber studies of Wilson and others. The basic idea is to begin with air saturated with water vapor and then adiabatically expand the volume, decreasing the temperature and increasing the supersaturation in the process. Calculating the final state is a standard problem in the statistical mechanics of ideal gases, but creating a well-controlled expansion with minimal heat flow to the walls is not an easy task. To my knowledge, there have been no significant studies of snow crystal growth using this technique, although its uncontrolled variant yields the expansion nucleator described above. Here again, however, adiabatic expansion may find gainful employment in future snow-crystal investigations.

Huang and Bartell used an extreme form of gas expansion to examine the nucleation of water clusters of ~5000 water molecules, observing the preferred formation of ice cubic ice Ic in the process [1995Hua]. One can imagine using similar experimental methods to explore the nucleation and subsequent growth of "nano" snow crystals from water vapor, although the technical challenges in such studies would be substantial.

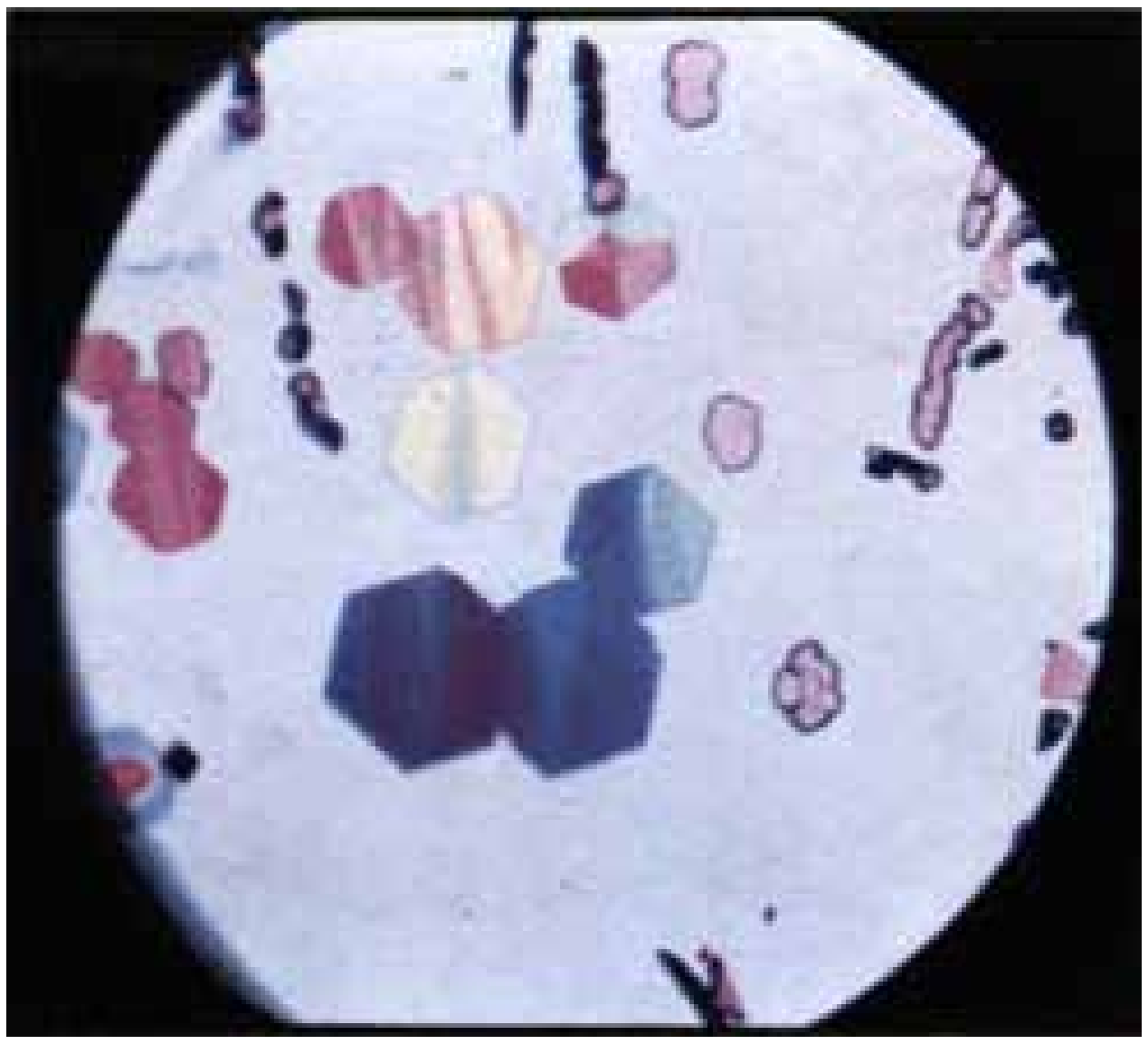

**Figure 6.39: Thin ice-crystal plates grow epitaxially on a cleaved covellite crystal. Note the alignment of the different ice prism facets, indicating the orientation of the underlying substrate crystal. Colors arise from the interference of reflections off the ice/air and ice/CuS surfaces, with the overall color depending on the thicknesses of the ice. (Image from [1959Bry].)**

## Epitaxial Growth

If two crystals exhibit similar lattice structures, then it is often possible to grow one crystal on the other *epitaxially*, meaning that the two lattices match one another at the interface. In the case of ice Ih, crystals of silver iodide, lead iodide, and the mineral covellite (CuS) have similar structures in their basal planes, and ice has been observed to grow epitaxially on all three [1959Bry, 1965Kob, 1984Cho], as illustrated in Figure 6.39.

Although this is an intriguing technique for growing oriented ice crystals, the initial nucleation has been difficult to control, usually resulting in numerous closely spaced ice crystals, with Figure 3.39 being one example. Nevertheless, it may be possible to pattern an otherwise hydrophobic substrate with isolated epitaxial growth sites [1999Aiz], or perhaps an epitaxial substrate could be covered with a hydrophobic coating except in an array of tiny holes. In either case, one can imagine nucleating a large array of microscopic seed crystals and performing simultaneous experiments in a highly controlled manner. This could represent a sizable technological step forward from wafting ice-crystal-laden air from a seed-crystal generator over a waiting substrate.

## Negative Snow Crystals

Pulling a vacuum through a thin capillary tube embedded in a block of single-crystal ice can yield "negative" snow crystals like the one shown in Figure 6.40. Faceting arises from the strong nucleation barrier in the sublimation kinetics (see Chapter 3, Figure 3.40), and this effect could potentially be put to use in experiments aimed at understanding the attachment/sublimation kinetics. Here again, one can imagine an array of microscopic holes in a substrate contacting a block of ice, allowing one to perform hundreds of simultaneous sublimation experiments. Nothing of this ilk has been attempted to date, but it presents another avenue that might be explored someday.

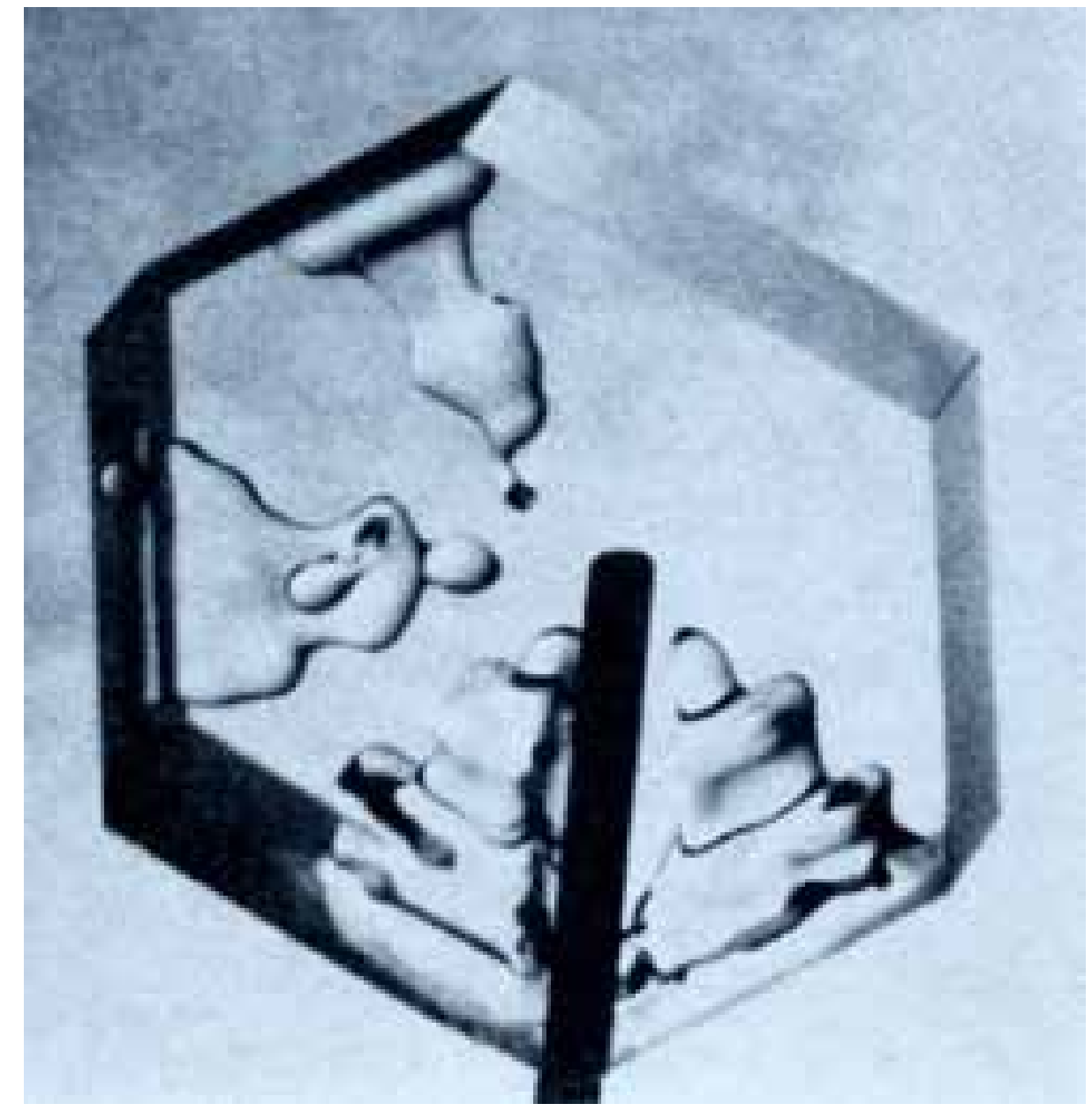

**Figure 6.40: A "negative" snow crystal growing at -14 C in a block of single-crystal ice, as a vacuum pump removes water vapor via a 0.45-mm-diameter capillary tube. The morphology of the void is determined from thermal diffusion effects together with a strong nucleation barrier in the sublimation kinetics. As with normal snow crystal growth, surface energy effects here are negligible compared to anisotropic sublimation kinetics. (Image from [1965Kni].)**

## Targeted Experiments

Although I have focused much of this chapter on some of my favorite proven snow-crystal growth technologies, the broader field remains largely unstudied. There is much parameter space left to survey, and novel experimental techniques could lead to important breakthroughs now unimagined. But while general-purpose experiments covering large swathes of temperature and supersaturation have been instrumental in the past, the future will likely be dominated by targeted experiments aimed at investigating specific physical processes in detail. This appears to be the case especially with the attachment kinetics, where the underlying molecular dynamics seems to depend on temperature, supersaturation, background gas, and even the width of faceted surfaces, as I discussed in Chapter 4.

To this end, I delve more deeply into some specific experimental techniques in Chapter 7 (measuring the growth rates of small ice prisms for investigating the attachment kinetics), Chapter 8 (growing larger specimens on electric ice needles for direct comparisons with computational snow crystals) and Chapter 9 (creating designer snow crystals mainly as an artistic pursuit). Where all this leads is anyone's guess, but clearly there are many possible routes leading to future scientific and artistic progress, using creative hardware solutions that further explore the dynamics of snow crystal formation.

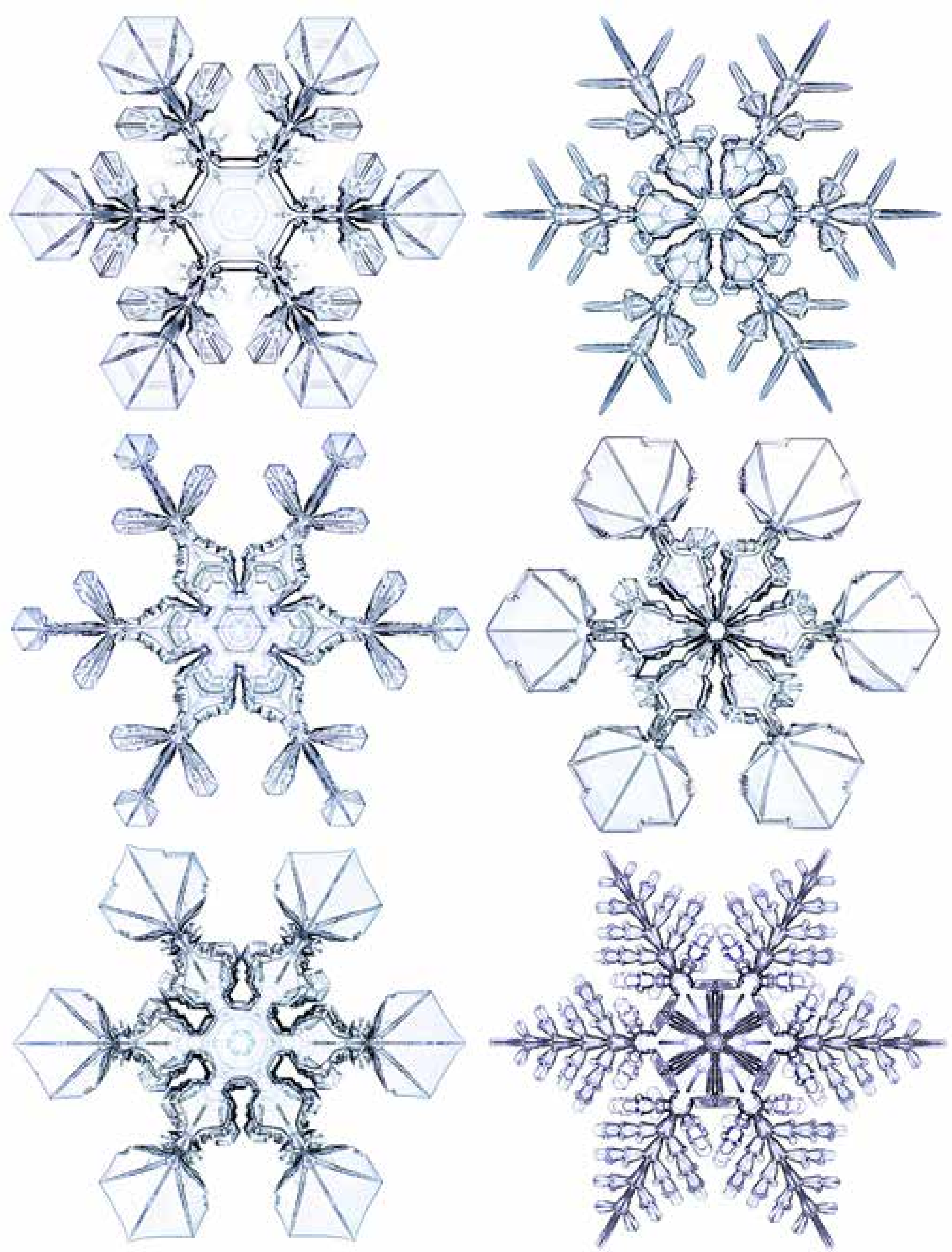

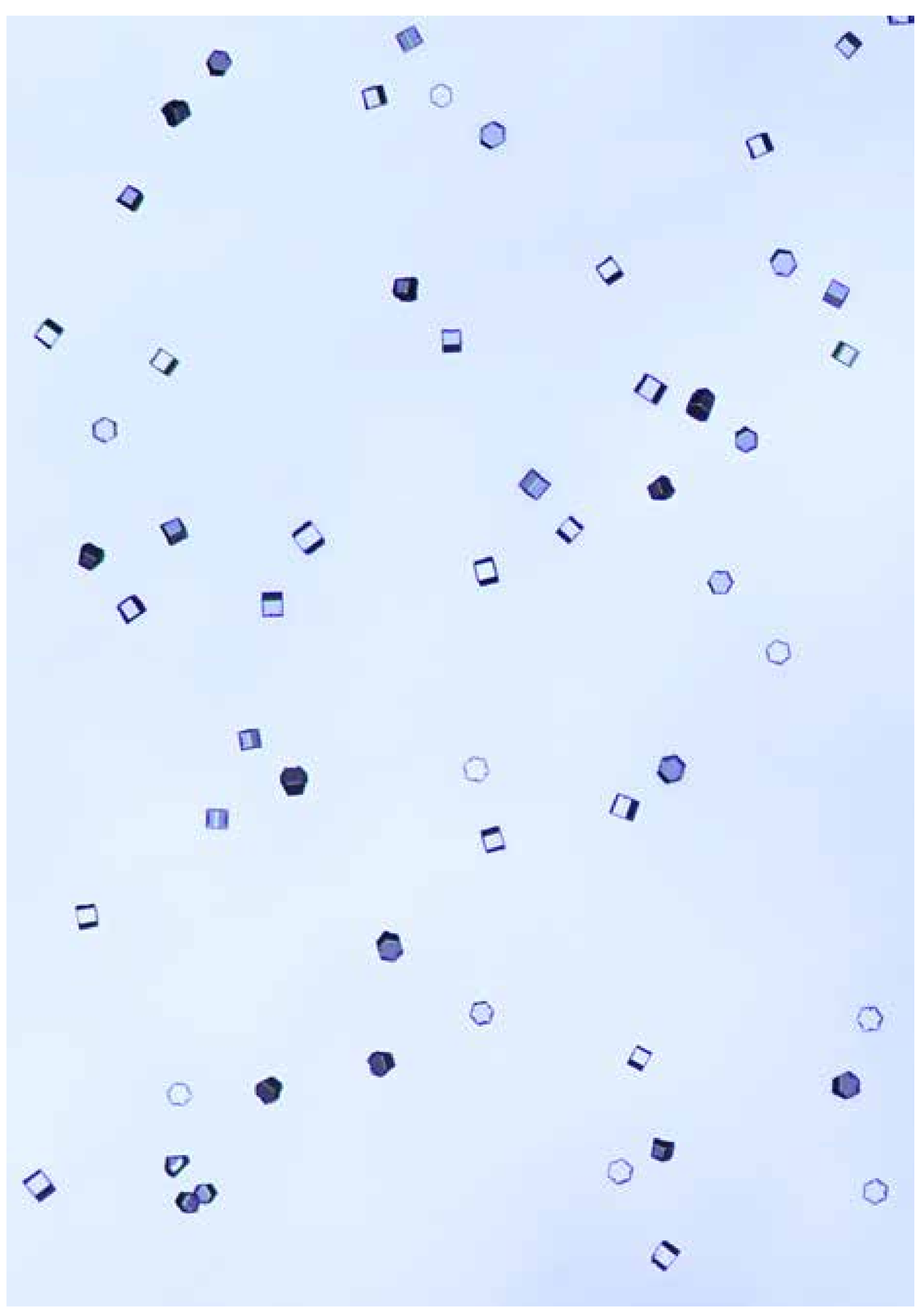

# Chapter 7

# Simple Ice Prisms

*To see a World in a Grain of Sand,*
*And a Heaven in a Wild Flower.*
*Hold Infinity in the palm of your hand,*
*And Eternity in an hour.*
– William Blake
*Auguries of Innocence, ~1803*

As I have stressed throughout this book, the morphology of a growing snow crystal is greatly influenced by the molecular attachment kinetics that takes place on its basal and prism facets. The formation of faceted surfaces is itself a direct result of anisotropies in the attachment kinetics (see Chapter 4), and the well-known morphological transitions between plate-like and columnar growth reflect corresponding changes in anisotropy with temperature (see Chapter 3). If we wish to create computational snow crystals (Chapter 5) that faithfully reproduce laboratory and natural specimens, it is essential that we begin with an accurate model of the anisotropic molecular attachment kinetics.

**Facing page: A collection of simple ice prisms growing in near vacuum at -7 C on a sapphire substrate. The size of typical crystal in this image is about 30 μm.**

As we saw in Chapter 4, however, our present understanding of the attachment kinetics is remarkably primitive. Experimental measurements indicate that the attachment coefficient depends strongly on many factors, including temperature, supersaturation, background gas pressure, facet width, and possibly surface chemistry. The prevailing theories of crystal growth cannot explain all these observations, suggesting that there are numerous shortcomings in our basic molecular picture of how water vapor molecules attach to ice surfaces.

Past experience suggests that the best way to make progress in this area is with ever-improving experimental investigations, especially those involving precise measurements of the growth of tiny ice prisms under carefully controlled conditions. Essentially everything we currently know about the attachment kinetics has been obtained from observations of these simplest snow crystals, as studying small prisms reduces the complicating effects from particle diffusion and complex growth morphologies.

This chapter focuses on how one goes about measuring the growth of simple ice prisms as a means to better understand the attachment kinetics on faceted surfaces. I will examine some experimental strategies that have worked in the past, analyze techniques for extracting attachment coefficients from growth data, and take a careful look at several systematic errors that can easily lead to erroneous conclusions if one is not careful. My overarching goal is to describe the current state-of-the-art for making direct measurements of the ice attachment kinetics, and to examine how one might create improved experiments for future investigations.

The principal methodology in precision ice-growth measurements is straightforward enough – measure the growth velocity $v_n$ (normal to the surface) of a faceted surface, determine the water-vapor supersaturation $\sigma_{surf}$ just above the growing surface, and use that information to extract the attachment coefficient using the usual growth equation $v_n = \alpha v_{kin} \sigma_{surf}$. Repeat this procedure to map out the attachment coefficient $\alpha(\sigma_{surf}, T, P)$ for both the basal and prism facets as a function of growth temperature $T$, near-surface supersaturation $\sigma_{surf}$, gas pressure $P$, and perhaps other factors.

With a sufficient quantity of accurate empirical data, hopefully one can develop a sensible theoretical picture of the underlying physical processes that determine the attachment kinetics. Progress has been made toward this final theoretical goal, but molecular dynamics is a nontrivial subject, and only the first steps have been made in what might be quite a long journey.

The reason for measuring small crystals can be seen from the spherical-growth analysis presented in Chapter 3. If the crystal is so large that $\alpha_{diff} \ll \alpha$, then the growth rate will be limited primarily by particle diffusion, making it difficult to extract useful information about $\alpha$ from growth measurements. In this case, when the growth velocity is strongly diffusion-limited, the attachment kinetics may still play an important role in guiding the overall crystal aspect ratio and morphology; but disentangling kinetics from diffusion becomes extremely tricky in practice, with many opportunities for an erroneous analysis. Working with smaller crystals increases $\alpha_{diff}$ and thus reduces the relative importance of diffusion effects. For this reason, the highest-quality $\alpha$ measurements are usually obtained from observations of small, simple, faceted ice prisms.

Measuring ice growth rates in near-vacuum conditions also greatly reduces the effects of particle diffusion, again making it easier to extract the attachment kinetics from the data. As described in Chapter 3, $\alpha_{diff}$ is inversely proportional to the background air pressure, so $\sigma_{surf}$ will be closer to the experimentally determined $\sigma_\infty$ far from a growing crystal at low background pressures. Low-pressure data have been instrumental in creating the comprehensive model of the attachment kinetics presented in Chapter 4, but observations also suggest that $\alpha$ depends on air pressure, at least at temperatures near -5 C. So making additional progress requires that we obtain quantitative data as a function of $\sigma_{surf}$, $T$, and $P$, which is a challenging experimental problem.

Prior experience tells us that obtaining accurate ice-growth measurements, and inferring the attachment kinetics from the data, can be influenced by a host of sometimes-subtle systematic errors. Particle-diffusion, substrate interactions, crystal dislocations, and possible chemical effects must all be carefully considered. Much of this chapter, therefore, focuses on identifying and reducing systematic measurement and analysis errors, so future experiments will be better able to make progress in this important aspect of snow-crystal formation.

## 7.1 Analysis of a Basic Ice-Growth Experiment

When researchers began exploring the ice attachment kinetics in earnest [1972Lam, 1982Kur, 1982Bec2, 1983Bec, 1984Kur1, 1989Sei], it soon became apparent that growth measurements at low background gas pressure were the best way to isolate surface kinetics from bulk diffusion processes. Unfortunately, the results presented in these papers showed a great deal of variation, and there was no easy way to make sense of the discrepancies between different experiments. It was later found that there were a number of important systematic errors that could affect ice-growth data, and essentially all the early experimental papers suffered from these errors to some degree. I surveyed many of the early results in [2004Lib], so I will not repeat that discussion here, in part because the data are not sufficiently reliable to inform our discussion of the attachment kinetics.

Instead, I will begin by performing a careful analysis of a simple ice growth experiment. Doing so allows us to identify most of the important systematic errors in detail, and hopefully this discussion can serve to guide the design of future experiments. The basic layout of the experiment is shown in Figure 7.1, showing a single test crystal on a substrate at temperature $T_{substrate}$, with a nearby ice reservoir at temperature $T_{reservoir}$. The crystal will grow as long as $T_{reservoir} > T_{substrate}$, and the goal of the experiment is to determine the attachment coefficient $\alpha(\sigma_{surf})$.

### Crystal Crowding Effects

If we remove the test crystal for a moment, then it is straightforward to determine the temperature $T(\vec{x})$ and water-vapor number density $c(\vec{x})$ as a function of position $\vec{x}$ throughout the chamber. Solving the heat-diffusion equation yields a simple linear temperature gradient profile

$$T(\vec{x}) = T_{substrate} + \Delta T \cdot (z/L) \qquad (7.1)$$

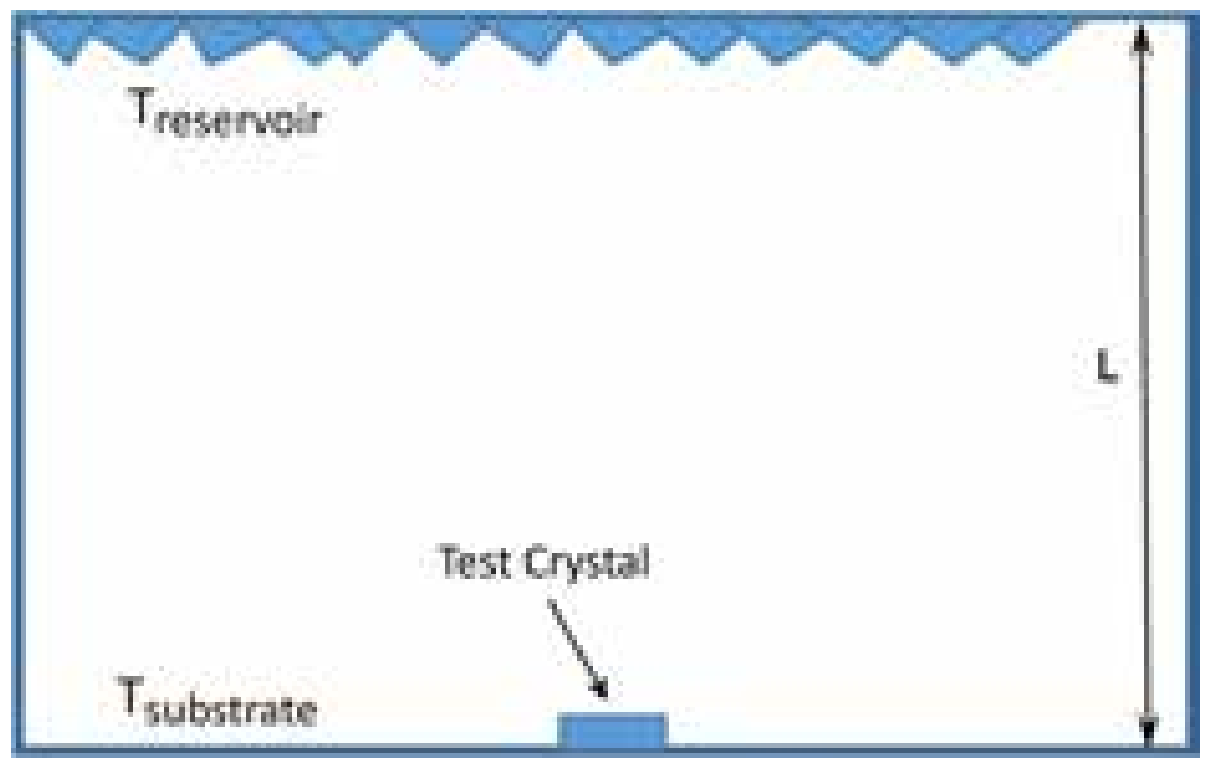

**Figure 7.1: A schematic layout of a basic ice-growth experiment. The single test crystal being measured rests on a isothermal substrate at temperature $T_{substrate}$, and no other ice crystals are present on the substrate. Water vapor is supplied by a large collection of frost crystals on the top of the chamber at temperature $T_{reservoir}$. The walls are free of ice crystals and have a relatively low thermal conductivity.**

where $z$ is vertical distance above the substrate, $L$ is the chamber inner height, and $\Delta T = T_{reservoir} - T_{substrate}$. This simple solution is valid in the presence of nearby walls, provided their heat conductivity is relatively low.

Likewise, solving the particle diffusion equation (see Chapter 3) yields the simple solution $c(\vec{x}) = c_{sat}(T_{reservoir})$ throughout the chamber. Putting these two solutions together yields the supersaturation

$$\sigma_1 = \frac{c_{sat}(T_{reservoir}) - c_{sat}(T_{substrate})}{c_{sat}(T_{substrate})} \qquad (7.2)$$

at the surface of the substrate, and $\sigma = 0$ at the surface of the reservoir. Note that this solution only applies when there are no ice crystals present on the substrate or the walls of the chamber.

If we now place a single test crystal on the substrate, as illustrated in Figure 7.1, and the size of the crystal is much smaller than $L$, then it will begin growing as if surrounded by a far-away boundary condition $\sigma_\infty = \sigma_1$. Unfortunately, it is not always easy to drop a single, isolated test crystal at the bottom of a growth chamber. More often, several crystals may end up on the substrate, as illustrated in Figure 7.2. In this case, we need to consider how the entire set of crystals affects our calculation of the supersaturation.

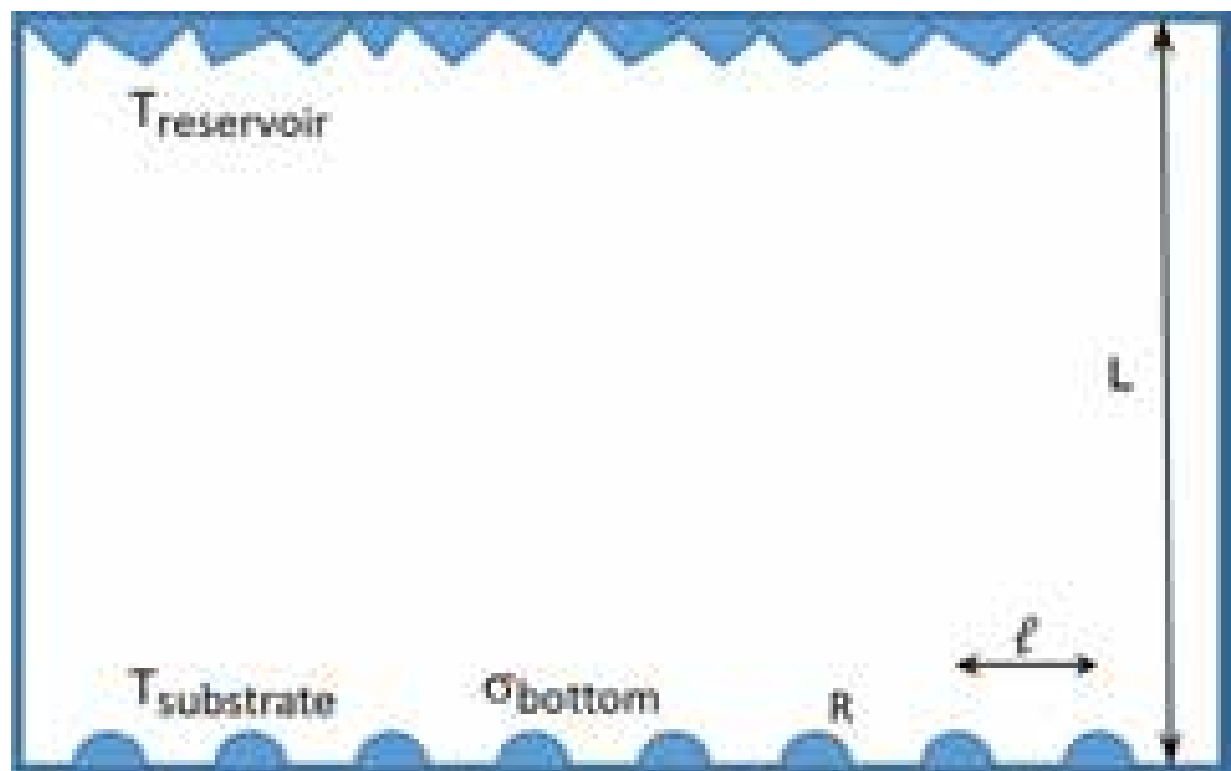

**Figure 7.2: If several test crystals are present at the bottom of the growth chamber, they collectively reduce the supersaturation $\sigma_{bottom}$ near the bottom of the chamber. As described in the text, even a small number of ice crystals, covering just a tiny fraction of the substrate area, can have a sizable effect on $\sigma_{bottom}$.**

To aid with the analysis, I have drawn Figure 7.2 with a simple square array of ice crystals on the substrate, all hemispherical in shape with a uniform radius $R$, each spaced a distance $\ell$ from its nearest neighbors. The presence of these crystals will not affect the temperature solution in Equation 7.1, but they will alter the water-vapor field $c(\vec{x})$, so our first task is to calculate the supersaturation $\sigma_{bottom}$ at the bottom of the chamber. We will assume $R \ll \ell \ll L$.

Using the spherical solution for diffusion-limited growth (see Chapter 3 or Appendix B), the substrate crystals will all grow at a rate

$$v = \varepsilon \alpha_{diff} v_{kin} \sigma_{bottom} \tag{7.3}$$

where

$$\varepsilon = \frac{\alpha}{\alpha + \alpha_{diff}} \tag{7.4}$$

and $\alpha$ is the usual attachment coefficient. This implies an overall downward flux of water molecules equal to

$$F = \frac{2\pi R D c_{sat} \varepsilon \sigma_{bottom}}{\ell^2} \tag{7.5}$$

where $c_{sat} = c_{sat}(T_{substrate})$. Equating this to the downward diffusion flux $F = D\nabla c$ then gives our result

$$\begin{aligned} \sigma_{bottom} &\approx \left(1 + \frac{2\varepsilon\beta L}{R}\right)^{-1} \sigma_1 \\ &\approx \left(1 + \frac{2\pi\varepsilon R L}{\ell^2}\right)^{-1} \sigma_1 \end{aligned} \tag{7.6}$$

where

$$\beta = \frac{\pi R^2}{\ell^2} \tag{7.7}$$

is the filling factor, equal to the fraction of the substrate surface area that is covered with ice crystals. Note that Equation 7.6 does not depend on $D$, so is independent of the background gas pressure in the chamber. This result will hold as long as the background gas pressure is greater than the water vapor pressure, and the mean free path of gas molecules is substantially less than $L$.

If we put in some typical numbers, with a filling factor of $\beta = 0.01$, a chamber height $L = 1$ cm, a crystal radius $R = 10$ μm (giving $\ell \approx 180$ μm), and simple diffusion-limited growth $\varepsilon = 1$, we find $\sigma_{bottom} \approx 0.05\sigma_1$. Looking down at such a collection of small, widely spaced ice crystals, covering just a small fraction of the substrate area, one might naively think that $\sigma_{bottom} \approx \sigma_1$ would be reasonably accurate. But this naïve assumption would overestimate $\sigma_{bottom}$ by a factor of twenty!

Many ice-growth measurements described in the literature [for example: 1972Lam, 1982Bec1, 1989Sei, 2014Asa] have used growth chambers with geometries roughly like the one illustrated in Figure 7.3, which turns out to be a somewhat poor choice. This is topologically similar to the geometry shown in Figure 7.1, but the effective $L$ is quite large because the ice reservoir is located far from the growing crystals. The authors in these papers generally assume $\sigma_{bottom} \approx \sigma_1$ without presenting a rigorous diffusion analysis to verify this assumption. As I described in [2004Lib], the neglect of this chamber-diffusion effect might easily explain many of the discrepancies between reported growth data. It is possible to overestimate $\sigma_{bottom}$ by even a factor of 100 if one is not careful [2015Lib], and it has become my impression that people have not always been careful. For this reason, I have found that most of the earlier growth measurements are generally unreliable for determining attachment coefficients [2004Lib].

Going back to Figure 7.2, a related analysis shows that one can use the substrate growth observations directly to estimate the systematic error in determining $\sigma_{bottom}$, provided one can view the extraneous crystals. The basic idea is to watch the crystals growing on the substrate and use the measured crystal sizes over time to estimate $\dot{V}$, equal to the change in ice volume per unit time and per unit area on the substrate. Because this ice is supplied by a downward flux of water vapor from the reservoir, one can show

$$\sigma_{bottom} \approx \sigma_1 - \frac{c_{ice}}{c_{sat}} \frac{H}{D} \dot{V} \qquad (7.8)$$

Importantly, one can estimate an uncertainty in $\dot{V}$ from the observations, and Equation 7.8 then turns this into an uncertainty in $\sigma_{bottom}$.

The moral of this story is that it is difficult to produce precision measurements of $\alpha$ without limiting the number of crystals on the substrate to exceedingly low values, preferably just a single, isolated test crystal. Achieving $\sigma_{bottom} \approx \sigma_1$ requires

$$\beta \ll \frac{R}{2\varepsilon L} \qquad (7.9)$$

or, equivalently, a spacing between test crystals of

$$\ell \gg \sqrt{2\pi\varepsilon RL} \qquad (7.10)$$

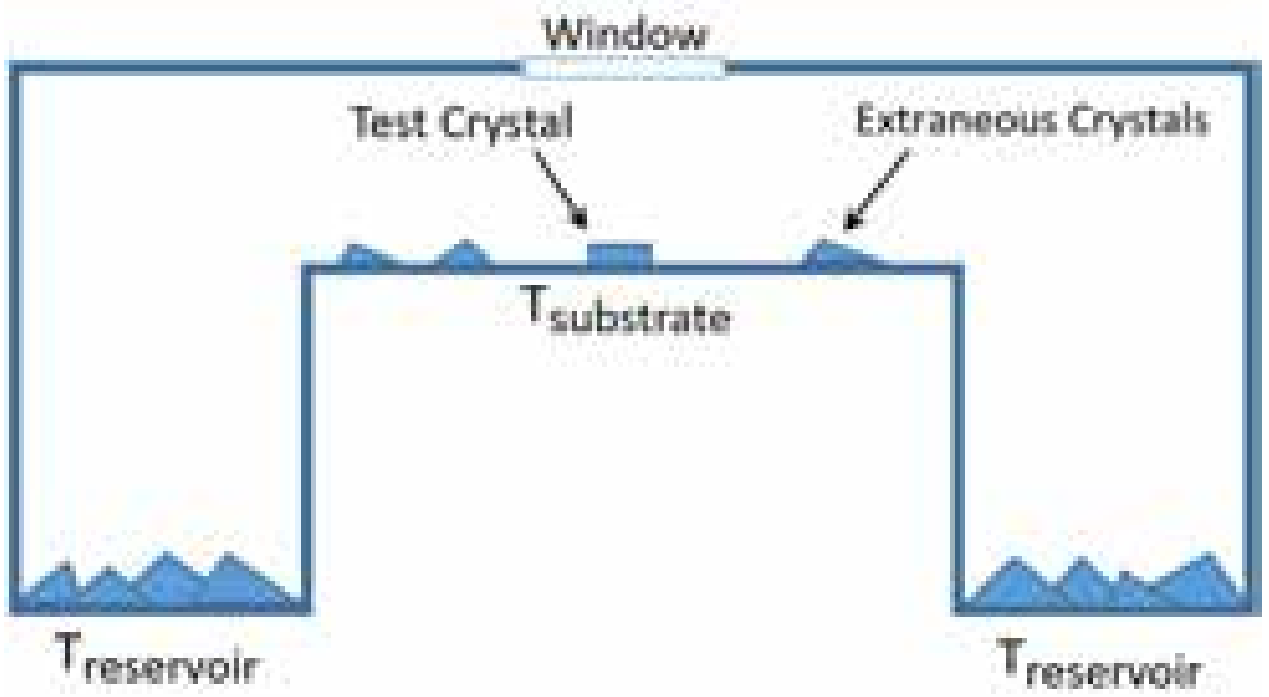

**Figure 7.3: An example of a poor chamber design for making precision ice-growth measurements. The presence of extraneous ice crystals on the substrate, coupled with the ice reservoir being far from the test crystal (making $L$ large in Equation 7.6), make it nearly impossible to determine the supersaturation near the test crystal with reasonable accuracy.**

and that turns out to be a pretty high bar to clear. Reducing the background gas pressure to the point that the chamber contains essentially only water vapor is one way to get around this diffusion issue. But achieving a pure water-vapor environment is itself a nontrivial task, as pulling a hard vacuum tends to sublimate away much of the ice in the chamber. In my studies to date, I have found that the best way to avoid these crowding problems is to produce what amounts to a single, isolated test crystal on the substrate.

## The Monopole Approximation

Proceeding with our analysis, let us now assume that we have managed to place a single, isolated test crystal on the substrate, thus achieving the ideal experimental geometry illustrated in Figure 7.1. Assuming that the test-crystal size is much less than $L$, this means that the crystal is surrounded by a supersaturation given by in Equation 7.2. Put another way, the test crystal will grow as if it were immersed in a uniform supersaturation environment with $\sigma_\infty \approx \sigma_1$. Starting with this far-away boundary condition, we must then determine $\sigma_{surf}$, the supersaturation at the crystal surface, in order to extract $\alpha$ using the growth equation $v_n = \alpha v_{kin} \sigma_{surf}$.

For a faceted test crystal, the near-surface supersaturation will vary with position on the surface, so numerical modeling of the diffusion equation is likely the only way to determine $\sigma_{surf}(\vec{x})$ over the crystal surface with high accuracy. But we can obtain a useful first approximation by using the analytic solution for the growth of a spherical ice crystal (see Chapter 3 or Appendix B).

For the spherical case, the analytic solution gives us both the crystal growth rate and the supersaturation field $\sigma(r)$ at all radii, and we can write the latter as

$$\sigma(r) = \sigma_\infty - \frac{dV/dt}{4\pi r X_0 v_{kin}} \qquad (7.11)$$

where $dV/dt$ is the rate of change of the volume of the growing crystal. For a hemispherical crystal on a substrate, this changes by a factor of two, giving

$$\begin{aligned} \sigma(r) &= \sigma_\infty - \frac{\dot{V}_s}{2\pi r X_0 v_{kin}} \\ &= \sigma_\infty - \frac{\dot{V}_s c_{ice}}{2\pi r D c_{sat}} \end{aligned} \qquad (7.12)$$

where $\dot{V}_s$ is the time derivative of the volume of the substrate test crystal.

When $r$ is much larger than the size of the crystal, the true solution will look much like this spherical solution, as Equation 7.12 is essentially a monopole approximation to the real solution. For a nearly isometric test crystal, the monopole solution is likely a good representation of the real solution nearly all the way in to the crystal surface. Thus, Equation 7.12 provides a useful estimate of the surface supersaturation [2012Lib]

$$\sigma_{surf} \approx \sigma_\infty - \frac{\dot{V}_s c_{ice}}{2\pi R_{eff} D c_{sat}} \qquad (7.13)$$

where $R_{eff}$ an effective radius derived from the size and morphology of the test crystal. This expression also indicates to what degree the growth is limited by diffusion. Of course, Equation 7.13 is less accurate for highly anisometric crystals, but it can be used as a reasonable first approximation.

## 7.2 Ice Growth at Low Background Gas Pressure

The most precise set of ice-growth measurements to date were made by Libbrecht and Rickerby [2013Lib], and these data played a central role in developing the comprehensive model of the attachment kinetics I described in Chapter 4. In this section I examine these measurements in detail, focusing on the experimental apparatus, ice-crystal handling, measurement methods, and data analysis. Special attention is given to the identification and minimization of potential systematic errors. My motivation for this section in two-fold: 1) I feel that this experiment demonstrated a substantial improvement over previous efforts, and so deserves some attention and scrutiny, and 2) I believe that this discussion presents a valuable case study when considering future precision ice-growth experiments.

### The VIG Experiment

Figure 7.4 illustrates the Vacuum Ice Growth (VIG) apparatus used in [2013Lib], which was designed to approximate the ideal growth-chamber geometry shown in Figure 7.1, at least to the degree that this is possible in an actual laboratory experiment. The easiest way to describe this device is simply to walk through the steps used during its operation. This serves to explain the apparatus details seen in the diagram, and it stimulates a discussion of various design choices and their consequences.

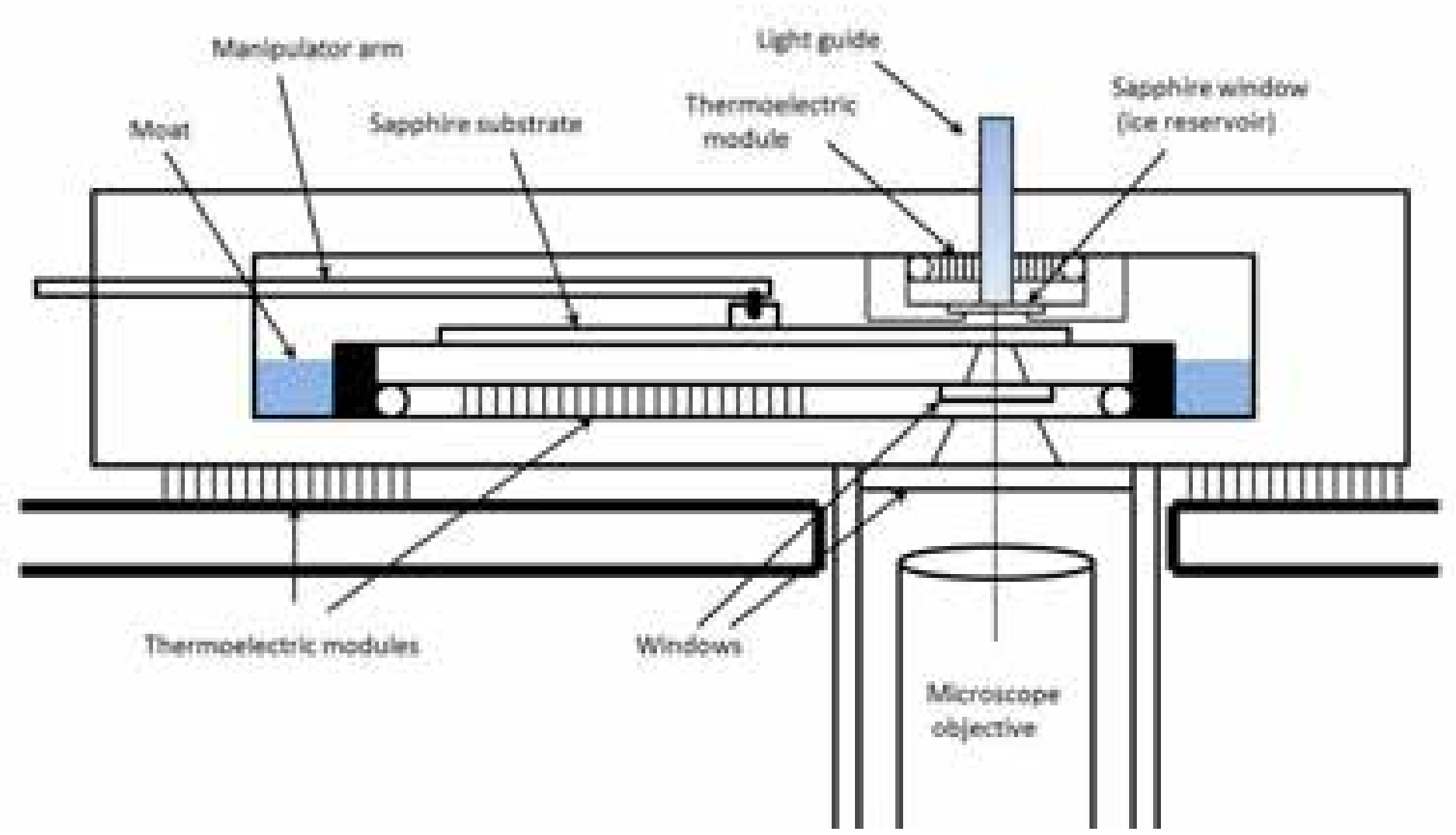

**Figure 7.4: (Left) A schematic diagram of the dual-temperature chamber used in [2013Lib] to measure the growth of small ice crystal in near vacuum. Figure 7.5 shows a closer view of the ice-growth region.**

**Vacuum Chamber.** The large outer box in Figure 7.4 depicts a short cylindrical vacuum chamber with a diameter of 7.5 cm, machined out of aluminum. The chamber is black anodized to seal the aluminum surfaces, and it includes a lid that bolts to the lower box, sealed using a silicone O-ring. The high thermal conductivity of aluminum keeps the chamber walls at a uniform temperature, and the silicone O-ring retains its pliability (and vacuum seal) at low temperatures. A digital temperature controller maintains a constant chamber temperature using a thermistor sensor embedded into the aluminum using thermally conducting epoxy together with thermoelectric modules on the bottom of the box. The chamber is opened, cleaned, and baked between each run to minimize chemical vapor contaminants. The substrate is also removed and thoroughly cleaned between runs to remove dirt and chemical residues. The substrate is given a final rinse with deionized water before being installed in the chamber, again to reduce remaining solvent residues.

**The Sub-Chamber.** Figure 7.5 shows a close-up of the sub-chamber where the test crystal grows. A key feature in this region is the spacing $L = 1$ mm between the ice reservoir and the test crystal. This spacing is as small as practical to minimize the difference between $\sigma_{bottom}$ and $\sigma_1$, as given in Equation 7.6. The 0.25-mm gap between the substrate and inner walls (see Figure 7.5) is large enough to allow free movement of the substrate, but small enough to isolate the sub-chamber somewhat from the main vacuum chamber, reducing perturbations to the supersaturation in the sub-chamber. The geometry of the sub-chamber, together with the procedure for placing a single test crystal within it (described below), are quite important for creating a well-known supersaturation in the vicinity of the test crystal, as quantified in the chamber analysis above.

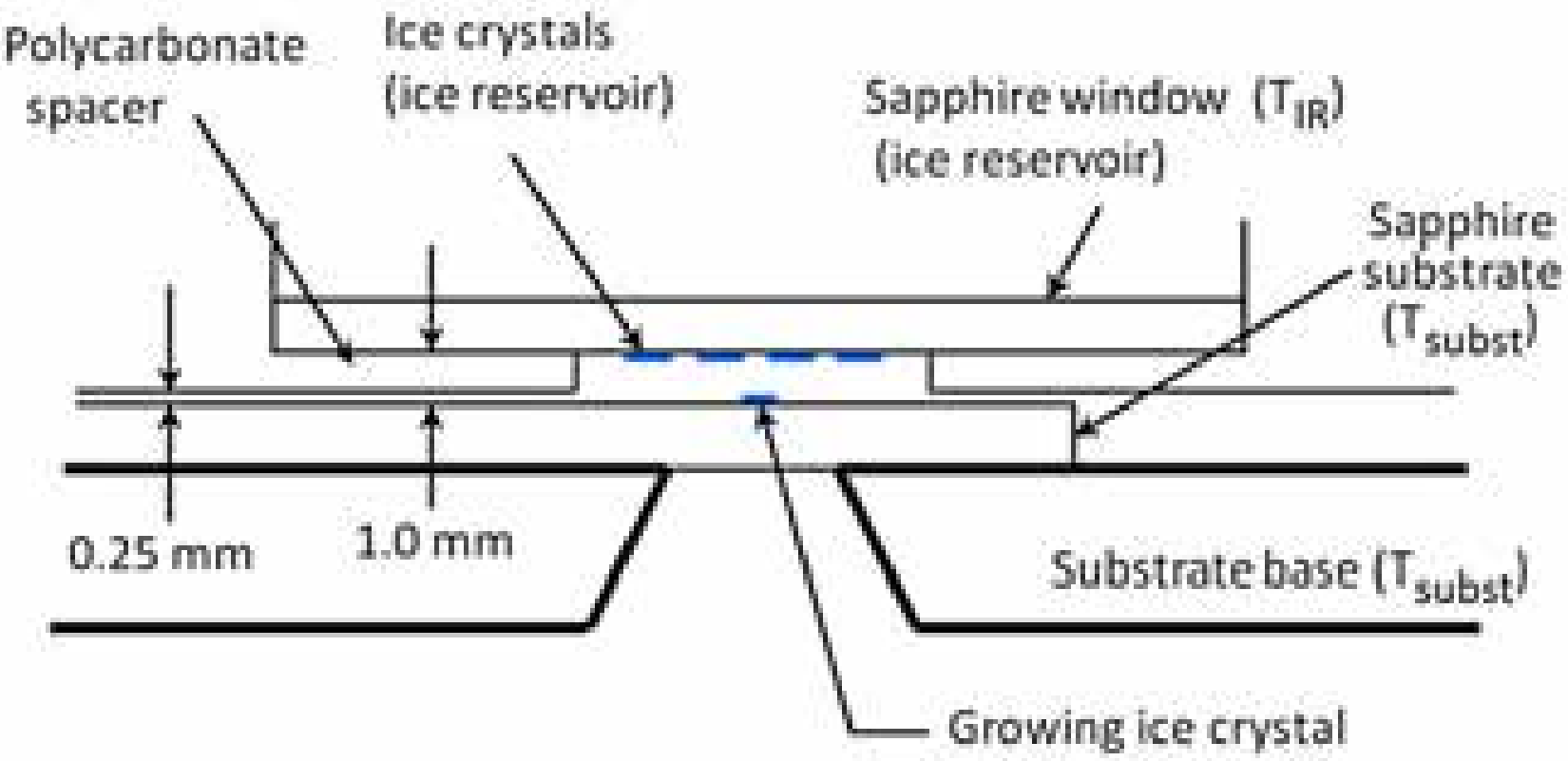

**Figure 7.5: (Left) A close-up view of the ice-growth region of the apparatus in Figure 7.4, showing the ice reservoir and test crystal. Note the similarities to the ideal growth-chamber geometry shown in Figure 7.1.**

**Temperature Control.** The aluminum chamber, substrate base, and ice reservoir are all independently temperature regulated using thermistor sensors and thermoelectric heating/cooling. The sapphire window that defines the ice reservoir is bonded using thermal epoxy to a small copper plate that contains a small thermistor for temperature sensing. The high thermal conductivity of copper, sapphire, and thermal epoxy makes for a well-defined ice-reservoir temperature. The sapphire substrate can slide freely over the anodized aluminum substrate base, and only the latter is temperature regulated. However, the large-area flat-on-flat contact between the substrate and base, along with little heat load on the substrate, keeps the substrate and base at essentially the same temperature.

The ice reservoir servo uses a home-built controller that does not regulate $T_{reservoir}$ directly, but instead regulates $\Delta T = T_{reservoir} - T_{substrate}$. This allows one to control $\Delta T$ with high accuracy, minimizing effects from any substrate temperature drifts. In normal operation, $T_{substrate}$ and $T_{chamber}$ are essentially equal, while $\Delta T$ is quite close to zero, yielding a nearly isothermal environment within the vacuum chamber.

**Temperature Calibration.** The thermistor response is known from the manufacturer's specifications, and the chamber and substrate temperatures are set using calibrated temperature controllers. The value of $\Delta T$ is especially critical, so the $\sigma = 0$ point is measured for each test crystal by adjusting $\Delta T$ (by adjusting $T_{reservoir}$) until the crystal is between growing and sublimating. This can be accomplished with especially high accuracy at low pressures, when the response to temperature changes is swift, allowing one to locate the $\sigma = 0$ point with an absolute uncertainty of about $\delta\sigma \approx 0.001$. This is significantly better than one can do with normal temperature controller calibration, as it is equivalent a temperature accuracy for $\Delta T$ of about 0.01 C.

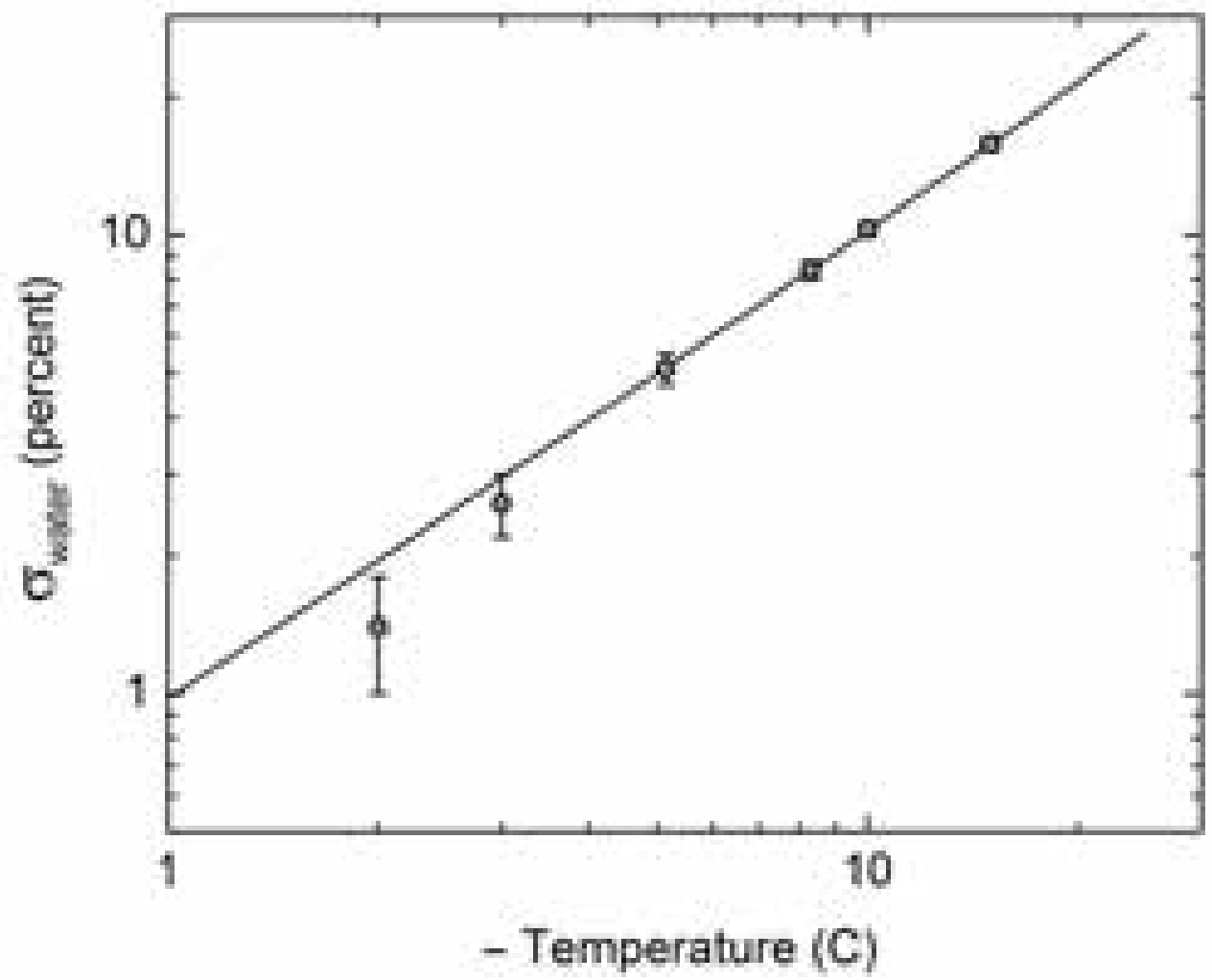

**Figure 7.6: Measurements of $\sigma_{water}$ at the substrate, as determined by the temperature difference $\Delta T_{stable}$ at which water droplets are stable. The line is not a fit to the data, but shows the theoretical expectation with no adjustable parameters.**

**Supersaturation Verification.** Observing the condensation of water droplets on the substrate provides an excellent method to verify that the supersaturation is equal to that given by Equation 7.2. With the chamber evacuated and no test crystals present, one can increase $\Delta T$ until water droplets appear on the substrate, and then slowly adjust $\Delta T$ until the droplets are neither growing nor shrinking. This value, $\Delta T_{stable}$, at which the droplets are just stable, must be producing a supersaturation equal to $\sigma_{water}$ at the substrate surface. Calculating $\sigma_{water}$ from the measured $\Delta T_{stable}$ yields the measurements shown in Figure 7.6, along with the theoretical expectation. The excellent agreement between theory and measurements, which includes no adjustable parameters for fits to the data, confirms the validity of Equation 7.2.

Note that the error bars in Figure 7.6 reflect the accuracy with which it is possible to measure $\Delta T_{stable}$. This measurement becomes more difficult at higher temperatures, as the value of $\Delta T_{stable}$ goes to zero as one approaches the melting point. The larger error bars at higher temperatures are indicative of the fact that it becomes generally more challenging to determine supersaturation accurately in a dual-temperature chamber at higher temperatures.

**Seed Crystal Generator.** The vacuum chamber shown in Figure 7.4 rests on the bottom of a much larger refrigerated chamber filled with ordinary air, about one meter in height. The larger chamber serves as a continuous seed-crystal generator (see Chapter 6), producing a constantly replenished cloud of small ice crystals that slowly drift down all around the aluminum vacuum chamber. These crystals grow in free fall, and any particular crystal grows for just a few minutes before settling out. New crystals are nucleated every ten seconds using an expansion nucleator, thus yielding a steady supply of pristine seed crystals. As described in Chapter 6, the seed-crystal chamber is somewhat self-cleaning in that the cloud of fresh ice crystals tends to continuously remove residual chemical impurities from the air in the chamber.

**Preparing the Ice Reservoir.** At the beginning of a run, after the system has reached its operating temperature and the vacuum chamber is stably temperature controlled, a butterfly vacuum valve on top of the aluminum vacuum chamber (not shown in Figure 7.4) is opened, allowing some seed crystals to fall onto the large sapphire substrate. A small electric motor rotates the substrate about a central pivot point so seed crystals land at all points around the circumference of the substrate. The vacuum pump, controlled by a variable needle valve, slowly draws air and seed crystals from the seed-crystal chamber into the vacuum chamber during this process.

With an ample supply of ice crystals on the substrate (still a low filling factor), the ice-reservoir temperature is set lower than the substrate temperature. The butterfly valve is closed and a vacuum is drawn inside the chamber while the substrate slowly rotates. Some ice on the substrate then sublimates and deposits on the sapphire ice-reservoir window. The microscope is focused on the latter surface to verify that a thick coating of frost appears on the window, forming the ice reservoir. After this loading process is complete, the substrate is further warmed to drive off any remaining ice, and air is let back into the chamber.

With the ice reservoir prepared, $T_{substrate}$ and $T_{chamb}$ are set to the desired operating temperature while $\Delta T$ is set to zero. The chamber is then ready for the main experimental session to commence. Additional ice can be added to the ice reservoir during the run as needed.

**Positioning an Isolated Test Crystal.** When the substrate is ice-free and a test crystal is desired, the butterfly vacuum valve is opened once again, allowing some seed crystals to fall onto the rotating sapphire substrate. A needle valve to the vacuum pump is opened slightly to draw air down from the seed crystal chamber, facilitating the transfer of seed crystals to the substrate. This process is continued for some tens of seconds to yield a low density of seed crystals on the substrate.

With the butterfly valve closed, a live video view through the microscope objective (see Figure 7.4) is scrutinized while the substrate slowly rotates to search for a suitable seed crystal. The substrate's central pivot can also be translated using the manipulator arm shown in Figure 7.4, allowing a 2D sweep of the substrate surface for test crystals. The crystal density is low, and not every seed crystal has an ideal prism morphology, so it often takes some searching to locate a suitable test crystal.

Typically, one looks for a well-formed ice prism with either a basal or prism facet lying flat on the substrate. Polycrystalline forms, malformed crystals, or poorly oriented crystals are all rejected. Crystals with nearby neighbors are also rejected. If need be, the substrate can be heated to remove all the crystals, so it can be reloaded with new seed crystals for another attempt. Finding high-quality, isolated seed crystals is actually a fairly laborious process,

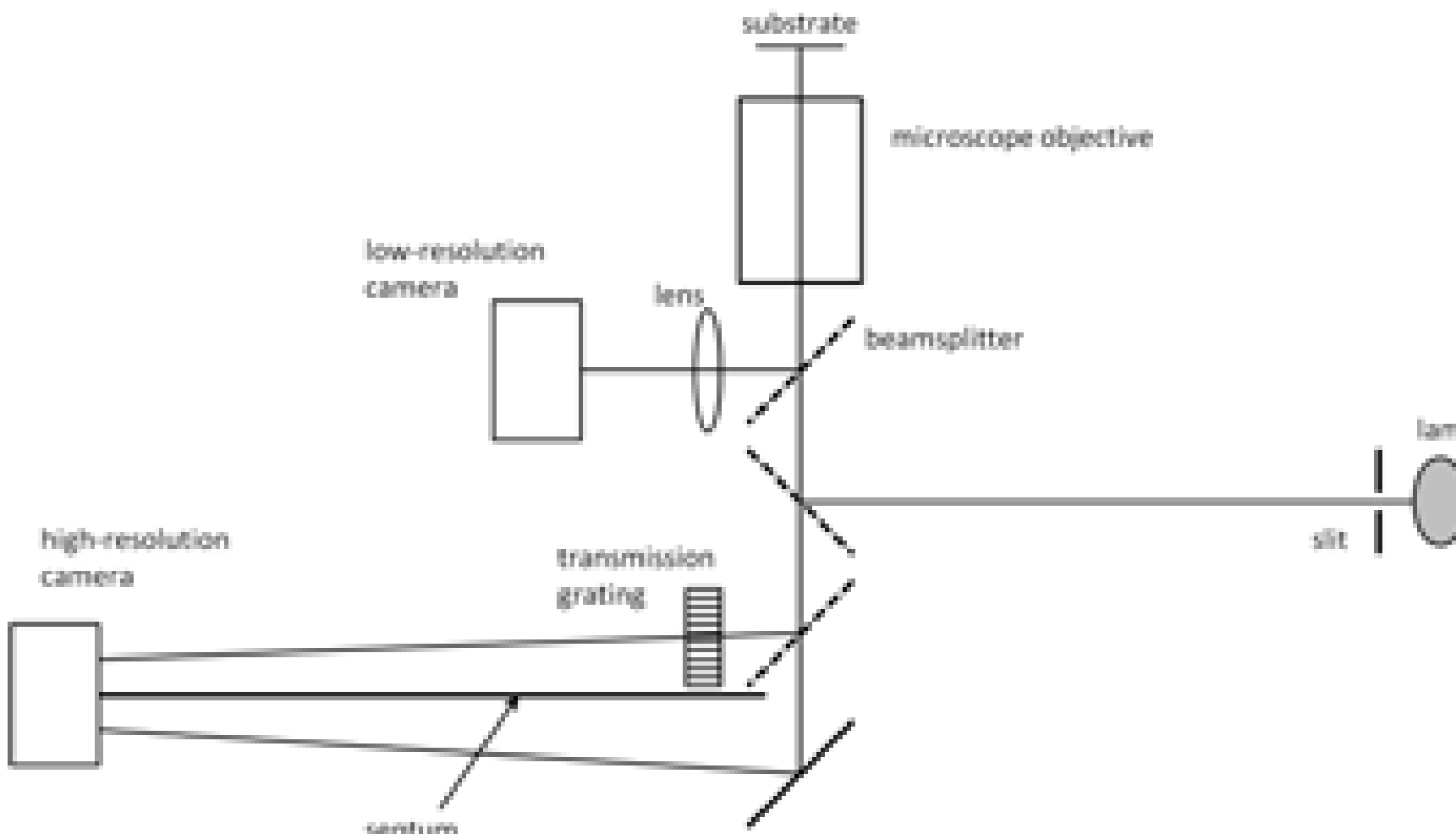

**Figure 7.7: The optical layout for observing the growth of a test crystal. The substrate and microscope objective are the same as in Figure 7.4. As described in the text, the low-resolution camera is used for finding and positioning a test crystal, while the high-resolution camera records a direct image of the test crystal along with a broad-band interferometer signal used to measure the crystal thickness.**

sometimes taking 10-20 minutes to locate a suitable specimen.

**Pumping Out the Chamber.** Once a high-quality test crystal has been positioned at the center of the microscope field of view, the variable needle valve is opened slowly to begin the pump-down. This is a somewhat tricky step when the pressure gets low, as it is relatively easy to sublimate away the test crystal with overzealous pumping. The operator carefully watches the test crystal and adjusts $\Delta T$ and the pump-out speed appropriately to make sure that the test crystal neither grows nor sublimates appreciably during the pump-down. Once the pressure has been reduced to about 20 mbar, the test crystal is ready for a growth measurement.

**The Optical System.** Figure 7.7 shows the optical layout for measuring the growth of a test crystal. The main illumination is from above, with light passing through the ice reservoir as shown in Figure 7.4. Calculations show that this light is far too weak to affect the temperature of crystals on the ice reservoir. The low-resolution camera is used for finding and positioning a suitable test crystal, and to verify that the test crystal has no nearby neighbors. A second image is projected onto the high-resolution camera for direct imaging of the test crystal.

White-light interferometry is used to measure the crystal thickness (see Chapter 6), and an image of the fringe pattern is also projected onto the high-resolution camera. A septum (see Figure 7.7) prevents these two images from overlapping on the camera sensor. A small prism located next to the high-resolution camera sensor (not shown in Figure 7.7) projects a third small image of a voltmeter onto the same image plane. Figure 7.8 shows two typical images from the high-resolution sensor. A shutter blocks the slit light periodically, allowing a better direct view of the crystal than is shown in Figure 7.8.

Projecting these three images onto a single camera sensor provides a convenient, low-budget method for data acquisition. The lateral size of the test crystal can be measured from the direct image, the crystal thickness can be extracted from the interferometer fringe pattern, and the voltmeter reading can be used to determine $\Delta T$. With this optically generated split-screen image, a single video recording simultaneously provides all of these measurements as a function of time, avoiding the need for synchronizing separate imaging data streams. The audio channel of the video is used to record verbal notes from the operator as the crystal is being grown. The relevant information is all transcribed from the video during data analysis. Figure 7.9 shows the VIG experiment in the lab.

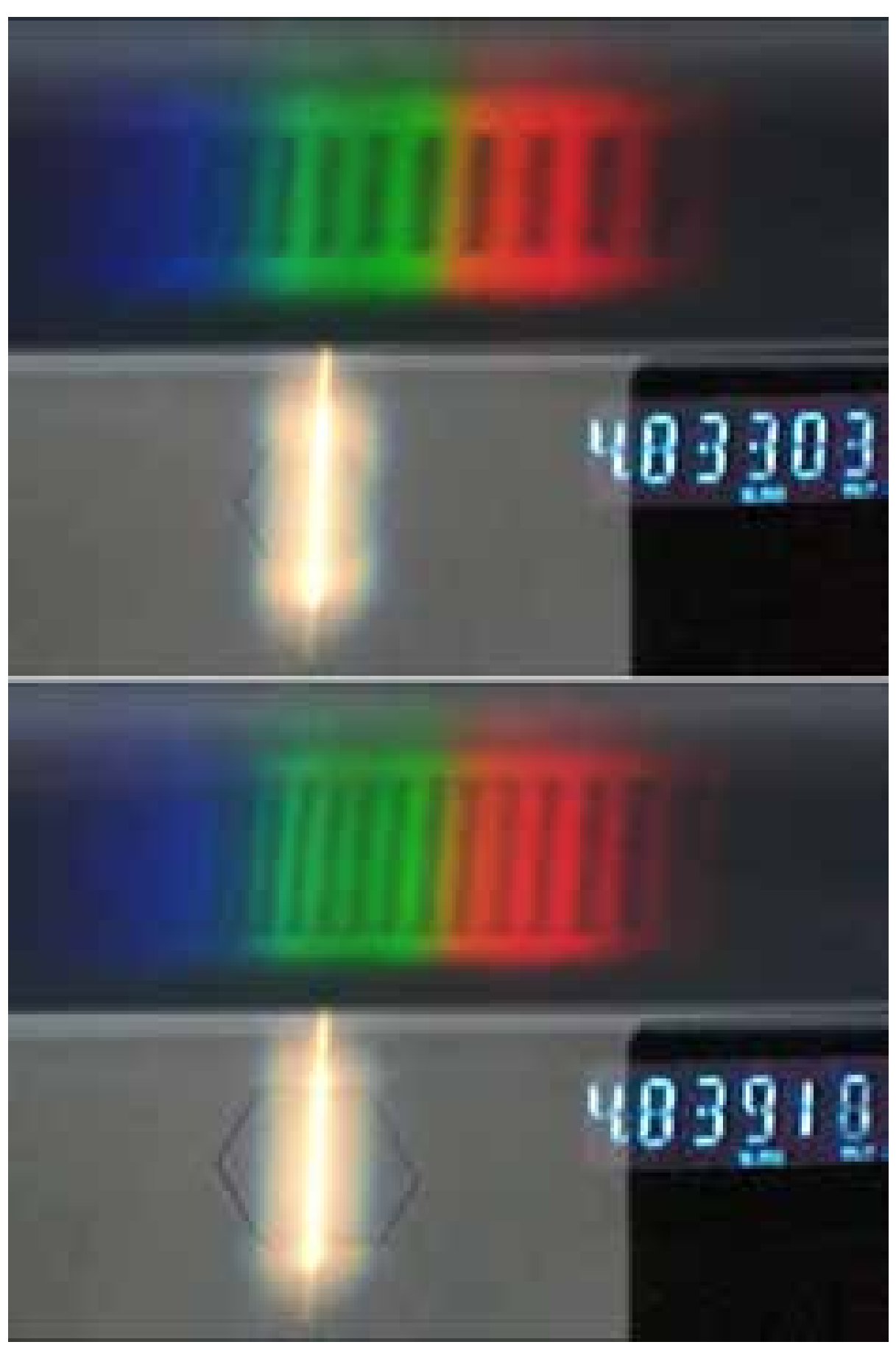

**Figure 7.8: (Left) Two still images from a video showing the growth of a test crystal. The top parts of both images show interferometer fringes in true colors, while the bottom parts of the images show a direct image of the crystal including illumination from the slit. The numerical section in the images show a voltage from which Δ*T* can be determined. The upper image was taken near the beginning of the growth cycle. The lower image shows the same crystal after it had grown larger and thicker. The bright slit light is periodically blocked to provide a better direct image of the crystal.**

**Figure 7.9: (Below) The Vacuum Ice Growth experiment in the lab. The aluminum vacuum chamber is a small black package at the bottom of the much larger copper-walled seed-crystal chamber. A heat lamp is baking the system in this photo. The optics are covered in black panels below the large chamber, and the high-resolution camera is contained in the white styrofoam box at the lower left.**

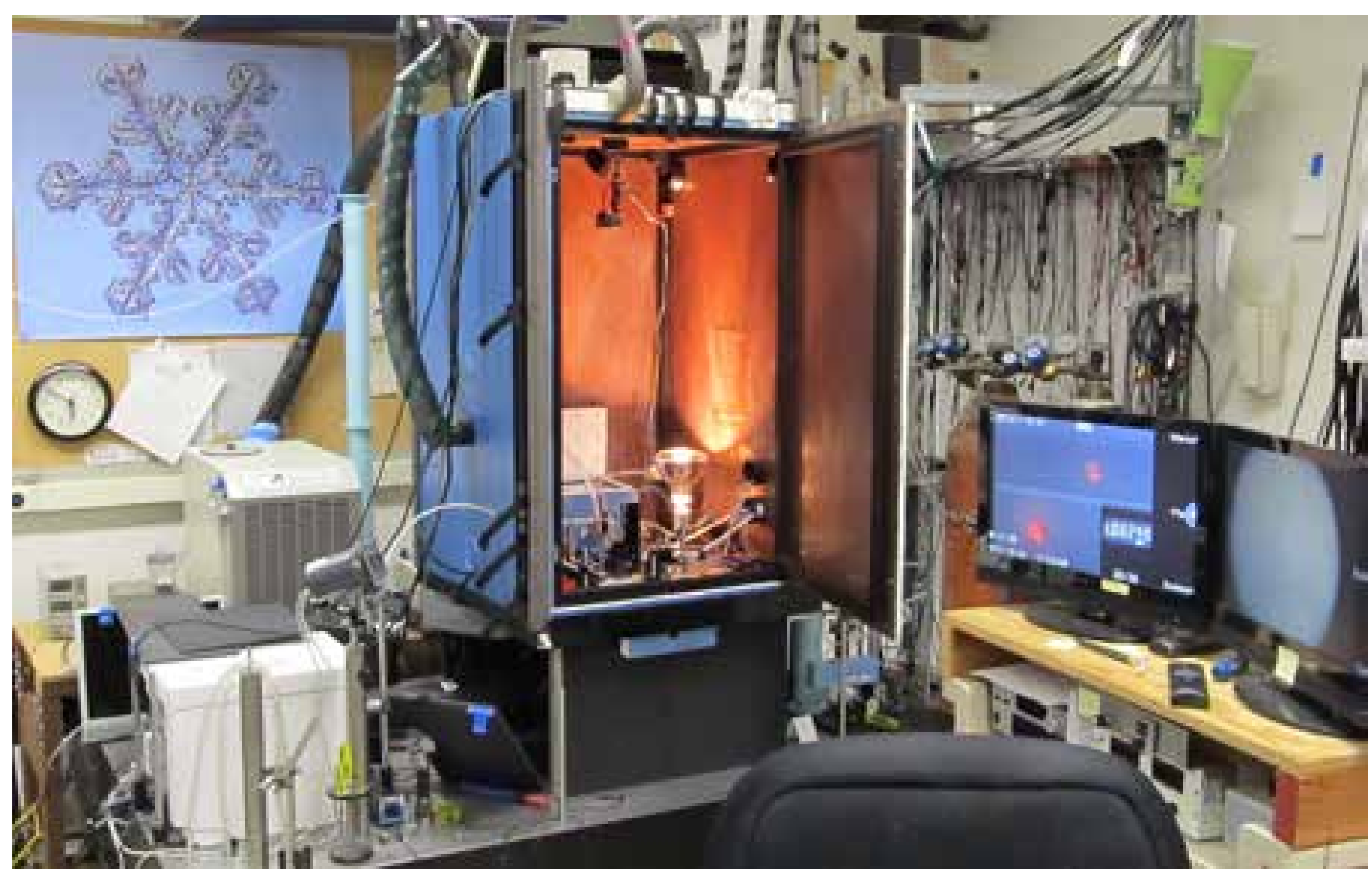

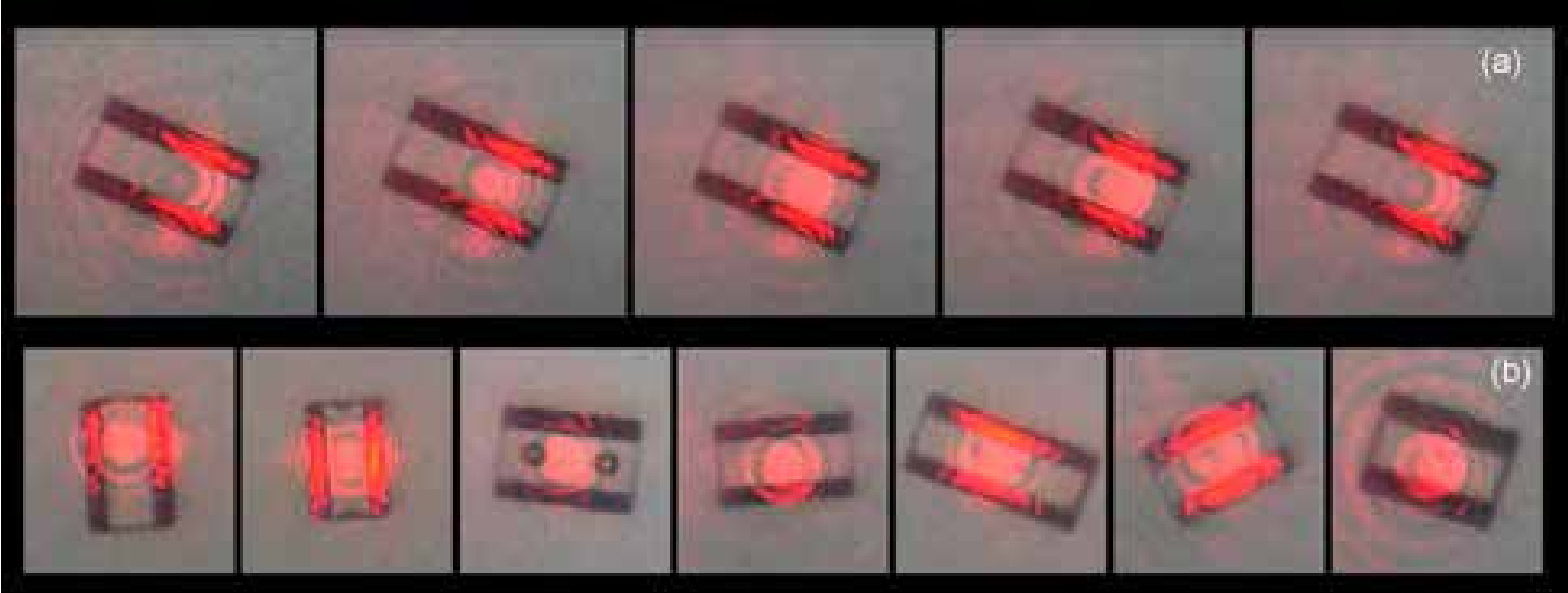

**Figure 7.10: (a) Several images of a columnar ice crystal showing the interference of two reflections from an incident Helium-Neon laser beam, one from the substrate/ice interface and one from the ice/vacuum interface. As the crystal grew (left to right) the spot brightness oscillates from dim (first image) to bright (middle image) and back to dim (last image). (b) Snapshots from several other example crystals.**

**Basal and Prism Growth.** As implemented in this experiment, the white-light interferometric measurement of the crystal thickness works only for thin, plate-like crystals, as demonstrated in Figure 7.8. As the crystal thickness increases, the fringes become closely spaced and the fringe contrast diminishes. Thus the technique is well-suited for measuring growth of basal facet surfaces using thin plates, but it cannot be used when measuring the growth of prism facet surfaces using columnar crystals.

When performing a run measuring columnar test crystals, the lamp and slit in Figure 7.7 are replaced by a Helium-Neon laser, giving direct images like those shown in Figure 7.10. As with light from the slit, the He-Ne beam reflects off both the substrate/ice and ice/vacuum interfaces, and these two reflections interfere with one another. The brightness of the reflected spot then depends on crystal thickness, oscillating between bright and dark as the crystal grows. This method is not as precise as white-light interferometry, nor does it yield an absolute measurement of the crystal thickness. With care, however, it can yield acceptable growth measurements of columnar crystals.

**A Growth Sequence.** Once a suitable test crystal has been found and positioned, and the chamber pumped down to about 20 mbar, the $\sigma = 0$ point is then determined by adjusting the ice reservoir temperature until the test crystal is neither growing nor sublimating. This procedure also gives the chamber a few minutes to equilibrate with $\sigma = 0$. A growth sequence then commences by slowly increasing $\sigma$ by increasing the ice-reservoir temperature while monitoring $\Delta T$. The substrate and chamber temperatures remain constant as $\Delta T$ increases. The top panel in Figure 7.11 shows typical data during a growth sequence, where $\sigma$ here is defined from $\Delta T$ using Equation 7.2.

The small size of the ice reservoir, and the high thermal conductivity of copper and sapphire, give the reservoir a fast temperature response while keeping the overall temperature equal to that indicated by its thermistor sensor. Thus the voltage number shown in Figure 7.8 gives an accurate indication of $\Delta T$ and thus $\sigma$ during the sequence. The temperatures of the substrate and vacuum chamber are kept fixed during this time, as they have a longer thermal response time.

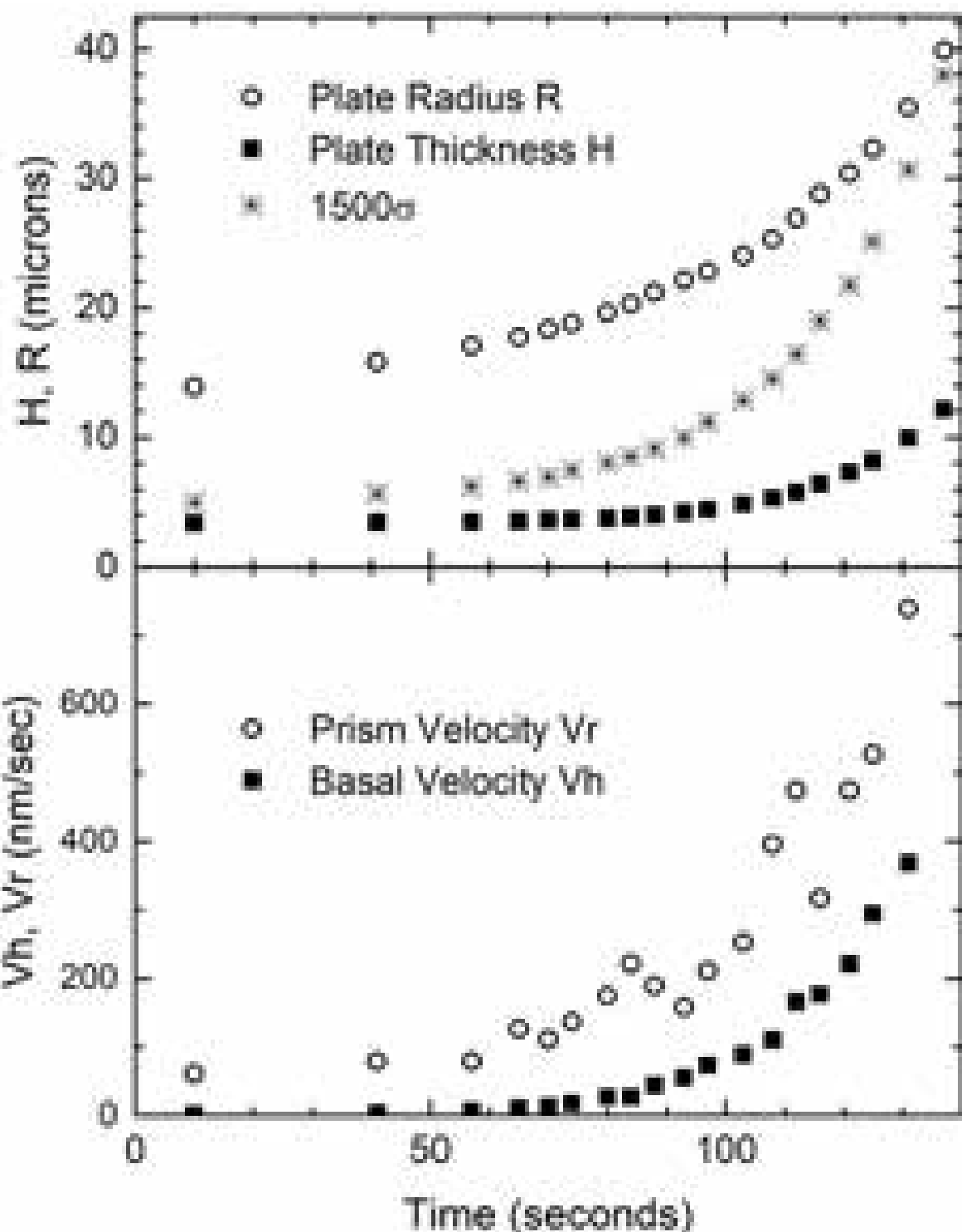

**Figure 7.11: (Top) A typical growth sequence for a single thin-plate crystal at -12 C, like the one shown in Figure 7.8. As the supersaturation is slowly increased by increasing $\Delta T$, the plate thickness H is measured using white-light interferometry while the plate "radius" R (here defined as half the distance between opposing prism facets) is measured from direct imaging. Note R>H, indicating a thin, plate-like crystal. (Bottom) Growth velocities of the basal and prism facets derived from the size data above. The higher accuracy of the interferometric measurements yields a lower noise in the basal radius and velocity data.**

At the end of the growth sequence, the test crystal is discarded by heating the substrate, which removes all the seed crystals. We have found that "recycling" a test crystal – sublimating it back down to a smaller size and growing it out again – generally leads to unreliable results. Surface impurities become concentrated during large-scale sublimation, and this often seems to corrupt the subsequent regrowth of the crystal. Sometimes a recycled crystal behaves normally, but this is not always the case. To avoid any potential problems from crystal recycling, we typically use a test crystal only once.

**Chemical Contamination Tests.** Surface contamination from unwanted chemical vapors is always a concern in any ice-growth experiment, as one is never sure how clean is clean enough. Opening the entire system and baking it between runs is a first line of defense again chemical contaminants, as they tend to bake out after numerous thermal cycles. Another plus is using a continuous seed crystal generator (see Chapter 6), as fresh seed crystals are produced every few minutes, and the crystals themselves absorb contaminants from the surrounding air and pull them to the bottom of the chamber as they fall. This self-cleaning feature keeps the air in the seed crystal generator chamber quite clean during a run.

As a test crystal is being grown (like that shown in Figure 7.11), the growing ice surface is also somewhat self-cleaning in regard to chemical contaminants. As a crystal grows, each expanding ice terrace edge tends to push surface chemicals ahead of it, as few chemicals are readily incorporated into the ice lattice. Surface contaminants are thus swept aside as a crystal grows, cleaning the faceted surface in the process.

One can test this process in a single growth sequence by first increasing $\Delta T$ with time and then decreasing it back to zero. After a short period of rapid growth, when $\Delta T$ is high, the fresh ice surface should be especially free of contaminants. Quickly bringing $\Delta T$ back down then allows a growth measurement of this pristine surface. If the growth velocity $v(\sigma_{surf})$ looks the same whether $\Delta T$ is increasing or decreasing, then this suggests that the initial ice surface was reasonably clean.

## Data Analysis

To analyze a growth sequence, the video is first transcribed to produce time-dependent measurements of $H$, $R$, and $\Delta T$ as a function of time, as shown in Figure 7.11. Then the $H$ and $R$ data are used to extract growth velocities, also seen in the figure, while $\sigma_{bottom}$ as a function of time derives from the $\Delta T$ data. Because the chamber design keeps the correction in Equation 7.6 quite small, the $\Delta T$ data give $\sigma \approx \sigma_{bottom} \approx \sigma_1$ directly, where the latter is given by Equation 7.2. Plotting the velocities $V_H$ and $V_R$ versus $\sigma$ then yields the data shown in Figure 7.12. Note that the growth sequence of a single test crystal yields growth velocities as a function of supersaturation. Thus every test-crystal growth sequence can be used to extract $\alpha(\sigma_{surf})$ for $\sigma_{surf}$ ranging from zero to some maximum value.

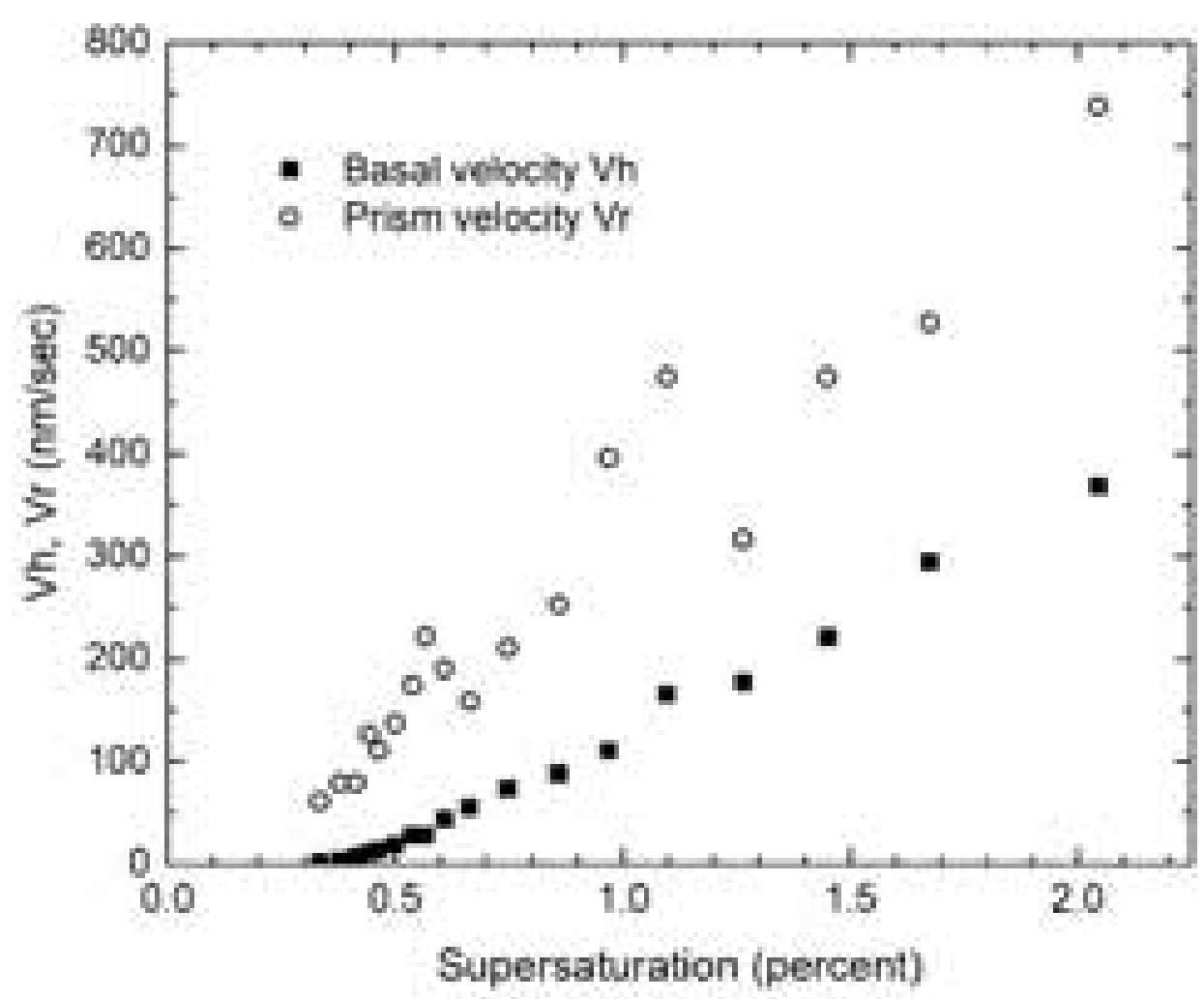

**Figure 7.12: The growth sequence in Figure 7.11 after being converted to velocities versus supersaturation $\sigma_{bottom}$ just above the substrate surface. Note that $\sigma_{bottom}$ is not generally equal to the supersaturation $\sigma_{surf}$ at the crystal surface.**

**Substrate Interactions.** As I described in Chapter 6, substrate interactions can have serious detrimental effects on ice crystal growth measurements [2012Lib]. Especially important is that a low ice/substrate contact angle may cause spurious nucleation of new terraces, thereby increasing growth rates compared to surfaces that do not contact the substrate. This effect can be seen directly in Figures 7.11 and 7.12, as the prism growth velocity $V_R$ is much larger than the basal growth $V_H$ at early times, when the supersaturation is low. In these data, the basal growth is suppressed by a large nucleation barrier, while the prism growth is not.

Measurements of prism facets that are not contacting the substrate, however, reveal a similarly large nucleation barrier for prism growth [2013Lib]. This behavior suggests that the $V_R$ data in Figures 7.11 and 7.12 are distorted by substrate interactions, especially at low supersaturations. The hypothesized model of substrate interactions described in Chapter 6 [2012Lib] provides a sensible explanation for this overall growth behavior.

In general, we have found that substrate interactions can be both significant and somewhat unpredictable. For example, the ice/substrate contact angle is sensitive to surface chemical residues, so may vary with position on the substrate. In addition to terrace nucleation, surface chemical effects may either increase or decrease growth rates if the substrate is not sufficiently clean. We have found that the best solution to this problem is simply to discard growth velocity data for all surfaces that contact the substrate directly. Interestingly, Beckmann et al. [1983Bec] observed similar substrate interactions to those just described, but chose to discard data from the facets that were not contacting the substrate, keeping data from the those that did, rather than the other way around.

In Figure 7.12, for example, we discard the $V_R$ data for this plate-like crystal, but retain the $V_H$ data, as the latter came from measurements of the top basal surface, which was parallel to the underlying substrate. For columnar crystals like those in Figure 7.10, we retain only growth data from the upper prism surface for the same reason.

**Diffusion Correction.** The next step in the data analysis is to recognize that while the $\Delta T$ data give $\sigma_{bottom}$ with good accuracy (for a suitably isolated test crystal), this is not generally equal to $\sigma_{surf}$ at the surface of the growing crystal. Even at a pressure of 20 mbar, the diffusion correction is significant and must be addressed. Ideally, one could use a computational growth model (see Chapter 5) to analyze the velocity data, but this approach is time-consuming and more rigorous than is needed, given other measurement uncertainties.

A substantially faster, simpler analysis method is to use the monopole approximation described in Equation 7.13 to convert $\sigma_{bottom}$ to $\sigma_{surf}$. (In this notation, $\sigma_{bottom} = \sigma_\infty$) The volume derivative $\dot{V}_s$ can be calculated from the crystal size and velocity data [2012Lib], using both the $V_H$ and $V_R$ data, so the $\delta\sigma$ correction contains no adjustable parameters. The correction is only as good as the data used to derive it, but the underlying physics is well understood.

Figure 7.13 shows this diffusion correction being applied to the data in Figure 7.12. The correction changes $\sigma_\infty$ to $\sigma_{surf}$, which moves a given point to the left. The same point moves vertically upward as $\alpha_{uncorrected} = v/v_{kin}\sigma_\infty$ changes to $\alpha = v/v_{kin}\sigma_{surf}$. The filled points in Figure 7.13 then give the desired function $\alpha(\sigma_{surf})$ for this test crystal.

Clearly the correction is quite large for the highest-velocity points, as $\sigma_{surf}$ is only about half as large as $\sigma_\infty$ for the final point in Figure 7.13. Because the diffusion correction is not extremely precise, one must expect that the resulting high-velocity data will be noisy and may exhibit some systematic errors. For the lower-velocity points, however, the correction is substantially smaller, giving one greater trust in the final corrected data.

**Heating Correction.** As described in Chapter 3, the generation of latent heat at the growing ice crystal produces another small correction factor. For growth on a substrate, this heat is readily conducted through the ice to the large substrate below, which can be considered an infinite heat reservoir at fixed temperature. From a calculation of heat flow through the ice,

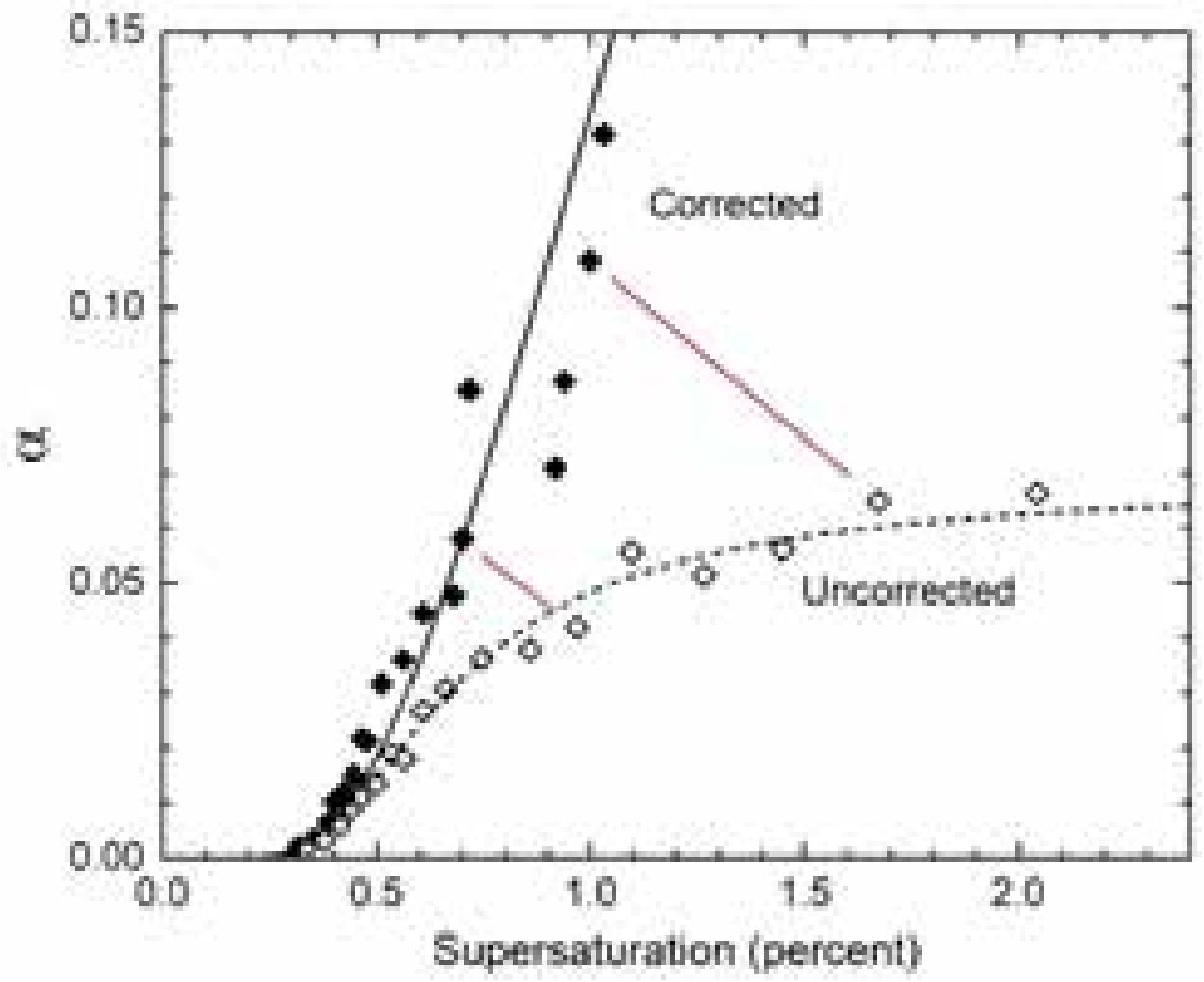

**Figure 7.13: The basal growth data from Figure 7.12 converted to the basal attachment coefficient $\alpha$ as a function of supersaturation. The open points show the uncorrected $\alpha_{uncorrected} = v/v_{kin}\sigma_\infty$ plotted as a function of $\sigma_\infty$, while the filled points show the corrected $\alpha = v/v_{kin}\sigma_{surf}$ plotted as a function of $\sigma_{surf}$. The red lines show how two individual points transformed from uncorrected to corrected.**

**A nucleation-limited growth model $\alpha = A\exp(-\sigma_0/\sigma_{surf})$ (solid line) provides a good fit to the corrected data using $A = 1$ and $\sigma_0 = 2$ percent. Including diffusion gives the dashed curve, equal to $\alpha\alpha_{diff}/(\alpha + \alpha_{diff})$ with $\alpha_{diff} = 0.075$ (see Chapter 3). At low $\sigma_{surf}$, the basal growth is strongly limited by a nucleation barrier, so $\alpha \ll \alpha_{diff}$ and the diffusion correction is small. At higher $\sigma_{surf}$, $\alpha \gg \alpha_{diff}$ and the diffusion correction is large.**

the perpendicular growth velocity $V_H$ can be written (see Appendix B)

$$V_H \approx \frac{\alpha\alpha_{heat}}{\alpha + \alpha_{heat}} v_{kin}\sigma_{surf} \qquad (7.14)$$

with

$$\alpha_{heat} \approx \frac{G\kappa_{ice}}{\rho_{ice}L_{sv}v_{kin}\eta H} \qquad (7.15)$$

where $H$ is the crystal thickness and $G$ is a dimensionless geometrical factor of order unity. For a broad, thin-plate crystal, $G = 1$. At a temperature of -5 C, this becomes

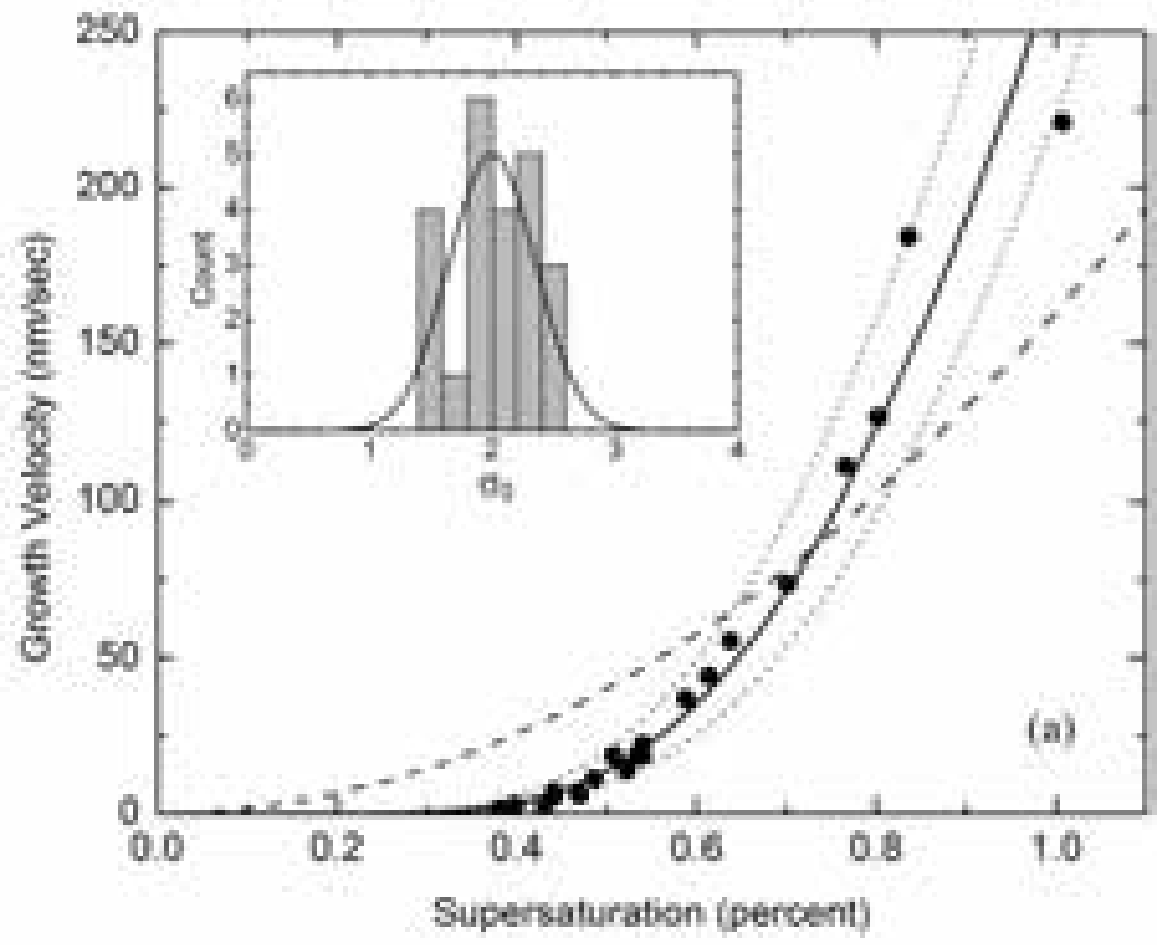

**Figure 7.14: Sample corrected measurements showing the growth velocity of the basal surface of a single ice crystal as a function of $\sigma_{surf}$ at $-12$ C, where data points were taken as the supersaturation was slowly increased. The line through the points gives the model $v_{basal} = \alpha_{basal} v_{kin} \sigma_{surf}$ with $\alpha_{basal}(\sigma_{surf}) = \exp(-\sigma_0/\sigma_{surf})$ and $\sigma_0 = 2.3 \pm 0.2\%$. The dashed line shows a spiral-dislocation model with $v \sim \sigma_{surf}^2$, which gives a poor fit to the data. The inset graph shows an unweighted histogram of measured $\sigma_0$ values for 23 crystals. A weighted fit to these data gives an estimated mean $\langle \sigma_0 \rangle = 1.95 \pm 0.15\%$.**

$$\alpha_{heat} \approx 22 \left( \frac{1\ \mu m}{H} \right) G \qquad (7.16)$$

When measuring basal growth with thin, plate-like crystals (e.g., in Figure 7.8) the thickness $H$ is typically just a few microns, so $\alpha \ll \alpha_{heat}$ and the heating correction is always quite small. The correction becomes larger at higher growth temperatures, however, when $v_{kin}$ becomes larger, and when using columnar crystals, because then $H$ is also larger. For most of the VIG data, the heating correction is negligible. But, as I describe below, heating effects may be distorting the measurements of $\alpha_{prism}$ at high temperatures when $\alpha$ is also high.

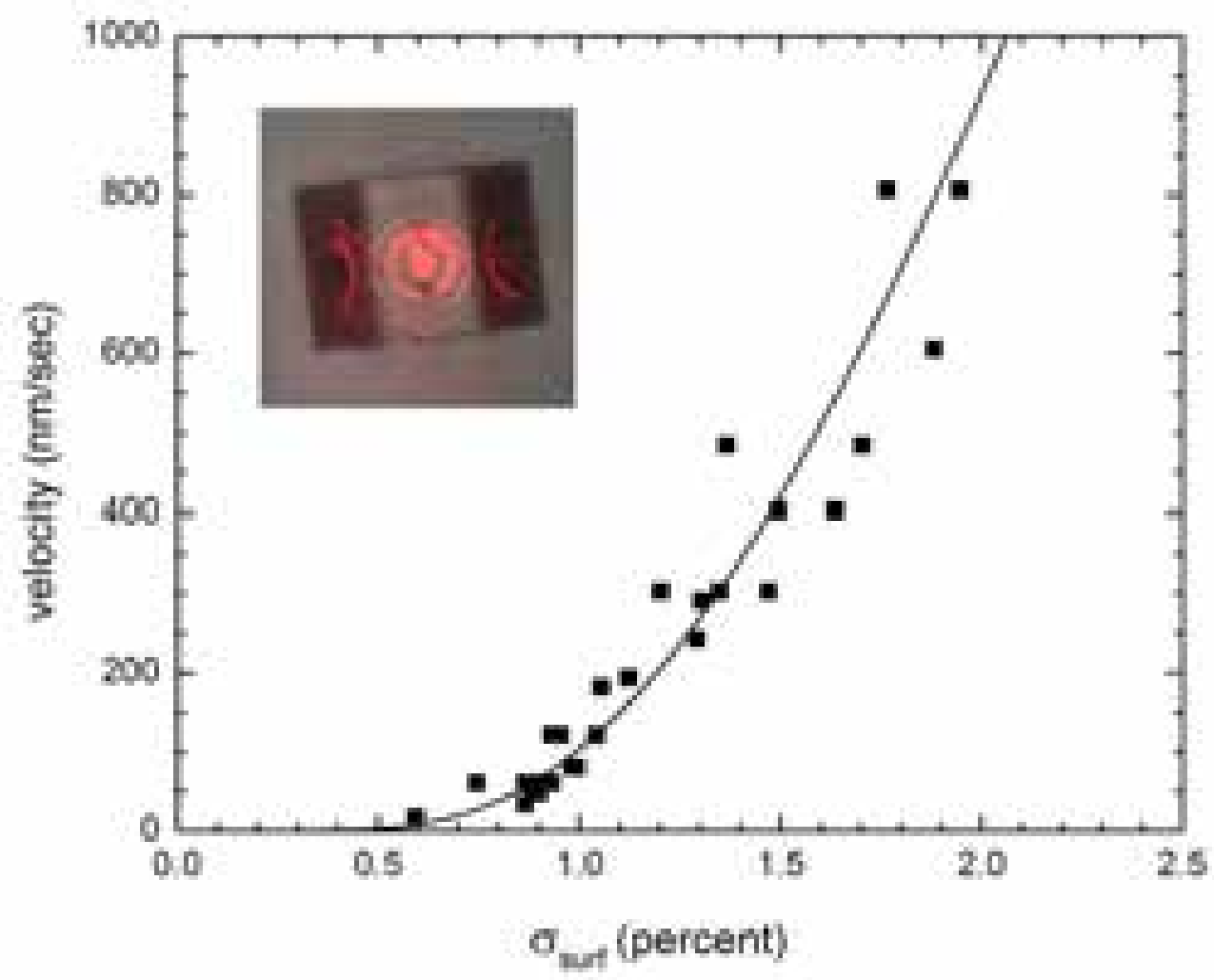

**Figure 7.15: Sample measurements of the prism growth velocity as a function of supersaturation at the facet surface. Data points were taken at -15 C with a background air pressure of 20 mbar, using oscillations in the brightness of the laser spot to interferometrically measure the growth velocity [2013Lib]. The line shows $v_n = \alpha v_{kin} \sigma_{surf}$ with $\alpha(\sigma_{surf}) = \exp(-\sigma_0/\sigma_{surf})$ and $\sigma_0 = 3$ percent. Note that the data fit this model well at growth velocities up to nearly 1 micron/second. The inset photo shows a test crystal being measured.**

**Nucleation-Limited Growth.** Once corrected for diffusion effects, the VIG data strongly suggest that the attachment kinetics are primarily limited by terrace nucleation. Figure 7.14, for example, shows that the data are well fit by a nucleation model, while a spiral-dislocation model does not fit the data (see Chapter 4). This conclusion applies to both basal and prism growth data, and Figure 7.15 shows one example for prism facet growth at a temperature of -15 C.

Displaying $\alpha$ versus $1/\sigma_{surf}$ in a semi-log plot provides a good view of the nucleation-limited growth behavior seen in the data, as the functional form $\alpha(\sigma_{surf}) = Ae^{-\sigma_0/\sigma_{surf}}$ appears as a straight line in such a graph. Figure 7.16 shows measurements of $\alpha_{basal}$ from 12 separate crystals, both before and after the diffusion correction. The uncorrected data clearly show a behavior indicative of diffusion-limited growth, rolling off to an effective $\alpha_{basal}$ equal to $\alpha_{diff}$ at high supersaturations. The corrected data, however, indicate $\alpha_{basal} \to 1$ as $\sigma_{surf} \to \infty$, which is the asymptotic behavior one would expect for nucleation-limited attachment kinetics. Figure 7.15 shows additional measurements of $\alpha_{basal}$ plotted as a function of $1/\sigma_{surf}$ for a variety of temperatures.

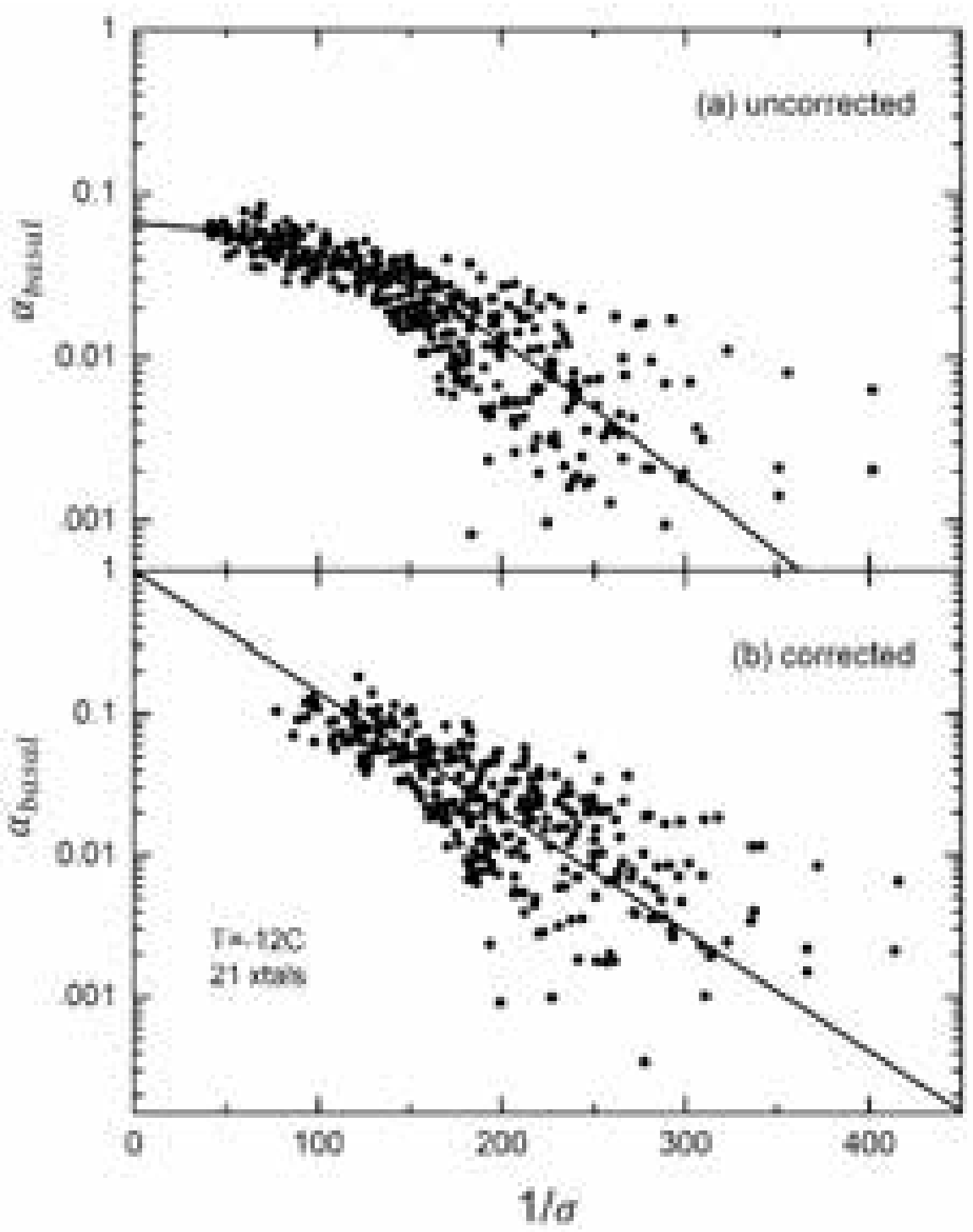

**Figure 7.16: A set of basal-facet growth measurements taken using 23 ice crystals at $T = -12C$, both before (a) and after (b) correction for residual diffusion effects. The straight line shows $\alpha(\sigma) = \exp(-\sigma_0/\sigma)$ with $\sigma_0 = 0.02$, while the curved line shows the same $\alpha(\sigma)$ combined with a diffusion-limited growth model. As with Figure 7.13, the uncorrected data show $\alpha_{uncorrected}$ plotted versus $1/\sigma_\infty$, while the corrected data show $\alpha_{corrected}$ plotted versus $1/\sigma_{surf}$. Each data point was shifted horizontally using a diffusion model that calculated $\sigma_{surf}$ from $\sigma_\infty$ using Equation 7.13 [2013Lib].**

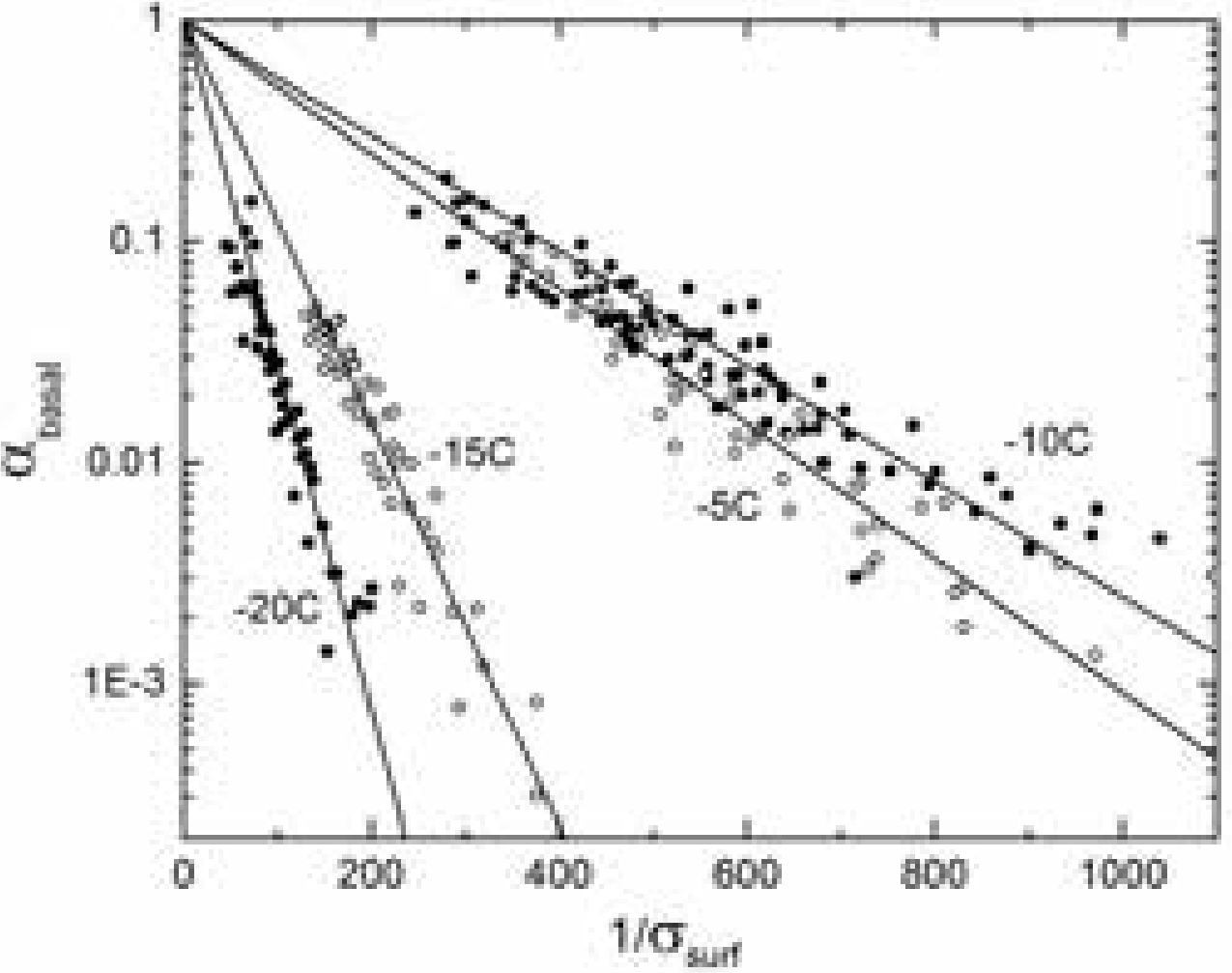

**Figure 7.17: Experimental data showing the attachment coefficient $\alpha_{basal}$ plotted as a function of $1/\sigma_{surf}$, including measurements taken at four different temperatures. Plotted this way, data exhibiting a nucleation-limited growth behavior with $\alpha_{basal}(\sigma_{surf}) = Ae^{-\sigma_0/\sigma_{surf}}$ appear as straight lines. The data at each temperature extrapolate to $\alpha \approx 1$ at large $\sigma_{surf}$, indicating rapid kinetics in the absence of a nucleation barrier. The values of $\sigma_0$ at different temperatures can be extracted from the slopes of the lines [2013Lib, 2017Lib].**

Note that the growth data were not constrained in any way to produce $\alpha_{basal} \to 1$ at high supersaturations; the analysis could have equally well yielded either higher or lower values of $\alpha_{basal}$ as $\sigma_{surf} \to \infty$. The fact that the data naturally yielded the theoretical expectation for fast growth suggests that the experiment is working well and that the corrections are quite accurate. I find that the convergence to $\alpha_{basal} \to 1$ at high supersaturations, over a broad range of temperatures, is an especially satisfying result from this experiment, as no previous ice-growth experiment had achieved the overall level of measurement precision to observe this theoretically pleasing result.

**Precision Measurements.** One important lesson from the VIG experiment is that obtaining high-quality measurements of the attachment kinetics requires a great deal of attention to detail regarding apparatus design, systematic errors, and data analysis. With the VIG experiment, this included:

1) The chamber was designed specifically to produce a well-defined supersaturation near the test crystal, following a careful diffusion analysis. The correction factor described by Equation 7.6 is especially important. Even at low pressures and with a small filling factor of crystals on the substrate, diffusion effects from crystal crowding can still be quite significant.
2) Several measures were taken to avoid chemical vapor contamination. The entire system was baked between runs, a self-cleaning seed crystal generator was used, and the vacuum chamber was purged with fresh air throughout each run.
3) We were careful to select only test crystals with the highest visual quality. Only near-perfect ice prisms with no nearby neighbors on the substrate were chosen.
4) Crystals were not sublimated and then regrown. A new test crystal was selected for each growth sequence.
5) Much time was spent analyzing correction factors and characterizing possible systematic errors in the measurement process.
6) We spent a great deal of time perfecting the apparatus, data acquisition procedures, and analysis methods. Over 200 crystals were grown and analyzed, which allowed many consistency checks and redundancies.

## Attachment Coefficients

We found that all of the VIG data could be well represented using attachment coefficients having the functional form $\alpha(\sigma_{surf}) = Ae^{-\sigma_0/\sigma_{surf}}$, thus reducing the entire attachment kinetics problem to the functions $\sigma_0(T)$ and $A(T)$ shown in Figure 7.18. The data clearly favor a nucleation-limited model for ice crystal growth from water vapor over a broad range of environmental conditions, and I discuss the physical implications of this result in Chapter 4.

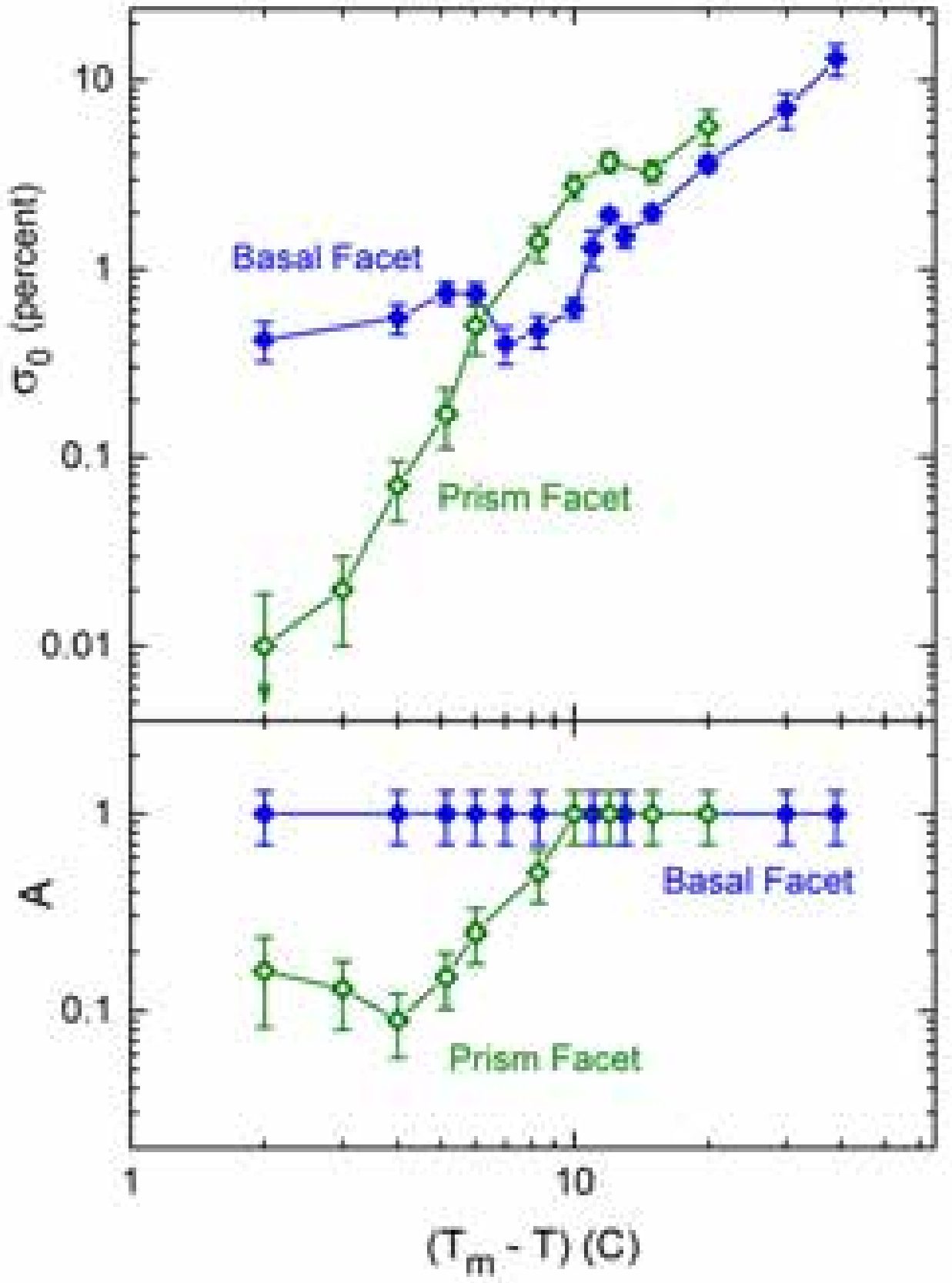

**Figure 7.18: Experimental data showing the attachment coefficient fit parameters $\sigma_0(T)$ and $A(T)$ for the basal and prism facets, assuming a functional form $\alpha(\sigma_{surf}) = Ae^{-\sigma_0/\sigma_{surf}}$ for both $\alpha_{basal}$ and $\alpha_{prism}$. The data were consistent with $A = 1$ for all the basal data and for the prism data when $T \leq -10C$, so this constraint was applied when fitting $\sigma_0$ over these temperature ranges [2013Lib]. The prism data with $T > -10C$ were better described with $A < 1$ as shown.**

To date, these are the best measurements of ice growth in a low-background-pressure environment, substantially surpassing previous efforts in overall precision. Remarkably, the resulting data suggest that ice crystal growth rates on the basal and prism facets are largely determined by the terrace step energies as a function of temperature, which are fundamental equilibrium properties of the ice lattice structure, as discussed in Chapter 2. If these step energies can be independently determined by molecular dynamics simulations, this would provide a major step forward in solving the full problem of snow crystal growth dynamics.

## Constrained A=1 Model

As seen in Figure 7.17, the basal growth data suggest a model with $A = 1$ at all temperatures, and fits to the data were consistent with this result. Theoretically, $A = 1$ means that $\alpha \rightarrow 1$ as $\sigma_{surf} \rightarrow \infty$, and such a behavior is expected in a nucleation-limited growth model. Physically, this is essentially equivalent to having $\alpha = 1$ on rough (non-faceted) surfaces, which is itself essentially equivalent to saying that any water vapor molecule striking a rough surface is immediately indistinguishable from other surface molecules.

Being biased by such a physically pleasing picture, we constrained all the basal data by assuming $A = 1$, then fitting the data to an attachment coefficient with the functional form $\alpha_{basal}(\sigma_{surf}) = e^{-\sigma_0/\sigma_{surf}}$ to determine $\sigma_0(T)$. The prism data were similarly consistent with $A = 1$ when $T \leq -10$ C, so this same fitting procedure was applied at these low temperatures, as shown in Figure 7.18. The higher temperature prism data were better described with $A < 1$, and a two-parameter fit gave the results shown in Figure 7.18.

Since publishing these data in [2013Lib], I have begun to suspect that $A = 1$ might actually provide a better description of the prism attachment kinetics over the entire temperature range $-40C < T < -2C$ (and perhaps beyond that range). My reasons for this change of thinking include:

1) On purely theoretical grounds, having $\alpha \rightarrow 1$ as $\sigma_{surf} \rightarrow \infty$ provides the most sensible picture for ice growth dynamics, regardless of the model specifics. I can think of no good theoretical reason that $A = 1$ should apply over a broad range of growth conditions, but then not on prism facets when $T > -10$ C.
2) The VIG growth measurements of prism facets are more susceptible to unmodeled systematic errors than with basal facets. The basal measurements are made using thin plates with typical thicknesses of 2-4 microns, while the prism measurements are made using columnar crystals with diameters of 20-40 microns. Systematic effects from both diffusion and heating are greater, and more difficult to subtract out, with thicker crystals.
3) The possibility of unmodeled systematic errors are also greater at higher temperatures, when the supersaturations are smaller than at lower temperatures.
4) Similarly, accurate growth measurements are especially difficult when $\alpha$ is high, again increasing the risk of unmodeled systematic errors in determining $A$ with precision.

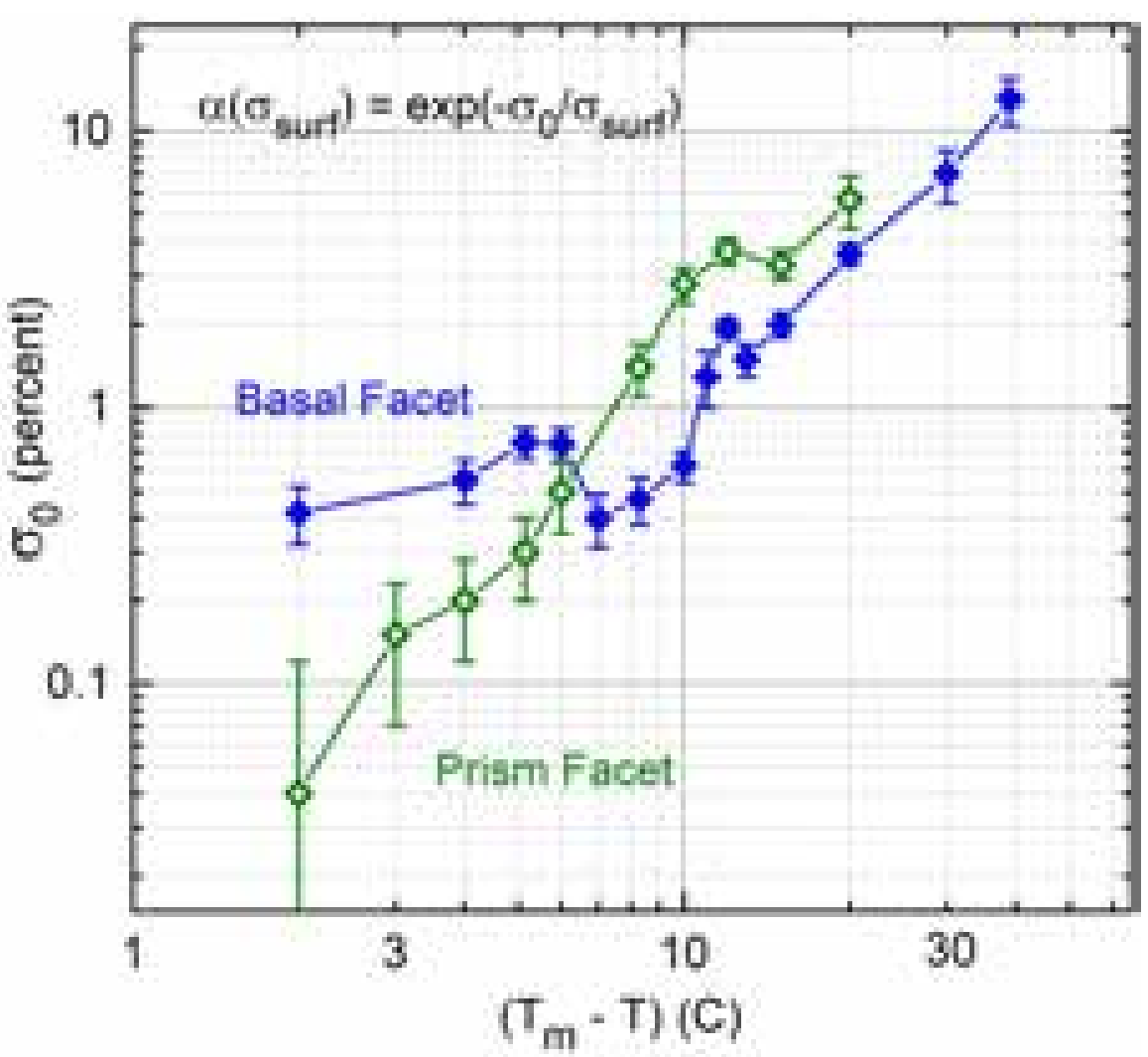

**Figure 7.19: A reanalysis of the data in Figure 7.18, now assuming $A = 1$ for both the basal and prism facets, yields the measured $\sigma_0(T)$ shown above. This analysis reflects a theoretical bias toward a purely nucleation-limited growth model with $A = 1$, plus the suspicion of possible unmodeled systematic errors in the prism data at high temperatures.**

Upon further analysis of the VIG data, I have not found any systematic effects that are clearly large enough to produce $A = 1$ for all the data, so I stand by our original results shown in Figure 7.18. Nevertheless, some of the diffusion and thermal effects are too large for comfort, and I now believe that having $A = 1$ over the entire data set is not definitely excluded.

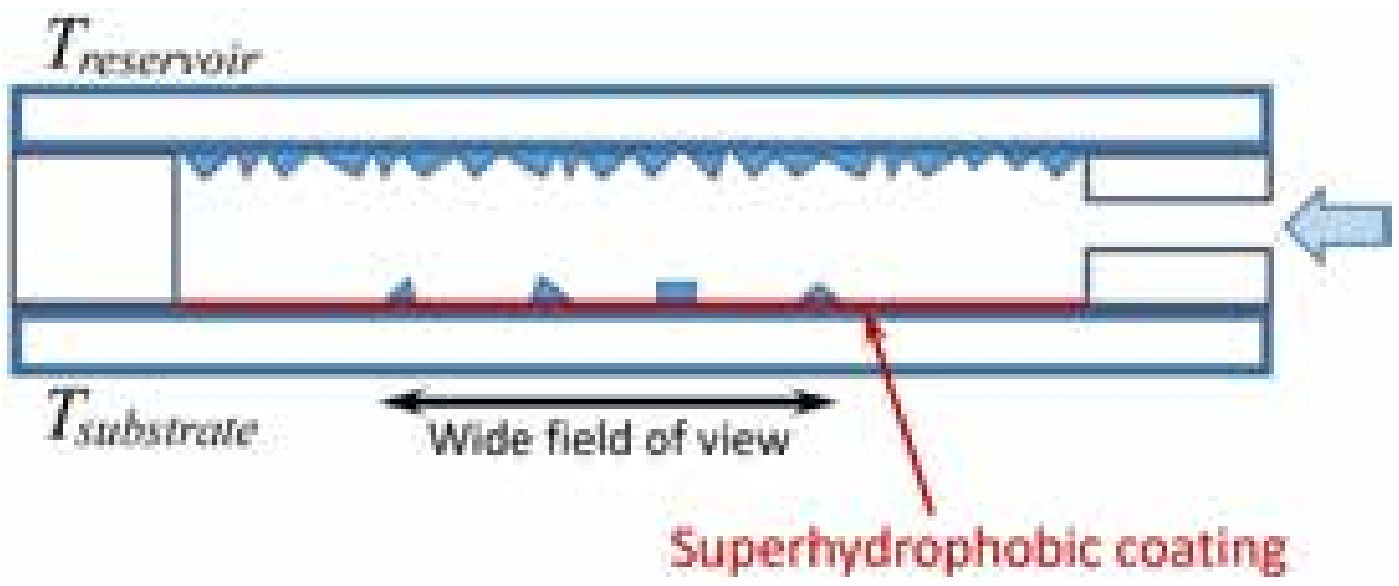

**Figure 7.20: A possible experimental setup for observing the growth of small ice crystals on a superhydrophobic surface. Once the temperatures are stable, an expansion nucleator (blue arrow) injects a pulse of nano-ice-crystals into the test chamber, where some land on the coated sapphire substrate at random positions with random orientations, where a wide-field camera records their subsequent growth. All diffusion-related corrections are reduced when the crystal sizes are in the few-micron range, allowing accurate measurements even at fairly high gas pressures. The compact chamber size allows rapid turn-around between nucleation pulses for efficient experimental throughput.**

Following this reasoning, I have reanalyzed the VIG data assuming $A = 1$ throughout, and the resulting fits for $\sigma_0(T)$ are shown in Figure 7.19. The values of $\sigma_0(T)$ for a faceted prism surface are somewhat higher than in Figure 7.18, but the overall trends remain unchanged. With this simplified model, the attachment kinetics at low background-gas pressures are reduced to a single function $\sigma_0(T)$ for each of the basal and prism facets.

A principal advantage of the reanalysis in Figure 7.19 is that it provides a simpler, one-parameter model for faceted growth. It may be an oversimplified model, but for now it gives a reasonable representation of the facts that is easier to think about than the two-parameter model. Moreover, snow crystal growth is mainly limited to quite low supersaturations, so the value of $\sigma_0$ is more important than the value of $A$. Reducing $A$ mainly affects fast-growth scenarios that do not generally apply in most realistic circumstances. For much of the remaining discussion, therefore, I will assume $\alpha(\sigma_{surf}) = e^{-\sigma_0/\sigma_{surf}}$ and the data in Figure 7.19 as my default model for ice growth at low background gas pressures.

## The Superhydrophobic Future

As transparent superhydrophobic coatings become available for sapphire substrates, it becomes possible to develop improved techniques for measuring ice growth over a broader range of conditions, building upon what was learned from the VIG experiment just described. Figure 7.20 illustrates one possibility that would allow: 1) measuring many crystals in parallel; 2) observing growth as a function of background gas pressure and species; 3) use of the smallest crystals that can still be observed optically; 4) avoiding substrate interactions using a superhydrophobic coating; 5) minimal heating effects for ice crystals on a sapphire substrate; 6) observations over a wide temperature range; and 7) rapid turnaround allowing measurements of many crystals.

One noteworthy feature of this potential experimental set-up is the rapid thermal diffusion time within the growth chamber. Assuming a 1-mm separation between the reservoir and substrate surface, the supersaturation profile will equilibrate in a diffusion time $\tau \approx L^2/D \approx 50$ msec at one atmosphere, and much faster at lower pressures. Thus, the nucleation pulse will produce only a very temporary change in growth conditions within the chamber, quickly settling to give the calculated supersaturation near the substrate.

If superhydrophobic coatings live up to expectation and largely eliminate substrate interaction effects, this and other experiments could allow observations of substantially smaller crystals than the VIG experiment, thus reducing diffusion effects and allowing quantitative measurements of the attachment kinetics as higher gas pressures.

## 7.3 Ice Growth in Air

Taken at face value, the VIG measurements in Figure 7.19 are in direct conflict with the well-established snow-crystal morphology diagram. Comparing $\alpha_{basal}$ and $\alpha_{prism}$ from the measured $\sigma_0(T)$ in Figure 7.19 predicts plate-like growth at -5 C and columnar growth at -15 C, which is opposite from what is seen in the morphology diagram. Although this discrepancy might be attributed to some major flaw in the VIG experiment, the more likely explanation is that the attachment kinetics is strongly affected by a background gas of air at one atmospheric pressure. This opens up a whole new dimension of pressure-dependent kinetic effects, and I discuss this topic at length in Chapter 4. For the present discussion, however, I want describe potential measurements of the growth of small ice prisms in air and how such data can contribute to our understanding of the molecular attachment kinetics.

The most obvious approach to measuring ice growth rates in air is to observe small prismatic crystals resting on substrates. The basic idea is similar to the VIG experiment described above, except with an added partial pressure of air or some other gas in the chamber. Such measurements have been attempted [1984Kur1, 1982Bec1, 1982Bec2, 1983Bec3], but I remain skeptical that they are sufficiently free of systematic errors to be reliable [2004Lib]. As described above (also see [2012Lib]), the corrections from diffusion effects and substrate interactions are already substantial at 20 mbar, and they become progressively worse at higher pressures. Perhaps a suitable superhydrophobic surface will greatly reduce the deleterious effects from substrate interactions, but such a magic surface has not yet been demonstrated. Without such a technological solution, I fear it may be nearly impossible to separate the attachment kinetics from other processes affecting ice growth.

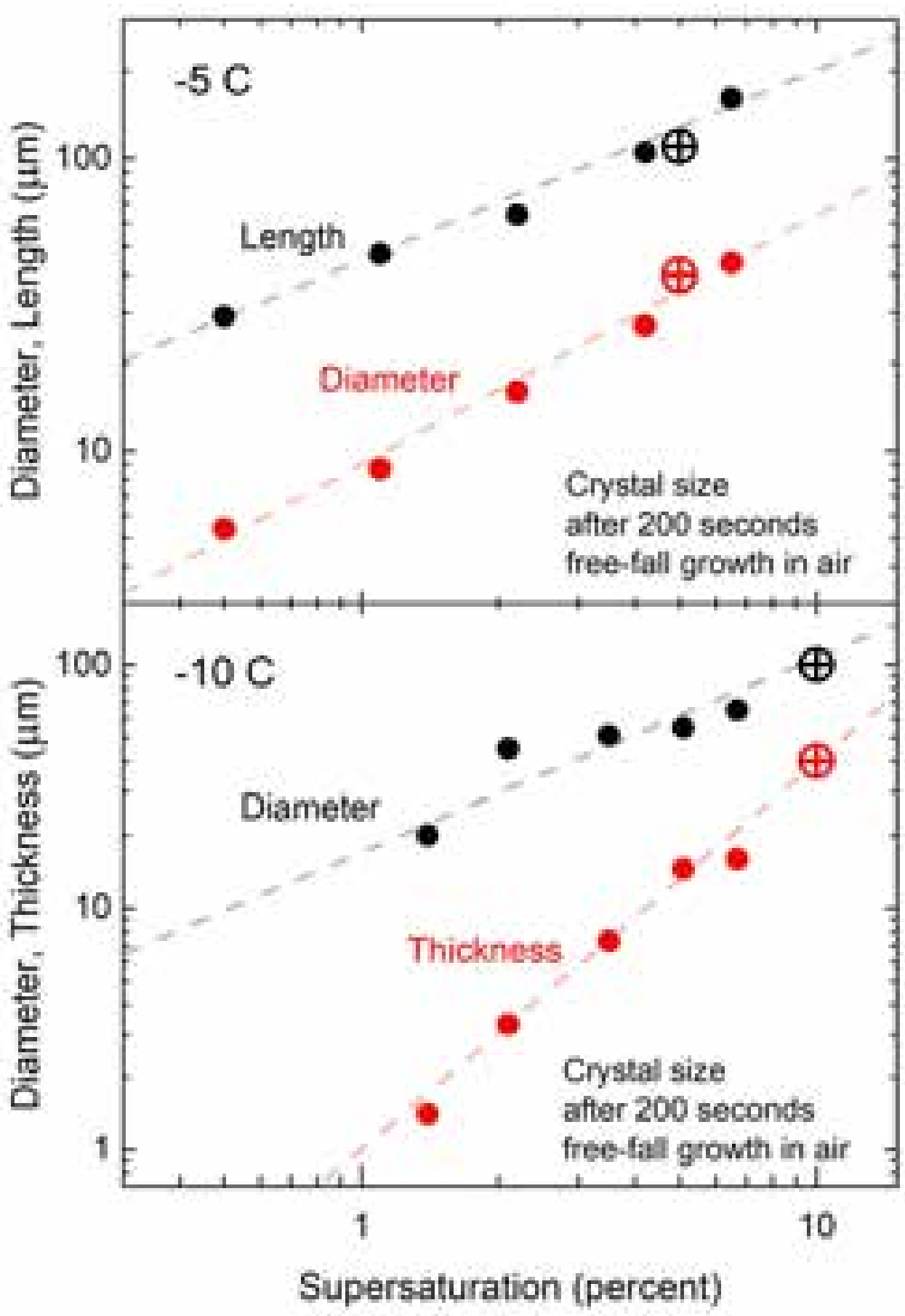

**Figure 7.21: Sizes of ice crystals after 200 seconds of growth in free fall through normal air, as a function of the background supersaturation $\sigma_\infty$. (not equal to $\sigma_{surf}$). The solid points are from [2009Lib], and the cross-circle points are from Yamashita in [1987Kob] (reproduced in Figure 6.22). Note that for both columnar crystals at -5 C and plate-like crystals at -10 C, the forms become more isometric with increasing supersaturation.**

### Free-Fall Growth Data

Another experimental approach is to avoid the use of a substrate altogether and observe ice crystals that are levitating or have experienced growth in free fall, as described in Chapter 6. The diffusion and heating effects are still quite large in these experiments, but the necessary corrections are manageable as long as $\alpha < \alpha_{diff}$, as shown in Chapter 3. Even with quite small crystals, satisfying this inequality limits the regions of parameter space that can be gainfully explored, but that is the nature of the beast.

There are essentially no useful levitation measurements to date, and remarkably little free-fall data either, but Figure 7.21 shows some examples. The basic idea is to nucleate some ice crystals at $t = 0$ in a chamber filled with air at a known supersaturation, let them grow for a fixed amount of time, and then sample some crystals and measure their sizes. Crystal-to-crystal size variations are typically a factor of two, which probably reflects some spatial variation in supersaturation within the growth chamber, averages being shown in Figure 7.21. As we will discuss below, the supersaturation is difficult to determine precisely, and is subject to a host of possible systematic errors, but for now we take the data in this figure at face value.

The Yamashita data probably have the most reliable supersaturation, as these crystals were grown in a cloud chamber containing a fog of cloud droplets in thermal equilibrium with the air. As long as the droplet number density is much higher than the crystal number density in the air, these conditions should yield $\sigma_\infty \approx \sigma_{water}$ with reasonable accuracy. The supersaturation in the [2009Lib] experiment was determined by differential hygrometry, which is more prone to systematic errors, especially at lower supersaturations.

One immediate take-away from these data is that the aspect ratios of the crystals become more extreme at lower supersaturations. That is, the diameter/thickness ratio of the plates at -10 C is largest at the lowest supersaturations observed, and so too is the length/diameter ratio of the columns at -5 C (although the supersaturation dependence is greater at -10 C). This behavior is often displayed in snow crystal morphology diagrams, which (incorrectly) show blockier crystals at low supersaturations. However, it is consistent with a kinetics model having $\alpha(\sigma_{surf}) = e^{-\sigma_0/\sigma_{surf}}$ for both the basal and prism facets (with a different $\sigma_0$ for each), as described above. This lends support to the notion that this simple parameterization in terms of $\sigma_{0,basal}(T)$ and $\sigma_{0,prism}(T)$ may provide a reasonable picture of ice growth in air as well as in near vacuum.

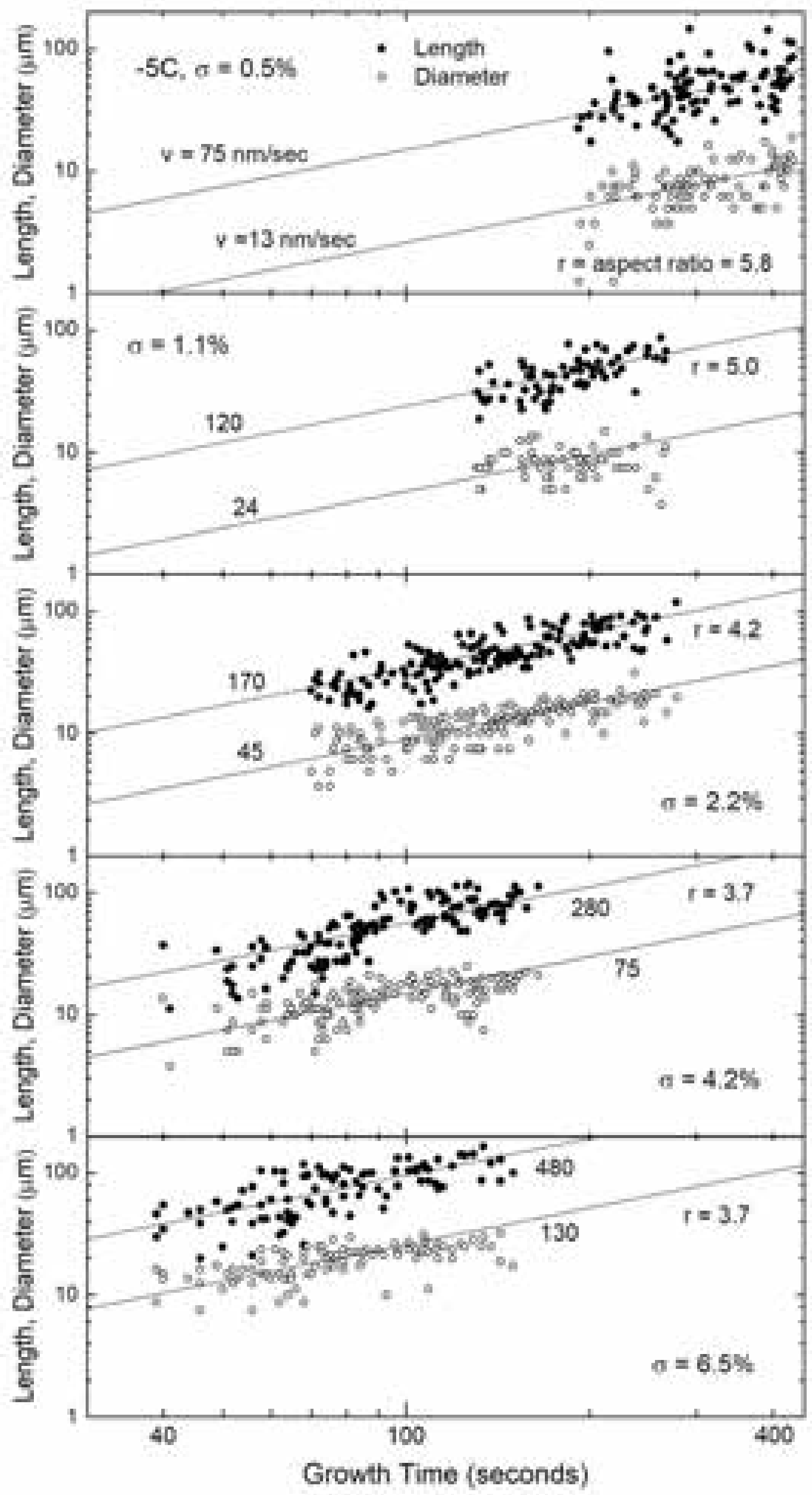

**Figure 7.22: Free-fall growth data at -5 C from [2009Lib], showing crystal sizes as a function of fall times for several different supersaturations (here $\sigma = \sigma_\infty$, which is generally not equal to $\sigma_{surf}$). Each pair of points (length and diameter) represent one observed crystal. Lines show constant-velocity trajectories, and the aspect ratio is the ratio of these velocities. Note that the aspect ratio tends toward unity with increasing supersaturation.**

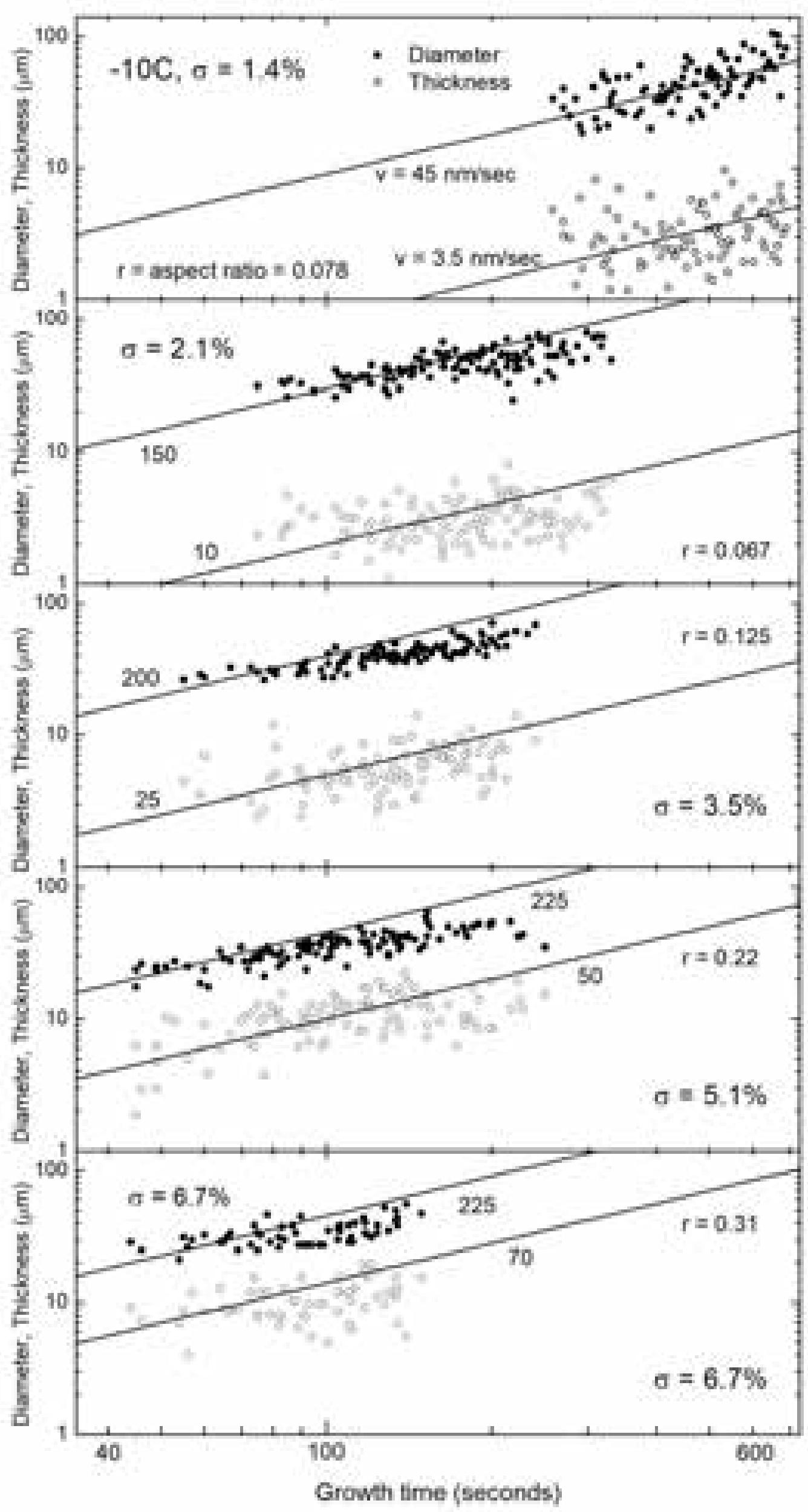

**Figure 7.23: Free-fall growth data at -10 C from [2009Lib], showing crystal sizes as a function of fall times for several different supersaturations (here $\sigma = \sigma_{\infty}$, which is generally not equal to $\sigma_{surf}$). Each pair of points (diameter and thickness) represent one observed crystal. Lines show constant-velocity trajectories, and the aspect ratio is the ratio of these velocities. Note that the aspect ratio tends toward unity with increasing supersaturation.**

The [2009Lib] data in Figure 7.21 are shown in expanded form in Figures 7.22 and 7.23, illustrating the full time dependence of the growth behaviors.

## Diffusion Corrections in Air

The growth data in Figure 7.22 provide a good example of the severity of the diffusion correction in air at one atmosphere. Using the monopole approximation described above, the supersaturation at the crystal surface is estimated as $\sigma_{surf} = \sigma_{\infty} - \delta\sigma$ with

$$\delta\sigma = \frac{dV/dt}{4\pi R_0 X_0 v_{kin}} \qquad (7.18)$$

where $dV/dt$ is the rate of change of the volume of the growing crystal, $X_0 \approx 0.15\ \mu m$ in air at one bar, and $R_0$ is an effective radius of the crystal. The value of $R_0$ is not well defined for a non-spherical crystal, but a reasonable approximation is to set the total crystal volume (known from measurements) to $V = (4/3)\pi R_0^3$. For the $\sigma_{\infty} = 0.5\%$ data (top panel in Figure 7.22), the monopole correction gives $\delta\sigma \approx 0.2\%$ at a growth time of 200 seconds. This is a substantial correction, given that the uncertainly in the nominal value of $\sigma_{\infty}$ is at least 0.1% [2009Lib].

Plunging forward and applying the monopole correction gives $\sigma_{surf} \approx 0.3\%$, and using the measured velocities in Figure 7.22 then yields $\alpha_{basal} \approx 0.05$ and $\alpha_{prism} \approx 0.009$ at $\sigma_{surf} \approx 0.3\%$,, consistent with the result from the modeling analysis in [2009Lib].

Moreover, the $\alpha_{basal}$ result is about equal to that expected from the vacuum ice growth measurements, $\alpha_{basal,vacuum} \approx \exp(-0.75\%/\sigma_{surf}) \approx 0.08$. Of course, the measured $\alpha_{prism}$ here is much smaller than $\alpha_{prism,vacuum}$, as the vacuum measurements do not even indicate columnar growth at -5 C.

The corrections become much larger as $\sigma_\infty$ goes up, however. Calculating the same monopole correction using the $\sigma_\infty = 6.5\%$ data yields $\delta\sigma \approx \sigma_\infty$, making it impossible to accurately estimate $\sigma_{surf}$. This simply tells us that the growth at $\sigma_\infty = 6.5\%$ is so strongly diffusion limited that even small uncertainties in $\sigma_\infty$ yield large changes in the corrected $\sigma_{surf}$, and thus wild uncertainties in the extracted $\alpha$ values. We did a direct analysis in [2009Lib], but a simple monopole analysis now tells us that this analysis was likely strongly influenced by small systematic errors in determining $\sigma_\infty$.

The moral in this story is one needs to be extremely cautious in analyzing ice growth data when the diffusion corrections are large, as I have stressed throughout this book. We used a proper numerical diffusion analysis in [2009Lib], but did not do a proper analysis of possible systematic errors from uncertainties in $\sigma_\infty$. My current thinking is that the data in [2009Lib] are fine, but the analysis to produce $\alpha_{basal}$ and $\alpha_{prism}$ values for the higher $\sigma_\infty$ data was flawed. I have reached a similar conclusion regarding essentially all of the early ice-growth measurements in air: the data are fine, but it is not possible to extract useful information about the attachment coefficients from a direct diffusion analysis, because the corrections are too large.

This conclusion follows as well from the simple spherical analysis presented in Chapter 3. When $\alpha_{diff} \ll \alpha$, the growth is strongly diffusion limited, and this means that the growth velocities are essentially independent of $\alpha$. As a result, one cannot extract much information about $\alpha$ from growth data when $\alpha_{diff} \ll \alpha$. Taking data at low pressures ameliorates this problem by increasing $\alpha_{diff}$, but this is not an option if one wishes to explore the attachment kinetics as a function of background gas pressure. Using the smallest possible ice crystals is desirable, as this also increases $\alpha_{diff}$. The technique used by Huang and Bartell [1995Hua] has great potential in this direction, but creating a well-known supersaturation will be a challenge.

## Anisotropy Analysis

Given the data at hand, we can obtain some useful information by analyzing the aspect ratios in the growth measurements. The aspect ratio of a crystal is largely determined by the ratio of the attachment coefficients, $\alpha_{basal}/\alpha_{prism}$, and this is true even when the overall growth is strongly diffusion limited. For the case of a simple ice prism, the anisotropy is defined simply by the length/diameter ratio, which is easily measured. (I often take the diameter to mean the distance between opposing prism facets, as this is a reasonable approximation. Doing a full hexagonal-prism analysis is unwarranted at present, given that the large measurement uncertainties in supersaturation, and supersaturation corrections, are vastly more important than the small geometrical correction that arises from treating a hexagonal prism as a simple cylinder.)

To see how one might apply an anisotropy analysis, consider once more the data in Figure 7.22, and assume that $\alpha_{basal}(\sigma_{surf})$ is already known from another source. With this assumption, one can use the basal surface as a "witness surface" to extract $\sigma_{surf}$ from the growth velocity using $v_{basal} = \alpha_{basal}(\sigma_{surf})v_{kin}\sigma_{surf}$. With this extracted value of $\sigma_{surf}$, the prism attachment coefficient is then simply obtained using $\alpha_{prism}(\sigma_{surf}) = v_{prism}/v_{kin}\sigma_{surf}$. Generally, this analysis only works well for small, simple prisms, where the diffusion corrections are not exceptionally large or complex, but I believe it can provide some useful insights to guide further experimental efforts.

## The “Well-Behaved-Basal” Model

Having examined quite a lot of ice-growth data over the years, I believe that the evidence supports two hypotheses: 1) the attachment kinetics depends on air pressure, and 2) the prism attachment coefficient changes more dramatically with pressure than the basal attachment coefficient. Moreover, the basal growth in both normal air and at low pressures might be described by the same $\sigma_{0,basal}(T)$ function, equal to that shown in Figure 7.19. The data certainly do not exclude some change in $\sigma_{0,basa}\ (T)$ with pressure, but it might not be a large change. At the same time, the change in $\sigma_{0,prism}(T)$ with pressure appears to be quite large. I will not expound at length on the evidence supporting these claims, because it is not a strong case. But there appears to be some truth in these statements, so perhaps the reader will allow me to continue thinking in this direction.

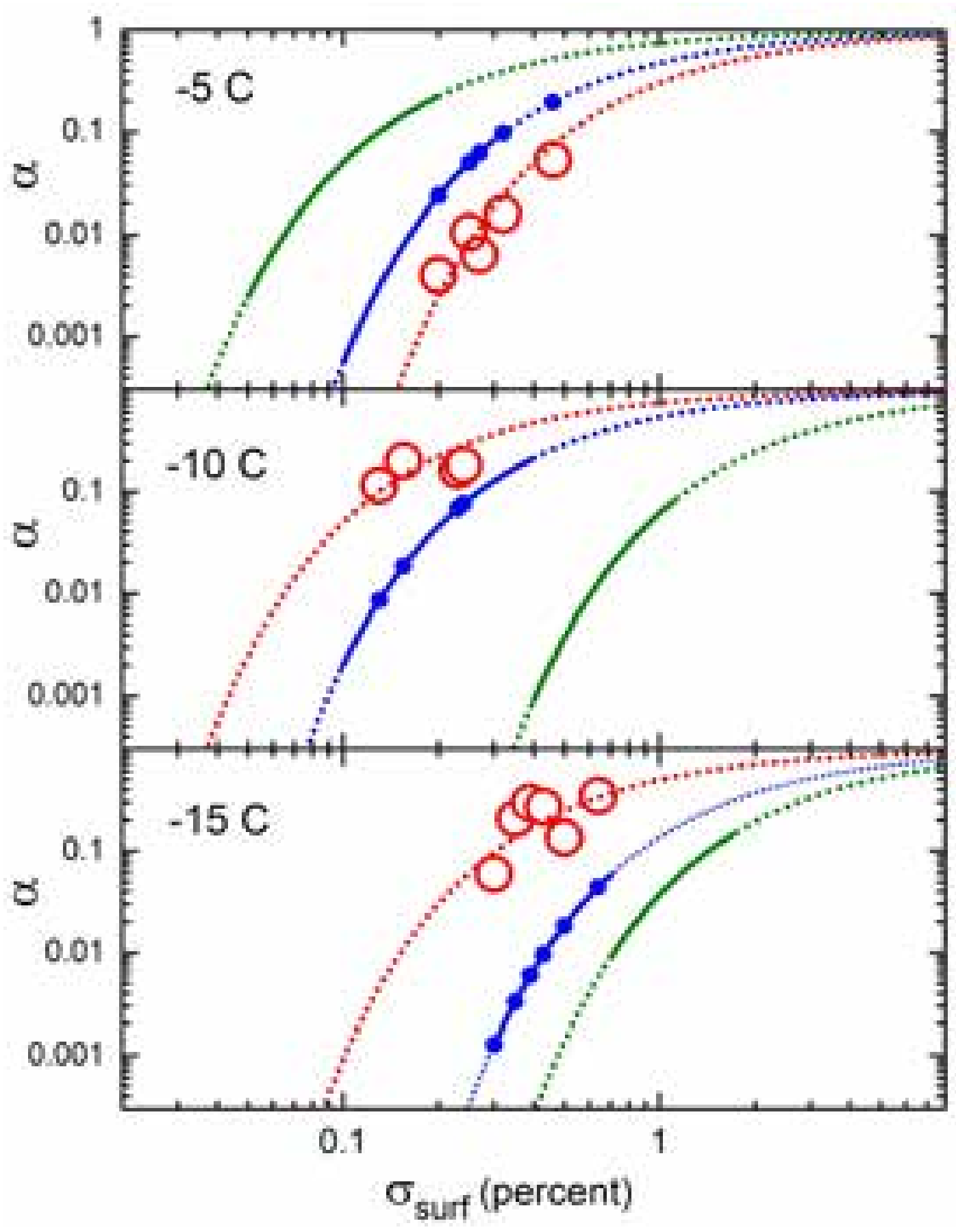

**Figure 7.24: Inferences (red circles) of $\alpha_{prism}(\sigma_{surf})$ from free-fall growth data, assuming the “Well-Behaved-Basal” model described in the text. The blue lines show the assumed $\alpha_{basal}(\sigma_{surf})$ from Figure 7.19 that were used to compute $\sigma_{surf}$ from basal velocity data. The green lines show $\alpha_{prism}(\sigma_{surf})$ from Figure 7.19, representing growth at low pressure, while the red lines show growth in air.**

The next step is to define a “Well-Behaved-Basal” (WBB) model, where I simply assume that the function $\sigma_{0,basal}(T)$ shown in Figure 7.19 applies independent of air pressure. Clearly this is just a rough approximation, but making this assumption allows us to draw some concrete inferences using the anisotropy analysis of ice-growth data in air.

The basic idea is as described above – use the basal growth velocity to determine $\sigma_{surf}$ and then use the prism growth velocity to determine a corresponding $\alpha_{prism}$ at that value of $\sigma_{surf}$. Applying this procedure using the growth data in Figure 7.22 then yields the results in the top panel in Figure 7.24. Other data from a variety sources (notably [2009Lib] and [2008Lib1] were used to determine $\alpha_{prism}(\sigma_{surf})$ values at other temperatures.

Putting all analysis together yields the comprehensive model of the attachment kinetics shown in Figure 7.25. This model assumes a purely nucleation-limited growth model, so $\alpha(\sigma_{surf}, T)$ is defined solely by $\sigma_0(T)$ for both facets. It further assumes that $\sigma_{0,basal}(T)$ does not depend on air pressure, which is a somewhat sketchy assumption. Although clearly just a first step toward a final, pressure-dependent model of the attachment kinetics, this model already suggests rather substantial changes in $\sigma_{0,prism}(T)$ with air pressure are required in any future model.

Because the WBB model is so speculative at this point, its main function is to direct future experiments toward the most informative areas of phase space. For example, Figure 7.25 suggests that measuring $\sigma_{0,prism}(T)$ as a function of background gas pressure may be especially fruitful near -15 C, where the changes are likely most dramatic. More importantly, the model indicates that many additional growth measurements as a function of background pressure will generally be needed to further our understanding of snow crystal growth dynamics.

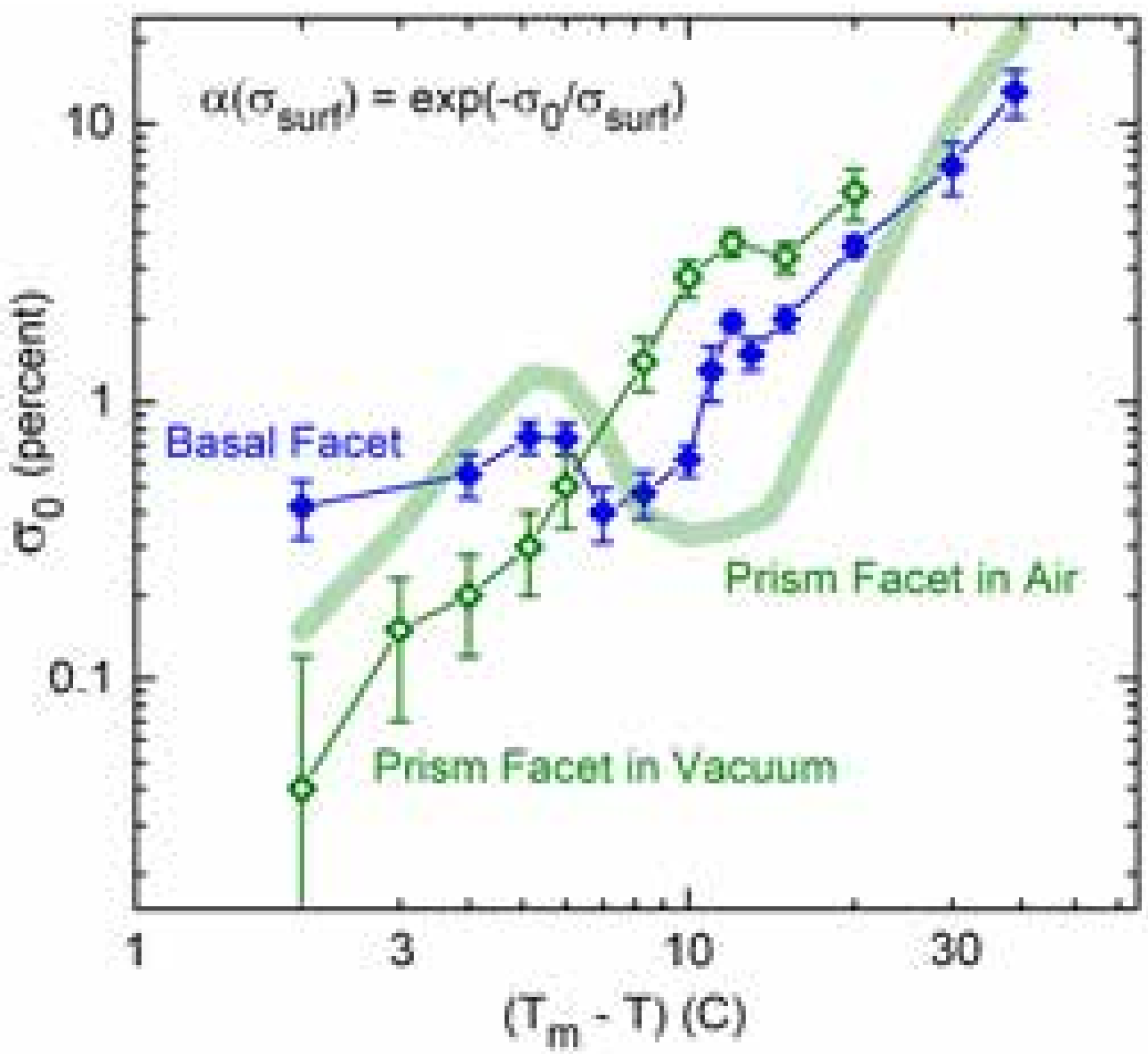

**Figure 7.25: A comprehensive model of the ice attachment kinetics in both air and near vacuum. This model assumes that the $\alpha_{basal}(\sigma_{\text{surf}})$ (given by the blue data) are independent of air pressure. The green data points then give $\alpha_{prism}(\sigma_{surf})$ at low pressure, while the thick green line shows $\alpha_{prism}(\sigma_{surf})$ in air at one atmospheric pressure. The physical origin of the large change in $\alpha_{prism}(\sigma_{surf})$ with air pressure is discussed in Chapter 4.**

## 7.4 Analysis of a Free-Fall Ice-Growth Experiment

As described above, there are remarkably few measurements of ice growth rates that can be used to determine the attachment coefficients as a function of background gas pressure. There is much room for improvement in this area, and Figure 7.26 illustrates one possible route to obtaining useful growth data in reasonably well controlled conditions. As an example of experimental design and the identification of systematic errors, this section will examine the kinds of measurements one could obtain with this apparatus.

To begin, a free-fall growth experiment avoids unwanted systematic effects from substrate interactions, which can be quite detrimental for obtaining accurate measurements on small ice crystals. A linear-gradient diffusion chamber is advantageous for this task because the temperature and supersaturation can be calculated with good accuracy, especially at low supersaturation values. Although both $T$ and $\sigma_\infty$ vary with position within the chamber, these inhomogeneities are probably tolerable for a basic experiment aimed a surveying overall trends. Uniform growth conditions would be more desirable, of course, but realizing this in an actual apparatus is not a trivial task.

Another beneficial feature of the linear-gradient chamber is that the air inside is stable with regard to convection, so seed crystals inserted near the drop point in Figure 7.26 will quickly reach terminal velocity and drift slowly downward as they grow, until some land on the substrate at the bottom of the chamber for observation. The resulting data will thus resemble those shown in Figures 7.22 and 7.23. As described in Chapter 6, the final crystal size will be roughly proportional to $H^{1/4}$, where $H$ is the fall distance, yielding sizes in the 10-20 μm range for $H = 20$ cm, depending on crystal morphology. This is a reasonable goal, as the crystals need to be large enough for optical imaging, but smaller crystals are better suited for reducing both diffusion and heating effects, as we will see below.

### Large-Scale Diffusion

As described in Chapter 6, the temperature profile inside a linear-gradient diffusion chamber is given by

$$T(\vec{x}) = T_{bottom} + \Delta T \cdot (z/L) \qquad (7.19)$$

where $z$ is vertical distance above the bottom of the chamber, $L$ is the chamber inner height, and $\Delta T = T_{top} - T_{bottom}$. If the chamber walls were moved outward to infinity while keeping $L$ constant, the water-vapor number density in the chamber would likewise be

$$c(\vec{x}) = c_{sat}(T_{bottom}) + \Delta c \cdot (z/L) \qquad (7.20)$$

where $\Delta c = c_{sat}(T_{top}) - c_{sat}(T_{bottom})$. With closer chamber walls as shown in Figure 7.26, $c(\vec{x})$ can be calculated from numerical modeling, yielding a broad maximum in $\sigma_\infty(\vec{x})$ near the center of the chamber (see Chapter 6). Interestingly, the modeling results are all independent of background gas pressure, as long as the molecular mean-free-path remains small compared other lengths in the problem. Thus $\sigma_\infty(\vec{x})$ will also be pressure-independent, making the linear-gradient diffusion chamber well-suited to observing growth rates as a function of pressure.

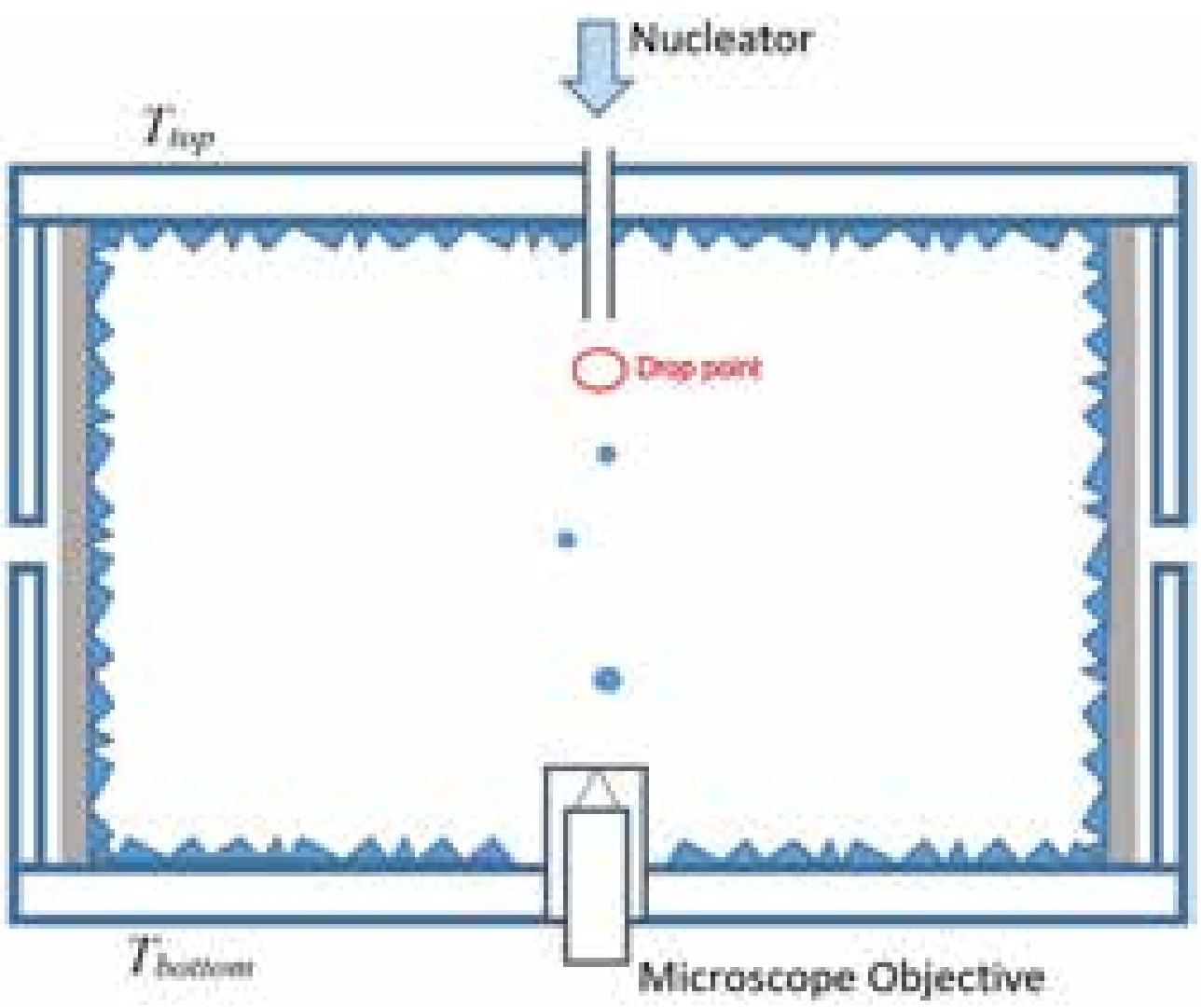

**Figure 7.26: A linear-gradient free-fall growth chamber (see Chapter 6) for observing ice growth rates as a function of temperature, supersaturation, and background gas pressure. A nucleator carefully places crystals near the "drop point" shown, and from there they grow and slowly fall onto the substrate for observation. To date, even basic trends in snow-crystal growth morphologies as a function of background gas pressure have been characterized, presenting a sizable gap in our knowledge of the underlying attachment kinetics.**

The flux of water vapor from the top of the chamber to the bottom does depend on pressure, being equal to $F = D\nabla c = D\Delta c/L$. Converting this flux to a sublimation velocity of ice from the top surface yields

$$v_{sublimation} \approx \frac{D\Delta c}{Lc_{ice}} \qquad (7.21)$$

If we take $L = 30$ cm, then $v_{sublimation}$ is about 0.2 nm/sec at 1 bar, meaning that the ice layer on the top surface will hardly change during the course of an experimental run. However, because $D$ is proportional to $P^{-1}$, this increases $v_{sublimation}$ to about 10 nm/sec at 20 mbar, so a 1-mm-thick ice layer will depleted after 24 hours. This should not be a problem, as long as the top surface is coated with a sufficiently thick ice layer at the beginning of a run.

The diffusion chamber also needs time to relax to its steady-state condition, which takes about $\tau \approx L^2/D \approx 1$ hour at 1 bar. Because a nucleation pulse releases crystals that absorb water vapor and reduce $\sigma_\infty$ in their vicinity, the chamber will have to be re-equilibrated for a time $\tau$ between successive nucleation events. Fortunately, $\tau$ is proportional to $P$, so the recovery times are substantially shorter at lower pressures. The recovery will also be quicker between runs if the nucleator carefully places just a small number of seed crystals at the drop point (easier said than done, alas).

## Crystal Crowding

The absorption of water vapor by a multitude of growing ice crystals can present an important perturbation of the supersaturation within a free-fall growth chamber. The crystals remove water vapor in a time that is short compared with the equilibration time $\tau$, so the actual $\sigma_\infty(\vec{x})$ within a cloud of growing crystals may be substantially less than that derived from steady-state modeling, which assumed that no crystals were present.

To see the magnitude of this perturbation, consider dropping a single crystal through a free-fall chamber, where it falls a distance $H$ and grows to an ice volume equal to $V$, at which point it lands on a substrate and is measured. After all the crystals in the chamber have grown and fallen, assume that there is one crystal per area $A$ on the detector on average.

Each observed crystal contains $Vc_{ice}$ water molecules, while the air column above it contains $HAc_{sat}$ molecules (slightly more, because the air is supersaturated, but this additional amount is negligible). The growing crystals thus removed water molecules from the air column and lowered its supersaturation by an amount

$$\delta\sigma \approx \frac{Vc_{ice}}{HAc_{sat}} \tag{7.22}$$

Taking $V \approx (25\mu m)^3$ and $H \approx 20$ cm, then keeping $\delta\sigma < 1$ percent means we must have roughly $A > (1\text{ mm})^2$. Having just a single ice crystal per square millimeter of substrate makes it difficult to accumulate data, especially when one must wait a time $\tau$ between nucleation events. This perturbation of the supersaturation becomes progressively more problematic as the temperature and supersaturation are lowered.

One consolation is that $\delta\sigma$ can be estimated for each nucleation event by adding up volume of crystals per unit area that fall onto the substrate. If the estimated $\delta\sigma$ correction becomes comparable to the modeled $\sigma_\infty$, then one must assume that the actual supersaturation is much lower than expected from the steady-state calculations. This effect need not be a show-stopper, as one can take care to produce a small number of crystals in each nucleation event. But it is certainly a serious concern, similar to the crowding issues discussed in Section 7.1.

This depletion effect may have influenced the data in Figure 7.23, as it was not carefully considered at the time of the experiment. If so, then the actual $\sigma_\infty$ for those data will have been substantially smaller than that indicated, especially for the lower-$\sigma_\infty$ measurements. Identifying systematic errors in ice growth measurements remains an ongoing battle.

## Small-Scale Diffusion

Assuming $\sigma_\infty$ is well-characterized within the diffusion chamber, one must still contend with how particle diffusion contributes to the overall growth behavior. From the spherical-crystal analysis presented in Chapter 3, we know that the growth is largely diffusion-limited if $\alpha_{diff} \ll \alpha$. In this regime, it is essentially impossible to determine $\alpha$ accurately, as small uncertainties in $\sigma_\infty$ translate to large uncertainties in $\alpha$. For the case of ice growth in air, this means that we can only determine kinetic coefficients when

$$\alpha < \alpha_{diff} = \frac{X_0}{R} \approx \frac{0.15\ \mu m}{R} \tag{7.23}$$

which leaves us with a rather restricted region of parameter space to explore. For example, if $R = 10$ μm and an air pressure of 1 bar, then we can only measure attachment coefficients with $\alpha < 0.015$. As seen in Figure 7.24, this means we can likely measure $\alpha_{prism}$ near -5 C and $\alpha_{basal}$ near -15 C, provided $\sigma_{surf}$ is quite low in both cases. The restriction is somewhat less severe at lower pressures, however, because $X_0 \sim P^{-1}$, and the numbers will change in other gases also.

## Heating Effects

Freely falling crystals may also be subject to substantial heating effects, as there is no nearby substrate to absorb the latent heat generated by growth. As with the particle-diffusion problem, the spherical-crystal analysis presented in Chapter 3 and Appendix B allows a reasonable estimate for how heating affects the growth behavior. At low pressures, when $X_0$ is large and particle diffusion is no longer important, heating effects kick in, limiting our ability to measure the attachment kinetics to the regime

$$\alpha < \alpha_{hea} = \frac{\kappa_{air}}{\eta L_{sv}\rho_{ice}v_{kin}R} \approx \frac{0.5\ \mu m}{R} \quad (7.24)$$

Again, taking $R = 10$ μm means we are limited to $\alpha < 0.05$, and again we see from Figure 7.24 that this is a serious limitation.

Because $\kappa_{air}$ is nearly independent of air pressure, we see that there is a cross-over in the diffusion-limited growth of freely falling crystals at a pressure of around $P_0 \approx 300$ mbar. Particle diffusion dominates above $P_0$, while heating effects kick in below $P_0$.

From this analysis, we see that there is essentially no way to measure high $\alpha$ kinetics from growth observations using freely falling ice crystals. The situation improves for crystals on a substrate, as $\kappa_{ice}$ is about 100x larger than $\kappa_{air}$. This allowed us to measure up to $\alpha \approx 0.1$ in the VIG experiment described above, and one could do even better by observing smaller crystals. In the end, it is likely that many different experimental approaches will be needed to create an accurate picture of the ice attachment kinetics.

## Pressure-Dependent Attachment Kinetics

Stepping back and looking at the big picture, we see that understanding the physics of snow crystal growth will require adding a third dimension of pressure to the morphology diagram. In terms of the attachment kinetics, this means obtaining accurate measurements $\alpha(\sigma_{surf}, T, P)$ for both the basal and prism facets. Doing so in different background gases may reveal interesting chemical effects as well, adding yet another dimension to the problem.

For now, looking growth in air, the free-fall experiment just described may provide some important insights. Focusing on measurements at low $\sigma_\infty$, it should be possible to determine $\alpha(\sigma_{surf}, T, P)$ directly from growth velocities, when the heat-diffusion and particle-diffusion corrections are manageable. From that starting point, it should be possible to expand to higher $\sigma_\infty$ measurements using the anisotropy analysis described in connection with Figure 7.24.

Over time, it should be possible to further develop a comprehensive model for $\alpha(\sigma_{surf}, T, P)$, building upon the discussion in Chapter 4. Even if the model is largely empirical in nature, it will provide necessary input for computational models (see Chapter 5), which can then be compared with quantitative observations covering the full range of snow-crystal morphologies (see Chapter 8). Hopefully the molecular physics underlying snow crystal formation will become better known along the way. There is much left to do before we have a good understanding of this fascinating and enigmatic phenomenon.

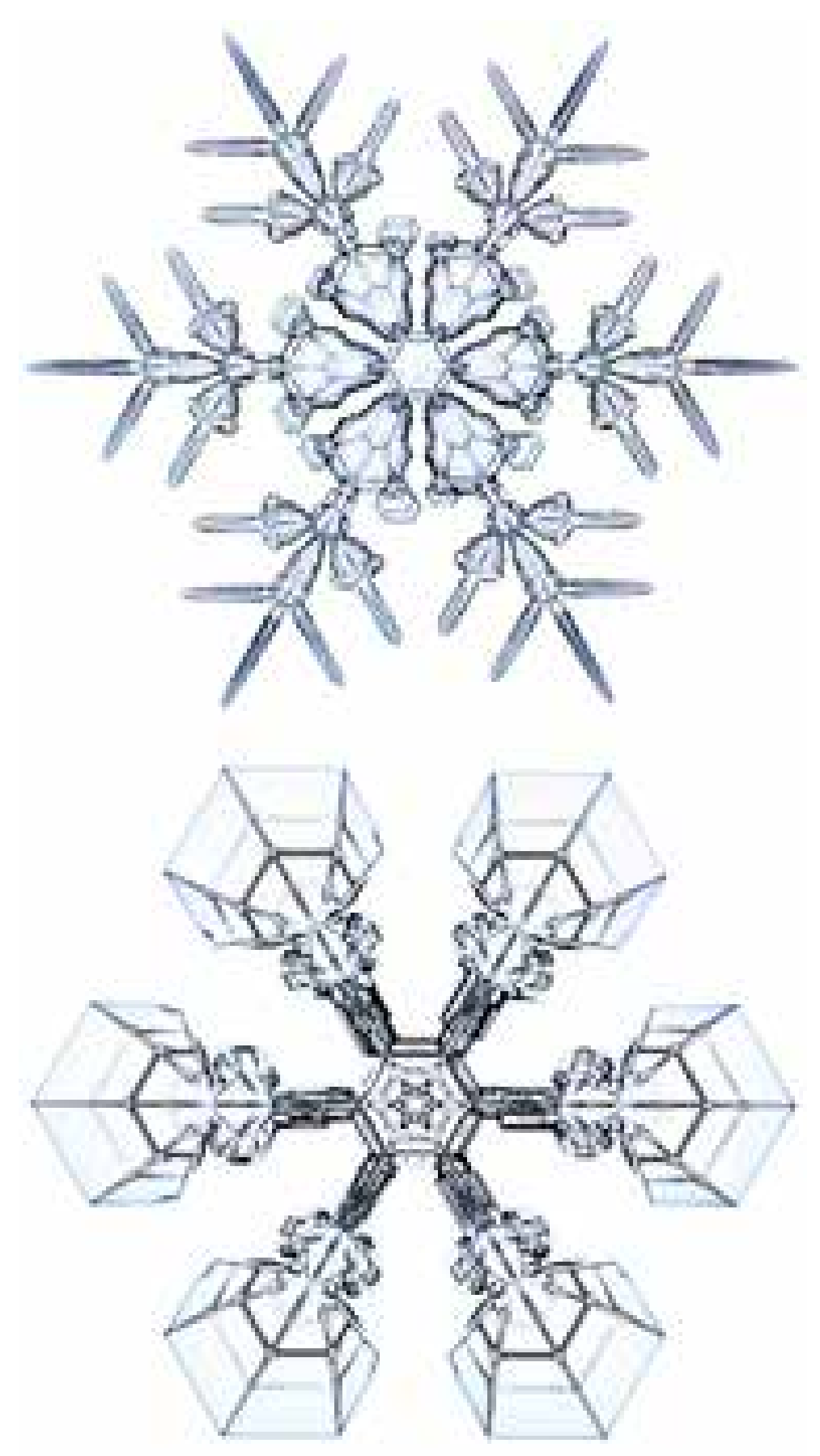

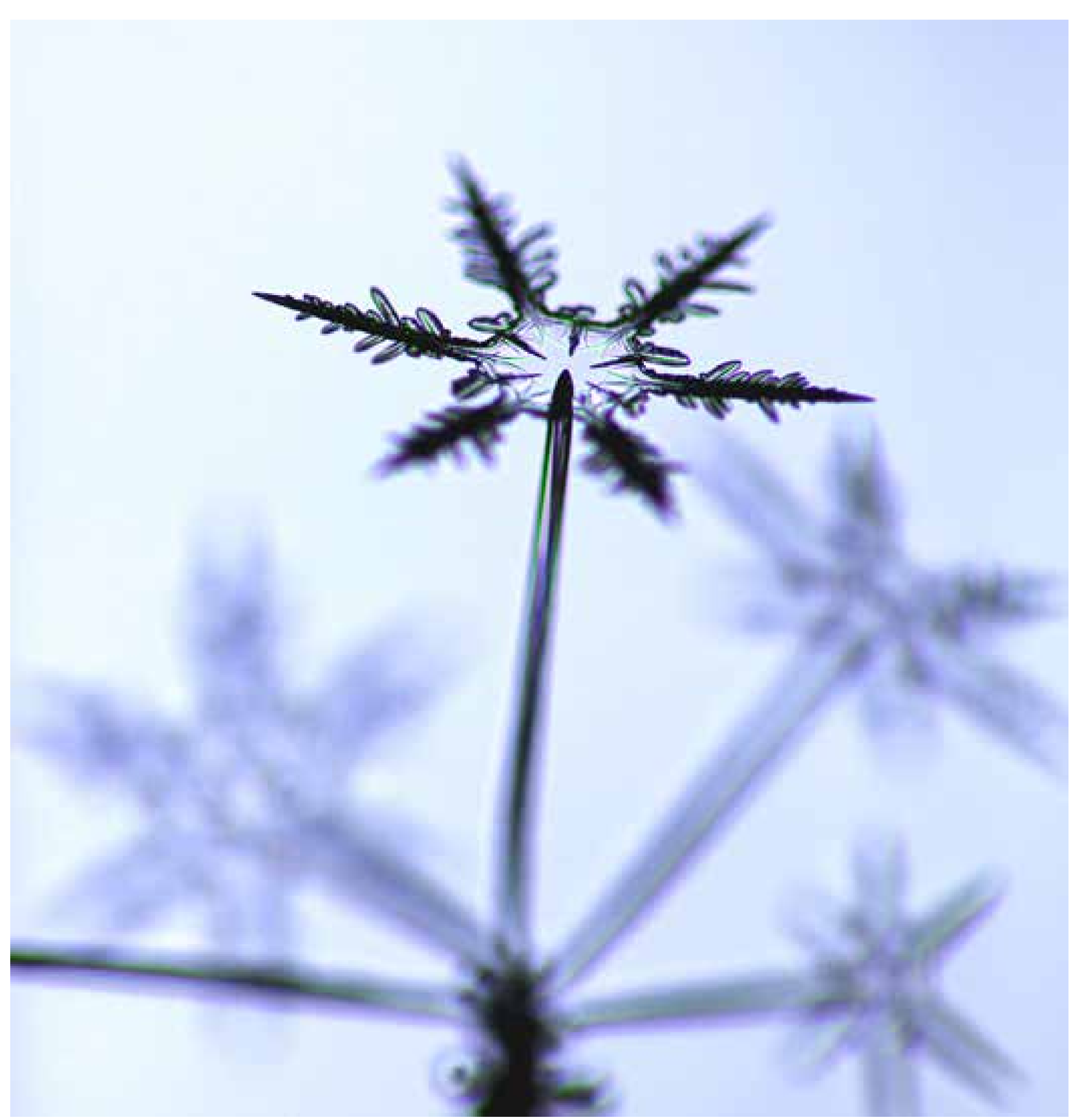

# Chapter 8

# Electric Ice Needles

*The universe is full of magical things*
*patiently waiting for our wits to grow sharper*
– Eden Phillpotts
*A Shadow Passes, 1919*

While the previous chapter looked at growing smaller, simpler snow crystals for investigating the molecular attachment kinetics, this chapter examines a particularly useful technique for producing and studying larger specimens with more complex structures. A primary goal in making these larger crystals is to examine the science of pattern formation in diffusion-limited growth, comparing synthetic snow crystals with their computational counterparts. Producing quantitative models that reproduce both morphologies and growth rates over a broad range of environmental conditions would be a substantial milestone in our understanding of the science of snow crystals.

**Facing page: A collection of stellar snow crystals growing on the ends of slender ice needles. The c-axis needles were created using high electric fields to accelerate normal crystal growth using the techniques presented in this chapter.**

A second goal is purely artistic – to create beautiful ice structures simply to watch them grow and develop, using the rules of snow-crystal growth to sculpt whatever nature allows. Both goals offer much appeal, and both require a nontrivial technical acumen to accomplish. This chapter is about developing a method that allows the formation of high-quality individual snow crystals over a broad range of growth conditions.

## 8.1 A Tool for Creating Isolated Single Crystals

Figure 8.1 shows a typical set of c-axis electric ice needles, which I also call *e-needles*. One begins with a metal wire exposed to highly supersaturated air in a diffusion chamber, so the wire quickly becomes covered with frost crystals. In Figure 8.1, one end of the wire is seen at the bottom of the photograph, covered with a collection of small frost crystals growing on its surface. In this example, the temperature surrounding the wire tip was near -6 C, so the frost crystals grew in random orientations with a generally columnar morphology, as expected from the snow crystal morphology diagram (see Chapter 1). The wire extends down to the bottom of the growth chamber where it exits and is connected to a high-voltage power supply.

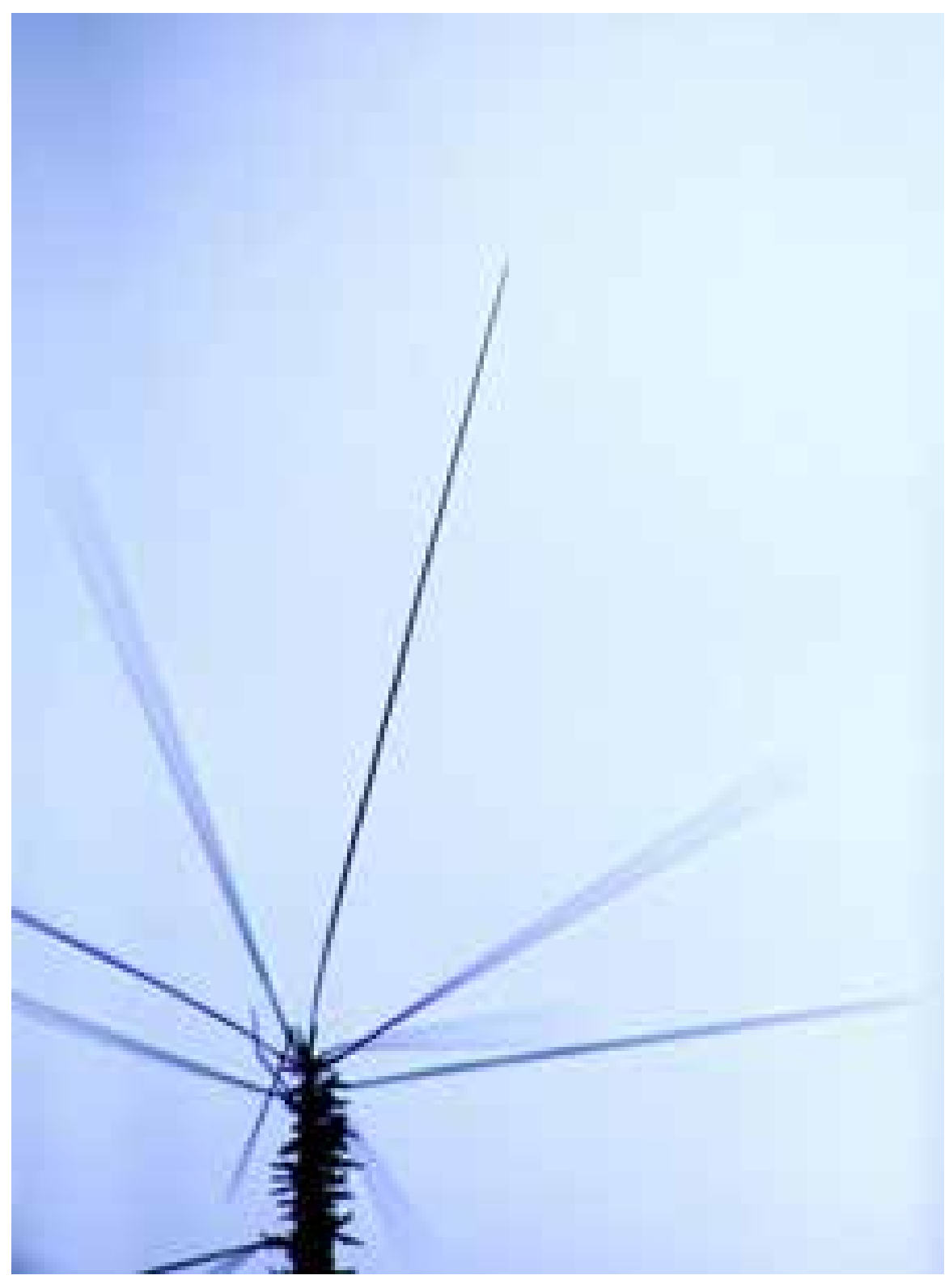

**Figure 8.1. A set of slender "electric" ice needles (e-needles) growing on the end of a frost-covered wire. The e-needle centered in the image is about 3 mm in length. The needles formed when +2000 volts was applied to the frost-covered wire at the bottom of the photo, which was simultaneously exposed to highly supersaturated air at a temperature near -6 C. The e-needle phenomenon is the result of an electrically induced growth instability.**

When a high voltage (typically +2000 volts DC) is applied to the wire, slender ice e-needle crystals spring forth from the frost crystals in just a few seconds, provided that the conditions are right. When the temperature at the wire tip is close to -6 C and the water-vapor supersaturation is near 100 percent, and the air contains trace quantities of acetic acid vapor, then the e-needles will typically emerge growing along the crystalline c-axis with tip velocities of around 100-150 microns/second

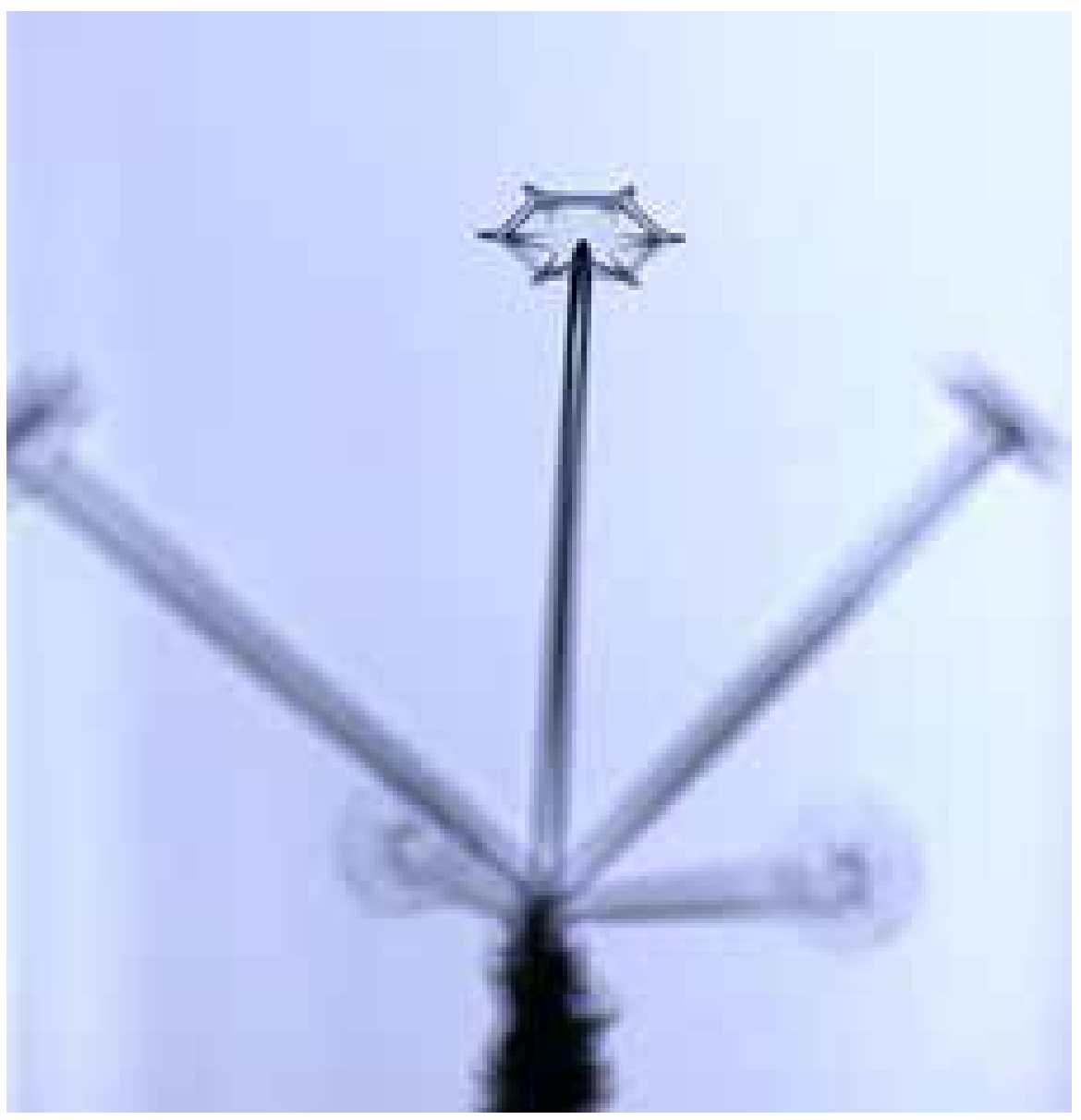

**Figure 8.2. Thin, plate-like snow crystals growing on the ends of c-axis electric ice needles. This example illustrates how e-needles can be used to cleanly support isolated snow crystals as they growth and develop.**

[2002Lib]. Thus the 3-mm-long e-needles shown in Figure 8.1 can be grown in under a minute.

While electrically modified growth is an intriguing phenomenon in its own right, c-axis e-needles are also quite useful as seed crystals for a broad range of snow crystal investigations. Normal (non-electrified) growth commences as soon as the applied high voltage is removed, yielding well-formed single-crystal specimens growing on the tips of the e-needles. In Figure 8.2, for example, plate-like crystals emerge and shield the ice needles upon which they grow, so the plate structures are only moderately perturbed by the presence of their supporting needles.

The e-needle method becomes an especially versatile tool for studying snow crystal growth in the dual-chamber apparatus described in detail below. The first chamber is optimized for rapidly and reproducibly growing high quality c-axis electric needles. Once created, the e-needles are then quickly transported to a second growth chamber that is optimized for observing the subsequent normal growth under a variety of carefully controlled conditions. With such a dual-chamber set-up, the entire morphology diagram can be explored using quantitative growth measurements. The challenge then becomes creating realistic computational models that can reproduce both the observed growth rates and morphologies. Figure 8.3 illustrates just a few of the possibilities.

Figure 8.4 shows a more quantitative example investigating the formation of a thin, plate-like crystal on the end of an electric needle via the Edge-Sharpening Instability (see Chapter 4). In focused studies like this one, comparisons with computer models can greatly inform our understanding of the molecular dynamics underlying snow crystal growth. It is likely that many additional studies covering a broad range of environmental conditions will be necessary to fully comprehend the enigmatic origins of the morphology diagram.

If there is one thing I have learned with great certainty in this field, it is that ice growth exhibits a mystifying variety of complex growth behaviors that will require much additional work to appreciate. A first important step is to continue developing a broad range of experimental techniques that can be used with numerical models of diffusion-limited growth to better understand the attachment kinetics. From there, molecular dynamics simulations can address the fundamental many-body molecular processes involved in ice crystal growth at the nanoscale. It will be a long journey, and I believe that experiments using electric ice needles will play an important role in advancing the science of snow crystal formation.

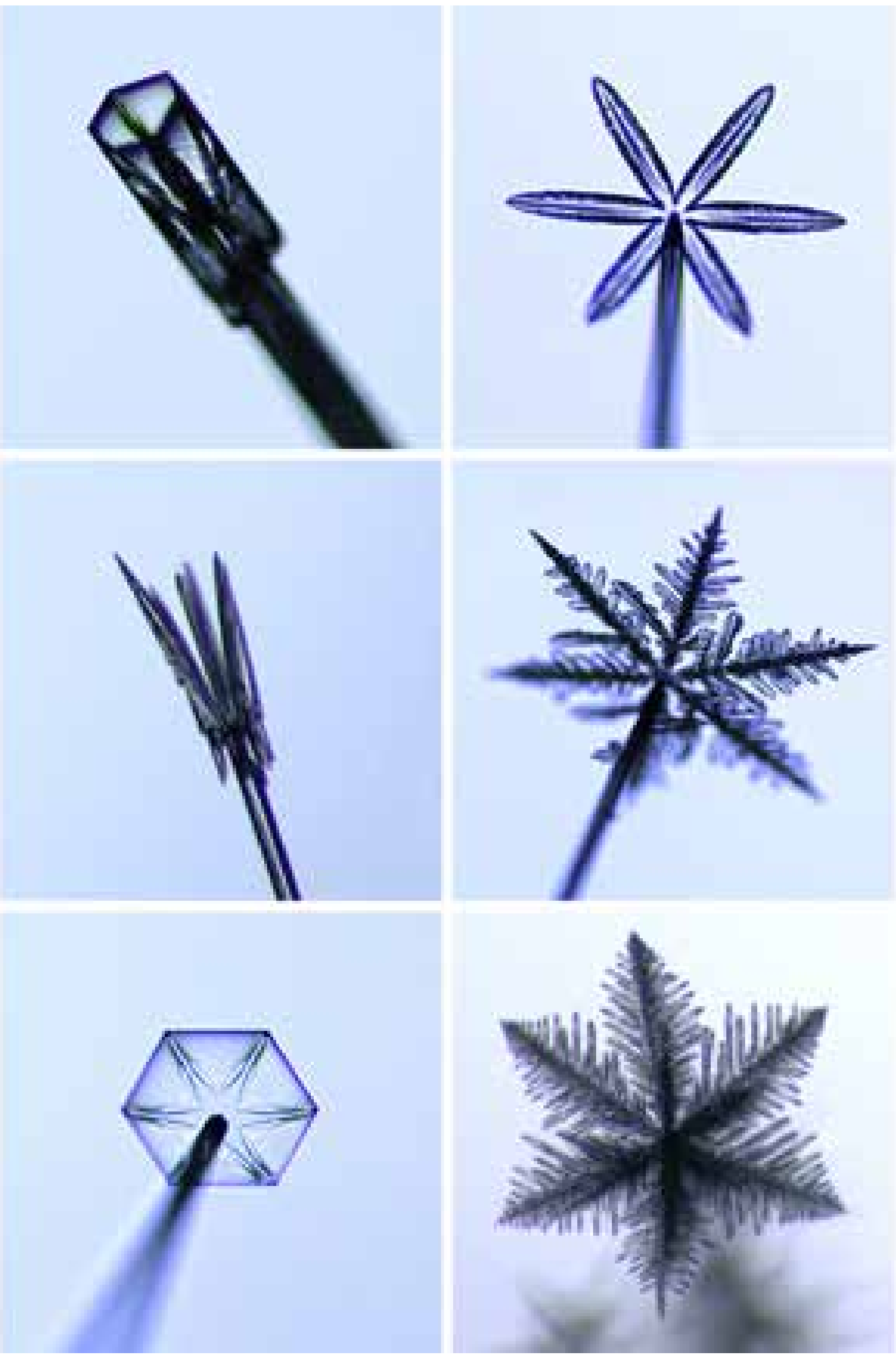

**Figure 8.3. Electric ice needles make excellent seed crystals for studying the development of complex snow-crystal morphologies, as shown in these examples. Once a c-axis e-needle has grown to a desired length, the high voltage is removed and normal growth commences on the needle tip. Each single-crystal structure shown above was grown in air at a constant temperature and supersaturation, and time-lapse photography can record the full growth history as desired. Observations like these are wonderfully suited for comparing quantitative growth measurements with detailed numerical models, hopefully leading to a better understanding of the physical dynamics of snow crystal formation.**

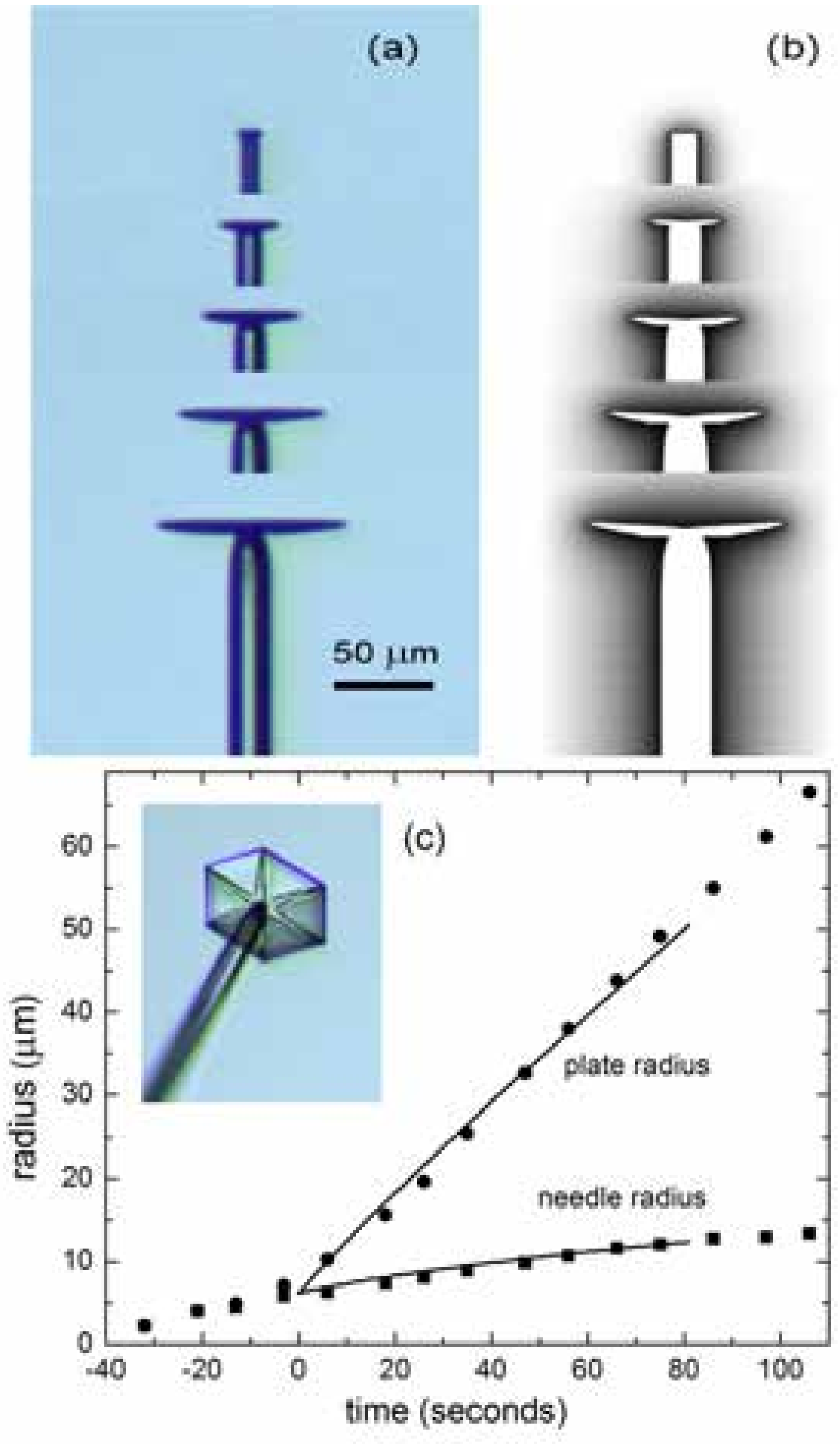

**Figure 8.4. These figures show one example of a quantitative comparison of laboratory and computational snow crystals [2008Lib, 2013Lib]. (a) A composite image made from five photographs showing the growth of a plate-like snow crystal on the end of an electric ice needle, viewed from the side. (b) A computational model of the same crystal, showing the water-vapor diffusion field around the crystal. (c) A quantitative comparison of experimental data (points) and the computational model (lines). The inset photo shows the crystal in (a) from a different angle.**

## Advantages…

I believe it is instructive, at this point, to examine what makes the e-needle method especially useful as a scientific tool, and to compare it with other experimental techniques for creating snow crystals for quantitative study.

**Single-crystal specimens.** One normally thinks of a seed crystal as being a tiny crystalline speck, for the case of ice a minute hexagonal prism. A single-crystal specimen is almost always desirable, as polycrystalline samples are unnecessarily complicated and less well suited for investigating the underlying crystal growth dynamics. But being small in all three dimensions is not an essential requirement, and a slender e-needle, small in two dimensions only, can still be considered a seed crystal. Moreover, with a well-defined crystal orientation and a sub-micron tip radius (at least while it is growing) [2002Lib], a c-axis e-needle embodies many qualities one seeks in an ideal seed crystal.

**Versatile support.** Once a seed crystal has been created, the question of supporting it while it grows must be addressed. In the case of e-needles, this question answers itself, as the support is already provided by the wire from which the needle originally grew. This support is robust and is easily manipulated, so the e-needle can be moved to a second growth chamber and positioned in front of a camera lens with relative ease. Moreover, a complex snow crystal growing on the end of a long, slender e-needle is well isolated from other parts of the apparatus, so its surrounding environment can be carefully manipulated and controlled.

**No substrate interactions.** When a seed crystal is supported by a non-ice surface

(e.g., a supporting surface, filament, capillary tube, etc.), there is a good chance that the substrate will influence the ice growth rates. I discuss the topic of substrate interactions in some detail in Chapters 6 and 7, because this has been a significant issue in many of my own ice growth experiments. Moreover, the substrate-interaction problem was often not fully appreciated by other researchers, sometimes yielding growth data of dubious quality [2004Lib]. With crystals growing on the tips of e-needles, there are no non-ice surfaces present, so substrate interactions are completely absent. With the entire structure made of ice, it all becomes part of the same numerical modeling problem.

**No water condensation.** Another problem with non-ice surfaces in a supersaturated environment is water condensation. If the supersaturation is above the dew point, then water droplets tend to condense on all available surfaces, greatly perturbing the surrounding supersaturation field. With seed crystals placed on a planar substrate, for example, droplet condensation readily occurs around the crystals, and this effectively precludes any useful studies at high supersaturation levels. E-needles avoid this issue, as the ice surfaces lower the nearby supersaturation so water condensation cannot occur. With e-needle support, therefore, one is free to explore quite high supersaturations with relative ease compared with other techniques.

**Witness surfaces.** Another persistent problem in studying ice growth is that the supersaturation level is difficult to know with high accuracy, as I discussed in Chapter 6. E-needles provide a way around this problem, at least partially, as the columnar body of the needle can serve a "witness surface" for determining the surrounding supersaturation. In many circumstances, the columnar growth satisfies $\alpha_{diffcyl} \ll \alpha_{prism}$ (see Chapter 3), and in this case the radial growth of the needle is determined to a good approximation by just the needle radius and the far-away supersaturation (see Chapter 3). In practice, this means that the measured growth of the body of the needle can be used to effectively measure the far-away supersaturation.

In Figure 8.4, for example, the outer-boundary supersaturation was adjusted in the computer model so the computed needle radius $R_{needle}(t)$ was a good fit to the data. The value of $\alpha_{prism}$ did not matter in this part of the calculation, because $\alpha_{diffcyl} \ll \alpha_{prism}$ was a good approximation. A good fit to $R_{needle}(t)$ thus constrains the supersaturation, as this is the only variable that significantly affects the radial growth rate $dR_{needle}/dt$. Having constrained the supersaturation in this way, the value of $\alpha_{prism}$ was then adjusted in the model to fit the plate growth.

This fitting procedure may sound like a circular argument at first blush, but its usefulness (and correctness) becomes apparent as one starts exploring quantitative modeling in detail. It soon becomes clear that a suitable witness surface allows for much more accurate numerical modeling, thus allowing a better understanding of the desired growth processes. I discuss examples of this in more detail later in this chapter.

**Rapid turnaround.** To make any real progress toward understanding the physics of snow crystal growth, one must measure a lot of crystals. The underlying molecular processes are complex and difficult to isolate, plus everything changes substantially with temperature and supersaturation (and perhaps other variables also). One of my favorite features of the e-needle method, therefore, is that it is possible to grow a lot of crystals in a short period of time, while still examining each one as it grows.

Because several e-needles typically form simultaneously on a wire tip (for example in Figure 8.2), the observer can select the best of several specimens, while examining the others to gauge the overall variability in growth and morphological development. The wide spacing between needle tips results in only minor interactions between the growth of the different crystals in a cluster.

Equally important, a new cluster of e-needles can be created in about a minute's time, allowing a large number of observations in a single observing session. By comparison, many other techniques look good in a first demonstration experiment, but soon lose their luster when it ends up taking all day to produce just two or three useful measurements.

**Simple geometry.** Another surprisingly beneficial feature of the e-needle method is that a needle tip contains only one exposed basal facet, compared to two basal facets on a small hexagonal prism. Looking at Figure 8.4, for example, a similar experiment performed with a hexagonal-prism seed crystal might yield a double plate during subsequent growth, and the competition between the two plates could then lead to one plate dominating over the other. Using an e-needle seed as in Figure 8.4, however, there is no possibility of a double plate forming.

Often the presence of two basal surfaces complicates the analysis, as numerical models must somehow deal with this double-plate competition effect. This is a nontrivial issue that may preclude the use of mirror symmetries that can otherwise be used to simplify the analysis (see Chapter 7). When comparing observations with numerical models, I have found that the end of a slender e-needle often offers a significant geometrical advantage when compared with a more traditional hexagonal-prism seed crystal.

**Useful asymmetry.** The geometry of a needle tip and its surroundings also brings with it a degree of symmetry-breaking that can be valuable when examining morphologies and growth rates in detail. For example, a plate-like crystal growing on the end of an e-needle typically exhibits a slightly concave upper basal surface and a slightly convex lower basal surface (depending on the detailed growth conditions).

This built-in asymmetry means that ridge structures form only on the lower basal surface while inwardly propagating macrosteps are found only on the upper basal surface. In fact, quite a large number of morphological features are similarly isolated in the needle-tip geometry, and this fact often turns out to be surprisingly helpful when trying to decipher the structures and physical origins of these features.

## …AND DISADVANTAGES

While the use of e-needles as seed crystals has numerous experimental advantages when examining complex snow crystal structures, there are some disadvantages associated with the technique as well.

**Larger diffusion effects.** Because $\alpha_{diffcyl} < \alpha_{diff}$ at fixed $R$ (see Chapter 3), the diffusion effects arising from an e-needle of radius $R$ are substantially larger than the diffusion effects of a hexagonal prism of overall size $R$. This is bad news for the e-needle method, as it means that e-needles are not especially well suited for making quantitative measurements of the attachment kinetics. As presented in the previous chapter, the smallest possible seed crystals are better suited for this purpose. I have used e-needles for some useful measurements of attachment kinetics, as I describe below, but these were done in somewhat special circumstances. In general, the e-needle method shines best when growing crystals with complex morphologies for comparison with numerical growth models.

**Weight restrictions.**

Although e-needles can support their own weight along with some build-up of material on their tips, there are limits. The contact point at the base of an e-needle is especially weak, and the needle will fall if it becomes sufficiently top-heavy. I have often found it odd that e-needles rarely crack and break the way one might expect from a crystalline structure. The e-needle itself behaves like a rigid structure, but its base support does not. When it becomes top-heavy, an e-needle usually pivots about its support point, slowly falling like a stick with its bottom end held in chewing gum.

**Complicated construction.** A final disadvantage with the e-needle method is that the apparatus is complex and therefore nontrivial to build. I describe my dual-chamber set-up in some detail below, but it is impossible to list every nuance of its construction and operation in a book of this nature. A great deal of trial-and-error experimentation was necessary to produce satisfactory crystals and growth measurements, and much of this effort would likely have to be repeated (to some degree) with a new e-needle venture. While this is a nontrivial disadvantage, I certainly hope that the results presented in this chapter stimulate at least some interest in developing the e-needle technology to greater heights.

## Comparisons with Other Laboratory Methods

To complete this discussion of the advantages and disadvantages of the e-needle method, we next compare e-needles with some alternative experimental methods that have been explored for supporting single snow crystals when studying their growth dynamics.

**Planar substrates.** Many experiments have been done observing the growth of snow crystals on flat substrates, and this technique is well suited for extremely small crystals, as discussed in the previous chapter. Two characteristics make this method somewhat ill-suited for studying complex snow crystals, however – substrate interactions and water condensation.

Water condensation covers a flat substrate with a field of small water droplets whenever the supersaturation rises above the water supersaturation level. The problem makes it essentially impossible to perform quantitative studies at higher supersaturations, a region of phase space that is especially interesting when investigating complex morphologies. In this regard, therefore, e-needles provide a better means of supporting snow crystals than flat substrates.

The problem of substrate interactions is subtler in nature, as the contact line between the substrate and ice surfaces can be an unwanted source of layer nucleation. I discuss this issue in more detail in Chapter 6. Here again, growth on e-needles involves no non-ice surfaces, so is immune from substrate interactions.

The Plate-On-Pedestal method described in the next chapter provides a novel way to create complex stellar snow crystals for artistic photography, but this technique is rather poorly suited for precision quantitative studies.

**Filament support.** Ukichiro Nakaya created the first laboratory snowflakes in the 1930s by suspending individual crystals from fine rabbit hairs, as I described in Chapter 1. The choice of rabbit hair was not accidental but arose only after Nakaya had examined a great many other filamentary materials. In nearly every case, a multitude of frost crystals appeared on the filaments, whereas isolated single crystals were desired for study. Desiccated rabbit hair yielded isolated seed crystals with reasonable reliability, and Nakaya employed this simple and ingenious method to first observe and quantify the snow crystal morphology diagram [1954Nak].

Supporting snow crystals on thin filaments thus has a long and illustrious history, and simplicity is a highly desirable feature of this method. On the other hand, producing a single initial seed crystal on a fiber was problematic for Nakaya, and it remains an issue for any studies involving filamentary support.

Interestingly, it appears that Nakaya's rabbit hairs exhibited relatively minor substrate interactions compared to other filamentary materials. This may be because natural oils on the hairs, combined with their complex microscopic structure, created a somewhat superhydrophobic surface that increased the ice/surface contact angle and thus suppressed the nucleation of new terraces.

Although the use of rabbit hair seems a bit archaic for modern use (to my knowledge, Nakaya's method has never been reproduced by other researchers), there may be superhydrophobic filamentary materials that provide better substrates for single-crystal ice-growth studies. Unfortunately, one must first find such exemplary materials and then figure out a way to create suitable seed crystals on them. Solving these two problems remains a challenge.

**Capillary tubes.** Nelson and Knight replaced Nakaya's rabbit hair with fine glass capillaries, providing an interesting new method of filamentary support [1998Nel]. One advantage is that a capillary tube is hollow, so ice can grow from liquid water in the tube until it becomes exposed at its tip for subsequent growth from water vapor. This solves the nucleation problem and yields a single seed crystal at the capillary tip, although perhaps with a random orientation of the seed crystal axes [1998Nel].

The problem of substrate interactions remains, however, making quantitative comparisons with numerical modeling a challenge. This issue can be seen in [1998Nel], as the growth rates of basal surfaces in contact with the substrate were generally higher than isolated basal facets, which is a commonly seen feature in substrate interactions (see Chapter 6).

Another issue is that the set-up time between crystals can be quite lengthy, especially if freezing water up the capillary tube is the nucleation method, as was described in [1998Nel]. In my opinion, the long turn-around time is perhaps the biggest factor that has kept capillary support in the demonstration phase. Water condensation on the capillary support is likely to be another problem that is difficult to surmount.

In many important ways, the e-needle method is superior to all filamentary supports, including capillary support. E-needles can be grown easily and quickly, with no issues relating to seed-crystal nucleation, substrate interactions, and water condensation. Nevertheless, filamentary support remains a simple and robust method for supporting snow crystals in circumstances where the e-needle method may be problematic, for example exceptionally large or heavy crystals, or perhaps in different background gases and at different pressures.

**Freely falling crystals.** I discussed free-fall growth chambers in Chapter 6, and this method has many advantages for studying small snow crystals, notably experimental simplicity and the multitude of crystals produced simultaneously. Also, one can start with extremely small seed crystals created by expansion nucleation, and substrate interactions are absent. However, studying complex morphologies in a simple free-fall chamber is problematic, as large crystals require long growth times and therefore long fall distances. The inability to view crystals as they grow is another distinct disadvantage of this method.

**Laminar flow levitation.** Takahashi and Fukuta observed freely falling complex snow crystals without long fall distances by levitating individual specimens in an upwardly moving column of air [1988Tak, 1991Tak, 1999Fuk]. This remarkable method allowed growth times of over an hour, and the authors produced snow crystals that were directly comparable to natural specimens. From a materials-science standpoint, however, the technique suffers several drawbacks that make direct numerical modeling difficult. For example, small crystals cannot be easily investigated, plus the supersaturation is either difficult to determine or limited to $\sigma = \sigma_{water}$.

The levitation method is also not easily adaptable to *in situ* observations, making it somewhat poorly suited for observing the growth of a single crystal as a function of time. And the apparatus is far from simple to build. None of these problems is insurmountable, but updraft levitation is generally less desirable (in my opinion) than the e-needle method for making quantitative studies.

**Electrodynamic ion traps.** Substrate interactions can also be avoided by suspending charged ice crystals in microparticle ion traps, as was pioneered by Swanson et al. [1999Swa]. This intriguing method is well suited for observing small crystals at low supersaturations, although the ion-trap electrodes may hinder efforts to produce a well-defined supersaturation around the growing crystal. Water condensation is also clearly an issue. Moreover, the charge/mass ratio changes rapidly as a levitated crystal grows, so ion trapping seems to be a poor contender for studying the growth of large, complex specimens. While early growth studies show promise [2016Har], many technical issues must be resolved before electrodynamic trapping becomes a workhorse for quantitative snow crystal studies.

**Broad applicability.** No single technique for growing snow crystals is optimal for all purposes, but I have found that growing crystals on the tips of electric needles has many desirable features. The e-needle method is especially well suited for scientific investigations involving complex snow crystal morphologies in air.

In the dual-chamber apparatus described below, it is possible to grow snow crystals over broad ranges of temperature and supersaturation, beginning with a particularly simple and reproducible initial geometry, free from substrate interactions and water condensation. Observations can be made *in situ*, and the set-up time between samples is quite fast. I believe that this technique hits a scientific "sweet spot" for observing the growth of complex snow crystals and making detailed, quantitative comparisons with numerical models.

## 8.2 E-needle Formation

Electrically enhanced ice growth was discovered in 1963 by Bartlett, van den Heuvel, and Mason [1963Bar], who observed the spontaneous formation of fast-growing electric needles when large, positive DC voltages were applied to ice crystals growing at high supersaturations. Little work was done to understand or apply e-needles for several decades until Libbrecht and Tanusheva explained the underlying physical cause as an electrically induced growth instability 35 years later [1998Lib, 1999Lib1, 1999Lib2, 2002Lib]. These authors also discovered the importance of chemical influences on the crystalline orientation of e-needle growth, and they developed the reliable technique for growing high quality c-axis needles described in this chapter.

### Basic Theory

The physical mechanism that produces electric ice needles can be understood by first considering the equilibrium vapor pressure of a charged ice sphere. From basic electrostatics, and assuming that the sphere has nonzero conductance, the static charge must all reside on the surface of the sphere, while the electric field inside the sphere is zero. Pulling a neutral water molecule off the sphere reduces its radius but not its charge, and this brings the surface charges closer together than they were before the water molecule was removed.

Because like charges repel, it requires some energy to reduce the size of the sphere and pull the surface charges closer together. It follows that pulling a water molecule off a charged sphere requires slightly more energy than pulling a water molecule off an uncharged sphere. For this reason, the equilibrium vapor pressure of a charged sphere is slightly lower than that of an uncharged sphere. A high voltage applied to the sphere has the same effect.

The argument is essentially the same as with the Gibbs-Thomson effect presented in Chapter 2, and the math is similar also. Adding in the electrostatic self-energy term, the equilibrium vapor pressure of a charged sphere of radius $R$ becomes

$$c_{eq}(R) \approx c_{sat}\left(1+\frac{2d_{sv}}{R}-\frac{R_{es}^2}{R^2}\right) \qquad (8.1)$$

where

$$R_{es}^2 \approx \frac{\varepsilon_0 \varphi_0^2}{2c_{ice}kT} \qquad (8.2)$$

and $\varphi_0$ is the applied electrical potential (aka voltage), while $\varepsilon_0$ is the vacuum permittivity in SI units (see Appendix A).

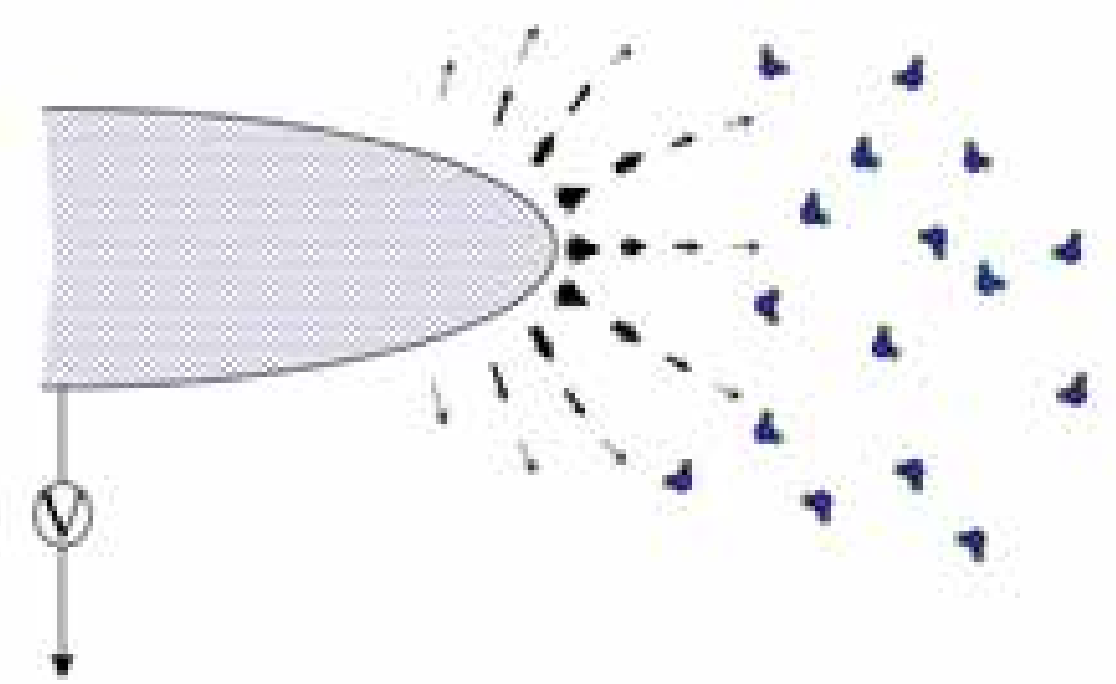

**Figure 8.5. When a high voltage is applied to an ice needle, strong electric fields (arrows) are concentrated at the tip. As described in the text, these fields decrease the equilibrium water-vapor pressure near the tip, increasing its growth rate. This produces a positive feedback effect: faster growth yields a sharper tip, which increases the electric fields near the tip surface, making the growth faster still. The result is a runaway growth instability that produces a slender, fast-growing e-needle.**

With this additional electrostatic effect, the equation describing the growth of a spherical crystal (see Chapter 3) becomes

$$v \approx \frac{X_0}{R} v_{kin}\left(\sigma_\infty - \frac{\sigma_\infty}{\alpha}\frac{X_0}{R}+\frac{R_{es}^2}{R^2}\right) \qquad (8.3)$$

where we have neglected surface tension for the reasons described in Chapter 3. Extending the discussion in that chapter, the corresponding equation for the tip velocity of a growing parabolic crystal becomes

$$v_{tip} \approx \frac{2X_0 v_{kin}}{BR_{tip}}\left(\sigma_{far} - \frac{\sigma_{far}}{\alpha}\frac{2X_0}{BR_{tip}}+\frac{GR_{es}^2}{R_{tip}^2}\right) \quad (8.4)$$

where $R_{tip}$ is the radius of curvature of the parabola at the tip and $G$ is a dimensionless geometrical factor.

Examining the individual terms in Equation 8.4 gives one a picture of the essential physics underlying the e-needle growth instability. The first term in the parentheses gives the constant tip velocity of a parabolic crystal when its growth is entirely diffusion-limited. This is the Ivantsov solution discussed in Chapter 3, arising solely from the solution to the particle diffusion equation.

The second term is rather small compared to the first, but it reduces the growth velocity as $R_{tip}$ becomes smaller. This term, albeit small, plays an essential role in stabilizing the normal growth of a parabolic crystal. It "selects" the final $R_{tip}$ for the growing Ivantsov parabola via solvability theory, as I discussed in Chapter 3. Together, the first two terms in Equation 8.4 describe the growth of a normal ice needle or dendrite that has an approximately parabolic shape near its tip.

The third term in Equation 8.4 tends to destabilize the normal parabolic growth, and this is the term that drives the e-needle instability. As $R_{tip}$ becomes smaller, this term increases the tip velocity relative to the normal growth, and the $R_{tip}^{-2}$ dependence means that this term eventually dominates over the kinetic term in the equation as the tip sharpens.

Including this third term in an extension of solvability theory and following the algebra through, the resulting equation for the parabola tip radius can be written in the form of a quadratic equation [2002Lib]

$$R_{tip}^2 - R_0 R_{tip} + AR_{es}^2 \approx 0 \qquad (8.5)$$

where $R_0$ is the tip radius in the absence of an applied electrical potential (the normal solvability-theory result) and $A$ is a dimensionless constant.

Solving the quadratic equation gives $R_{tip} = R_0$ when there is no applied potential, which is the normal solvability result. As the potential is turned on slowly, at first the solution yields a tip velocity that is only slightly larger than the normal velocity. In this regime, the normal solvability solution is only slightly perturbed by the applied potential, decreasing $R_{tip}$ and increasing $v_{tip}$ as $\varphi_0$ becomes larger. Thus, there is no dramatic effect when a small voltage is applied, as one would expect.

This "perturbative" regime remains in effect as long as $R_{tip} < 2R_0$, or, equivalently, as long as $v_{tip}$ is no greater than about twice its normal-growth result. Under typical ice-growth conditions, the perturbative regime holds as long as the applied voltage is less than about 1000 volts. Beyond that point, the quadratic equation no longer has any real roots, meaning that the second term can no longer stabilize the growth as described by solvability theory.

Physically, the destabilizing electrostatic term eventually brings about a full-blown growth instability. Above a threshold voltage $\varphi_{thresh} \approx 1000$ volts, the physical influence of the third term in Equation 8.4 exceeds that of the second term, providing a positive feedback effect that leads to runaway growth. Reducing $R_{tip}$ makes the tip electric fields higher, which turns up the growth rate and reduces $R_{tip}$ still more, further increasing the tip electric fields. All this quickly leads to an abrupt increase in $v_{tip}$ and the formation of an electric ice needle.

Figure 8.6 shows a direct comparison of experiment measurements with the theory described above [1998Lib], illustrating the initial perturbation of the solvability solution followed by a runaway instability leading to the formation of an electric needle. Consistent with the solvability model, the dendrite tip growth increases in a well-behaved fashion until reaching about twice its normal (zero applied potential) value, at which point the e-needle forms and the tip velocity increases abruptly.

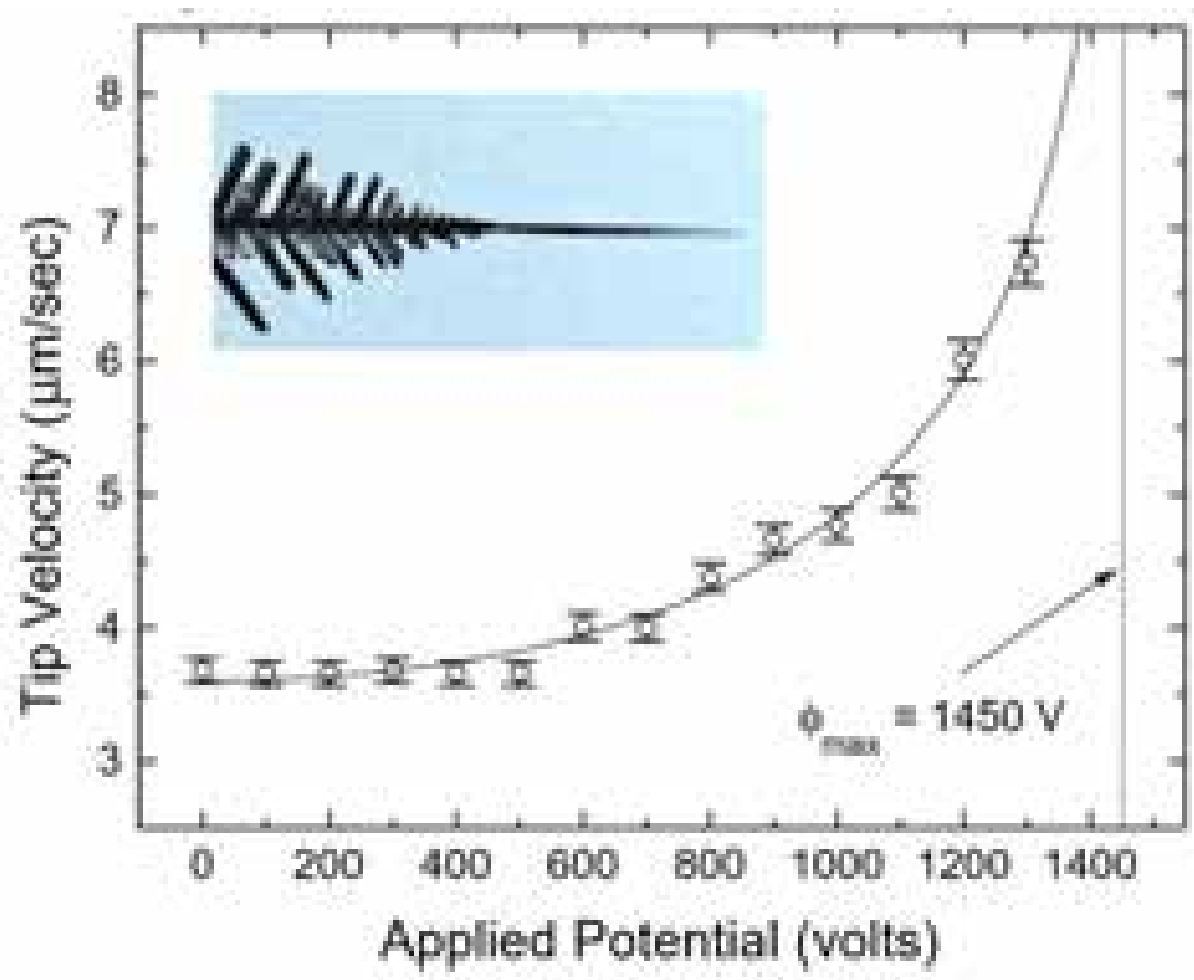

**Figure 8.6. Experimental measurements (data points) showing a gradual increase in the tip growth velocity of an ice dendrite at -15 C as the applied voltage was increased. The line through the data points comes from the theory contained in Equation 8.5, with $A$ adjusted to fit the data. Note that the dendrite morphology continued to exhibit sidebranching as the voltage was initially increased. Once a threshold voltage was exceeded, however, normal dendrite growth gave way to the formation of an electric needle. The inset image shows a similar run in which the normal dendrite growth transformed into a fast-growing a-axis e-needle above threshold. From [1998Lib].**

Figure 8.7 shows measurements of tip velocities near -5 C for normal needle growth, e-needle growth along axes other than the c-axis, and for c-axis e-needles. Although the tip radius $R_{tip}$ was often too small to measure optically in these measurements, it is reasonable to estimate $R_{tip}$ from the Ivantsov solution. This dictates that $R_{tip}$ is proportional to $v_{tip}^{-1}$, and the inferred tip radius $R_{tip}$ falls to values below 100 nm for the fastest growing c-axis e-needles [2002Lib].

The above electrostatic theory is a natural extension of solvability theory, and it clearly fits the observations reasonably well. But it only describes the tip behavior in the perturbative regime, when the tip radius is still stabilized by the kinetic term in Equation 8.4. Above the voltage threshold, there must be some other stabilization mechanism that selects the final e-needle tip velocity.

This necessary stabilization cannot be provided by vapor-pressure effects stemming from either surface energy or attachment kinetics, as these effects both go only as $R_{tip}^{-1}$, as seen with the second term in Equation 8.4. These terms, therefore, cannot compete with the $R_{tip}^{-2}$ term once the tip radius becomes small. At the 100-nm scale, surface tension may provide sufficient mechanical force to halt additional tip sharpening, but this is just a guess. At present, the e-needle tip stabilization mechanism is not known.

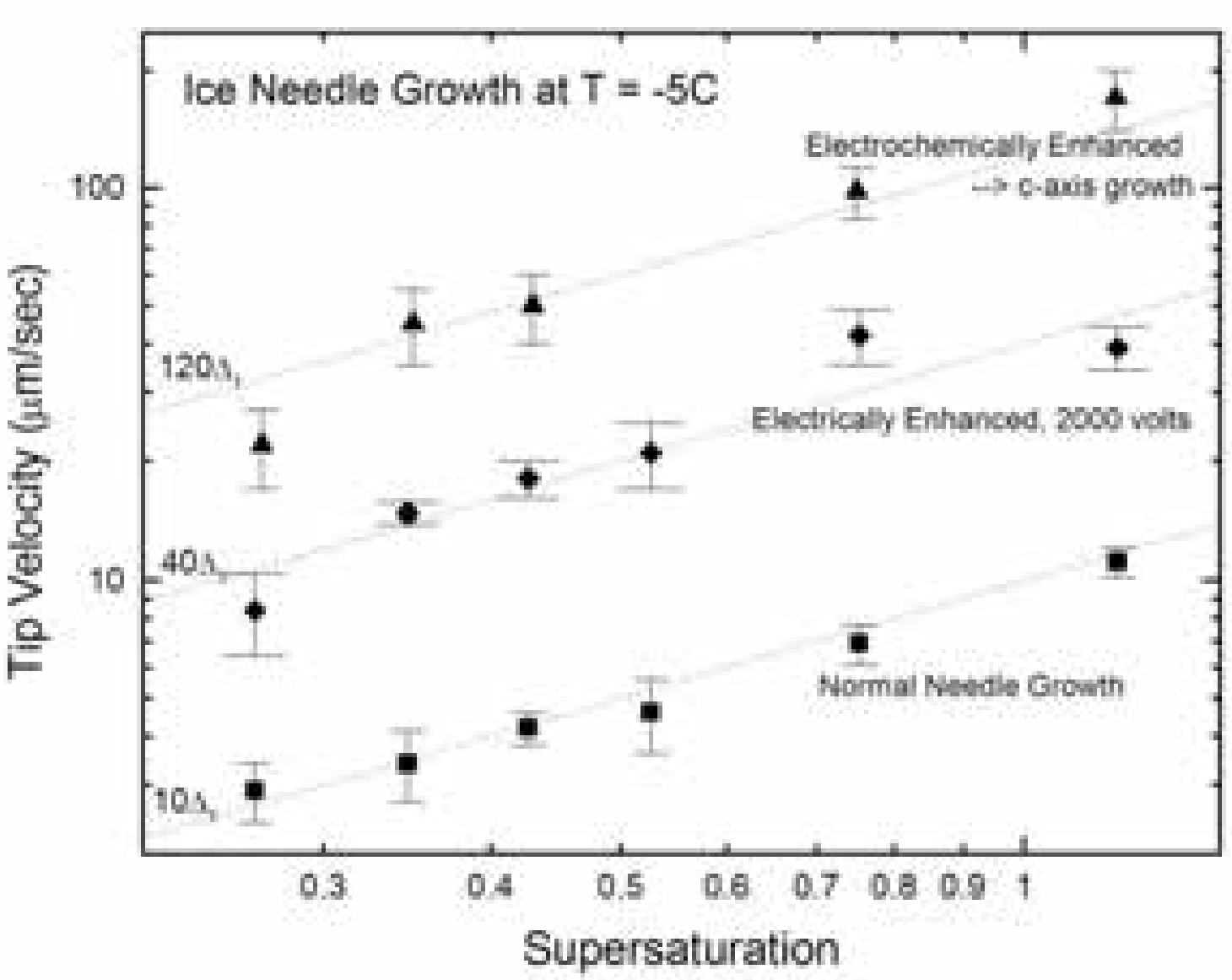

**Figure 8.7. Experimental measurements (data points) showing the tip velocities for normal needle growth (lower points), e-needles growing along crystalline axes that are not the c-axis (middle points), and c-axis e-needles (upper points). The data were all taken at a temperature of -5 C, and in all cases the tip growth velocity is linearly proportional to the supersaturation, as predicted by solvability theory (see Chapter 3). Growth along the c-axis was stimulated using chemical vapor additives as described in this chapter. From [2002Lib].**

## Polarizability Effects

The astute reader will note that the above theory does not involve the polarizability of the highly polar water molecule in the electric field near the needle tip. It turns out that molecular polarizability brings about two electrical effects that nearly cancel one another in the theory. First, the vapor pressure of a charged sphere is increased by the polarizability because removing a molecule from the zero-field region inside the sphere to the high-field region outside releases energy. Second, the water vapor density in the high-field region is increased as polarized water molecules are preferentially drawn into this region. The theory becomes somewhat complex at this point, but the final result is that the polarizability of the water molecule can be ignored to first order [1999Bre, 2002Lib]. It appears that molecular polarizability contributes somewhat to the energetics, but it is not as important as the electrostatic effect for creating e-needles.

The electrostatic effect is somewhat universal in that it does not depend much on the characteristics of the vapor molecules in the problem, including the polarizability. Because of this universality, one expects that the e-needle phenomenon should be observed in other high-vapor-pressure material systems besides ice. Indeed, Libbrecht, Crosby and Swanson [2002Lib] demonstrated a similar e-needle effect in iodine crystal growth, even though this simple dipole molecule has a very low molecular polarizability.

## E-needle Crystal Orientation

Electric ice needles can be persuaded to grow with a variety of crystalline orientations, depending on growth conditions, as is demonstrated in Figure 8.8. Near -15 C, e-needles often prefer to grow along the a-axis of the ice crystal, but sometimes they grow preferentially along the $[1\bar{1}00]$ axis, as shown in the top and middle images in Figure 8.8. I have not fully explored the causes of the different e-needle orientations, and the precise conditions needed to produce growth along the a-axis or the $[1\bar{1}00]$ axis are not presently known. The preference for growth along the $[1\bar{1}00]$ axis is especially puzzling, as this axis seems to play little role in other aspects of ice crystal growth.

At temperatures near the needle peak around -6 C in the morphology diagram, the e-needle growth is usually not along a well-defined crystal axis. Instead the growth axis appears to be roughly the same as the growth direction of fishbone dendrites, which is somewhat temperature and supersaturation dependent. However, we found that c-axis electric needles could be reliably produced near -6 C by adding trace quantities of vaporous chemical additives to the air in which the needles grew [2002Lib].

A variety of chemicals were found to promote c-axis e-needles near -6 C, including hydrocarbons (e.g. gasoline vapor), various alcohols, and other solvent vapors. After some trial-and-error investigations, I found that acidic acid vapor is especially effective, with concentrations as low as 1ppm readily promoting needle growth along the c-axis. However, the best vapor I have found for promoting c-axis e-needles is that emitted from GE Silicone II caulk. Acetic acid is the primary solvent used in this caulk, but the vapor appears to include additional proprietary volatile organic compounds at low concentrations.

The use of vaporous chemical additives to promote c-axis e-needle growth was a largely serendipitous discovery made by Tanusheva and Libbrecht in 1998, as our growth chamber was assembled using GE Silicone II caulk, which produced some residual vapor before the chamber was thoroughly baked. In the next section I outline how this chemical-vapor trick can be used to produce c-axis e-needles with nearly 100 percent efficiency. The underlying reason why trace chemical impurities promote c-axis e-needle growth so effectively, however, remains a mystery.

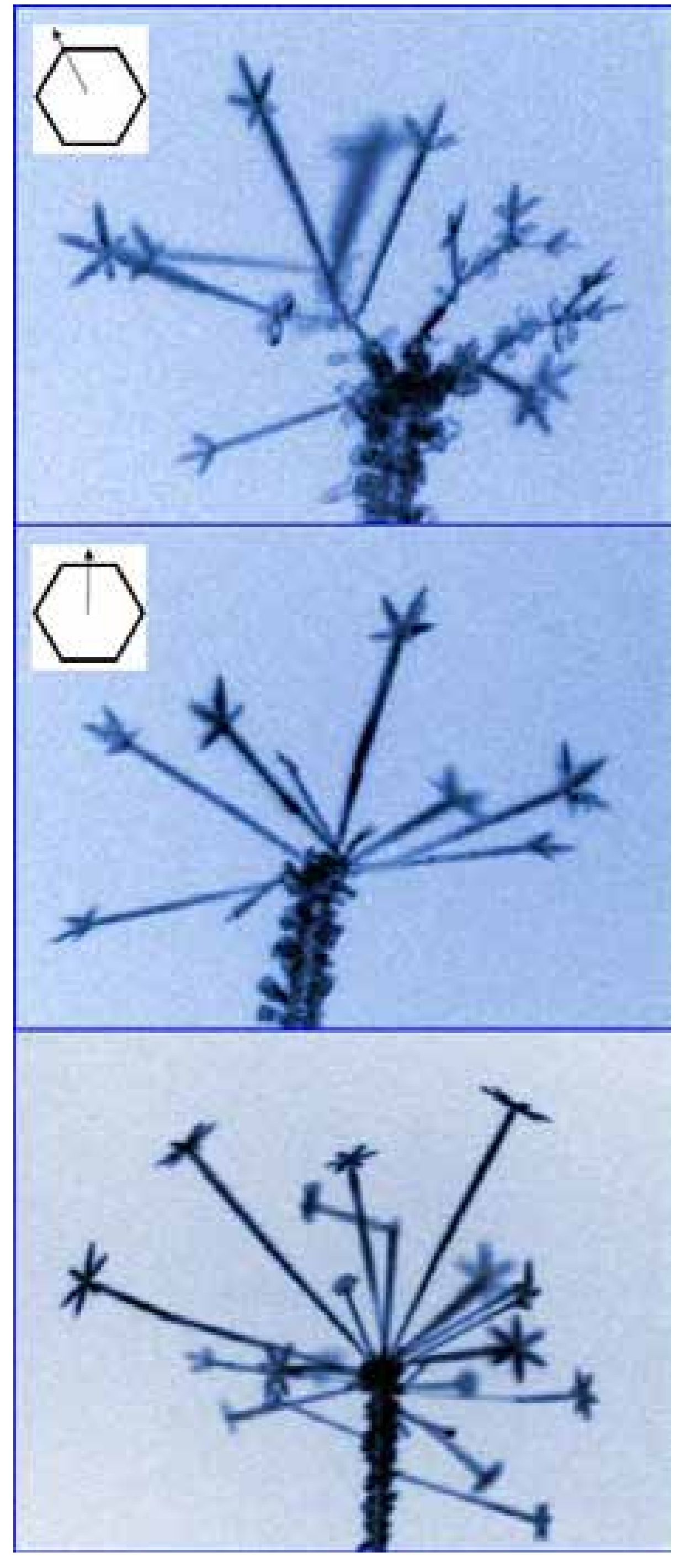

**Figure 8.8. (Above right) Under different experimental conditions, e-needles can grow preferentially along the a-axis (top photo), the $[1\bar{1}00]$ axis (middle photo), or the c-axis (bottom photo). In many conditions, e-needle growth is not along a well-defined crystalline axis but appears to be somewhat random and likely dependent on the orientation of the seed crystal from which the e-needle formed. In these photos, a period of e-needle growth was followed by a period of normal growth, during which there was no applied voltage. The orientation of the e-needle could then be determined by the orientation of the subsequent normal growth.**

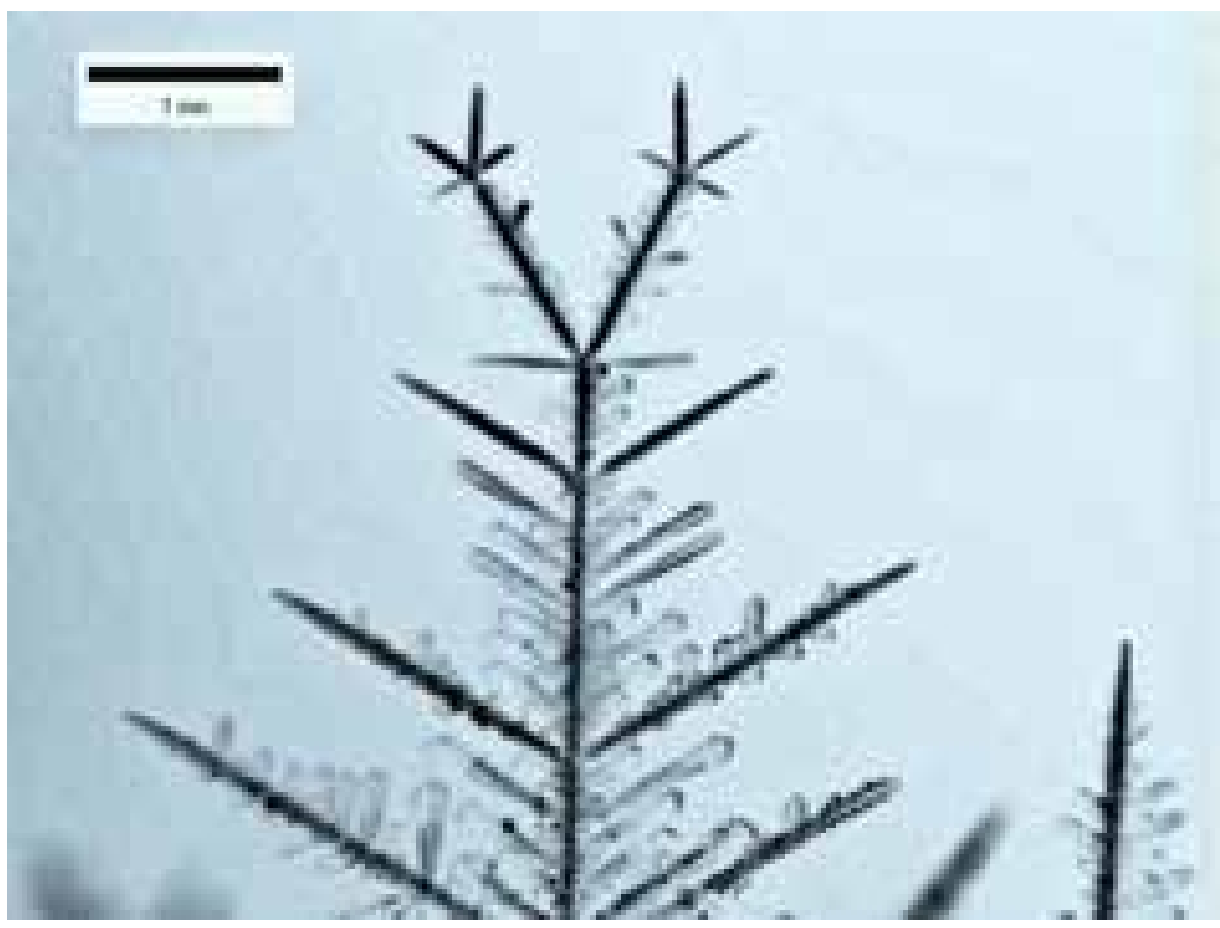

**Figure 8.9. The preferential growth of e-needles along the $[1\bar{1}00]$ axis sometimes yields a peculiar tip splitting phenomenon seen in dendrite growth near -15 C, illustrated in this photo. When the applied voltage is just slightly above the e-needle threshold, an $[1\bar{1}00]$ axis e-needle apparently begins to form, but the growth halts before the structure can turn into a full-fledged e-needle. Instead the reorientation of the crystal axis brings about a tip-splitting that yields two primary dendrite branches and two secondary branches, the latter perpendicular to the original branch axis. This split-tip structure lowers the electric fields sufficiently (at constant applied voltage) that normal growth commences from the split tip. In this example the two primary branches then grew farther apart until the electric fields passed through threshold again, so the two branch tips each underwent an additional splitting. This applied potential was not changed after the first tip splitting occurred.**

## Some Remaining Questions

Although the above theory is probably correct at some level, the e-needle phenomenon is nevertheless largely unexplored, both experimentally and theoretically. Some remaining questions and ideas for further research include:

- What stabilizes the e-needle growth above the instability threshold? Solvability theory is no longer adequate for this problem, so some new theoretical ideas seems to be required.
- Why do c-axis e-needles grow about four times faster than e-needles growing along less preferred lattice directions?
- What mechanism is responsible for chemical impurities promoting the growth of c-axis e-needles?
- How would the e-needle phenomenon change in different gases or as a function of gas pressure? No experiments along these lines have ever been performed, to my knowledge.
- Is it possible to grow individual c-axis e-needles in a more controlled fashion, for example producing c-axis e-needles that are oriented perpendicular to a fixed substrate? To date, I have only been able to grow c-axis e-needles from a frost covered wire, which leads to somewhat random spatial orientations.
- Why do e-needles sometimes prefer growth along the $[1\bar{1}00]$ axis near -15 C, while other times selecting the a-axis at this same temperature?
- Are there other preferred e-needle orientations under different growth conditions? The available parameter space has not been fully explored, leaving open the possibility of new discoveries.
- What other materials exhibit the e-needle growth instability? Iodine exhibited some e-needle-like behavior, but little work has been done exploring this phenomenon in other materials that grow from the vapor phase.

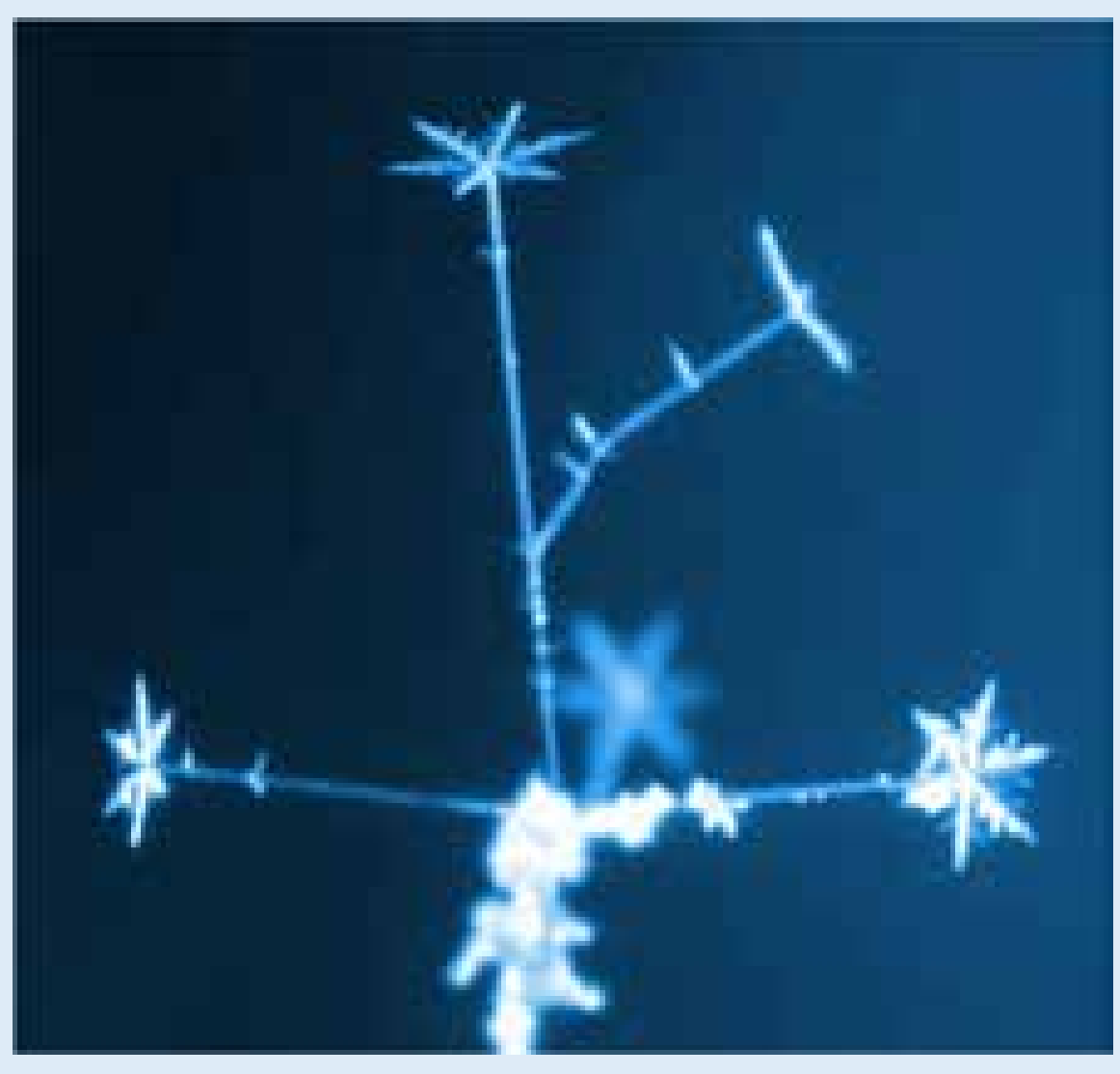

## A Serendipitous Discovery

In 1997, summer student Victoria Tanusheva captured the above photograph during our early studies of electric ice needles. The image shows five beautifully formed stellar snow crystals growing on the tips of the five e-needles. Notably, all five stars are perpendicular to their respective e-needles, indicating that all five e-needles had grown along the crystalline c-axis.

We had witnessed c-axis e-needles previously, but these were rare occurrences. Most of the time, the e-needle axes were somewhat randomly oriented with respect to the crystal axes, which was not ideal for growing snow crystals on the needle tips. Turning off the voltage and setting the growth conditions to produce stellar crystals on these e-needles yielded mostly lopsided, rather malformed stars. Seeing this image, with all five e-needles growing along the c-axis, we realized that there must be some recipe for reliably making c-axis e-needles.

During the weeks and months after seeing this lone photo, we were unable to reproduce the high yield of c-axis e-needles. We carefully explored growing e-needles at different temperatures, supersaturations, voltages, and other parameters in our apparatus, but nothing worked, and the desired recipe eluded us during many frustrating tests and trials.

Having excluded many other possibilities, we began to think that unwanted chemical vapors of our apparatus may have been affecting our results. The diffusion chamber was constructed from aluminum, styrofoam, glass, and other materials, and much of it was held together with silicone caulk, which does emit a characteristic odor.

Removing the contaminating vapors entirely was impossible, but baking the chamber would slowly reduce the contaminant levels. So we heated the chamber to about 50 C and left it alone for several weeks, focusing our attention on other projects for the duration.

When the time seemed right, and the odors had clearly subsided substantially, we turned off the bake and tried our luck once more. And, lo and behold, now we saw no c-axis e-needles whatsoever; the bake had reduced the yield to effectively zero.

At that point the light bulb went off and we realized that vapor contaminants were not the problem, but rather an essential part of our desired recipe. Adding a bit of caulk vapor was straightforward, and in short order we were producing superb c-axis e-needles reproducibly and with a nearly 100 percent yield.

After additional tests, we found that many chemical vapor additives could bring about the formation of c-axis e-needles. Just about anything with a significant odor seemed to do the trick. Acetic acid (a.k.a. vinegar) worked especially well, and it was entirely fortuitous that this was a main constituent in the caulk we had been using all along (G.E Silicone II caulk). In fact, the caulk vapor ended up being slightly better than pure acetic acid, and better than any other chemical additive we tested.

We soon developed a highly reproducible procedure for creating copious c-axis e-needles, which is presented in this chapter. As with many empirical recipes, however, we still do not understand why it works!

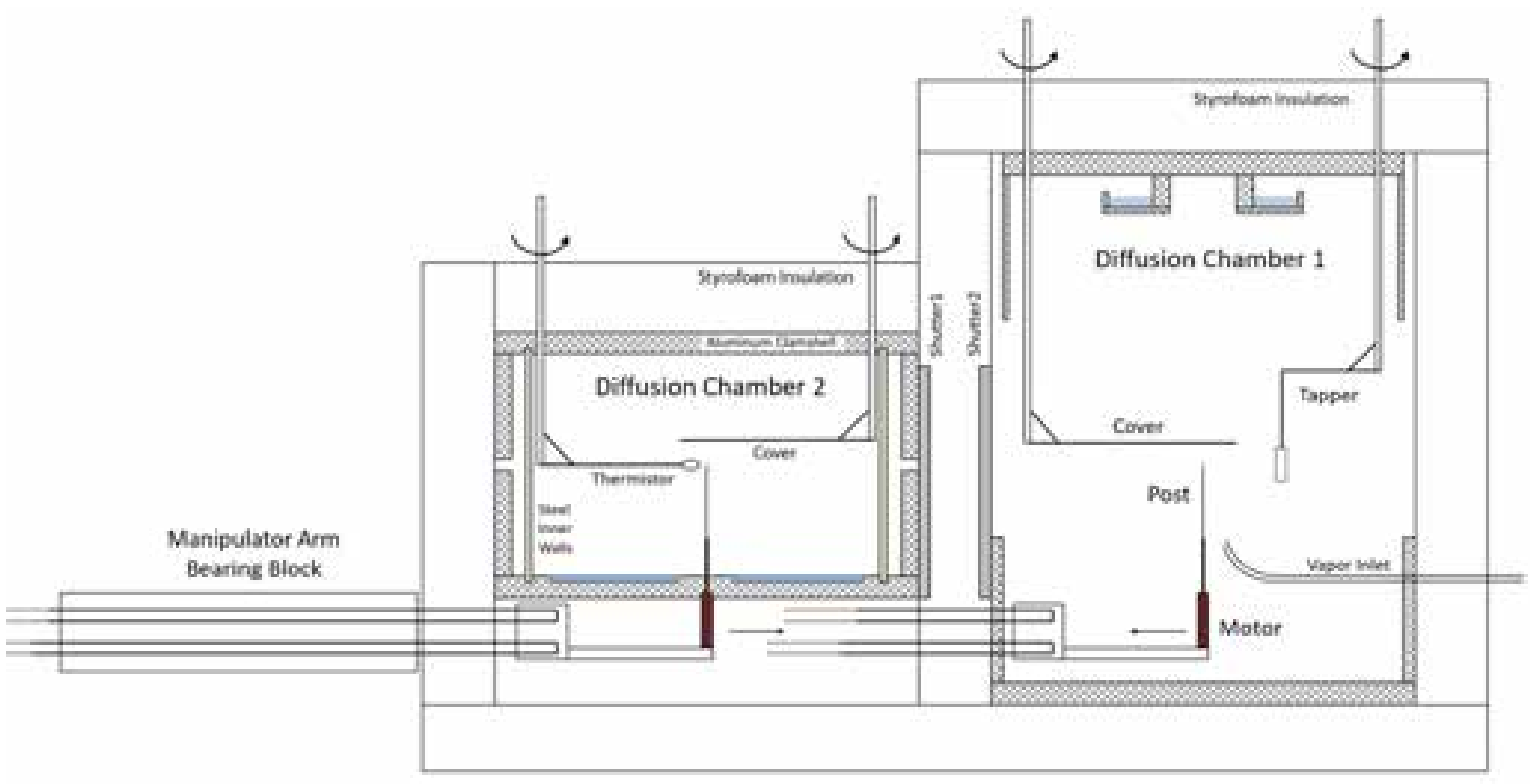

**Figure 8.10. A dual diffusion chamber apparatus for observing snow crystal growth on electric ice needles. Diffusion Chamber 1 (DC1, on the right) provides the necessary conditions for creating c-axis electric needles easily and quickly. The needles are then transported to Diffusion Chamber 2 (DC2, on the left), which provides a well-controlled environment that can achieve a broad range of temperatures and supersaturation levels. The inside height of DC2 is 10 centimeters [2014Lib].**

## 8.3 An E-needle Dual Diffusion Chamber

In an effort to exploit the electric ice needle method for additional studies of snow crystal growth, I constructed the dual-diffusion-chamber apparatus shown in Figures 8.10 and 8.11 [2014Lib]. The basic idea here is to grow c-axis e-needles in one diffusion chamber and then move the e-needles to a second diffusion chamber where their subsequent normal growth can be observed over a broad range of conditions. The first diffusion chamber can then be optimized for the task of creating c-axis e-needles quickly and reliably, while the second chamber is designed to produce a well-controlled environment with a precisely known temperature and supersaturation level.

In many respects, this is a next logical step in the morphological studies begun by Nakaya [1954Nak] and advanced by Mason [1958Hal, 1963Mas], Kobayashi [1961Kob], Bailey and Hallett [2004Bai, 2009Bai] and others. The main difference is that we can now make precise measurements of single, isolated crystals growing *in situ*, avoiding the complications of substrate effects and crystal crowding. This enables the next substantial step in the scientific progression – making detailed, quantitative comparisons with modern computational models that can yield new insights into the underlying physical processes that govern snow crystal growth.

### Diffusion Chamber 1

Referring to Figure 8.10, DC1 was designed to produce c-axis electric needles quickly, reliably, and easily. Its basic construction is a partial clam-shell diffusion chamber (see Chapter 6) with a top temperature of 59 C and a bottom temperature of -35 C [2014Lib]. These temperatures, along with the dimensions of the aluminum clam-shell walls, were adjusted (somewhat by trial-and-error) to yield a high supersaturation ($\sigma \approx 100\%$) and a temperature of -6 C at the location of the wire tip of the support post.

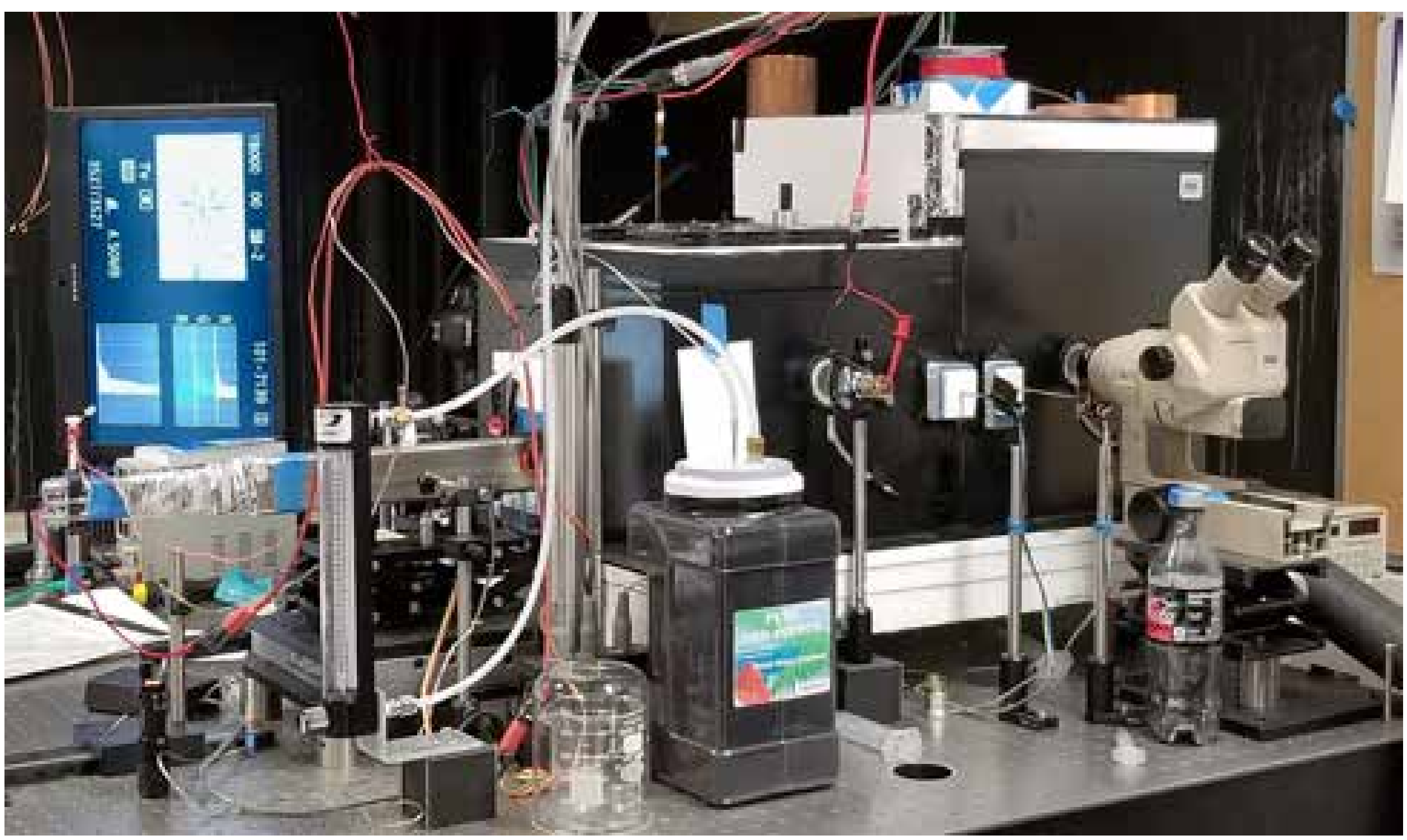

**Figure 8.11. A laboratory photograph of the dual diffusion chamber apparatus depicted in Figure 8.10. The recirculating chiller that cools the chambers, as well as several temperature controllers and other pieces of electronic hardware, are not visible in this picture.**

The base of the apparatus is cooled using a recirculating chiller that circulates methanol at -35 C to a 12x18-inch aluminum base plate upon which the rest of the hardware is assembled. Four 1x1-inch copper bars thermally connect this base plate to the top surface in DC2. Three sets of thermoelectric modules provide adjustable temperature control for the bases of both DC1 and DC2, set by independent electronic temperature controllers. These and other components of the cold-chamber design are described in Chapter 6.

A key feature in DC1 is that the supersaturation must be high enough to reliably produce c-axis e-needles, which do not readily form when $\sigma < 100\%$. The clam-shell diffusion-chamber design is well suited to this task, although it is difficult to calculate the supersaturation *a priori* from the design parameters. Some trial-and-error reckoning was necessary, therefore, to achieve the desired environmental conditions in DC1.

The temperature profile in a clam-shell diffusion chamber is typically nonlinear, and Figure 8.12 shows the vertical profile for the center of DC1. This profile varies with horizontal distance from the walls, so the air in the chamber is not stable against weak convection currents. The resulting slow air circulation in the chamber complicates any attempt to calculate the supersaturation using diffusion modeling. Even quite slow air currents are important to consider, as the time necessary to establish the final supersaturation profile is of order the diffusion time $\tau = L^2/D$, which is about 10 minutes in this chamber. It is not necessary, however, that the air be perfectly still, or that the temperature profile be precisely known. All that really matters in DC1 is that the temperature be near -6 C and the supersaturation be at or above 100 percent at the position of the wire tip.

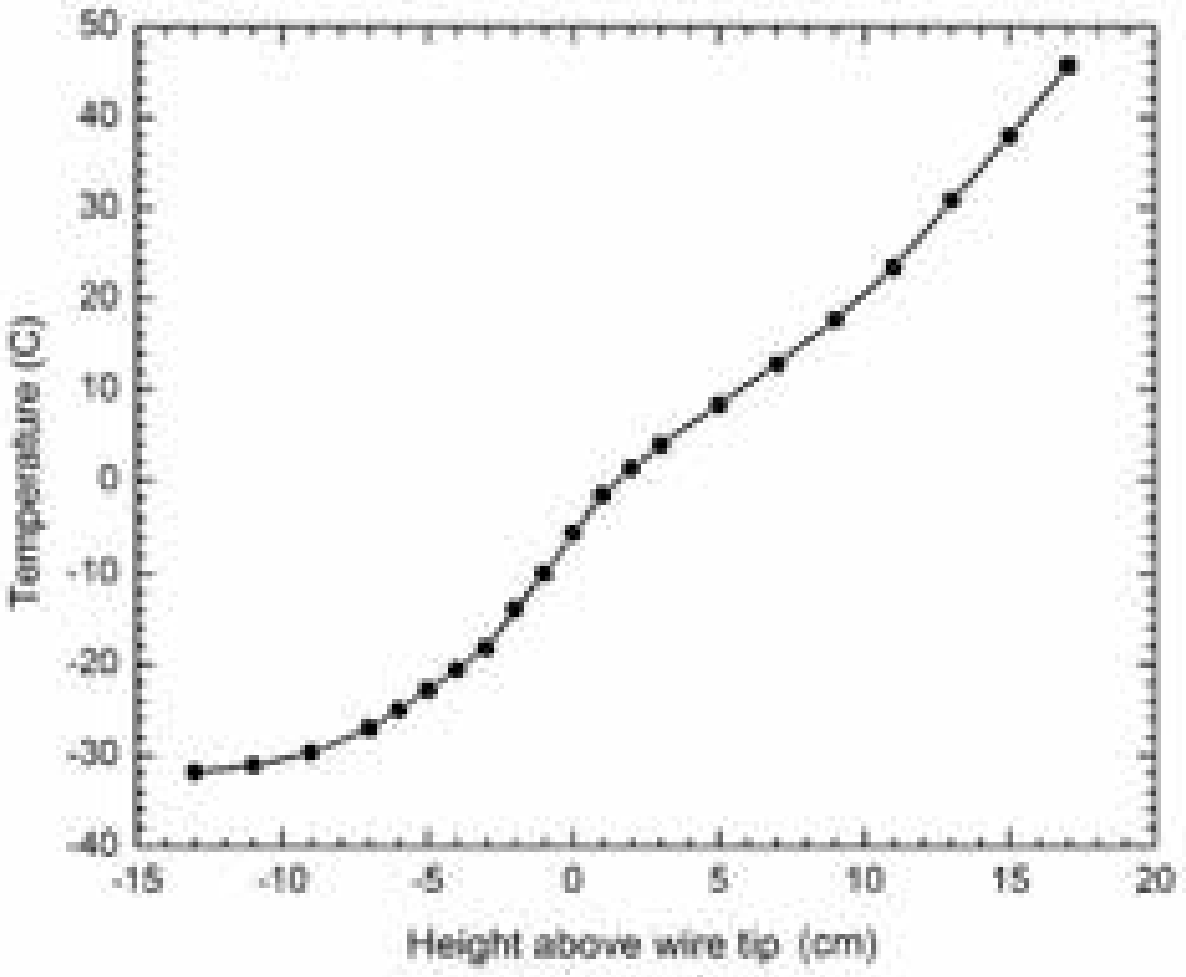

**Figure 8.12. A measurement of the temperature profile along the central vertical axis in DC1. The parameters of the DC1 design, particularly the lengths of the upper and lower clam-shell walls, were adjusted to produce an especially high supersaturation and a temperature of -6 C at the location of the wire tip ($z = 0$ in the graph). Note the especially steep $dT/dz$ at this position.**

The top plate in DC1 includes a water reservoir that holds about 100 mL in a shallow pan to facilitate evaporation. The plate, clam-shell walls, and water reservoir are all made from copper, soldered together for good heat conduction. The top-plate assembly is heated using a sealed resistive heating element, using a temperature controller to maintain a well-defined plate temperature throughout the duration of an experimental run.

At the beginning of a run, and for other testing purposes, it is often convenient to hang a weighted length of thin monofilament fishing line down the center of DC1 to observe the resulting ice growth at the position where the wire tip will be placed. Two cylindrical observing ports (see Chapter 6) are included in the DC1 walls for this purpose, and the long-distance viewing microscope can be seen in Figure 8.11. When DC1 is operating correctly, "fishbone" dendrites appear at temperatures near -5 C on the filament, as shown in Figure 8.13. These fast-growing dendrites serve as an effective measurement of the chamber temperature, and they indicate that a high supersaturation level has been achieved. This filament is removed during normal operation of the chamber, so as not to interfere with the e-needle growth on the wire tip (at the end of the Post in Figure 8.10).

Some amount of chemical vapor is needed to produce c-axis electric ice needles, as they do not readily form in clean air. At the same time, too much chemical vapor might contaminate the subsequent growth in DC2, which would clearly be undesirable for quantitative analysis. Therefore, the vapor inlet tube shown in Figure 8.10 is included in DC1 so only a minute amount of chemical vapor is used to create c-axis e-needles.

The following methodology was found to work quite well in practice. First, a good amount of GE Silicone II caulk is deposited into a half-liter soda bottle. The caulk has quite a strong odor, mostly from acetic acid, and capping the bottle nicely traps this vapor for weeks at a time. When the wire tip is in place in DC1, a syringe is inserted into the bottle through a small hole in the cap, and 2 ml of odoriferous air is drawn into the syringe. Note that the caulk itself remains untouched at the bottom of the bottle; only the air above it is used. The air in the syringe is then ejected through the inlet tube into the region surrounding the wire tip. A high voltage (typically +2000 volts DC) is quickly applied to produce c-axis e-needles. The injected air quickly disperses in DC1, and the quantity is too low to significantly contaminate the air in DC2.

Surprisingly little caulk vapor is needed to stimulate the formation of c-axis e-needles. It is often sufficient to draw 2 ml of caulk-bottle air into the syringe, eject this air out of the syringe (not into DC1), then draw another 2 ml of normal lab air into the syringe and eject that air into DC1. The residual caulk vapor coming from whatever is stuck to the inner walls of the syringe is enough to produce c-axis e-needles in DC1 with nearly 100 percent efficiency. However, completing these same steps with no initial caulk-bottle draw does not work, yielding almost entirely non-c-axis e-needles. Why this chemical vapor is needed to produce c-axis e-needles remains a mystery, and this procedure was developed almost entirely from an initial serendipitous observation followed by trial-and-error experimentation. As discussed in Chapter 4, our overall understanding of chemical vapor influences on snow crystal growth is extremely poor.

The swing-in Cover in DC1 (shown in Figure 8.10) is made from a strip of 0.1-mm-thick plastic sheet mounted horizontally, rigid enough to maintain its flat shape, about five centimeters in width. The cover quickly becomes covered with frost crystals, and it is normally kept near the chamber walls, where it does not perturb the supersaturation to a great extent. Swinging the cover into place, so it is positioned above the wire tip, quickly reduces the supersaturation at the tip by a substantial (but not well measured) factor. With this cover in place, the voltage can be turned off and the simple e-needle structure remains stable for some tens of seconds. Without the cover over the e-needles, turning off the high voltage produces normal fishbone growth that greatly broadens the tip structure in just a few seconds.

I have also found that judicious use of the swing-in cover greatly improves the transfer of e-needles from DC1 to DC2. The best procedure is to first get the e-needles started with the cover removed, as this initiation step works best with the highest available supersaturation. After the needles grow to about 1 mm long, swing the cover into place while leaving the high-voltage on. This slows the e-needle growth by about a factor of 2-3, but their sharp morphology remains. In this state, let the e-needles growth another 1-2 mm, then turn off the high voltage (leaving the cover in place) and pull the wire tip into DC2. Following this procedure, the transfer efficiency is nearly 100 percent; the e-needles mostly survive the journey into DC2. Without the slower growth step with the cover in place, however, the e-needles frequently break off from the wire tip before they make it into DC2, which can be quite frustrating in practice.

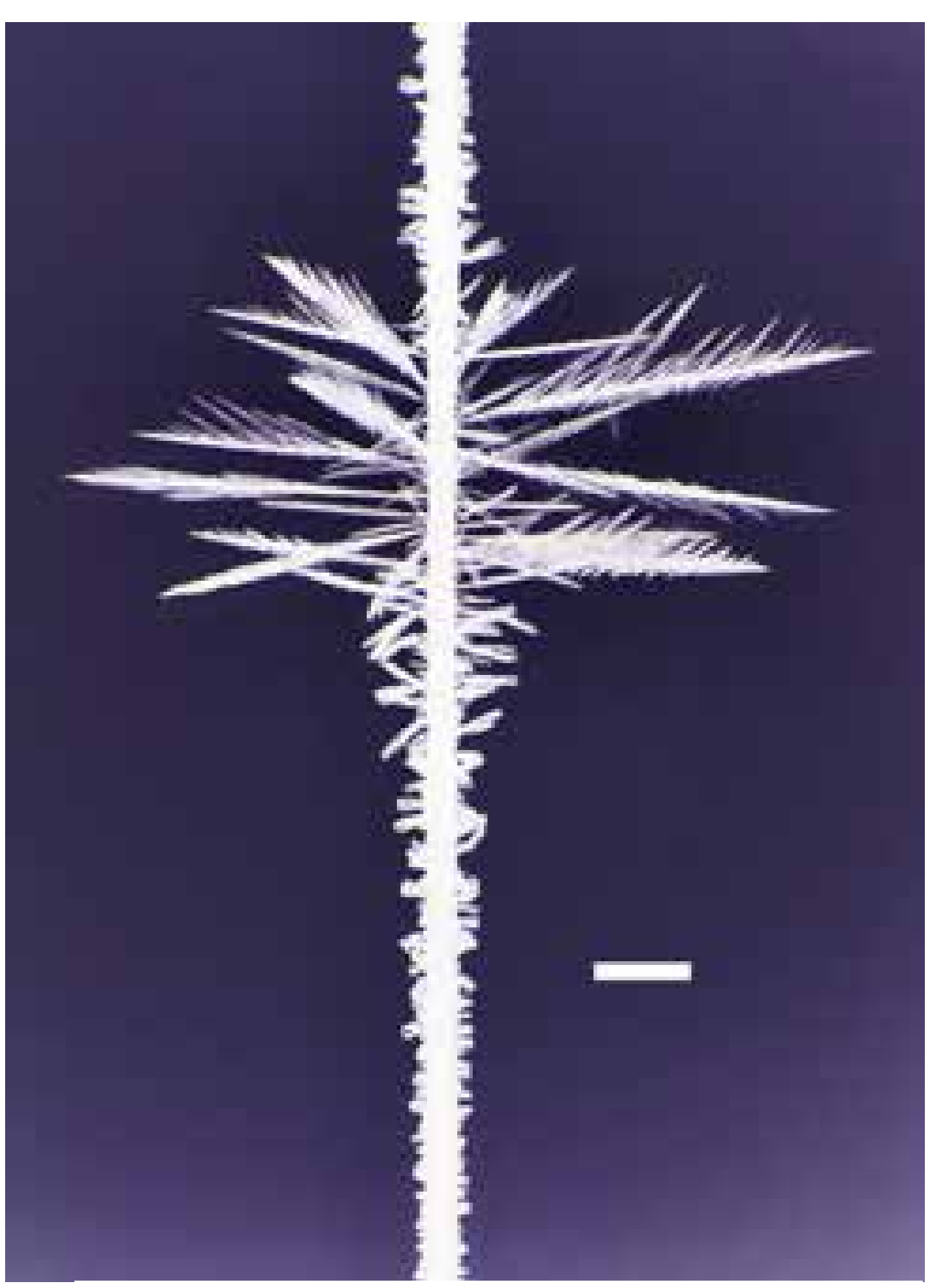

**Figure 8.13. An image showing ice crystals growing on a segment of 200-micon-diameter nylon fishing line hanging in the center of DC1. The scale bar is 1 mm long, and the crystal growing time was 19 minutes. [2014Lib]. The crystal morphology is strongly temperature dependent, with the fast-growing "fishbone" dendrites appearing near −5 C. The best wire tip location for producing c-axis electric needles is just below the fishbone peak, where the temperature is near -6 C. Hollow columnar crystals appear at this position.**

The "Tapper" in Figure 8.10 is a convenient tool for removing built-up frost from the wire tip. The tapper consists of a small cylindrical weight on a string that swings into place like the cover plate. The weight swings from its string and effectively knocks crystals off the wire, readying it for making more e-needles.

Over time, the frost buildup on the wire tip becomes so high that it cannot be removed effectively with the tapper. When this happens, the wire tip can be cleaned by inserting a long plastic rod into the top of the chamber. A plastic rod with a two-cm-long wire end seems to work best. Touching the thicker, room-temperature wire to the thin wire tip immediately melts and removes the ice buildup. The frost-covered tapper is then brought in to tap the wire tip a few times to nucleate freezing of the remaining water. (Without this nucleation step, only liquid water condenses on the wire tip, as the temperature is not cold enough to quickly nucleate spontaneous freezing.)

In summary, after a fair bit of experimentation, the following procedure was found to produce c-axis e-needles with nearly 100 percent yield:

1) With the nylon monofilament removed from DC1, and the Cover and Tapper (see Figure 8.10) out of the way, position the wire tip (on the end of the Post in Figure 8.10) in DC1.

2) Use the Tapper to remove frost from the wire tip if needed. If there is a lot of frost on the wire, insert a wire-tipped room-temperature rod in from the top of the chamber and touch the rod to the wire to melt the frost, leaving a nearly bare wire tip. Then touch the wire with the Tapper to nucleate ice on the wire.

3) Extract 1-2 ml of air from a bottle containing a caulk vapor and inject the air through the Inlet Tube that has one end near the base of the post. This sends a pulse of chemical vapor into DC1, which quickly disperses so some additive reaches the space near the wire tip.

4) Apply +2000 volts DC to the wire tip to start the growth of c-axis electric needles. (This voltage is supplied via a wire running through the manipulator arm, as described below.)

5) View the electric needle growth using a long-distance microscope (see Figure 8.11). When the needles are about 1 mm long, swing in the Cover to lower the supersaturation above the needles, while leaving the high voltage on. Let the e-needles grow another 1-2 mm to reach their desired final length.

6) Turn off the high voltage, immediately open the two shutters separating DC1 and DC2, and then pull the post assembly, with the electric needles, into DC2.

Although the above procedure sounds somewhat involved, the end result is an efficient electric-needle "factory" that works remarkably well and nearly always yields a set of c-axis e-needles in DC2 like those shown in Figure 8.1. Moreover, the dual chamber has a rapid sample turnaround that is especially important for turning a proof-of-principle demonstration into a workhorse experiment that produces valuable scientific data.

## Diffusion Chamber 2

DC2 was designed to be a linear-gradient diffusion chamber, optimized to allow accurate modeling of the interior supersaturation. As shown in Figure 8.10, a pair of aluminum clamshells provide a cold barrier around the chamber, while stainless-steel inner walls conduct heat vertically to establish a linear vertical temperature gradient along the walls and throughout the chamber interior. The thickness of the stainless-steel walls (1.6 mm) was chosen to provide sufficient conduction to define a linear temperature gradient, but not so thick that the resulting heat conduction is difficult to sustain. Figure 8.14 shows a measured temperature profile at the center of the chamber.

With a linear temperature gradient and frost-covered walls to produce the boundary condition $c_{wall}(z) = c_{sat}(T_{wall}(z))$ at the chamber walls, it becomes possible to accurately model the interior temperature and supersaturation. In the limit that the width of DC2 is much larger than the height, the heat and particle diffusion equations yield linear profiles for both the temperature and particle

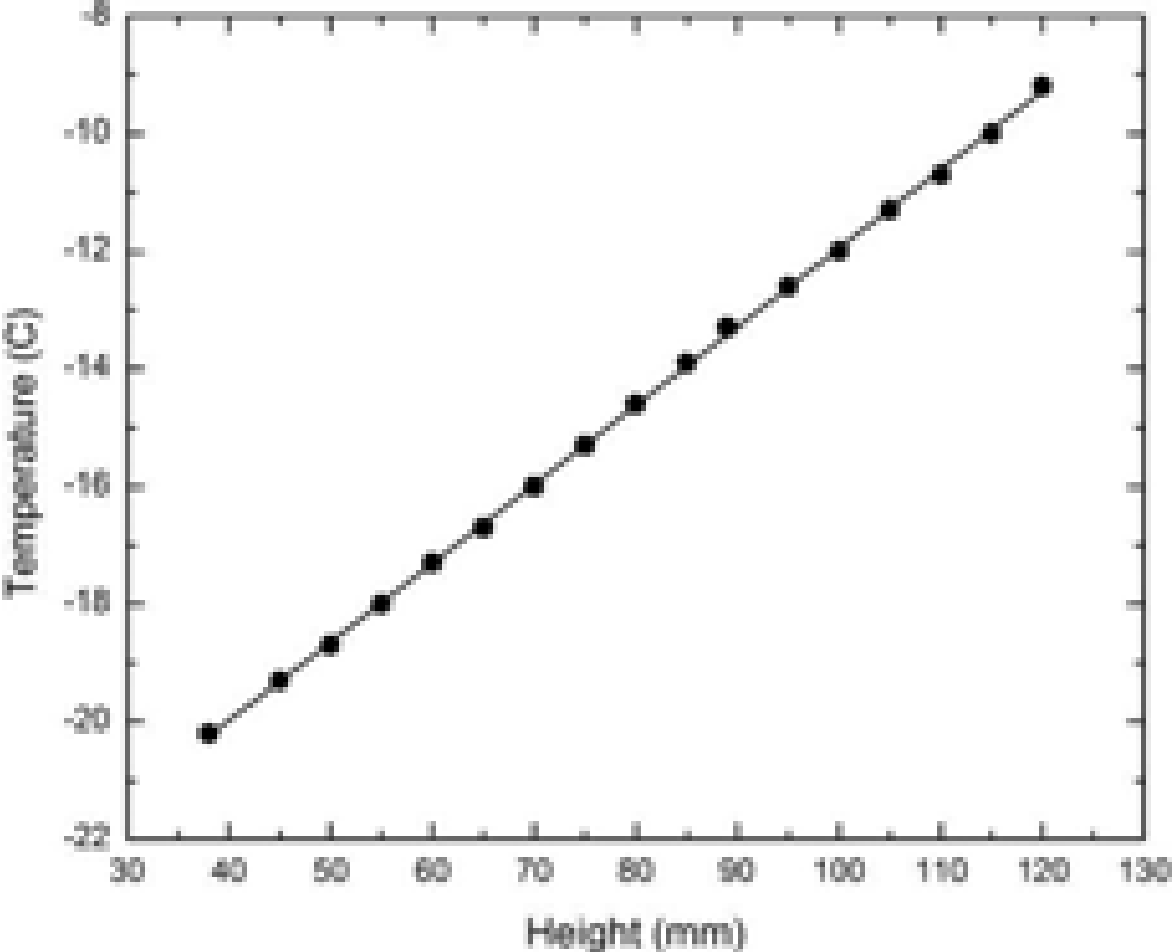

**Figure 8.14. Measurements of the vertical temperature profile in DC2. Unlike in DC1, this temperature profile holds near the walls and throughout the interior of the chamber. As a result, there are no horizontal temperature gradients inside the chamber, so the air is stable against convection. This stability allows precise modeling of the supersaturation within the chamber.**

density inside the chamber. From this, one obtains the supersaturation at the chamber center

$$\sigma_{center} = \frac{(c_{top} + c_{bottom})/2 - c_{sat}(T_{center})}{c_{sat}(T_{center})}$$

$$\approx \frac{1}{2}\frac{1}{c_{sat}(T_{center})}\frac{d^2 c_{sat}}{dT^2}(T_{center})(\Delta T)^2 \quad (8.6)$$

where $T_{center} = (T_{top} + T_{bottom})/2$ and $\Delta T = T_{top} - T_{bottom}$.

This far-away-walls approximation is not too far off for DC2, but the walls do reduce the supersaturation somewhat. Using a finite-element diffusion analysis to calculate the effects of the walls and the post supporting the crystals yields the supersaturation model shown in Figure 8.15 [2016Lib]. This model predicts that, over a broad range of growth conditions, the center supersaturation is well described by

$$\sigma_{center} \approx \frac{1}{2}\frac{G_{mod}}{c_{sat,center}}\frac{d^2 c_{sat}}{dT^2}(T_{center})(\Delta T)^2 \quad (8.7)$$

where $G_{mod} \approx 0.72$ is a correction factor arising from the walls and post. Note that the model shows very nearly the same $\sigma_{center} \sim \Delta T^2$ dependence as the limiting theoretical expression.

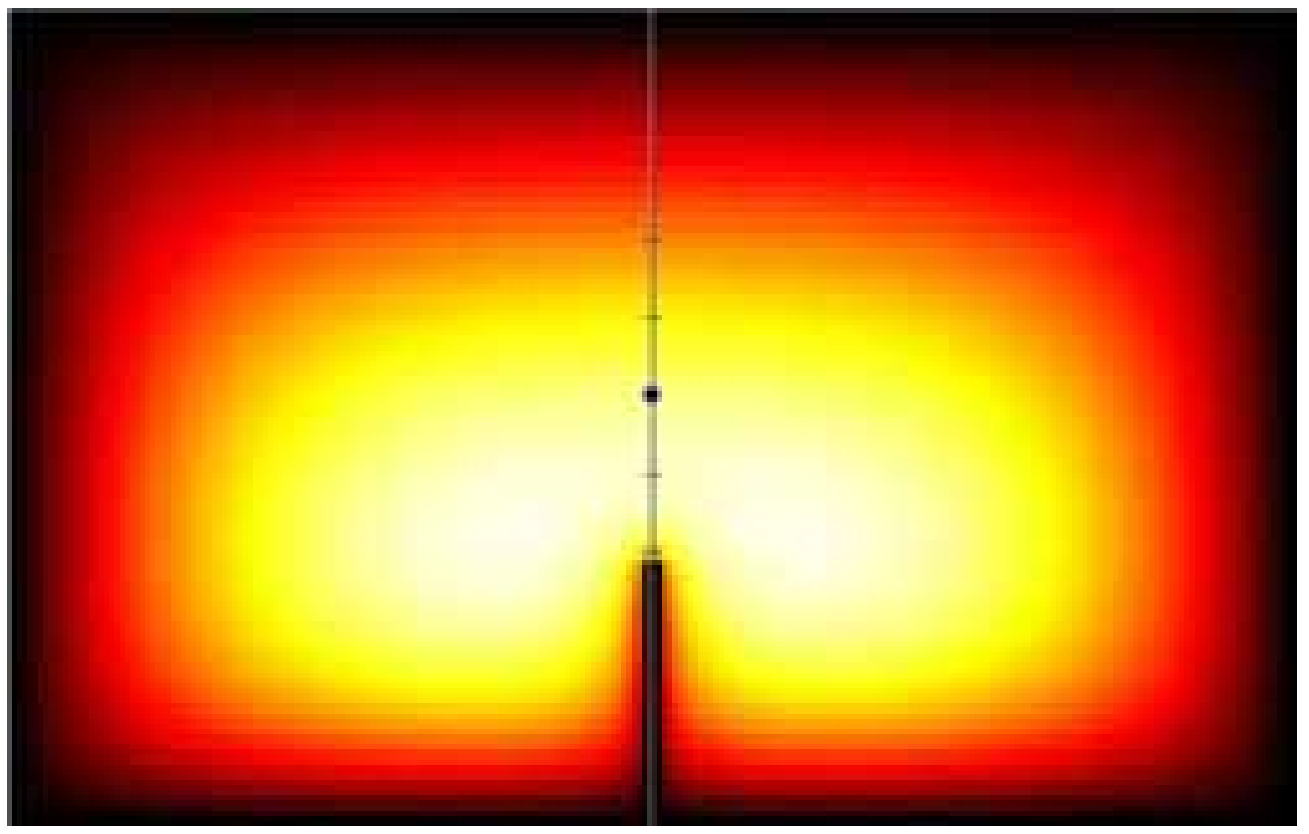

**Figure 8.15. A finite-element diffusion model of the supersaturation field in DC2 at a fixed ΔT. The supersaturation goes to zero (black) at the frost-covered walls and exhibits a broadly peaked maximum (yellow/white) below the chamber center. The black dot indicates the center of the chamber, which is the location of the growing e-needles. By running the model with different top and bottom temperatures, one can determine the supersaturation at the center of the chamber as a function of ΔT.**

DC2 also includes a swing-in thermistor to monitor the temperature at the chamber center, and a swing-in cover to lower the supersaturation if desired. The calibrated thermistor has an absolute accuracy of $\pm 0.1$ C and can be rotated in to measure the temperature at the location of the growing crystals. In principle, the center temperature is the average of the top and bottom temperatures, but small perturbations come from the imaging optics, optical viewports, and other factors.

The DC2 Cover consists of a 0.1-mm-thick, 1-cm-wide plastic strip mounted horizontally. This swings in to a position just above the center of the chamber, nearly turning off the supersaturation seen by the growing crystals. The cover is typically put in place before the e-needle transfer from DC1, which then allows time to position the needles and focus the camera under conditions of low supersaturation. Swinging the cover away then restores the normal supersaturation in a time of roughly $\tau \approx L^2/D_{air} \approx 5$ seconds, where $L \approx 1$ cm and $D_{air} \approx 2 \times 10^{-5}$ m$^2$/sec. The impact of the cover on the supersaturation can be verified by direct measurements of the crystal growth.

During the several-hour-long cool-down of the system, the bottom surface of DC2 is heated to produce strong convection within the chamber. Evaporation from bottom water reservoirs (see Figure 8.10) produces water vapor that is then transported upward to deposit as frost on the walls and upper surface of the chamber, and this frost provides the water vapor source for normal operation of the diffusion chamber. At the start of cooldown, the DC2 bottom temperature is set to +35 C, and it remains at this temperature for about an hour, producing a substantial evaporation rate for icing the upper surfaces. The DC2 bottom temperature drops as the chiller cools the base plate during cooldown. This temporary inverted temperature profile during cooldown provides an ample supply of ice on the upper surface and walls of DC2, thus maintaining the assumed boundary condition that all DC2 surfaces are covered with ice.

Starting during cooldown, clean air is sent slowly into the top of DC2 to reduce effects from chemical contamination. Air is first sent through an activated charcoal filter to remove residual chemical contaminants, and the clean air is then injected into the top of DC2 at a rate of 60 ccm via a flow meter. This slow trickle of clean air continually replaces the air in DC2 about once per hour without significantly affecting the temperature or supersaturation profiles. The clean-air purge is not essential for operating DC2, but it does seem to improve the overall reproducibility of the crystal growth observations.

## The Manipulator Arm

Moving the e-needles reliably from DC1 to DC2 is a nontrivial challenge, and Figure 8.10 shows the manipulator arm that was constructed for this task. The lateral motion is guided by a pair of precision-polished stainless-steel rods moving through linear-motion bearings, providing a smooth enough ride that the e-needles are not shaken off the frost-covered wire tip upon which they formed. The drawing in Figure 8.10 shows the post assembly in both DC1 and DC2, while there is actually only one post assembly that shuttles back and forth between these two chambers.

The post assembly consists of a set of telescoping stainless-steel capillary tubes that produce minimal perturbation of the supersaturation field while still providing the necessary support and rigidity. The top of the post, extending out from the smallest stainless-steel capillary tube, is a sharpened, 120-micron-diameter stainless-steel acupuncture needle (J type). The base of the post is connected, via an insulating coupler, to a 6-mm-diameter DC motor that rotates the entire post assembly about its vertical axis. Wires for the motor and the high-voltage brush connection to the post pass through a tube that runs along the entire length of the manipulator arm and out the back end.

A pair of insulating sliding-plate shutters (see Figure 8.10) are used to open and close a keyhole-shaped passage between DC1 and DC2. These shutters are normally kept closed to maintain the temperature profiles in the two chambers, and they are opened only briefly to allow passage of the post assembly. A narrow slot at the base of DC2 is not shuttered, as the temperature below the DC2 bottom plate is colder so the air below does not mix with the air in DC2.

### Optical Microscopy

Imaging of the growing e-needles in DC2 is done using a 3X Mitutoyo Compact Objective, with 0.07 Numerical Aperture and a 2.5-micron resolving power. The objective is built into the back wall of DC2, and the front surface of the objective has a fixed distance of 69 mm from the chamber center. A short tube placed over the front of the objective keeps frost from forming on the optical surface. As with all optical microscopy, this objective was chosen as a compromise between resolution, depth-of-field, and working distance.

A full-frame D-SLR camera at room temperature is positioned behind the objective, separated by a three-window cylindrical viewing port. No additional optics are placed between the microscope objective and the camera sensor, while extension tubes minimize scattered light from the room lighting. Focusing is done by moving the manipulator arm slightly (perpendicular to its main line of travel) and by sliding the camera back and forth on an optical rail.

Illumination is provided by an LED lamp positioned outside the chamber, and another viewing port near the lamp completes the optical path. The manipulator arm is held only by the bearing block, so the wire tip exhibits several microns of shake when in normal operation, brought about largely from unavoidable coupling to vibrations from the recirculating chiller. A camera shutter speed of $1/8000^{th}$ second effectively freezes the crystal motion to provide sharp imaging.

### Turn-Key Operation

This dual-chamber apparatus has become something of a workhorse for my ongoing investigations of snow crystal growth. The hardware has evolved to where it has a nearly turn-key operation, able to churn out observations of a broad range of single-crystal structures on the tips of e-needles. The results are already providing new insights into the attachment kinetics and the morphology diagram, and there is much potential for additional scientific progress using e-needles.

## 8.4 The Morphology Diagram on E-needles

Using the dual-diffusion-chamber apparatus just described, one can examine ice growth over a substantial range of temperatures and supersaturations, thus exploring a broad parameter space in the snow crystal morphology diagram. Figure 8.16 shows an array of photographs of normal growth on e-needles as a function of temperature from -0.5 C to -21 C as a function of supersaturation from 8 to 128 percent.

Each tile in this collection of photos shows a representative example using fixed values of temperature and supersaturation that remained constant as the crystals grew. In these images, the smallest needle-like structures have diameters of about 30 microns, while the largest dendritic plates have diameters of about 1.5 mm. The image scale and cropping were separately adjusted for each image. Growth times ranged from about 5 minutes at the highest supersaturations to 30 minutes at lower supersaturations.

One problem with earlier versions of the morphology diagram is that they were created from less-than-ideal snow crystal observations. Filamentary support influenced the growth behavior, as did competition from multiple closely spaced crystals. Conclusions regarding detailed morphological effects could not be obtained from these non-ideal observations, and detailed comparisons with numerical models were not practical.

In contrast, e-needle crystals can be used to grow isolated single crystals with no substrate interactions and in well-defined environmental conditions. The resulting crystals exhibit nearly flawless morphologies with excellent six-fold symmetry. Because each crystal begins as a simple ice needle, the subsequent tip growth behavior is remarkably reproducible over the entire observed range of environmental conditions. E-needle observations are thus nearly ideal for examining detailed morphological features.

Importantly, each of the photos in Figure 8.16 presents an opportunity for quantitative comparisons between observed growth behaviors and numerical models. Properly calibrated images of snow crystals growing on e-needles allow a broad investigation into the detailed physical processes that determine all the various growth behaviors. Thus e-needle observations open up new opportunities for better understanding the physical processes underlying the broad diversity of snow crystal growth behaviors.

## Robust Features

The e-needle observations shown in Figure 8.16 exhibit a variety of robust morphological features. In this context, I use "robust" to indicate a specific behavior that is easily generated in the lab, distinctive, and can be reliably found over a well characterized range of environmental conditions. Numerical models that cannot readily reproduce these robust features are clearly incomplete or incorrect in some way.

A first robust feature is one that has long been part of the morphology diagram, namely the increased degree of complexity in crystals grown at higher supersaturations. E-needles grown at low supersaturations often develop as simple columns, simple blocky structures, or perhaps thick plates on stout columns. At even lower supersaturations than those shown in Figure 8.16, the norm is growth as simple hexagonal columns.

As the supersaturation increases, branching often begins with six primary branches with little or no sidebranching. Sidebranching eventually develops at the highest supersaturations shown in Figure 8.16, although some temperatures are more prone to copious sidebranching than others.

Near -15 C, the dendritic branching is mainly confined to a nearly planar structure, as growth outside the plane is limited by strong basal faceting. On e-needles, these dendritic plates are typically slightly conical in overall shape, as the top basal surface grows faster than the bottom surface. The cone angle depends rather strongly on growth conditions, and these morphological trends tell a story about how $\alpha_{prism}/\alpha_{basal}$ varies with temperature and supersaturation.

Near -5 C, the six primary branches develop into fishbone dendrites at the highest supersaturations. Over the entire temperature range in Figure 8.16, the morphology of a single dendritic branch at high supersaturation (see Chapter 3) defines the shapes of the six primary branches. Unfortunately, reaching high supersaturations at low temperatures is experimentally difficult, so this region of phase space has yet to be explored. At -0.5 C in Figure 8.16, melting prevented the formation of snow crystals at $\sigma = 128$ percent.

Simple stars are a robust feature near (T, $\sigma$)=(-14C, 32%), and the spike-like primary branches are observed to grow stably with no sidebranches to substantial lengths. I suspect that this morphology will be difficult to reproduce in 3D modeling without the Edge-Sharpening Instability, but that question remains to be investigated. A long-armed example of a simple star, along with examples of several other robust morphological features, can be found at the end of this chapter.

**Figure 8.16. (Following five pages) The snow crystal morphology diagram depicted by crystals growing on the ends of slender ice needles, as a function of temperature and supersaturation.**

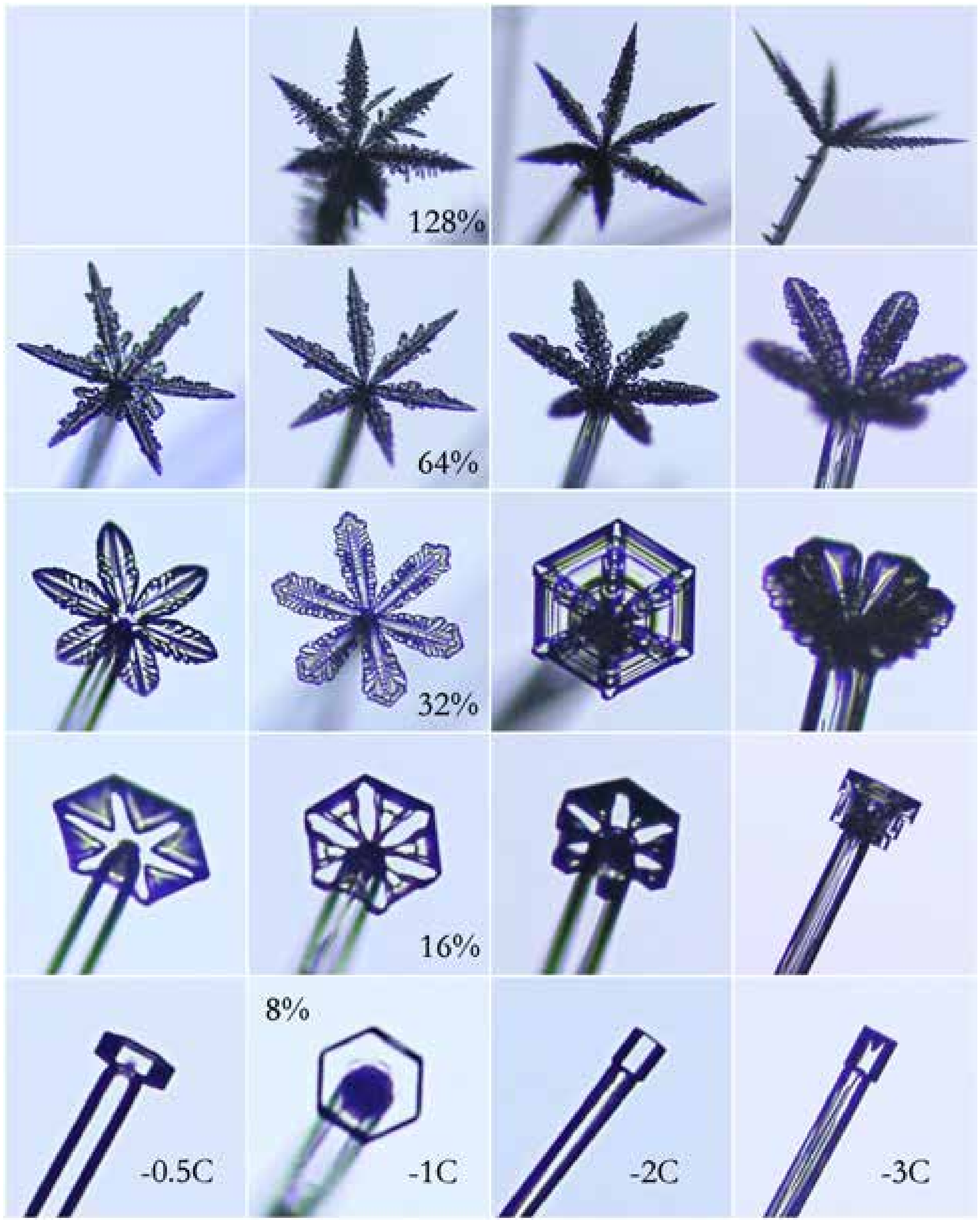

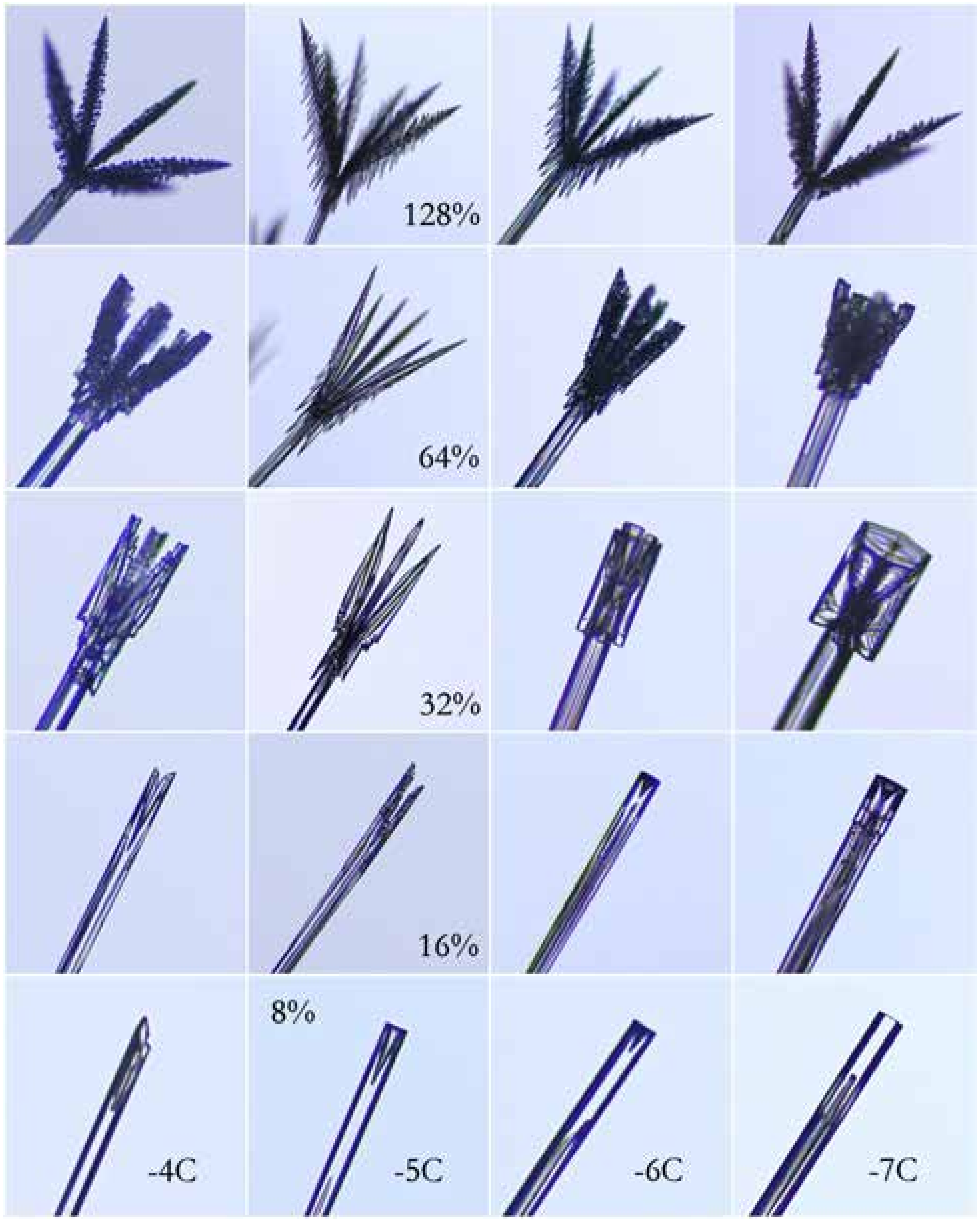

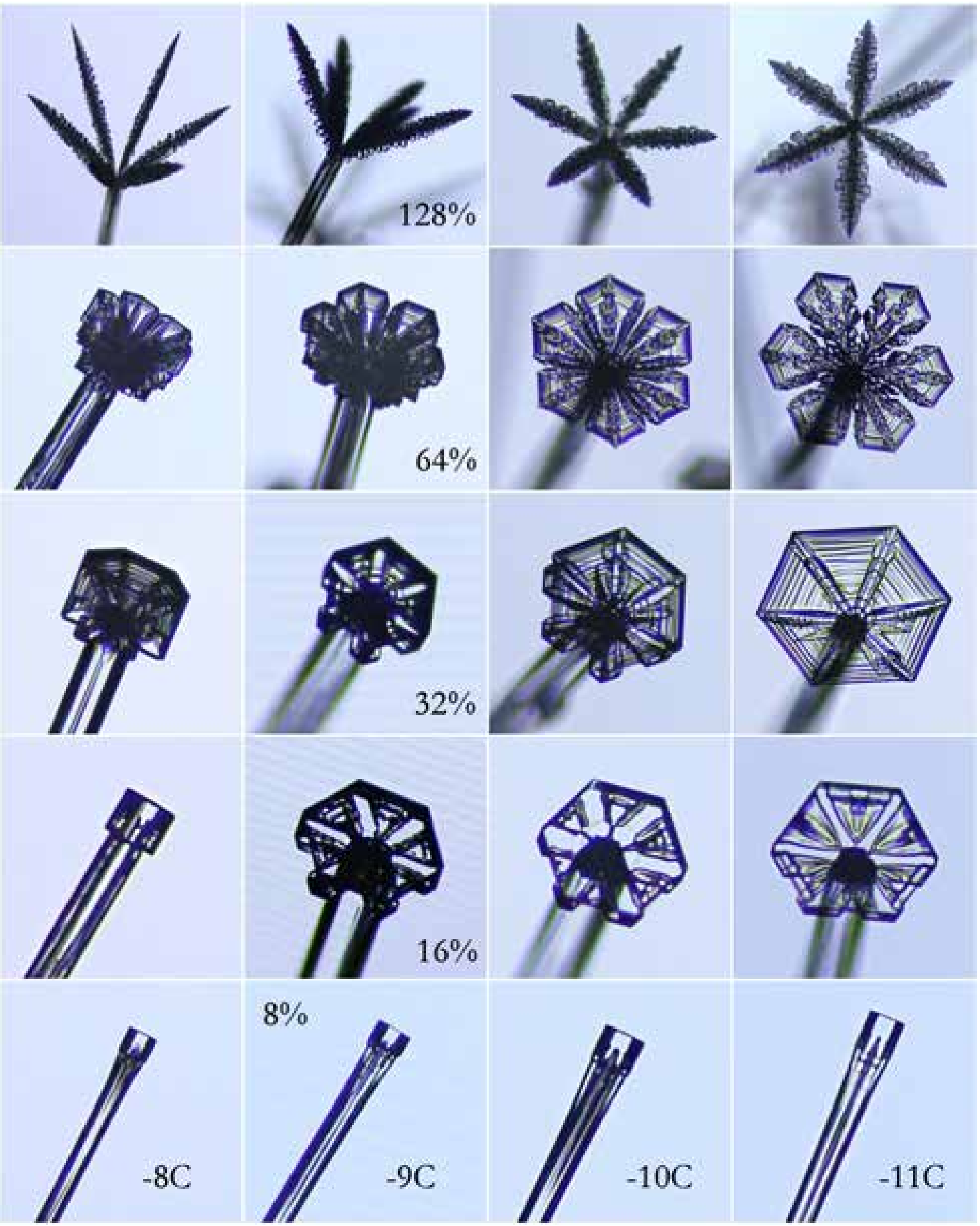

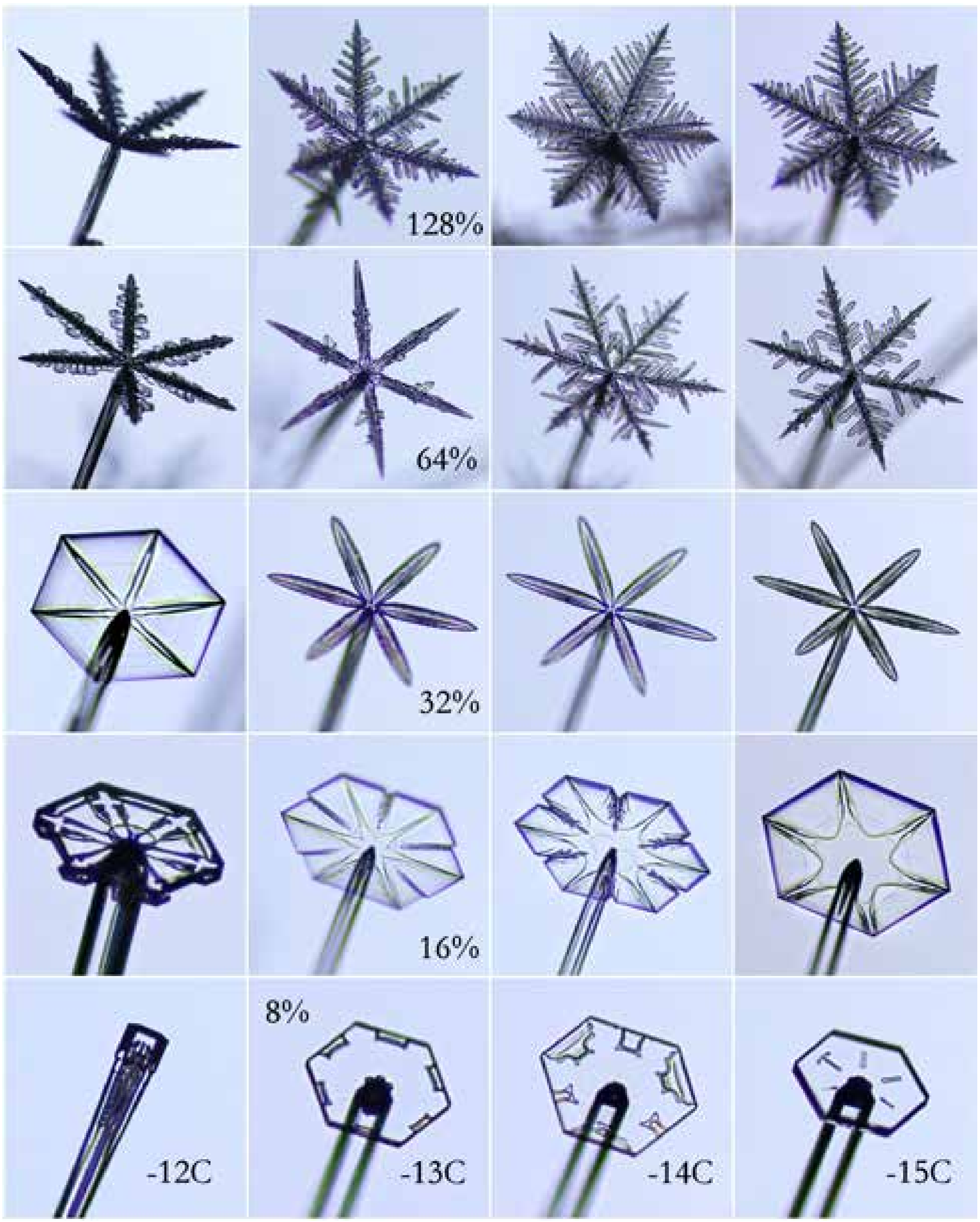

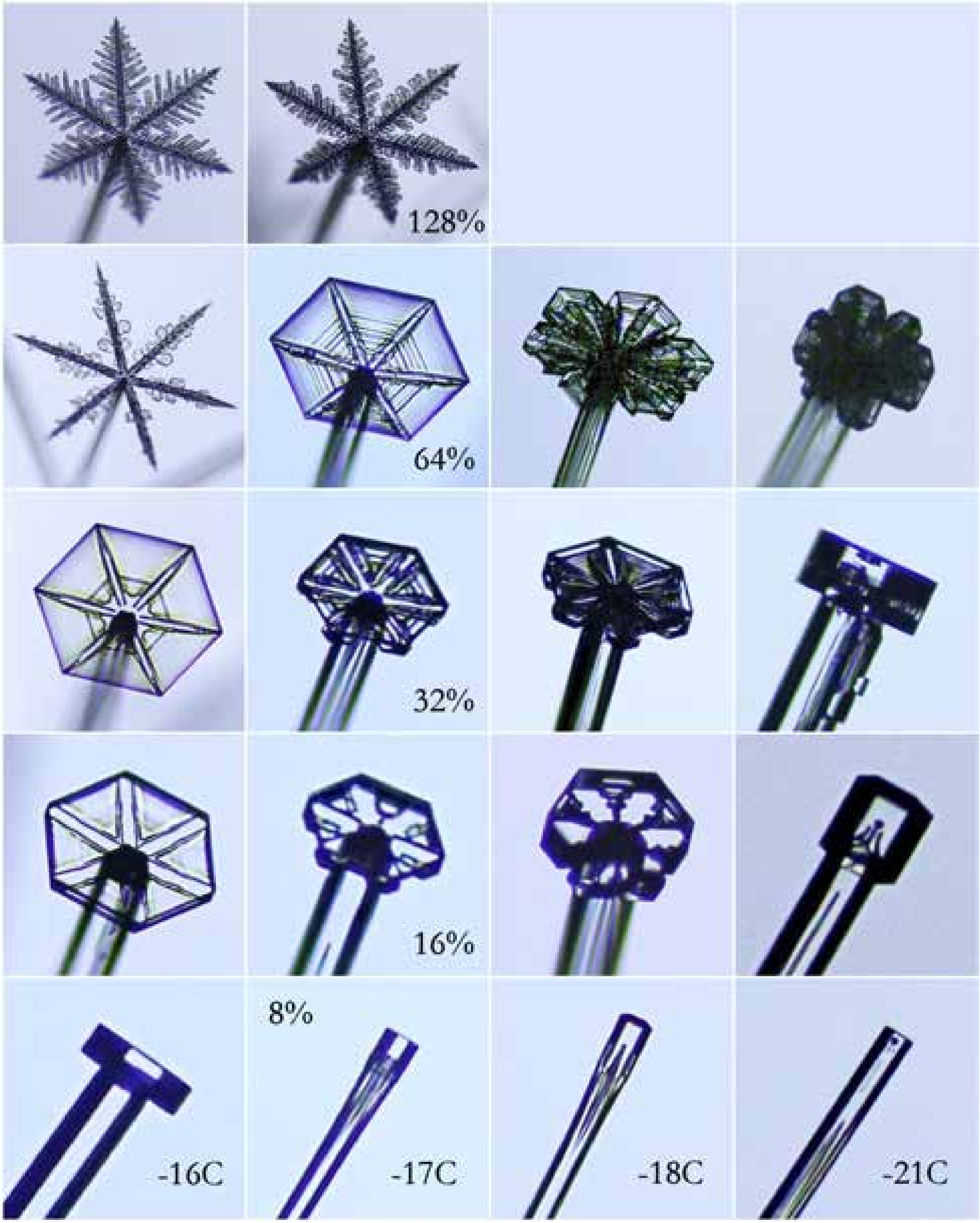

Hollow columnar structures appear on the ends of e-needles near -5 C, but they occupy a rather small region of parameter space. Hollows do not readily form if the supersaturation is too low, and the walls break up into a cluster of needle-like structures if the supersaturation is too high. Deep cups form near (-7 C, 32%), with the cup opening angle depending strongly on temperature. When growing on e-needles, the cups are typically flanked by straight "fins" on the outside edges.

Ridges on the six corners of hexagonal plates appear over a broad range of growth conditions, making these especially robust features. Their morphology depends quite strongly on the cone angle of the growing plates, and often the ridges develop an "I-beam" structure, for example seen clearly at $(T, \sigma)$=(-9C, 32%). Ridges generally become more pronounced and more structured as the cone angle of the plates increases, eventually yielding the fins described in the previous paragraph.

The diversity and widespread appearance of ridge-like structures is quite remarkable in e-needle growth. In part, this is because the supersaturation gradient around an e-needle tip yields slightly conical plates that exhibit especially distinctive ridge-like structures. As mentioned above, this built-in supersaturation gradient can be quite beneficial in that it accentuates these morphological features and facilitates their detailed investigation.

The formation of exceedingly thin plates on e-needle tips near -15 C is another noteworthy feature of the morphology diagram. As seen in Figure 8.16, remarkably thin, nearly featureless hexagonal plates form at several locations in the $(T, \sigma)$ plane near -15 C. Ridges are sometimes absent at low supersaturations, but delicate ridging is present in the largest, thinnest plates. Understanding why these thin plates form so dramatically near -15 C, in such a narrow temperature band so far from the freezing point, remains an enduring mystery in the science of snow crystals.

## The Next Grand Challenge

The obvious next step in all this is to quantify the e-needle observations over a broad range of conditions and then compare the observations with 3D numerical models. At least the first part of this statement is relatively straightforward, as the technology for creating and exploiting e-needles is already quite mature. Unfortunately, the theory side of this research program now substantially lags the experimental side, as I described in Chapter 5.

Full 3D cellular-automata models have yielded structures that nicely resemble many aspects of snow crystals, and this puts them ahead of other numerical modelling techniques. Nevertheless, the models to date have not progressed much beyond their demonstration phases. They have not yet incorporated realistic parameterizations of the attachment kinetics, so their morphological successes do not always reflect a good understanding of the underlying crystal-growth physics. Plus, the 3D models have not yet reached the kind of turn-key operation needed for churning out dozens and hundreds of models for direct comparison with observations of growth rates as well as morphologies.

There appears to be no obvious roadblocks to developing suitable numerical models at this point, however, so it appears likely that researchers will begin making quantitative 3D comparisons between observations and numerical models in the not-too-distant future. When this happens, I expect it will lead to rapid progress as the attachment kinetics and surface diffusion effects are adjusted to provide good quantitative agreement between observations and theory over a broad range of environmental conditions. And this, in turn, will spark new insights into the molecular processes that underlie the best-fit model parameters. At some point we may finally achieve a fundamental understanding of how snow crystals form.

## 8.5 Simplest E-needle Cylindrical Growth

Examining the most basic columnar growth of electric needles provides a good validation of the supersaturation model for DC2, and it supports our overall understanding of ice growth from water vapor [2016Lib]. It tests the accuracy of Equation 8.7, confirms our understanding of the 1D cylindrical growth model (see Chapter 3), and generally supplies a "reality check" that our basic picture of diffusion-limited growth is quantitatively correct. It is important to perform these kinds of model-validation experiments, in my opinion, if one expects to realize an accurate quantitative understanding of more complex ice-growth phenomena.

The general idea in this basic experiment is to start with a set of e-needles in DC2, like that shown in Figure 8.1, rotate the support post to bring a single e-needle into focus perpendicular to the imaging axis, and then measure the radial growth of the chosen needle as a function of time. The growth is subsequently compared with the 1D cylindrical model for different temperatures and supersaturations. If everything is working properly, then we should find good quantitative agreement between theory and measurements.

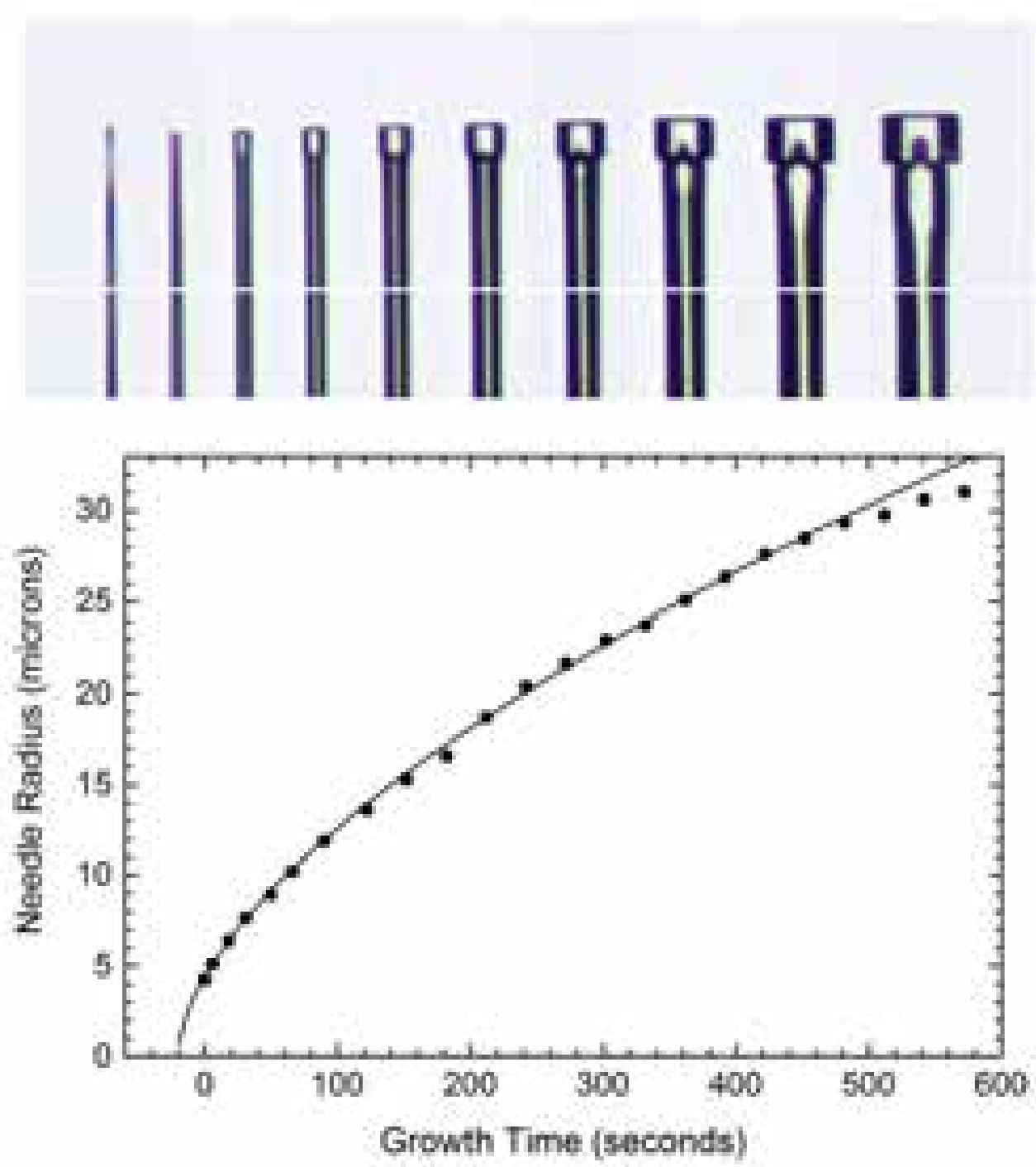

**Figure 8.17. Example observations at -2 C of the growth of a single electric ice needle after being transported into DC2. The composite image shows the needle at several different times, and the graph shows measurements of the needle radius as a function of time, measured at the position of the white horizontal line in the image, which is approximately 100 μm below the needle tip [2016Lib].**

### Infinite Cylinders

The theory of cylindrical growth is straightforward and very well understood, at least for the limiting case of an infinitely long cylinder. The mathematical details are presented in Chapter 3 and Appendix B, as well as in [2016Lib]. In most circumstances of interest in this section, $\alpha_{diffcyl} \ll \alpha_{prism}$ is a good approximation, so the growth is not substantially limited by attachment kinetics, but rather is determined solely by particle and heat diffusion around the cylinder. The theory then depends only on well-known physics, and, importantly, it is independent of the not-so-well-known molecular attachment kinetics.

Assuming Equation 8.7 provides an accurate model of the supersaturation around the growing ice needles in DC2, the radial growth velocity of an infinite ice cylinder is then given by

$$v \approx \frac{1}{1+\chi_0} \frac{G_{mod}}{B} \frac{X_0}{R} \frac{1}{2c_{sat}} \frac{d^2 c_{sat}}{dT^2} (\Delta T)^2 \quad (8.8)$$

where $R$ is the radius of the cylinder, $B = \log(R_{out}/R)$ derives from the cylindrical boundary conditions of the diffusion problem, and $\chi_0$ is a thermal parameter that derives from latent heating of the crystal during growth.

Sample data illustrating the measurement process are shown in Figure 8.17. Clearly the morphology of the growing e-needle is not that of an infinite cylinder, but it does approach that ideal case to some reasonable approximation at early times, and the cylindrical approximation works best at positions far from the end of the needle. Therefore, the measurements of the cylindrical radius $R_{needle}$ in Figure 8.17 were obtained at a distance 100 $\mu m$ from the needle tip.

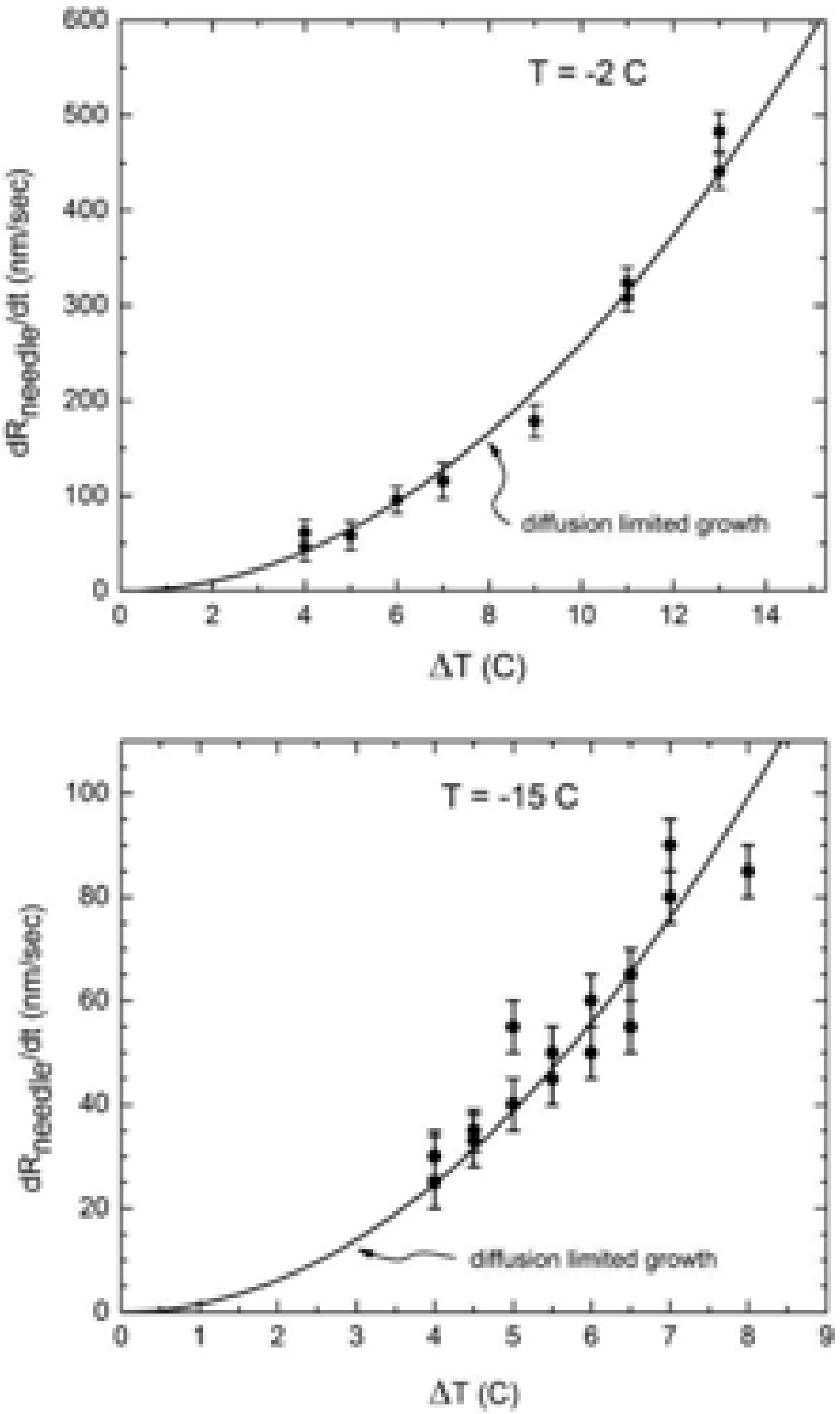

**Figure 8.18. Example measurements at -2 C and -15 C showing the radial growth velocity of cylindrical ice needles when the needle radius was equal to 5 μm. The data are shown as a function of $\Delta T = T_{top} - T_{bottom}$ in DC2, which is proportional to the supersaturation according to Equation 8.7. Fitting the data reduces it to a single parameter $A(T)$, with the growth velocity given by $dR/dt = A(t)\Delta T^2$ when the needle radius is equal to 5 μm [2016Lib].**

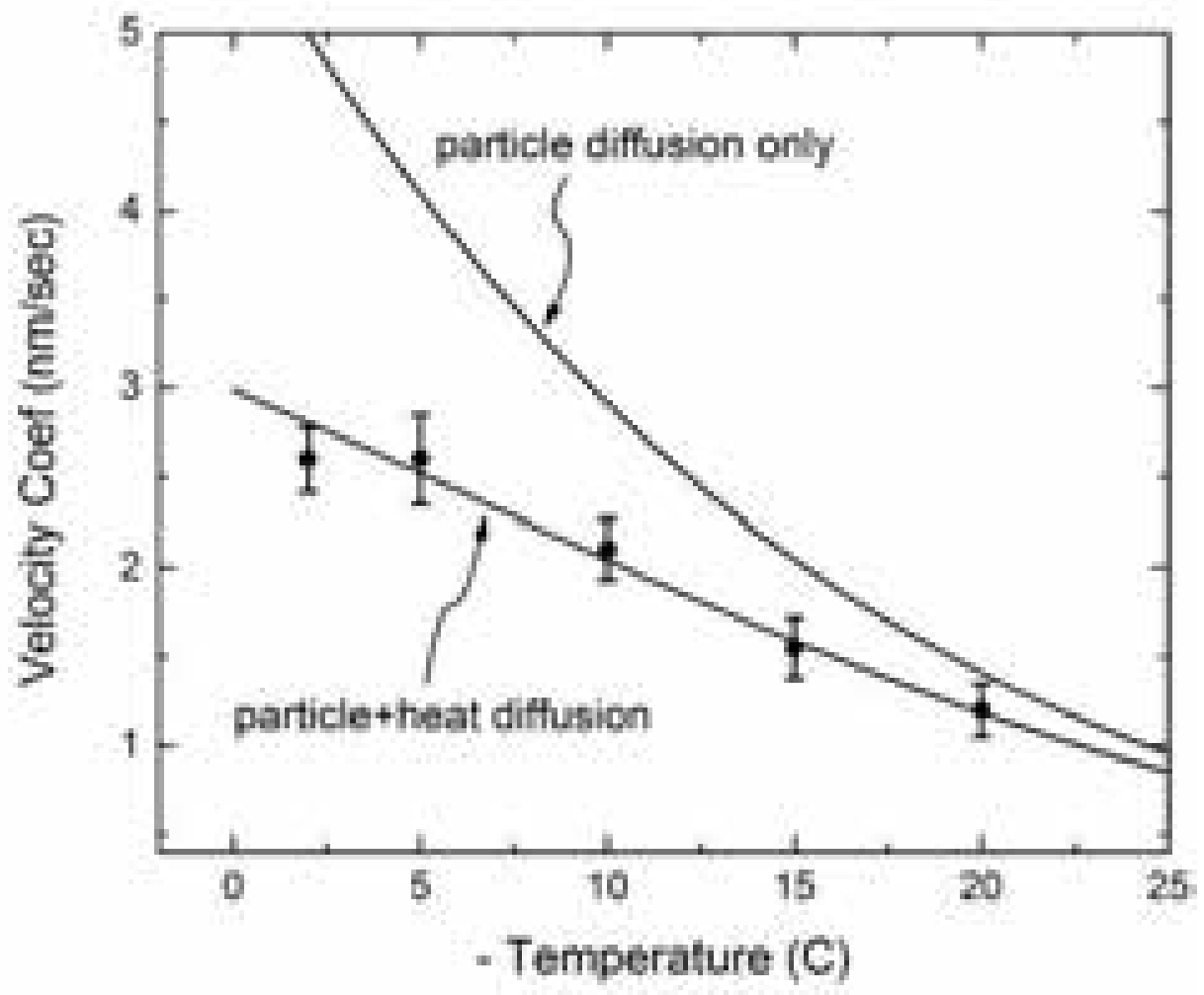

**Figure 8.19. The cylindrical growth parameter $A(T)$ extracted from the observations of needle growth in DC2, as a function of temperature. The top theory line is that predicted when the growth is limited entirely by particle diffusion (setting $\chi_0 = 0$ in the theory), while the lower theory curve includes both particle and heat diffusion. From [2016Lib].**

Drawing a line through these data, I then used the fit line to determine the growth velocity $dR_{needle}/dt$ when the needle radius was just $R_{needle} = 5\ \mu m$, before the blocky structure appeared at the tip of the needle. Analyzed in this way, producing a single $dR_{needle}/dt$ value at early times and at a position quite far from the needle tip, this value can be taken as a reasonable proxy for the analogous measurement of the growth velocity of an infinite cylinder with $R_{needle} = 5\ \mu m$.

Figure 8.18 shows a series of growth measurements reduced in this way, from measurements of $R_{needle}(t)$ on individual needles at different supersaturations. We see that $dR_{needle}/dt$ (when $R_{needle} = 5\ \mu m$) is proportional to $\Delta T^2$ to reasonable accuracy, meaning that the growth velocity is proportional to supersaturation at the growth region in DC2, as is expected from theory.

The proportionality constant $A(T)$ depends only on the growth temperature, and it is known directly from the analytical theory describing cylindrical growth. The only unknown parameter in the theory is $B$, which varies only logarithmically with $R_{out}$ and can be estimated to an accuracy of about 20 percent [2016Lib]. Figure 8.19 shows $A(T)$ as determined by the experimental measurements, after adjusting $B$ slightly to fit the data.

Figure 8.19 shows clearly that the measurements agree well with the full cylindrical theory that incorporates both particle and heat diffusion. In contrast, leaving heat diffusion out of the theory (by setting $\chi_0 = 0$) leads to a systematic discrepancy that is small at low temperatures but becomes increasingly important at higher temperatures.

This relatively simple experiment provides a nice demonstration that our basic understanding of particle and heat diffusion in snow crystal growth is indeed correct. To my knowledge, this is the first experiment that has clearly observed the simultaneous effects of both particle and heat diffusion in snow crystal growth. As such, it a step forward in the quest to make *quantitative* observations that can be compared with theoretical models of snow crystal growth. The experiment also demonstrates the inherent accuracy that can be obtained with careful modeling of linear diffusion chambers.

This series of measurements also illustrates a point I have made earlier in this chapter, that the radial growth of slender needle crystals provides a valuable calibration of the supersaturation surrounding a growing crystal. Because the radial growth is usually diffusion limited ($\alpha_{diffcyl} \ll \alpha$), it can be easily calculated from the supersaturation in the chamber. Turning this around, the sides of the column serve as "witness surfaces", whose growth can be used as an indirect measurement of the supersaturation that is usually better than any direct measurement we can devise. We will return to this idea again later in this chapter.

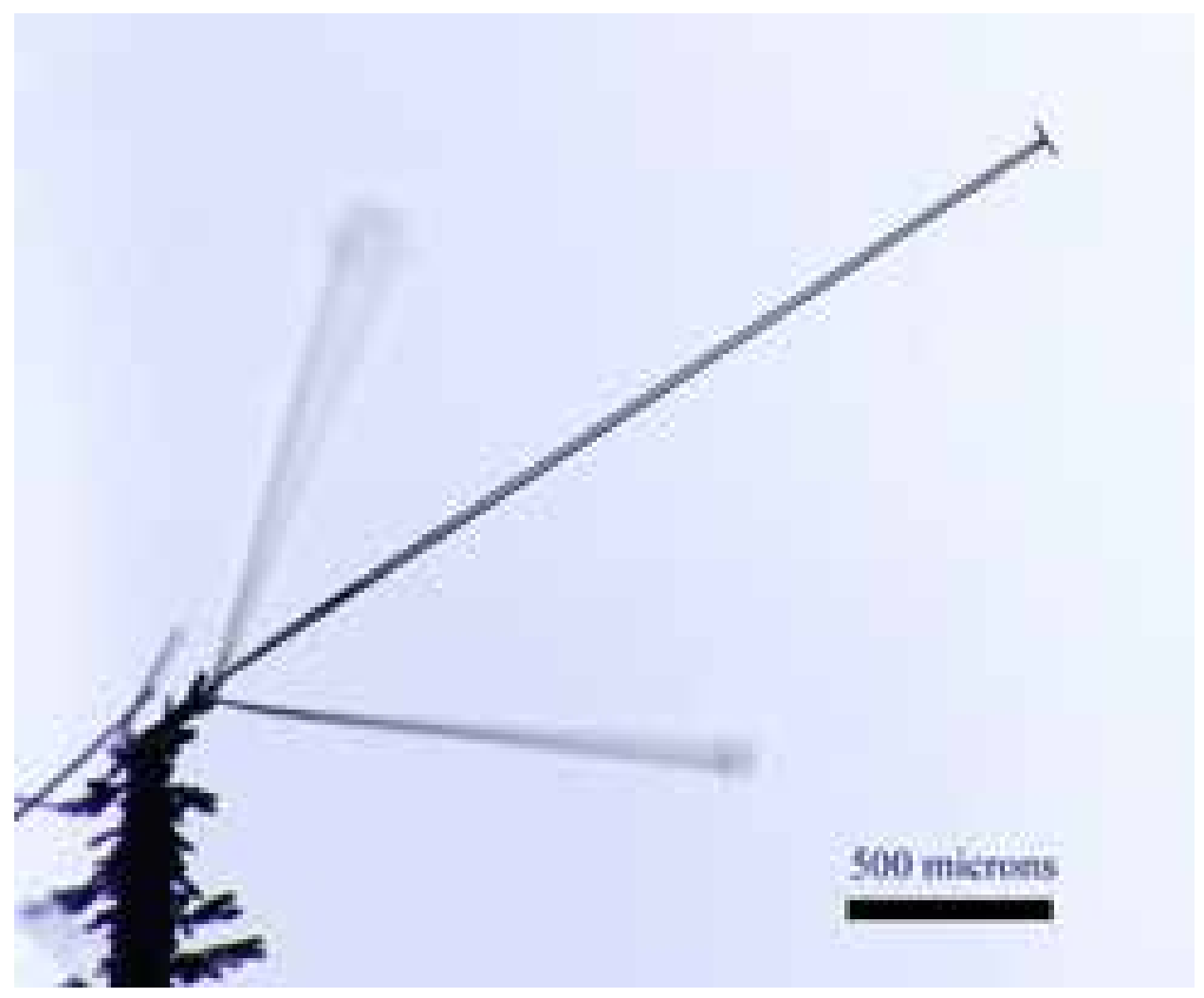

**Figure 8.20. Example of a thin snow crystal plate growing on the end of a slender ice needle. The supporting wire was rotated so the entire needle was in the focal plane of the image.**

## 8.6 Thin Plates on E-needles near -15C

One aspect of the snow crystal morphology diagram I have found particularly intriguing is the formation of thin, plate-like crystals at temperatures near -15 C. At the lower supersaturations shown in Figure 8.16, we see that the thinnest plates appear on e-needles in only a narrow temperature range between about -13 C and -15 C, and the question immediately arises as to why this behavior is observed, and what is so special about this narrow temperature band. In this section I describe a focused investigation of the formation of thin plates on ice needles [2015Lib2].

A notable feature in this investigation is that the phenomenon of thin-plates on e-needles is especially amenable to quantitative analysis, as the morphology of a thin hexagonal plate growing on a slender hexagonal e-needle is quite simple in its overall structure. As described in Chapter 5, these morphologies can be modeled with reasonable accuracy using a cylindrically symmetrical 2D numerical model, which is substantially easier to create and run than a full 3D model. With a 2D model, therefore, one can quickly run dozens of models using a wide range of parameters for comparison with experimental measurements.

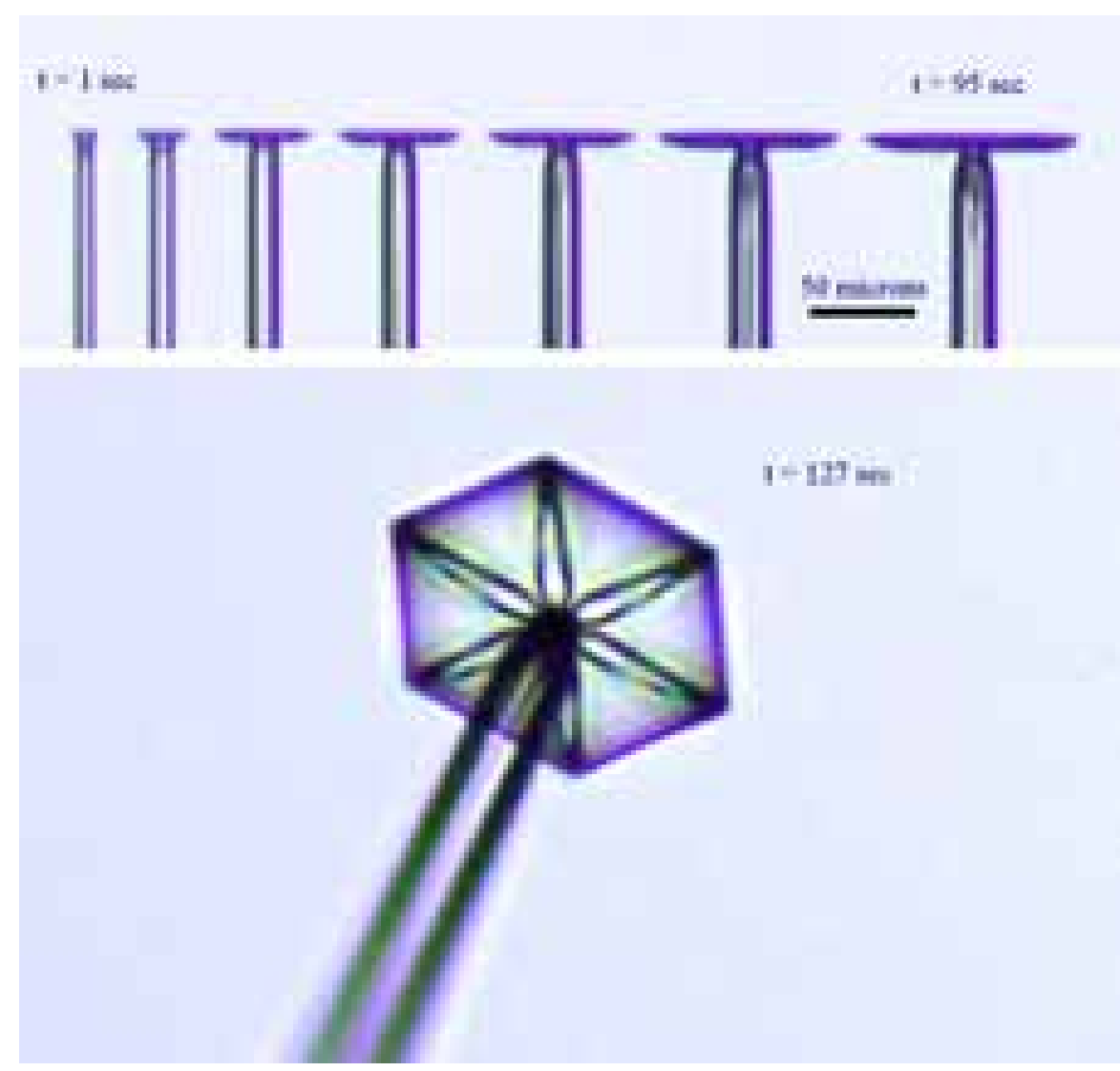

**Figure 8.21. Observations of a thin plate-like snow crystal growing on the end of an ice needle, as seen from the side (top set of images) and from a more face-on direction (bottom). The crystal grew at a temperature of -15 C with a supersaturation at the center of DC2 of about 11 percent.**

## An In-Depth Example

On the experimental side, I have found that selecting an especially well-formed, representative crystal specimen and analyzing it in detail tends to give better results than trying for form averages over many crystals. With this in mind, the present observations began with the creation of e-needles in the dual-chamber apparatus described above and photographing the normal growth of the crystals in the linear diffusion chamber (DC2). Figure 8.20 shows a single image selected from a series of photographs recording the growth of a thin plate on an ice needle.

In a typical run, the overhead cover in DC2 was rotated into place to reduce the supersaturation during the e-needle transfer from DC1. After orienting the ice needle and bringing it into focus on the camera, the cover was then rotated away, restoring the supersaturation to its normal level in a few seconds time. The growing crystal was then photographed at regular intervals, and the images were subsequently analyzed to give quantitative measurements. Figure 8.21 shows several images from a series, and Figure 8.22 shows the data extracted from the entire set of images. The data plot also includes several numerical models of the growing crystal.

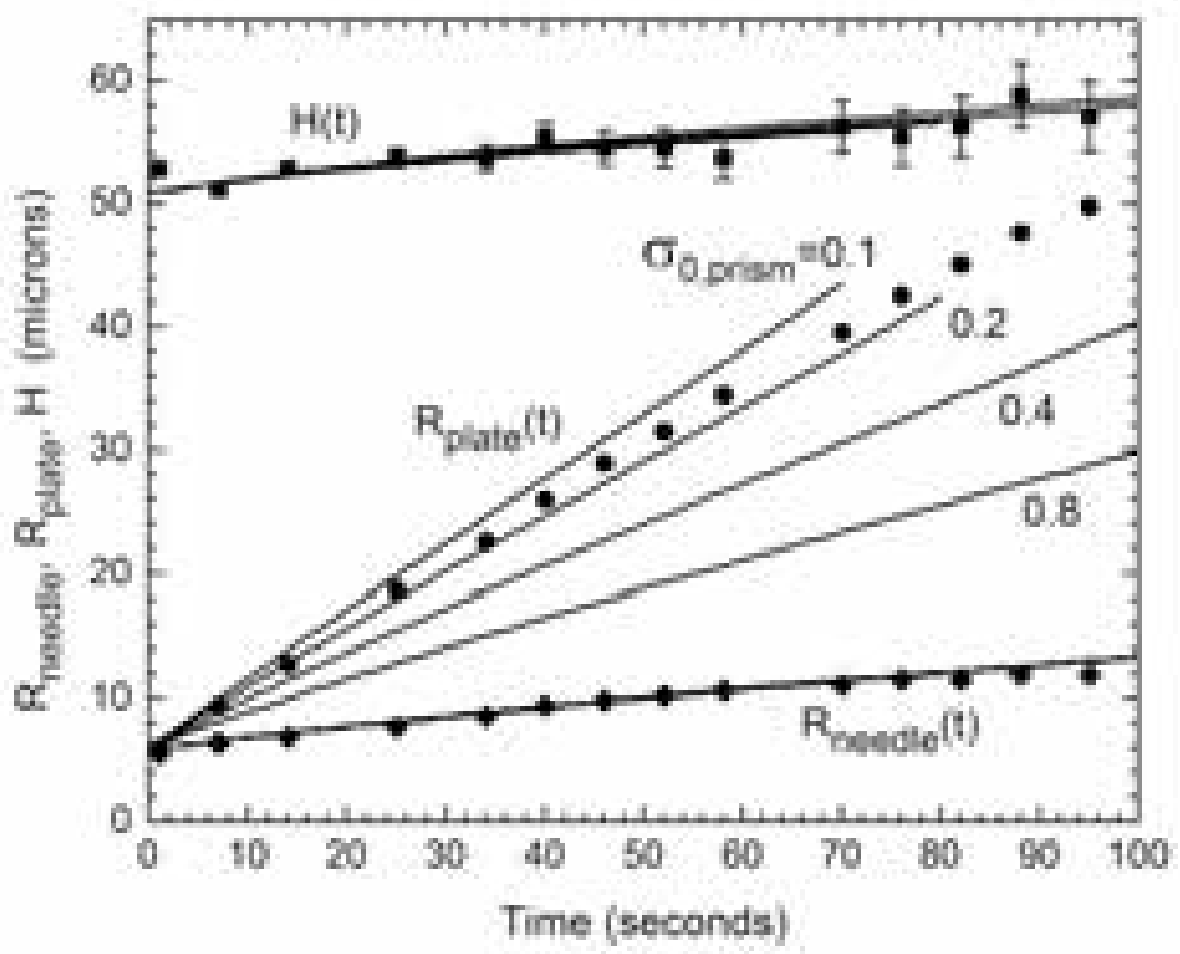

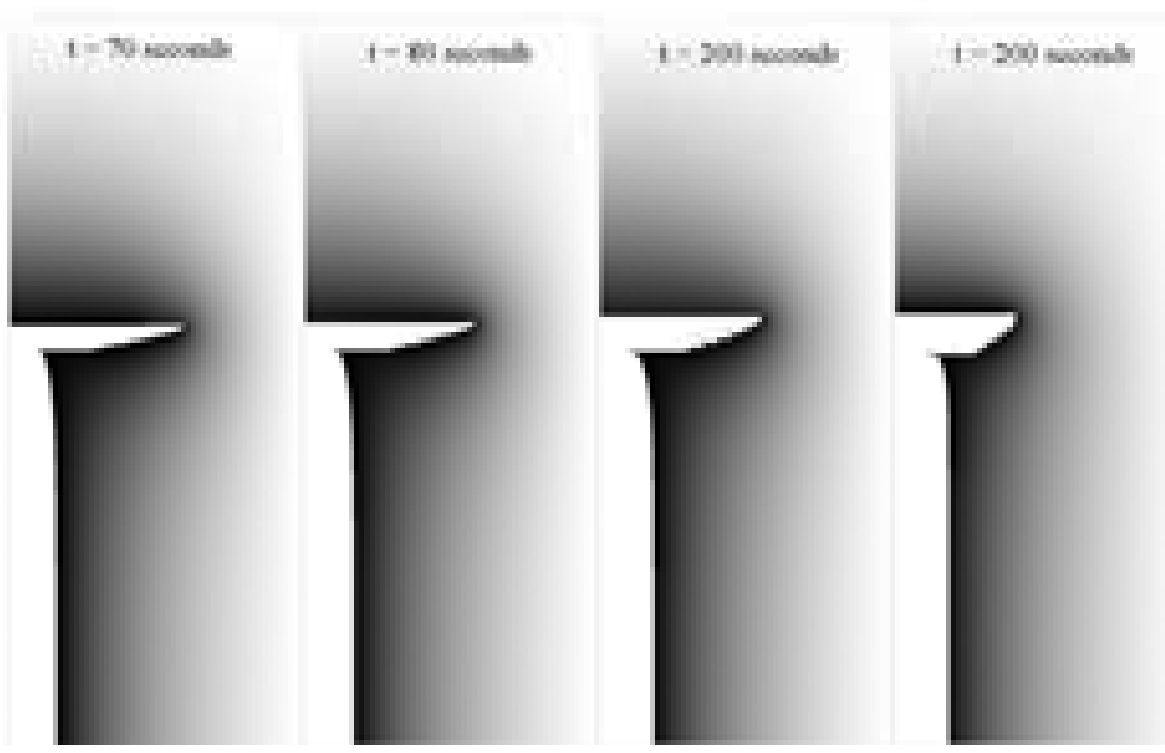

**Figure 8.22. (Upper) Measurements of the plate radius, needle radius, and axial growth as a function of time, from the data shown in Figure 8.21 (plus other images not shown). The plotted height is after subtracting a constant value, as the overall needle length was much larger, as illustrated in Figure 8.20. Lines through the data are from numerical models of the growing crystals, as described in the text. Note that changing $\sigma_{0,prism}$ strongly affects $R_{plate}(t)$ but has little effect on $R_{needle}(t)$. (Lower) Computer-generated cross sections of the four model crystals shown in the upper graph. From [2015Lib2].**

The first step in the modeling process was to determine the correct value of $\sigma_{far}$, the supersaturation at the outer boundary of the model space (Chapter 5 describes the 2D cylindrically symmetric cellular automata used in this study). Because the model boundary is quite close to the growing crystal (for reasons of computational expediency), $\sigma_{far}$ will be substantially smaller than $\sigma_{\infty}$, the latter being the background supersaturation at the center of DC2 quite far from the crystal. While there are ways of extending the model out to large distances, a better approach is to simply adjust $\sigma_{far}$ in the model so that it gives a good fit to $R_{needle}(t)$ in the measurements.

This procedure is an example of the "witness surface" idea in practice. The needle growth $R_{nee}$ $(t)$ depends roughly linearly on $\sigma_{far}$, as seen from model calculations in Figure 8.23. But $R_{needle}(t)$ is essentially independent of $\alpha_{prism}$, as can be seen by the model calculations in Figure 8.22 (where changing $\alpha_{prism}$ in the model changes $R_{plate}(t)$ substantially but has little effect on either $R_{needle}(t)$ or $H(t)$).

That $R_{needle}(t)$ in the model calculation depends only on $\sigma_{far}$ and not at all on $\alpha_{prism}$ may seem counterintuitive, but it follows directly from the analytical theory for cylindrical growth presented in Chapter 3, because $\alpha_{diffcyl} \ll \alpha_{prism}$ is a good approximation for the needle growth.

The take-away message from this first step in the analysis is that the needle growth can be used to determine $\sigma_{far}$ with quite good accuracy, even if $\sigma_{\infty}$ and $\alpha_{prism}$ are not so well known. As a bonus, the actual position of the outer boundary is not terribly important in the analysis, so one can use numerical models that do not extend to great distances from the growing crystal. The witness-surface method is thus a powerful analysis technique that works remarkably well in the case of snow crystals growing on the ends of e-needles.

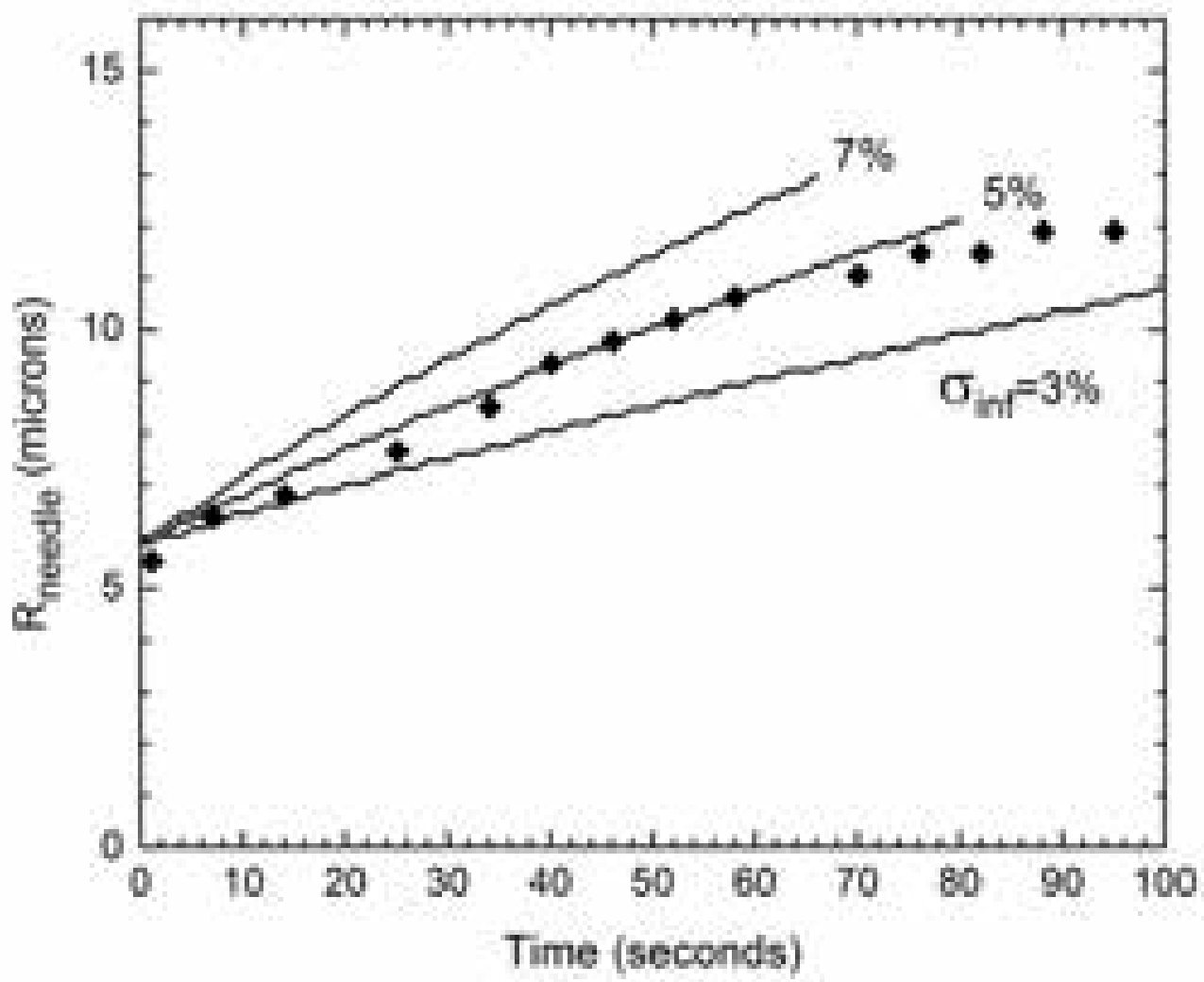

**Figure 8.23. The needle growth data from Figure 8.22 along with several models using different values of $\sigma_{far}$. As long as $\alpha_{diffcyl} \ll \alpha_{prism}$ (which is true in most circumstances), then the modeled $R_{needle}(t)$ is essentially independent of $\alpha_{prism}$, while being approximately proportional to $\sigma_{far}$, as seen in this graph. Thus, the measured $R_{needle}(t)$ allows one to determine $\sigma_{far}$ with quite good accuracy, even when $\alpha_{prism}$ is not well known. This result is somewhat counter-intuitive, but it follows directly from the analytical theory of the diffusion-limited growth of ice cylinders discussed in Chapter 3.**

Having determined that $\sigma_{far} = 5$ percent is a good fit to $R_{needle}(t)$ in Figure 8.23, the next step in the analysis is to look at the plate growth as a function of $\alpha_{prism}$ in the model. Now things get a bit model-dependent, as $\alpha_{prism}$ is not a known quantity, and, in fact, the whole exercise is aimed at determining $\alpha_{prism}$ from a comparison of model and measurements. In these calculations, I assumed a model with $\alpha_{prism} = \exp(-\sigma_{surface}/\sigma_{0,prism})$, as the attachment kinetics on faceted surfaces are typically well described by a layer-nucleation model (see Chapter 4).

Looking at the models in Figure 8.22 (again, skipping over some of the details presented in [2015Lib2]), we see that the models indicate that we must have $\alpha_{prism} \approx 1$ to fit the plate growth data $R_{plate}(t)$. And, as expected, changing $\alpha_{prism}$ mainly affects the calculated $R_{plate}(t)$ in the models but has little effect on either $R_{needle}(t)$ or $H(t)$.

The result indicating $\alpha_{prism} \approx 1$ agrees with the fact that the hexagonal plates forming at this supersaturation level are just barely stable against branching. The morphology data in Figure 8.16 indicates that plates at -15 C soon give way to complex dendritic structures at higher supersaturations. Faceted prism growth is generally quite stable when $\alpha_{prism} \ll 1$, but becomes unstable to branching when $\alpha_{prism} \rightarrow 1$, as was discussed in Chapter 3. Thus it is not surprising that the models indicate $\alpha_{prism} \approx 1$ for these data, because we know that the faceted plate growth is close to sprouting branches.

A first conclusion, therefore, is that this thin-plate experiment agrees well with our overall picture of snow crystal growth. The formation of thin plates on the e-needle tips means we must have $\alpha_{basal} \ll \alpha_{prism}$, and the fact that the plate growth is about to become unstable to branching indicates that $\alpha_{prism} \approx 1$. The witness-surface growth allows us to determine $\sigma_{far}$ with good accuracy in the models, and we can find a final model that reproduces all the growth rates as well as the overall plate-on-needle morphology. Looking at just this one crystal (which is representative for growth at this temperature and supersaturation), the models make sense and it all paints a reasonably self-consistent picture.

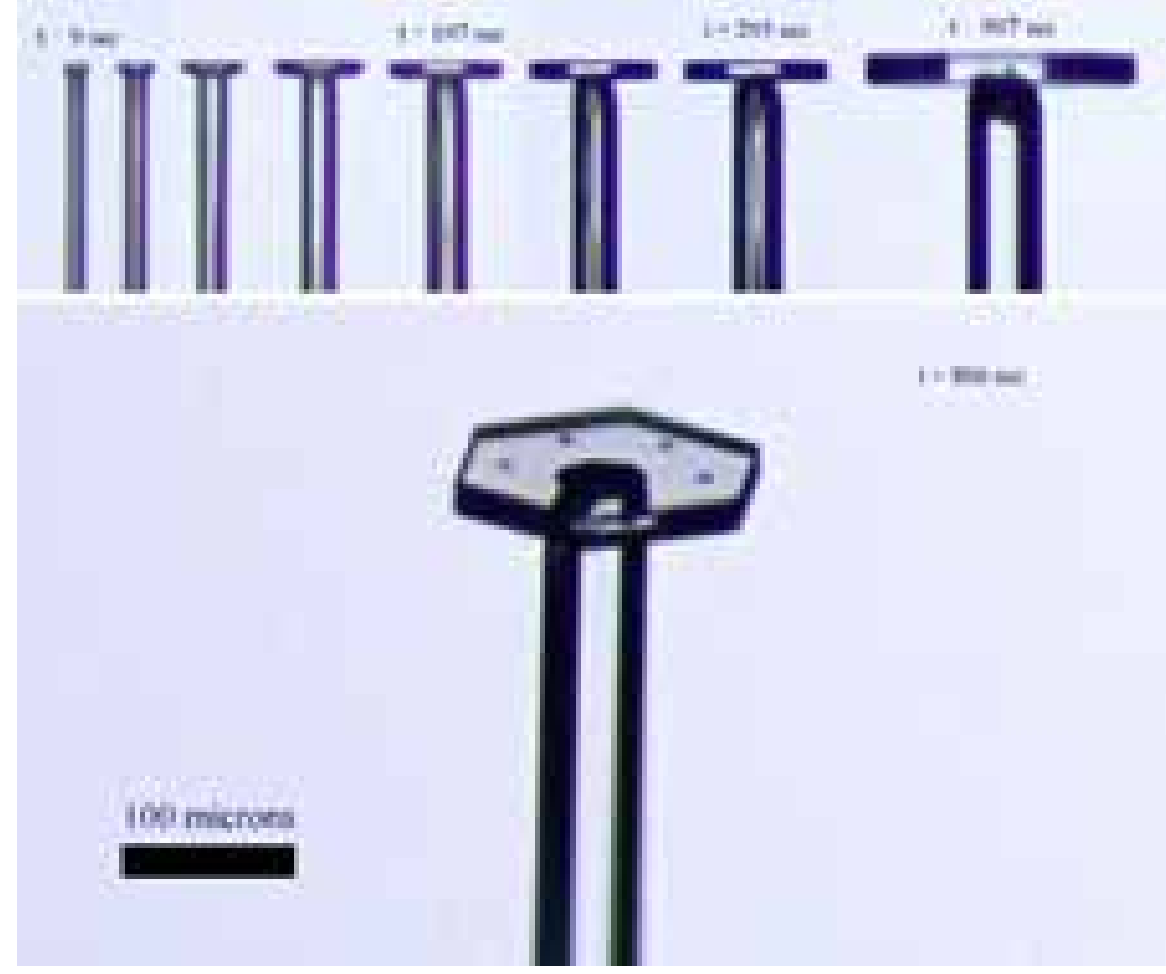

**Figure 8.24. Observations of a thicker plate-like snow crystal growing on the end of an ice needle, as seen from the side (top set of images) and from a more face-on direction (bottom). The crystal grew at a temperature of -15 C with a far-away supersaturation of about 7 percent.**

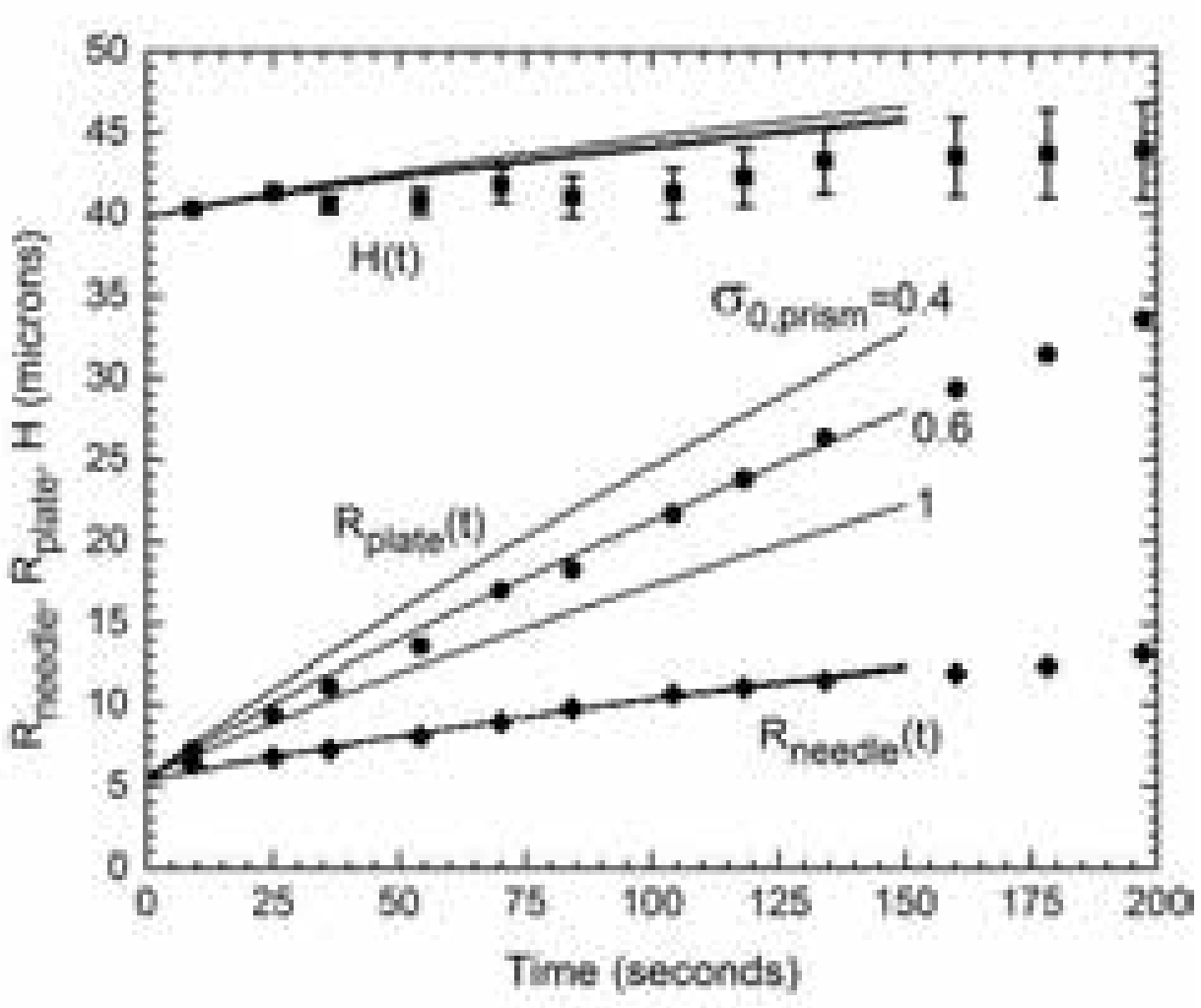

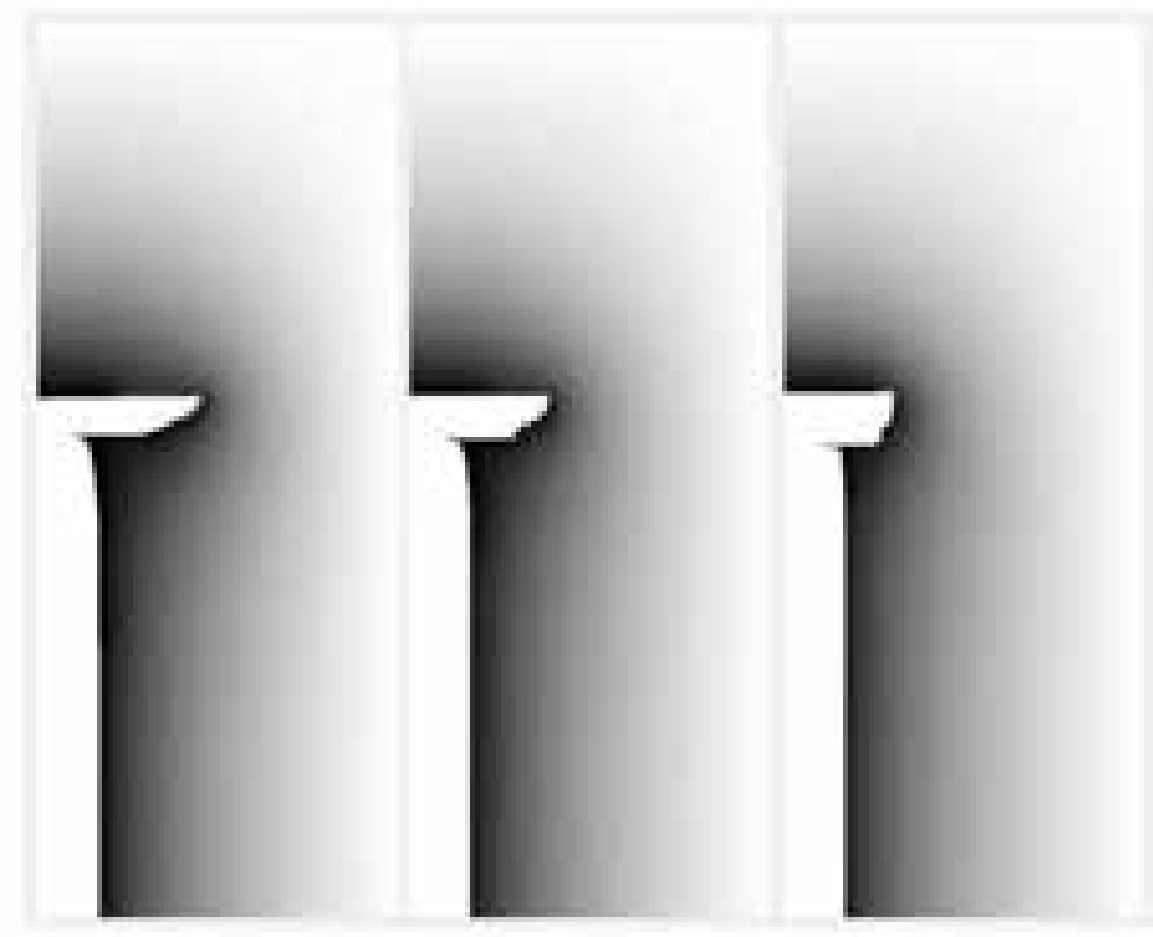

**Figure 8.25. Measurements of the plate radius, needle radius, and axial growth as a function of time, from the data shown in Figure 8.24. Lines through the data are from numerical models of the growing crystals, and the lower images show computer-generated cross sections of the three model crystals shown in the upper graph. From [2015Lib2].**

## Varying the Supersaturation

Things become a bit more interesting when one starts varying the parameters, and Figures 8.24 through 8.27 show two more growth measurements at different supersaturation levels [2015Lib2]. As $\sigma_\infty$ is lowered from 11 percent to 7 percent to 4.6 percent, the tip growth changes from a thin plate to a thicker plate to a more blocky crystal, which agrees with the overall behavior we see in the morphology diagram.

Once again, we can produce numerical models that fit the growth measurements and reproduce the growth morphologies reasonably well, but only after making substantial changes in $\alpha_{prism}$. Importantly, contrary to what one might have expected, no model with a single $\alpha_{prism}(\sigma_{surface})$ can be found to fit all the data.

As described in [2015Lib2], this set of experiments leads one to a conclusion that a relatively simple model of snow crystal growth does not explain the transition from blocky crystals to thin plates at -15 C. Solving the diffusion equation using a reasonable model for the attachment kinetics, and assuming a single-valued function $\alpha_{prism}(\sigma_{surface})$, does not fit the data. The growth data are likely fine, so we conclude that something is wrong with the model assumptions.

Having looked careful at this discrepancy, I have not found an obvious resolution. Surface energy effects appear to be too small to have much influence. Surface diffusion may play a role, but this is not obviously the answer either. I believe that the Edge-Sharpening Instability described in Chapter 4 could be the correct explanation, and I have found that incorporating the ESI does yield a single model that fits all the data reasonably well [2015Lib2]. Of course, additional study is needed to either verify the ESI or find some other explanation, and this research is a work in progress. It does seem clear, however, that the transition from blocky crystals to thin plates at -15 C is an interesting part of the snow crystal puzzle.

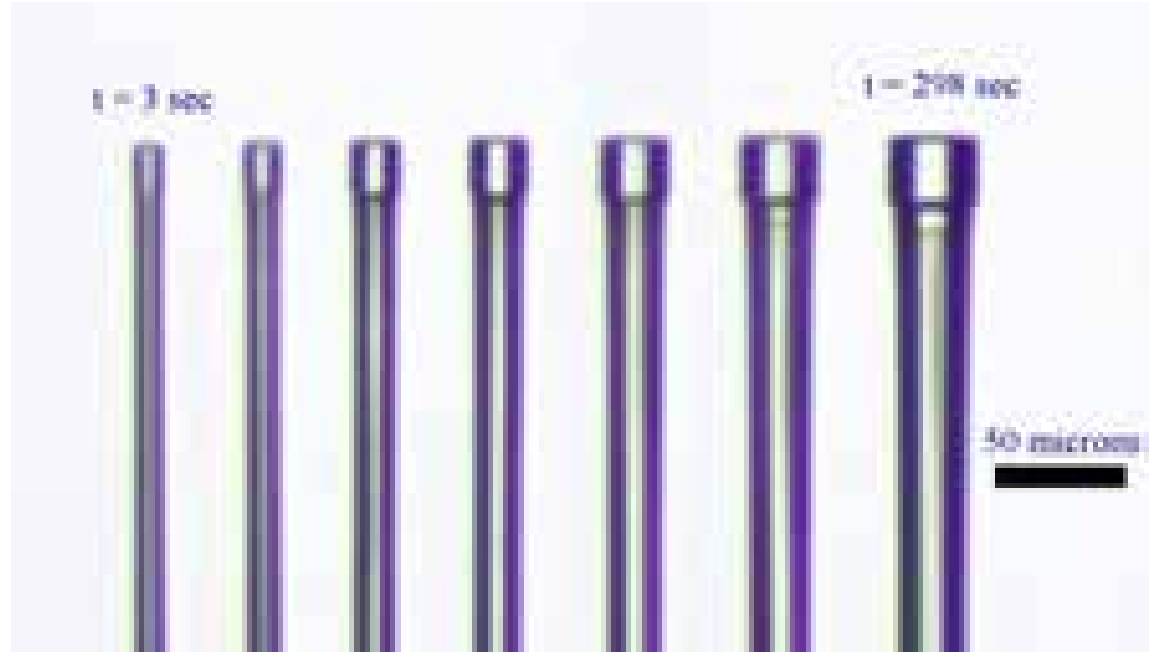

**Figure 8.26. Observations of a blocky snow crystal growing on the end of an ice needle at a temperature of -15 C with a far-away supersaturation of about 4.6 percent.**

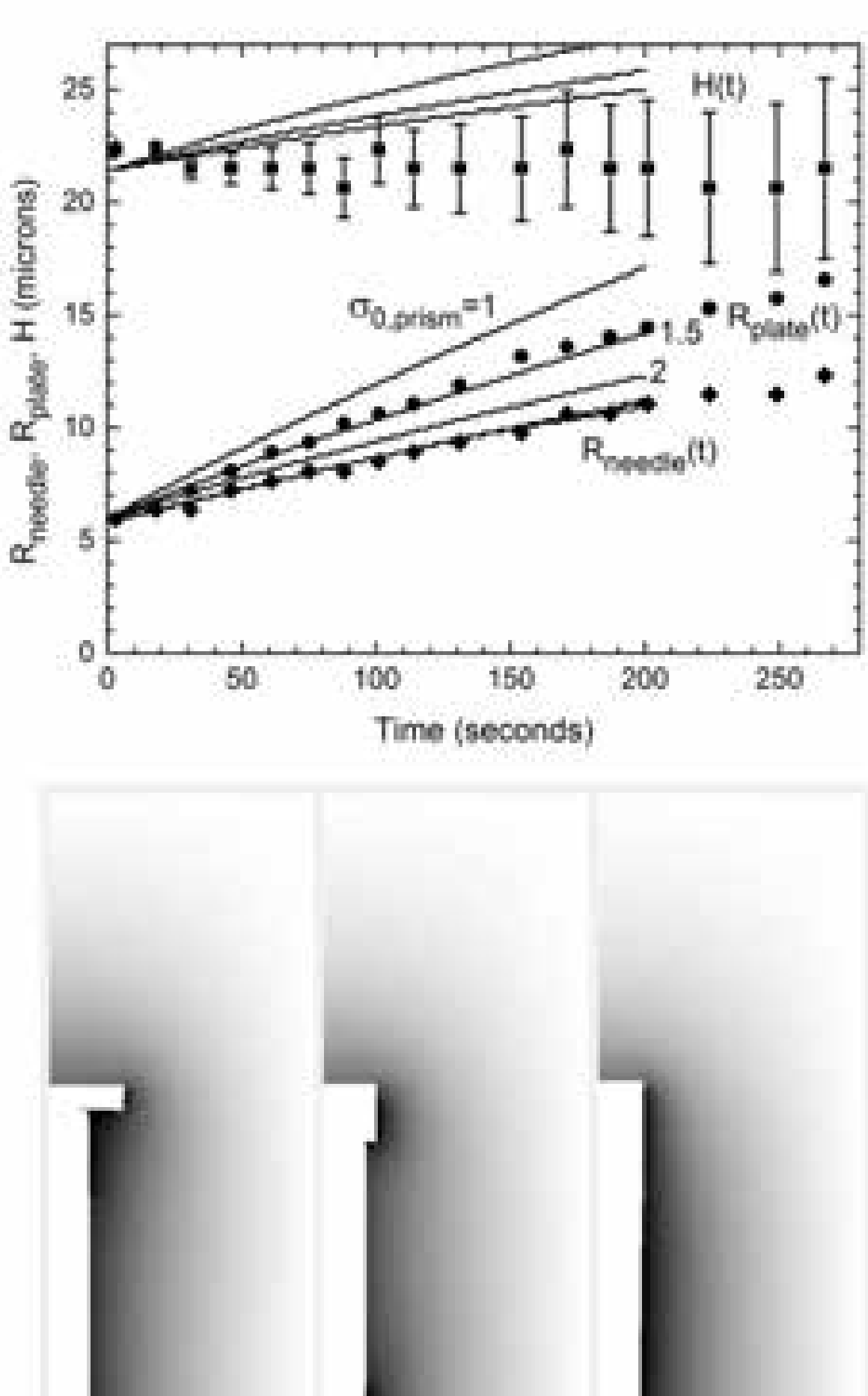

**Figure 8.27. Measurements of the plate radius, needle radius, and axial growth as a function of time, from the data shown in Figure 8.26. Lines through the data are from numerical models of the growing crystals, and the lower images show computer-generated cross sections of the three model crystals shown in the upper graph. From [2015Lib2].**

## Near-Surface Supersaturation

From the above measurements and models, we can also glean some additional insights by examining the supersaturation levels found at the crystal surfaces, which is a byproduct of the numerical modeling. Two factors give us confidence in this exercise: 1) the best-fit models reproduce both the morphologies and growth rates with reasonable fidelity in all three experiments, and 2) the numerical solutions to the diffusion equation is likely accurate, as the underlying physics is well understood. It stands to reason, therefore, that the model determinations of the supersaturation fields around the crystals, in particular at the crystal surface, are fairly accurate.

One interesting result that we obtain from the models is that the supersaturation at the outermost prism edges in all three crystals is about 0.5 percent. In particular, the value of $\sigma_{edge}$ appears to be nearly the same for the fast-growing thin plate in Figure 8.22 as it is for the slow-growing block in Figure 8.27 [2015Lib2]. This result seems counterintuitive, but it begins to make sense when you look at some contour plots of supersaturation around growing crystals, as shown in Figure 8.28.

For the case of the blocky crystal in Figure 8.28, the prism facet and basal facets grow at roughly the same rate, as this is required to produce an approximately isometric morphology. At distances more than a few crystal radii away, therefore, the contour lines in the supersaturation field are nearly indistinguishable from those around a growing sphere. As one approaches the faceted surfaces, the surface supersaturation is highest near the corners, which is known as the Berg effect.

In contrast, the thin-plate crystal in Figure 8.28 requires that the prism surfaces grow at a much higher rate than the basal surfaces. This means that the flux of material at the basal facets is much lower than the flux at the prism facets. Diffusion theory tells us that flux is proportional to the gradient in the supersaturation field, so the contour lines near the edges of the thin plate in Figure 8.28 are much more closely spaced than the contour lines at the basal facets. This also means that the surface supersaturation at the prism surfaces must be substantially lower than at the basal surfaces.

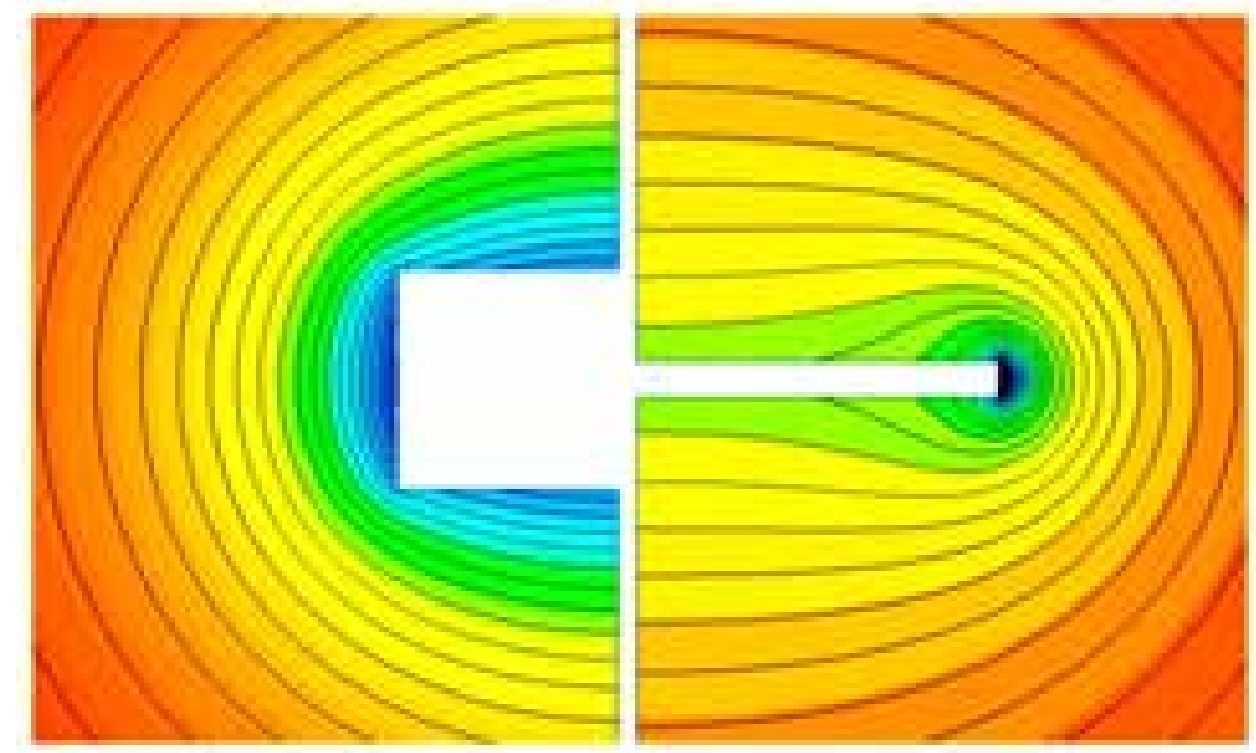

**Figure 8.28. Calculated contour plots of supersaturation levels around growing ice crystals, shown here in (r,z) coordinates. Around a blocky crystal (left), the supersaturation is highest near the corners of the faceted prism, a phenomenon called the Berg effect. Around a thin-plate crystal, the contour lines are tightly bunched at the plate edge. This is because the fast-growing edge requires a steep gradient in the supersaturation to supply a sufficient flux of water vapor molecules.**

These calculations assumed nothing about the attachment kinetics, but only that the basal and prism facets grew at a fixed rate across each facet (and this assumption follows from the fact that the morphology remains faceted as it grows). The contour lines, and the surface supersaturations that result, arise solely from diffusion theory, including mass conservation at the crystal surface. The diffusion equation can be solved knowing only the far-away supersaturation (the outer boundary condition) and the growth velocities (the surface boundary conditions). These quantities, therefore, are sufficient to determine the entire supersaturation field surrounding the crystal.

Looking at the calculations in Figure 8.28, it begins to make sense that the surface supersaturation at the edge of a thin plate growing at high $\sigma_\infty$ can be similar to the supersaturation at the surface of a blocky crystal at substantially lower $\sigma_\infty$. For this to happen, however, $\alpha_{prism}$ must be much higher on the edge of the thin plate as compared to $\alpha_{prism}$ on the edge of the blocky crystal. Once again, we find that the model conclusions make some amount of sense, while examining the supersaturation fields surrounding these crystals helps build one's intuition regarding diffusion-limited growth.

## Lessons Learned

A primary conclusion from this focused study is that one cannot find a single, physically reasonable expression of the prism-facet attachment kinetics – that is, a simple function $\alpha_{prism}(\sigma_{surface})$ – that fits all the growth data. This is an important result, because it says that there must be some additional physics lurking somewhere in the problem. Combining what we know and expect from diffusion-limited growth, surface-energy effects, and attachment kinetics, we still come up short.

This is the primary puzzle presented by this set of experiments. Looking at the faceted prism surfaces on the thin plate and the blocky crystal, we see that they grow at the same temperature and with the same near-surface supersaturation. So if these faceted surfaces are exposed to essentially the same growth conditions, why do they grow at such different rates?

One way to explain the data is to postulate that the attachment kinetics on a prism facet changes when the facet width approaches atomic dimensions. This postulate replaces a simple function $\alpha_{prism}(\sigma_{surface})$ with a more complicated function $\alpha_{prism}(\sigma_{surface}, w)$, where $w$ is the width of the prism terrace. I described this theory in more detail in Chapter 4, and it produces the Edge-Sharpening Instability (ESI) I described in that chapter.

This postulate is new physics, in that it does not fit nicely into conventional theories of crystal growth. Putting such an effect into the numerical model does reproduce the observations reasonably well, as we described in [2015Lib2]. However, it must be noted that other nonlocal phenomena, such as peculiarities in surface diffusion or other effects, may provide a viable alternative explanation. Additional work is needed to decide if the ESI theory is correct.

Another conclusion from this study, and perhaps the more important take-away message, is that there is much to be learned by combining careful, quantitative observations of snow crystal growth under well known conditions with precise numerical models. The devil is in the details, and quantitative studies like this one allow one to identify the most interesting growth phenomena and hopefully develop unequivocal models to explain the underlying physical processes in detail. Morphological studies are not enough, and the underlying molecular physics is far too complex to simply intuit from first principles. I have become firmly convinced, therefore, that detailed quantitative comparisons between measurements and models are the only viable route to making real scientific progress.

Another conclusion is that the "witness surface" method is remarkably useful during data analysis, especially when considering the growth on the tips of e-needles. Adjusting $\sigma_{far}$ to fit the observed $R_{needle}(t)$ constrains the theory nicely, even when $\sigma_\infty$ is not well known experimentally, and it allows some flexibility in choosing the outer boundary in the model. This greatly reduces one of the biggest problems in the quantitative study of snow crystal growth – the inherent difficulty in determining the supersaturation in laboratory measurements. In my opinion at least, using the witness-surface method with e-needle seed crystals opens up many new opportunities for detailed quantitative studies of snow crystal growth dynamics over a broad range of conditions.

Finally, this study demonstrates that the big picture will require making lots of measurements on lots of snow crystals over a broad range of environmental conditions. Growing a few crystals is a start, but a few measurements will not be sufficient to fully understand the remarkable richness of snow crystal growth phenomena. Only by examining crystals growing at different temperatures and supersaturations, with different morphologies, and in a variety of background gases at different pressures, will we begin to fully understand the underlying molecular processes. The snow crystal story is only beginning to unfold.

## 8.7 Air-Dependent Attachment Kinetics Near -5C

Another nice result I obtained with the dual-chamber e-needle apparatus described in this chapter was a clear demonstration of how $\alpha_{prism}$ near -5 C depends strongly on air pressure [2016Lib1]. The tale begins with a series of observations of growing needle crystals, including the data shown in Figure 8.29.

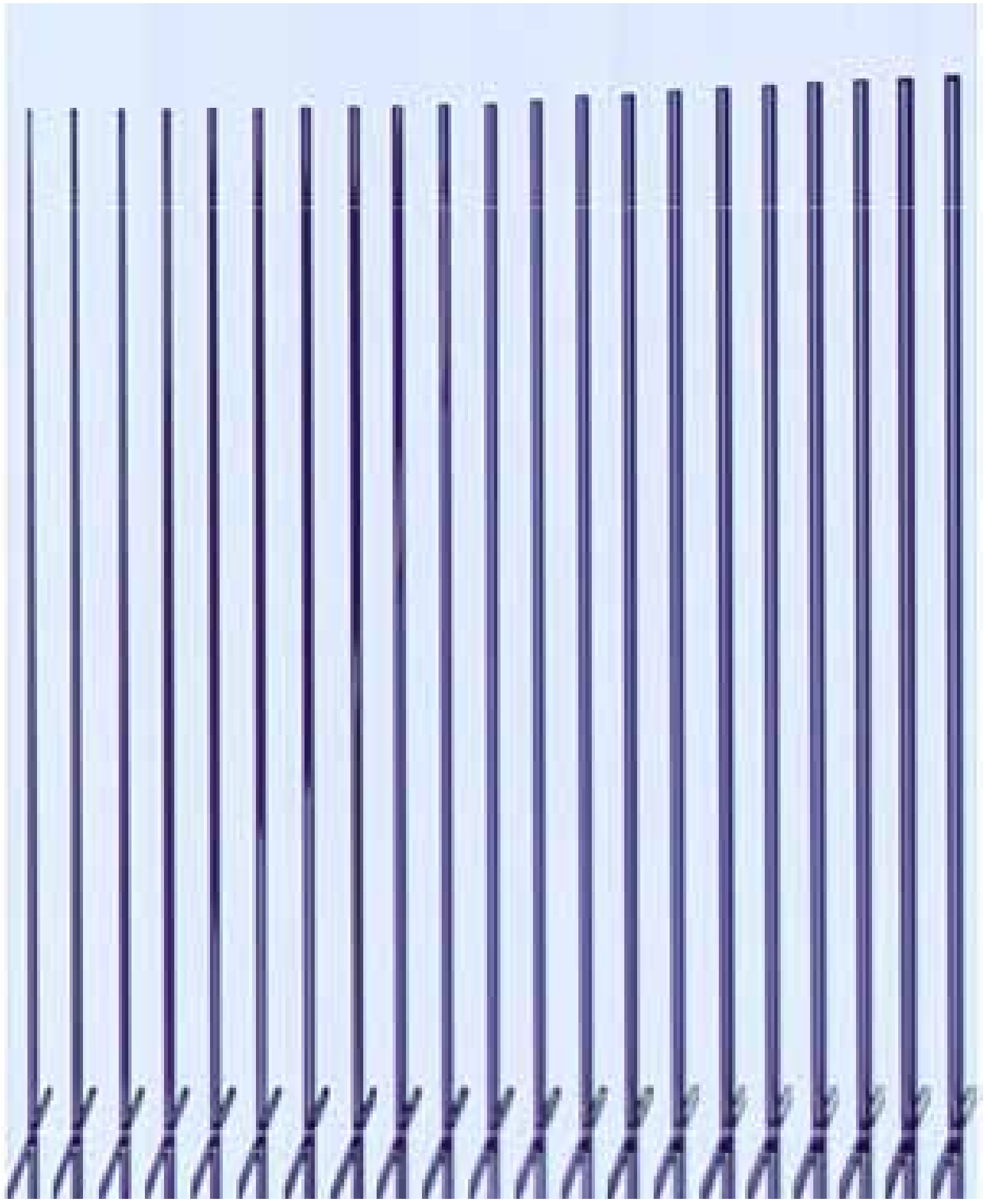

**Figure 8.29. A composite image showing the growth of an e-needle at -5 C in a supersaturation of $\sigma_\infty$=1.8 percent. The near-constant structure near the base of the needle was used to align the images, allowing an accurate measurement of the axial growth rate. The radial growth was measured from direct imaging of the needle widths. Note how the axial growth is slow at first, but then transitions to a faster rate. The transition occurs just as the needle goes from slightly tapered to fully faceted.**

The first image was taken soon after the high voltage was turned off and the e-needle was moved to DC2. The subsequent images were taken periodically as the needle grew, allowing a measurement of the axial and radial growth as a function of time. The temperature and supersaturation $\sigma_\infty$ were constant during the observations. Both $R_{needle}(t)$ and $H_{needle}(t)$ were then obtained directly from the images, and the results are shown in Figure 8.30. Immediately the data show a peculiar transition in growth behavior that happens at a time near $t = 130$ seconds.

### A Growth Transition

A key to understanding this peculiar behavior is realizing that the transition at $t = 130$ seconds occurs just as the needle morphology changes from a slightly tapered needle to a fully faceted column. At early times, when the needle is tapered, the attachment coefficient on the columnar sides is that of a vicinal surface, which we write as $\alpha_{vicinal}$. Because there is a continuous train of molecular steps on a vicinal surface, their growth is not limited by layer nucleation, and we expect $\alpha_{vicinal}$ to be quite high, approaching that of a rough surface ($\alpha_{rough} \approx 1$).

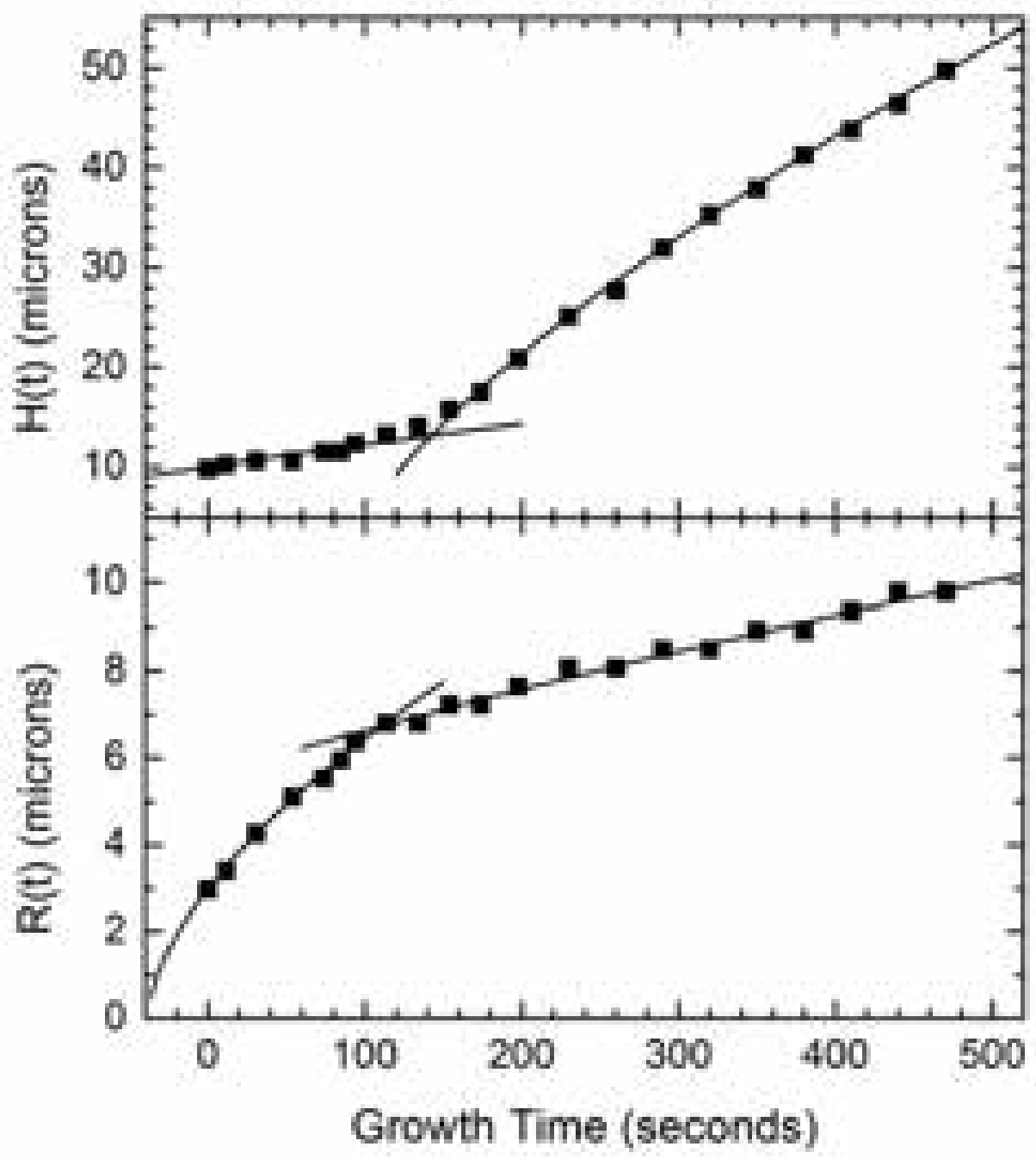

**Figure 8.30. Measurements of $R_{needle}(t)$ and $H_{needle}(t)$ obtained from the image data in Figure 8.29. Here $H_{needle}(t)$ is the needle length minus a constant offset. Lines were drawn through the data to guide the eye. The needle growth shows two distinct behaviors: 1) rapid radial growth and slow axial growth at times t<130 seconds, and 2) slower radial growth and faster axial growth for t>130 seconds. The images reveal that the transition occurred when the needle morphology changed from slightly tapered to fully faceted.**

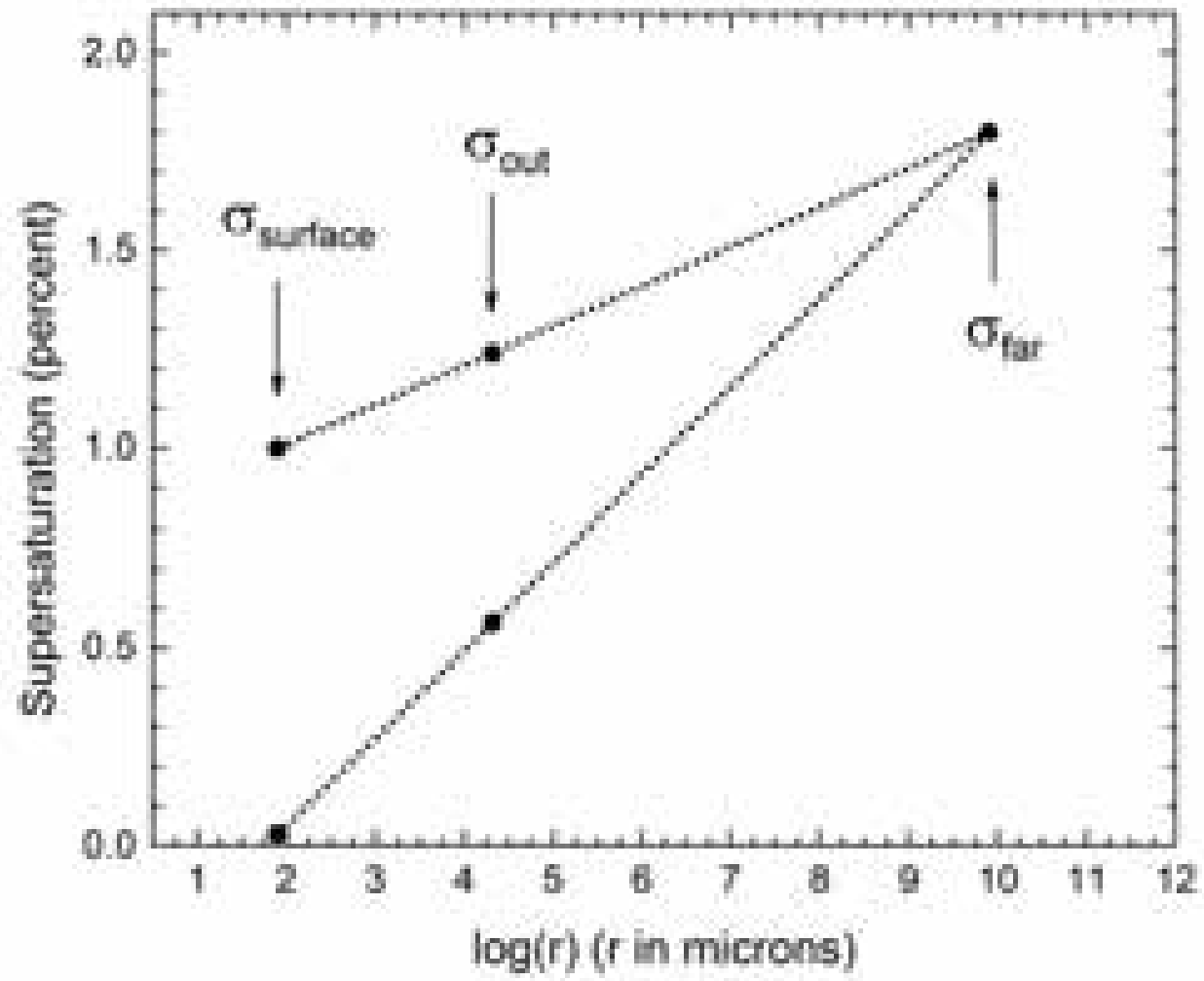

**Figure 8.31. A simple analytical model showing the supersaturation field σ(r) surrounding an infinite cylinder. The bottom line assumes a fairly large $\alpha_{vicinal}$ at the surface, while to top line assumes a much smaller $\alpha_{prism}$ at the surface. Changing the value of the attachment coefficient has a large effect on the supersaturation near the crystal surface.**

The fact that $\alpha_{vicinal}$ is high means two things. First, the high attachment coefficient results in a relatively fast radial growth, as is seen in Figure 8.30. Second, the high attachment coefficient means that the supersaturation in the vicinity of the vicinal surface is relatively low, as excess water vapor is rapidly depleted by the growing surface.

Importantly, once we have $\alpha_{vicinal} \gg \alpha_{diffcyl}$, the exact value of $\alpha_{vicinal}$ does not matter much when calculating the radial growth. Moreover, solving the diffusion equation for cylindrical growth (see Chapter 3) reveals that the surface supersaturation is $\sigma_{surface} \approx (\alpha_{diffcyl}/\alpha_{vicinal})\sigma_{far}$, which is quite low because of the above inequality. In this mainly-diffusion-limited regime, the radial growth rate is essentially independent of the exact value of $\alpha_{vicinal}$.

Once the prism surfaces become faceted, however, the attachment coefficient drops abruptly from $\alpha_{vicinal}$ to $\alpha_{prism}$, and for this crystal $\alpha_{prism} \gg \alpha_{diffcyl}$ is not a good assumption. After this transition, the low value of $\alpha_{prism}$ brings about two changes. First, the radial growth drops, as seen in Figure 8.30. And second, $\sigma_{surface}$ increases because of the slower radial growth, and this brings about the observed increase in the axial growth rate.

Figure 8.31 shows a simple model of the supersaturation field around the column that gives one a good picture of what is going on. Solving the diffusion equation for the growth of an infinite cylinder gives the supersaturation field $\sigma(r) = A_1 + A_2 \log(r)$ around the cylinder, where $A_1$ and $A_2$ are constants determined by the boundary conditions.

The lower line in Figure 8.31 shows the supersaturation around the tapered needle in this model. The value of $\sigma_{surface}$ is low, and a simple line (in this semi-log plot) connects $\sigma_{surface}$ and $\sigma_{far}$, the latter being 1.8 percent in this case. Once the needle becomes faceted, $\sigma_{surface}$ goes way up while $\sigma_{far}$ stays the same, yielding the upper line in Figure 8.31.

A closer examination of all this with real models and numerical simulations is presented in [2016Lib1], and the overall conclusion is that this picture holds together nicely after more careful scrutiny. However, the models insist that we must have $\alpha_{prism} < \alpha_{diffcyl}$ after the transition at $t = 130$ seconds, and some pretty basic math gives $\alpha_{diffcyl} \approx 0.003$. After examining the numerical models, a value of $\alpha_{prism} \approx 0.002$ seems to give a pretty good fit overall.

What makes this measurement interesting is that there is simply no way to explain any of this if $\alpha_{prism}$ on a faceted prism surface is much higher than 0.002 after the transition. The vicinal surface prior to $t = 130$ seconds provides an effective witness surface in the analysis, and this confirms the overall picture. The slower radial growth after the transition, accompanied by the faster axial growth, gives a strong indication that $\alpha_{prism} < \alpha_{diffcyl}$ after the transition.

This experiment demonstrates the use of a time-dependent witness surface, as the radial growth before the transition is essentially independent of $\alpha_{vicinal}$, while the growth is mainly limited by $\alpha_{prism}$ after the transition. Without the data taken before the transition, it would have been difficult to draw any meaningful conclusions about $\alpha_{prism}$. When $\alpha_{prism}$ is close to $\alpha_{diffcyl}$, as it is in this case, then it is pretty much impossible to tell the difference between diffusion-limited growth and kinetics-limited growth from measurements of $R_{needle}(t)$ alone. But the witness surface that was present before the transition removed this problem and allowed a solid measurement of $\alpha_{prism}$.

This transitional behavior appears to be limited to a very small region of the morphology diagram – only at low supersaturations near -5 C. Everywhere else, $\alpha_{prism}$ is high enough that $\alpha_{prism} > \alpha_{diffcyl}$ is a good approximation. Thus the transition seen in Figure 8.30 is indeed unusual, and is not seen in other columnar growth data.

## Prism Attachment Kinetics In Air and Vacuum

The result from this rather simple experiment takes on some importance when comparing it to other growth data. In particular, ice growth measurements at low air pressures clearly indicate that $\alpha_{prism}$ is much larger than 0.002 for crystals growing at the same temperature and with the same $\sigma_{surface}$ (see Chapter 4). The almost inescapable conclusion, therefore, is that $\alpha_{prism}$ is smaller in the presence of air at one atmosphere than $\alpha_{prism}$ on similar surfaces in near vacuum.

This conclusion is supported by the fact that no one has ever (to my knowledge) observed the growth of slender columns in near-vacuum conditions near -5 C. If it were true that $\alpha_{prism} \ll \alpha_{basal}$ at low pressures, then we would expect to see columnar growth at low pressures. The *lack* of any definite observations of slender columns in any low-pressure experiments lends support to the notion that the prism attachment kinetics at -5 C is affected by air pressure.

Others have suggested that the presence of air affects the attachment kinetics in some circumstances, but the measurements were always plagued by a poor understanding of the surface supersaturation levels in the experiments [1984Kur1]. And this, in turn, makes it difficult to clearly separate kinetics effects from diffusion effects. As a result, the previous experiments were, in my opinion at least, not very convincing. The combination of e-needle measurements and low-pressure measurements, however, makes a much stronger case for air-dependent attachment kinetics near -5 C.

Once again, we find that careful, quantitative measurements of ice growth, accompanied by careful diffusion analyses, are absolutely necessary to make definitive conclusions regarding the physical processes that govern snow crystal growth. In addition, we again see that e-needles make excellent seed crystals for undertaking many focused studies of the attachment kinetics and complex diffusion effects over a broad range of environmental conditions.

## 8.8 E-needle Vignettes

This final section presents simply a collection of different growth behaviors observed on the ends of e-needles. This is by no means a complete accounting of morphological features observed on e-needles, but rather a sampling of phenomena I have observed to date. As with any scientific endeavor, this is a work in progress, and the full morphology diagram offers considerable opportunity for further investigation and discovery. The scientific potential will become vastly increased once 3D numerical modeling has been developed to a suitably advanced state, but that may have to wait for the next generation of snow-crystal researchers.

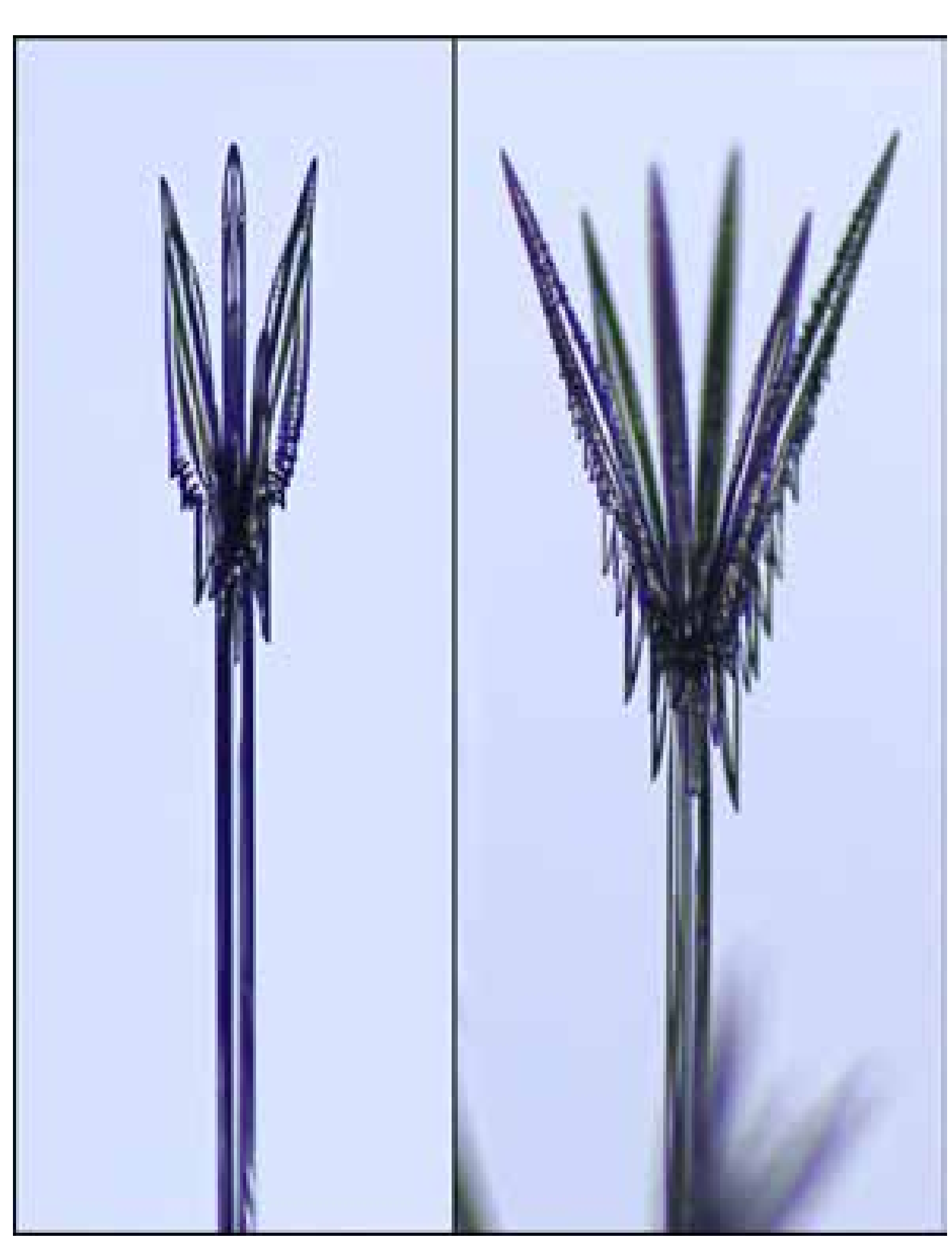

**Tridents and Witch's Brooms.** **This composite image shows two e-needle crystals both grown at -5 C, the crystal on the left at a supersaturation of 32 percent, the other at 64 percent. At the lower supersaturation, the smaller opening angle between the primary branches results in a strong competition for available water vapor. The result is that alternating branches stunt the growth of their immediate neighbors, yielding a three-branched crystal that resembles a *trident*. The phenomenon is related to the formation of triangular snow crystals, which is discussed in Chapter 3. At the higher supersaturation, the opening angle is larger, resulting in less competition and crystals with all six branches resembling a *witch's broom*.**

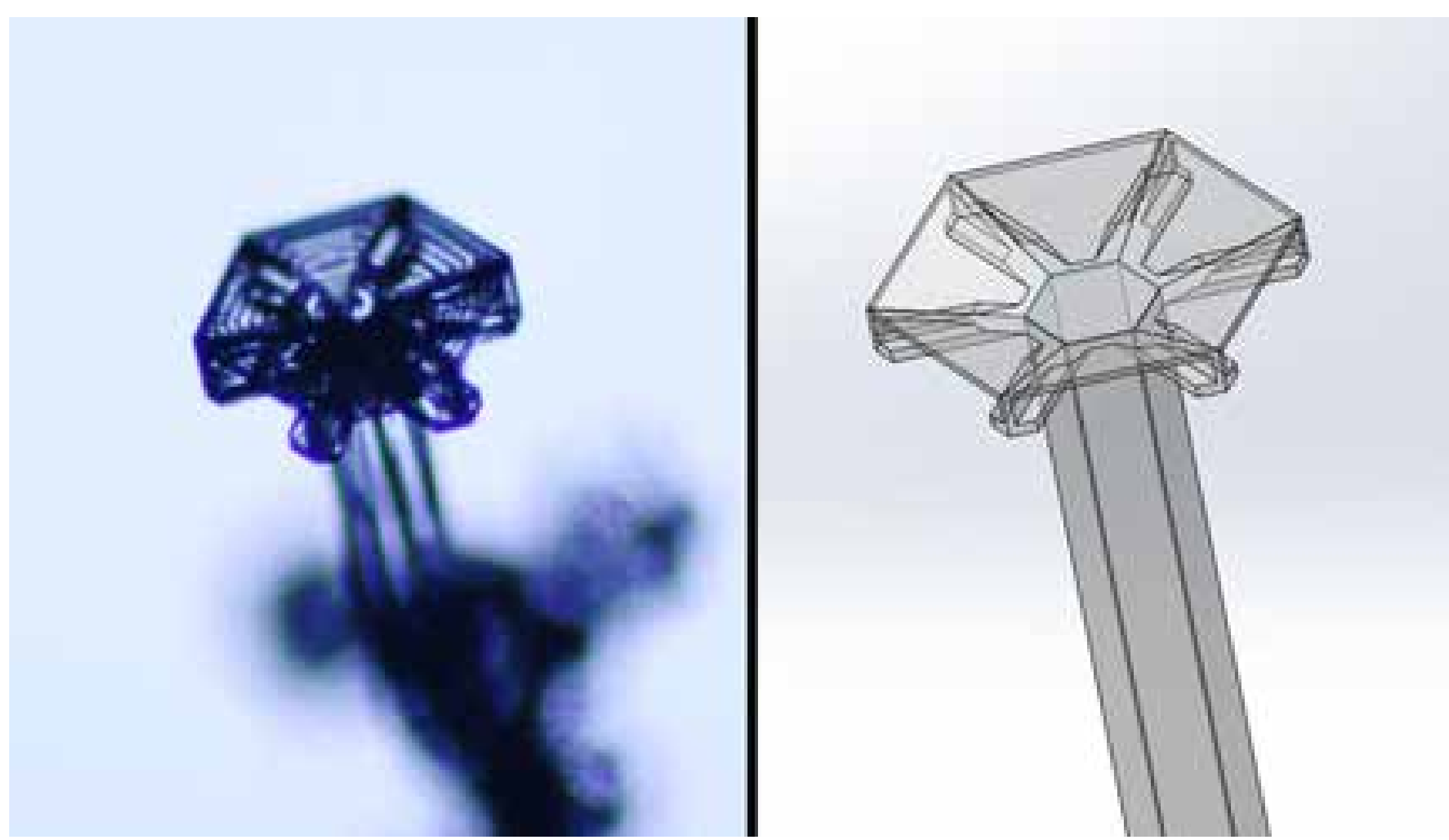

**I-beam Ridges. (Above)**
**This common e-needle morphology exhibits a slightly conical plate, pronounced ridges on the undersurface of the plate, and plate-like extensions growing from the bottoms of the ridges. The ridge structure thus has an "I-beam" like appearance. The 3-D rendering is meant to clarify the structure, which can be difficult to discern from a single photograph. As seen in Figure 8.16, variations of the I-beam ridge structure can be found in many regions of the e-needle morphology diagram.**

**Cups with Fins. (Below)**
**Near $(T, \sigma)$ = (-7 C, 32%), hollow cup-like crystals form, flanked by six plate-like "fins" on their sides. As the temperature is lowered, the opening angle of the cup increases, eventually transforming this morphology into a slightly conical plate with I-beam ridges.**
**[SolidWorks drawings by Ryan Potter.]**

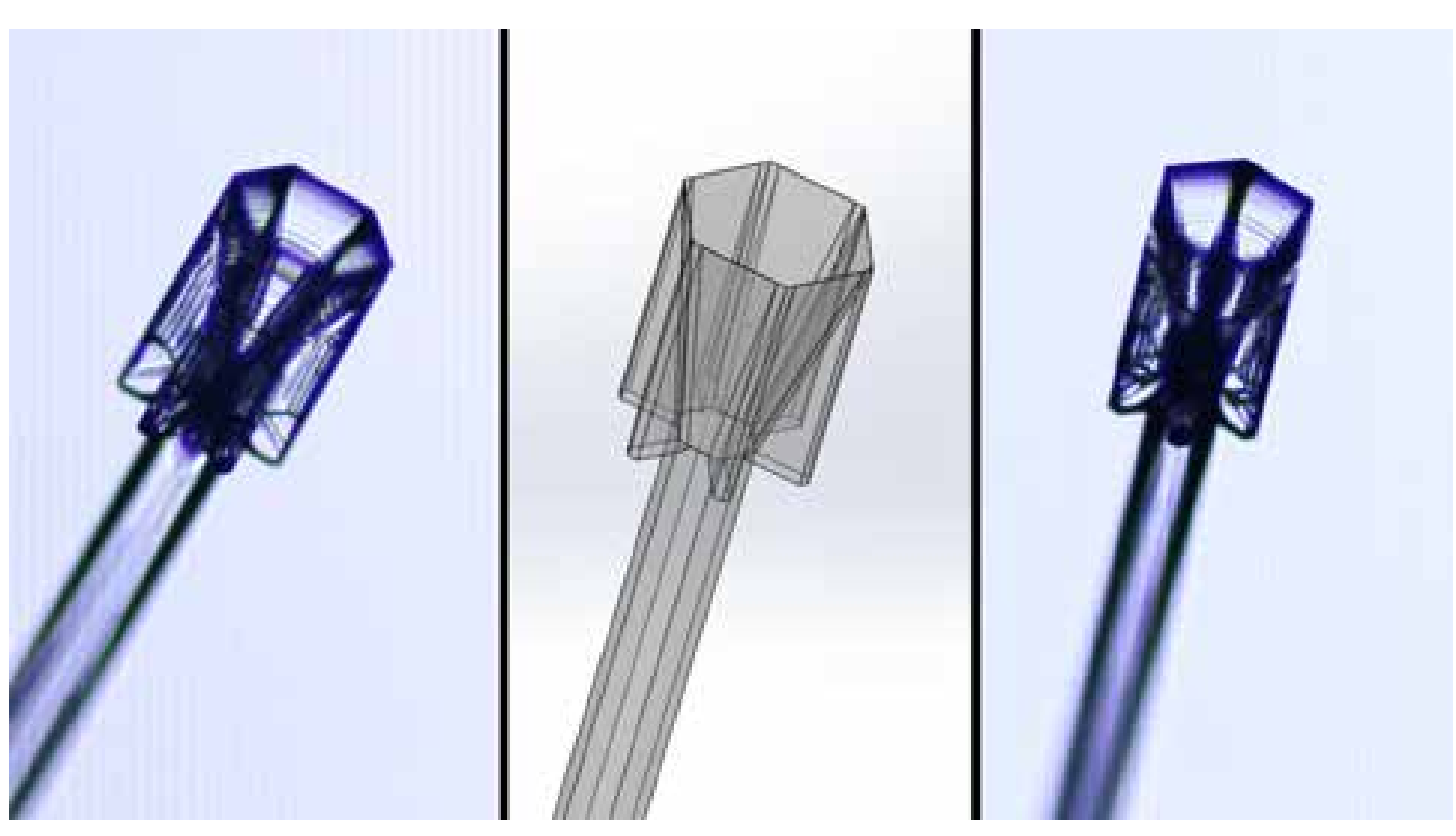

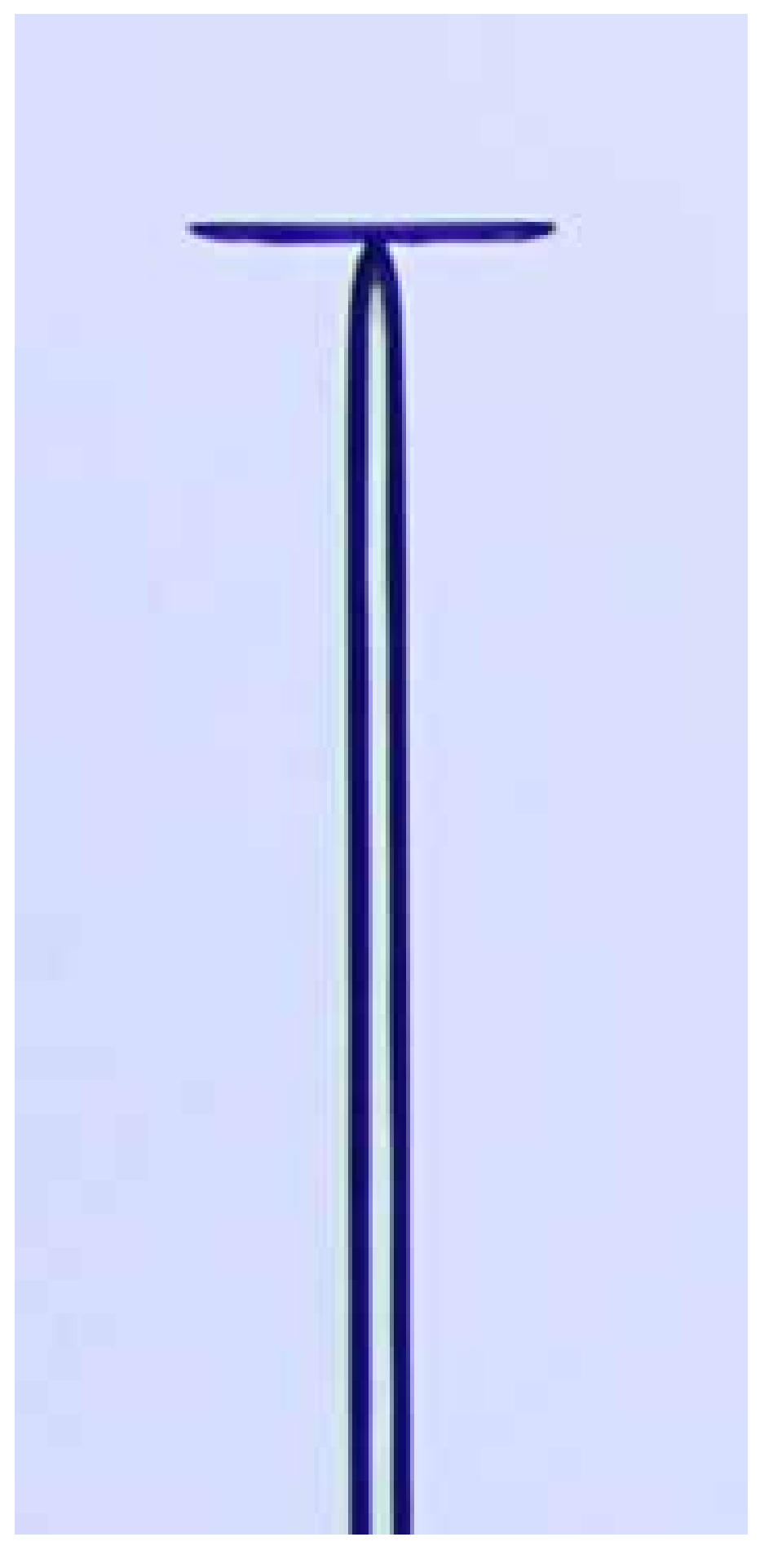

**Isolated Plates. (Above)**
**This crystal grown near -15 C shows a thin plate (in side view) on an e-needle. The broad plate shielded the growth of the tip of the needle, while the column below it was less shielded and grew thicker. The result is a plate-like crystal perched rather precariously upon the tapered tip of the column.**

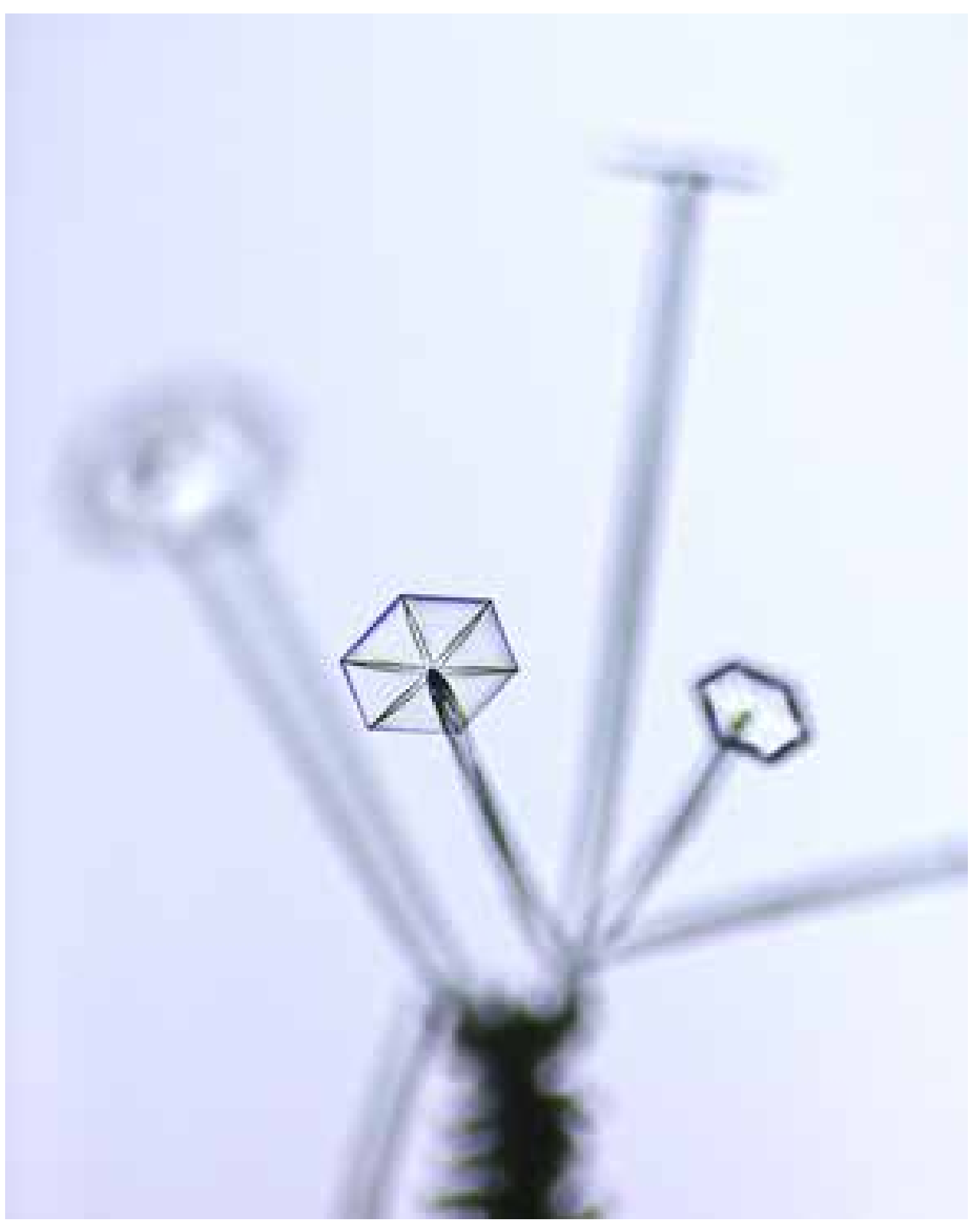

**Crystal Bouquet. (Above)**
**Thin, near-perfect hexagonal plates readily grow on the ends of c-axis e-needles, making it possible to obtain some delightful group shots.**

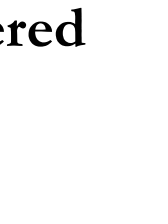

**Hollow Plates. (Right)**
**This crystal grown on an e-needle at $(T, \sigma)$ = (-16 C, 16%) exhibits deep hollows in the six prism facets. Reducing the supersaturation would cause the outer edges to fill in, leaving thin trapped bubbles in the plate. Bubbles in plates are remarkably common (Chapter 3), and sometimes display interference colors in snow crystal photographs (Chapter 11).**

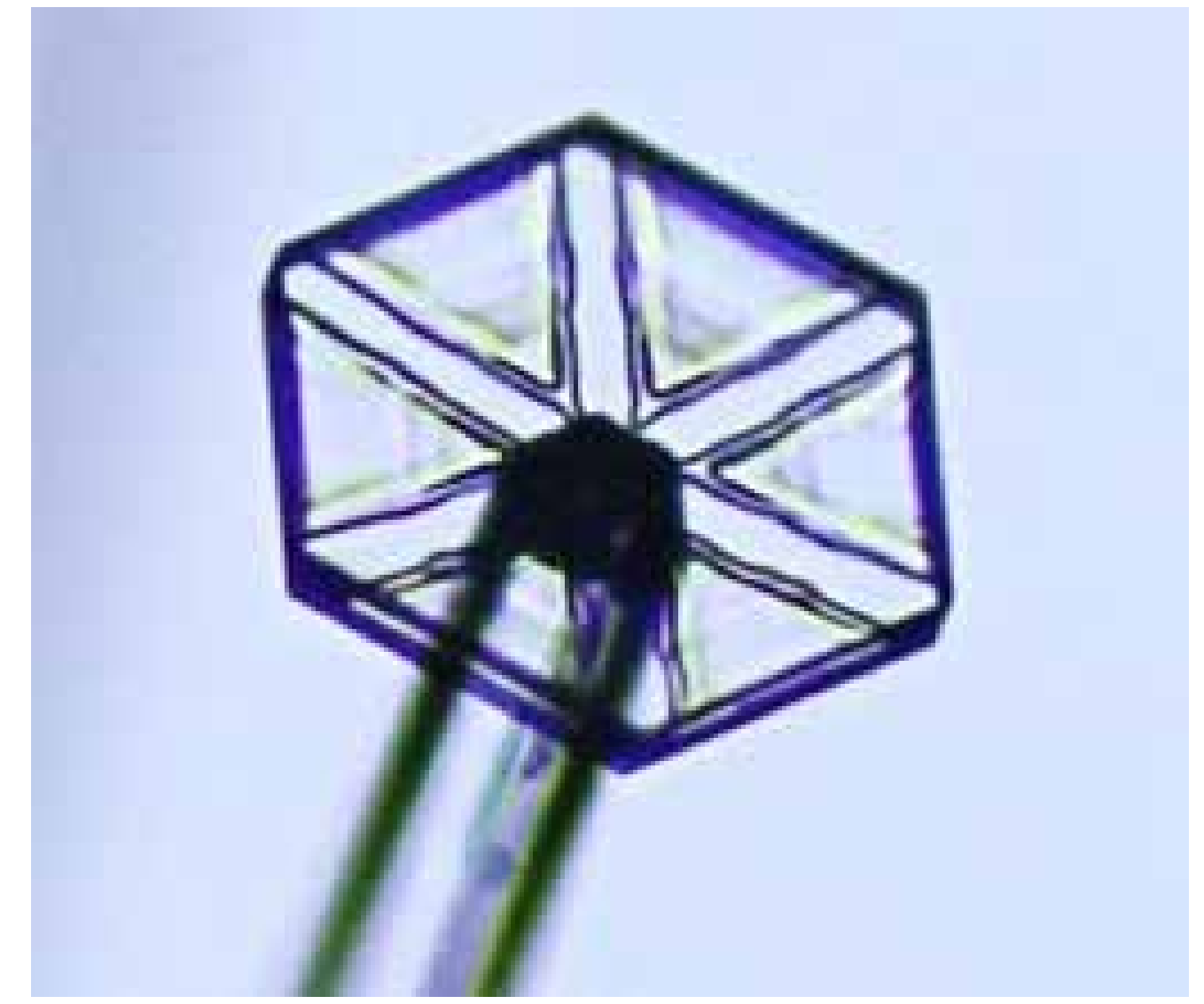

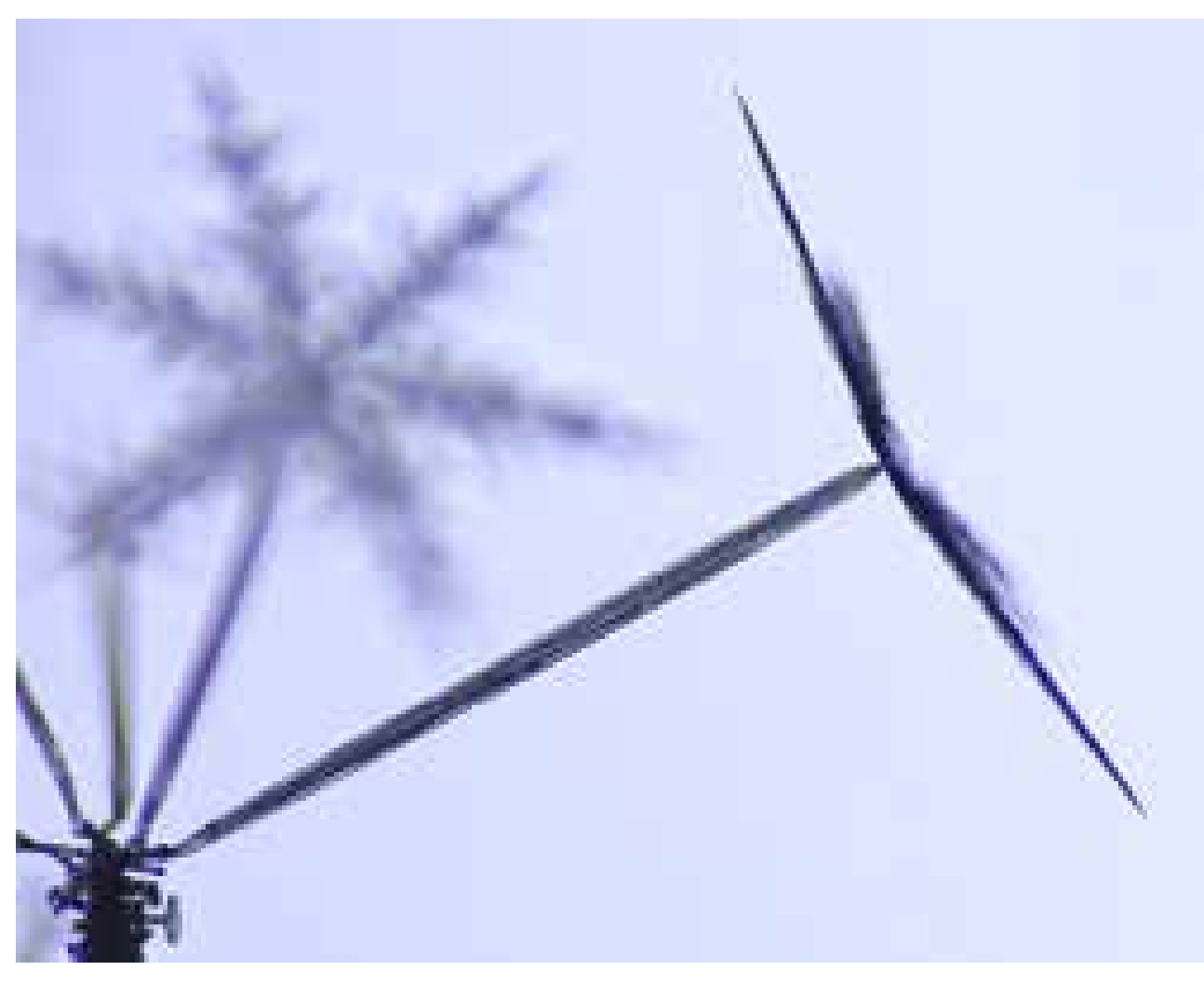

**Slightly Conical Plates.**
When plate-like crystals grow on the ends of e-needles, including dendritic structures like these, the outer edges typically growth slightly upward (away from the e-needle), because of a slight gradient in the supersaturation, which is lowered by the presence of the e-needle. As a result, plate-like crystals are often not flat, but exhibit a slightly conical overall shape.

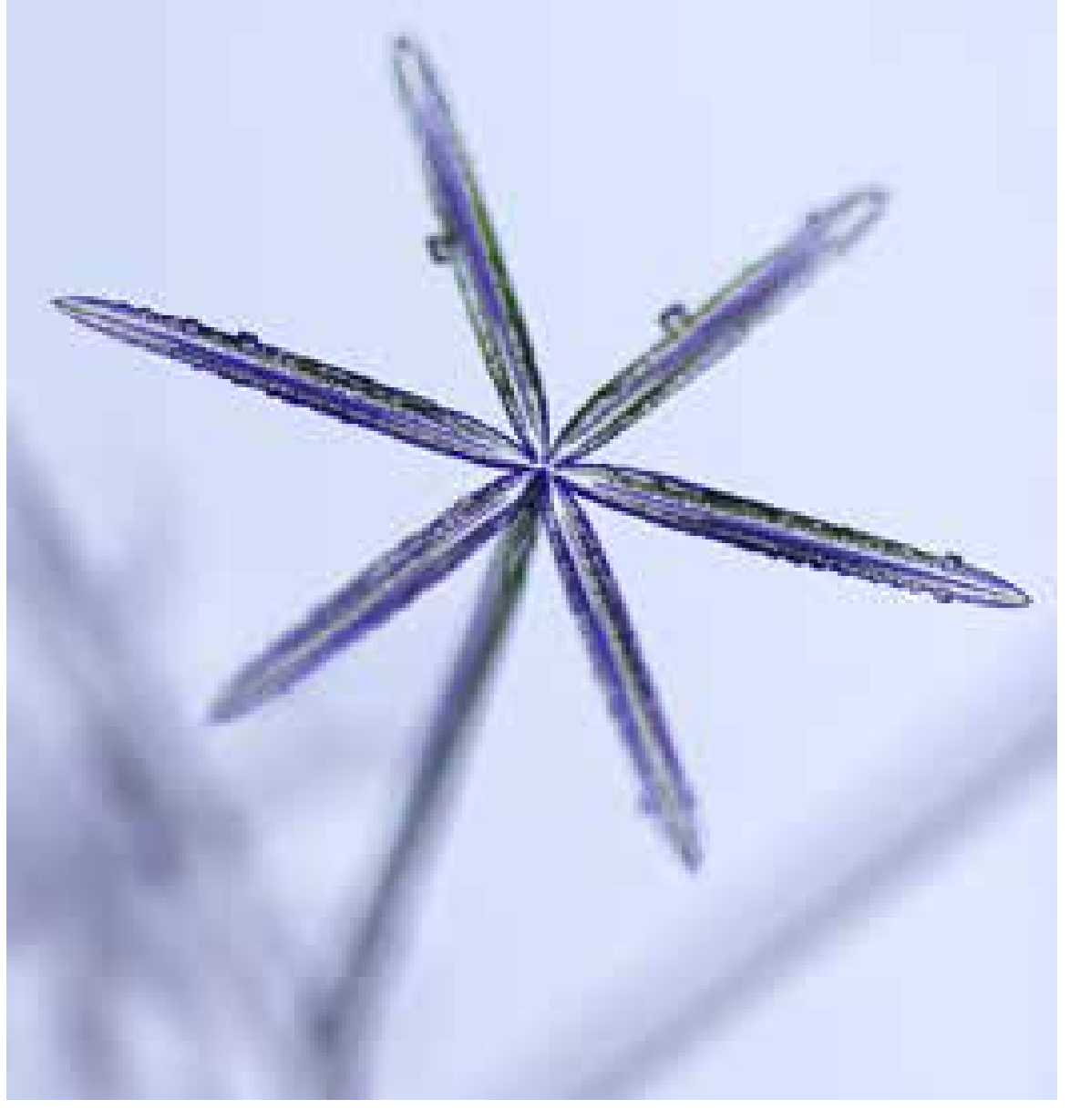

**Simple Stars.**
At conditions near $(T, \sigma)$ = (-14 C, 32%), stellar crystals appear with simple primary branches and essentially no sidebranching. These straight branches can grow quite long with no change in overall morphology; in this example the crystal is nearly 2mm in diameter. Modeling these stars may require the Edge-Sharpening Instability, as $\alpha_{prism}$ must be high at the tips but substantially lower on the sides of the branches.

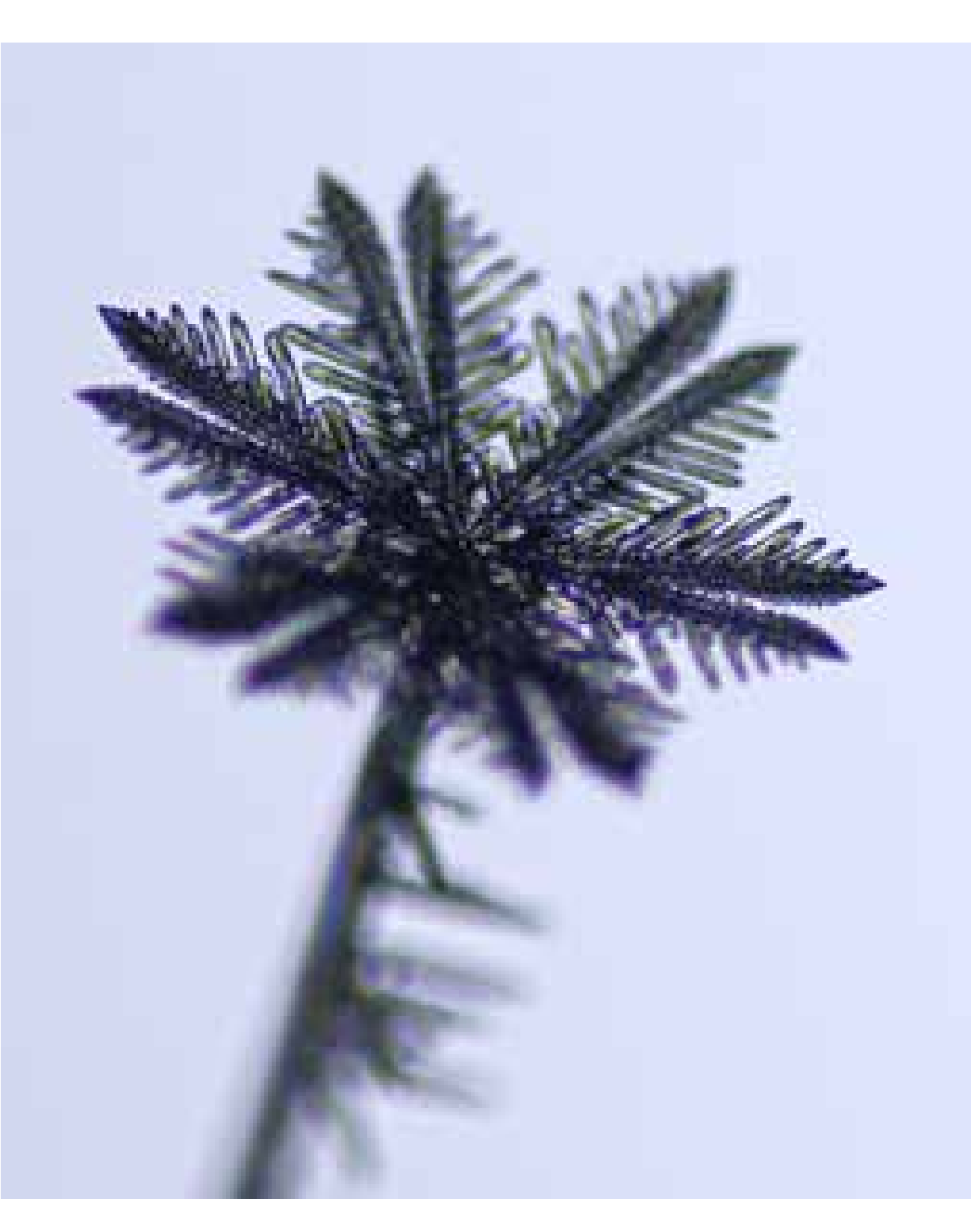

**Tip Splitting. (Left)**
This photograph shows a crystal grown on the end of an e-needle under the conditions $(T, \sigma)$ = (-15 C, 128%). Early in this crystal's development, when the crystal was small and the near-surface supersaturation especially high, all six branches split, yielding six sets of split branches. As discussed in Chapter 3, tip splitting indicates nearly isotropic prism attachment kinetics, in this case because $\alpha_{prism}$ was very close to unity. As the crystal grew larger, diffusion lowered the surface supersaturation, which in turn lowered $\alpha_{prism}$, so the branches experienced no additional tip splitting. Tip splitting is fairly common at these extreme conditions around -15 C, although symmetrical tip splitting like in this example is unusual.

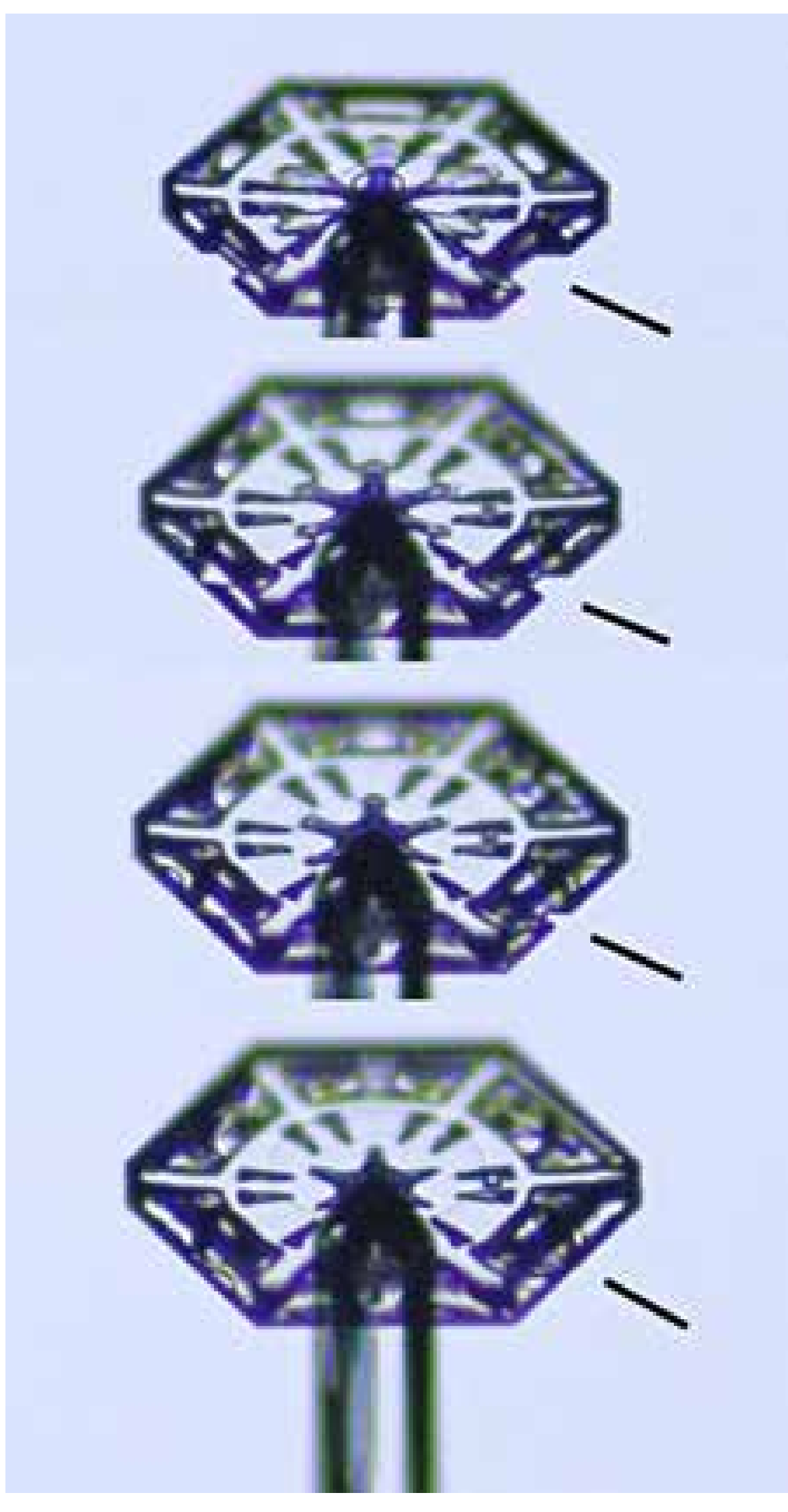

**Merging I-beams. (Above)**
**The top image in this series shows a thin plate with I-beam ridges that formed at $(T, \sigma)$ = (-12 C, 16%). As the plate continued to grow, the I-beam extensions grew together to form a deep hollow in the thick plate.**

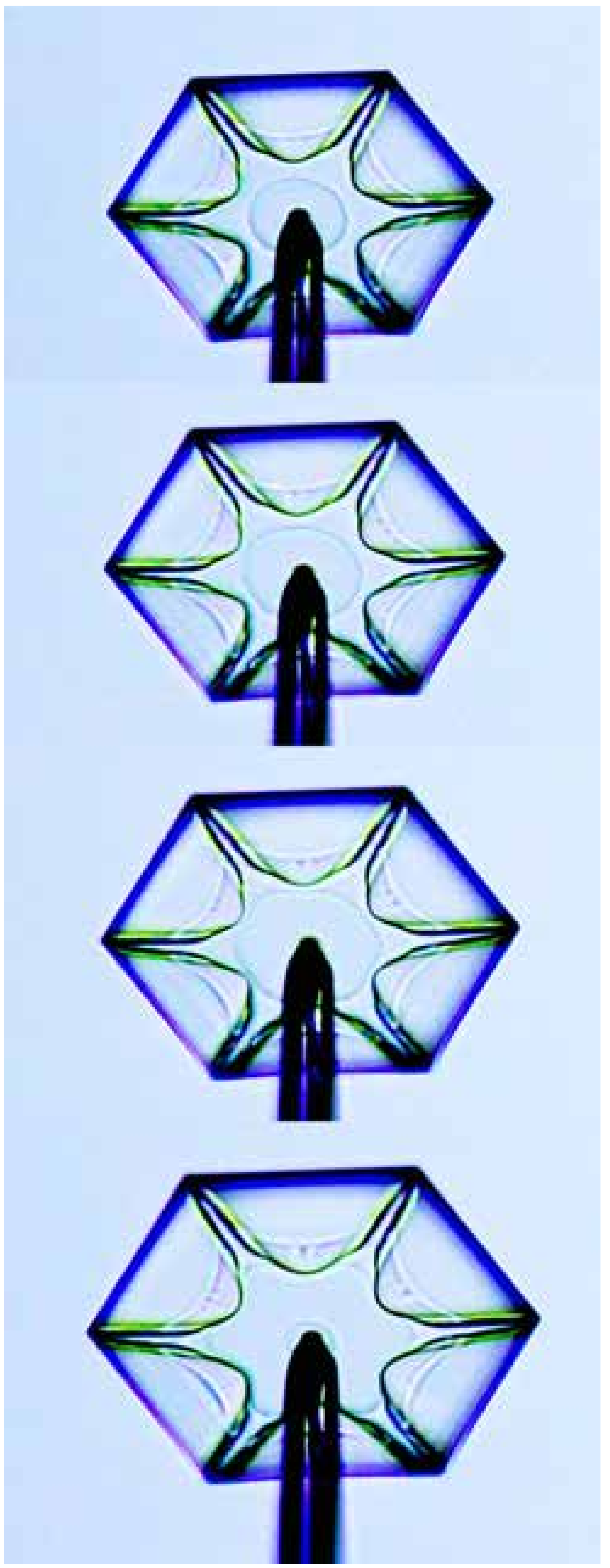

**Outward Propagating Macrosteps. (Right)**
**This series of images shows, from top to bottom, an outwardly propagating macrostep on the bottom surface of a thin plate, observed at $(T, \sigma)$ = (-15 C, 16%). The steps originated from the contact point with the e-needle on the underside of the plate. In the final image, the first macrostep has reached the edge of the faceted region, and a new small circular macrostep has appeared.**

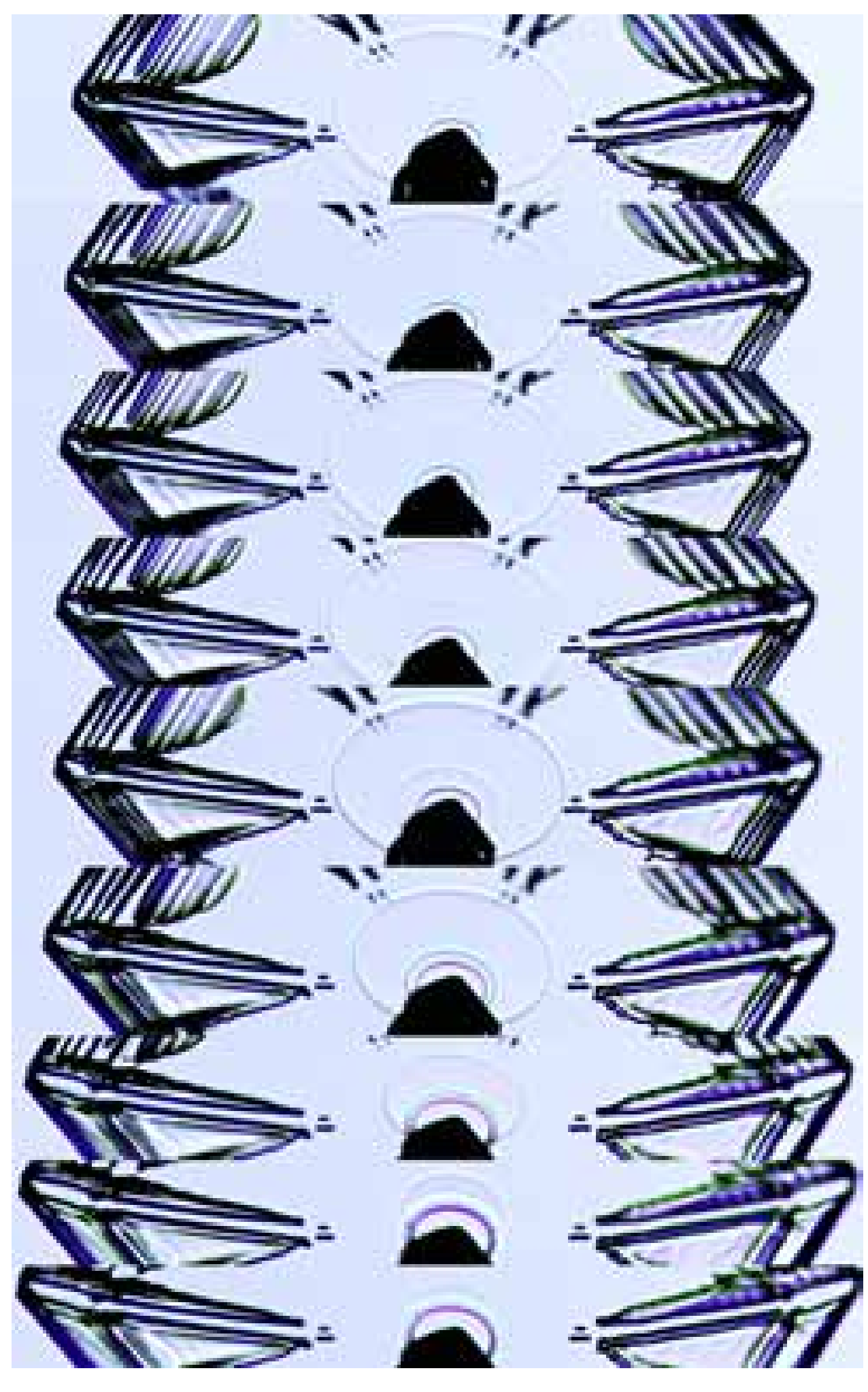

**Multiple Macrosteps. (Left)**
This series of images shows several macrosteps propagating on both the top and bottom surfaces of a hexagonal plate crystal atop an e-needle. Unfortunately, the macrostep growth dynamics are complex and not readily amenable to quantitative analysis on these crystals. In photos like these, even deciding whether a particular macrostep resides on the top or bottom surface can be challenging.

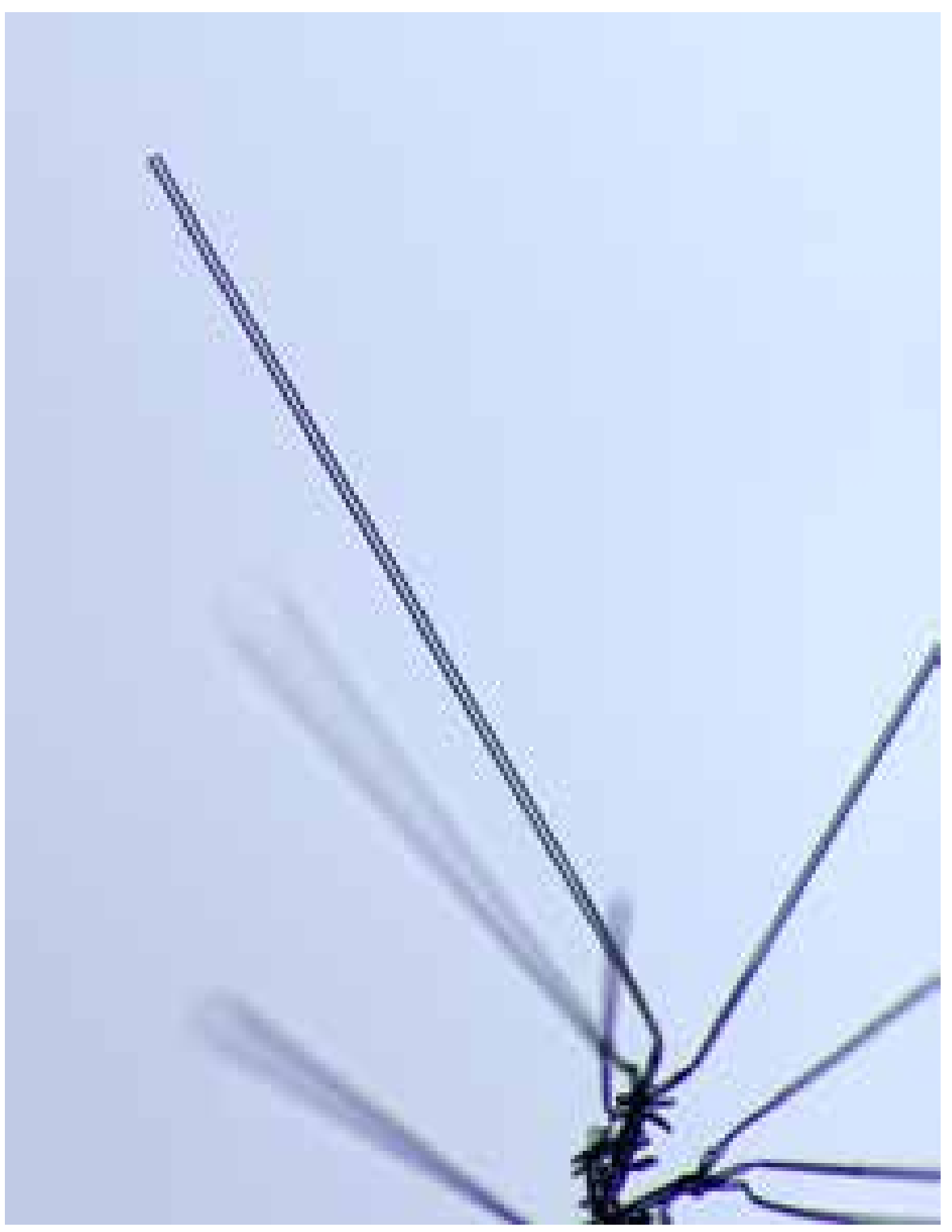

**Simple Columns. (Above)**
When the supersaturation is especially low at any temperature, e-needles typically grow into simple columnar forms.

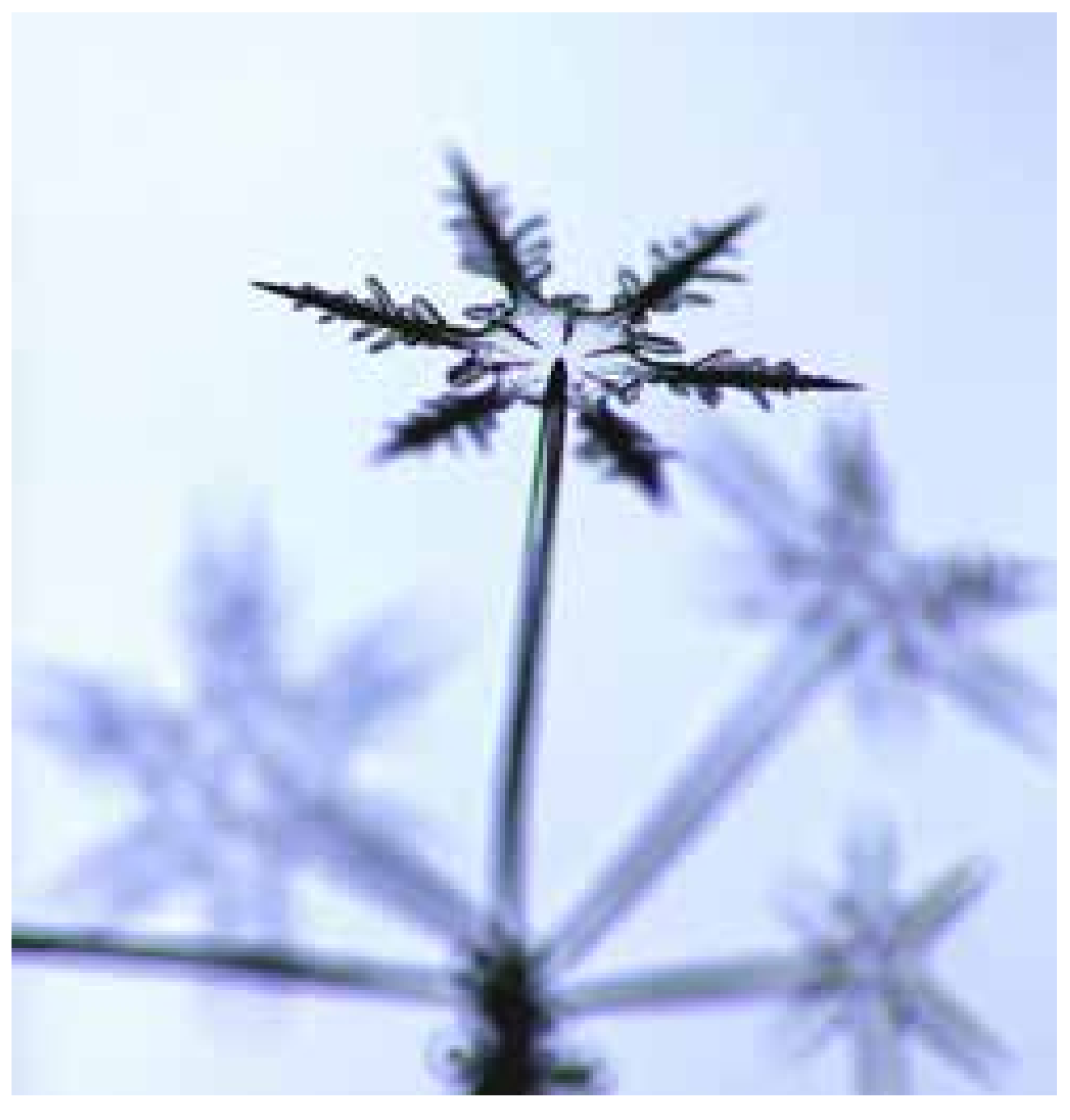

**Changing Conditions. (Left)**
Abrupt changes in growth behavior usually occur only when there is a causative change in growth conditions. In this example, a thin plate was first created on an e-needle at a relatively low supersaturation near -15 C. Then the supersaturation was suddenly increased, resulting in the formation of dendritic branches on the six corners of the hexagonal plate.

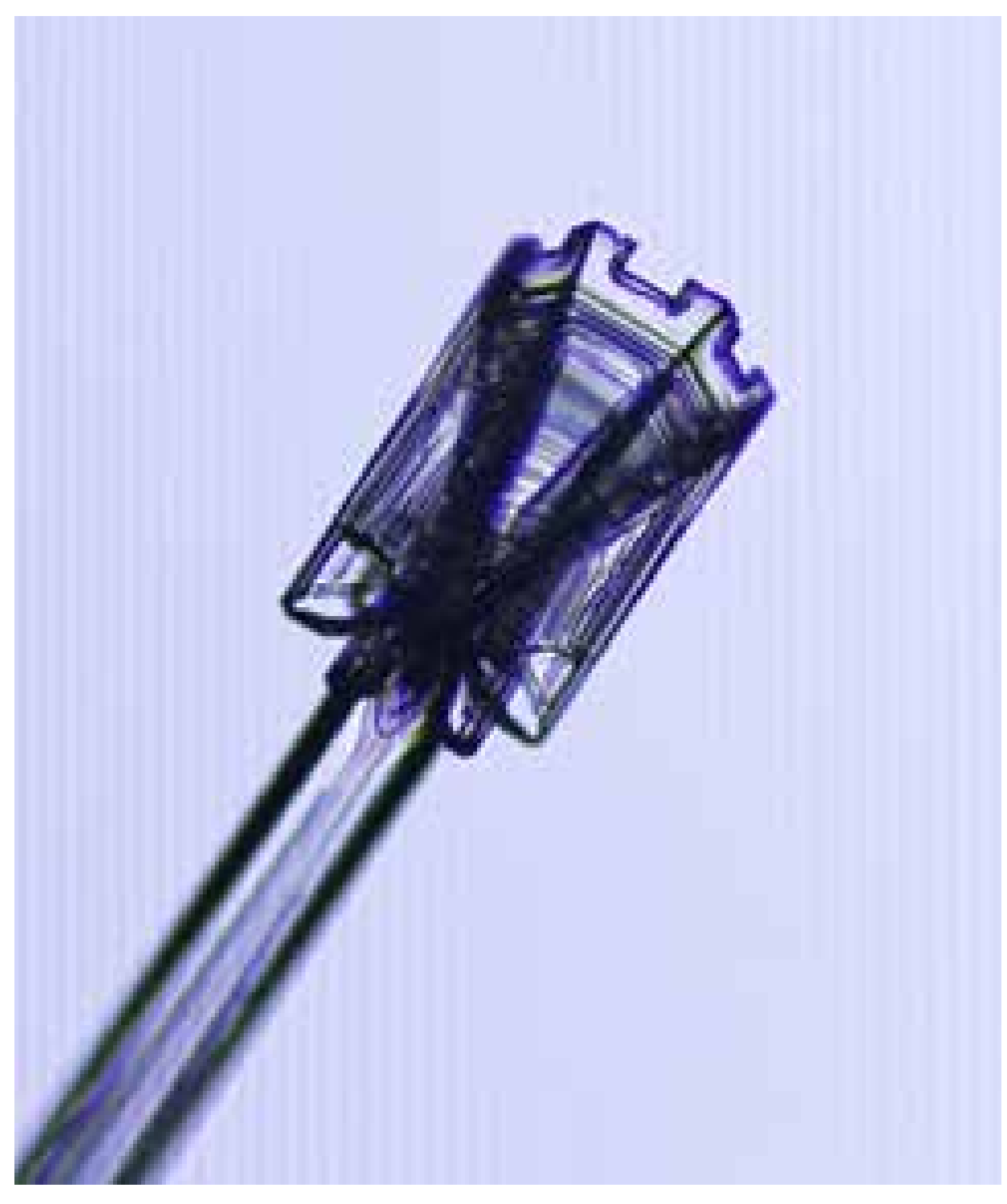

**Crystal Battlements.**
This unusual structure emerged when a cup-with-fins crystal grown near $(T, \sigma)$ = (-7 C, 32%) was subsequently exposed to a higher supersaturation, causing the rim of the cup to break up into six sections.

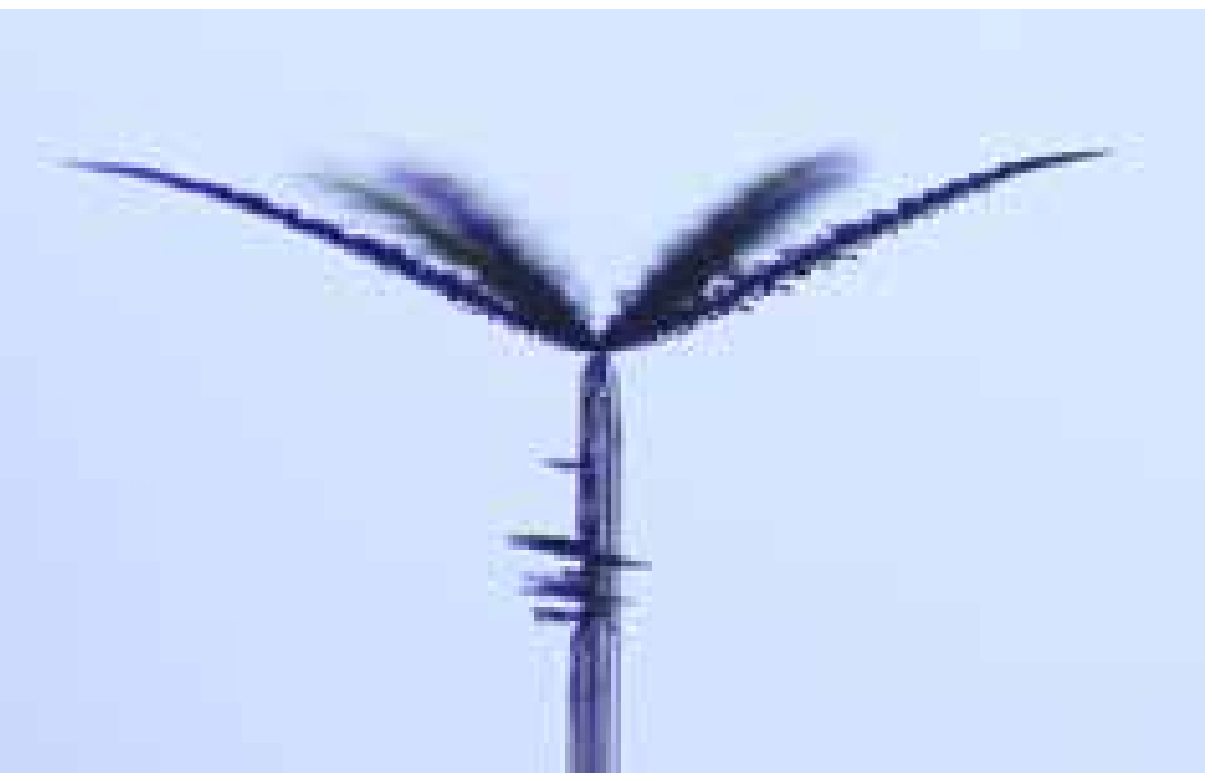

**Curved Branches.**
This e-needle was first exposed to $(T, \sigma)$ = (-12 C, 128%), growing a somewhat conical star with modest dendritic sidebranching (see Figure 8.16). Near the end of its growth, the supersaturation was lowered to 64% and the branch direction changed as basal faceting became more pronounced. In general, lower supersaturations yield flatter plates when the temperature is near -14 C.

**Bubble Evolution. (Below)**
This series of images shows the formation and evolution of two bubbles inside a columnar crystal grown over several hours near -7 C. The large markings running down the length of the column are surface features, while the bubbles grew from hollows at the end of the column.

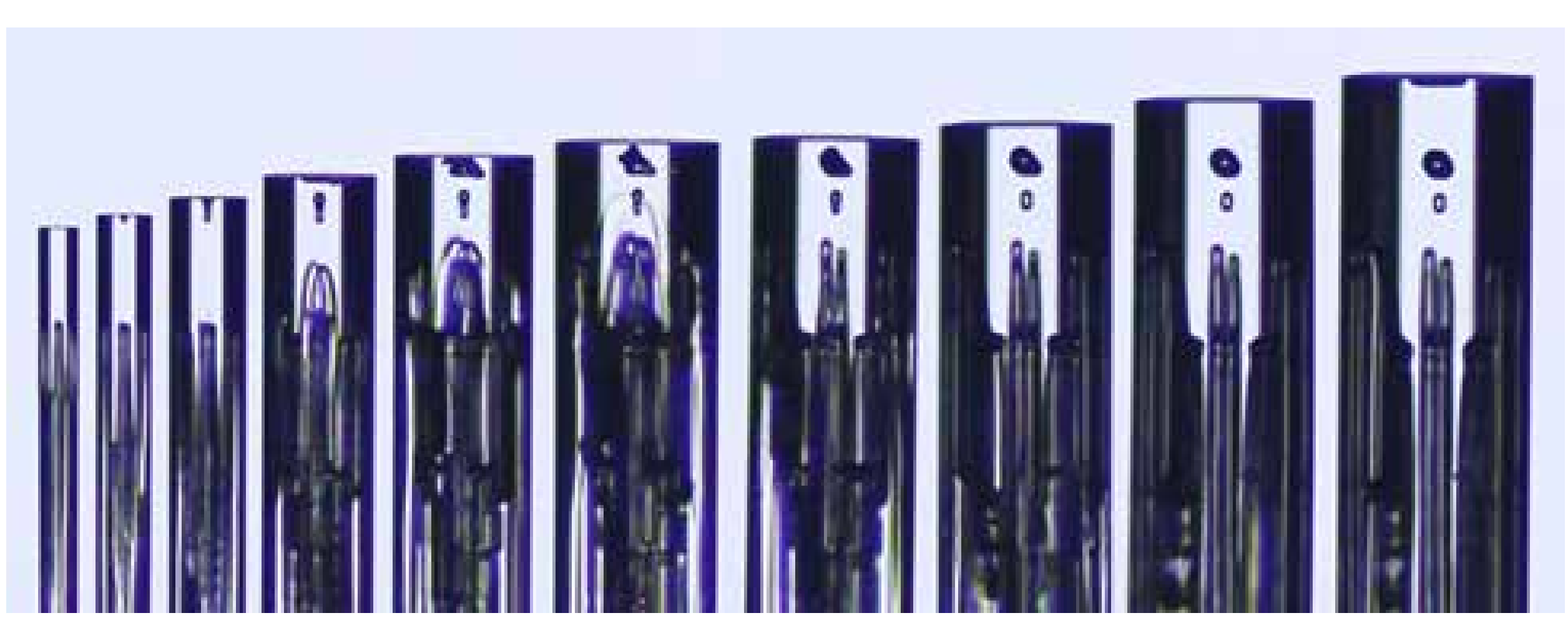

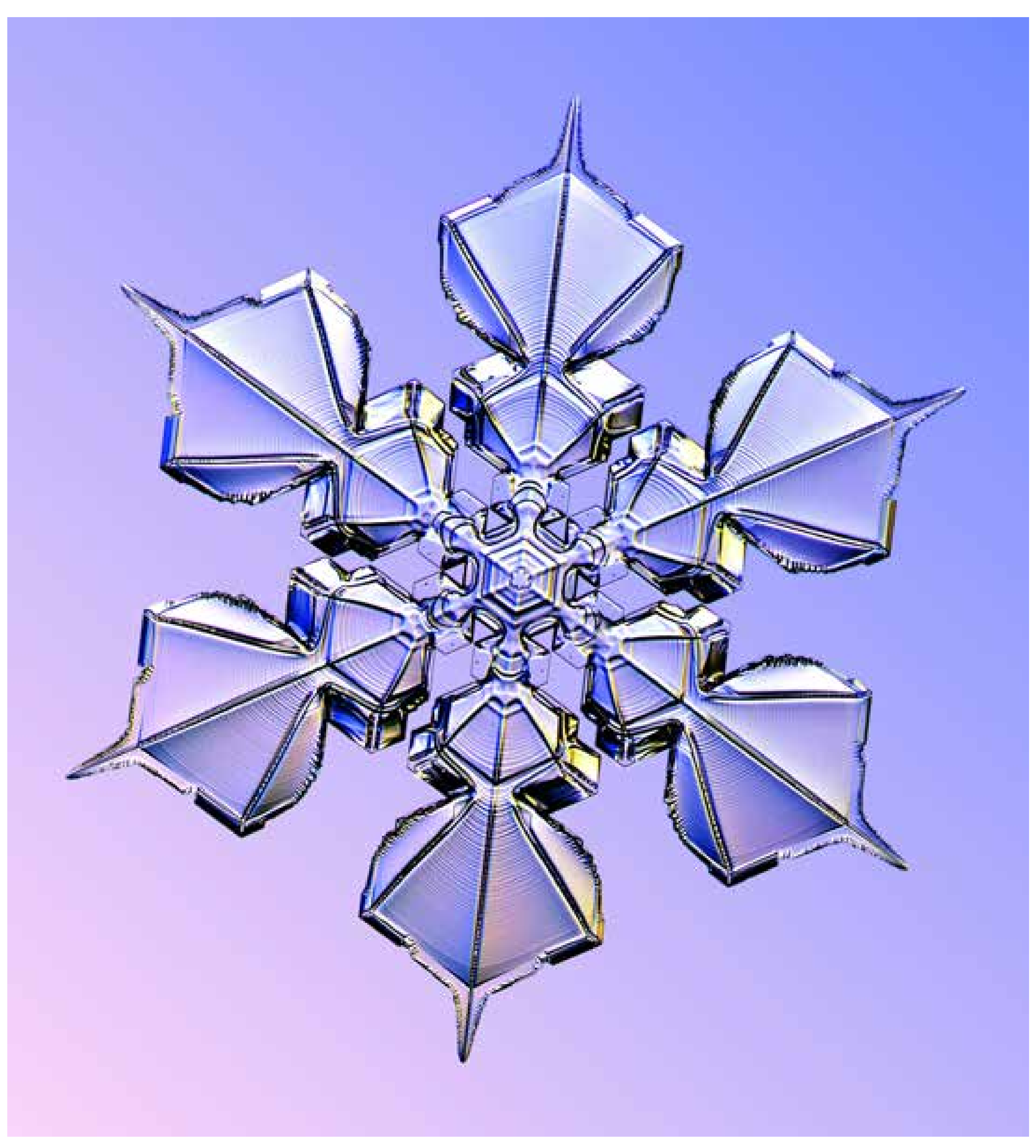

# Chapter 9

# Plate-on-Pedestal Snow Crystals

*Build a better snowflake,*
*And the world will shovel a path*
*to your door.*
– KGL

While creating synthetic snow crystals is an essential endeavor for investigating the physical processes governing their growth, the occupation also presents an excellent opportunity for artistic expression. Beyond photographing natural specimens falling from the clouds, it is possible to fashion synthetic snow crystals that are even more spectacular than the best nature has to offer. I like to call these *designer snowflakes*, as one can create a specific growth morphology simply by adjusting the applied temperature, humidity, and other environmental factors as a function of time. Moreover, in contrast to the natural variety, once can photograph designer snowflakes as they form, allowing time-lapse recordings of the entire growth process. The activity becomes a novel type of ice sculpture, discarding the chisel in favor of using molecular self-assembly and the laws of crystal growth to create beautiful, lacy, symmetrical crystalline structures.

**Facing page: A laboratory-grown Plate-on-Pedestal snow crystal, measuring 3.3 mm from tip to tip, grown by the author. Remarkably, most of this stellar plate is not in contact with the transparent substrate, but lies above it, balanced atop the small ice nub seen at the center of the crystal.**

Designer snowflakes have lagged behind their natural counterparts as photographic subjects, in large part because it had not previously been possible to grow synthetic snow crystals of a quality that compared to the best natural specimens. Just as synthetic diamonds have only recently begun to rival their quarried counterparts in size and quality, designer snowflakes are only now surpassing those found in nature, exhibiting sharper facets and more precise symmetries.

In this chapter, I describe in detail a *Plate-on-Pedestal* (PoP) technique that I developed for creating beautiful stellar snow crystals like the one shown on the facing page. The PoP apparatus is relatively straightforward to construct and operate, plus it was specifically engineered for capturing high-resolution photographs of designer snow crystals as they form and develop. Both the growth temperature and supersaturation can be separately adjusted as a function of time, allowing the manufacture of a near-infinite variety of complex morphologies. As an added bonus, one need not wait for a suitable snowfall, or brave the frigid weather, to photograph these icy creations. All the cold parts of the apparatus are conveniently shielded from the room-temperature environment, allowing one to explore the artistic side of snow-crystal growth on one's own schedule while remaining in quite comfortable surroundings.

## 9.1 The Plate-on-Pedestal Method

Creating a PoP snow crystal begins by producing a cloud of small ice prisms in a free-fall growth chamber (see Chapter 6) and letting a few fall onto a transparent substrate held at a temperature near -12 C. Some of the prisms will land with one basal facet resting flat against the substrate surface, as illustrated in Figure 9.1. Next expose this small crystal to a moderately high supersaturation in air, and a thin ice plate will commence growing out horizontally from the top surface of the prism, as shown in Figures 9.1 and 9.2. Because the upper plate is supported above the substrate by the central ice prism, I refer to this as a Plate-on-Pedestal geometry.

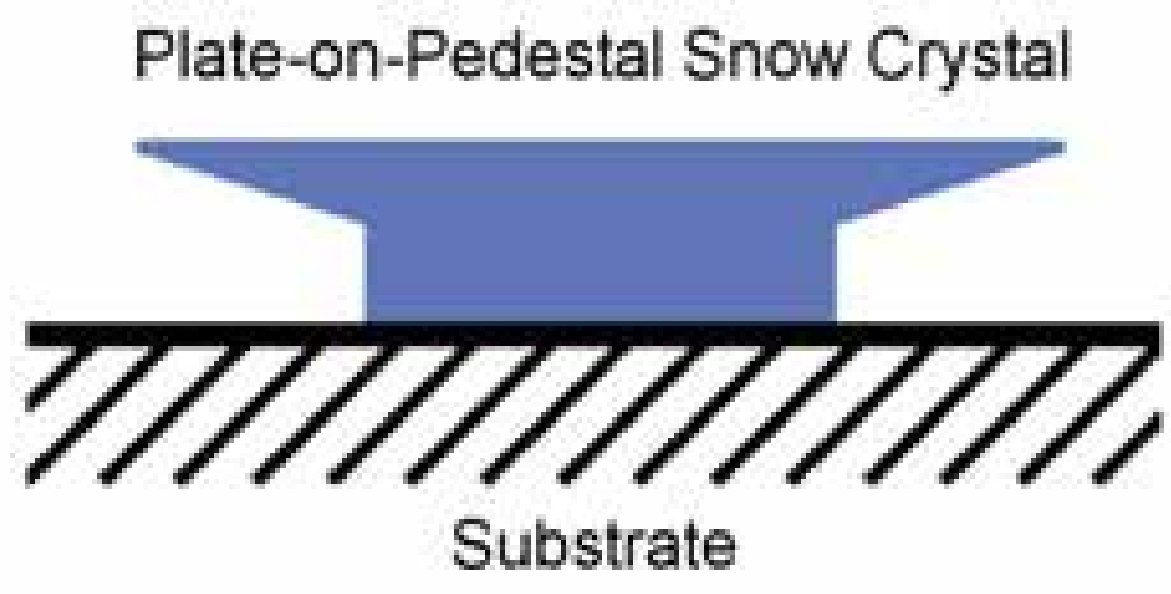

**Figure 9.1: A side view of the basic Plate-on-Pedestal (PoP) snow-crystal geometry. A thin plate-like crystal grows outward from the top edge of a small hexagonal prism, while one basal face of the prism rests on a transparent substrate.**

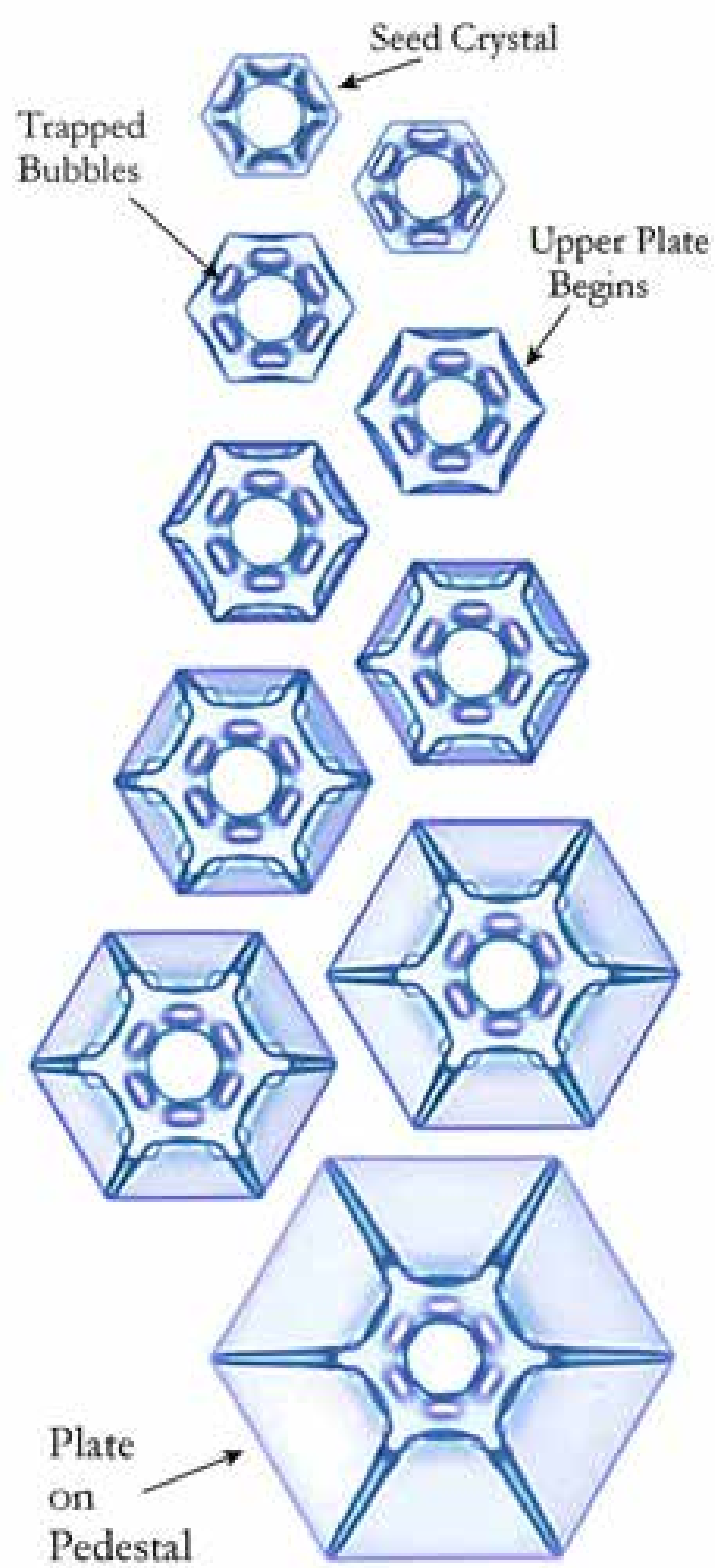

**Figure 9.2: Multiple images showing the formation of a PoP snow crystal from a seed crystal, here growing from about 50 μm to 170 μm in diameter. The seed crystal exhibited some non-faceted structure, and its early growth trapped six small bubbles between the ice and the substrate surface. Once formed, these isolated bubbles did not evolve substantially as the upper plate formed. Note that the final thin, hexagonal plate is not touching the substrate, but is supported above it, as illustrated in Figure 9.1.**

**Figure 9.3: (Right) The apparatus used to grow and photograph PoP snow crystals. Small, freely falling ice prisms in the seed-crystal chamber first pass through a shutter, and some of them fall randomly onto a waiting sapphire substrate (loading phase). The substrate is then moved to the growth region for subsequent PoP crystal growth.**

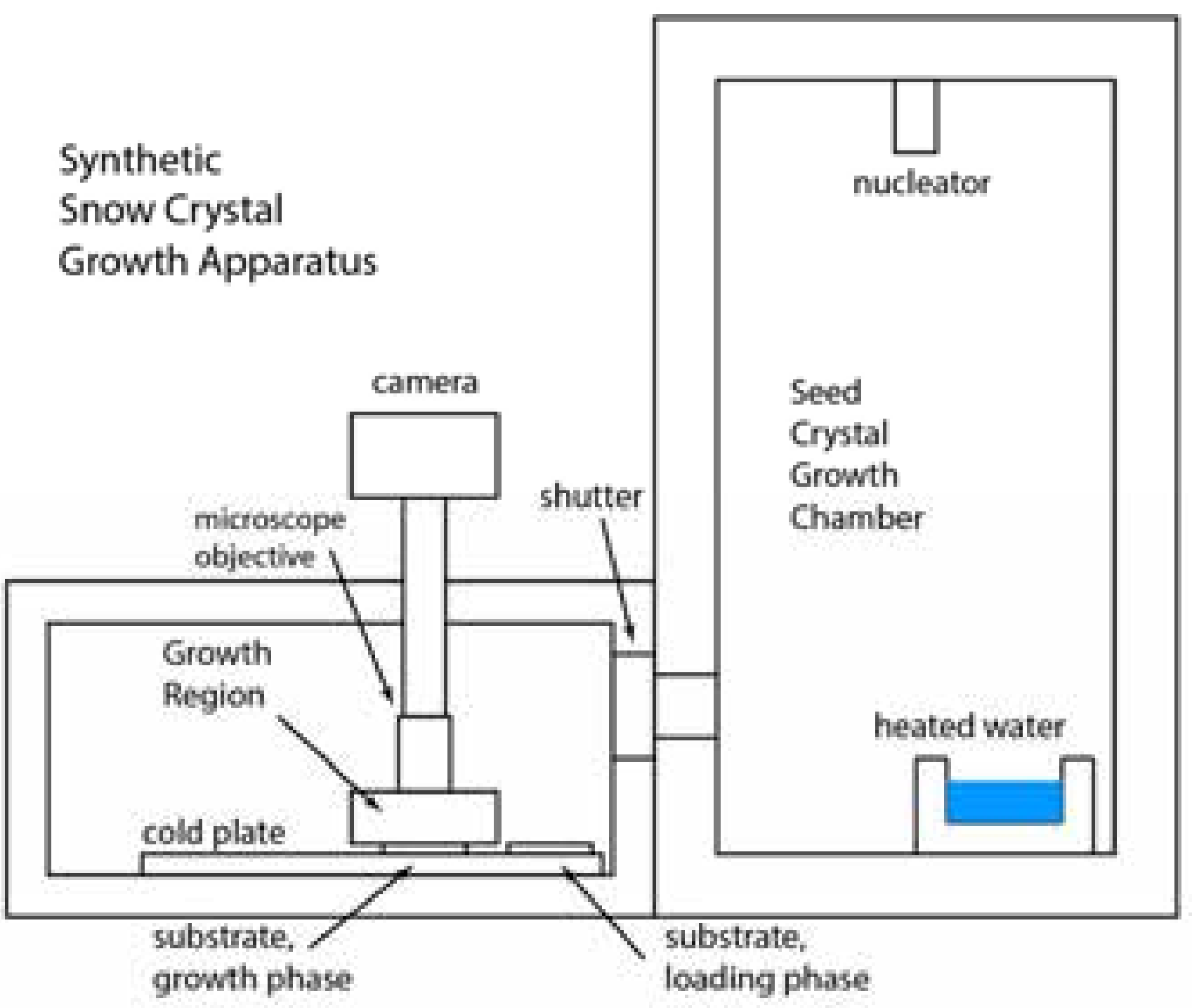

The physics underlying the formation of the PoP geometry arises from diffusion-limited growth and the Edge-Sharpening Instability (ESI) described in Chapter 4, which is further related to how the attachment kinetics on prism surfaces depends on background air pressure. None of this is completely understood at present, as the attachment kinetics is generally a subject of current research (see Chapters 3 and 7). But, like many aspects of engineering, one need not understand a phenomenon perfectly to put it to good use. Figure 9.2 shows the formation of a small PoP crystal, which is remarkably easy to achieve in practice.

Once the PoP structure has been established, it will continue so long as the subsequent growth occurs within several degrees of -15 C. Even as the ice plate becomes quite large, it grows entirely above the substrate, balanced atop the initial small prism. I have used this technique to engineer a remarkable variety of large, beautiful stellar snow crystals, many of which are shown throughout this book. Being stationary and supported above a transparent substrate, PoP snow crystals can be photographed easily as they form, allowing high-resolution imaging and striking time-lapse videos of their developing structure.

It appears that Gonda, Nakahara, and Sei grew snow crystals using a similar technique in the 1990s [1990Gon, 1997Gon], although these papers do not explicitly describe the plate-on-pedestal structure. Little subsequent work appeared after these initial results, perhaps because the PoP technique is not especially well suited for scientific measurements. After recognizing the PoP geometry in my own studies, I continued its development as a means to grow snow crystals in a more artistic realm.

## PoP Hardware

Figures 9.3 shows a sketch of the apparatus I developed for creating and photographing PoP snow crystals [2015Lib3]. The tall chamber is essentially the seed-crystal generator described in Chapter 6, producing a continuous cloud of small ice prisms that slowly fall through the chamber as they grow. Upon opening a shutter connecting the seed chamber to the adjoined lower chamber, some of the seed crystals waft through the opening and fall randomly onto a waiting sapphire substrate, shown in its loading position in Figure 9.3. After loading crystals for a few seconds, the substrate is moved over to its growth position under a photomicroscope, shown in greater detail in Figure 9.4. While watching the microscope image displayed on a TV monitor, the substrate is moved around using a pair of manipulator arms to search for a well-formed, isolated hexagonal prism that can be grown into a PoP snow crystal.

Once a suitable seed crystal has been positioned in the microscope field, moist air is blown gently down onto the crystal in the growth region. With proper temperature and supersaturation conditions begin applied to the seed crystal, the early growth produces the PoP geometry, which can then be grown further into a large stellar plate.

It typically takes 20-60 minutes to produce a large PoP snow crystal, during which time one can monitor its progress using the photomicroscope. Changing the substrate temperature $T_1$ (see Figure 9.4) changes the growth temperature of the crystal, while changing the heat-exchanger temperature $T_2$ adjusts the effective supersaturation around the crystal. Increasing the air flow though the heat exchanger increases the supersaturation as well. Changing the growth conditions frequently and abruptly tends to produce especially complex growth morphologies with a high degree of six-fold symmetry.

With a bit of experience, one can, at least to some degree, plan the structural features of a PoP snow crystal in advance, or improvise its morphological development as it grows. Each change in the temperature and supersaturation alters the growth behavior, and these parameters become the tools needed to create a wide variety of snow-crystal forms. As one begins to understand the rules of snow-crystal growth, the process becomes a unique and quite satisfying form of additive ice sculpture. In working with this apparatus to date, I have typically observed the crystal formation in real time and made temperature and air-flow adjustments without a great deal of pre-planning. However, one could easily add computer control to these inputs and develop specific algorithms for different growth behaviors.

## The Seed-Crystal Generator

The seed crystal growth chamber in Figure 9.3 has inside dimensions of approximately 40x40x100 cm, and is made from a frame of aluminum T-rail (see Chapter 6) covered with 1/8” thick aluminum panels. Methanol coolant from a recirculating chiller flows through central holes in the four vertical T-rails, and heat conduction through the aluminum rails and panels is sufficient to cool the remainder of the chamber, which is well insulated from the room by Styrofoam panels.

An insulated container containing one liter of ordinary tap water rests on the bottom of the seed-crystal chamber, as shown in Figure 9.3, and the water temperature is kept constant by an electronic regulator using an immersed water heating element and water temperature sensor. The top of the container is open to the air, so water evaporates and the vapor is carried by convection throughout the rest of the chamber. The continuous evaporation and convection maintains a steady-state supersaturation within the seed-crystal chamber that depends strongly on the water temperature, as described in Chapter 6.

**Figure 9.4: (Left) An expanded schematic view of the PoP growth region shown in Figure 9.3. In operation, saturated air at temperature $T_2$ blows gently down onto the substrate with its growing PoP crystal at temperature $T_1 < T_2$. Not shown is a white-light LED lamp placed underneath the color filter in this sketch.**

Water vapor is continually removed from the air by the growing ice crystals and by frost depositing on the walls of the chamber, and this water vapor is continually replenished by evaporation from the water reservoir. The air temperature is typically kept near -15 C, as measured by a thermistor near the center of the chamber, as this yields small plate-like seed crystals. The supersaturation is more difficult to determine, but can be inferred to some degree by the morphology of the growing crystals. The chiller temperature is typically set to -19 C (giving a -15 C air temperature) and the water temperature to 17 C, as this yields a continuous supply of thin, hexagonal plate-like seed crystals with diameters in the 20-50 micron range. Higher water temperatures yield somewhat branched morphologies, which is not desired for PoP seed crystals.

**Figure 9.5: A photograph of the apparatus depicted in Figure 9.3. The TV monitor displays a live view from the camera, here showing a growing PoP snow crystal surrounded by a field of water droplets. The cold chambers are covered in sealed Styrofoam panels for thermal insulation and to prevent condensation from the room air.**

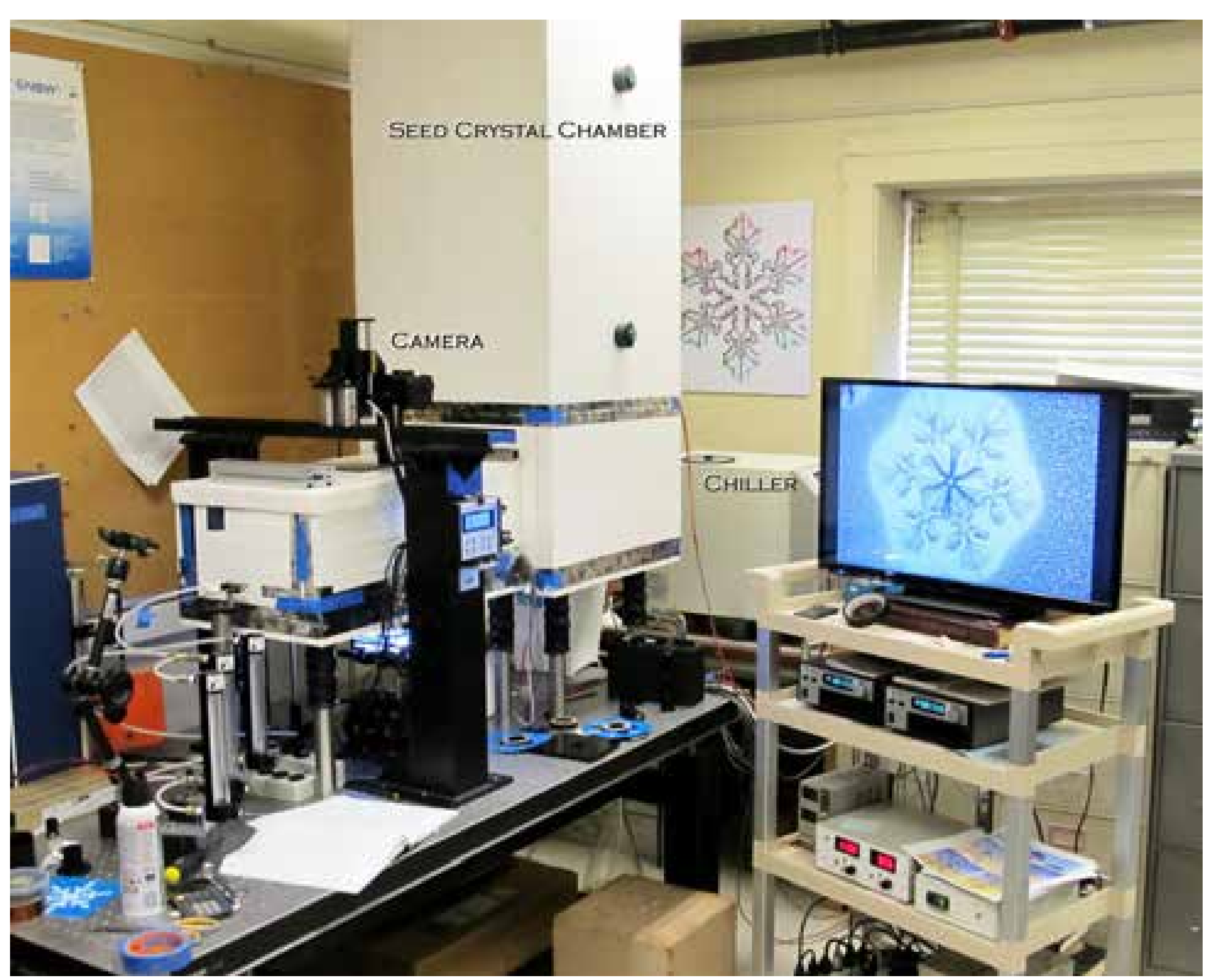

The expansion nucleator at the top of the seed-crystal chamber consists of a standard 1.33-inch Conflat vacuum nipple with an interior volume of about 25 cubic centimeters that is connected to a solenoid valve on the output side. Pressurized room air flows into the nucleator through a needle valve that constricts the rate of input air flow. The nucleator assembly is placed inside the growth chamber so its temperature is approximately -15 C during operation. The overall flow rate is slow enough that the air temperature becomes roughly equilibrated inside the nucleator.

During operation, frost condenses on the inner walls of the nucleator, so the water vapor content of the air is near the saturated value. Every ten seconds the solenoid valve is pulsed open, causing the pressurized air to rapidly expand and enter the growth chamber. The rapid expansion produces a small amount of air that is sufficiently cooled to nucleate ice crystals. Air pressures as low as 15 psi will usually nucleate some crystals, while 30 psi produces many thousands per pulse. Water buildup inside the nucleator is removed after each run by operating it for several hours when the chamber is at room temperature. With no initial ice buildup, the nucleator can run continuously for at least ten hours without difficulty.

The nucleated ice crystals float freely as they grow, until they eventually settle to the bottom of the chamber. The fall times are typically a few minutes, depending on temperature and supersaturation. Pulsing the nucleator valve open every ten seconds thus produces a steady-state in which roughly a million seed crystals are growing inside the chamber at any given time (this number being determined by a visual estimate of the typical spacing between crystals floating inside the chamber during operation). Shining a bright flashlight into the chamber reveals sparkles caused by reflections off the crystal facets, and this is a convenient way to verify that seed crystals are present.

Compressed air for both the nucleator and the crystal growth region is supplied by an ordinary oil-free workshop air compressor with a built-in storage tank and regulator, which automatically maintains the required 30 psi air pressure. The compressed air is passed through an oil filter and then an activated charcoal filter (containing coconut-husk charcoal) to remove remaining chemical contaminants from the air, and then a fine-pore fiber filter to remove any remaining charcoal dust.

A 50-mm-diameter hole in the side of the seed crystal growth chamber connects it to the adjoining main growth chamber seen in Figure 9.3. The cold plate at the bottom of the growth chamber is cooled using the same circulating coolant that flows through the walls of the seed chamber. To grow a POP crystal, the ice-free substrate is first moved to its loading position (see Figure 9.3) and a simple plate shutter is slid open between the two chambers. The convective air currents in the seed chamber cause a slight air flow between the chambers that carries a small number of seed crystals into the growth chamber, and some of these fall onto the substrate, a process that takes a few seconds. The shutter is then closed and the substrate is moved to a covered region within the main growth chamber, and a suitably isolated seed crystal is centered under the microscope for subsequent growth and observation.

## The Growth Chamber

As illustrated in Figure 9.4, the substrate is an uncoated sapphire disk, 50 mm in diameter and 1 mm thick, with the sapphire c-axis perpendicular to the disk surface. The principal advantages to using sapphire in this application are its high thermal conductivity and its resistance to scratching. Using c-axis sapphire avoids birefringence issues that can interfere with optical imaging.

The substrate slides on a smooth anodized aluminum plate with its temperature $T_1$ maintained by a temperature controller

(Arroyo Instruments model 5310) using a thermistor temperature sensor (Cole-Parmer Digi-Sense 08491-15) in the aluminum plate, together with thermoelectric heating/cooling modules beneath it.

The thermistors have an absolute accuracy better than $\pm 0.1$ C, and the temperature regulation is stable to better than $\pm 0.01$ C under normal operation. However, the temperature of the substrate and especially the air immediately above it are not precisely equal to the aluminum plate temperature, and this adds some uncertainty to the ice crystal growth temperature.

The heat exchanger above the substrate is an aluminum plate at a temperature $T_2$ maintained by a separate temperature controller. Filtered room air from the air compressor first passes through a baffled precooler kept near $T_{precool} = 0$ C, which reduces the heat load in the main heat exchanger and removes a large fraction of the water vapor it contains to prevent ice clogging the main heat exchanger.

The precooled air then passes through a series of serpentine channels in the heat exchanger before blowing down onto the substrate and the growing snow crystal. The air flow rate $F$ is typically 200-300 cubic centimeters per minute (ccm), measured using a tapered-tube flow meter and controlled with a simple needle valve. This flow rate replaces air in the guide tube (between the heat exchanger and the substrate, as shown in Figure 9.4) about once per second, which is comparable to the time needed to equilibrate the air temperature to that of the guide tube.

This growth-chamber design provides three adjustable parameters that can be used to control the crystal growth behavior: $T_1$, $T_2$, and $F$. The crystal temperature is nearly equal to $T_1$, which is typically kept within a few degrees -15 C to grow stellar-plate snow crystals. The quantity $\Delta T = T_2 - T_1$ mainly determines the supersaturation, which can be at most $\sigma_{max} = [c_{sat}(T_2) - c_{sat}(T_1)]/c_{sat}(T_1)]$, a quantity that is roughly proportional to $\Delta T^2$.

The interior diameter of the guide tube is 1.6 cm, and its overall length (from the bottom of the heat exchanger to the substrate) is approximately 2.3 cm. Air flows into the guide tube via four channels in the heat-exchanger plate, arranged symmetrically around the circumference of the top of the guide tube.

The equal flow rates through the four input channels, along with the cylindrical geometry of the guide tube assembly, were engineered to produce a nearly cylindrically symmetric downward flow pattern within the guide tube, with the flow axis centered on the growing crystal. The guide tube temperature is kept near the substrate temperature $T_1$, and the guide tube is thermally isolated from the heat exchanger by a short section of thin-walled plastic tube.

An important consideration in the heat exchanger design is the uniformity and symmetry of the air-flow pattern around the growing snow crystal. If the temperature, supersaturation, or air flow are substantially nonuniform across the face of a stellar crystal, this will compromise the symmetry of the final morphology, creating a lopsided crystal. For the pursuit of artistic snow-crystal perfection, therefore, engineering carefully uniform environmental conditions in the growth chamber is quite critical.

During a cooldown of the apparatus, air is passed through the heat exchanger for 30 minutes to deposit ice on its inner surfaces. The temperature is set to $T_2 < -20$ C during this time to make sure ice (and not supercooled water) is deposited inside the heat exchanger. Once the heat exchanger has been preconditioned in this way, air passing through it will exit at temperature $T_2$ and be saturated with water vapor relative to ice at $T_2$. As it approaches the substrate, the air cools to near $T_1 < T_2$ and thus becomes supersaturated.

Modeling the temperature and supersaturation at the growing ice surface is problematic with this apparatus for a number of reasons. The Reynolds number of the flow is approximately 10, so the flow is probably not perfectly laminar, and the timescale for the air in the guide tube to become equilibrated with the guide-tube walls via diffusion is comparable to the time it takes air to flow through the tube. Moreover, a stagnation point in the flow occurs where the flow axis of the system intercepts the substrate surface, at the position of the growing crystal, further complicating the air-flow and thermal analysis. In general, however, a higher $\Delta T$ and a higher $F$ produce a higher supersaturation around the growing snow crystal.

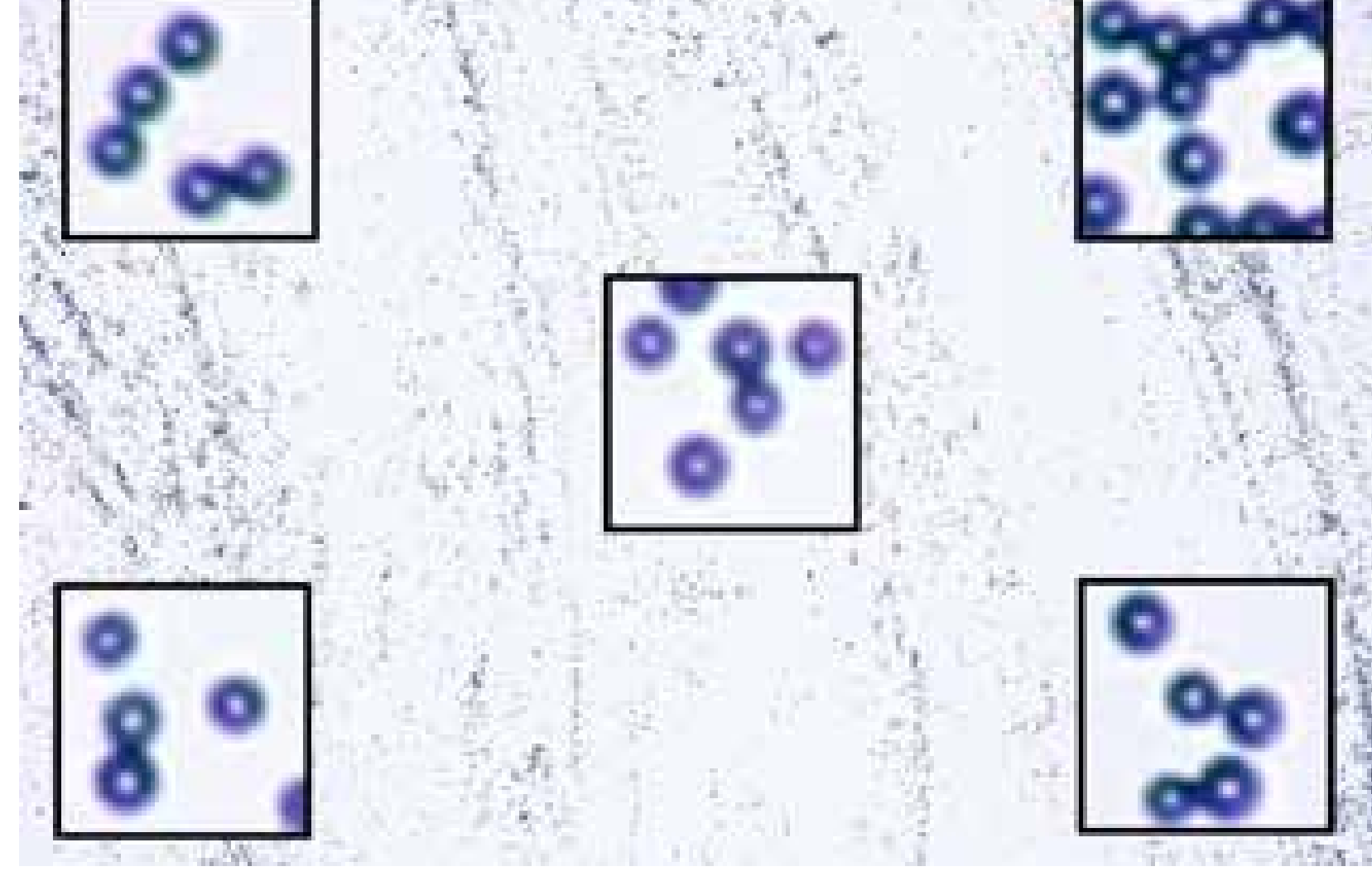

**Figure 9.6: The background image here shows a dusting of 10-μm-diameter glass beads spread over the sapphire substrate, for testing image quality across the 56x37 mm field of view. Sub-images near the center and four corners were magnified to gauge image quality. The sharpness in the corners of the frame was noticeably improved in this image by using focus stacking of five images.**

To complicate the supersaturation analysis even further, water droplets often condense on the substrate near the crystal, as I describe below. The presence of liquid water substantially alters the supersaturation field, and the amount of water condensation changes substantially with changes in $\Delta T$ and $F$. The thermal connection between the edge of a growing PoP crystal and the underlying substrate is also difficult to determine accurately, given the pedestal geometry.

For all these reasons, I do not expect that the apparatus described here will ever be well suited for performing precision measurements of ice growth rates under known conditions. It is better suited for more qualitative studies examining ice crystal morphologies and growth behaviors, as well as for creating snow crystals purely for artistic objectives.

## Optical Imaging

The microscope objective shown in Figure 9.4 is part of the heat exchanger assembly, but it is kept a few degrees warmer than $T_2$ using a heater dissipating 1-2 Watts into the objective body. This elevated temperature is necessary to keep fog from condensing on the glass face of the objective, which would interfere with optical imaging.

A Mitutoyo 5X Plan Apo objective with a 250-mm focal length achromatic reimaging lens immediately behind it works well for single-frame imaging. The infinity-corrected objective has a working distance of 34 mm, a numerical aperture of 0.14, resolution of 2.0 μm, and a depth of focus of 14 μm, yielding excellent image quality. Focusing is done by moving the camera body (on a StackShot rail), and some amount of focus stacking (see Chapter 11) is typically needed for optimal imaging of large crystals, owing to the shallow depth of focus. The image projects to about 1 μm per pixel of the 36x25-mm, 5616x3744-pixel sensor in a Canon EOS 5D camera.

Small glass beads are a convenient tool for testing the focus quality of the optical system, as shown in Figure 9.6. With the 5X objective, even a slight tilt of the substrate relative to the focal plane can degrade the image sharpness across the field of view, owing to the small depth of focus. This can be nicely corrected using focus stacking, which is also quite helpful for PoP crystals that are slightly concave in shape, which is common for larger specimens. Obtaining high-quality microscope imaging is always a challenge with a home-built apparatus, so some kind of image testing like this is highly recommended.

**Figure 9.7: (Right) A simplified ray diagram of the optical layout shown in Figure 9.4. Note that the microscope objective images the snow crystal onto the camera sensor ($A \rightarrow A'$), while the field lens images the color filter onto a pupil within the microscope objective ($B \rightarrow B'$).**

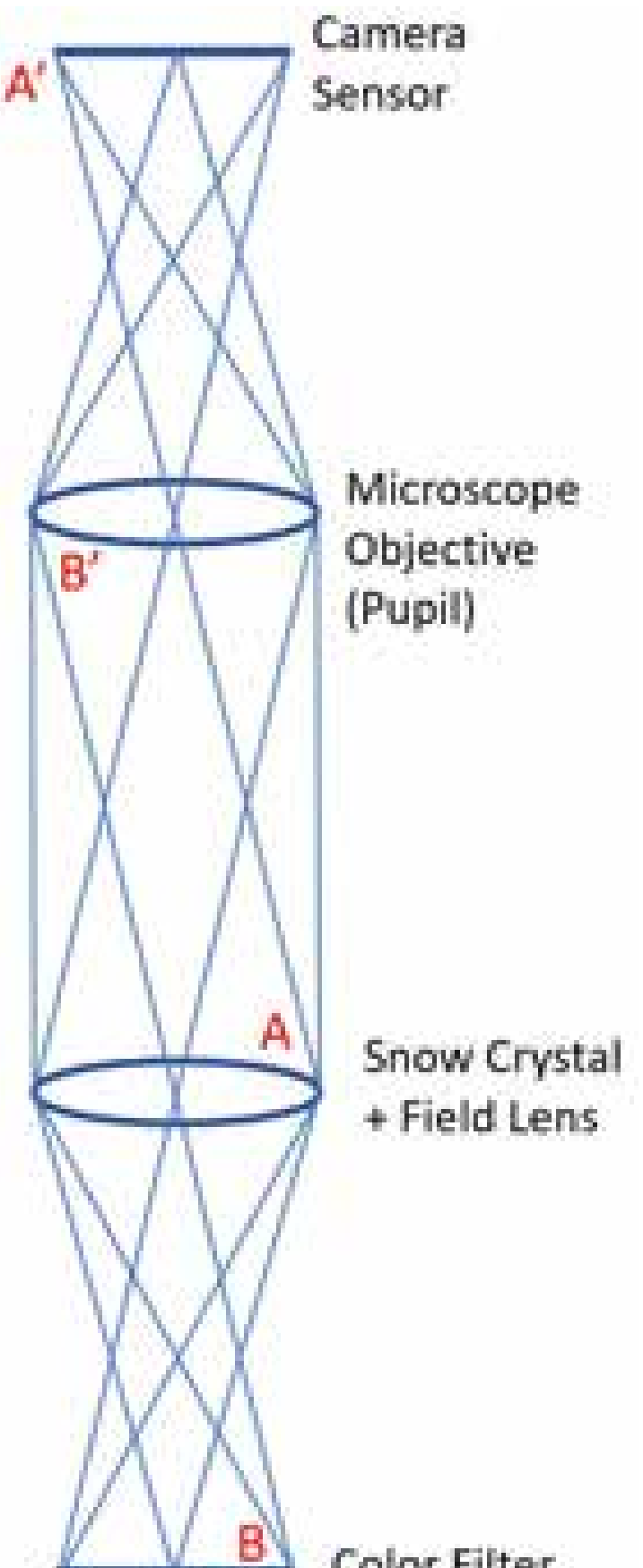

When shooting still images for making time-lapse videos of growing PoP crystals, a 3X Mitutoyo Compact Objective is preferred, with a resolution of 2.5 microns and a depth of focus of 23 microns. The 3X images are noticeably less sharp, but focus stacking is no longer needed, simplifying video production from the series of time-lapse stills. Also, the reduction in image sharpness is hardly noticeable in a video of a growing snow crystal, as the edges are constantly moving.

The field lens shown in Figure 9.4 reimages a color filter onto a pupil within the objective for achieving a variety of illumination effects, which are discussed below. Figure 9.7 gives a ray diagram that shows how the microscope objective creates an image of the snow crystal on the camera sensor, while the field lens images the color filter onto the pupil inside the microscope objective. Understanding this ray diagram, particularly the importance of the pupil plane, is most useful for creating desirable illumination effects that yield especially eye-catching photographs.

A rectangular field stop placed over the field lens blocks all incident light that does not transmit to the field of view of the camera, thus reducing the amount of unwanted scattered light in the optical system. The lens tube above the objective is also baffled and lined on its interior with a highly light-absorbing material to further reduce scattered light. The window cell (see Chapter 6) provides thermal insulation between the room and the cold plate.

Often a pellicle beamsplitter is placed right after the microscope objective to send an additional image to a second camera not shown in Figure 9.4. This is useful when collecting images rapidly from the main camera, as its live view seen on the TV monitor experiences a substantial dead time each time an image is being recorded. During these times, when the crystal growth behavior is being changed quickly, the second camera live view (seen on a second TV monitor) can be used to inform choices of $T_1$, $T_2$, and $F$ that will achieve the desired morphological effects. Several specific examples are described later in this chapter.

## Choosing a Seed Crystal

Finding a well-formed, isolated seed crystal on the substrate is perhaps the most difficult step in using this apparatus. Seed crystals fall randomly during loading, and their surface density on the substrate is adjusted by how long the shutter remains open with the substrate in the loading position. Also, many seed crystals are malformed or do not lie flat on the substrate, and are therefore not suitable candidates for further growth.

If the surface density of loaded seed crystals is too low, it may not be possible to locate a well formed specimen. If the density is too high, then it may not be possible to obtain a sufficiently isolated crystal. Often several tries are needed to find a suitable specimen, with the substrate being heated between attempts to evaporate away the existing crystals. The search process can be laborious and may end up taking anywhere from 1-30 minutes.

A heating trick helps somewhat in finding larger and better formed seed crystals on the substrate. After warming the substrate to clean it, the substrate temperature controller is set to the desired loading temperature $T_1$, and then the loading process is begun before $T_1$ is reached, so the substrate temperature is still warmer than $T_1$. With proper timing, the warmer substrate tends to evaporate away smaller crystals as they land, while larger ones survive long enough for the substrate temperature to reach $T_1$, at which point they remain stable. This process tends to yield a greater fraction of well-formed seed crystals on the substrate.

When growing large, plate-like ice crystals, the ideal seed crystal is a simple, well-formed hexagonal plate with its basal surfaces parallel to the substrate, and with no additional ice within at least several millimeters from the chosen crystal. The subsequent growth phase typically lasts 20-60 minutes and is recorded via the imaging system. The temperatures $T_1$ and $T_2$ are adjusted with time (it requires about a minute for each to stabilize), along with the flow rate, to obtain the desired growth behaviors. At the end of the growth phase, the substrate is heated to just below 0 C so that the ice crystals sublimate away, so the cycle can be started once again.

After a typical day-long run growing crystals, the entire system is warmed to room temperature and baked to remove water. The base plate is typically heated to 40 C via its temperature controller, while the seed crystal chamber is heated via an internal heat lamp, after emptying the water reservoir. Following about a day of baking, the entire system is clean and dry, reducing the presence of residual chemical contaminant vapors.

Because a growing crystal is surrounded by air that has passed through the heat exchanger, special care is taken to reduce chemical contaminants in that air. The charcoal filter in the air stream removes contaminants coming from the air compressor, and the fiber filter downstream from the charcoal filter contributes little odor emission. Moreover, the heat exchanger is baked at 40 C overnight while clean air is passed through it before each run to remove residual contaminants. That fact that thin plates grow readily near -15 C is a good indication that the air flowing into the growth region is quite clean, as chemical contaminants readily inhibit thin-plate growth at this temperature [2011Lib].

## 9.2 Illumination and Image Post-Processing

A variety of illumination techniques can be explored using the relatively simple optical imaging system illustrated in Figure 9.4. The fixed microscope position and orientation are mainly optimized for photographing plate-like stellar snow crystals, achieving a high imaging resolution while capturing crystals as they grow and develop. In addition to changing the crystal morphology, the type of illumination used can greatly affect the overall character of a snow-crystal photograph. Ice is an almost perfectly clear material, so the choice of illumination is substantially more important than one might be accustomed to from photographing opaque subjects. Digital post-processing can also be used for creating a variety of novel artistic effects, and, as with illumination techniques, there is considerable opportunity for enriching the overall look of a PoP snow-crystal photograph.

### Uniform Illumination

The most straightforward illumination method is to apply uniform white light from behind the crystal, replacing the color filter in Figure 9.4 with a simple round aperture. Uniform illumination tends to produce the sharpest microscopic details, and Figure 9.8 gives one example. This photo shows off the bitingly sharp facets and corners that are a special characteristic of most PoP snow crystals.

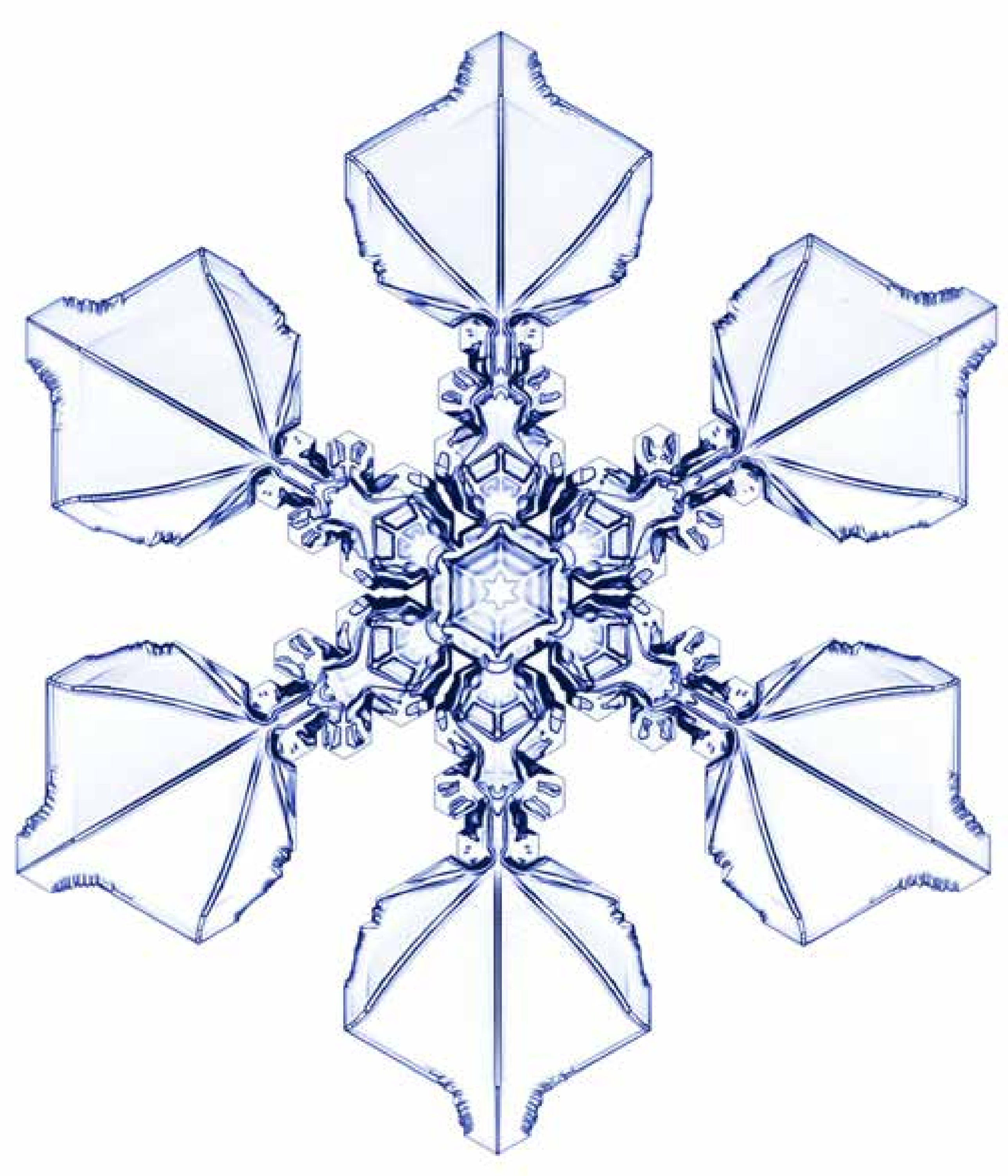

**Figure 9.8: Plain white-light illumination from behind yields a particularly high-resolution image, revealing exceptionally sharp facets and fine details in the surface structures. Five images were combined to make this image, using focus stacking to improve the resolution slightly. Minimal post-processing was applied, mainly just adjusting to background to full bright white and applying a slightly bluish hue to the image.**

In contrast, natural snow crystals usually experience some sublimation after they fall out of the clouds, which gives their features a generally softer, "travel-worn" appearance (see Chapter 11). This is not the case with PoP snow crystals, because they are being recorded as they are growing, when their facets and corners are especially sharp. Thus, photographs of PoP crystals often reveal features that are rarely, if ever, seen in natural snow crystals. You need to be something of a snowflake connoisseur to notice the difference, or perhaps to care, but it does give PoP snow crystals a unique crispness in their finer details.

The specific appearance of the PoP snow crystal in Figure 9.8 can be understood from how light is transmitted through the clear ice. A flat pane of ice reflects some light incident on its surfaces, just as a flat pane of glass reflects some light. But not much is reflected, so the overall appearance of the flat areas of the crystal in Figure 9.8 are quite bright, almost as bright as the background.

In contrast, the edges of a crystal refract transmitted light, diverting it away from its initial path. If the curvature of the edge is high, some of this light is diverted to such large angles that it does not enter the microscope objective, and this gives the edges a darker appearance. More generally, the clear ice acts like a complex lens that refracts some of the transmitted light through a variety of angles. Edges tend to refract light to large angles, so the edges appear darkest in the image. Flat panes of ice refract the light less, so appear brighter. No light at all is reflected or refracted where there is no ice, so the background is the brightest part of this photograph.

I have created a large number of images like the one shown in Figure 9.8, which I call my "blue-on-white" collection. These have some commercial value because they look nice on a white piece of paper and can be used to fill extra space in nearly any document. As the reader may have noticed, I have sprinkled blue-on-white images throughout this book. In general, white-light illumination is good for showing off the detailed structure in a PoP snow crystal, but the resulting images tend to be a bit "flat" in character, as they give little sense of the snowflake's rich three-dimensional structure.

When using a simple round aperture to produce white-light illumination, one soon finds that the size of the aperture affects the character of the resulting photograph. When the aperture is small, the resolution of the image is decreased, yielding generally fuzzier edges. The reason comes from how the field lens images the aperture onto the microscope objective, specifically onto the pupil plane, as shown in Figure 9.7. A close look at this ray diagram reveals that a smaller illumination aperture is essentially equivalent to reducing the aperture of the objective, because now no light enters the outer part of the objective. This reduces the resolution because imaging is generally diffraction-limited in microscopy. A smaller input aperture means a lower resolution in the diffraction limit, and using a small illumination aperture produces the same effect.

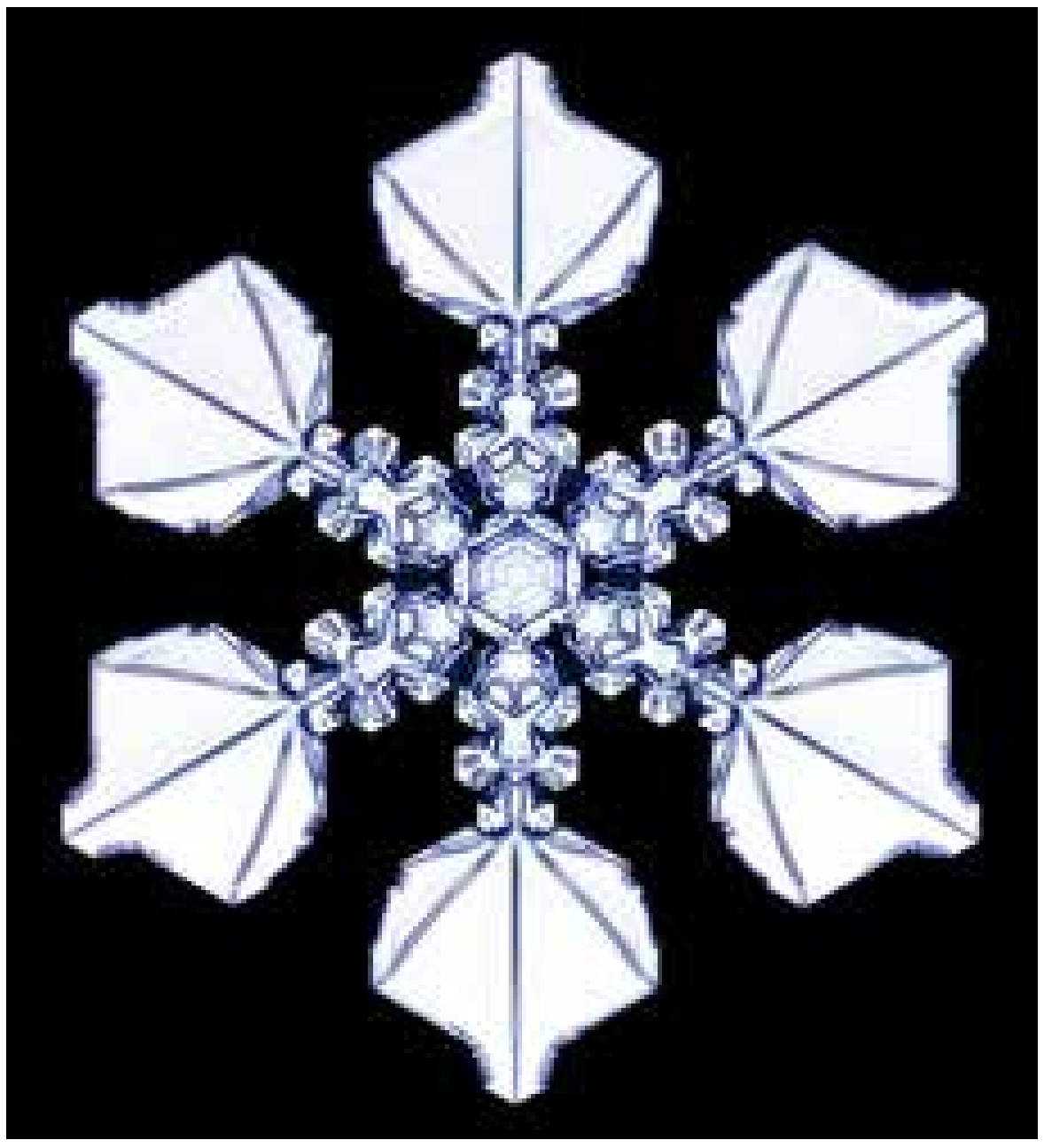

**Figure 9.9: Setting the background black in Figure 9.8 yields this image, looking similar to snow crystal photographs taken by Wilson Bentley (see Chapter 1).**

If the illumination aperture is especially large, then the edges will not be as dark as they would otherwise be, which reduces the contrast of the final image. In the limit that white-light is incident from behind the crystal at all angles, covering a full $2\pi$ steradians, the contrast would drop to nearly zero. In this extreme case, even large-angle refraction from the edges would not reduce the amount of light entering the objective (which can be seen by considering light rays running in reverse, from the image plane to the illumination source).

The optimal illumination aperture size is that where the image of the aperture just fills the entrance of the microscope objective. This gives the maximum resolution, as the full objective is being used. And it produces a high contrast as well, as even small-angle scattering from the crystal edges will reduce the amount of light entering the objective.

As can be seen from this exercise, a good understanding of the principles of optics is most helpful for taking photographs of snow crystals. Commercial cameras and microscopes are not optimized for this purpose (as snow-crystal photography is certainly not their primary market), so some DIY design effort, and a fair bit of trial-and-error, can be quite valuable for obtaining high-quality photographic results.

Figure 9.9 shows this same photo after applying a "Bentley blocker" that digitally sets the background color to black. Wilson Bentley modified nearly all his photographs this way (see Chapter 1), although he did it the hard way by scraping the background emulsion off his glass photographic plates with a razor blade. Digital image processing reduces this task to a few clicks, but I am not a fan of the flat, high-contrast look. The main advantage of the Bentley blocker is that the crystal now appears white, which many people feel is a more natural look for snow, even though individual snow crystals are actually not white (see Chapter 11).

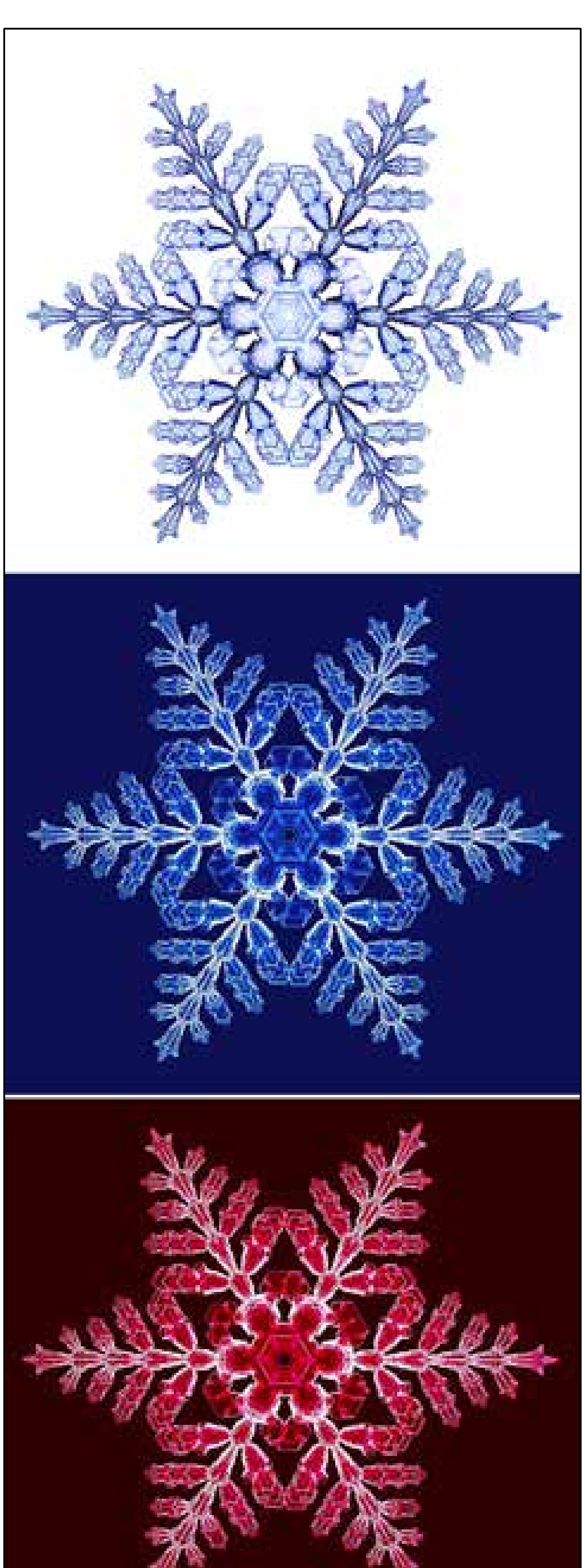

**Figure 9.10: (Right) Beginning with a PoP snow crystal illuminated with white light, these modified images show a blue-on-white version from minimal post-processing (left), an inverted image after color modification to give a blue-on-dark appearance (center), and another inverted image modified to give a red-on-dark look (right).**

Figure 9.10 shows some additional image modifications that can be applied to a simple white-light PoP image to yield colorful effects. The various adaptations do not change the underlying snow crystal structure, but simply present it in different ways. Whenever photographing snow crystals, I like to explore a multitude of lighting and post-processing effects like these. Additional examples are presented at the end of this chapter.

## Dark-Field Illumination

Another approach using white-light is to replace the color filter in Figure 9.4 with a simple annulus that blocks the central light while letting a ring of light illuminate the crystal from an oblique angle. Figure 9.11 gives one example of this use of dark-field illumination in PoP snow-crystal photography. Here the color filter was replaced with an opaque disk on a clear glass holder. As shown in Figure 9.7, the field lens images the disk onto the microscope objective, so no light enters the microscope if no ice is present, giving the image a dark background.

The snow crystal again acts like a clear, complex lens, this time refracting some of the light coming from outside the disk in such a way that it does enter the objective. For the photograph in Figure 9.11, the opaque disk filter was moved around and placed slightly off-center, producing different amounts of refraction on different sides of the crystal.

Comparing Figures 9.9 and 9.11, one can see that off-center dark-field illumination gives the image a pleasing sense of depth, with an overall "glassy" look, as the brightness variations accentuate the three-dimensional structure of the snow crystal. The "flat" image in Figure 9.9 is much less vibrant by comparison, and it gives the viewer no sense of the full crystal structure. Using illumination to create a realistic sense of depth is one of the tricks of snow crystal photography (see Chapter 11), which applies as well to PoP crystals.

**Figure 9.11: (Left) An image of a PoP snow crystal using dark-field illumination. In the absence of any ice, no light enters the microscope objective, so the background is dark. The crystal appears bright because the ice refracts some light from oblique angles toward the objective. In this case, the central light was blocked using a slightly off-center opaque spot in place of the color filter in Figure 9.4, giving an asymmetry in the overall illumination of the crystal.**

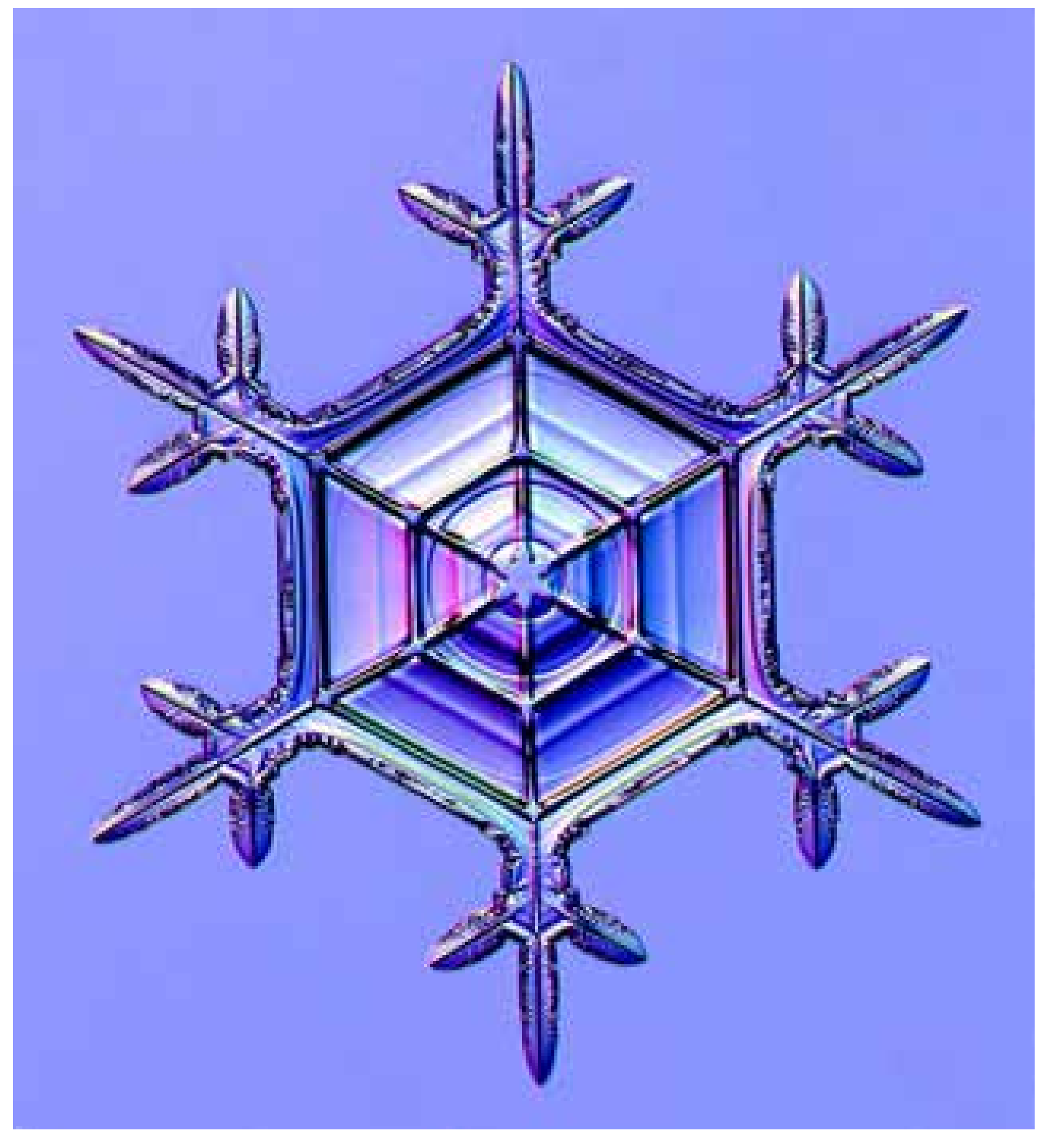

**Figure 9.12: (Left) Rheinberg illumination here accentuates surface structure, maintains high resolution, and adds some color to this PoP snow crystal image. Note that the background hue is quite uniform, while the snow crystal shows some red highlights introduced by the colorful Rheinberg filter.**

## Rheinberg Illumination

The technique of placing a patterned color filter in the pupil plane is a variation of dark-field illumination that was first described by microscopist Julius Rheinberg in 1896, and is now called *Rheinberg illumination*. Figure 9.12 shows one example of a PoP crystal photographed using this method. I especially like Rheinberg illumination because it provides an excellent sense of depth to snow crystal photographs, accentuating the full three-dimensional structure better than other types of illumination. Surface features remain sharp with high contrast, and it adds a new dimension of color to snowflake photography.

Filters presenting vibrant colors and strong patterns often yield good photographic results, and a few example filter designs are shown in Figure 9.13. Filters can be constructed from pieces of gel filters, or by creating designs digitally and simply photographing one's computer screen onto standard 2x2-inch color slides (if one still has access to a film camera, which is becoming less likely). Replacing the white LED light source and color filter with a small computer projector is perhaps a more modern solution when using Rheinberg illumination.

**Figure 9.13: (Below) A sampling of nine color filters I have used for Rheinberg illumination of PoP snow crystals. A filter design using bright colors with abrupt transitions often yields images that exhibit rich shading and vibrant highlights.**

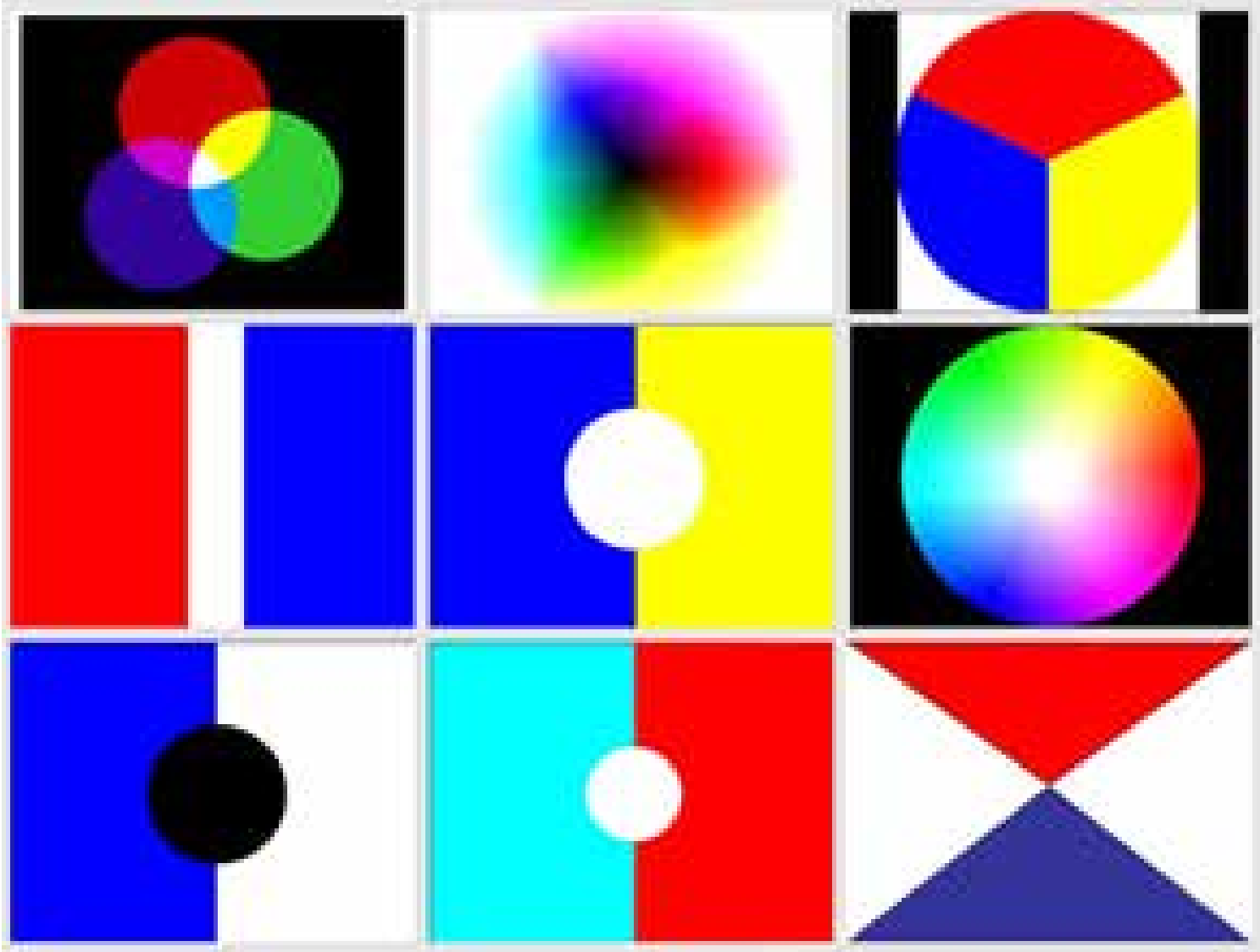

Because the color filter is placed at a pupil in the optical system (see Figure 9.7), the pattern in the filter itself is not seen in the background image. One way to think about this illumination method is that different colors of light are shining onto the crystal from different angles. Each point on the focal plane receives equal amounts of all the colors, so the background image has a uniform average color, as seen in Figure 9.12. With this technique, the background color in the image plane remains uniform regardless of the color filter used. This is typical of optical systems, as patterns in the pupil plane have little effect on what is seen in the image plane.

Note also that the colors seen in snow-crystal images like this do *not* result from any dispersion effects, like what you see from a glass prism. Color dispersion is negligible in snow crystals because the ice is simply too small and thin. It would take quite a large, thick ice prism to produce much color dispersion, and even then it would only be noticeable with careful lighting and large, flat prismatic surfaces. Ordinary glass objects like cups and plates also show negligible color dispersion, for the same reason.

With Rheinberg illumination, the colors all come from the color filter being used. As the light passes through the snow crystal, the ice again acts like a complex lens that refracts the light and changes its direction of travel, and this process is how color variations are produced. For example, if a bit of red light is shining on the image plane from an oblique angle, none of this light will normally enter the microscope objective, so none will make it onto the camera sensor. But if the ice bends some of that red light and sends it into the objective, and thus onto the camera sensor, then some red highlights will appear on the snow-crystal image. And this is how the red highlights in Figure 9.12 were created.

When photographing PoP crystals, I try quite a few color filters on a single crystal, moving each around while observing the live view on the TV monitor, looking for pleasing effects. If the crystal is growing slowly, as is often the case, one has plenty of time to experiment with different lighting effects. Moreover, a PoP crystal is constantly changing as it grows, and each new morphological development provides what is essentially a new subject to record. With natural snowflakes, a crystal falls to earth, and that is what you have to work with. But photographing a PoP snow crystal is something of a continuous process, as each stage of its development presents a new photographic opportunity. After attaining some level of proficiency with the hardware, a single day of crystal growing can yield a bounty of excellent photographs.

## 9.3 PoP Growth Behaviors

Having described the PoP hardware, optics, and photography, I would now like to step back and discuss how one goes about growing a PoP snow crystal. Having worked with this apparatus for some years, I have developed a number of strategies for producing different morphological features under different growth conditions, and these have basically become a set of "recipes" for designing and fabricating different types of PoP snow crystals. As we will see, this hardware is quite versatile in that it can be used to grow a great variety of highly symmetrical stellar-plate snow crystals.

### Simple Hexagonal Plates

Loading and positioning a seed crystal is usually done by setting the substrate temperature to -12.2 C, the heat-exchanger temperature to -12.0 C, and the air flow at about 250 ccm. This establishes a slight supersaturation and a correspondingly modest degree of ice growth on the substrate, allowing some time to search for a suitable seed crystal. Upon opening the shutter (see Figure 9.3), the substrate is moved into its loading position, where it stays for just a few seconds before being pulled back to the growth region. Then the substrate is moved around in a 2D raster pattern to search for a well-formed seed crystal, one that exhibits a clean hexagonal-prism morphology and has no nearby neighboring crystals. If none can be found, the substrate is heated to -6 C for several minutes to drive off the seed crystals, and then back to -12.2 C for another attempt.

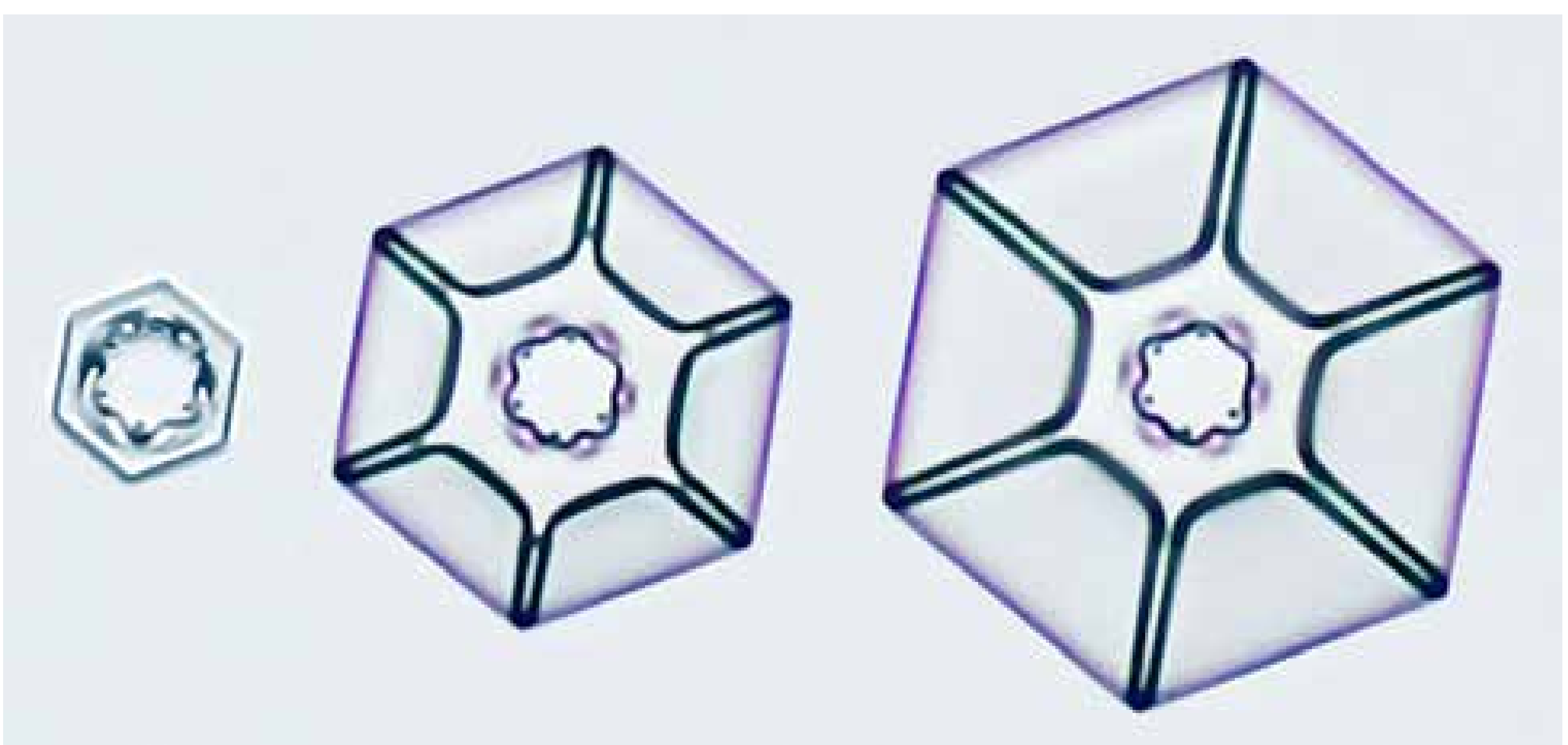

**Figure 9.14: Starting with an initial seed crystal (left), a Plate-on-Pedestal geometry develops. In this example, the initial hexagon grew out slightly before the upper plate formed. The odd central patterning reflects some surface structure on the bottom basal surface that became trapped by contact with the substrate. In contrast, the top basal surface soon grew into a nearly prefect faceted form, filling in any surface structure that had been present on the seed crystal. The seed crystal shown measures 90 μm from facet-to-facet, while the final thin plate measures 300 μm.**

Once a suitable seed is in position at the center of the microscope field, the substrate temperature is lowered slightly to -12.5 C to commence the initial growth of a hexagonal PoP geometry. At lower growth temperatures, the plate becomes unstable to branching, quickly yielding a small stellar crystal, which may not be desired. At higher temperatures, however, the PoP geometry might not appear, if a thin plate cannot emerge from the upper edge of the initial hexagonal prism. Figure 9.14 shows an example of this process, which takes just a few minutes.

If a large plate-like crystal is desired, it is best at this point to move to a somewhat higher substrate temperature, perhaps as high as -10 C, while keeping the supersaturation fairly low. A thin upper plate, similar to that shown in Figure 9.14, may not readily grow out from a small seed crystal at such high temperatures, but it will likely continue growing once it has formed. This type of hysteresis seems to be a feature of the Edge-Sharpening Instability (ESI, see Chapter 4). As seen in Chapter 8, exceedingly thin plates will grow out directly from columnar forms near -15 C, as this where the ESI is most effective. Closely spaced double plates are most likely to grow near this temperature also, for the same reason. Because the initial seed crystal in Figure 9.14 is quite thin, it is necessary to be near -15 C for the PoP geometry to appear, and -12.5 C has been found to work well for growing plates up to around 0.5 mm in size.

It is quite a challenge, however, to grow a large hexagonal plate using the PoP method when the temperature is less than a degree or two away from -15 C. The attachment coefficient on the plate edge is so close to unity that the plate becomes unstable to branching, and reducing the supersaturation sufficiently to prevent this from happening is problematic. After some amount of trial-and-error, I found that beginning near -12.5 C and slowly transitioning to -10 C is a fairly good recipe for growing a large, simple hexagonal plate. Doing so requires quite a lot of patience, however, so the largest simple PoP plates I have made this way have measured only as large as 1.5 mm from facet-to-facet.

## Fog Droplets

If the supersaturation at the substrate ever exceeds $\sigma_{water}$, then water droplets will readily condense on its surface, as illustrated in Figure 9.15. As these small droplets continue growing, they will typically coalesce into larger droplets over time. The first image in Figure 9.16 shows a small PoP crystal surrounded by an array of individual droplets that recently nucleated, while the second image shows the same crystal after much additional growth. As the crystal grows outward, it "pushes" the $\sigma \approx \sigma_{water}$ perimeter out in front of it, as nearby droplets evaporate to provide water vapor for the growing crystal.

The supersaturation near the droplets is essentially clamped at $\sigma_{water}$, making it nearly impossible to achieve supersaturations substantially above this level using the PoP method. As a result, one cannot grow true fernlike stellar dendrites using this apparatus, as these crystals require a higher supersaturation level (see Chapter 8).

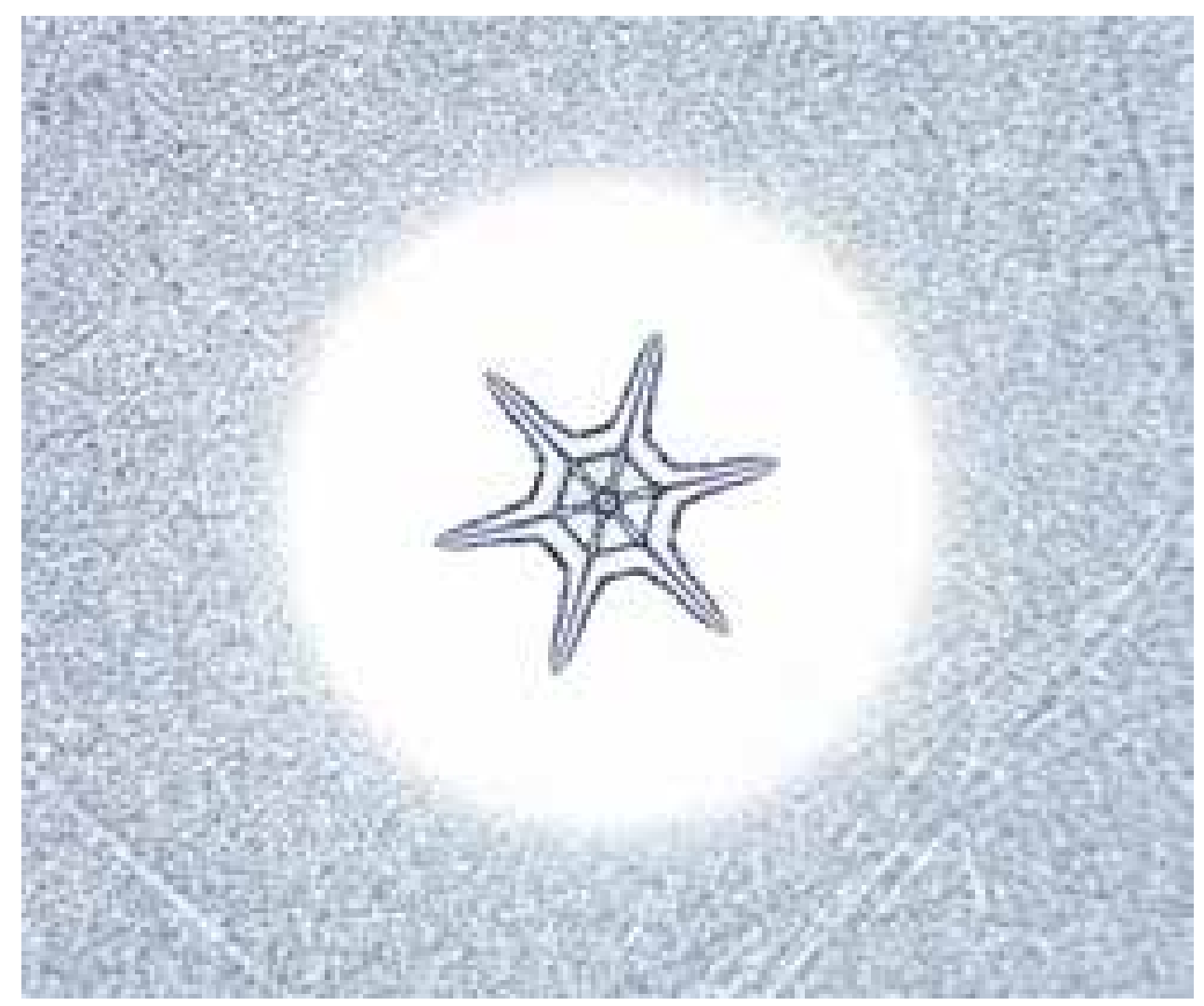

**Figure 9.15: (Above) When the supersaturation exceeds $\sigma_{water}$ near the substrate, water droplets will condense on it, here appearing as a fog around a small PoP crystal. The growing crystal absorbs water vapor in its vicinity, keeping the supersaturation below $\sigma_{water}$ nearby. The circular transition region indicates where $\sigma \approx \sigma_{water}$ at the substrate.**

**Figure 9.16: (Below) At high magnification (left), one can see individual droplets that recently nucleated on the surface in this photo. At lower magnification (right), the droplets take on the appearance of a continuous fog. In this case, the $\sigma \approx \sigma_{water}$ contour is not circular, but follows the overall shape of the PoP crystal.**

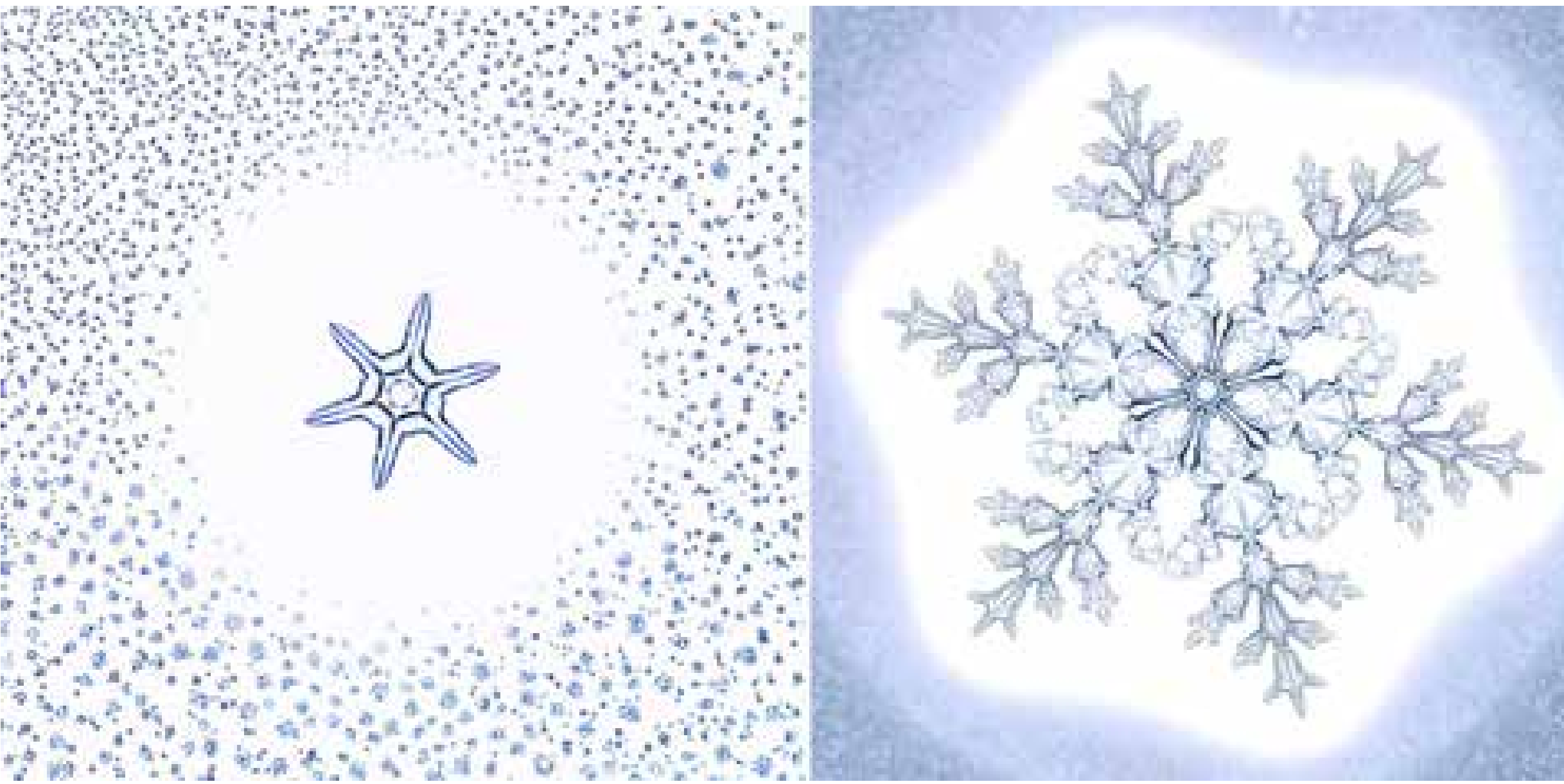

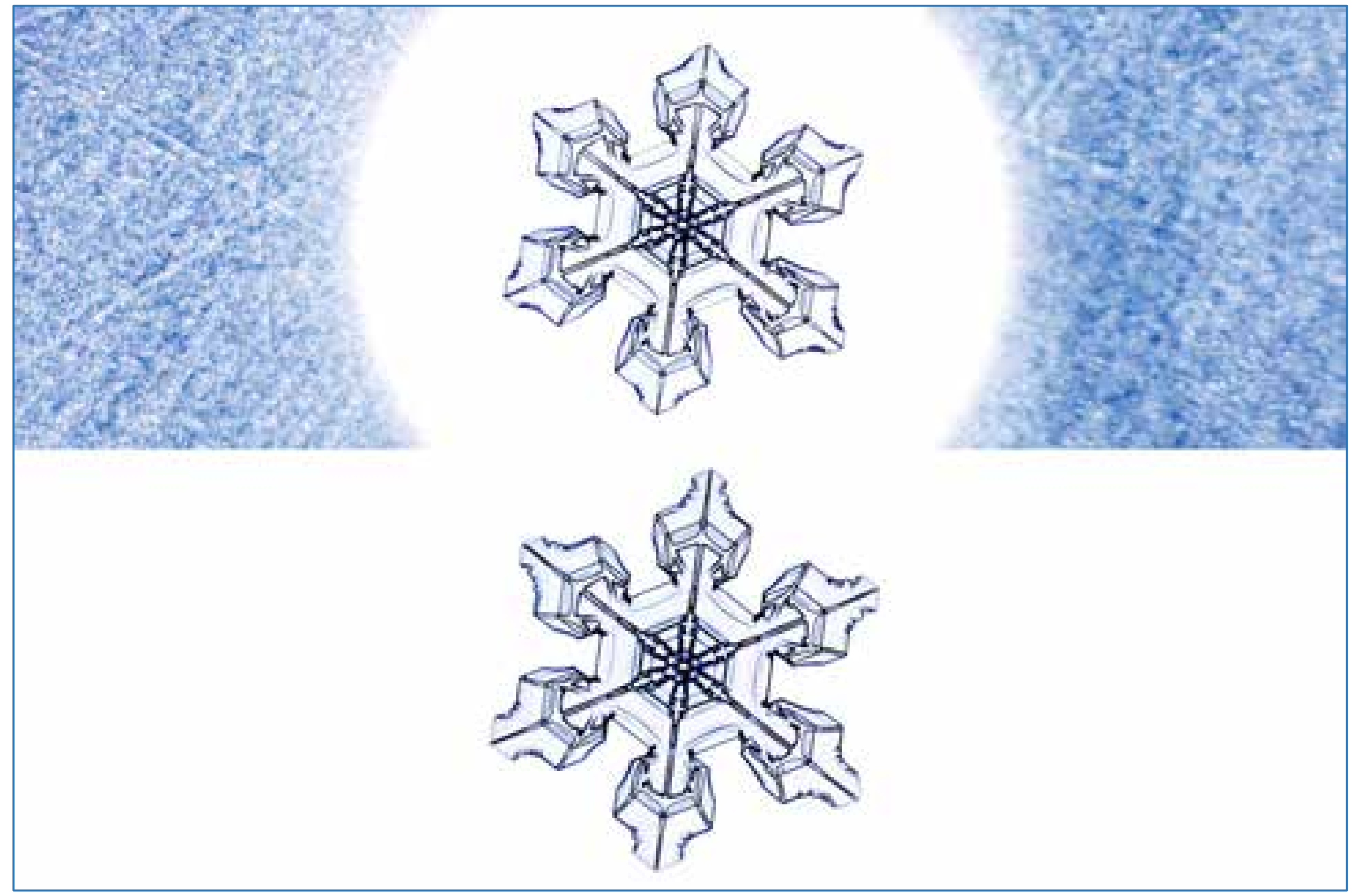

**Figure 9.17: After growing the PoP crystal in the upper image, surrounded by a fog of water droplets, the supersaturation was reduced to a value $0 < \sigma < \sigma_{water}$. Under these conditions, the water droplets soon evaporated away, while the crystal continued growing slowly, yielding the lower crystal surrounded by a clean substrate, free of droplets.**

It is possible to remove the condensed droplets around a crystal by reducing the supersaturation below $\sigma_{water}$, and Figure 9.17 shows an example of this process. For a relatively large crystal like this one, it may take 5-10 minutes before all the droplets are completely gone, depending on the supersaturation level and droplet sizes. The water evaporation is largely diffusion limited and therefore quite slow for a large field of droplets.

Because the substrate temperature is invariably below 0 C, eventually the liquid-water droplets will freeze. They can remain in a metastable unfrozen state for quite some time, however, often more than an hour even at temperatures down to -20 C. In some cases, a rapidly growing ice branch will approach a droplet field faster than the droplets evaporate away, causing the ice to grow into the supercooled water. The first droplet touched by the ice freezes instantly, and the solidification front slowly migrates outward as the frozen droplets each grow outward toward isolated liquid droplets in their immediate vicinity. Within several minutes, typically, the entire droplet field becomes interconnected and frozen.

Once this happens, the supersaturation is then clamped near zero by the ice field, thereby greatly reducing the supersaturation over the entire substrate. The PoP growth is thus greatly slowed when the droplet field freezes, as now the PoP crystal must compete for water vapor with all the surrounding ice. Increasing $\Delta T$ and the air flow rate will cause the whole ice field to continue growing slowly, along with the PoP crystal. But the supersaturation is generally too low to allow significant branching, and the PoP symmetry will degrade with additional growth also. In some of my PoP photographs, Figure 9.8 being one example, I have digitally removed the droplet field surround the crystal, simply to de-clutter the image background.

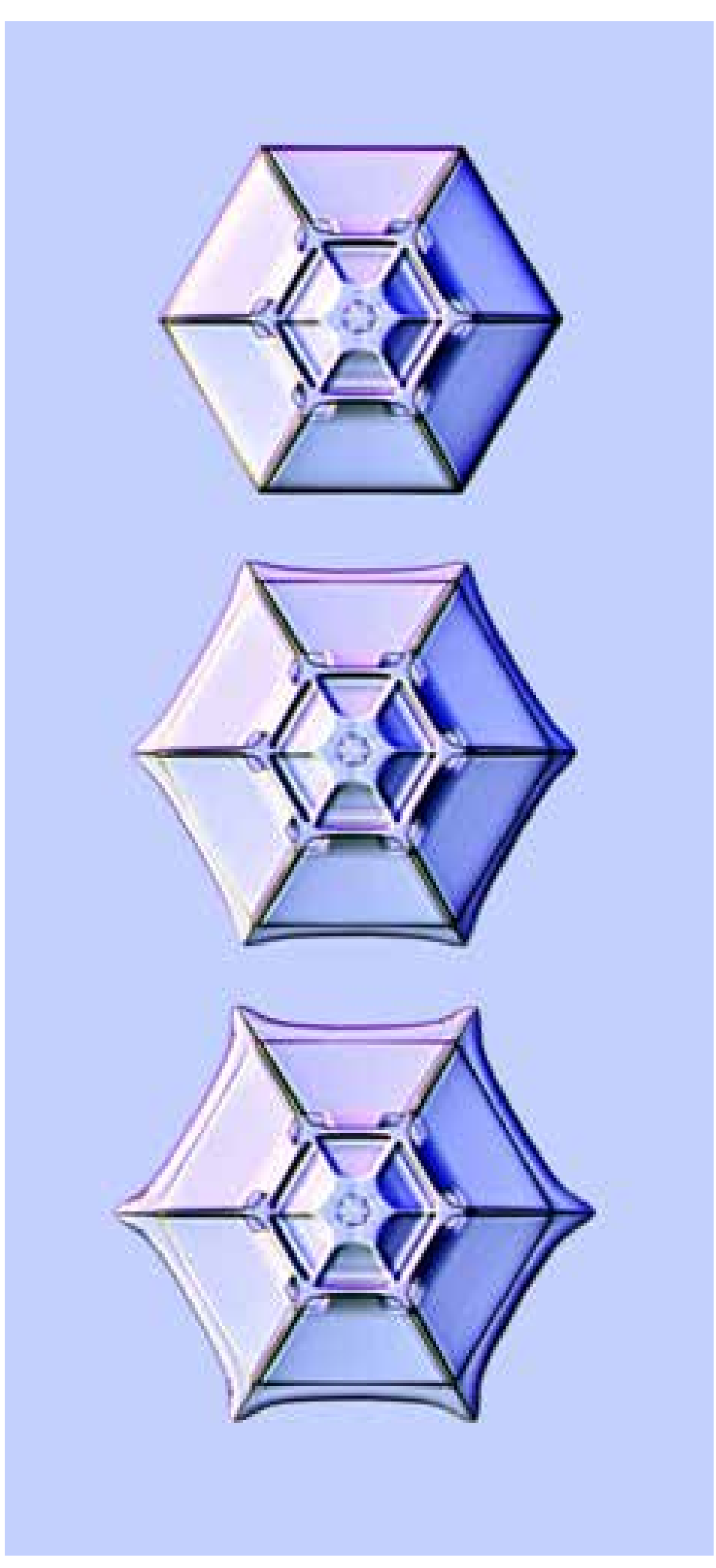

**Figure 9.18: After growing a simple hexagonal PoP crystal at -12.5 C (top), the substrate temperature was reduced to -15 C to stimulate branching. At first the outer edge of the plate became thinner (middle), and then branches sprouted from the six corners of the hexagon (bottom). Note that the plate is supported above the substrate by the central ice pedestal, and the observed surface patterning is all contained on the underside of the plate. The top basal facet, in contrast, is essentially perfectly flat over the entire crystal.**

## Branches, Wrinkles, & Spikes

Figure 9.18 illustrates the branching instability being initiated on the corners of a thin, faceted hexagonal plate. I use the word "initiated" because the branching process was brought about in this case by lowering the temperature of this crystal. Left at -12.5 C, the hexagonal plate would have grown much larger, and the transition to branching would have been weaker, resulting in broad, faceted protuberances. Reducing the substrate temperature to -15 C both changed the growth temperature of the crystal and also increased $\Delta T$, thus simultaneously increasing the supersaturation, and this initiated the sharp spikes seen in Figure 9.18.

The branching transition near -15 C is one of the most mysterious aspects of snow crystal formation, as the overall growth behavior is highly sensitive to temperature in this region. The attachment coefficient on a plate edge is quite high right at -15 C, yielding extremely thin plate edges and prism facets that are quite unstable to breaking up to form branches, as shown in Figure 9.18. Just a few degrees warmer or cooler, however, and faceted prism growth is much more stable, allowing the formation of quite large faceted prism edges on plates. None of these growth behaviors is difficult to comprehend at a specific temperature, but why there is such a strong temperature dependence so far from the melting point remains a substantial puzzle.

Figure 9.19 shows another example of the branching transition on a simple hexagonal plate. Once again, the abrupt sprouting of narrow branches was not a spontaneous occurrence, but was stimulated by bringing the substrate temperature closer to -15 C. And again the plate edge first became thinner, and then the prism facets became unstable to the formation of spike-like branches.

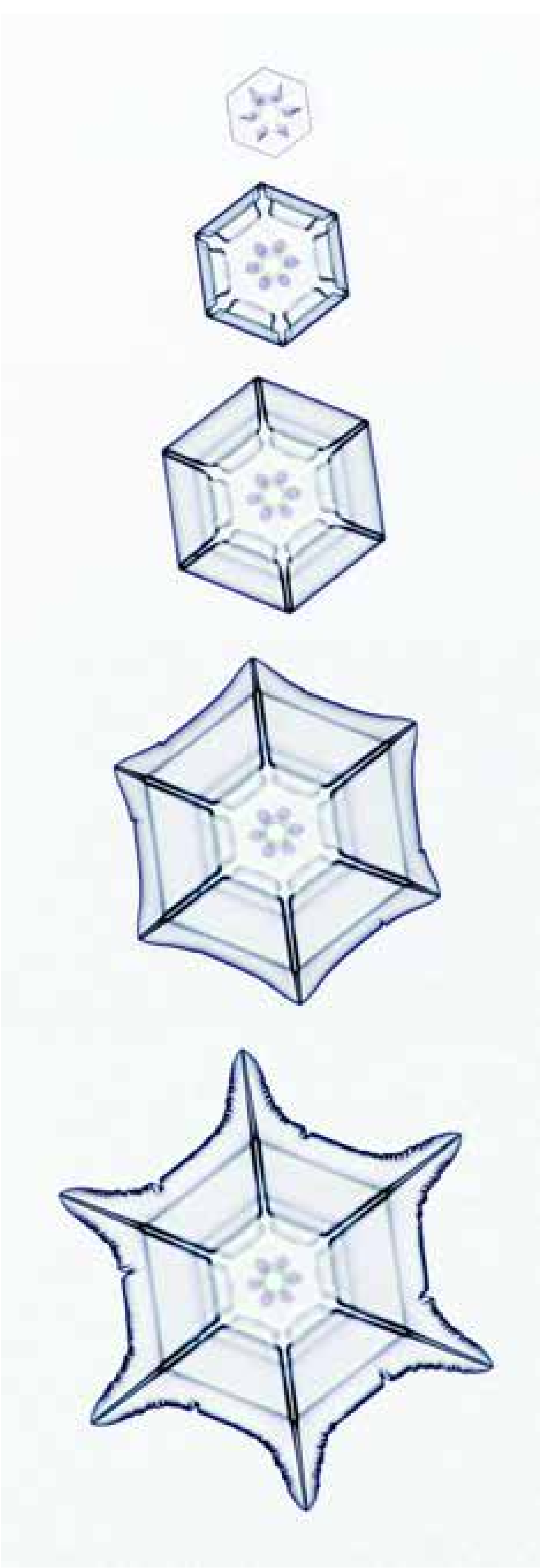

**Figure 9.19: (Left) Another example of the onset of branching on a simple hexagonal plate, brought about by lowering the temperature to near -15 C and increasing the supersaturation. First the edge of the plate first grew thinner, then the prism facet started to become curved, and finally the branching instability kicked in. The small "petals" at the center of the crystal are bubbles trapped at the base of the pedestal, and again all of the surface structure is located on the underside of the plate, while the upper basal surface is almost perfectly flat.**

As the branches develop in Figure 9.19, they leave behind a set of concave plate edges that experience a new kind of "wrinkling" instability, yielding the serrated edges seen in Figures 9.19 and 9.20. Near the branch tip, $\alpha$ is near unity and the growth is limited mainly by diffusion, resulting in a roughly parabolic tip shape, related to the Ivantsov solution to the diffusion equation (see Chapter 3). Farther from the tip, the supersaturation is lower, so prism faceting becomes important. In this region, the faceting process turns the smooth concave edges into a series of faceted segments, yielding the overall serrated contours observed. Similar features can be seen in natural snow crystals, albeit not as clearly.

**Figure 9.20: (Below) When the formation of branches results in a concave plate edge, the initially smooth edge is unstable to localized faceting that produces a serrated edge and a series of wrinkled surface features.**

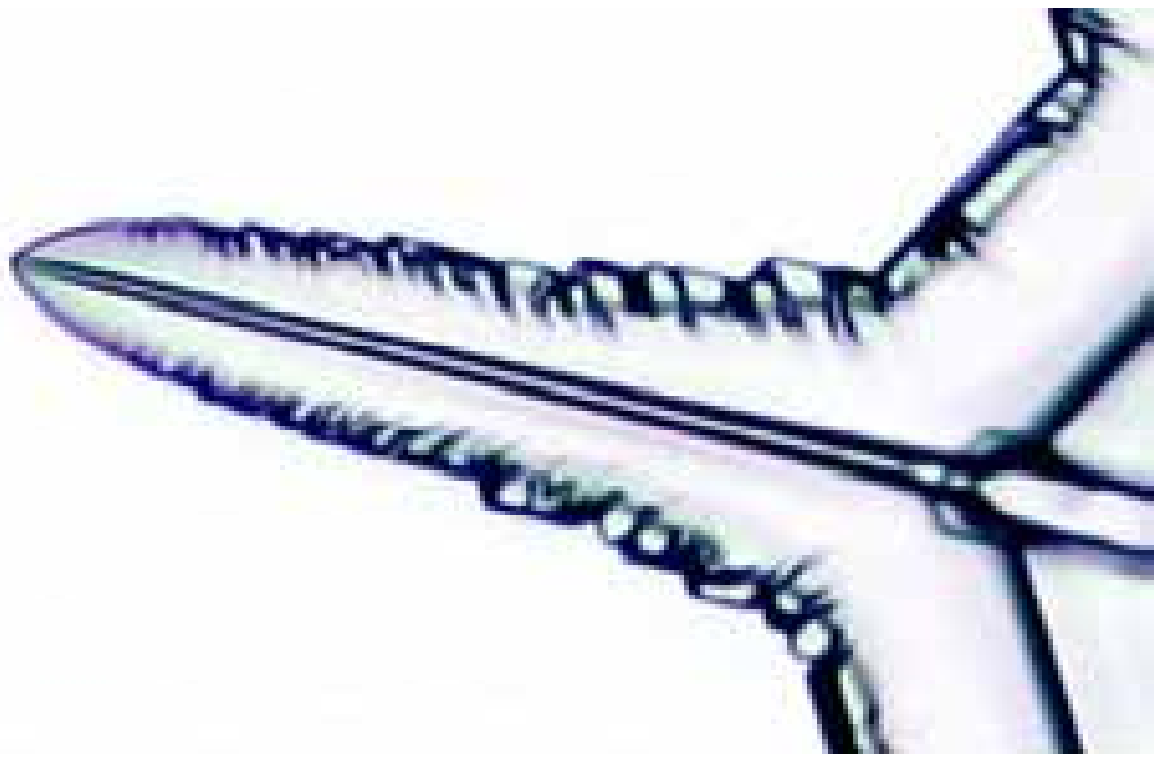

Figure 9.21 shows another example of the branching instability taking place, this time with a crystal growth temperature near -13 C and a relatively modest supersaturation. Under these conditions, the transition is weaker, yielding broad branches that form slowly and exhibit faceted tips. As a general rule, the branching instability is stronger, yielding narrower, faster growing branches when the temperature is closer to -15 C and the supersaturation is higher.

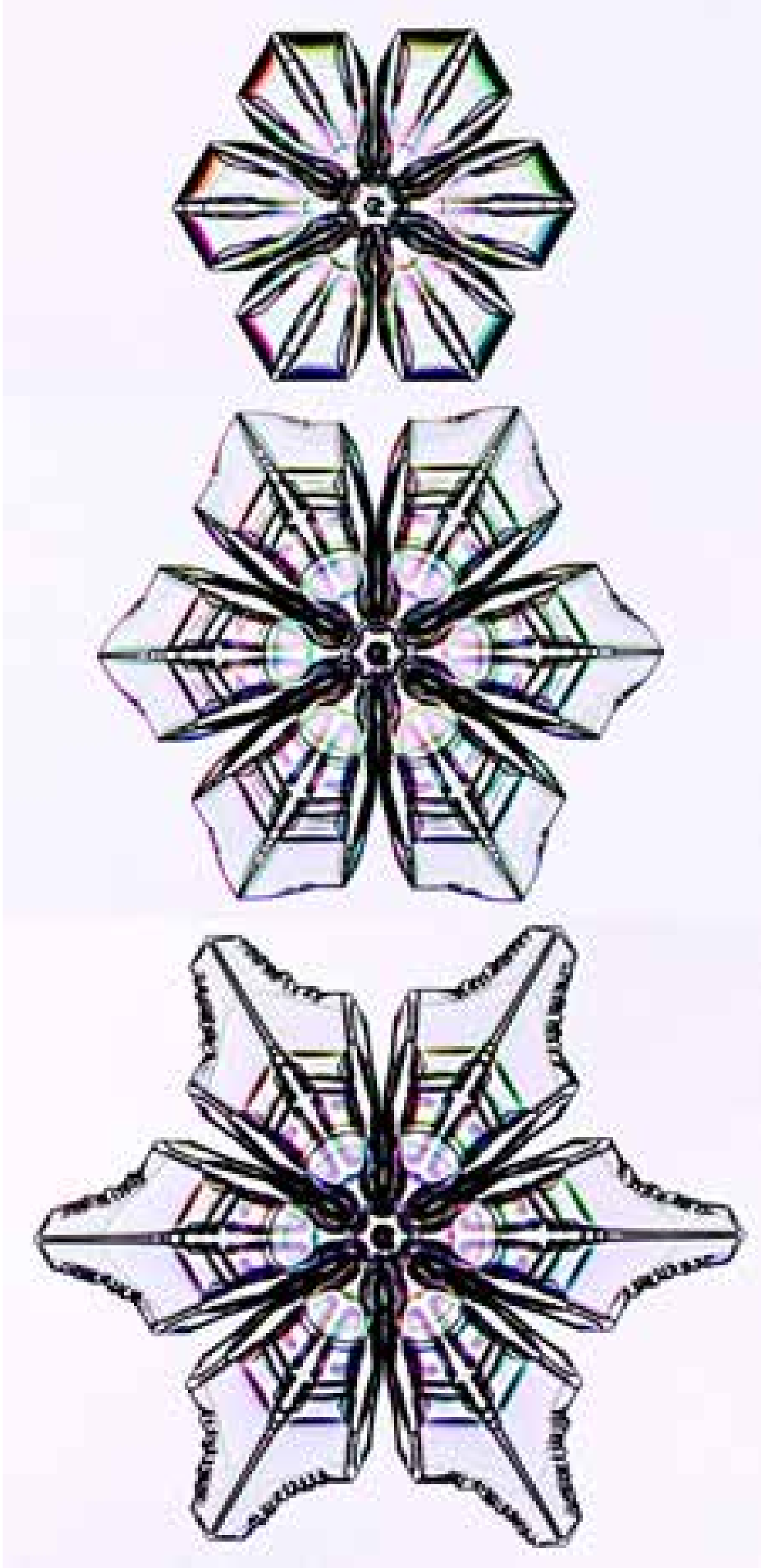

**Figure 9.21: This PoP crystal experienced a period of weak branching, resulting in broad outgrowths with faceted tips, in contrast to the narrower branches with rounded tips seen in Figure 9.20.**

## Induced Sidebranching

While it is straightforward to form six primary branches on a PoP snow crystal, the supersaturation is always too low to observe spontaneous sidebranching. As described above, the condensation of water droplets on the substrate prevents the supersaturation from attaining values much above $\sigma_{water}$, and this is too low for the type of dendritic sidebranching seen in fernlike stellar dendrites (see Chapter 3). Even near -15 C, when the branching instability is strongest, the best one can do under constant growth conditions is create a crystal with spike-like branches, as in the example shown in Figure 9.22.

Nevertheless, it is possible to create sidebranches by manipulating the growth conditions as a function of time, using a process of *induced sidebranching* that is illustrated in Figure 9.23. The basic idea is to first form the primary branches, then lower the supersaturation so the branch tips become faceted, and then increase the supersaturation abruptly to a high value, stimulating the formation of branches at all three of the exposed prism corners. Thus the sidebranch formation is "induced" by creating a faceted tip geometry and quickly exposing it to a high supersaturation.

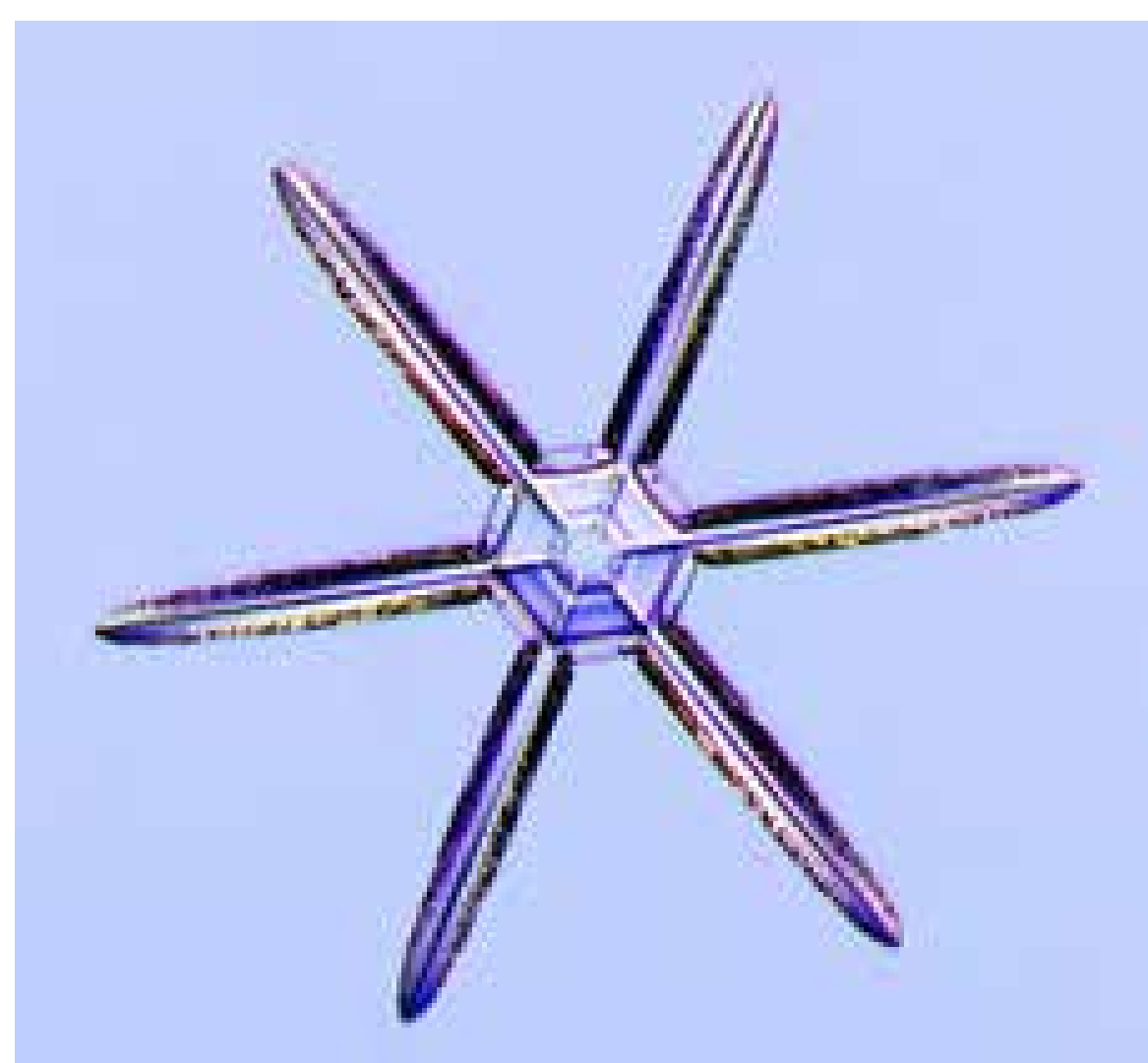

**Figure 9.22: Narrow PoP branches can grow quite stably near -15 C, yielding a spiked star like this one. Dendritic structures will not form under constant growth conditions, however, because droplets condensing around the crystal will prevent the supersaturation from becoming high enough to drive spontaneous sidebranching.**

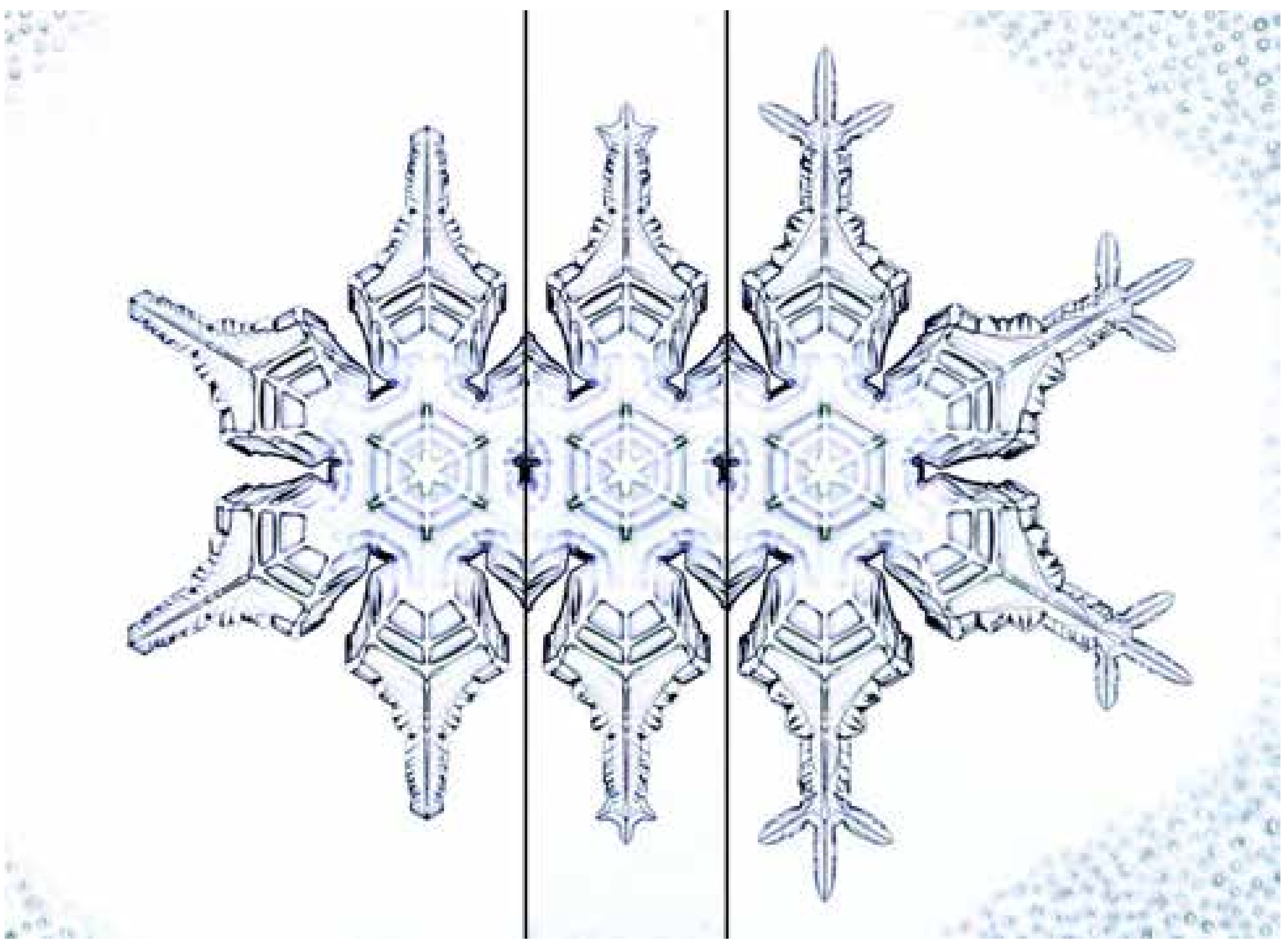

**Figure 9.23: The process of *induced sidebranching* is illustrated in these three photos of a growing PoP snow crystal. After growing a set of six primary branches, the supersaturation was lowered so the branch tips became faceted (left photo). Once the three outermost facet corners were well developed, the supersaturation was abruptly increased, causing branches to sprout from these corners (center). Keeping the supersaturation high, the primary and side branches all continued to grow outward at about the same rate (right).**

When spontaneous sidebranching happens in a fernlike stellar dendrite, the primary branch tips are always roughly parabolic in shape. The smooth, rounded tip profile makes it more difficult to initiate sidebranch formation, so doing so requires a high supersaturation. With induced sidebranching, one first prepares the tip by giving it a faceted profile. The sharp, faceted corners are then more susceptible to the branching instability, so a lower supersaturation level is needed to make it happen.

Another feature of spontaneous sidebranching is its overall random character. The primary branch tip always has a roughly parabolic shape, and sidebranches arise at essentially random times. The process is so haphazard that often even the opposing sides of a single primary branch exhibit sidebranches at different positions, reflecting their generally uncoordinated formation. Looking at photographs of a fernlike stellar dendrites (see Chapter 10), one sees an overall asymmetrical placement of sidebranches that reflects this random process.

With induced sidebranching, on the other hand, the process is highly coordinated by the timing of the events that caused it. Sidebranches form on all six primary branches simultaneously, and on both sides of each primary. This results in an overall crystal structure that is both complex and symmetrical, a feature that is something of a defining characteristic of snow crystals. For both PoP and natural snow crystals, we see that the large-scale, complex symmetry does not result from any preordained crystal design or communication between the primary branches. Rather it arises simply from the time-dependent environmental conditions being applied to the growing crystal.

Of course, the process of induced sidebranching can be applied repeatedly, at varying temperatures and supersaturations, and with varying wait times between events. Figure 9.24 shows one example of how a series of growth transitions can be used to fabricate a complex, yet symmetrical, PoP snow crystal.

**Figure 9.24: This PoP snow crystal underwent a series of induced sidebranching events, each producing a set of symmetrical sidebranches flanking each of the primary branches. The final crystal displays a decidedly complex overall structure with a degree of six-fold symmetry that is rarely seen in natural snow crystals. Just before this photograph was taken, the supersaturation was increased to condense a fog of water droplets quite close to the outer perimeter of the crystal.**

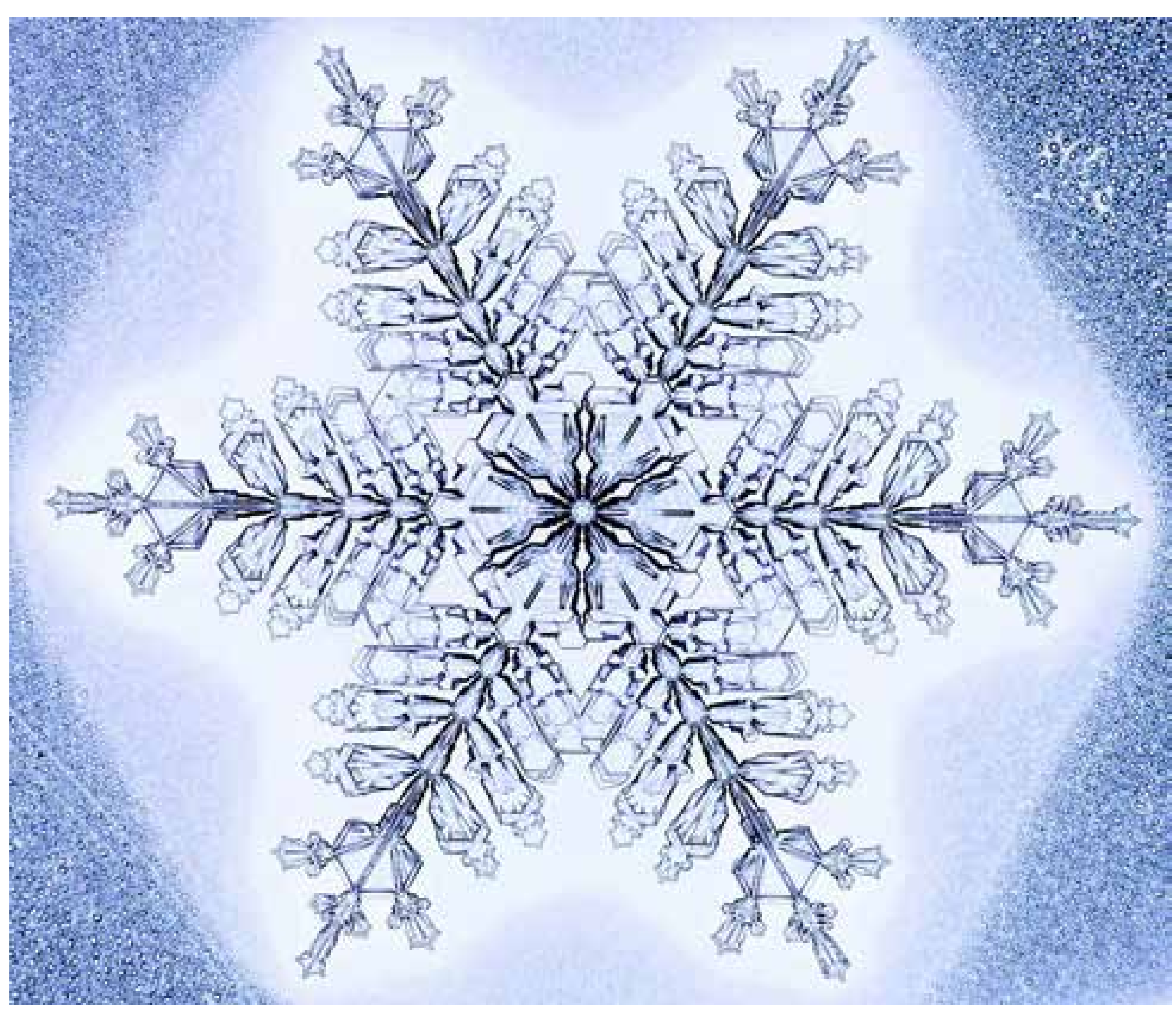

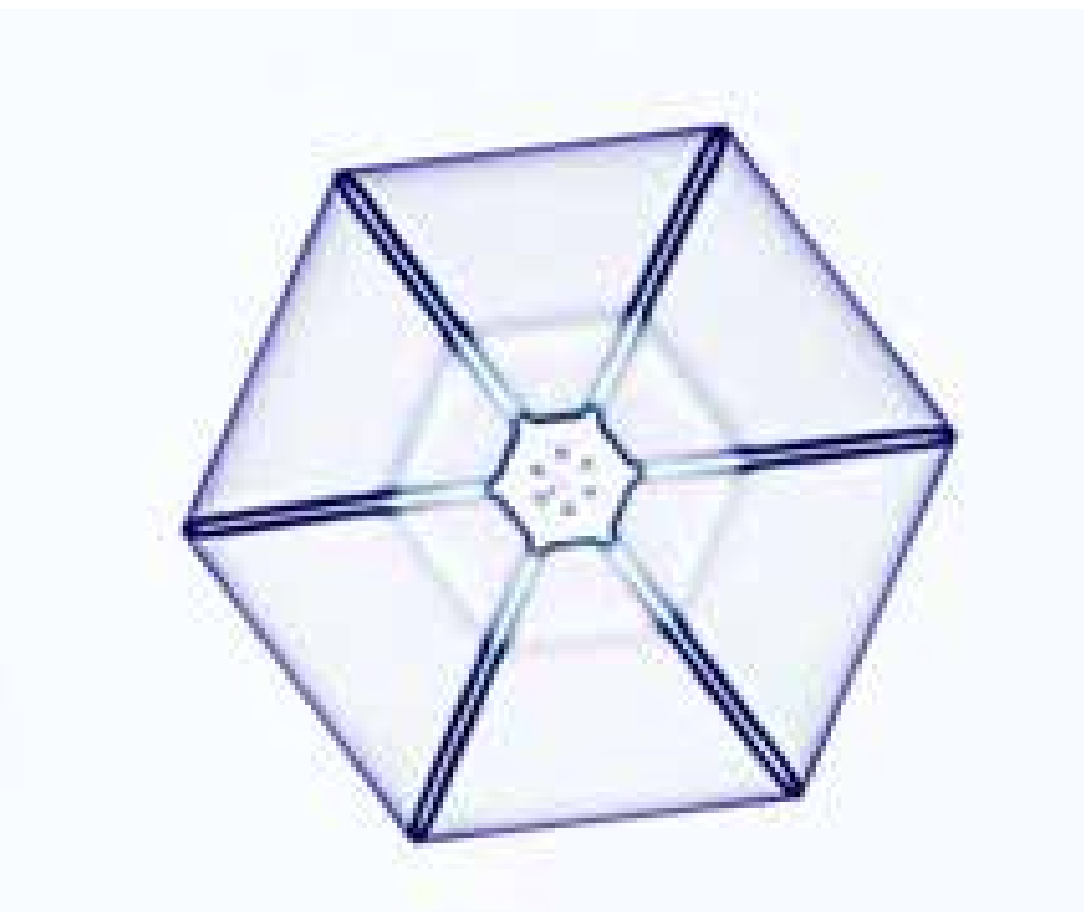

**Figure 9.25: Small PoP snow crystals often exhibit a set of six radial *ridges* that connect to the corners of the hexagonal plate. These ridges form on the underside of the supported plate, while the upper basal surface is essentially perfectly faceted. This simple sectored-plate snow crystal measures about 400 μm from facet to facet.**

## Ridge Growth

The initial PoP geometry is nearly always accompanied by the formation of a set of six *ridges*, as illustrated by the small hexagonal plate shown in Figure 9.25. Natural snow crystals exhibit many ridge-like structures also, but the phenomenon is especially clear in PoP snow crystals, with Figure 9.25 providing an especially simple example of a sectored plate snow crystal. The upper surface of this specimen is essentially a perfectly flat basal facet, with no surface structure of its own. The ridges are confined to the lower side of the plate, which is supported above the substrate by the central ice pedestal.

Figure 9.26 sketches the formation of a single ridge, which arises from a variant of the usual branching instability (see Chapter 3). Looking first at the top basal surface, it grows slowly because $\alpha_{basal}$ is low, even though this surface is exposed to a relatively high supersaturation. Under these conditions, the upper basal surface is essentially perfectly faceted and grows slowly upward. The lower surface of the plate, on the other hand, is so close to the substrate that it sees $\sigma \approx 0$. Away from its outer edges, therefore, the underside of the plate experiences almost no growth.

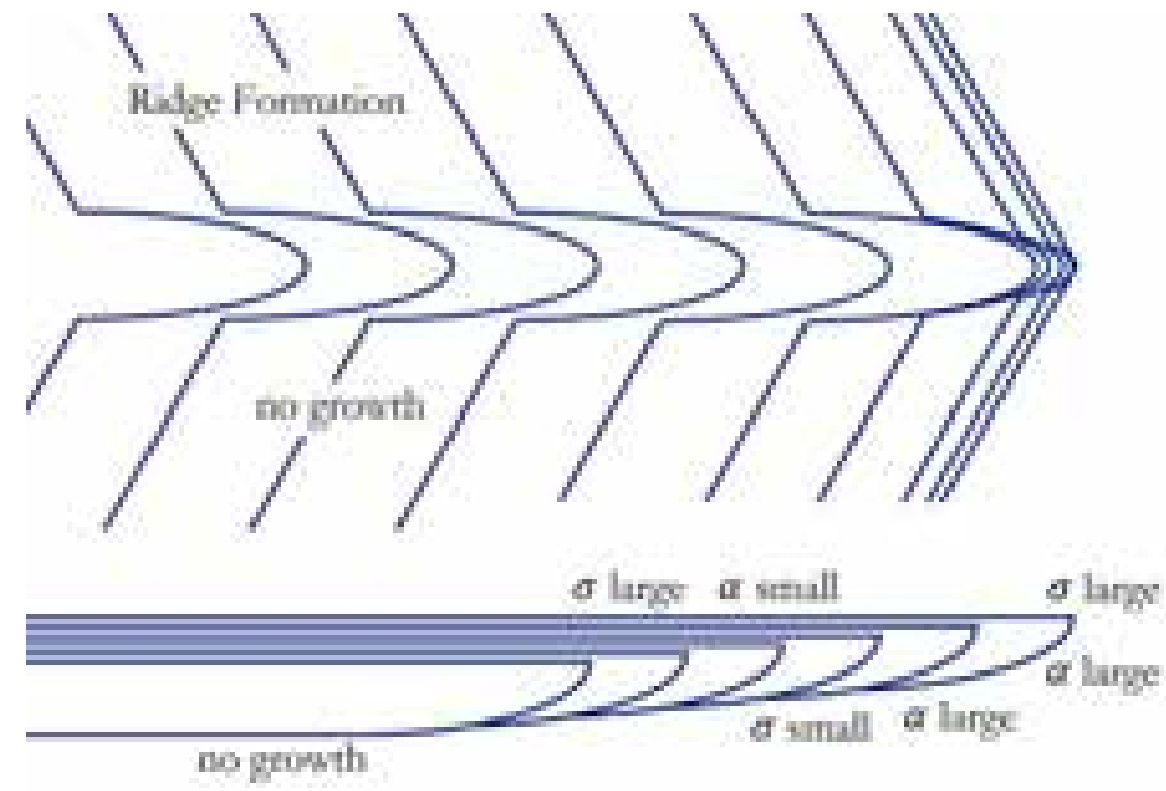

**Figure 9.26: Diagrams illustrating the formation of a snow-crystal ridge. In the top sketch, the lines represent molecular terrace edges on the underside of the plate. Each terrace has an overall hexagonal shape, but the corners see a slightly higher supersaturation and are thus unstable to the usual branching instability. The lower sketch shows a side view of the ridge structure. As ice is added to the top surface and the plate edge, the ridge is left behind. Away from the edge, neither the ridge or the surrounding surface experiences any additional growth.**

In terms of the overall plate structure, therefore, we have the upper basal surface growing slowly, the plate edge growing out quickly, and the underside of the plate hardly growing at all. Over time, this results in a shallow conical overall shape, with the top surface flat and the lower surface slightly convex in shape. Referring to Figure 9.26, the contour lines in the top sketch show the convex shape of the lower surface, and each line can be thought of as the edge of a single molecular terrace.

Far from the outer edge of the crystal, these terraces are essentially static, because they see $\sigma \approx 0$ in their vicinity. Nearer the edge, the supersaturation is a bit higher, especially near the corners. In this case each terrace experiences a variant of the usual branching instability, causing the corners to grow out faster than the terrace edges farther from the corners. This results in the spiked contours shown in Figure 9.26, which have the same overall shape as the spiky branches in Figure 9.19. The full prism edge of the crystal is still stable against branching, so it keeps its faceted shape. But the molecular terraces sprout one-layer-thick primary branches as illustrated. Looking at these edges as contour lines, the overall shape becomes that of a ridge in the ice.

The lower sketch in Figure 9.26 shows why the ridge has a nearly constant width over the entire crystal. Soon after the branching instability creates the ridge near the crystal edge, the edge moves outward, because the plate edge is growing rapidly. Once the ridge is left far behind on the underside of the plate, away from the edge, it sees $\sigma \approx 0$ and develops no further. Thus the ridge structure originates when it is near the growing corner and remains static thereafter.

I believe this growth process explains the simple ridge structure seen in Figure 9.25 quite accurately and naturally, requiring only a slight variation on the usual branching instability. Moreover, it makes a prediction that ridges like this can only form on slightly convex basal surfaces. This a simple model that appears to fit all the available observations, and computational models support it somewhat as well (see Chapter 5). Comparing controlled PoP or e-needle observations with corresponding computational models would provide direct verification, and may provide additional insights into this commonly observed ridge phenomenon.

On simple hexagonal plates like the one shown in Figure 9.25, the ridges grow radially outward in a hexagonal pattern, and this shape represents the simplest example of a sectored-plate snow crystal (see Chapter 10). Ridges also readily form on thin plates at the ends of branches, often called sectored-plate extensions, and Figure 9.27 shows a particularly minimal PoP example.

Unlike with a simple hexagonal plate, here the plate-like extensions become crowded together as they grow, thus making them compete for the available water vapor in their vicinity. The crowding distorts the development of each plate, and this often gives rise to the formation of curved ridges. Because each ridge originates from one corner of the plate, the ridge lines trace out where the plate corners were at earlier times.

## Induced Rib Structures

Ribs are another common surface patterning feature seen in natural snow crystals, and again we can create especially clear examples in PoP crystals. Figure 9.28 shows a small hexagonal plate that was created as an example of rib formation, exhibiting a series of concentric hexagonal ribs looking something like a spider's web.

The essential recipe for creating a rib structure on the underside of a PoP plate is illustrated in Figure 9.29. The first step is to create the PoP geometry as described previously, initiating the Edge-Sharpening Instability by applying a sufficiently high supersaturation at a temperature near -15 C. This yields a relatively fast-growing plate with a thin edge, as shown in the first sketch in Figure 9.29.

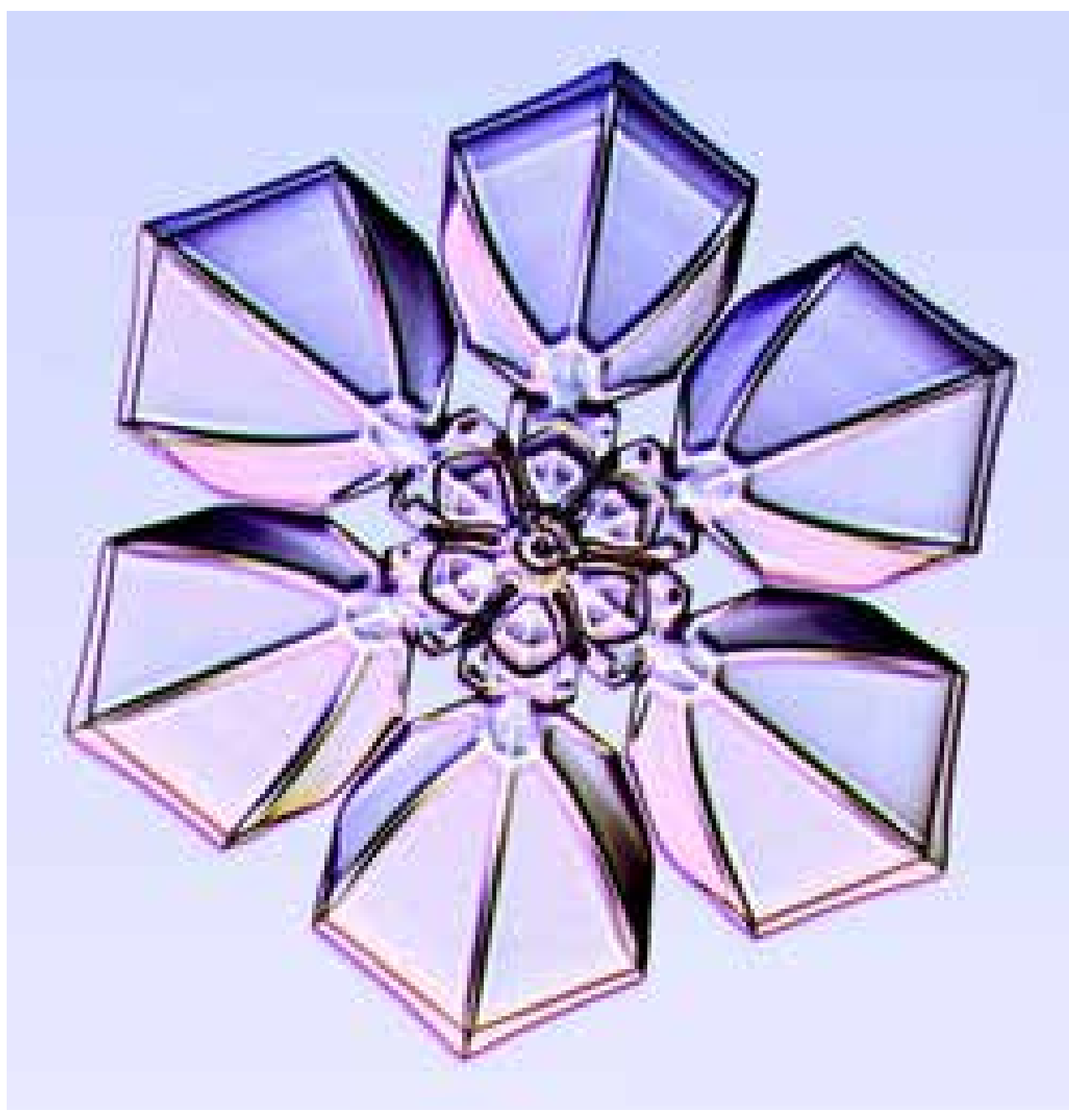

**Figure 9.27: (Left) Sectored plate extensions. Curved ridges often form on broad-branched sectored plates, looking a bit like "duck's feet" in this PoP example. Because ridge formation originates at the faceted corners of a plate, the ridges trace out the location of the corners as a function of time as the crystal developed.**

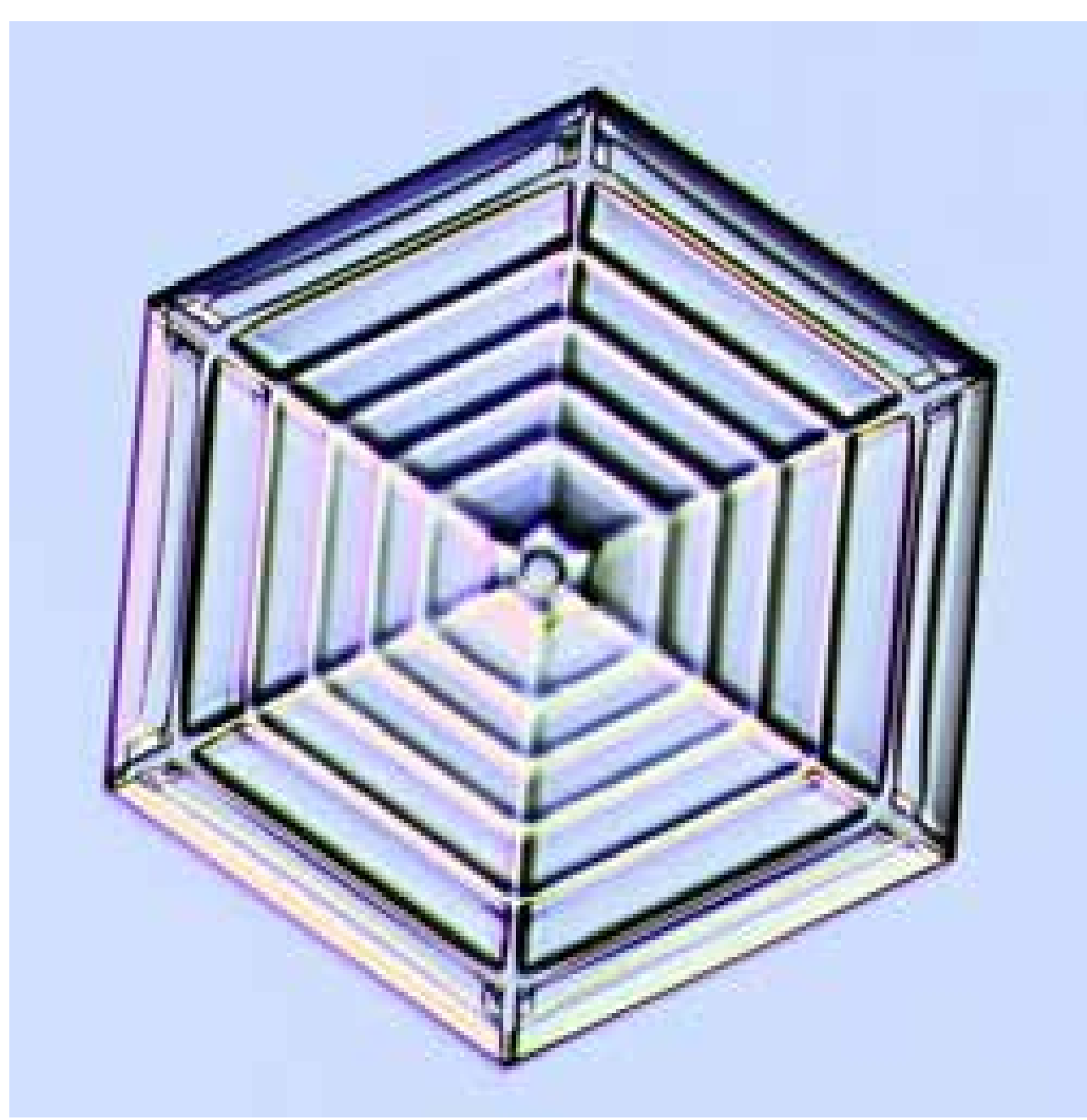

**Figure 9.28: The "spider-web" pattern on this small PoP crystal was created by inducing a series of evenly spaced rib structures. Each hexagonal rib was created by lowering $\sigma$ for about thirty seconds to thicken the plate edge, and then raising $\sigma$ back up again to continue the thin-plate growth. Both ribs and ridges form on the undersides of PoP plates, as shown in Figure 9.29.**

Next reduce the supersaturation, which turns off the ESI and yields a blockier edge on the plate, as shown in the second sketch in the figure. Note that the convex underside of the plate means there is no nucleation barrier on that surface, so most of the edge growth occurs on that surface. There is a strong nucleation barrier on the top basal surface, however, so that surface grows slowly and remains faceted. Finally, increase the supersaturation to its previous value, which again initiates the ESI and causes a thin plate to emerge from the top surface of the thick rib. Again we see that the rib structure, like the ridges, is confined to the underside of the PoP plate. This same mechanism can create ribs on plate-like branches, as illustrated in Figure 9.30.

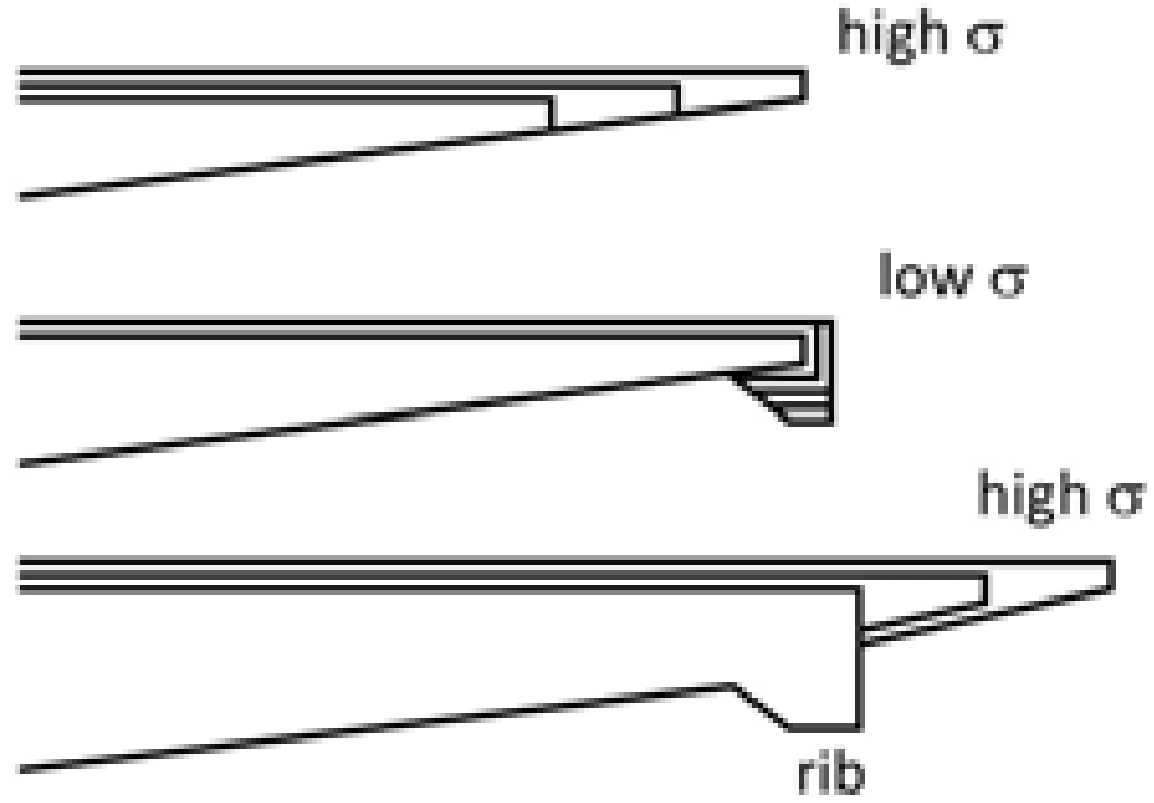

**Figure 9.29: The above series of sketches shows a side view of the formation of a rib structure on a PoP snow crystal. The rib forms on the edge of the plate, but is subsequently left behind when the ESI brings about a thinner plate from the top edge of the ridge. Applying this sequence of supersaturation changes several times yielded the multi-rib pattern seen in Figure 9.28.**

**Figure 9.30: (Below) A series of ribs were added to the plate-like extensions of this PoP snow crystal.**

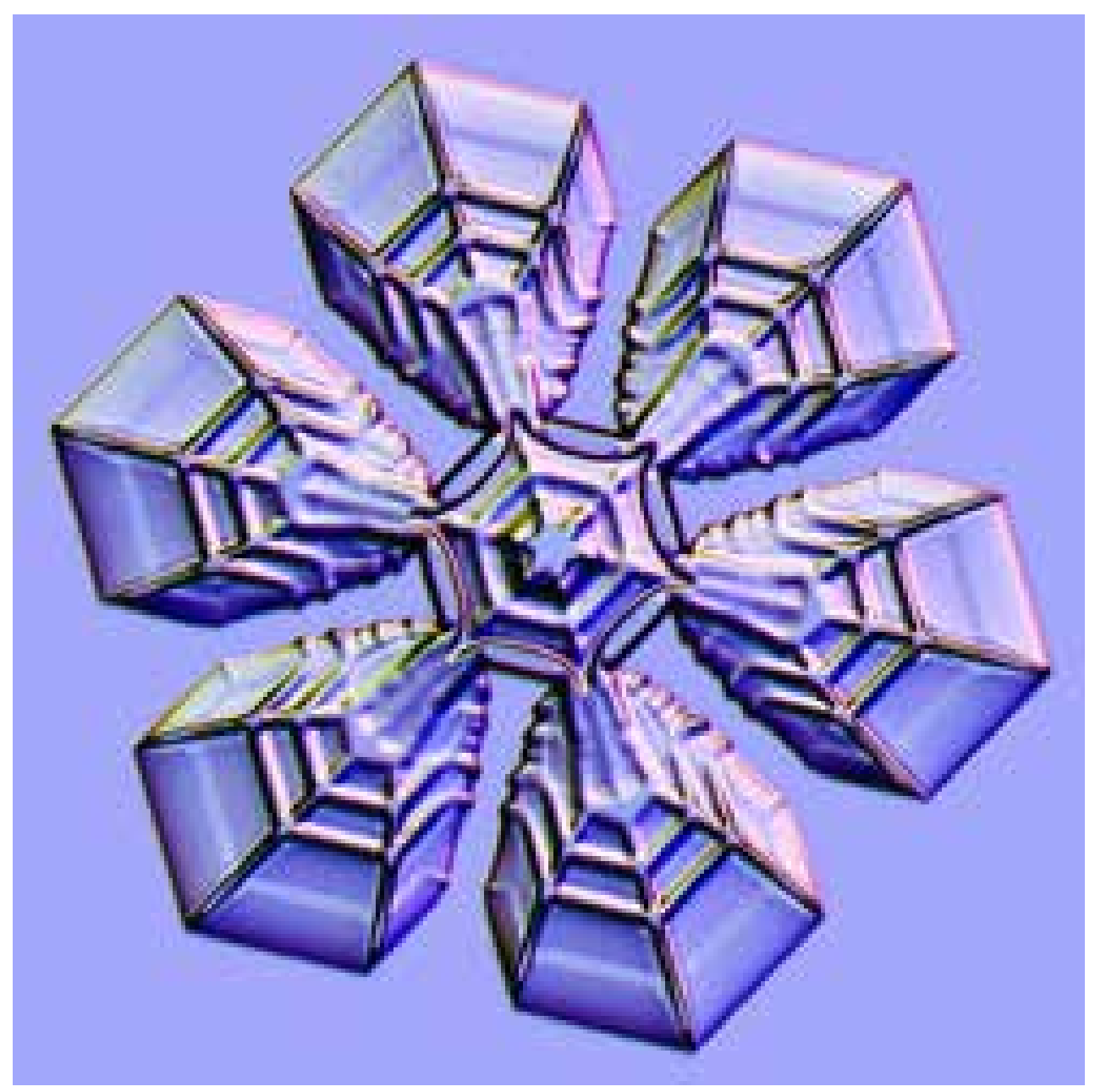

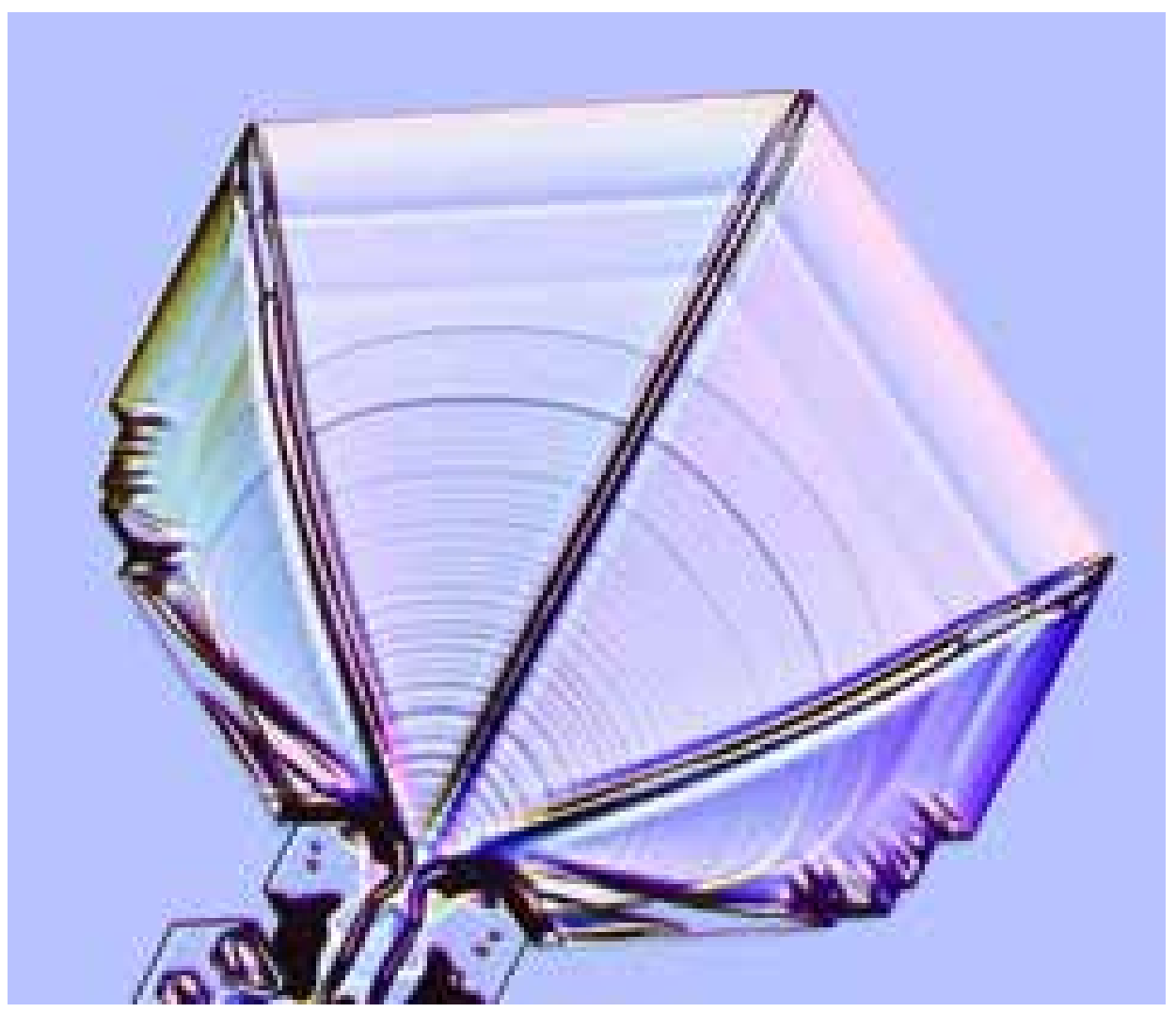
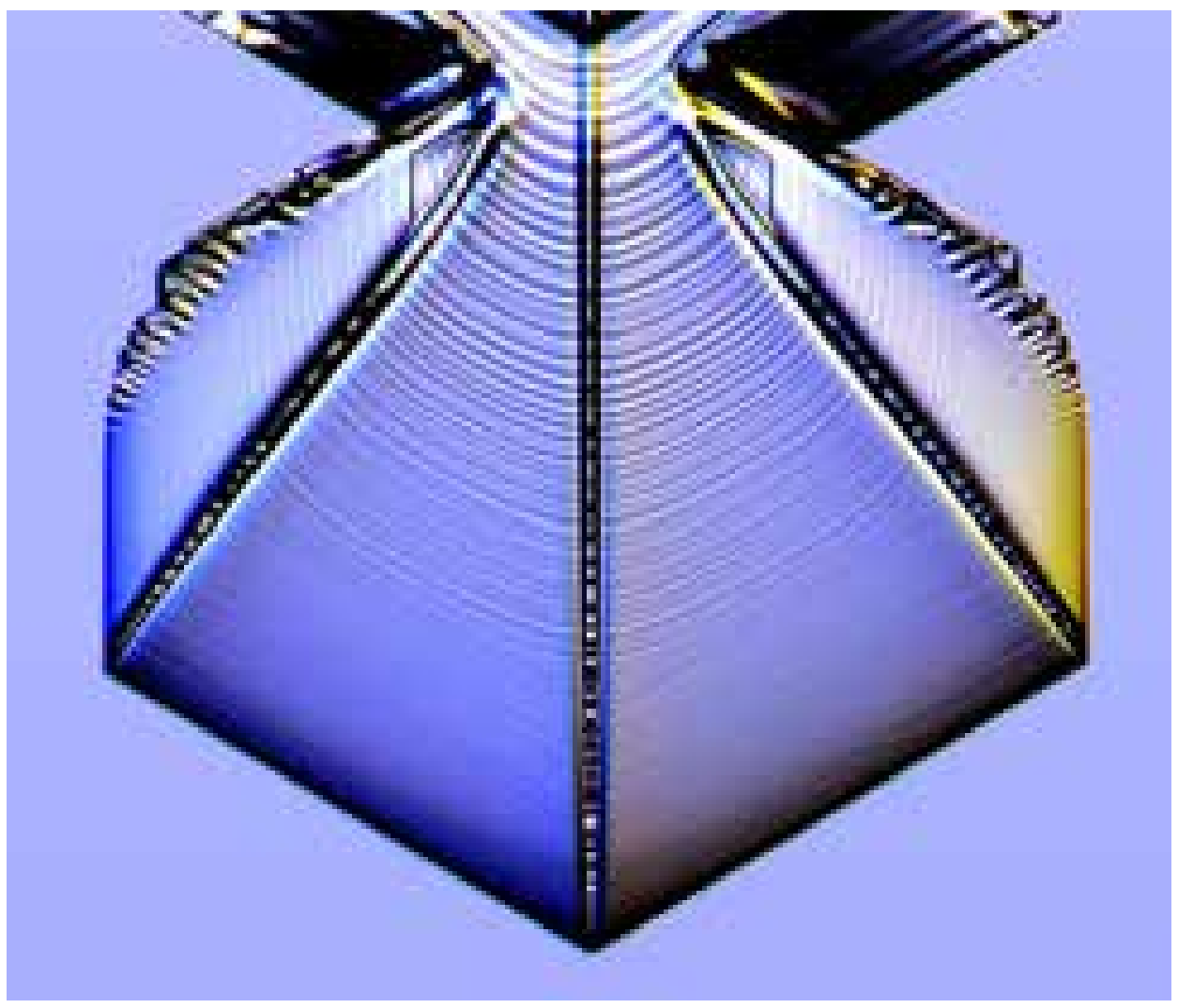

**Figure 9.31: Large sectored-plate extensions on PoP snow crystals often exhibit concentric-ring patterns like those seen in these two examples. The rings propagate inward slowly as the plates grow, and they are located on the top basal surfaces, as sketched in Figure 9.32. Ridges and faint ribs can also be seen, and both these features are located on the lower plate surfaces.**

## Inwardly Propagating Rings

Figure 9.31 shows two examples of inwardly propagating rings, which are another commonly observed phenomenon in PoP snow crystals. The rings are a result of step bunching (see Chapter 3), which occurs on basal surfaces when they become slightly concave in shape. I have been remiss in not growing a simple hexagonal PoP crystal with a suitably large, concave shape, as this would present the cleanest example of this phenomenon, illustrated in Figure 9.32.

Similar ring-like structures can be found in natural snow crystals, but the built-in asymmetry of PoP crystals makes the rings especially noticeable. Because water vapor is supplied from above the substrate, a large plate grows outward and slowly upward, as illustrated in Figure 9.32. This gives the upper basal surface a slightly concave morphology that includes a series of concentric molecular terrace steps.

The outermost edges of the crystal generally experience the highest supersaturation, and this means that the terrace steps nearest the plate edges will grow inward the fastest. This scenario brings about the phenomenon of *step bunching*, which transforms a plane of evenly space molecular steps into a set of *macrosteps* that are large enough to be seen in Figure 9.31. The distribution of macrostep sizes will depend on bulk diffusion, surface diffusion, the slope of the vicinal surface, and other factors, making it difficult to calculate with confidence.

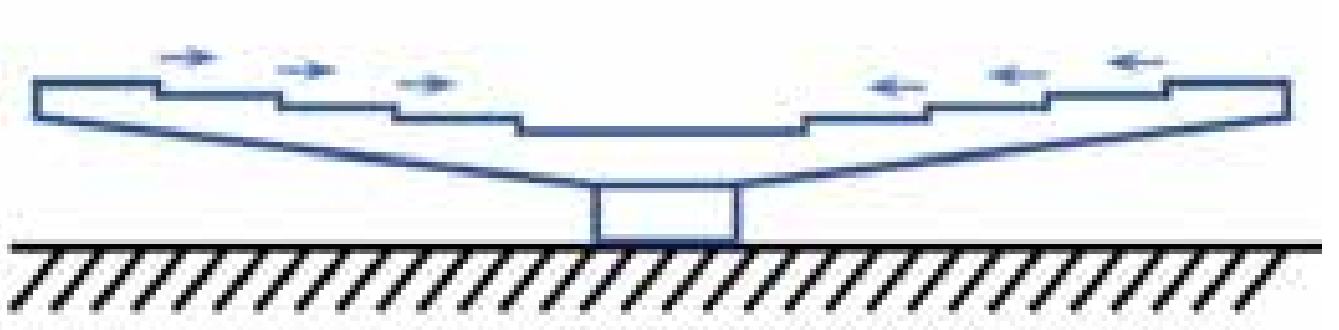

**Figure 9.32: A sketch illustrating the side view of a (hypothetical) simple PoP crystal exhibiting inwardly propagating steps. Because water vapor is supplied from above, the plate grows outward and slowly upward, giving the upper basal surface a slightly concave morphology.**

The above examples were grown at a temperature of -17 C, in a moderate supersaturation, surrounded by a field of water droplets. these conditions are especially conducive to the formation of large sectored-plate extensions on PoP snow crystals.

## Columnar Forms

While the PoP apparatus was designed for creating stellar-plate snow crystals, columnar forms can also be observed. Figure 9.33, for example, shows an example of a small column that fell onto the substrate and soon grew some short, sheath-like extensions on the ends of the column. Figure 9.34 shows the peculiar shape of a twinned crystal (see Chapter 2).

As I am usually growing large stellar snow crystals in the PoP chamber, errant columns that land on the substrate usually develop into half-double plates, as shown with several examples in Figure 9.35. In all these cases, the crystals obviously cannot grow into the substrate, and substrate interactions may distort their growth in unpredictable ways. The presence of the substrate thus makes the morphologies of these half-double plates a bit challenging to interpret.

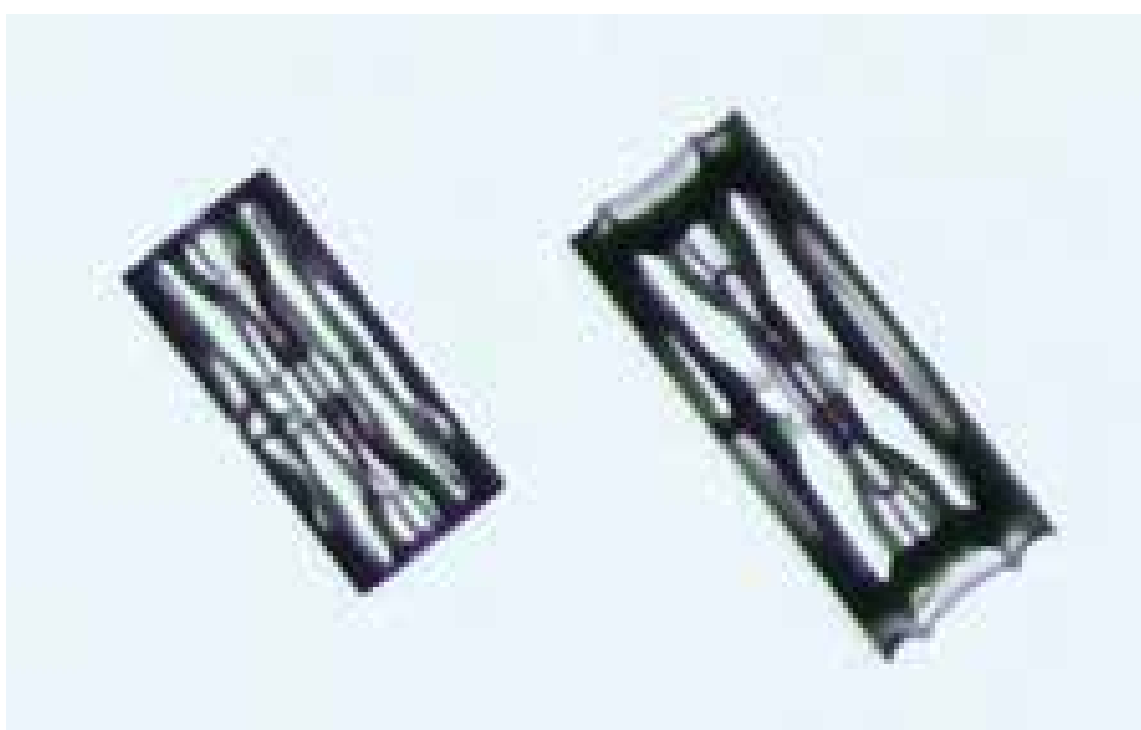

**Figure 9.33: Sheath-like growth of a small columnar crystal observed in the PoP apparatus.**

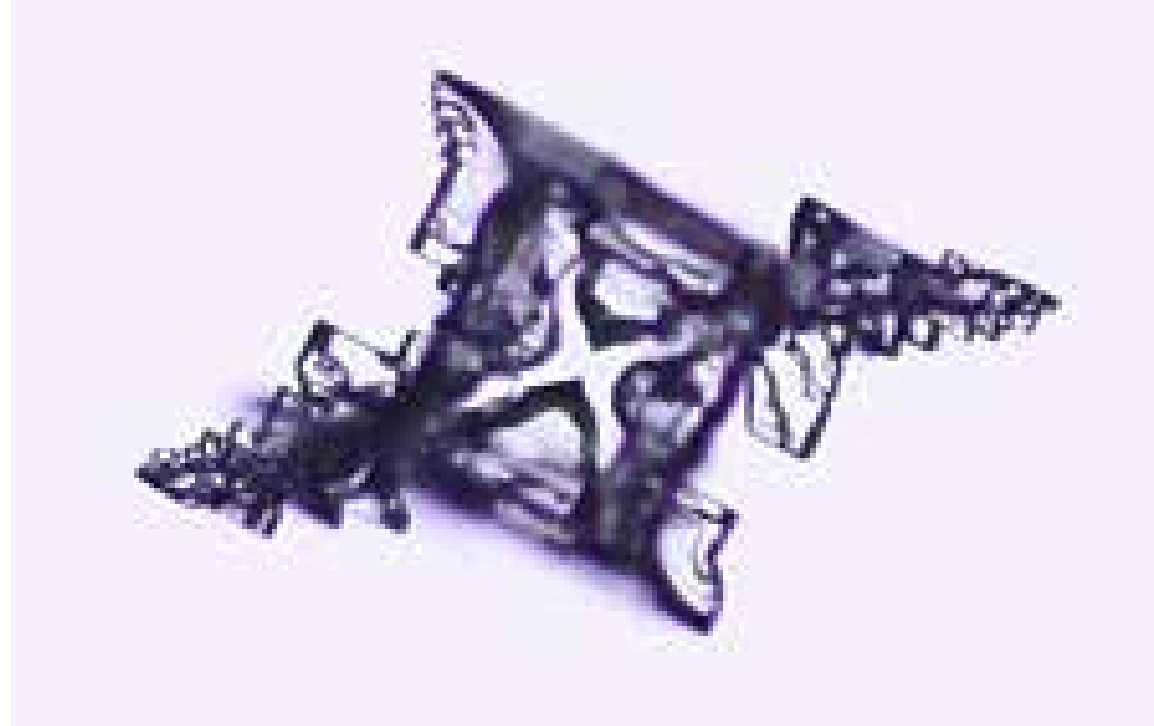

**Figure 9.34: A small twinned snow crystal that happened to fall on the PoP substrate.**

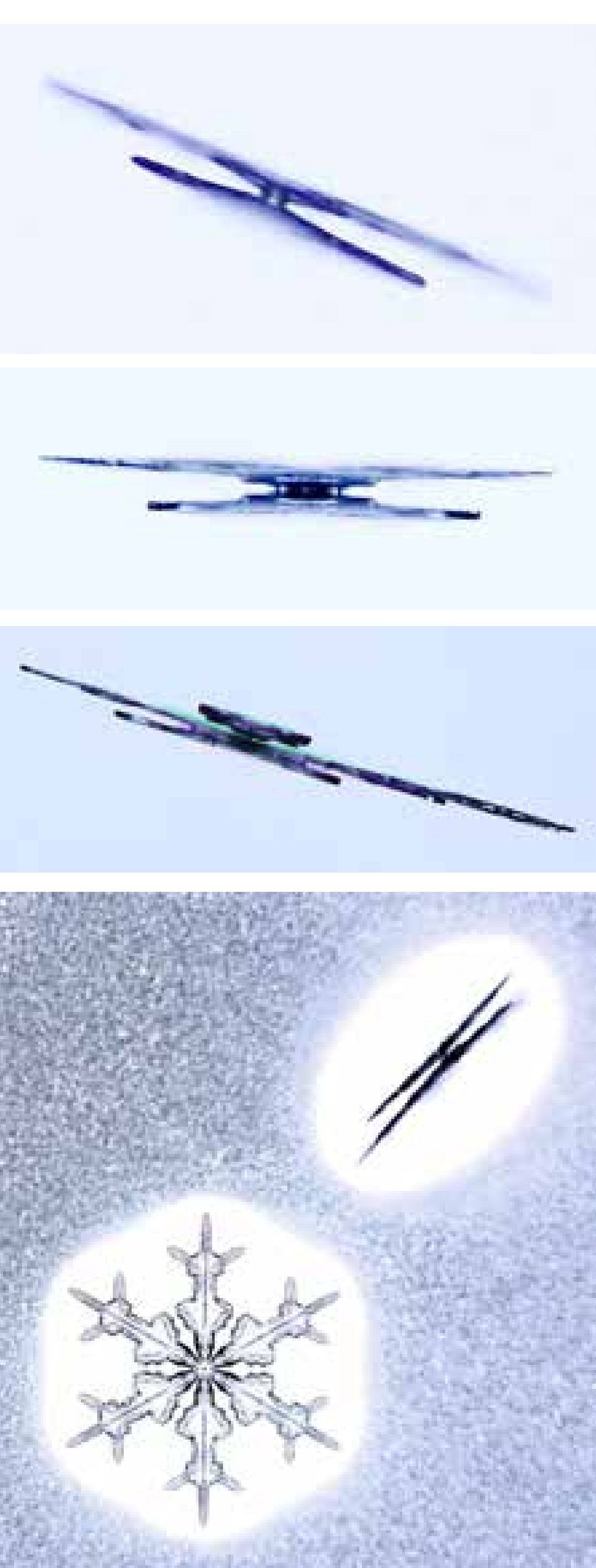

**Figure 9.35: When grown alongside plate-like PoP snow crystals near -15 C, columnar seed crystals usually develop into what are essentially "half" double plates, with the perpendicular plate structures only growing above the substrate.**

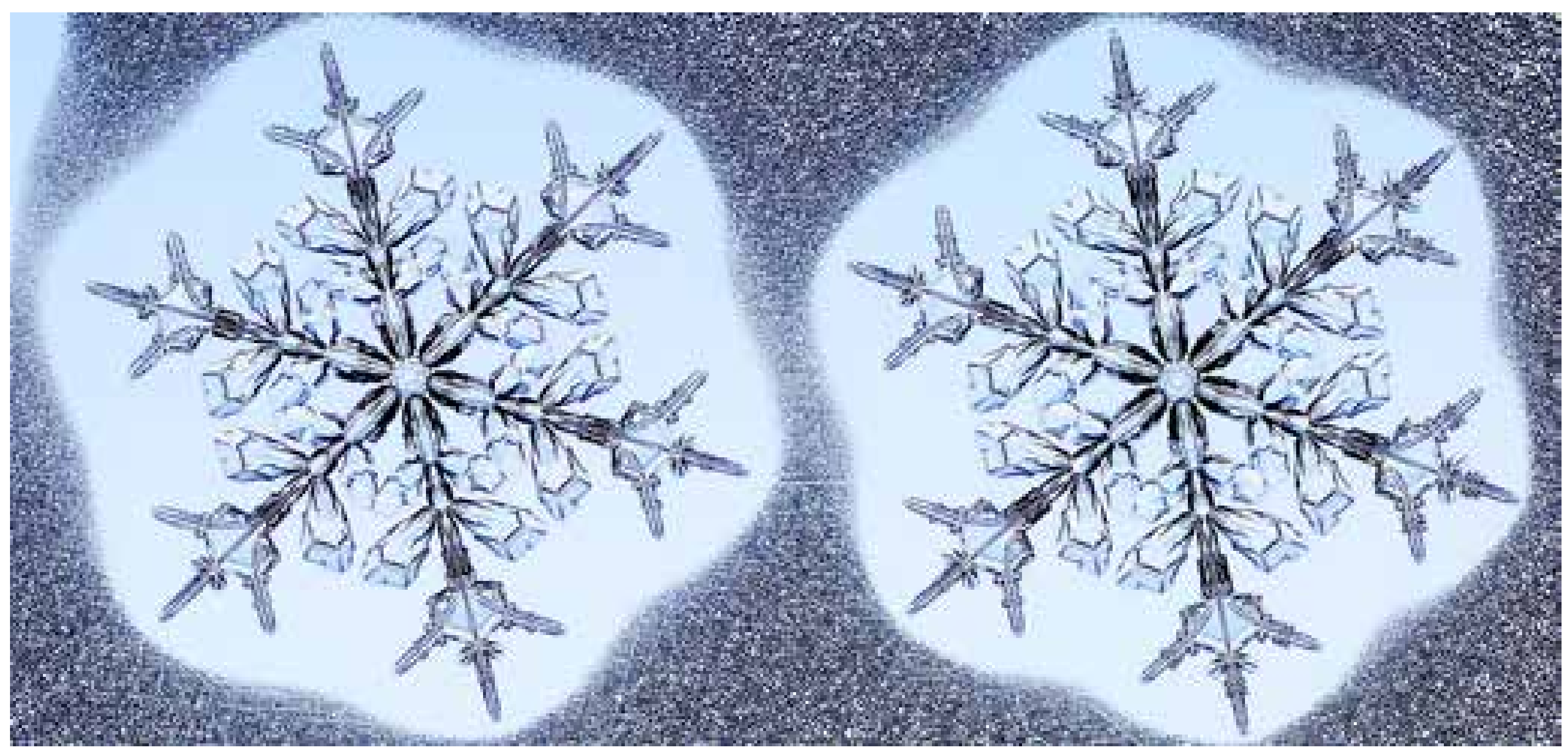

**Figure 9.36: A pair of "identical-twin" snow crystals, grown side-by-side in the PoP apparatus.**

## 9.4 Identical-Twin Snow Crystals

The old adage that no two snowflakes are alike appears to have had its origin with the photographs of Wilson Bentley. In his book with William Humphreys, Bentley presented pictures of nearly 2000 stellar snow crystals, selected for their beauty and symmetry, and each was clearly different from all the others. Since then, the notion of snowflake uniqueness seems to be something we all learn at a young age (at least in North America), probably while participating in the near-universal craft of cutting snowflakes out of paper. No one I have asked can remember when they first learned that no two snowflakes are alike, so I am guessing before the age of ten.

As I describe in Chapter 1, there is good reason to believe that no two complex, natural snow crystals will ever look quite the same. Because each snowflake follows a different path through the clouds, guided by the motions of a turbulent atmosphere, each experiences different growth conditions during its journey. The number of possible variations is vast, so the probability of finding two identical snowflakes is vanishingly small.

But this discussion changes when you consider growing PoP snow crystals. Now the growth conditions are not determined by random paths through a turbulent atmosphere. Now the temperature and supersaturation are controlled by precision temperature controllers that can be set and changed according to a prescribed schedule. In principle, one might engineer a precision snow-crystal factory that would produce a continuous flurry of essentially identical snowflakes. (Of course, one has to be careful about how you define the word "identical", as I discussed at some length in Chapter 1.) I have not had any great urge to create such a factory, but it not outside the realm of the possible.

Instead of growing identical PoP snowflakes one after another, a much easier approach is to grow two at the same time, side by side, as illustrated in Figure 9.36. This photo, unmodified except for cropping and small global brightness/contrast/color adjustments, shows two snow crystals that I grew simultaneously in the PoP apparatus. (I have demonstrated the process of growing similar crystal pairs in person to numerous colleagues and reporters, just to have some witnesses, but I am also hoping you will trust me on this.) I like to call these "identical-twin" snow crystals because they are not perfectly identical, and small differences can easily be found if you look carefully at the photo. But, like identical-twin people, they are clearly much more similar than one might ever expect to see under normal circumstances. In the spirit of full disclosure, Figure 9.37 shows the original, unmodified image.

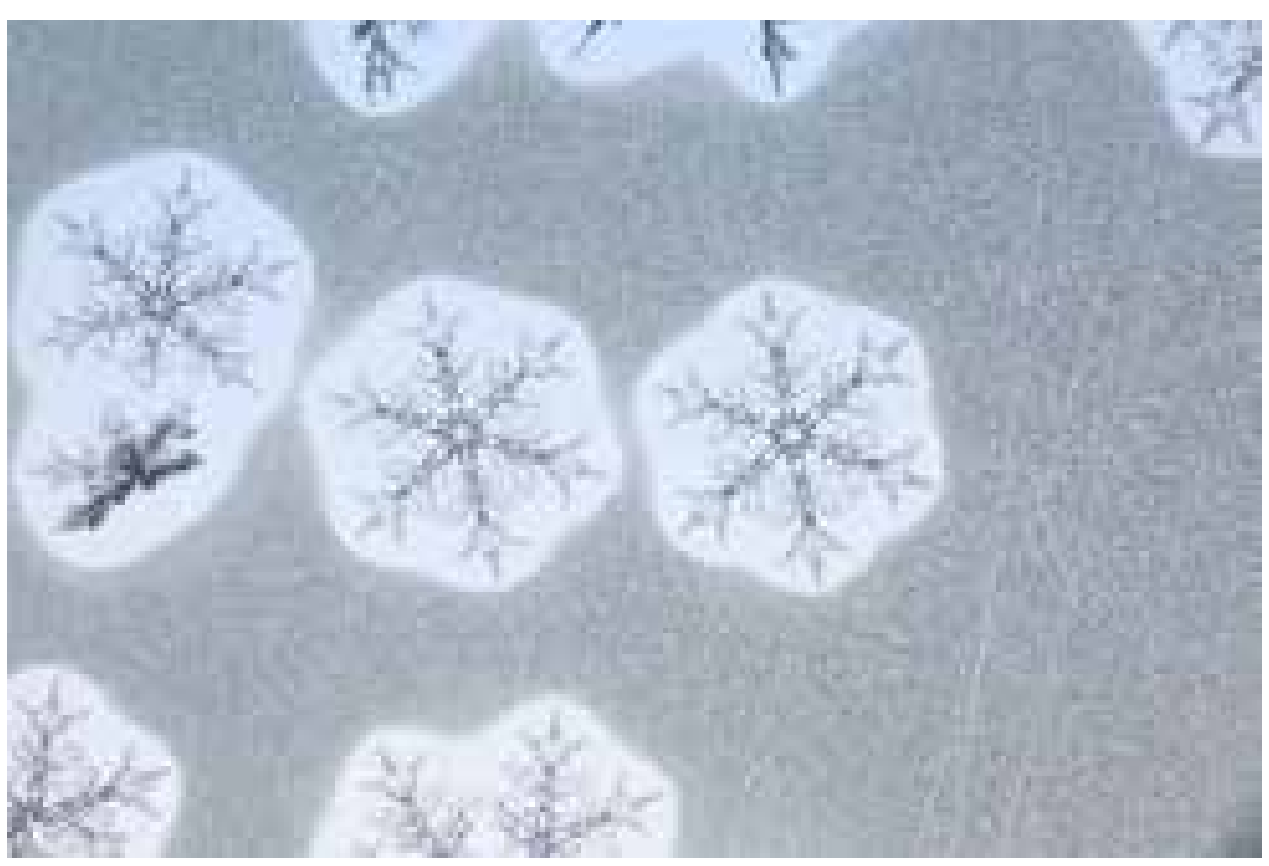

**Figure 9.37: The completely unretouched, uncropped, straight-out-of-the-camera original version of the image shown in Figure 9.34.**

To create this pair of identical-twin snow crystals, I first had to locate a pair of well-formed seed crystals, close to one another, but not too close, and reasonably isolated from any additional crystals. This is a high bar to clear, and Figure 9.37 shows that there was quite a bit going on over the entire substrate, a fact that the cropped image does not convey. Pretty much every photo in this chapter was cropped in order to focus the viewer's attention on the primary subject.

With a suitable pair of seed crystals in place, I then proceeded to grow the dual PoP crystals by applying branching, faceting, and other effects at various times. I typically have no predetermined strategy for creating a specific large-scale morphology; usually I just make it up as the crystal develops. Someday I will put everything under computer control and take a more systematic approach to growing PoP snow crystals, but that is a task for a future date.

A key trick for growing identical-twin snow crystals is to make lots of large, abrupt changes in growth conditions. Rapidly changing the temperature by a substantial amount, for example, causes a correspondingly large and abrupt change in the growth behavior. Smaller, gradual changes typically yield less perfect symmetry between the two crystals, or even among the six branches of a single crystal. There are inevitably some weak temperature gradients in the growth region, and the neighboring crystals perturb the local environment to some extent also. Making large, abrupt changes tends to mask these weaker effects, thus improving the overall symmetry of the growing crystals.

Another trick is to create a field of water droplets around the two crystals early on, and to maintain a well-defined droplet perimeter around both crystals, always with a "barrier" of droplets separating the crystals, as shown in Figure 9.37. The droplets provide a stabilizing influence on the supersaturation, greatly reducing the perturbations that arise from neighboring crystals. This makes sense because the droplets hold the supersaturation at $\sigma \approx \sigma_{water}$ in their immediate vicinity, and the boundary condition of having a clean droplet perimeter around both crystals tends to improve the overall symmetry of the two crystals.

The final trick is just knowing when to stop. In Figure 9.37, for example, the line of droplets between the crystals will soon evaporate away as the crystals grow larger, and then the crystals will begin to interfere with one another's growth. Figure 9.38 shows what happens when the droplet barrier disperses and the crystals continue to grow. The branches growing between the two crystals compete for the available water vapor, thereby stunting their growth relative to the outer branches, which are still supplied by the nearby droplets. Viewing a video showing first the nearly identical growth of nearby crystals, followed by the stunted growth of the branches between them, gives one a much better appreciation of what affects the overall growth process.

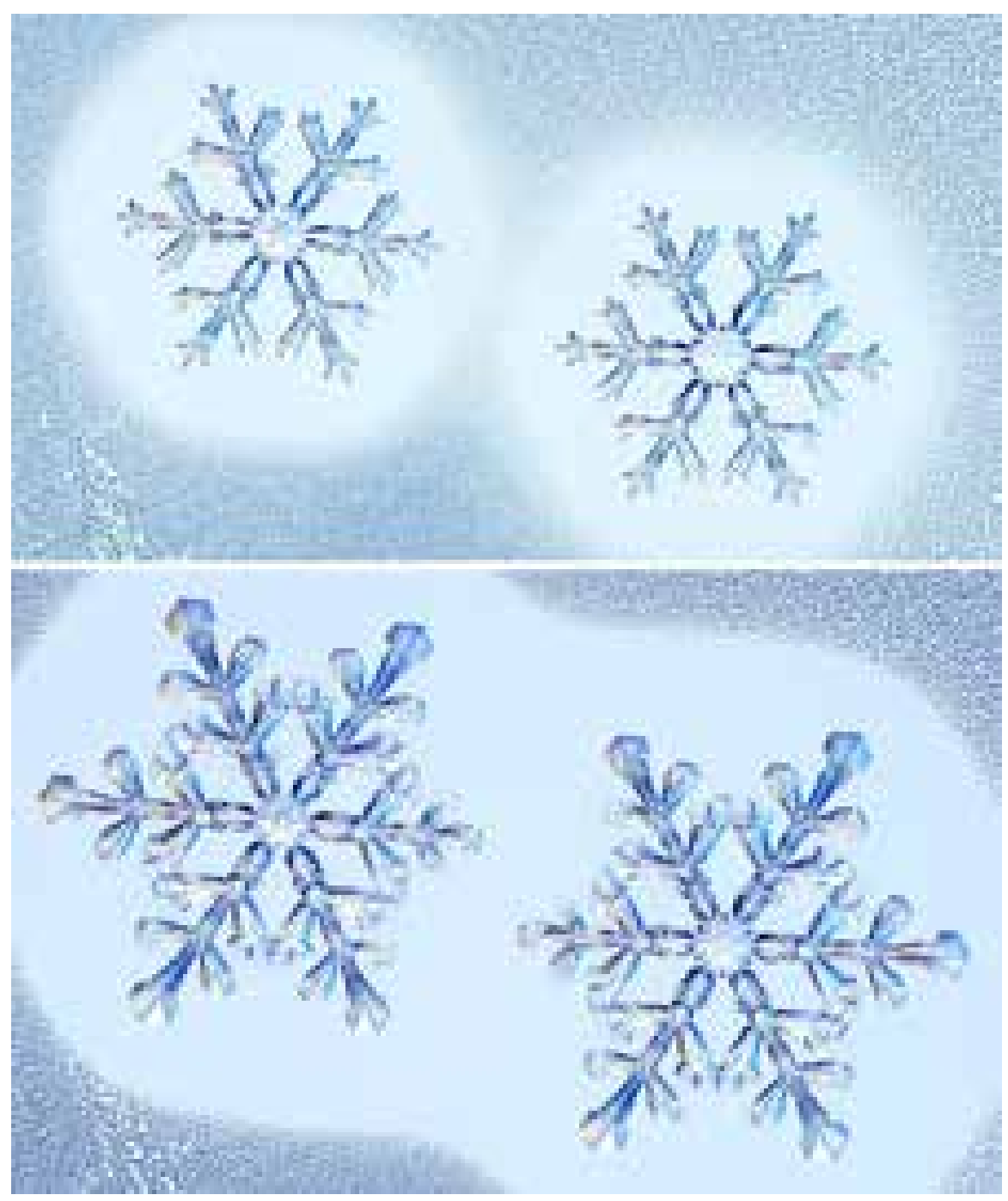

**Figure 9.38: Top: A field of water droplets surrounding a PoP snow crystal tends to stabilize its growth, providing the same level of supersaturation for all twelve branches. Bottom, same crystals: Once the droplets disappear between the crystals, the facing branches grow more slowly, yielding asymmetrical snow crystals.**

Note that the phenomenon of twelve branches growing in synchrony is really no different than with six branches. For a standard natural snow crystal, only its six conjoined branches experience the same growth conditions as a function of time. But in the PoP apparatus, this limitation is removed, as all crystals are attached to the same fixed substrate. There is no need to stop at twelve either, and Figure 9.39 shows larger groups of similar-growing PoP snow crystals.

Overall, observing PoP snow crystals is not ideal for investigating the underlying science, as the growth conditions are rather poorly known. But studying the formation of these crystals in detail, especially in video form, gives one many useful insights into the underlying physical processes.

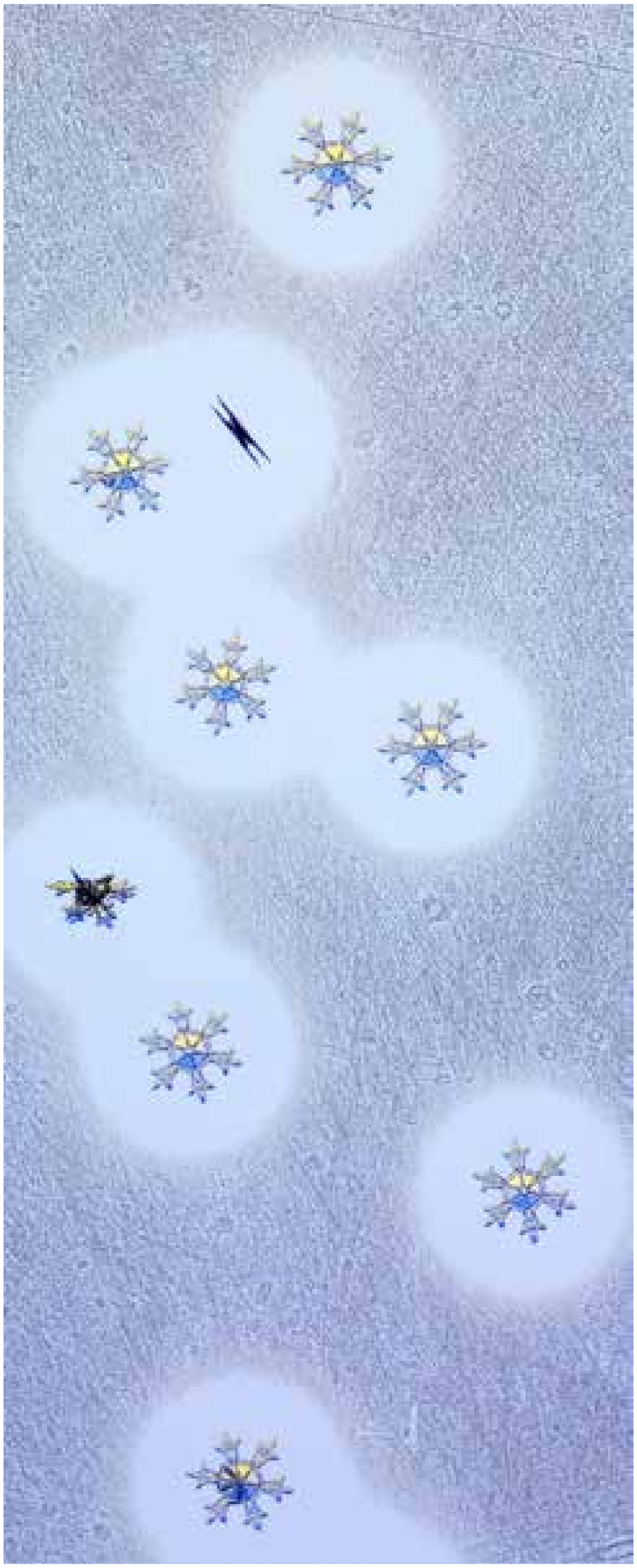

**Figure 9.39: (Above, and facing page) Clusters of small PoP crystals all exhibiting about the same growth conditions as a function of time, and thus all growing into similar shapes.**

## 9.5 PoP Art

The remainder of this chapter presents a gallery of synthetic PoP snow crystals that I grew using the apparatus and techniques described in the previous sections. The images that follow show real PoP snow crystals, with essentially no digital modifications of the overall crystal structures. However, I did use a fair bit of artistic license when adjusting brightness, contrast, cropping, sharpness, and a host of color effects. In some images I also removed droplets and/or other distractions from the background around the growing crystals. My overarching goal in this gallery is to examine the growth of synthetic snow crystals as a novel art form, rather than a tool for scientific discovery.

While science and art are normally quite distinct endeavors, they come together beautifully in snow crystal growth. My understanding of the science allowed me to engineer the PoP apparatus, leading to the creation of high-resolution images and videos showing details not observable in natural snow crystals. I believe there is much left to explore in this novel, additive form of ice sculpture.

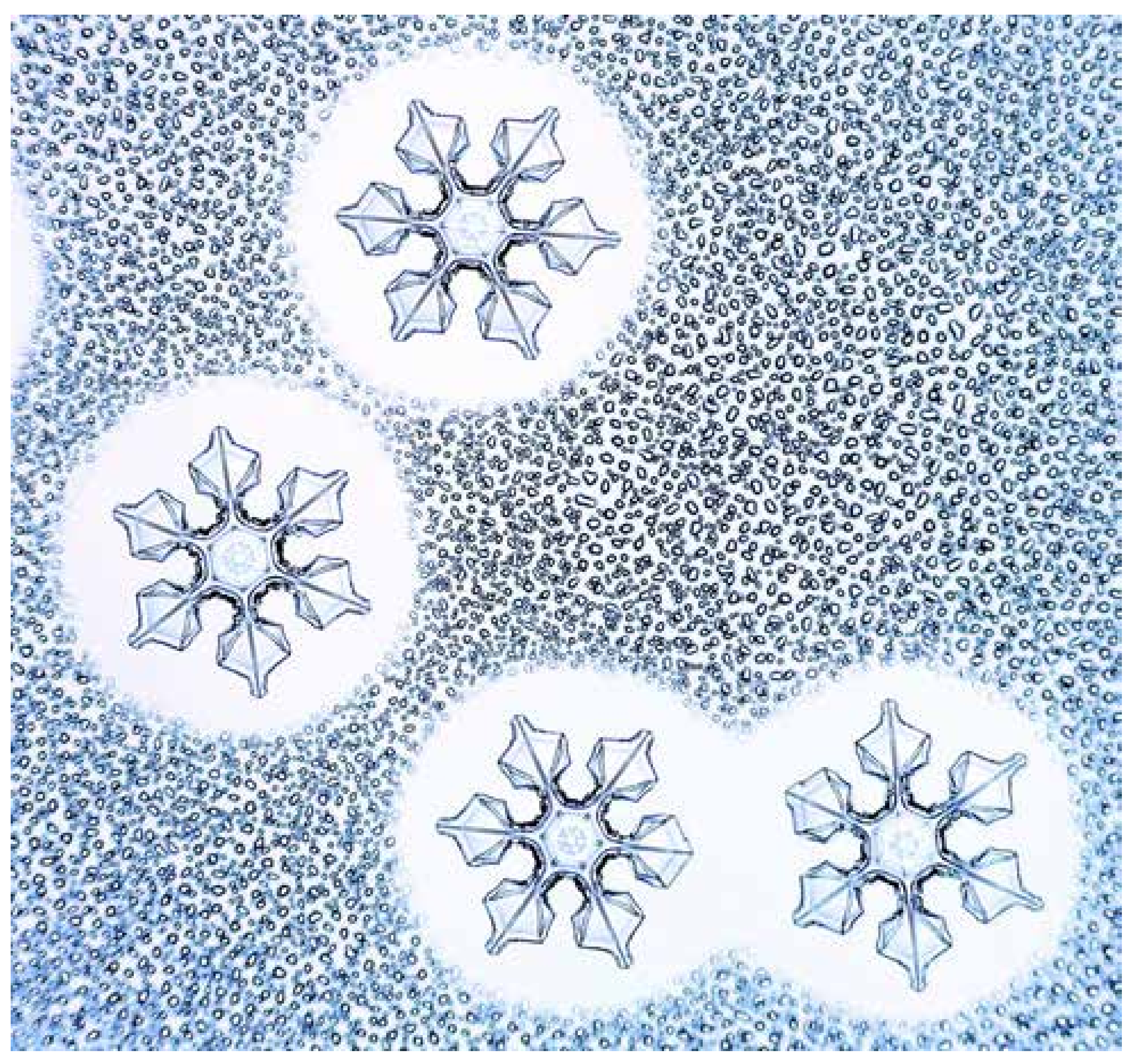

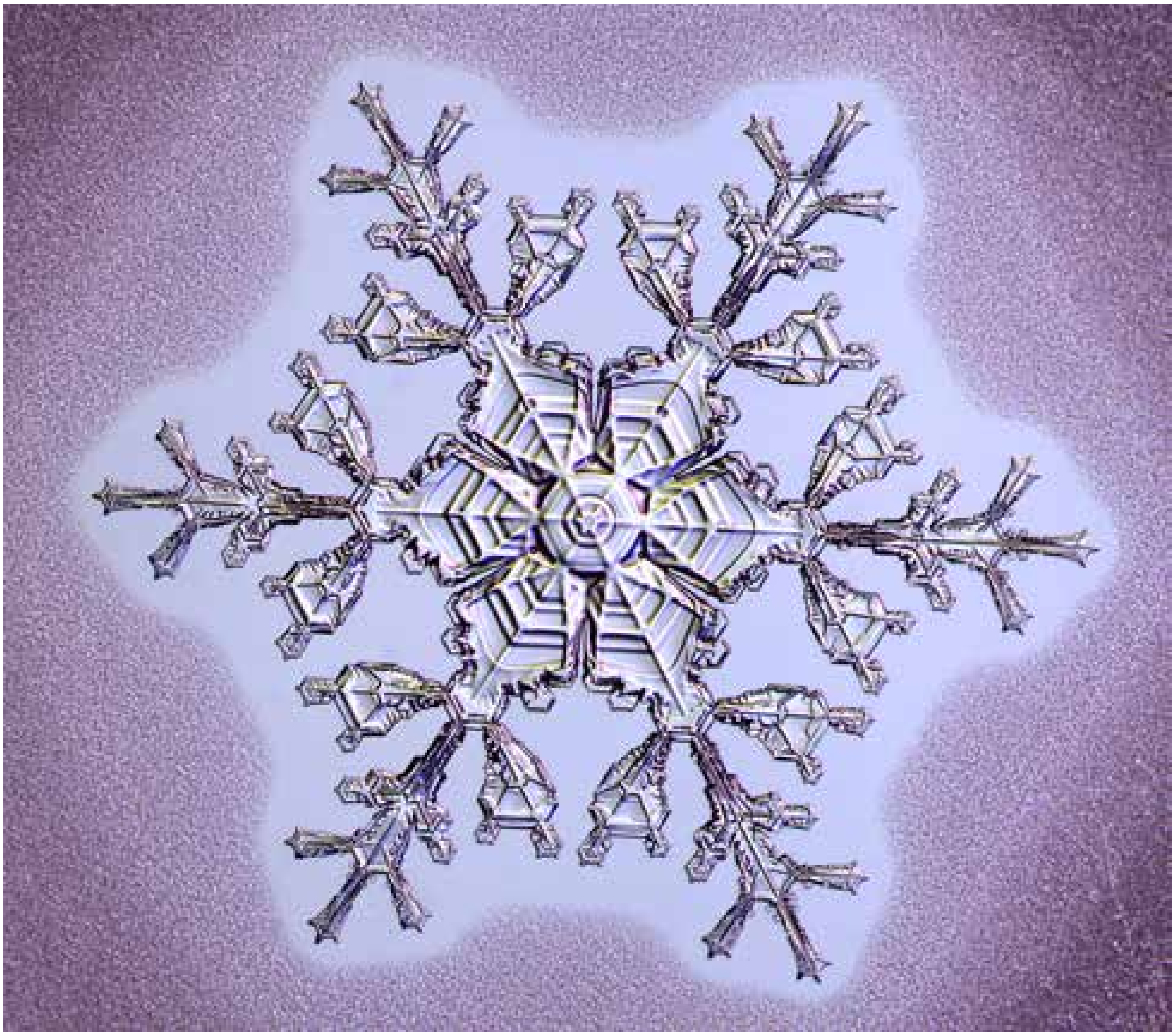

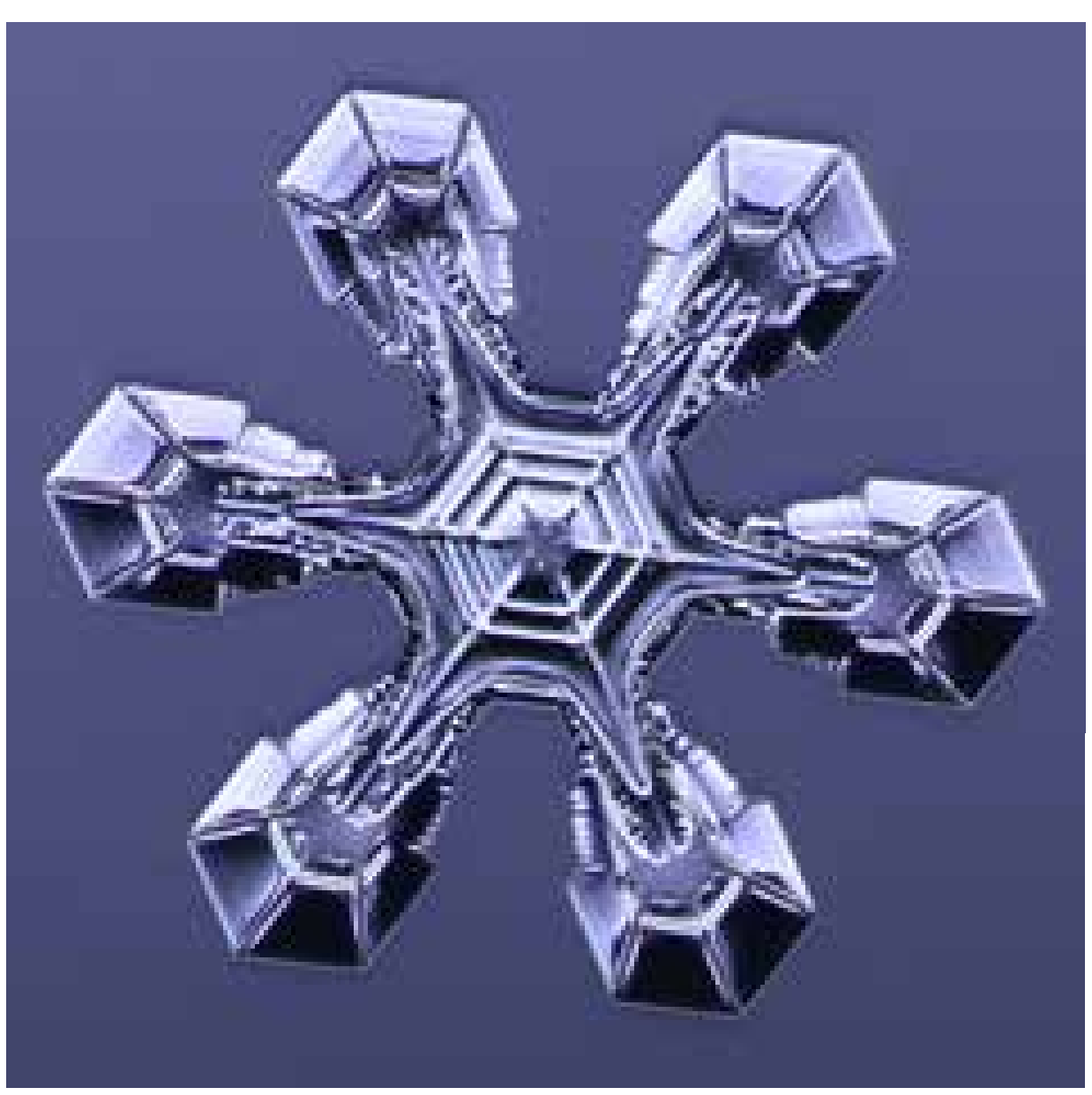

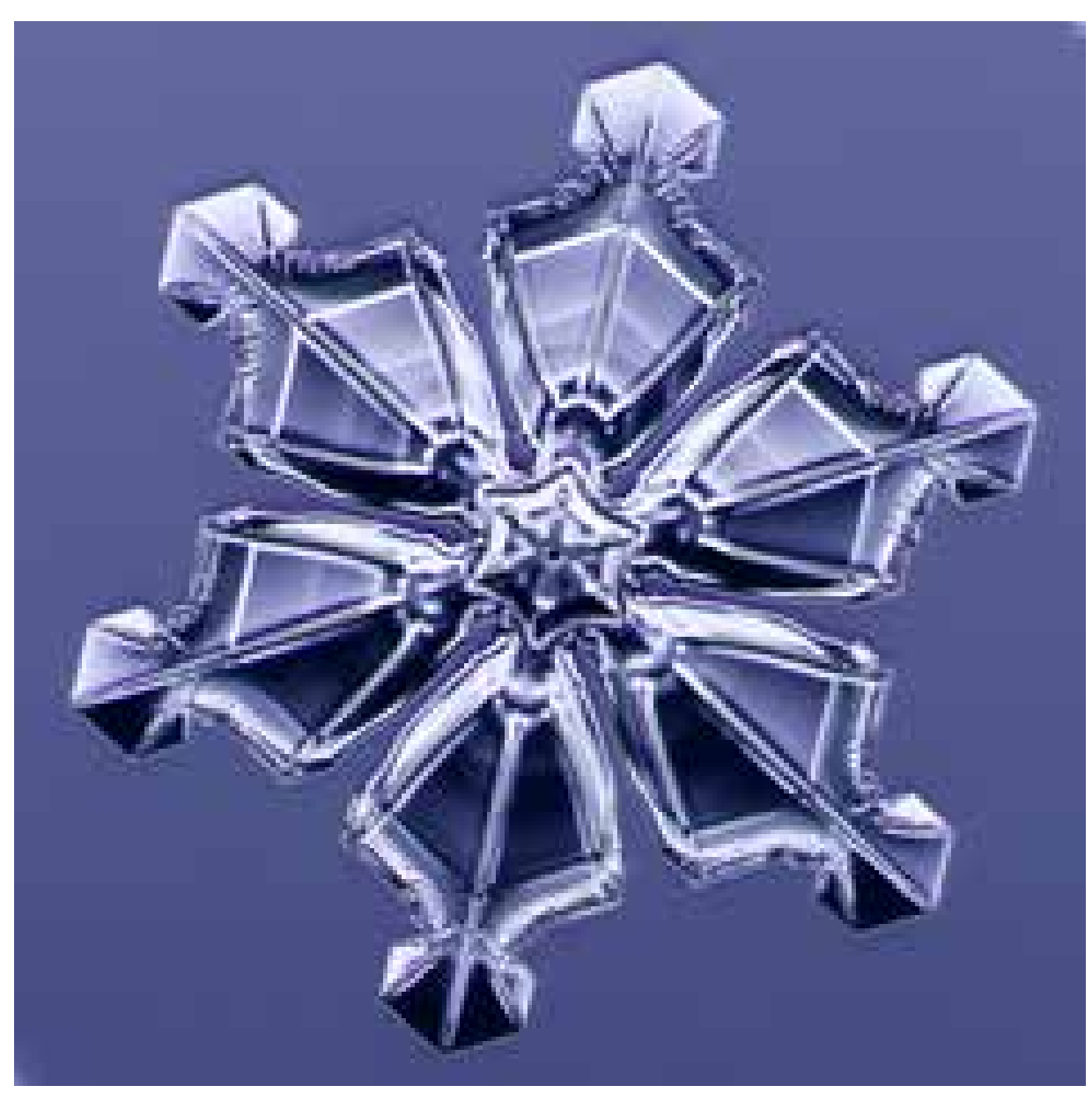

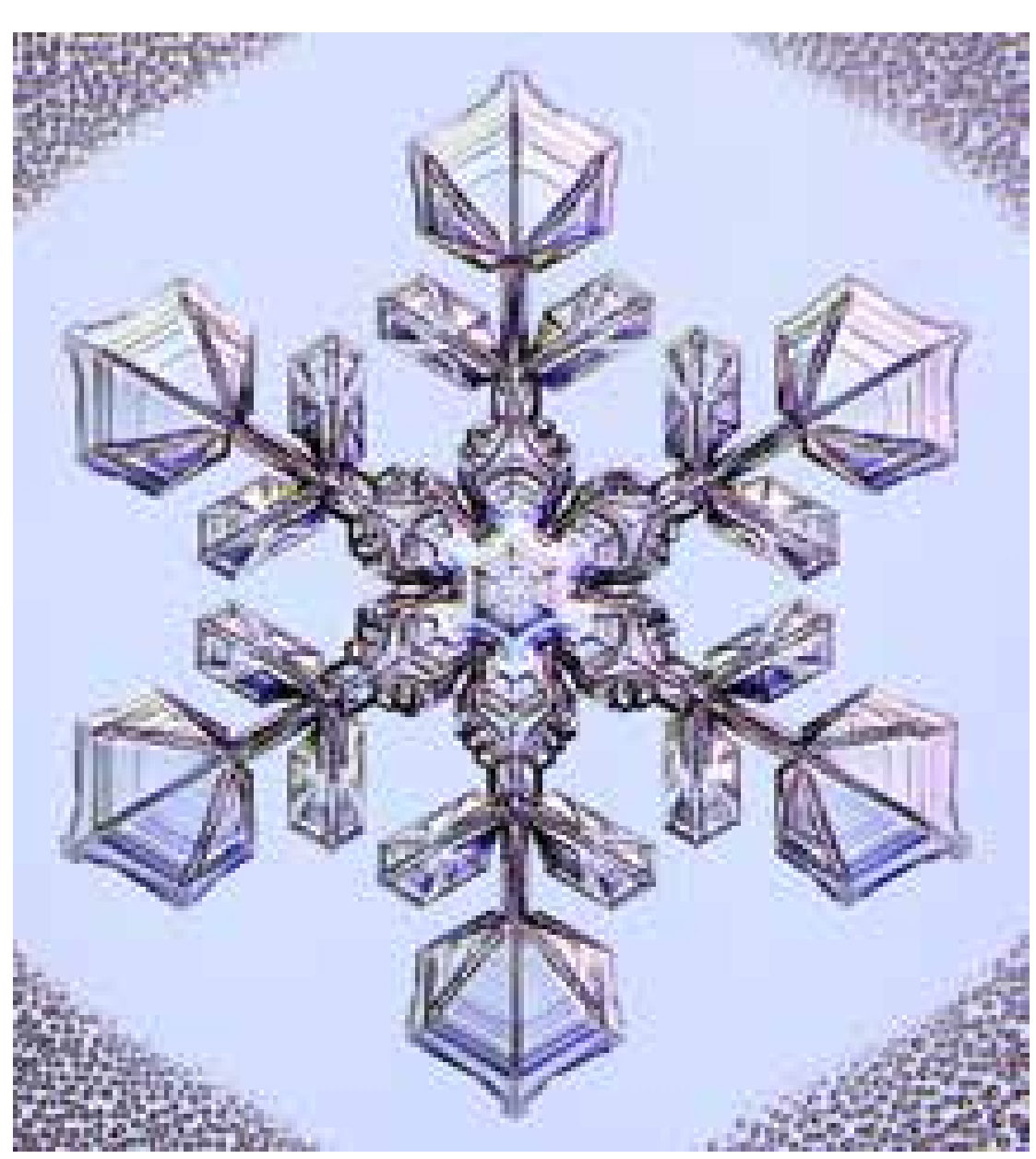

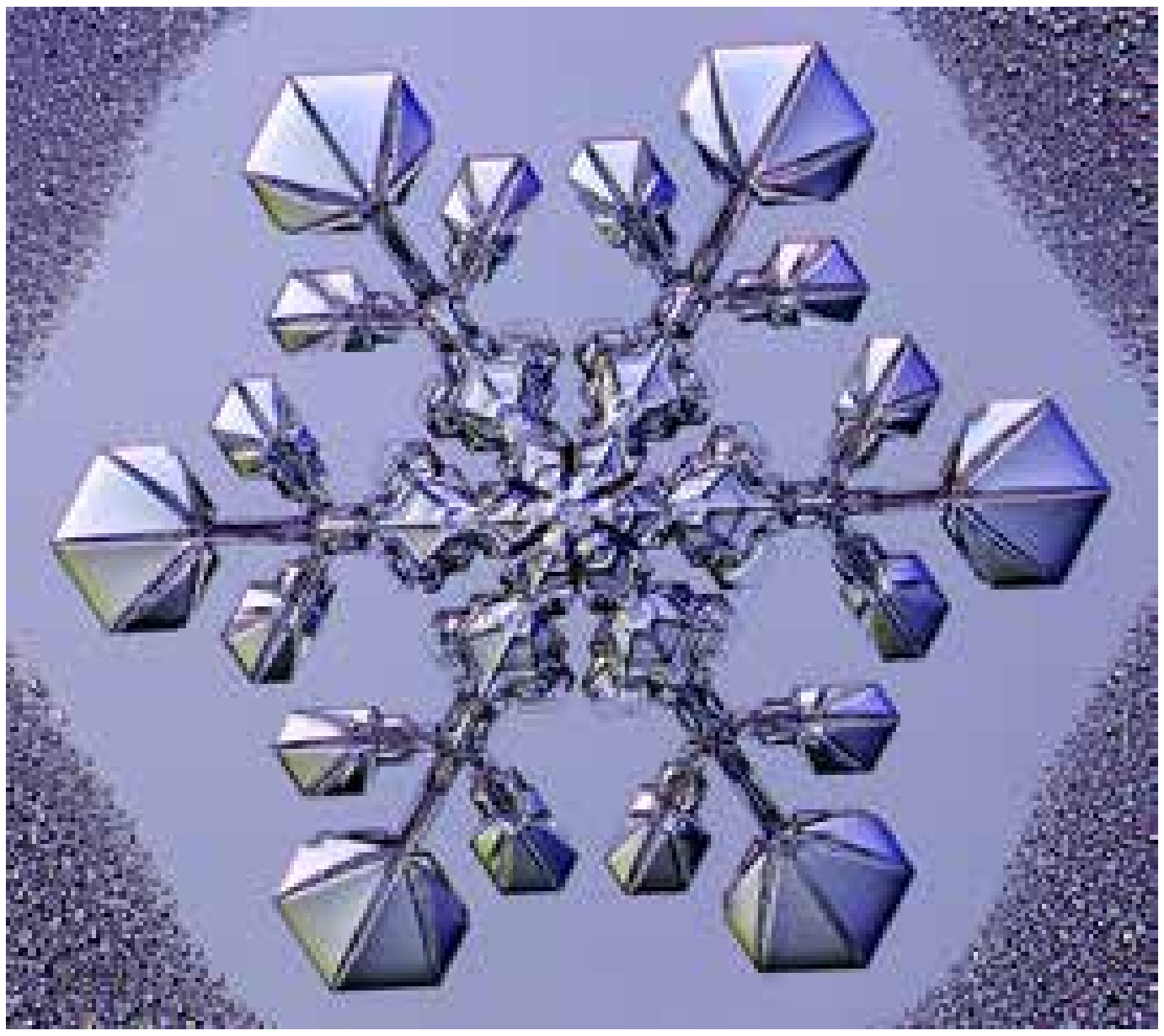

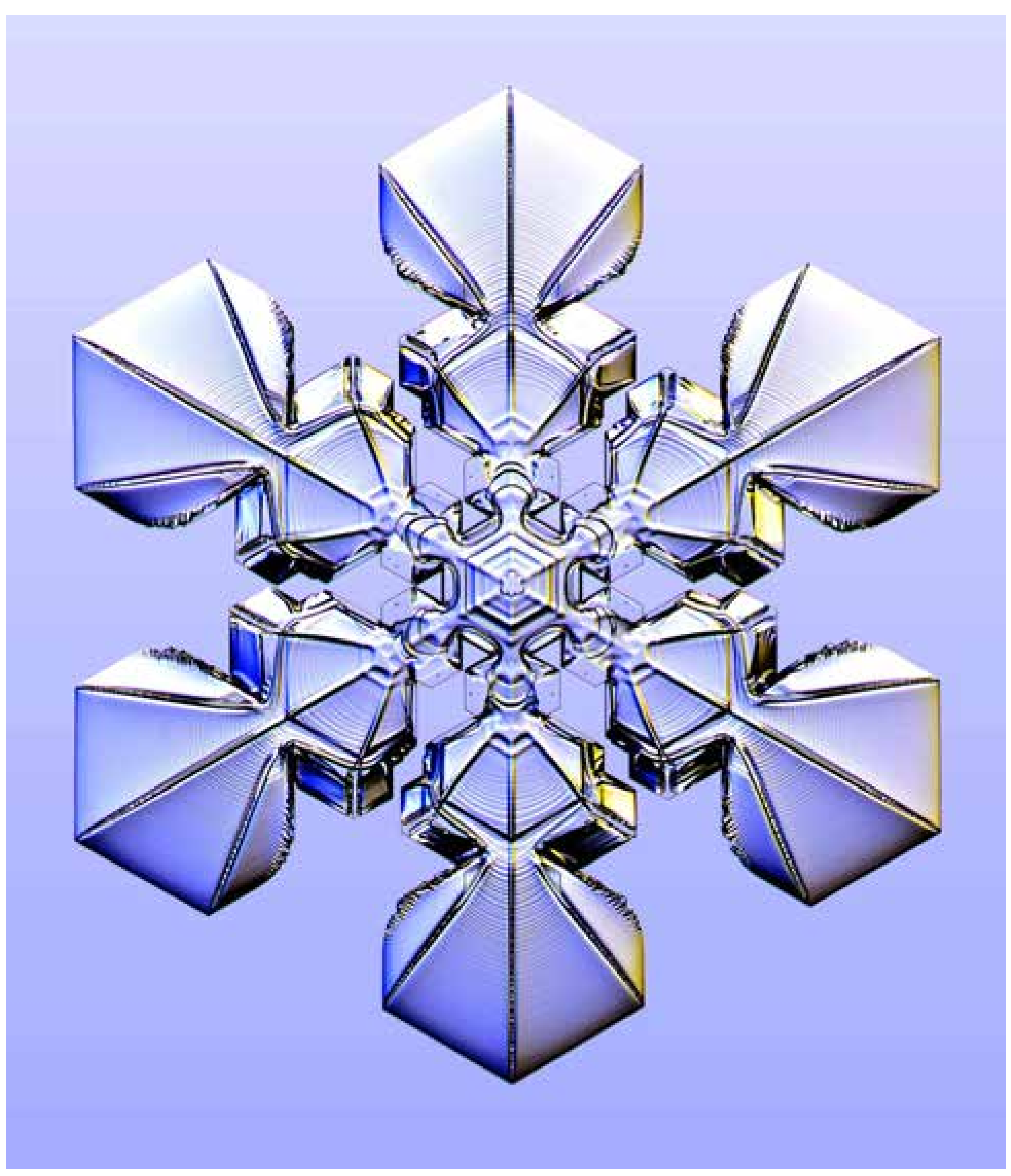

**Figure 9.40: (Above and facing page) Snow crystal portraits. The PoP technique is well suited for creating large stellar crystals with complex designs and nearly flawless symmetry. When photographing natural snowflakes, beautiful specimens like these are hard to find! Moreover, PoP snow crystals often exhibit razor-sharp facets and exceptionally crisp surface features, because they are photographed while they are being grown. In contrast, natural snow crystals usually experience a bit of sublimation after they leave the clouds, rounding their features and generally giving them a bit of a "travel-worn" appearance.**

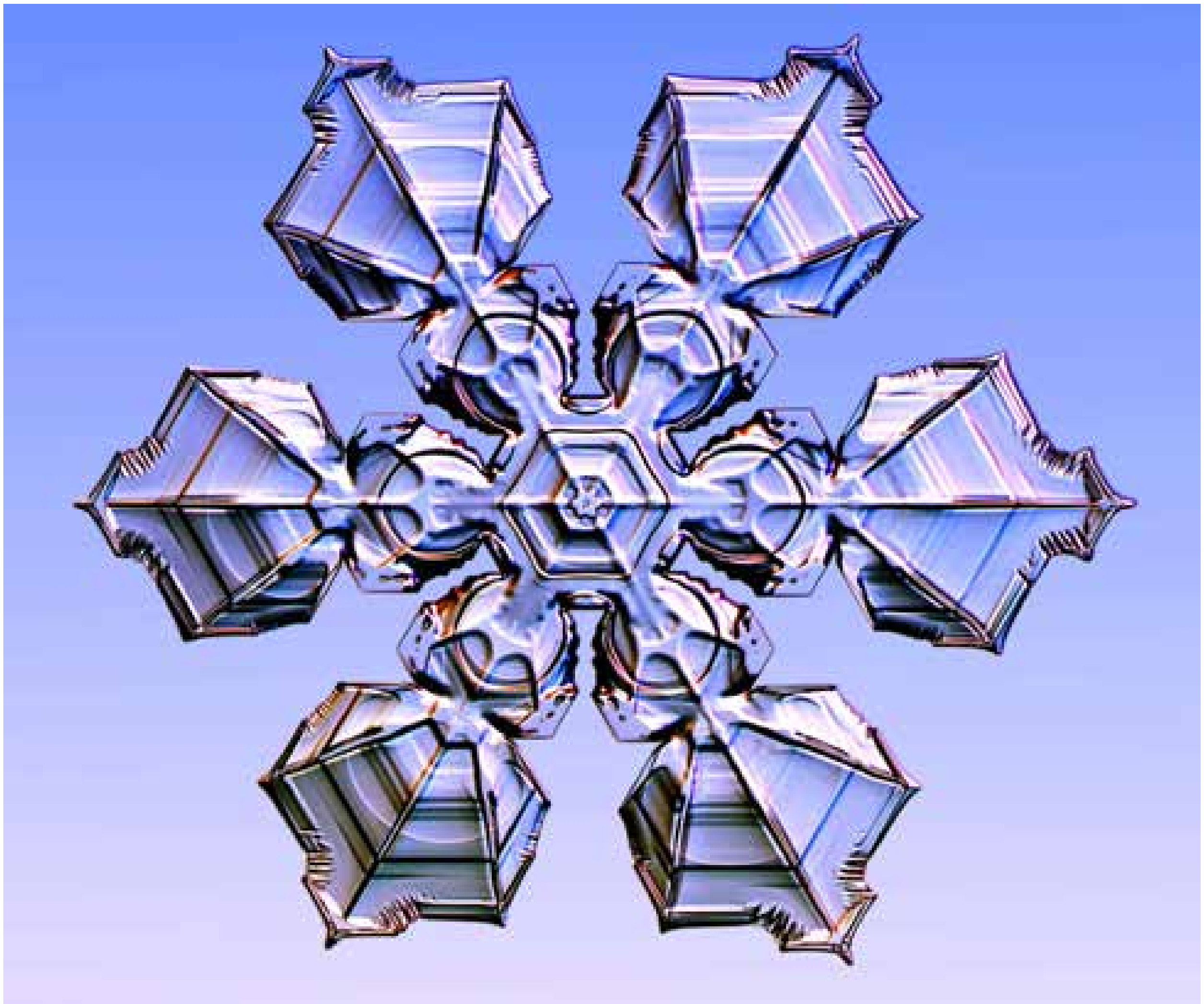

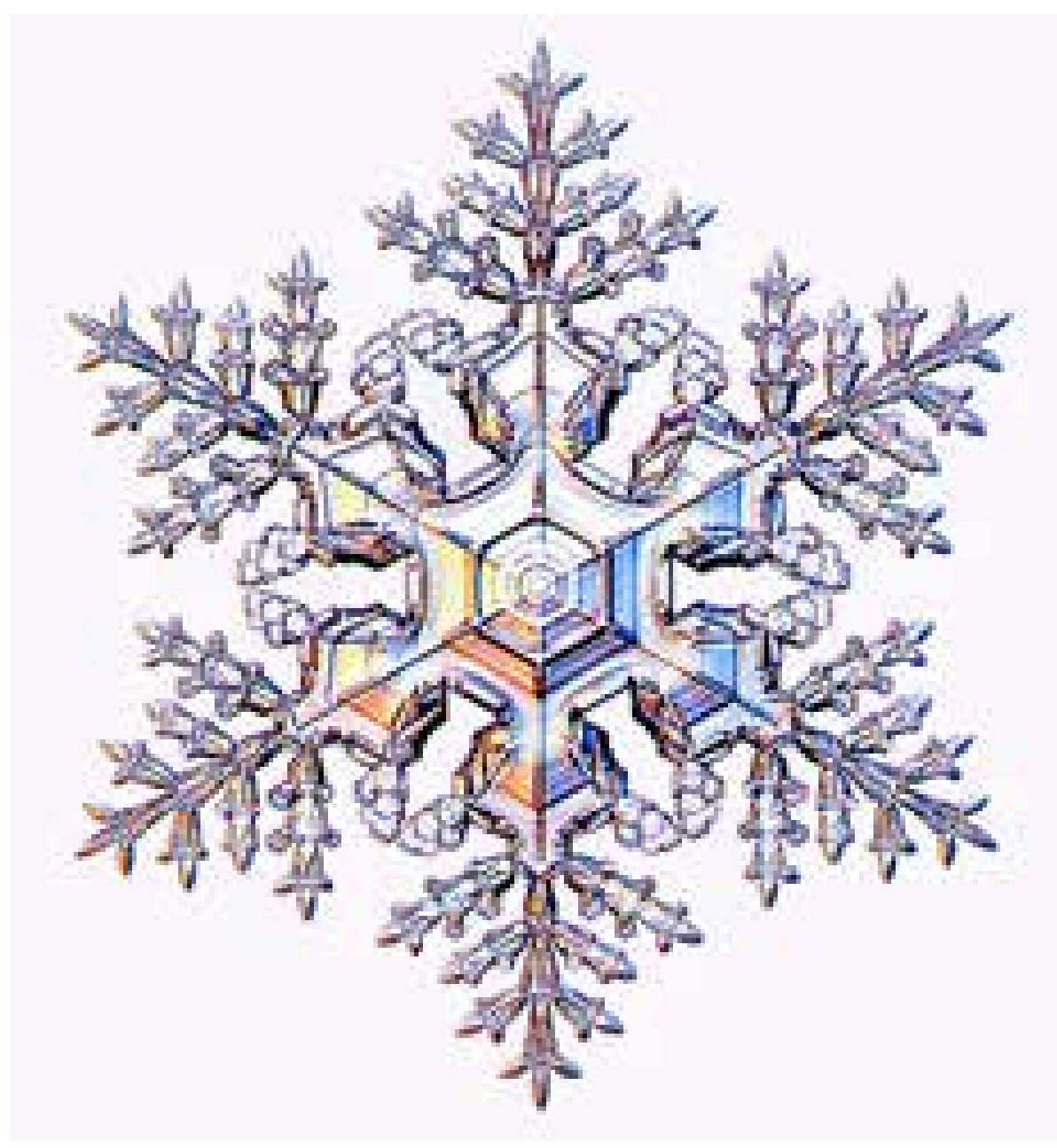

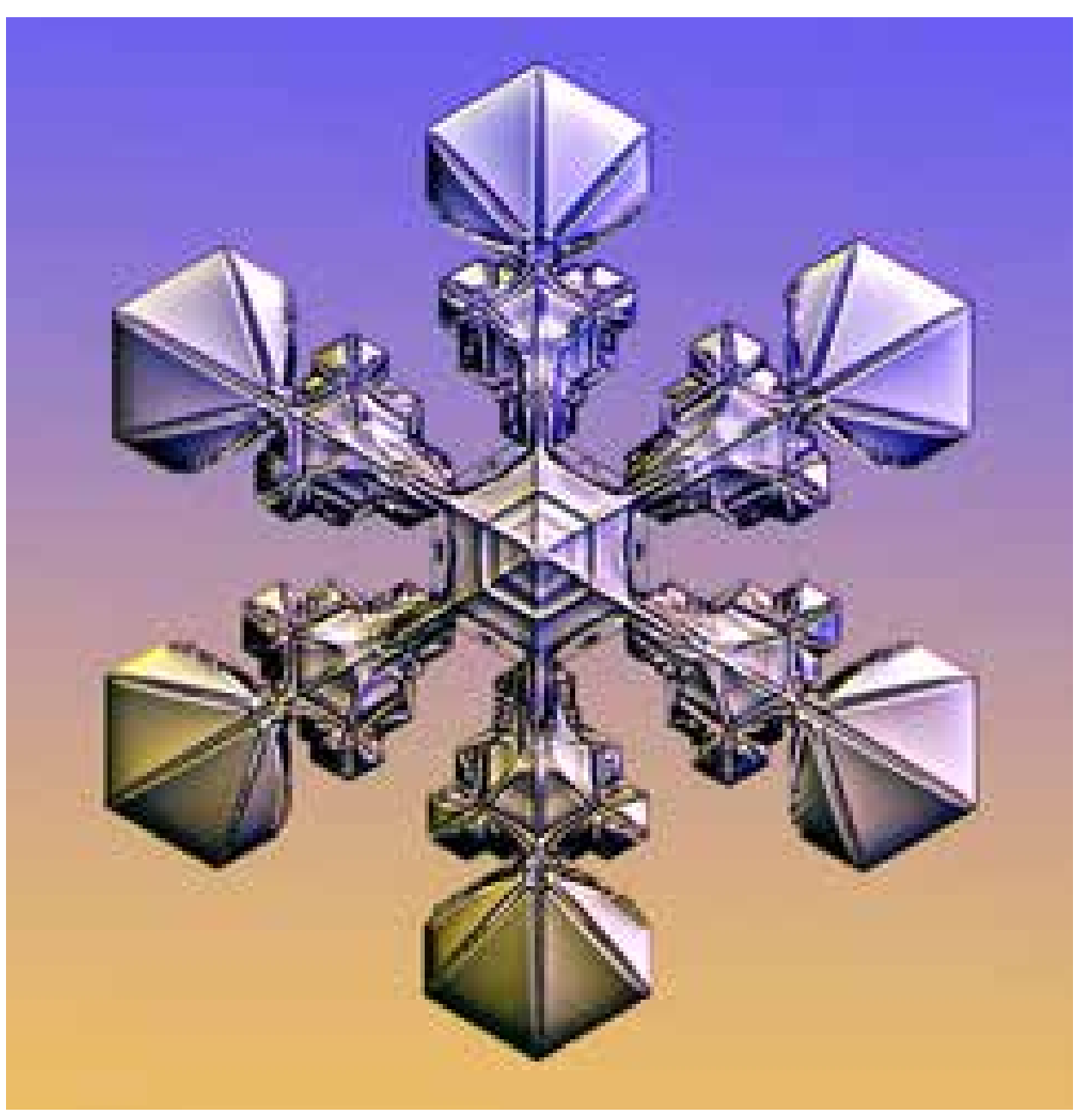

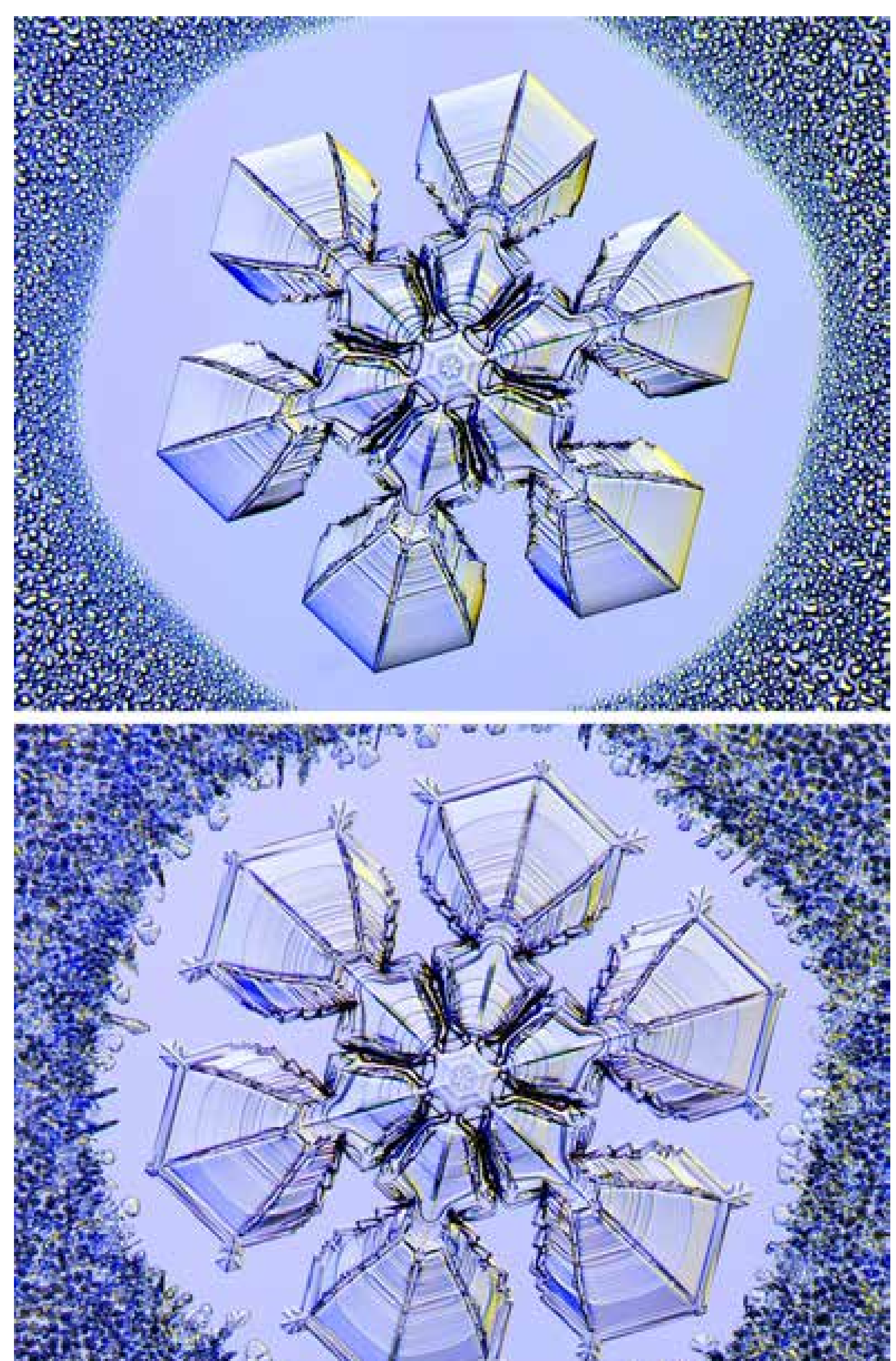

**Figure 9.41: (Above and facing page) In these sets of images, the top photos show PoP crystals surrounded by fields of water droplets, taken soon before the droplets froze. The lower images show further development of the crystals, now framed by frost, after the air flow was substantially increased to boost the supersaturation.**

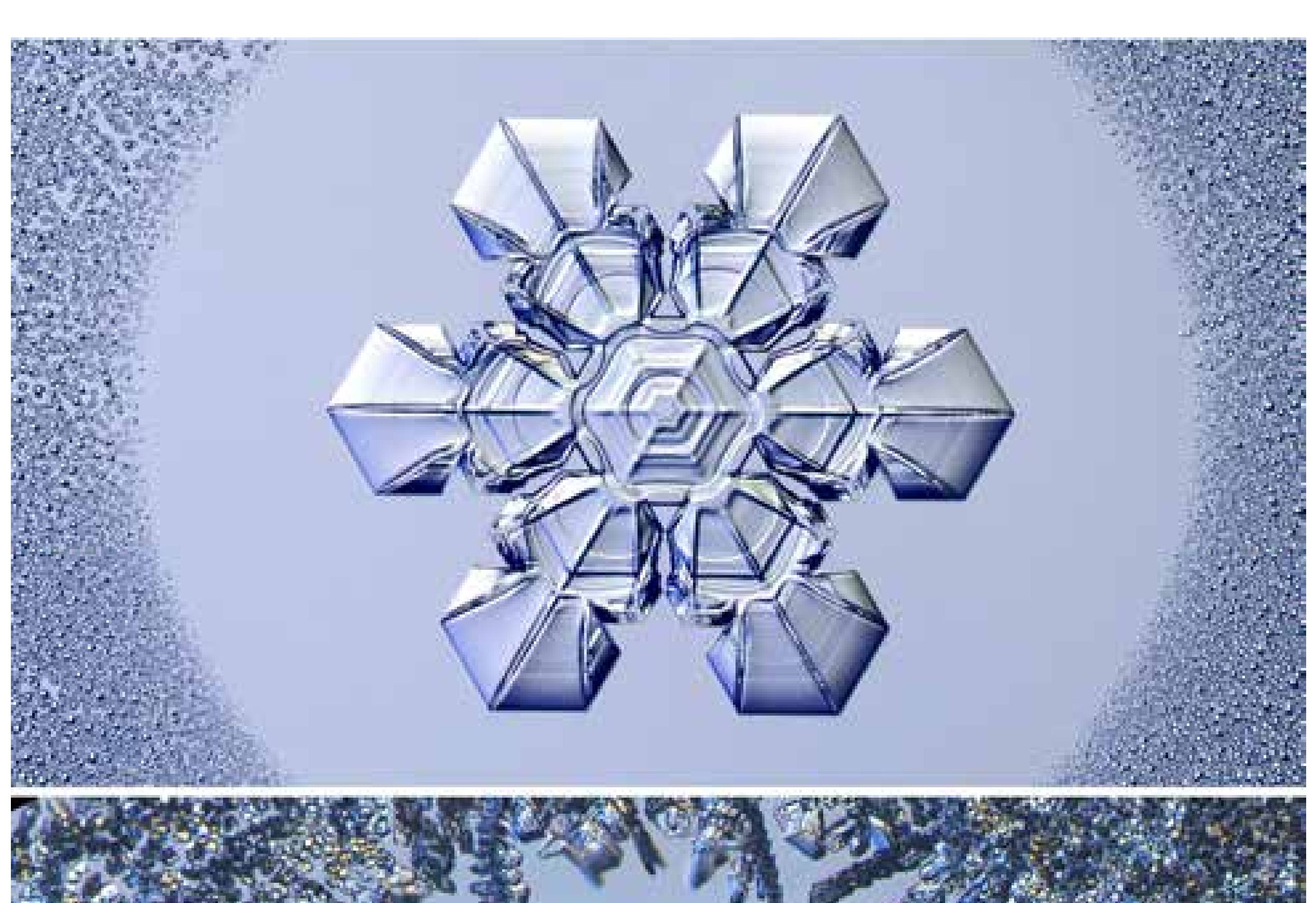

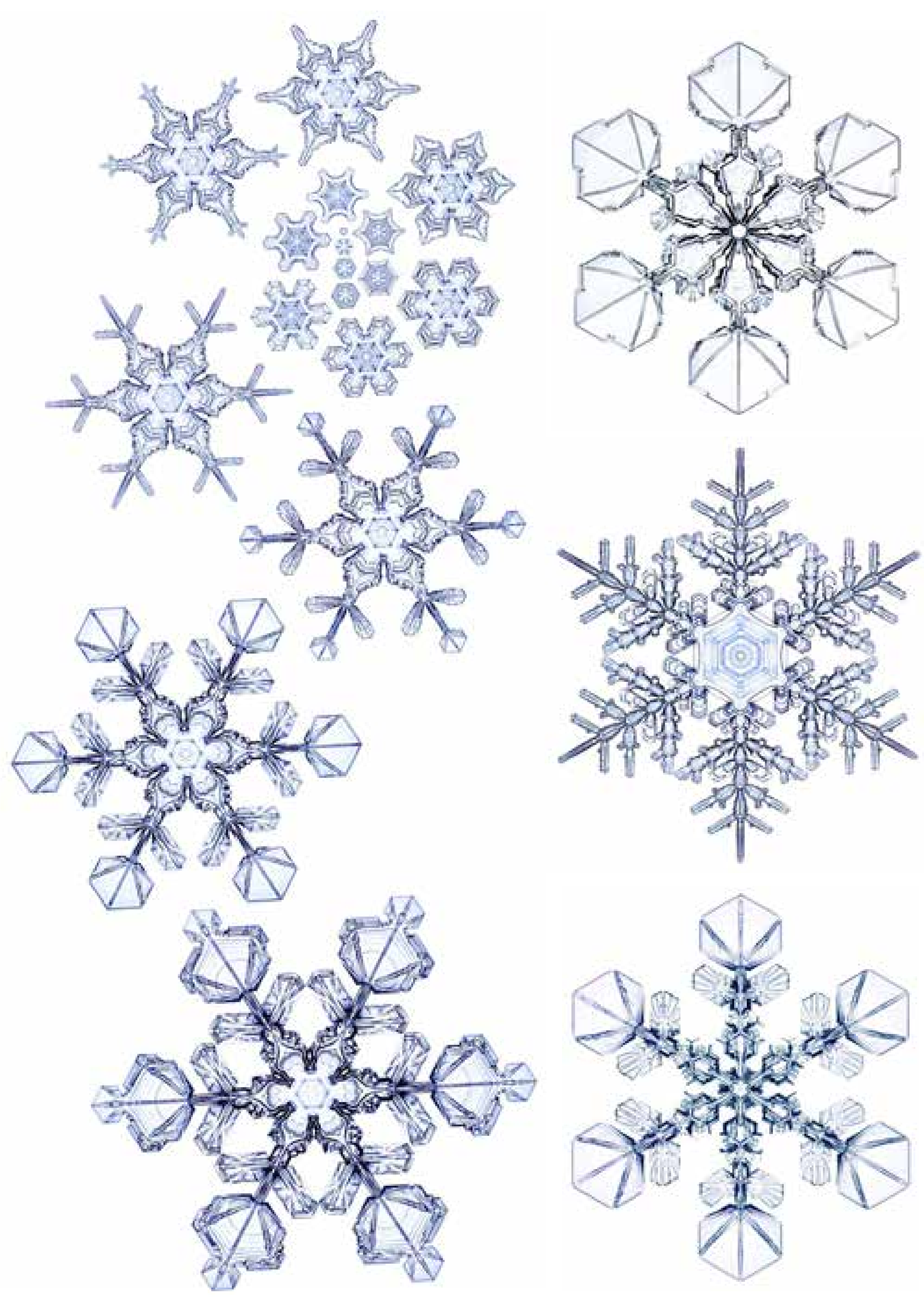

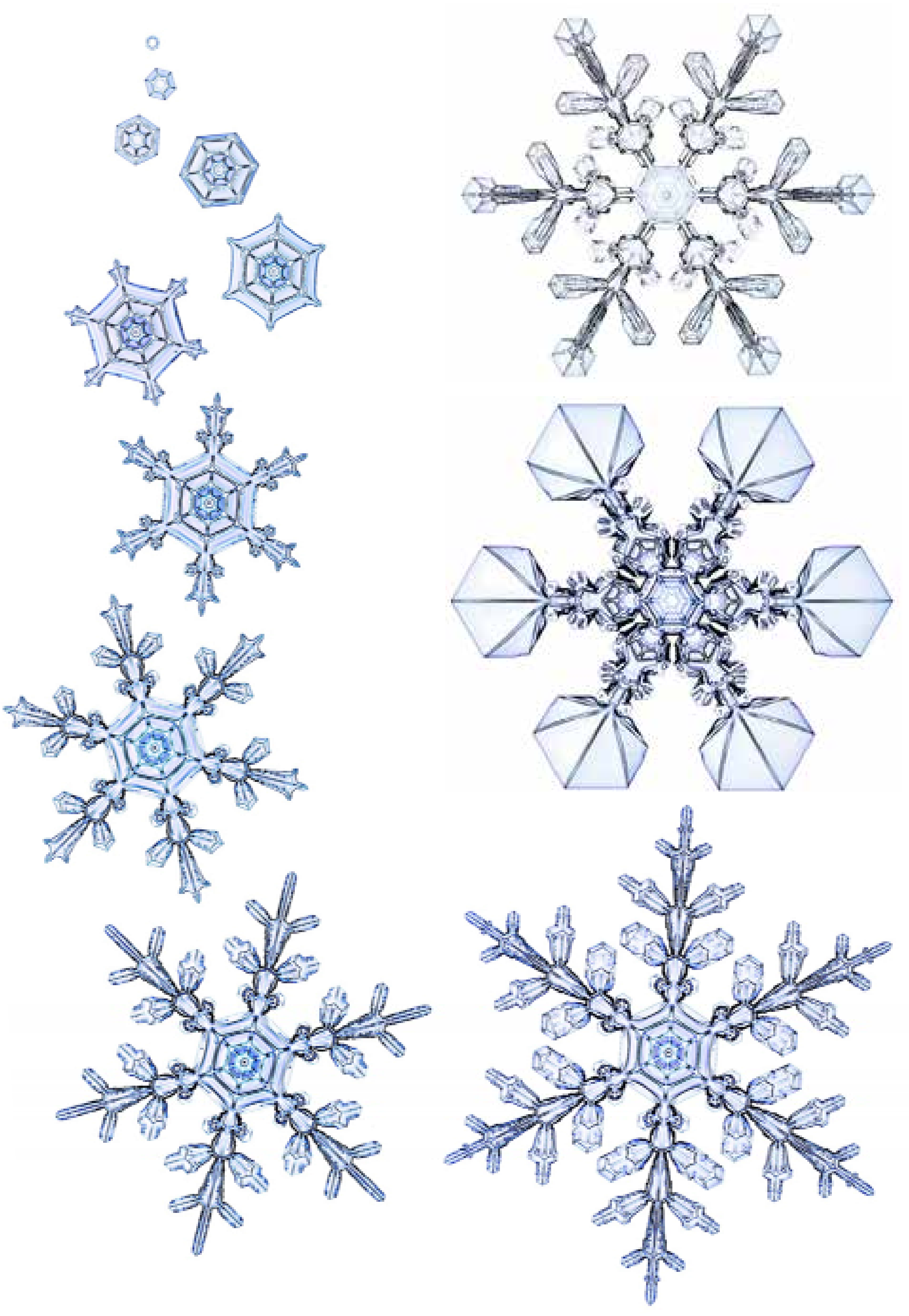

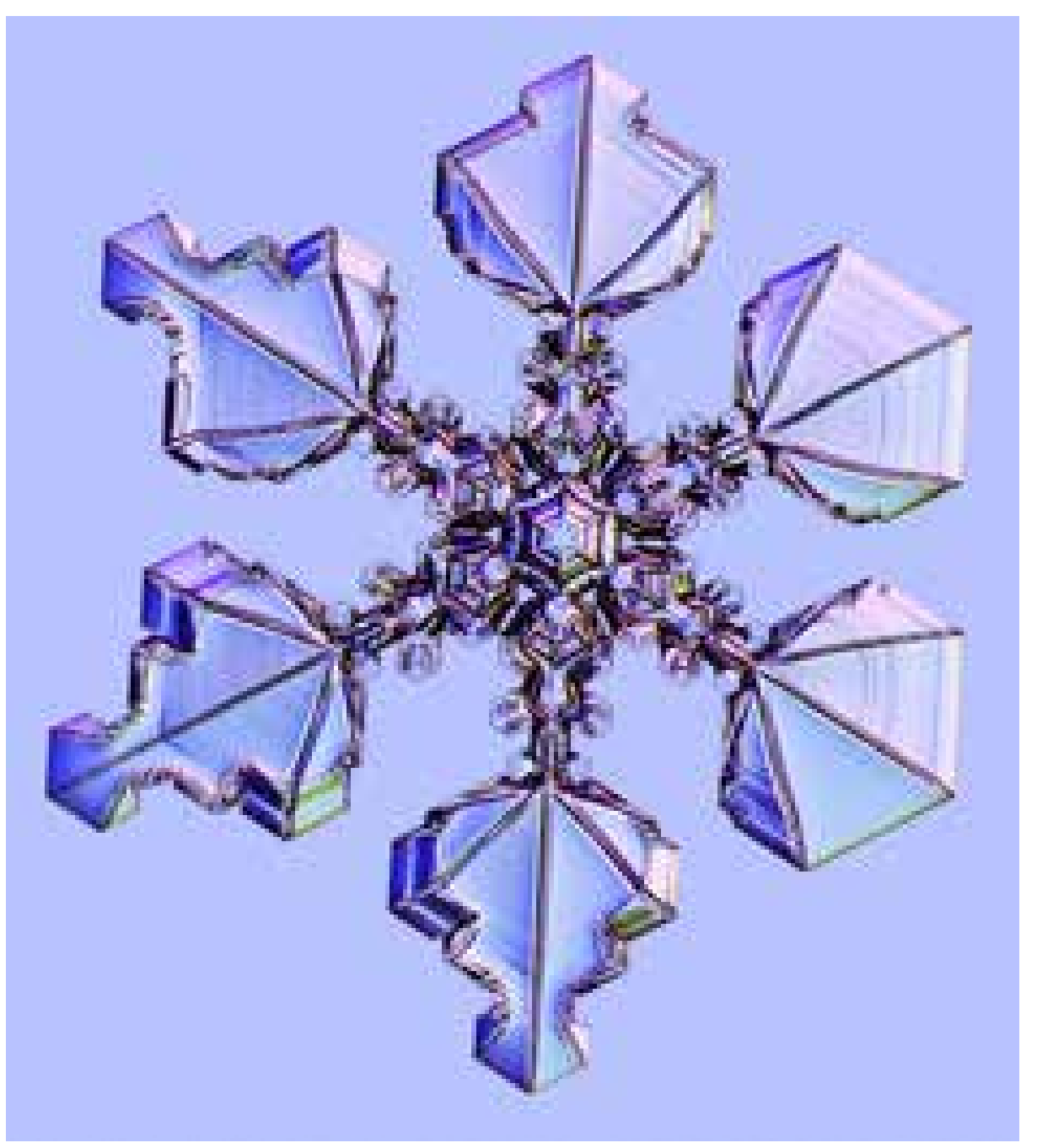

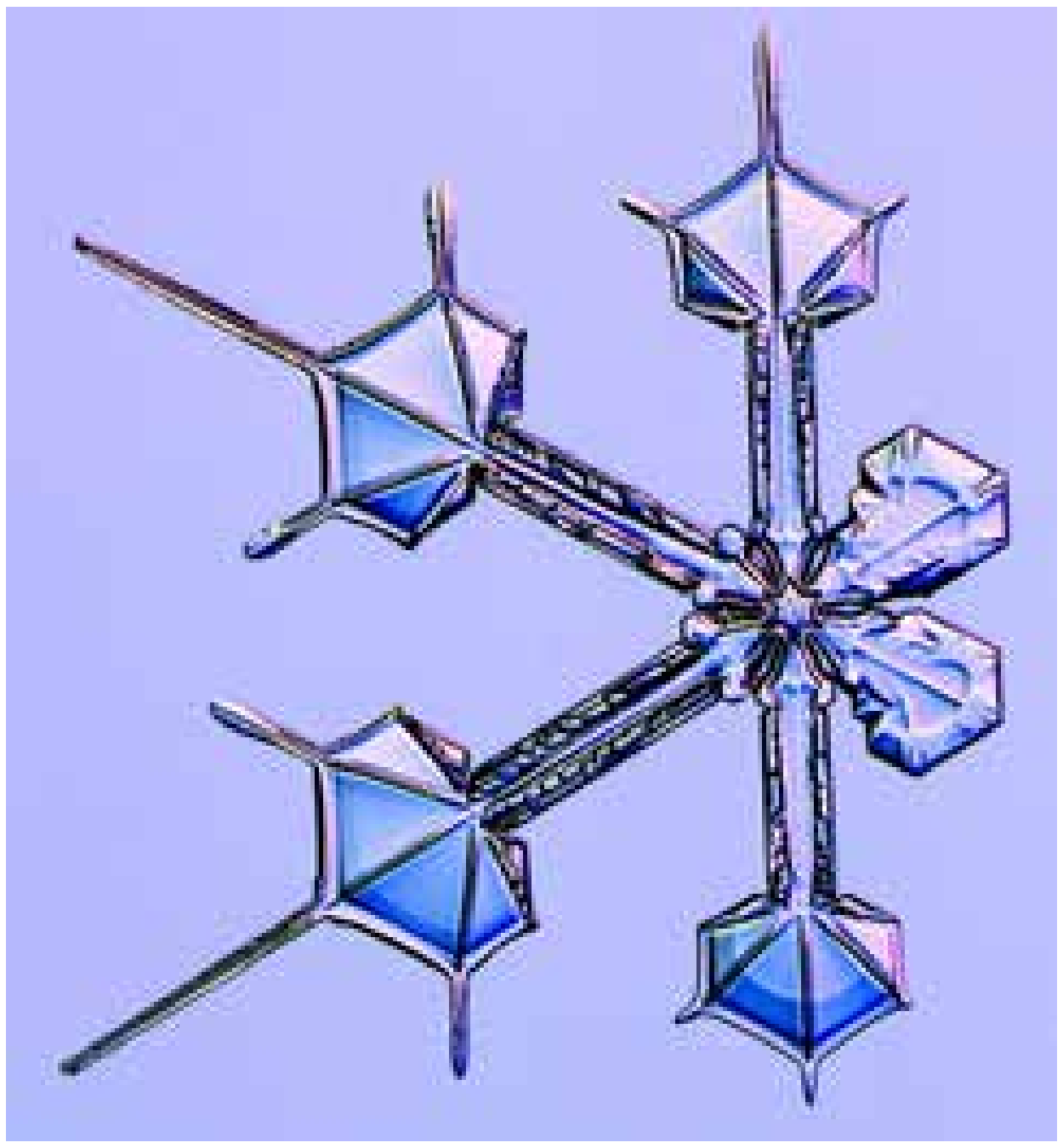

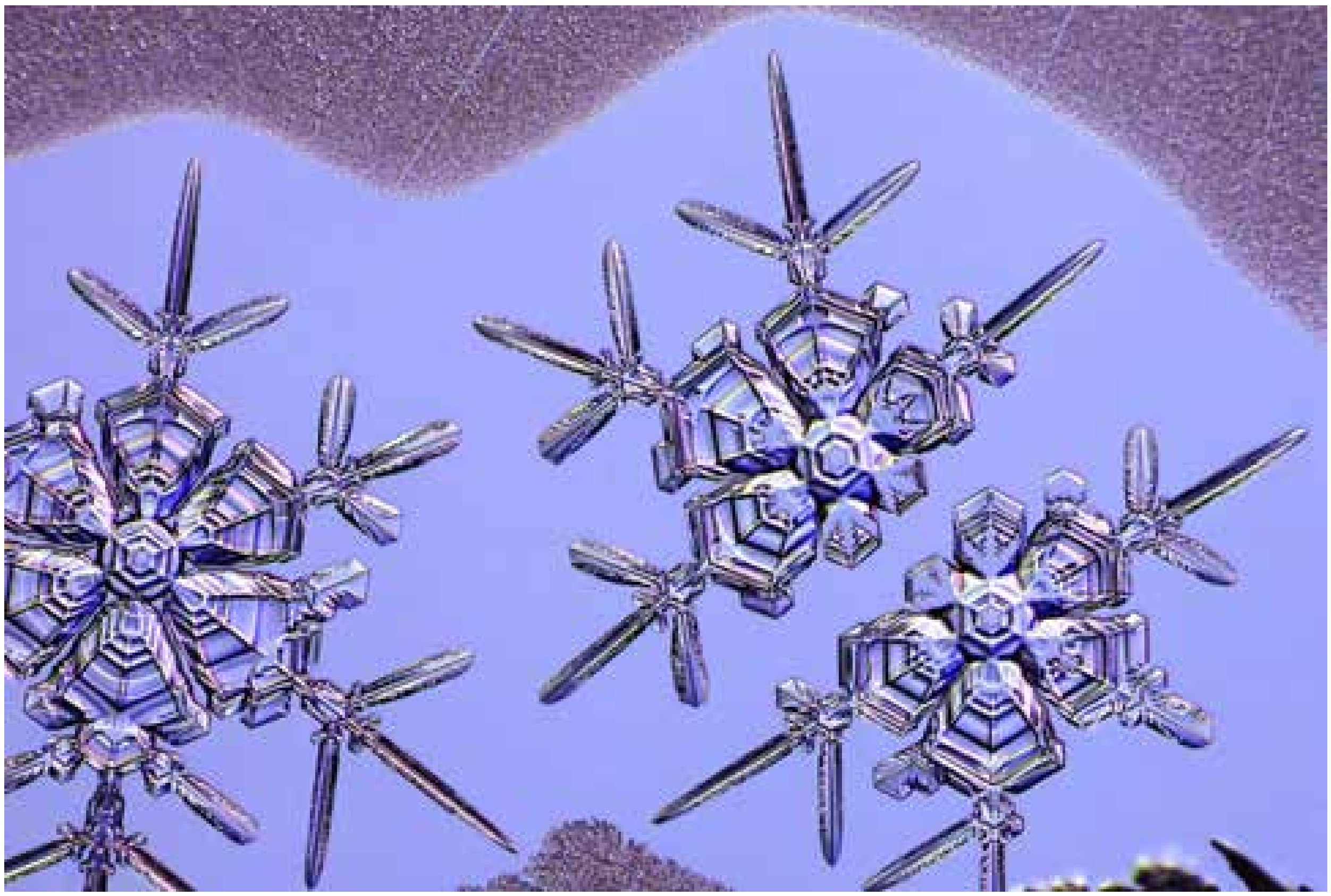

**Figure 9.42: (Above and facing page) Asymmetry all around. When two or more PoP snow crystals grow in close proximity, they compete for the supply of water vapor in their vicinity, which can lead to quite a variety of odd asymmetrical crystals. For most of these photos, I focused my attention on an isolated crystal at the center of the camera's field of view, which grew with good symmetry. When finished with that crystal, I then looked outside the original field to see what else looked interesting.**

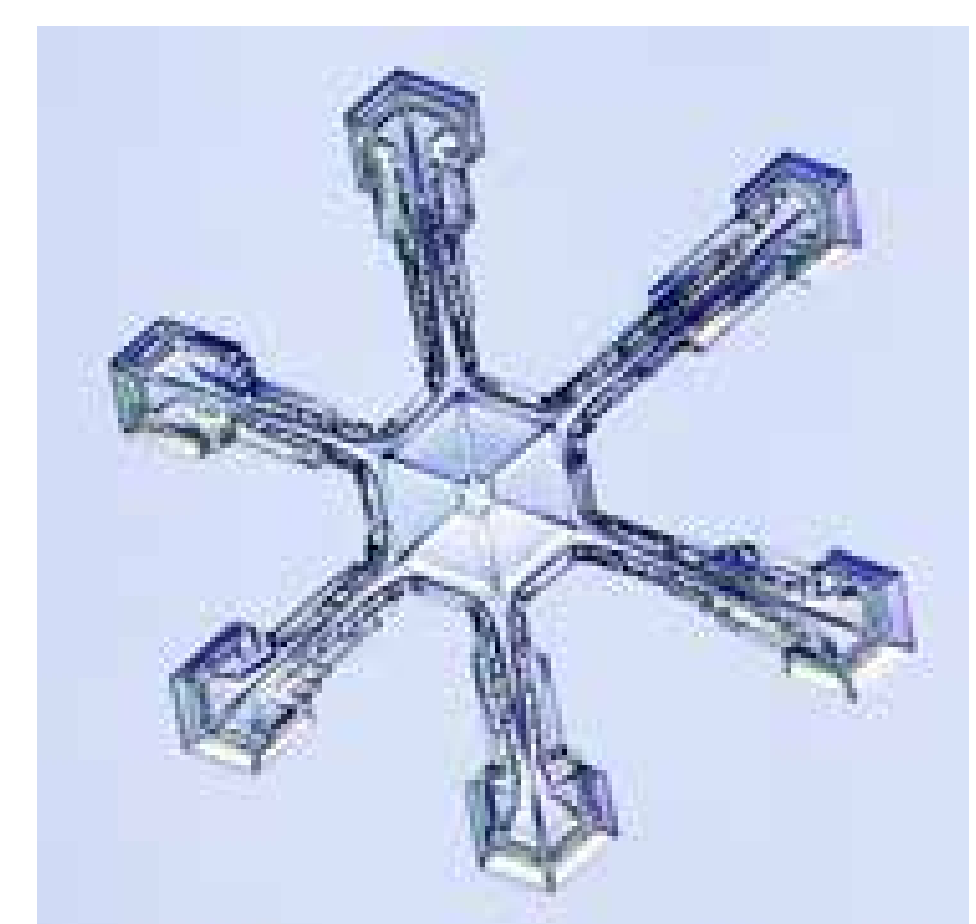
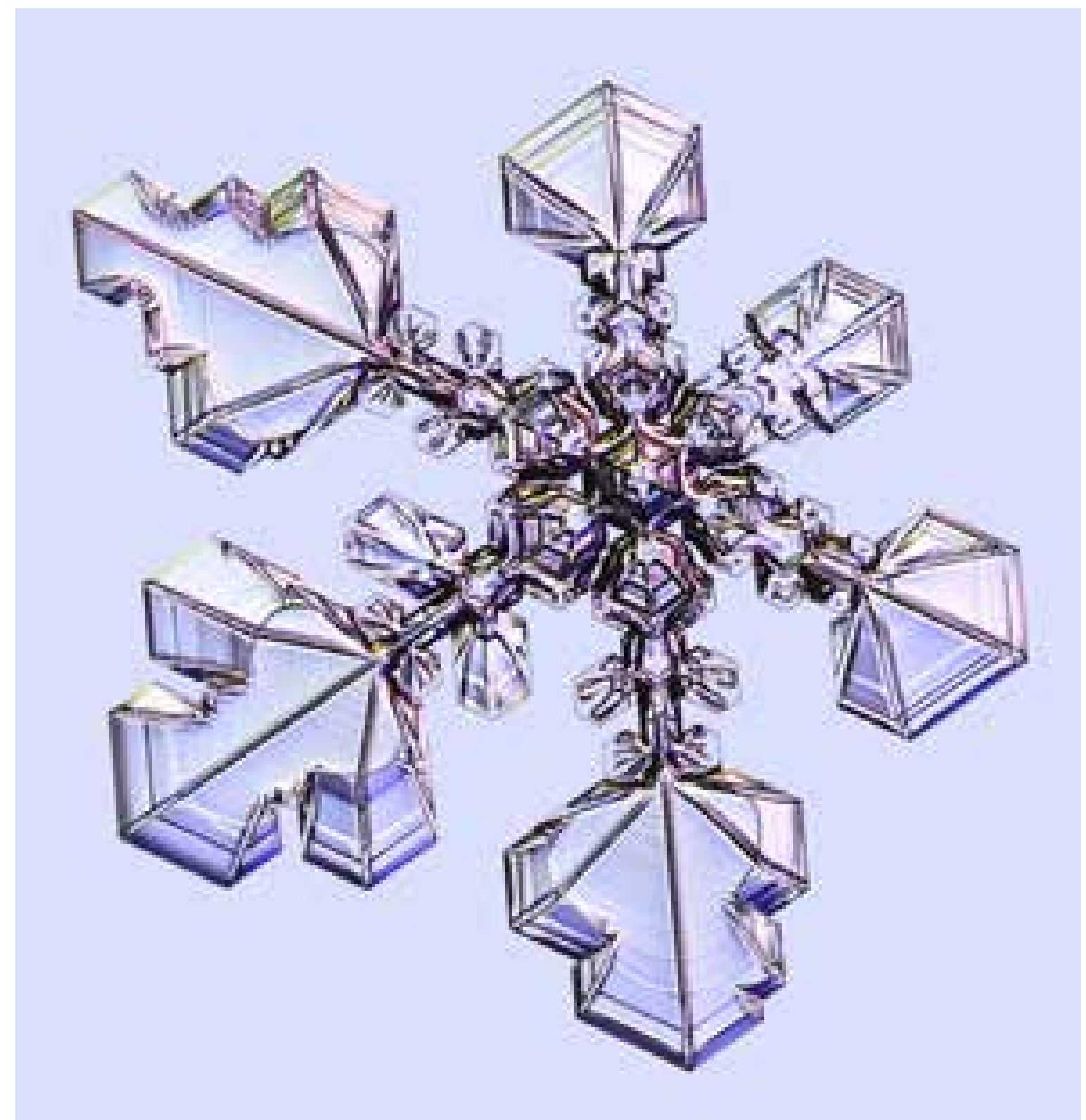
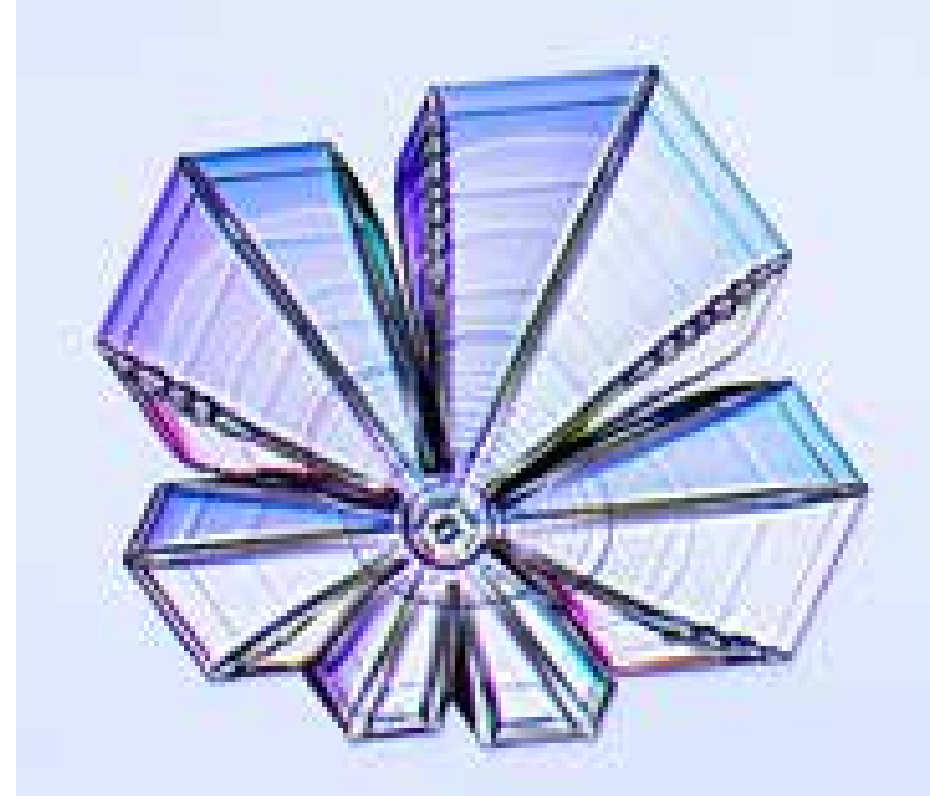
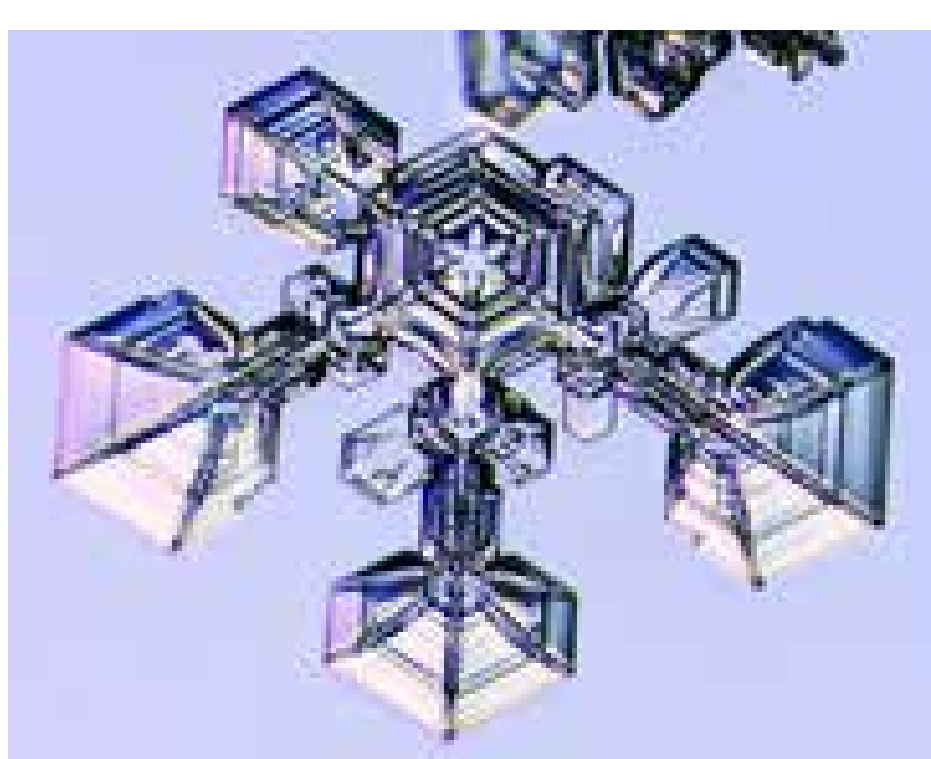
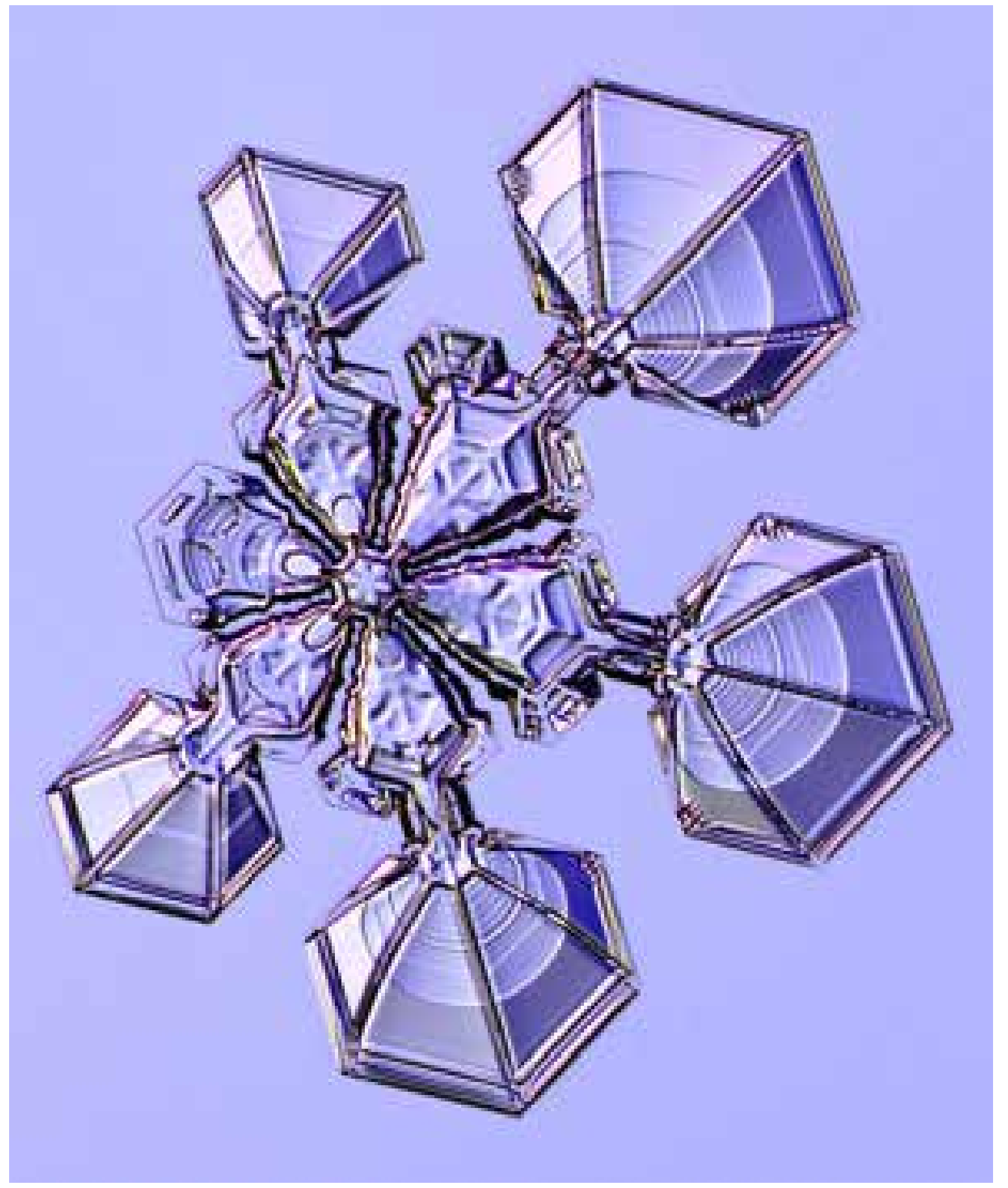
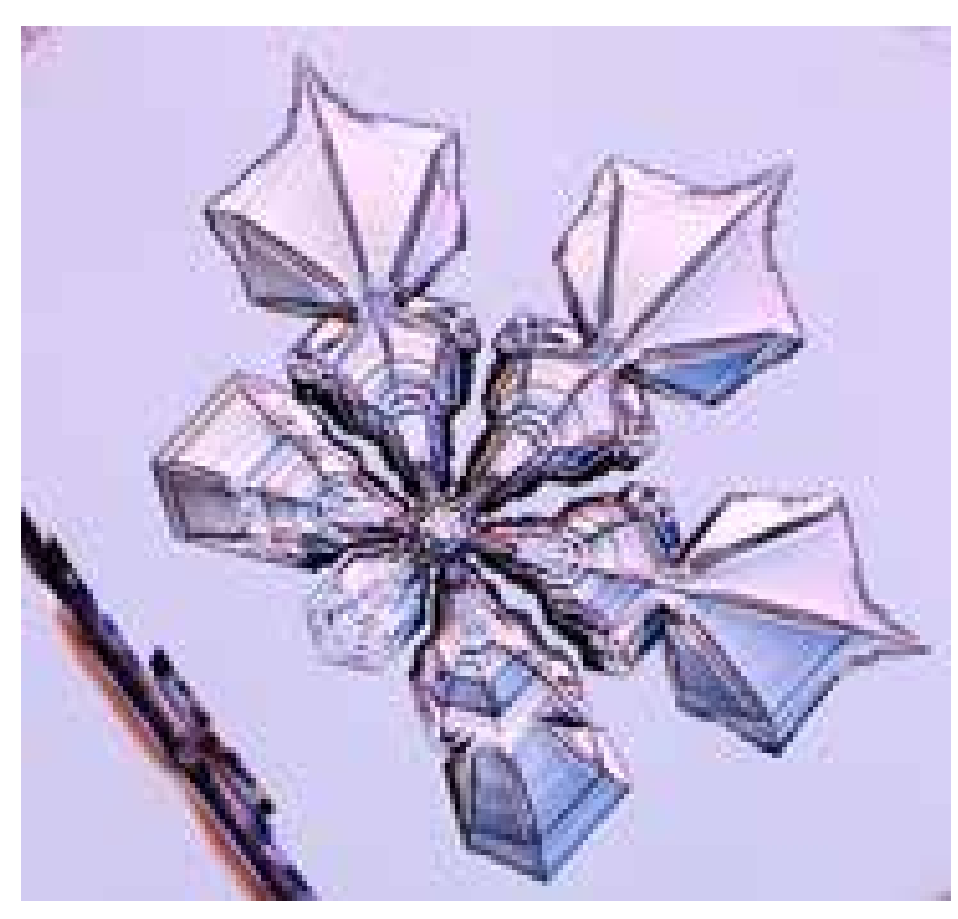

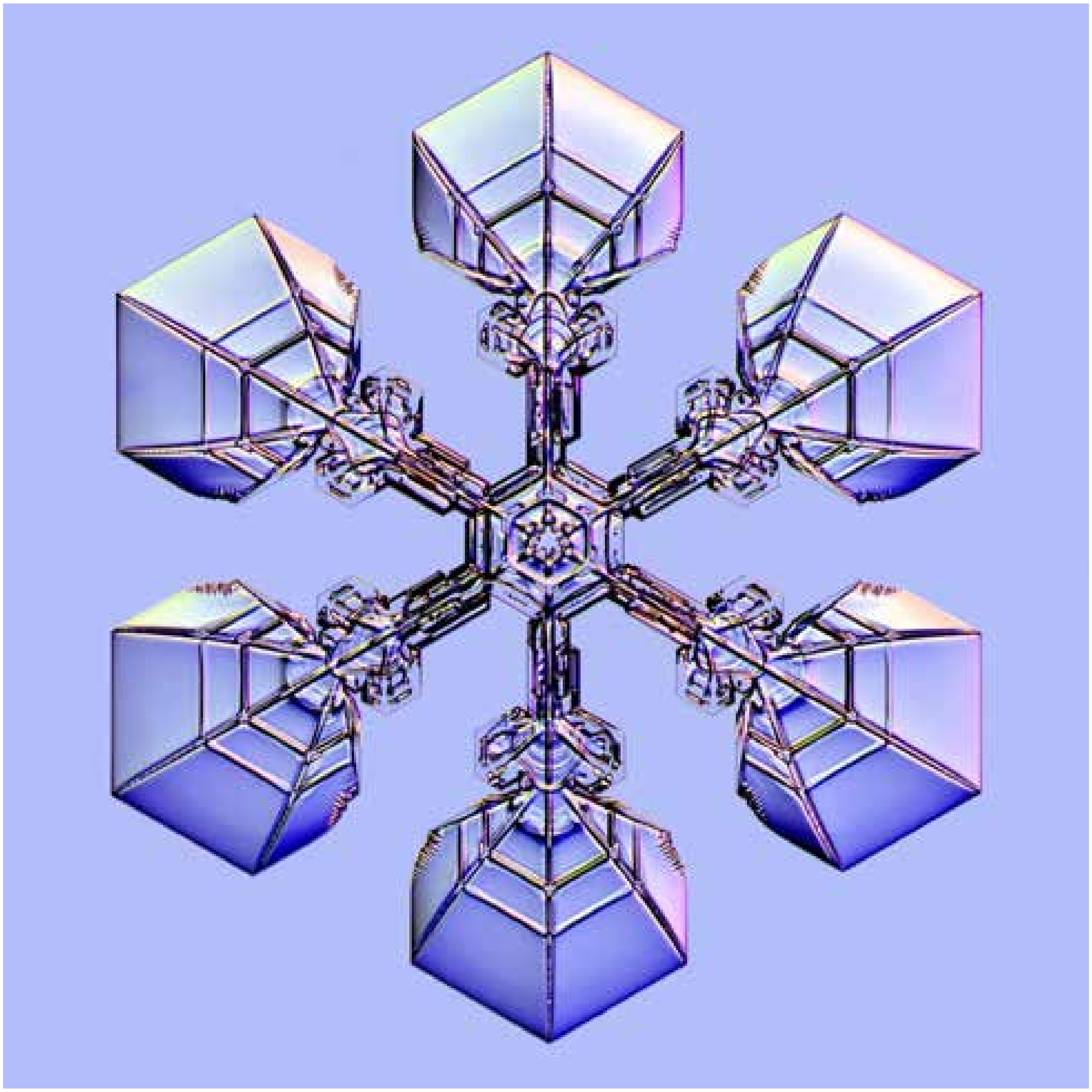

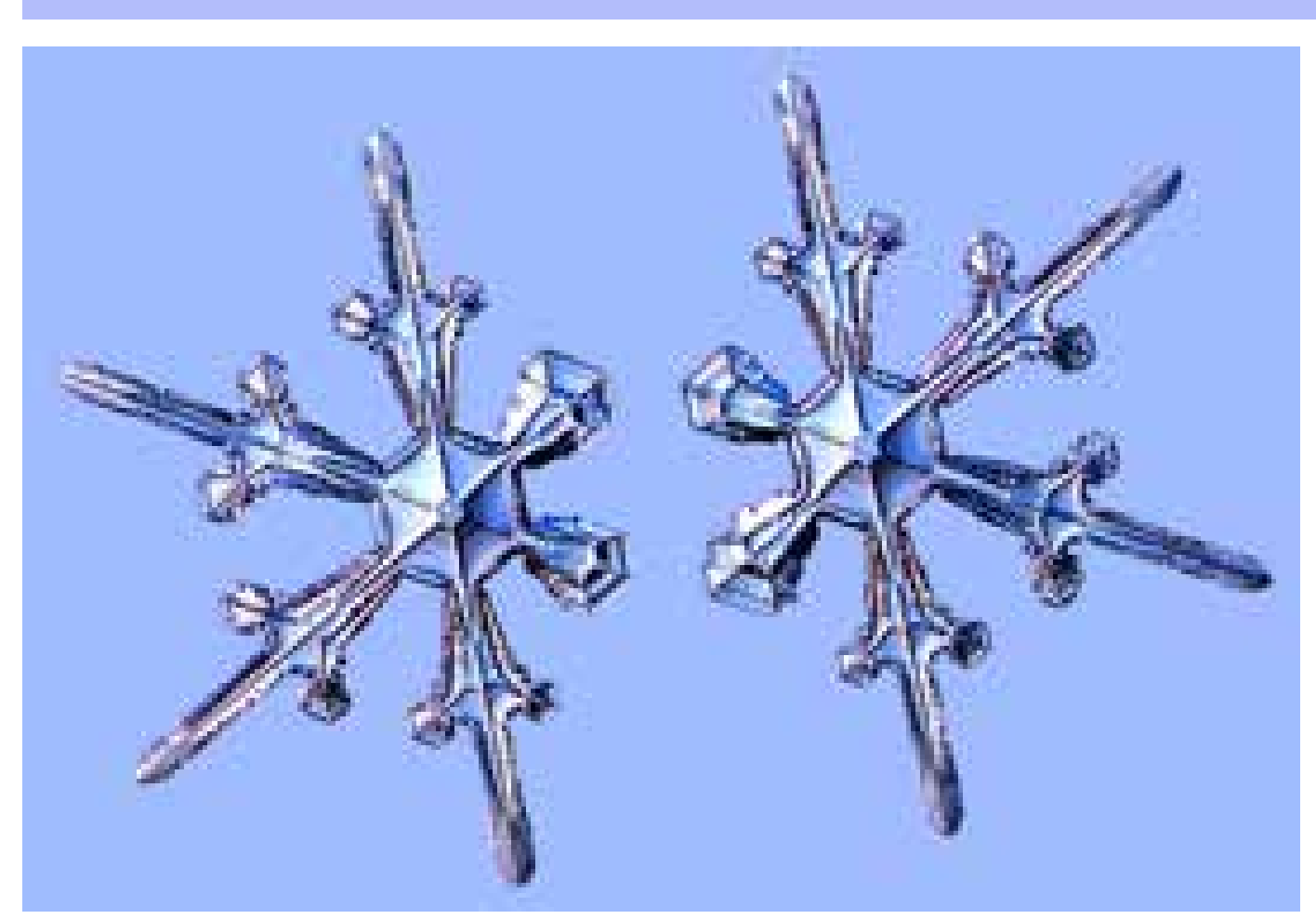

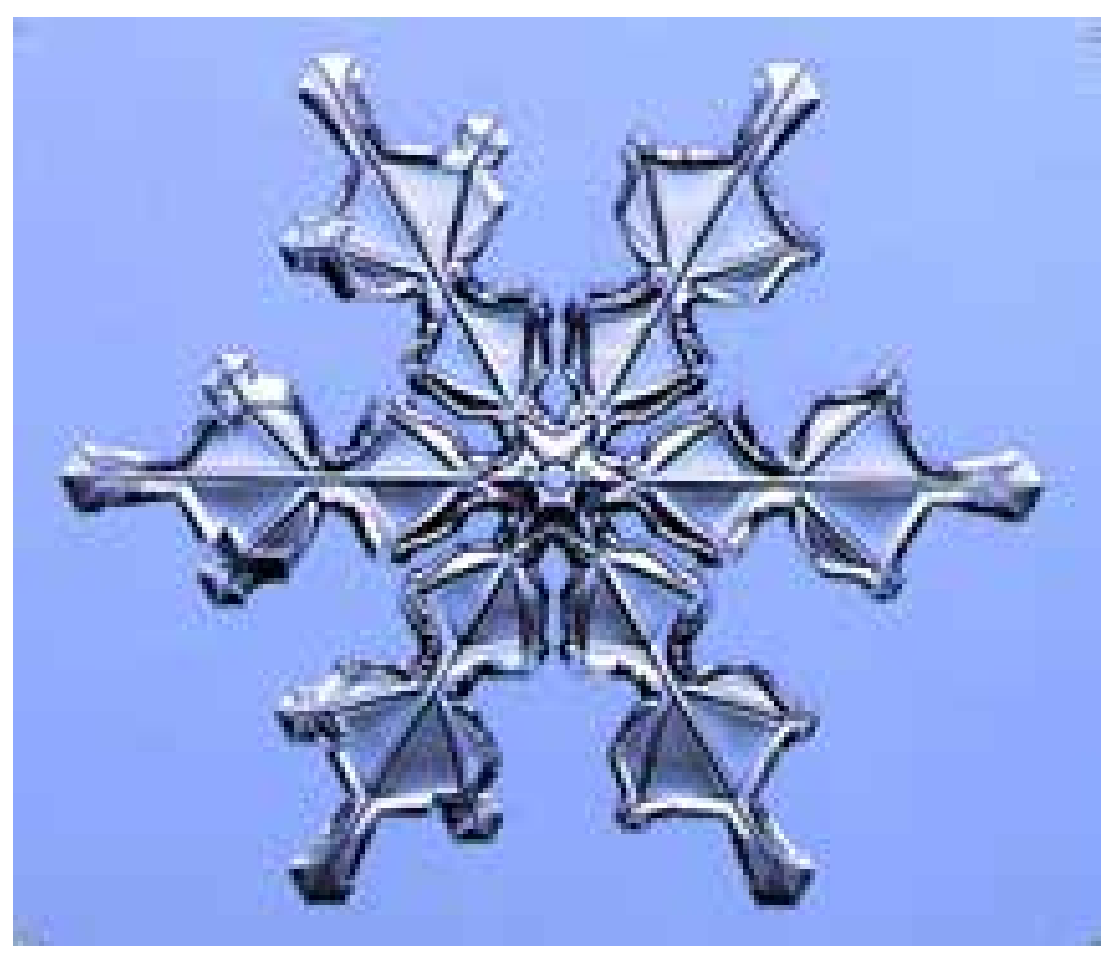

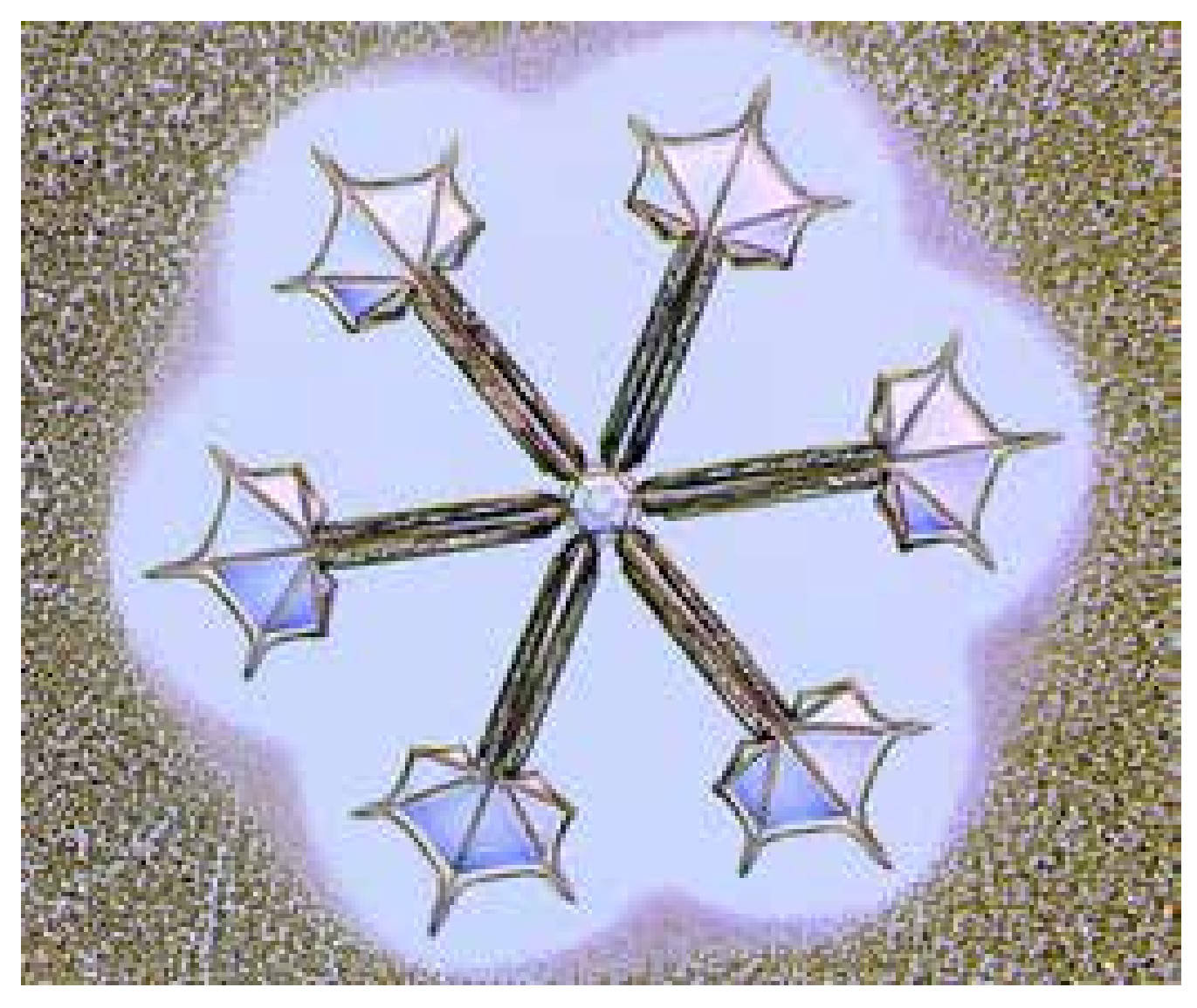
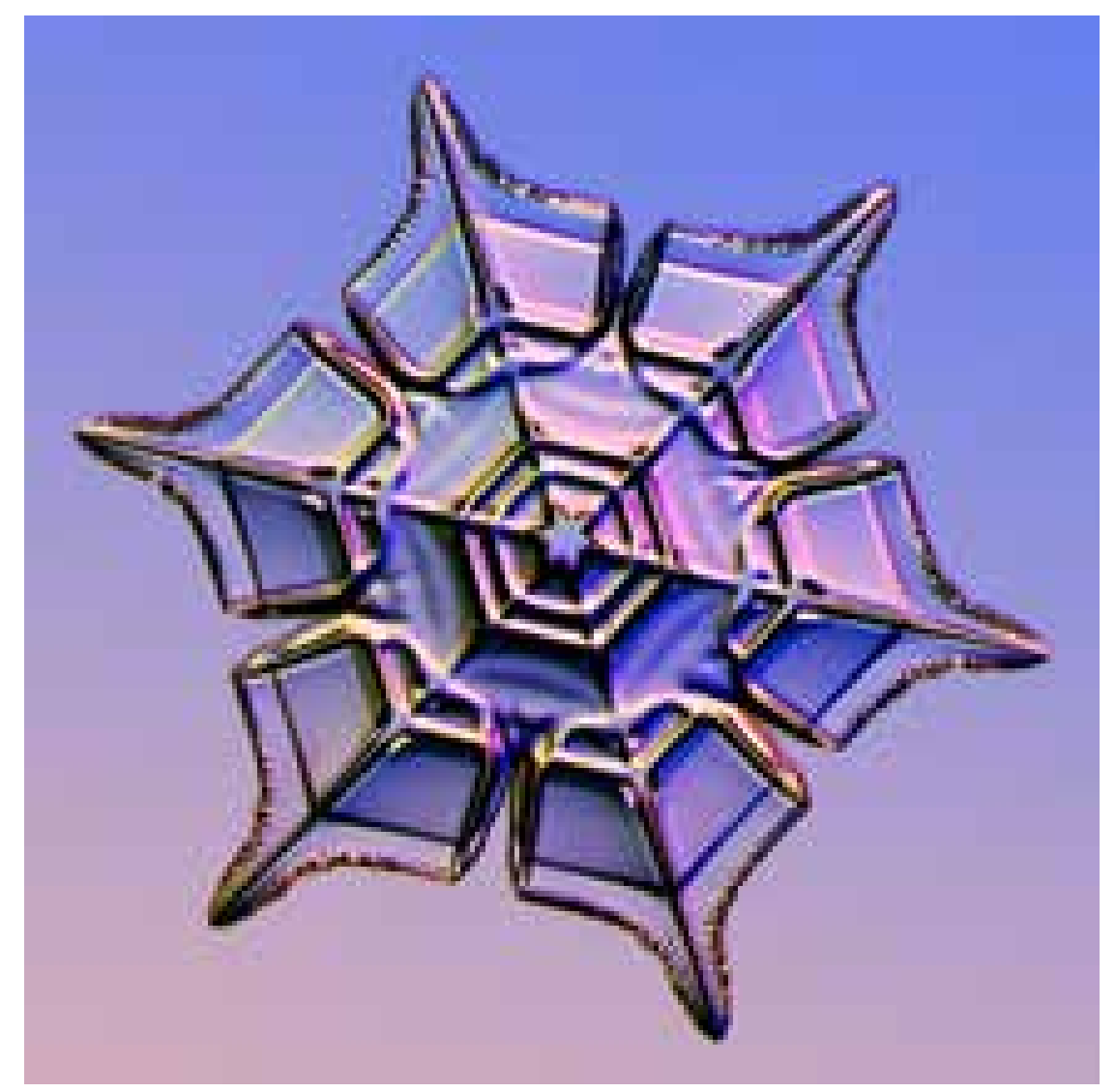
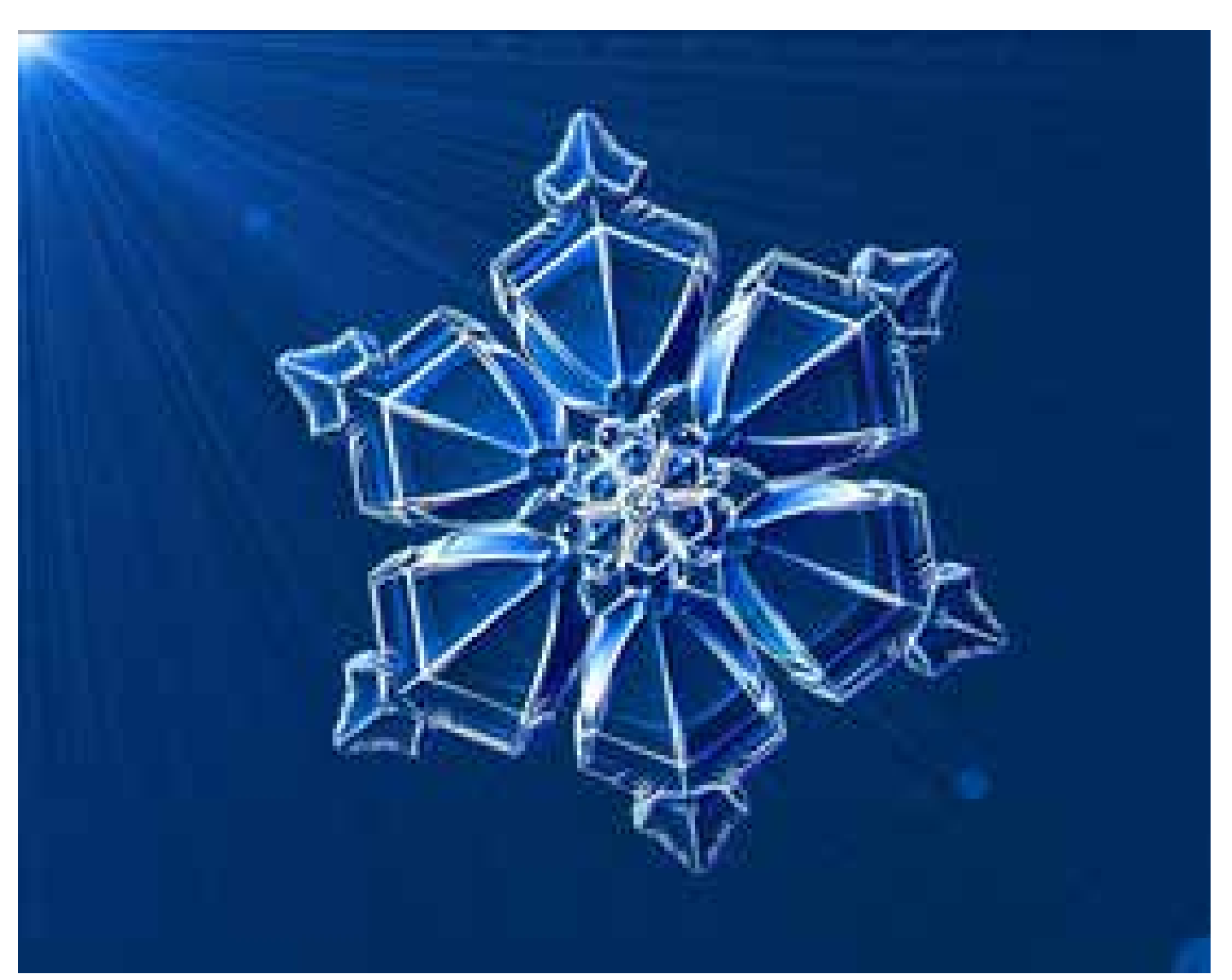
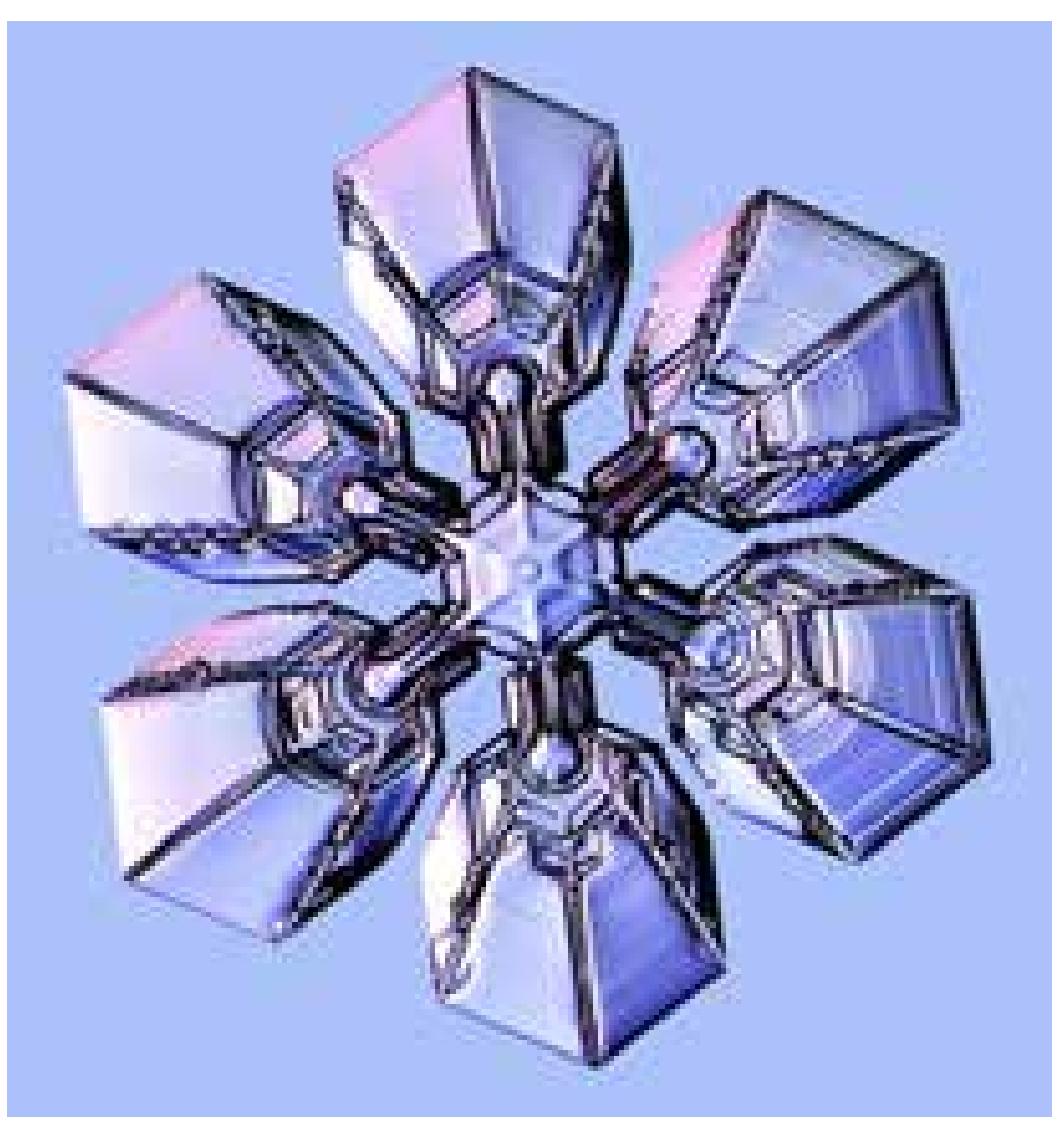
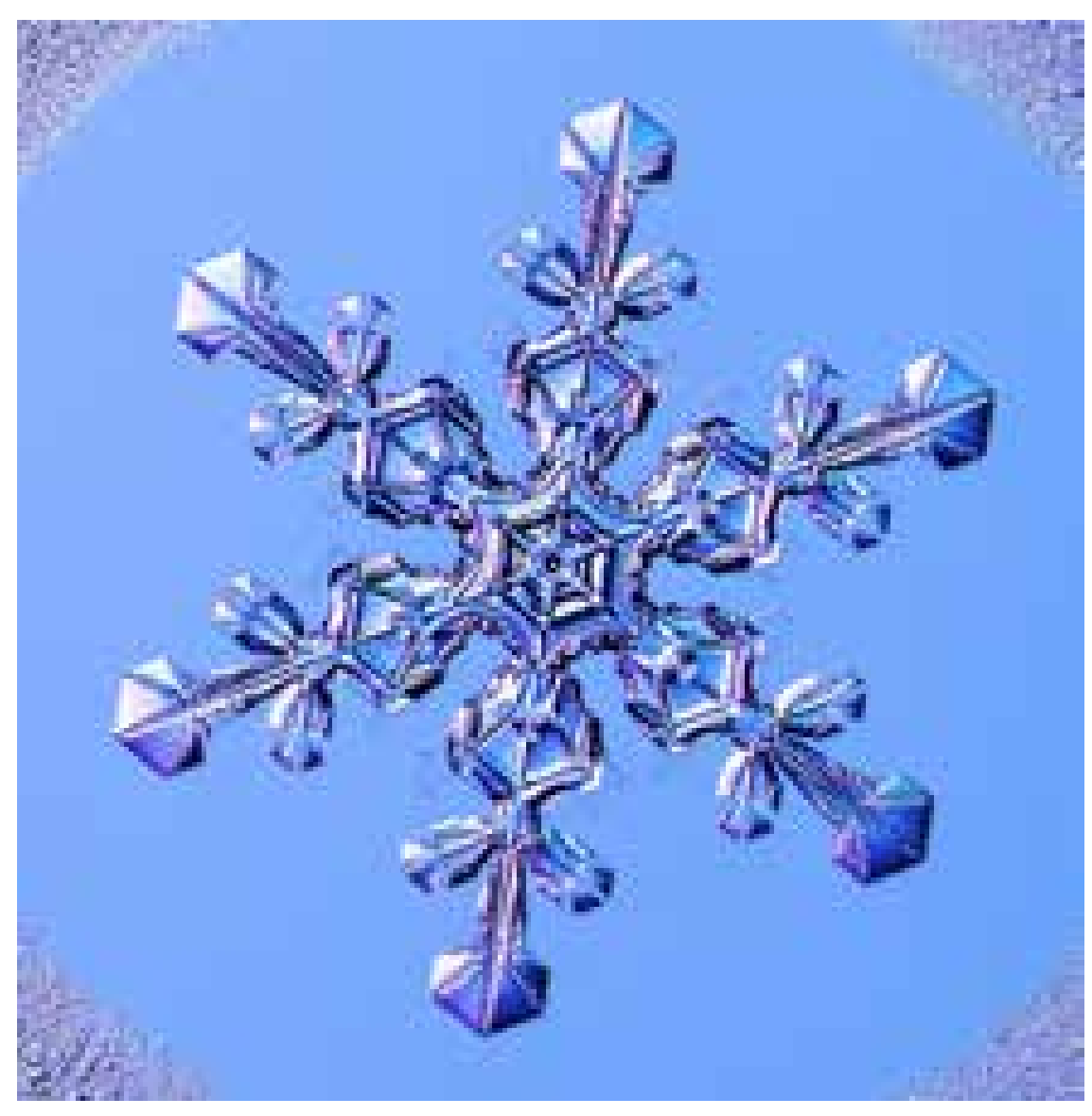
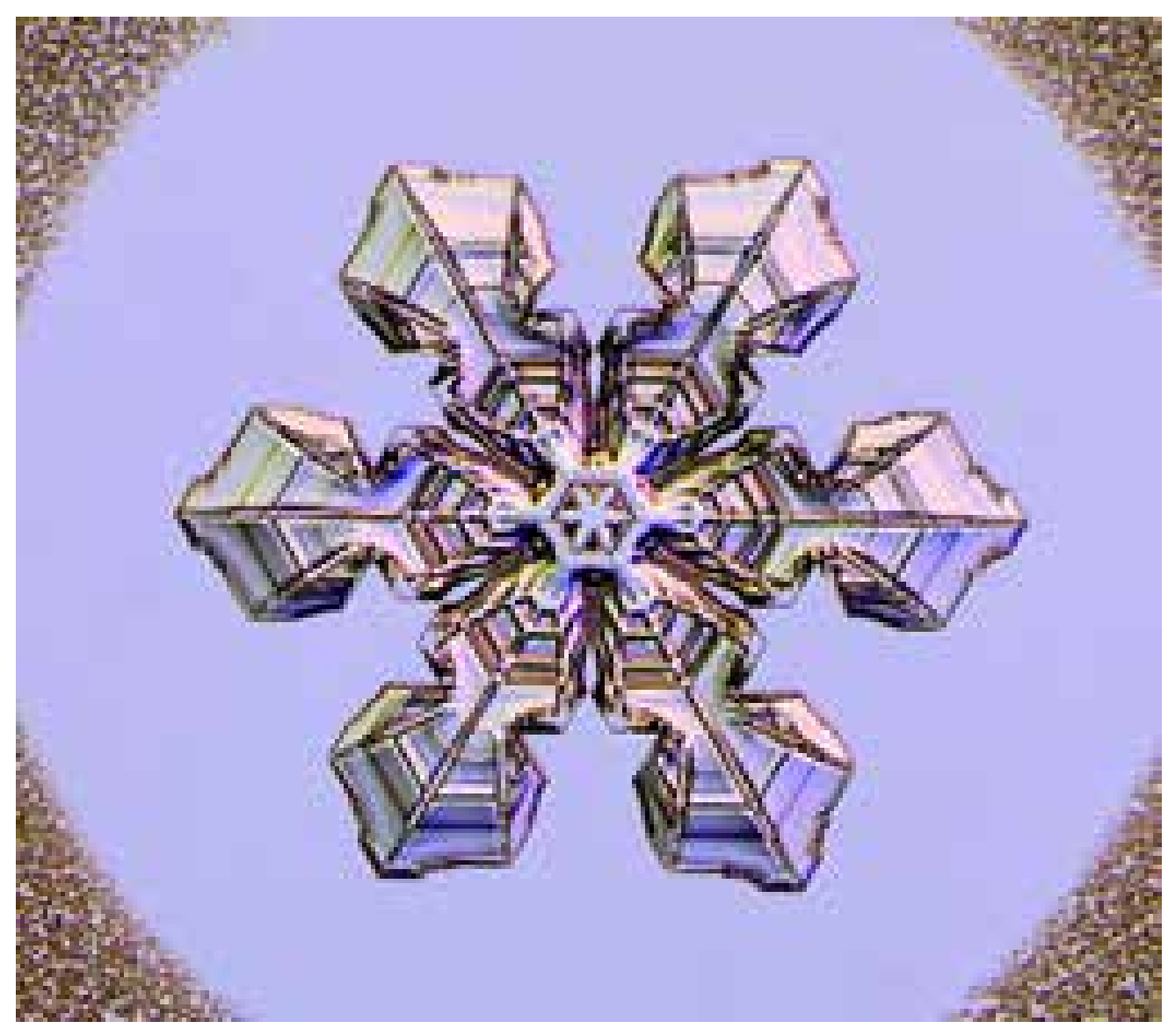

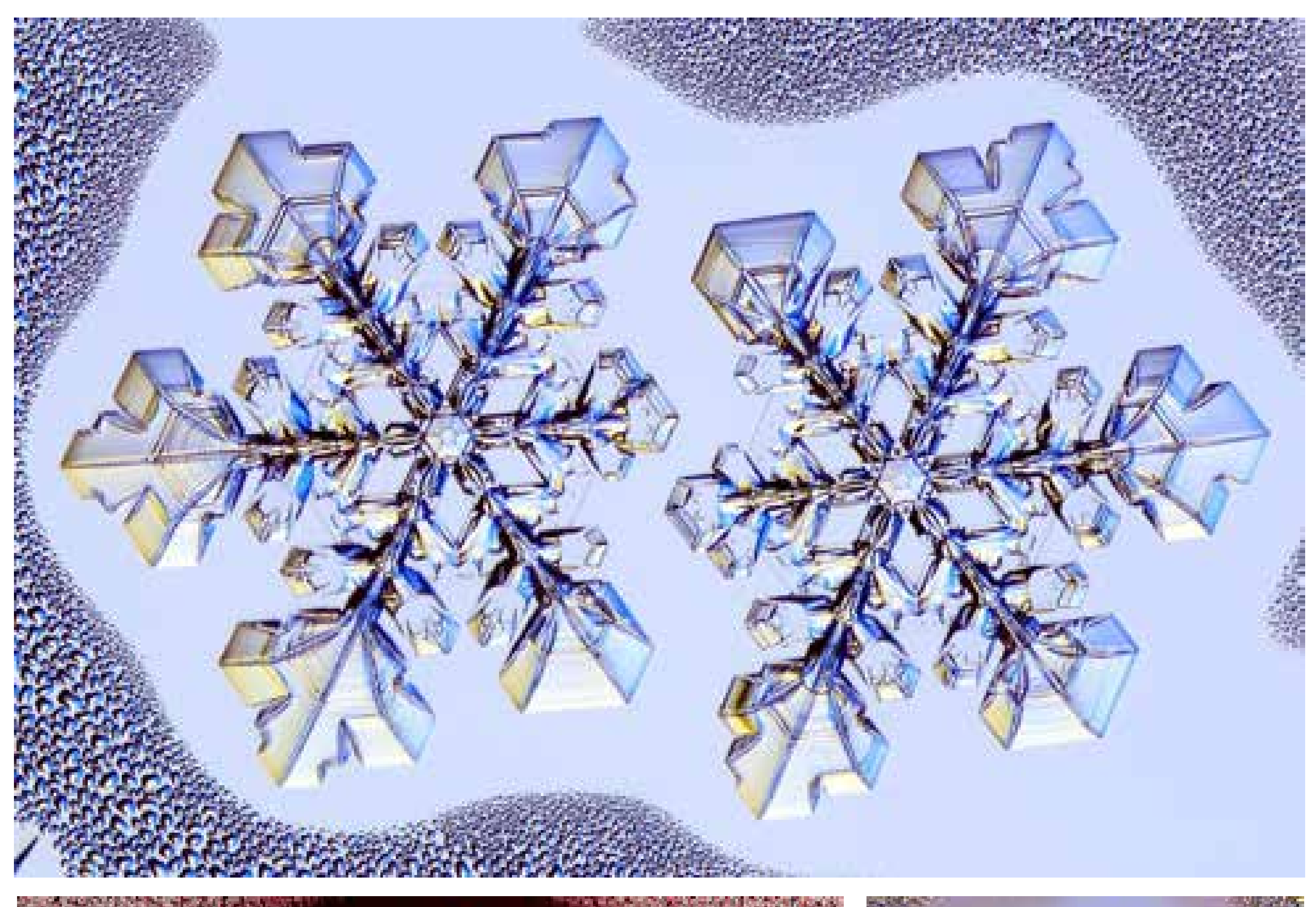

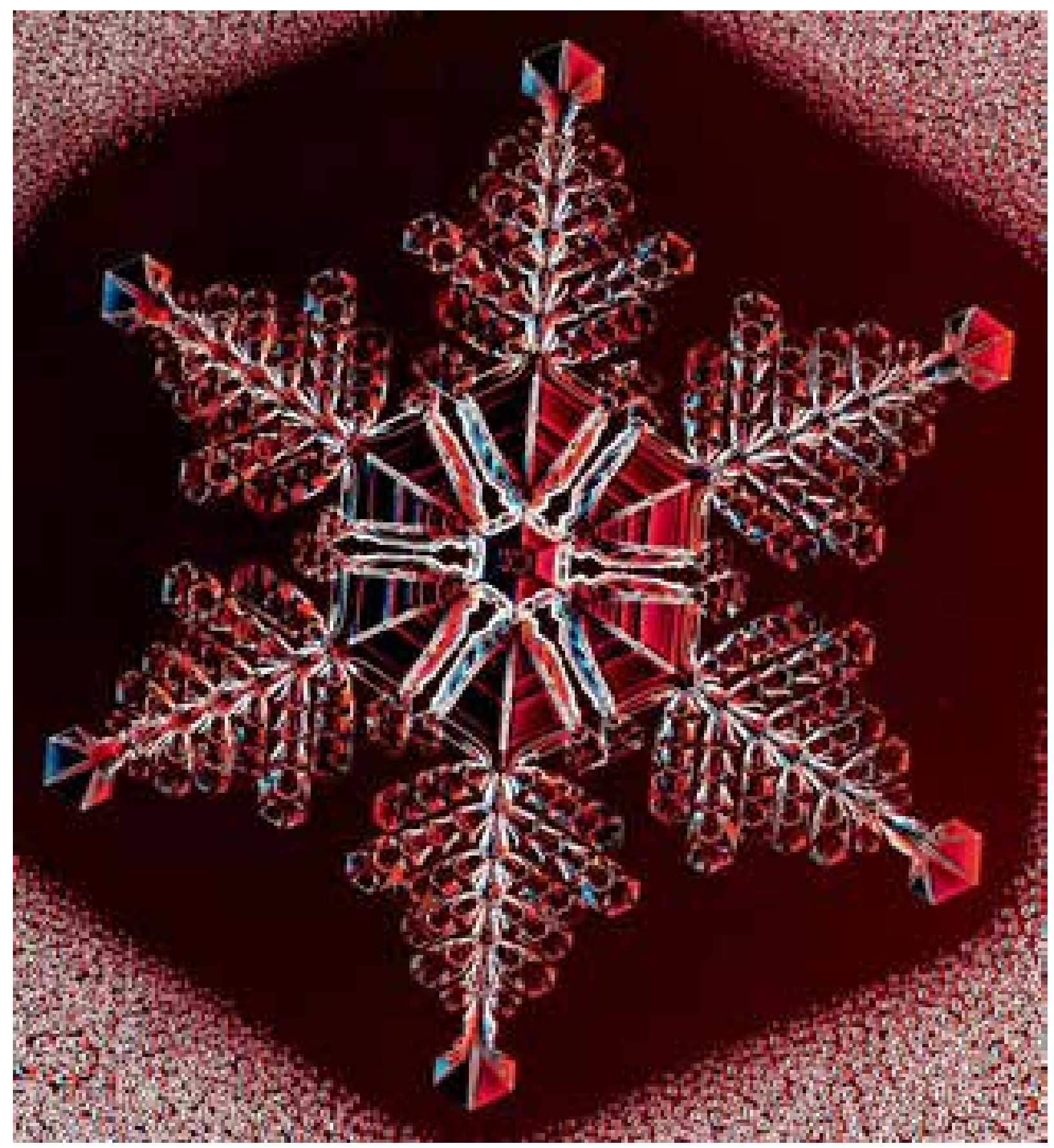

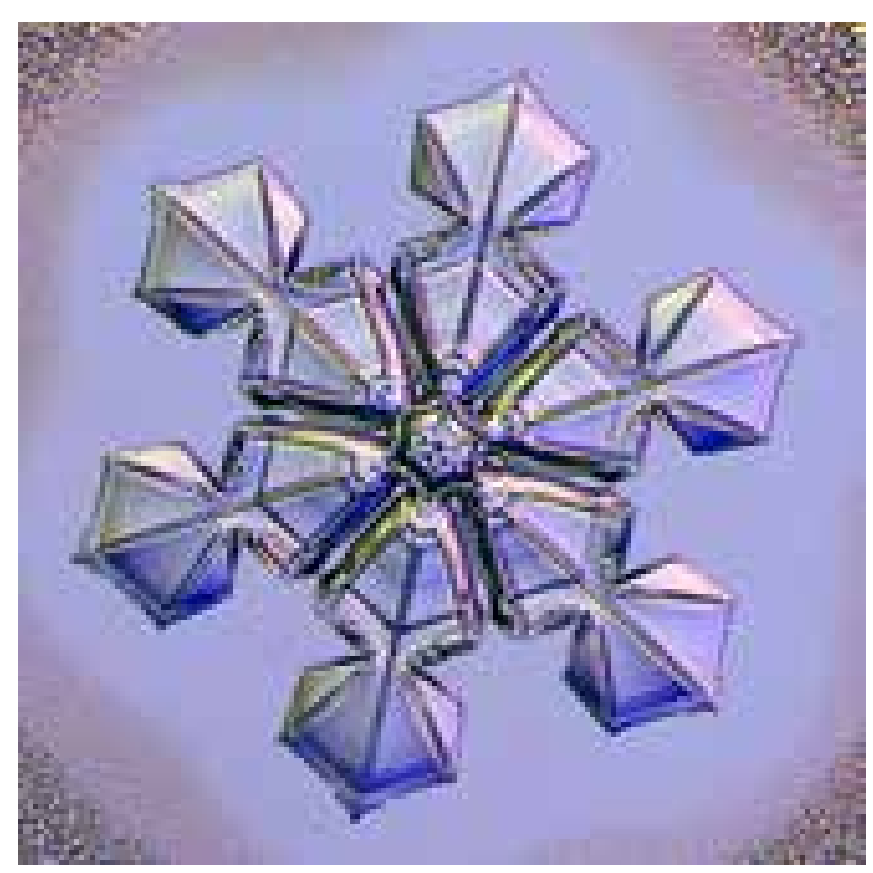

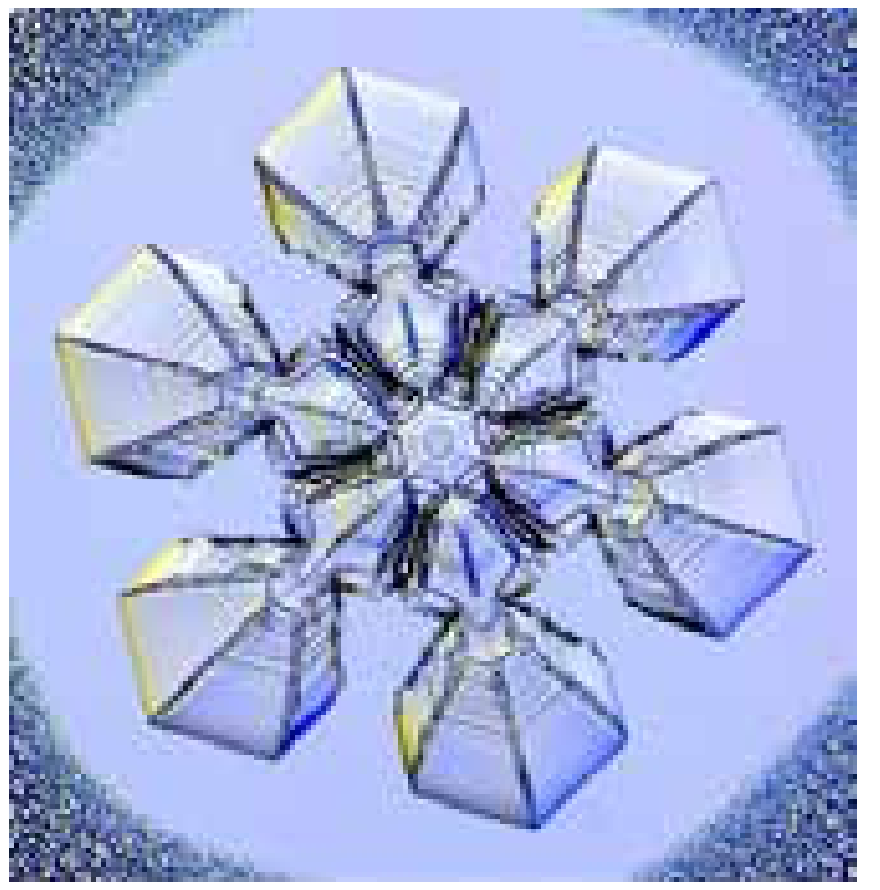

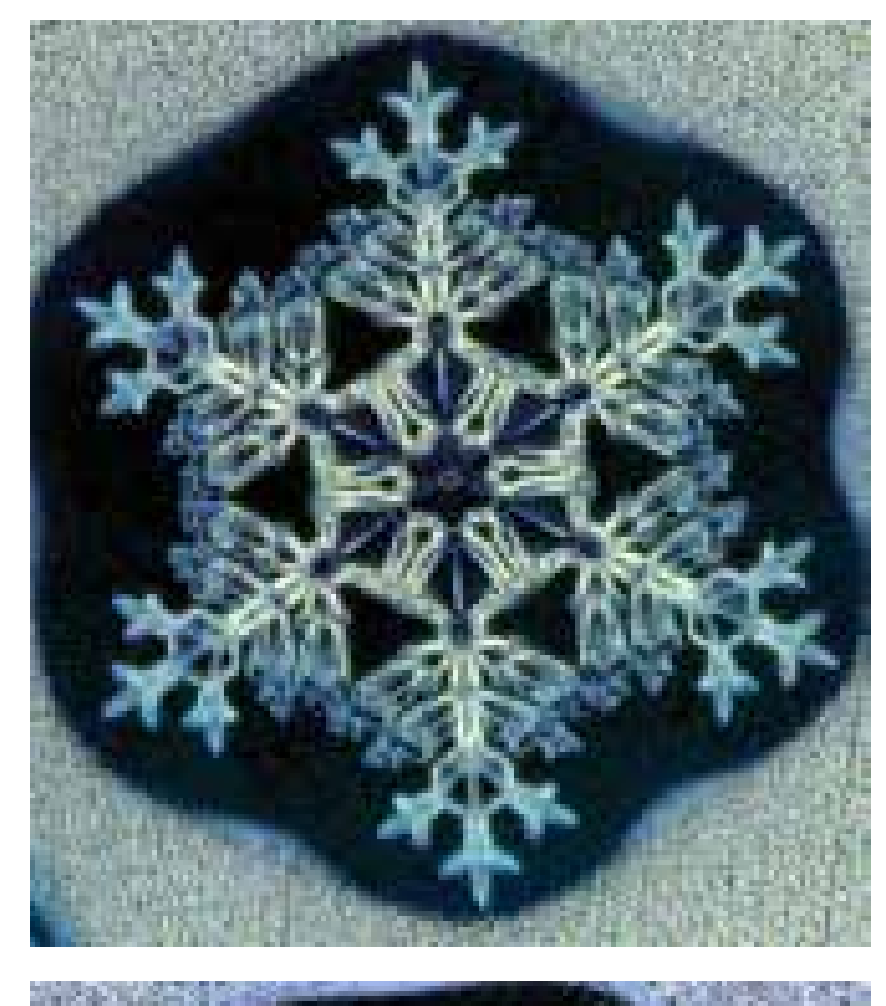
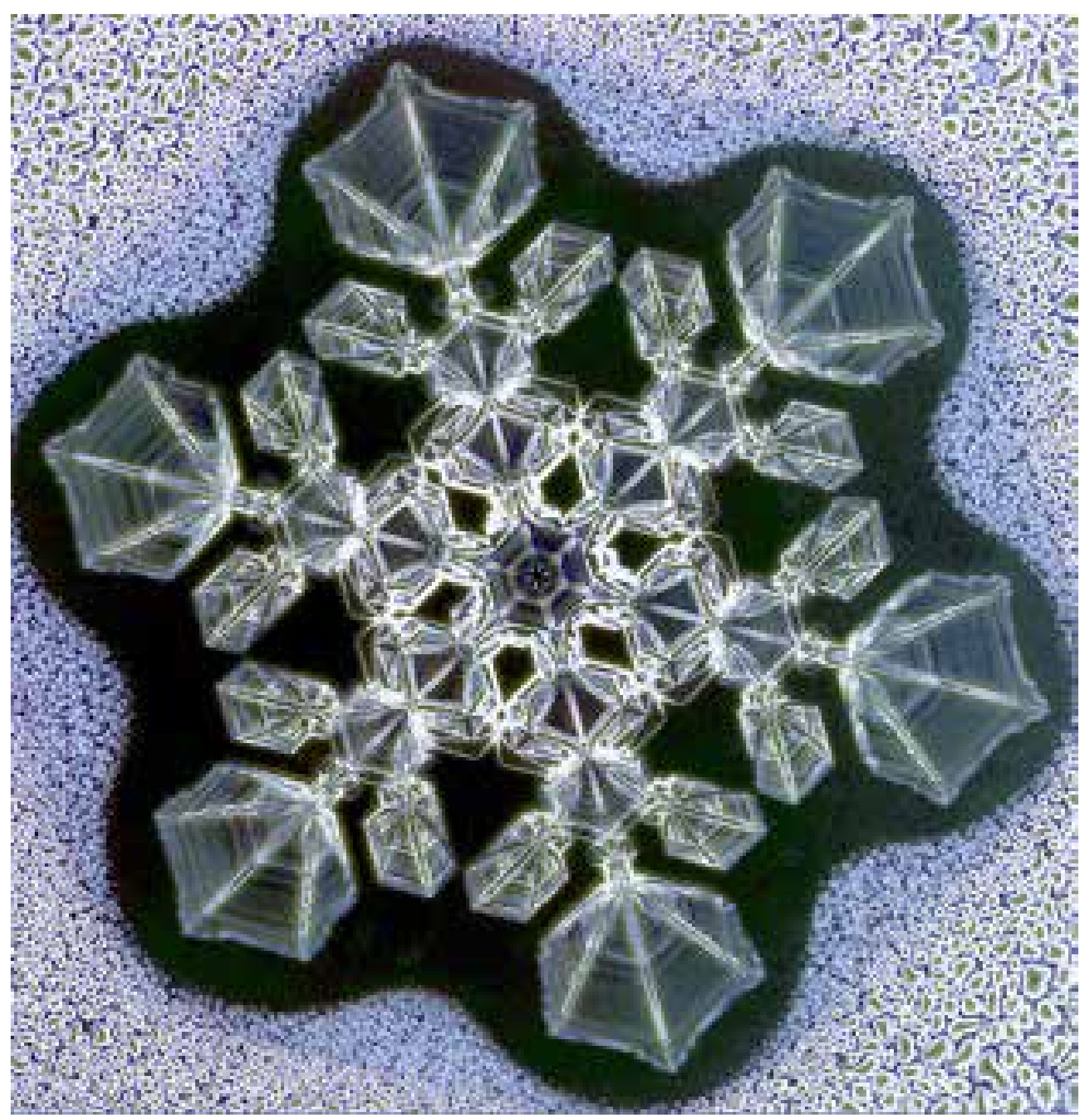
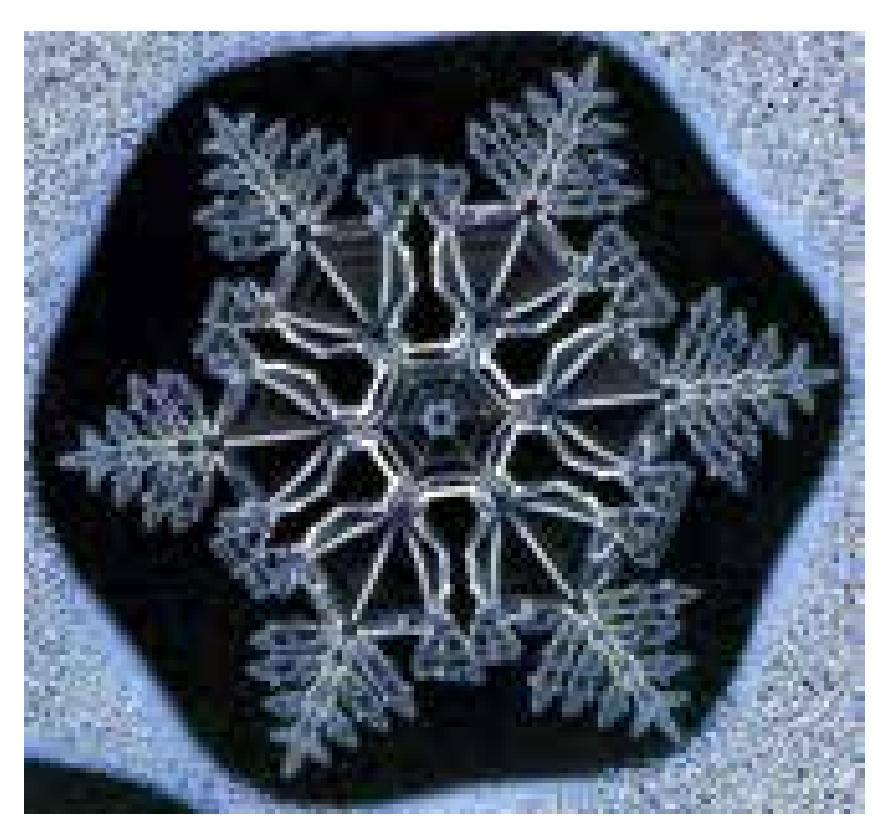
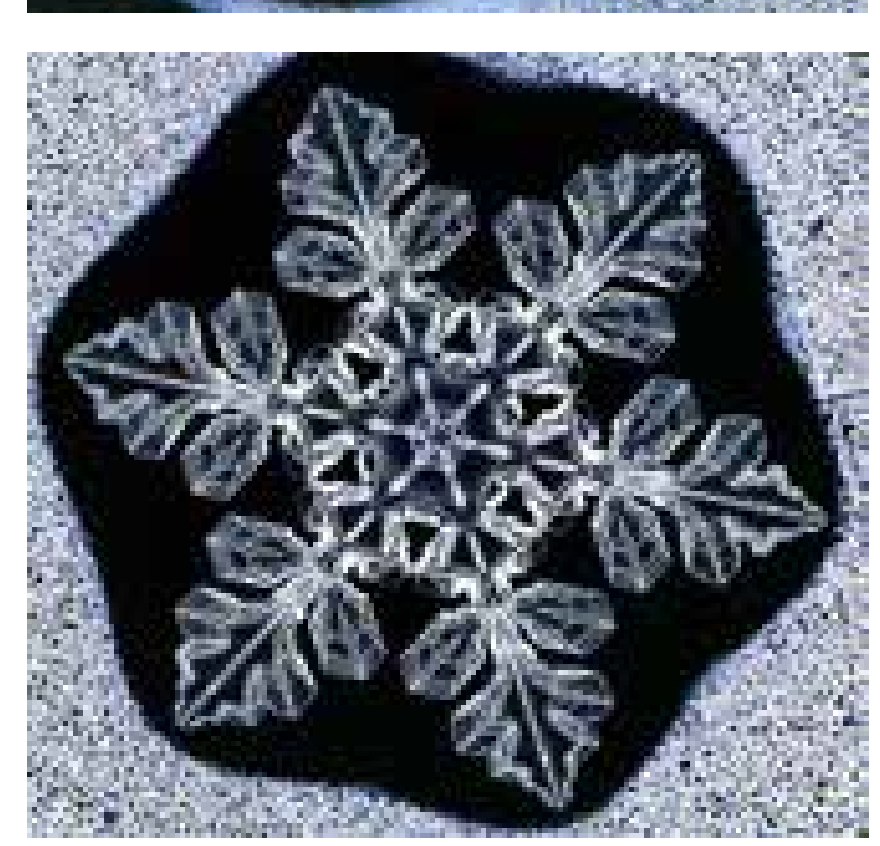
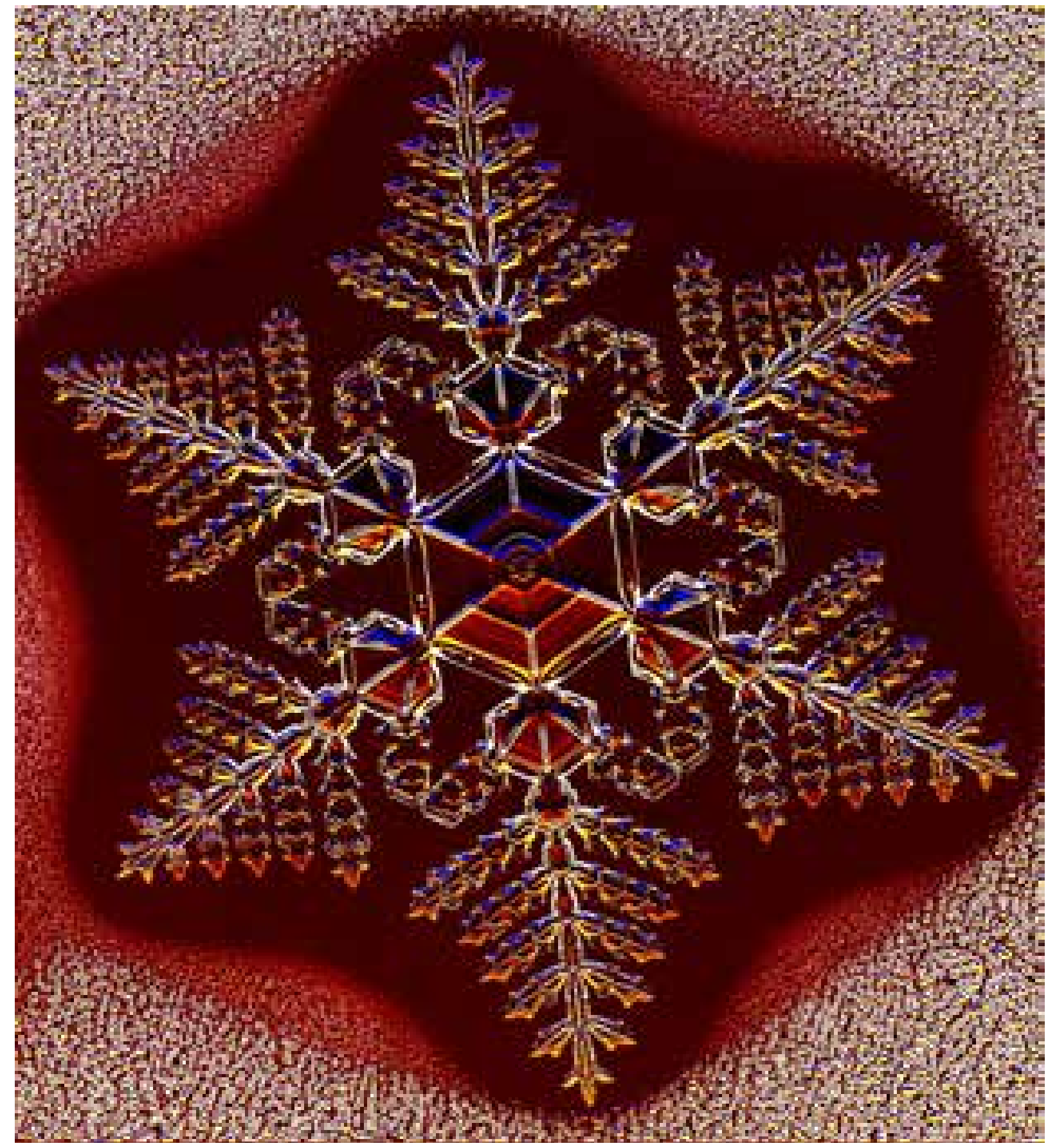
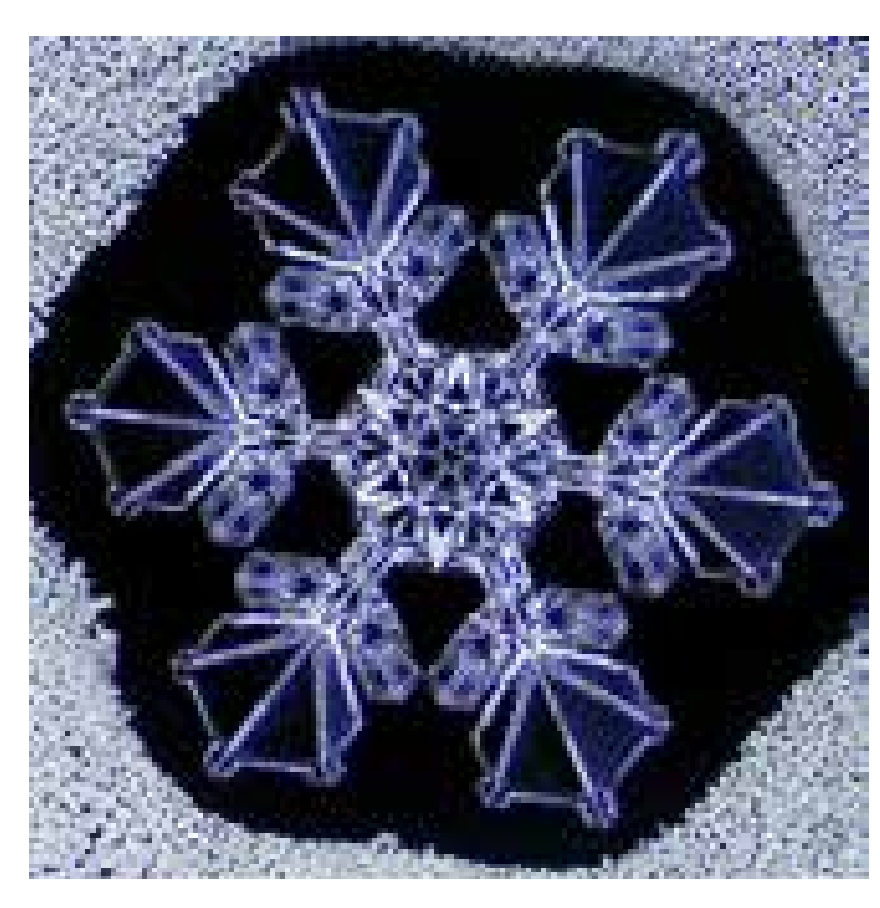

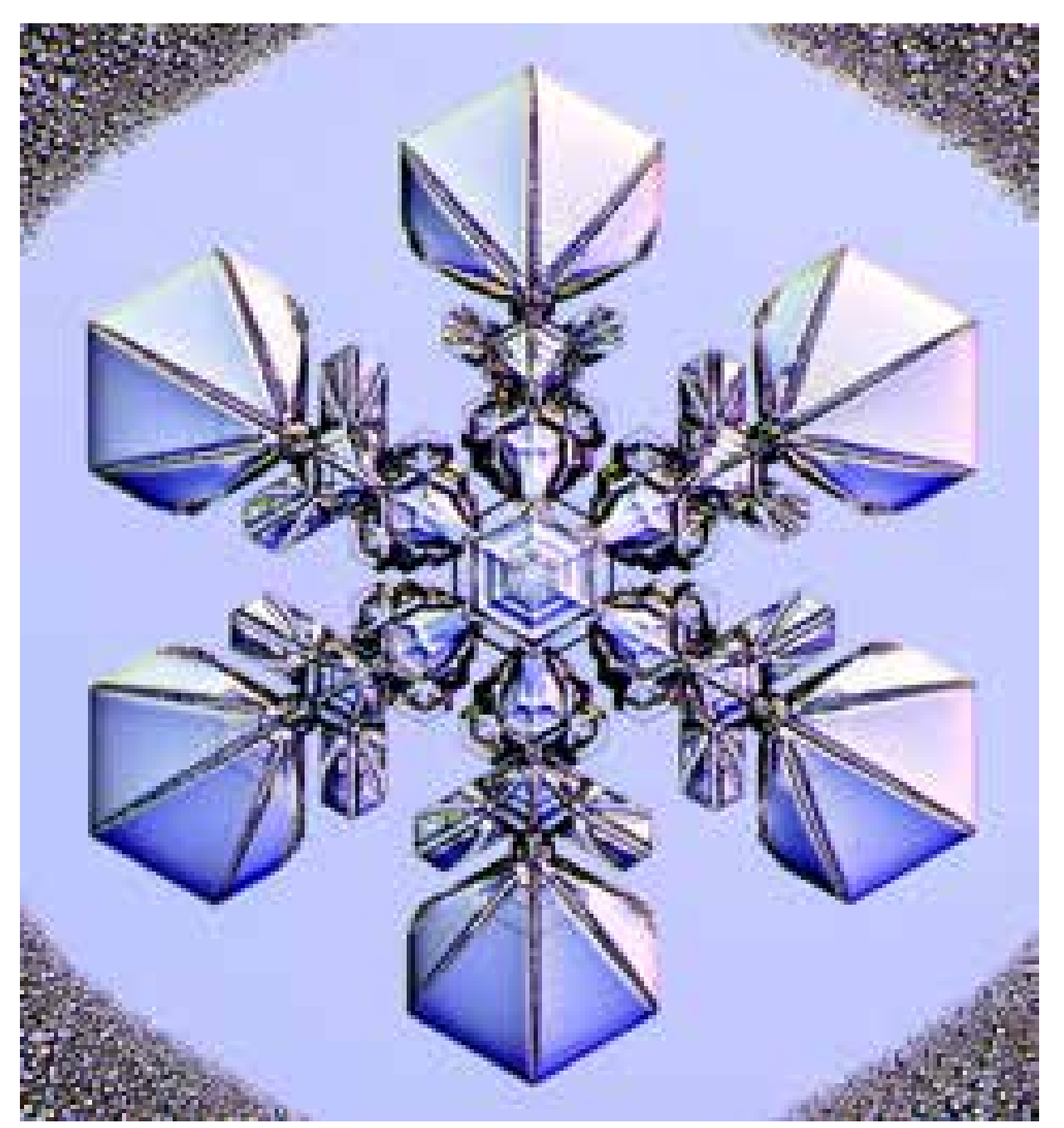

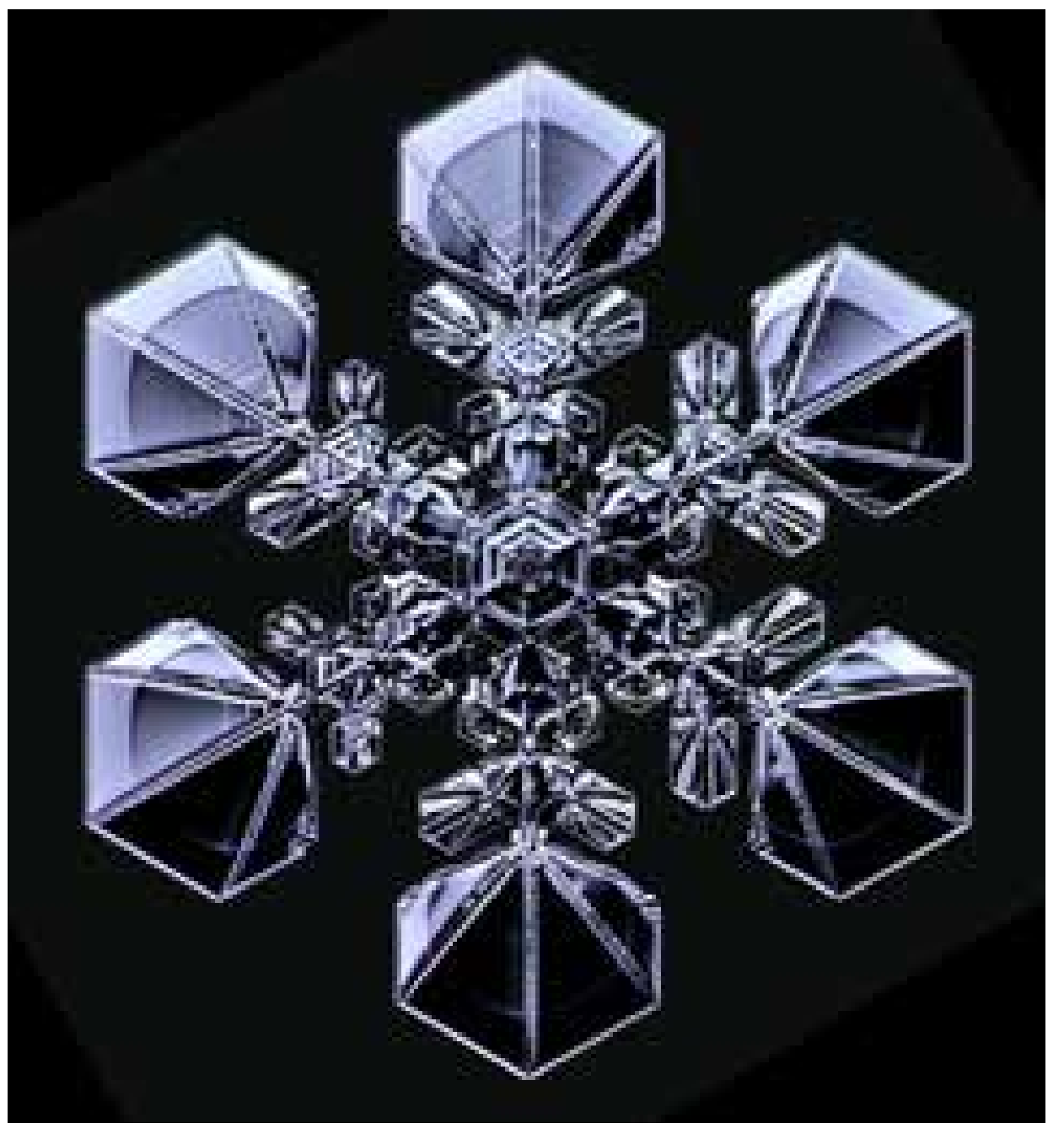

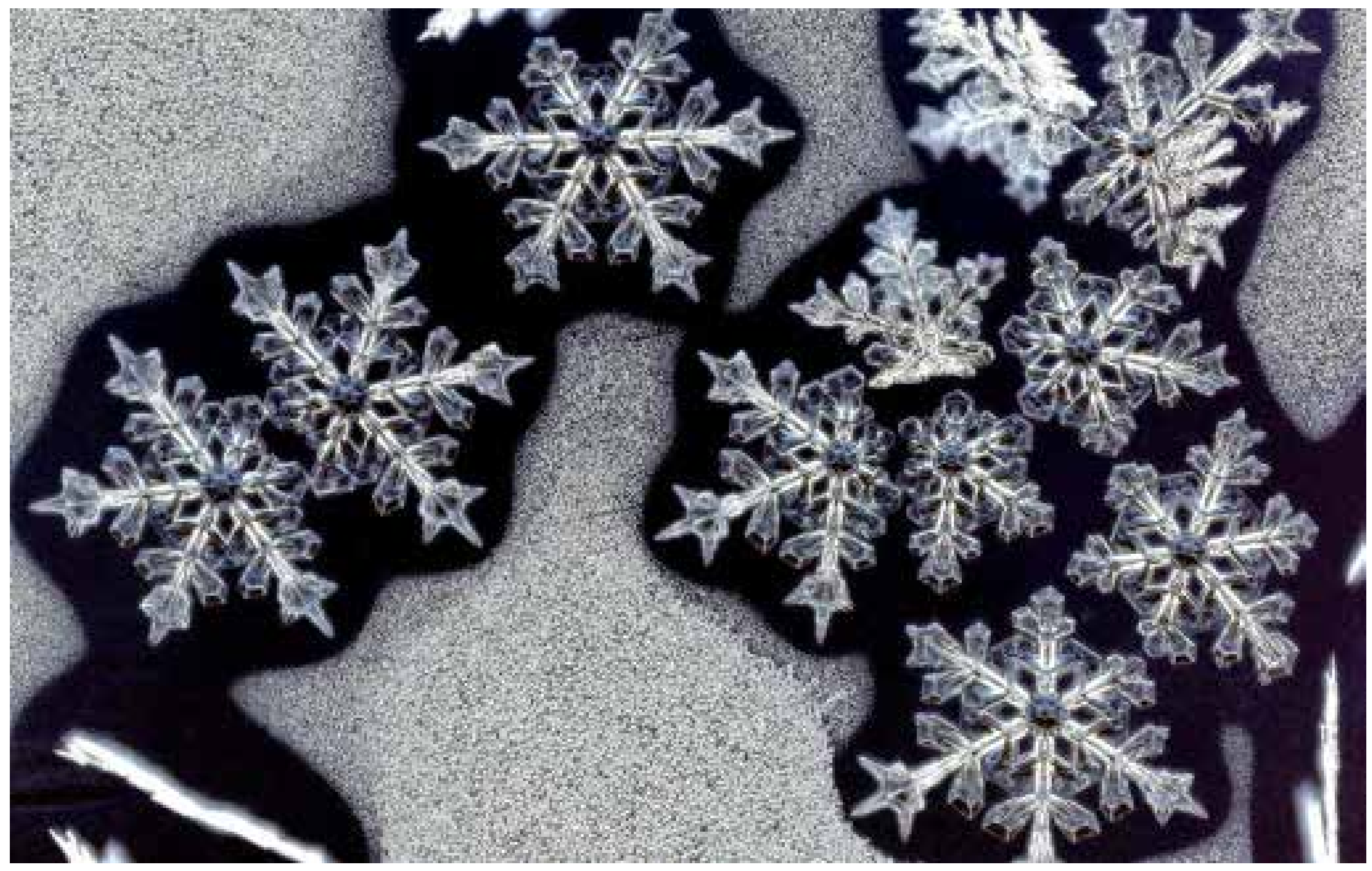

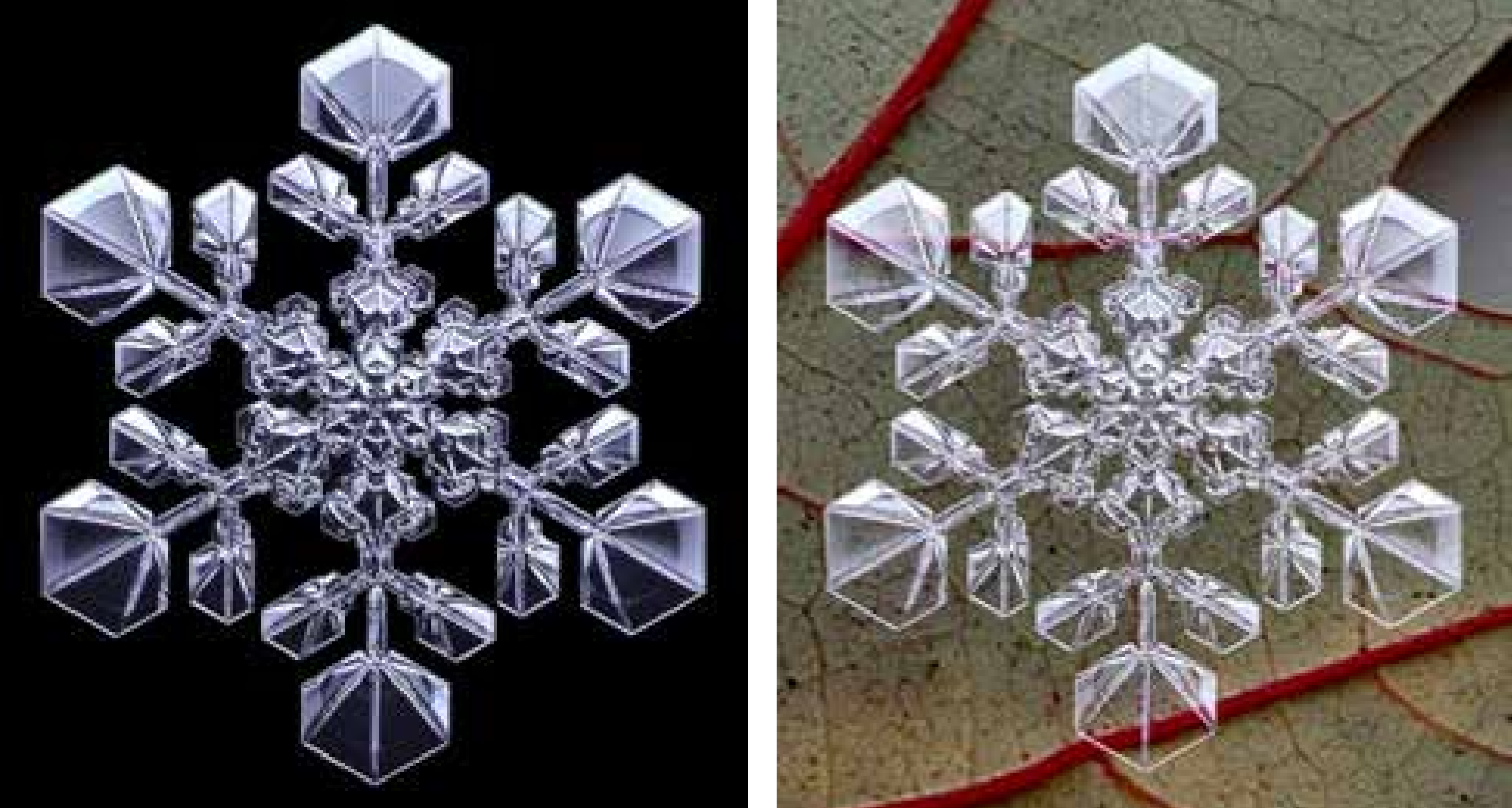

**Figure 9.43: Post-processing with a dark-field photo. The left image shows a PoP snow crystal photographed using dark-field illumination, yielding a bright crystal on a dark background. For the image on the right, the dark-field image was digitally superimposed with a close-up image of a leaf, (taken by Damon Taylor, posted at Flickr.com), with the composite image equal to (1-S)*B+S, where S is the snowflake image (normalized to a maximum brightness of 1) and B is the background image.**

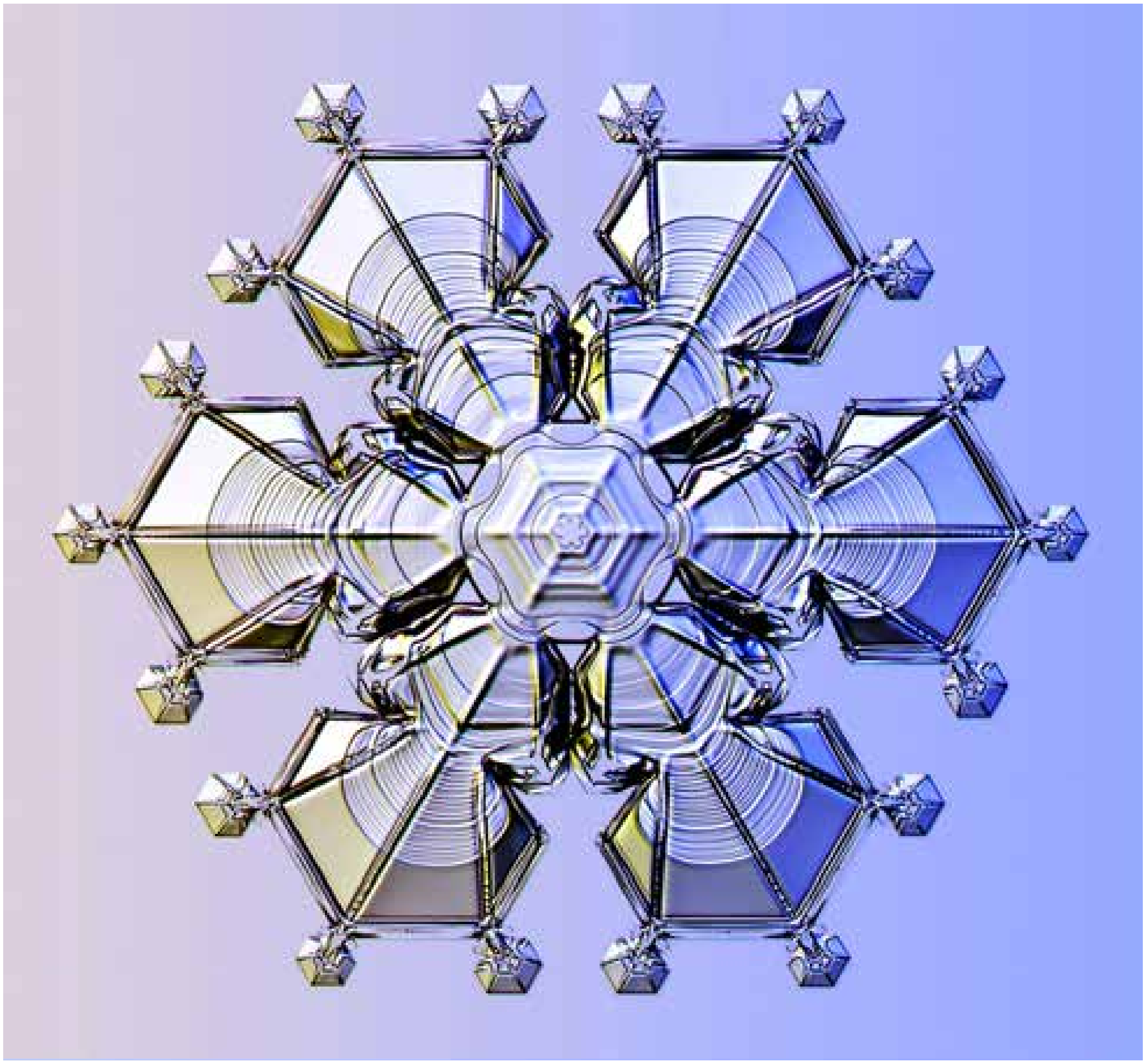

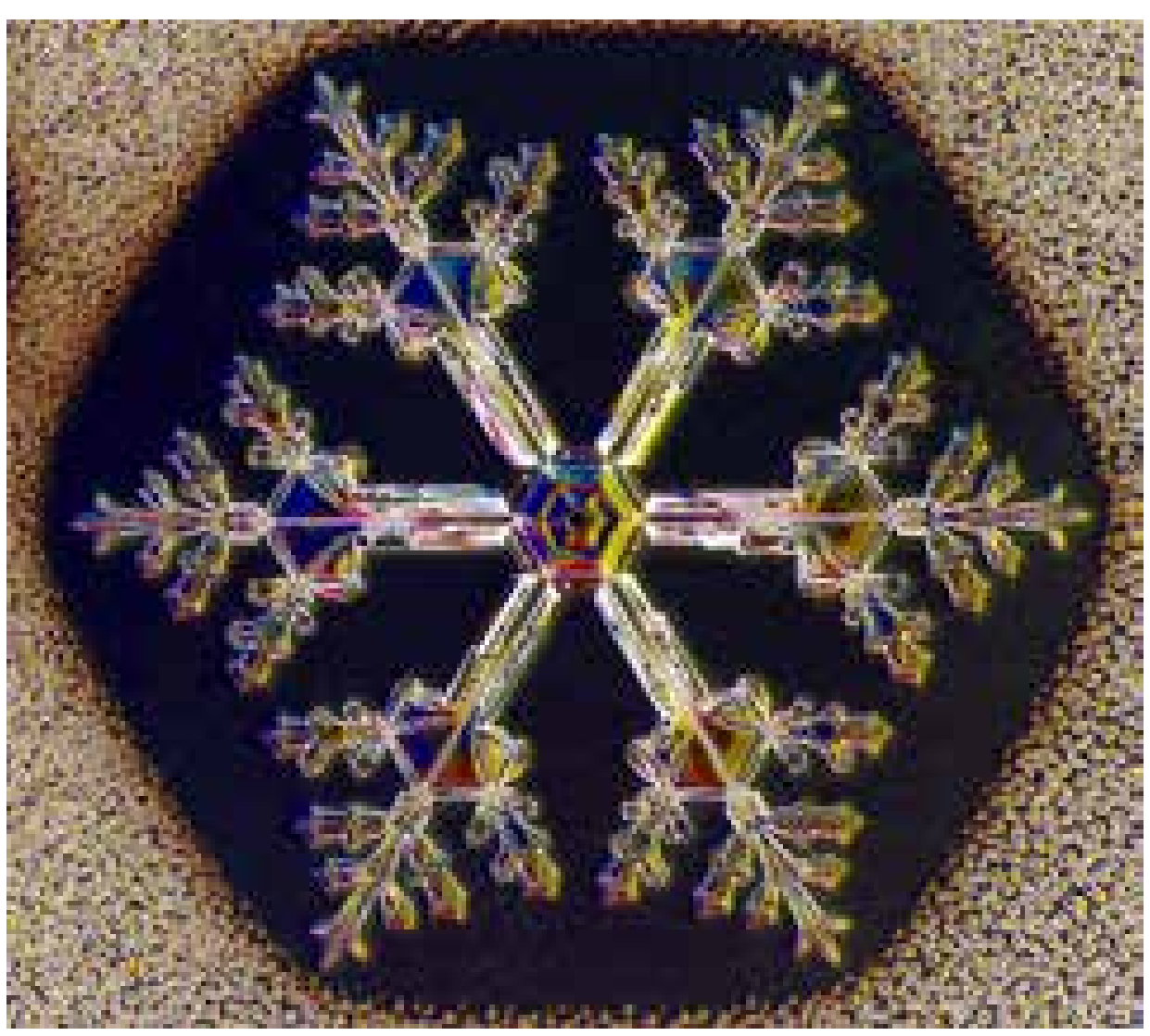

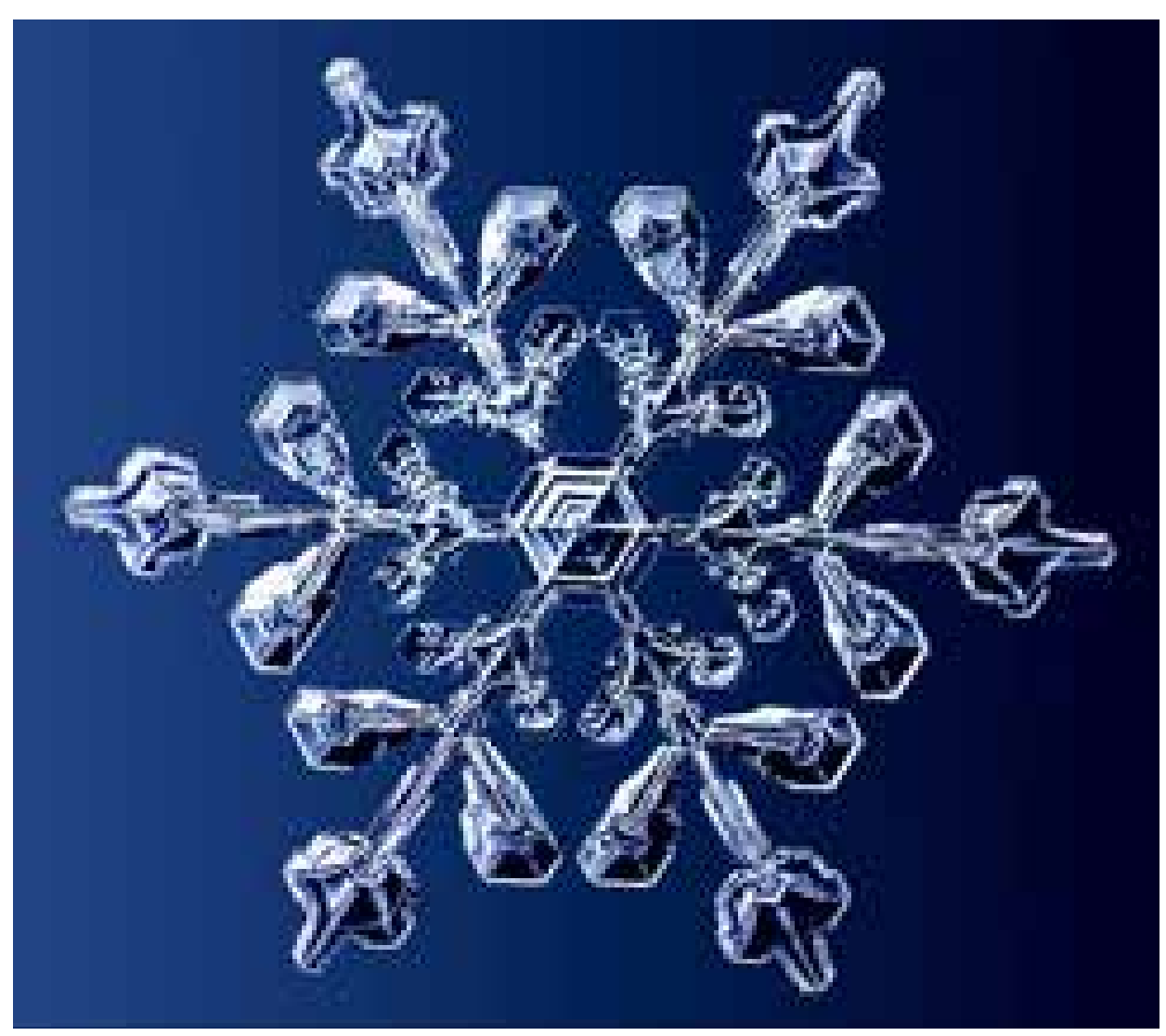

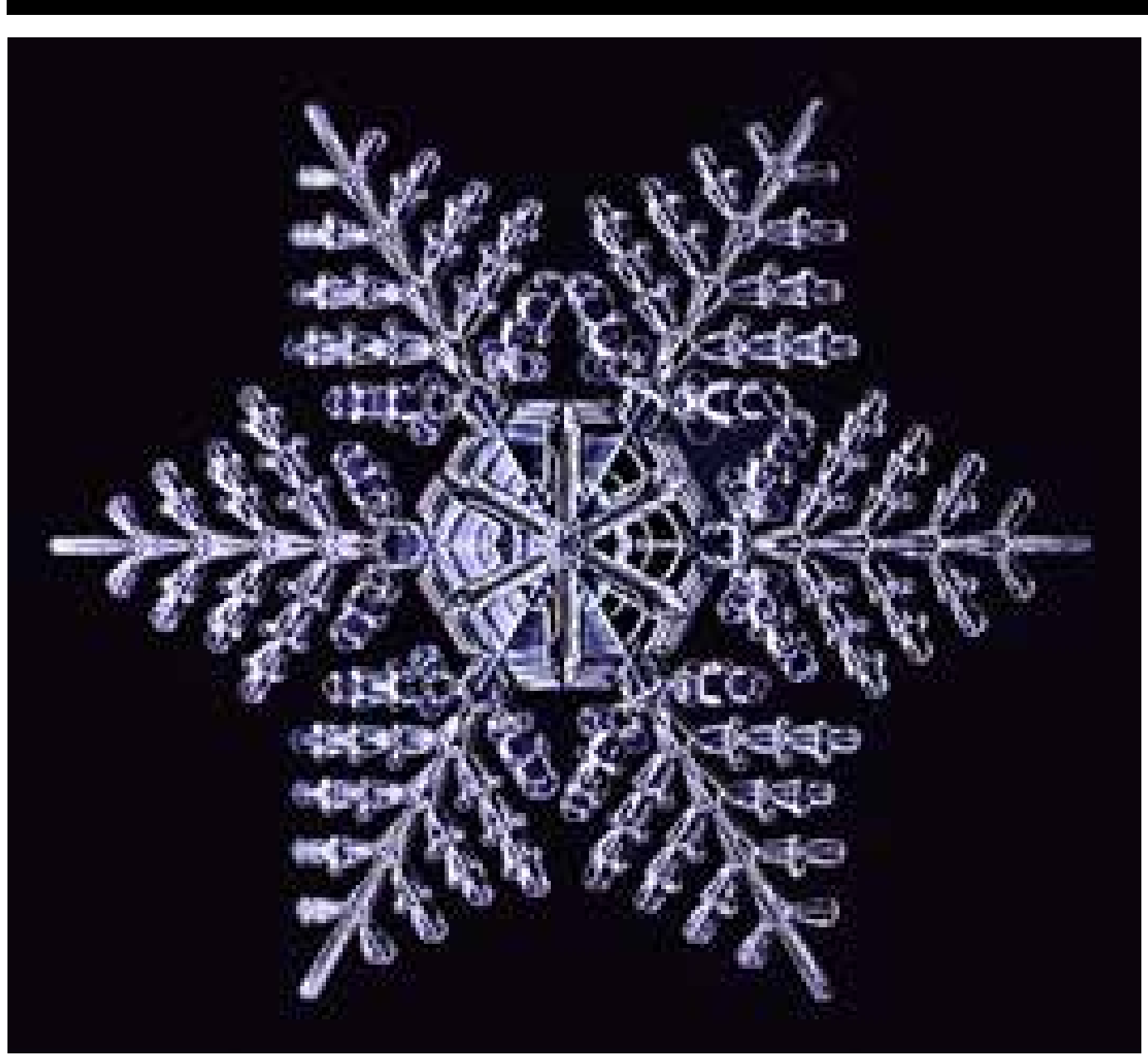

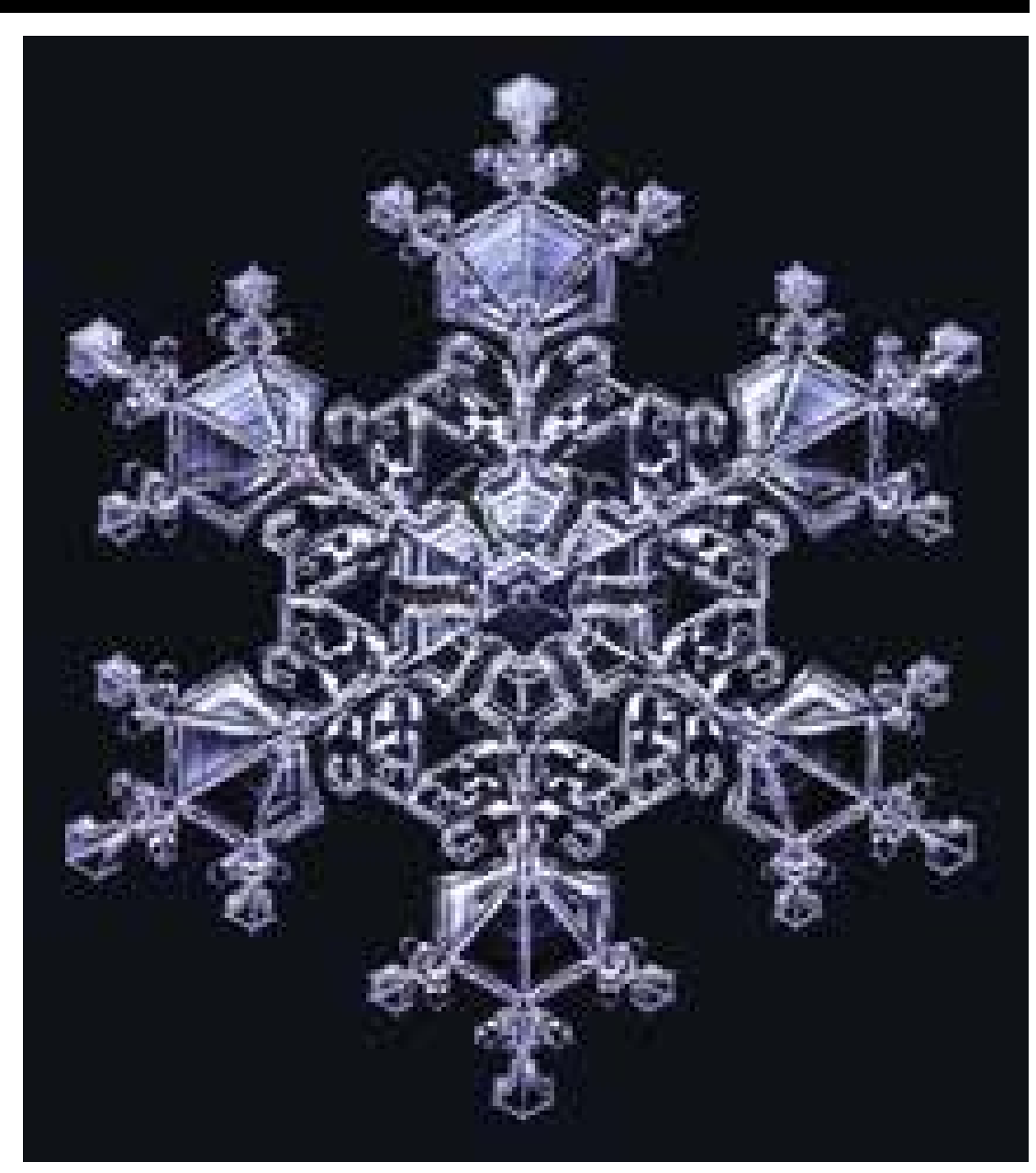

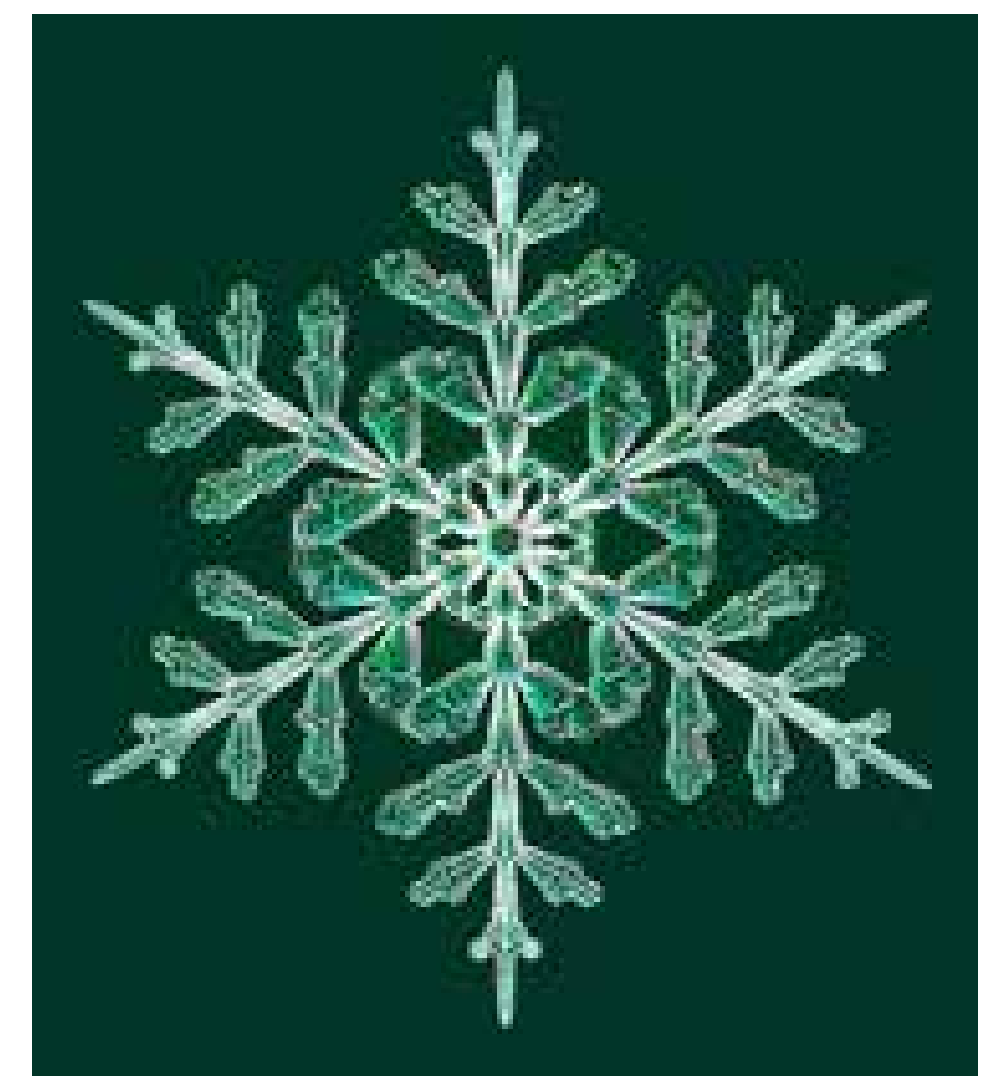
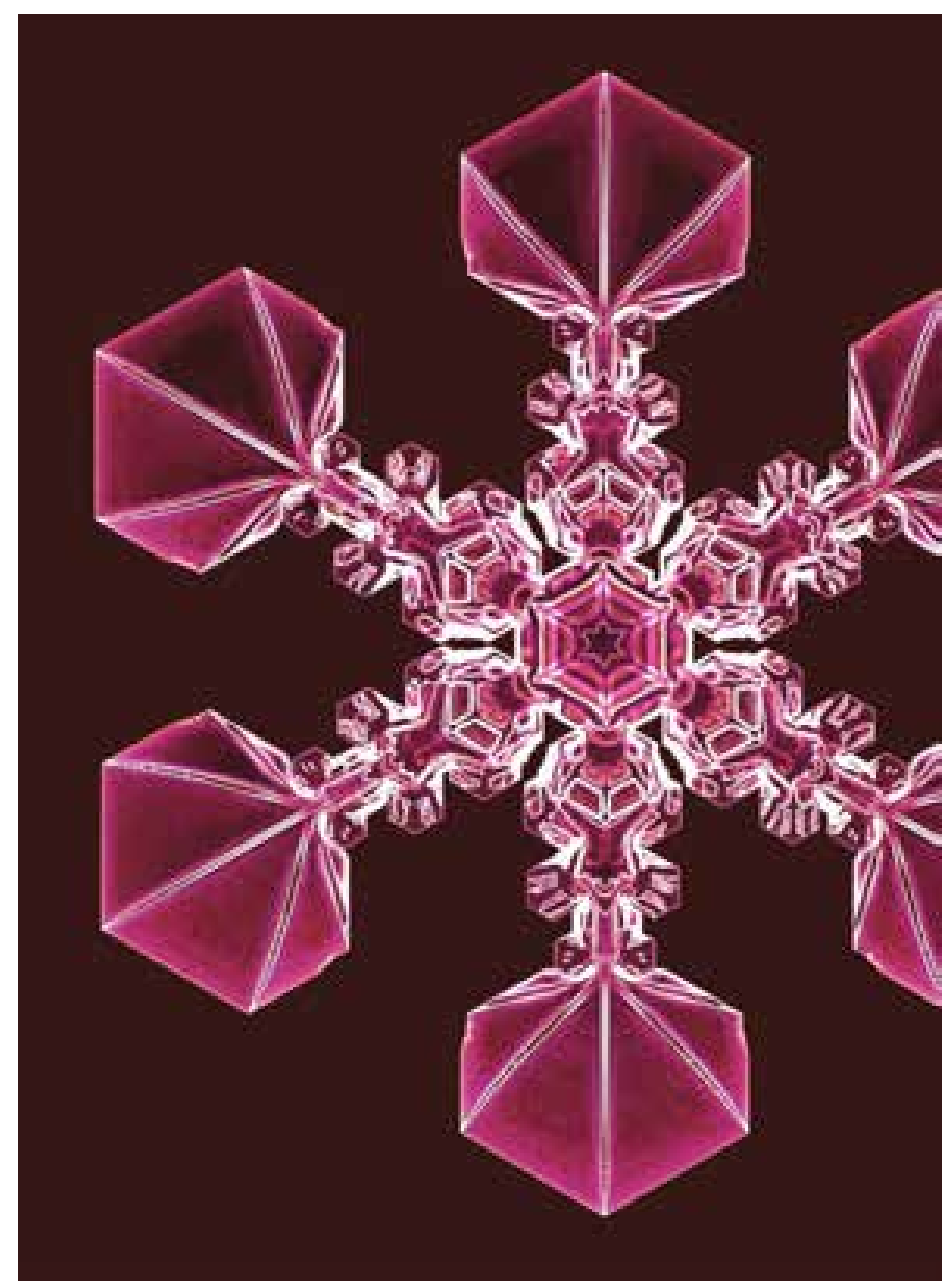
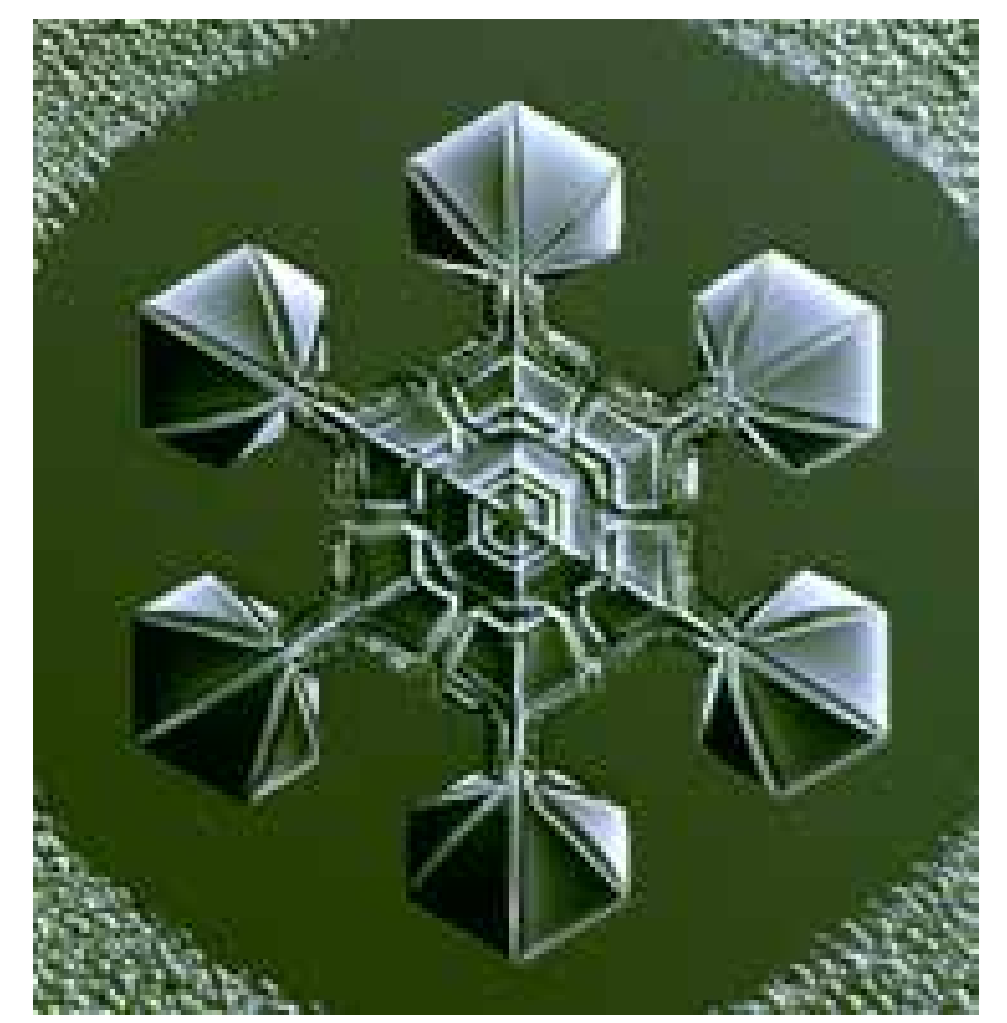
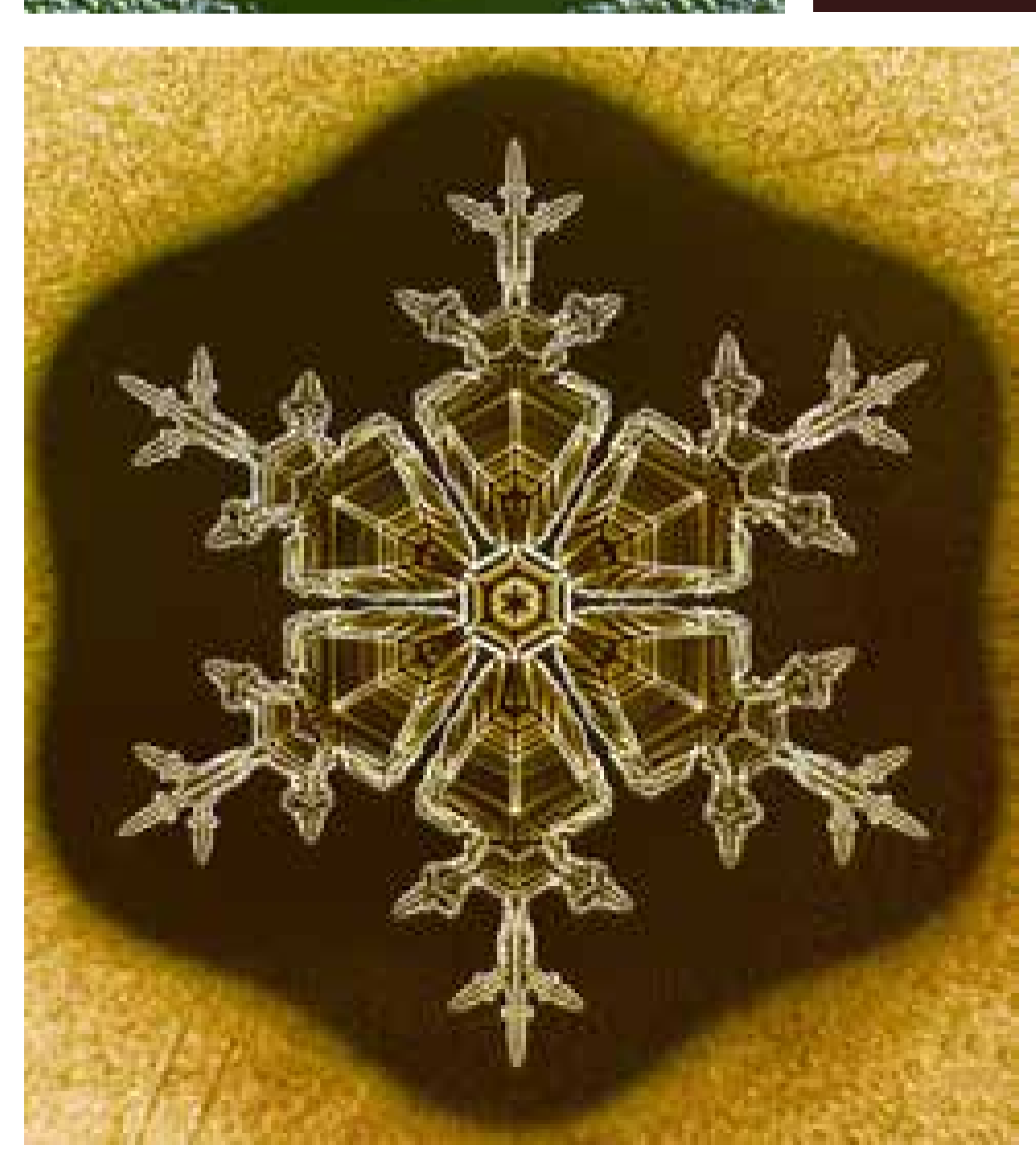
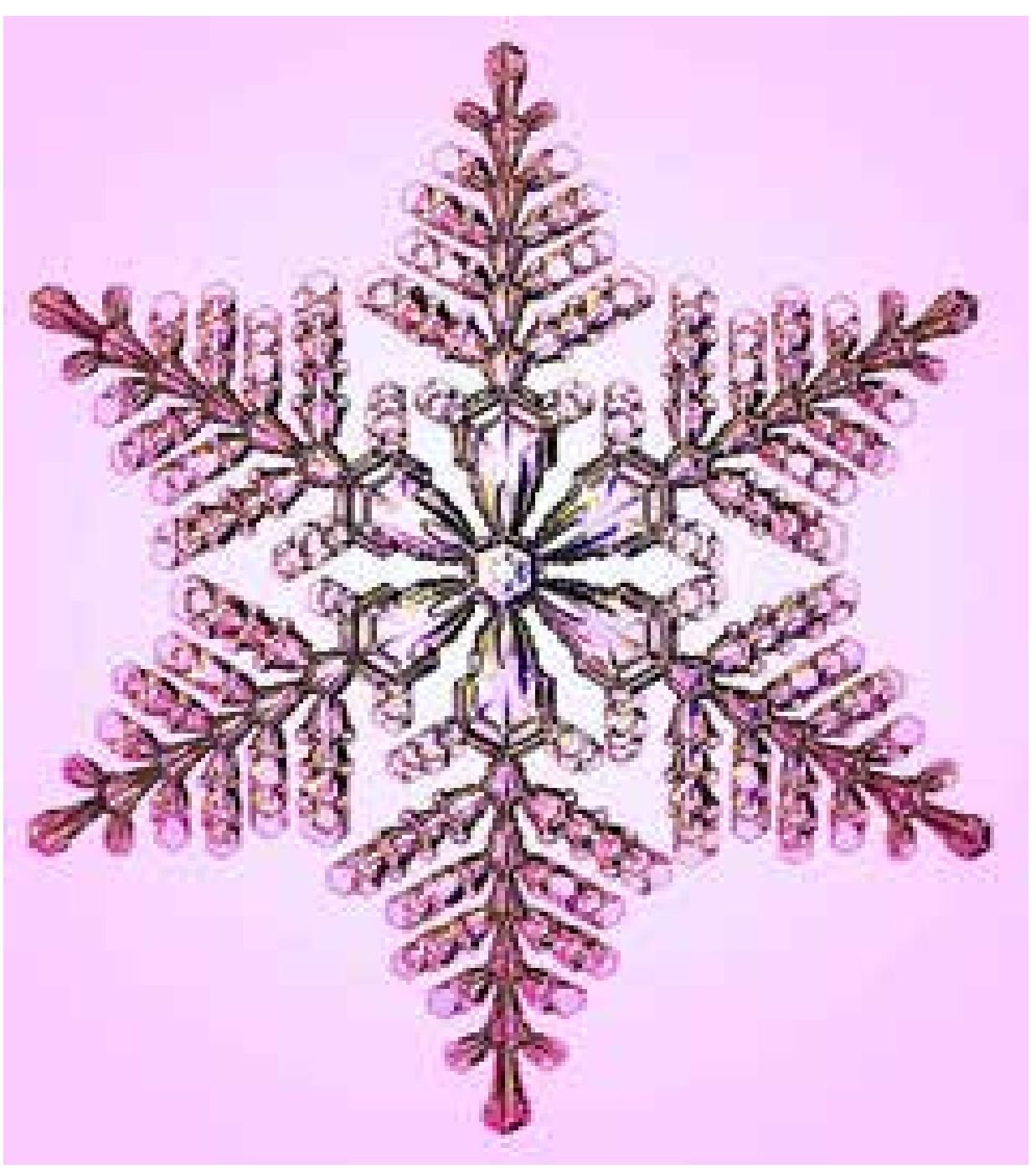

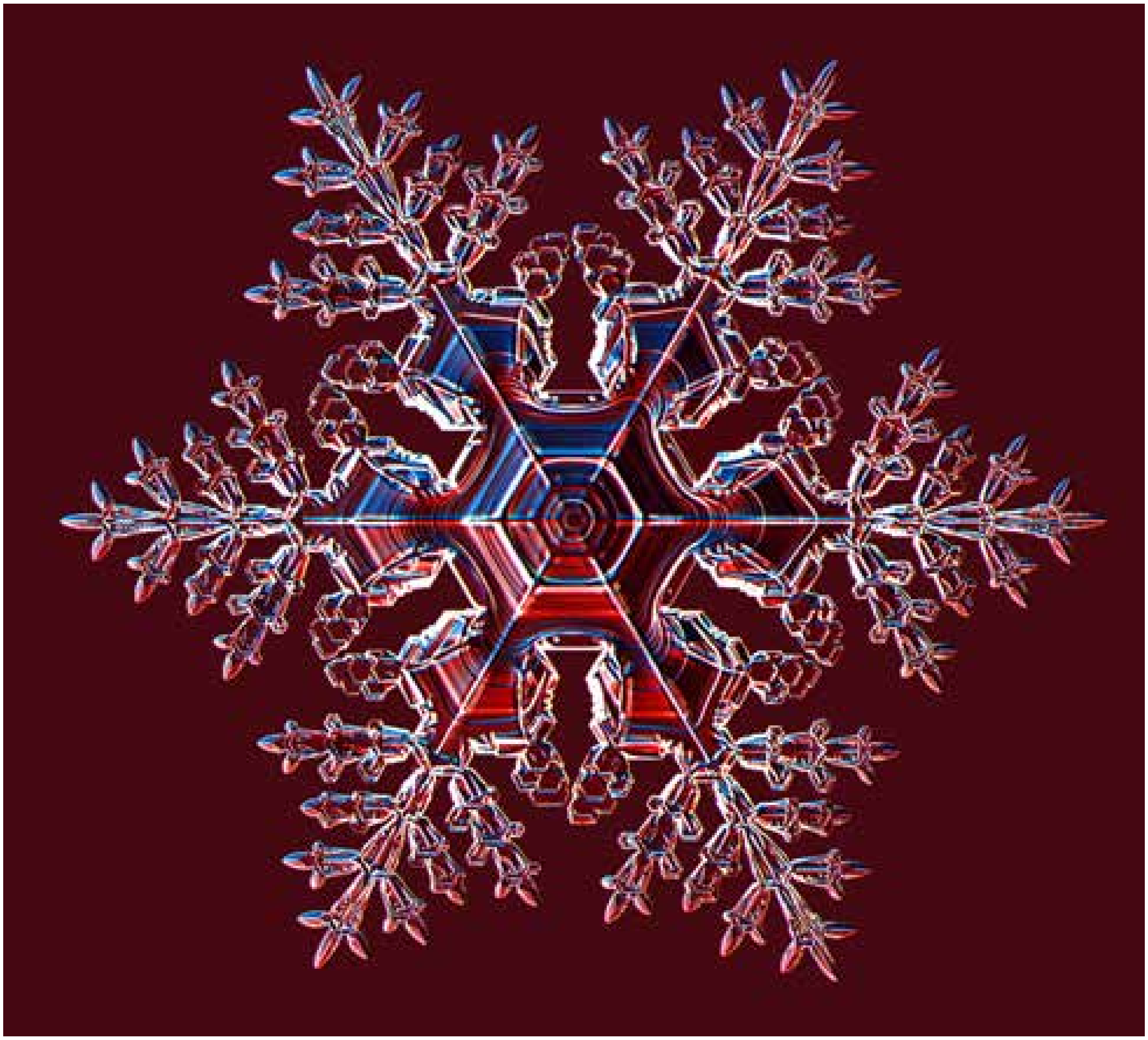

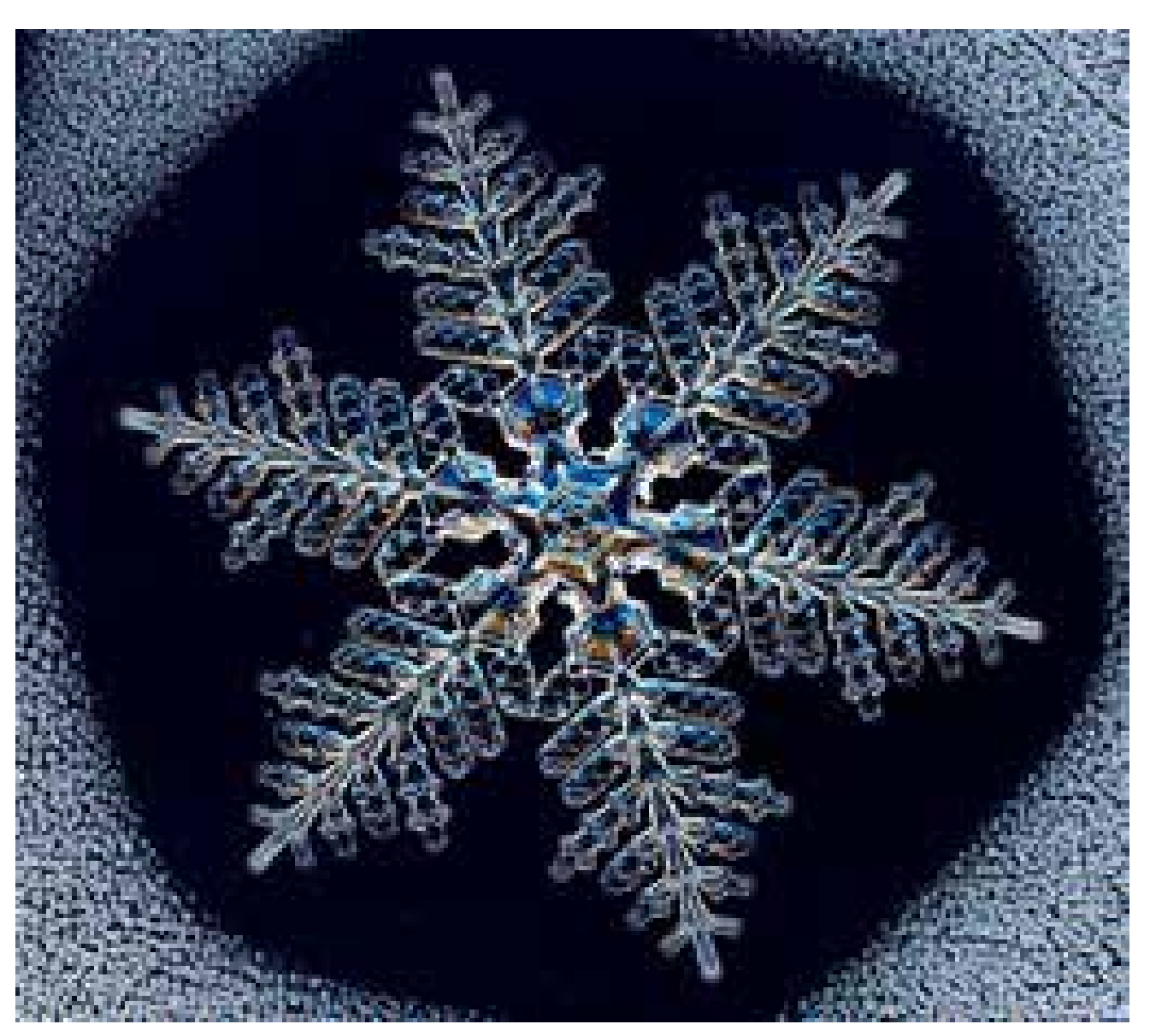

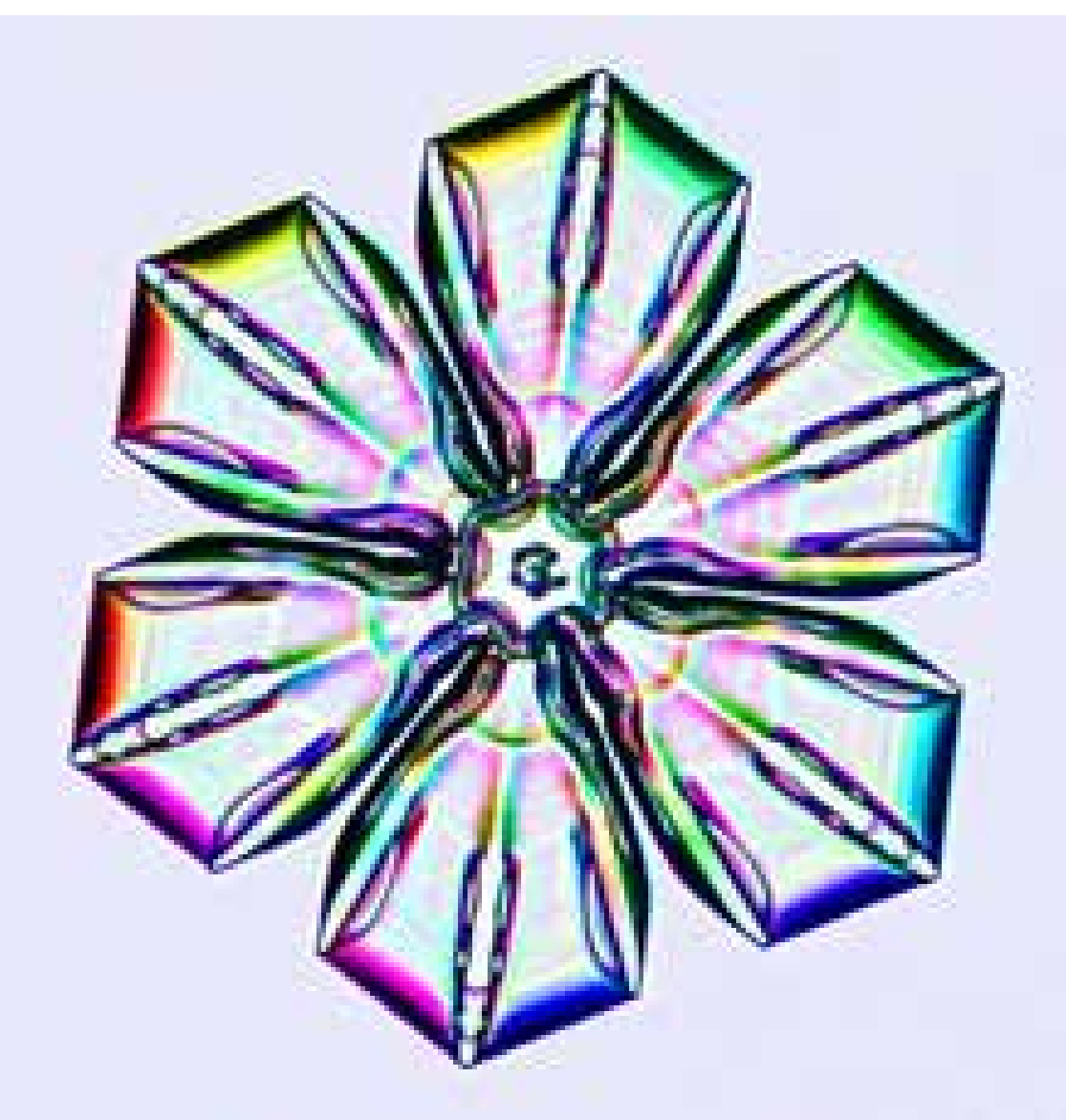

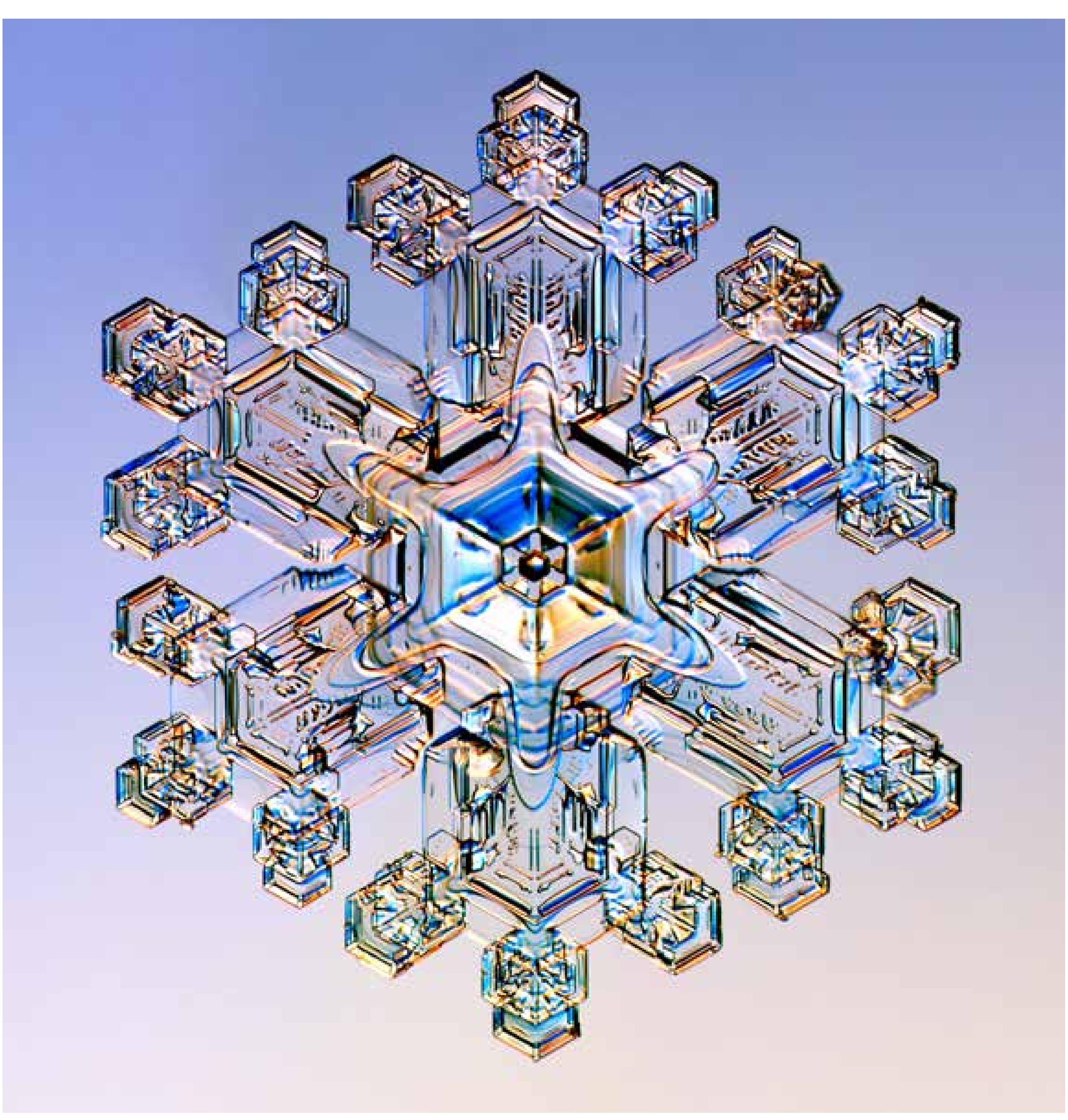

# Chapter 10

# A Field Guide to Snowflakes

*One cannot fix one's eyes on the commonest natural production without finding food for a rambling fancy.*
– Jane Austen, *Mansfield Park*, 1814

Nature provides a marvelous laboratory for examining the morphological diversity of snow crystals. With a simple magnifier and a robust tolerance for cold weather, one can observe a remarkable variety of crystal forms falling from the winter clouds. The possibilities range from simple plates and prisms to hollow columns, sectored plates, fernlike stellar dendrites, capped columns, and a host of rare and exotic varieties. Each snowfall has its own character, and there is always something new to discover.

Before embarking on a personal quest for snowflake sightings, however, it is helpful to have a handbook that describes what others have observed and documented over the years. To this end, the present chapter examines different types of snow crystals and defines a nomenclature that can be employed to describe one's own observations. Much of this material can also be found in *Ken Libbrecht's Field Guide to Snowflakes* [2006Lib1], which is still available as of this writing. The book is inexpensive and sized to fit into the ample pocket of a typical winter coat, so it may provide the best option if you want something to carry out into the field. With either the book or this chapter, my main objective is to increase snowflake awareness by whatever available, and perhaps to persuade a few hearty souls to venture out into the cold to have a look for themselves at this marvelous and often-unappreciated natural phenomenon.

**Facing page: A large stellar-plate snow crystal with complex surface markings, measuring just over three millimeters from tip to tip, captured by the author in Burlington, Vermont.**

## Hieroglyphs from the Sky

Ukichiro Nakaya described snow crystals as "hieroglyphs from the sky" [1954Nak] because the form of each crystal can be interpreted, at least to some approximation, to reveal the atmospheric conditions it experienced as it grew and developed. There is some truth to this supposition because the process of snow-crystal growth is largely deterministic, meaning that two crystals experiencing the same conditions as a function of time will grow into nearly identical shapes. The "identical-twin" snow crystals presented in Chapter 9 provide a direct confirmation of this deterministic behavior.

A primary tool for deciphering the shape of a specific snow crystal is the snow crystal morphology diagram shown in Figure 10.1 (see also Chapter 1 and Figure 8.16). The morphology diagram provides a valuable overview that connects the seemingly disparate observations of falling snow into a generally coherent picture of what is happening up in the clouds. Of course, there are limitations to how much one can say about the growth history of a snowflake by examining a single photograph of its final state; but usually one can envision a plausible scenario to explain an observed morphology.

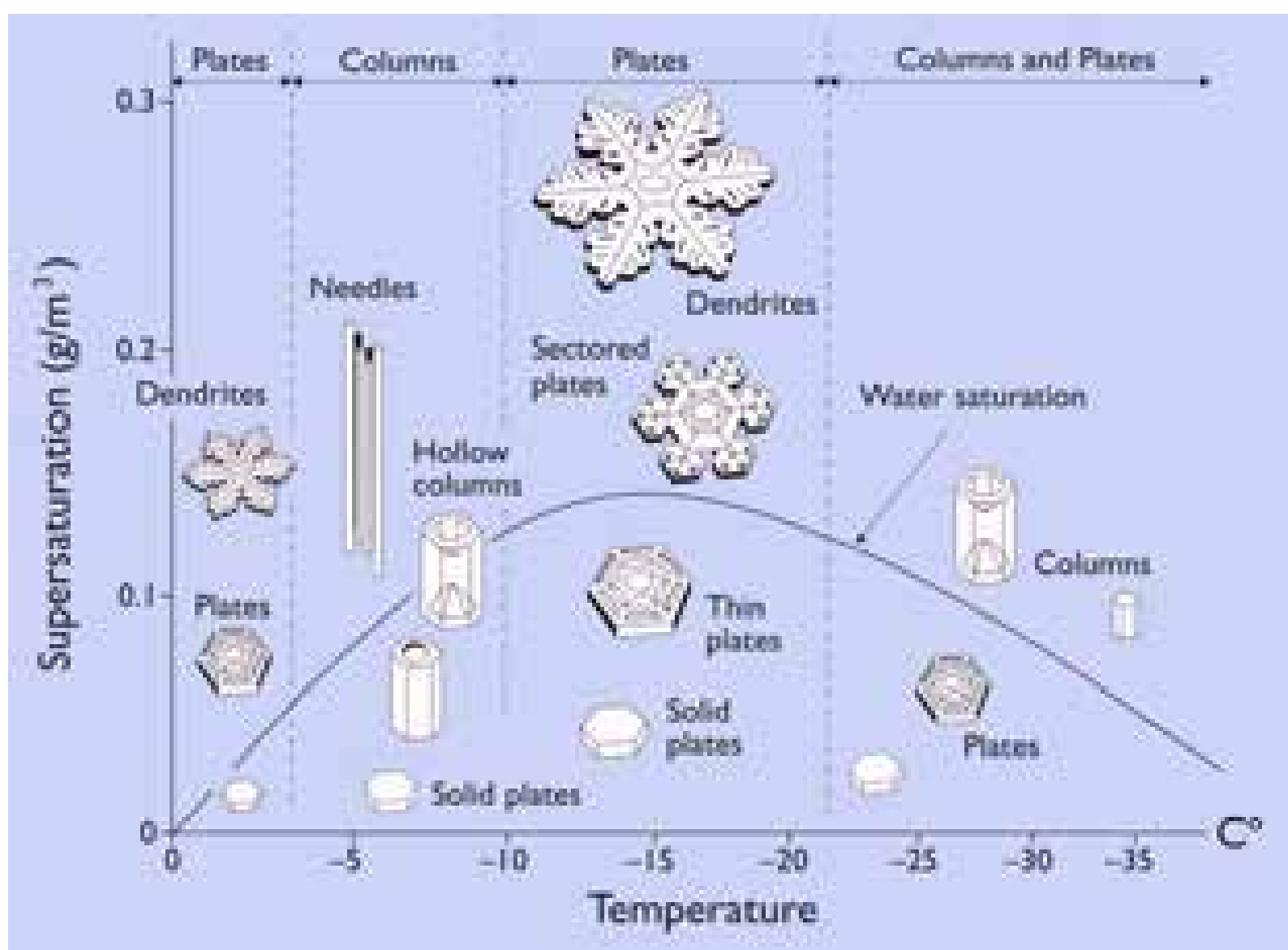

**Figure 10.1: The snow crystal morphology diagram illustrates what types of crystals form in air as a function of temperature and water-vapor supersaturation. Note that this chart provides just a rough approximation of the different morphologies, plus it applies only if the growth conditions are constant in time. The "water saturation" line shows the supersaturation found in a dense cloud of liquid water droplets.**

Another use of the morphology diagram is to predict what types of crystals will appear in different weather conditions. For example, if one wishes to find large, well-formed stellar snow crystals (a popular photographic goal), it is useful to know that such crystals can only be found when the cloud temperatures are around -15 C. (Although the morphology diagram indicates that plate-like crystals also appear near -2 C, these higher-temperature specimens are typically quite small.)

## 10.1 Snowflake Watching

Most of this chapter presents an extensive catalogue of photographic examples showing different types of snowflakes and snow crystals, along with a discussion of their various identifying features and characteristics. Unless otherwise identified, all are photographs of natural snow crystals taken by the author. The presentation is in the form of a field guide, aimed at assisting snowflake photographers or other readers who want to see for themselves what the winter clouds have to offer. Although such exploration could be done without any guidance, the venture is nearly always more rewarding when you set out knowing what to look for.

As an occasionally avid bird watcher, I like to think of snowflake watching as an entirely analogous activity. It can be entertaining, educational, and a surprisingly enjoyable leisure activity. Keeping an eye out for interesting crystals is a fine pursuit whenever you happen to be outside during a light snowfall. You could be riding the chair lift at your local ski area, taking a stroll through the winter woods, or just waiting in your car somewhere. If the snow is falling all around you, why not have a look from time to time to see what you can find?

Growing up on a farm in North Dakota, I experienced a lot of snow, and I saw my share of birds. But I never really noticed either until someone showed me what to look for, which happened long after my childhood days. Looking back on this particular aspect of my youth, these were lost opportunities. We all live in nature, but it takes some awareness to notice the natural phenomena around you. If you happen to live in a cold climate, I suggest that you think about snowflakes occasionally, and perhaps go outside to have a look for yourself. You never know what you might find.

## Snow Crystal Classification

We name snowflakes for the same reason we name most things – so we can talk about them. Certain morphologies are common and have a distinctive appearance, and those have fairly well-defined names. Stellar plates, stellar dendrites, fernlike stellar dendrites, hollow columns, and capped columns have all been part of the snowflake vernacular for some time. But there is no absolute classification system for snow crystals, and there never will be, because there is no clear way to divide snow crystals into distinct, non-overlapping categories.

Some things are intrinsically well suited to classification. Biological species, for example, cannot easily interbreed, so they mostly form well-defined, non-overlapping categories (aside from many exceptions, such as the mule). Naming chemical species works quite well also, as each name refers to a specific chemical formula (and perhaps a specific isomer). In these cases, naming conventions can be quite precise, making it reasonably straightforward to identify the named group that a given individual fits into.

Other groups of items are not so easily categorized. We can talk about different types of bread, cheese, cookies, breeds of dogs, types of hobbies, or musical instruments, but the names are generally human constructs with few natural partitions. People organize and catalog all these items, but different people have different lists, and the names often to change over time. As a specific example, skiers have many names for different types of snow on the ground, but again the categories are a bit arbitrary. Snow scientists have done the same for falling snow crystals.

Asking "what kind of snowflake is that?" is not an especially good question, because there may not be a well-defined answer. The shape of a snow crystal depends on its entire growth history, which is somewhat analogous to saying that the breed of a dog depends on its entire ancestry. If a specific snow crystal had an unusual history, then it may not fit well into any category, no matter how many categories one defines. There is no way to avoid these ambiguities, so classifying snow crystals is a practice with somewhat limited usefulness. Nevertheless, some taxonomy is necessary to guide the conversation, so classification systems have been devised for this purpose.

Nakaya first recognized the need for nomenclature and constructed the first classification system shown in Figure 10.2, containing seven primary categories that branch out into a total of 41 snowflake types [1954Nak]. This was later expanded by Magono and Lee to 80 categories [1966Mag] and recently expanded again by Kikuchi and Kajikawa to include 121 distinct snowflake types [2013Kik, 2011Kik], as shown in Figure 10.2.

For the purposes of everyday snowflake watching, I prefer the somewhat simpler chart shown in Figure 10.3. This is essentially a modernized version of Nakaya's original list, placing a greater focus on the physical processes that underlie the different morphological types. With 35 different named categories, this chart includes the common designations that have evolved over the years, and it describes most of what can readily be found in the wild. I have found this chart quite useful for identifying and describing natural snow crystals, so I continue to promote its use. There is no definite, ideal method for classifying snow crystals, but Figure 10.3 is the chart I prefer.

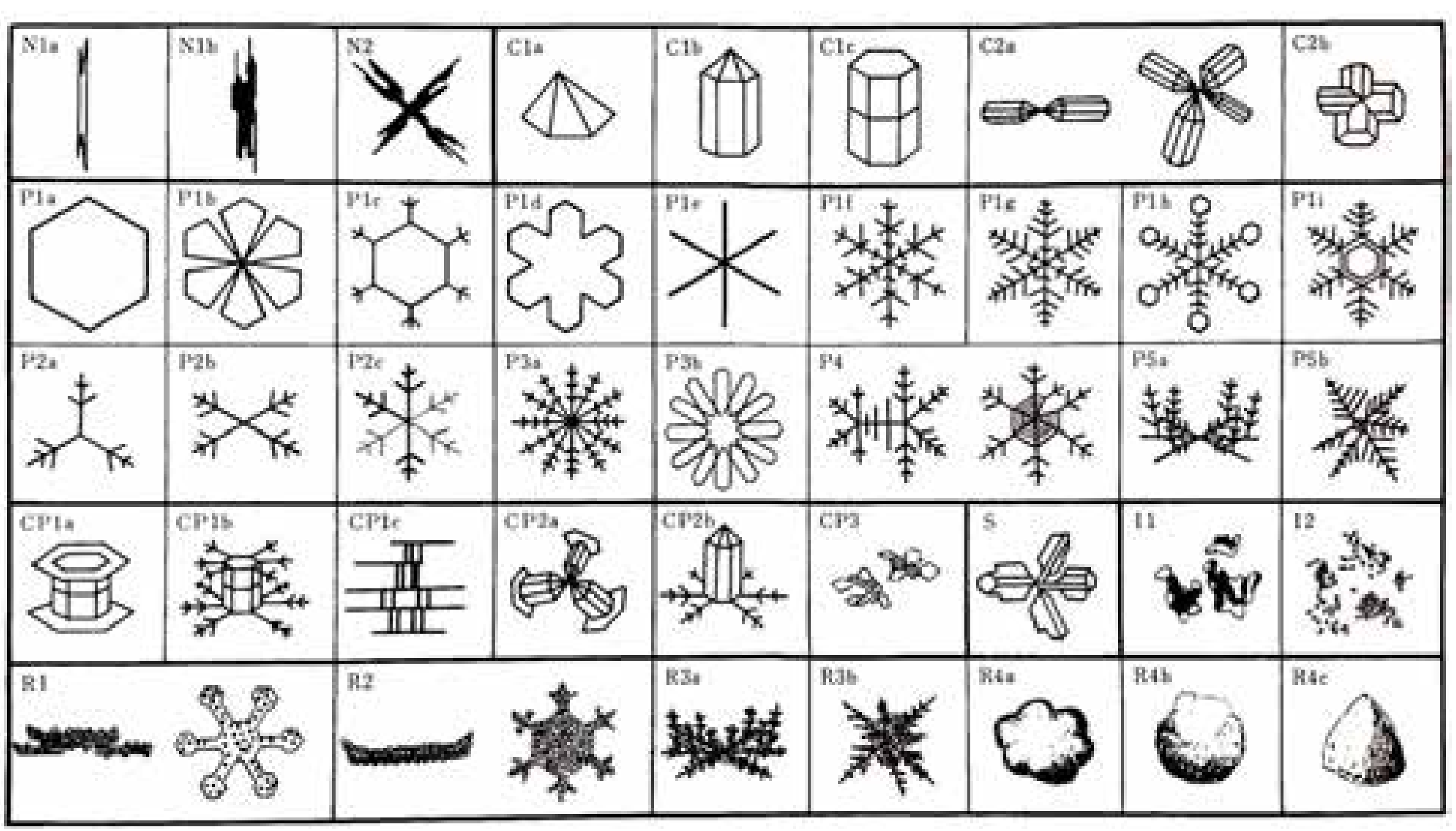

|  |  |
|---|---|
| N1a | Elementary needle |
| N1b | Bundle of elementary needles |
| N1c | Elementary sheath |
| N1d | Bundle of elementary sheaths |
| N1e | Long solid needle |
| N2a | Combination of needles |
| N2b | Combination of sheaths |
| N2c | Combination of long solid colmns |
| C1a | Pyramid |
| C1b | Cup |
| C1c | Solid bullet |
| C1d | Hollow bullet |
| C1e | Solid column |
| C1f | Hollow column |
| C1g | Solid thick plate |
| C1h | Thick plate of skeleton form |
| C1i | Scroll |
| C2a | Combination of bullets |
| C2b | Combination of columns |
| P1a | Hexagon plate |
| P1b | Crystal with sectorlike branches |
| P1c | Crystal with broad branches |
| P1d | Stellar crystal |
| P1e | Ordinary dendritic crystal |
| P1f | Fernlike crystal |
| P2a | Stellar crystal with plates at ends |
| P2b | Stellar crystal with sectorlike ends |
| P2c | Dendritic crystal with plates on ends |
| P2d | Dendritic crystal with sectorlike ends |
| P2e | Plate with simple extensions |
| P2f | Plate with sectorlike extensions |
| P2g | Plate with dendritic extensions |
| P3a | Two branched crystal |
| P3b | Three-branched crystal |
| P3c | Four-branched crystal |
| P4a | Broad branch crystal with 12 branches |
| P4b | Dendritic crystal with 12 branches |
| P5 | Malformed crystal |
| P6a | Plate with spatial plates |
| P6b | Plate with spatial dendrites |
| P6c | Stellar crystal with spatial plates |
| P6d | Stellar crystal with spatial dendrites |
| P7a | Radiating assemblage of plates |
| P7b | Radiating assemblage of dendrites |
| CP1a | Column with plates |
| CP1b | Column with dendrites |
| CP1c | Multiple capped column |
| CP2a | Bullet with plates |
| CP2b | Bullet with dendrites |
| CP3a | Stellar crystal with needles |
| CP3b | Stellar crystal with columns |
| CP3c | Stellar crystal with scrolls at ends |
| CP3d | Plate with scrolls at ends |
| S1 | Side planes |
| S2 | Scalelike side planes |
| S3 | Combination of side planes, bullets, and columns |
| R1a | Rimed needle crystal |
| R1b | Rimed columnar crystal |
| R1c | Rimed plate or sector |
| R1d | Rimed stellar crystal |
| R2a | Densely rimed plate or sector |
| R2b | Densely rimed stellar crystal |
| R2c | Stellar crystal with rimed spatial branches |
| R3a | Graupel-like snow of hexagonal types |
| R3b | Graupel-like snow of lump type |
| R3c | Graupel-like snow with nonrimed extensions |
| R4a | Hexagonal graupel |
| R4b | Lump graupel |
| R4c | Conelike graupel |
| I1 | Ice particle |
| I2 | Rimed particle |
| I3a | Broken branch |
| I3b | Rimed broken branch |
| I4 | Miscellaneous |
| G1 | Minute column |
| G2 | Germ of skeletal form |
| G3 | Minute hexagonal plate |
| G4 | Minute stellar crystal |
| G5 | Minute assemblage of plates |
| G6 | Irregular germ |

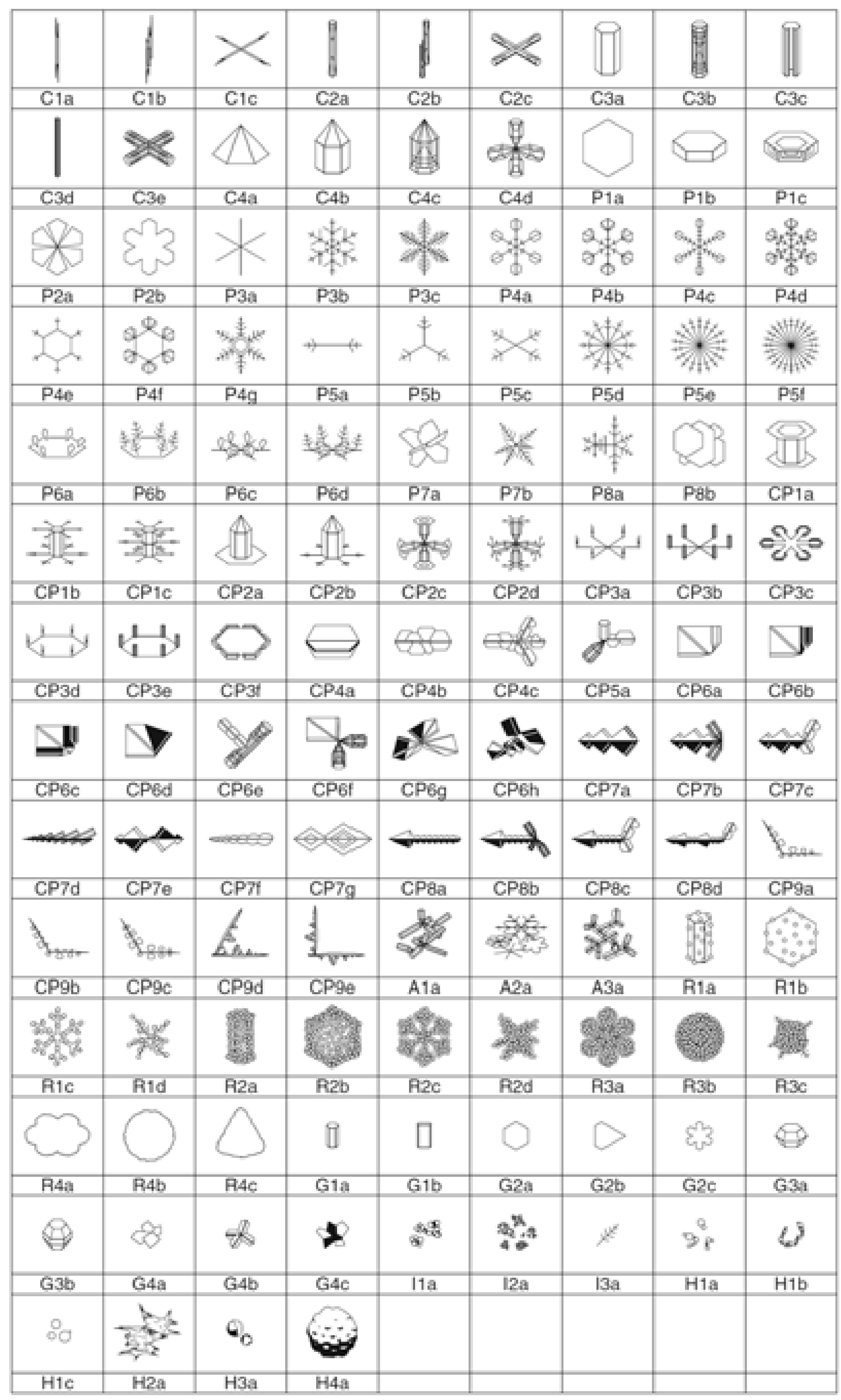

**Figure 10.2: (This and facing page) A progression of snow-crystal classification systems introduced by Nakaya (facing page, top) [1954Nak], extended by Magono and Lee (facing page, bottom) [1966Mag], and further extended by Kikuchi et al. (this page) [2013Kik].**

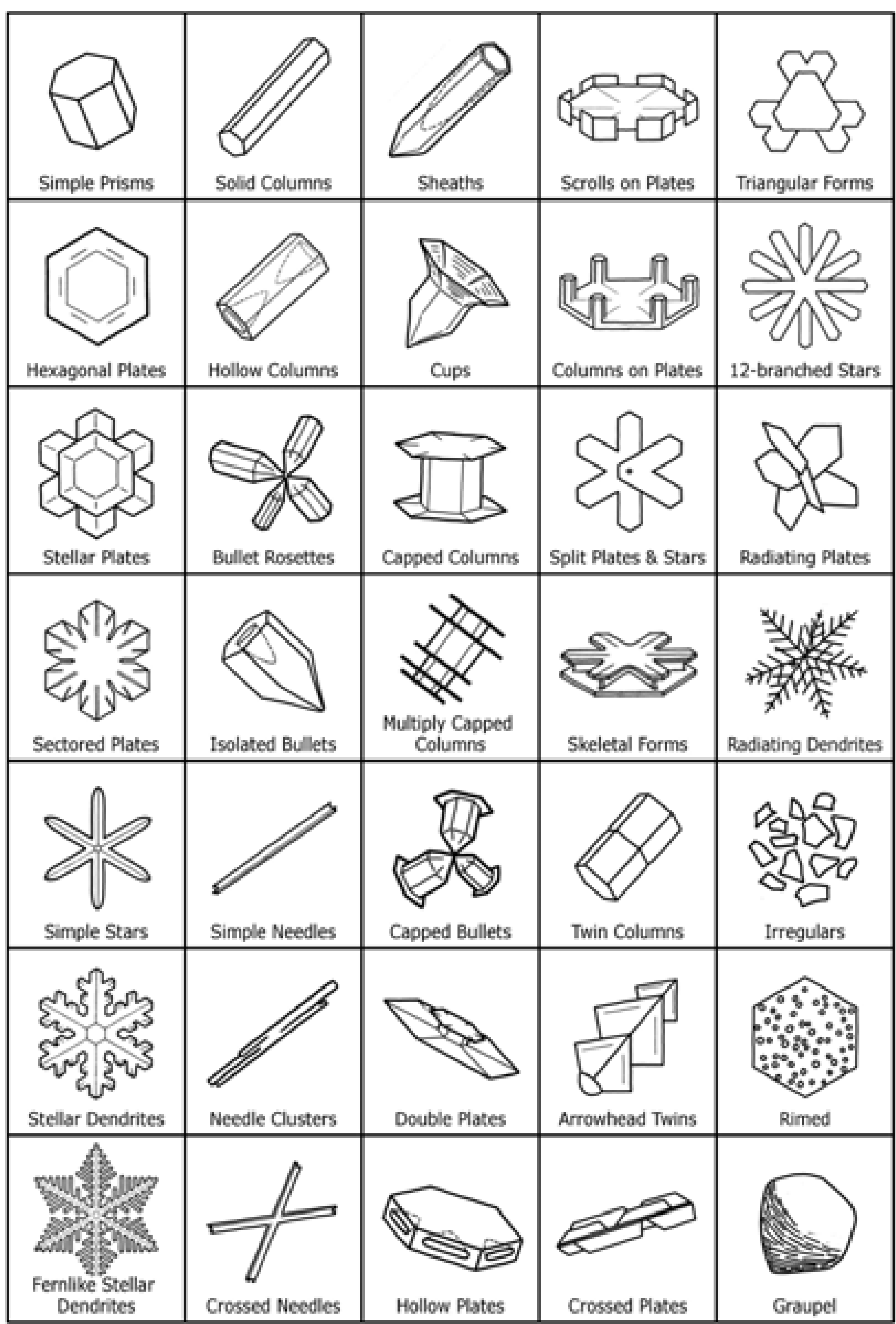

**Figure 10.3: I devised this simplified snow-crystal classification system for everyday observing [2006Lib1]. While there is no definitive method for dividing snow crystals into precise categories, these 35 types provide a reasonable overview of the morphological diversity found in natural snow crystals.**

## Biased Sampling

In the spirit of full disclosure, I like to point out that well-formed snow crystals like those illustrated in classification charts are not the norm, and most of the snowflake types shown are actually quite rare. You may not realize this from photos you have seen, because photographers invariably present a heavily biased sample of what falls from the clouds. We stand out in the bitter cold for hours on end, searching for especially photogenic examples that exhibit well-formed, strikingly symmetrical features. Exceptionally beautiful snow crystals are a delight to behold, so we work hard to find them. And because most people are not eager to buy a book or read an article showing unattractive snowflakes, those photos do not get published.

To witness an unbiased snowflake sample, you need only go outside during a light snowfall and have a look. Every snowfall has a different character, and certain weather conditions are conducive to producing photogenic crystals (see Chapter 11). But essentially all snowfalls bring many examples from the "Irregulars" category in Figure 10.3, and the crystals in Figure 10.4 provide a representative sample. These small, somewhat malformed plate-like crystals are extremely common, and some snowfalls deliver little else. I sometimes call this "granular snow" because the crystals look a lot like icy grains of sand.

Another common occurrence is when growing snow crystals collide with cloud droplets, creating what is called *rime* – basically collections of minute, frozen droplets. Figure 10.5 shows an example where a stellar snow crystal first developed normally in a region relatively free of cloud droplets, and then moved into a dense cloud and accumulated a thick coating of rime. If the rime coating becomes so thick that the entire structure is mostly an agglomeration of frozen droplets,

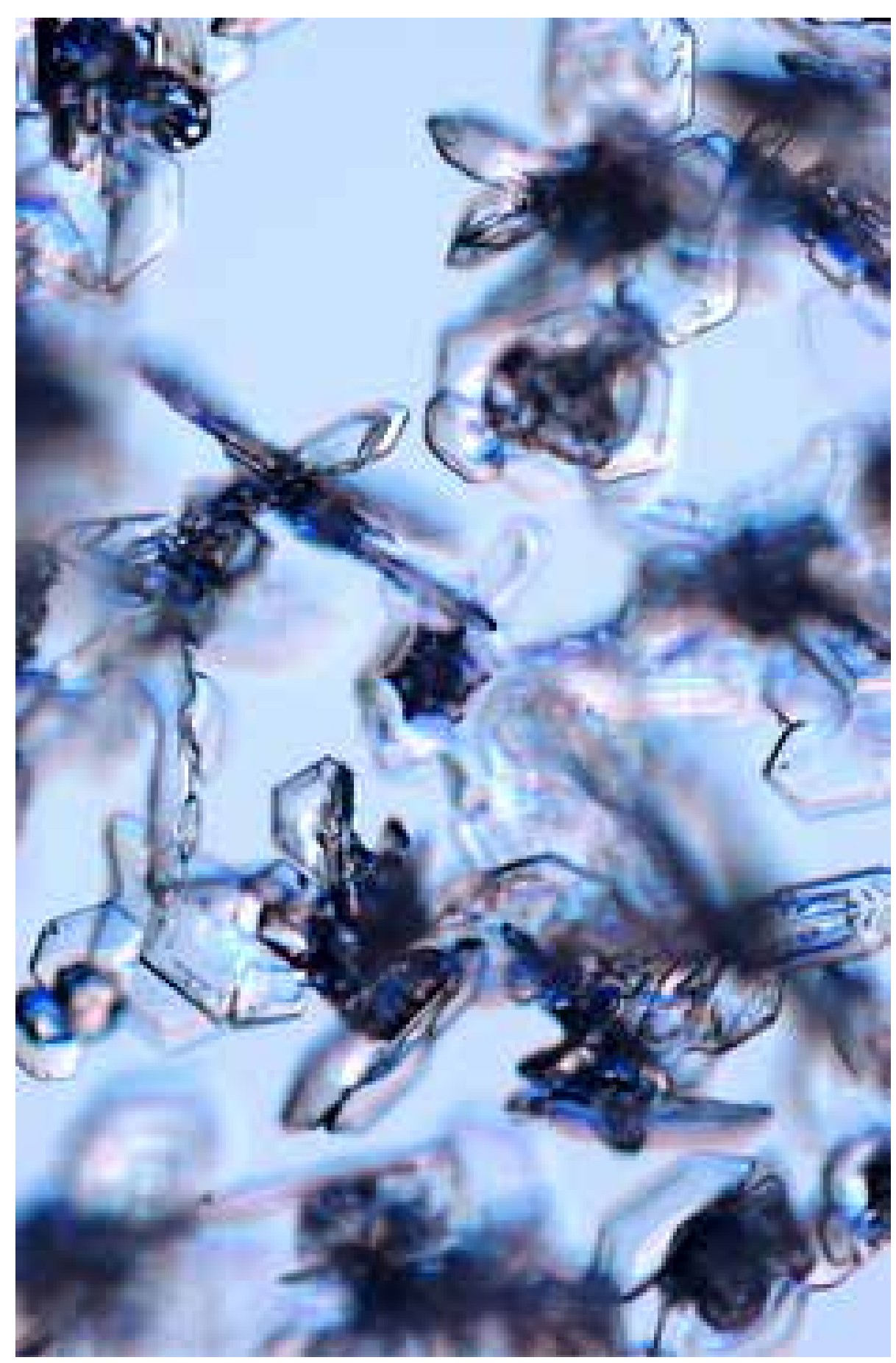

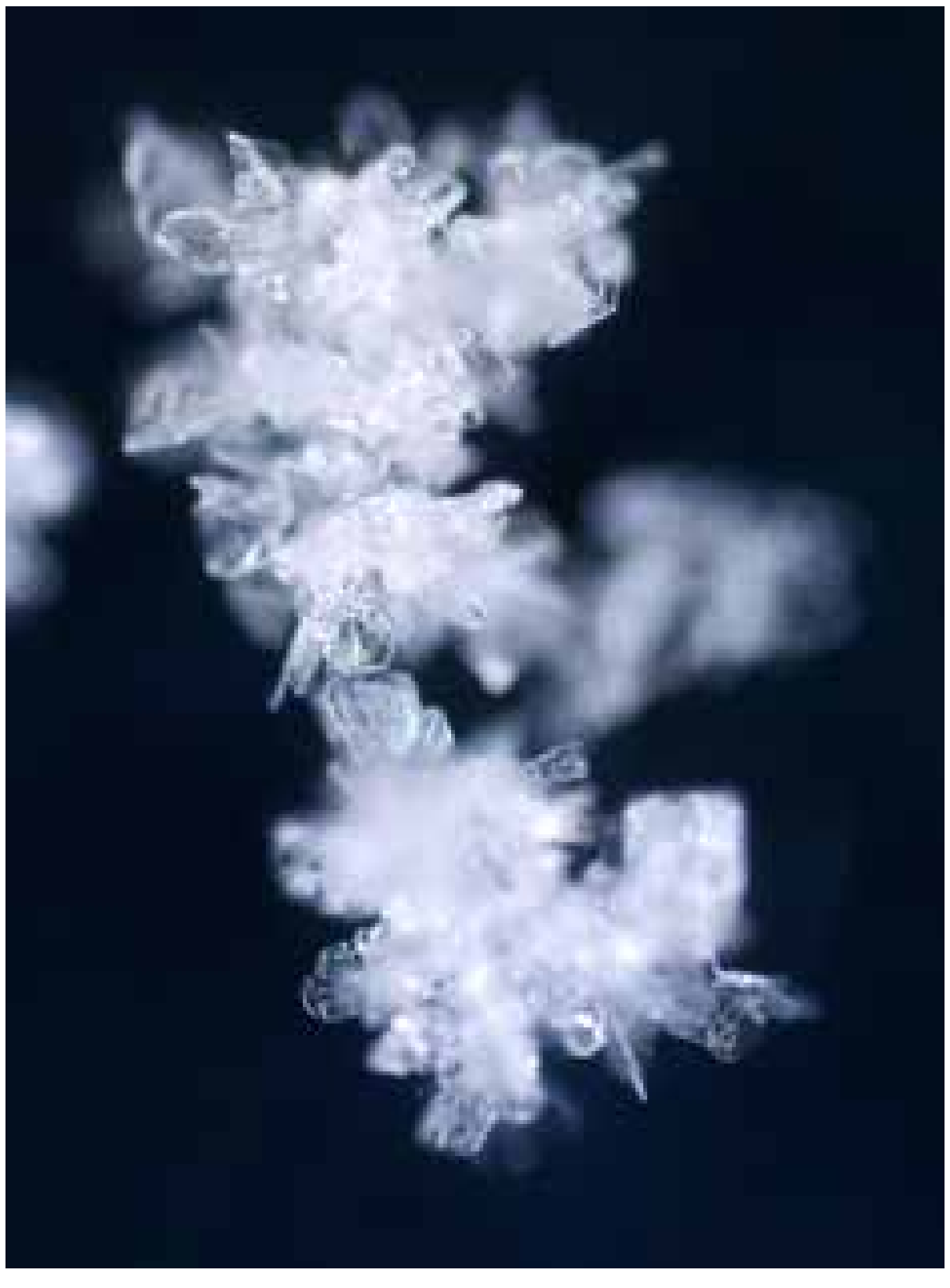

**Figure 10.4: Although this chapter focuses on different types of well-formed snow crystals, clumps of small "irregular" crystals like those shown in these two photos are far more common.**

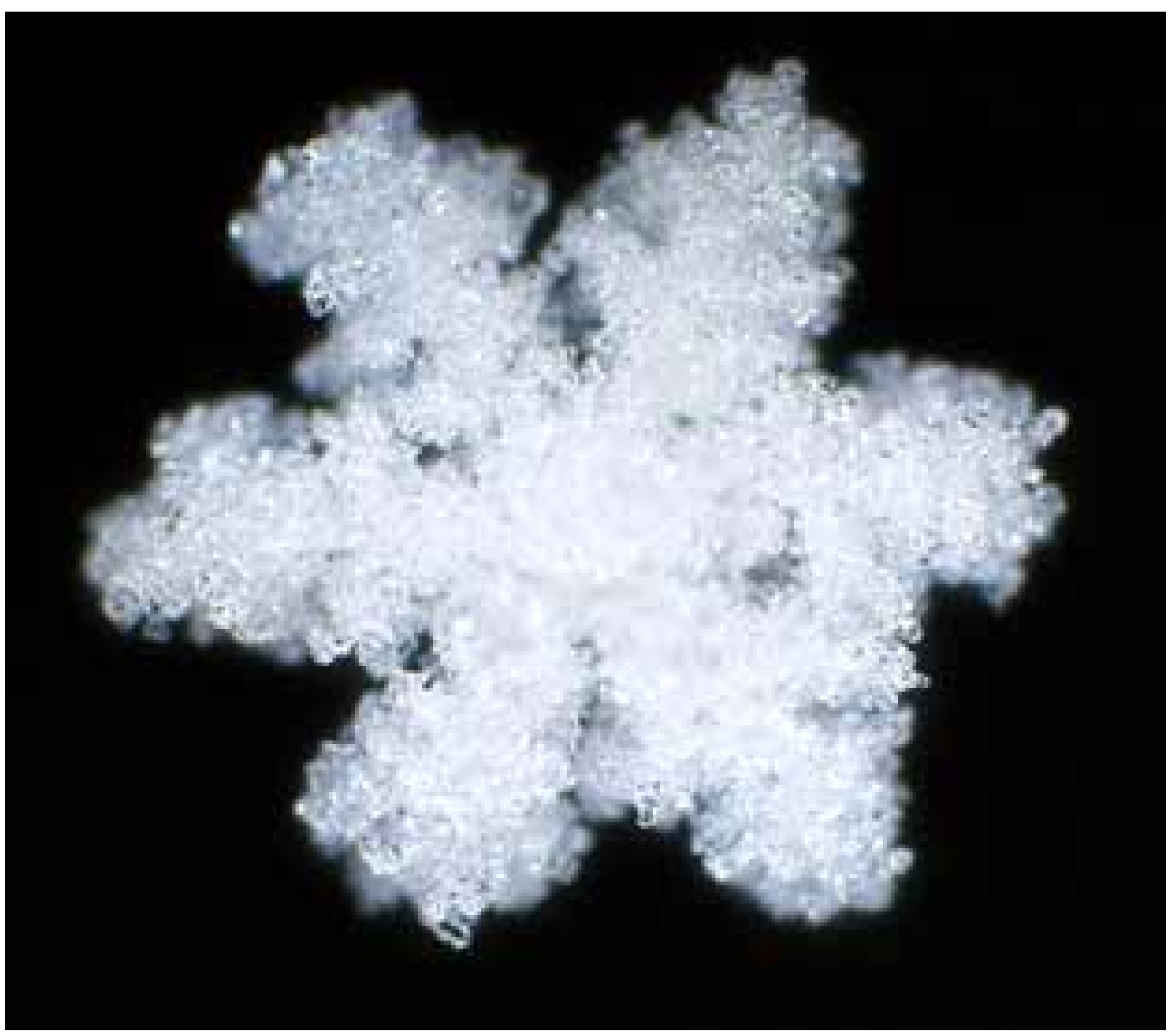

**Figure 10.5: As they are growing and falling, snow crystals often collide with water droplets from the surrounding clouds. The supercooled droplets immediately freeze onto the ice surface, and this example shows a thick coating of *rime* on a stellar crystal.**

then it is called *graupel*, or *soft hail*. Again, some snowfalls deliver mostly rimed crystals.

An unfortunate truth when snowflake watching is that granular snow and rimed crystals are especially prevalent when the temperature is near 0 C, which includes a lot of snowfalls. Because population centers tend to form in moderate climates, and -15 C (5 F) is considered bitter cold by typical standards, the laws of probability suggest that most people will rarely observe exceptionally beautiful snow crystals where they live, even when snow is fairly common. I discuss this and related problems further when considering snowflake photography in Chapter 11.

The goal and tenor of the field guide in this chapter is much like what you find in a guide to mineral crystals. Mineral books tend to focus on beautiful, single-crystal specimens, as these represent the basic mineral types. But if you go hiking up in the mountains to look for yourself, all you will likely encounter is rather ordinary rocks. Small mineral crystals can be seen if you look closely at many rocks, as any geologist will quickly point out; but large single-crystal mineral specimens are exceedingly rare.

The good news for snowflake watching is that finding high-quality snow crystals is much easier than finding quality minerals. Rock hounds have already removed nearly all the nice specimens that were easily retrievable, leaving few behind to discover. You can find large mineral crystals in museums, and for purchase, but not so much in the wild. New mining operations are among the best places to find quality mineral specimens, as they expose unexplored material.

With snow crystals, however, all are made anew for every snowfall, so you have a good chance of finding some outstanding examples. If the temperature is somewhere between -10 C and -20 C, and you know what to look for, you will almost certainly find some noteworthy crystals if you are persistent. Not every snowfall brings exquisite snow-crystal gems; but occasionally one can witness beautiful crystals falling to earth in large numbers. Patience and persistence are often needed, but those few magical snowfalls bringing exquisite crystals make up for all the granular snow and rimed crystals.

The remainder of this chapter presents many examples to illustrate the different snow-crystal morphologies presented in Figure 10.3, along with some discussion of the physical processes involved in their creation. If you go outside to look at the falling snow, magnifier in hand, you may find this chart useful for observing and identifying different crystals. Human nature being what it is, you are more likely to spot a triangular crystal, a bullet rosette, or a double plate if you know what to look for.

**Figure 10.6: (Right) Many snow crystals exhibit a somewhat "travel-worn" appearance, especially when the temperature is warm. In this example, the branch tips are rimed, and nearly all the crystal edges are rounded from sublimation.**

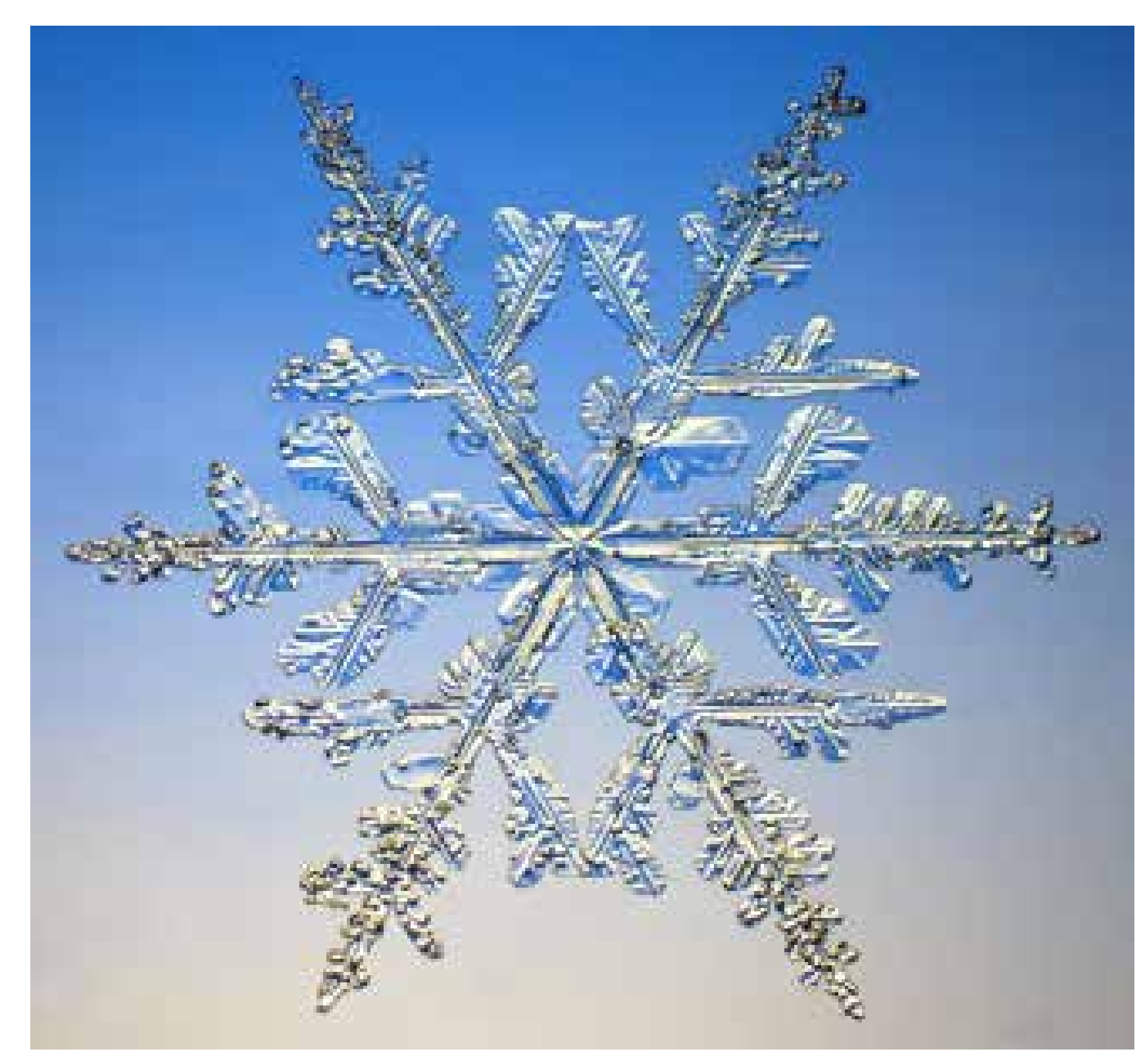

**Figure 10.7: (Below) Six-fold symmetry – the signature characteristic of a well-formed stellar snow crystal – is often over-represented in popular photographs. Most natural specimens exhibit some degree of asymmetry, as illustrated in these examples.**

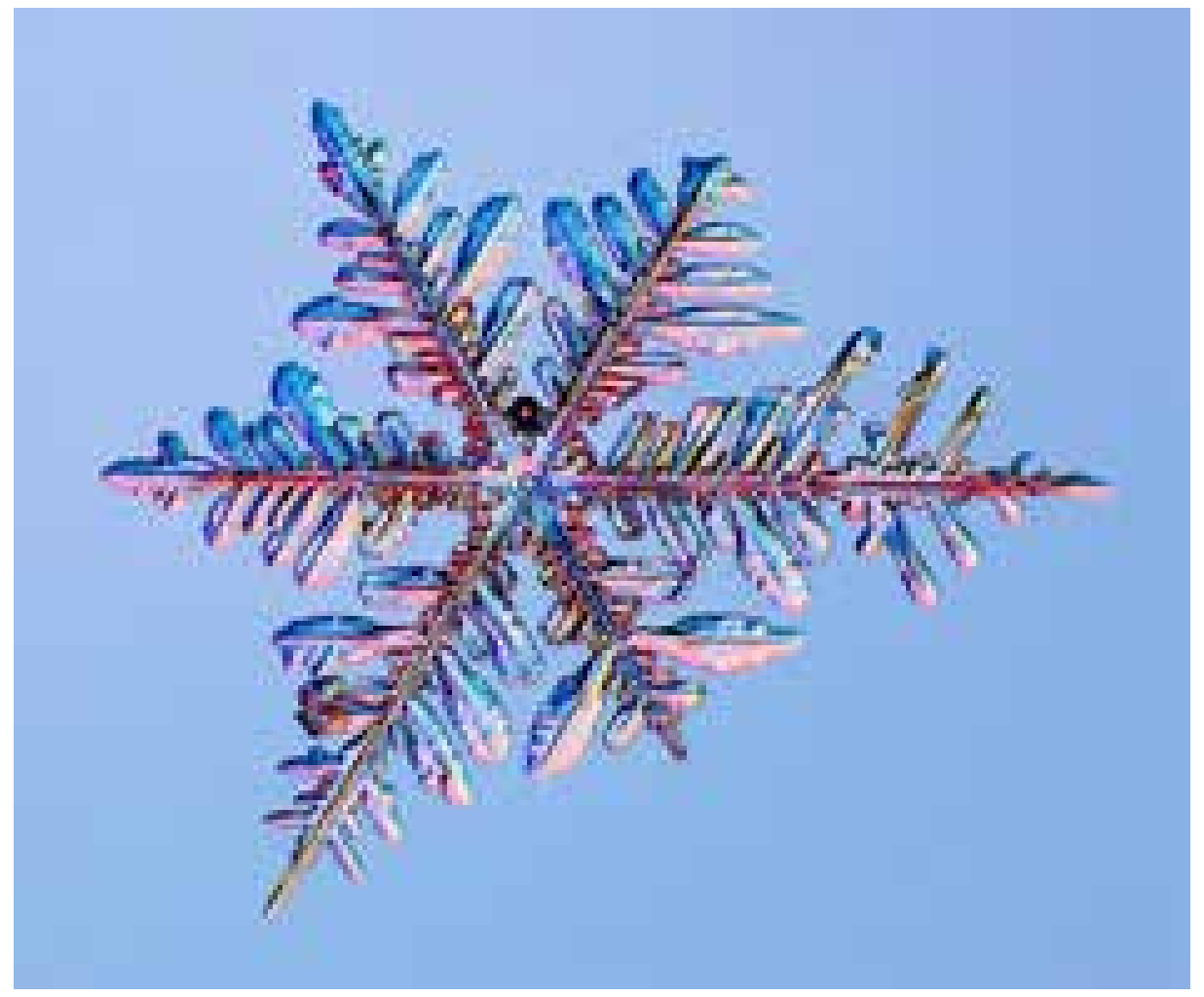

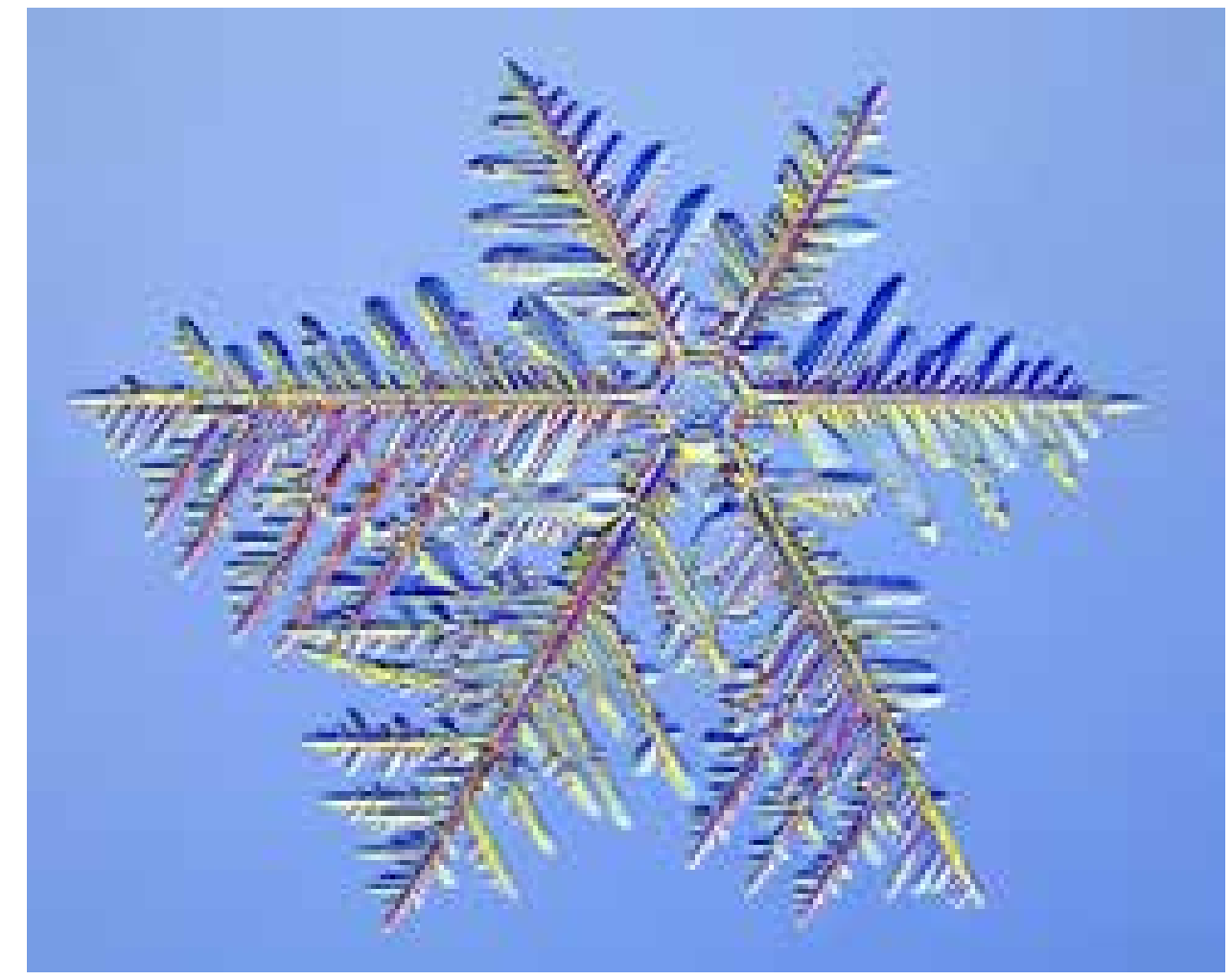

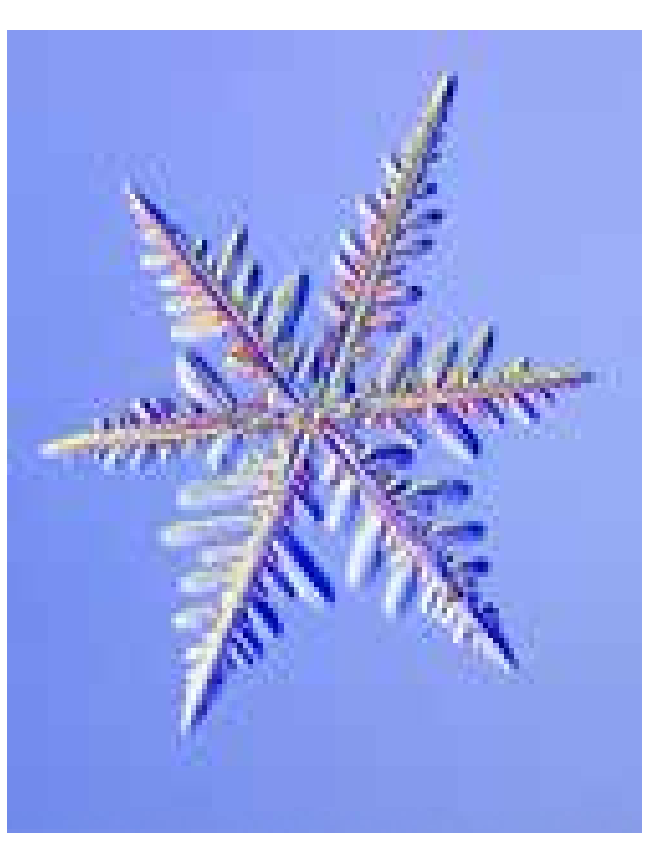

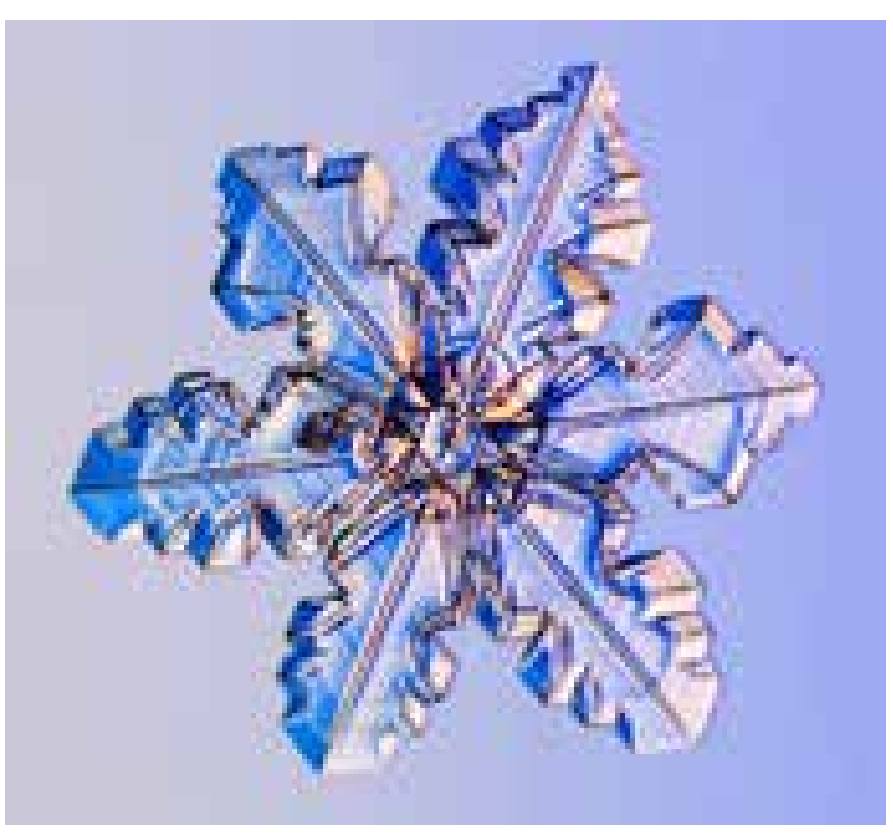

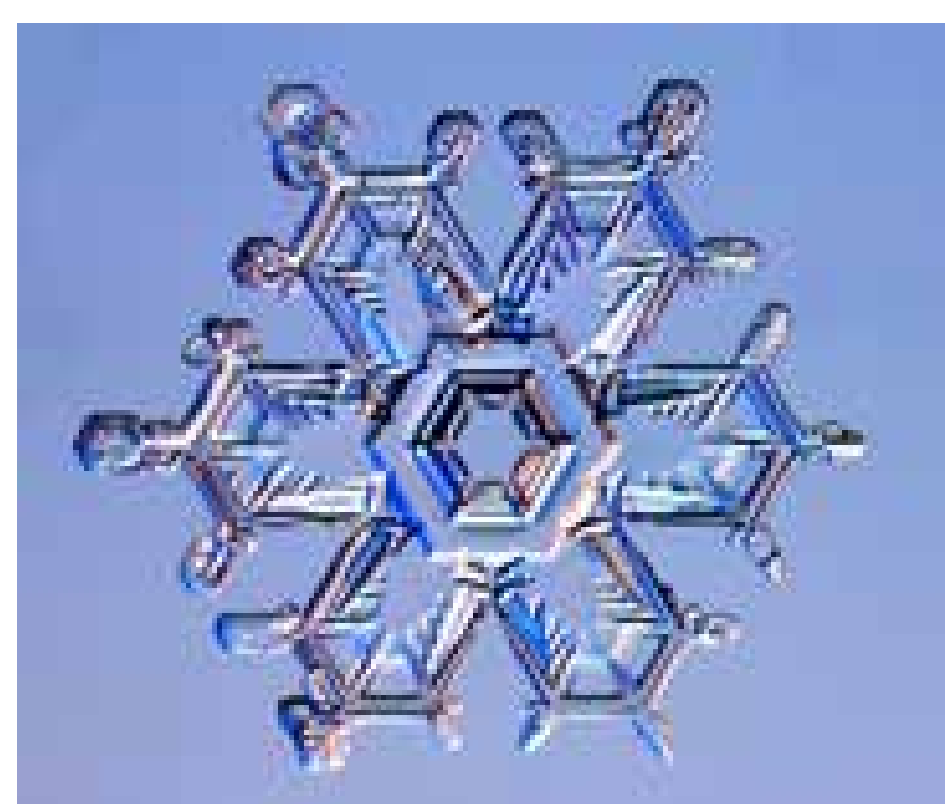

## SIMPLE PRISMS

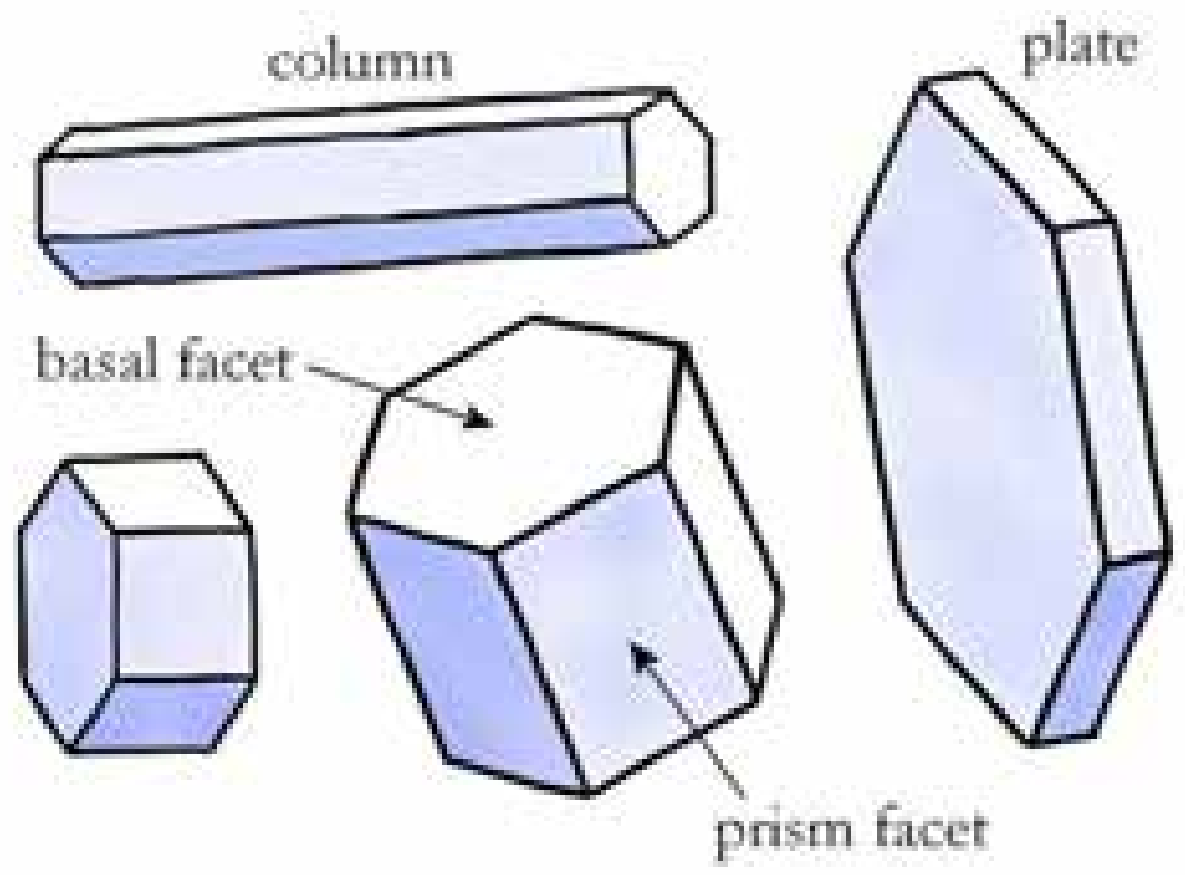

*Simple prisms are small, faceted snow crystals that range from plate-like to columnar in form. They have relatively plain shapes, with minor patterning and no branching. These minimalist snowflakes are common and can be found in most snowfalls, regardless of temperature. However, most simple prisms are so tiny that you need a microscope to see them clearly.*

Every snowflake has its beginning, and these small crystals are essentially young snowflakes that have not had time to grow into larger and more elaborate shapes. The examples shown on this page are roughly 0.3 mm in size, about as large as the period at the end of this sentence.

Faceting is a dominant force in the development of simple prisms because they are still small. The transition to branching has simply not yet had a chance to occur. A rough rule of thumb is that branching begins when a crystal grows to more than half a millimeter in size, although this rule is only approximate. If the humidity is especially high, branching can occur sooner. If the humidity is low, crystals will remain faceted longer.

Occasionally you can observe these small crystals on bitter cold days when the sun is out, so their mirror-like facets sparkle brightly as they tumble through the air. With that image in mind, you can see why these are also called *diamond dust* snow crystals.

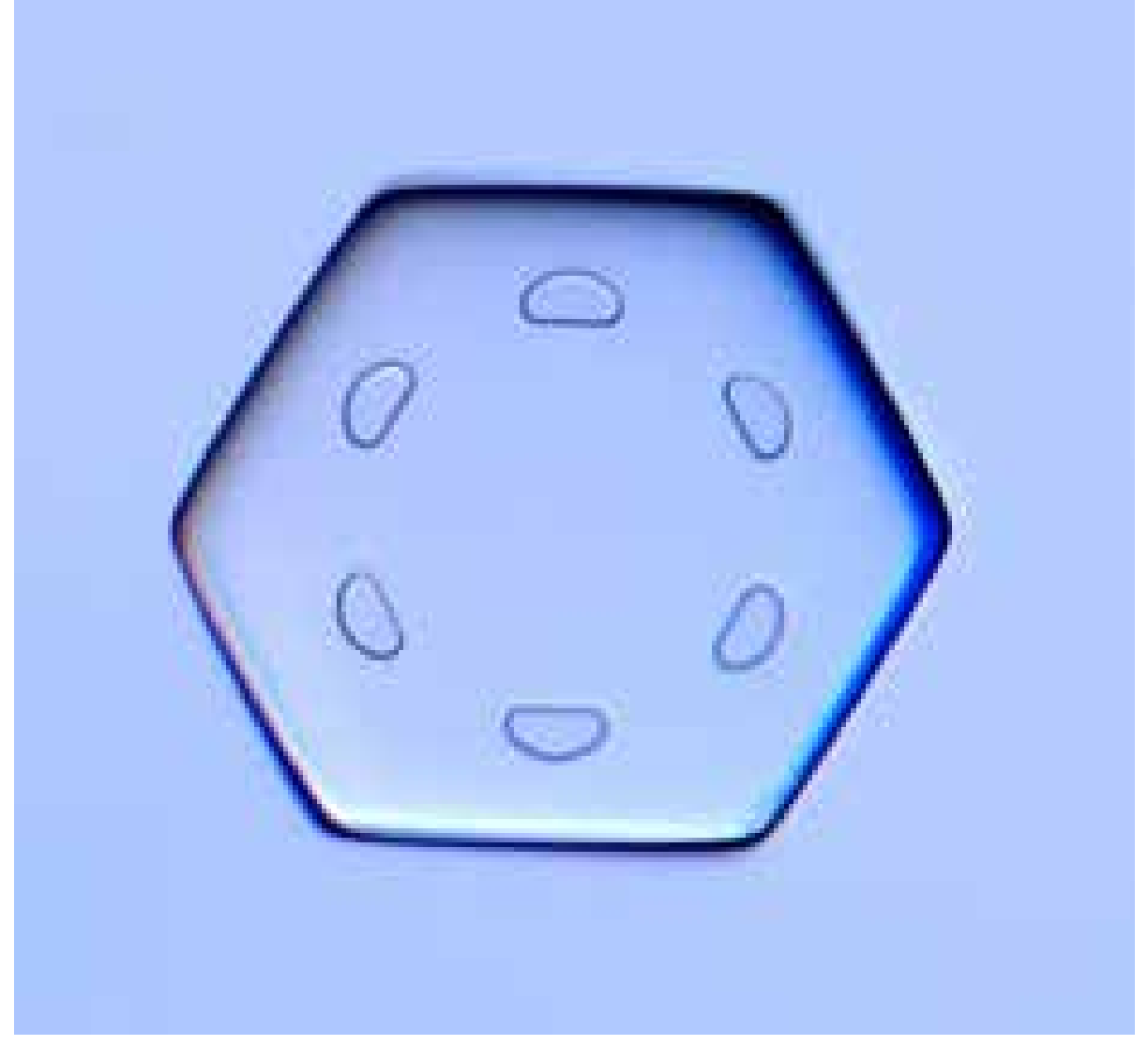

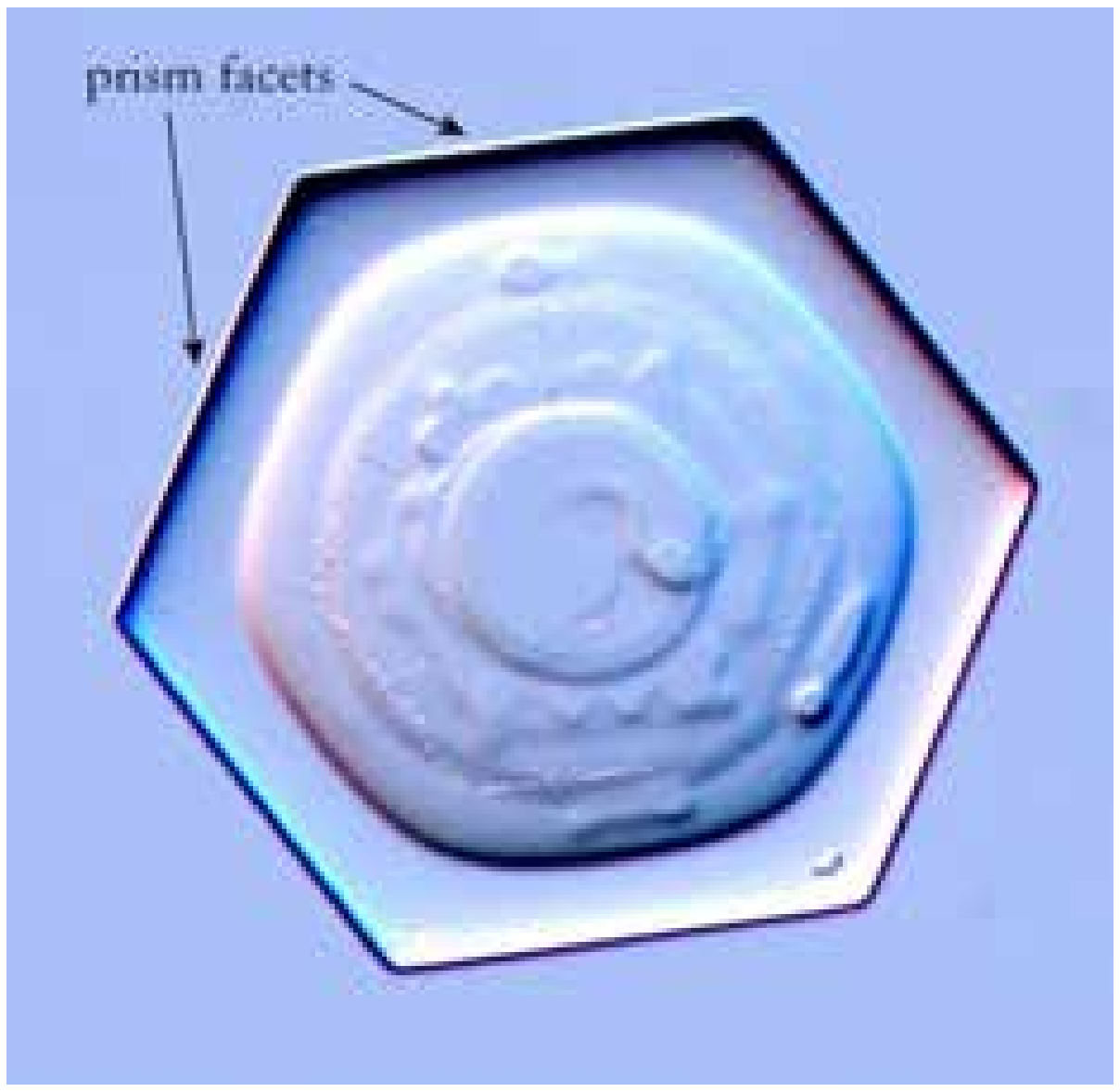

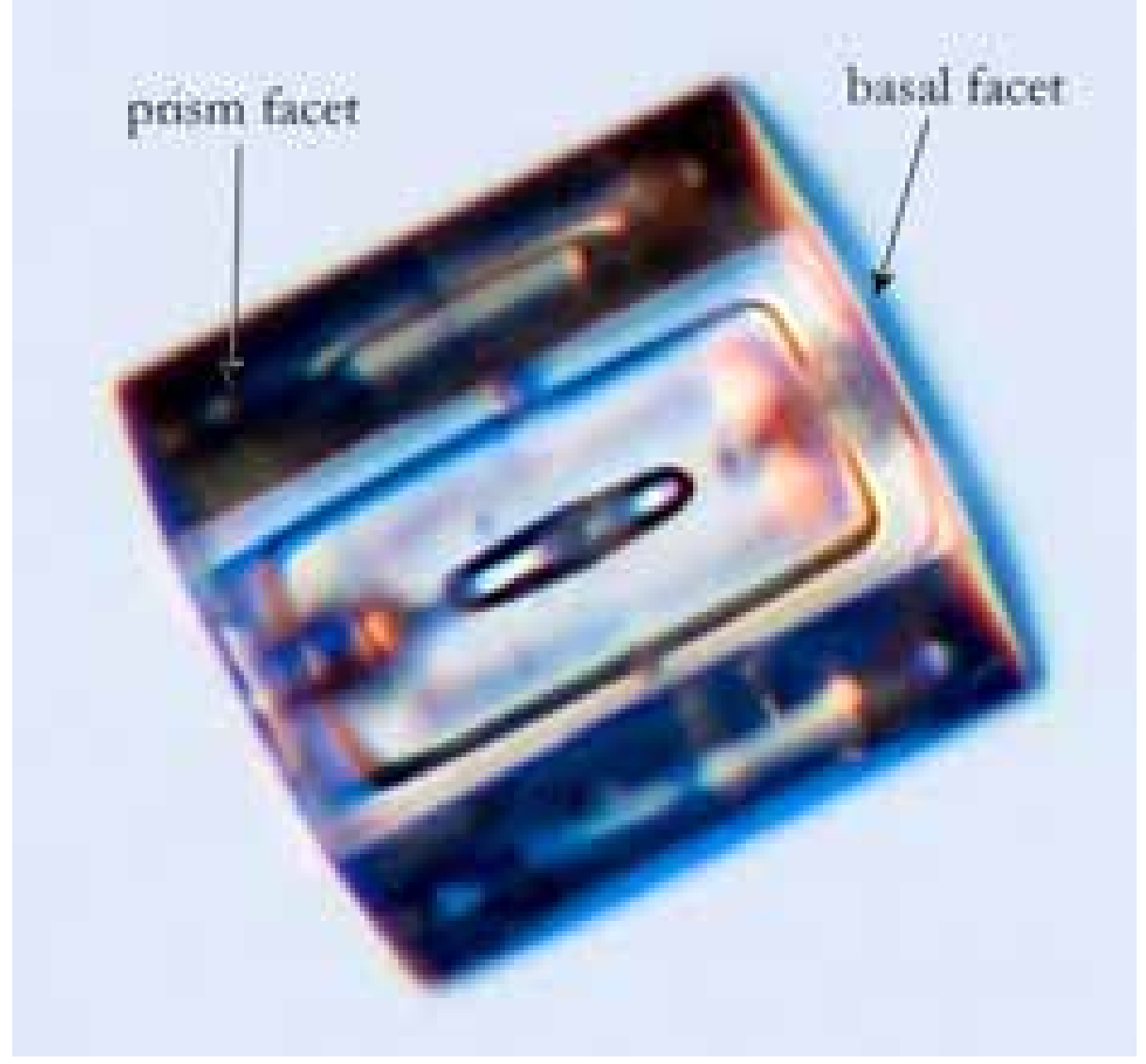

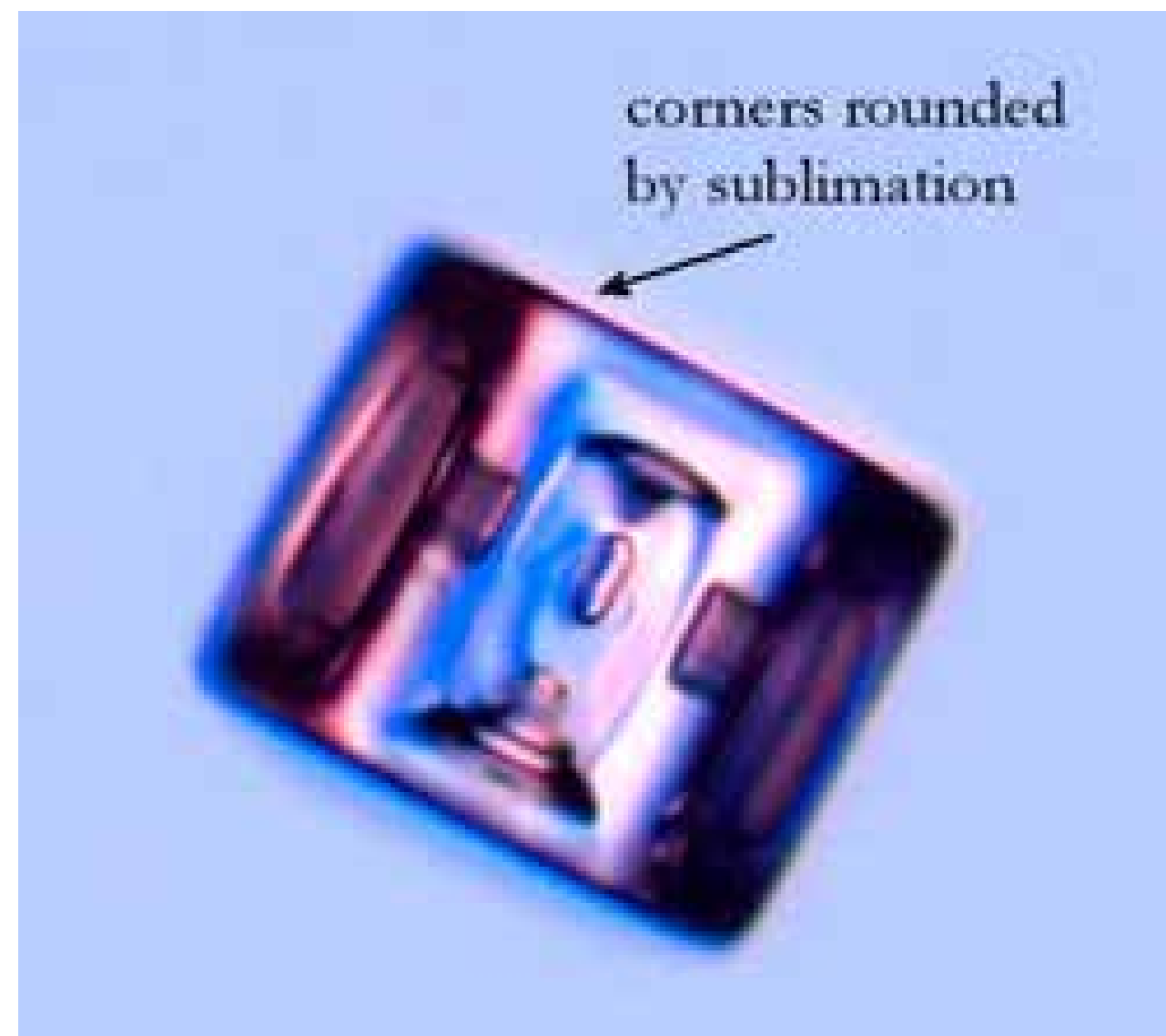

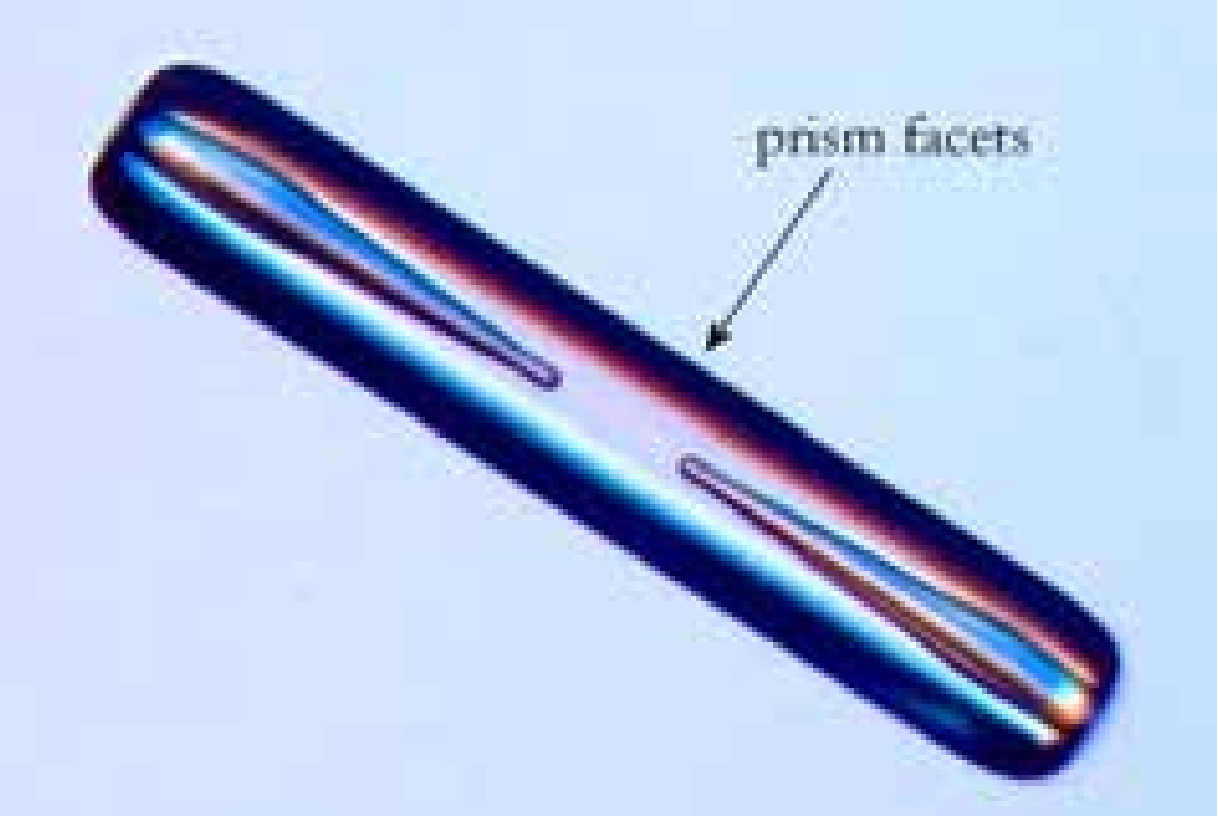

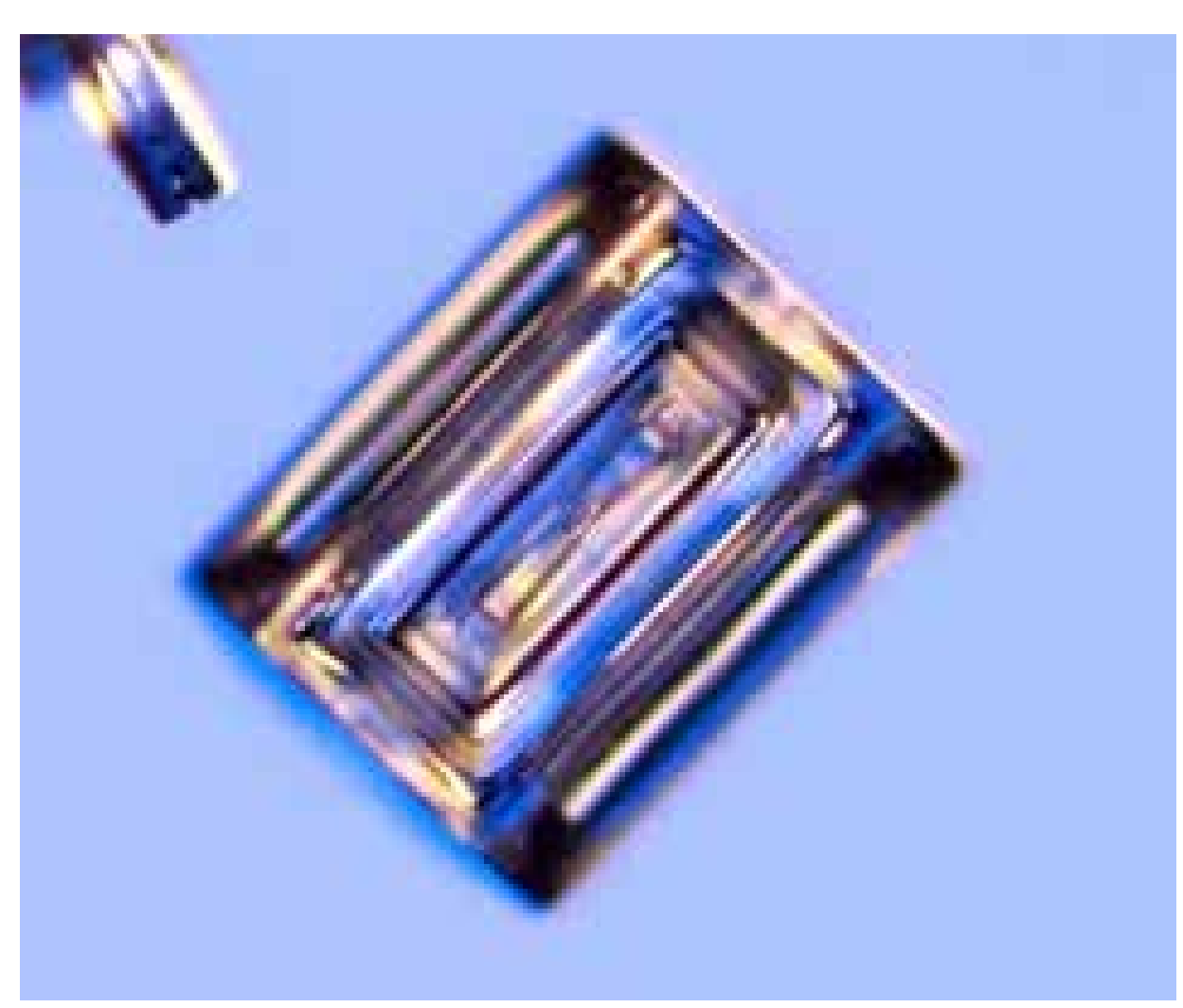

Well-formed crystal facets have razor-sharp corners during growth, but this is not always what you see in the pictures. Sublimation will often round the edges, as you can see with the small prism above. The rounding of sharp features is especially noticeable on smaller specimens, and when the temperature is warm. Sublimation is always an unknown factor when snowflake watching, because you don't know what conditions the different crystals have been through after forming. By the time it reaches the ground, a crystal may look quite different than it did when it was growing up in the clouds.

When photographing snow crystals, I usually illuminate them from behind with colored lights, giving my photos a bright background. In the photo at right, Canadian photographer Don Komarechka [2013Kom] used a ring flash to capture these bright, glass-like crystals on a dark background. See Chapter 11 for more about snowflake photography.

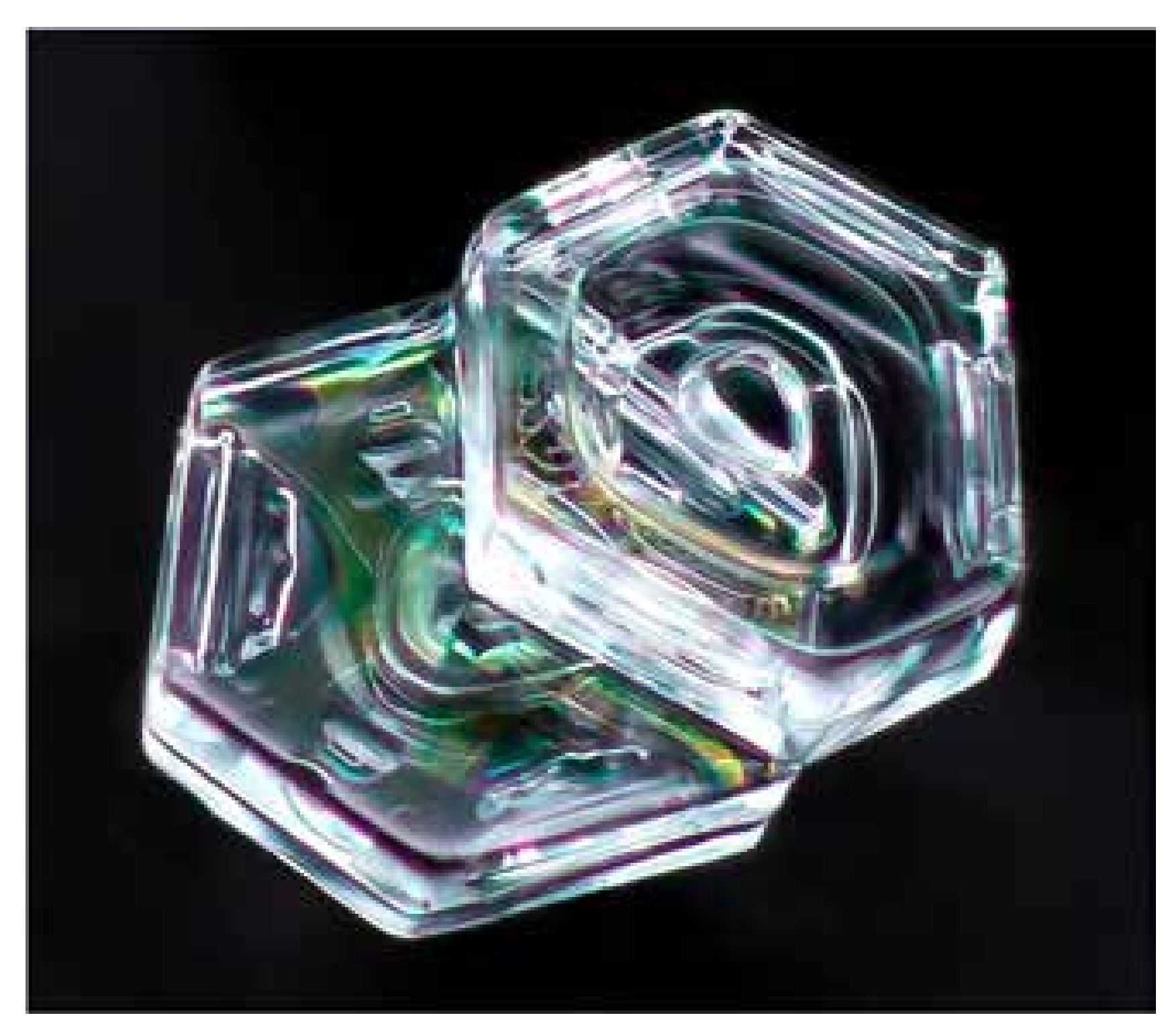

**Hollow Faces.** (Right) For this photograph I focused my camera on one face of a diamond dust prism, about 0.3 mm in size. I caught this crystal quickly on an especially cold day, so sublimation has not yet taken its toll; the corners are still distinct and sharp. I like this picture because it illustrates the hollowing sometimes seen in prism facets. During growth, diffusion gives the corners of the crystal a greater supply of water vapor. The facet centers receive less, so they accumulate material more slowly. Over time, the facet centers lag behind the growth of the edges, as shown in the accompanying sketch. This is a common growth behavior, and is a first step in the transition to branching.

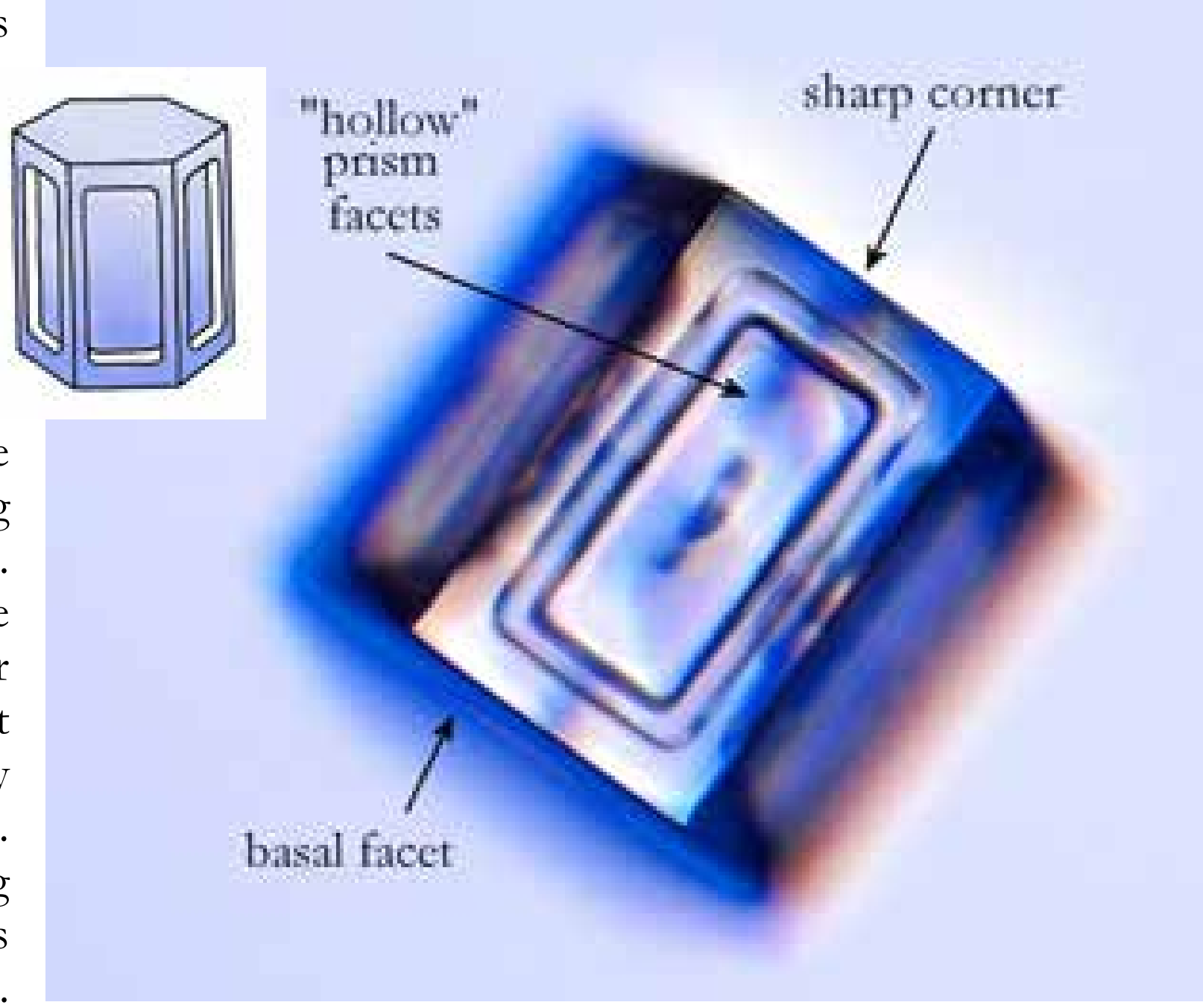

**Antarctic Snow Crystals.** (Below) These tiny crystals were photographed at the South Pole by Walter Tape [1990Tap]. In the dry, bitter cold arctic conditions, snow crystals often grow into simple, sharply faceted prisms like these.

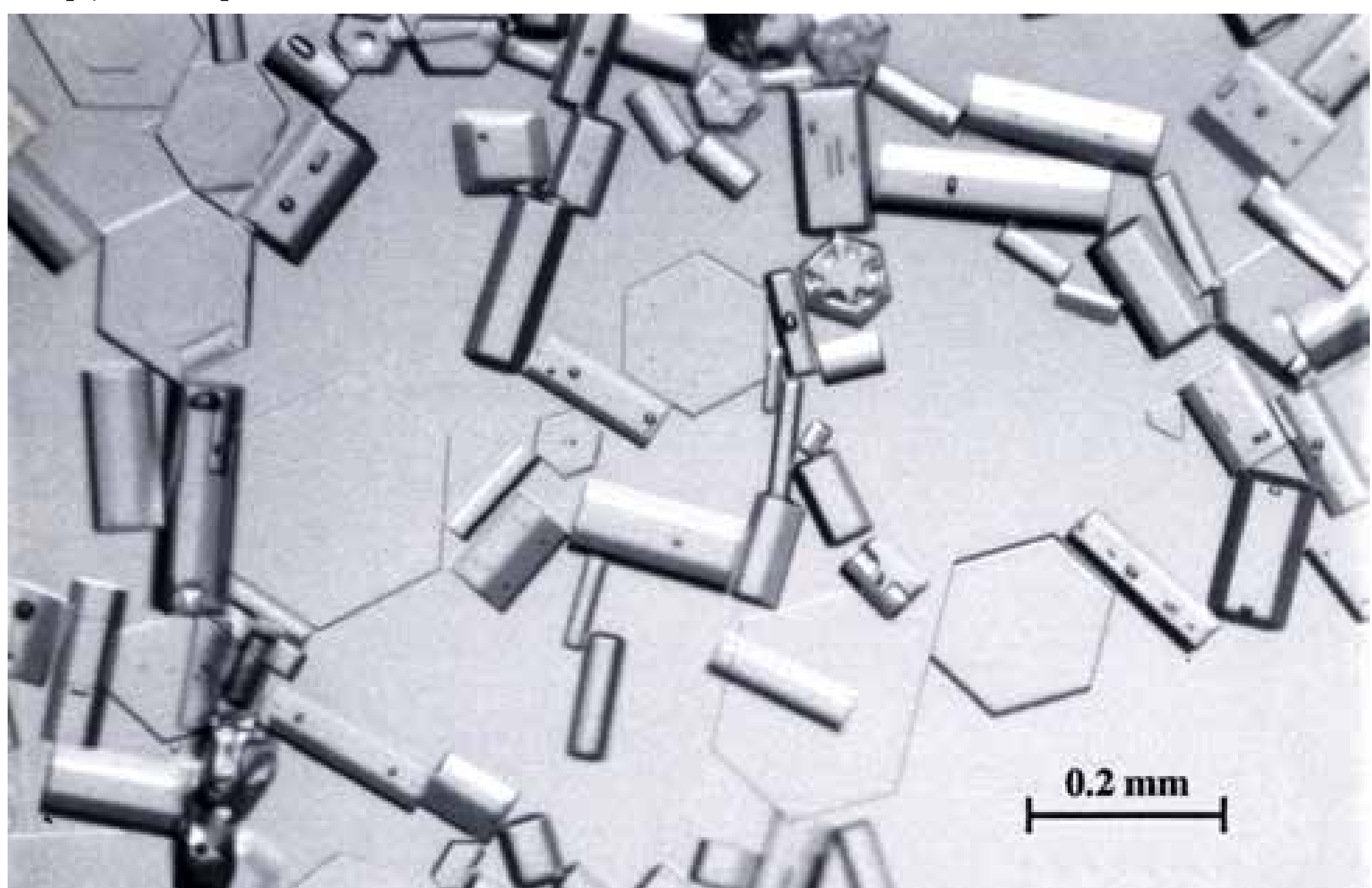

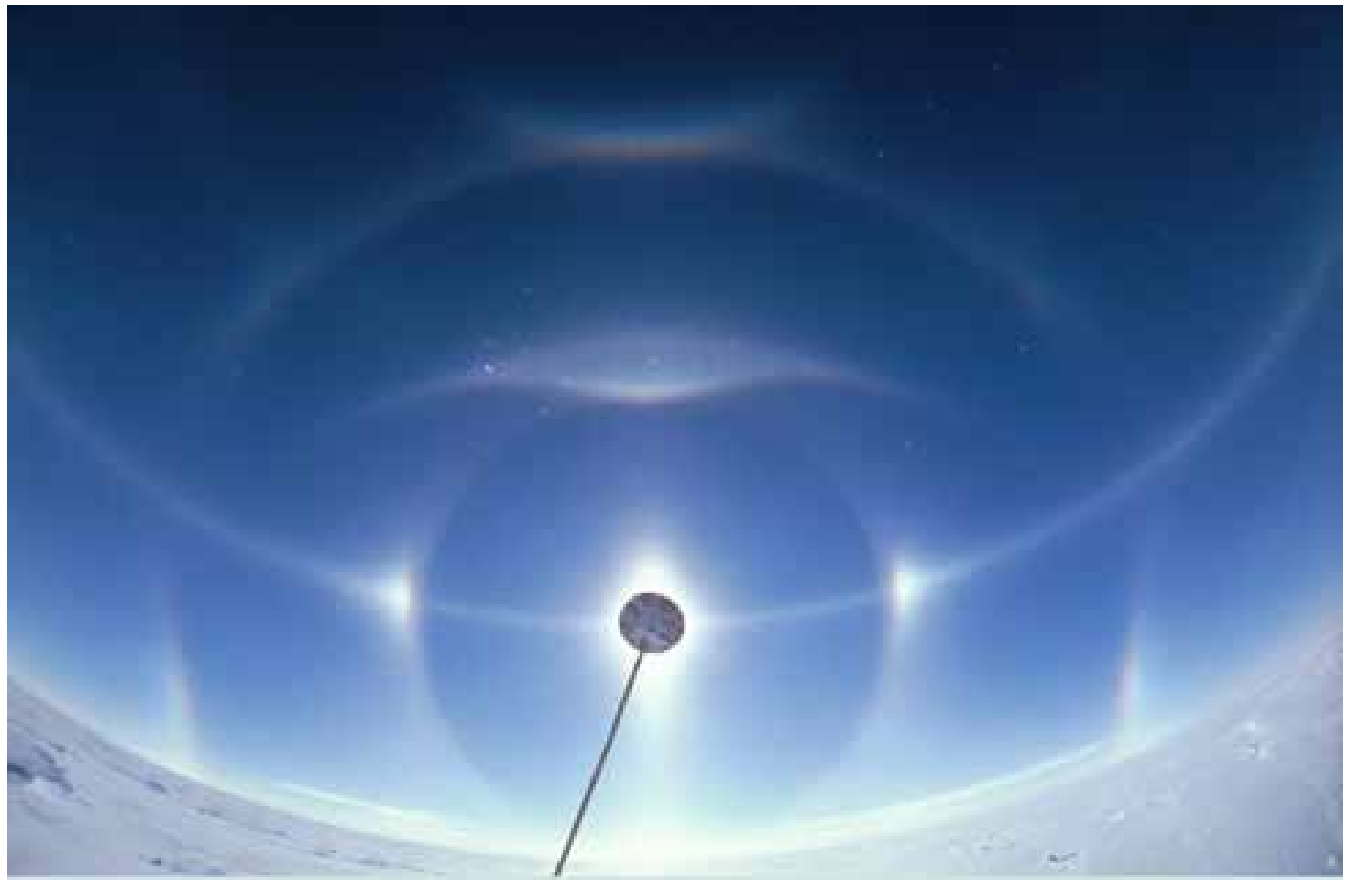

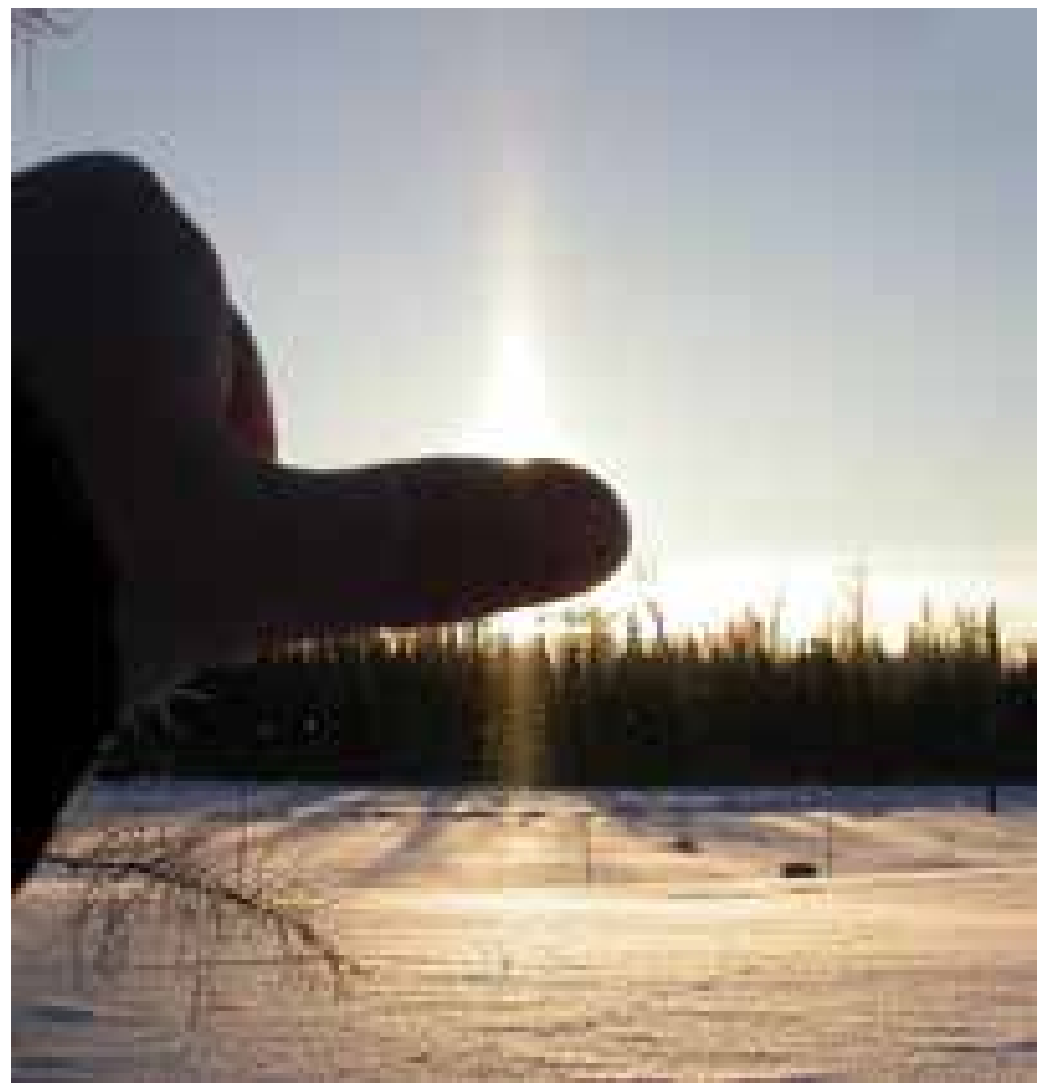

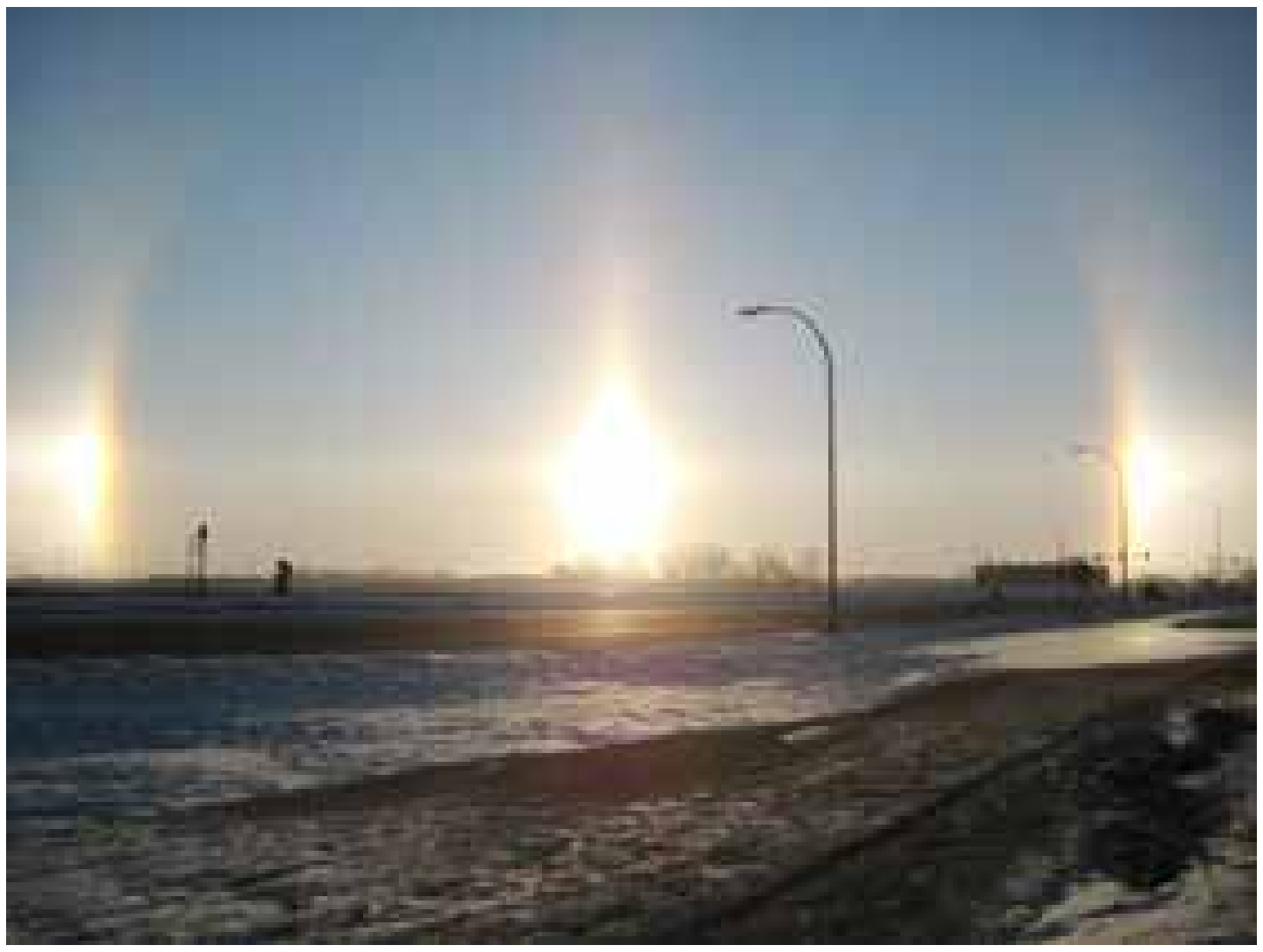

**Atmospheric Halos.** Simple-prism snow crystals are responsible for a variety of atmospheric halo phenomena. The top photo above shows a spectacular halo display captured at the South Pole by Walter Tape [1990Tap]. The lower-left image illustrates a simpler *light pillar* phenomenon that includes the author's thumb blocking the sun, photographed in Cochrane, Ontario. The lower right photo shows the sun flanked by a pair of *sundogs* captured in Fargo, North Dakota by Gopherboy6956. Much has been written about halo phenomena [1980Gre, 1990Tap], and complex reflection/refraction models are needed to explain how falling ice crystals can create such complex patterns of light. For many halo features, including sundogs, the falling crystals must be aligned relative to vertical by aerodynamic forces (see Chapter 3).

# Stellar Plates

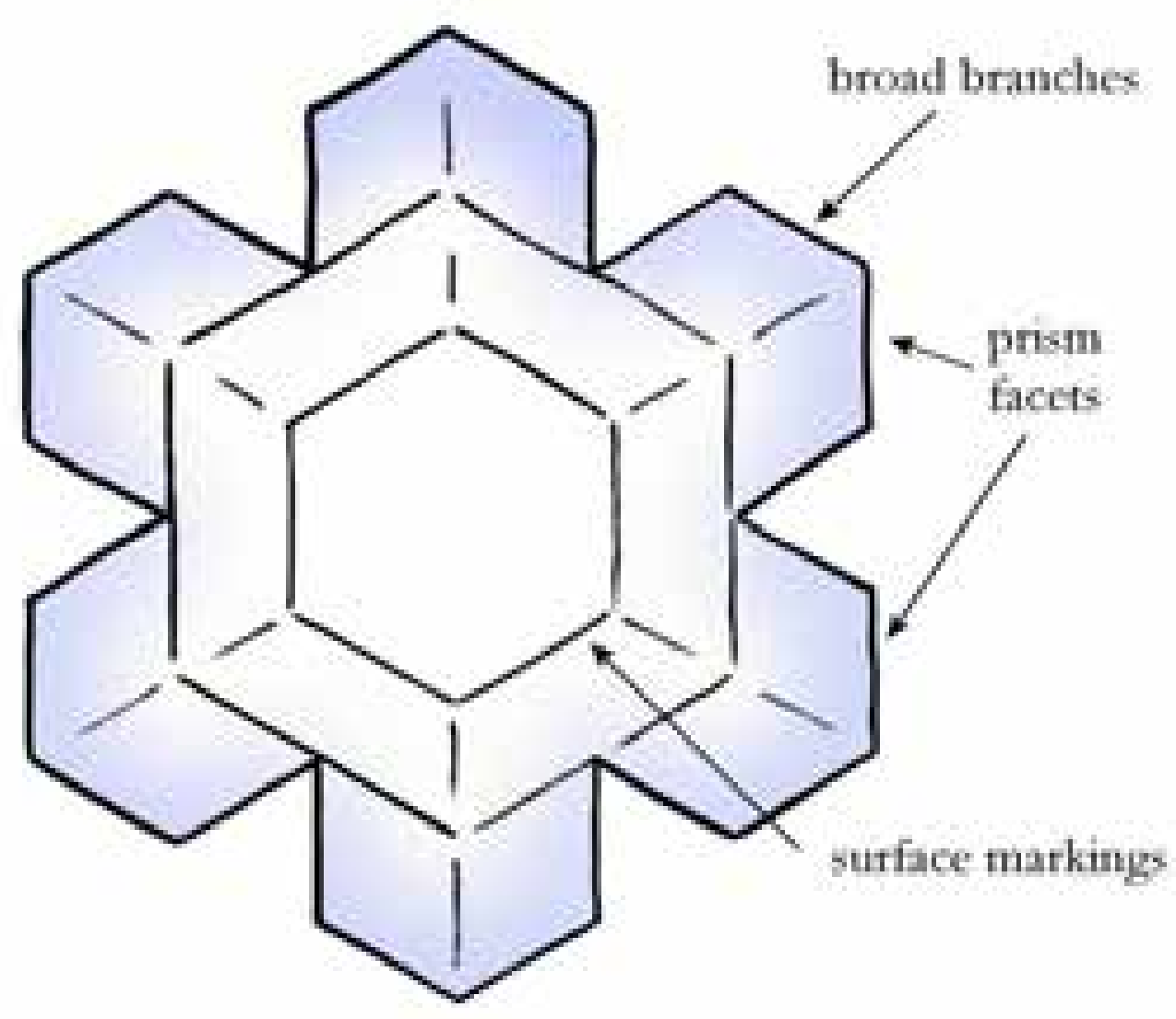

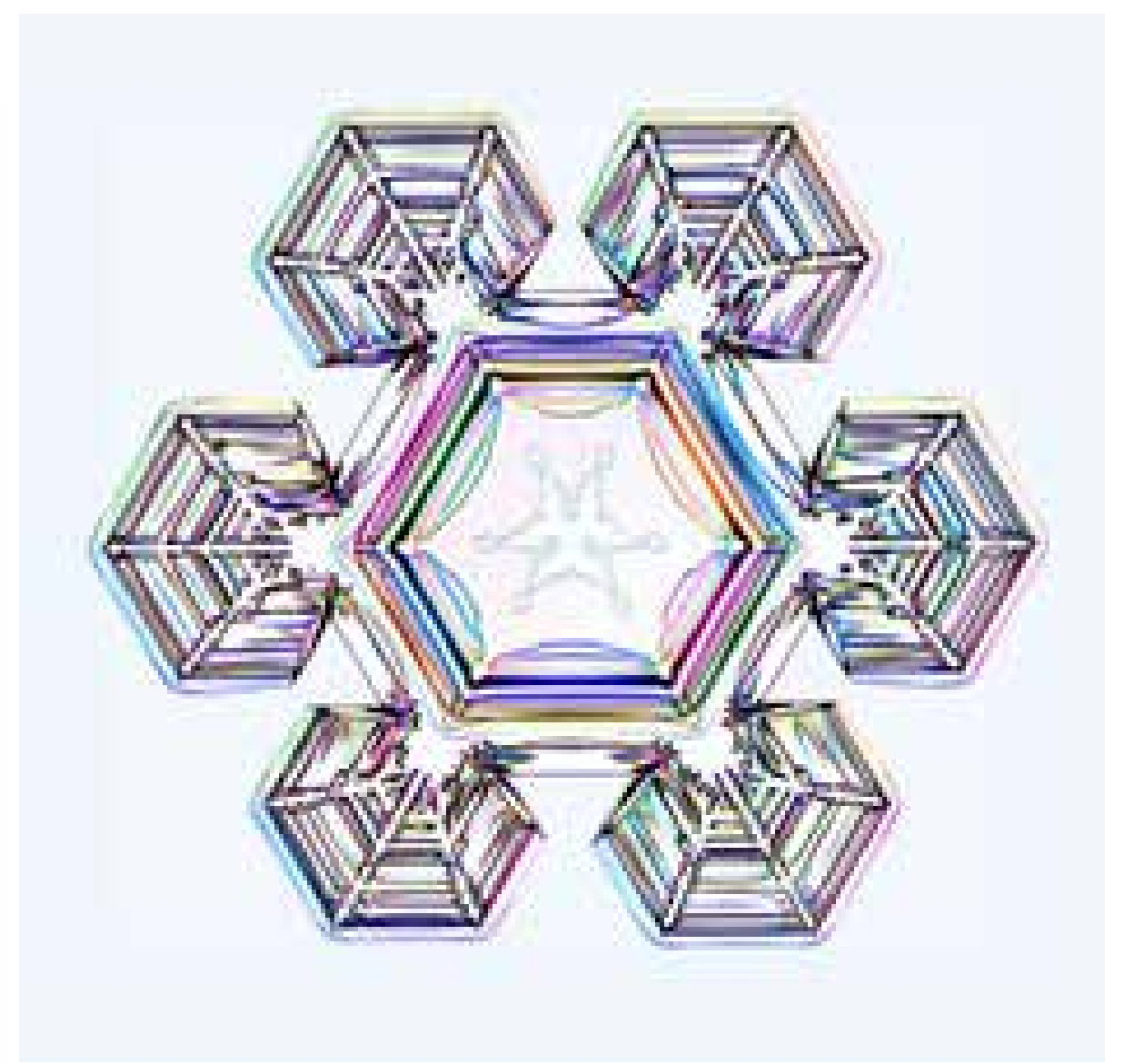

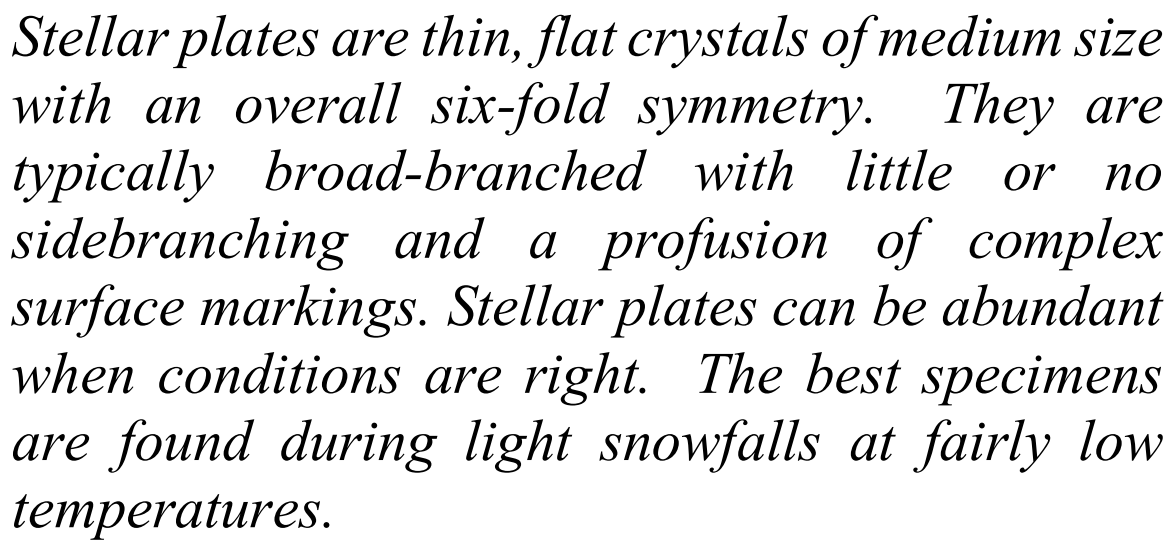

*Stellar plates are thin, flat crystals of medium size with an overall six-fold symmetry. They are typically broad-branched with little or no sidebranching and a profusion of complex surface markings. Stellar plates can be abundant when conditions are right. The best specimens are found during light snowfalls at fairly low temperatures.*

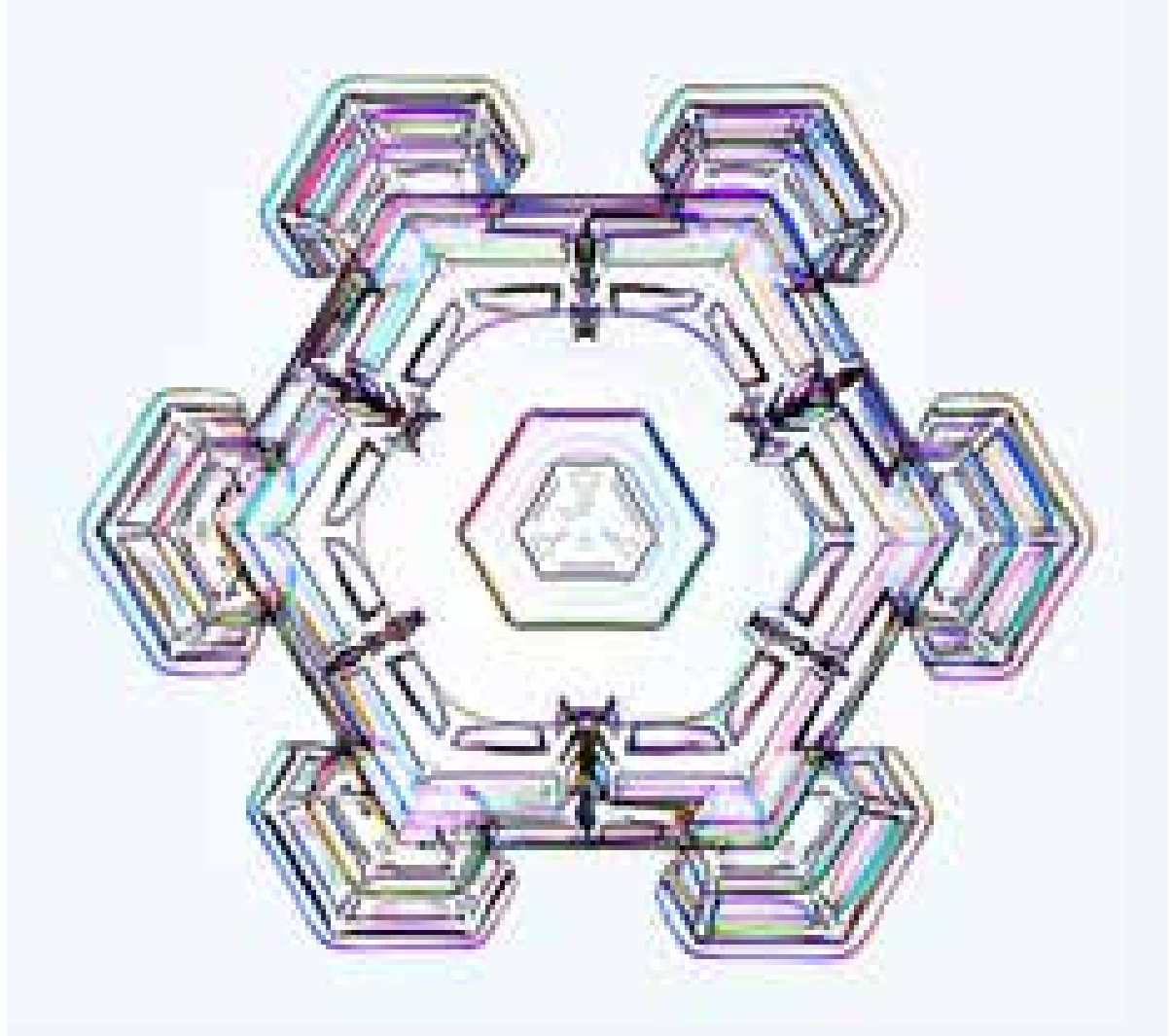

The sparkle you see in falling snow often comes from stellar plates, when their flat basal surfaces catch the light. These crystals are large enough that a simple magnifier gives you a pretty good view of their overall structure, as a good-sized specimen might be two millimeters in diameter. A microscope opens up a whole new realm of observing, however, allowing a detailed look at the intricate patterning on each crystal.

Stellar plates form over a narrow range in temperature, so are not present in all snowfalls. The morphology diagram tells us that large, plate-like crystals will grow when the clouds are near either -15 C or -2 C. At the higher of these temperatures, however, one does not generally find well-formed crystals, because of sublimation and other factors. Thus large stellar crystals mainly appear when the temperature is within a few degrees of -15 C. If you want to find some beautiful stellar plates, you have to wait for just the right conditions.

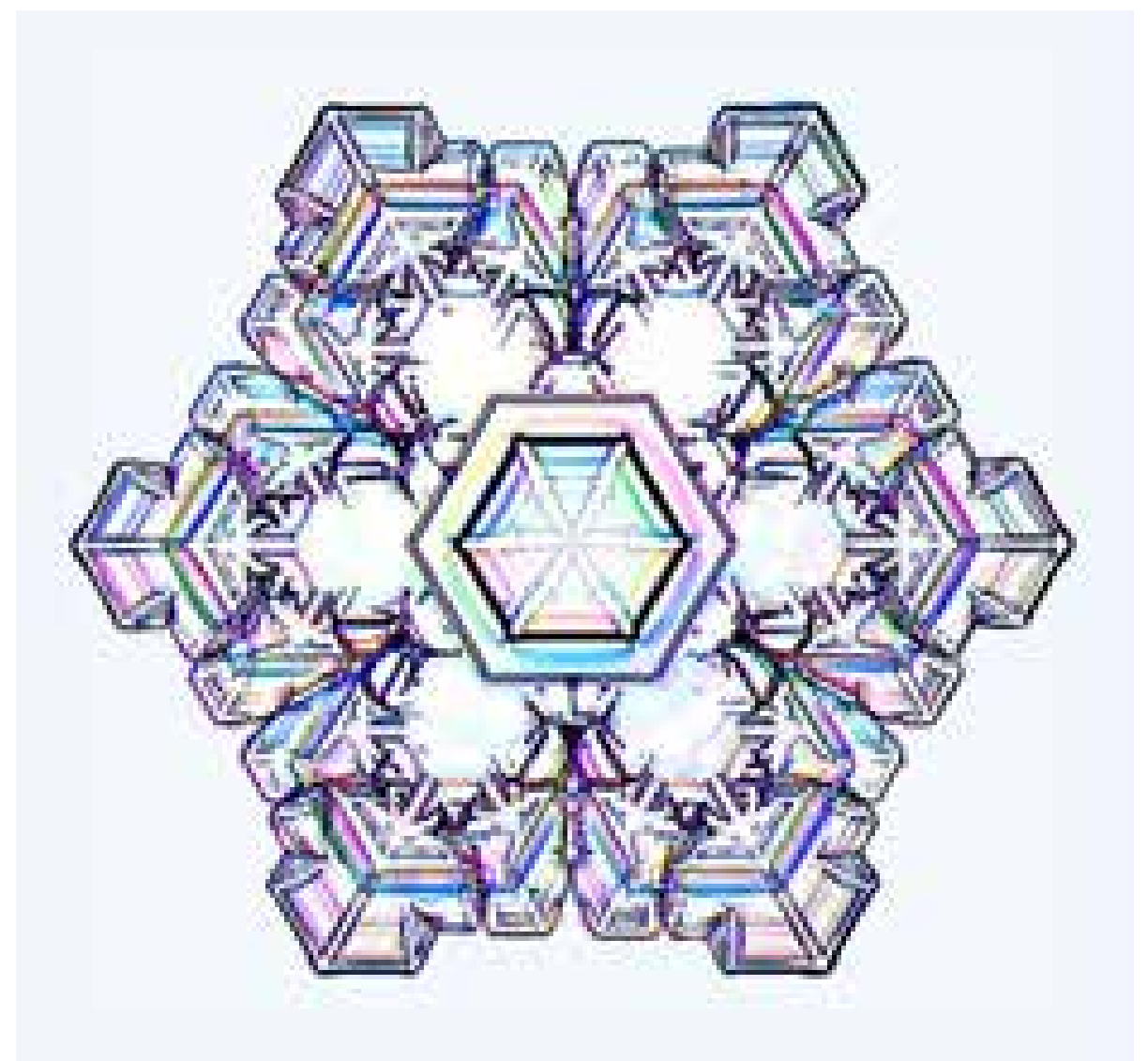

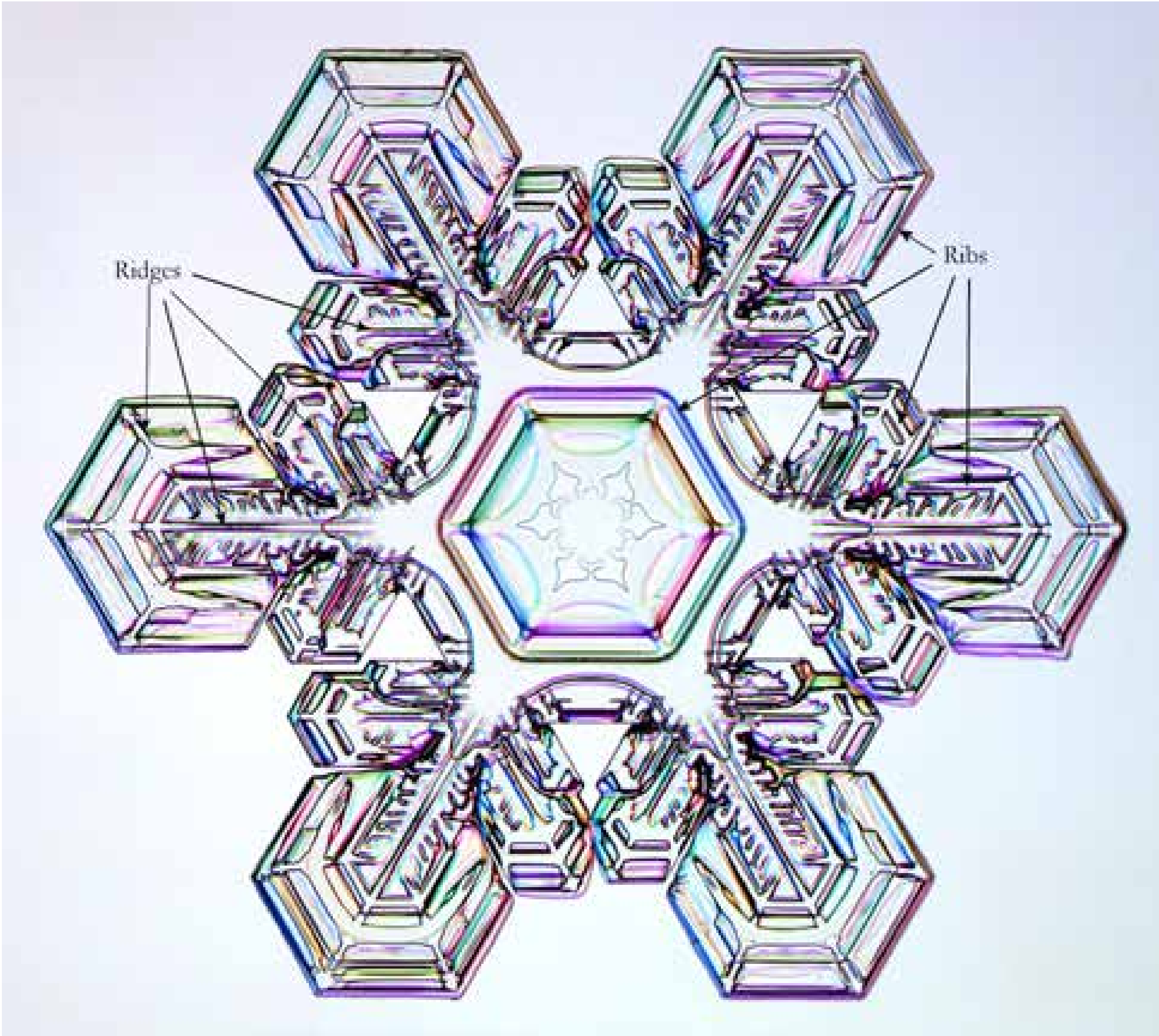

**Ribs, Ridges, and Rings.** The ribs, ridges, and inwardly propagating rings described in Chapters 4 and 9 are often prominent surface features on stellar-plate snow crystals. The complexity of the patterning reflects the ever-changing conditions experienced by each crystal, brought about by the convoluted path it followed through the atmosphere. The possible permutations are endless, and some crystals develop highly complex surface markings. Laboratory-grown snow crystals allow these different features to be analyzed in isolation, as described in Chapter 9.

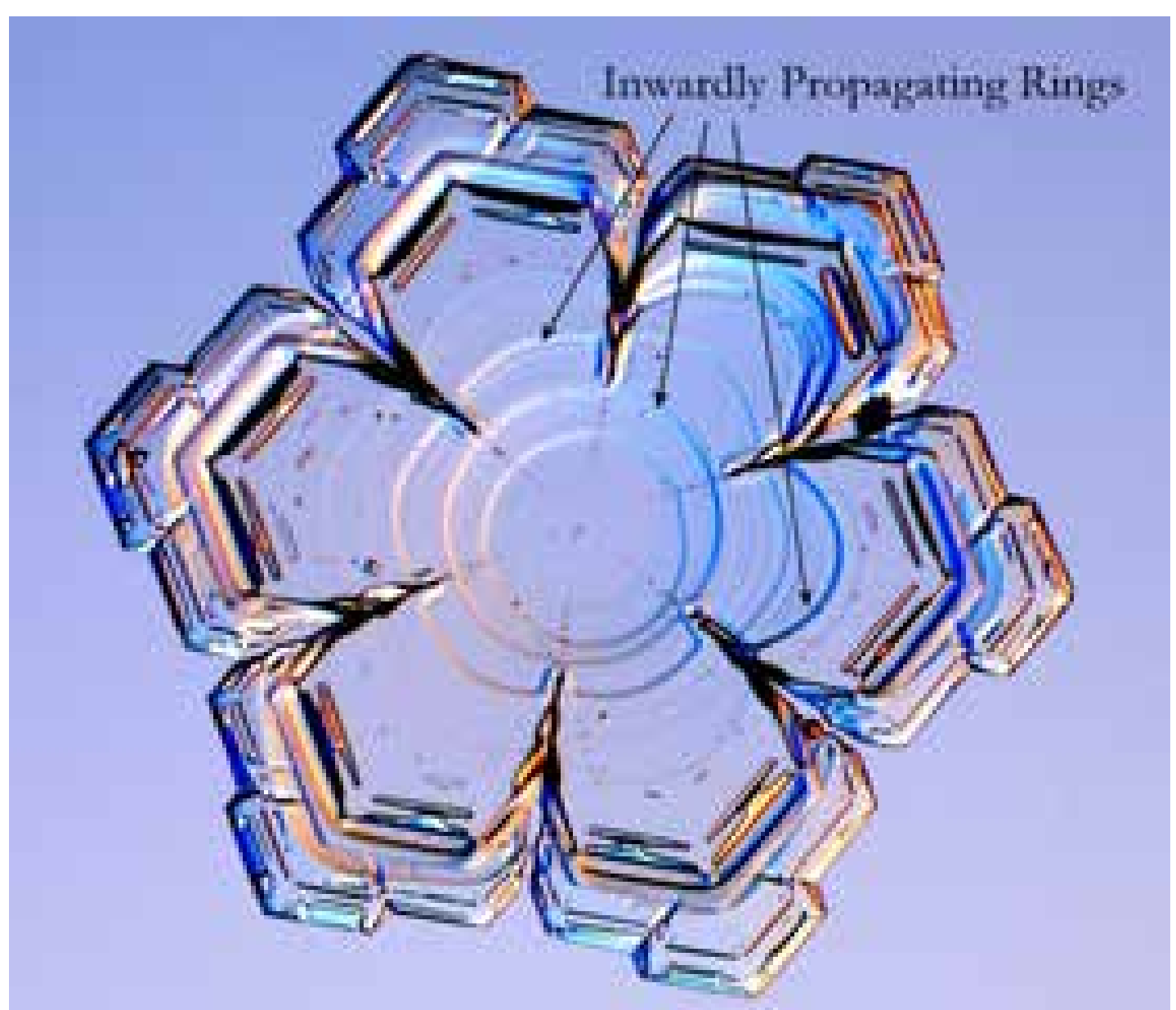

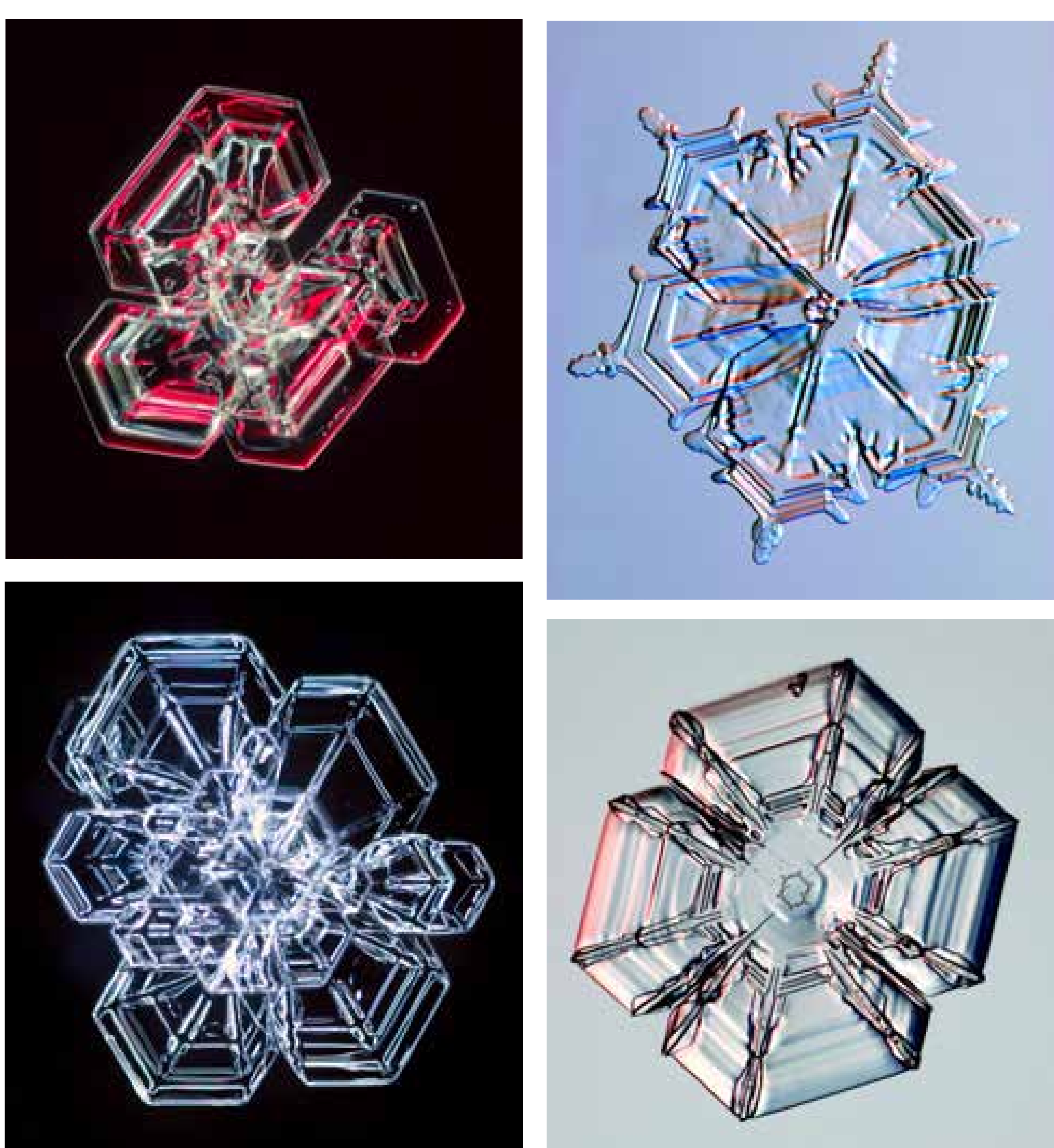

**Malformed Plates.** Most stellar plates are not beautifully formed and flawlessly symmetrical, as you can verify by spending ten minutes with a magnifying glass in any snowfall. The quintessential, well-formed snowflake is actually quite rare. The above pictures provide several examples of imperfect, somewhat malformed stellar plates. These are all single crystals of ice, as you can tell from the relative alignment of the various facets on each crystal. The facets reveal the underlying molecular order, which we see is the same throughout each plate. The odd shapes of these crystals came about because their growth was perturbed in some way. Perhaps they experienced some lattice defects during growth, or suffered collisions with rime particles or other falling crystals. There are many potential problems that can interfere with symmetrical growth.

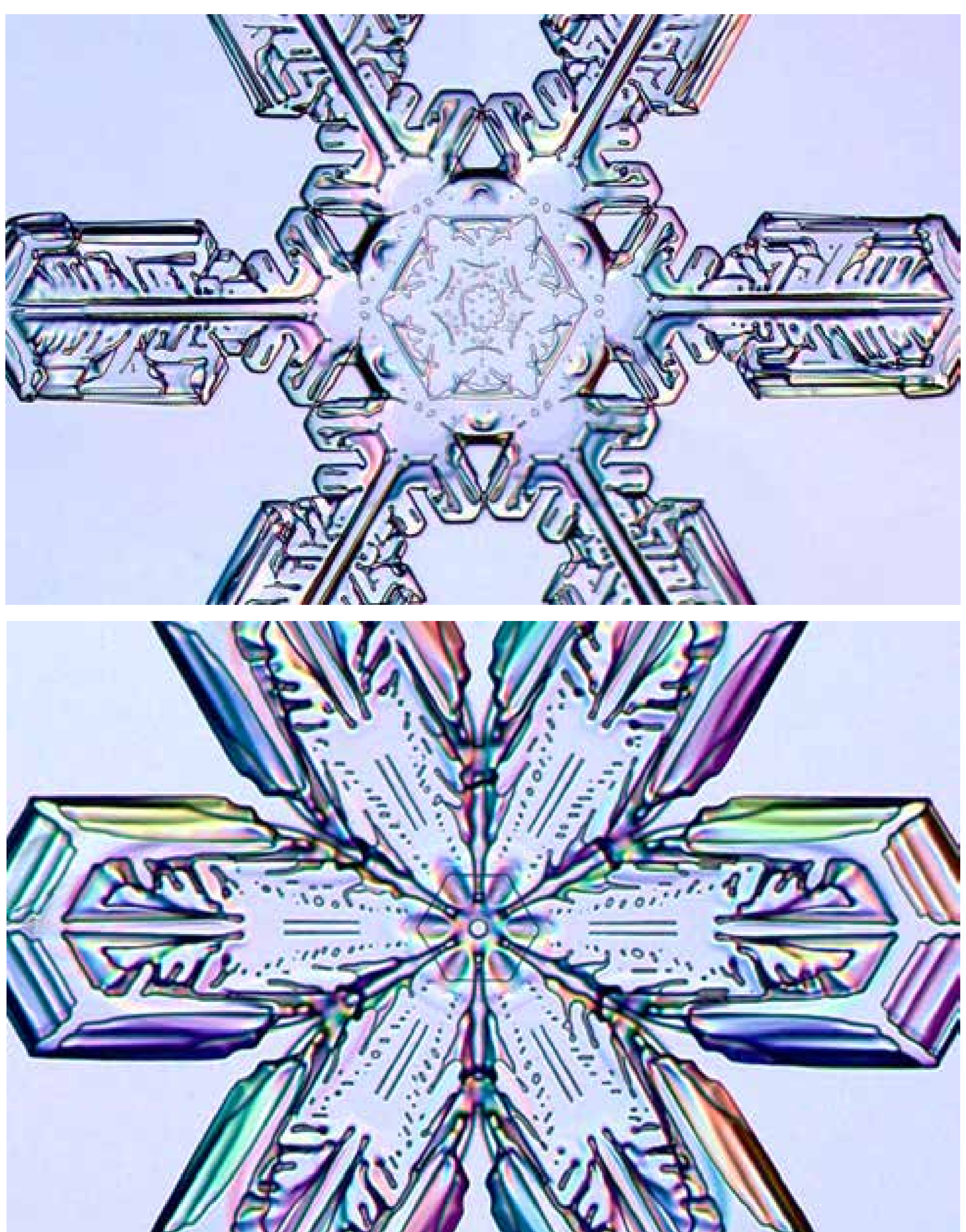

**Surface Patterns.** Some stellar plates exhibit remarkably complex and symmetrical surface markings, especially in their central regions. The structures are so small that a microscope is usually needed to see them.

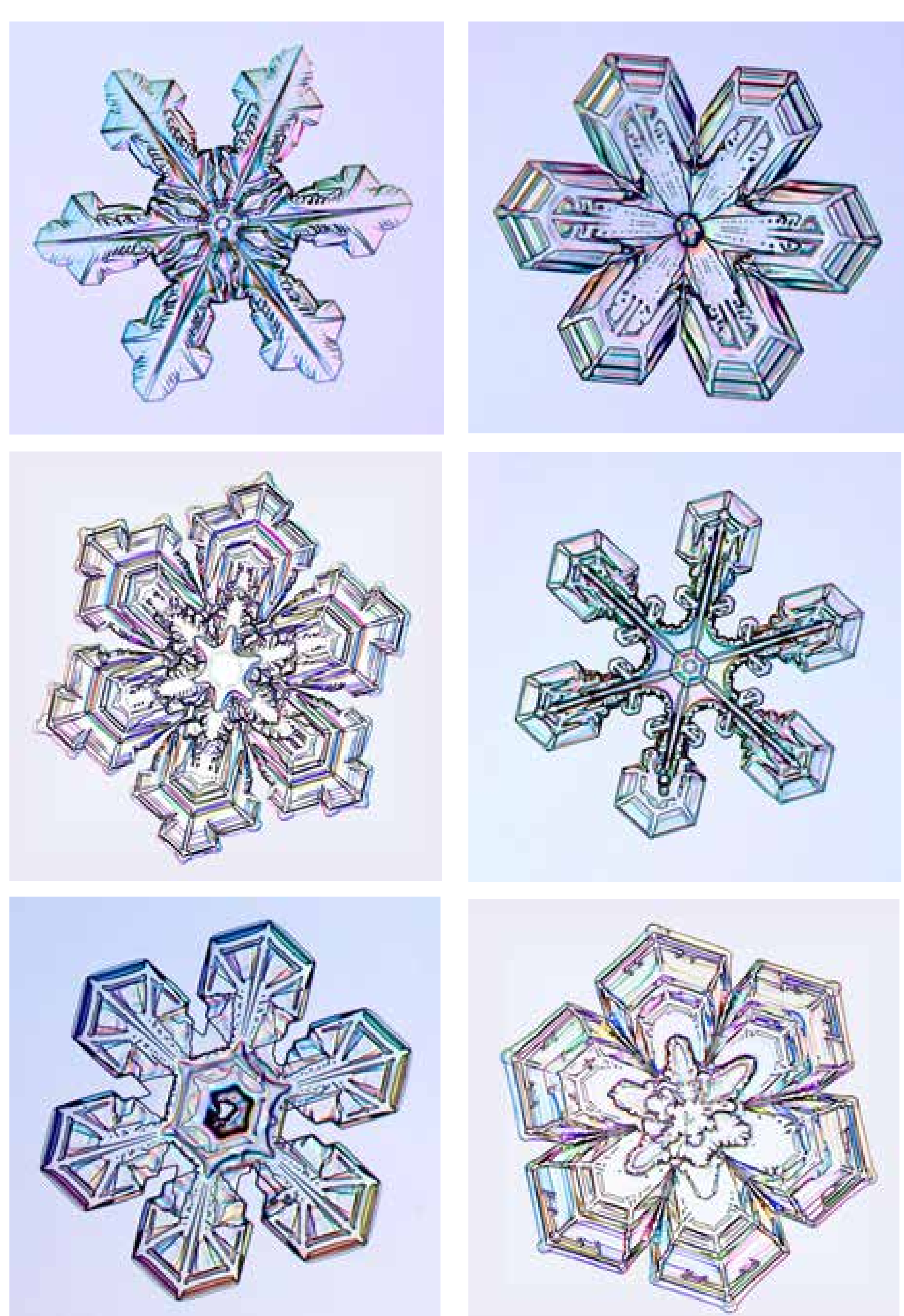

**Perpetual Variety.** (Two pages) Stellar-plates snow crystals exhibit an endless diversity of complex surface patterns.

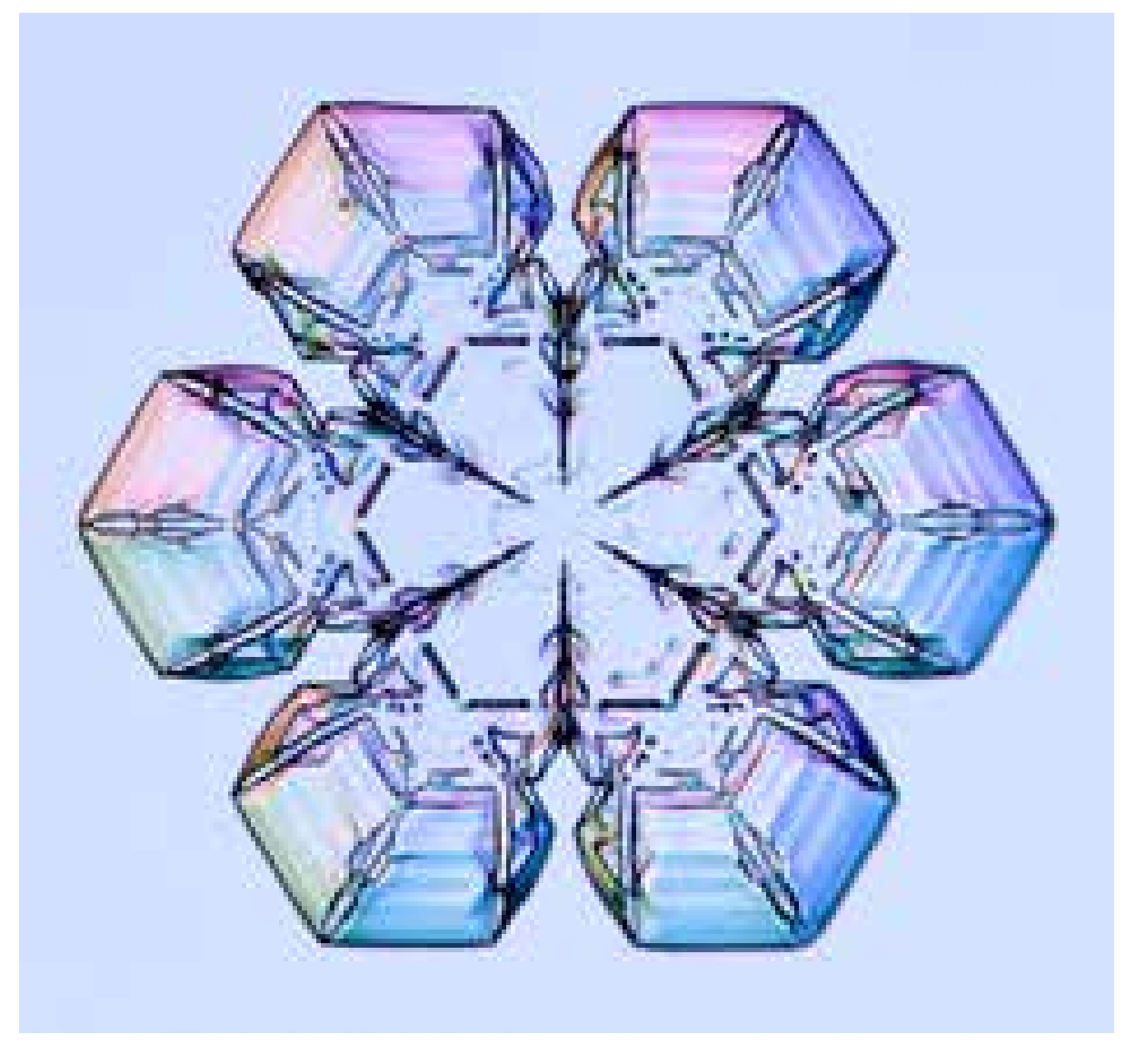
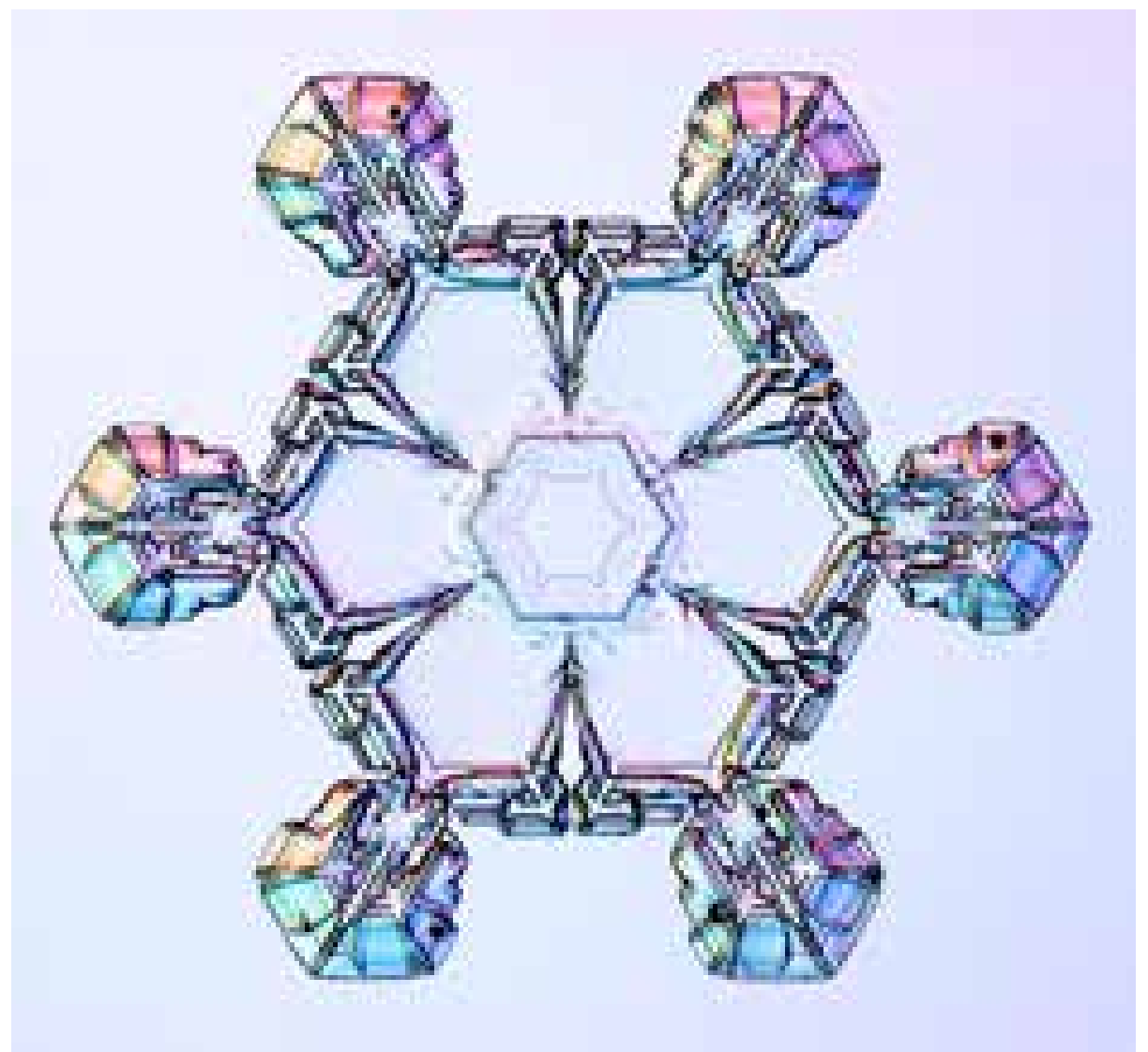
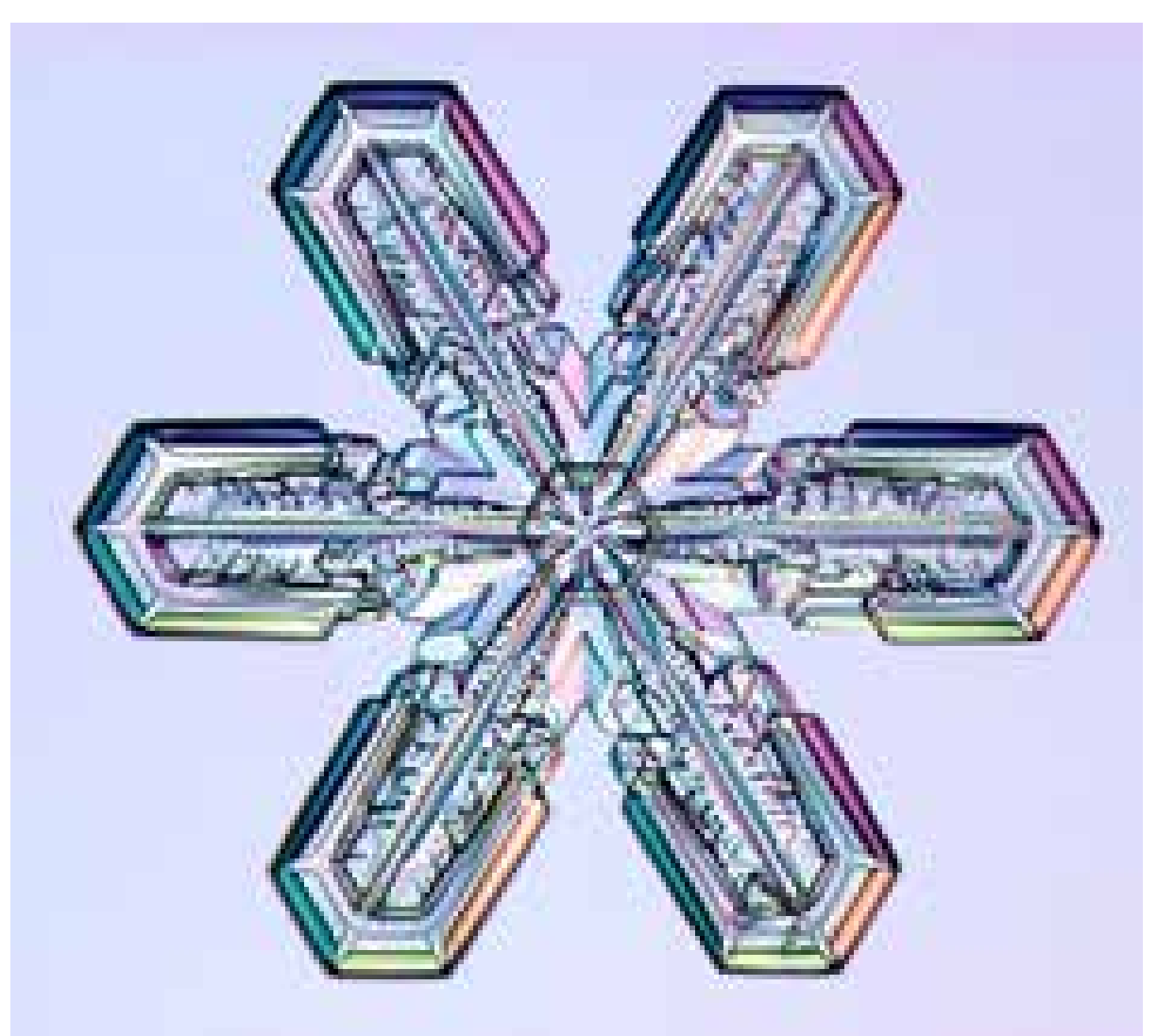
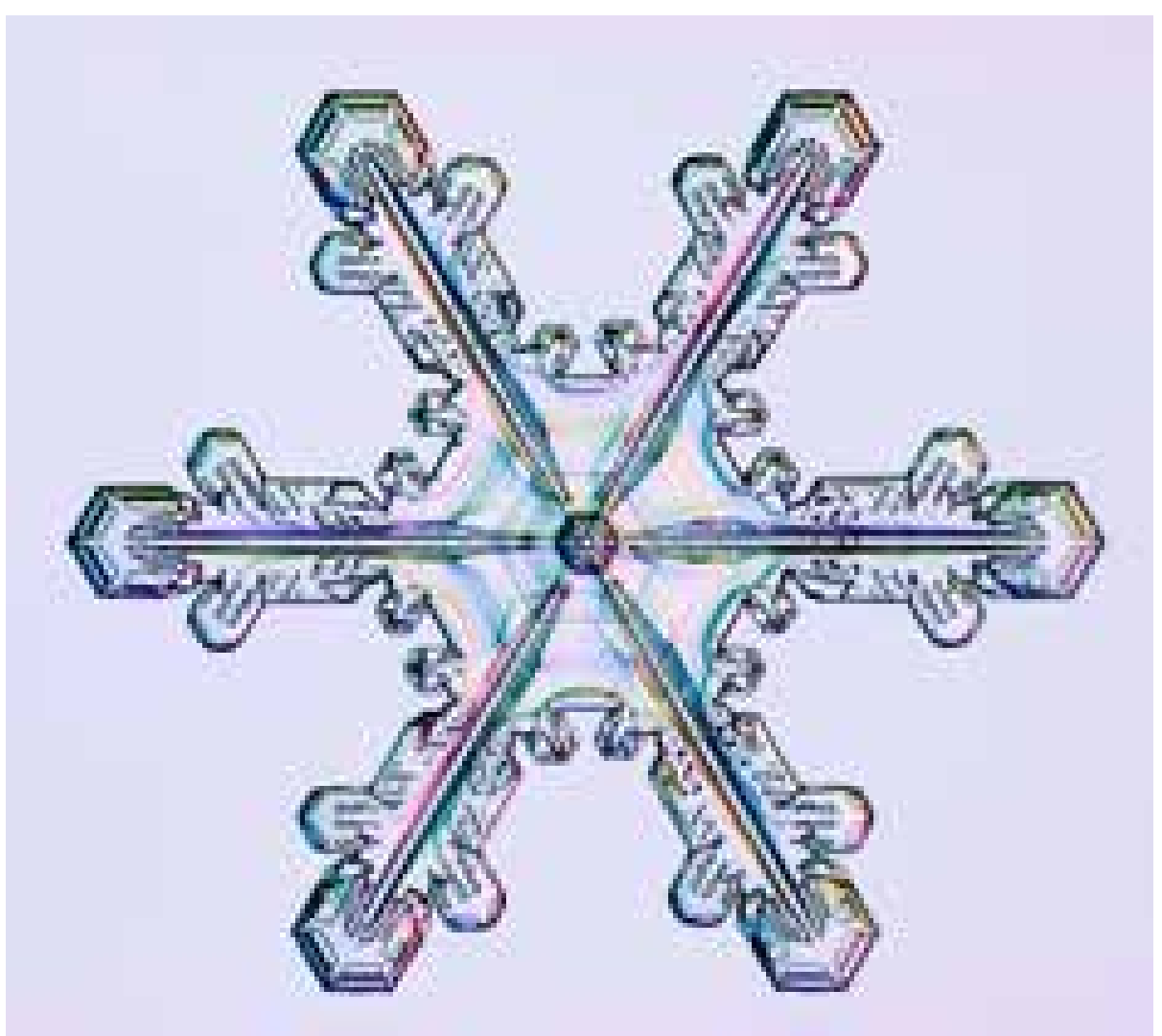
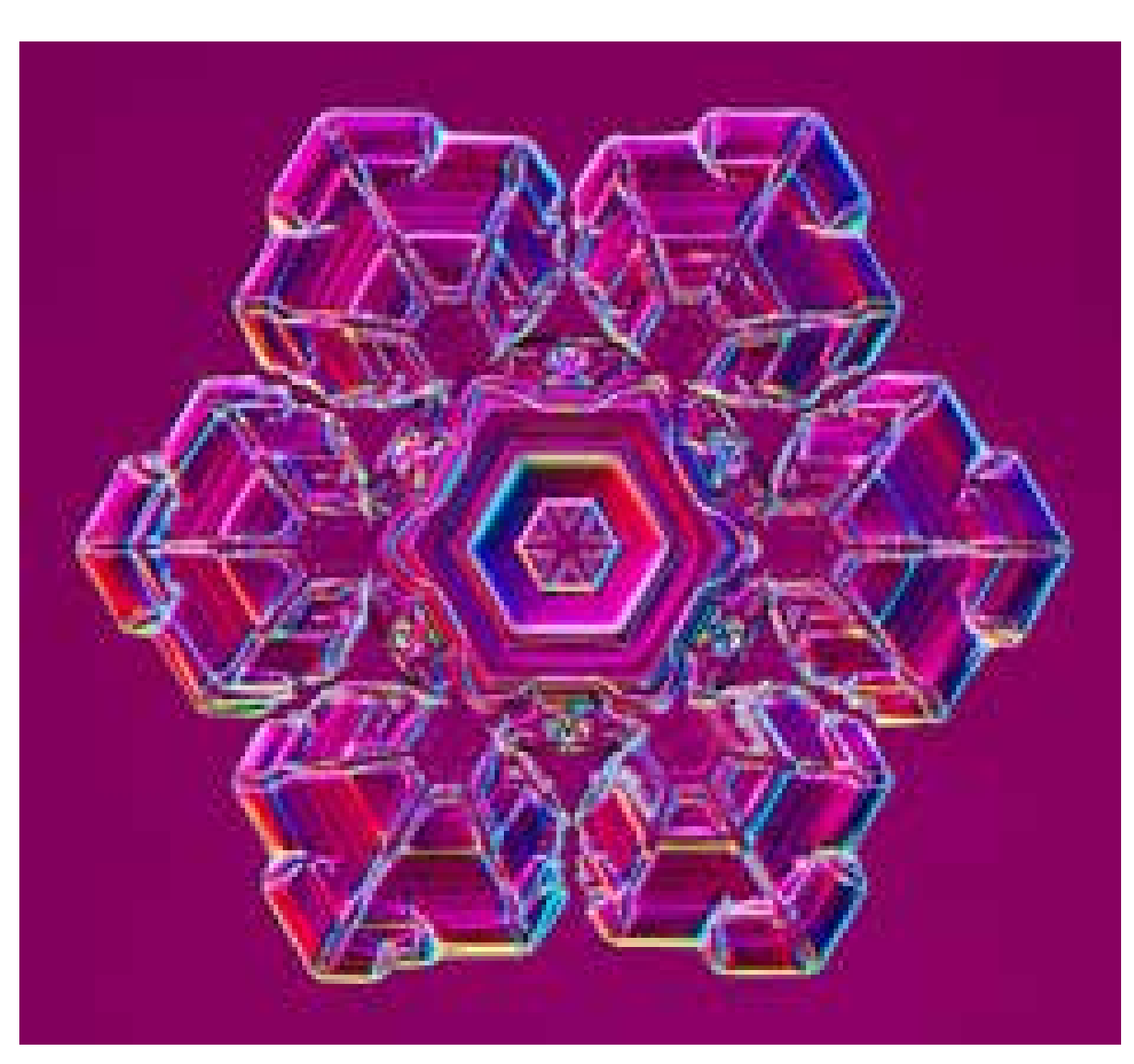
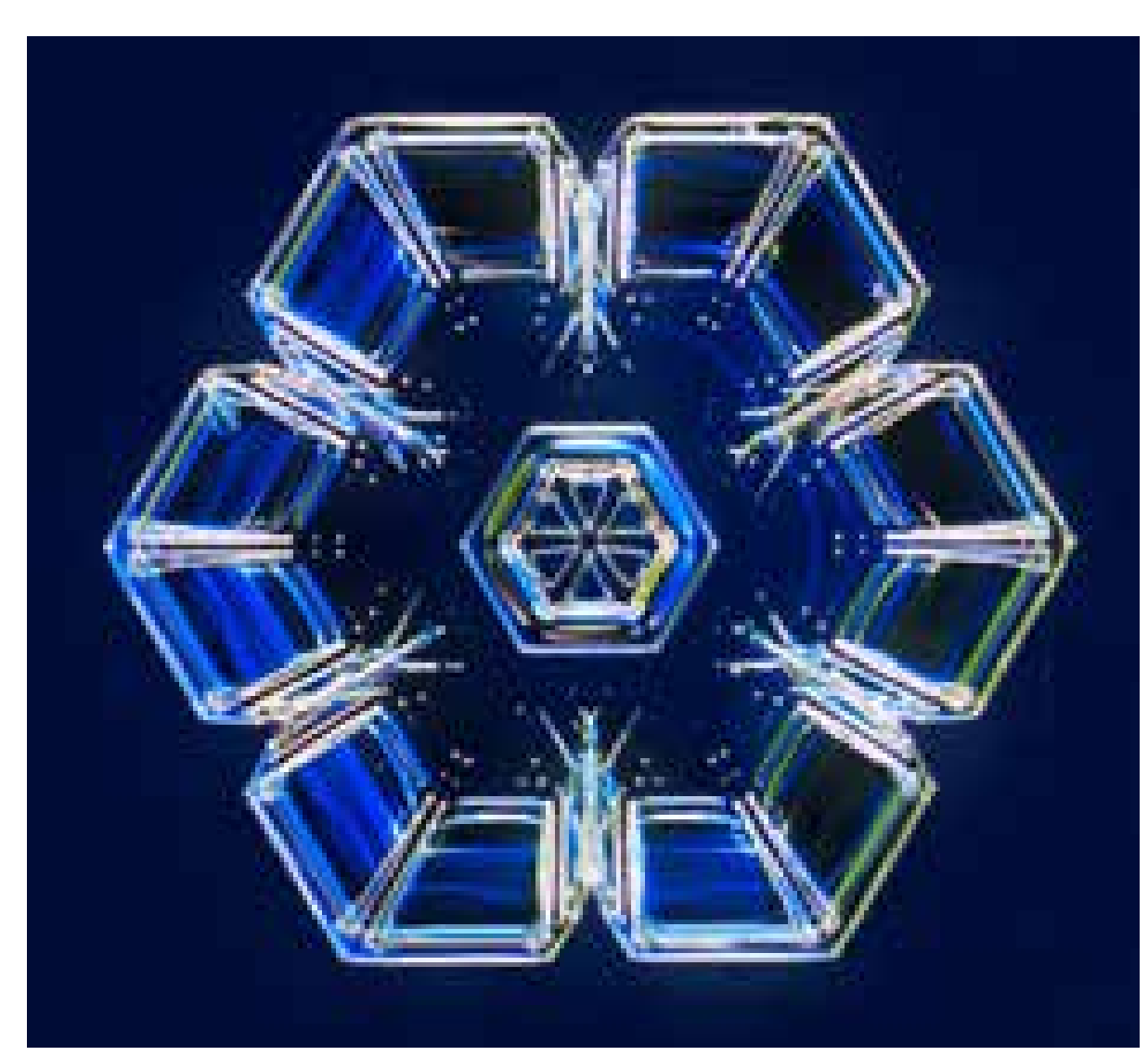

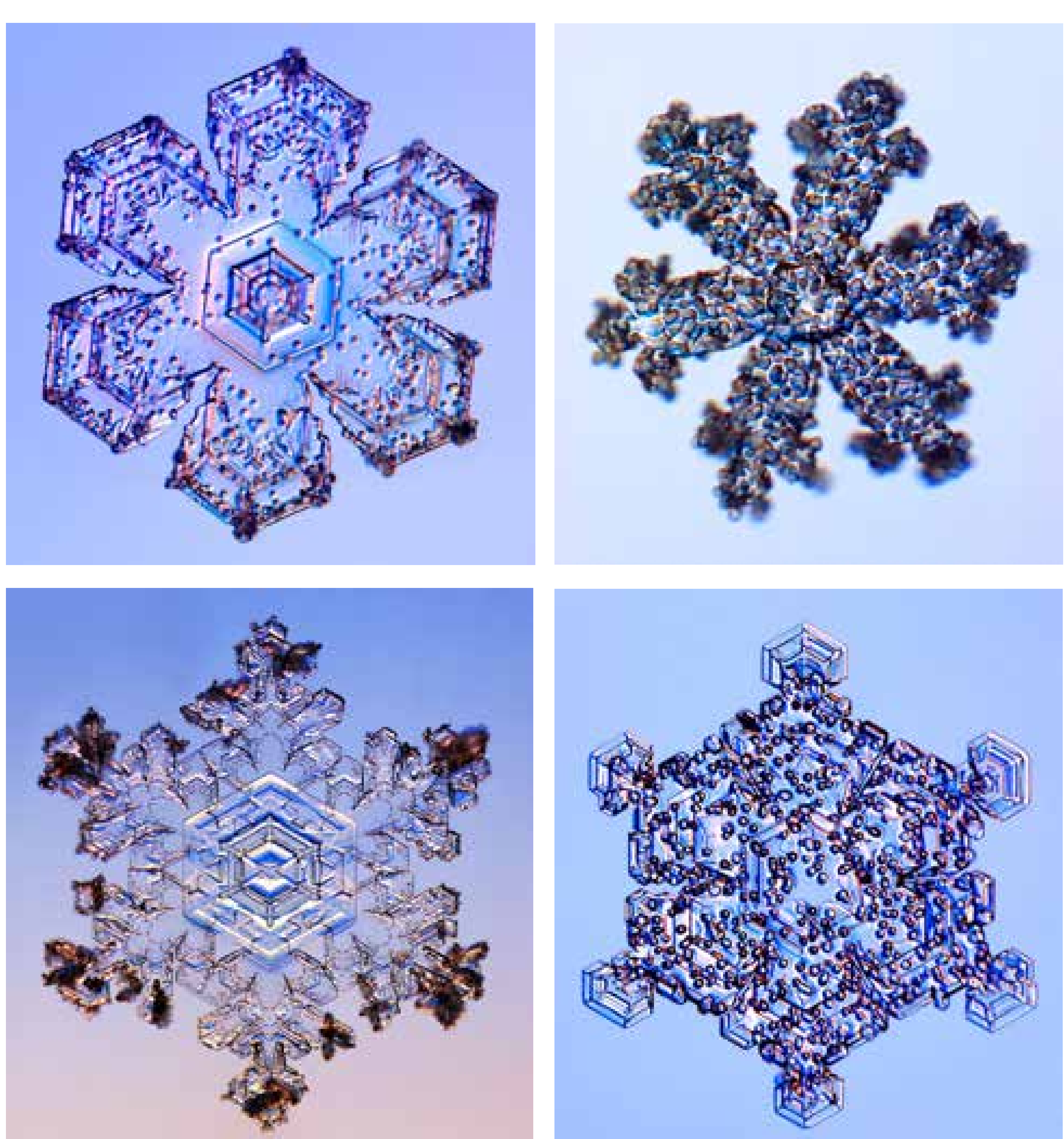

**Rimed Crystals.** Snow crystals are often decorated with rime particles, ranging in number anywhere from one to thousands. A typical droplet has a size of roughly 0.03 millimeters, which is half the diameter of a human hair. Large crystals can be especially prone to rime, as the high humidity necessary for their growth requires a high density of cloud droplets. Aerodynamic forces often deposit rime particles on the edges of large plates, as seen in the lower-left image. The lower-right crystal is unusual in that it picked up quite a bit of rime, but then it moved to a region with fewer cloud droplets, where the crystal tips grew out relatively unperturbed by the rime.

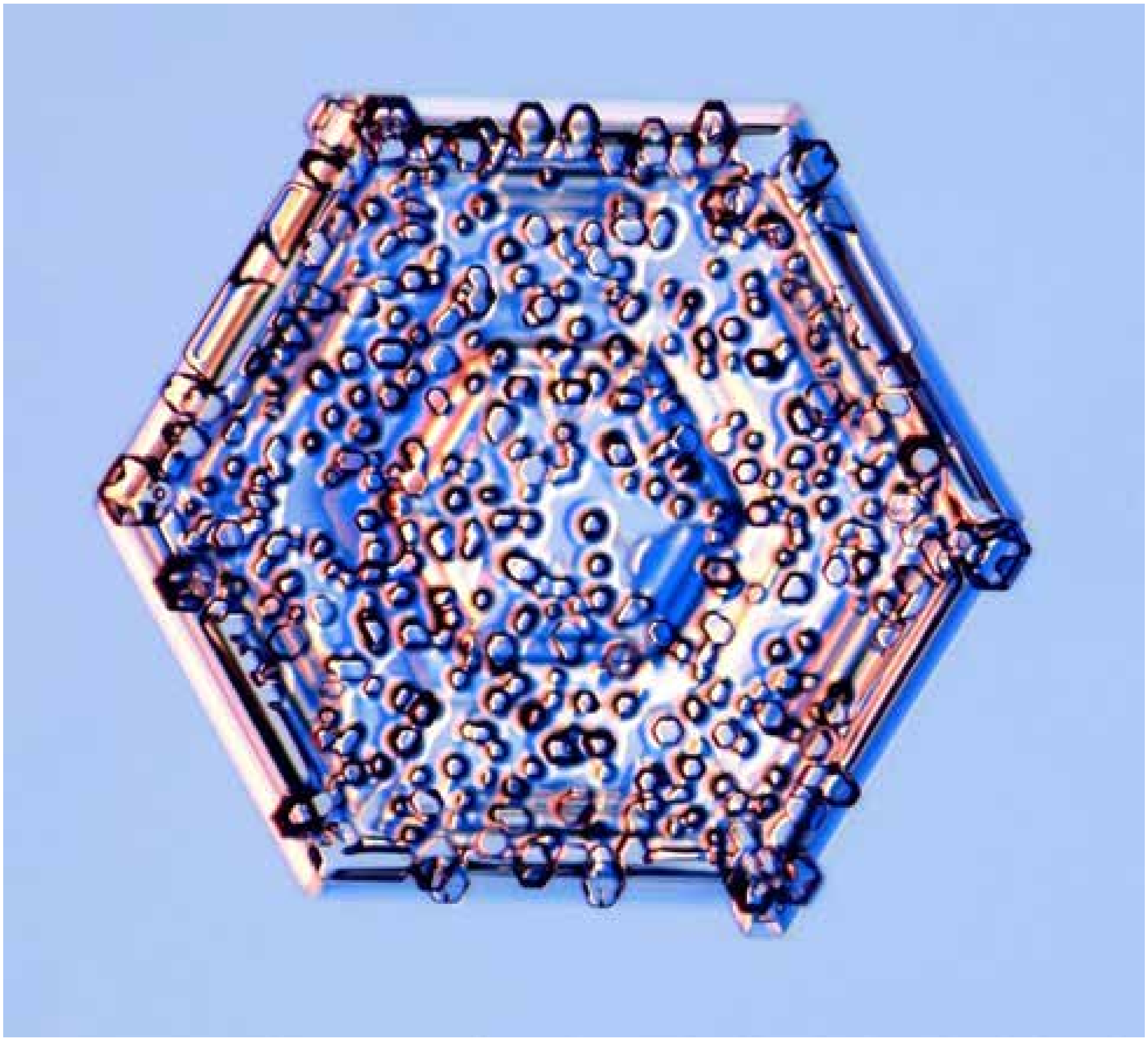

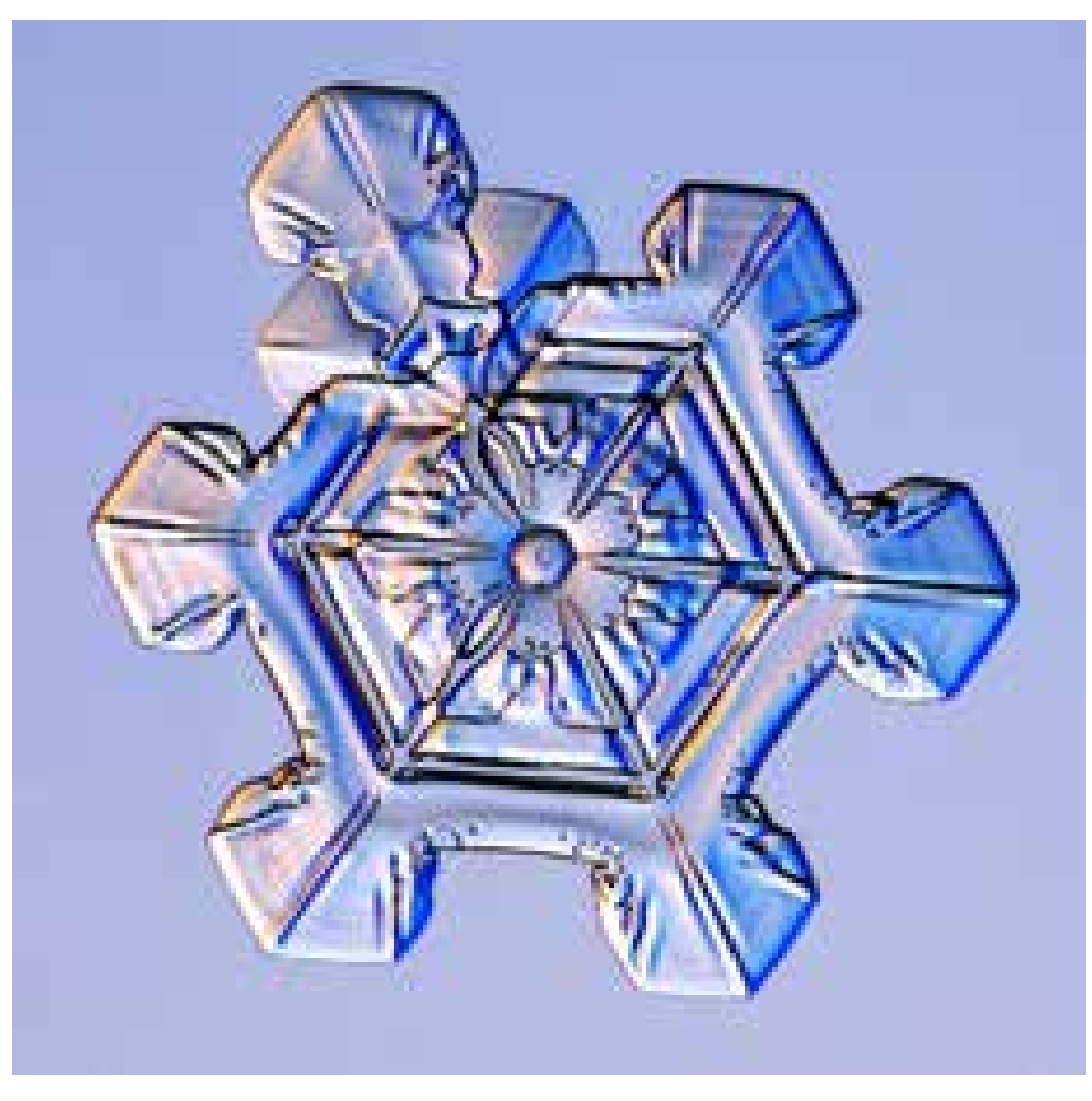

**Epitaxial Growth.** After growing into a small hexagonal plate, the above crystal wandered into a dense region of a cloud and picked up a good dusting of rime droplets. Each droplet froze upon contact with the ice, and, if you look closely, you can see that the facets on the frozen droplets are mostly aligned with the facets of the plate. This is an example of *epitaxial growth*, as the plate ice served as a template to guide the molecular orientation of the freezing liquid. The crystal on the right apparently acquired a single rime droplet when it was smaller, which froze epitaxially and then stimulated the growth of an errant branch that grew differently from its siblings.

# Sectored Plates

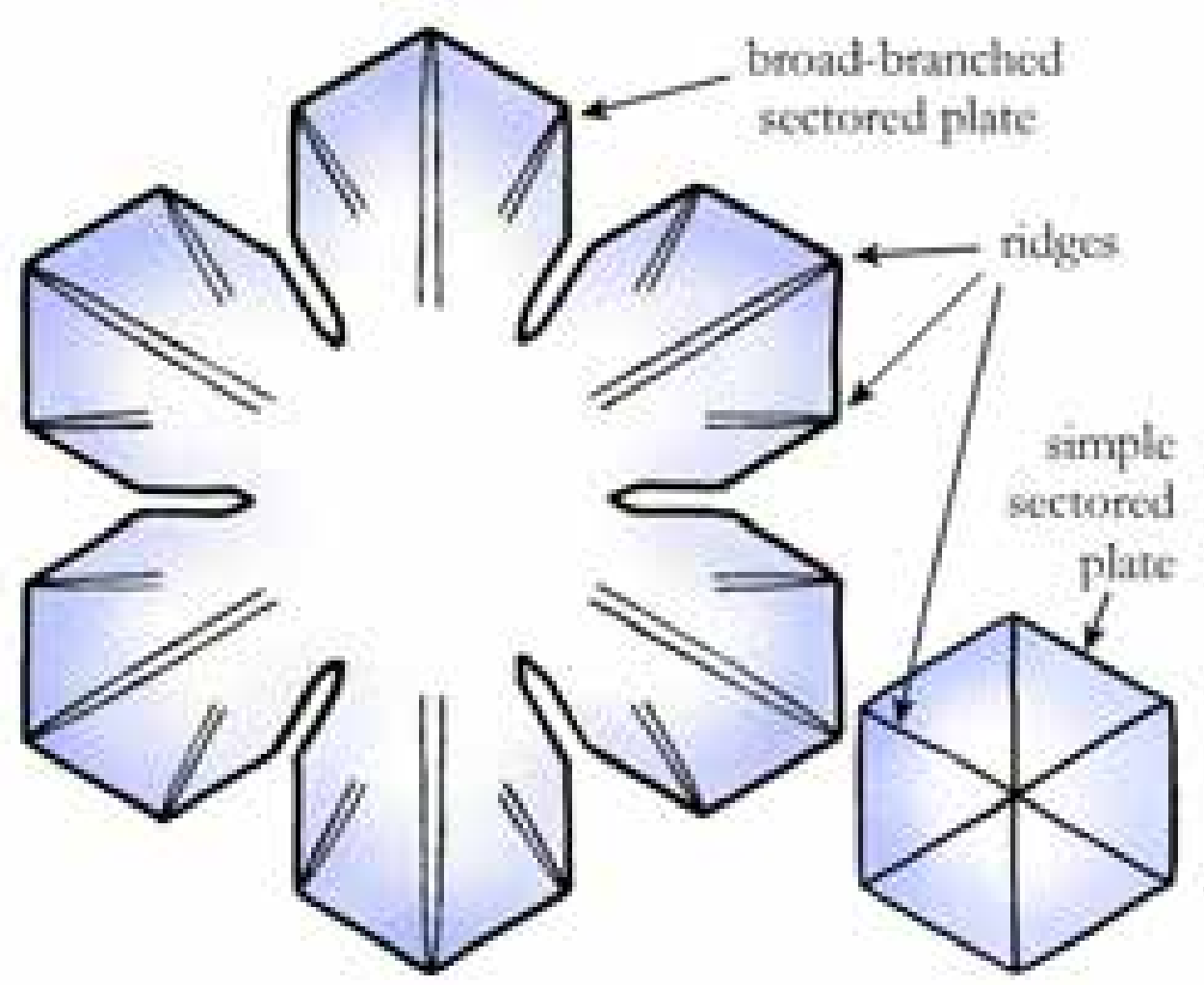

*Sectored plates are flat, broad-branched crystals decorated with pronounced radiating patterns of ridges. They get their name from the way the ridges seem to neatly divide the plates into sectors. At times these surface markings look like veins on a leaf, giving some snowflakes an almost plant-like appearance.*

The simplest sectored plates have a basic hexagonal shape divided into six sectors, like the example shown below-right (see also Chapter 9). Broad-branched crystals with sectored-plate extensions are more common. Sectored plates can be considered a sub-class of stellar plates, and there is no sharp dividing line between the two categories. If a stellar plate shows especially prominent ridging, and few other surface markings, then we tend to call it a sectored plate.

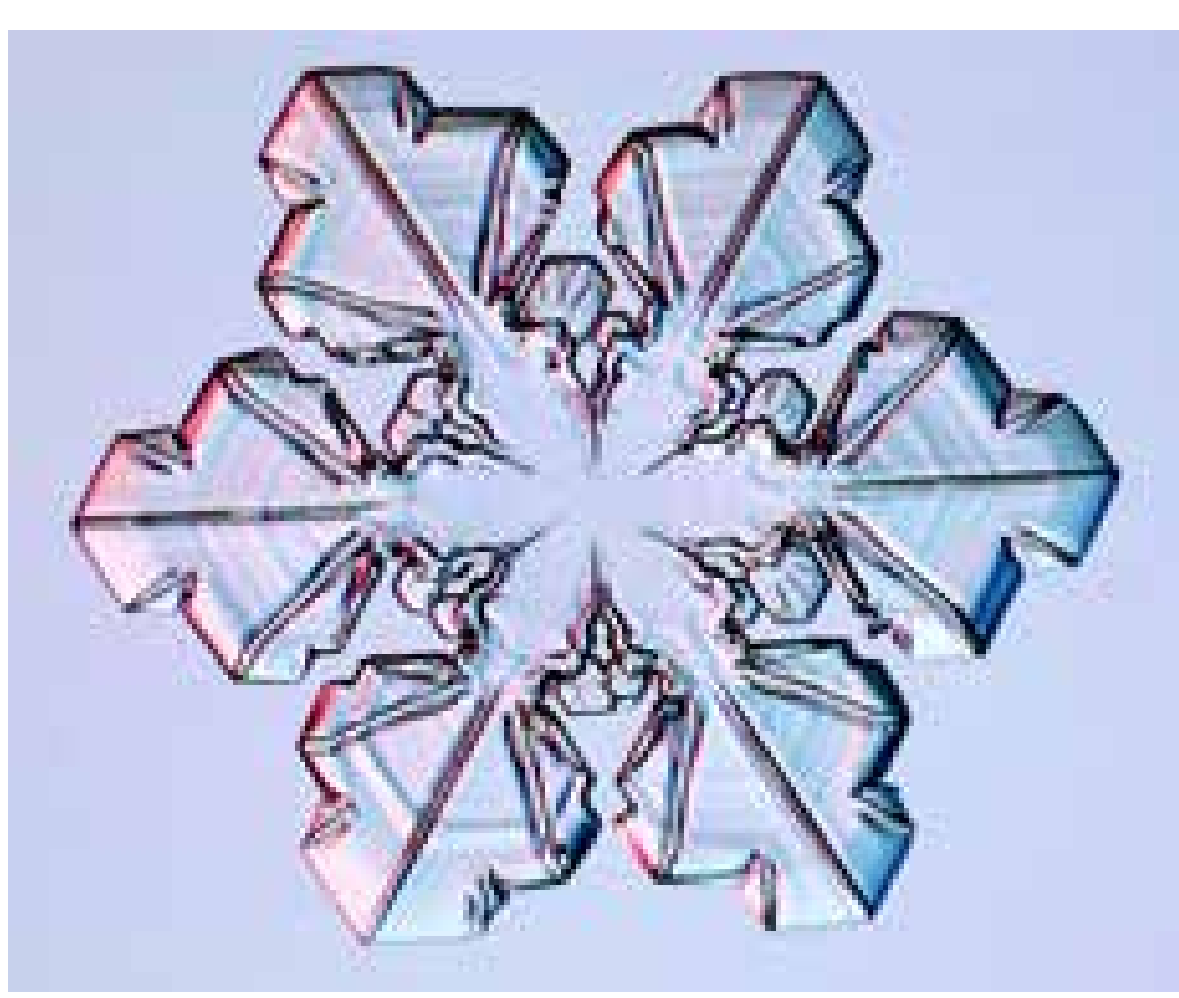

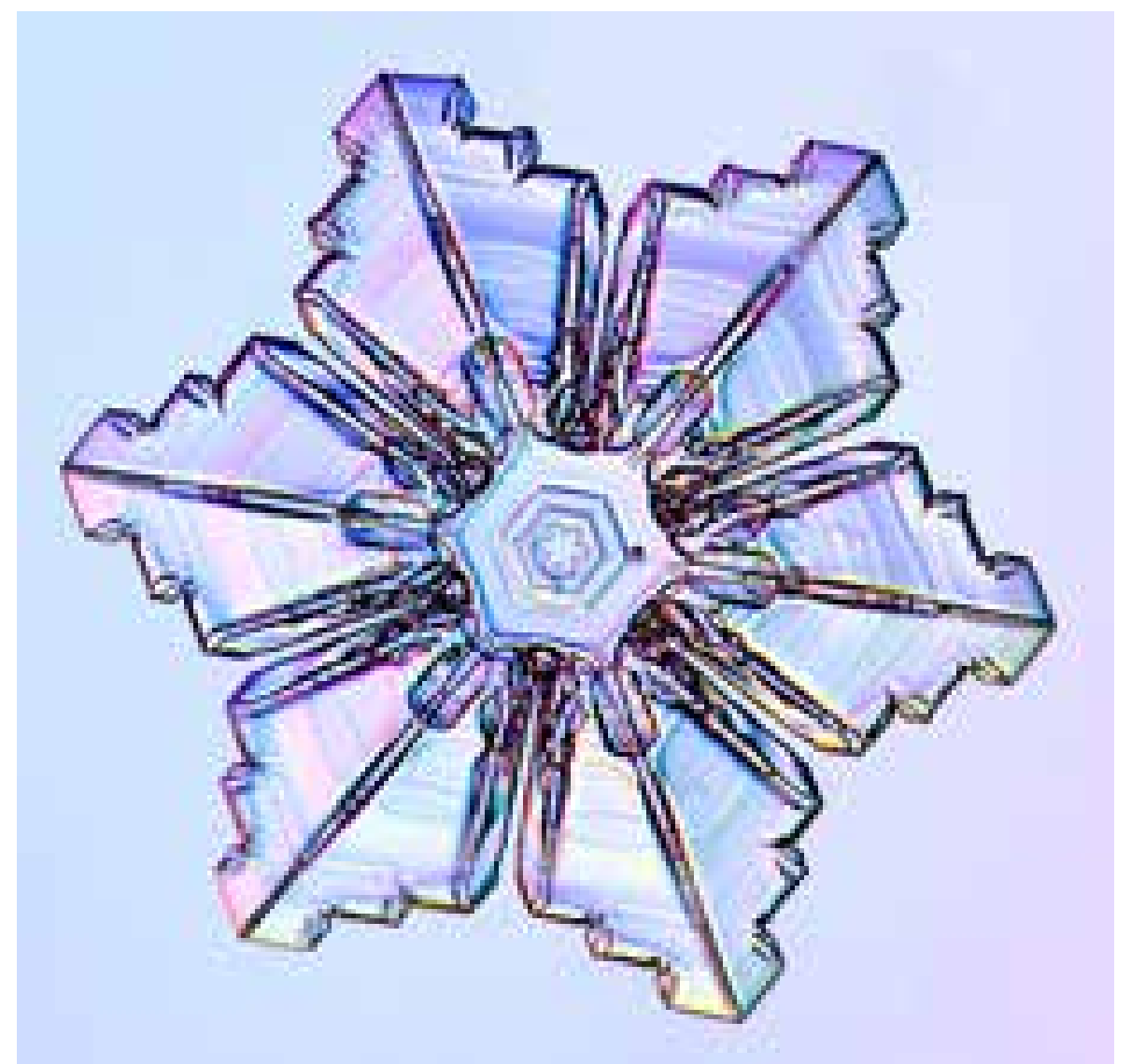

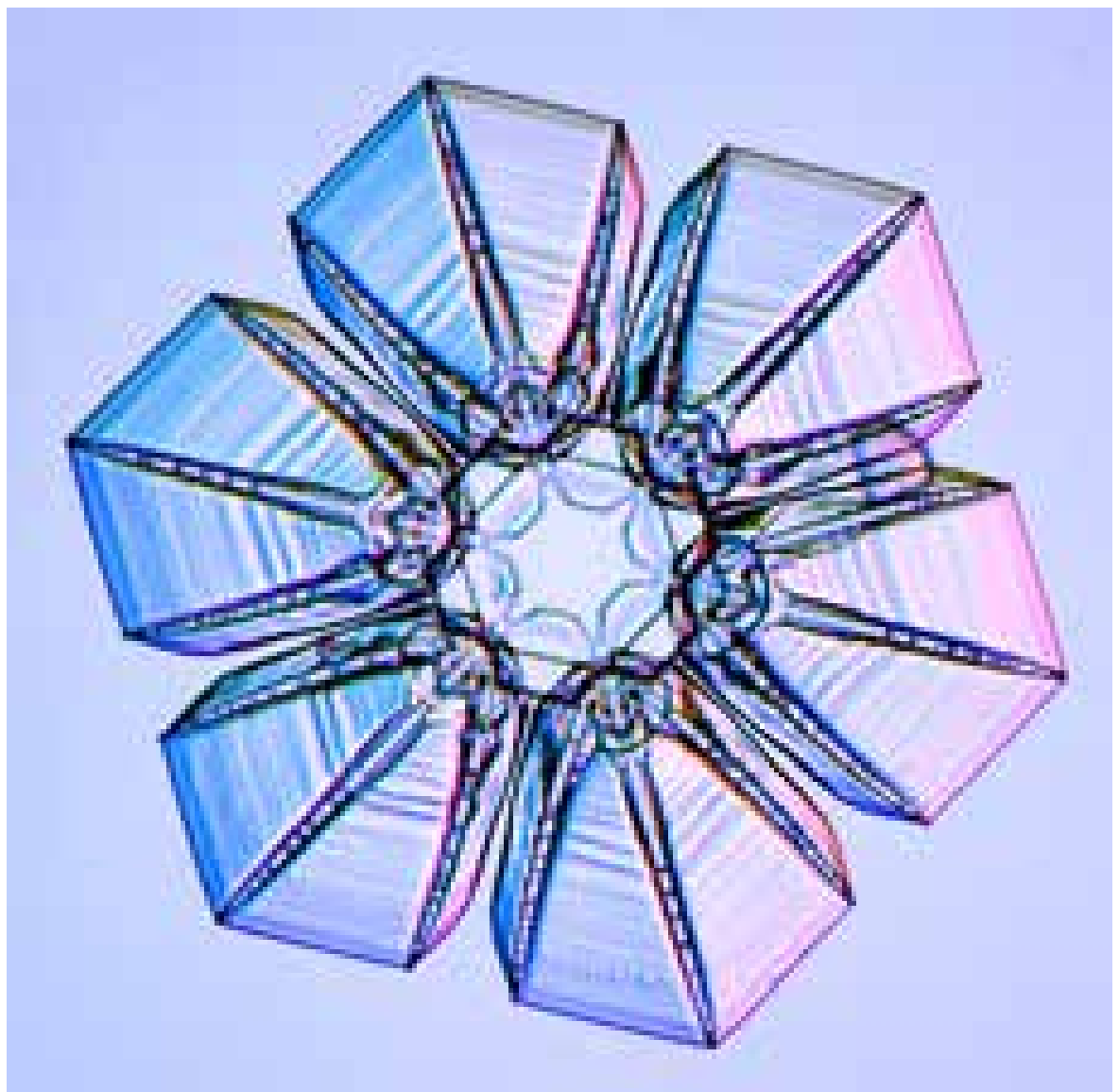

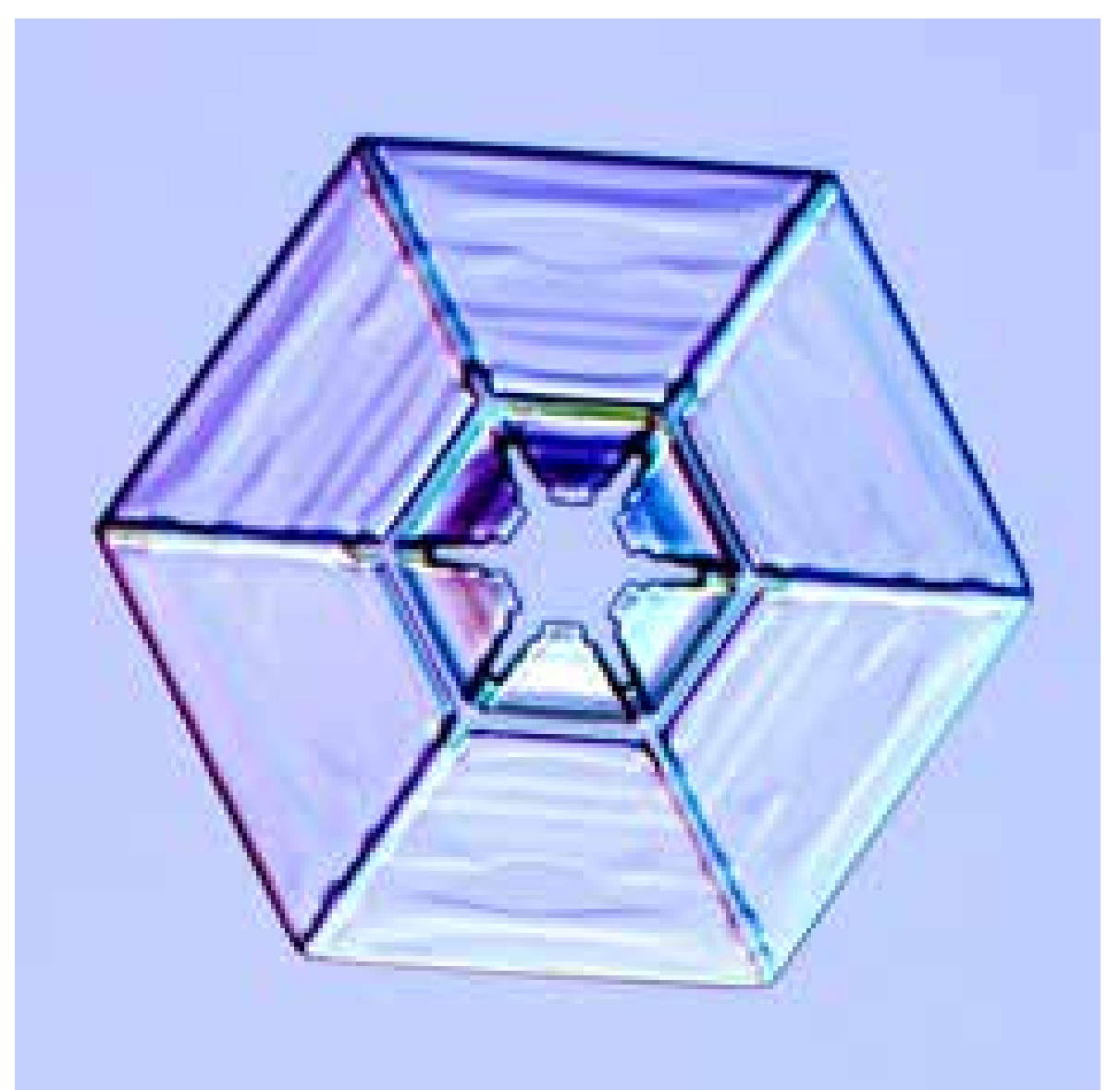

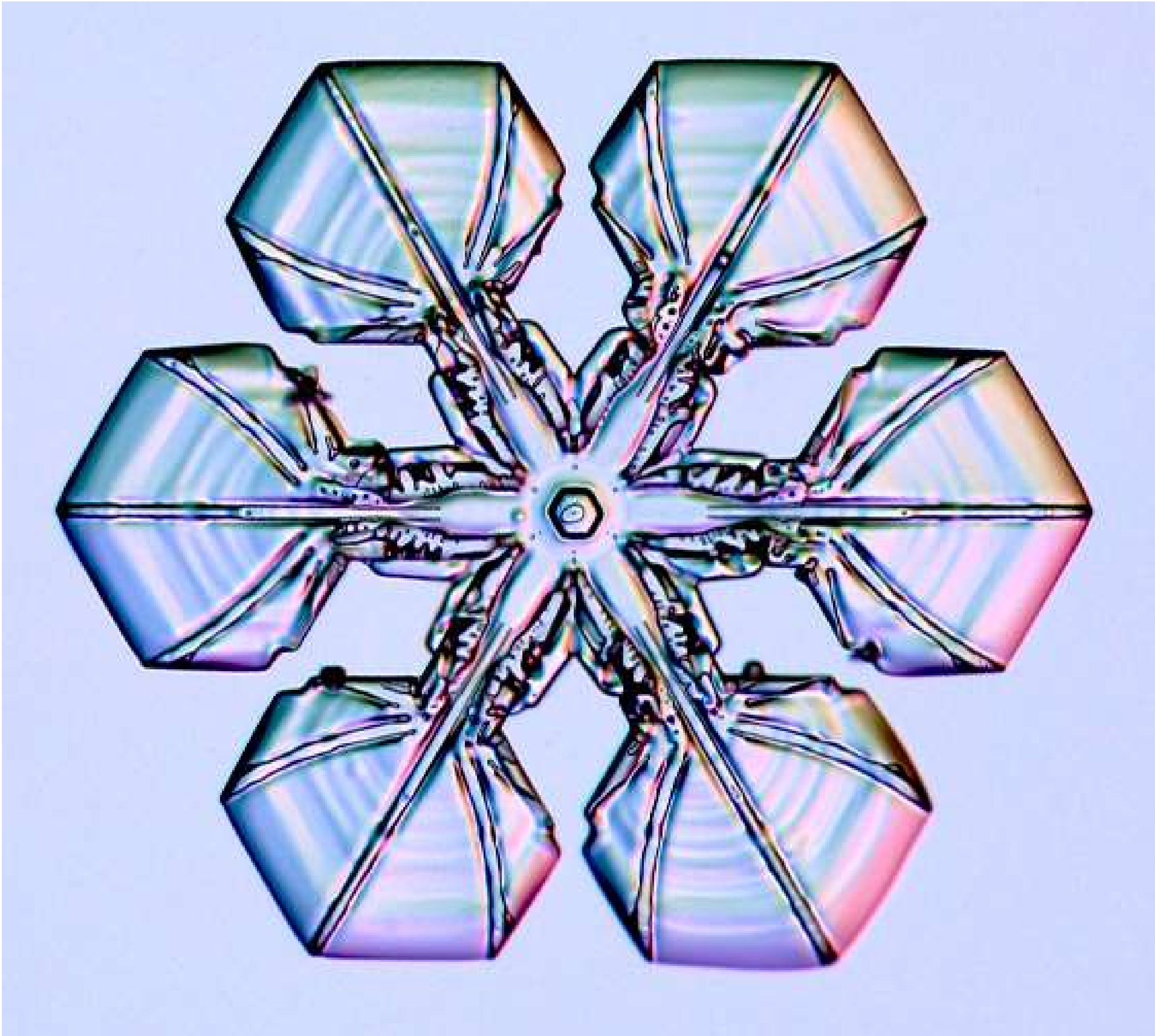

**Sectored-plate Extensions.** The relative simplicity of their surface markings indicates that sectored plates form in relatively constant conditions, without large swings in temperature or humidity. Thus the plate-like branches are generally flat and smooth (aside from the ridges) and the prism facets tend to be well-formed and large. Then they look a bit like duck's feet.

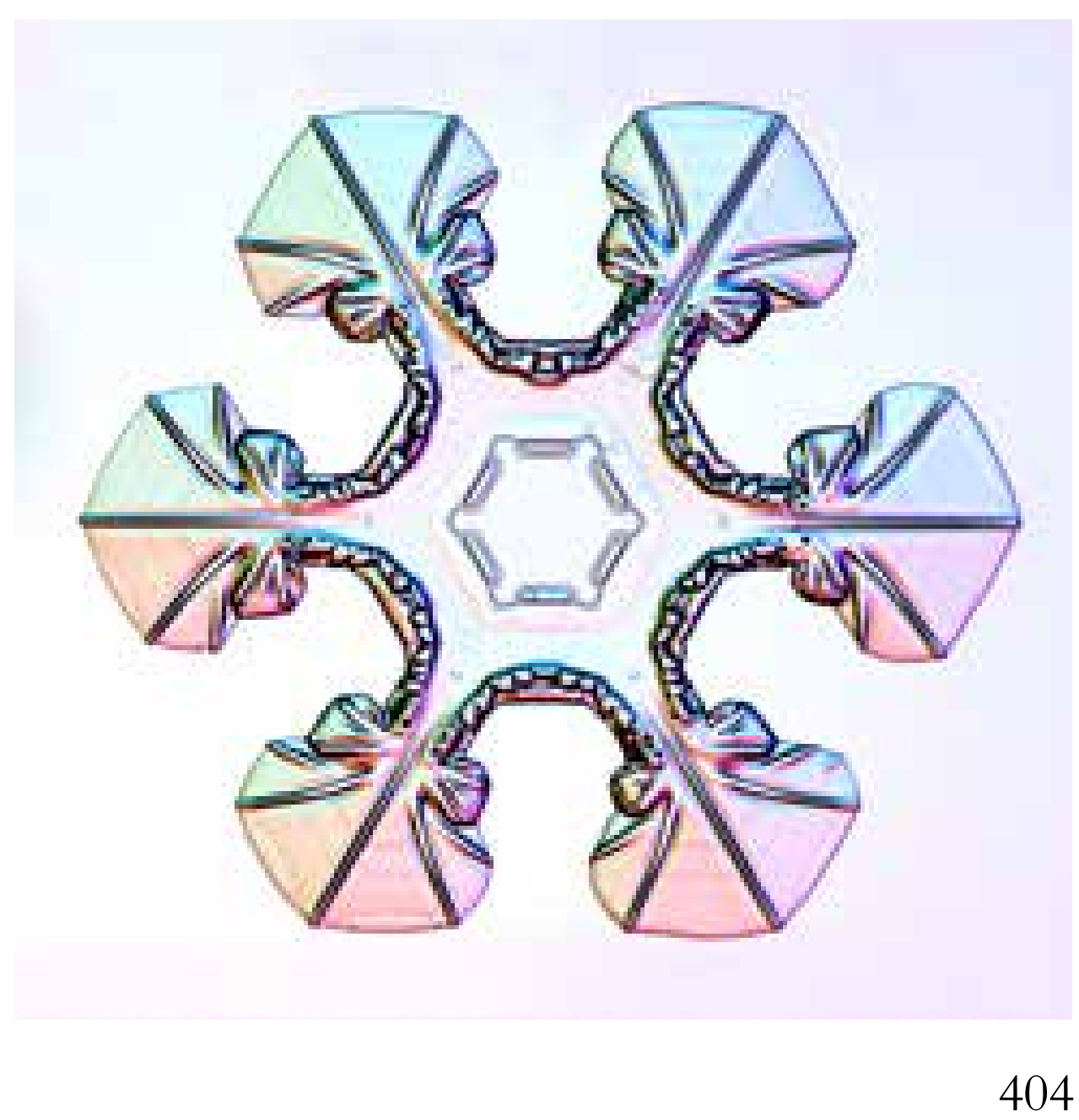

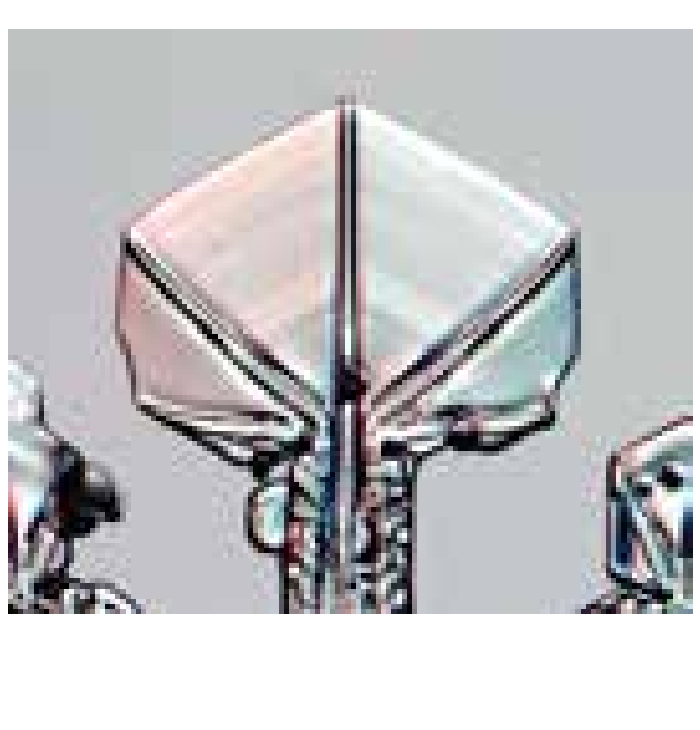

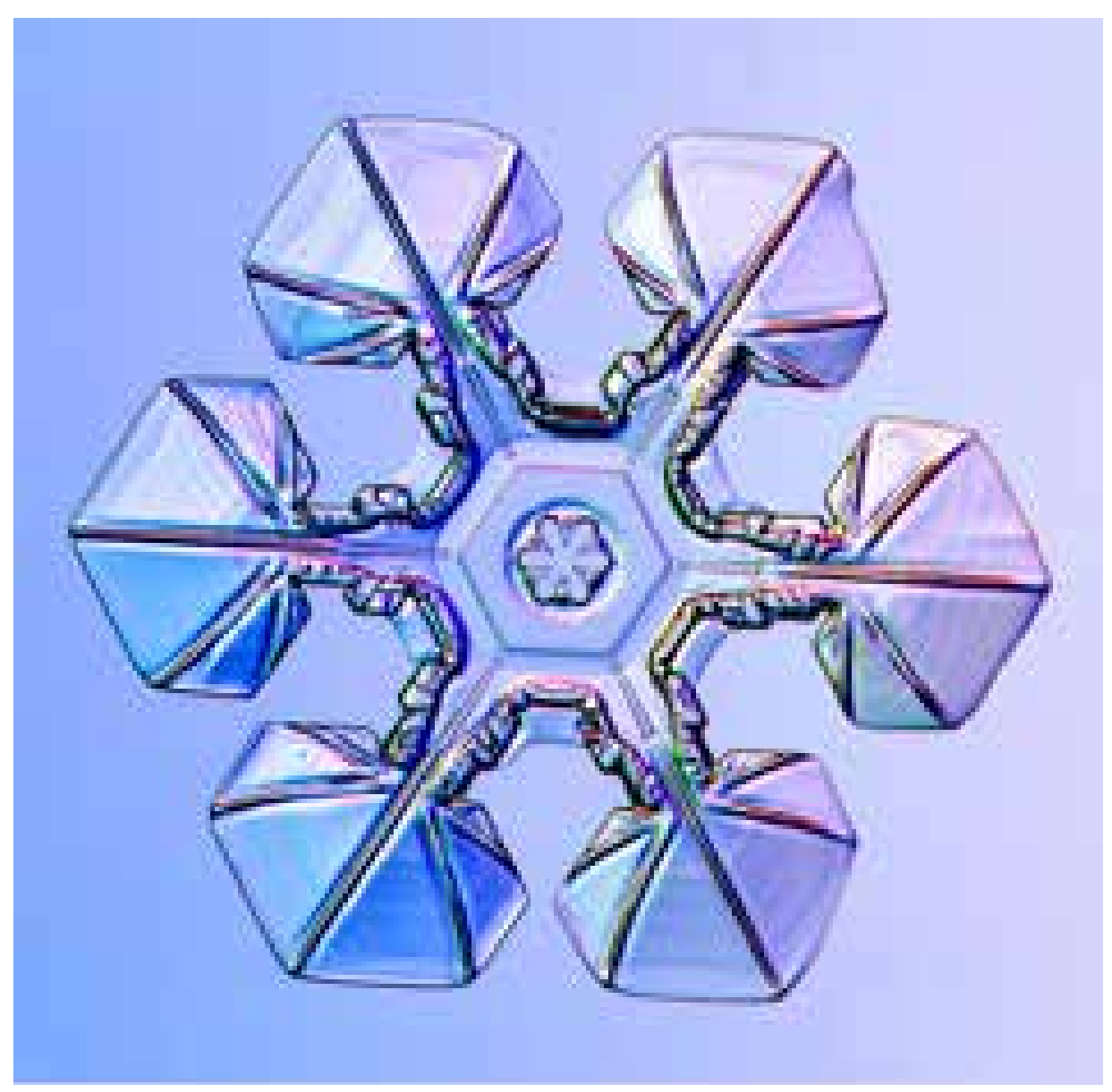

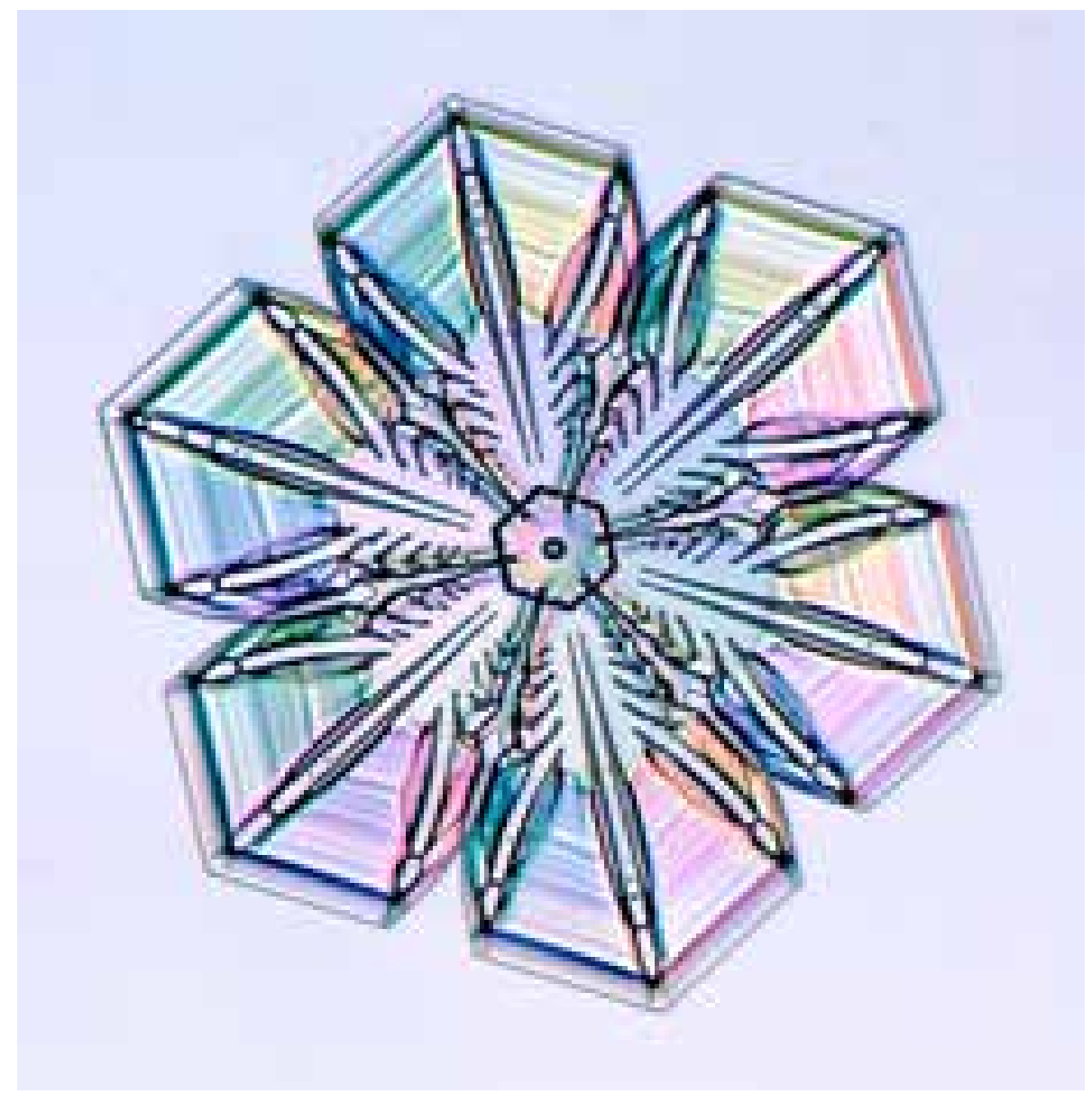

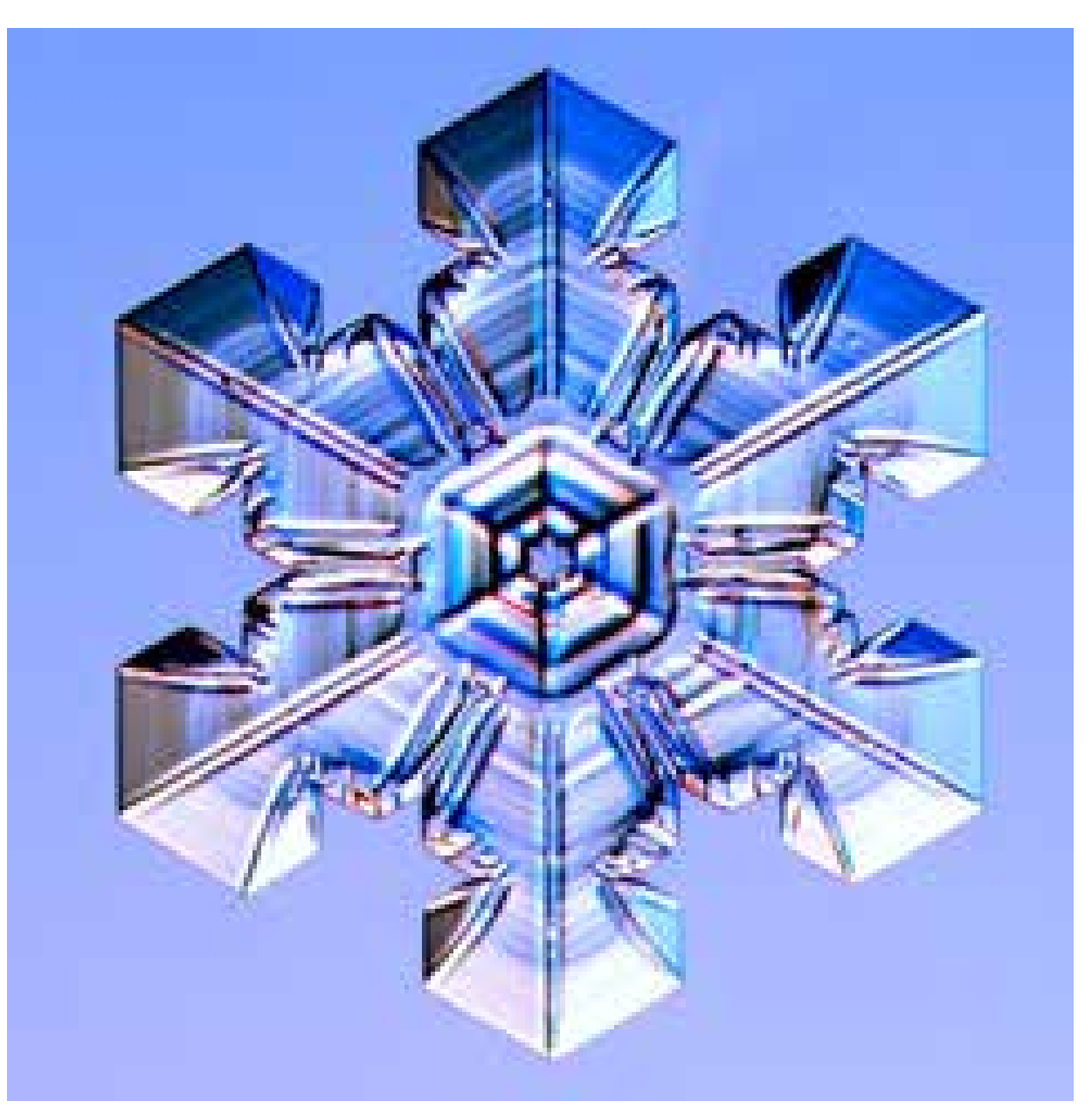

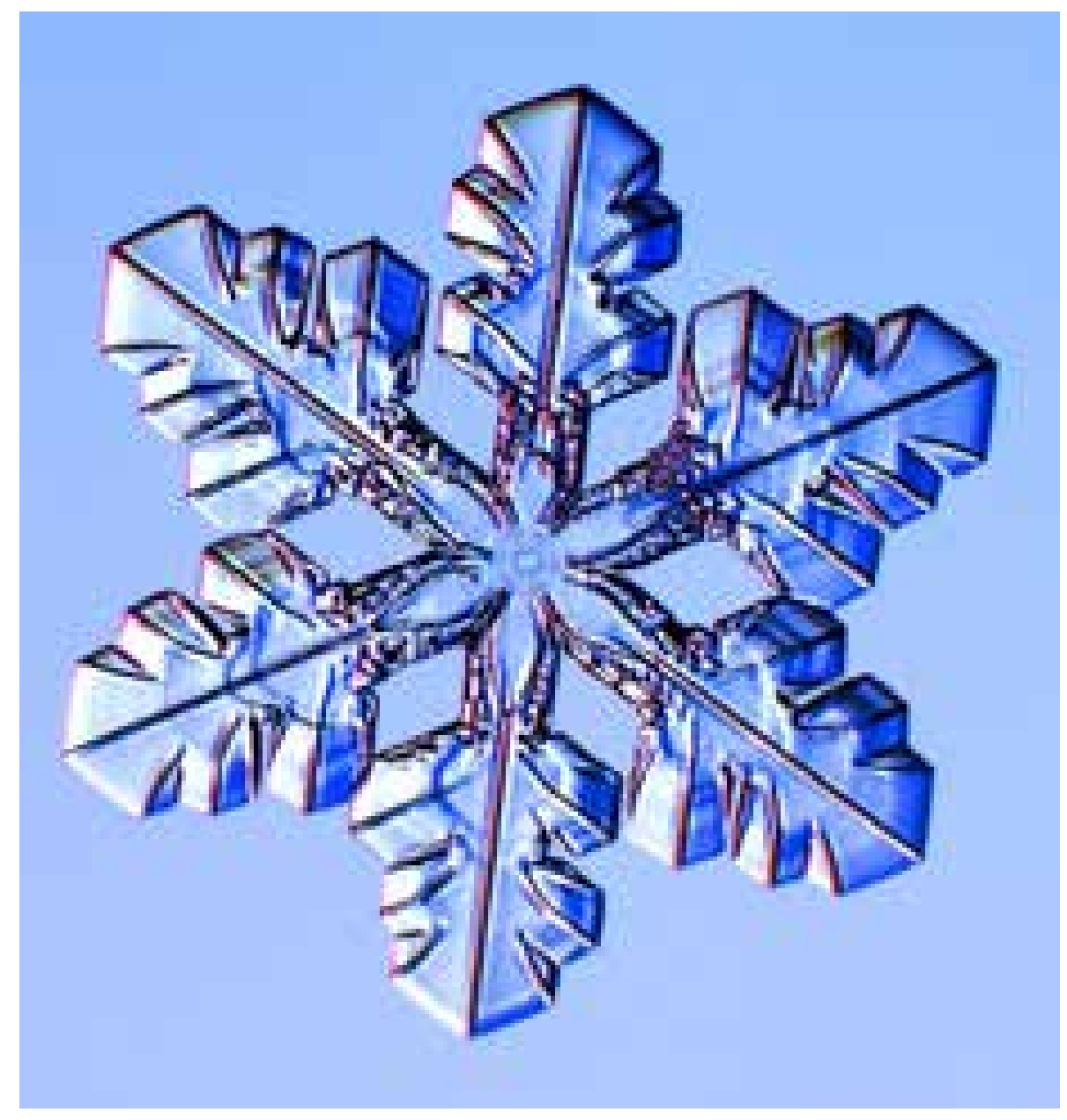

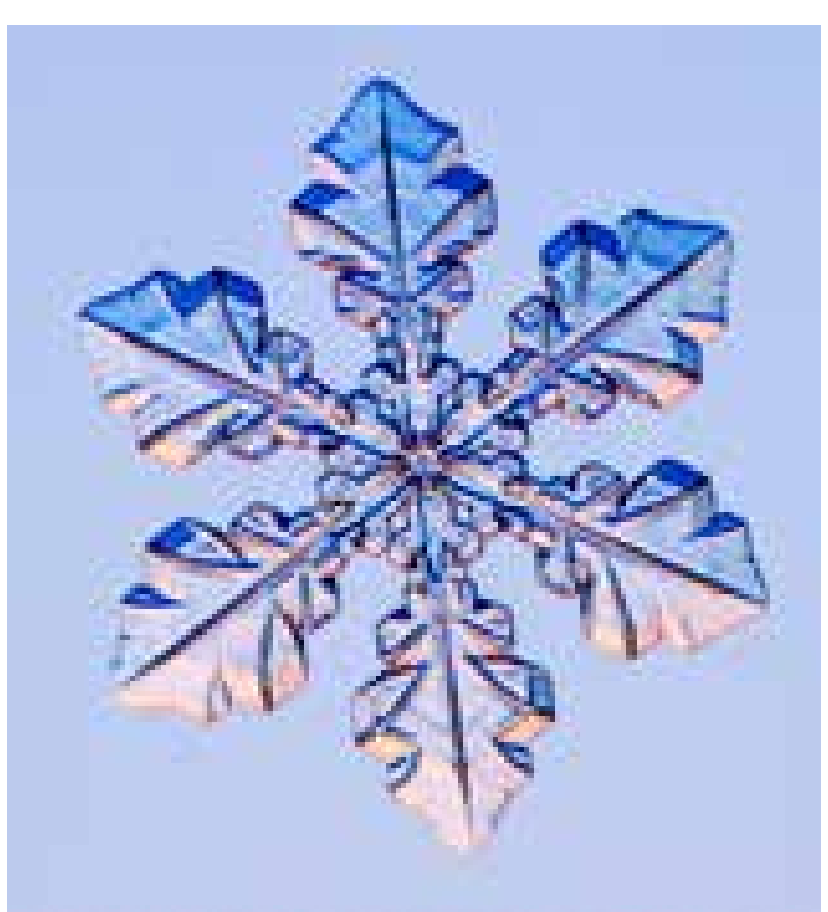

**Prominent Ridges.** Sectored plates merit a separate name because ridging is such a robust feature in snow-crystal growth. As described in Chapters 4 and 9, ridges form on slightly convex basal surfaces, owing to a patterning growth instability that is a two-dimensional version of the branching instability. The e-needle crystals in Figure 8.16 shows that ridging occurs over a broad range of temperatures and supersaturations, making it one of the most prevalent snow-crystal features. If you look carefully, you can usually find some evidence of ridge formation on most thin-plate crystals.

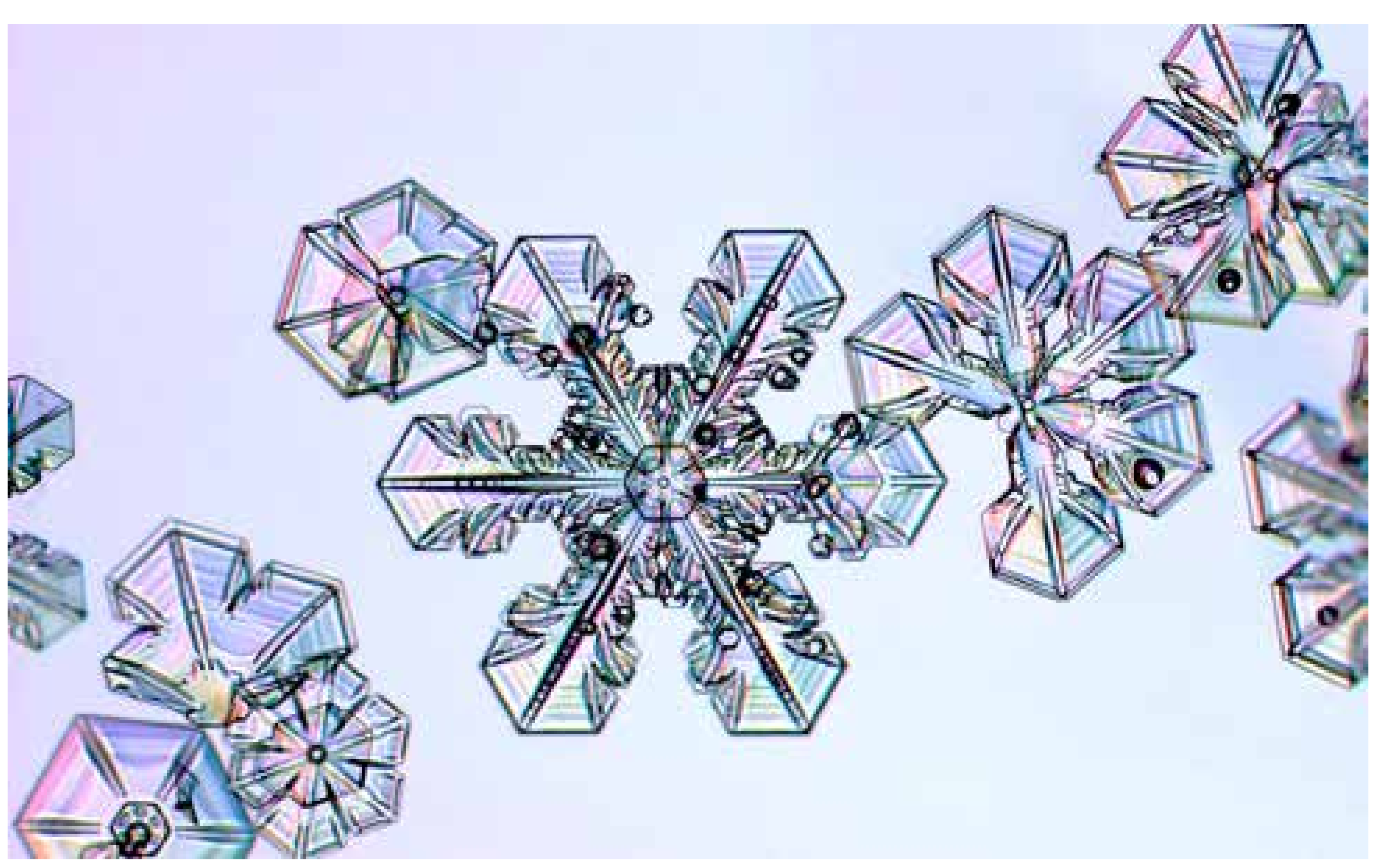

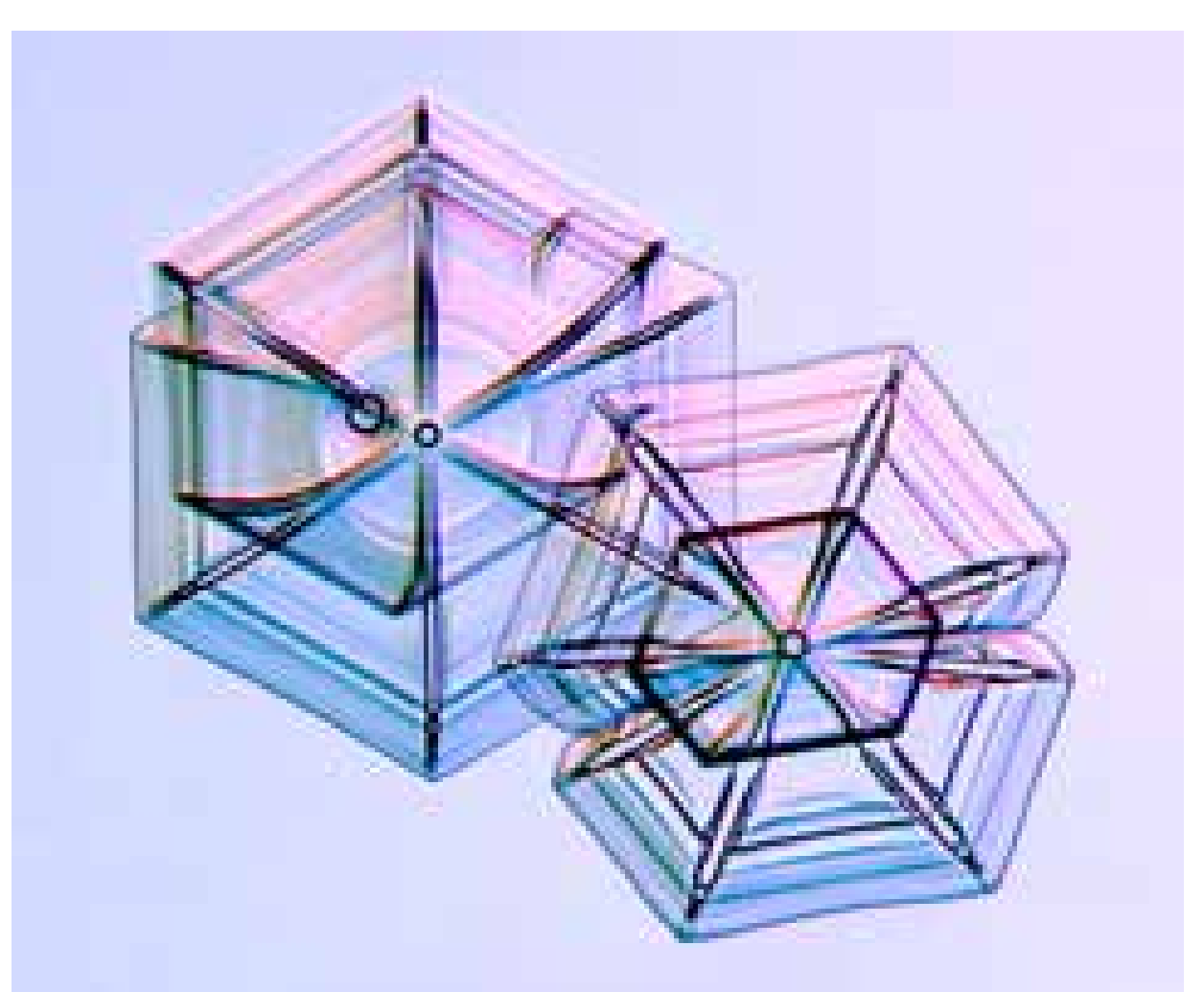

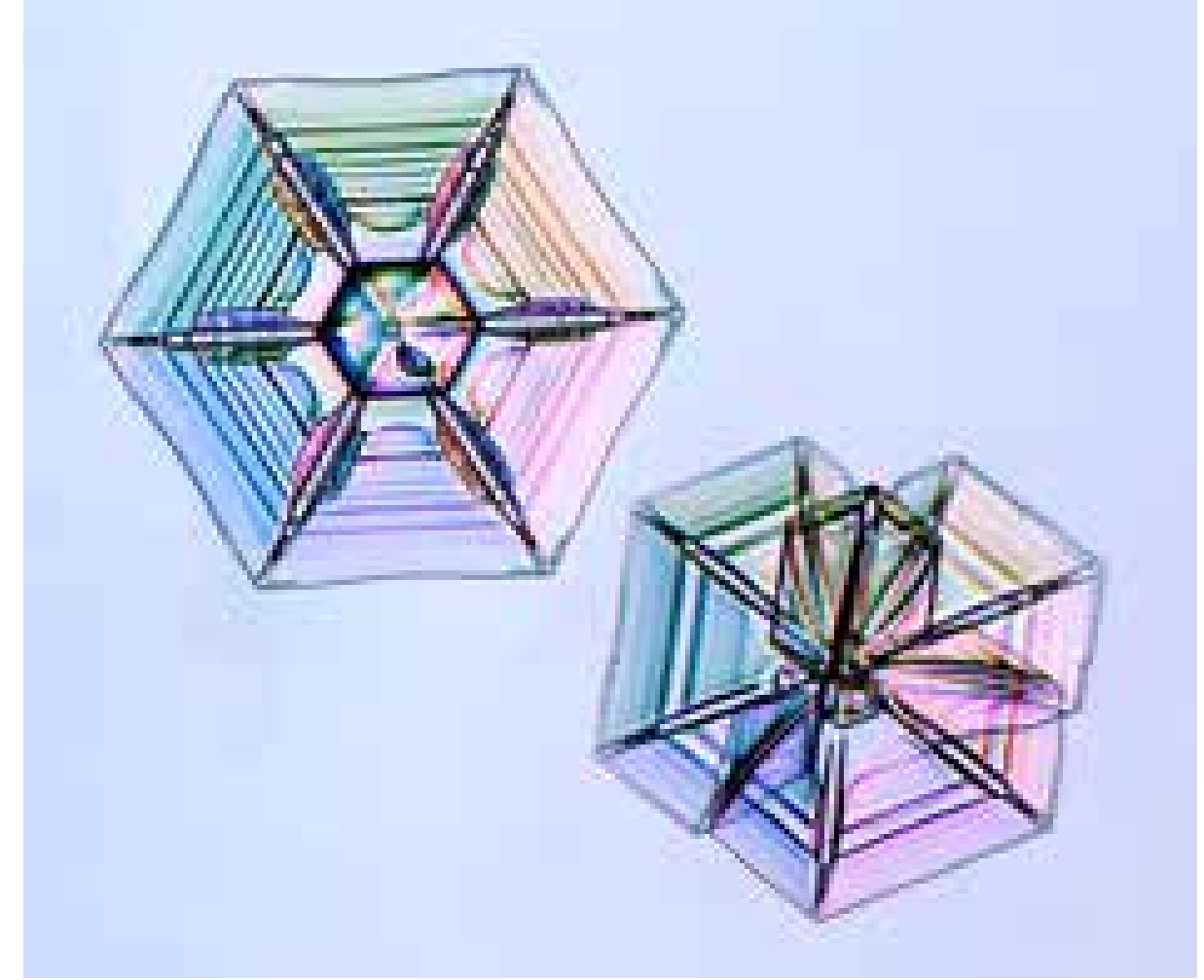

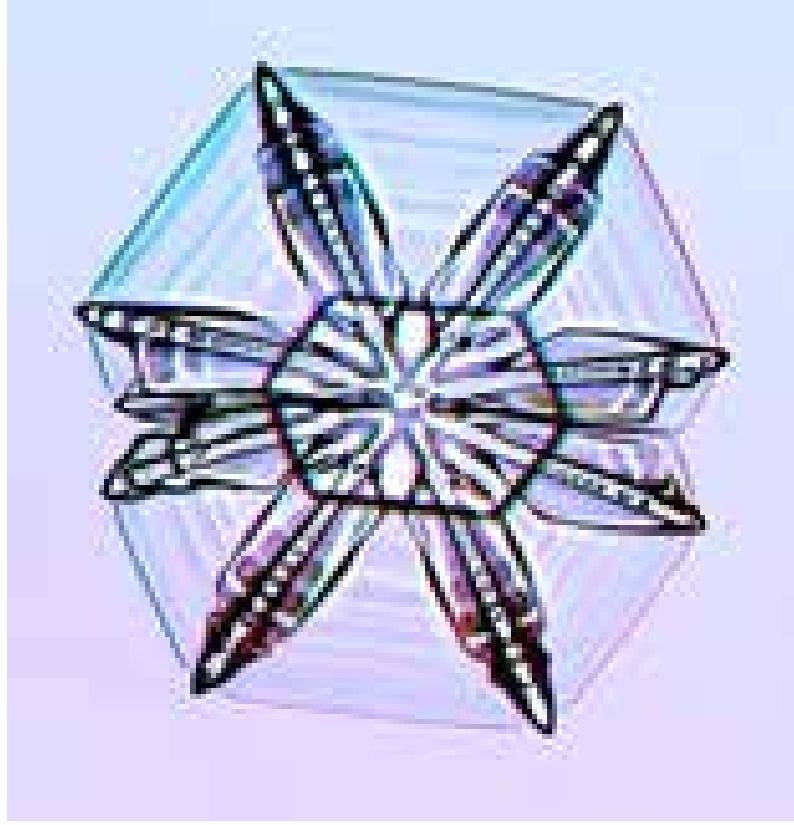

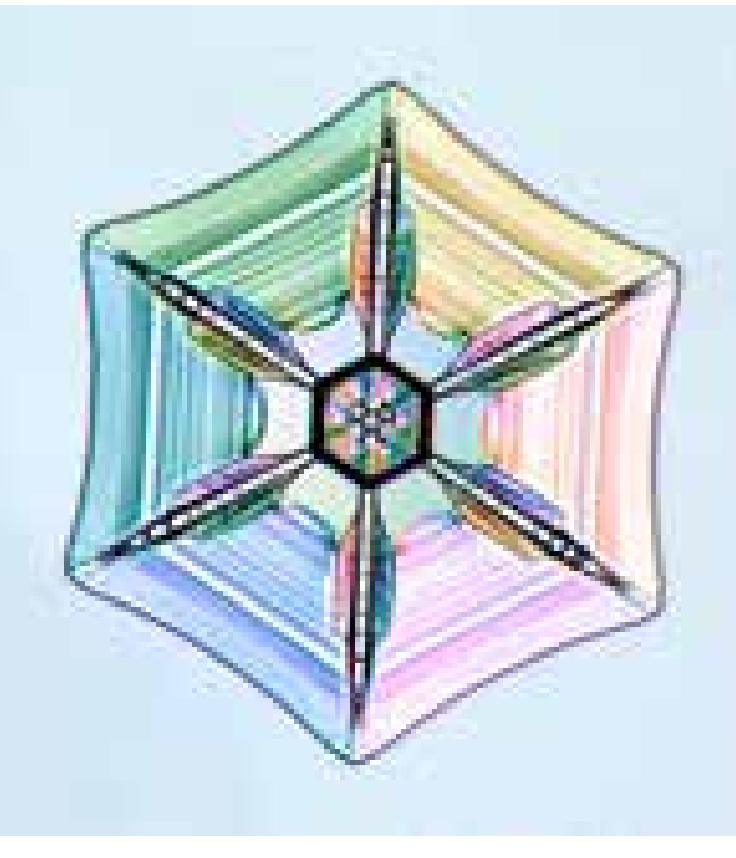

**Contemporaneous Crystals.** It is not atypical to see a cluster of comparable crystals all falling within a short period of time, as these did one day. When the conditions are right to form a particular crystal type, the clouds can release them in large numbers. As usual, some are well formed, but many are not.

# Stellar Dendrites

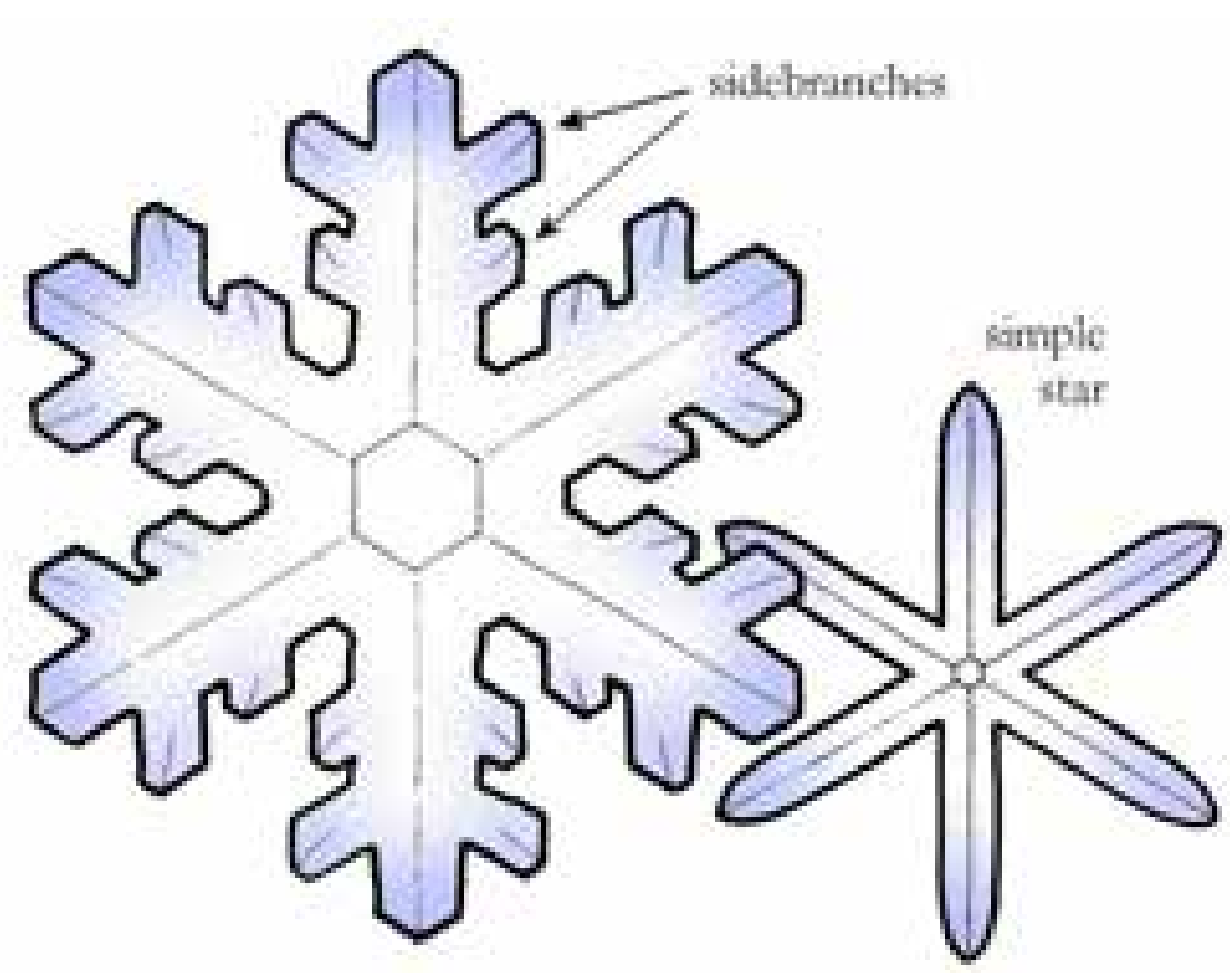

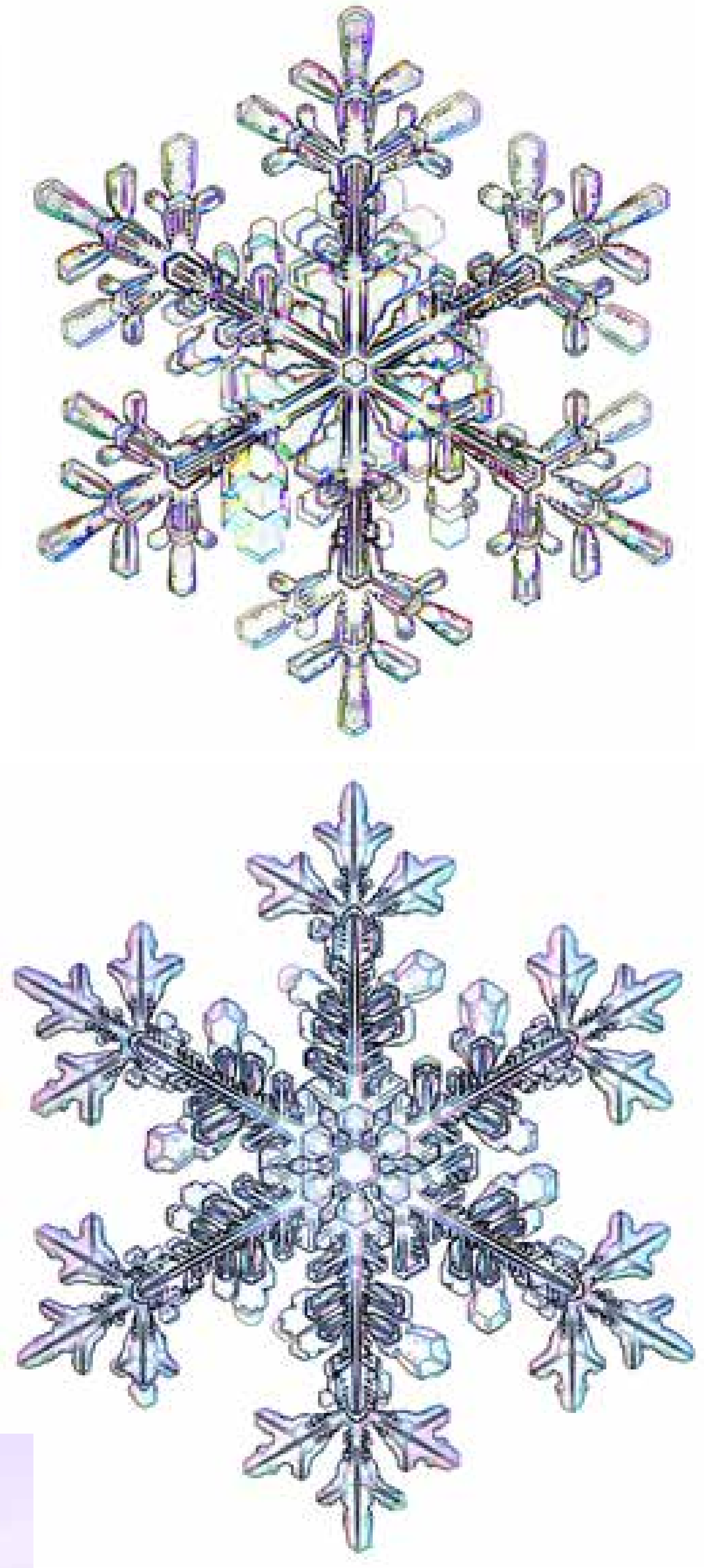

*Stellar dendrites are plate-like crystals with narrow branches decorated with numerous sidebranches. They tend to be larger than stellar plates, with generally less prominent faceting and more complex shapes. These crystals can be readily found with the naked eye, and considerable detail can be seen with a simple magnifier. Stellar dendrites are common in many snowfalls, often arriving in great numbers.*

The word dendrite means "tree-like," which is an apt description of these extravagant crystals. They form around -15 C when the humidity is fairly high. The ample water vapor supply drives the branching instability to produce numerous sidebranches. Stellar dendrites are often conspicuous, as a generously sized specimen might measure three millimeters from tip-to-tip. They are also quite thin and flat. Their ornate shapes with outstanding symmetry make stellar dendrites the much-celebrated canonical holiday snowflakes.

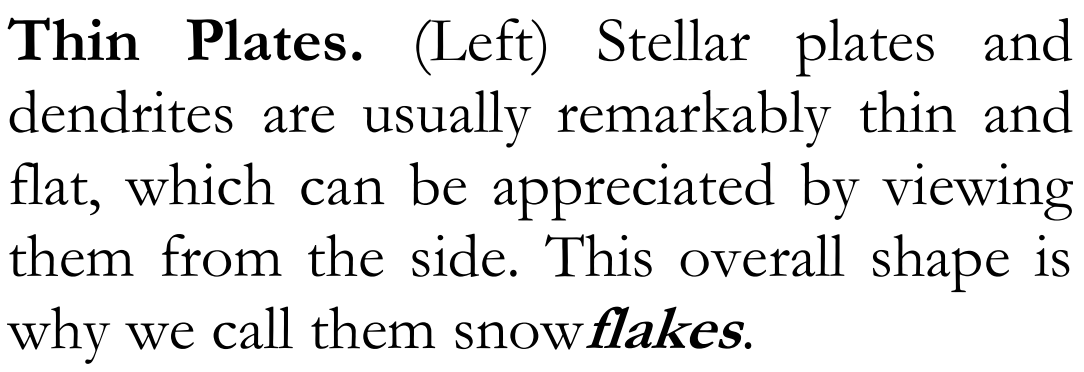

**Thin Plates.** (Left) Stellar plates and dendrites are usually remarkably thin and flat, which can be appreciated by viewing them from the side. This overall shape is why we call them snow***flakes***.

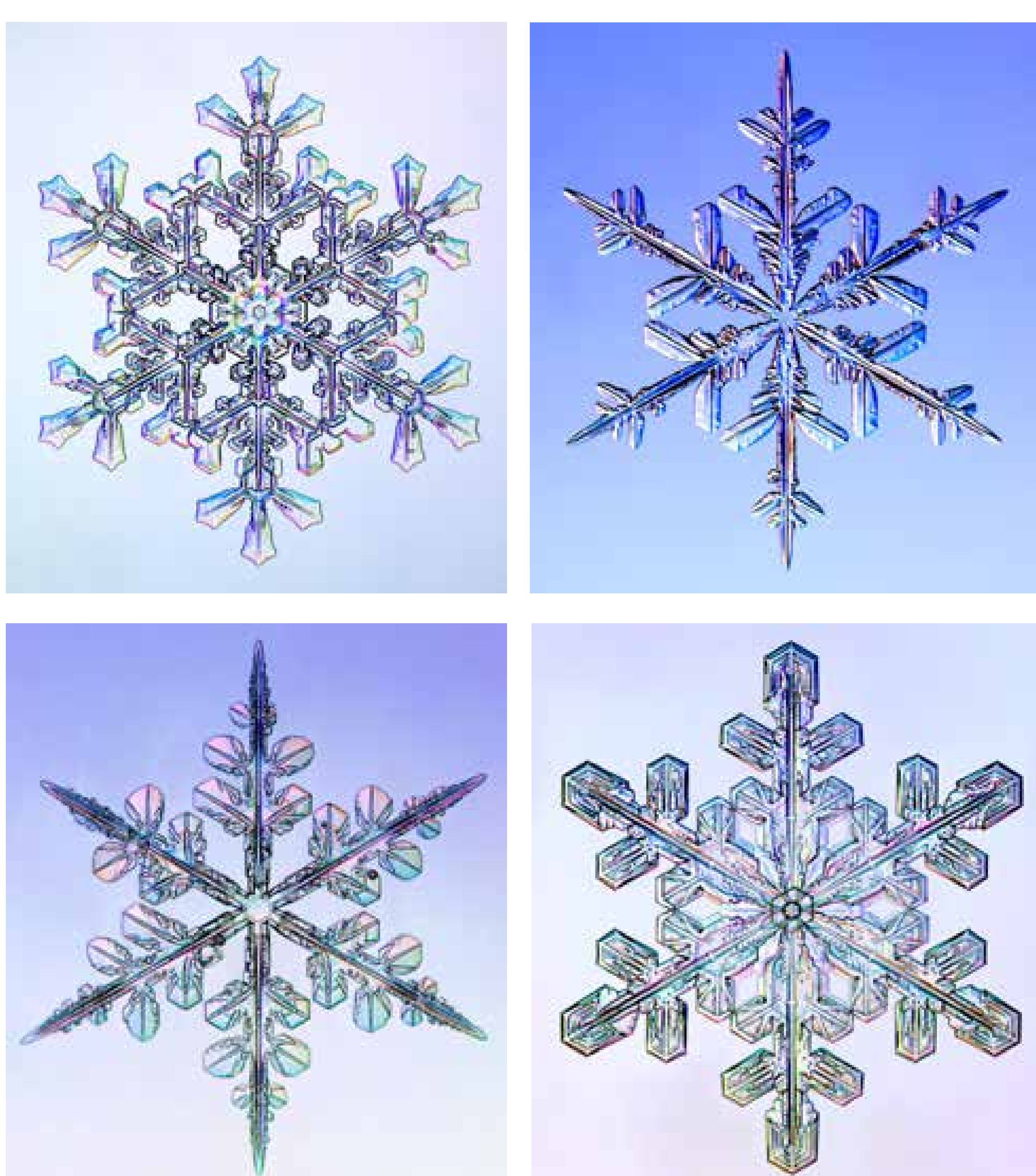

**Induced Sidebranching.** Symmetrical sidebranching on a stellar dendrite is typically brought about by induced sidebranching, as described in Chapters 4 and 9. This is the only known mechanism that will cause sidebranches to sprout synchronously from all six primary branches on a large stellar crystal. It requires a carefully orchestrated series of events to produce several sets of well-formed symmetrical sidebranches, which is why good examples are rare and difficult to find in nature. They are substantially easier to create in the lab (see Chapter 9) under controlled conditions.

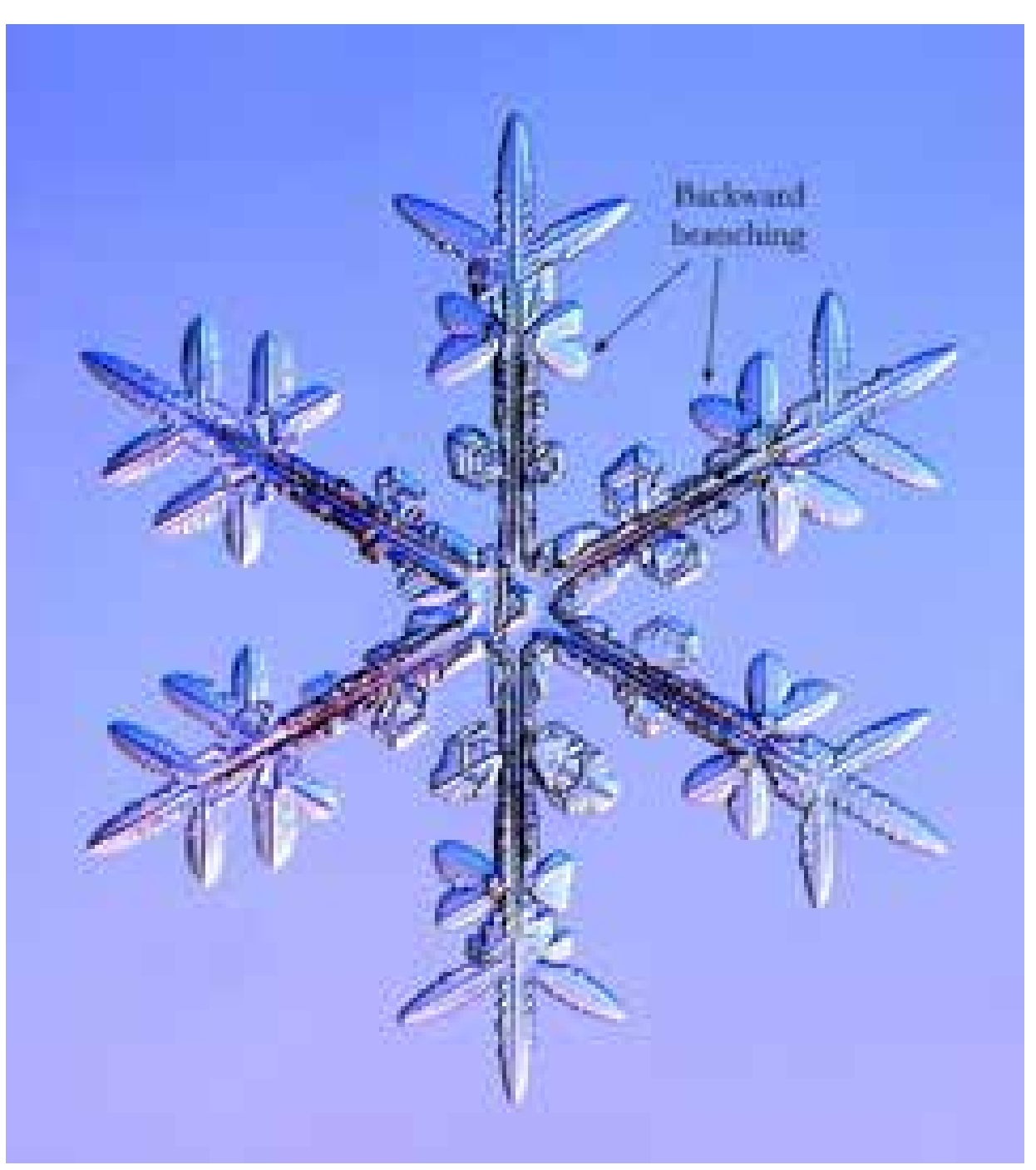

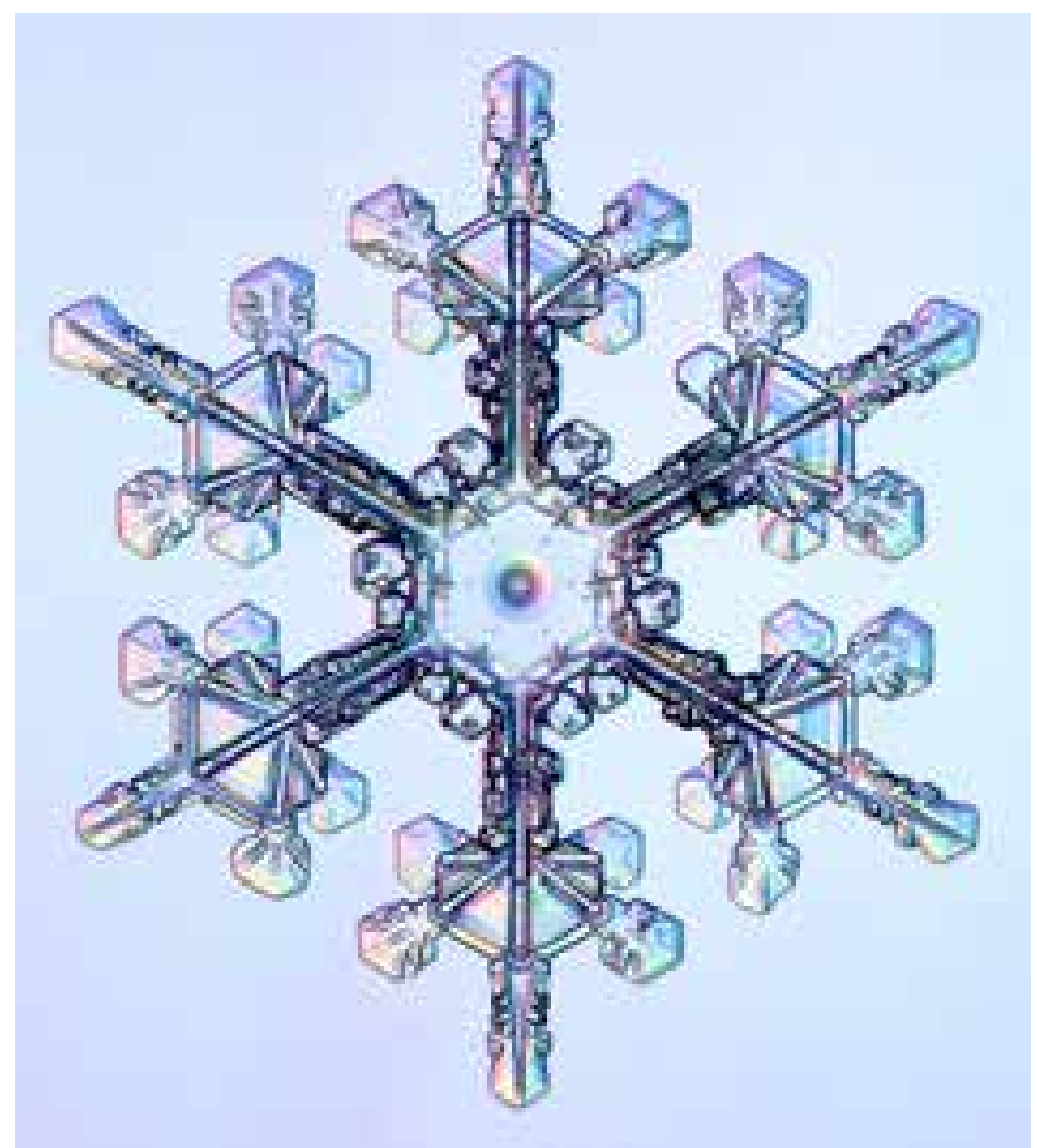

**Backward Branches.** On rare occasions, one can find sidebranches that appear to be growing 60 degrees off from the usual forward direction, so I call them "backward" sidebranches.

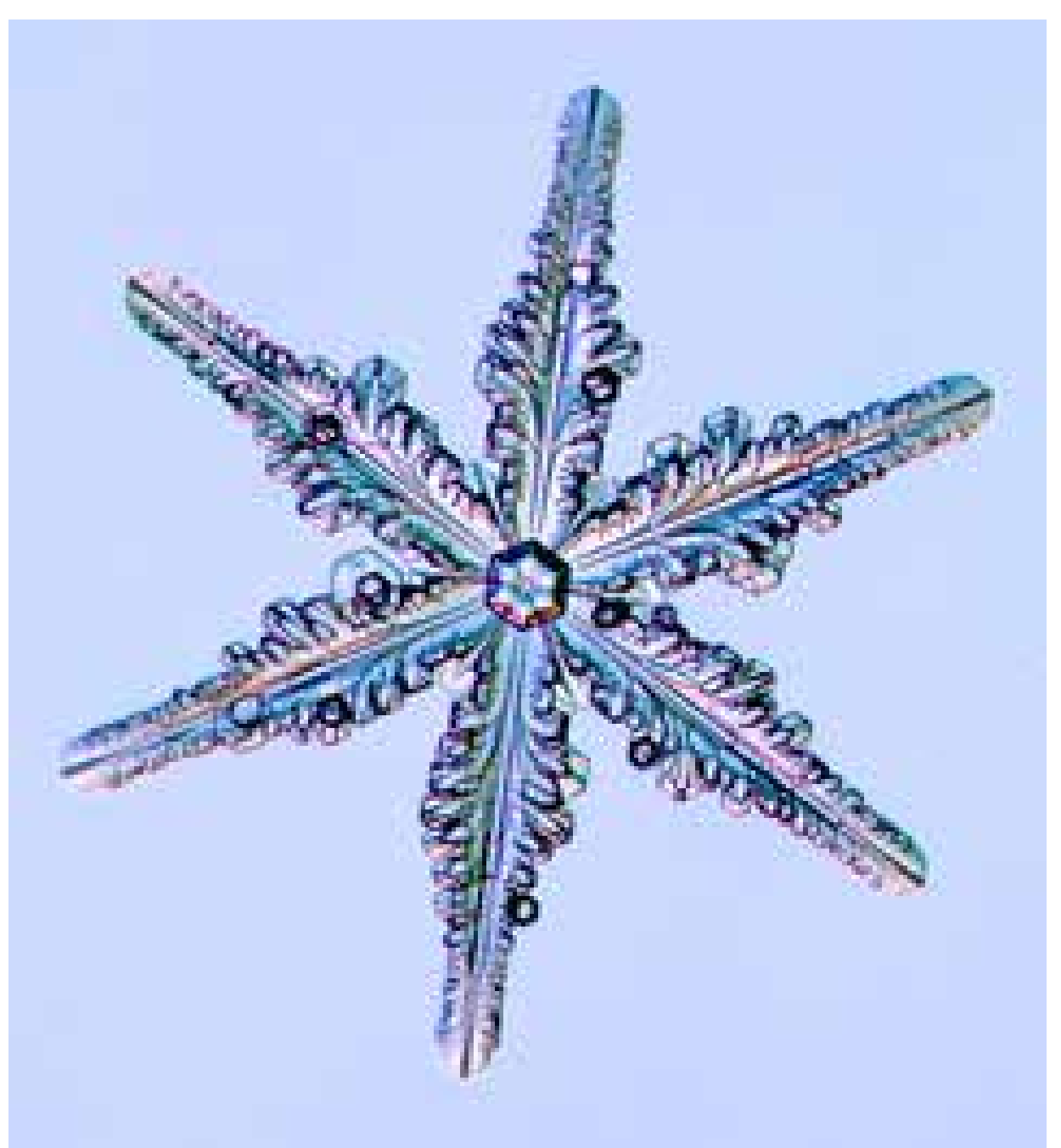

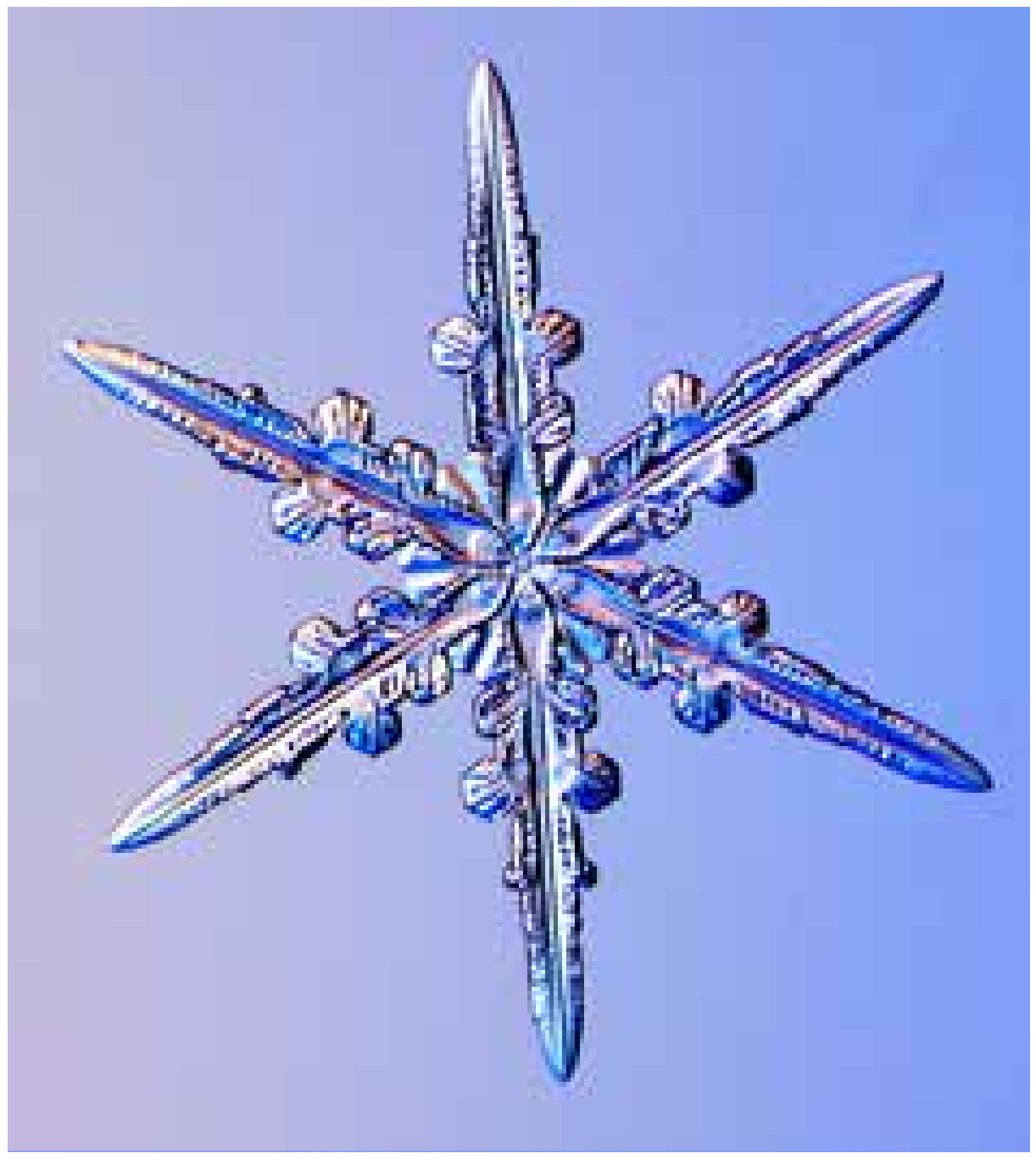

**Simple Stars.** These minimal stellar crystals appear when the supersaturation is high enough to produce narrow branched growth, but too low to create much sidebranching. Figure 8.16 reveals that these crystals grow only over a fairly narrow range of parameter space, so they are not especially common. They are also fairly small and easy to overlook.

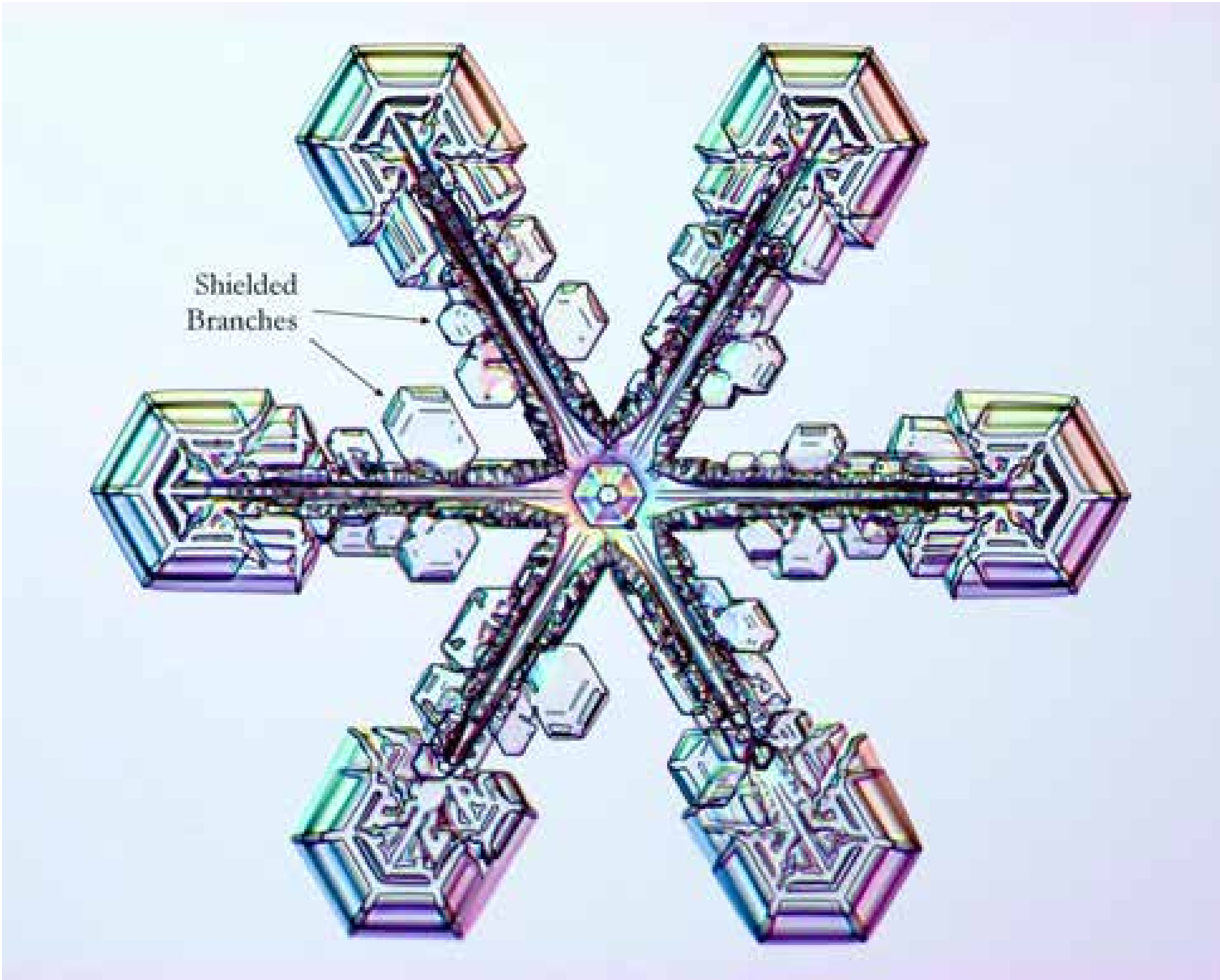

**Shielded Branches.** Snow crystals grow fastest at their outer edges, which have the greatest supply of water vapor. But sometimes the interior branches will grow substantially even after the outer branches have grown out and left them behind. Because the interior branches are shielded by the outer branches, they receive a reduced supply of water vapor. Such conditions often yield thin, rather featureless plate-like structures that are rather asymmetrically placed, as seen in these examples.

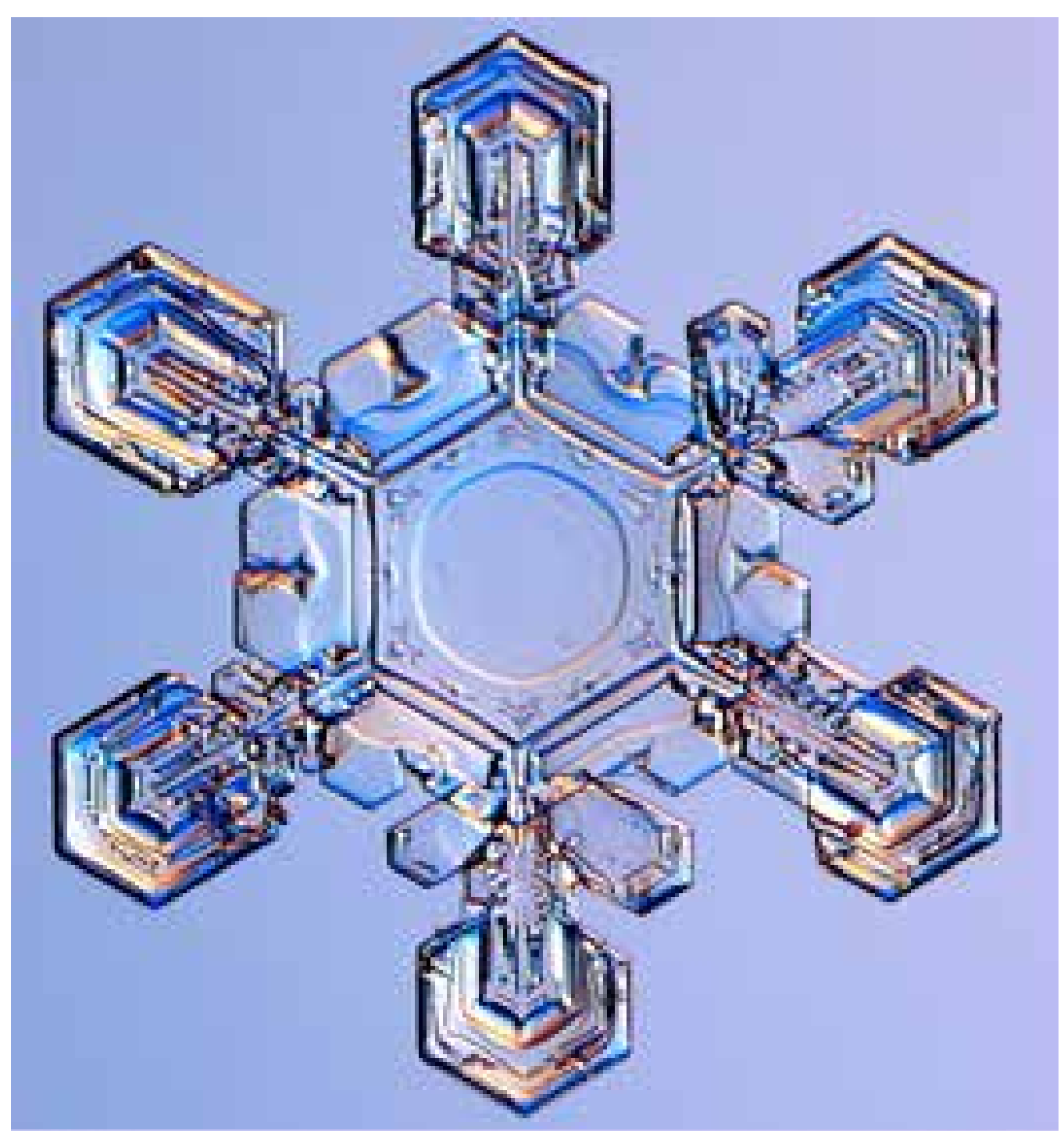

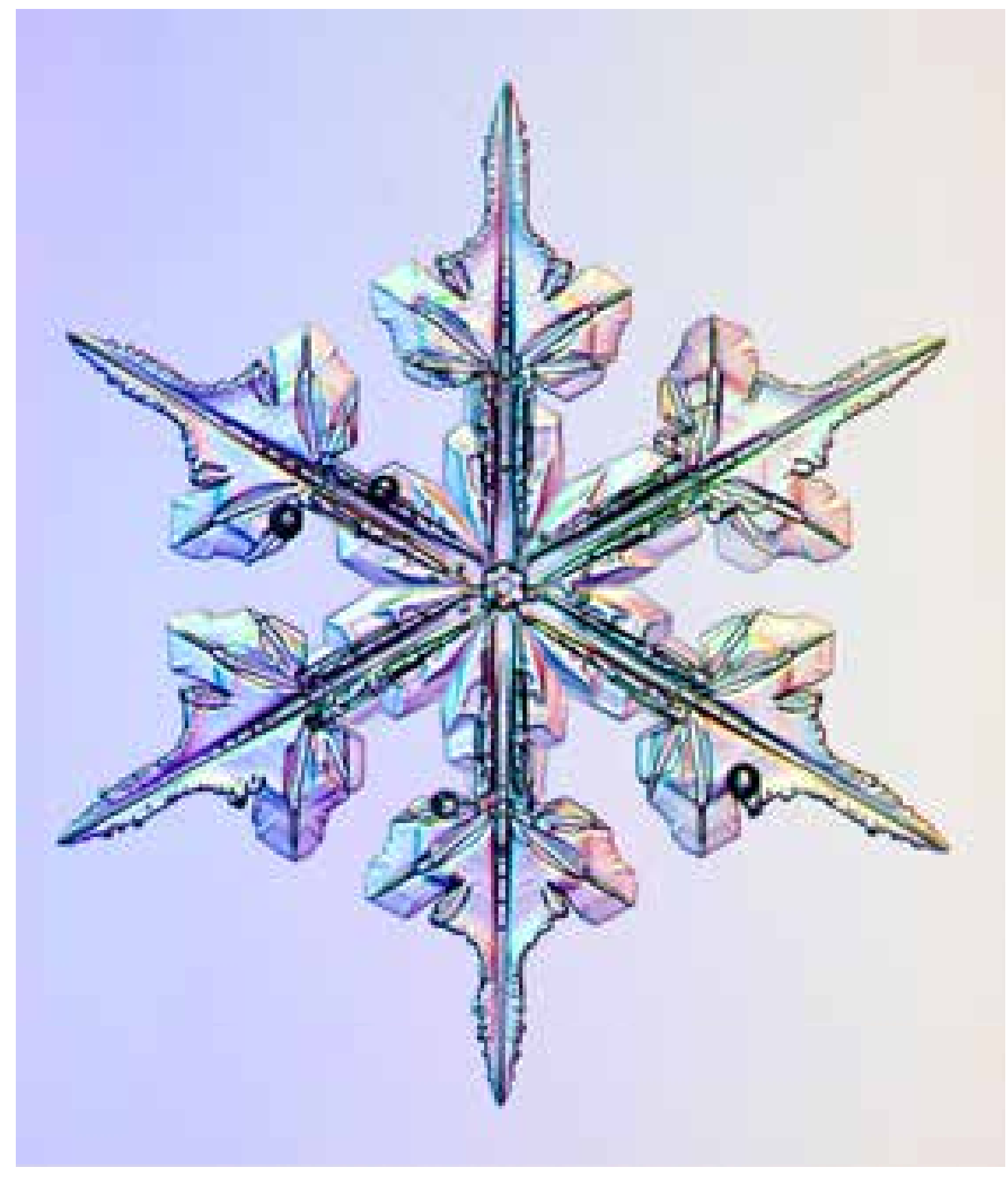

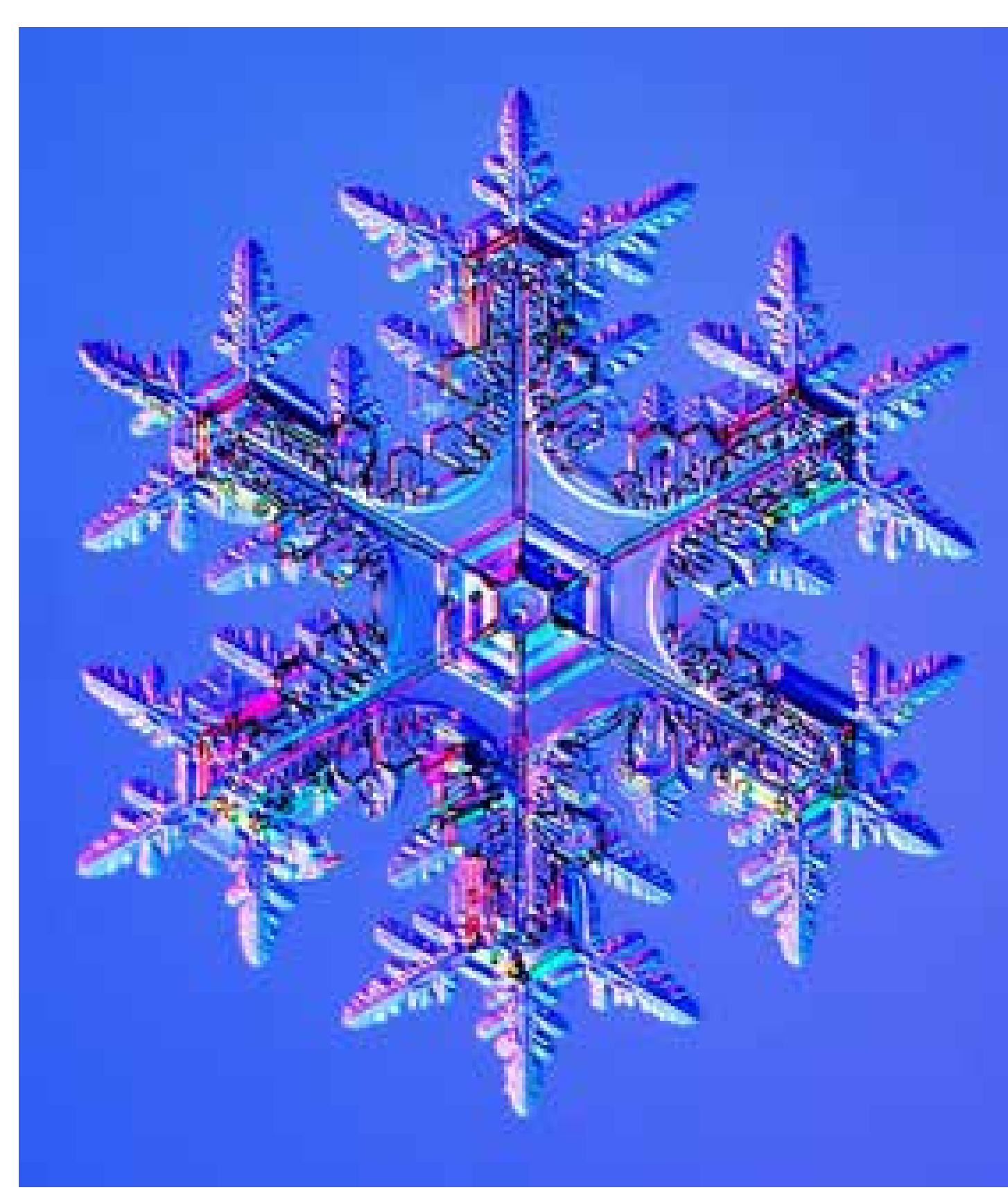

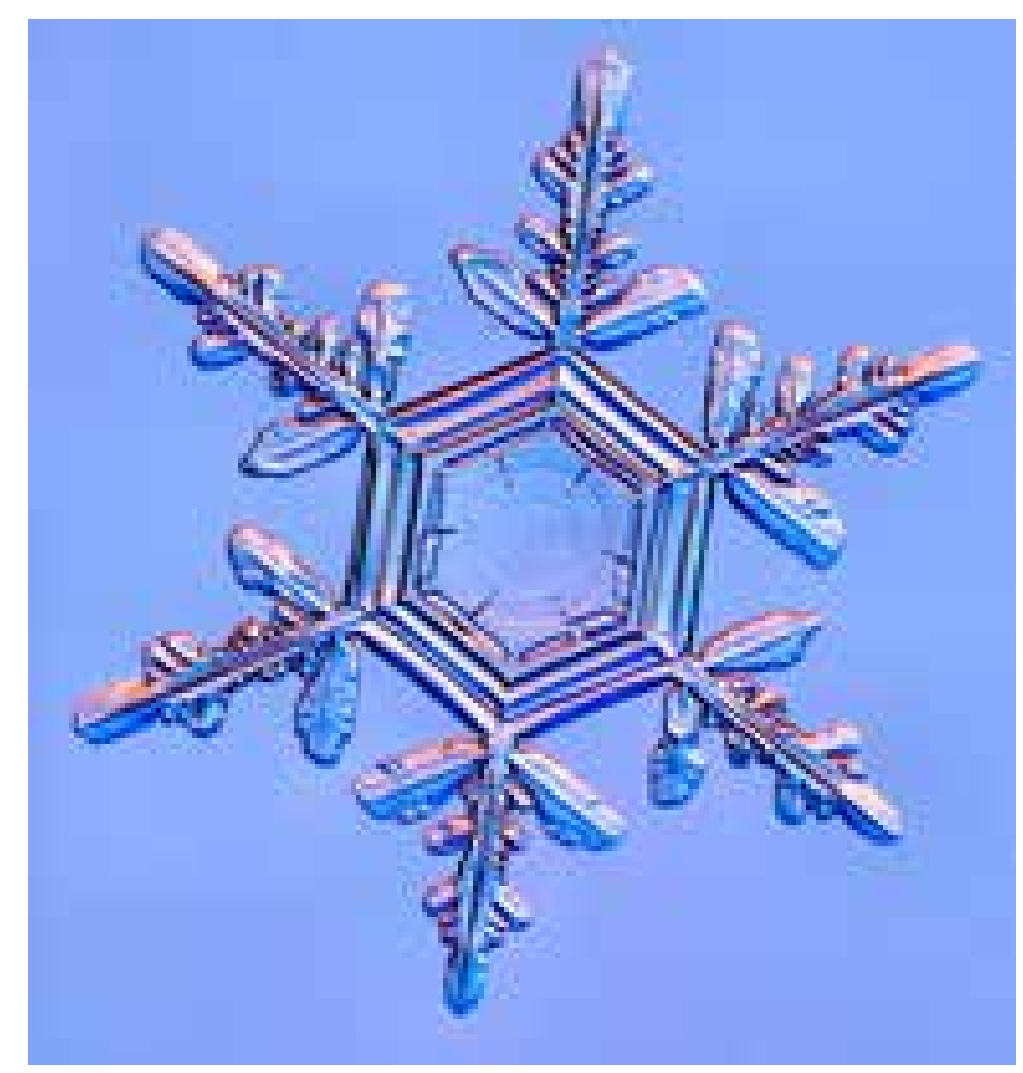

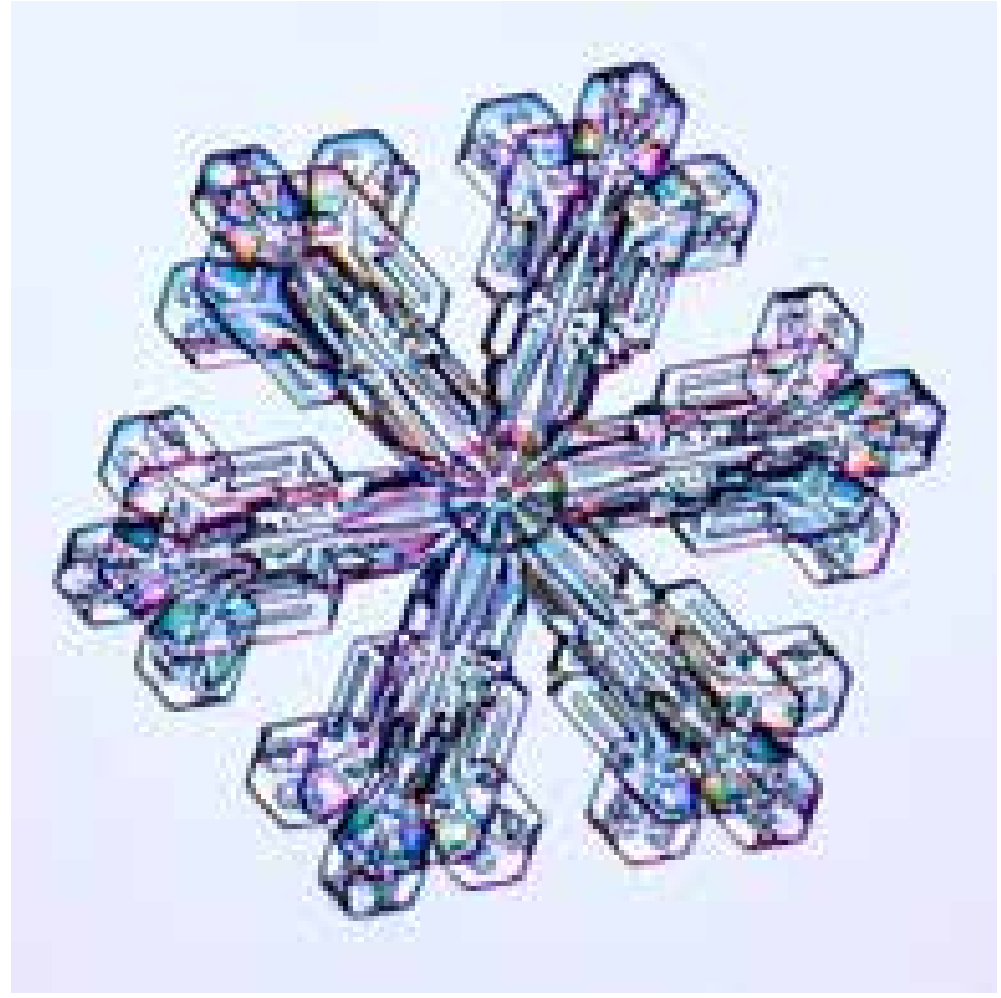

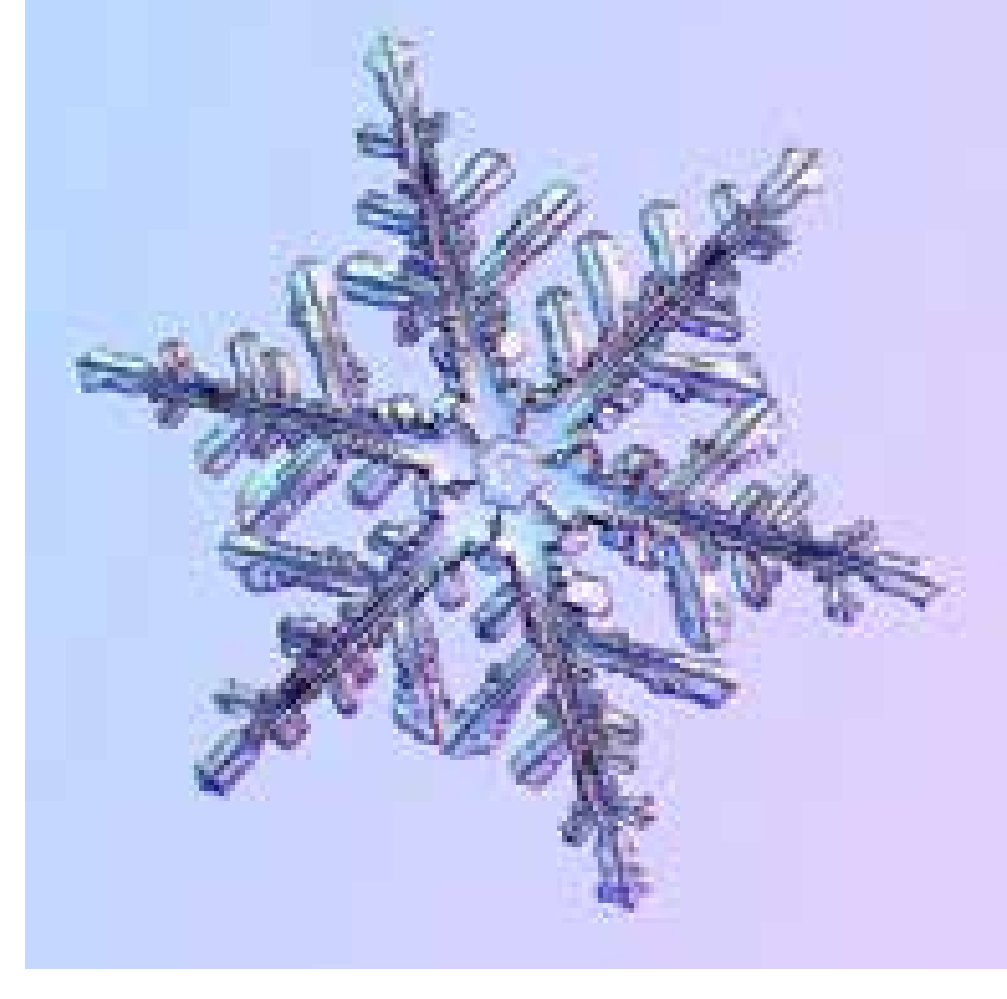

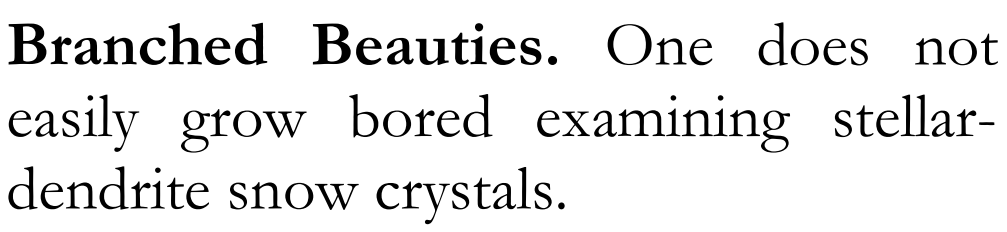

**Branched Beauties.** One does not easily grow bored examining stellar-dendrite snow crystals.

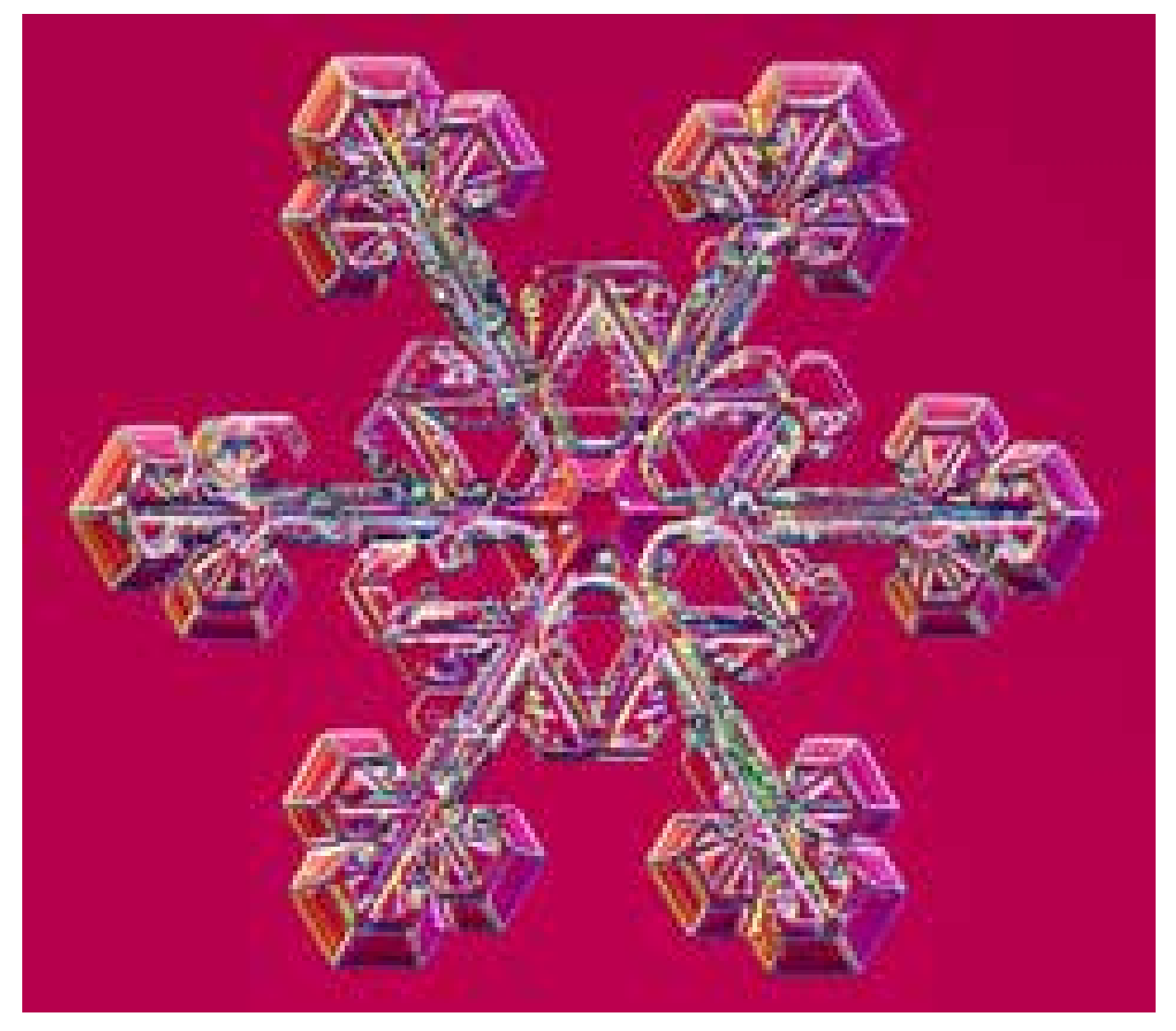
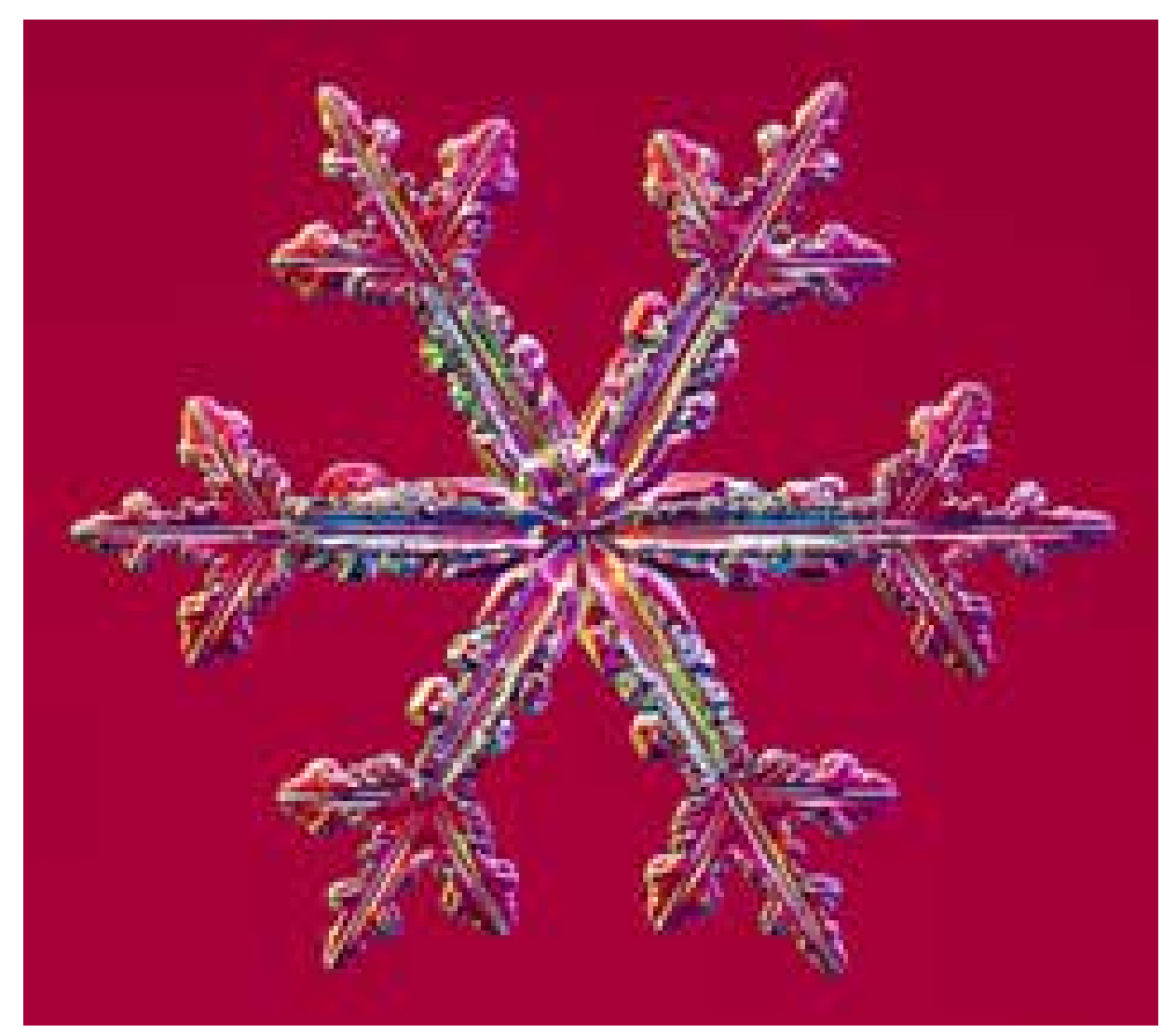
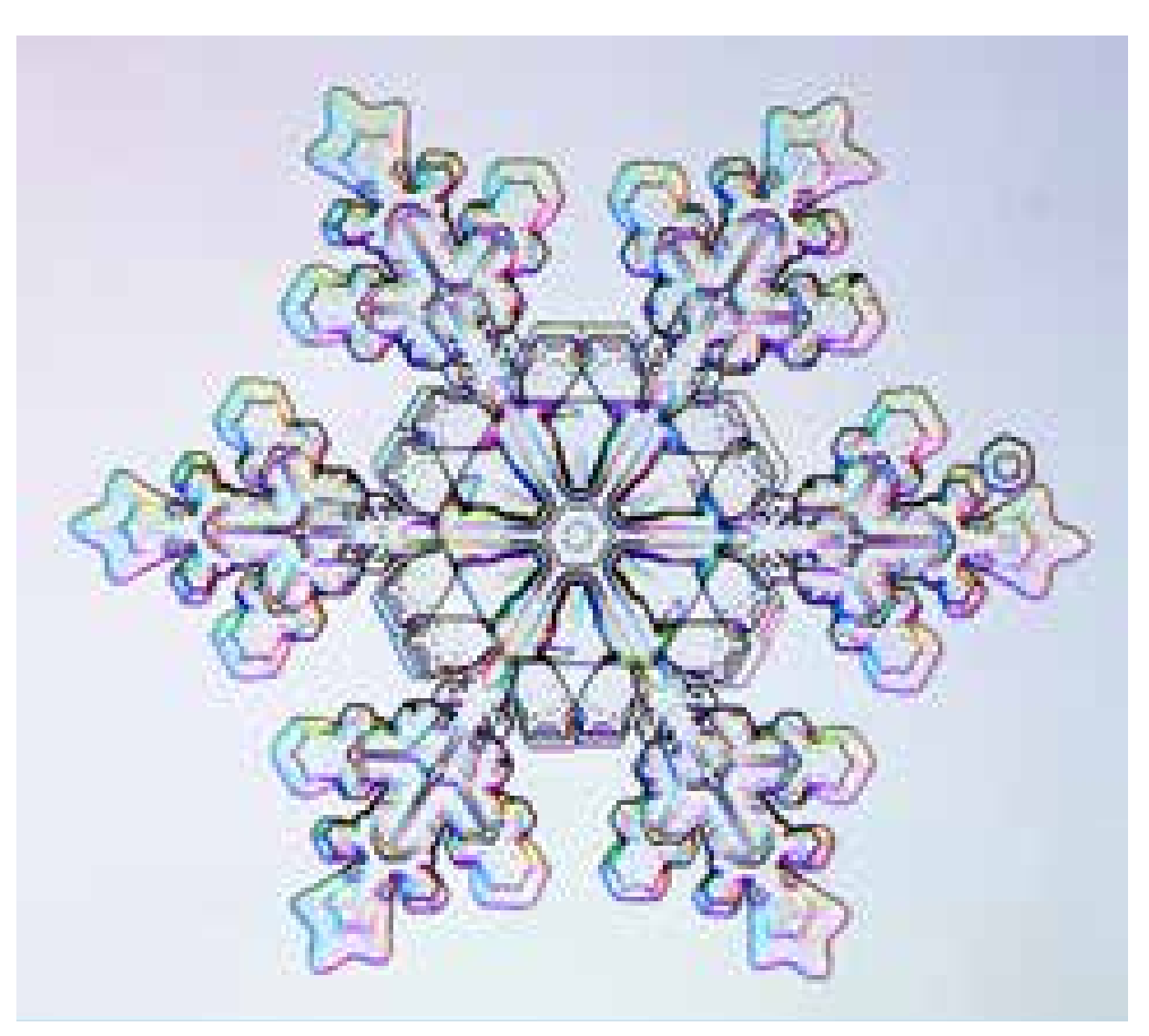
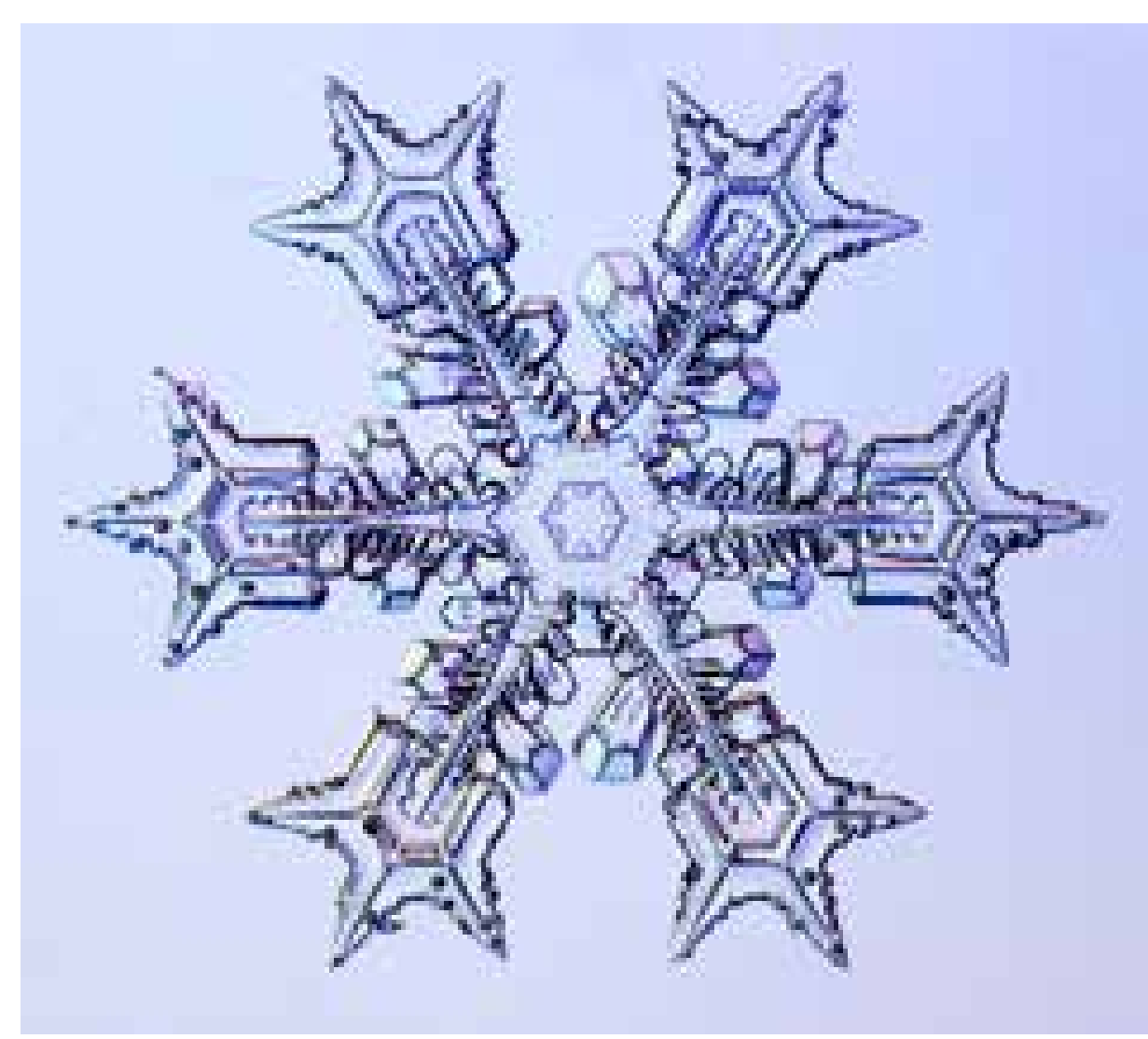
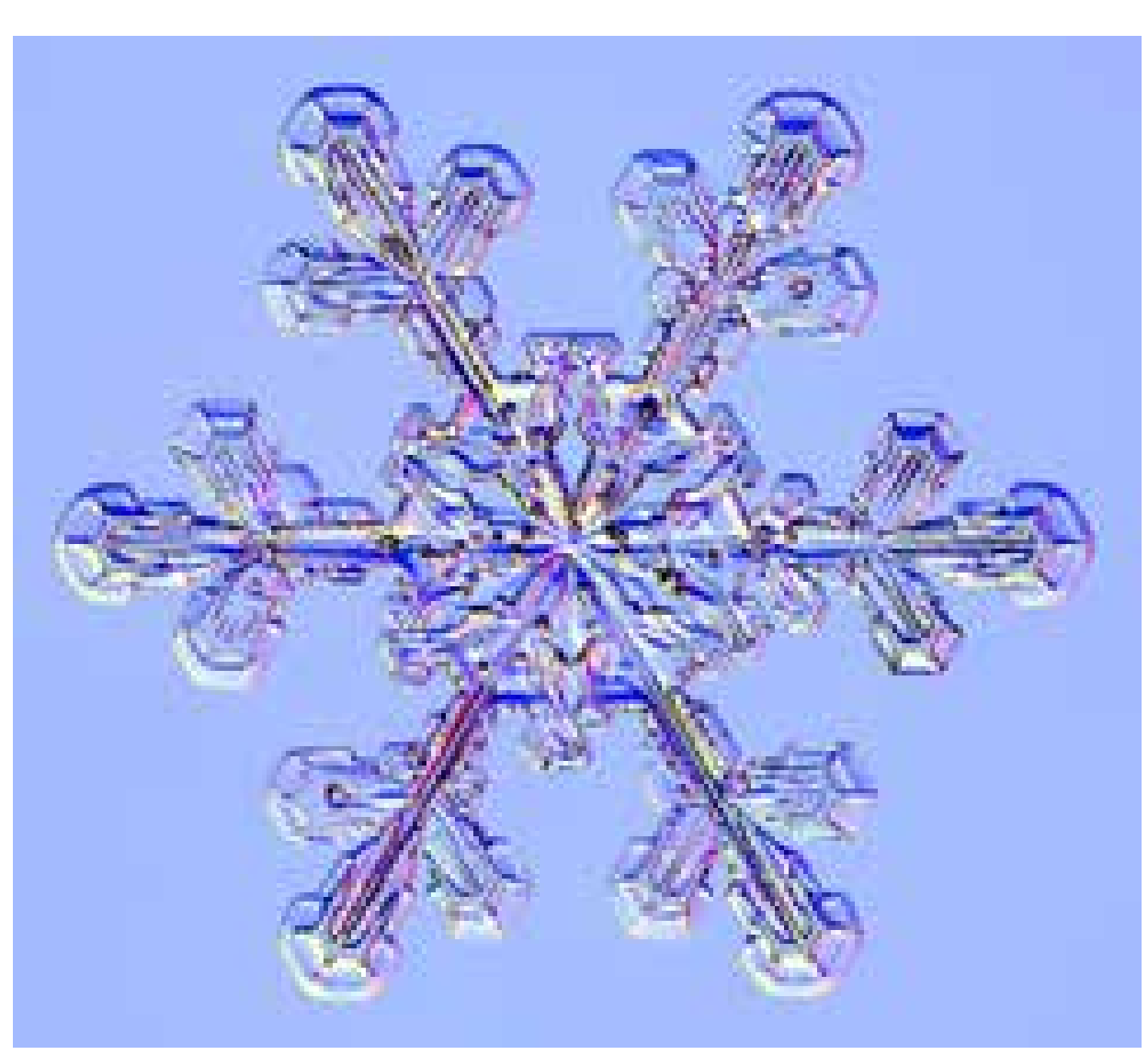
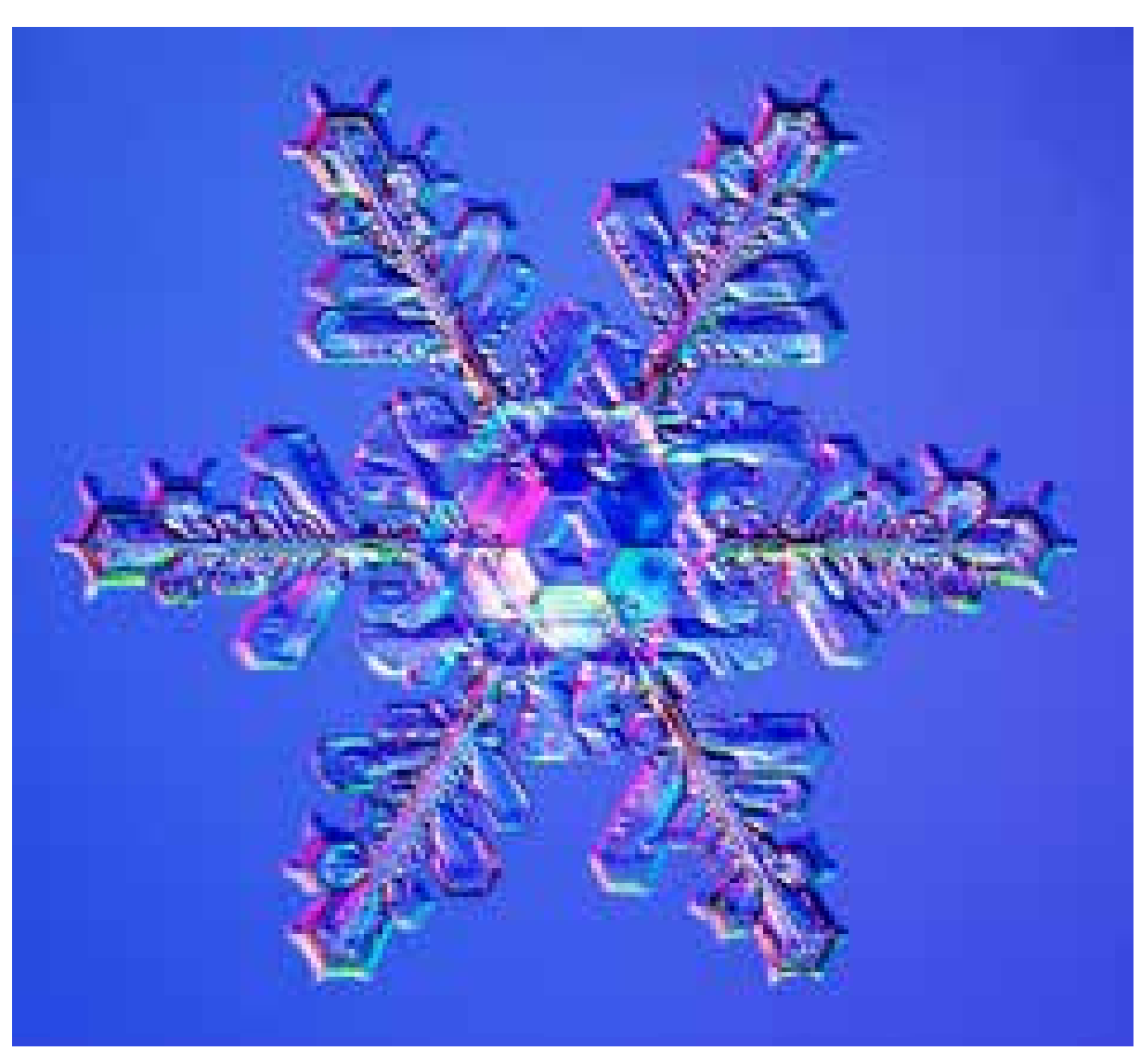

# Fernlike Stellar Dendrites

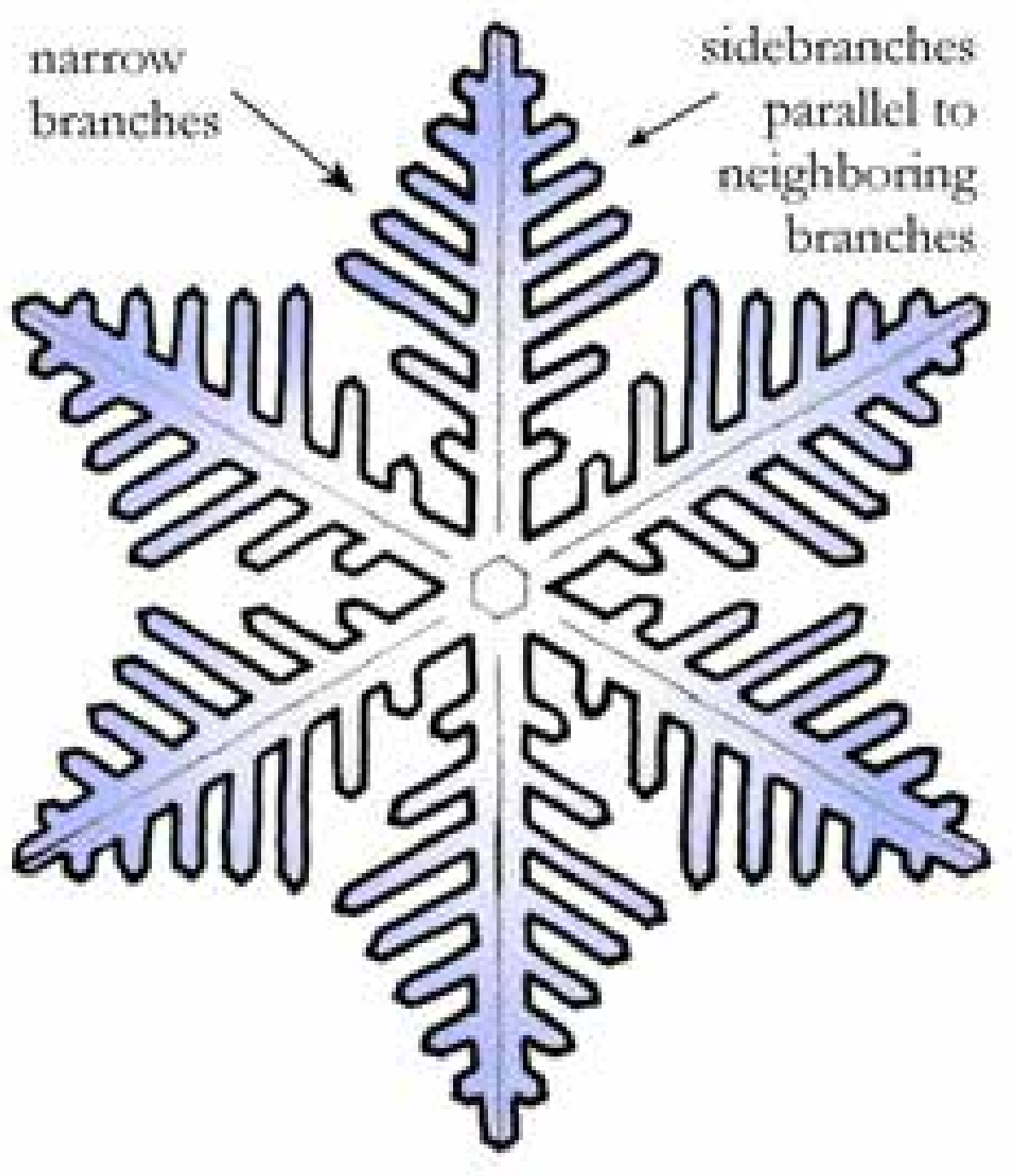

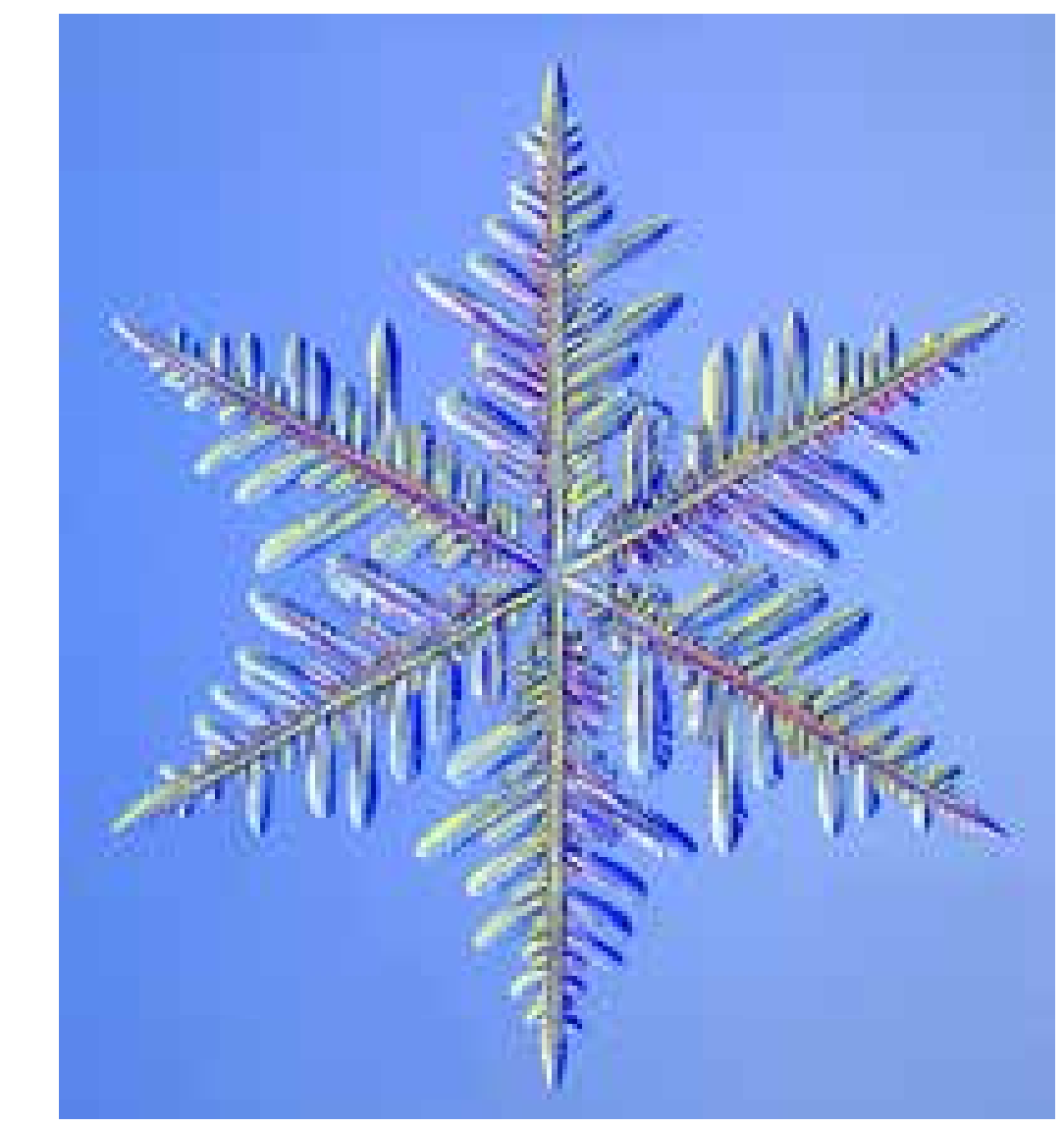

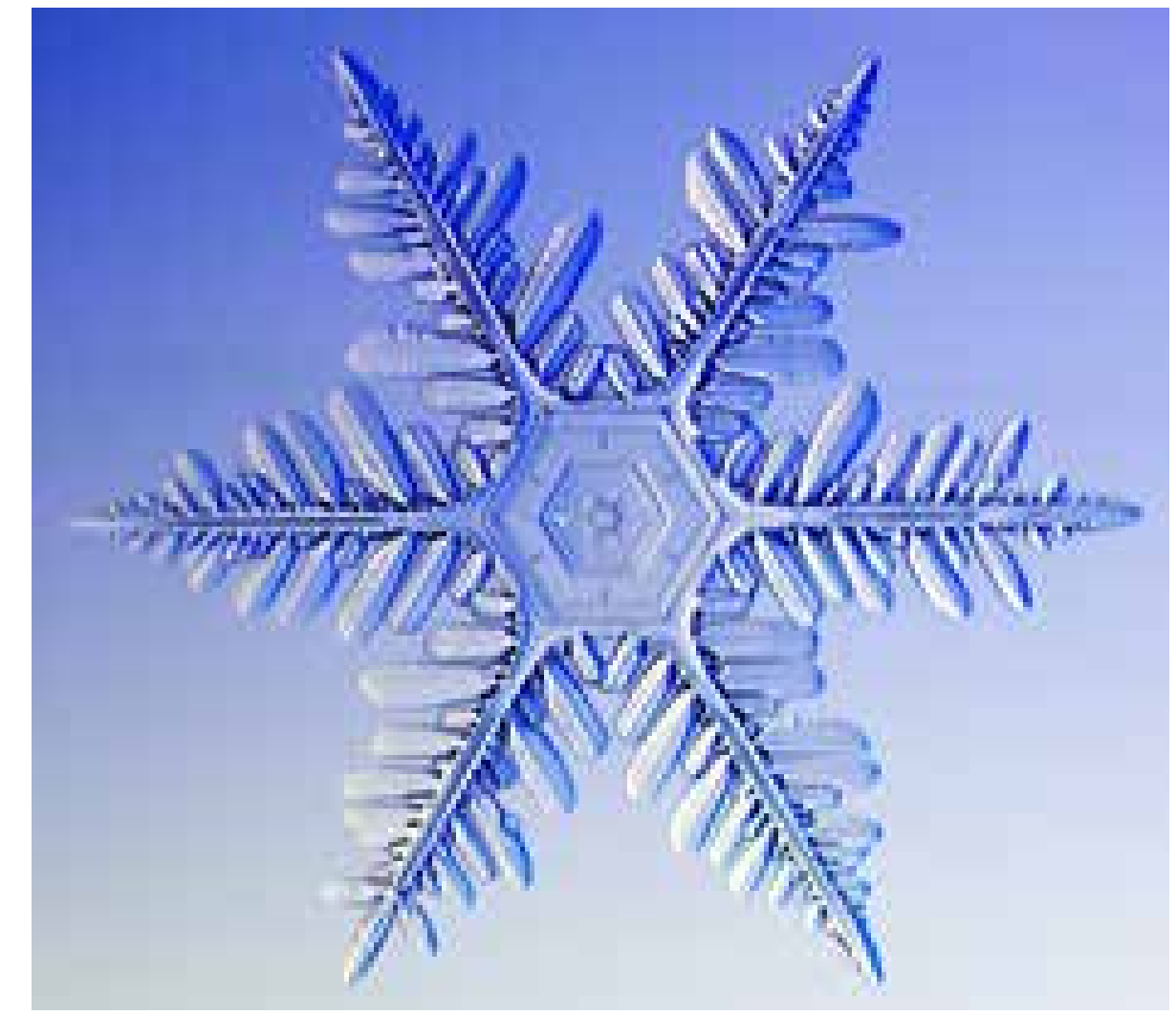

*Fernlike stellar dendrites are large, thin plates with narrow branches and sidebranches that looks similar to a fern. Sidebranches typically form at 60-degree angles relative to their primary branches. These crystals are common, and their exceptionally large size makes them easy to spot.*

Fernlike stellar dendrites are the largest snowflakes, on rare occasions measuring over 10 mm in size. Their thickness may be a hundred times less than this, however, making them extremely thin, flat, plate-like crystals. They only form near -15 C when the humidity is exceptionally high, which drives their rapid growth with copious sidebranching.

The well-defined 60º angles between the branches and sidebranches of fern-like stellar dendrites indicate that they are single crystals of ice. In spite of their complex shapes, the molecules are all lined up from one tip to the other.

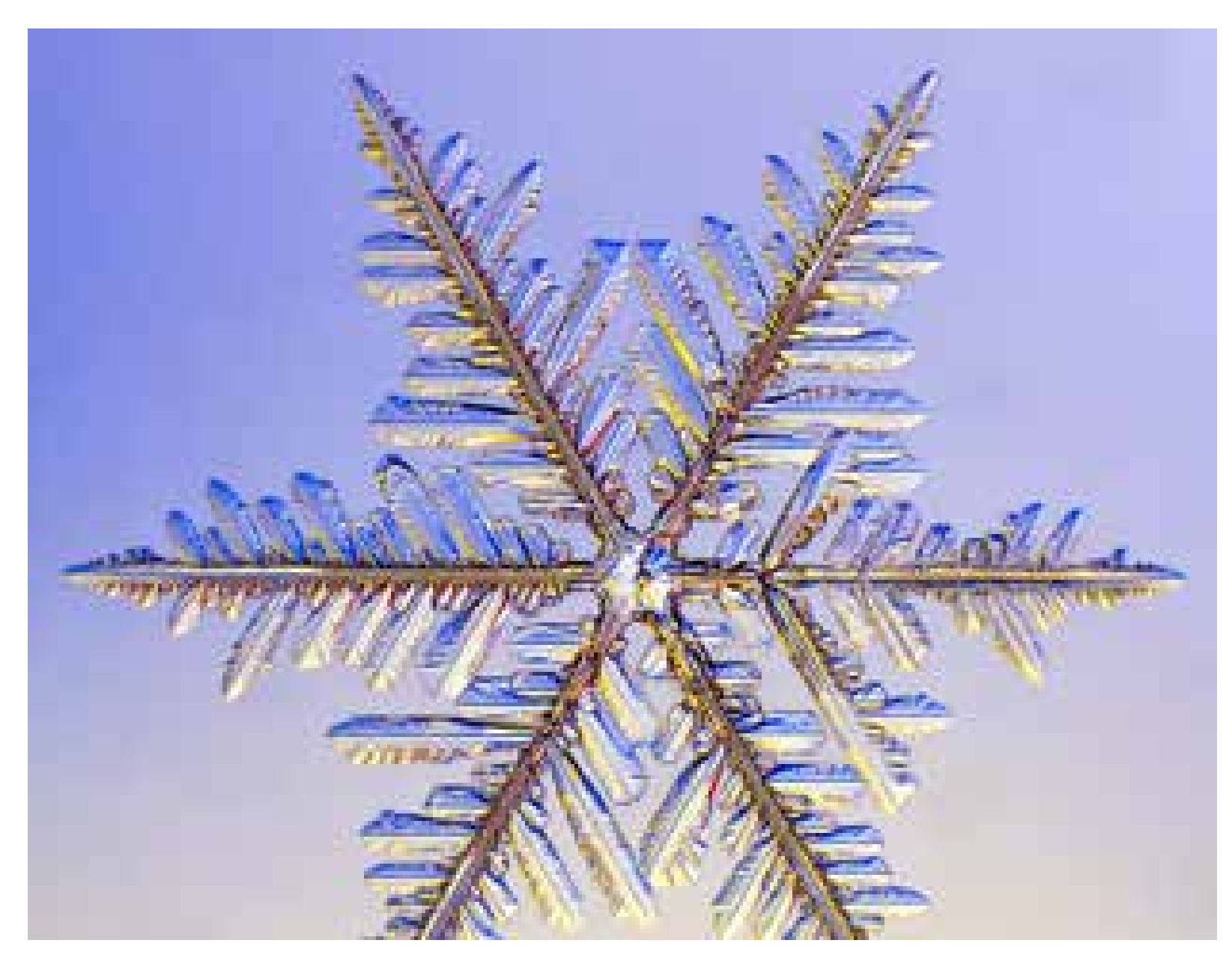

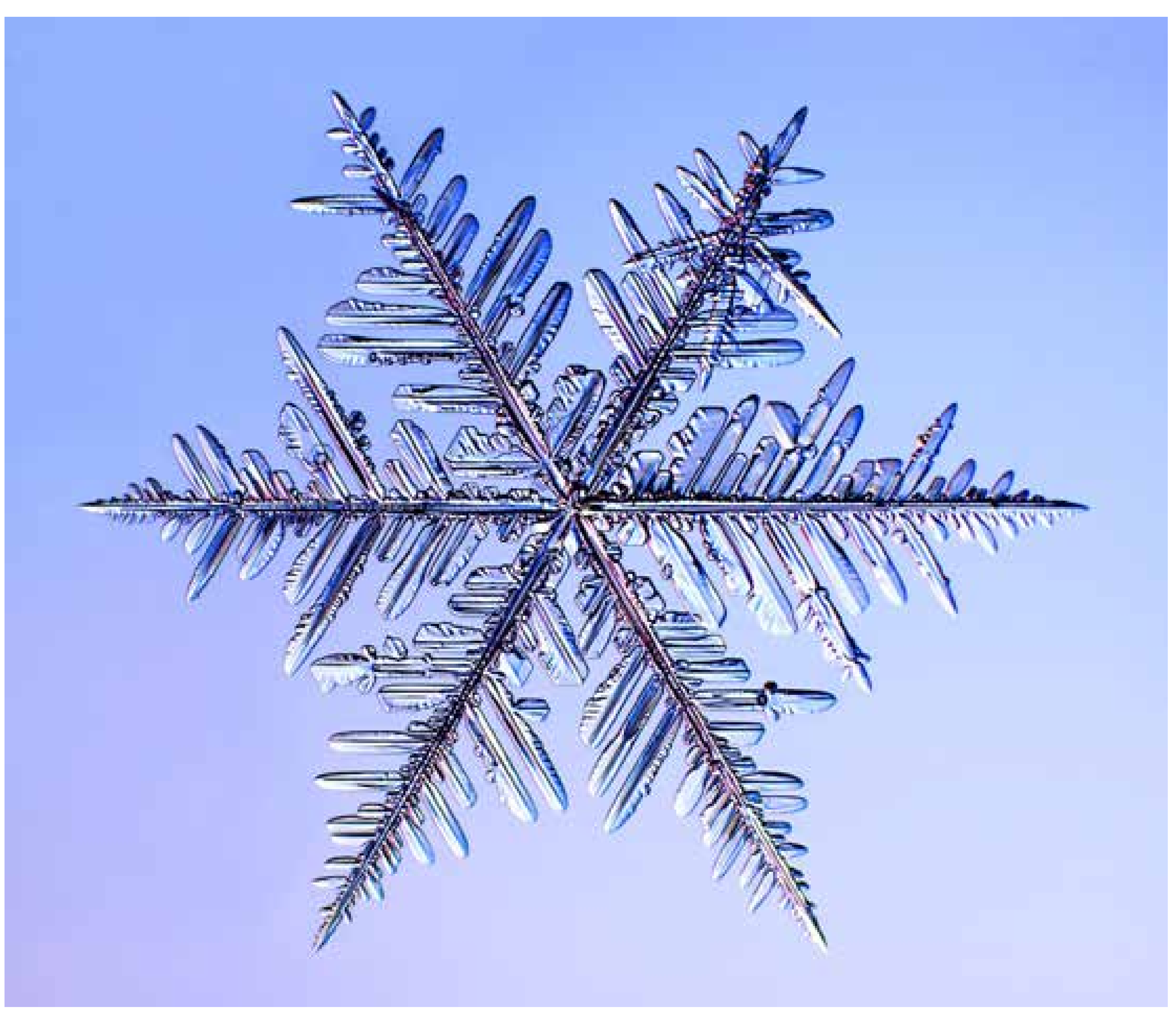

**Chaotic Branching.** The shape of this snow crystal reflects the humid environment in which it grew. As soon as it was born, the abundance of water vapor drove the branching instability hard, so the transition from faceted to branched growth occurred early. As a result, at its center there is little visible remnant of the crystal's initial faceted stage. Once the six principal arms were established, the high humidity resulted in narrow, closely spaced sidebranches with no prism faceting. The absence of faceting meant no induced sidebranching events, and thus no six-fold symmetry in the placement of the sidebranches. Even the sidebranches on opposite sides of a single primary branch are uncorrelated. In a sense, the growth of this crystal was too fast to be orchestrated. This is a medium-sized dendritic specimen, just over two millimeters from tip to tip, but it is also quite thin and flat. Basal faceting, with some assistance from the edge-sharpening instability, mainly restricted its growth to two dimensions. Because it stayed thin and light, the crystal made a slow descent through the clouds, never falling faster than about half a meter per second.

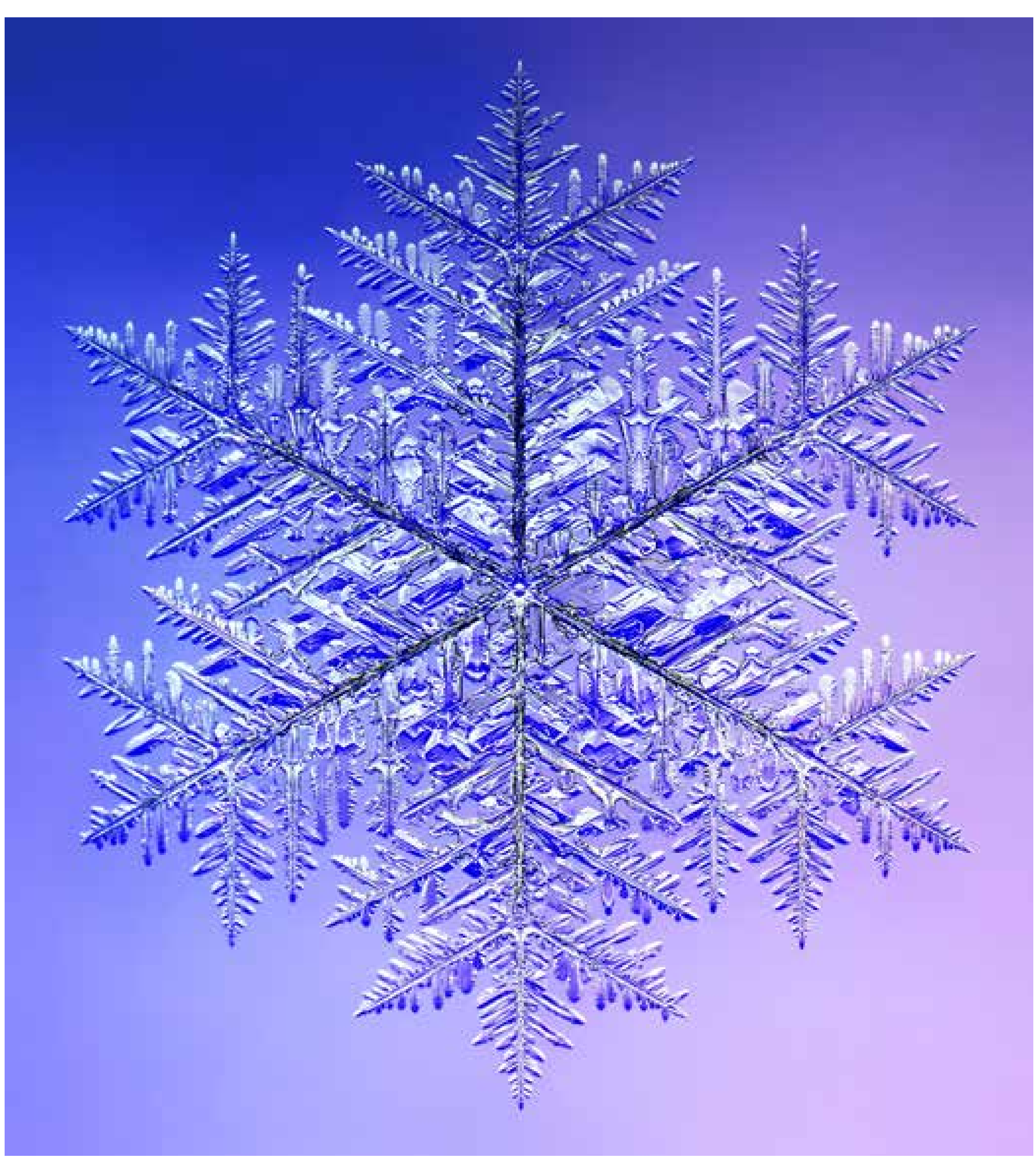

**Monster Snowflake.** To my knowledge, this is the largest snow crystal ever photographed – a fernlike stellar dendrite measuring just over 10 mm from tip to tip. I have witnessed such large crystals only twice, both times in Cochrane, Ontario, and both times for just a few short minutes. Each branch holds first-generation sidebranches along with second-, third-, and even fourth-generation sidebranches. Extensive higher-order sidebranching like this is rare in snow crystals.

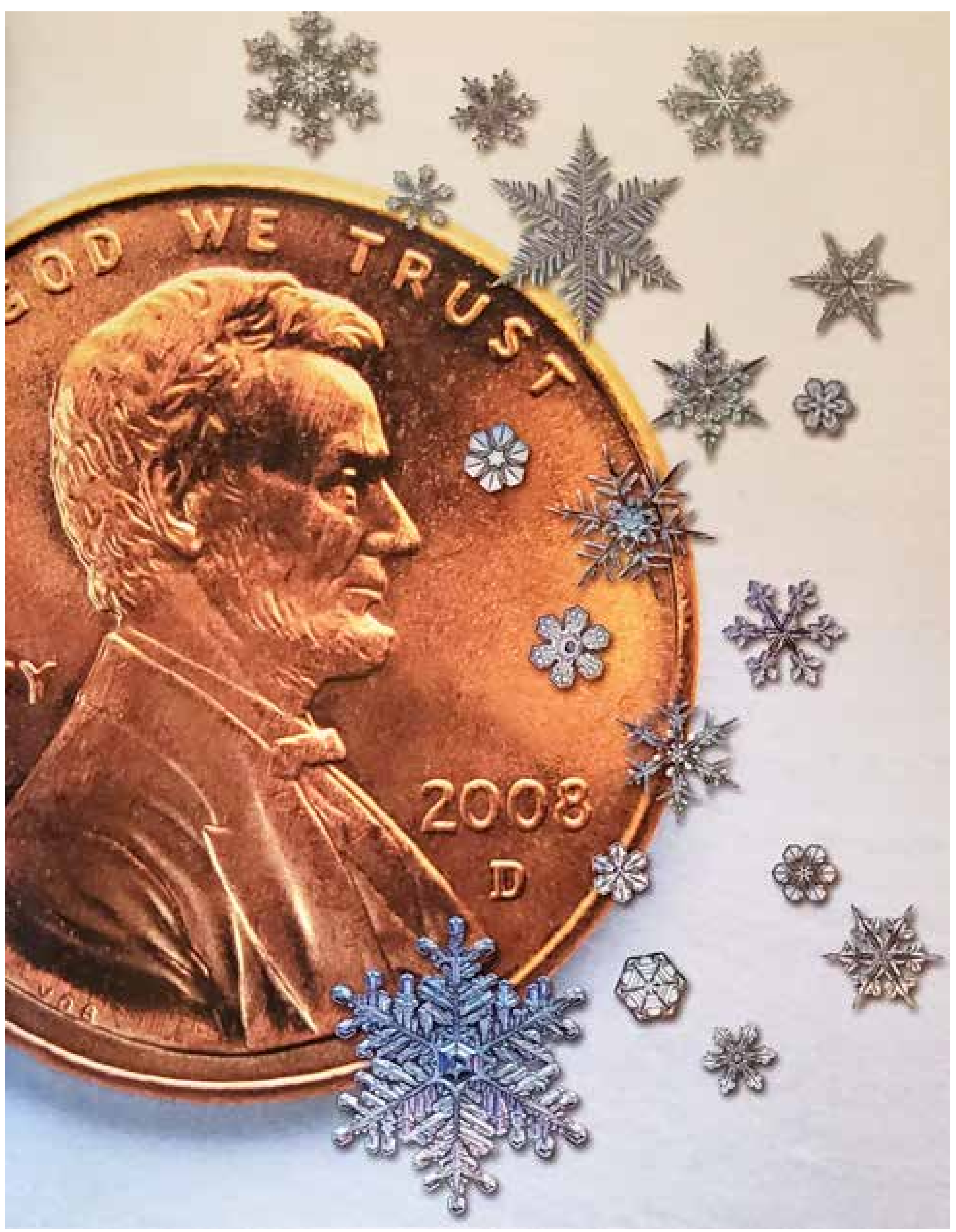

**Snowflake Sizes.** This true-to-size composite image shows several snow crystals next to a penny (19 mm in diameter). The monster snowflake on the preceding page is about as large as Lincoln's head.

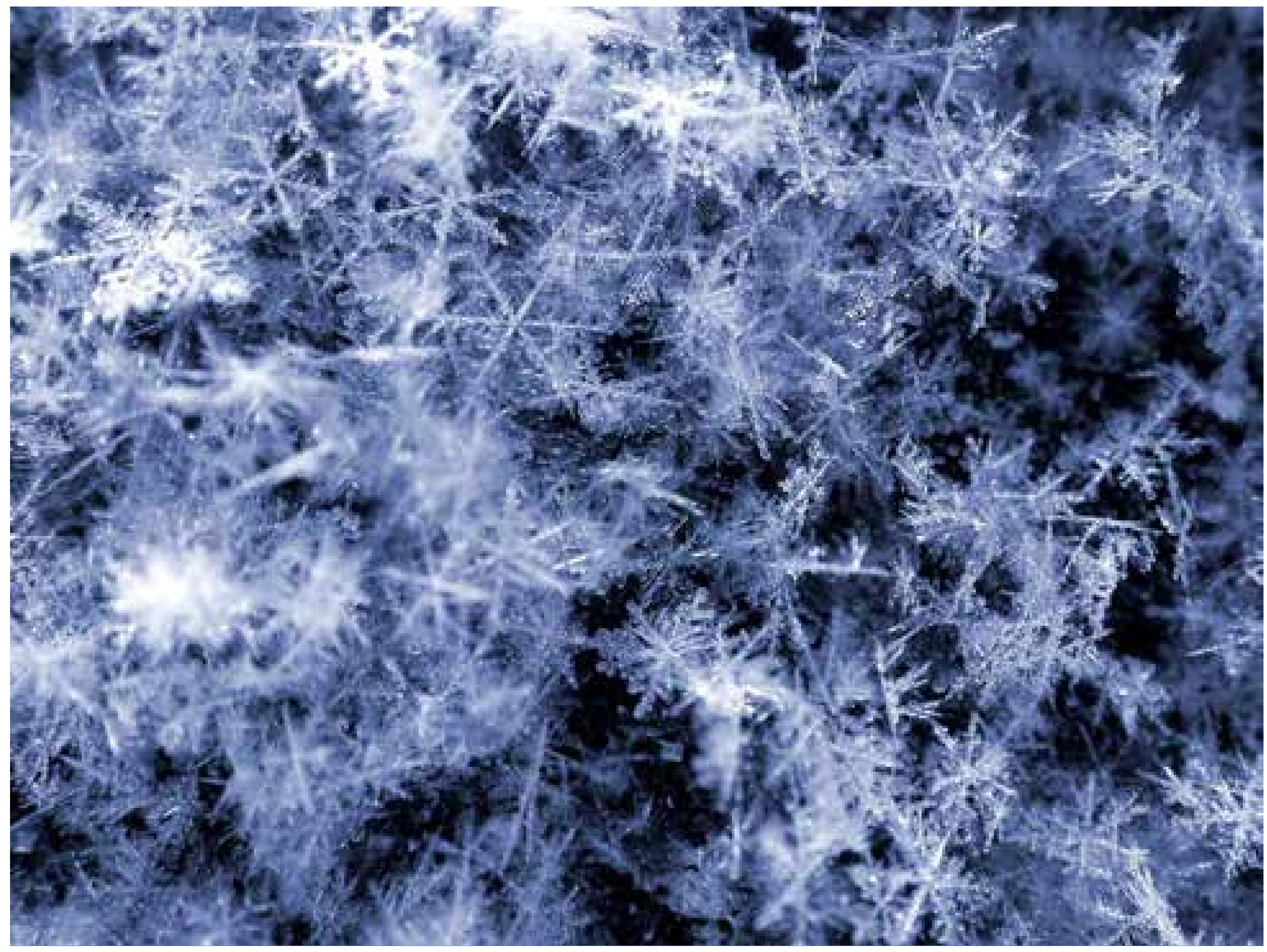

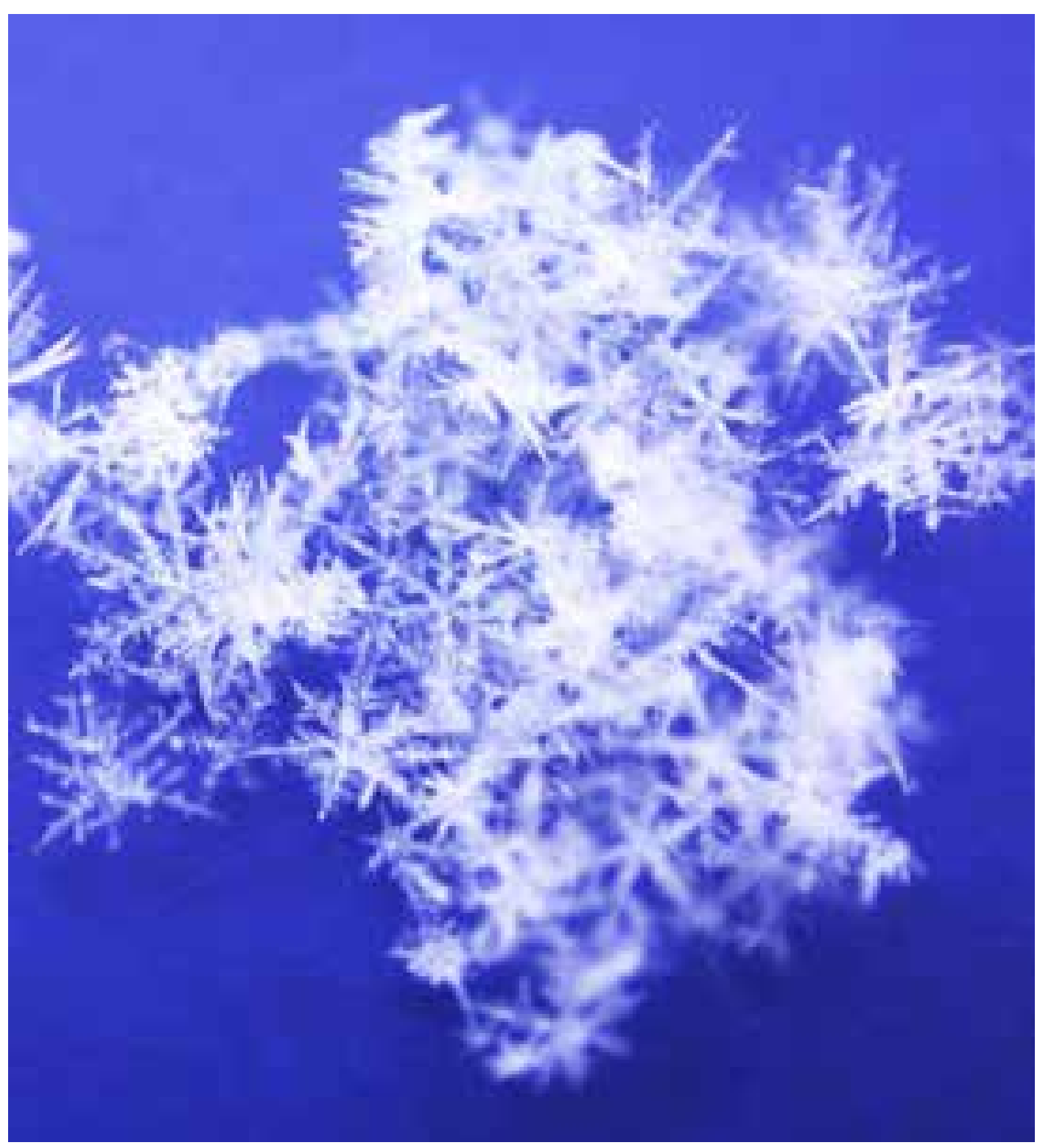

**Powder Snow.** When conditions are right for the formation of fernlike stellar dendrites, they can fall in abundance. The picture above shows a close-up view of the windshield of my car after a snowfall that dropped almost entirely large stellar snowflakes. You can see how the barbed branches locked together to form an exceptionally light, fluffy blanket of ice. On the ground, this kind of snow is called fresh powder, and the airy structure is so soft that a skier might sink waist-deep into it, skies and all. After being exposed to the sun and wind for a day or two, however, the snow packs down into a denser, less yielding composition.

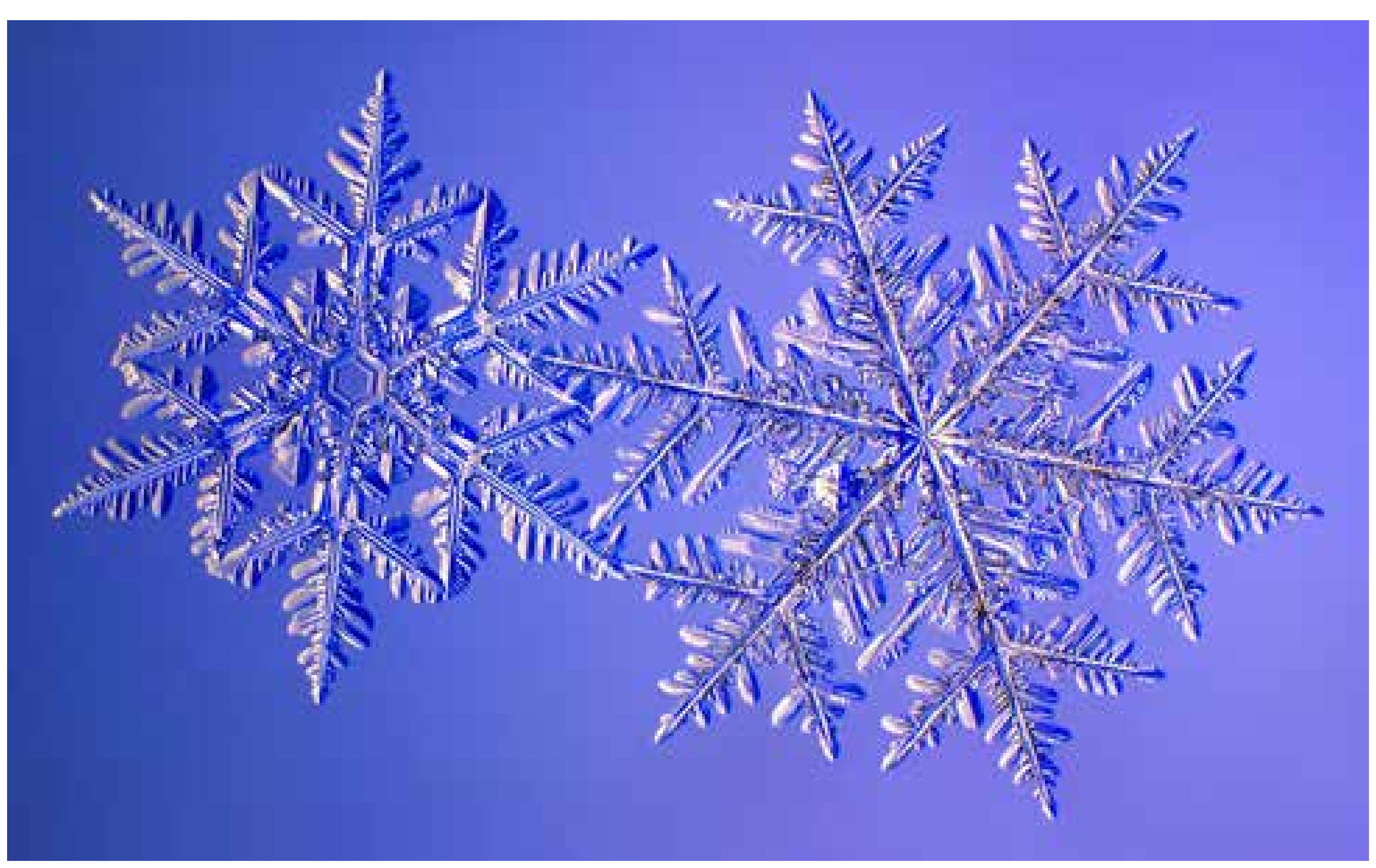

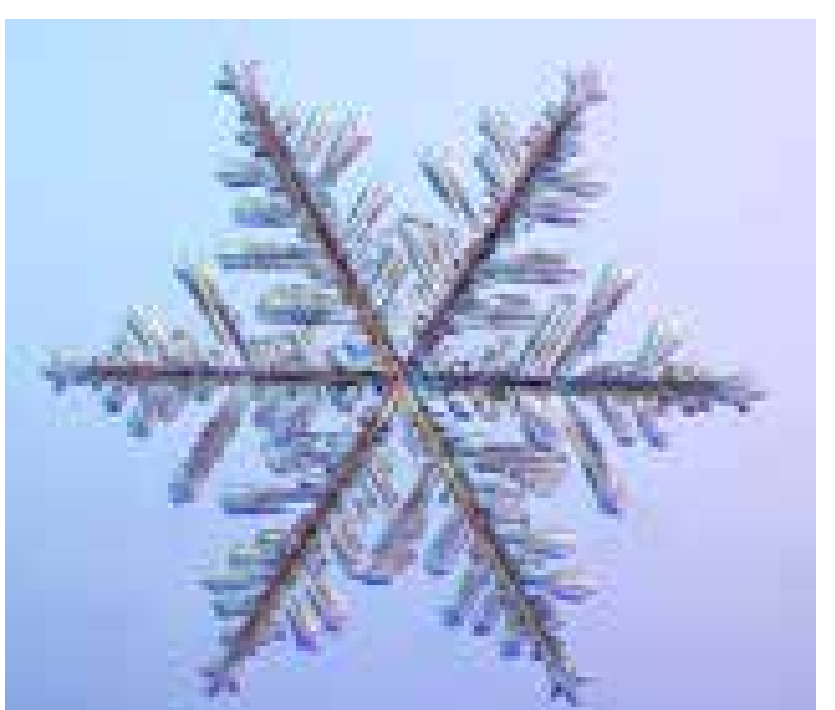

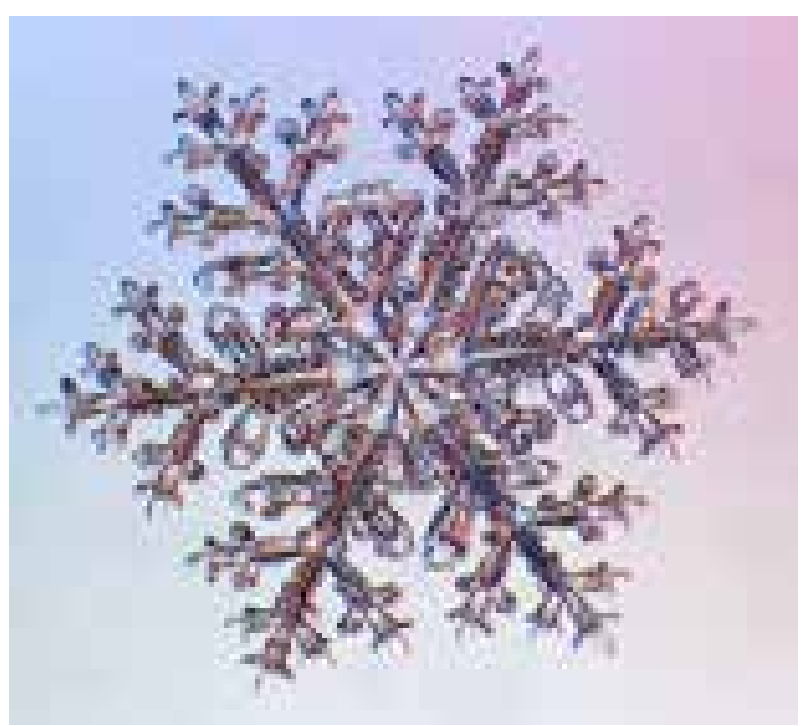

**Variations.** Once again, there is no sharp dividing line between the stellar-dendrite and fernlike-stellar-dendrite categories. Many crystals display aspects of both types, so they rightfully belong somewhere between these two classifications.

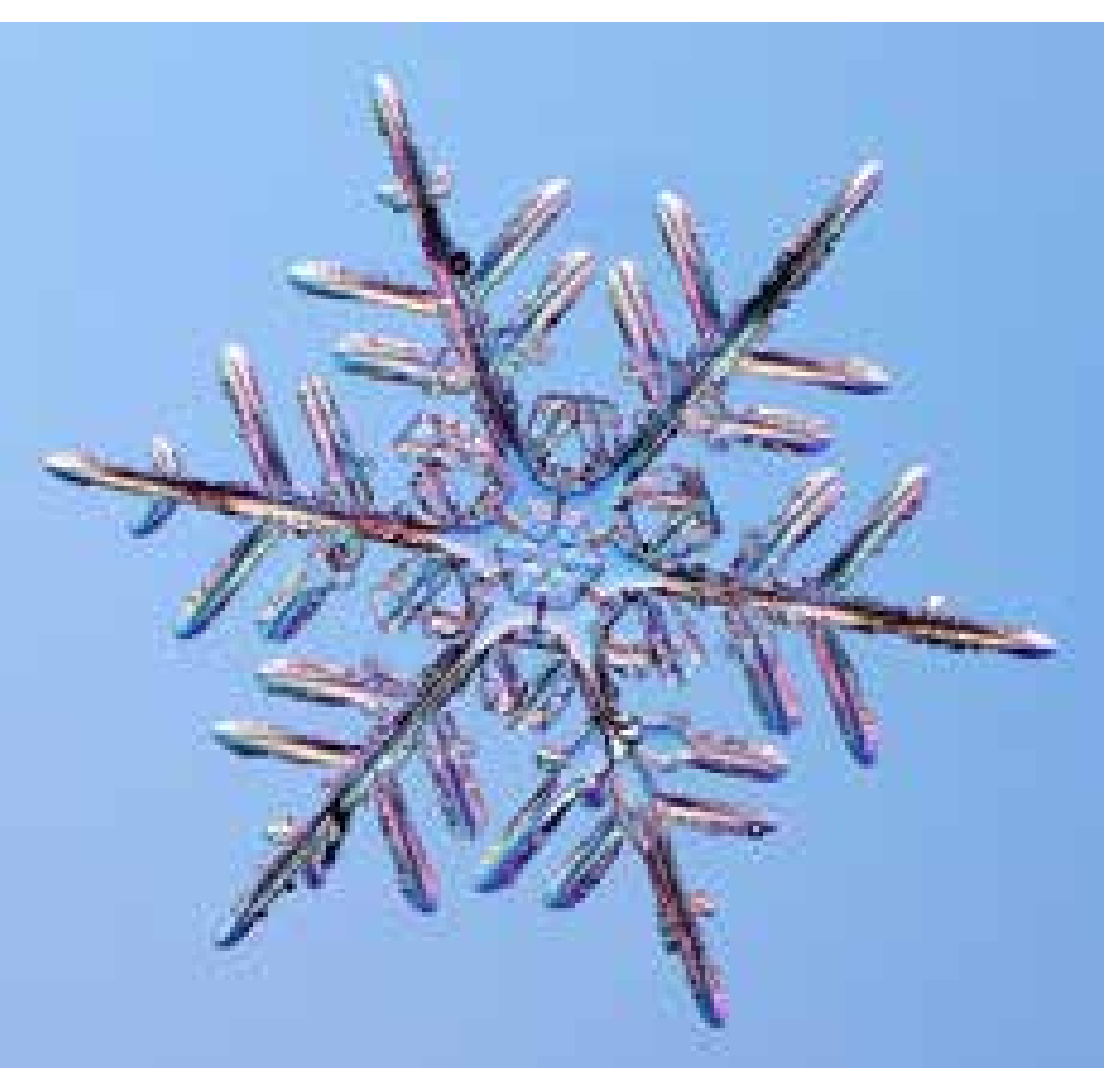

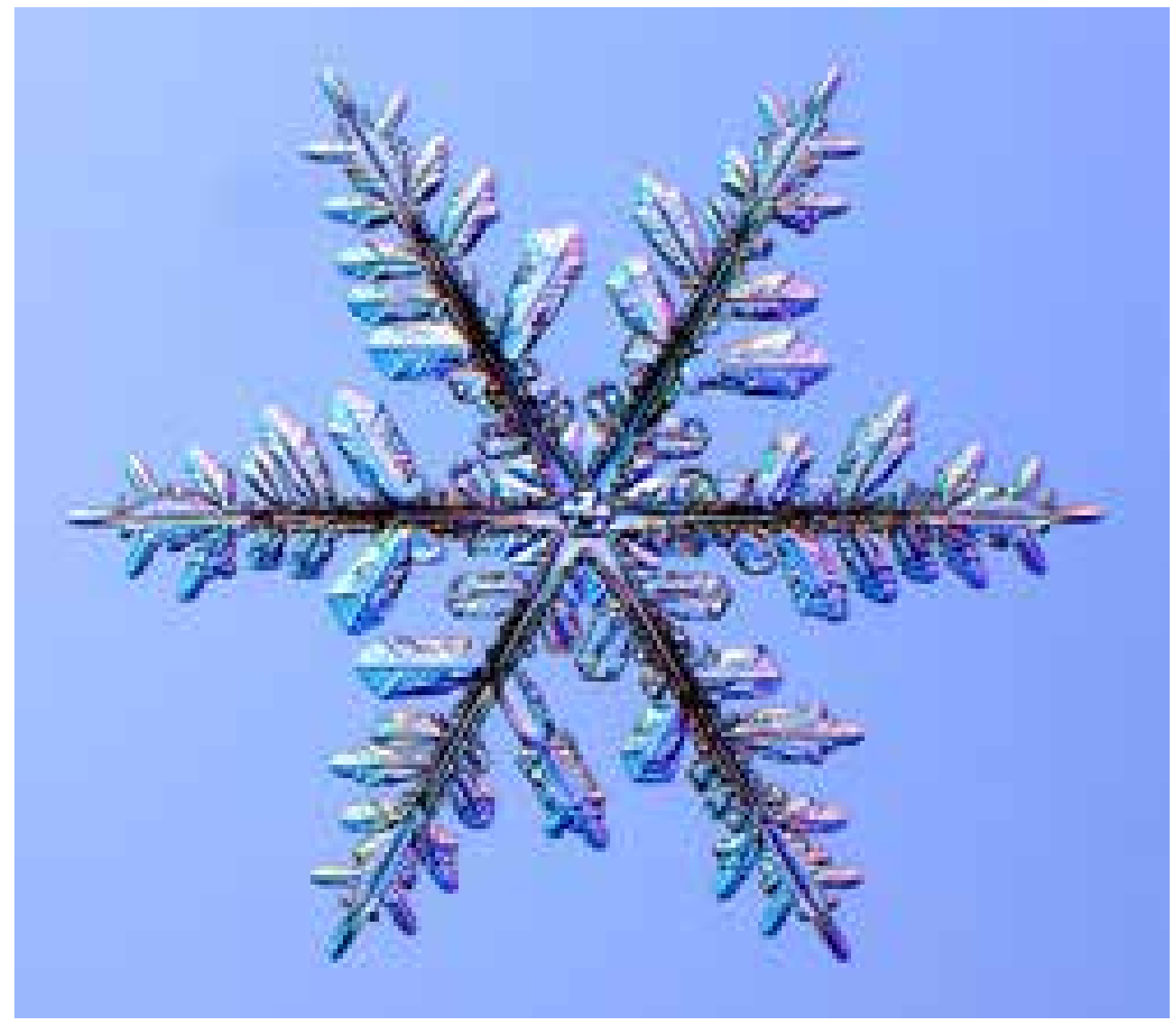

## Hollow Columns

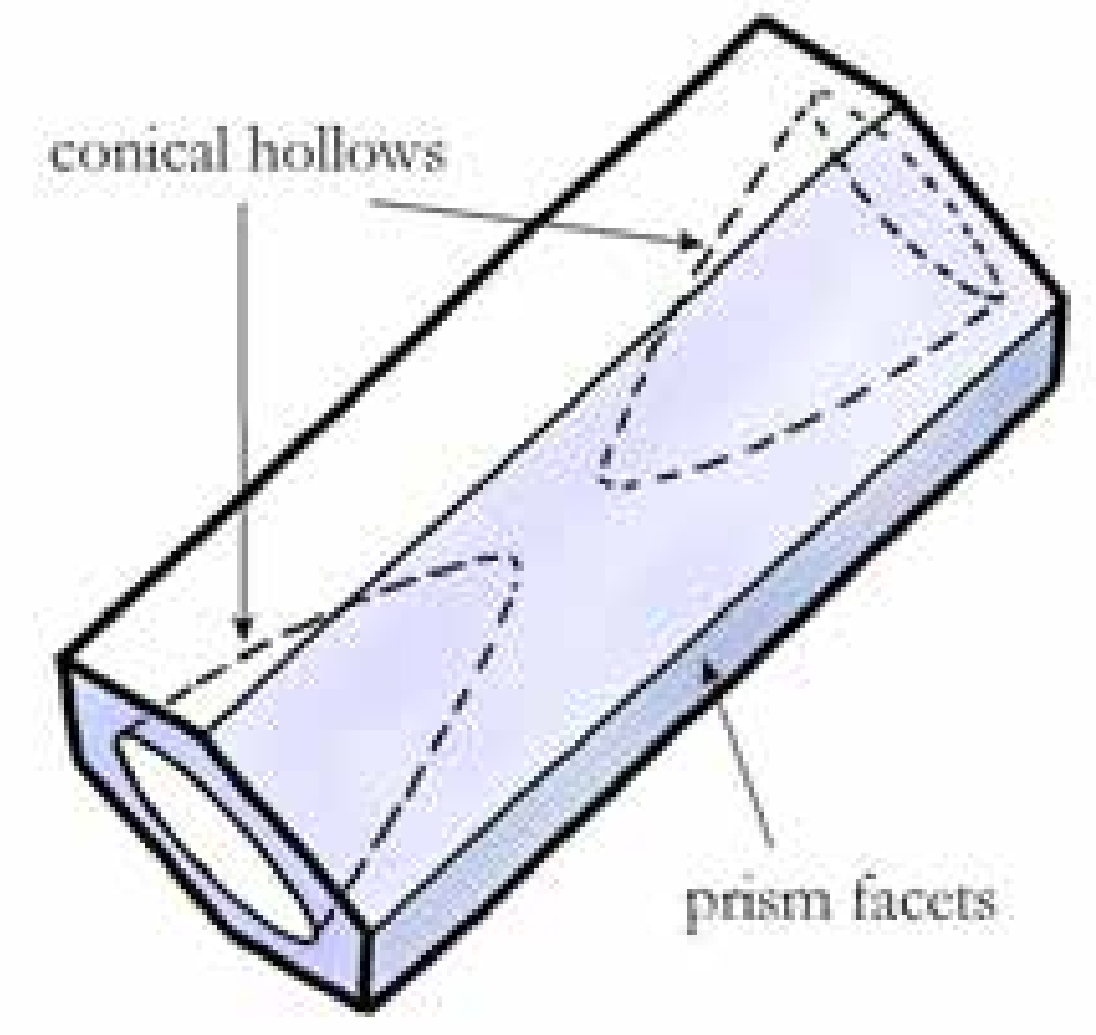

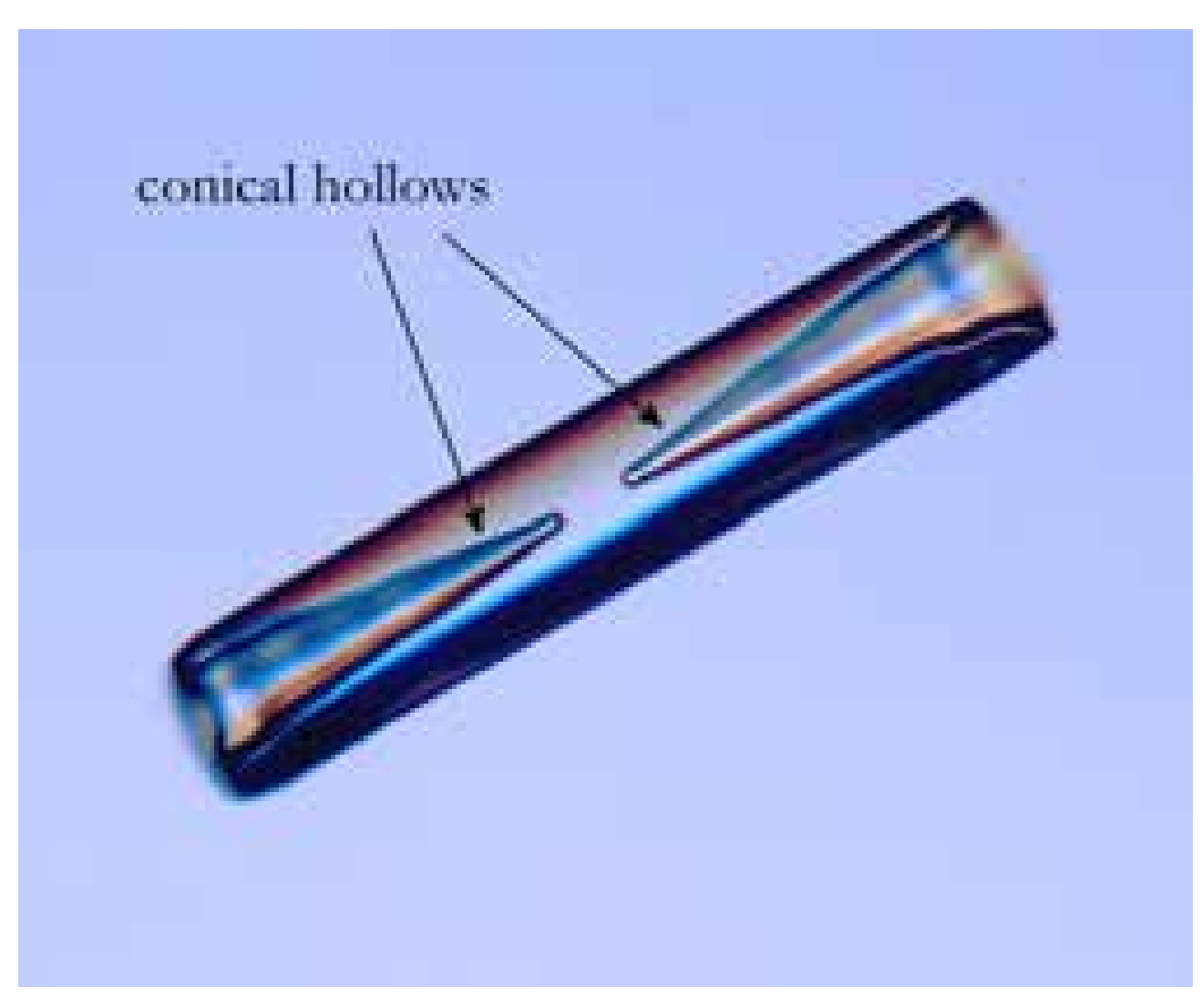

*Hollow columns are simple hexagonal ice prisms with conical voids extending down from their ends. The recesses typically appear in a symmetric pair running along the central axis of a crystal, with the tips nearly touching at the waist. Hollow columns are small and easy to overlook with the naked eye, and their internal structure is best viewed with a microscope. They are a relatively common columnar morphology and can frequently be found in warmer snowfalls.*

Hollow columns are most likely found when the temperature is near -5 C, as indicated in the morphology diagram. The overall hexagonal columnar structure is often not apparent because the prism corners have been rounded by sublimation, which is especially rapid at these warmer temperatures. Thus hollow columns may look more like round cylinders than hexagonal columns.

The crystals pictured here are fairly small, close to a millimeter in length; the one above is slightly shorter and the one below is slightly longer. The best way to find and view these crystals is to let some snow fall onto several glass slides and then view the slides under a microscope. When the temperature is high and granular snow is the norm, one can often find a few well-formed hollow columns in the mix. As is true with most snow-crystal types, finding well-formed examples can be a challenge.

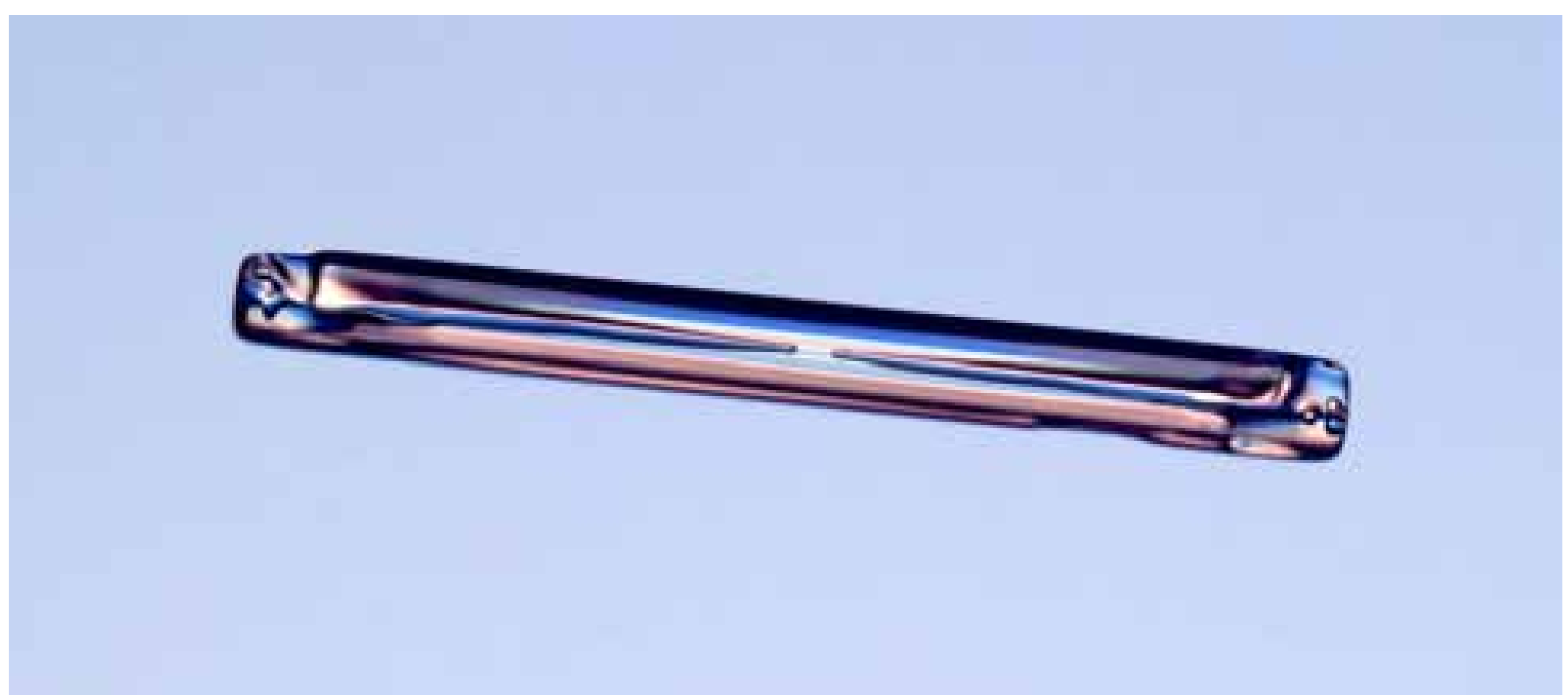

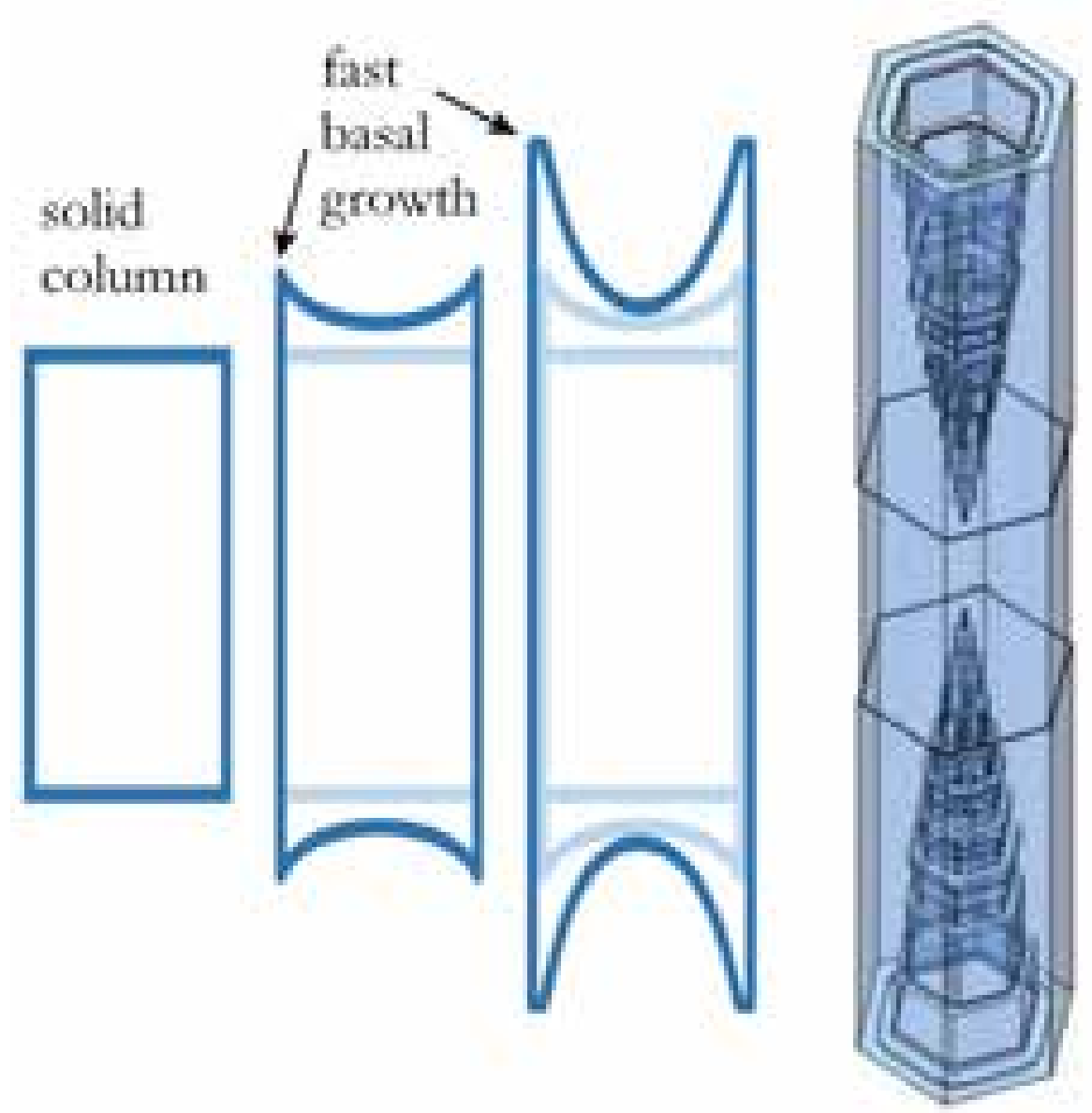

The formation of a hollow column is a manifestation of the familiar branching instability (see Chapter 3). The diagram above shows cross-sections of a hollow column at different times, together with a numerical model result [2009Gra]. Faceting initially yields a small solid column, then diffusion favors growth of the basal edges. This mechanism predicts that there can never be a fully hollow column (hollow like a pipe), and none has ever been observed. The initial seed crystal will always leave behind a solid central core.

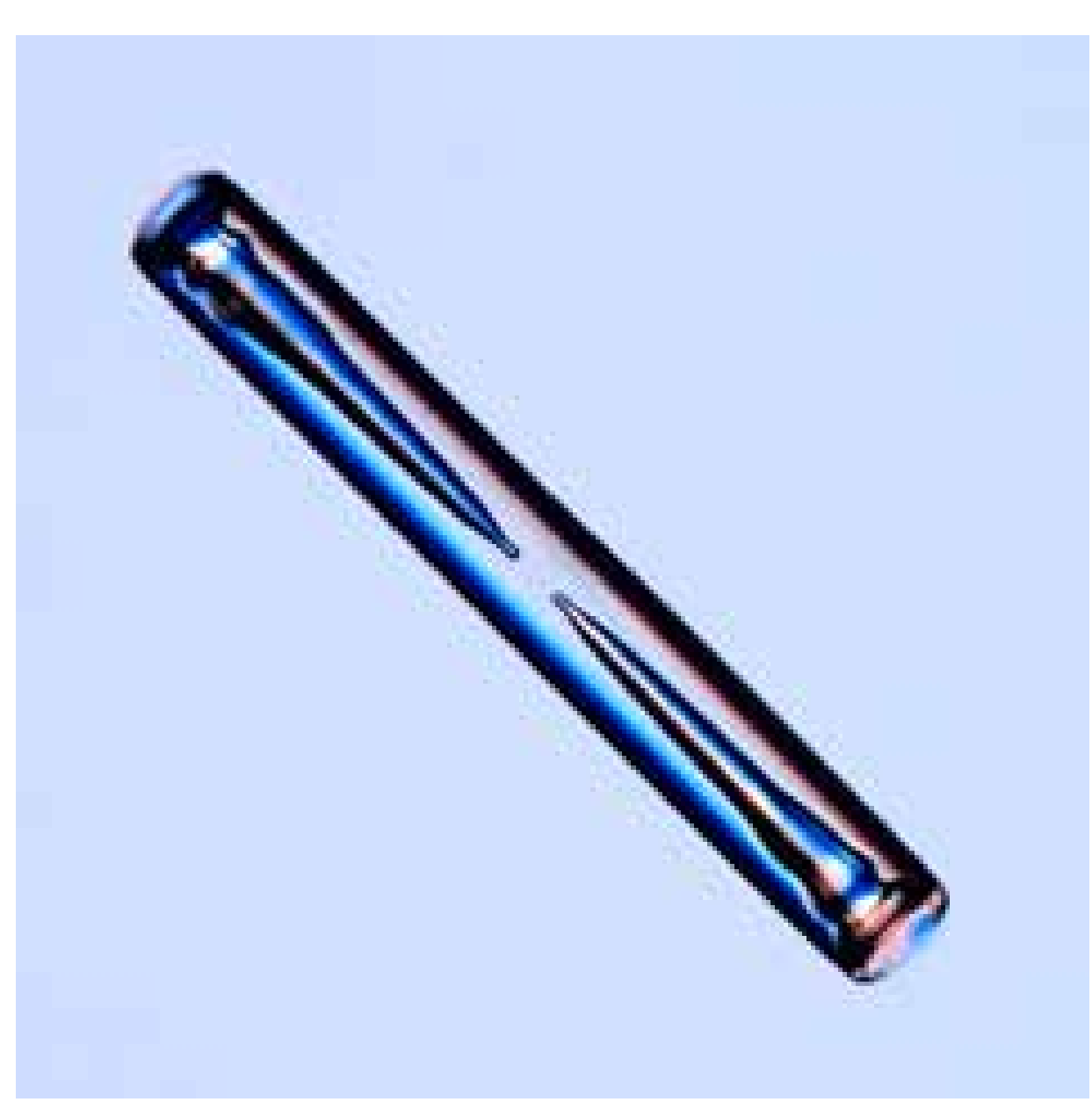

The hollow column shown at the lower left on this page exhibits nearly conical hollows near its center along with some distinctive hollow "wiggles" near the two ends of the column. These likely arose from changing conditions around the crystal during its development, which altered the growth of the hollows. Because both ends of the column experienced the same changes at the same times, the wiggles developed similarly on the two ends.

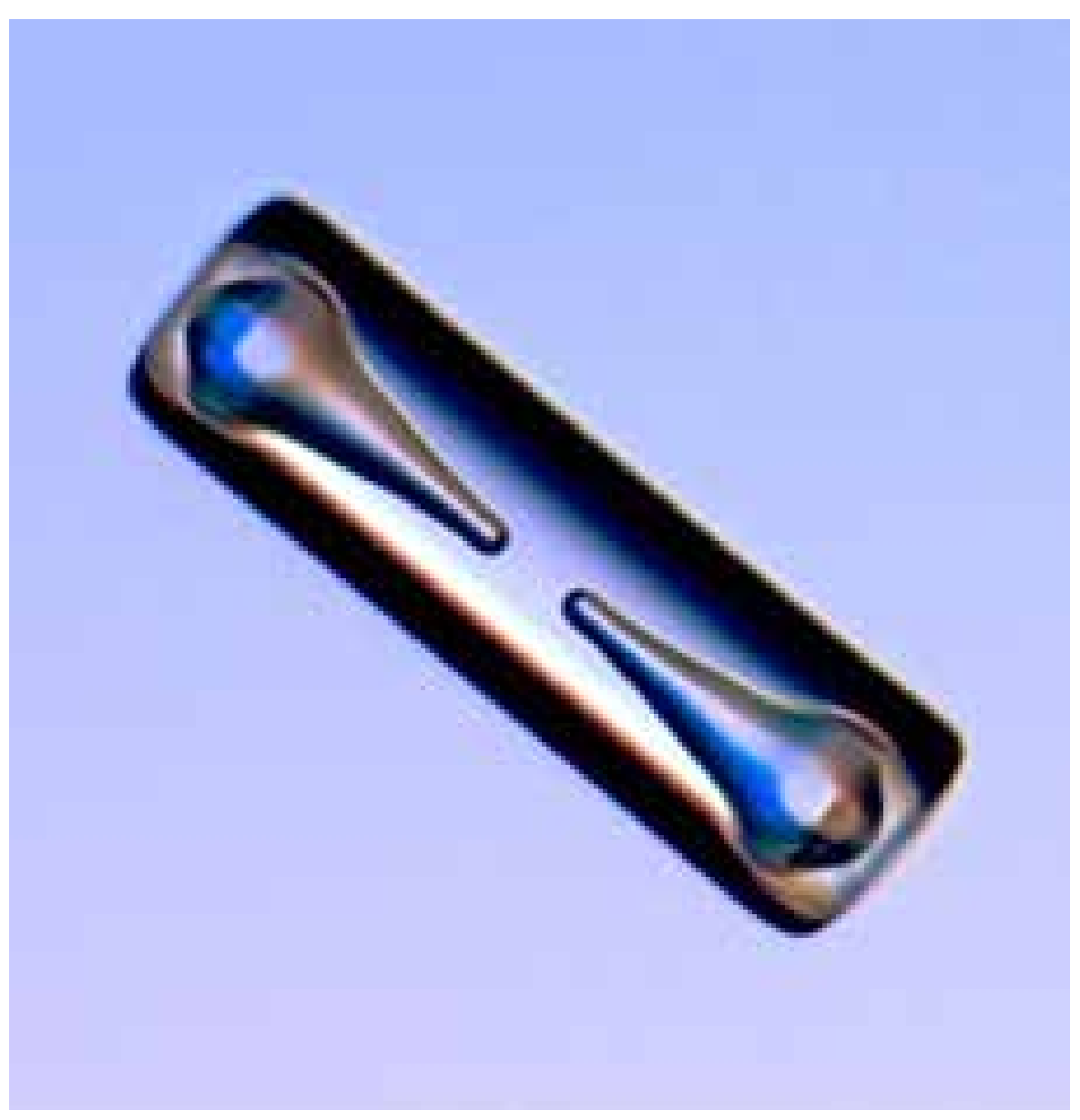

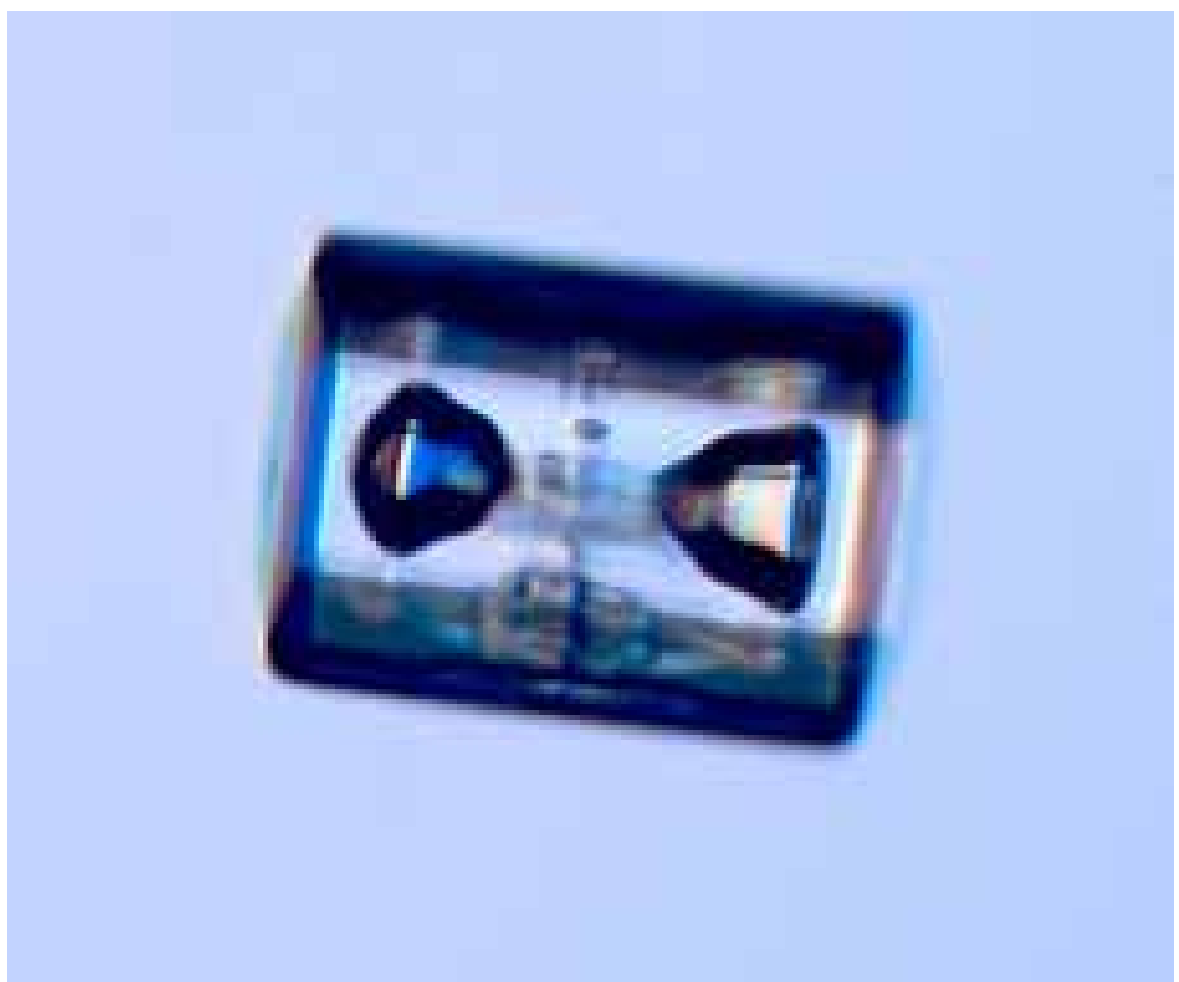

**Columnar Bubbles.** In some hollow columns, including the two examples above, the hollows close up when the supersaturation becomes low (see Chapter 3), leaving pairs of columnar bubbles in the ice.

# Needles

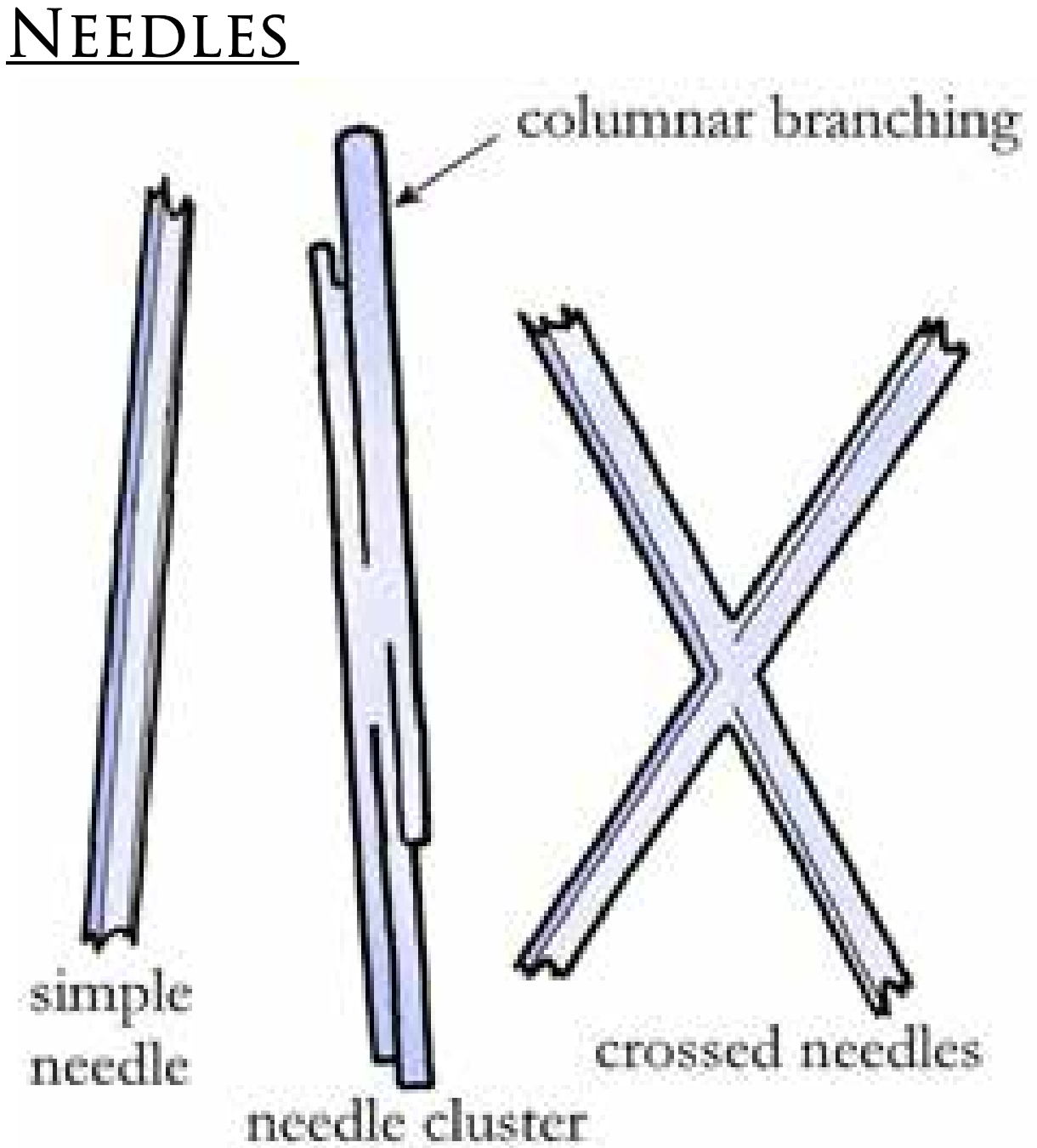

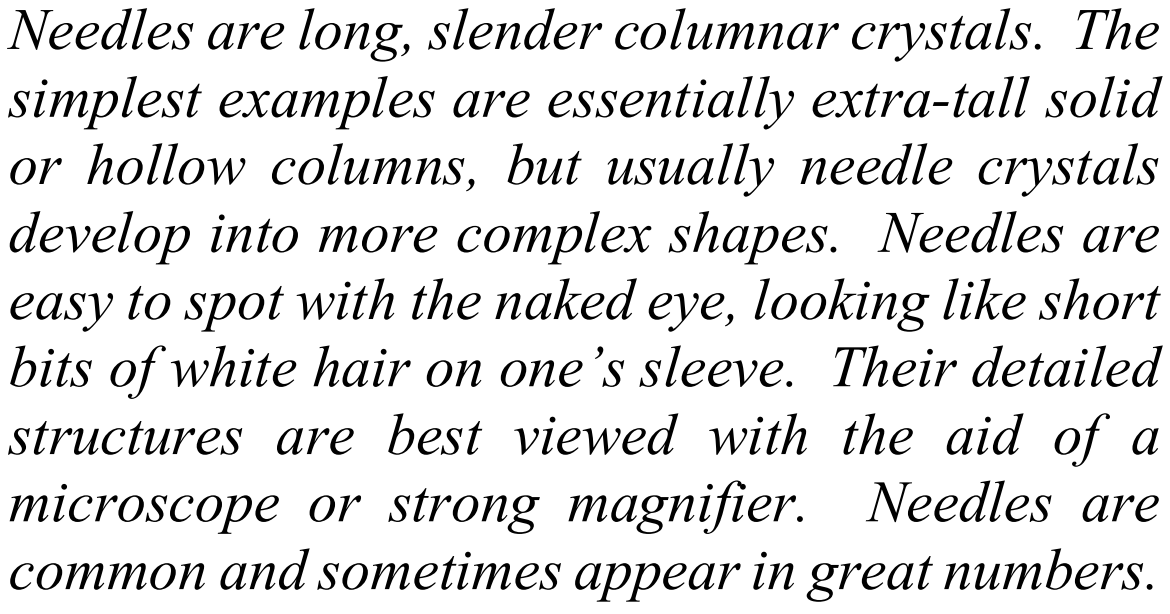

*Needles are long, slender columnar crystals. The simplest examples are essentially extra-tall solid or hollow columns, but usually needle crystals develop into more complex shapes. Needles are easy to spot with the naked eye, looking like short bits of white hair on one's sleeve. Their detailed structures are best viewed with the aid of a microscope or strong magnifier. Needles are common and sometimes appear in great numbers.*

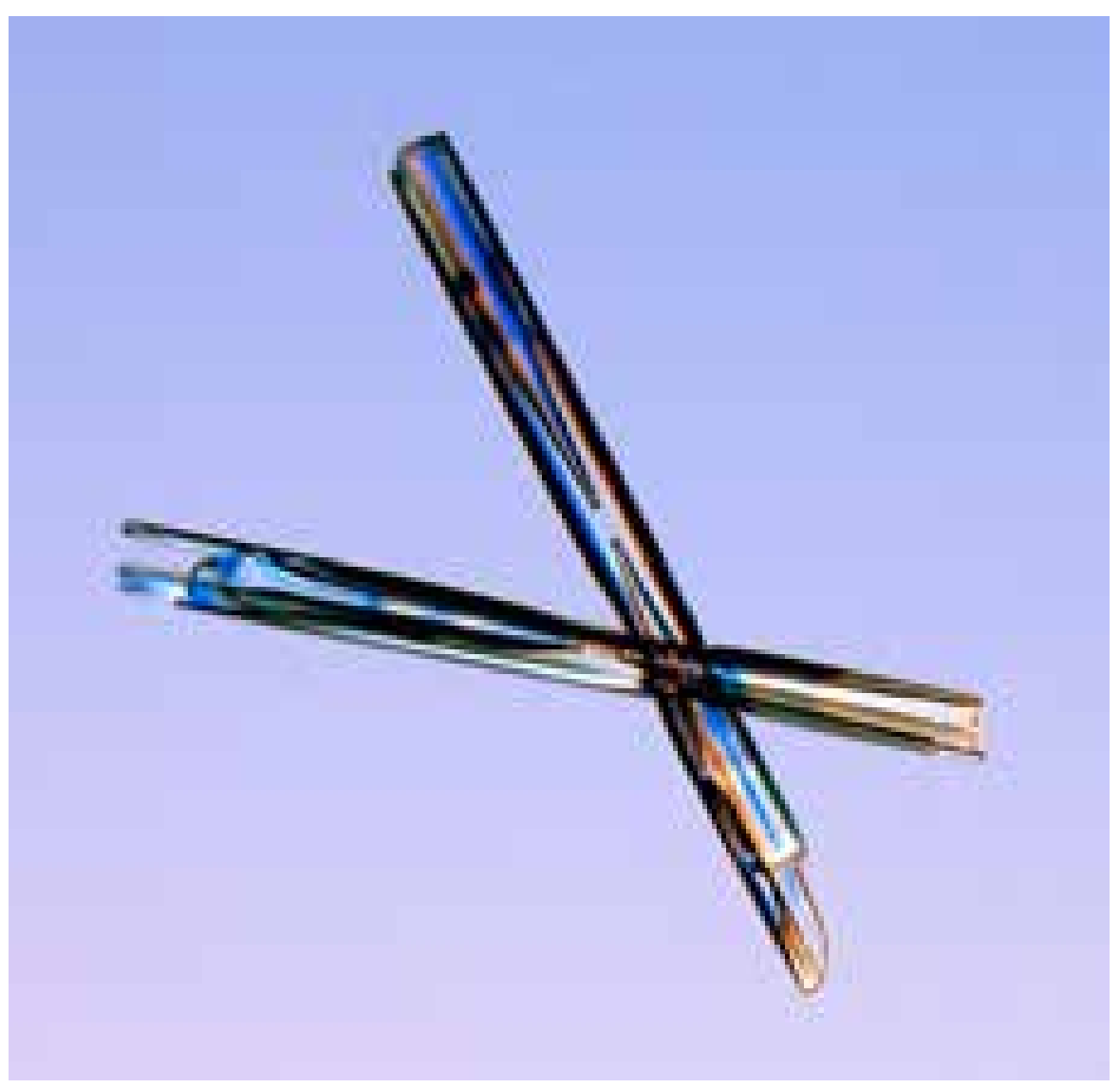

Needle crystals are the product of warm, wet snowfalls, forming when the temperature is close to -5 C and the humidity is high. With lengths often up to 3 mm, needles are the longest of the columnar snow crystals.

A crossed needle arises either from a polycrystalline seed crystal or from the mid-air collision of two simple needles.

Needle clusters are another result of the branching instability, as secondary needles sprout from the corners of a primary needle end. As with fernlike dendrites, fast growth yields complex structures.

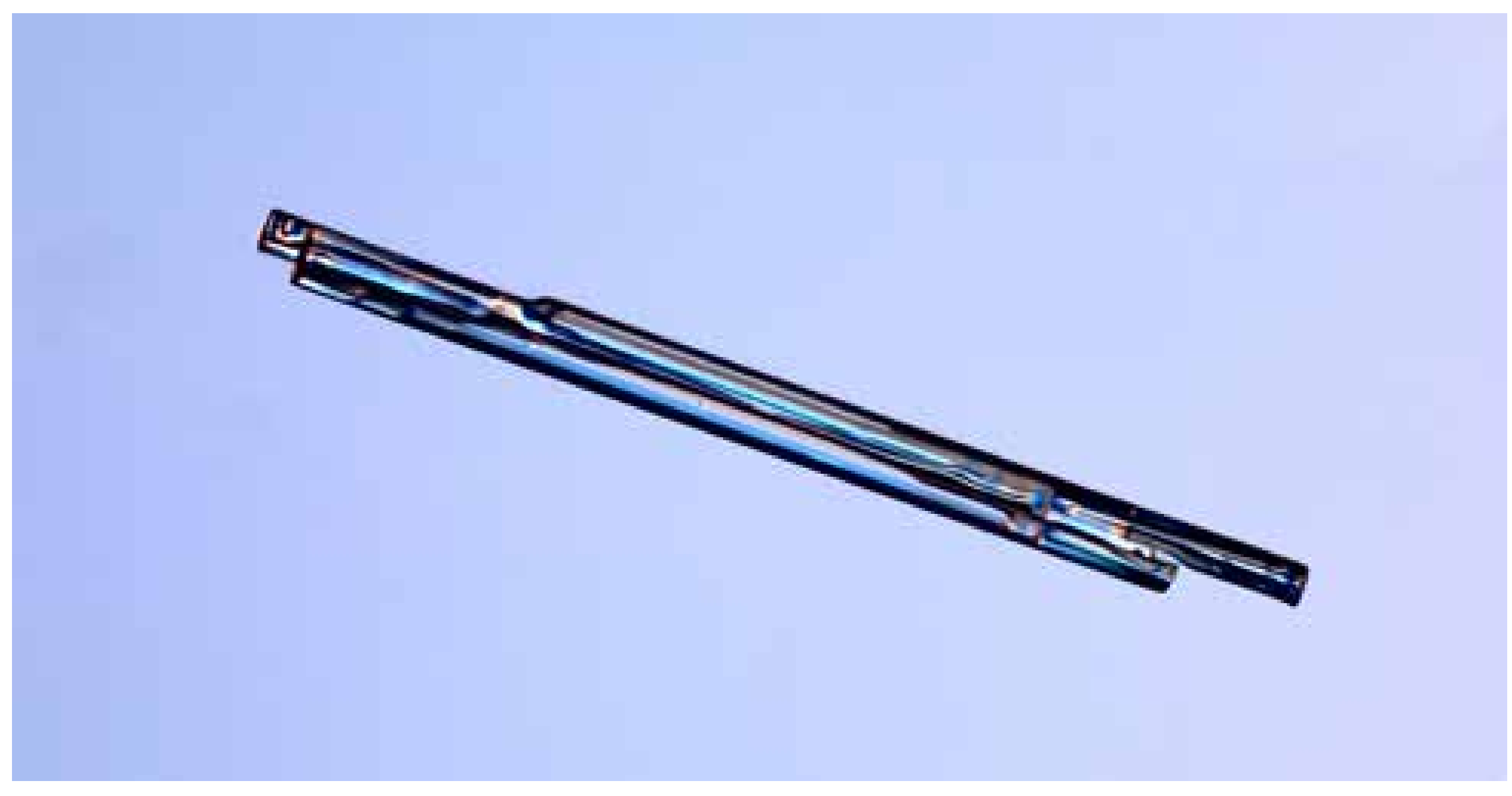

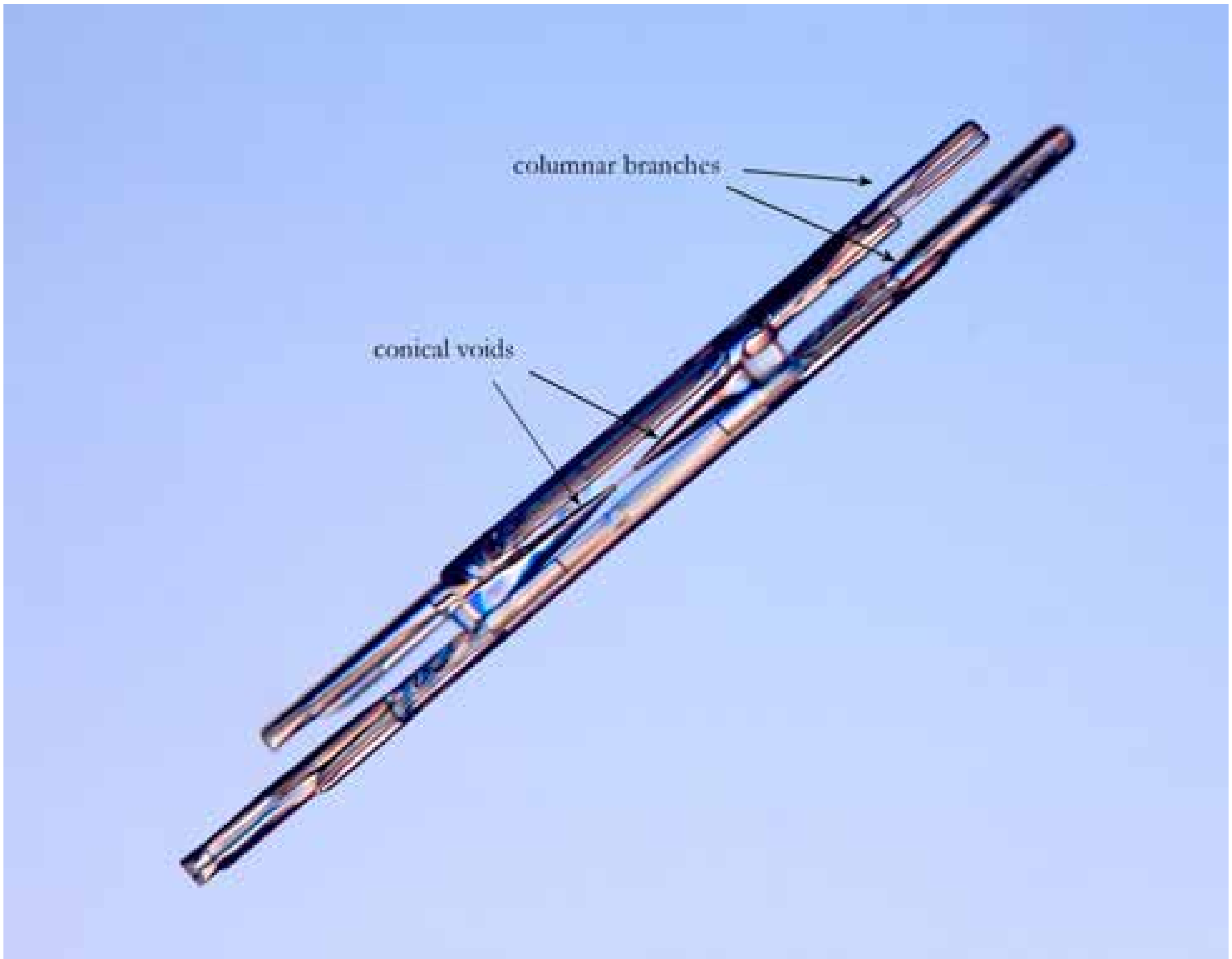

**Hollow column with needle extensions.** The complex needle crystal above began as a hollow column, as evidenced by the conical voids seen deep inside the structure. As the hollow column grew larger, needle-like branches sprouted from the corners of the columnar ends.

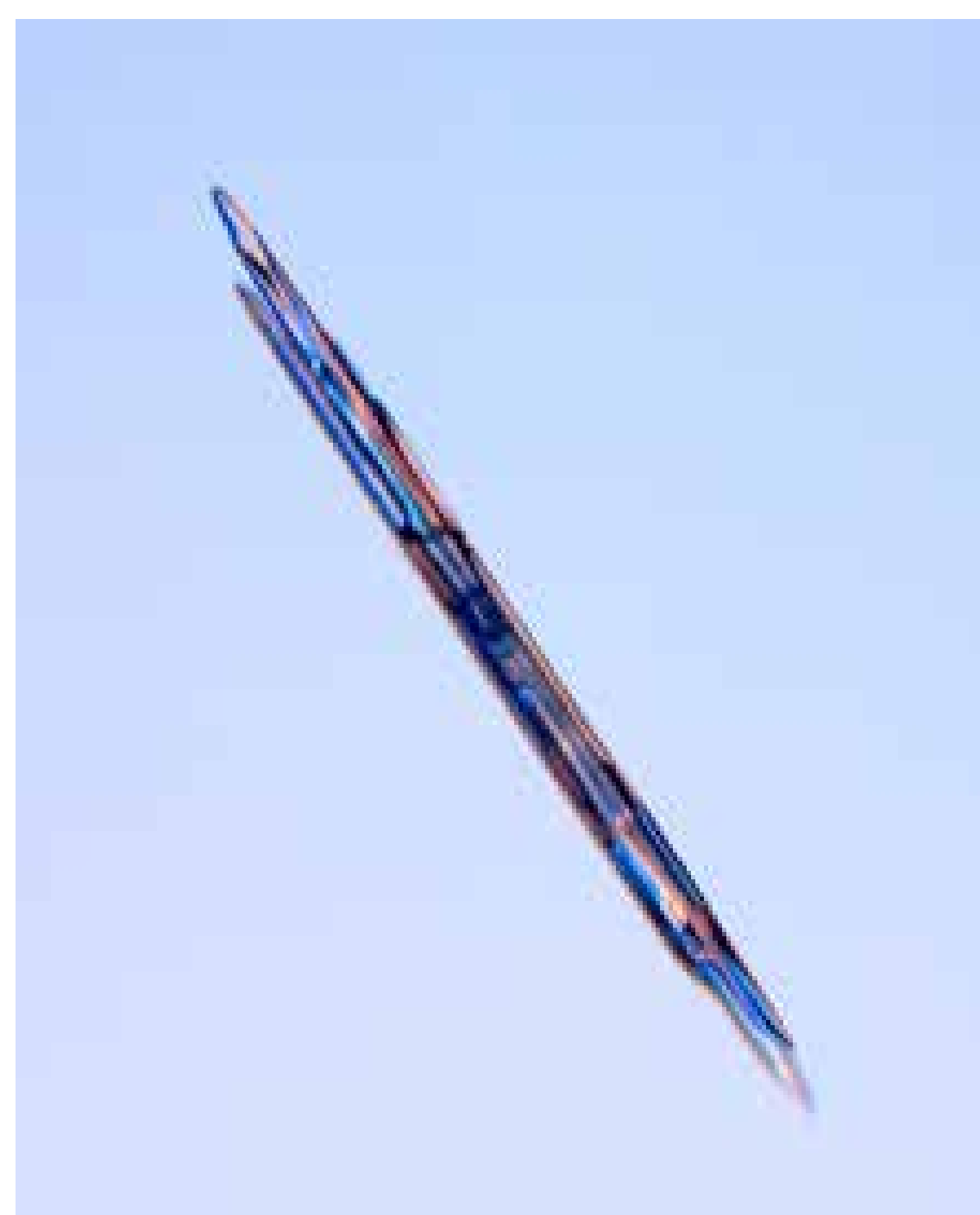

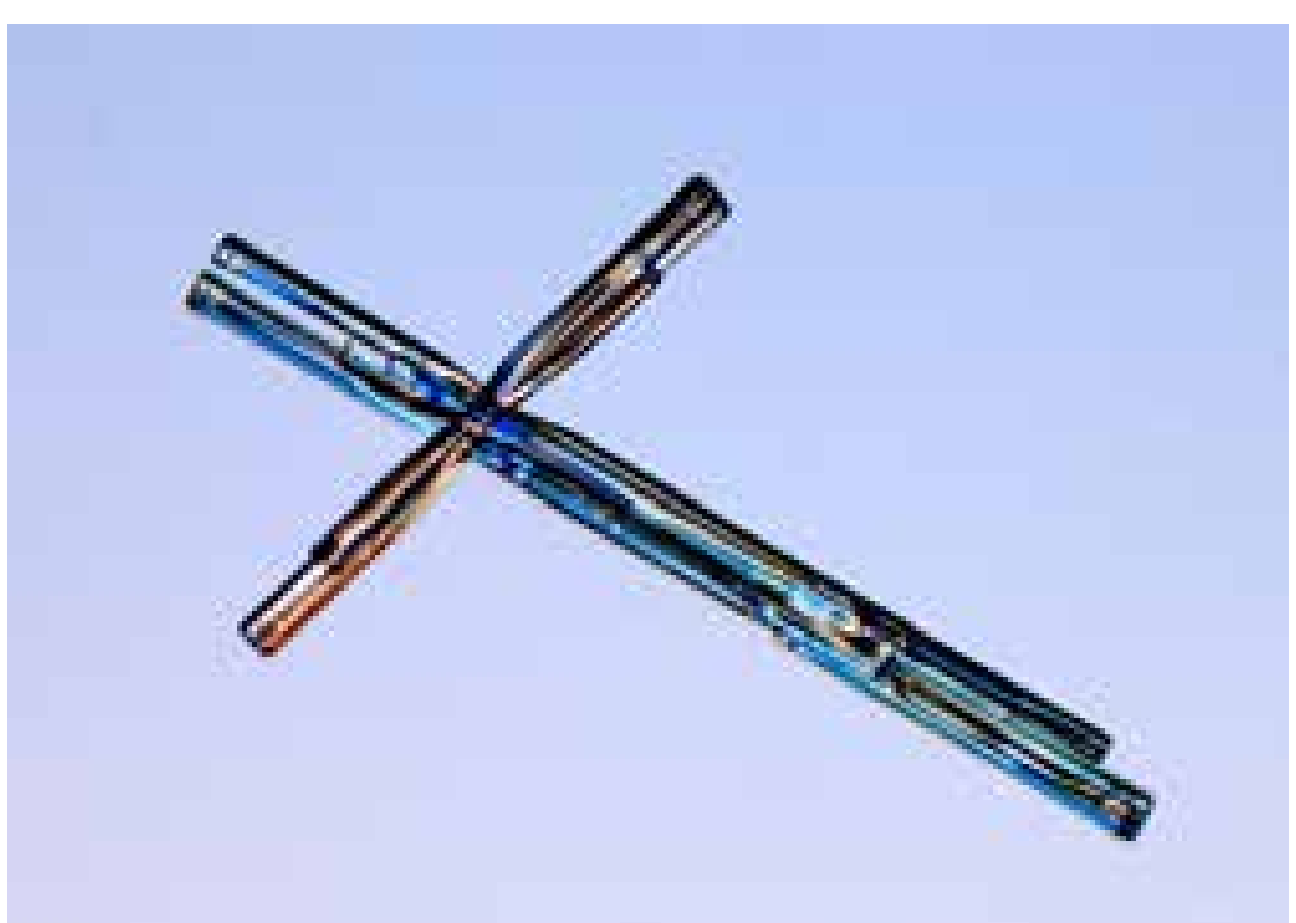

# Capped Columns

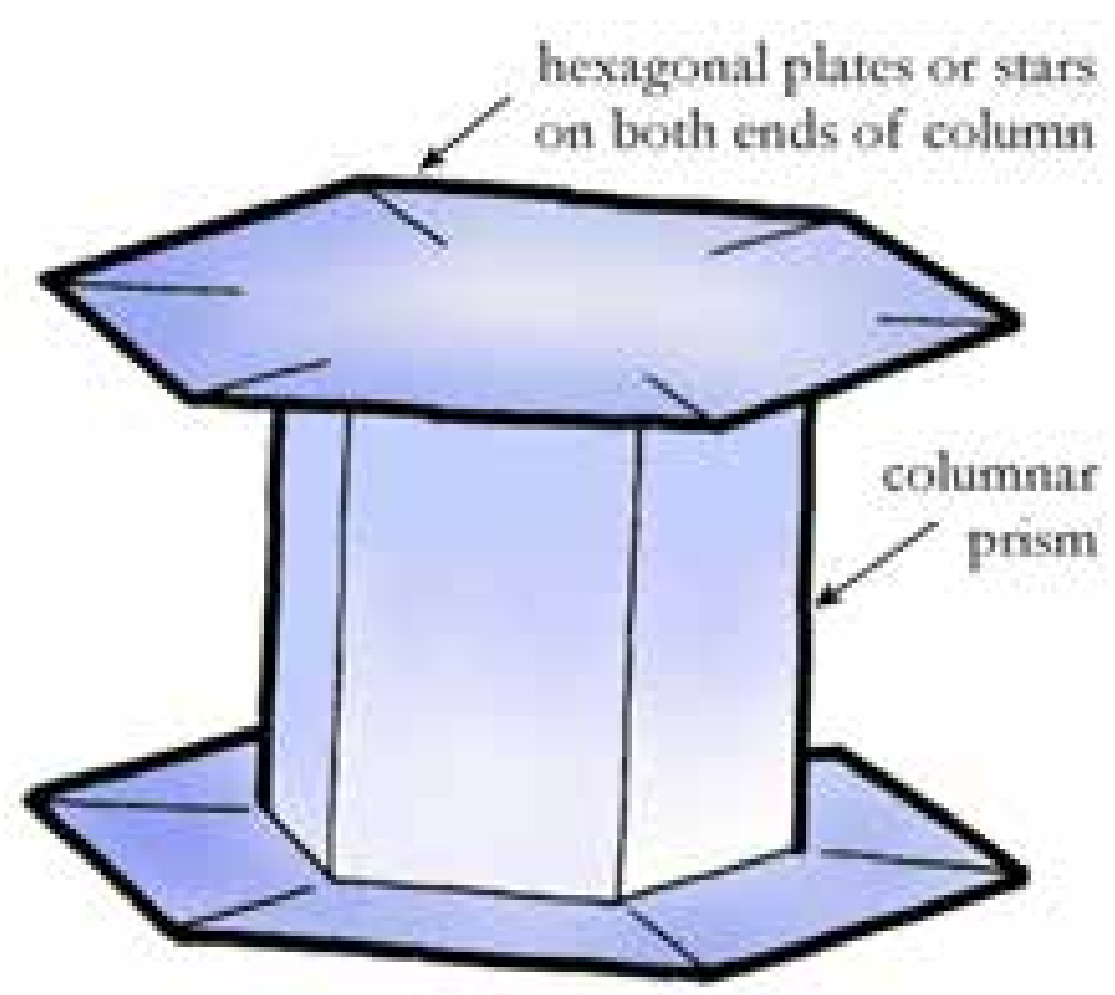

*Capped columns are columnar crystals with stellar plates on their ends. A typical specimen looks like a stubby axle flanked by two hexagonal wheels. Although these crystals are not especially common, a trained eye can often find a few mixed in with simple columnar crystals in warmer snowfalls. Capped columns are just large enough to be spotted with the naked eye, and their distinctive shape makes them easy to identify.*

A capped column forms when a snow crystal experiences its own style of midlife crisis, abruptly changing its growth behavior from columnar to plate-like. This can happen when a large mass of moist air is pushed upward by a passing storm front. The air cools as it rises, carrying its suspended cloud droplets along with it. When the temperature falls to around -6 C, some of the droplets freeze and begin growing into columns. If the air continues to rise, the temperature may drop to around -15 C, promoting plate-like growth on the columnar ends, yielding capped columns.

A common feature of capped columns is that the transition from columnar to plate-like growth is usually quite abrupt, owing to the edge-sharpening instability (see Chapter 4). This same physical effect allows the formation of Plate-on-Pedestal crystals described in Chapter 9.

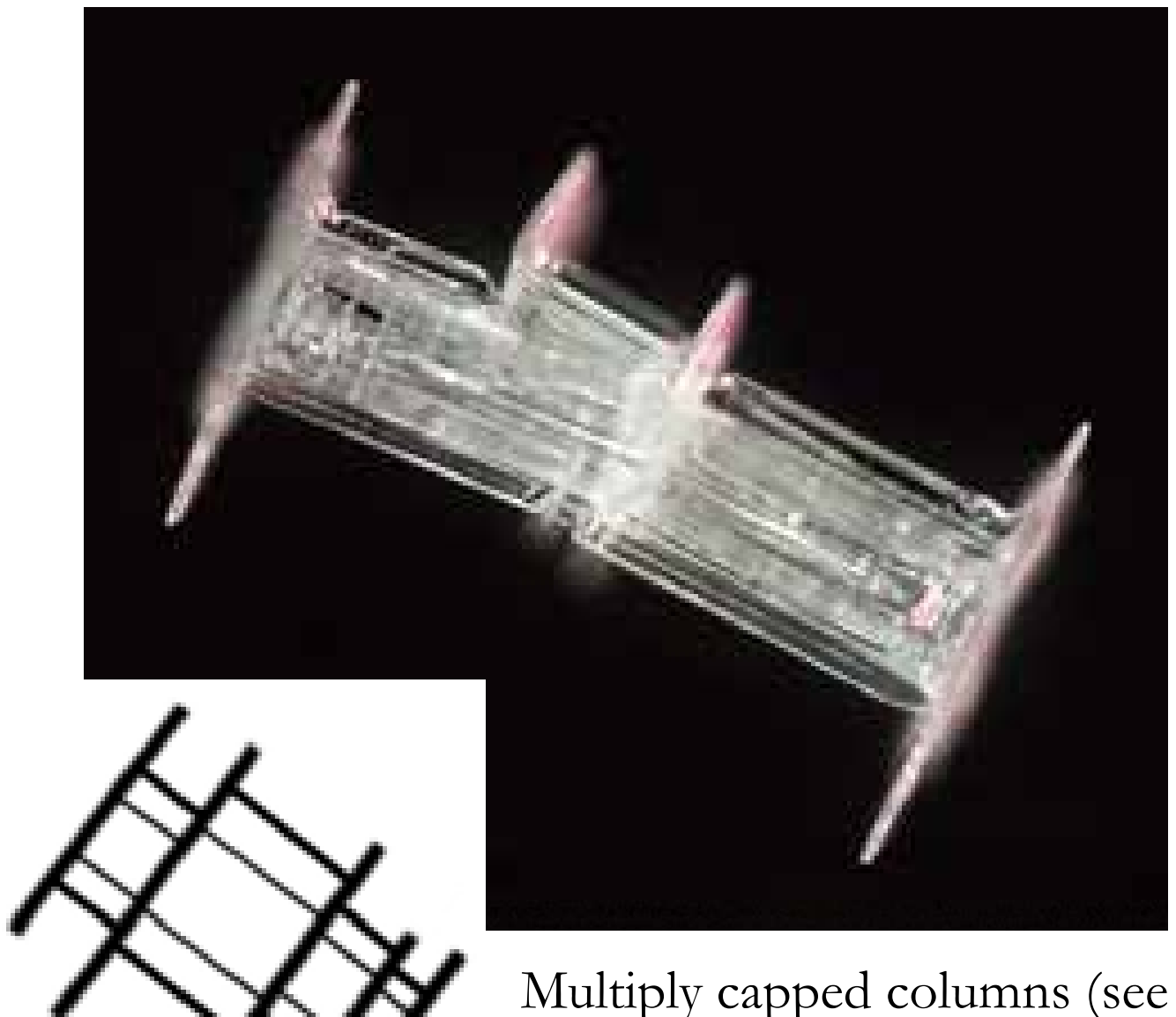

Multiply capped columns (see Figure 10.3) exhibit additional plate-like extensions, as shown in the example above. These extra plates sprout from the exposed ledges in needle clusters, so typically they are symmetrically placed on the crystal.

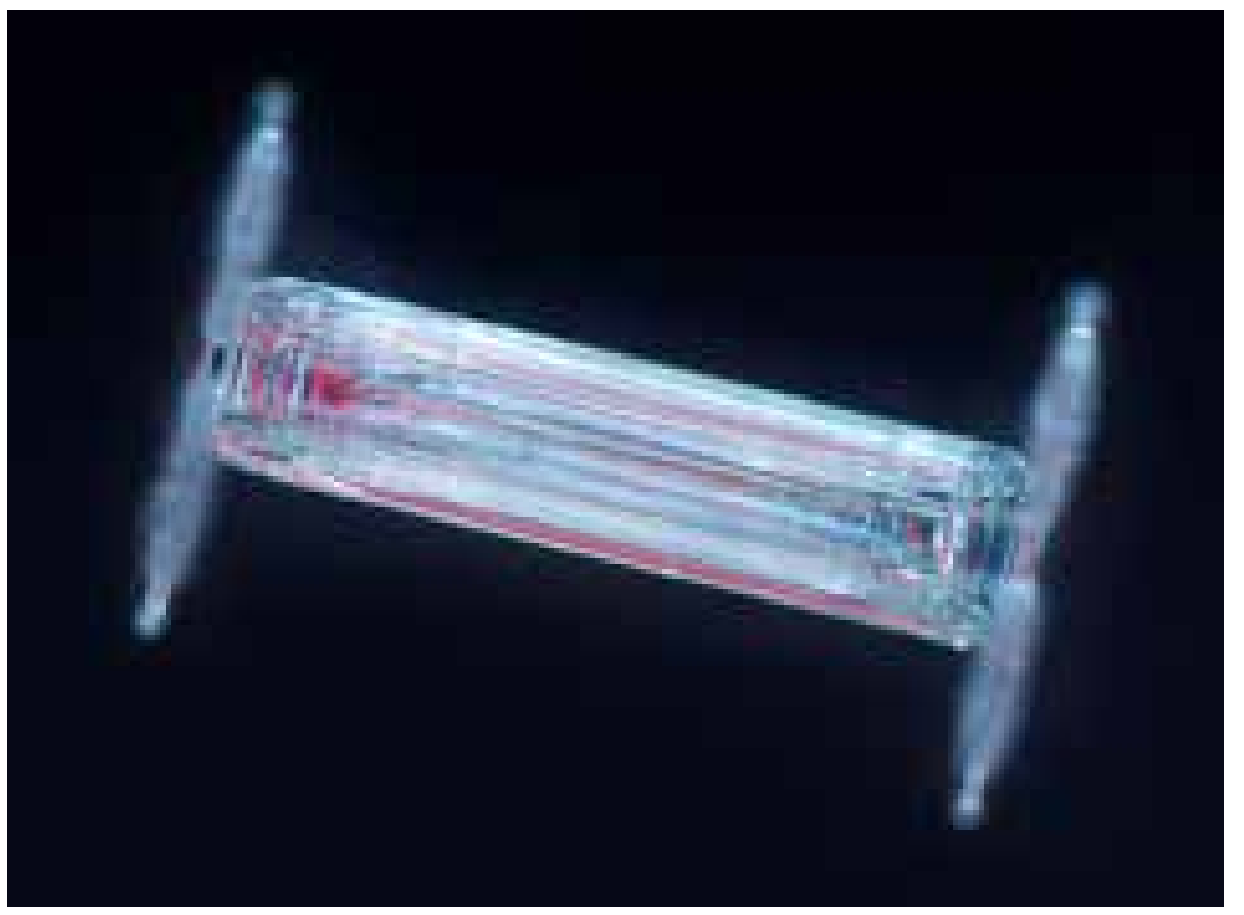

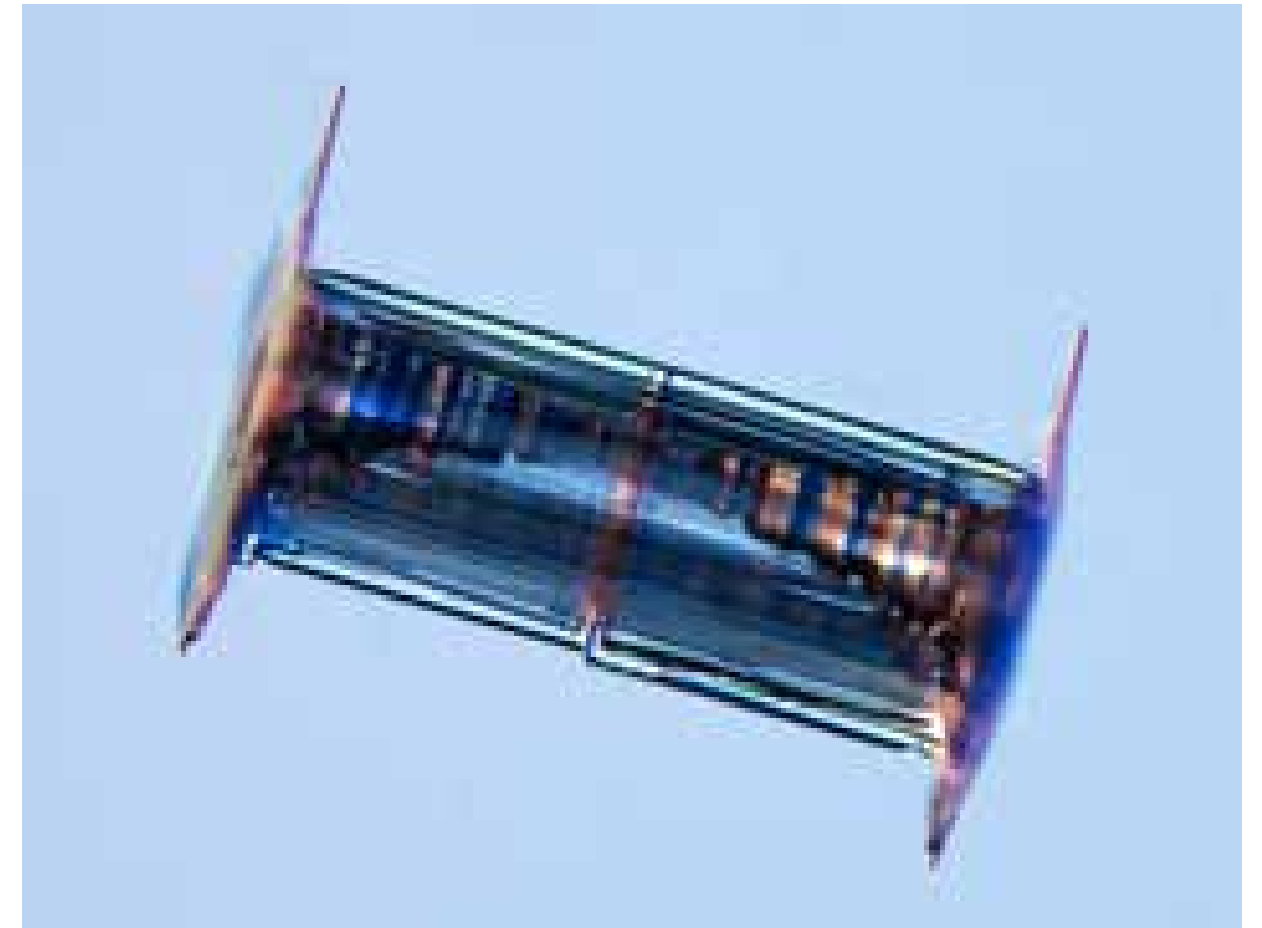

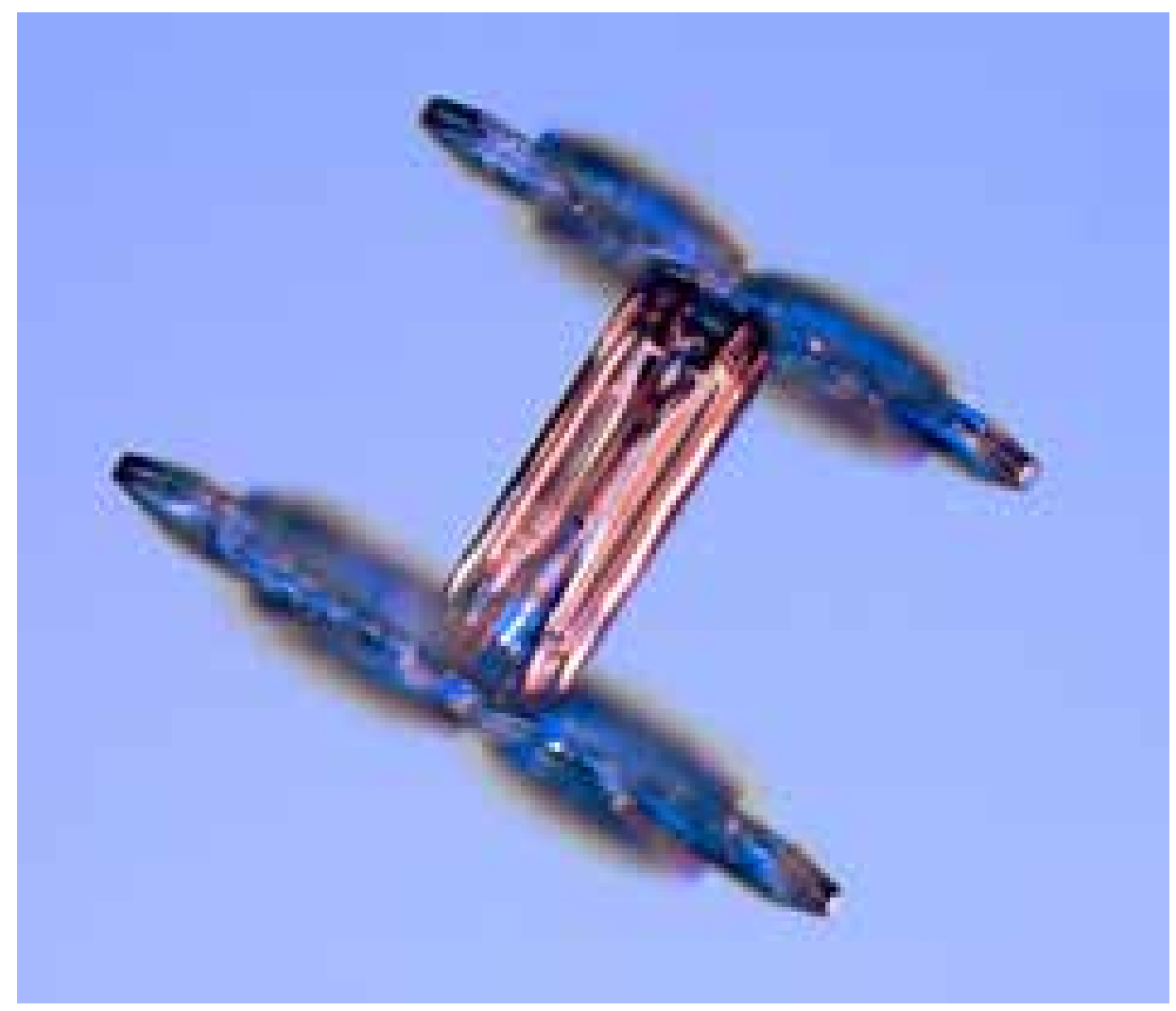

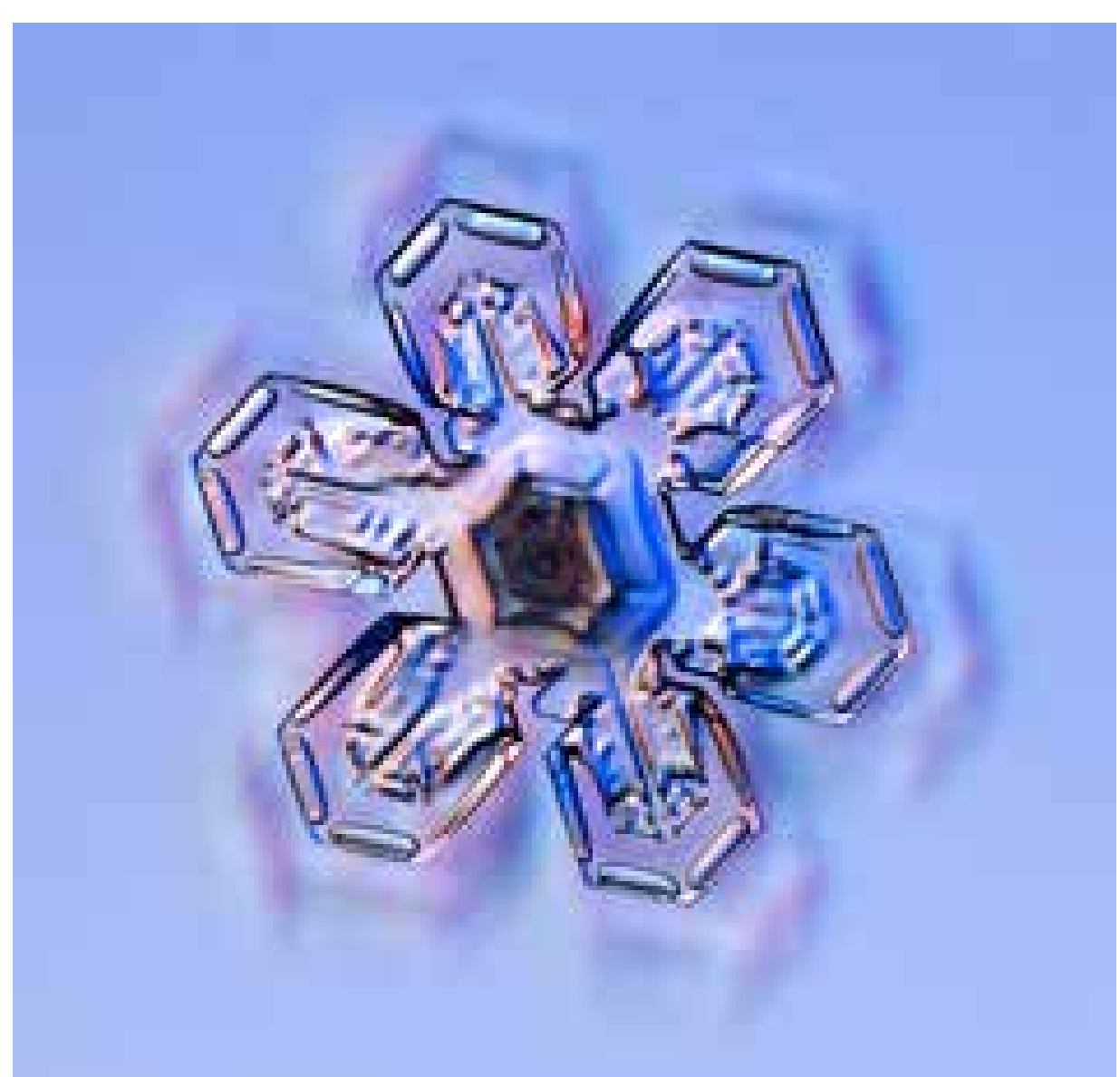

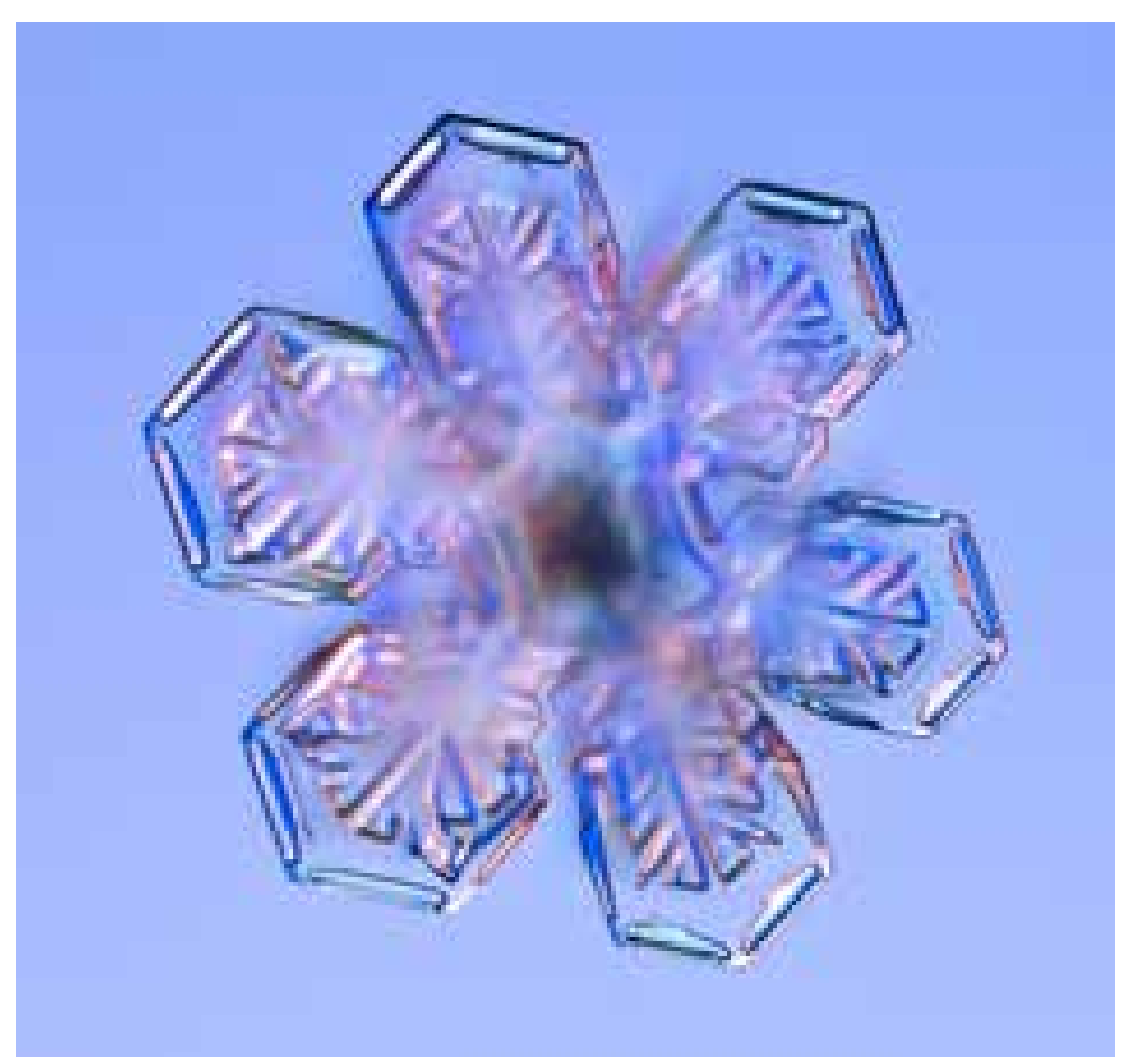

**Capped Column Close-up.** The three photographs at left are all of the same crystal, but with two different orientations and with different focal planes. The top picture shows the crystal in the orientation I found it, after it had fallen onto a glass slide. This shows a nice side view of the column upon which the plates formed. Some hollowing is present, so at one point this crystal must have looked like a simple hollow column. After photographing the crystal as it had fallen, I then used a fine paintbrush to flip it onto one face. Focusing my microscope on the smaller upper plate yielded the second picture, which looks like a typical stellar plate. The symmetry is subtly imperfect, and you can see a central dark spot where the column attaches. Without moving the crystal, I then re-focused on the lower plate to produce the third picture. Note that the blurry upper plate now obscures the lower plate to some degree. The lower plate looks a lot like the upper one, as you would expect because the two formed under nearly identical conditions.

The crystal below is a capped column with especially distinctive stellar plates on both ends. The image was captured in Moscow by Russian photographer Alexey Kljatov.

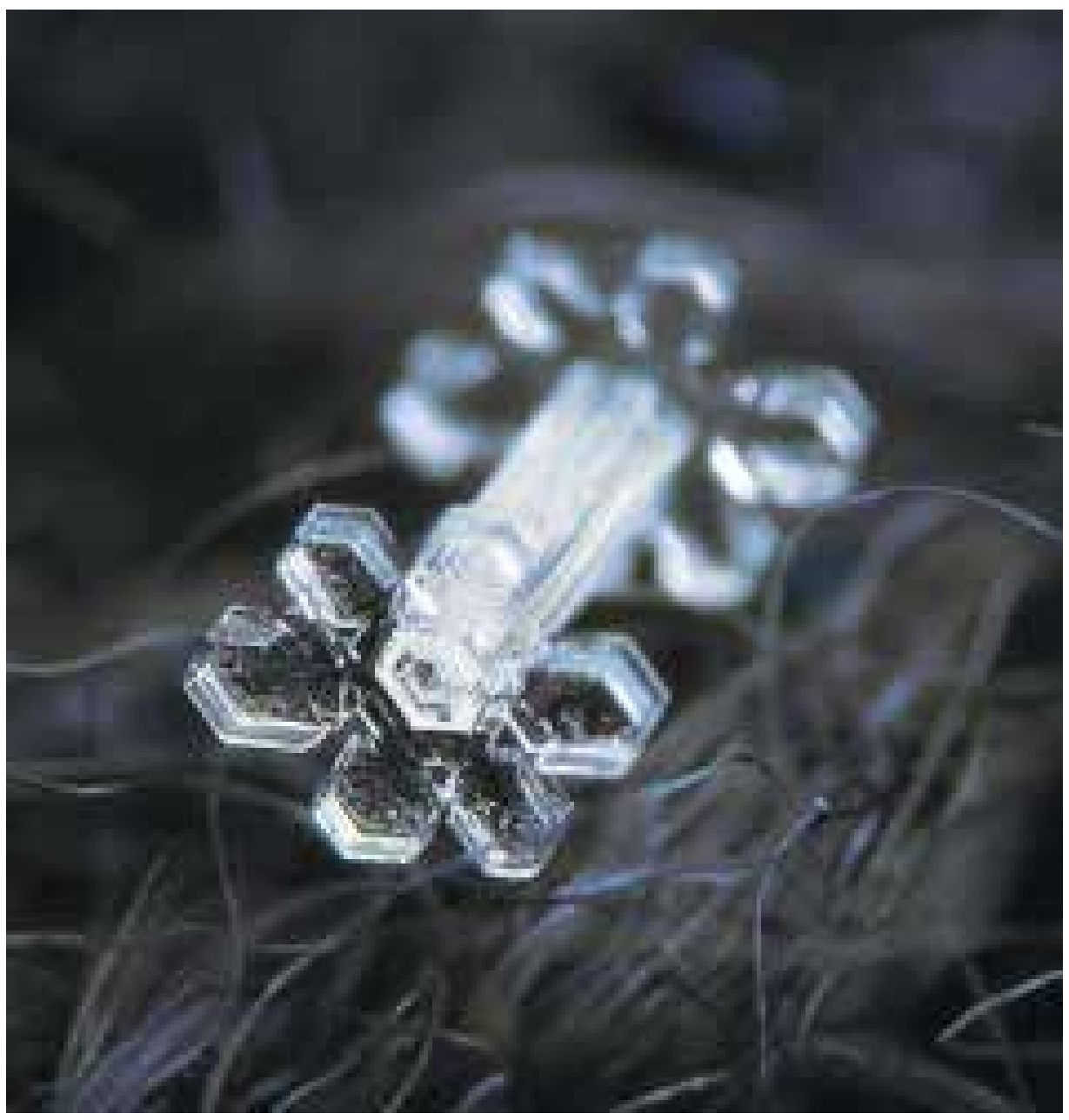

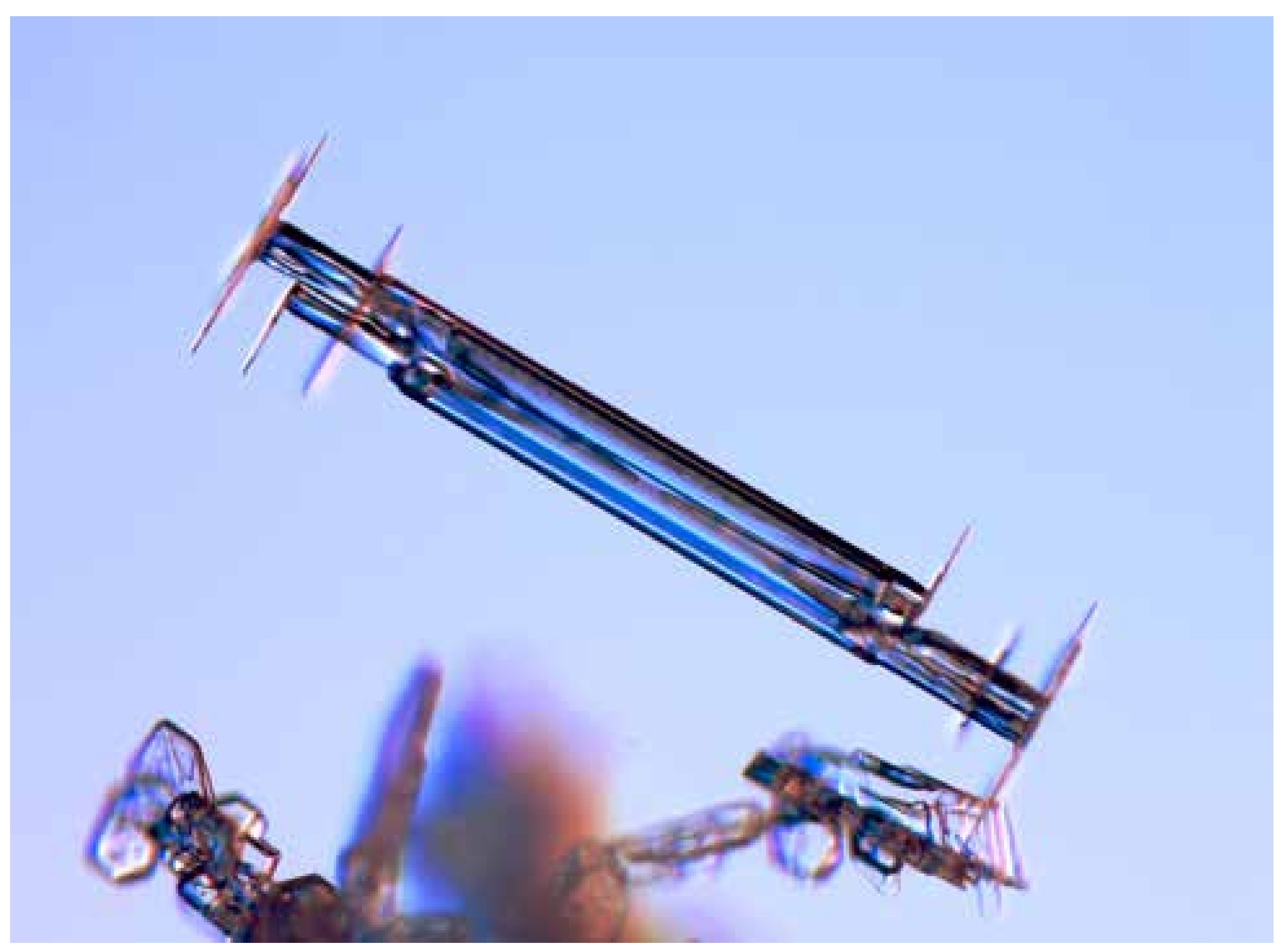

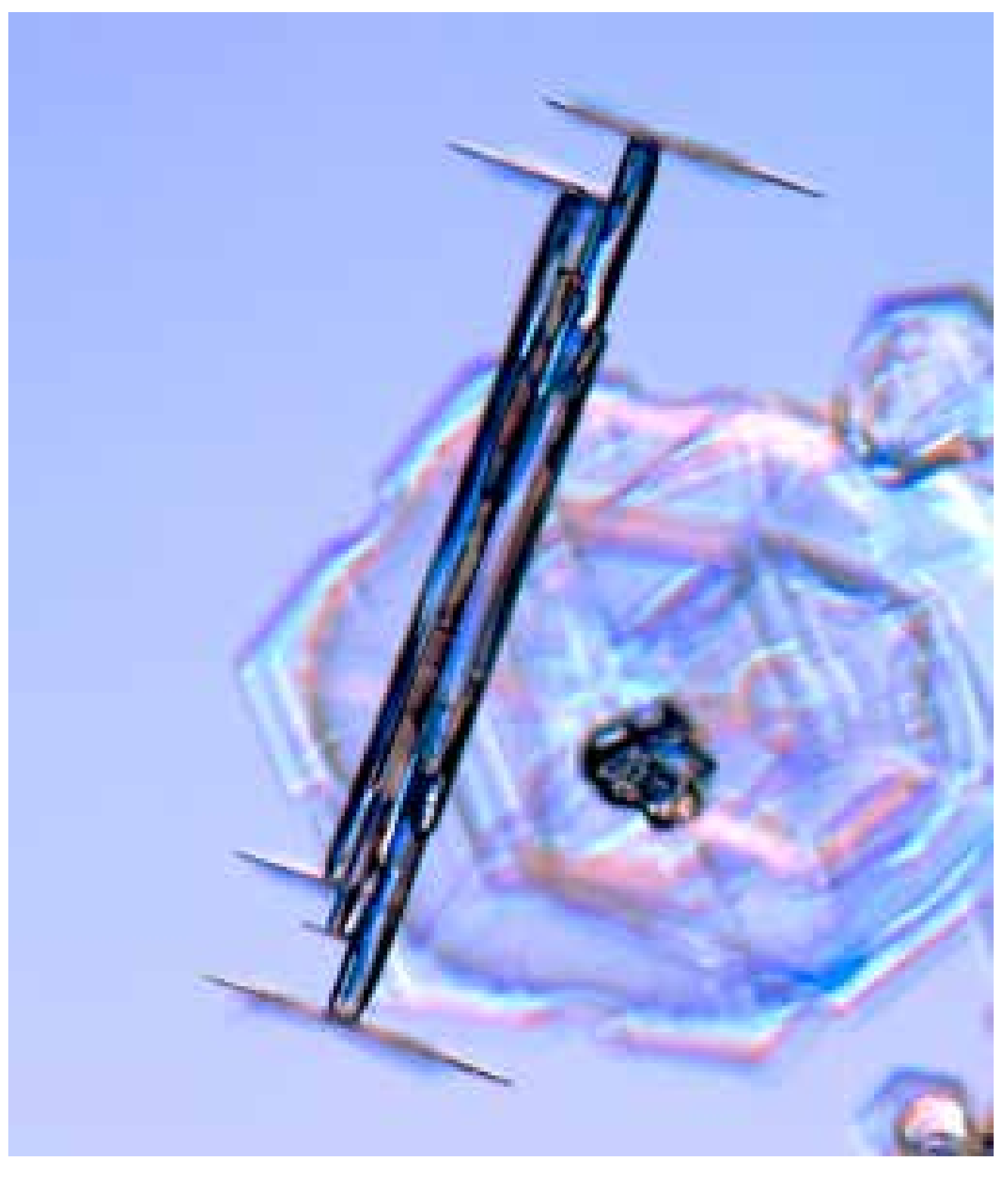

**Capped Needles.** These two excellent specimens are essentially capped columns, but might be more appropriately called capped needle clusters, as each has multiple plates growing from the ends of sizable needle clusters. Both are about 1.7 mm long, and the various plates (seen edge-on) are all amazingly thin, with razor-sharp edges. Moreover, the column-to-plate transitions are especially abrupt. Here again these crystals provide excellent demonstrations of the edge-sharpening instability in action. I have encountered large capped needles like this only once, on one remarkable day in the Michigan Upper Peninsula. These two crystals fell within a few minutes of one another, and I spotted several others like them as well. When the conditions are just right, rare snow crystals can fall in abundance.

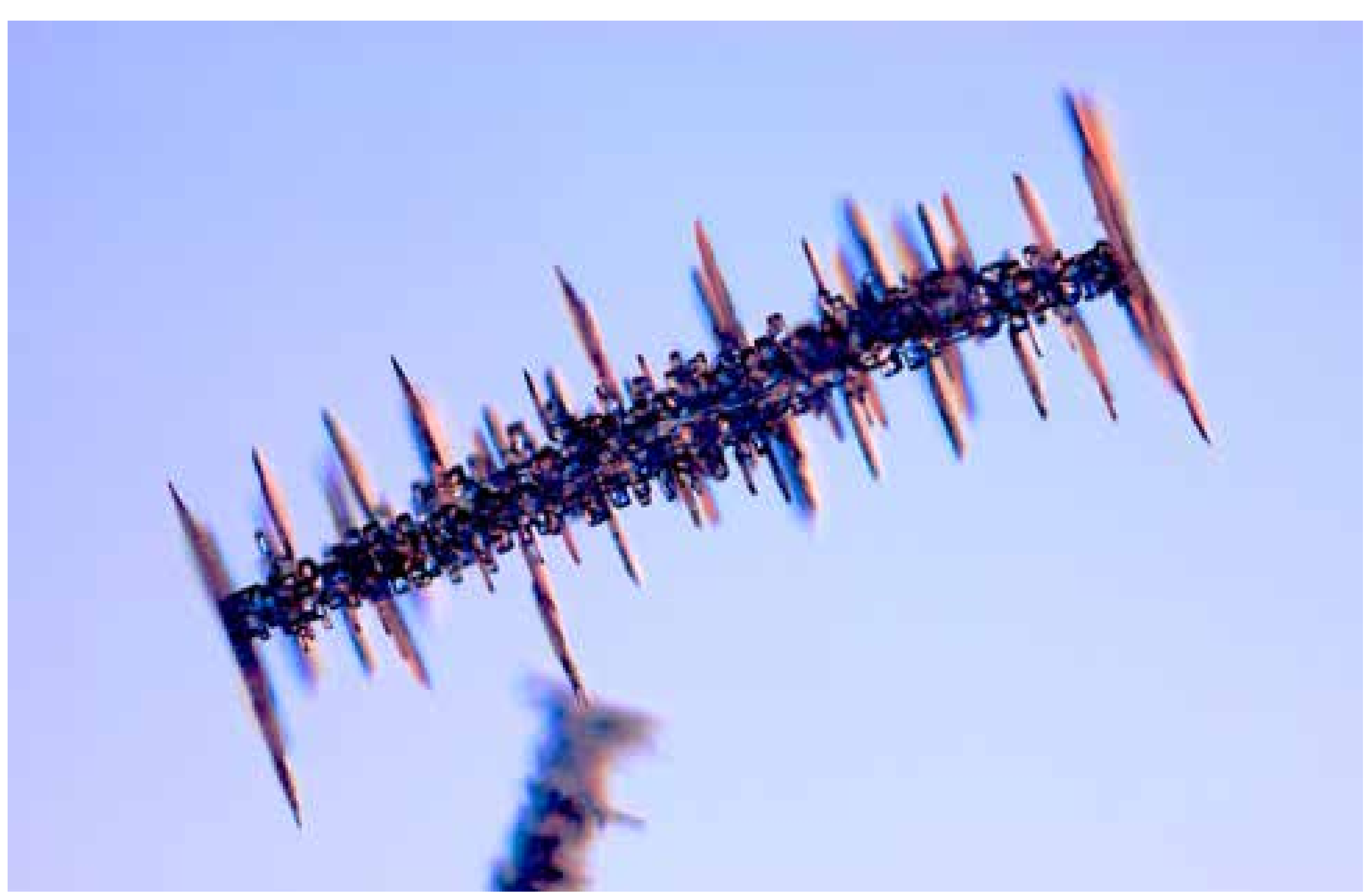

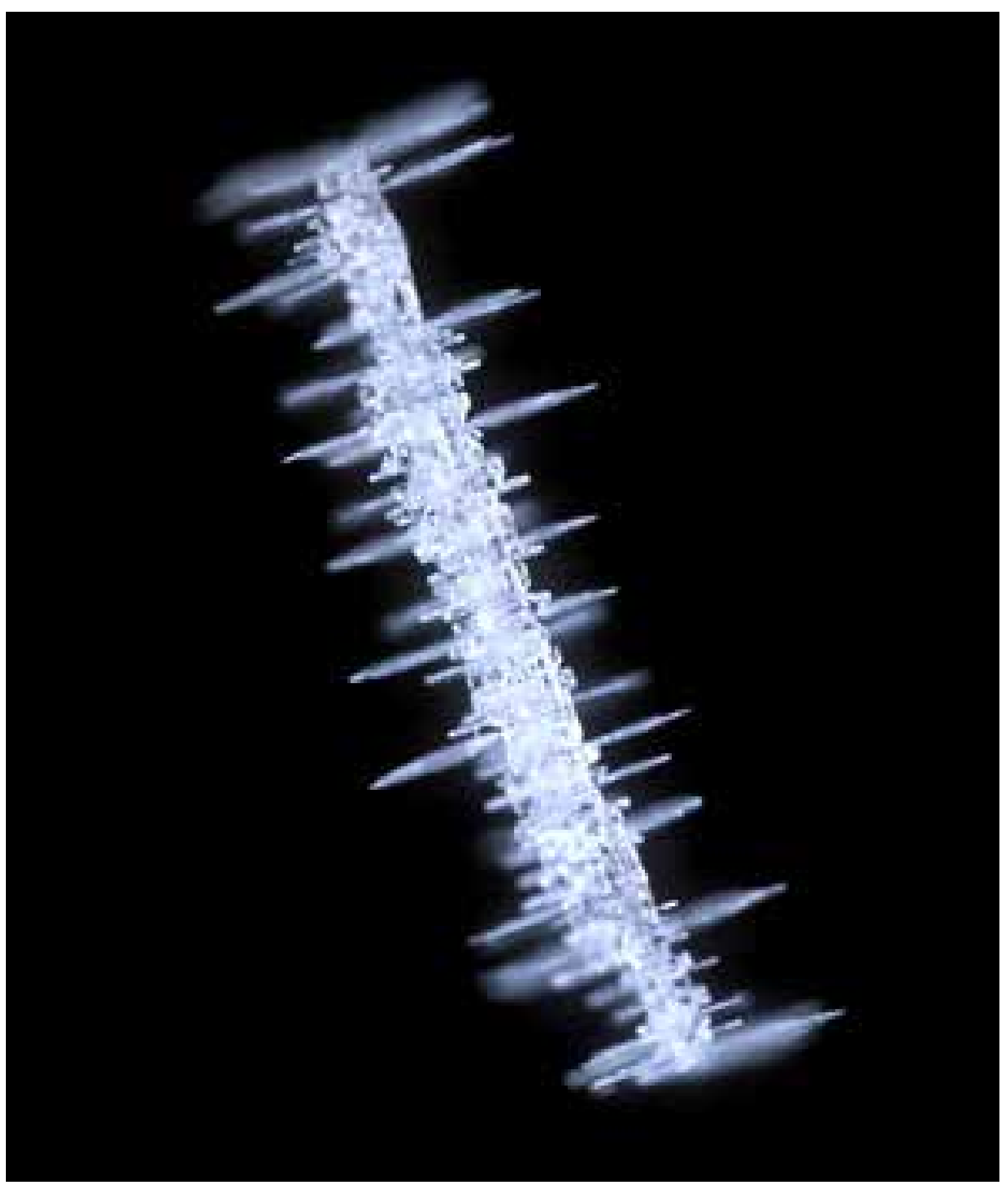

**Plates from Rime.** The world of multiply capped columns is inhabited by some exotic beasts, such as these two ice caterpillars. Both are relatively simple needle crystals festooned with copious side plates. (The plates are all seen edge-on in the pictures). Each of these crystals started out as a simple needle, which then became coated with rime. Next the temperature dropped and plates sprouted from many of the rime droplets. Note that the rime froze with the same lattice orientation as the underlying needles, so the side plates are all parallel to one another. Thus each of these seemingly disordered structures is in fact a single crystal of ice; the water molecules are aligned throughout. Both crystals were found in the Michigan Upper Peninsula, on the same day as the crystals on the previous page.

## Double Plates

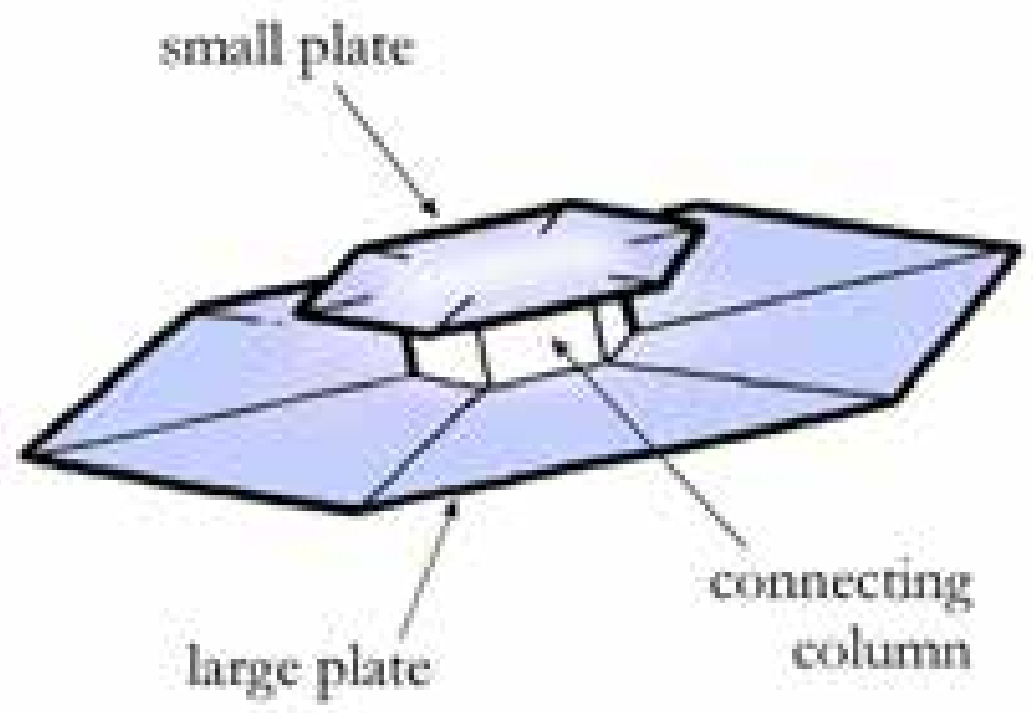

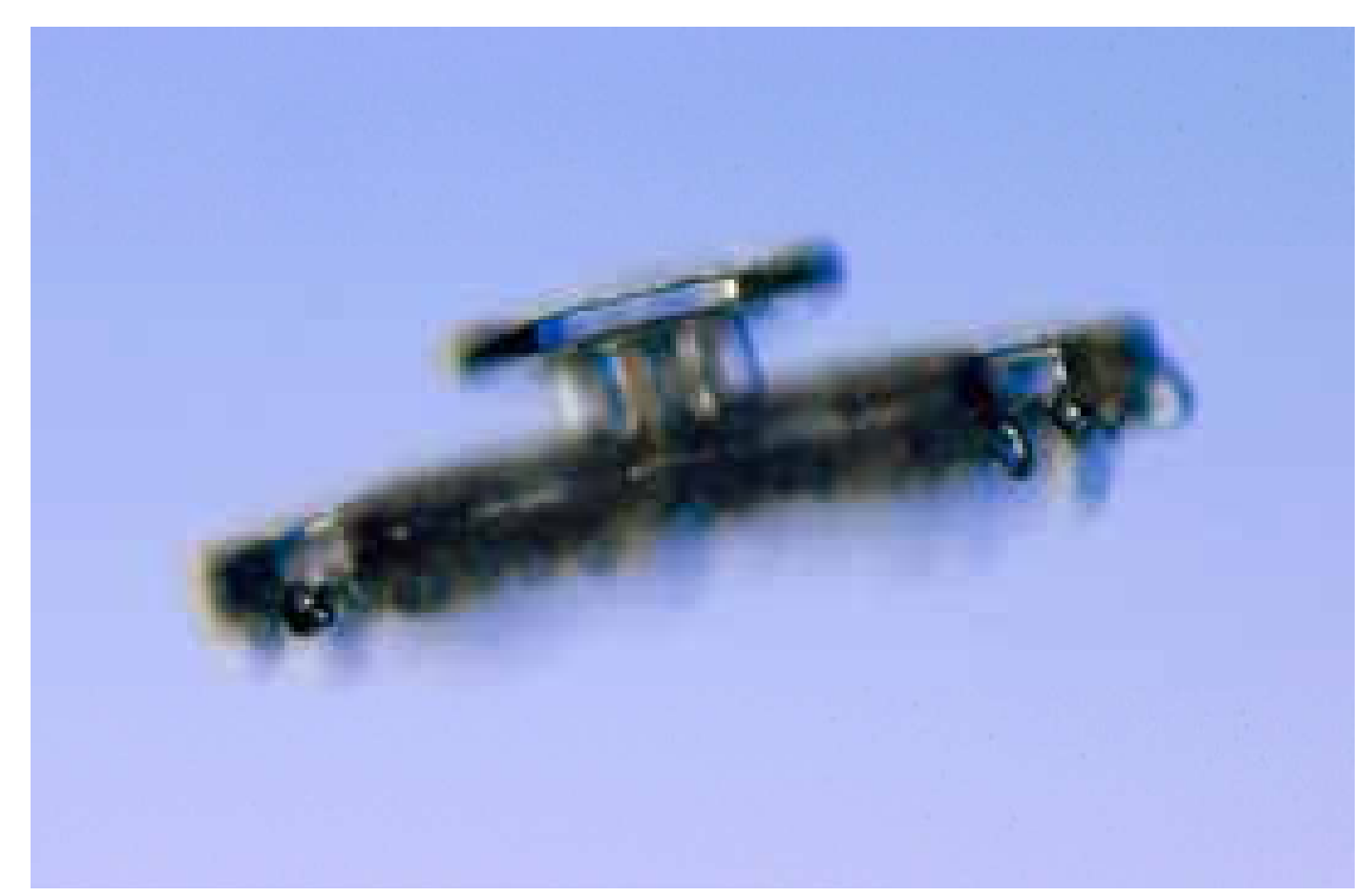

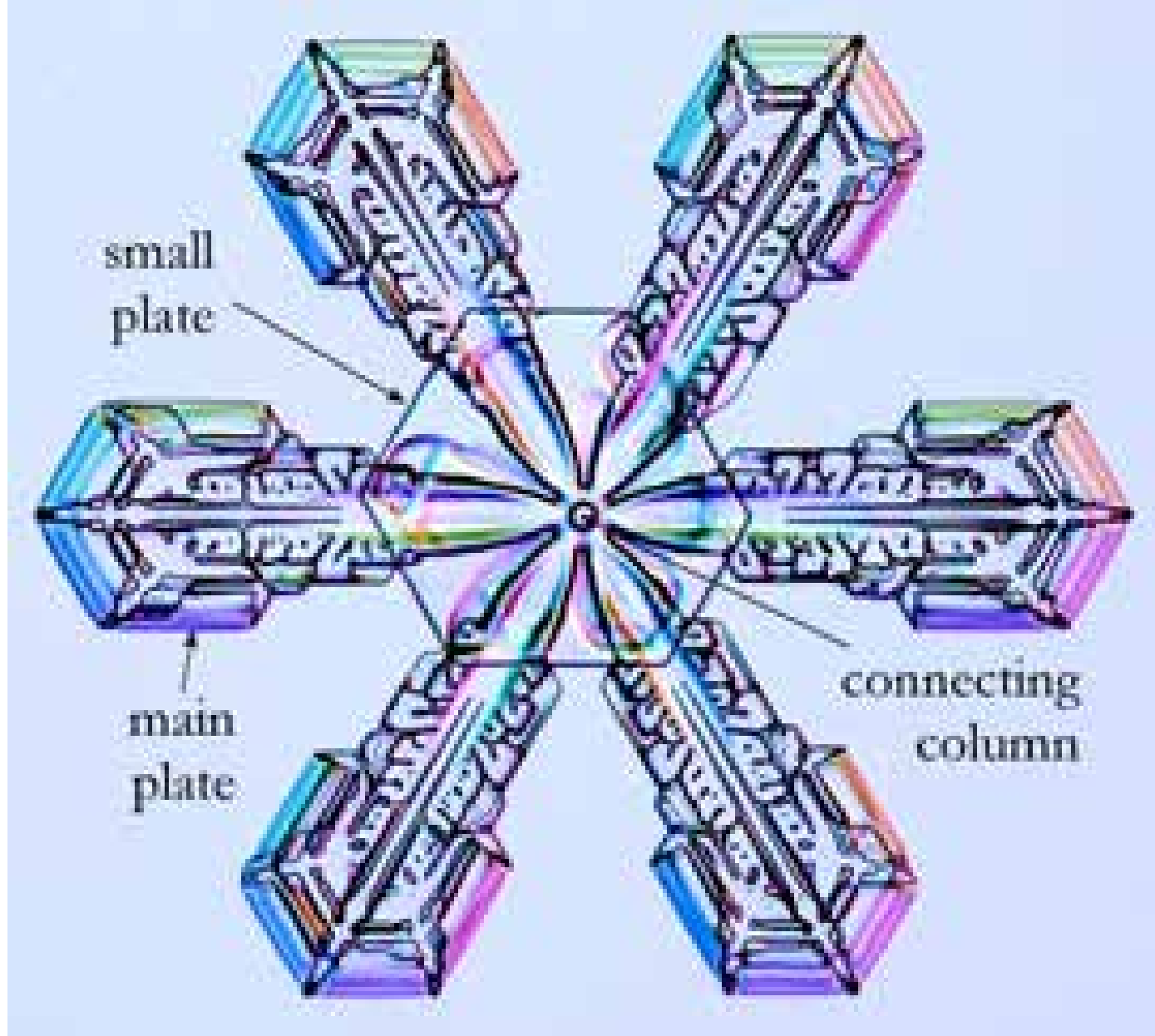

*Double plates are pairs of thin, plate-like crystals held together by a small connecting column. Often one side is a large stellar plate while the other is a smaller hexagon, although many other variations are possible. This phenomenon is relatively common, and many stellar crystals are actually double plates if you look closely.*

Double plates are basically extreme versions of capped columns that result in two closely spaced plates. The two plates compete for water vapor, leading to a growth instability: any slight perturbation can cause one plate to overshadow the other, yielding one dominant and one recessive plate. The two photos below show the same rimed crystal with separate focus on the top and bottom plates.

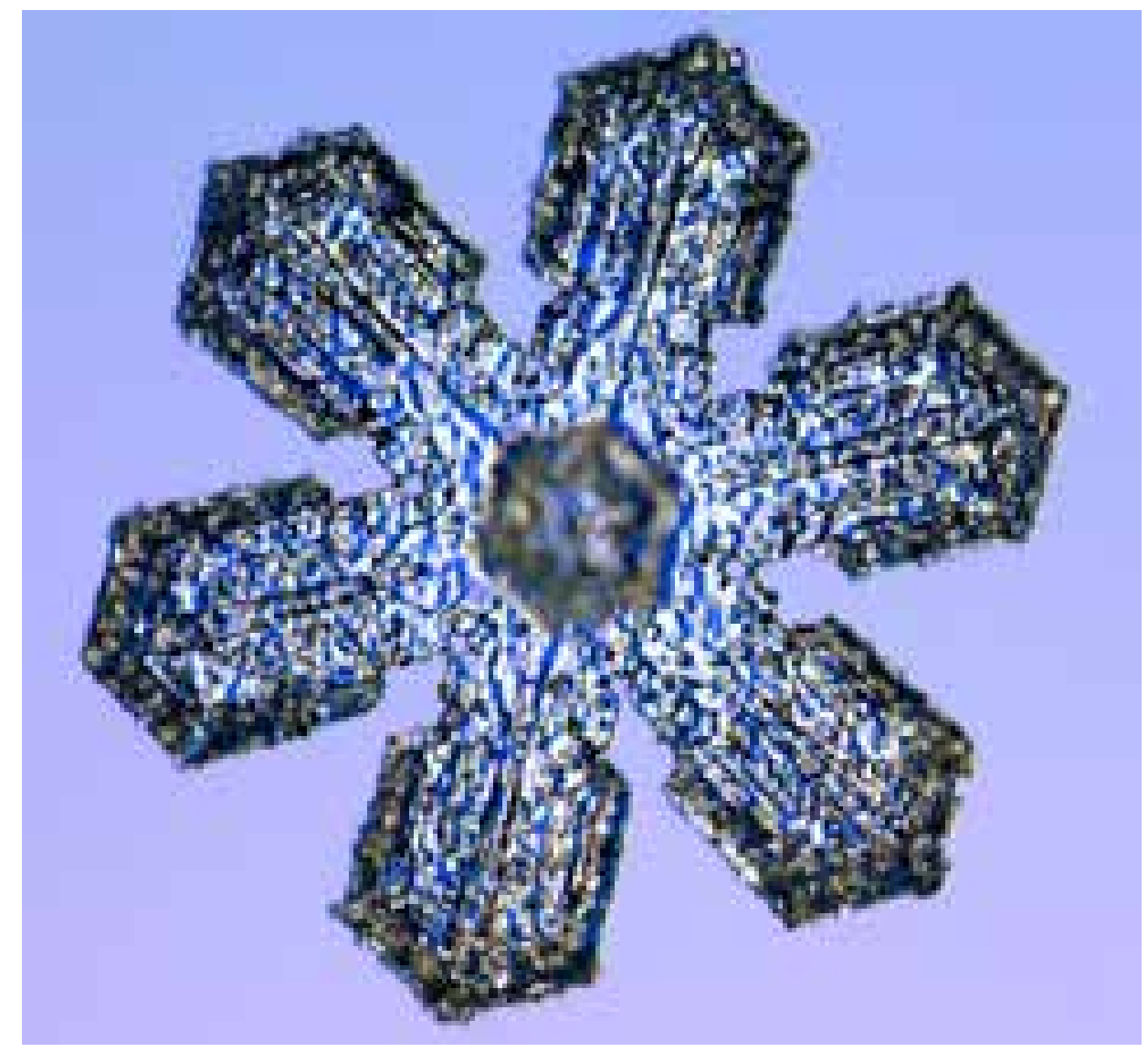

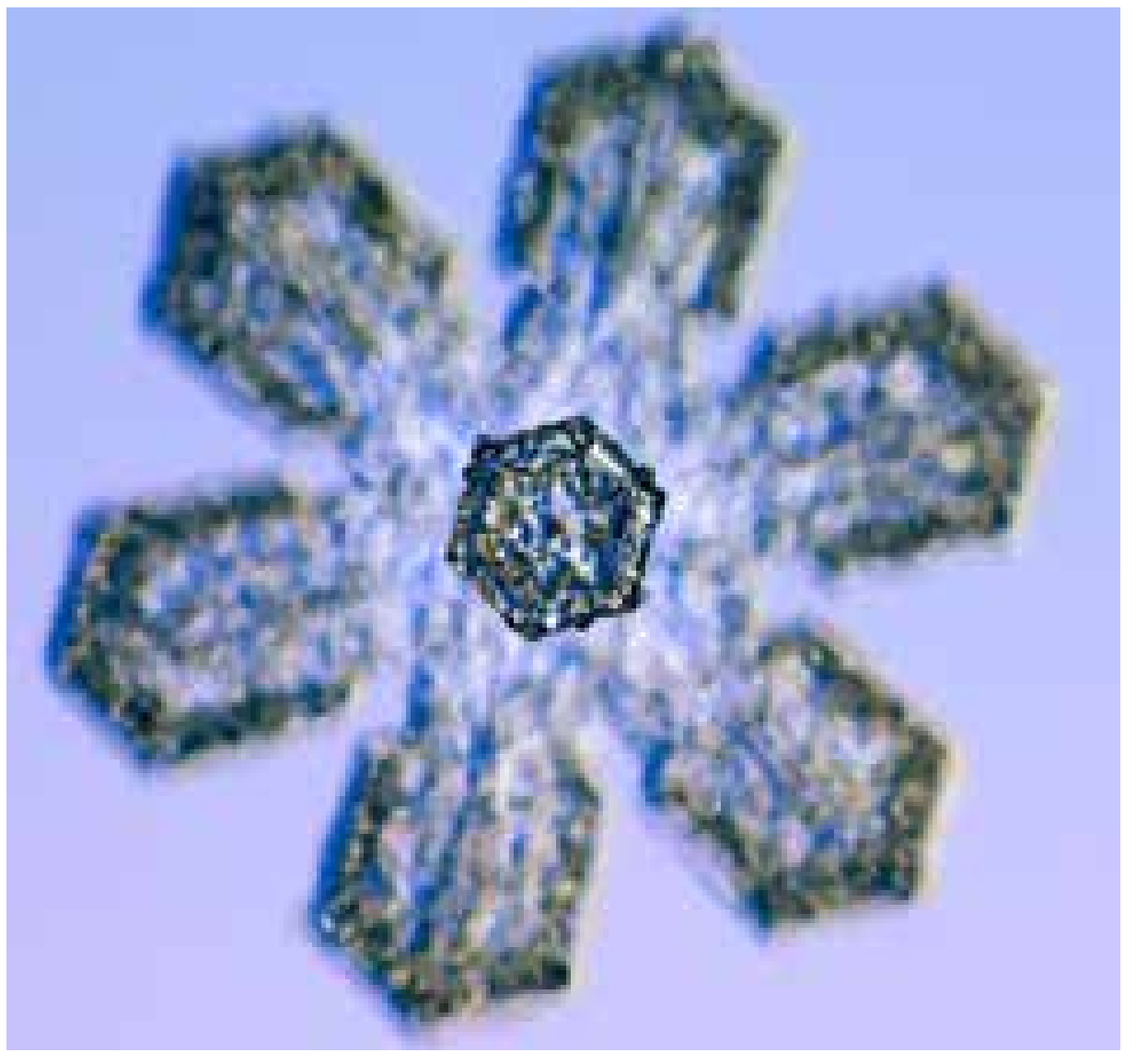

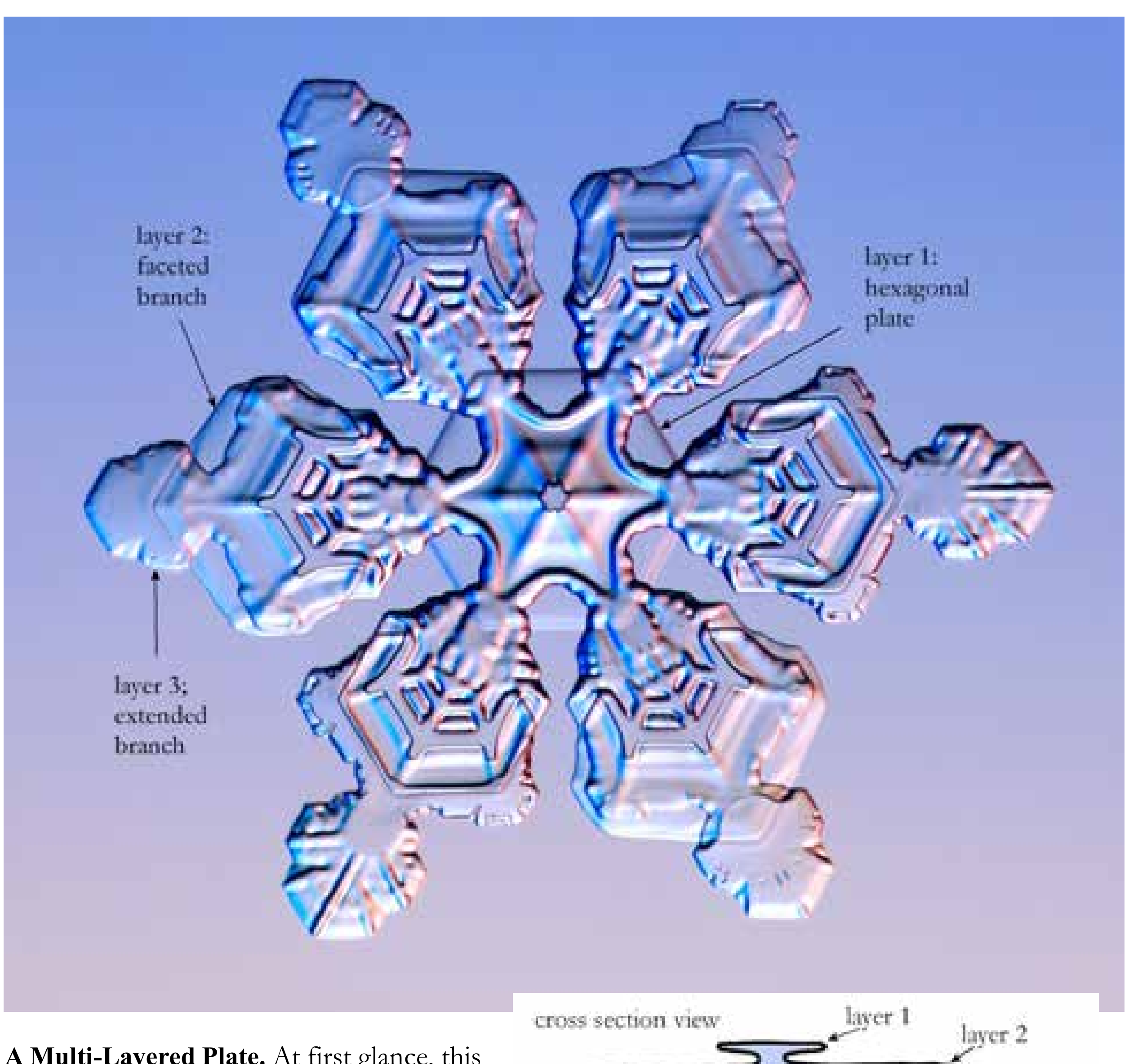

**A Multi-Layered Plate.** At first glance, this snow crystal may look like an ordinary stellar plate, but a closer inspection reveals three distinct layers, as shown in cross-section in the inset sketch. Note first the nicely formed hexagonal plate near the center of the crystal (layer 1), which is slightly out of focus in this picture. This hexagon was one half of a double plate when the crystal was small. The other half grew out faster and branched, and in doing so it deprived the hexagon of water vapor. Because it grew relatively slowly, the hexagon remained smaller and faceted. One often sees double plates where the larger sheet is branched and the smaller one is faceted for this reason. When the crystal was about half its final size, it ran into low humidity and the branches grew thicker. Later the humidity picked up again and the branches became double plates of their own (layers 2 and 3). Here again one plate was left behind growing slowly (layer 2) while the other grew out more quickly and became branched (layer 3). If you look carefully, many stellar crystals show multiple layers like this one.

## Split Plates and Split Stars

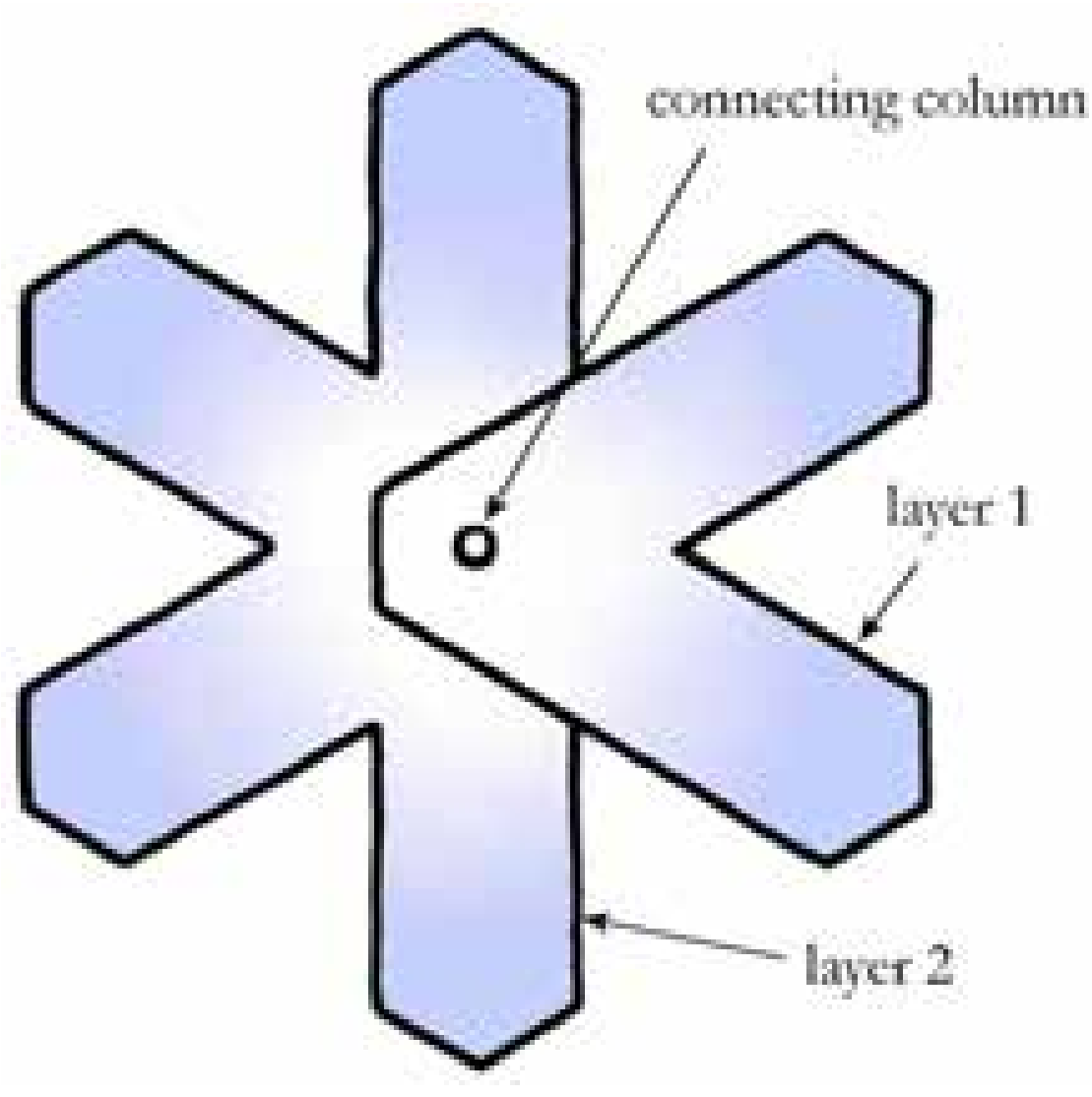

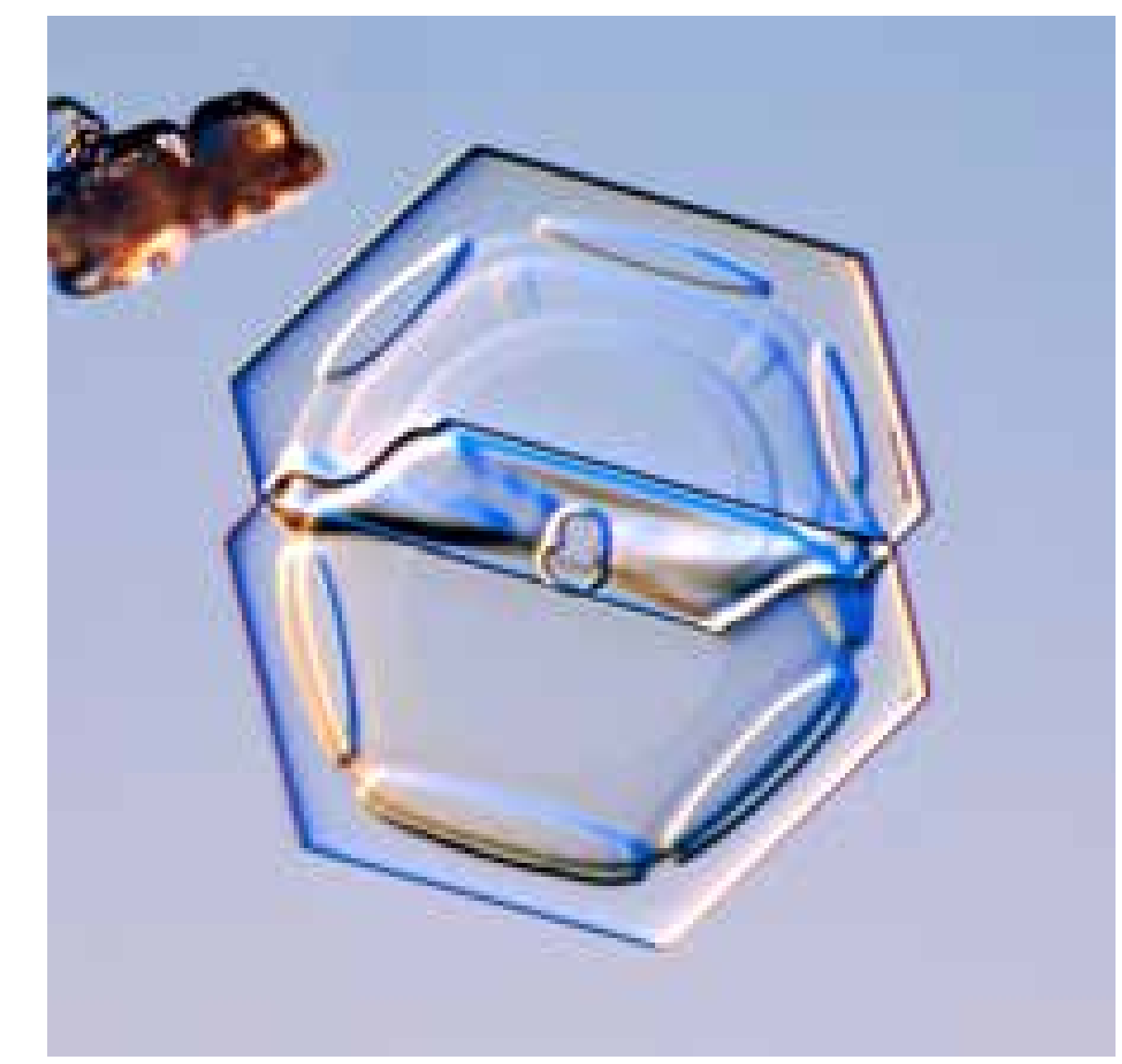

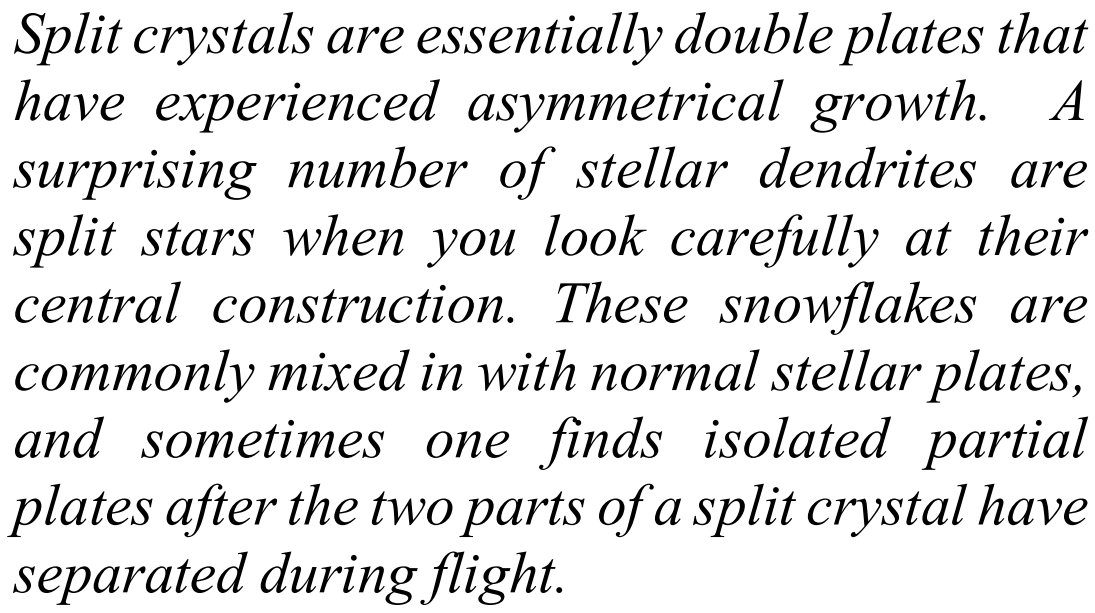

*Split crystals are essentially double plates that have experienced asymmetrical growth. A surprising number of stellar dendrites are split stars when you look carefully at their central construction. These snowflakes are commonly mixed in with normal stellar plates, and sometimes one finds isolated partial plates after the two parts of a split crystal have separated during flight.*

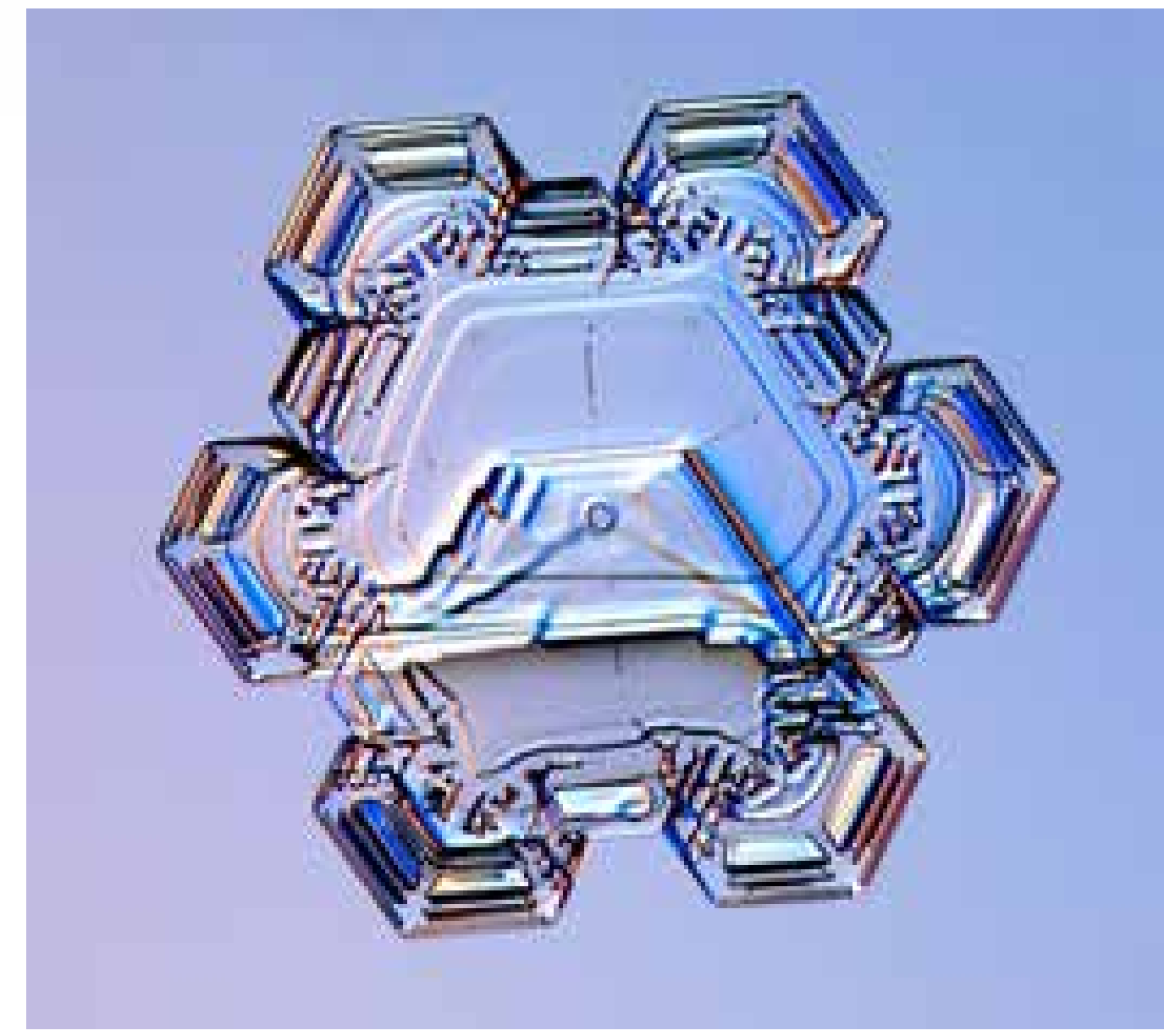

The formation of a split crystal is driven by a growth competition between the two members of a double plate. The pair starts out symmetrical, looking much like a short capped column. But if one branch or corner happens to edge ahead of its nearby sibling, then the growth of the latter is soon stunted from overshadowing. If one entire plate dominates over the other, then the result is a double plate. But if parts of both plates prevail, then the crystal will develop into a split plate or split star. If the split occurs early, the six dominant branches may grow into a surprisingly symmetrical stellar crystal.

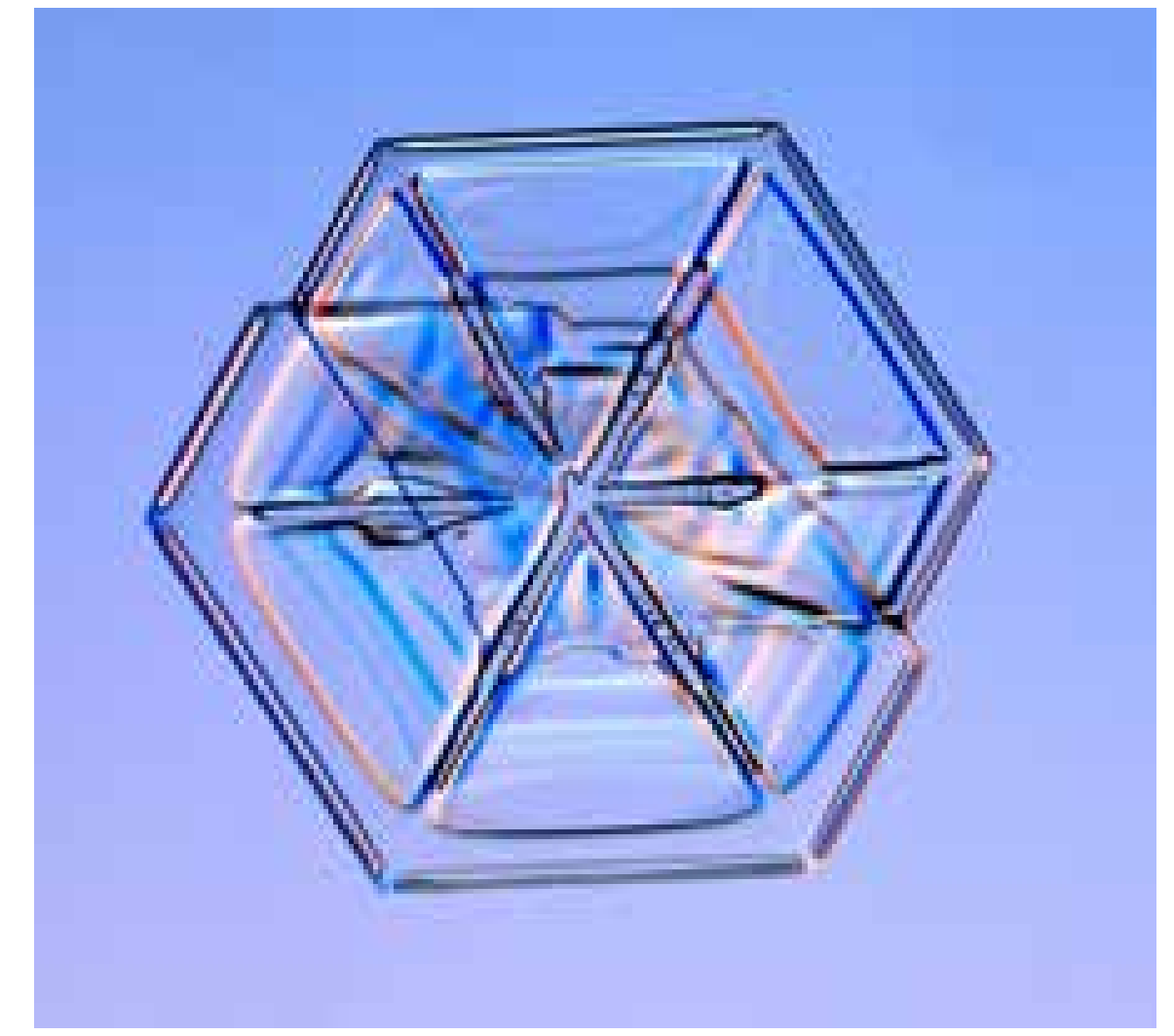

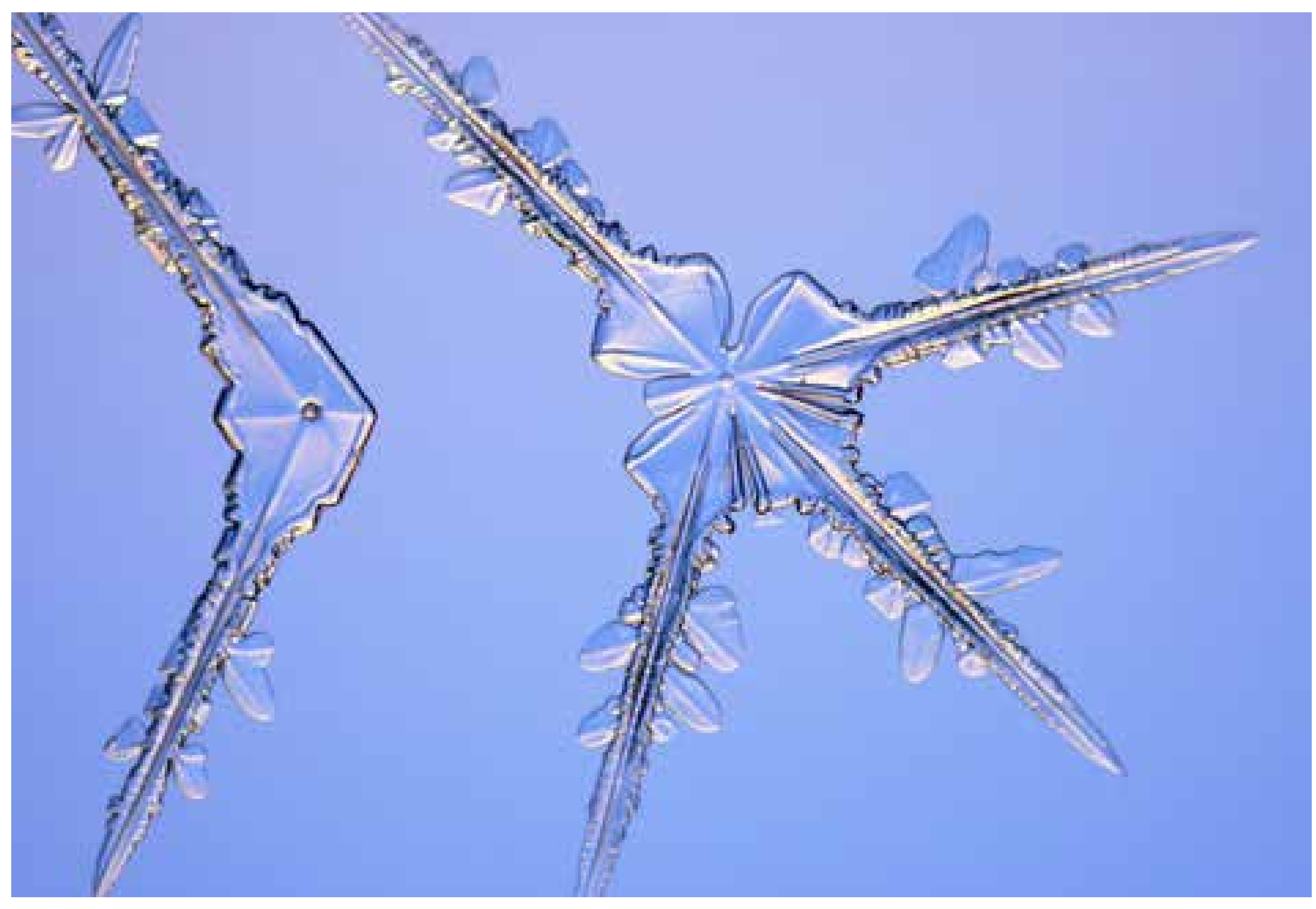

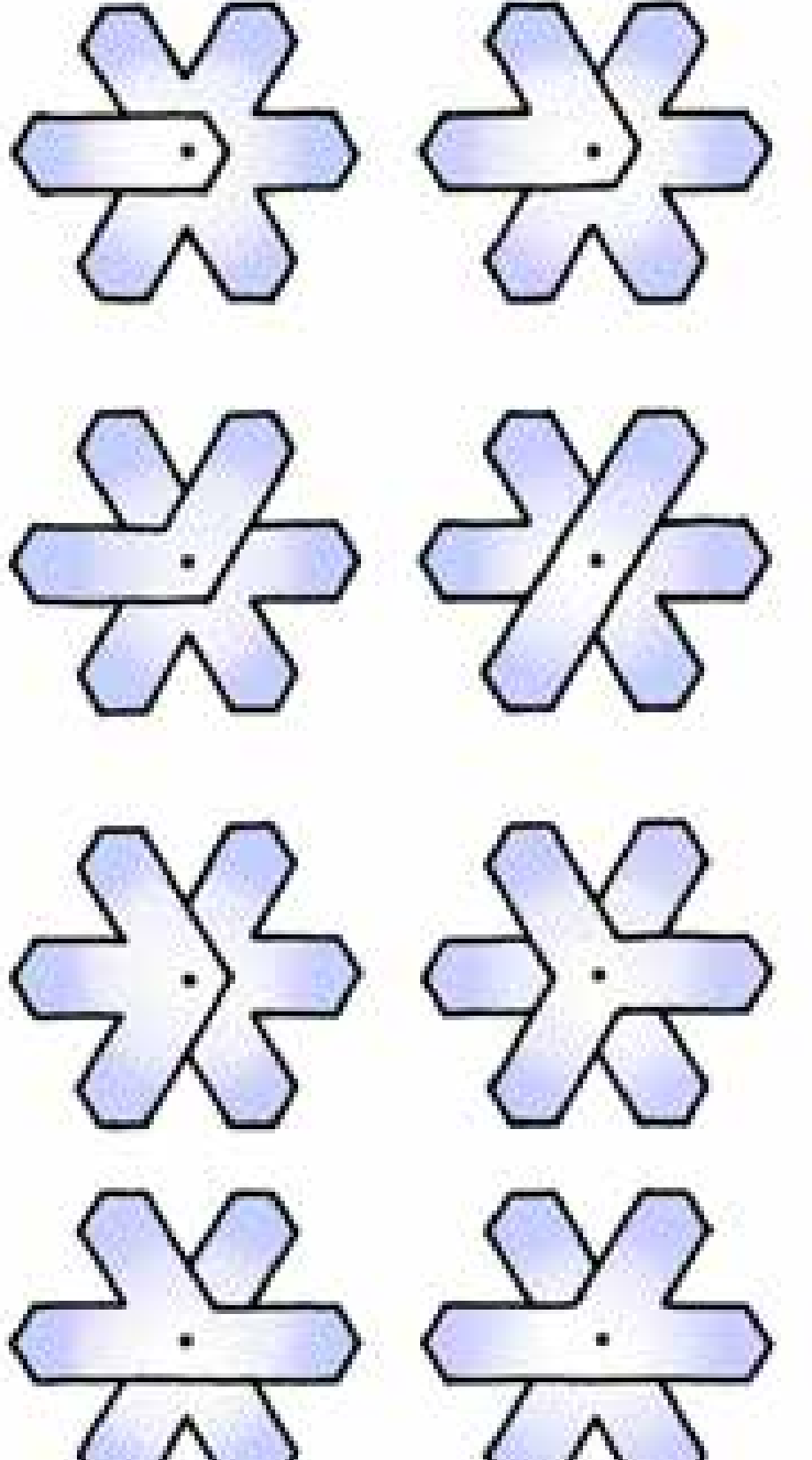

**Split Categories.** There are eight possible topologies for making a split plate or star, as illustrated in the sketches on the left. The crystal above is one of the three possible 4+2 variants, photographed by Patricia Rasmussen [2003Lib2]. Here the two parts of the crystal broke apart during handling, giving a nice look at a "disassembled" split star.

The photo below shows one piece from another 4+2 variant.

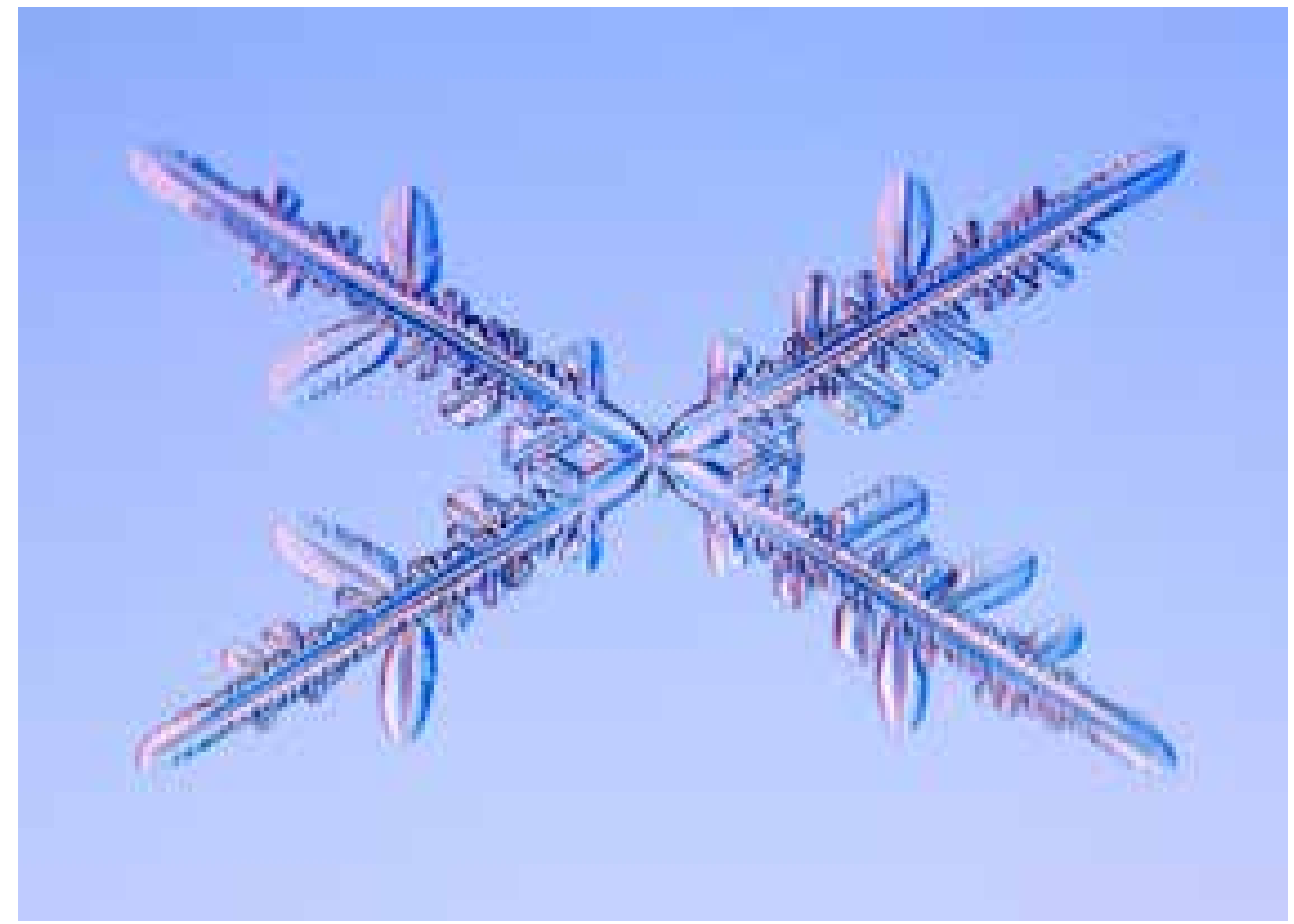

## Hollow Plates

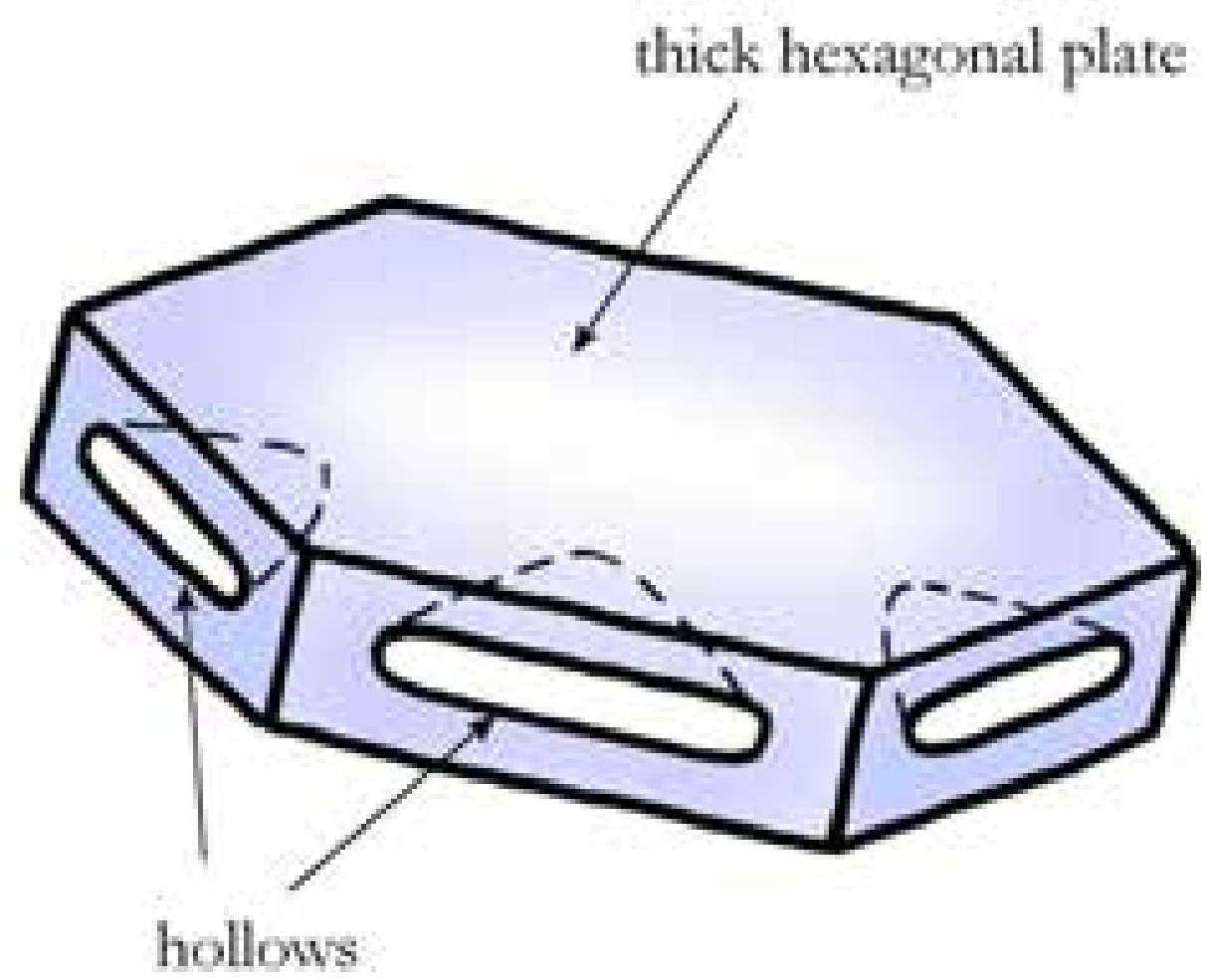

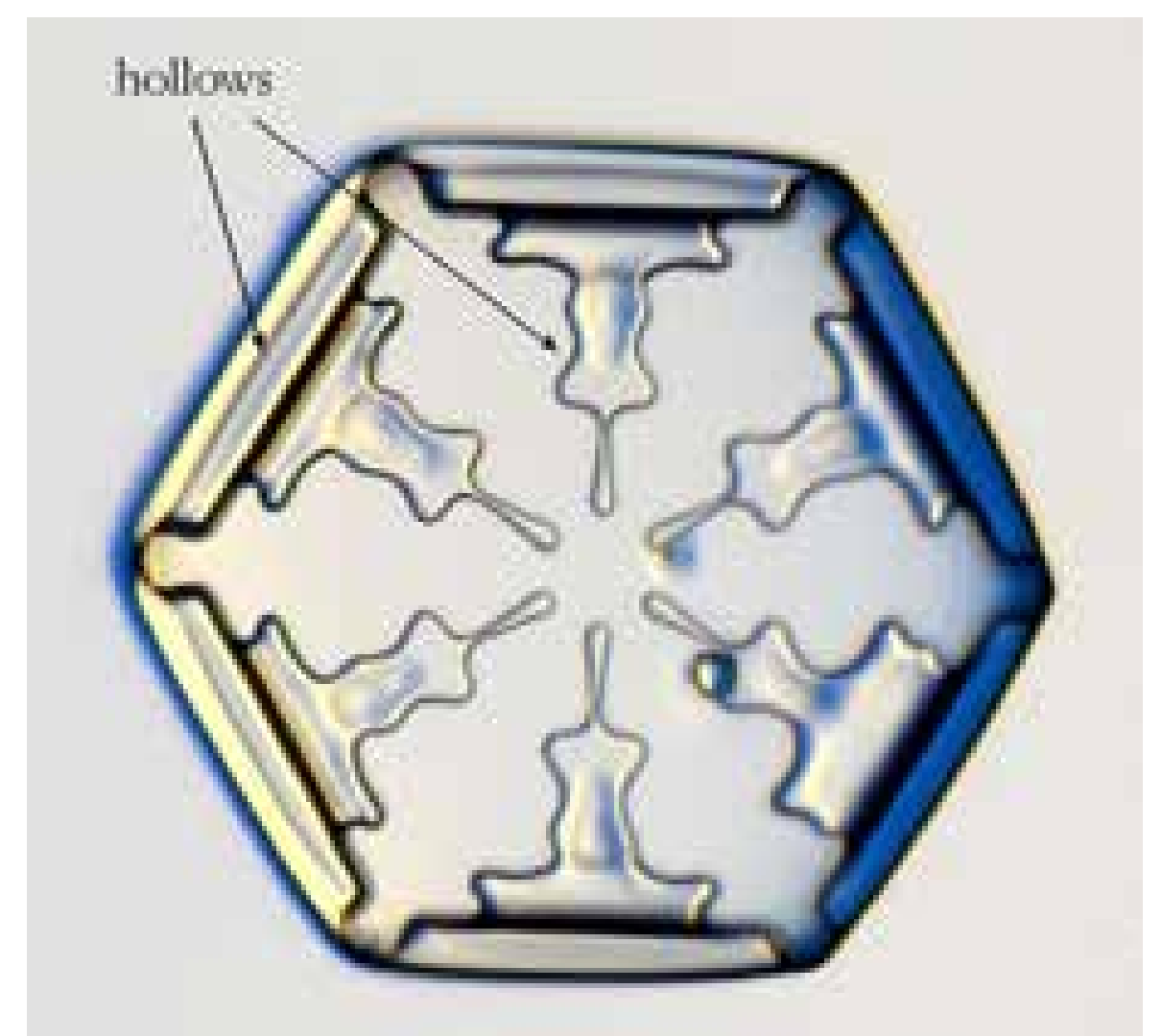

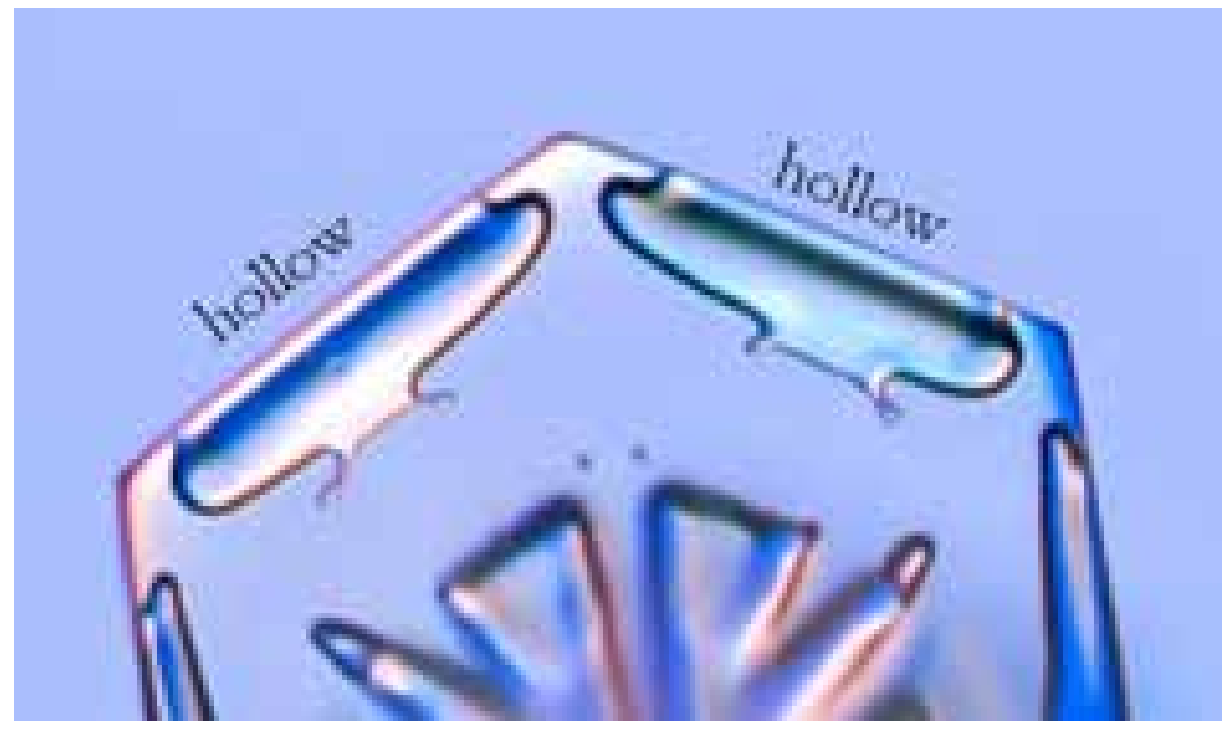

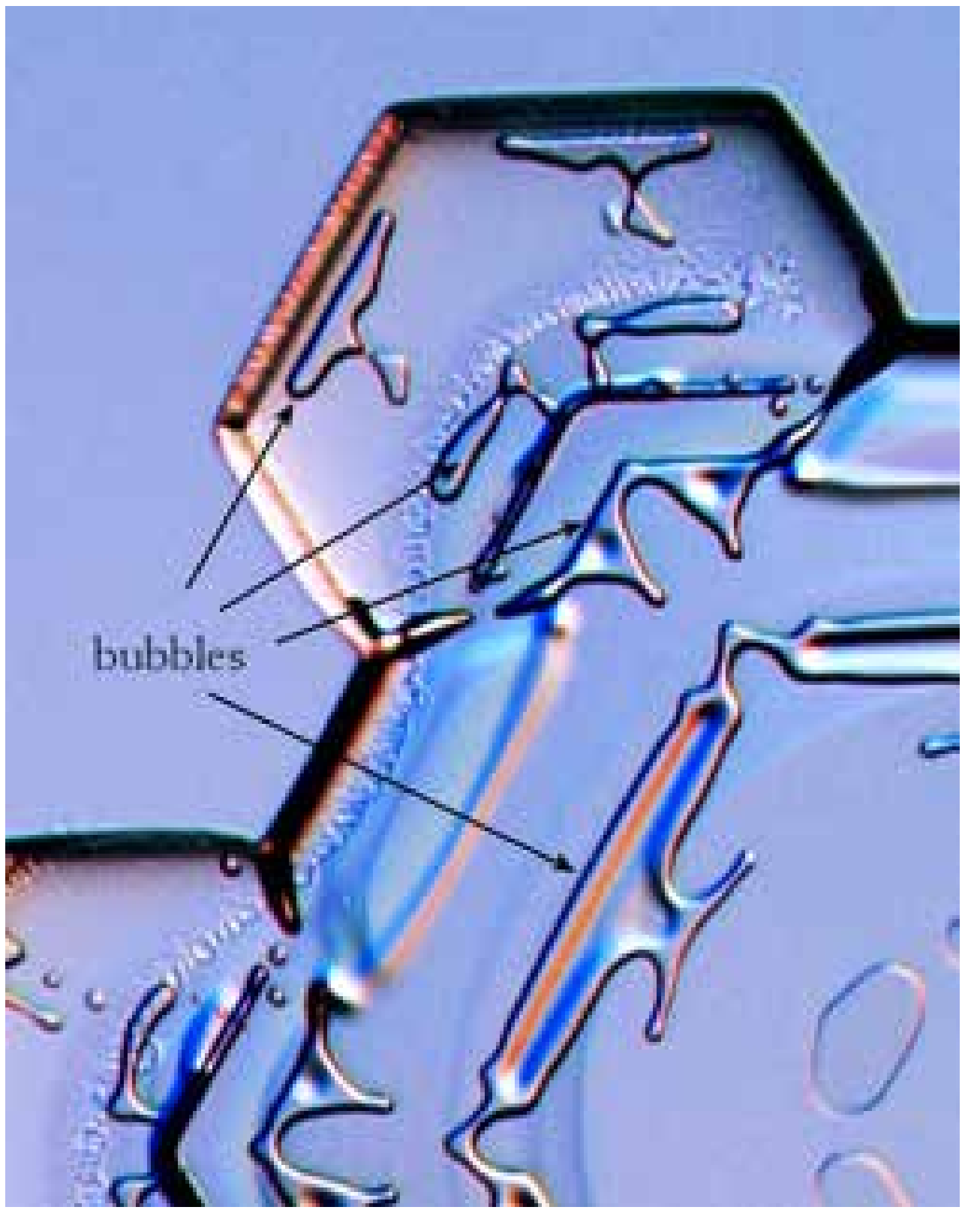

*Hollow plates are thick plates with voids extending down from their prism faces. Sometimes the faces grow over the voids to enclose thin bubbles in the ice. These features are occasionally found in small prisms and on the broad branches of stellar crystals, although it can be difficult to tell the difference between "dimples" that are depressions on basal surfaces and nearly enclosed "voids".*

Hollow plates are essentially the plate-like analog of hollow columns. One starts with a thick-plate crystal, and then the facet edges grow faster than the centers, eventually leaving behind hollows in the prism faces. The sketch above shows a hexagonal hollow-plate crystal, but the phenomenon is more of a structural feature than a snow-crystal type. Like ridges and ribs, small hollows are fairly common features in broad-branched stellar plates and other thick-plate crystals.

Hollow plates are most likely to grow when the temperature is either just above or just below -15 C, as can be seen in Figure 8.16. Fluctuations in temperature and humidity can yield rather oddly shaped voids, although broad, wide voids are more typical.

In reflected light, interference effects can make voids show up as colorful regions in photographs, which I will discuss further in Chapter 11.

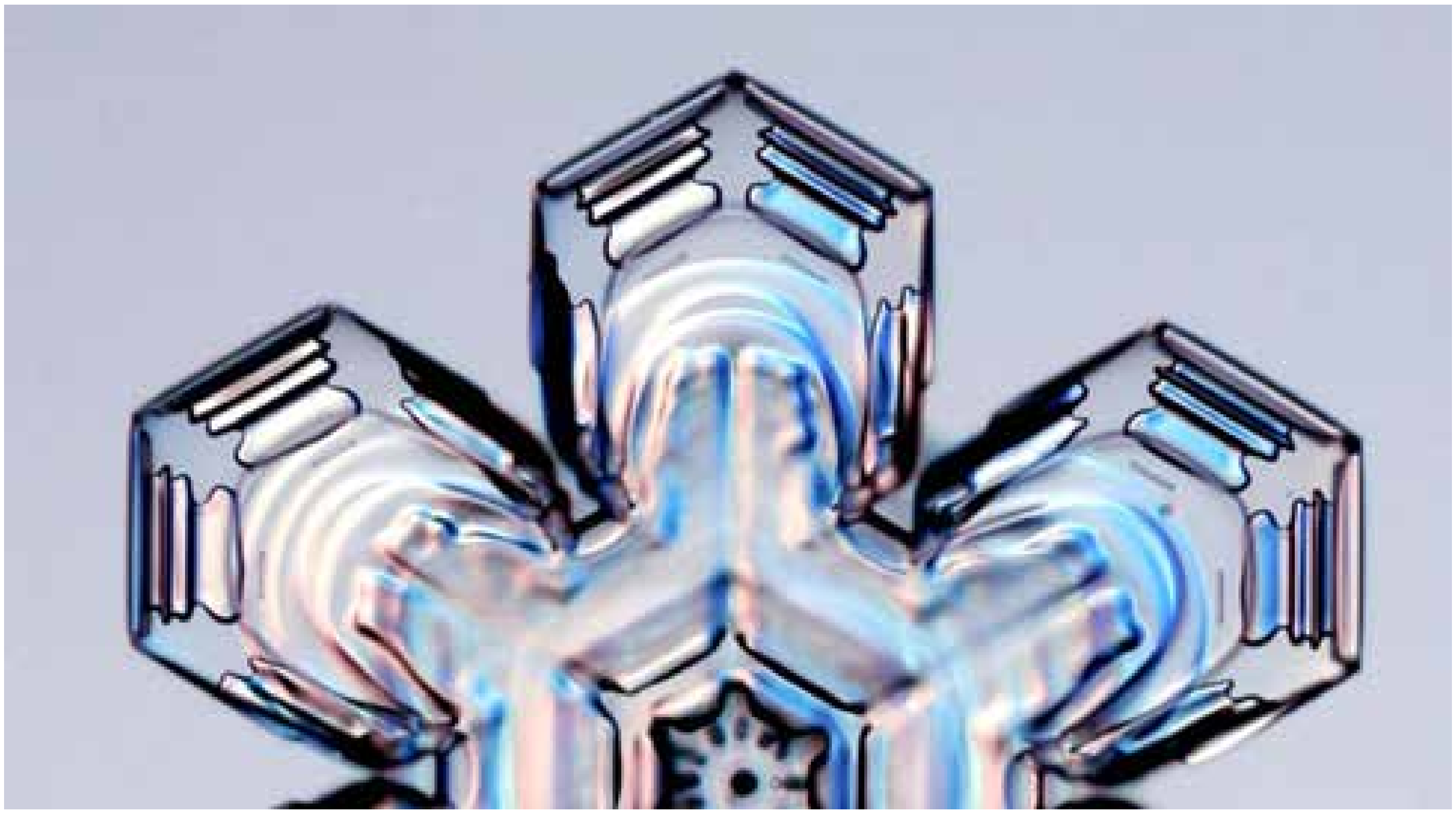

When you begin to look closely, hollows and bubbles can be found on many plate-like snow crystals, as in these two examples.

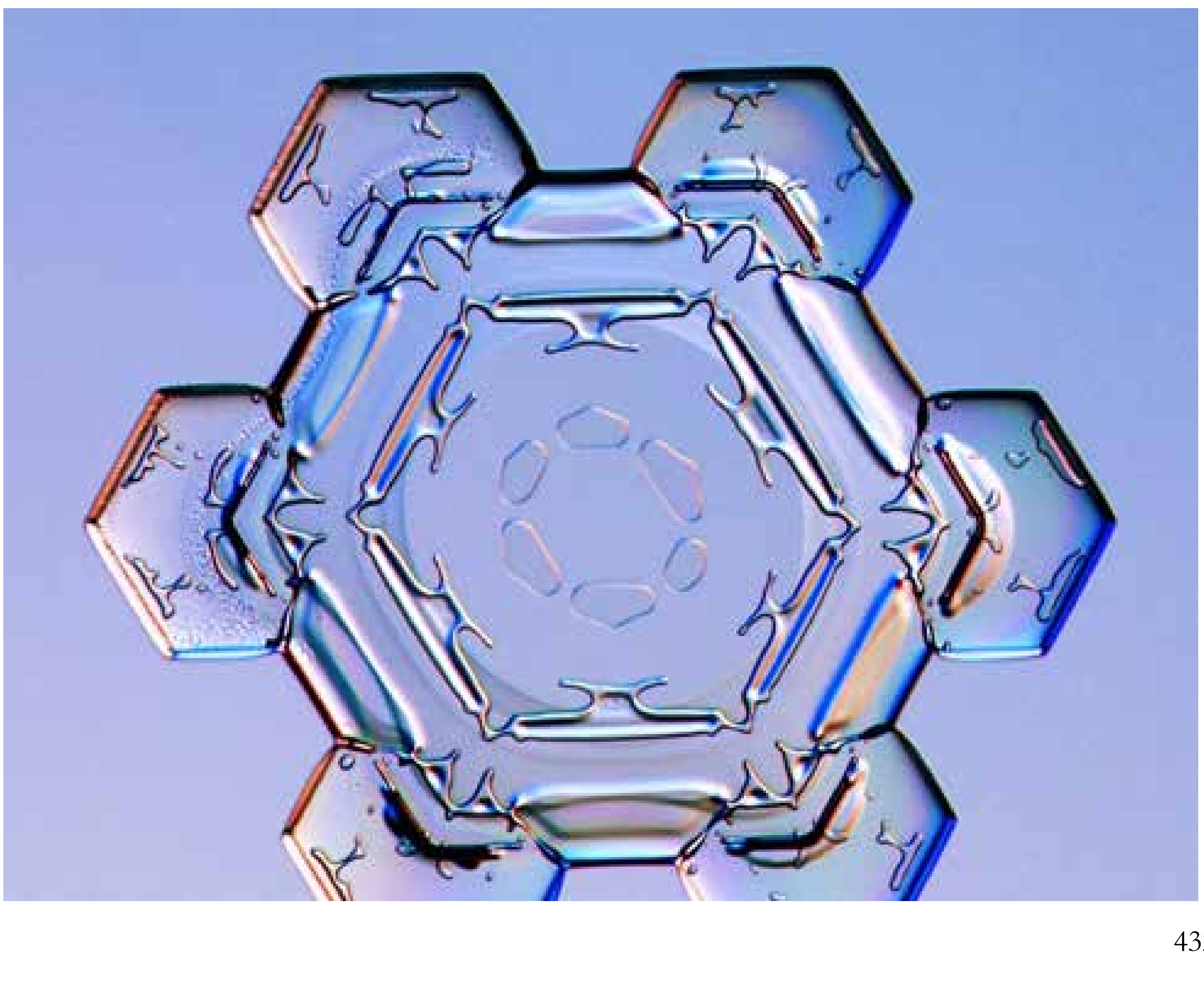

## Skeletal Forms

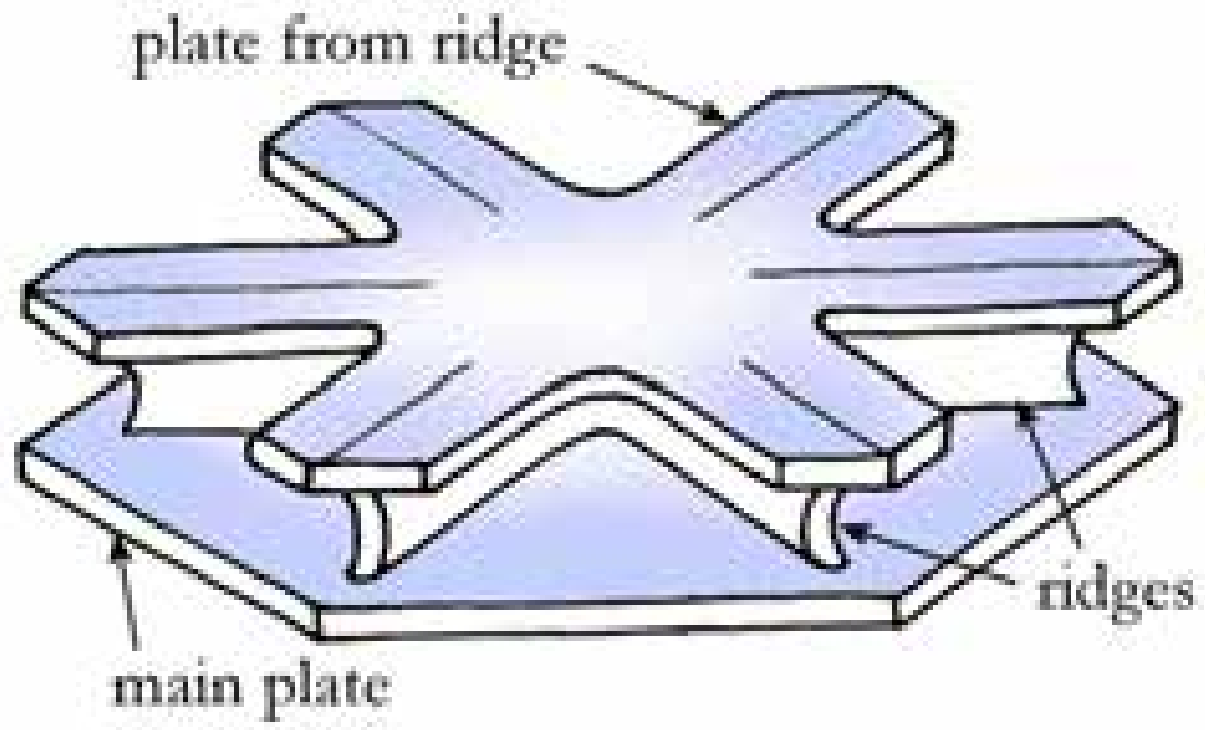

*This category refers to the formation of thick ridges on a basal surface followed by secondary plates growing out from the ridges. The quite distinctive "I-beam" structures that result are a fairly common morphological feature that can be found in many snow crystals.*

While thin ridges are the defining feature of sectored-plate snow crystals, thick ridges are the basis for skeletal forms. In both cases, the underlying physical phenomenon is the spontaneous appearance of ridge structures on convex basal surfaces, which is described in Chapters 3 and 8. Both thin- ridge and thick-ridge phenomena are clearly seen over a broad range of conditions in Figure 8.16, showing that they grow under constant growth conditions (in contrast to capped columns, for example, which cannot form in constant conditions).

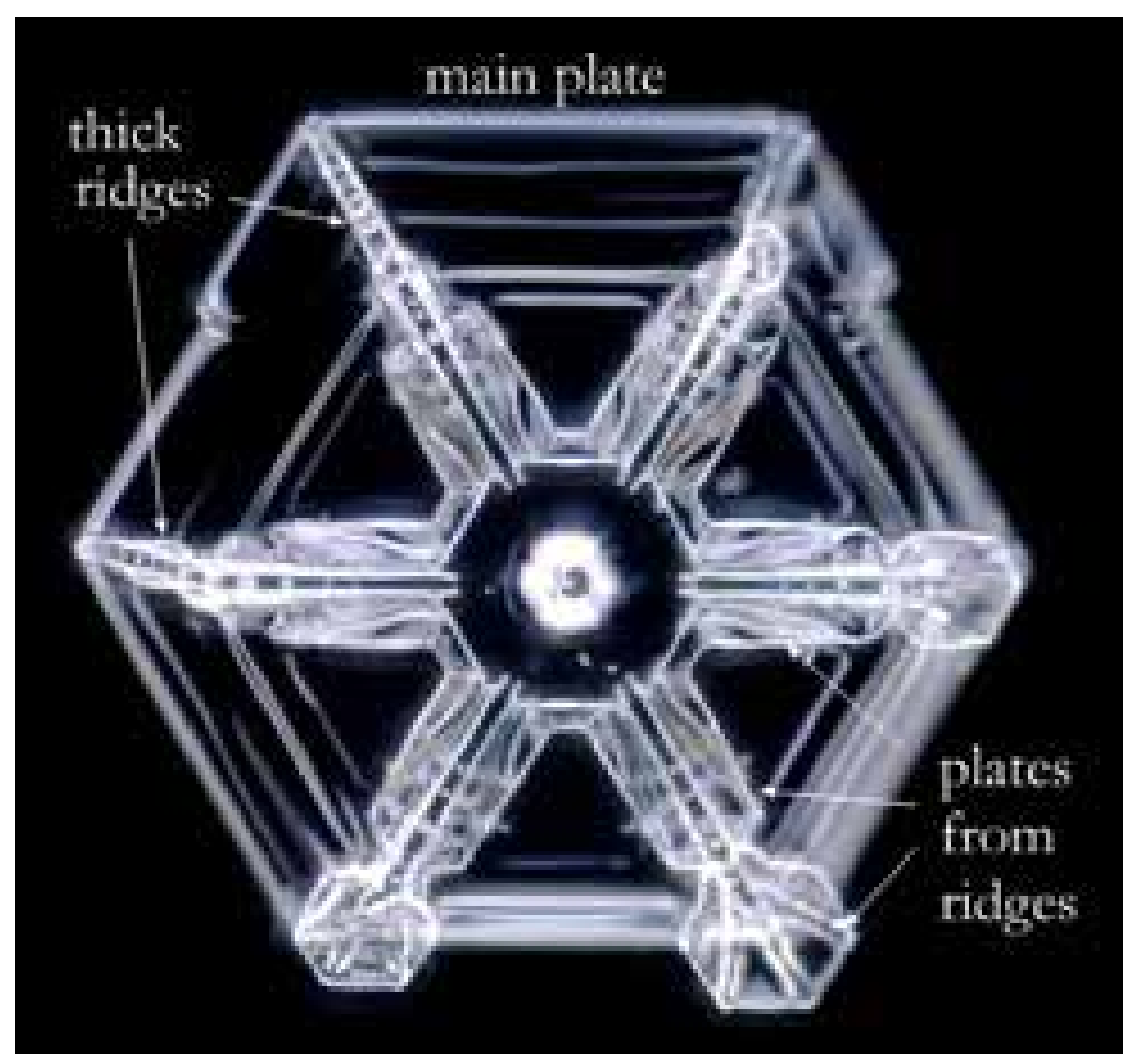

When circumstances are right, plate-from-rib skeletal forms can be remarkably common.

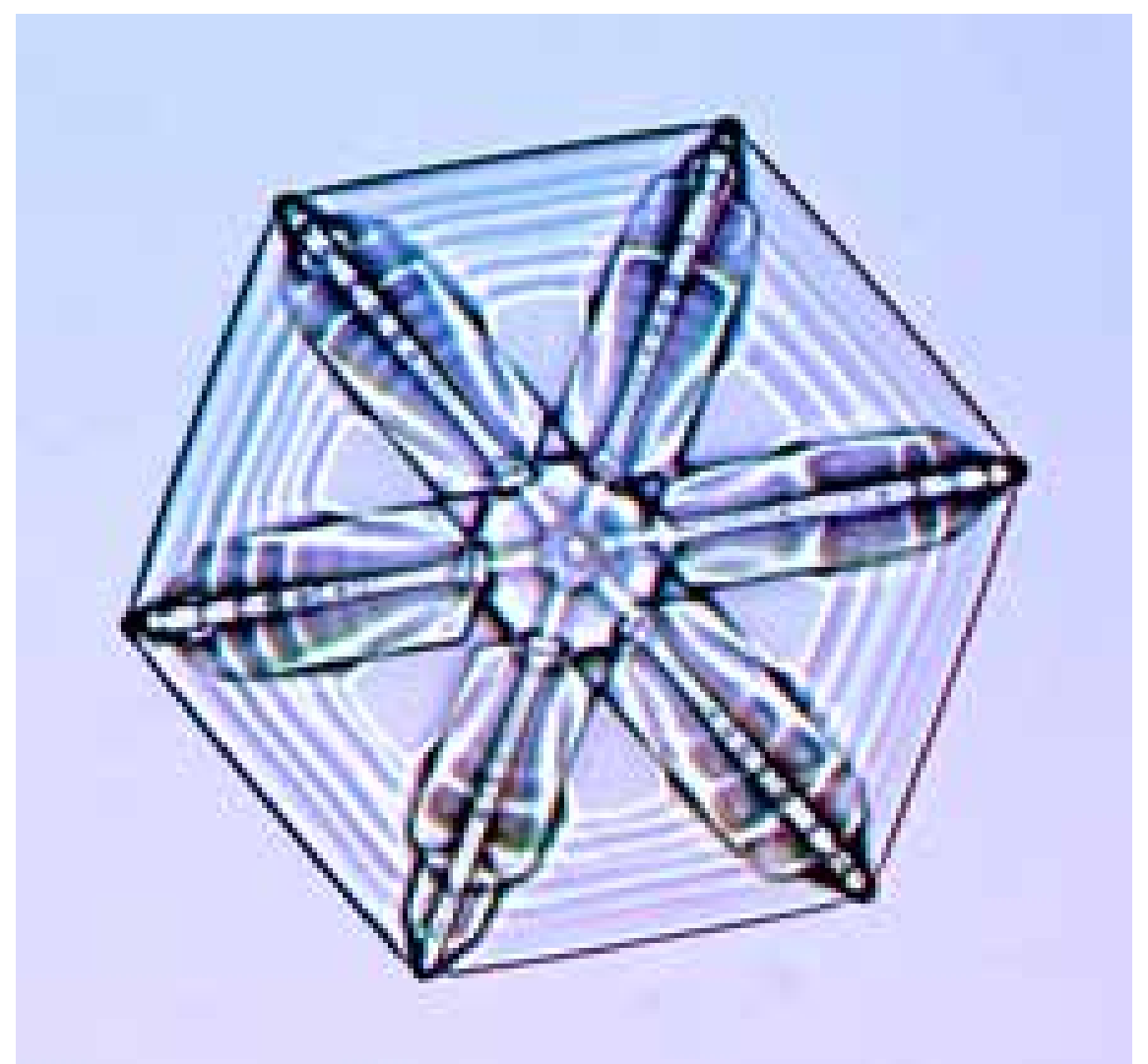

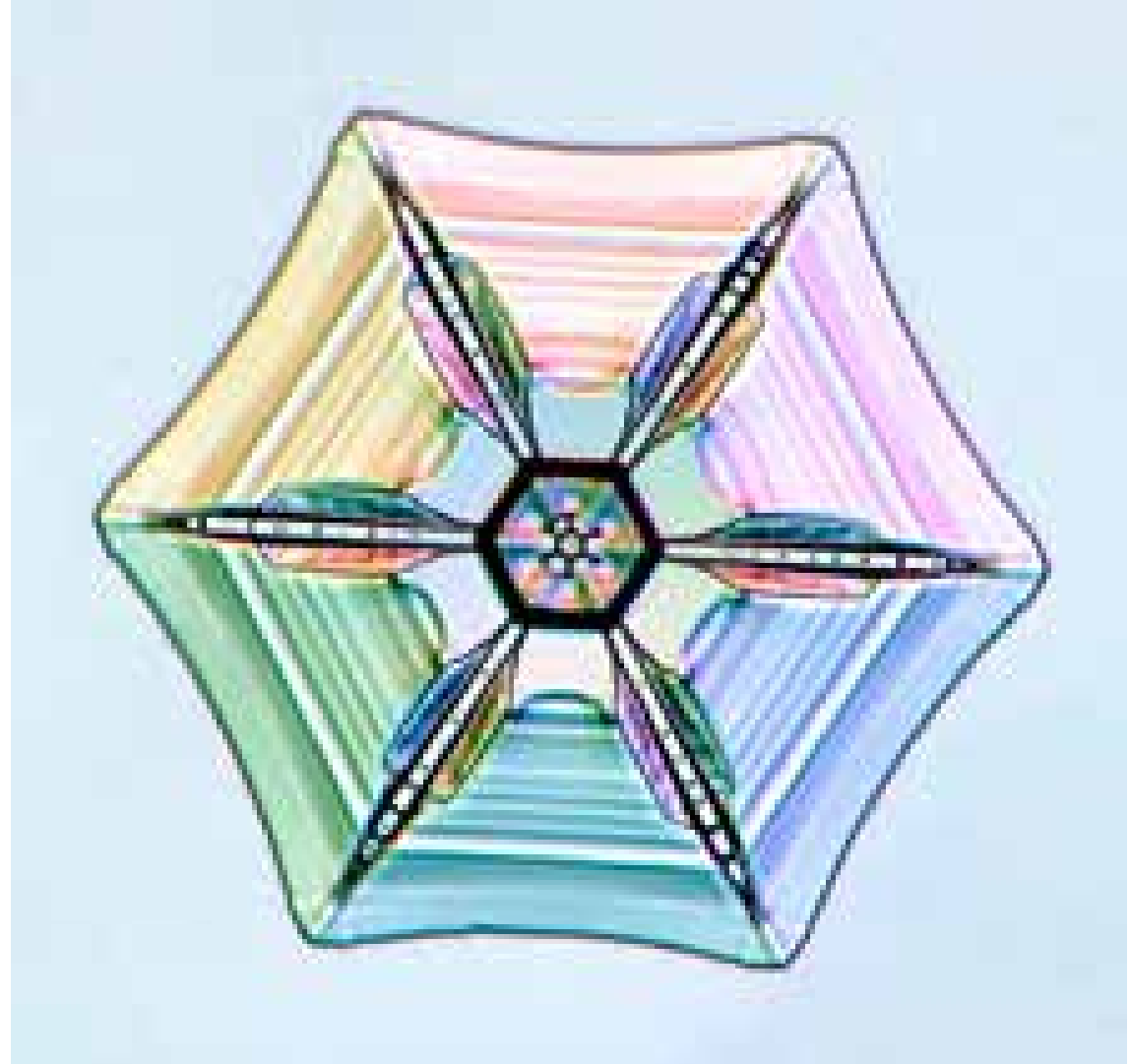

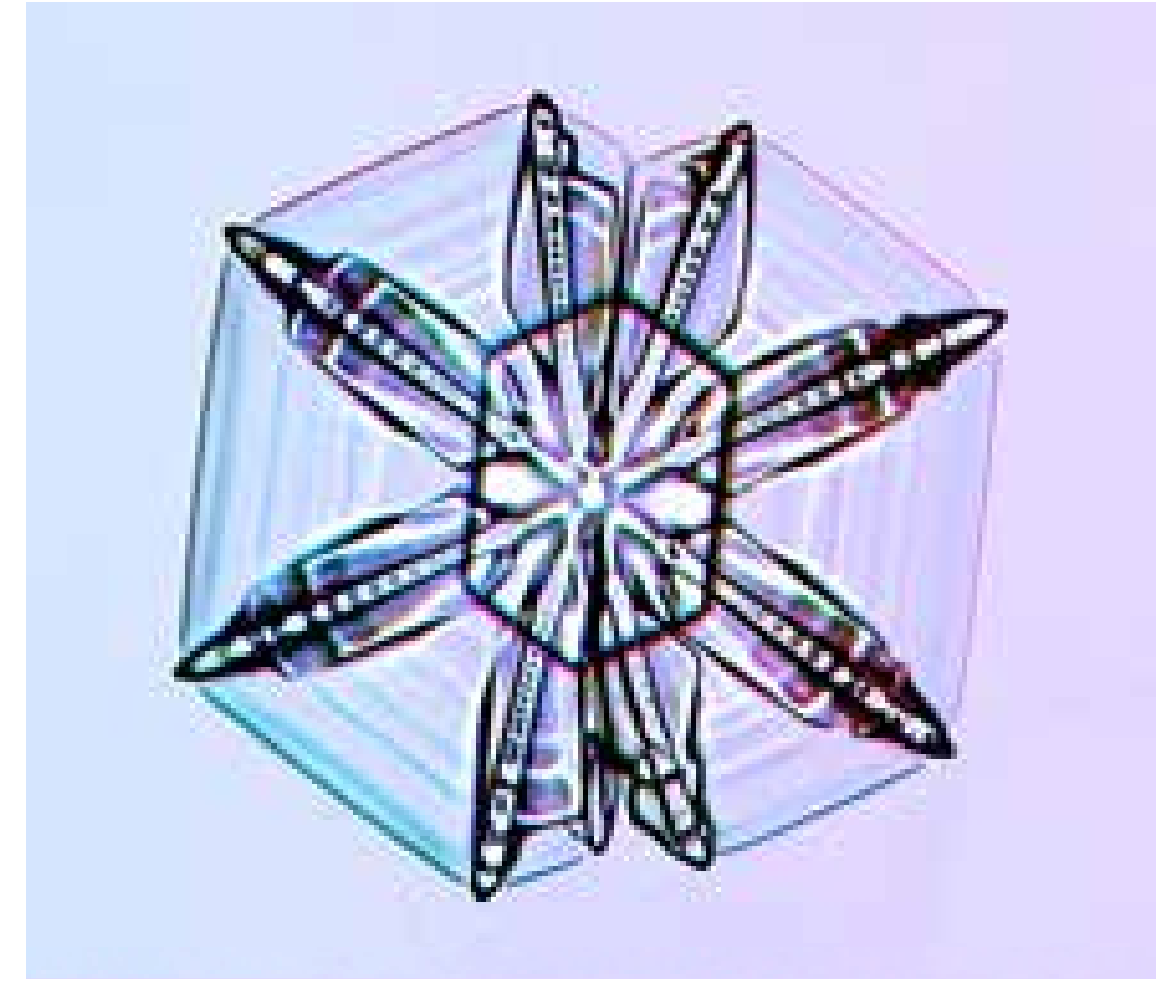

# Columns on Plates

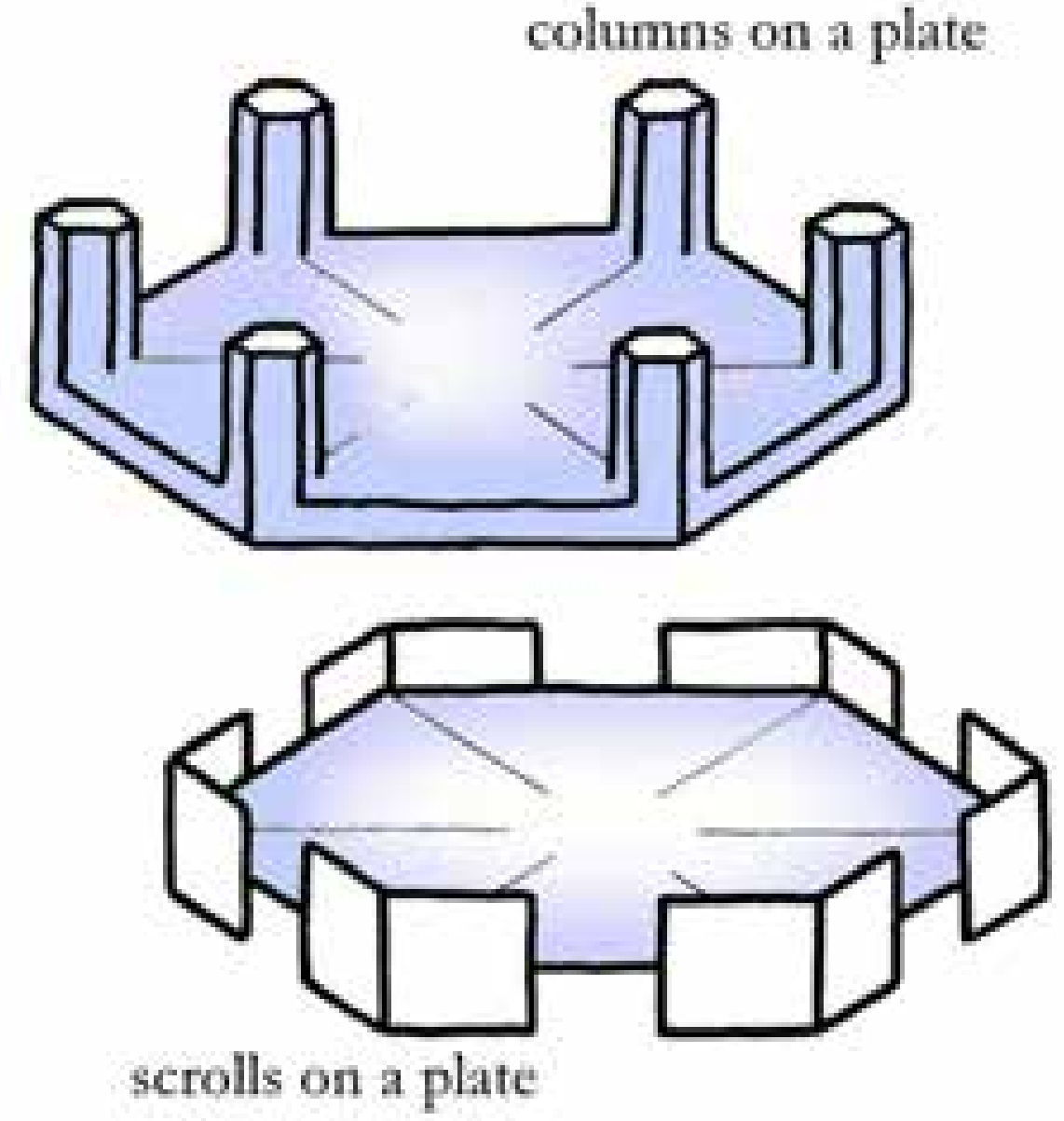

*This category includes crystals for which plate-like growth was followed by columnar growth, which is essentially the opposite of capped columns. As shown in the sketches, the columnar growth can take the form of simple columns (top), or sections of hollow columns called scrolls (bottom). Unlike capped columns, simple examples of these forms are exceedingly rare.*

A typical cooling cloud may transition through -2 C (not cold enough to freeze droplets) to -6 C (droplets start to freeze, columns form), to -15 C (plates form), and this common behavior can yield capped columns. Weather scenarios that produce columns after plates are unusual, so any kind of column-on-plate growth behavior is quite rare.

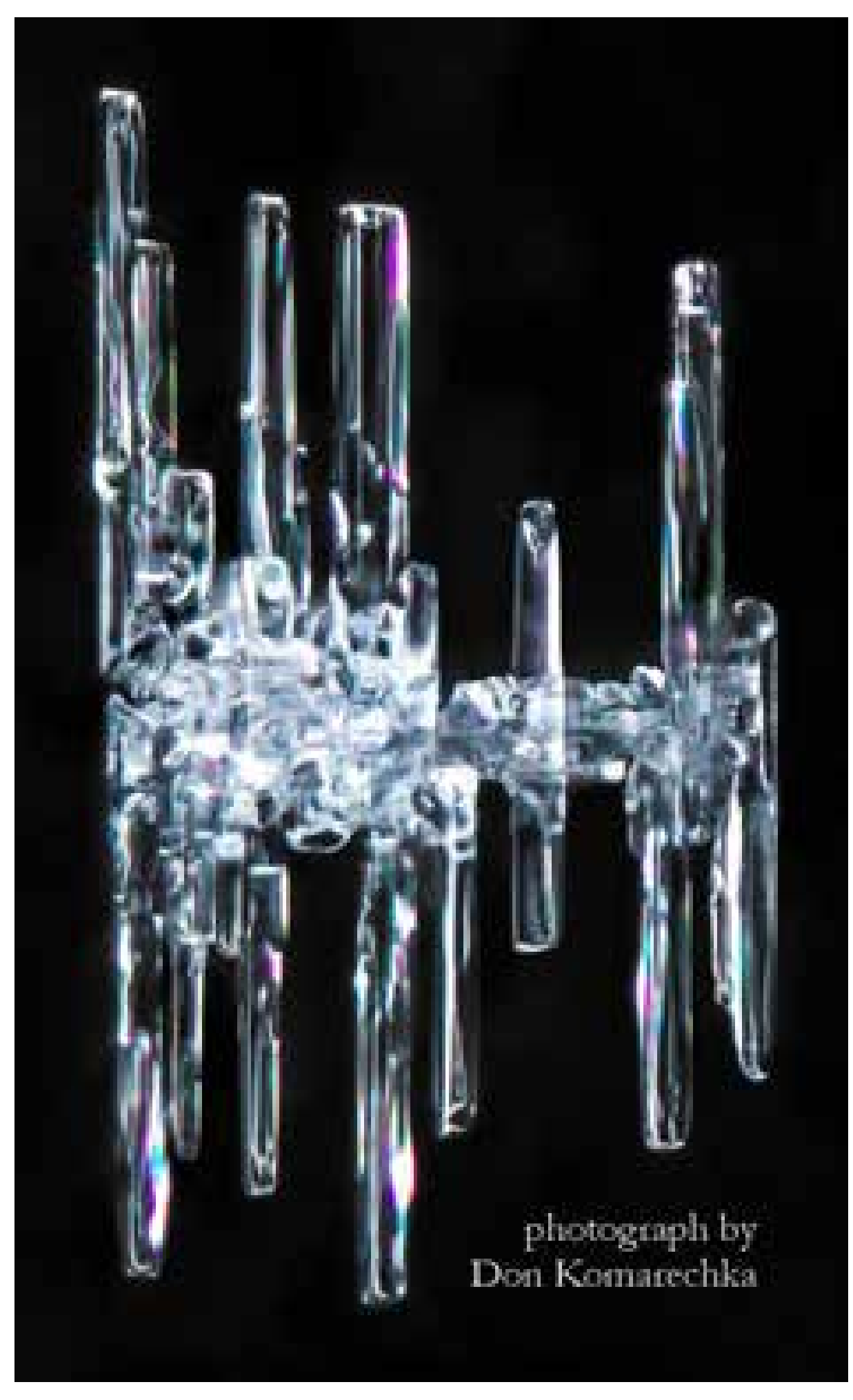

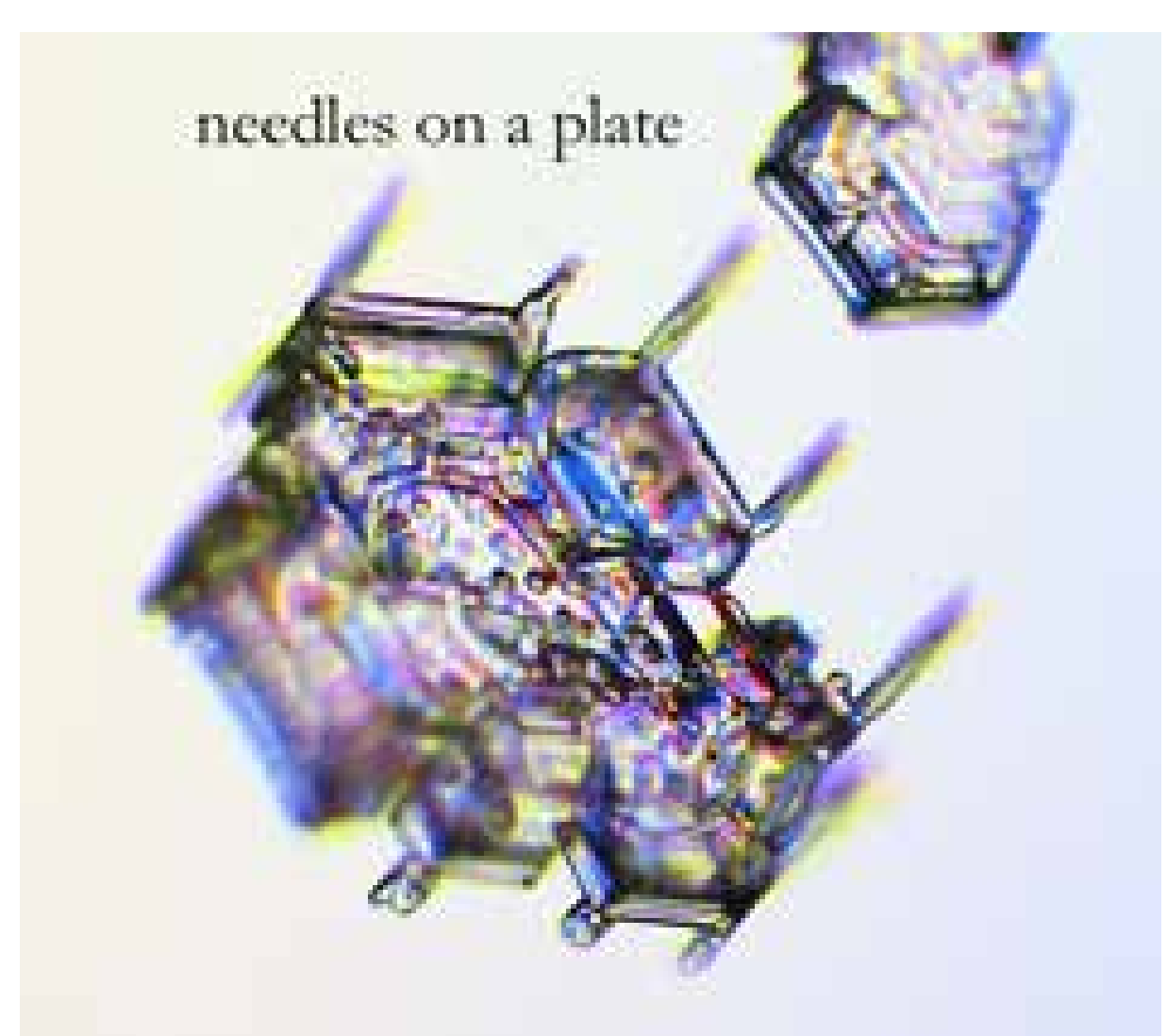

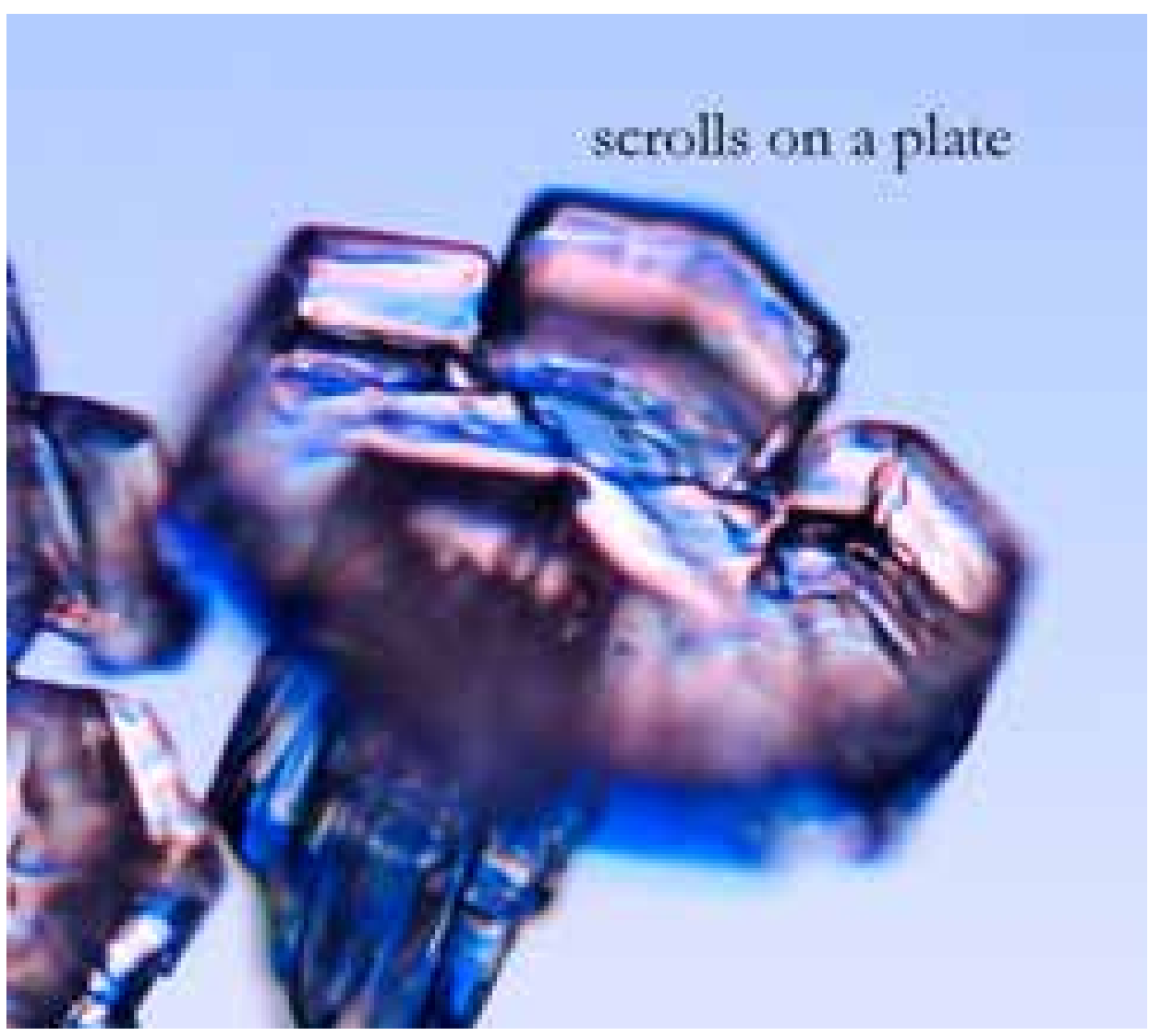

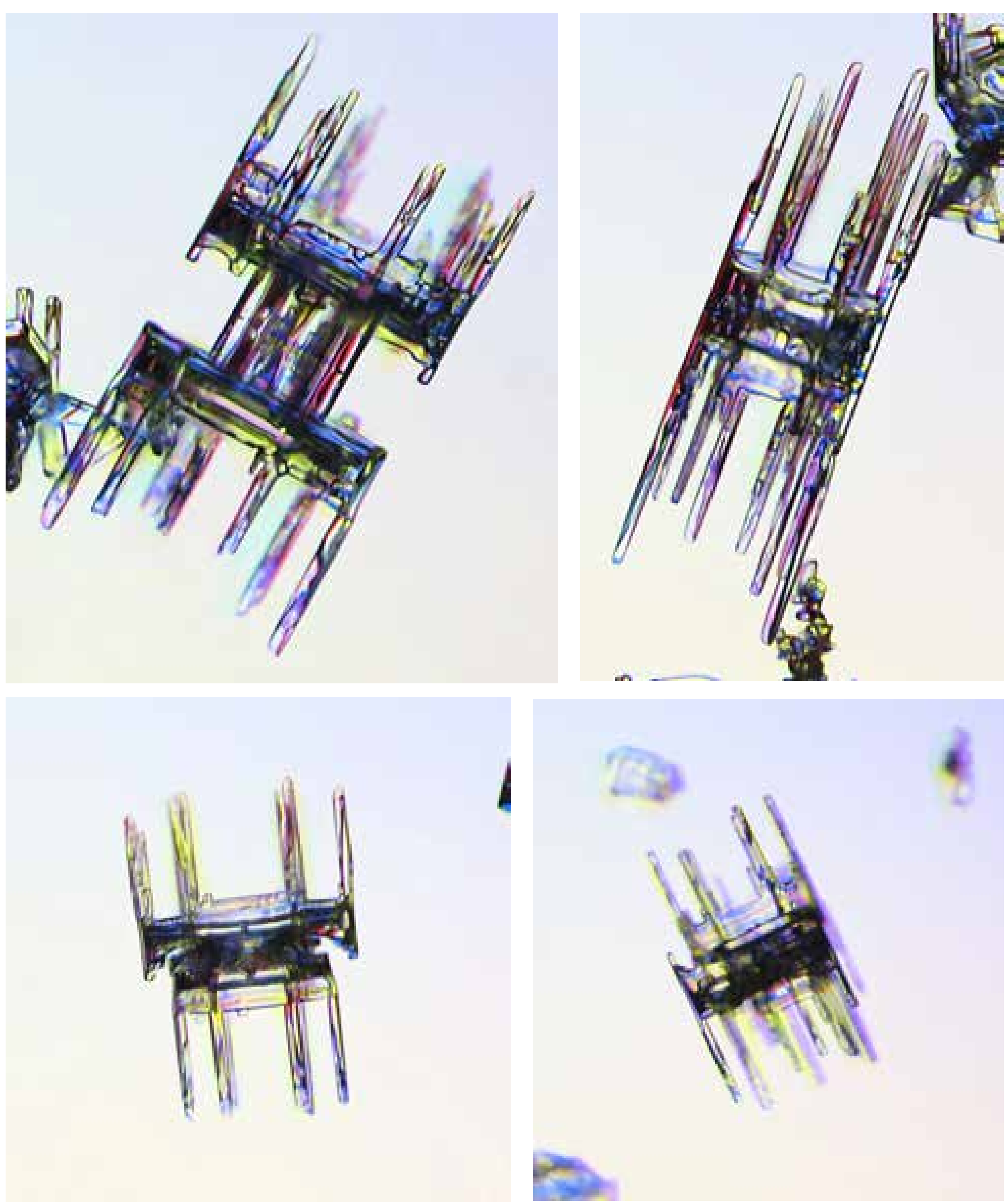

**Locally Abundant.** In birdwatching, rare but "locally abundant" birds are generally hard to find, except when they are all around you. Snow crystals can be the same way. When a snowfall produces just the right conditions, ordinarily rare crystals can be quite abundant, at least for a short while. I photographed all the odd column-on-plate crystals on this page during a 20-minute period in Fairbanks, Alaska, when the temperature was near -5 C. One never knows what the clouds may deliver.

# Triangular Crystals

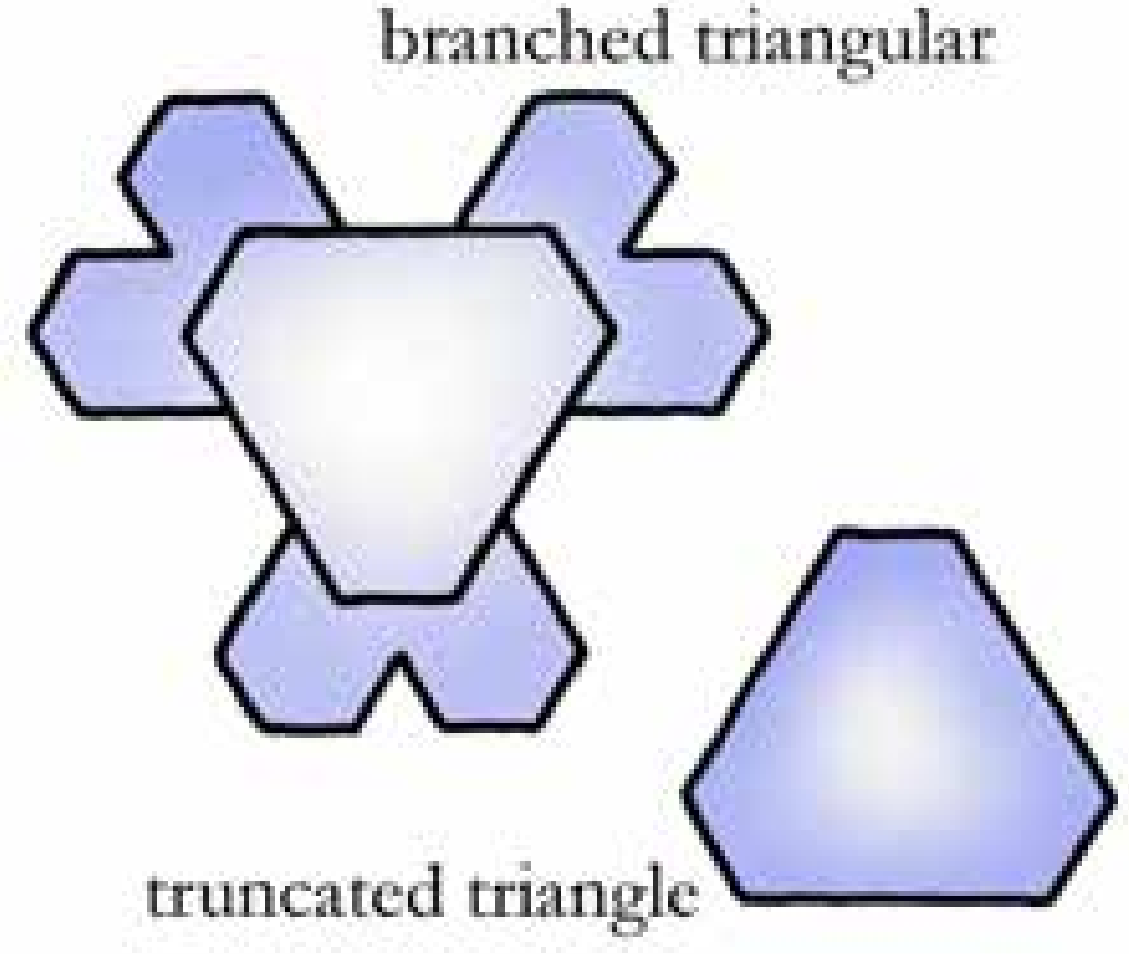

*Triangular snow crystals display an overall three-fold symmetry rather than the usual six-fold symmetry. The most common shape is a truncated triangular plate, sometimes with branching. Triangular crystals are relatively rare and usually small. They are most likely to be found in warmer snowfalls, mixed in with other small plates.*

As described in Chapter 3, there is a weak growth instability that can cause a hexagonal plate to transform into a triangular shape. A slight perturbation in that direction will be amplified by diffusion-limited growth, and, once begun, the transition from hexagonal to triangular is irreversible. Exactly how and when this instability is triggered, however, is not yet well understood.

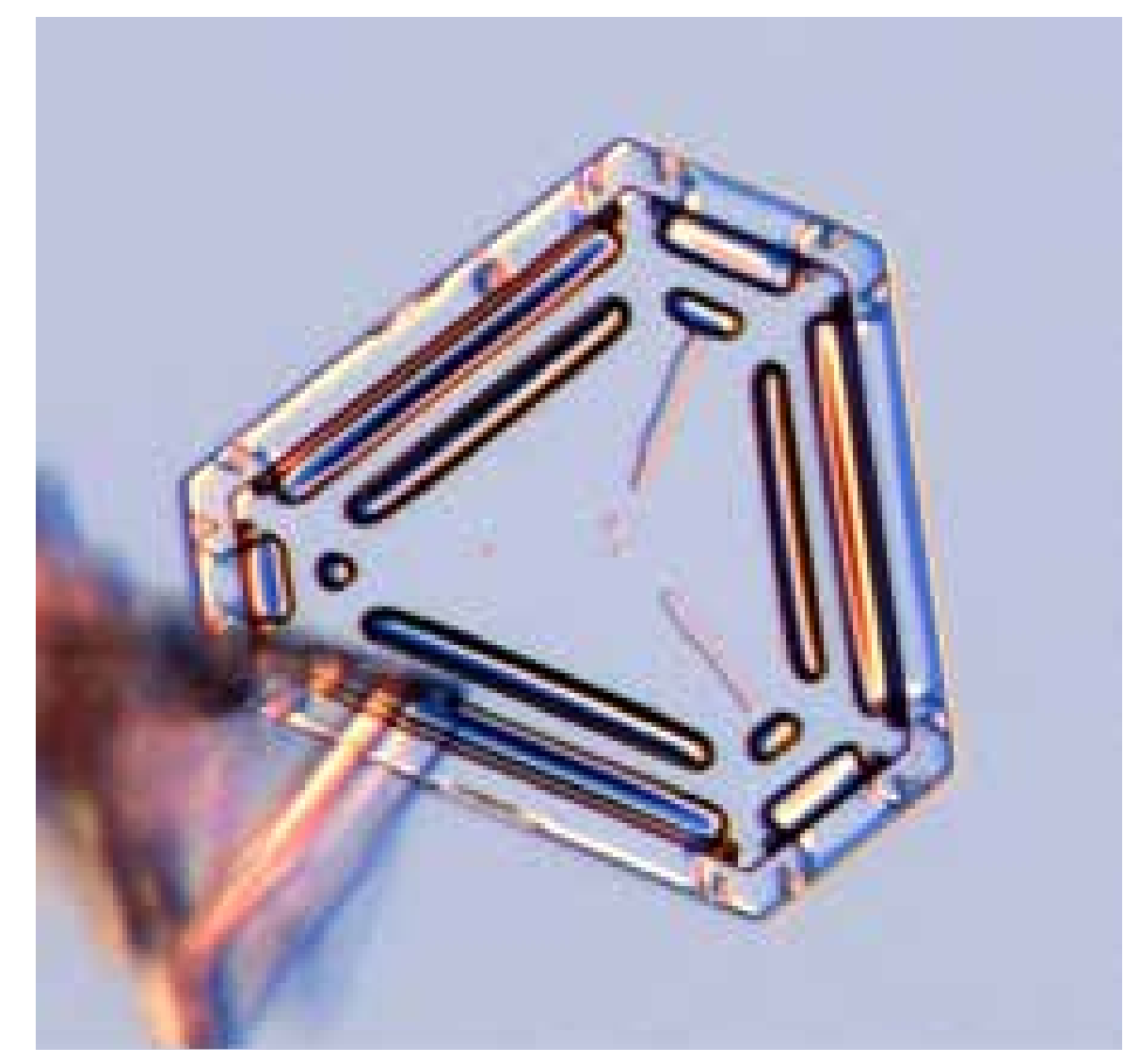

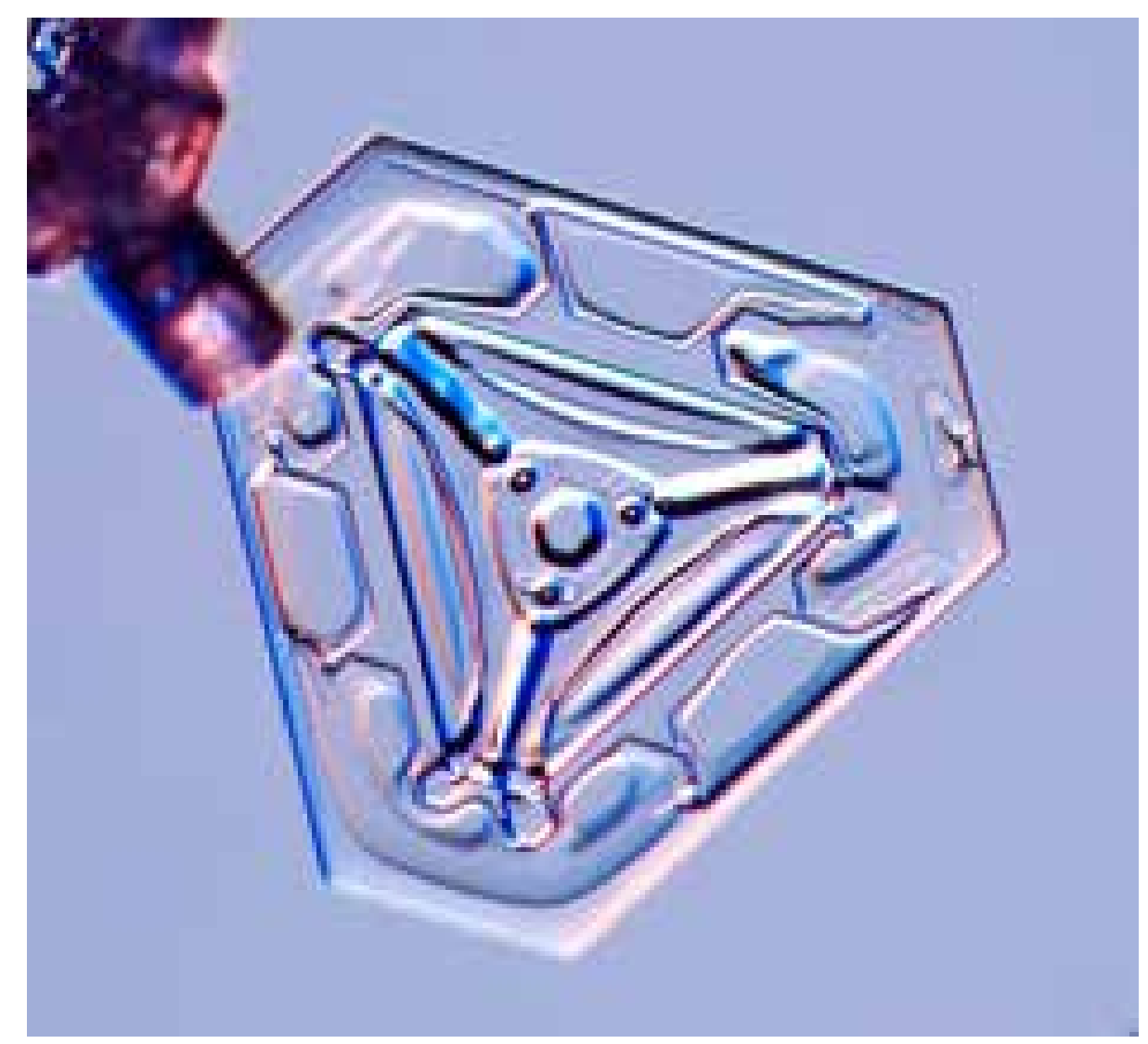

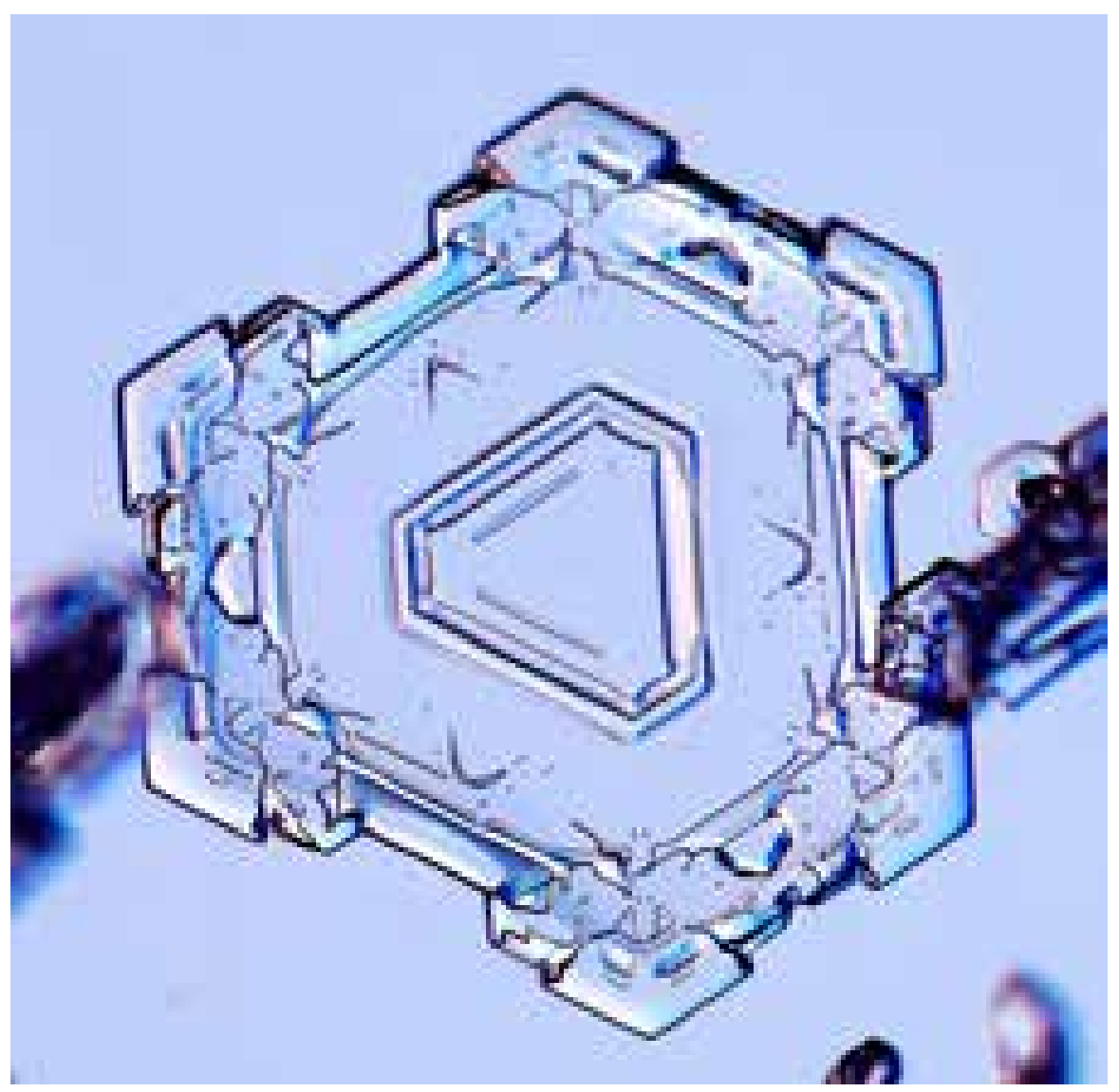

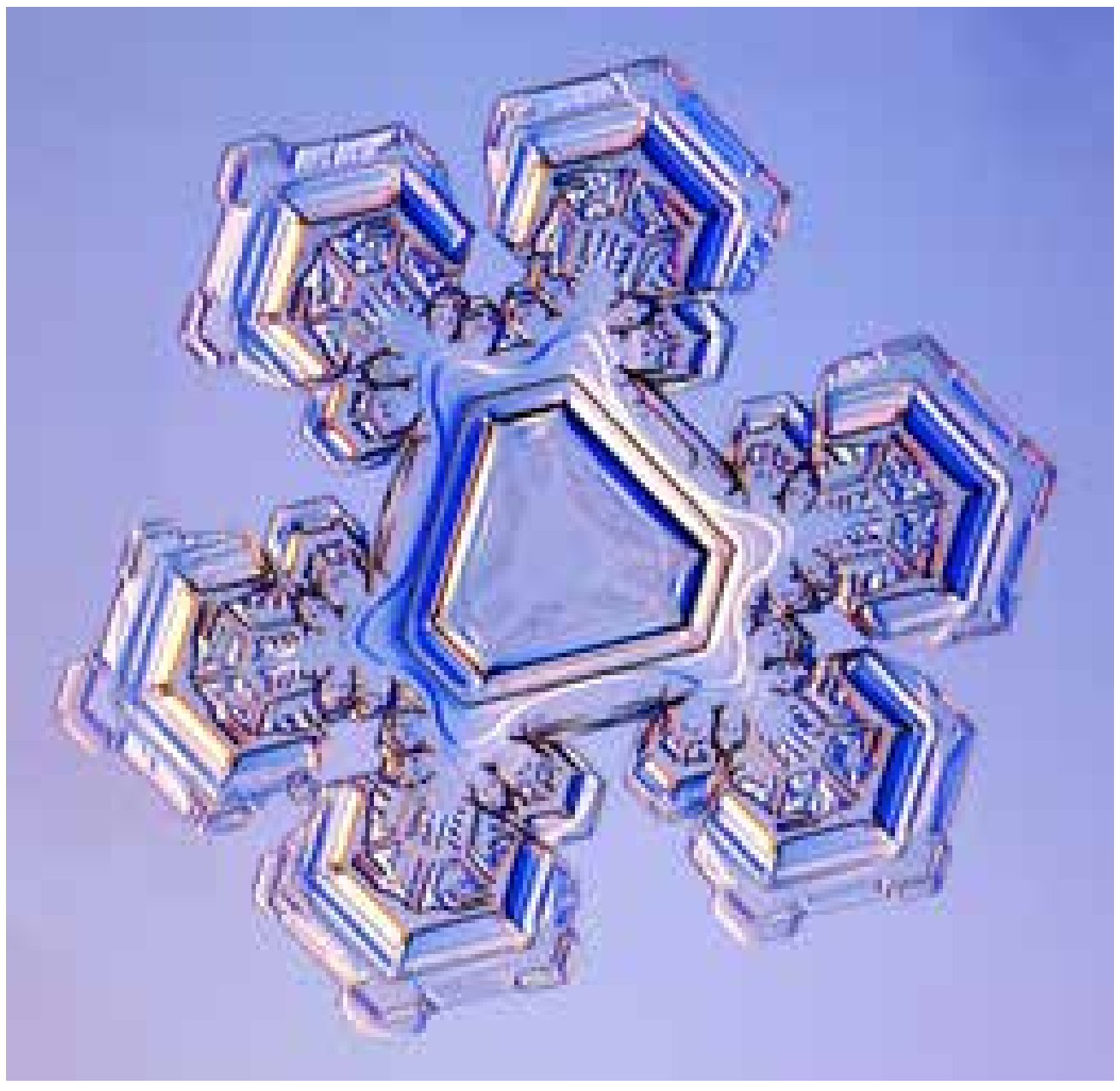

## Bullet Rosettes

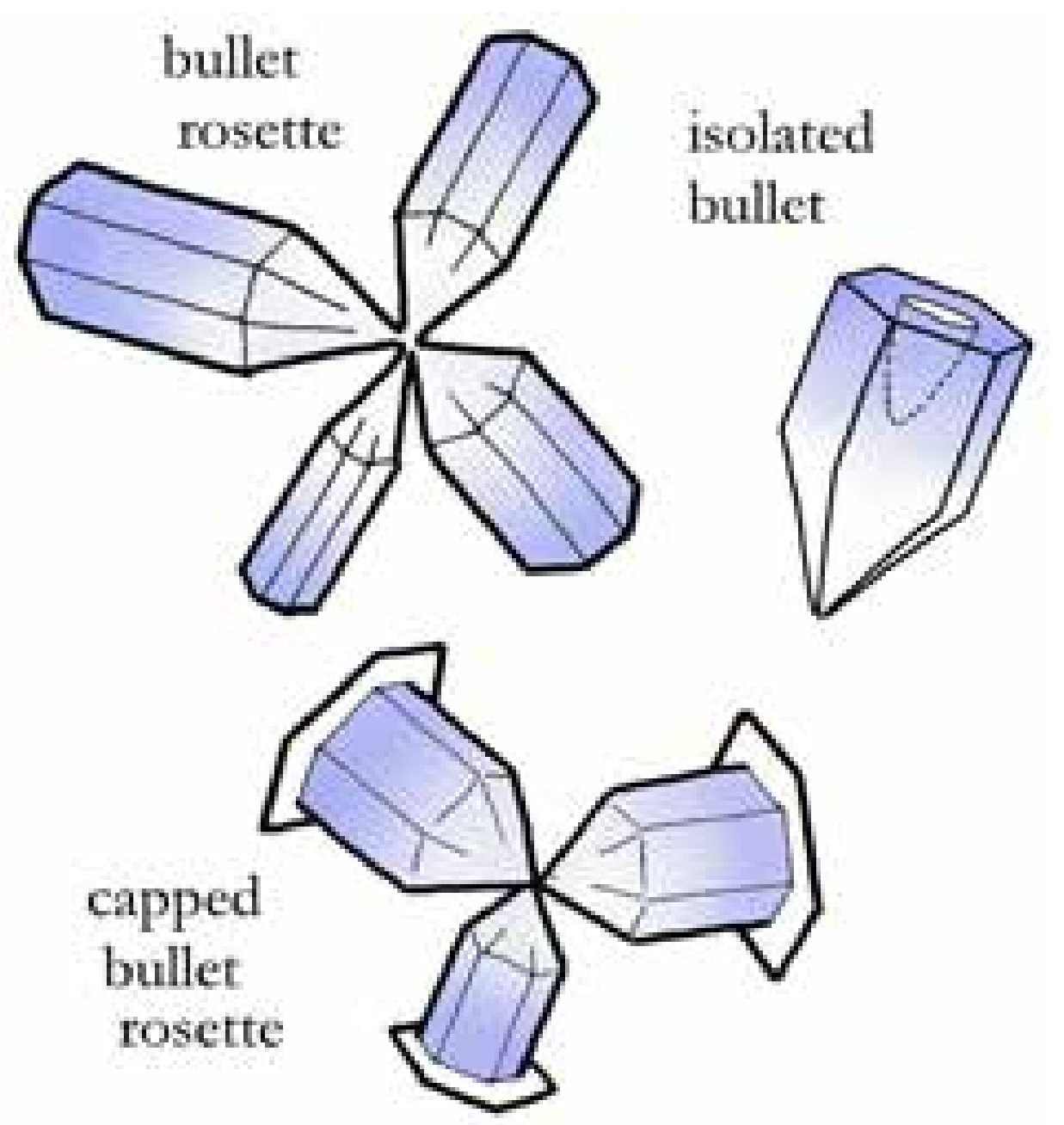

*Bullet rosettes are collection of columnar crystals that form together around a single nucleus. Competition for water vapor inhibits growth near the center, giving each column a bullet-like shape. Individual bullets come from the breakup of bullet rosettes. These snowflakes are typically found mixed with columnar crystals in warmer snowfalls. Bullet rosettes are polycrystalline forms, which means that the entire structure is made of several individual crystals that grew out from an initially polycrystalline seed.*

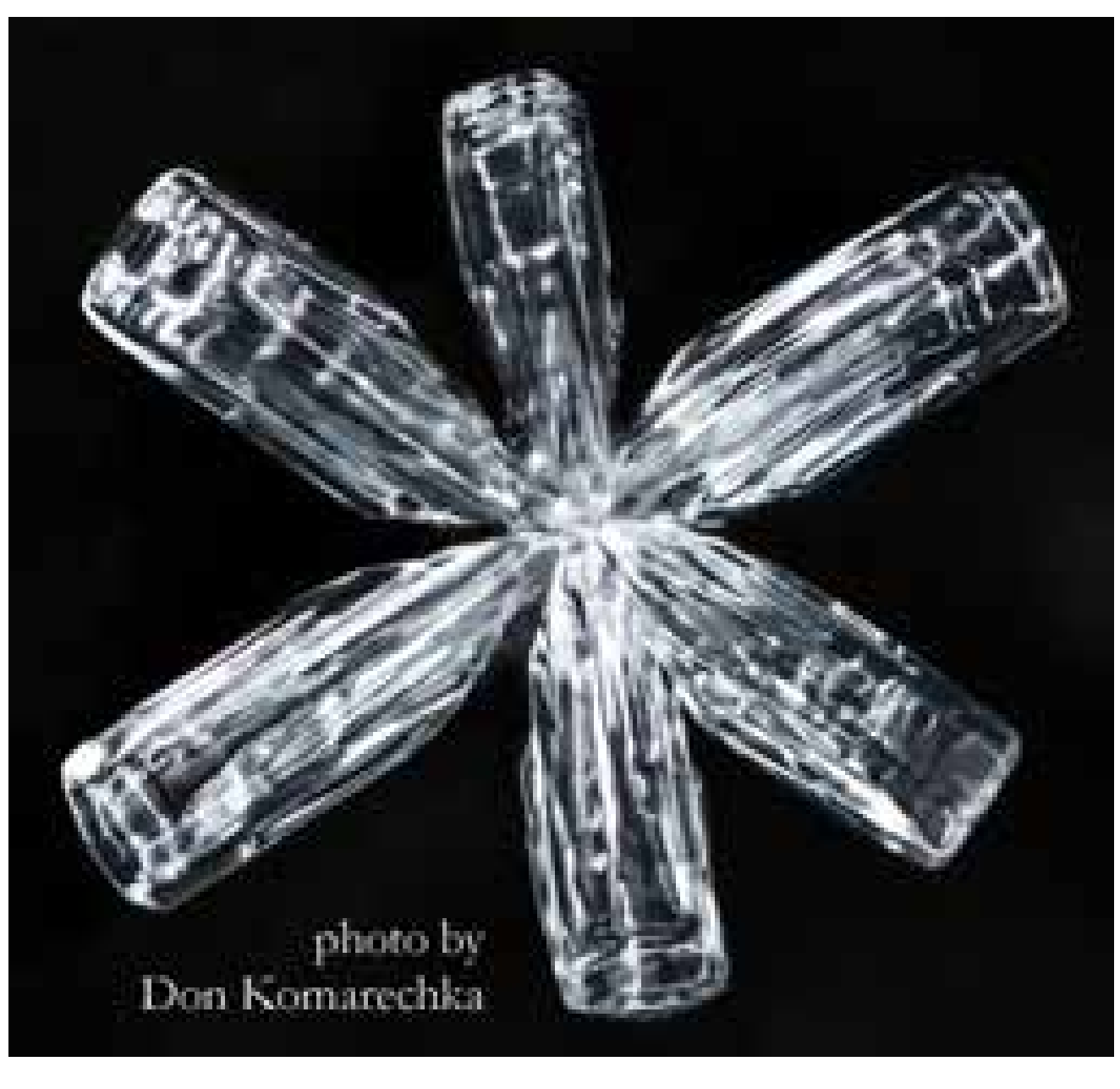

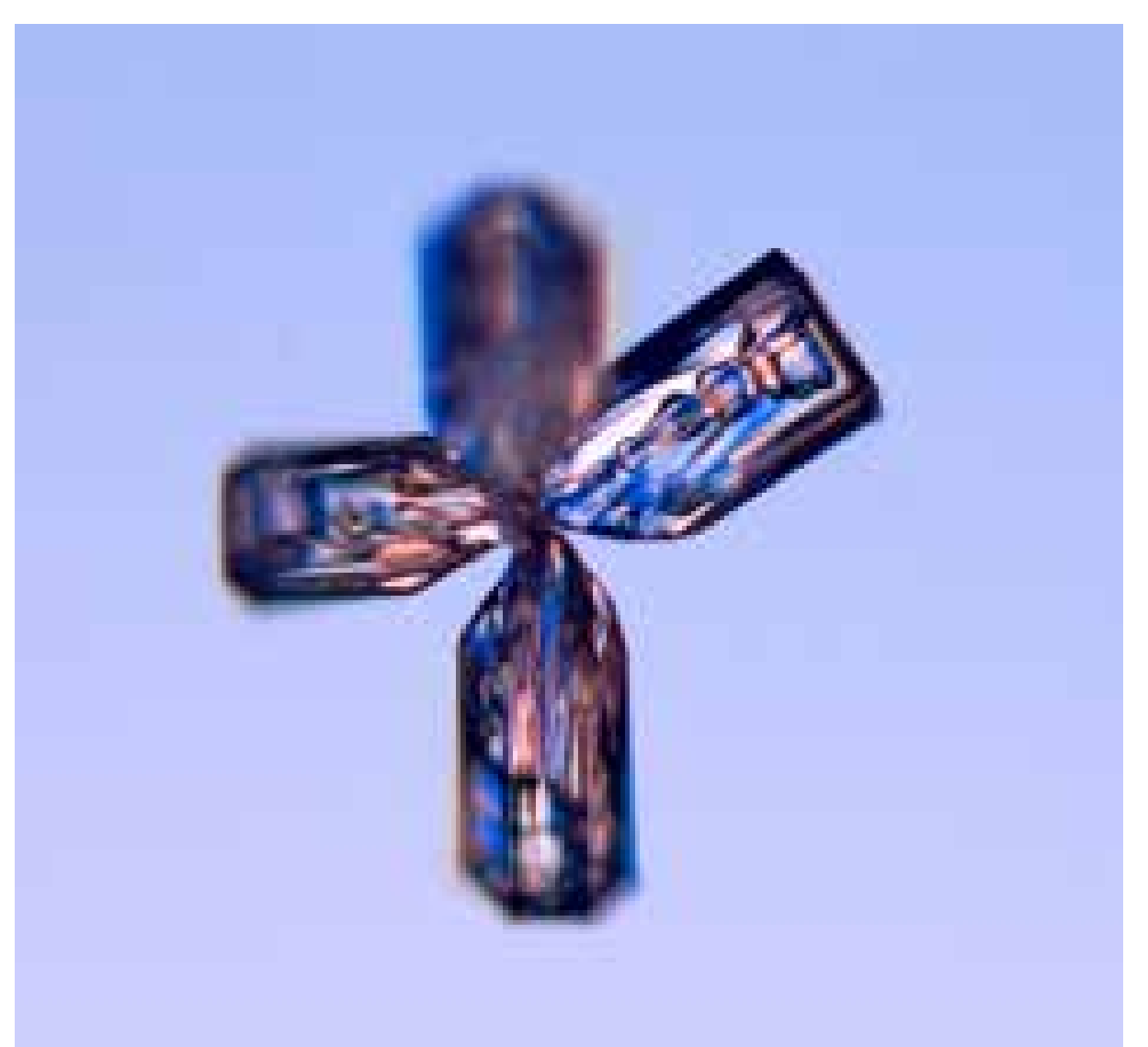

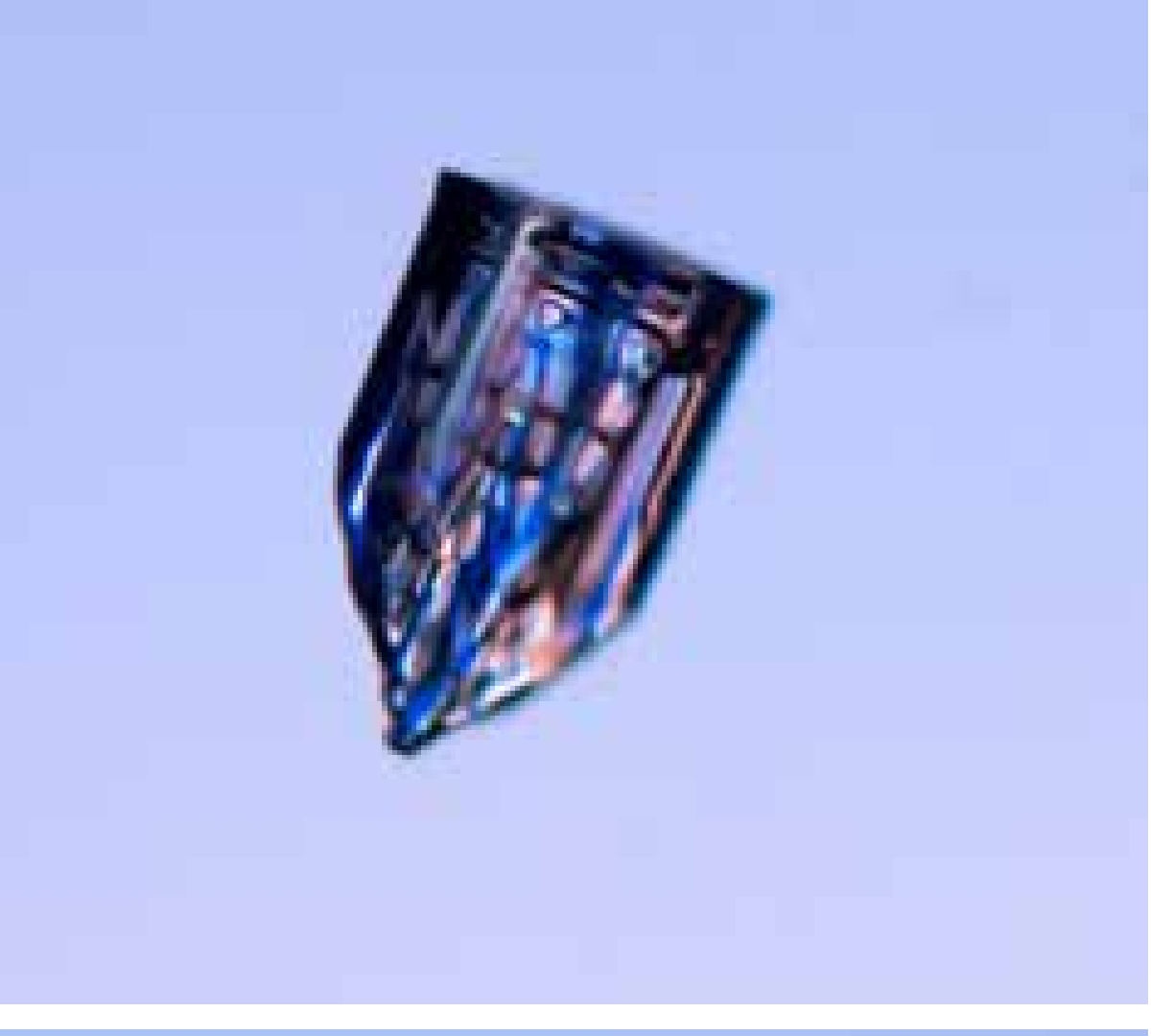

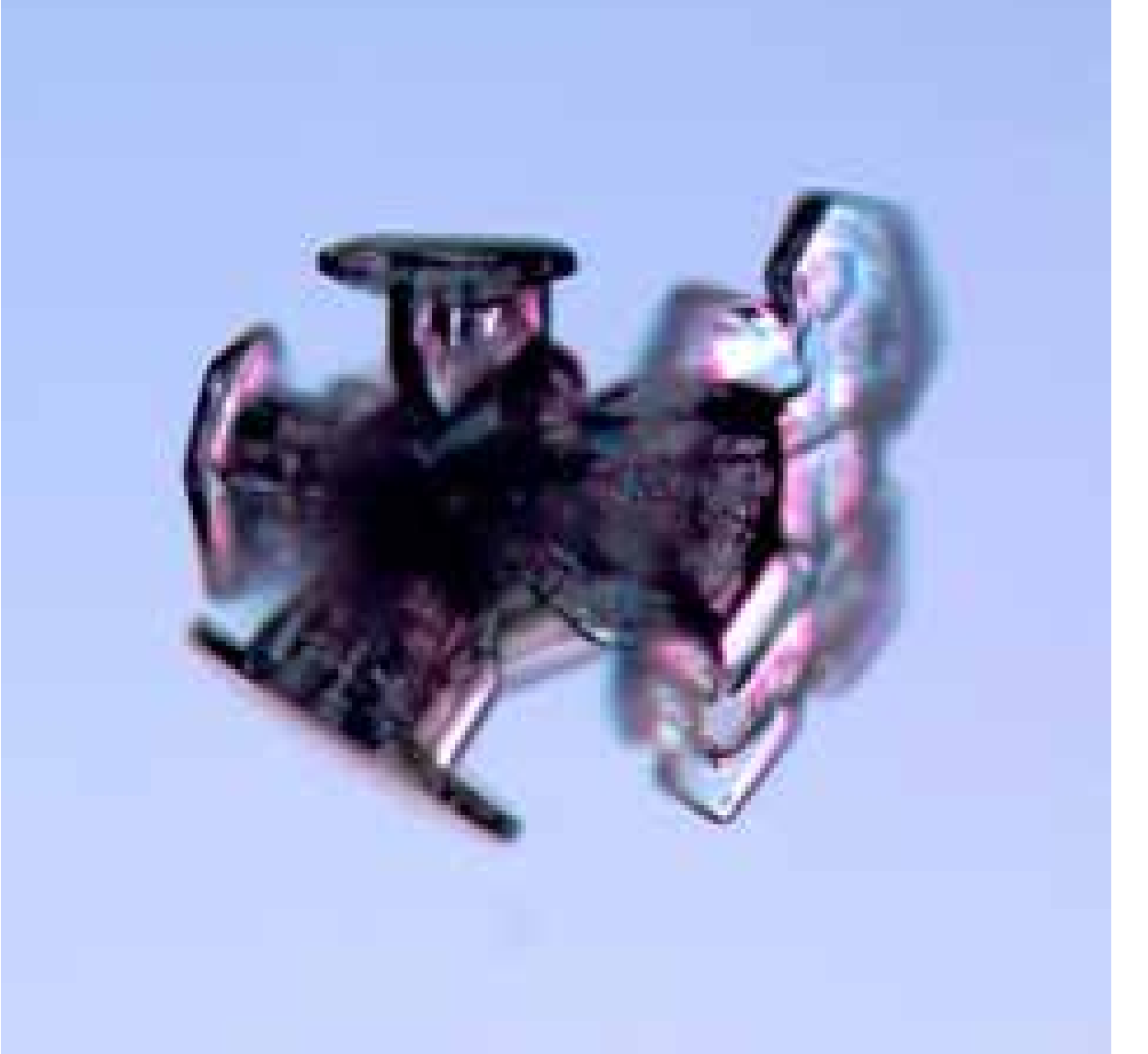

# Radiating Plates and Dendrites

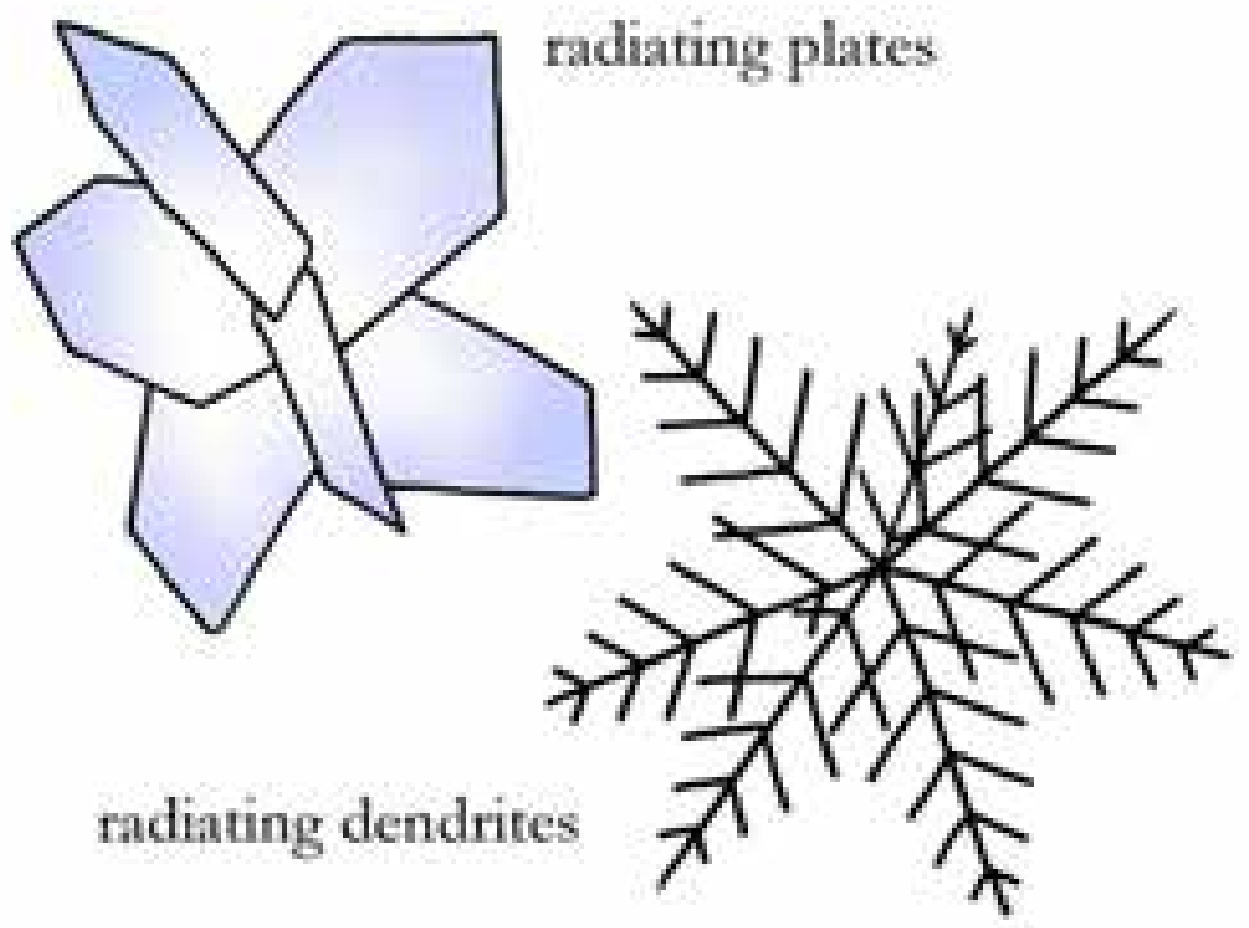

*Radiating plates and dendrites are polycrystalline forms much like bullet rosettes, except with a collection of plate-like crystals instead of columns. Typically, the different segments grow out from a common center, and their structure can be anything from simple faceted plates to fern-like dendrites. These composite structures are common and typically found mixed in with other plate-like crystals.*

Whether a cloud droplet freezes into a single ice crystal or a polycrystal depends on many factors. Larger droplets are more likely to become polycrystalline, as are highly supercooled droplets. Polycrystals can also form when particles collide and stick. The crystal at the bottom of this page probably picked up a rime droplet that froze with some random crystal orientation. This nucleated the formation of the additional branches you see growing out of the plane of the photograph.

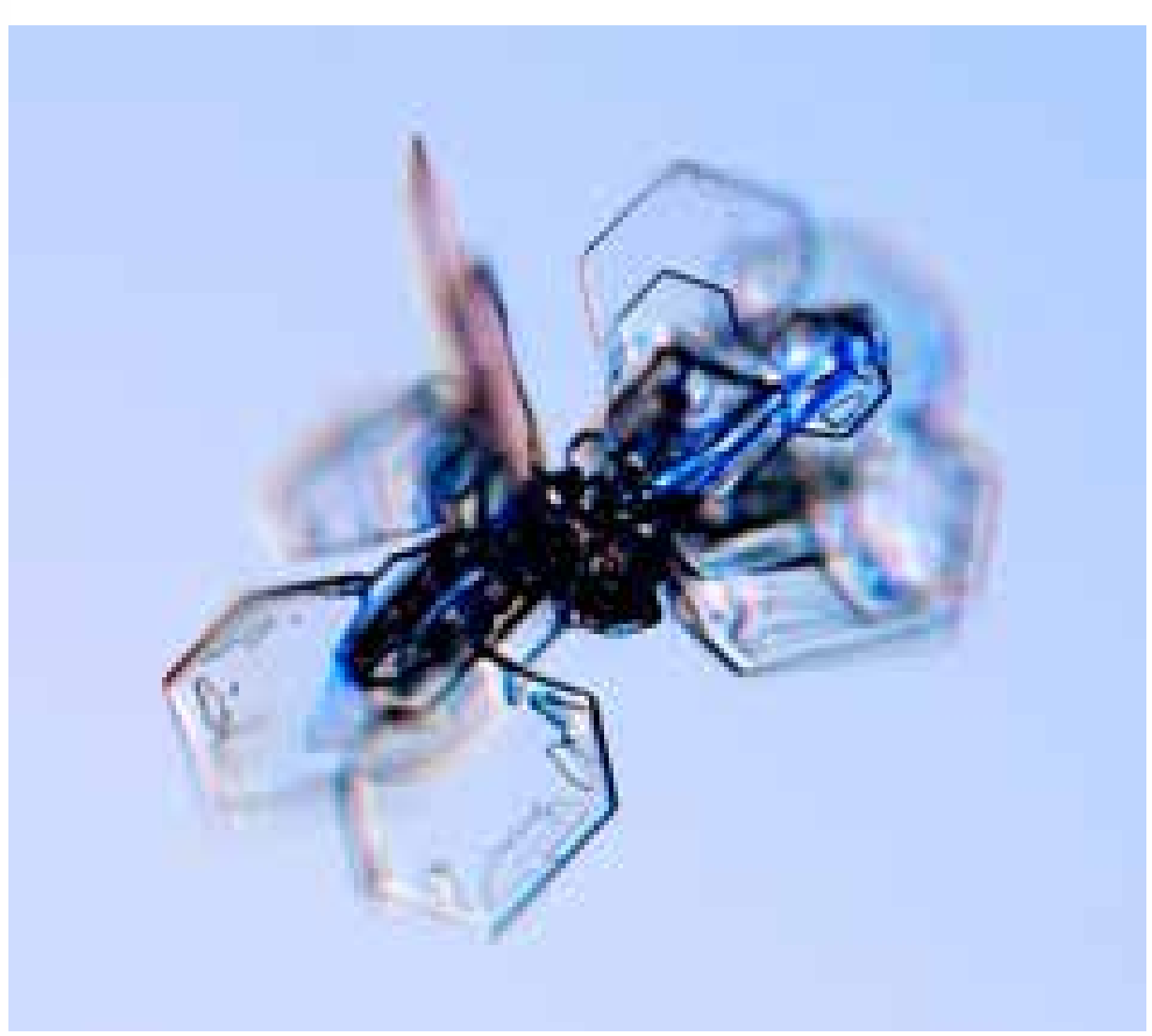

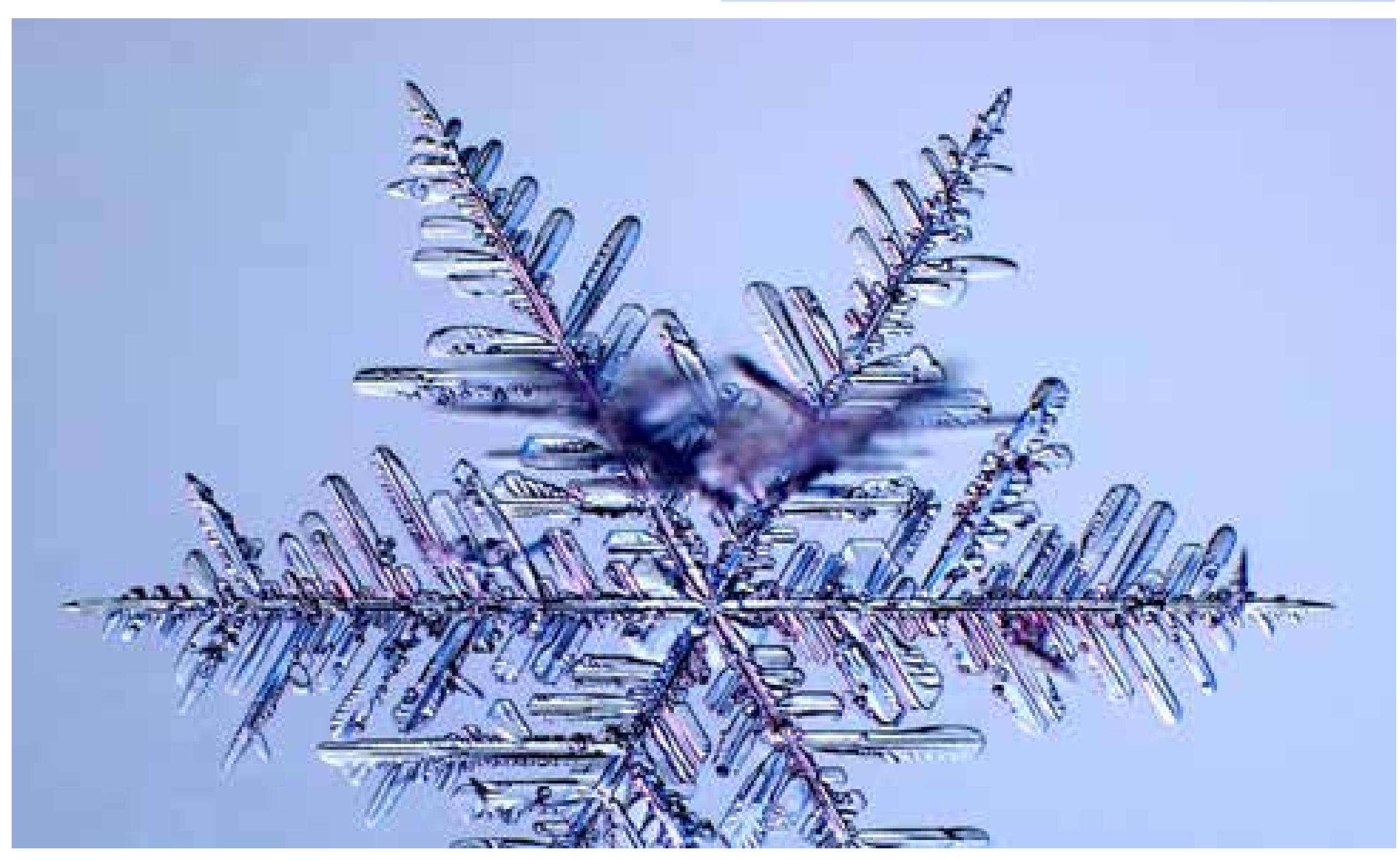

# Sheaths and Cups

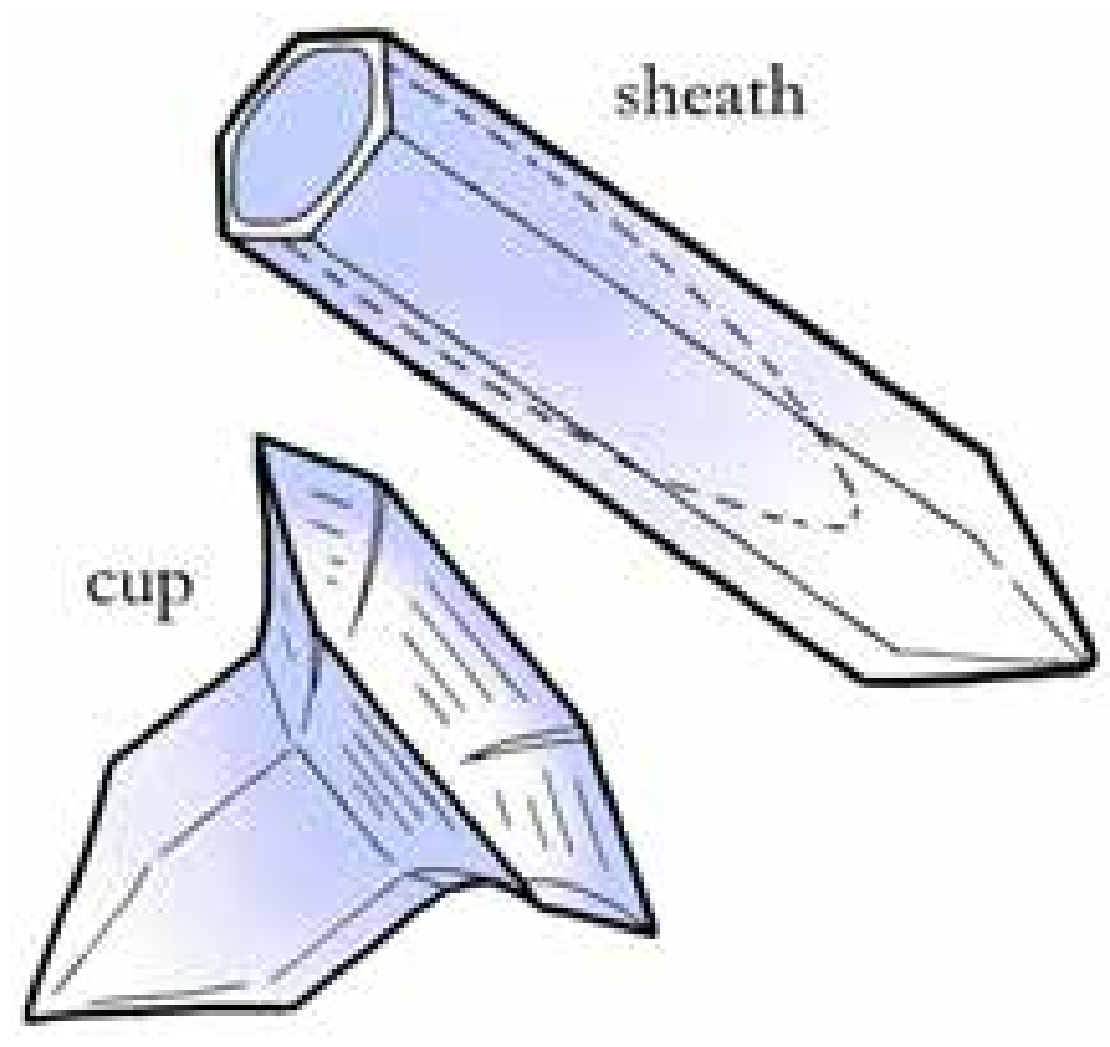

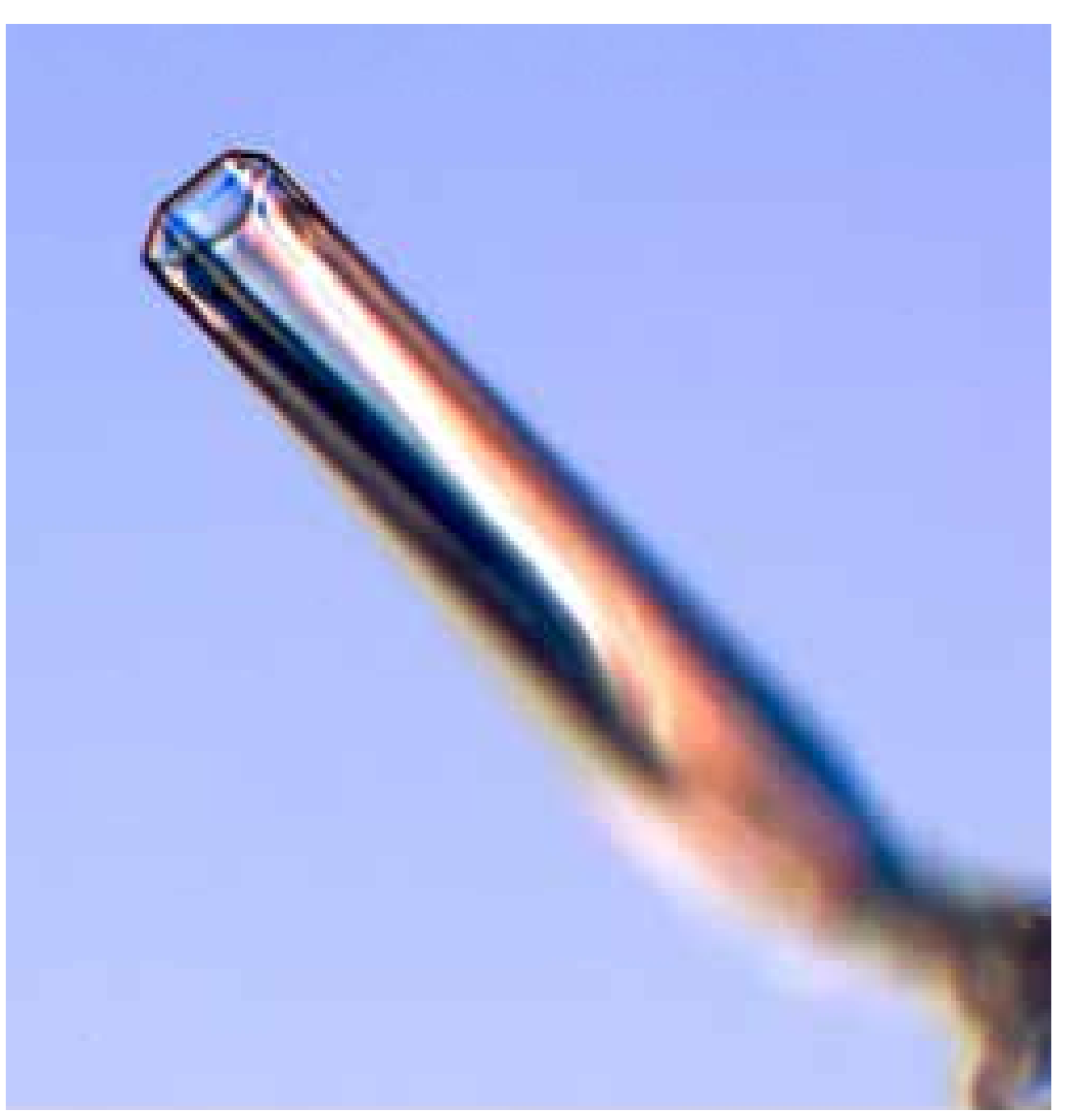

*Sheaths are exaggerated hollow columns with exceptionally thin walls and deep hollows. Cups are stout crystals with flared walls that resemble shallow hexagonal goblets.*

In terms of growth mechanisms, these crystals could be included in the hollow-column and capped-column categories; but both can be quite distinctive in appearance, so they have picked up their own names over the years. These crystals are generally small and rare, so are easily overlooked. I found these by scanning over collections of small irregular crystals that had landed on glass slides.

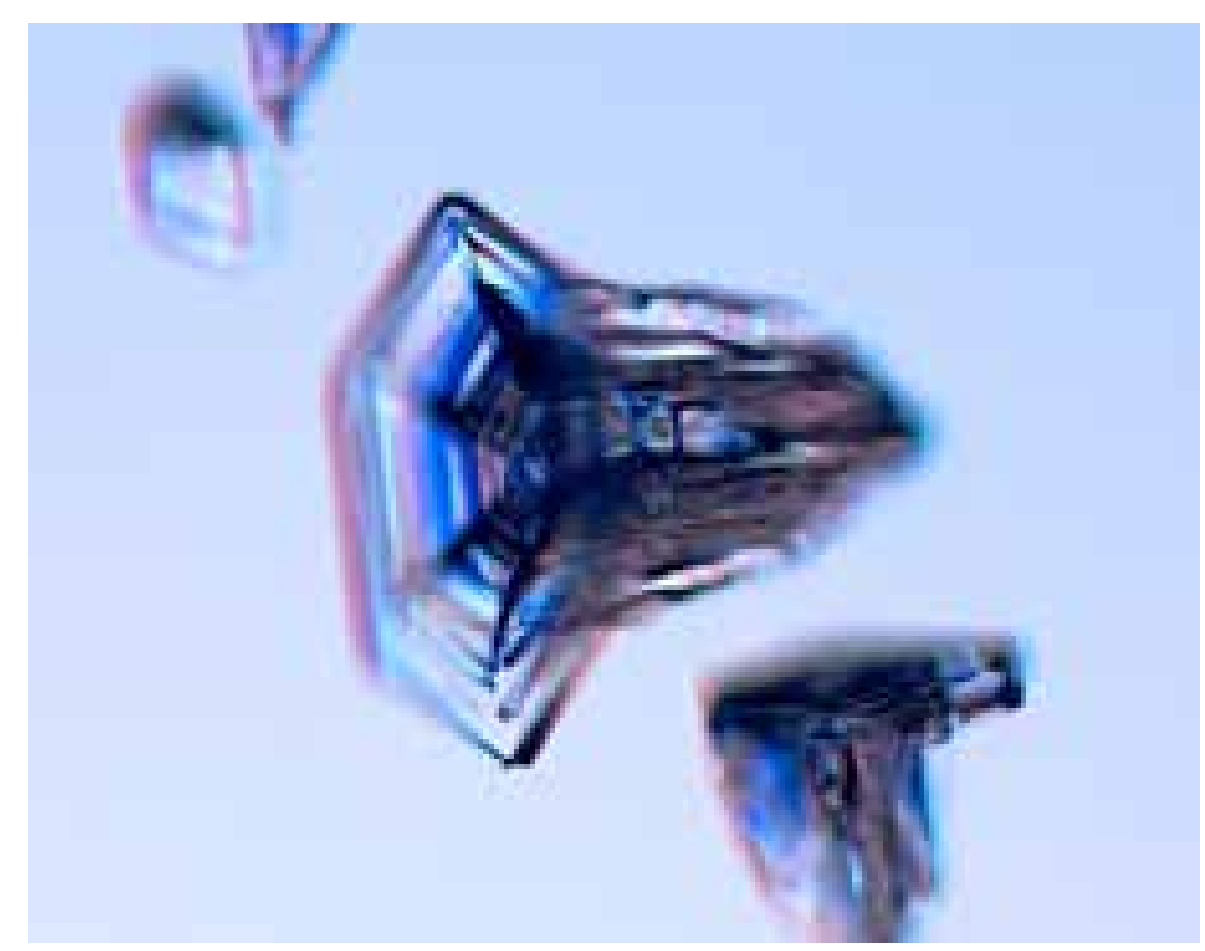

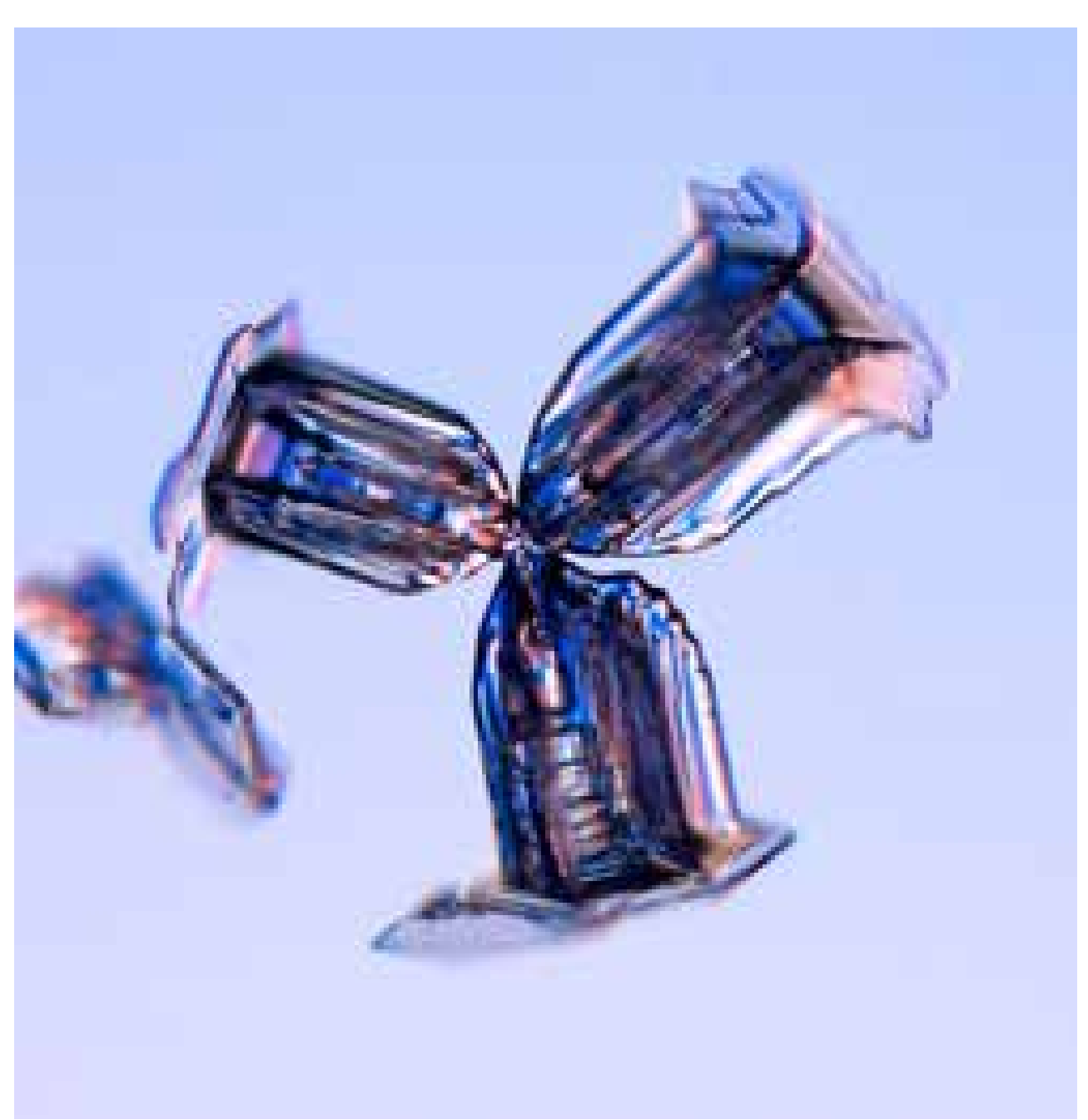

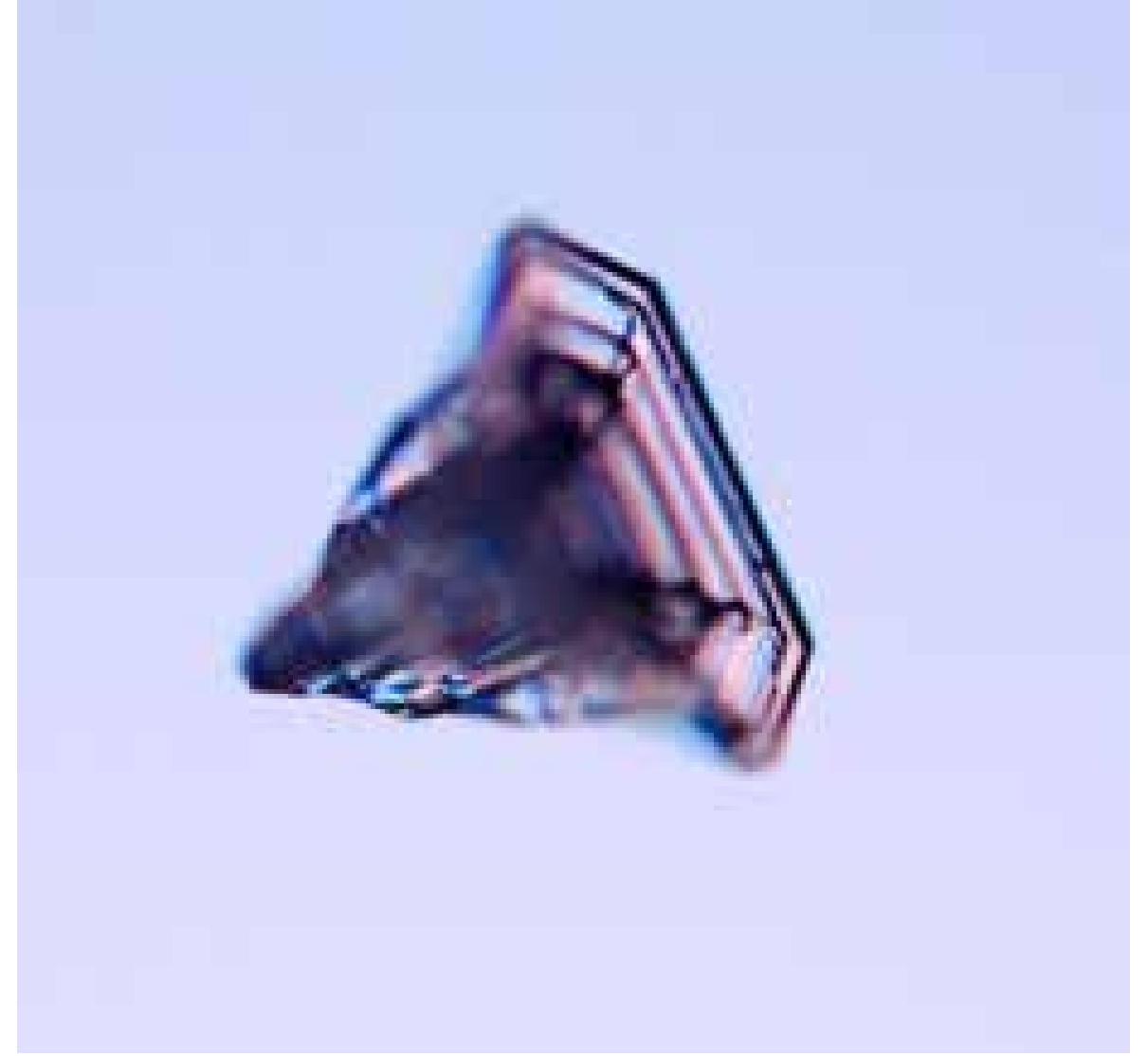

# Crystal Twins

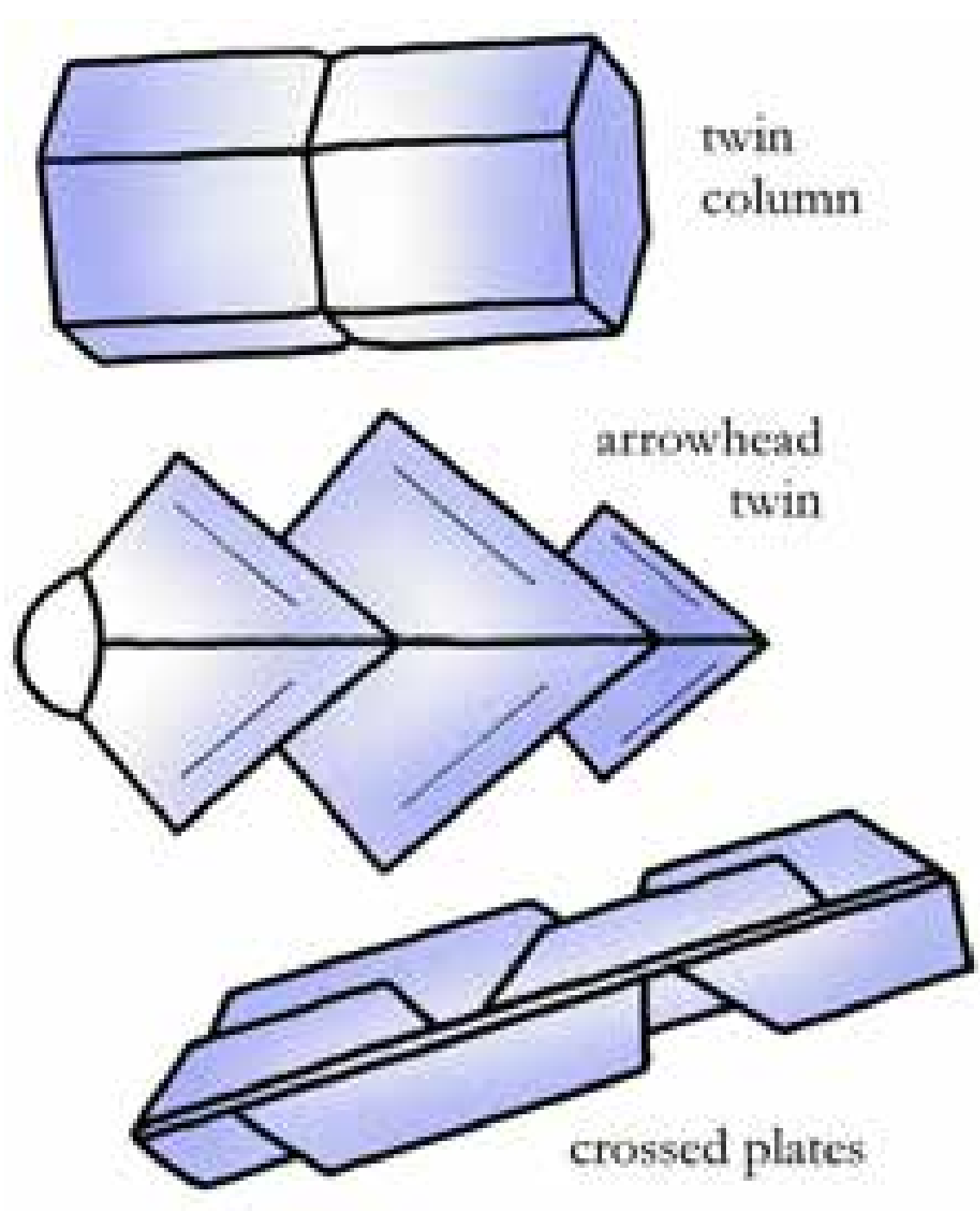

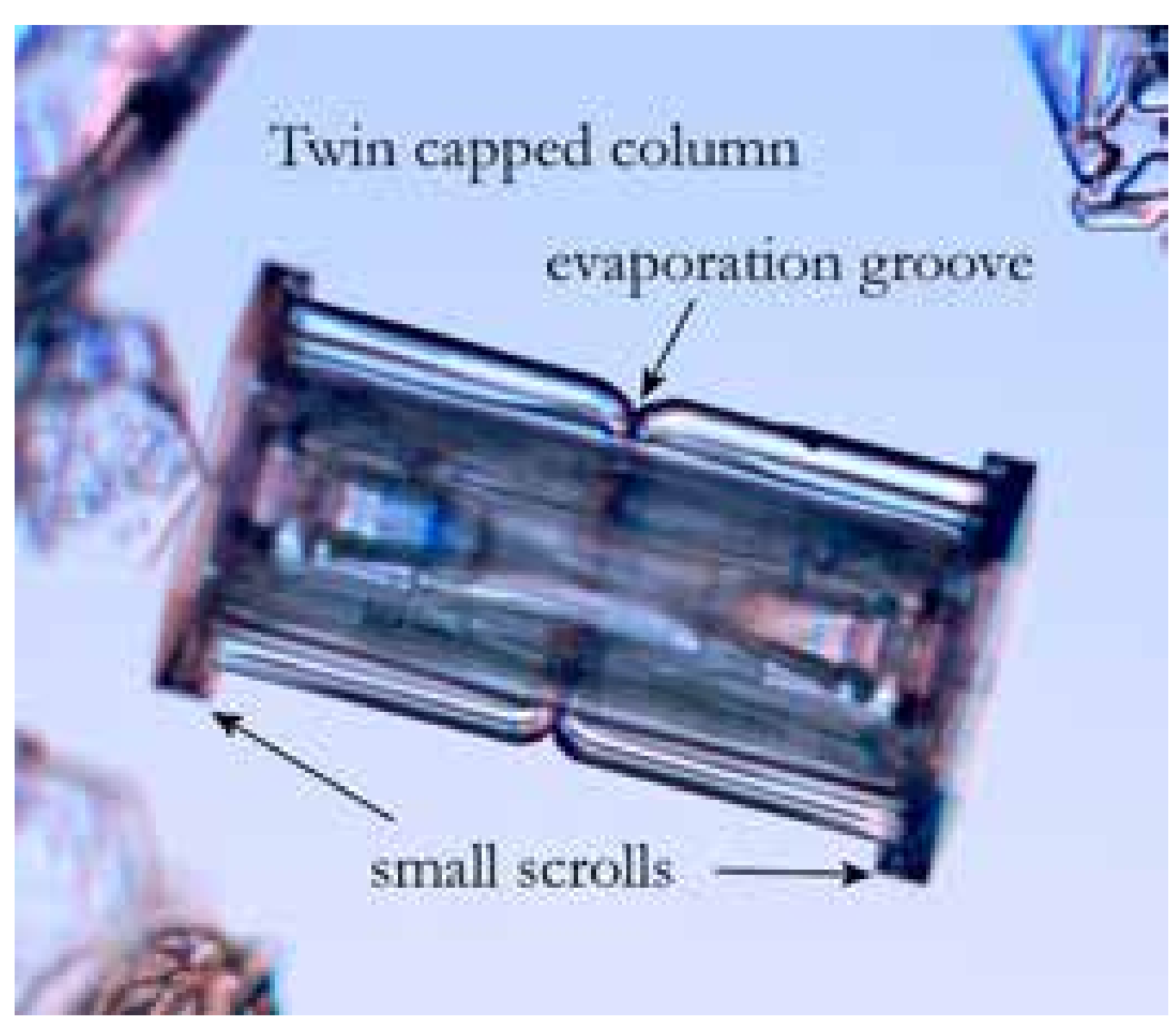

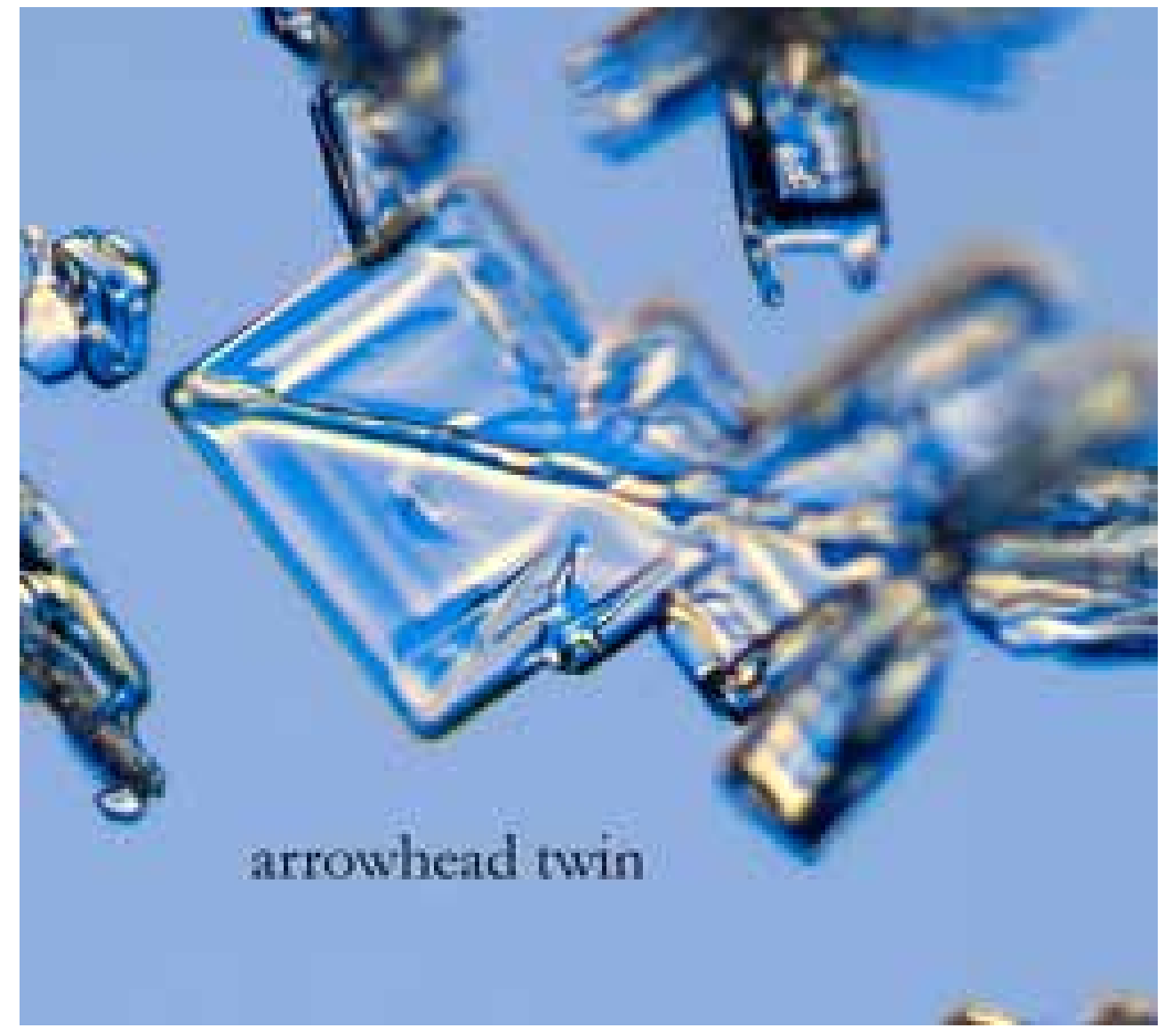

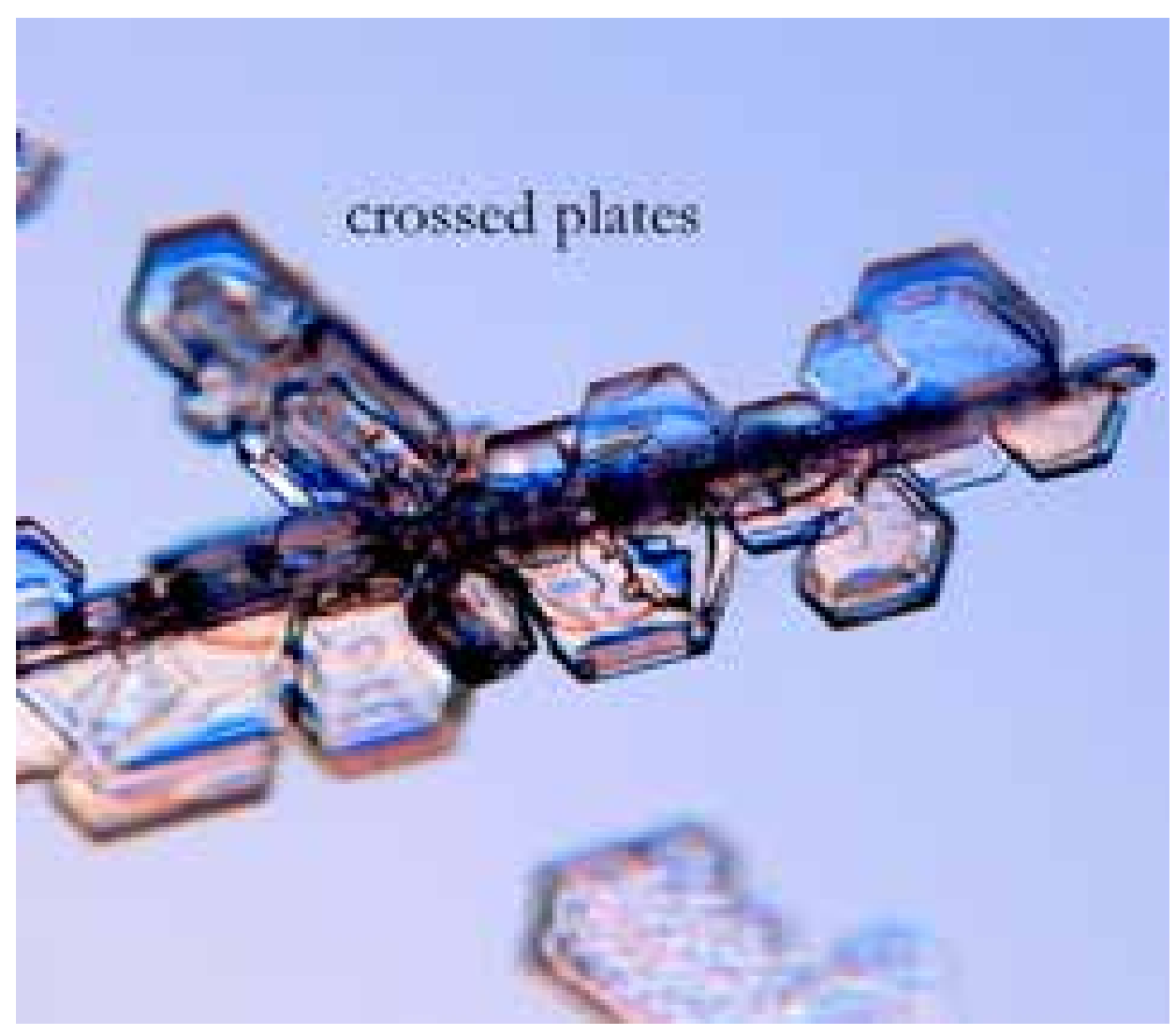

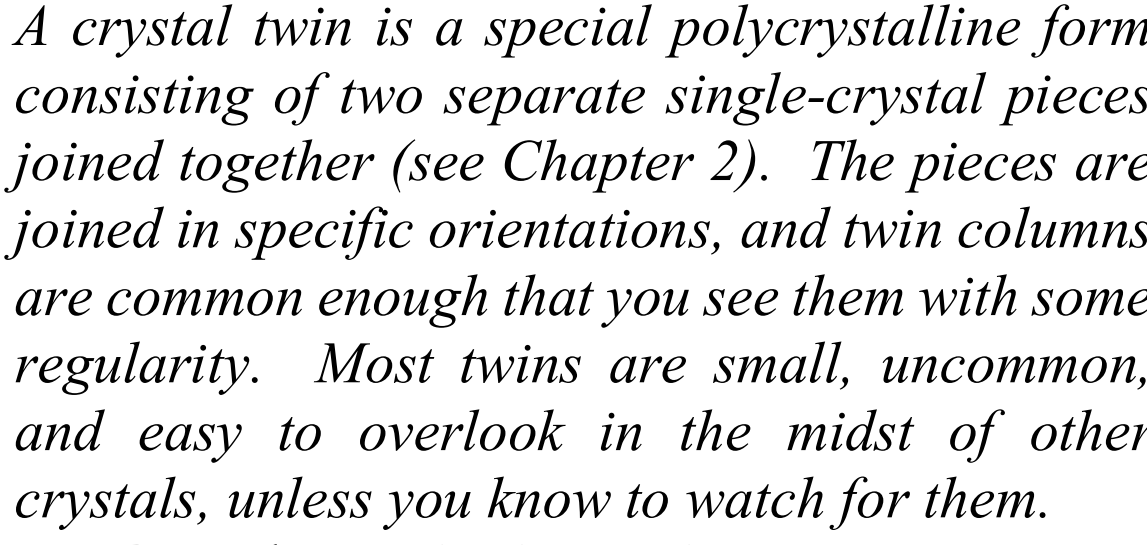

*A crystal twin is a special polycrystalline form consisting of two separate single-crystal pieces joined together (see Chapter 2). The pieces are joined in specific orientations, and twin columns are common enough that you see them with some regularity. Most twins are small, uncommon, and easy to overlook in the midst of other crystals, unless you know to watch for them.*

Crystal twinning is a common mineralogical phenomenon involving an initial molecular lattice mismatch that develops into a pair of co-growing crystals. The alignment of the pieces indicates the lattice construction of a twin crystal, and I described the known possibilities for snow crystals in Chapter 2. Crossed plates and arrowhead twins are both quite rare in the wild, although some variants are fairly easy to produce in the lab. Twin columns can be quite common; they look almost exactly the same as normal columns, but often one can see a distinct "evaporation groove" around the column's waist (see Chapter 2), indicating the weaker molecular bonding in that plane.

# Twelve-Branched Snowflakes

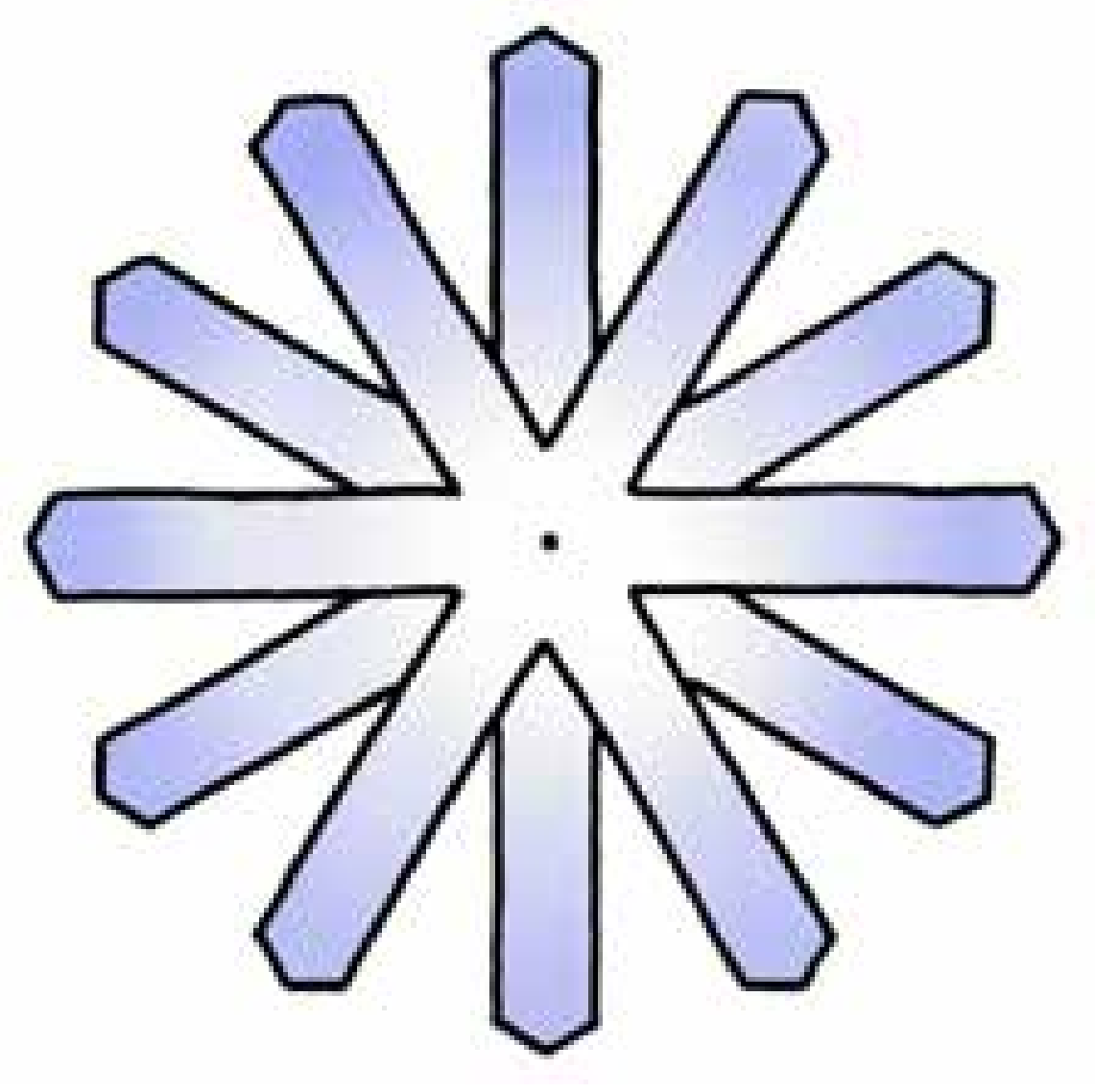

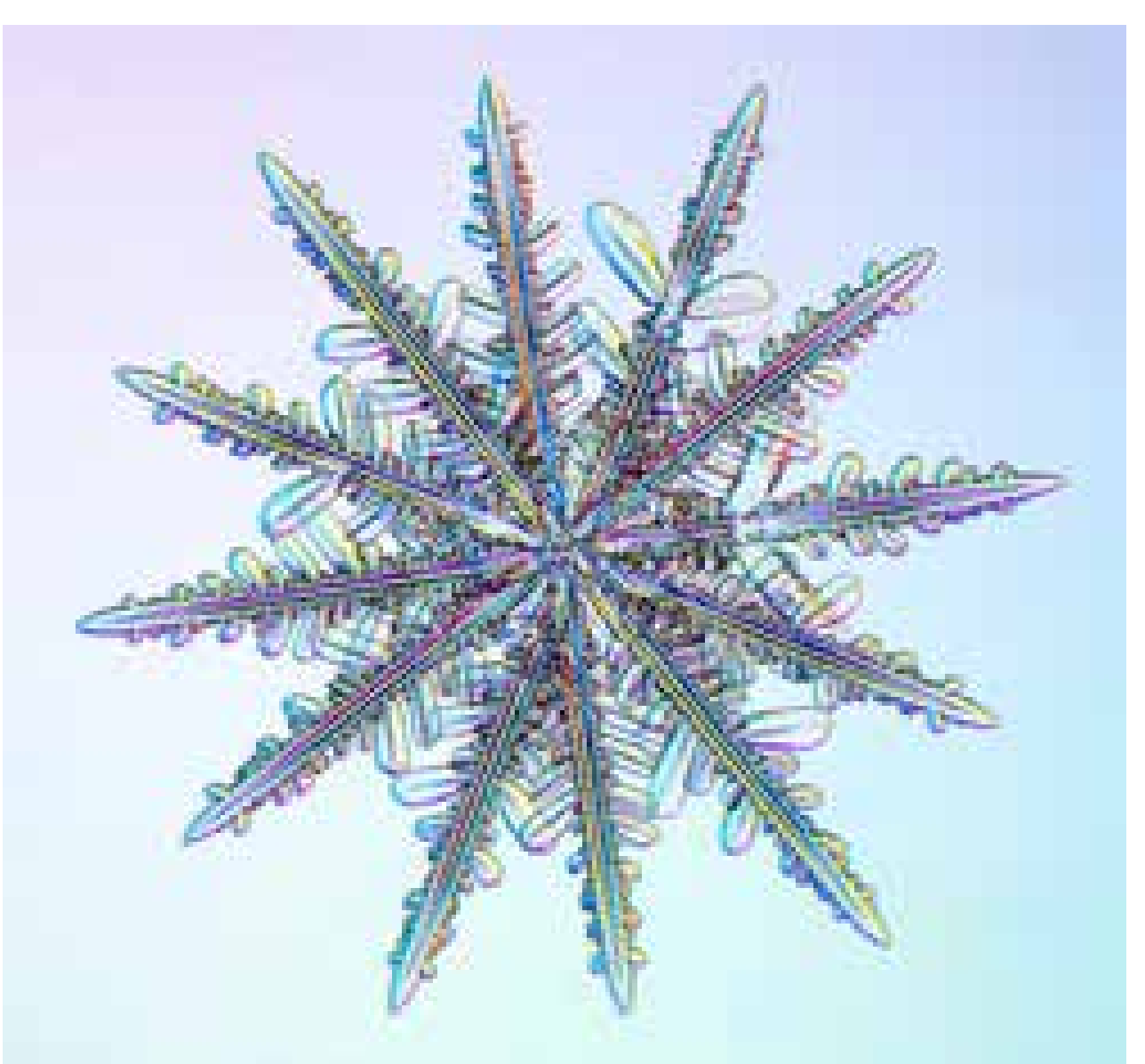

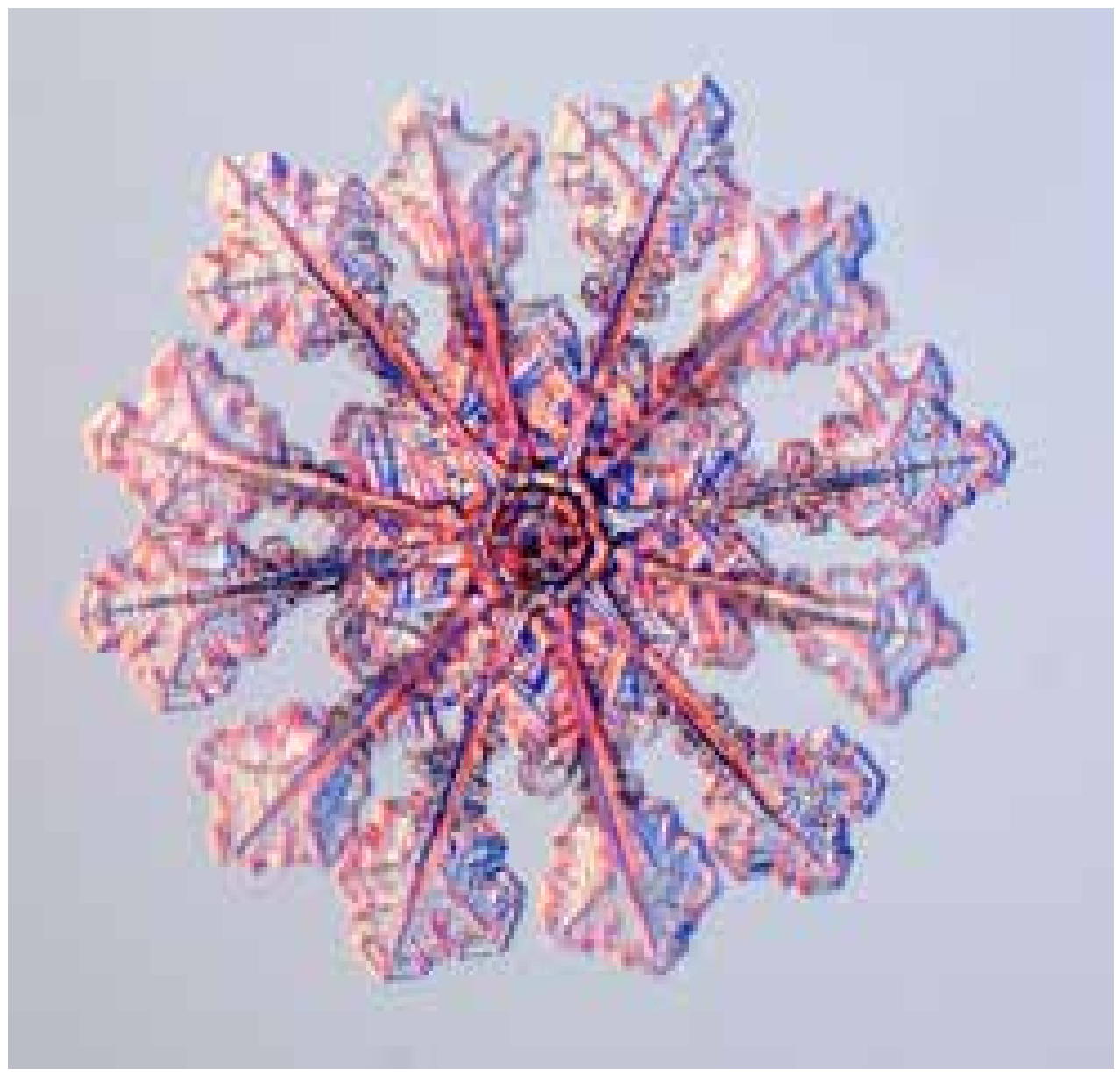

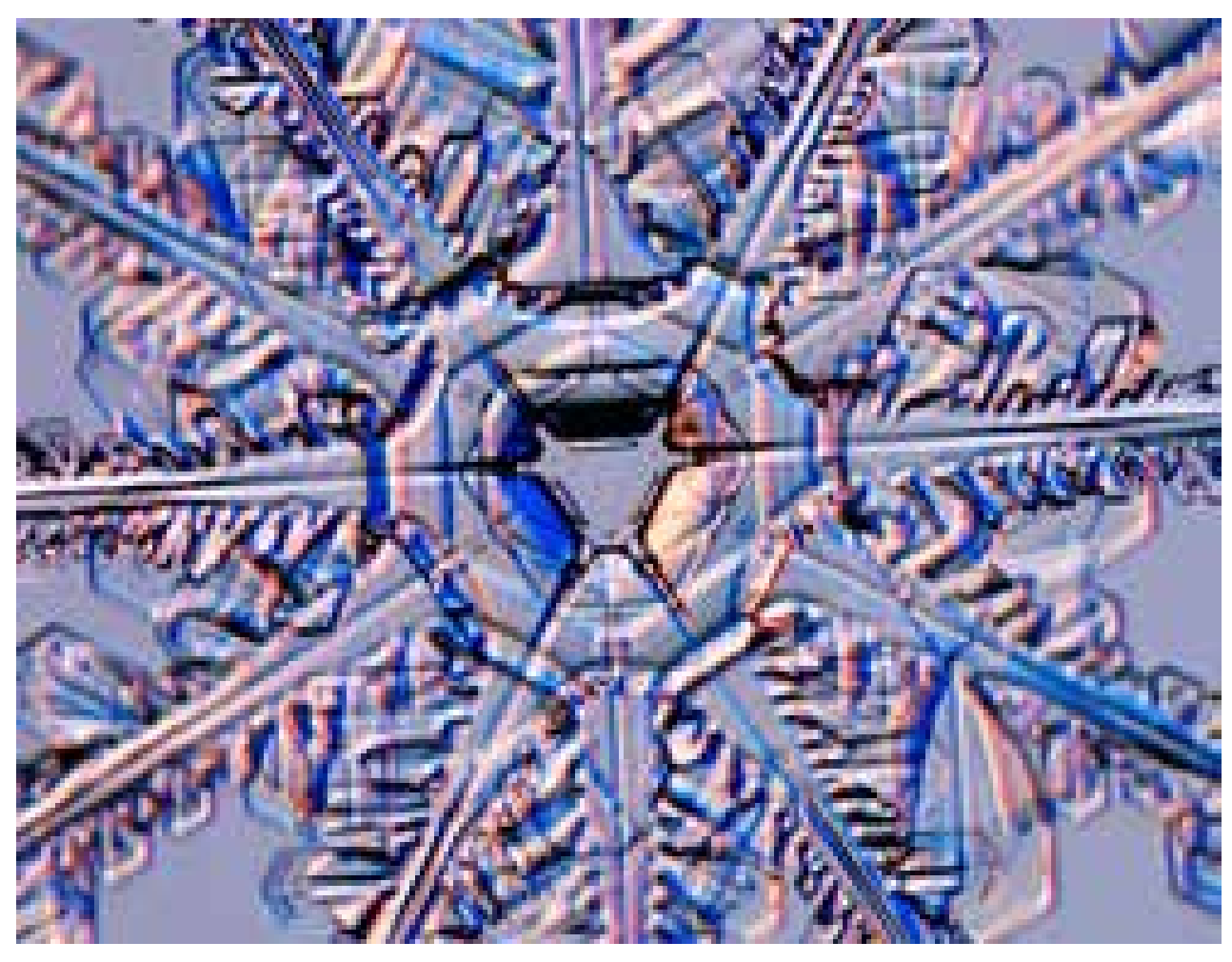

*A twelve-branched snowflake is essentially a matched pair of six-branched stellar crystals attached at their centers, with one rotated 30 degrees relative to the other. Twelve-branched snowflakes are uncommon, but they can be quite large and distinctive in appearance. Some snowfalls bring quite a few mixed in with normal stellar crystals.*

The evidence to date suggests that a twelve-branched snowflake is nothing more than two six-branched crystals that collided and stuck together when they were small. The near-perfect 12-fold symmetry in some examples appears to arise from a selection effect: if two tiny prisms experience a collision that bonds their basal faces together with close to a 30-degree rotation between them, then the pair will develop into a well-developed and easily spotted 12-branched crystal. However, if the collision is less ideal (which is far more likely), then the pair will develop into an inconspicuous radiating dendrite. A key to this model is that your brain is quite adept at noticing symmetrical snowflakes in the midst of a great deal of malformed clutter. The fact that many 12-branched crystals are not quite aligned, either in position or angle, supports this selection-bias hypothesis.

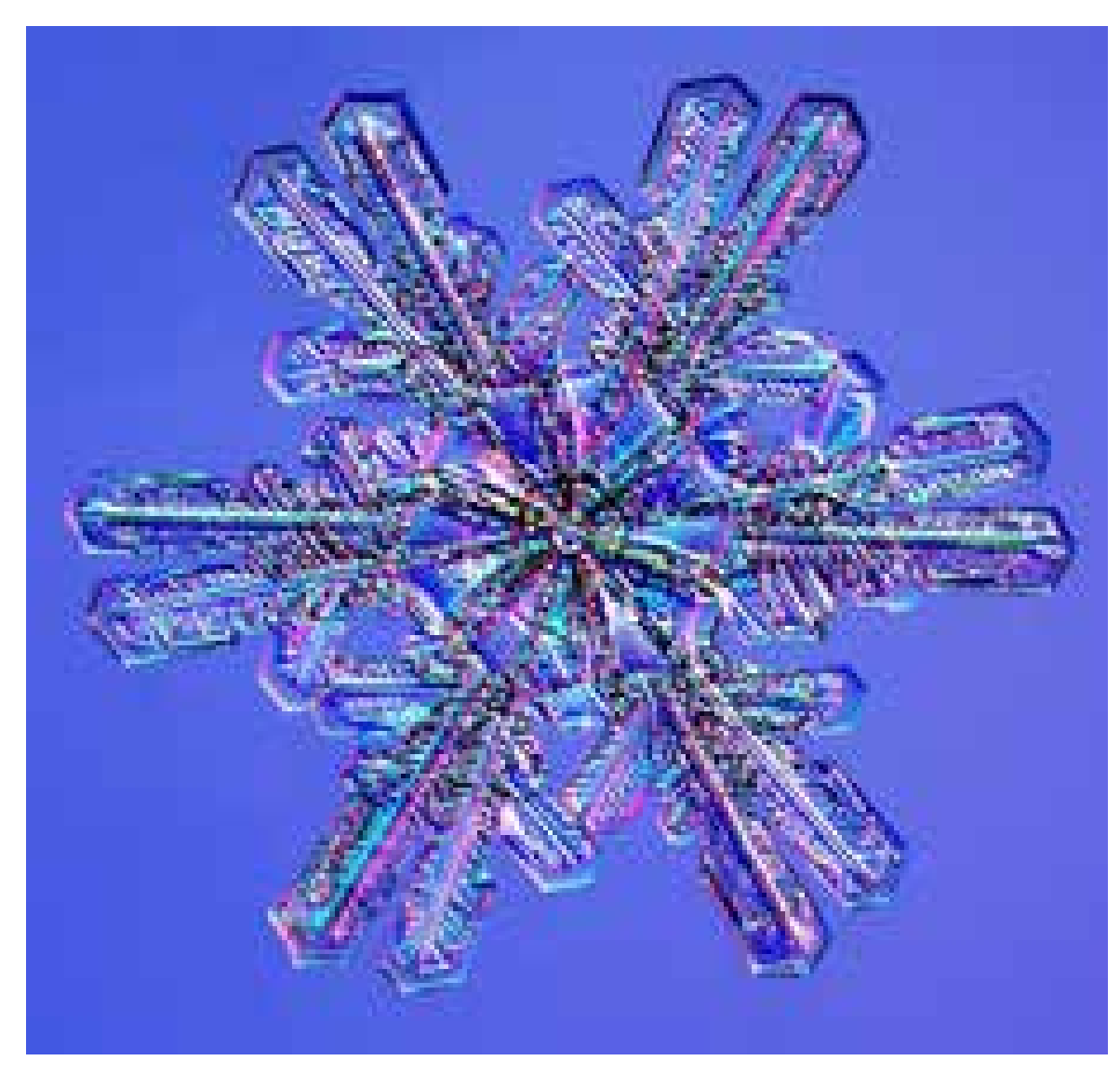
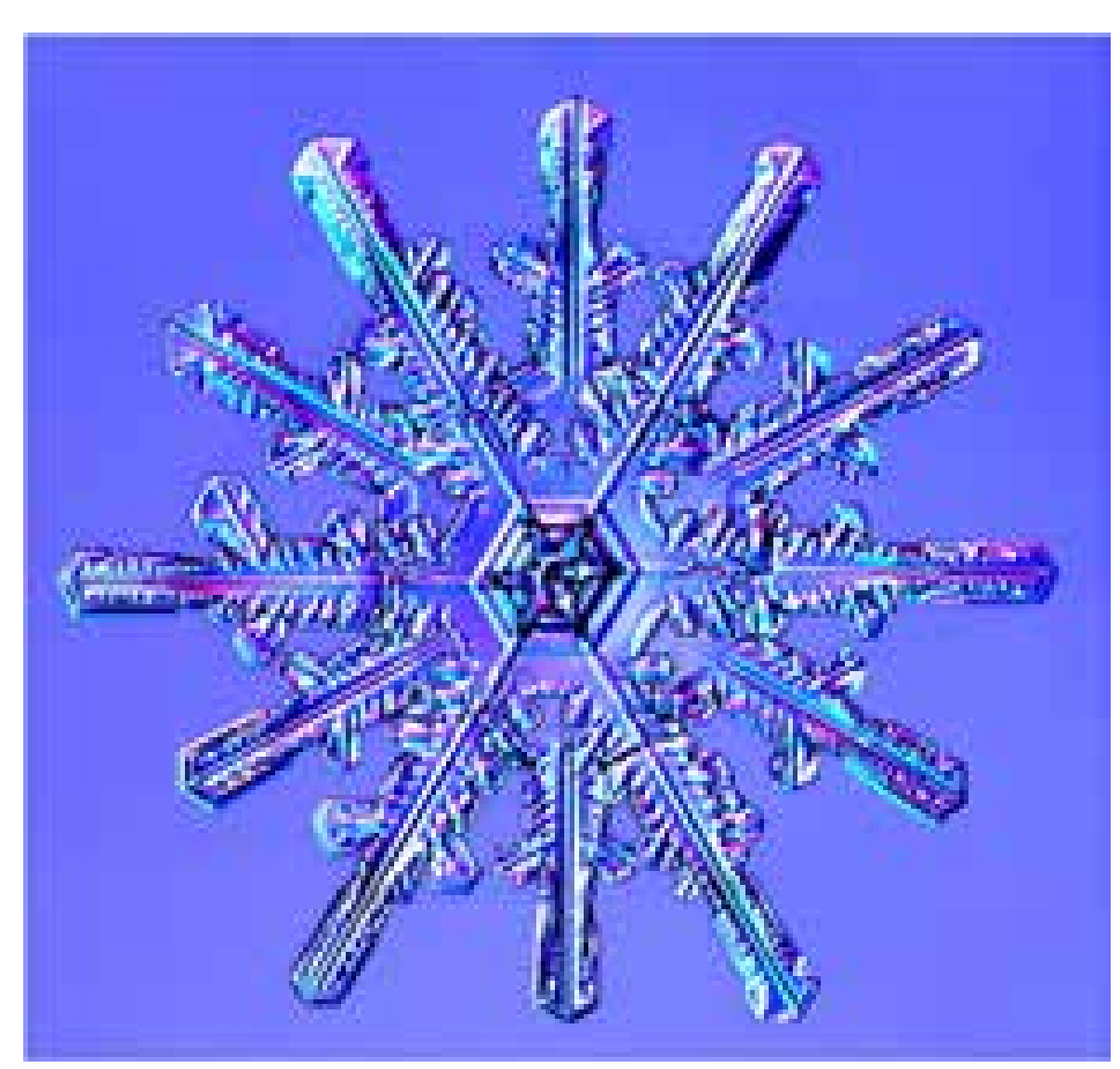
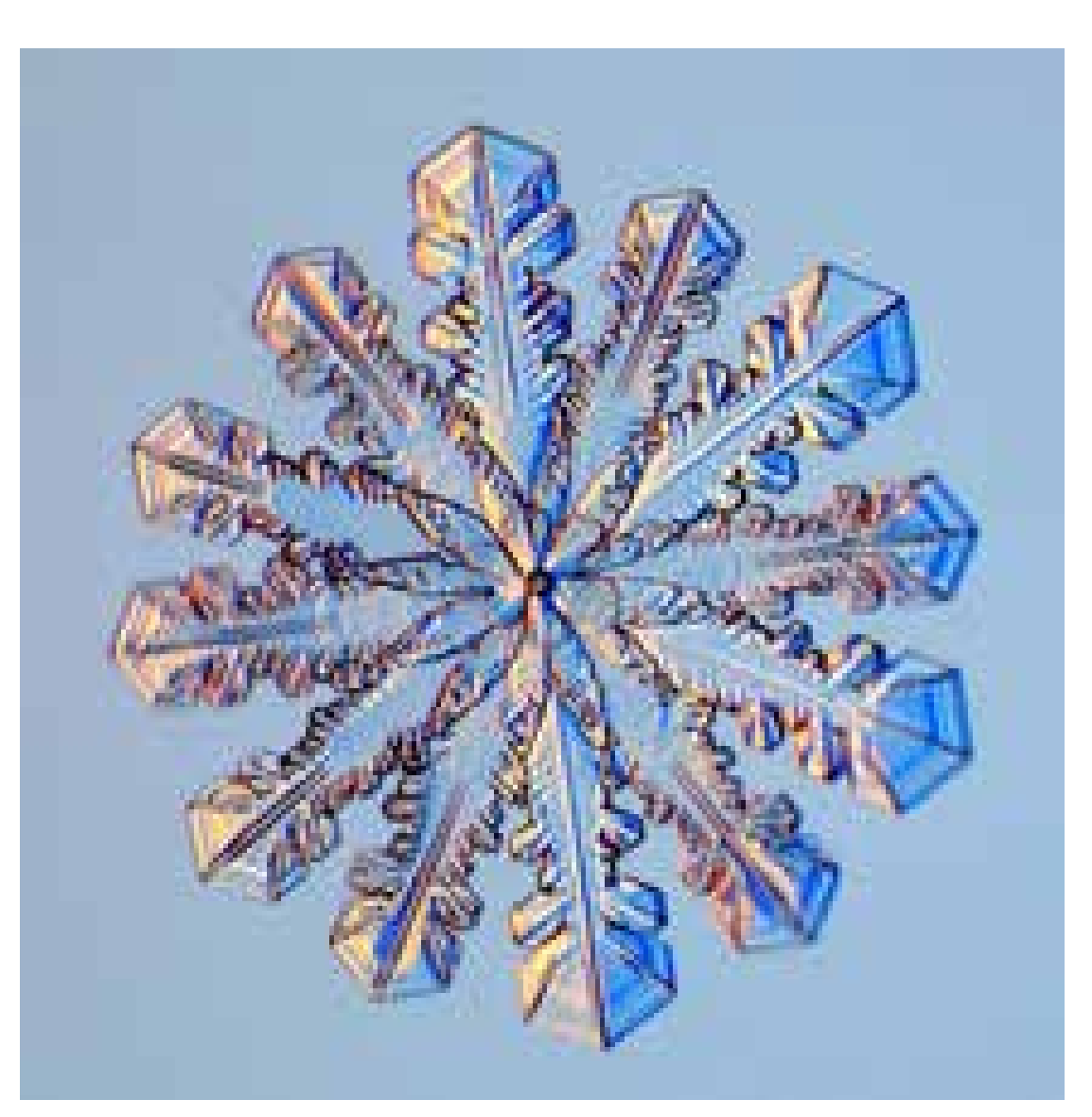
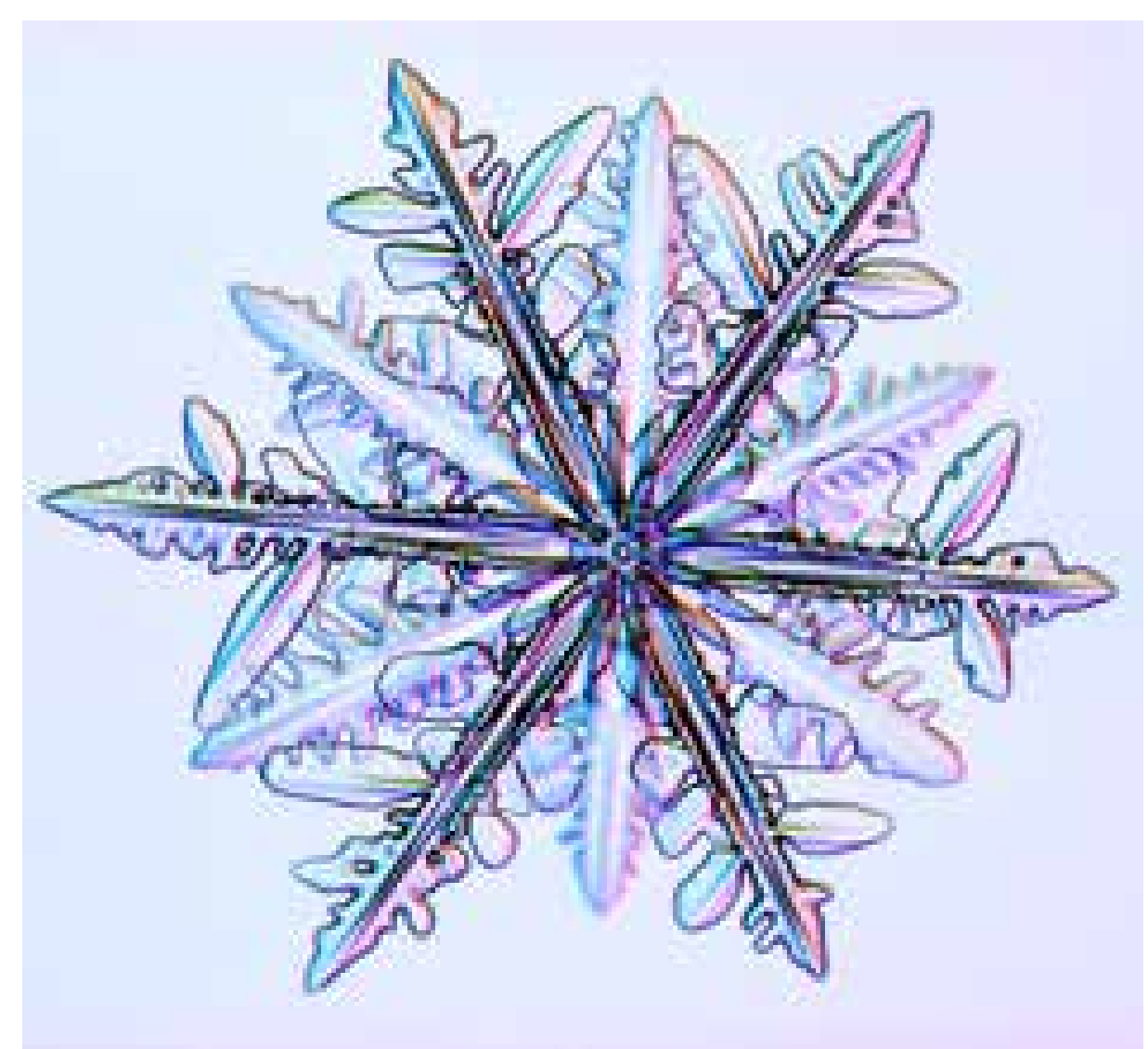
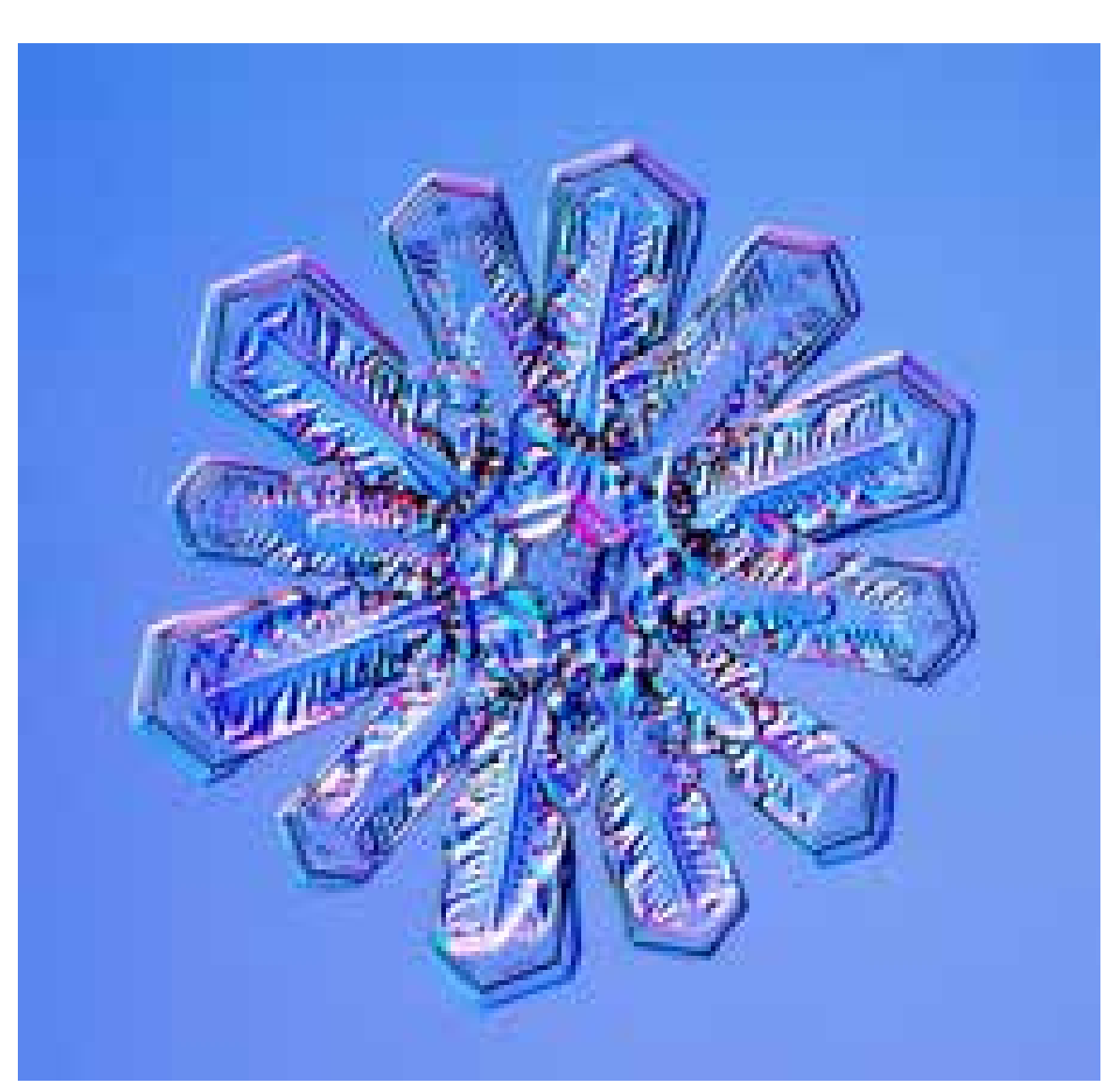
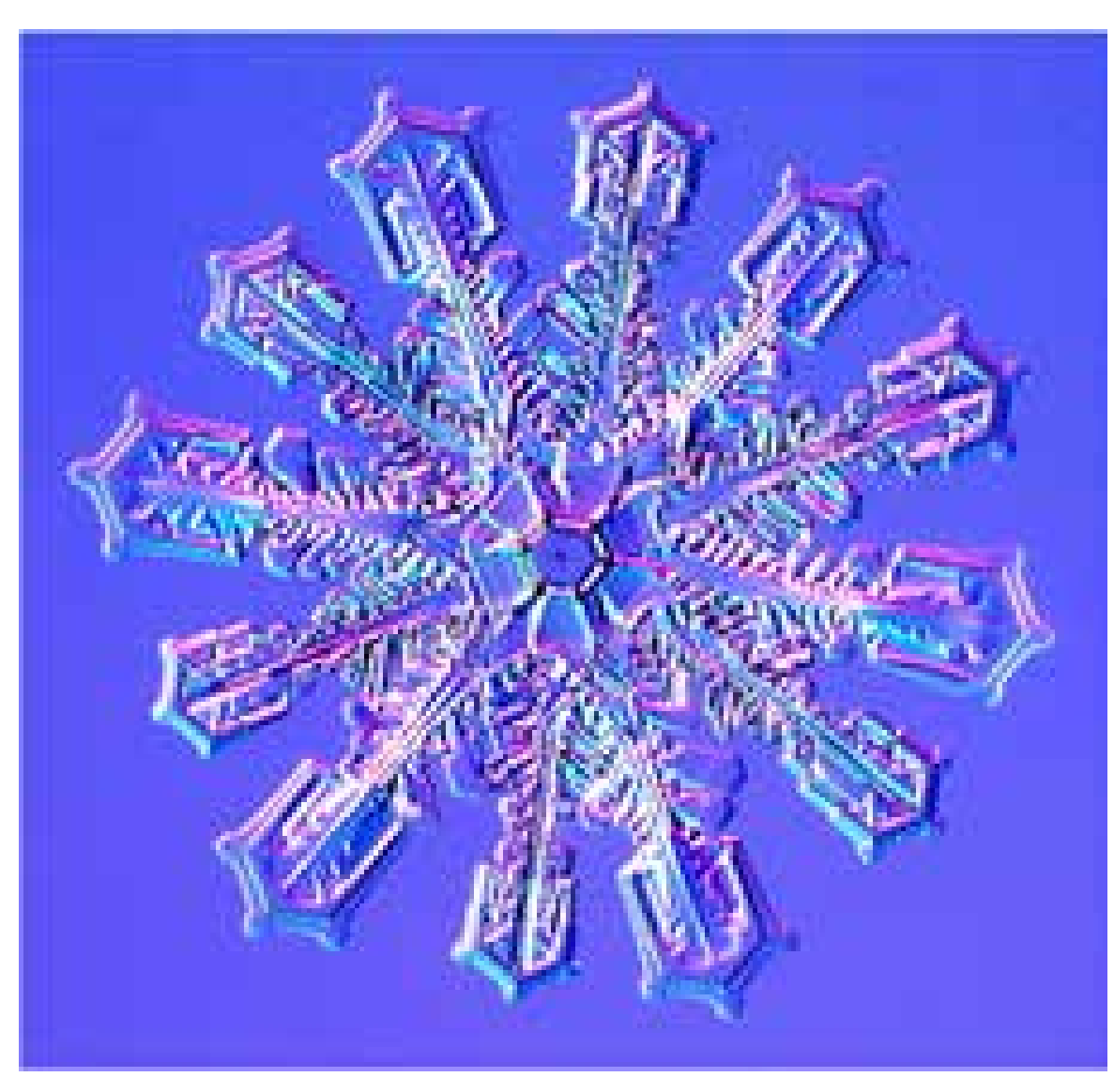

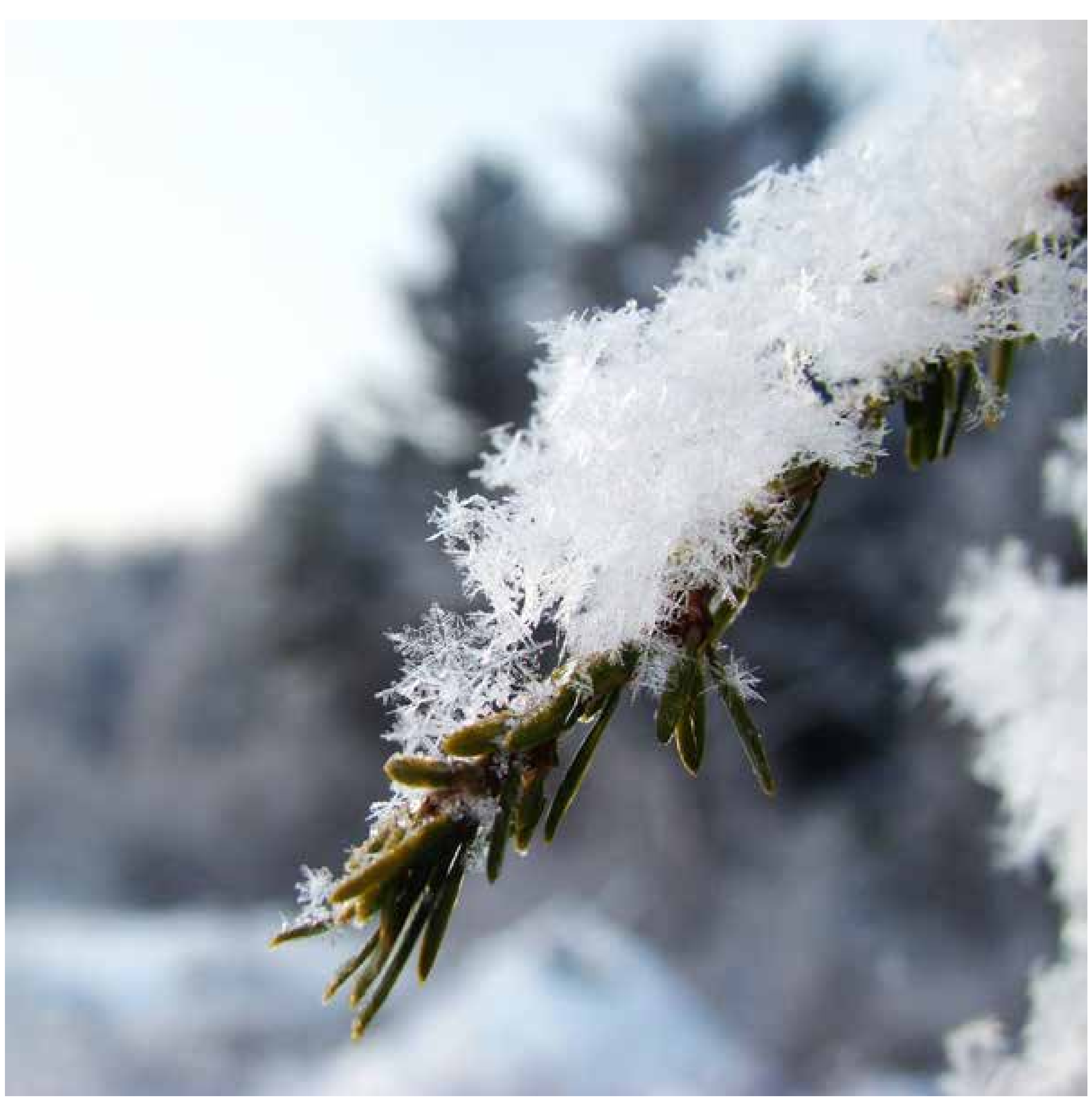

# Chapter 11

# Snowflake Photography

*It is extremely improbable that anyone has as yet found, or, indeed, ever will find, the one preeminently beautiful and symmetrical snow crystal that nature has probably fashioned when in her most artistic mood.*
– Wilson Bentley, *The Vermonter*, 1922

Snowflake photography has much in common with other forms of nature photography; it requires an artistic eye, some suitable optical gear, and a compelling desire to just go out there and take some pictures. The activity presents its own unique challenges as well, in that snow crystals are small, somewhat fragile, prone to evaporation and melting, and – as if that weren't enough – they need to be handled outside in the cold. As a semi-professional snowflake photographer for many years, I have managed to pick up a few tricks and techniques regarding lighting, handling, equipment, and other considerations that matter out in the field. Moreover, I have studied the subject fairly extensively and have tried to learn from other prominent snowflake photographers as well. When you take a deep dive into the subject, there are a substantial number of rather subtle issues involved in capturing quality images of these tiny slivers of ice. In this chapter, I attempt to document what I have learned about snowflake photography, in the hope that others can continue developing this fascinating craft.

**Facing page: Freshly fallen snow crystals perched on a branch of eastern hemlock in Vermont. Photo by Martha Macy.**

In my experience, three factors are of primary importance in snowflake photography: finding suitable subjects, using quality equipment, and developing a solid technique, especially regarding lighting. If any one of these factors is sufficiently lacking, the quality of the resulting photographs will suffer. Patience is a virtue as well, along with an artistic eye and a willingness to try different approaches. And, as with all types of nature photography, success sometimes requires just being in the right place at the right time.

## 11.1 Finding Snowflakes

Perhaps the most common difficulty one encounters in snowflake photography is simply a dearth of quality subjects. One cannot control what the clouds are producing, and not every snowfall brings superb crystals. As I described in Chapter 10, the most common bits of frozen precipitation are best classified as "irregular" or "rimed" (see Figure 10.3), and these are undoubtedly the least photogenic of all snow-crystal types. Crystals from more-desirable categories can be quite difficult to find, and they are usually mixed in with a sizable number of irregular specimens. The first step in snowflake photography, therefore, is learning how to find nice specimens.

To begin, proper snowflake photography can only be done with freshly fallen crystals. Once the flakes hit the ground, they will stick together and soon metamorphose into clumps of crystals with much changed morphologies. The character of ground snow is of considerable interest to skiers and people studying avalanches, but that takes us outside the scope of this book. Moreover, hoarfrost crystals can also be amazingly beautiful, and they too make worthy photographic subjects. But if you want to photograph snow crystals in all their glory, you have to catch them before they hit the ground.

The optimal strategy for photographing snowflakes will depend on what kinds of crystals are falling. When it begins to snow, my first step is to leave the camera behind and just go outside to have a look. My preferred tools at this point are a sheet of dark-blue foam-core cardboard and a small magnifier like the one shown in Figure 11.1. The foam-core provides a smooth matte surface that makes it easier to spot nice crystals, and the magnifier is handy for evaluating the quality of the crystals.

It is not unusual to observe a lot of small, grainy, gloppy, rimed, and generally undesirable crystals at this point; my generic name for this is "granular" snow, because the crystals look like small icy grains of sand. As described in Chapter 10, this type of snow offers little appeal for snowflake photography. If there is nothing falling from the clouds but granular snow, then one's best option is probably just to go back inside and try again later. Wishing there was something better to photograph is not especially helpful.

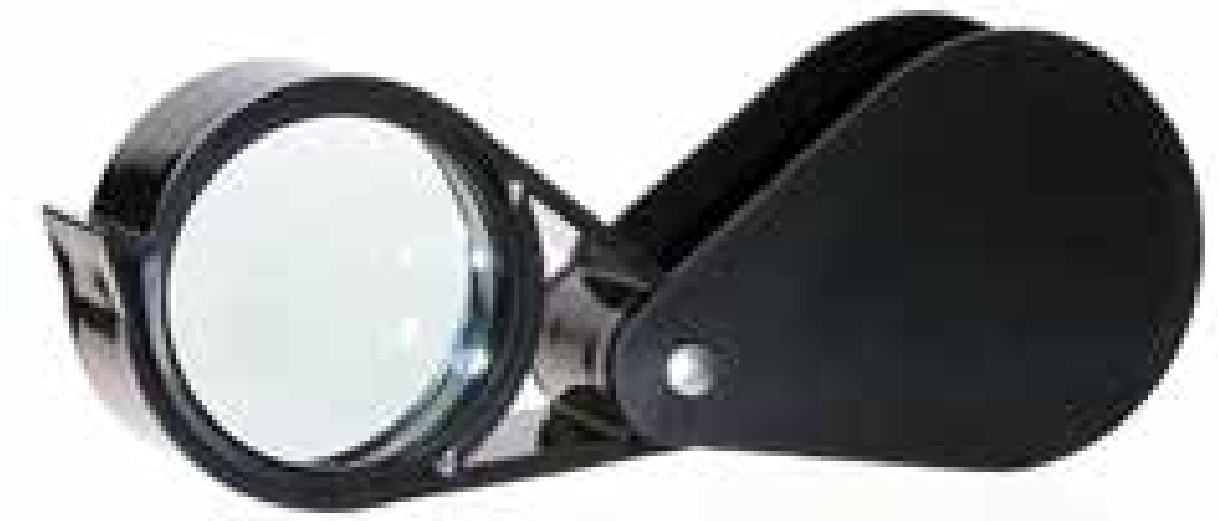

**Figure 11.1: An inexpensive fold-up magnifier, or *loupe*, is a convenient tool for appraising the overall quality of falling snow crystals. A magnification of 4X or 5X is about right, as this provides a reasonable amount of detail with a fairly wide field of view.**

I find it important, however, not to give up too quickly. Even when there is a lot of granular snow all around, there might be some interesting crystals in the mix. Moreover, some of the rare and quite captivating crystal types are usually quite small, and I like to photograph those almost as much as the canonical stellar variety. Capturing the full menagerie of snow crystal types is a worthwhile and often fascinating activity in its own right.

One useful trick I have learned is to hold the foam-core out under a bright light, perhaps a streetlight or a yard light that is essentially a single point of bright illumination from a distance. By moving the foam-core around under such sharp lighting, even small faceted ice surfaces will sparkle clearly, making them easier to spot when surrounded by unfaceted granular snow. As a general rule of thumb, if I can see some sparkle on the board, then there is a reasonable chance that some interesting crystals are present.

If the snow has been falling for a while, and it happens to be dark outside, another trick is to just look out a window and view the reflection of a bright streetlight off a nearby snowbank. Pure granular snow, especially heavily rimed snow, has little or no sparkle, and this gives a snowbank a flat, white appearance. A snowbank that shows some sparkle, on the other hand, suggests that there are some nice, faceted crystals falling.

It is also important to keep watch on the crystals throughout a snowfall, even when it appears that there will be little of interest to see. The character of the falling crystals can change dramatically with time, and you can miss some great pictures if you are not sufficiently diligent. There have been times when, as I was outside photographing, some exceptional crystals appeared only briefly, for perhaps 10-20 minutes. Granular snow does not usually change to great snow that quickly, so I typically check the crystals every 30 minutes or so. I have witnessed many snowfalls that started out as granular glop, then improved somewhat, then improved more, and then delivered some excellent photographic subjects for an hour or two, only to go back to granular snow as the snowfall waned. Like all other aspects of weather, snow crystal production can be highly variable and quite unpredictable.

I should also point out that snowflake photography is often best done at night. Partly that is simply because the nights are long in the winter, especially at high latitudes. As one ventures farther north in the dead of winter, working in the dark becomes a matter of statistical necessity, as the daylight hours are so short. Also, the temperature is typically lower at night, and lower temperatures are desirable at most locations. So a dedicated snowflake photographer can expect to spend long hours outside, alone, in the cold and dark. I suppose the hobby is not for everyone, but it does satisfy one's hermit-like tendencies.

## Weather & Climate

In principle, one could use the morphology diagram to predict what kinds of snow crystals fall in different weather conditions, and this works to some extent. For example, like most snowflake photographers, I am always keen to find large stellar crystals, and these occur almost exclusively when the temperature is near -15 C. More precisely, because the temperature is usually slightly higher on the ground than up in the clouds, around -13 C is close to an ideal ground temperature for finding good specimens. However, the weather is not as predictable as that sounds; in reality, well-formed stellar crystals might be found anywhere from -10 C to -20 C. But the probability falls off substantially outside that temperature range.

Warmer snowfalls often bring a great variety of snow-crystal types, including columnar crystals near -5 C, or perhaps capped columns and other exotic forms, as described in Chapter 10. These crystals tend to be on the small side, however, and they are nearly always accompanied by lots of granular snow. As I describe below, working a warmer snowfall is best done with high magnification and a different collection strategy compared with stellar crystals. But small can be beautiful, and I have captured many excellent photographs of unusual snow crystals in relatively warm (above -10 C) conditions.

While temperature is the most important parameter for predicting snow-crystal types, many other factors will influence quality. For example, wind can be quite detrimental, as the crystals can get beaten up by mid-air collisions. A heavy snowfall is not ideal for the same reason. From my experience, the best specimens can usually be found during calm, cold, light snowfalls, providing just a steady dusting of crystals drifting slowly downward.

Another meteorological phenomenon I have come to appreciate is low-hanging clouds. When the clouds are high in the sky, a kilometer or more above the ground, that usually yields what I call "travel-worn" snowflakes. The problem is that the crystals stop growing once they leave the clouds, and they can experience quite a bit of sublimation as they slowly descend. This rounds the faceted corners and yields somewhat shabby-looking crystals. When I see snow falling from especially high clouds, I know that finding extraordinary specimens will be unlikely.

I should stress, however, that while it can be relatively easy to predict low-quality snow crystals from the weather conditions, predicting high-quality crystals is almost impossible. I have experienced some snowfalls that checked off all the boxes for great crystals, yet brought nothing but granular snow. Moreover, often granular snow gives way to beautiful stellar crystals, or vice versa, with no obvious change in weather conditions. It is certainly true that hollow columns and needles generally form only around -5 C, and large stellar plates are restricted to around -15 C; but there is not much one can reliably say beyond that. The atmosphere is not a precisely controlled laboratory environment, so it is impossible to predict exactly what kinds of crystals will appear, at least not with any real accuracy. For the snowflake photographer, this means that waiting and watching are simply part of the process.

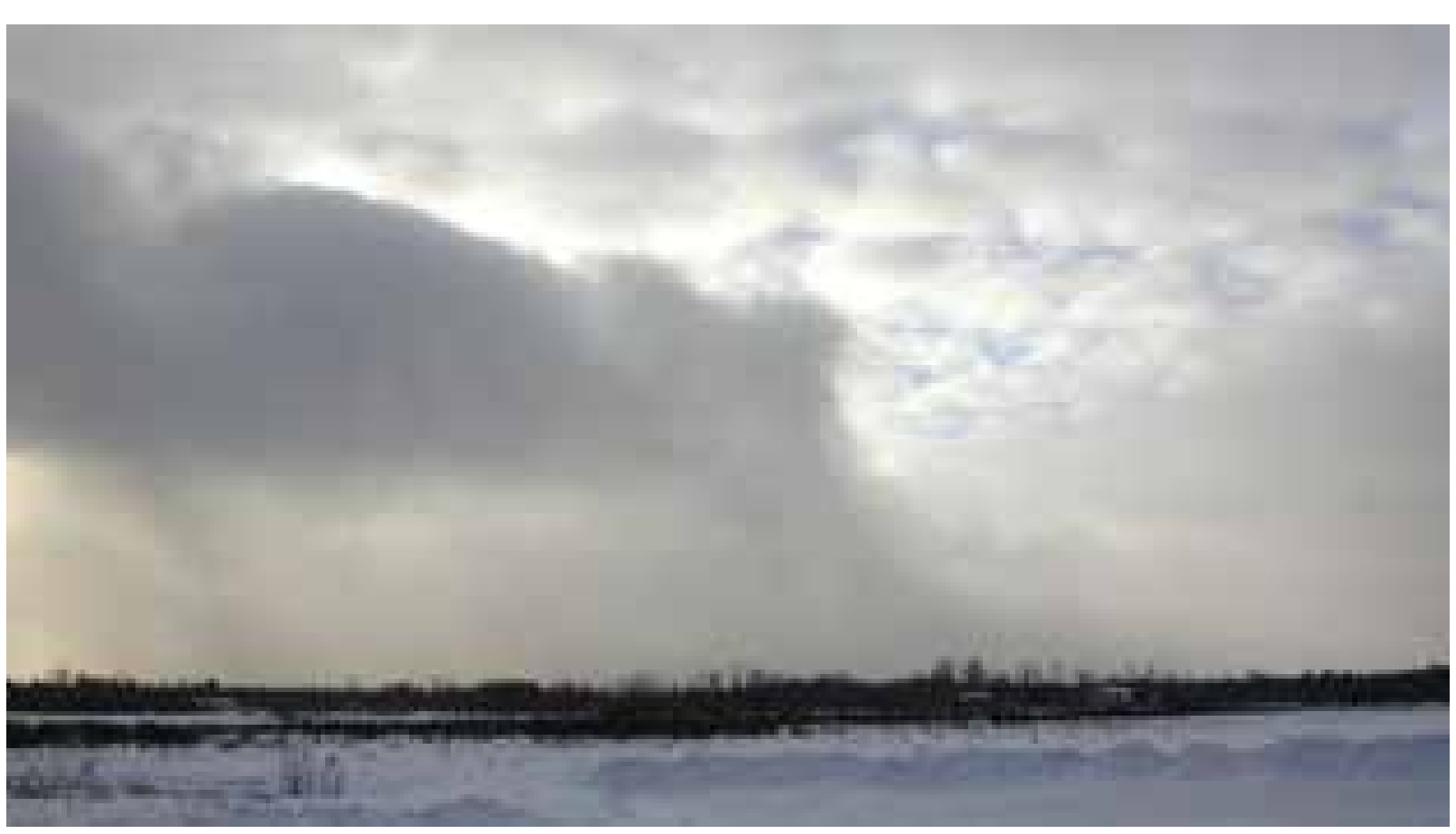

**Figure 11.2: The scene during a near-perfect storm for snow-crystal photography, taken by the author in Cochrane, Ontario. On rare occasions, the meteorological conditions seem to conspire to create the most beautiful snow crystal forms.**

That being said, I have been especially attentive when it comes to finding the best conditions for observing large stellar crystals, as these are such a delight to photograph. As a concrete example, Figure 11.2 shows a photo I took during a "perfect storm" that lasted about eight hours and gave me some of my best snowflake photographs. Looking back on the day, I noted several beneficial characteristics of the weather:

1) The temperature had hovered around -13 C all day, which is the ideal temperature for finding stellar crystals.
2) It snowed lightly all day, so the crystals did not much interfere with one another in the clouds, or on my collection board.
3) There was essentially no wind all day.
4) The clouds were hanging low in the sky, barely above ground level, so the crystals continued growing during most of their descent, yielding sharply faceted crystals.
5) The clouds were thin and patchy, so the varying conditions resulted in a good deal of morphological diversity in the falling crystals.

Even at a good location, one might encounter a high-quality snowfall like this maybe a few times during a winter season. As I mentioned above, some degree of patience is essential in snowflake photography.

## Location Matters

Quality snowflakes can appear anywhere, as long as the temperature and other weather conditions are favorable. Location is a factor only because the probability of experiencing such conditions varies from place to place. Being a snowflake photographer who happens to live in Southern California, I have tried to find locations that maximize the probability of finding high-quality snow crystals, especially large stellar dendrites, and I have studied this problem quite a bit over the years.

Temperature is the most important factor, as I mentioned above. One of my favorite locations is the small town of Cochrane, Ontario, where the average January temperature is -18 C and the average daily high is -12 C. This means that the probability of finding stellar crystals is reasonably high on average, notably in the daytime when being outside is most pleasant. The average January precipitation is a respectable 3.5 cm (water equivalent), and this arrives in frequent, light snowfalls. Wind speeds are generally low as well (7 mph), and overall I have found Cochrane to be an excellent location for photographing snowflakes. Many time zones away, the town of Kiruna, Sweden has comparable average conditions, and I have found some excellent crystals there also.

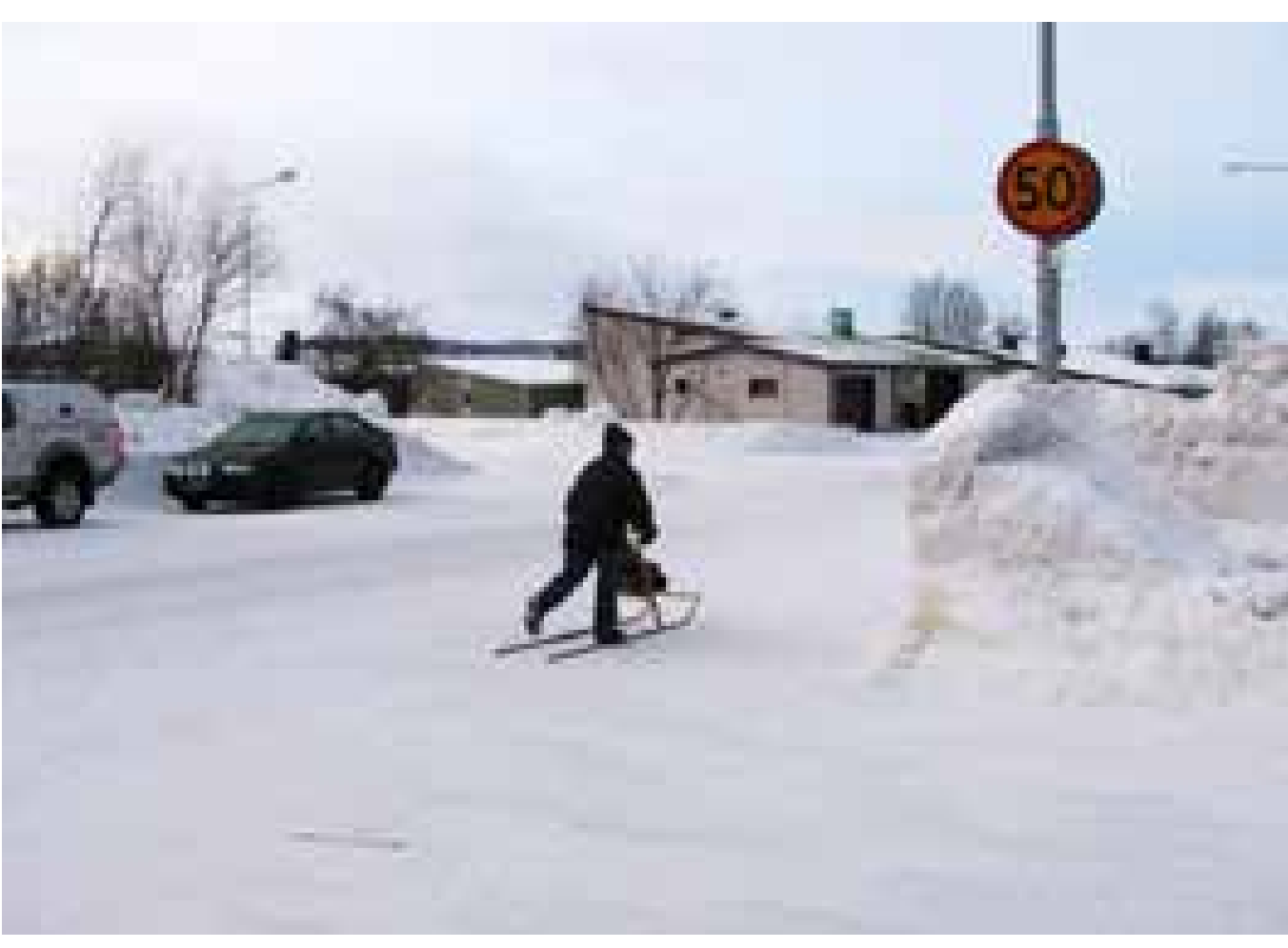

**Figure 11.3: Roads packed with accumulated snow often indicate a good location for snow-crystal photography. In Kiruna, Sweden, shown here, some exercise-conscious residents use sleds for their grocery shopping, using the snow-packed roads to good advantage. Photo by the author.**

One noteworthy characteristic of both Cochrane and Kiruna is that residential roads in January are almost always covered with packed snow, as illustrated in Figure 11.3. This is a good sign for snowflake photography, as it tends to indicate consistent low temperatures (as the snow does not melt) and plenty of falling snow.

My hometown of Fargo, North Dakota, provides a good example where a low average temperature is not the only parameter to consider. The average January temperature in Fargo is -13 C, which sounds good; but the average precipitation is only 0.7 cm, and the brisk winds can be quite incessant. Snowfalls are somewhat infrequent, and much of the winter precipitation comes during intense blizzards. Although Fargo has a favorable average temperature, it is not an ideal location for snowflake photography.

Ukichiro Nakaya lived in Sapporo, Japan, where the January average is a balmy -4 C, although conditions are better in nearby Asahikawa at -8 C. The January precipitation tops an impressive 10 cm water equivalent, with typically calm winds, so there is certainly no shortage of snow. Central Hokkaido is also well-known as an excellent location for snowflake photography, as evidenced by Nakaya, Katsuhiro Kikuchi, Yoshinori Furukawa, and others from that region.

Another good case-study is Barrie, Ontario, which is home to noted snowflake photographer Don Komarechka. The average January temperature in Barrie is -8 C, and the average low is -12 C, so overall I would rate this location as being a bit on the warm side. Nevertheless, Don has taken some of the world's best snowflake photographs in Barrie, so the site is obviously working for him. It helps that it snows a lot, bringing 4 cm on average in January, and the average wind is not too bad (9 mph).

Moscow is worthy of consideration also, as this location is the home of Alexey Kljatov, another renowned snowflake photographer. Here the average January temperature is -8 C with an average snowfall of 4 cm, and Moscow can boast a remarkably low average wind speed (3 mph).

Wilson Bentley, the founding father of snowflake photography, made his home in Jericho, Vermont, where the January average is -7 C, although this number was a bit lower in the 1880s. The precipitation and wind speeds are also both suitable, and Vermont remains a prime location for snowflake photography.

The climate data indicate that Barrie, Moscow, and Jericho are all quite similar in average January conditions, so certainly that says something regarding the availability of quality snow crystals. Personally, I would rate Cochrane a bit higher, with its colder average temperatures, but the statistics are thin all around; even the best locations deliver exceptionally well-formed crystals only rarely. Note also that the population density drops off rapidly with the average winter temperature. I suspect that slightly warmer conditions yield more snowflake photographers simply because so few people live in colder regions. Details notwithstanding, snowflake photography is a craft best practiced near the cold edges of human civilization.

Other than average weather conditions, there does not seem to be anything special about any of these locations; there is no "magic" behind producing quality snowflakes, other than the fact that favorable weather conditions are more likely in some places than others. I have never been especially fond of mountain locations, mainly because of generally high winds, and high temperatures are problematic in most highly populated areas.

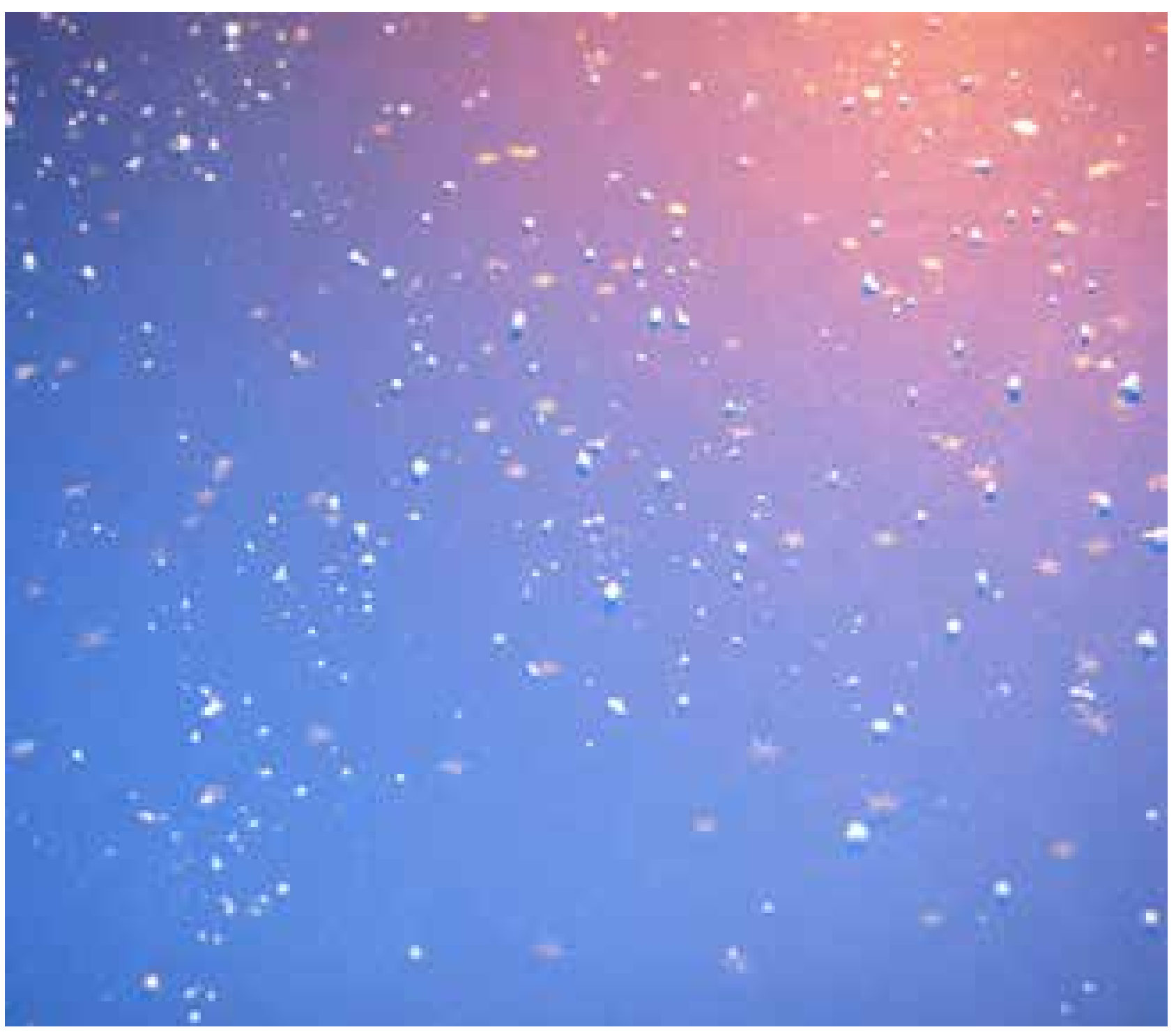

**Figure 11.4: A collection of freshly fallen snowflakes on a dark-blue foam-core collection board. The glow in the upper right comes from a bright lamp shining down on the crystals, producing strong reflections from smooth faceted surfaces.**

If you happen to live in a place that experiences sufficiently cold winters with plenty of snow, then you will likely find some excellent snow crystals if you look for them. The best way to find out is simply to go outside with a simple magnifier to have a look for yourself, preferably sampling multiple snowfalls at multiple times, as not every storm brings exceptionally photogenic crystals. If you like what you see, and you are willing to spend some time outside in the cold, then you might well enjoy snowflake photography.

## Handling Snowflakes

When I am photographing snowflakes, the handling technique I use depends on whether the interesting crystals are larger or smaller than about 2 millimeters. With larger specimens, I let the crystals fall onto a foam-core collection board, with the result looking something like what is shown in Figure 11.4. Especially photogenic specimens are rare, so a large foam-core collection board gives one a lot of crystals to look over, and the eye is remarkably adept at noticing especially nice crystals in a field of mostly granular snow. In Figure 11.4, I would say that over one percent of the crystals are reasonably well formed, and that is considered a pretty good yield. The overall average yield is much lower than one percent, as many snowfalls bring nothing but granular glop. With maybe a thousand snowflakes on the board (which is just a modest 32x32 array of crystals), one can scan around and find the best of the bunch in a minute or so, thus delivering a one-in-a-thousand snow crystal to photograph. With a quick brush of one's sleeve, the board is cleared for another round. Scanning over a board like this every few minutes, before long one can capture some exceptionally photogenic, one-in-a-million specimens.

When I spot what looks like a promising candidate on my collection board, I next pick it up using a small paintbrush and transfer it from the board to a glass microscope slide. This works surprisingly well, as the fine bristles will lift a typical stellar snow crystal with hardly any damage, at least most of the time. Best is to gently roll the bristles under the crystal, lifting it in the process. Of course, some breakage in handling is to be expected, and Figure 11.5 shows one example. A worse problem, in my experience, is carefully lifting a highly promising specimen onto the brush and then poof, a slight gust of wind sends it flying off, gone forever.

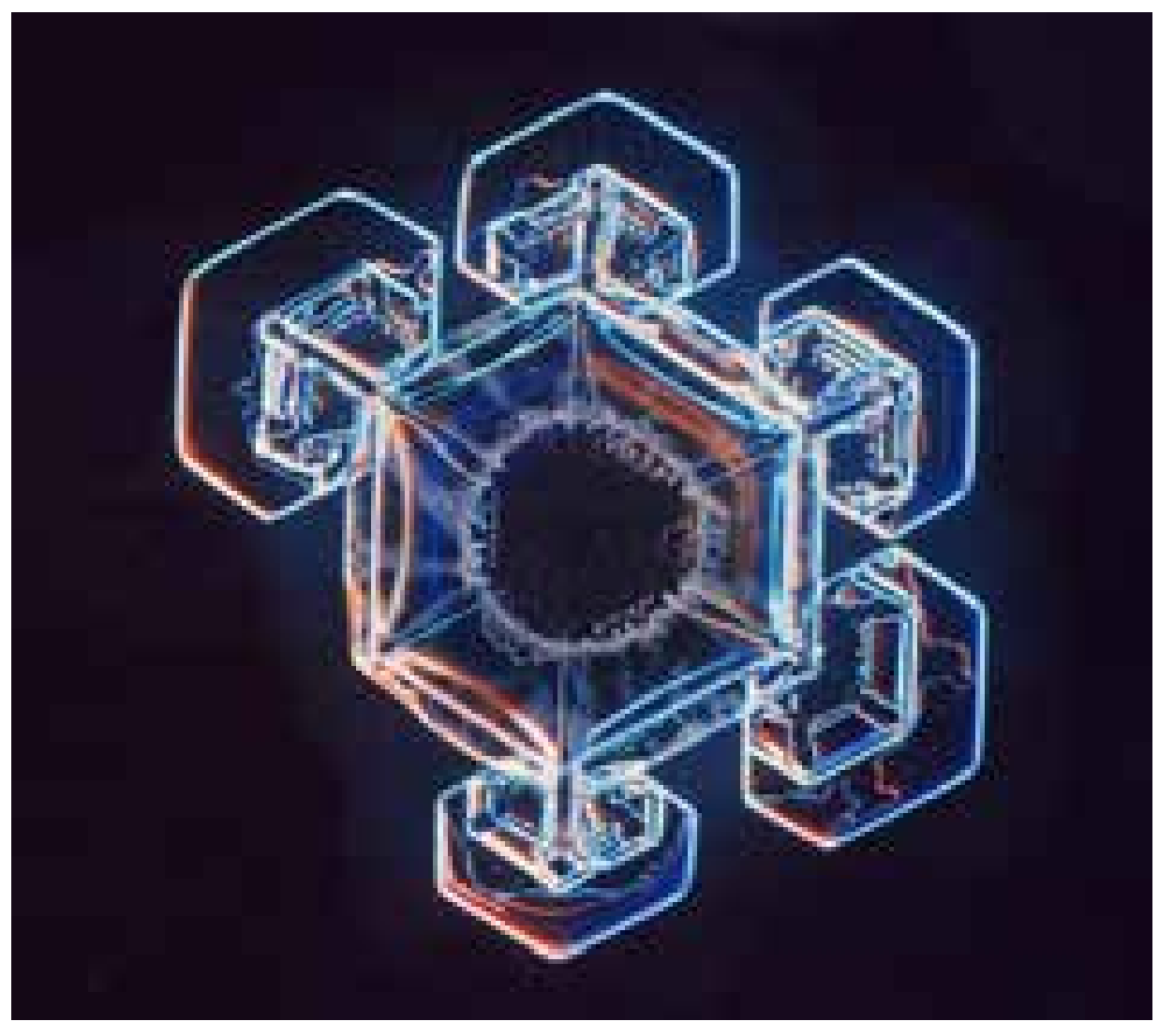

**Figure 11.5: A fine paintbrush works quite well for picking up and placing snow crystals, but damage is not uncommon. I broke this snow crystal when I tried to pick it up and move it onto a glass slide, losing a plate-like branch in the process.**

A glass microscope slide is certainly not the only destination for placing a snow crystal, and one might want to frame a photo in any number of ways. Regardless of how you want to proceed, a large foam-core surface and a small paintbrush can be used to scan through a large number of snowflakes quickly, thus allowing one to choose the nicest specimens. If you want to photograph rare crystals, including large, well-formed stellar crystals, then it is essential to scan over as much falling snow as possible.

Another secret to photographing exquisite crystals is to move fast. A good snowfall will not last forever, so it pays to capture as many crystals as possible while the clouds are being generous. Scan the collection board, find a subject, pick it up, place in on a slide, put it under the microscope, adjust the lighting, and take the shot … repeat. On a good day, I can do a crystal every minute or two this way, thus achieving a fairly large throughput. I have never met a photographer who managed to get terrific pictures with every single shot; taking lots of pictures is essential for yielding a much smaller number of outstanding photos.

Another good reason to hurry is to avoid sublimation. Figure 11.6 shows a nice example of a small stellar crystal that slowly evaporated away as it sat in the bright lights of my microscope. Figure 11.7 shows a further example of a snow crystal that melted as it was being photographed. Melting is mostly a problem only when the temperature is close to 0 C, and sublimation is not a huge issue at temperatures near -15 C. Nevertheless, speed is a virtue when photographing snowflakes.

**Figure 11.6: (Below) This snow crystal experienced quite a lot of sublimation during the two minutes that elapsed between the first photo and the last. You can see how the finer structural features on the crystal extremities are the usually first to disappear during sublimation.**

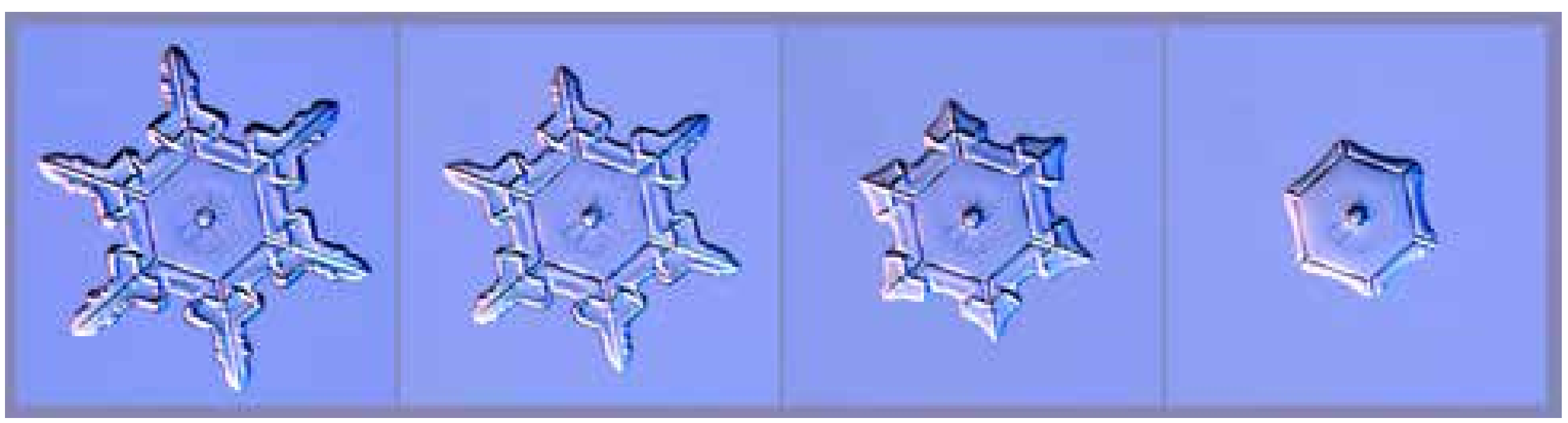

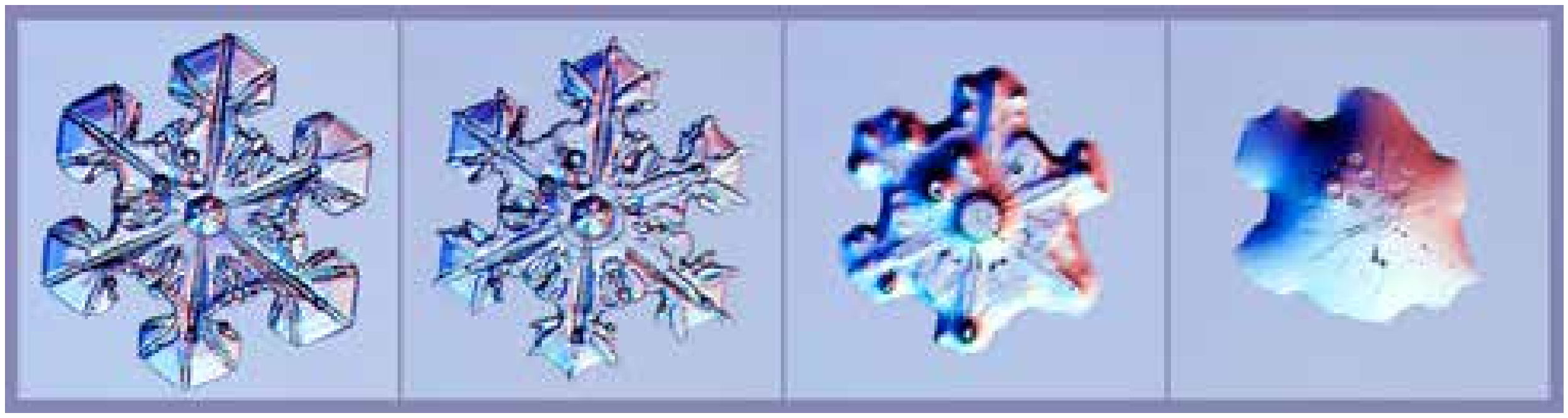

**Figure 11.7: This series of photos shows a snow crystal melting, with just 27 seconds elapsed between the first and last image. The temperature was just below 0 C during this series, illustrating how challenging it can be to photograph snow crystals at such warm temperatures.**

Although I am partial to a foam-core collection board and glass slides, this is by no means the only approach to snow-crystal photography. Many practitioners prefer to let the crystals fall onto some kind of dark-colored, wooly fabric for direct point-and-shoot photography, as illustrated in Figure 11.8. Nice looking specimens are often supported by a single cloth fiber, and the tangle of neighboring fibers provide an interesting backdrop for the photo. No collection board is needed when the crystals fall directly onto a piece of fabric, but searching through large numbers of specimens will be somewhat slower using this technique.

As I describe further below, my foam-core and glass-slide approach appeals to my science side, as it gives exceptional clarity and resolution, revealing find details in the crystals. But Alexey Kljatov's point-and-shoot technique is wonderfully pleasing from the artistic side, yielding a more natural view of these tiny slivers of ice. There are many different ways to photograph snow crystals!

## Granular Gems

Although photographing large, well-formed stellar snow crystals can yield some spectacular results, I have obtained many excellent pictures while focusing on small specimens, typically around 1-2 mm in size or even smaller. Many of the exotic snow-crystal varieties described in Chapter 10 are invariably quite small, and these tiny gems are worth pursuing. However, recording the full menagerie of snow-crystal types requires different techniques to find and photograph falling snow in this regime.

One big change when working with small crystals is that it is no longer possible to scan over a collection board to pick out promising specimens with the naked eye, at least not effectively. The crystals usually must be placed under a microscope just to see what you have. Although this sounds like something of a painstaking process, it is actually quite simple and enjoyable, and it can yield some remarkably interesting photos even when the clouds seem to be delivering little more than granular snow.

**Figure 11.8: (Left) A snow crystal supported by the fibers of a dark fabric, photographed by Alexey Kljatov in Moscow. Note how the out-of-focus fibers provide a pleasing background that adds a sense of depth and scale to the photo.**

The technique I like best is to lay out a set of glass microscope slides to catch the falling snow, and then just pick one up and look it over under my photo-microscope. If a crystal looks worthy of a photograph, I focus, adjust the lighting, and take the shot. If I find nothing worthwhile on the slide, I clean it off, set it back out to catch more snow, and pick up another slide to scan. Cycling through a half-dozen slides usually works quite well, allowing each slide to accumulate a new dusting of snow while one is looking through the others. I support the slides on a pair of knife-edge "rails" (made from tape) to keep them elevated. This keeps the bottoms of the slides clean and free of snow.

If the clouds are being unkind, nearly all the falling snow may consist of gloppy, granular, or rimed crystals, so at those times there is little one can do but try again later. Small hexagonal plates can usually be found even in quite retched conditions, but there are only so many photos one can take of those forms, as they all look pretty much the same. Surprisingly often, however, if one has some patience, there are interesting crystals waiting to be found in the mix, at least from time to time. Most of the photos of the smaller exotic crystals described near the end of Chapter 10 were taken by scanning over hundreds of glass slides when there were no large stellar plates to be had.

## 11.2 Optics and Lenses

There are many equipment options available for photographing snowflakes, depending on the image quality you seek and how much money you are willing to spend. At the low end, a smartphone with a $10 clip-on macro lens can yield some reasonably nice snowflake photos; not super-sharp, but good enough to capture the overall shape of stellar crystals, including some surface detail. Many people have been experimenting with this simplest form of snowflake photography, and a quick web search will yield many examples. This is a fine approach for getting started, just to see what kinds of crystals nature has to offer in your part of the world.

A next step up, if you already own some camera equipment, is to use a "reversed lens" at the end of an extension tube to make a relatively inexpensive macro lens. This technique is discussed in considerable detail on various photo blogs and websites, so again a web search will provide much more information than I care to write down here. By my estimation, a reversed-lens macro system can achieve an optical resolution of perhaps 10-20 microns, or even better if done with care using a high-quality lens. This is sufficient to take some excellent snowflake photos, and no one has demonstrated this better than Alexey Kljatov, who has captured many stunning snow-crystal photographs using a reversed-lens system. With quality crystals, an artistic eye, and some patience and effort, this technique can yield outstanding photos without spending a lot of money on fancy optical gear.

Given the scientific nature of this book, my main focus here will be on achieving exceptionally high optical resolution, with the overarching goal of revealing the finest details in snow-crystal structure, especially with smaller specimens. Obtaining resolutions of 2-5 microns is not an inexpensive undertaking, but the exceptional photos that result take one to a whole new level in snowflake photography. This kind of professional-grade hobby is clearly not for everyone, but such is the nature of this book.

To begin, it has been my experience that the choice of camera sensor is not especially important in snowflake photography. There are many high-quality, reasonably priced camera bodies on the market with sensors in the 20- to 30-megapixel range, and most would work well in this application. Sensors with larger physical dimensions tend to be better than smaller sensors, other things being equal, and the lens requirements are somewhat relaxed with a larger sensor as well; but this detail is probably not terribly important. As long as you have a reasonably modern digital camera, the imaging sensor will likely not be the limiting factor in obtaining quality photographs.

It is necessary to have at least two pixels for each real resolution element, so the optimal camera field-of-view is one that is matched to the optical resolution of the lens (described below). For example, if one wants to achieve a 2-μm optical resolution, the camera field-of-view should have about 1 μm/pixel. Put another way, a two-micron feature on the object snow crystal should image onto two sensor pixels, regardless of the actual physical size of the sensor pixels. Oversampling the image just wastes camera real estate, while undersampling will compromise the optical resolution. A 20-megapixel sensor imaged to produce 1 μm/pixel will give about a 4x5 millimeter field of view, and already this is larger than most snow crystals. If the physical size of the sensor pixels is 5 μm, which is typical for many cameras, this means that the lens should provide 5X magnification.

This exercise shows that a 100-megapixel sensor would not yield substantially better snowflake photos than a 20-megapixel sensor. A larger field-of-view is of little use, as very few snowflakes would fill it, and more pixels per micron would not help either, because the optical resolution of the lens is what usually limits the photo, as is described below. Occasionally one encounters a really huge crystal, and a field of view larger than 4x5 mm would be handy. But those situations are rare, and it is straightforward to just take two or more photos and stitch them together digitally in post-processing. When you consider the full parameter space, the camera is usually not the limiting factor for achieving high-quality snowflake photos.

While the choice of camera is not so important, the choice of lens is quite critical, especially when the goal is to obtain the highest possible resolution. Because resolution is of central importance in this chapter, I will use resolution as a starting point in the discussion of lens options. The usual definition of optical resolution (a.k.a. resolving power) is about what you would expect – the distance between two point-like objects that can just barely be resolved in an image. In practice, this means that a resolution of 2 μm will allow you to clearly distinguish features that are separated by at least 4 μm. Features that are 2 μm apart would be "barely resolved", which usually means they are almost completely blurred together, so not easily distinguished. Of course, one can provide a proper mathematical definition of resolution, but this rule-of-thumb is adequate for the present discussion.

In the case of snow crystals, the smallest structural features are about one micron in size, and this is a real physical limit imposed by surface tension and the Gibbs-Thomson effect (see Chapter 2). Any significant surface structure (like a rib, ridge, or sharp edge) that is substantially smaller than one micron would have such a high vapor pressure that it would soon sublimate away unless under extreme environmental conditions. Thus, unlike with most solid objects, one does not observe ever more detail in snow crystal structure by observing with ever higher resolution. There just isn't much to see beyond a resolution of about 1 μm. This is why electron microscope images do not generally reveal more structural details than optical images, as discussed in Chapter 6.

In my personal experience photographing snow crystals, I have found that using a lens with 2-micron resolution yields noticeably better photos than a lens with 4-micron resolution; the edges are crisper and overall the image has a sharper appearance. This can be seen fairly easily in side-by-side comparisons of a single snow crystal. Put into more-typical photographic language, I would say that a 4-micron-resolution lens yields a noticeably "softer" snowflake image than a 2-micron-resolution lens. This may sound obvious, but using a 1-micron-resolution lens generally does *not* yield substantially higher-quality images than a 2-micron-resolution lens, at least not when looking at snowflakes. Depth-of-focus is one reason for this (discussed below), but the main reason is simply that there are few additional structures to be seen in snow crystals at super-high resolution. The take-away message is that one gains little by going beyond using a 2-micron-resolution lens when photographing snow crystals. That has been my experience anyway.

## Macro and Micro Lenses

In terms of overall resolution, snowflake photography falls roughly between the usual regimes of macro photography and full-blown microscopy. Macro lens tend to yield resolutions in the 5- to 20-micron range, and a few exceptional lenses can do a bit better. Unfortunately, the optical resolution of most macro lenses is not listed in their specification sheets, even though high resolution is pretty much the main reason one purchases a macro lens. I have limited experience with the broad range of macro lenses that are available, but my experience has been that a 5-micron-resolution lens is quite good, and photography reviewers will speak of its bitingly sharp images. And it will cost around $1000. A 10-micron lens will likely cost less, but it will be reviewed as only okay, yielding somewhat "softer" images in high-resolution tests, lacking in their finer details.

For example, I have done some testing with the Canon MP-E 65mm lens, which is something of a high-resolution macro stalwart, well reviewed by many macro photographers. When set to its highest-resolution setting (5X), I measured an overall resolution of about 4 microns using this lens (discussed below). I imagine others have made similar measurements, but I have not found much resolution data online, neither from the manufacturer or lens reviewers. The world of macro photography is generally not a very quantitative place, and that can make it difficult to know what you are buying. It is hard to know what to make of adjectives like "soft" or "bitingly sharp".

In contrast, microscope objectives invariably list resolving power (resolution), numerical aperture, working distance, and depth of focus as part of their specs. These numbers can still be deceptive, as they refer only to on-axis viewing, and inexpensive microscope objectives can have dreadful optical quality even with supposedly good specs. However, for most reputable manufacturers (Zeiss, Olympus, Mitutoyo, etc.), the specifications provide a fairly reliable assessment of the quality of their objectives. In this respect, it is generally easier to purchase a quality microscope objective of known performance than a high-resolution macro lens, and microscope objectives are usually somewhat cheaper as well.

While most people equate microscope objectives with full-blown (and expensive) microscopes, Figure 11.9 shows how a simple microscope objective can be turned into a DIY photo-microscope. This configuration is identical to the usual reversed-lens setup, just replacing the reversed lens with a higher quality microscope objective.

The biggest drawback with this layout is scattered light, which sends unwanted light onto the camera sensor. Fortunately, this problem can often be ameliorated by carefully covering the inside of the extension tube with highly absorbing black flocking paper. A field stop in the object plane is also useful, as this prevents otherwise unused light from entering the objective and rattling around inside the extension tube.

The simple optics and fixed extension tube in Figure 11.9 means that focusing involves either moving the camera or moving the subject, usually with some kind of mechanical translation stage. This focusing method is the norm for both microscope objectives and high-resolution macro lenses. Moving lens elements within a lens (using a focusing ring) is generally not practical at high resolution, nor is in-lens auto-focus.

Personally, I tend to favor microscope objectives over macro lenses for several reasons:

1) There is a larger selection of microscope objectives available, at generally higher quality (in my opinion).
2) Microscope objectives have clear specifications that include their optical resolution, unlike macro lenses.
3) Microscope objectives are typically somewhat cheaper than macro lenses, for a given resolution.
4) Microscope objectives are far more compact than macro lenses, making the much easier to incorporate into snow-crystal growth chambers.
5) Microscope objectives are easily adaptable for use with different camera bodies.

## The Diffraction Limit

In the microscopy world, optics are nearly always diffraction limited, meaning that the wavelength of light is ultimately what limits the image resolution. This is not true with normal photography, but the diffraction limit will play a role in macro photography at the highest resolutions. As a general rule, if the overall image resolution is smaller than about $10\lambda$, where $\lambda \approx 0.5$ μm is the wavelength of visible light, then the diffraction limit will begin to be an important consideration. Because I am mainly concerned with high-resolution imaging in this book, I will assume that diffraction is one of the main factors limiting the overall optical resolution.

Because a 2-μm optical resolution is substantially larger than $\lambda$, the diffraction limit takes on a relatively simple mathematical form. Using the terms defined in Figure 11.10, and assuming an index-of-refraction of unity for imaging in air, we can assume a small-angle approximation with $\sin\theta \approx \tan\theta \approx \theta$, where $\theta$ is measured in radians. For $\theta = 0.15$, $\sin\theta = 0.149$ and $\tan\theta = 0.151$, so this is an excellent assumption.

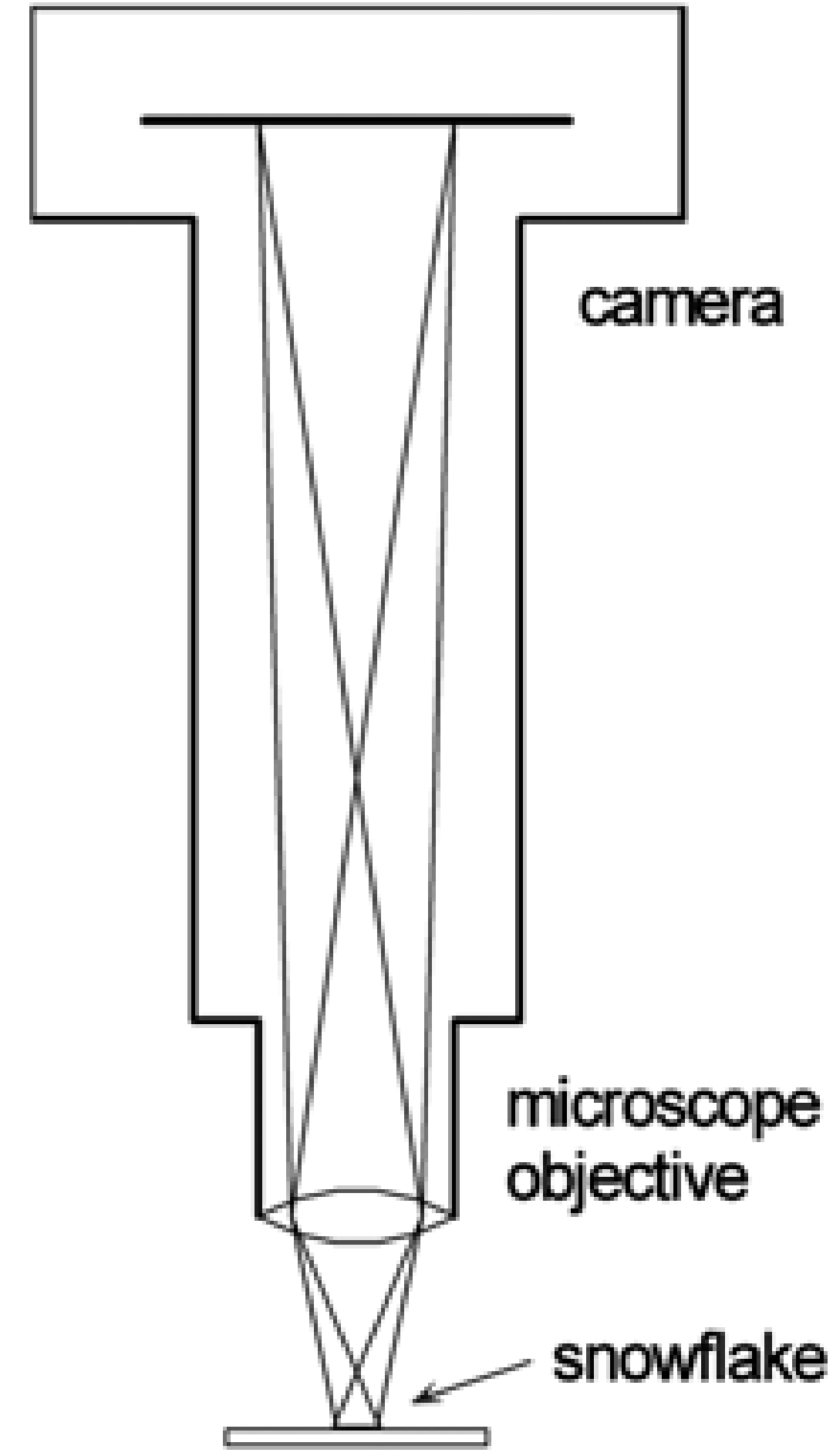

**Figure 11.9: A basic DIY photo-microscope consists of little more than a microscope objective, an extension tube, and a camera body. Viewing is done through the camera, for example displaying the image on a TV monitor via the live-feed camera output. The field of view of the camera can be set by choosing an appropriate length for the extension tube, which is the same as with reversed-lens setups.**

Microscope objectives are typically specified by a *numerical aperture*, $N_A$, and in our small-angle approximation this is given by

$$N_A \approx \theta \qquad (11.1)$$

In photography, a lens is specified by its *f-number*, $f_\#$, which can adjusted by changing the

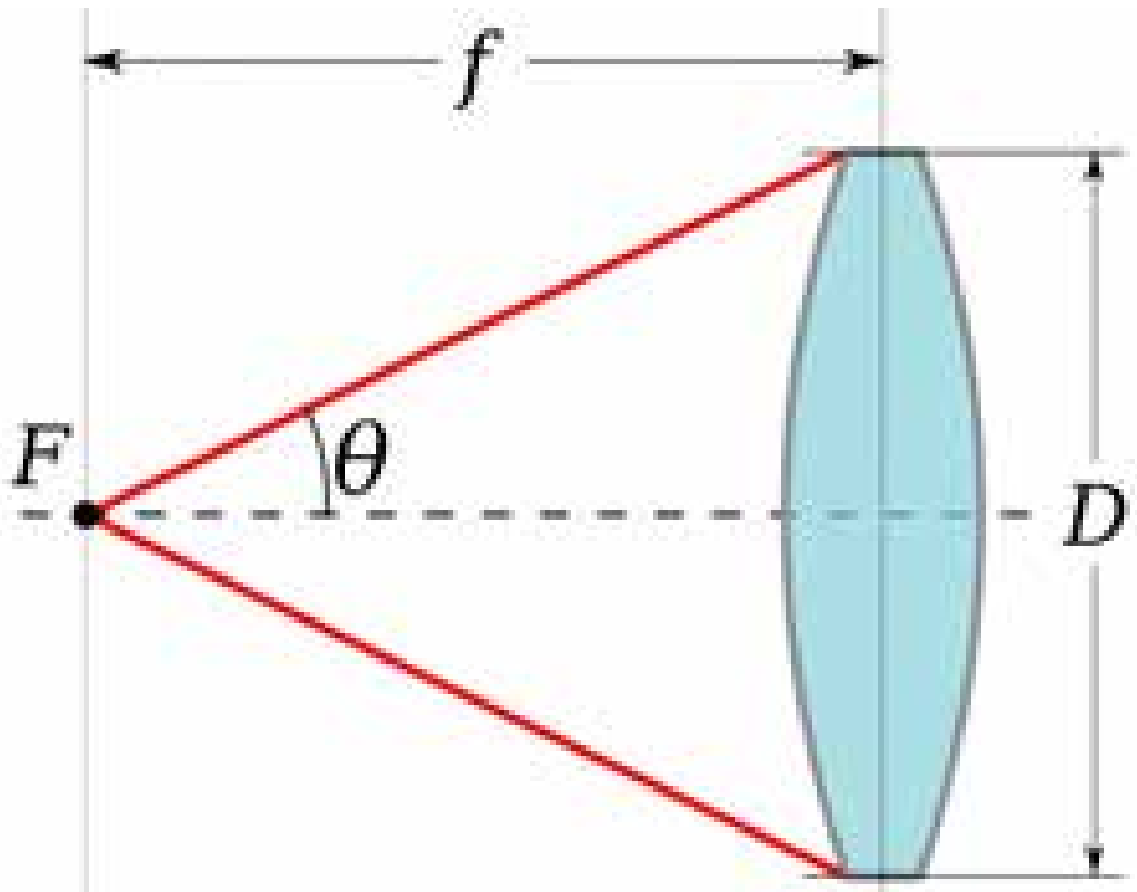

**Figure 11.10: This sketch defines the focal length $f$, the lens aperture $D$, and the half-angle $\theta$ for a simple microscope objective.**

aperture of the lens, and this is given by (in the small-angle approximation)

$$f_{\#} \approx \frac{1}{\theta} \approx \frac{1}{N_A} \quad (11.2)$$

Photographers and microscopists tend to use different nomenclatures, but the underlying optical physics is the same.

In the diffraction limit, the optical resolution is given by

$$R_{xy} \approx \frac{\lambda}{2N_A} \quad (11.3)$$

and I will typically assume $\lambda \approx 0.5$ μm.

Additionally, there is a corresponding resolution perpendicular to the image plane, $R_z$, which is given by

$$R_z \approx \frac{\lambda}{2N_A^2} \quad (11.4)$$

This is usually called the *depth-of-focus* or depth-of-field; parts of the object that are within $\pm R_z$ of the focus position will be essentially in-focus, while parts outside this range will be considerably out-of-focus.

For example, if we want an optical resolution of $R_{xy} = 2$ μm, then we need a microscope objective with a numerical aperture of at least $N_A \approx 0.125$, and this means the depth-of-focus will be a scant $R_z \approx 16$ μm. This latter number can be problematic, because most snow crystals are thicker than 16 μm, even plate-like crystals. This is an inescapable problem in snowflake photography – one cannot have both high resolution and a large depth-of-focus simultaneously. In normal photography, one closes down the aperture to increase the depth-of-focus, but that no longer works when the resolution is diffraction limited.

As another example, the Canon MP-E 65mm macro lens has an f/2.8 aperture, and the above equations give a corresponding numerical aperture of $N_A = 0.36$ with a theoretical resolution of $R_{xy} \approx 0.7$ μm, which would be awesome. However, this lens is not diffraction limited when used at is maximum resolution, and the measured resolution (see below) is about 4 μm. This is typical for most traditional camera lenses, even high-quality macro lenses. Any good microscope objective will be diffraction limited, or close to it, so the above equations are quite useful for evaluating the performance of quality objectives. Inexpensive objectives may have serious deficiencies, but quality objectives should always meet spec, at least on-axis. With camera lenses, however, it is often not possible to know the optical resolution unless you measure it yourself.

One straightforward way to estimate the optical resolution of a lens is to image a calibrated resolution target, as demonstrated in Figure 11.11 for several example lenses. In all cases, the camera sensor was not a limiting factor in determining the quality of the images. Although making an absolute measurement of $R_{xy}$ is difficult, comparisons between lenses are straightforward. By my reckoning, these images reveal that the 5X Mitutoyo objective seems to meet its spec of having a 2-μm

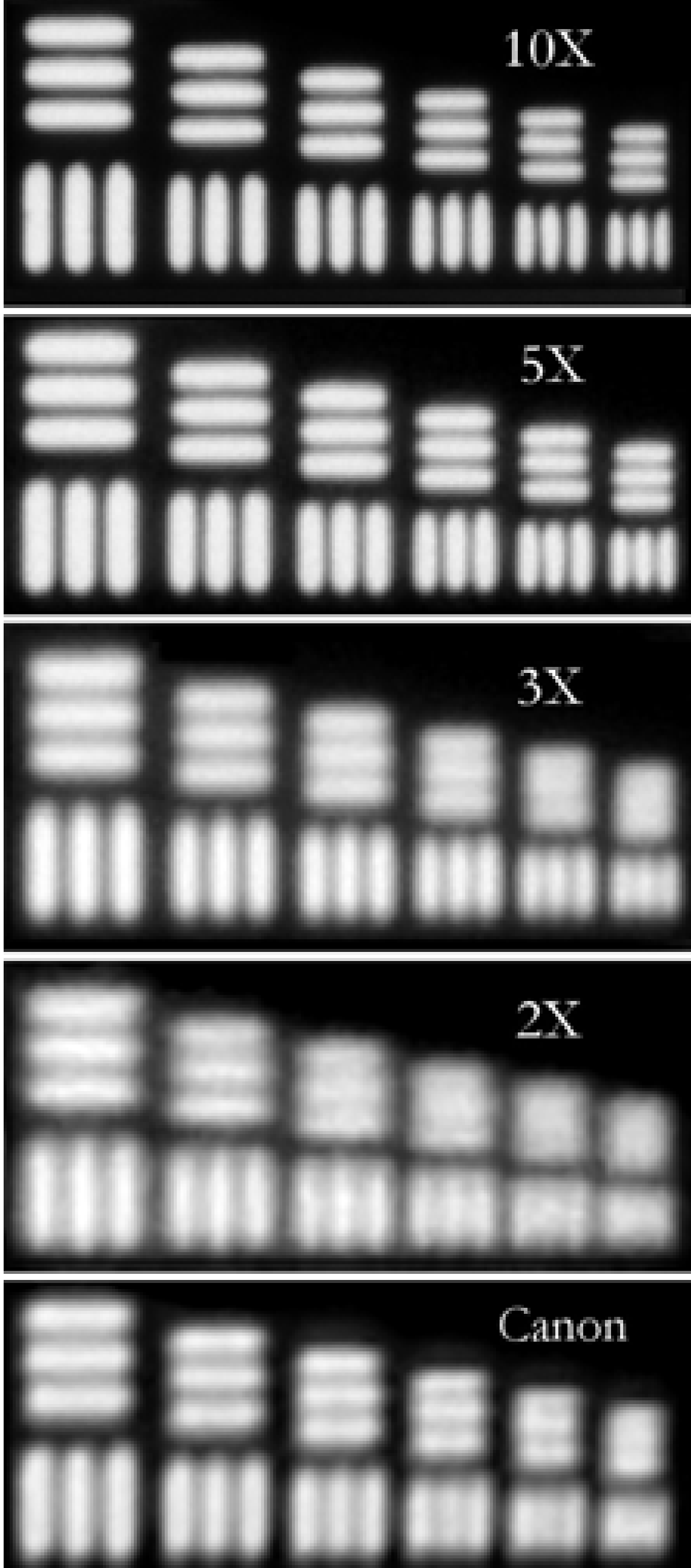

**Figure 11.11: Images of a calibrated resolution target using three Mitutoyo Plan APO Objectives (10X, 5X, and 2X), a Mitutoyo 3X Compact Objective (3X), and the Canon MP-E 65mm 1-5X Macro Lens set at 5X/f2.8 (f4.0 is similar). The Mitutoyo 2X Compact Objective (not shown here) yields a resolution-target image that is quite similar to the 3X objective. The spacing between the bars is 7.8, 7.0, 6.2, 5.6, 4.9, and 4.4 μm.**

resolution, and from this I obtained the measurements shown in Figure 11.12.

The microscope objectives mostly met spec, except for the Compact 3X, where the specified $R_{xy} = 2.5$ μm simply does not agree with the measured resolution of about 4 μm. However, the specified numerical aperture (0.07) gives a theoretical resolution of $R_{xy} \approx 3.6$ μm, which agrees reasonably well with the measurement. As far as I can tell, this is a specification error by Mitutoyo, which would be unusual for this company. The Canon lens has no resolution specification, and the 4-μm number is the best I could get using this lens. The Canon resolution-target images at 5X with f/2.8 and f/4.0 were similar, and the resolution rapidly deteriorated at lower magnification or higher f-numbers, as one would expect.

Another useful method for examining resolution is to sprinkle 10-μm beads onto a glass substrate, as shown in Figure 11.13. In this example, the Mitutoyo 5X objective was not properly infinity-corrected, which resulted in a slightly non-flat image plane. Thus the center beads are in focus while the corner beads are a bit out-of-focus. With a slight focus adjustment, the corner beads can be brought into focus while the center beads are then slightly blurry. The array of small beads makes it easy to evaluate the focus and resolution across the entire field of view, which is useful for optimizing an optical system.

## Focus Stacking

As I mentioned above, snowflake photography always involves a trade-off between resolution and depth-of-focus. When the resolution is high, the depth-of-focus is low, so only a thin plane is brought into sharp focus on the sensor. If the snow crystal is tilted with respect to that plane, or if the crystal is not thin and flat, then not all parts of the crystal can be brought into focus at the same time. This is a fundamental feature of diffusion-limited optics, so there is no way to avoid this trade-off.

Focus stacking, however, is an effective work-around that allows one to photograph complex snow crystals at high resolution over their entire structure. The basic idea is to take several pictures at different focus settings, with each photo bringing a different part of the crystal into focus. The images can then be combined digitally in post-processing to stitch together the in-focus pieces of each of the individual photos, thus creating a single image that appears to be in focus throughout. Several software packages are available to do the image reconstruction (for example, Helicon Focus), and much information about focus stacking can be found online. There are even hardware systems (such as from StackShot) that will automatically move the camera focus in programmable steps using a translation stage to acquire the desired series of images.

| Lens/Objective | NA spec | Rxy(μm) spec | Rz(μm) spec | Rxy(μm) meas | Rz(μm) calc | Rxy(μm) calc | Working dist(mm) | Price US$ |
|---|---|---|---|---|---|---|---|---|
| Mitutoyo Plan APO 10X | 0.28 | 1 | 3.5 | 1.5 | 3.2 | 0.9 | 34 | 880 |
| Mitutoyo Plan APO 5X | 0.14 | 2 | 14 | 2 | 13 | 1.8 | 34 | 700 |
| Mitutoyo Plan APO 2X | 0.055 | 5 | 91 | 5 | 83 | 4.5 | 34 | 930 |
| Mitutoyo Compact 3X | 0.07 | (2.5) | (23) | 4 | 51 | 3.6 | 78 | 500 |
| Mitutoyo Compact 2X | 0.06 | 4.6 | 76 | 4.5 | 70 | 4.2 | 92 | 460 |
| Canon MP-E 65mm @5X | | | 50 | 4 | | | 42 | 950 |

**Figure 11.12: Measurements and specifications of several microscope objectives and the Canon MP-E macro lens. For the Compact 3X objective, my measurements suggest that the specified numerical aperture is accurate, but the specified resolution (shown in parentheses) is incorrect.**

Nearly all skilled snowflake photographers use focus stacking to some extent, as this is a straightforward technique for effectively increasing the depth-of-focus while maintaining a high optical resolution. Large stellar snow crystals are intrinsically thin and flat, so photographing these crystals face-on usually requires minimal focus stacking even at high resolution. It is hard to avoid some tilt of the crystal relative to the image plane, however, so I often take 2-3 pictures while adjusting the focus to make sure all the branch tips are nicely in focus. This kind of minimal focus stacking is easy to apply and nearly always yields good results.

It is often desirable to tilt a flat crystal over quite a large angle to obtain specular light reflections (see the section on Specular Reflection illumination below), and in this case a great deal of focus stacking is needed at high resolution. Don Komarechka is the undisputed focus-stacking champion in snow-crystal photography, often combining 30-50 individual shots to obtain a single in-focus image, as I describe below [2013Kom].

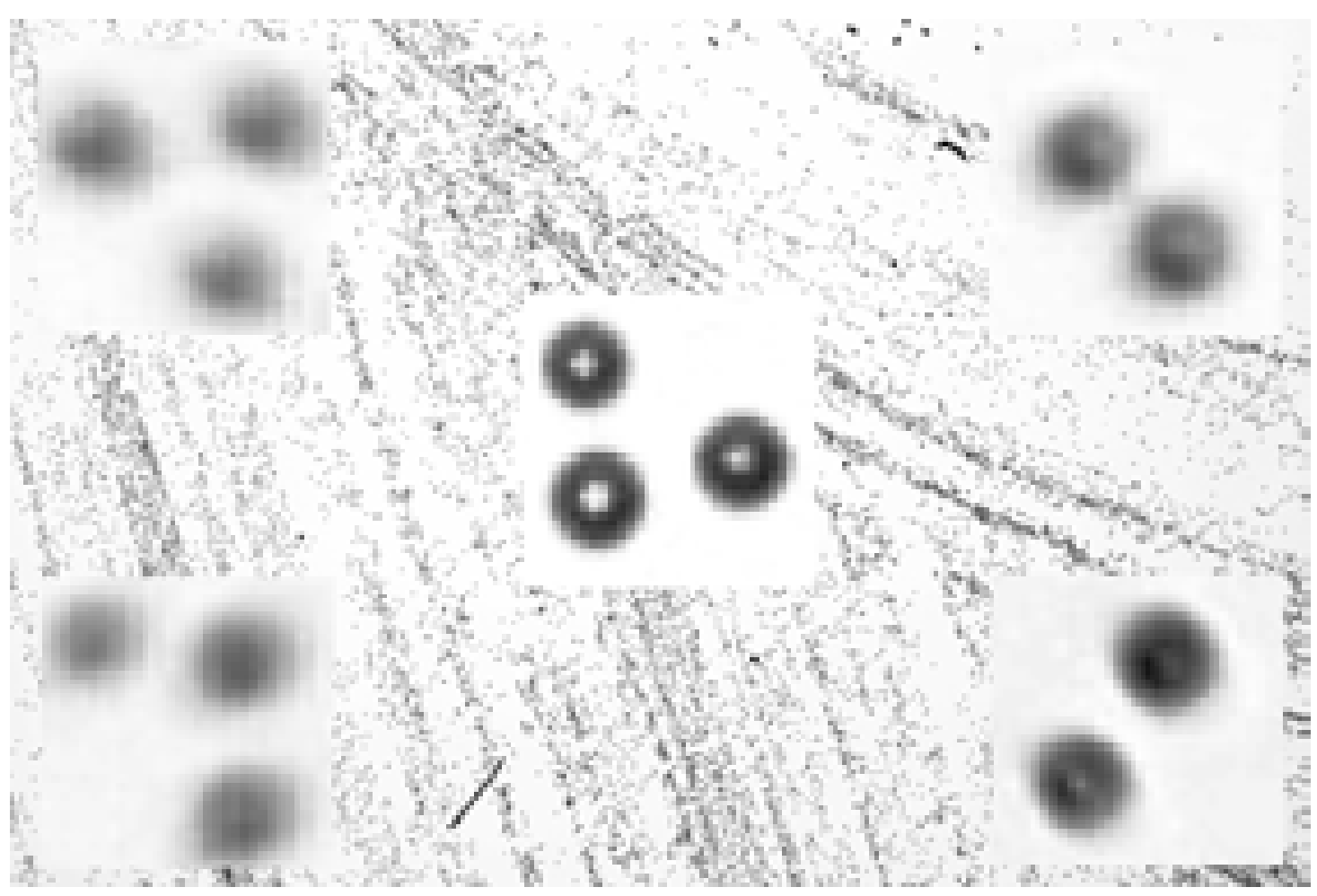

**Figure 11.13: (Left) A photograph of a glass substrate sprinkled with 10-μm glass beads, imaged with a Mitutoyo Plan APO 5X objective. The total field of view is about 5x4 millimeters, and the inset images show magnified snippets of the main image taken from the center and the four corners. The focus was set to best image the center beads, which resulted in somewhat out-of-focus beads in the corners. Combining several images using focus stacking yielded a sharp focus across the entire image.**

Like any photographic tool, focus stacking can be employed or not, depending on what is being photographed and what kinds of optical effects are desired. If one opts for a lower overall optical resolution, then perhaps a single image is sufficient. Moreover, having the extremities of a crystal appearing slightly out of focus often gives an image a pleasing sense of depth, and this type of optical illusion is often used by photographers (a version of *bokeh*). But if super-high resolution is desired over an entire crystal, then some focus stacking, and perhaps a lot of focus stacking, is usually required. Focus stacking is a nice trick that is both easy and inexpensive to use, so it has become a valuable addition to any snow-crystal photographer's toolkit.

## Point-and-Shoot versus Stable Mounting

In most circumstances, photomicroscopy is not performed in a point-and-shoot fashion using hand-held optics. Microscopes tend to be rigid structures where the camera, the optics, and the object being viewed are all solidly mounted. The reason is that photomicroscopy subjects are so tiny that it is nearly impossible to hold everything steady enough by hand to get good pictures. On the other hand, macro photography is often a point-and-shoot affair, as a hand-held camera+lens gives the photographer plenty of freedom to move around an object to get just the right angle for an artistic shot. Snowflake photography is somewhere between these two, as a resolution of 5-10 microns is quite low by microscopy standards, but is quite high for macro photography. Given this intermediate position, some snowflake photographers use the point-and-shoot method while other go with rigidly mounted hardware. Both can be made to work, but there are trade-offs for each.

For low-resolution imaging at 10-20 microns, point-and-shoot is relatively easy, inexpensive, and effective. Not as easy as normal photography, but doable. You let the snowflakes fall where they may, and then simply photograph them as you would anything else. It requires a steady hand, because the crystals are small, and a bright flash is useful to freeze any remaining camera motion. The point-and-shoot method is especially convenient in that there is no additional investment in mounting hardware. Even at relatively low resolutions, focusing is not accomplished by rotating a lens ring, but simply by moving the camera+lens in and out.

For high-resolution imaging, point-and-shoot becomes substantially more challenging, and the savings gained by not having to buy a lot of mounting hardware tends to be lost in the need for high-performance camera equipment and a great deal of image processing. A bright flash becomes an absolute necessary at high resolution, as it is practically impossible to hold a camera steady enough by hand unless you use a super-fast shutter speed. In addition, you probably want to take a lot of photos quickly, because the camera is moving around somewhat and changing the focus, and this means that both your camera and flash need to be capable of taking several pictures per second for best results. Don Komarechka has described his point-and-shoot methods in [2013Kom], and it involves some pretty high-end camera gear.

Because I like to achieve the highest possible resolution with relatively simple camera gear, I prefer to use a stable mounting platform, essentially like a traditional microscope. I usually make my own mounting setup using the basic single-lens optical layout shown in Figure 11.9, using a variety of hardware options described briefly in Chapter 6. There are many options for mounting hardware, but a tripod is one of the worst, as tripods are generally too unwieldy and unstable for microscopy.

Focusing is the most expensive part of a rigid mounting system, as focusing requires a mount that is both stable and movable. One option is to use a linear positioning stage (see Chapter 6), and I am partial to those available a Thorlabs. A focusing ring works also, as shown in Figure 11.14. Another option is to mount the camera on a motorized stage like those from StackShot. The StackShot option is somewhat expensive, but it provides the additional benefit of automated focus stacking, which quickly becomes an indispensable tool if you choose to build it into your system.

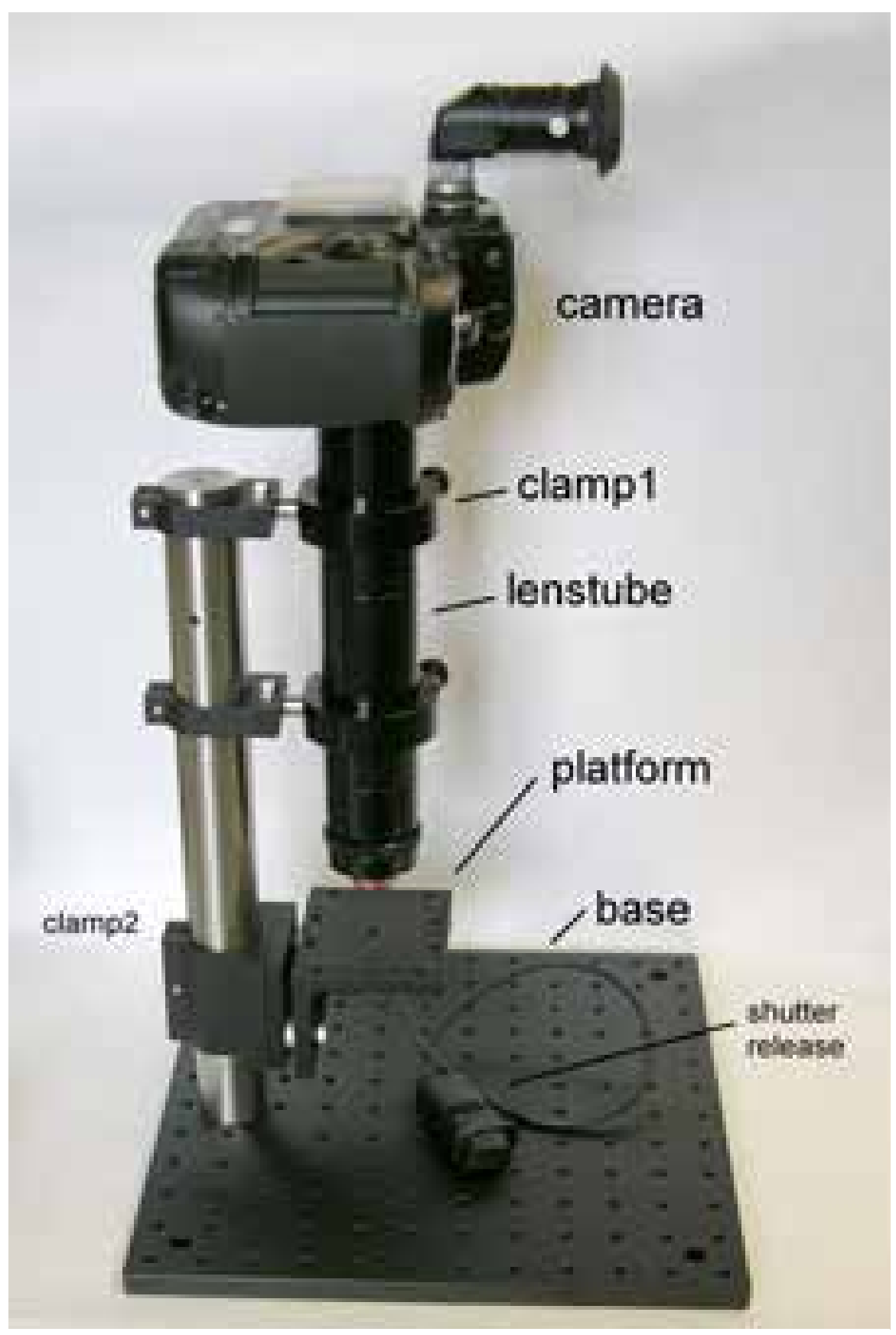

**Figure 11.14: One possible rigid mounting system for snowflake photography, using the basic optical layout shown in Figure 11.9. Here the microscope objective is mounted on a threaded ring that moves the objective up and down by rotating the ring. A right-angle eyepiece is used to view the image through the DSLR camera. The base plate, vertical bar, clamps, and other mounting components are commercially available from Thorlabs and Edmund Optics.**

One of the biggest advantages of a rigid mounting system is ease of use. You drop a snowflake onto a glass slide (for example), place it under the microscope objective, and it just sits there solid as a rock. You can move it around in the field of view, adjust the focus, adjust the lighting, and take the shot when everything looks good. No flash is needed, and slow shutter speeds are not a problem. Focus stacking is easy as well, because the crystal does not move laterally when you tweak the focus, at least not if you use a quality linear positioning stage. I clearly prefer the up-front costs of stable mounting hardware over the constant trials and tribulations of point-and-shoot photography out in the cold, but that is a matter of personal taste. Overall, rigid mounting tends to win over point-and-shoot at the highest resolutions.

## 11.3 Illumination Matters

One thing that clearly separates snow-crystal photography from other types of photography is that ice is transparent, like glass. With opaque materials, one can simply shine some light on an object and expect to get a reasonably good picture. Of course, lighting is important for taking excellent photos in any situation, but the type of illumination one uses can be particularly critical in producing high-quality snow-crystal photographs.

To understand why the type of illumination matters so much in snow-crystal photography, it is necessary to examine how transparent objects scatter and refract light. For example, a bank of snow looks white because it is made from a large number of transparent ice crystals. When light shines on these crystals, some light reflects off every air/ice interface. Only a few percent reflects from each surface, while the rest is transmitted, and very little light is absorbed in the process. But after encountering thousands of air/ice surfaces in the snow, the light is mostly scattered this way and that until it makes its way back out of the snowbank. The net result is that light striking the snow is scattered in all directions with little absorption, and this is exactly what being "white" means. Any pile of transparent grains appears white, as shown in Figure 11.15. And paper is white because it is made from tiny, transparent cellulose fibers. This answers that amusing question: when a snowbank melts, where does all the white go?

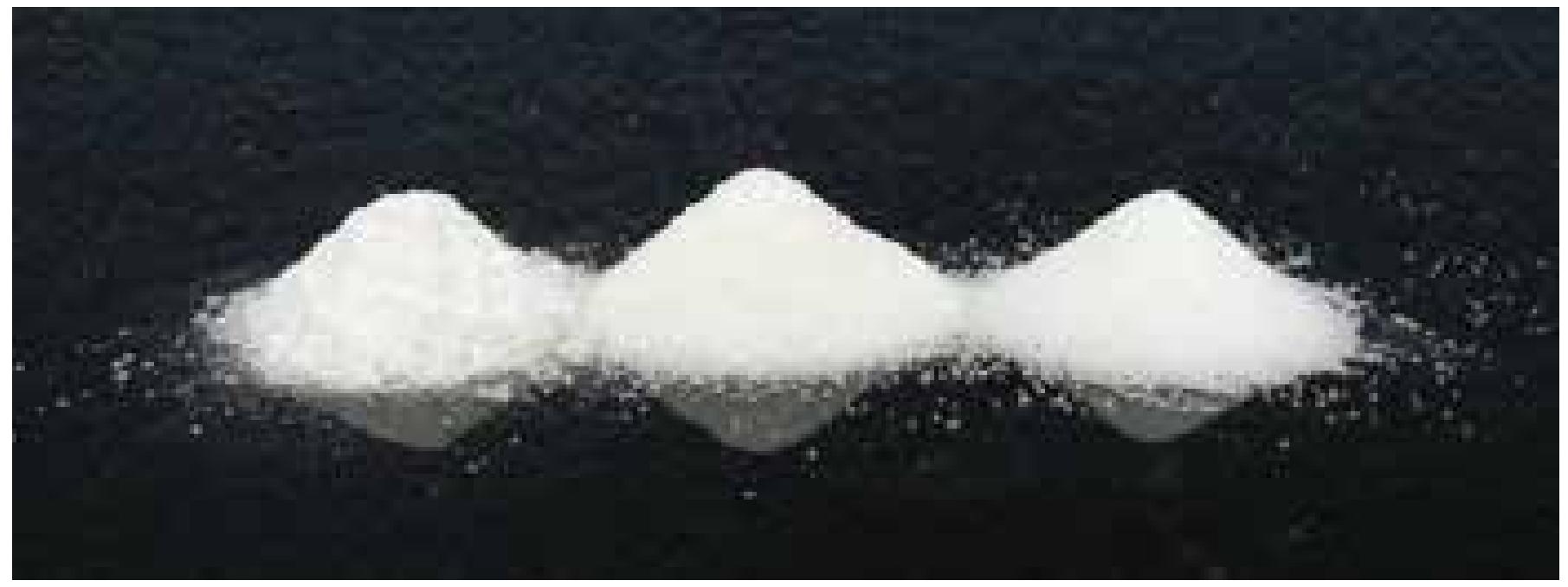

**Figure 11.15: (Left) This photo shows piles of crushed glass (left), sugar crystals (center), and salt grains (right). In all cases, the individual particles are transparent. The piles look white because light scatters off the countless small surfaces.**

A pane of glass looks transparent because it contains only two smooth, planar surfaces. When you look at the pane from certain angles, you can see the light reflecting like a weak mirror. But otherwise light transmits through the glass and it looks transparent. If you scratch up the surfaces with sandpaper, however, then the light scatters in all directions and the surface takes on a whitish appearance. It is not bright white because some light incident on the front surface of the scratched pane makes it out the back surface. With a pile of ground glass, the light mostly keeps scattering around until it comes back out the front; the probability that it makes it through to the back of the pile is extremely low.

In addition to reflecting light from it surfaces, transparent objects also bend light via refraction. This is how lenses work, and a lens-shaped piece of clear ice would behave similarly. The fact that ice is clear – transmitting, reflecting, scattering, and refracting light – brings an added dimension of lighting effects to snow-crystal photography. Even interference effects can be important in some circumstances, as I discuss below.

In some circumstances, a snowflake can be thought of as a complex lens that refracts light through various angles as it is transmitted through the clear ice. In other circumstances, the snowflake can be thought of as a small sliver of scratched-up glass that scatters light from its highly structured surface. And a heavily rimed snowflake begins to look like a small pile of crushed glass, because its surface is covered with a dense layer of frozen droplets.

One aspect of light transmission that is completely negligible in snowflakes is color dispersion. A beam of light transmitted through a glass prism will be dispersed into a rainbow of colors, and this would happen with ice as well. But it would have to be a very large block of ice for this to be even remotely noticeable. One does not normally observe color dispersion from glass bowls, pitchers, cups, or other glass objects. Likewise, color dispersion in tiny snow crystals is completely negligible.

Perhaps the easiest way to understand the different ways illumination affects snowflake photography is by example. People have been experimenting with different types of lighting for many years, and it is straightforward to categorize different photographs by the type of lighting used. The sections that follow focus on these lighting categories.

## Side Illumination

What I am calling "side illumination" could also be called using "ambient light" illumination. The basic idea is illustrated in Figure 11.16, and this is essentially the type of lighting you get when you simply photograph a snowflake resting on an opaque surface with a point-and-shoot camera. Light shines down on the crystal from all around, and some of that light enters the camera lens and is focused onto the sensor.

If you supply your own lighting, then there are an infinite number of possible variations of the side lighting method shown in Figure 11.16. For example, one might shine a bright light in from one side only. Or one might shine blue light in from one direction and red light in from another. The creative possibilities for using colored light in side illumination have received little attention, so I believe there are some interesting opportunities for branching out into new photographic directions here.

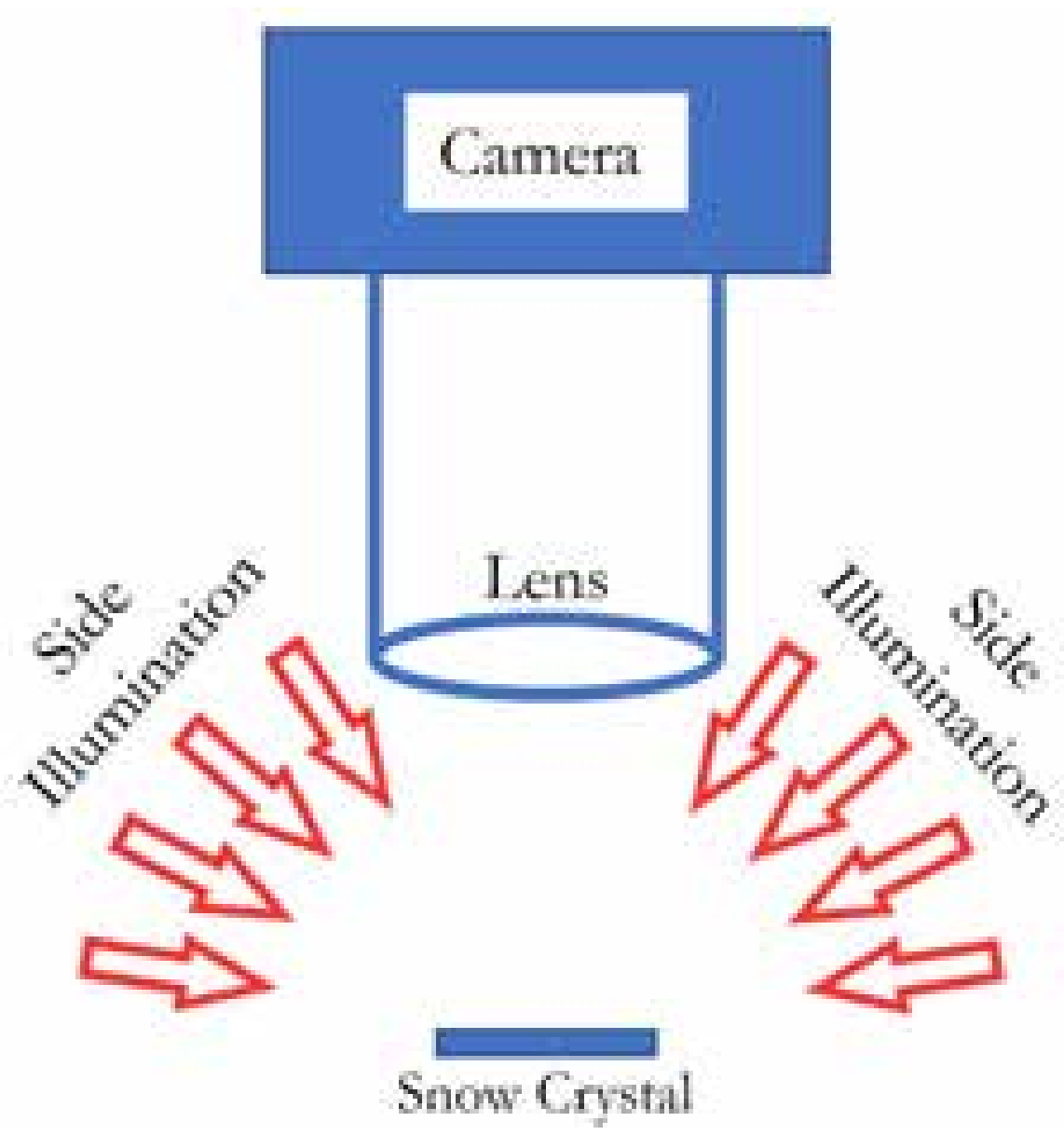

**Figure 11.16: Side illumination. In this straightforward method, light shines in from the side to illuminate the snow crystal, and scattered light enters the lens and is focused onto the camera sensor.**

One aspect of side illumination that I want to focus on is that only scattered light can contribute to a snowflake photograph taken using this method. Refraction is irrelevant because the transmitted light strikes the opaque substrate and is absorbed. What this means is that side illumination tends to accentuate the crystal edges and surface structures.

Another aspect is that a flat ice plate that lies perpendicular to the viewing angle will not reflect any light directly into the camera lens. When a flat plate is viewed face on, the direct reflection of side illumination will come out the other side of the crystal and be lost. Because no light shines down on the crystal from the position of the lens, no directly reflected light from the plate surface can enter the camera. This means that flat plates appear somewhat invisible, like a pane of glass, when using side illumination.

Alexey Kljatov is a master of side-illumination snowflake photography, and several examples from Alexey illustrate many features of this illumination method. A number of common features can be seen in Alexey's photos, including:

1) Thin, plate-like regions scatter little light, making them appear almost invisible in some photos (Side Illumination #1 (SL#1)).
2) Rimed structures appear bright white, like a pile of crushed glass (SL#2).
3) Crystal edges are generally quite bright, as they strongly scatter light into the camera (SL#1, SL#3).
4) Surfaces with a lot of structural detail tend to have an overall whitish appearance, like a small flake of etched glass.

**Side Illumination#1, by Alexey Kljatov.** (Below) In this photo, the flat, plate-like parts of the crystal scatter no light and are basically invisible, like small panes of glass. A small dimple marks the center of this crystal, surrounded by a well-formed hexagonal rib. The outer edges of the crystal scatter light strongly, giving them a bright white appearance. The crystal is supported by a few fibers from the underlying piece of cloth.

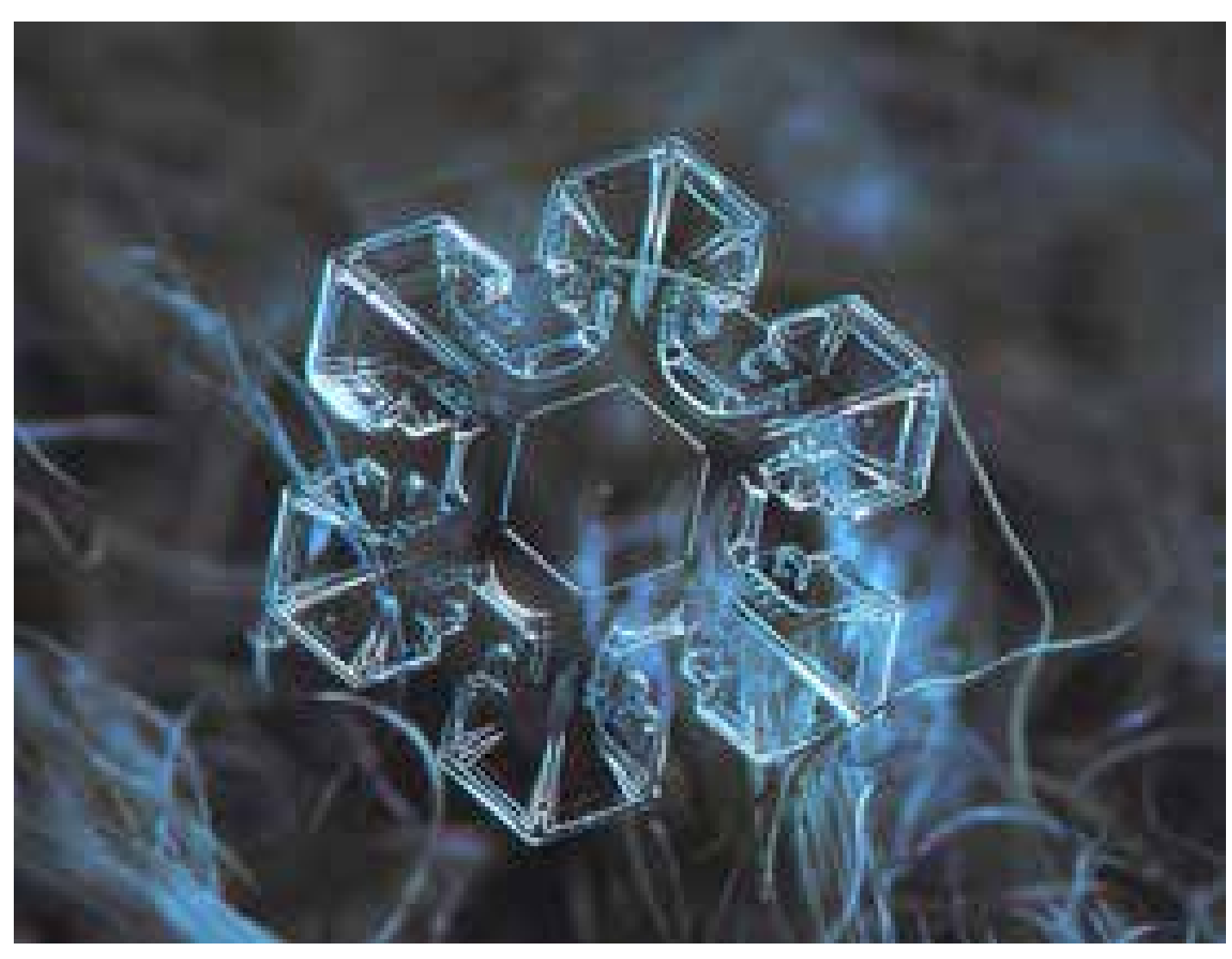

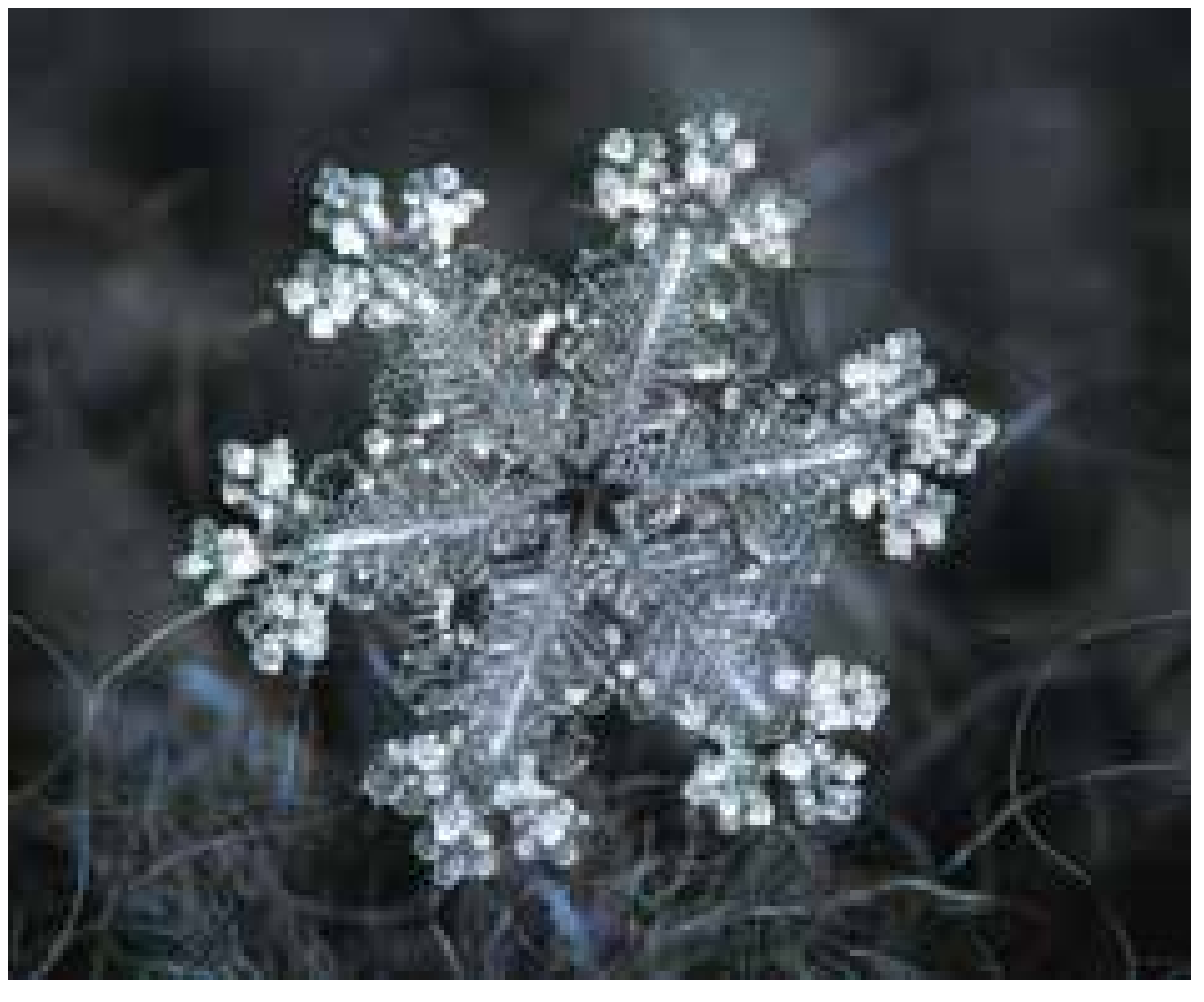

**Side Illumination#2, by Alexey Kljatov.** The outer branches of this stellar crystal are heavily rimed, and the rime particles scatter light strongly, giving these areas a bright-white appearance. Only the star-shaped central region is flat enough to appear transparent.

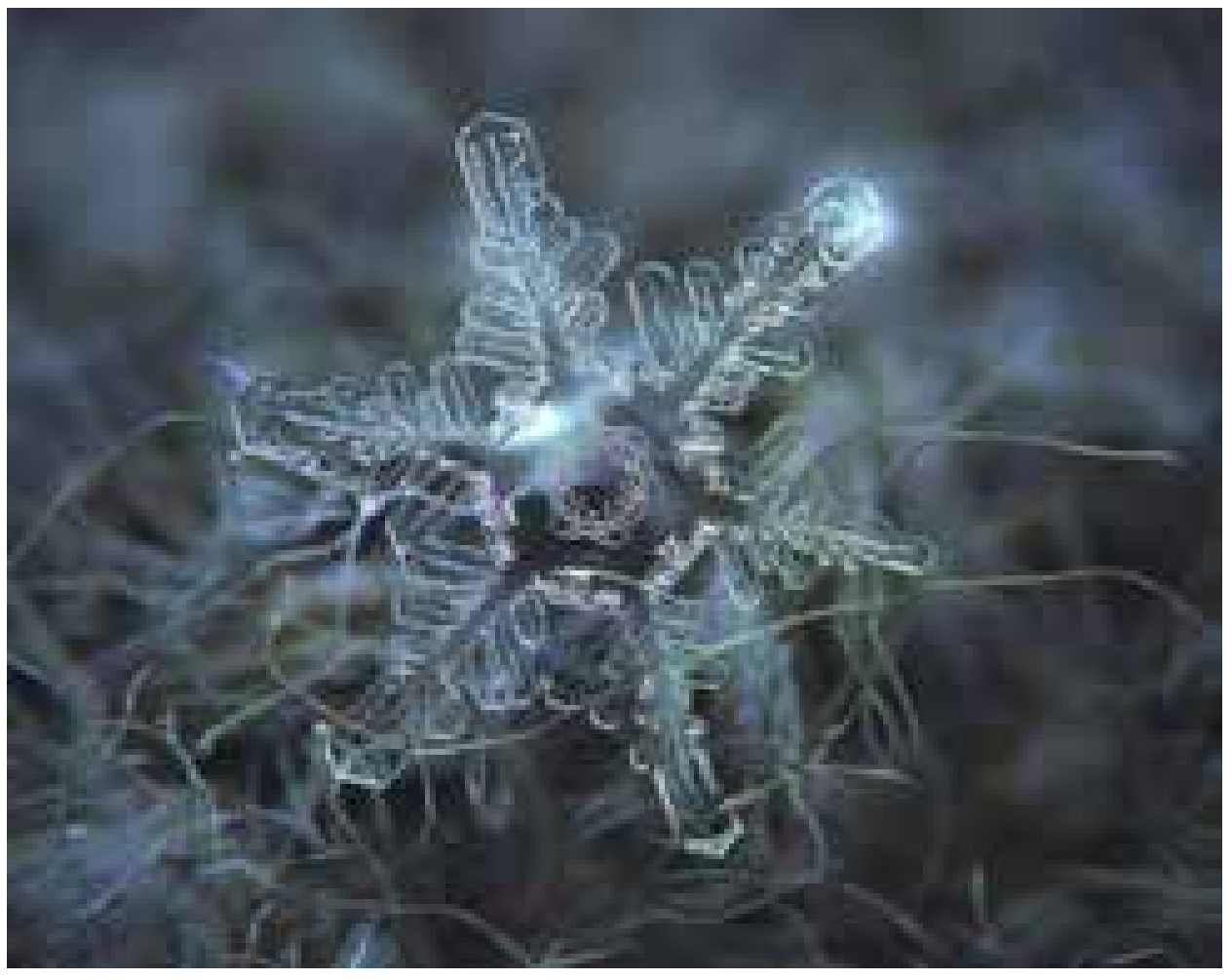

**Side Illumination#3, by Alexey Kljatov.** Some thin-plate snow crystals scatter so little light that they can become difficult to see using side illumination. The fact that one branch of this crystal is substantially smaller than the others is not immediately apparent.

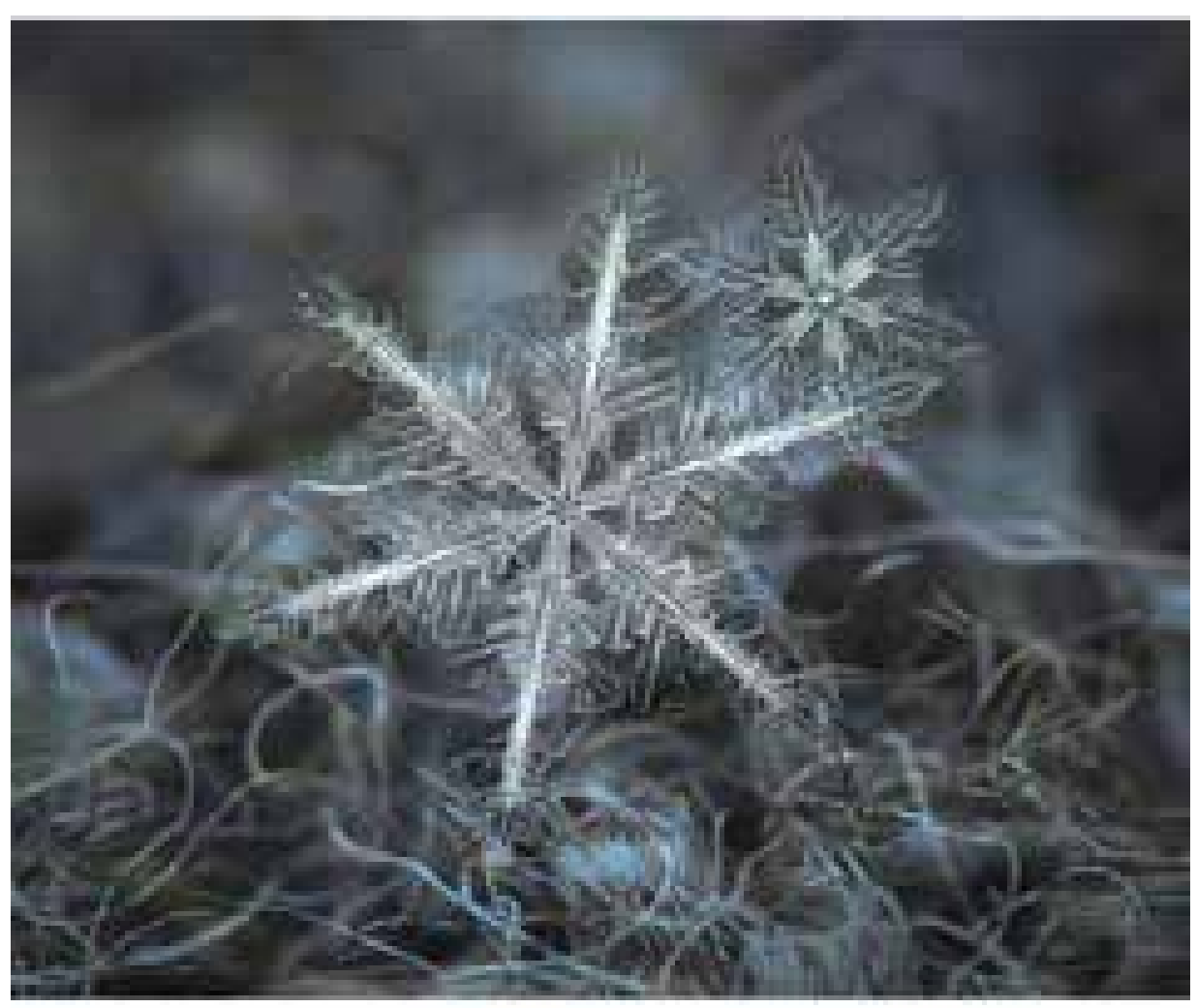

**Side Illumination#4, by Alexey Kljatov.** In many stellar dendrites, the axis of each branch is thick and structured, so these show up brightly using side illumination. Some of the outer, thin-plate regions of these crystals, however, can only be seen by virtue of their bright edges.

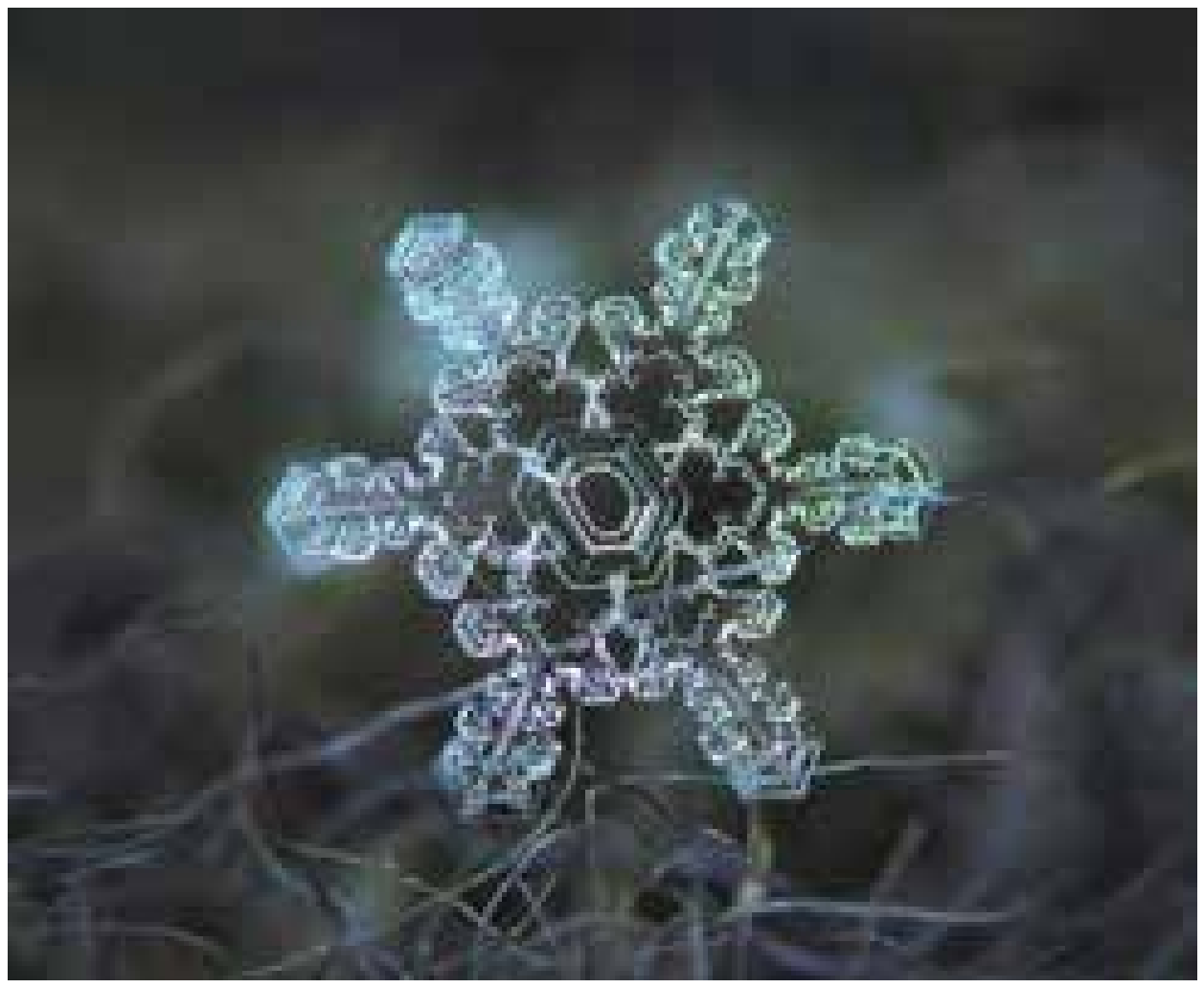

**Side Illumination#5, by Alexey Kljatov.** People sometimes report seeing snow crystals with "holes" in their centers. This is an optical illusion that comes from side illumination. The central region of this crystal is so thin and flat that becomes almost invisible.

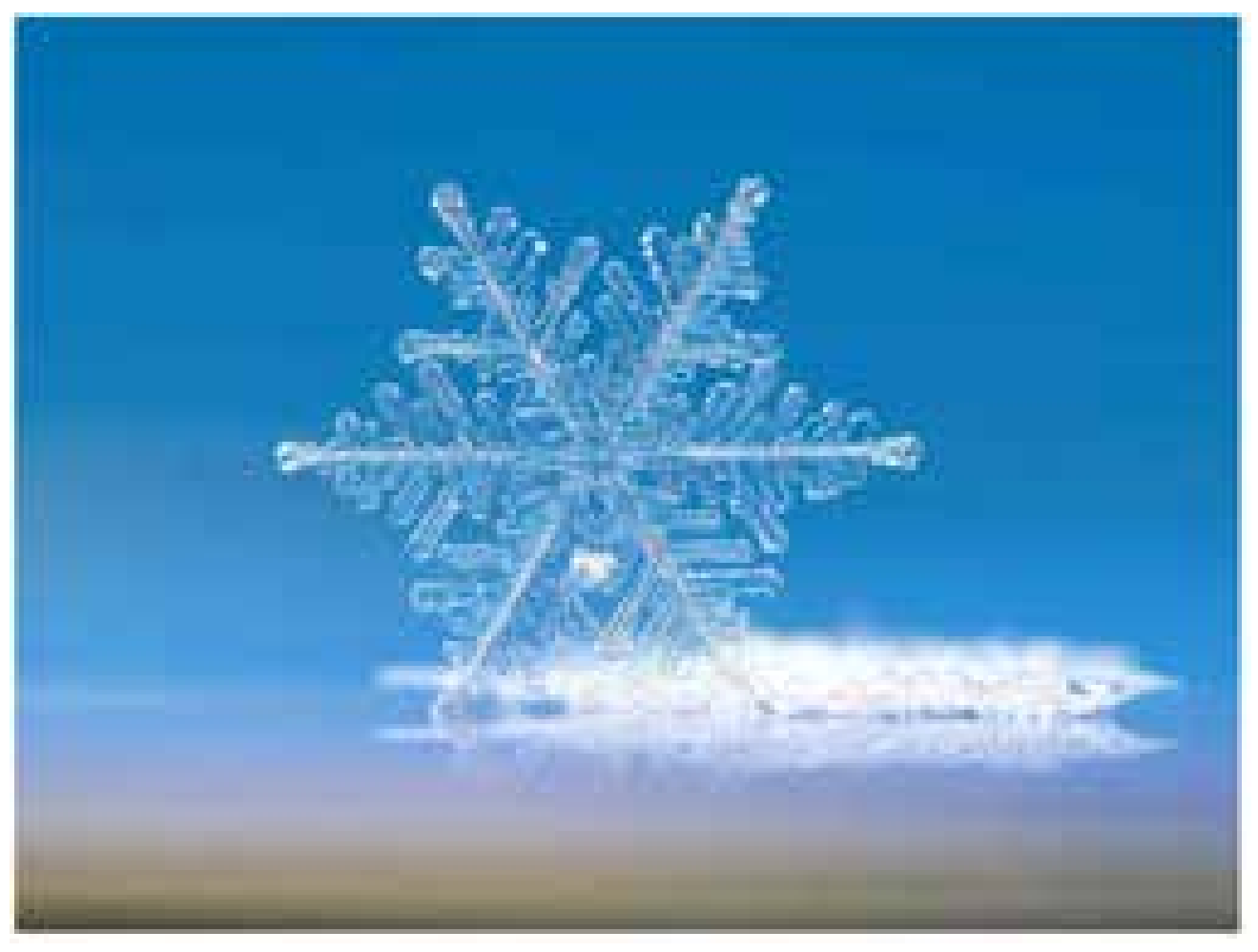

**Side Illumination#6, by Alexey Kljatov.** In this remarkable photo, Alexey dropped a bit of snow onto a plastic surface and then used this snow as support to balance a stellar crystal on its edge. Note the combination of white-light side illumination with colored back illumination.

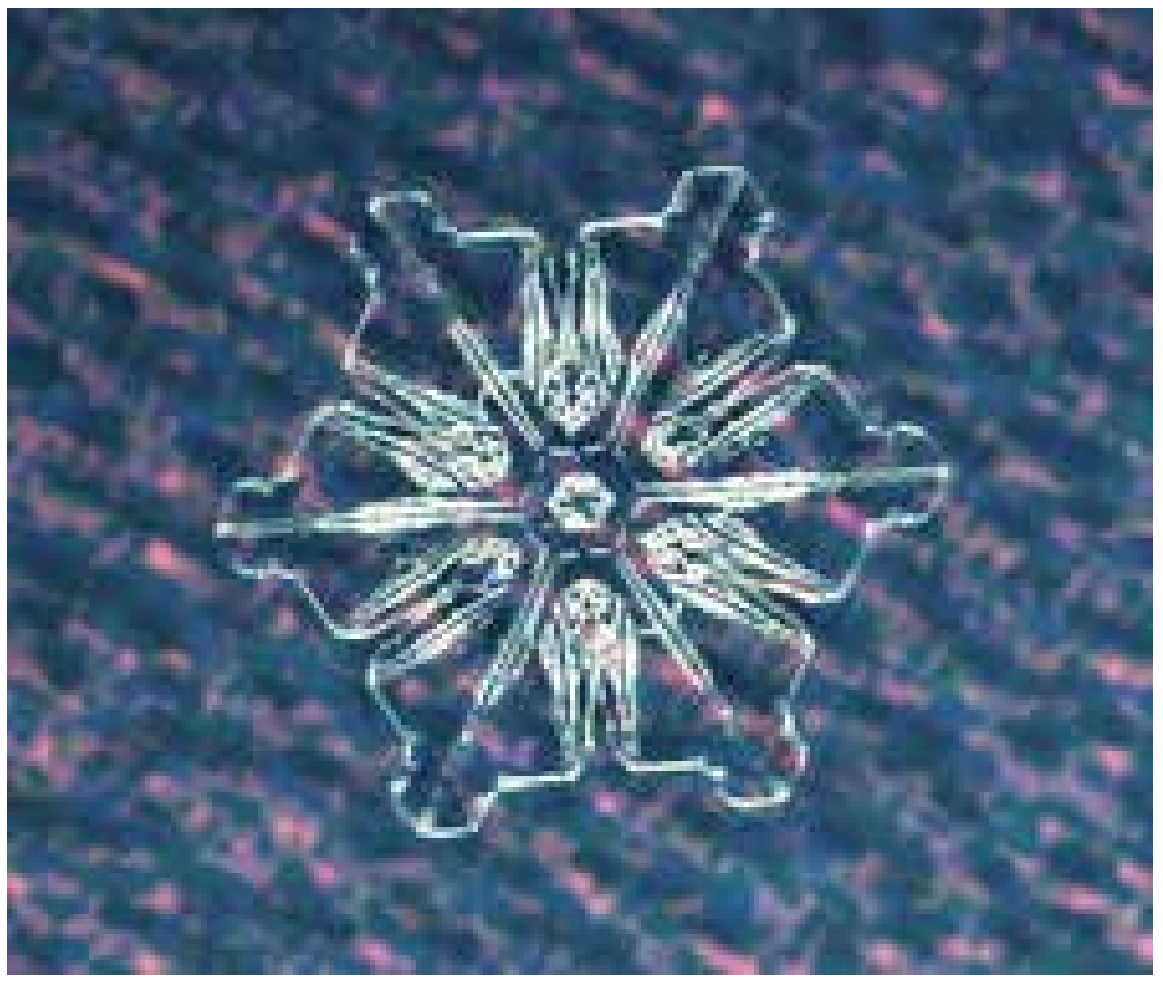

**Side Illumination#7, by the author.** Side illumination often yields a somewhat garish, high-contrast image because the edges appear so much brighter than flat regions of the crystal. The high contrast tends to obscure some of the subtler aspects of the crystal structure.

Although I have always been enamored with Alexey's beautiful photos, I have been generally dissatisfied with my own efforts using side illumination. In many photos (e.g. SI#7, SI#8, SI#9), the crystal edges scatter strongly and appear bright white, while flat regions scatter no light at all. This often results in quite a high-contrast image that tends to wash out the finer structural details within crystal. The camera's auto-exposure software tends to accentuate this problem as well. In my own experience, I have found that taking high-quality photos with side illumination is a rather difficult skill to master, and it does not work well with all types of snow crystals.

While plate-like crystals are especially problematic to photograph using side illumination, columnar and capped columnar crystals tend to yield impressive results quite reliably. While thin plates are nearly invisible except for their garishly bright edges, thicker crystals provide more varied light scattering that nicely reveals many internal structural details, as can be seen in SI#11, SI#12, and SI#13.

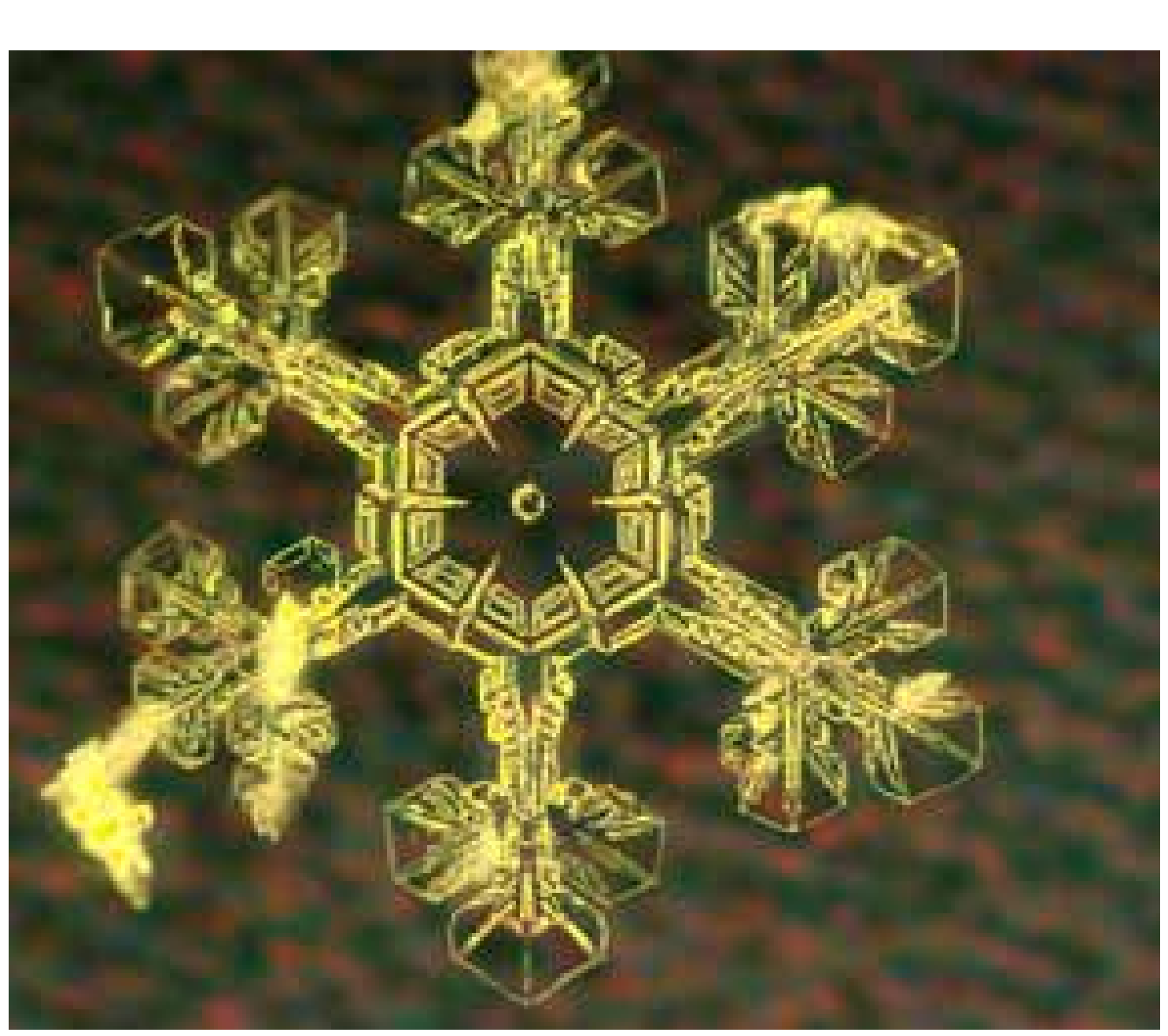

**Side Illumination#8, by the author.** (Left) Both the central region and the outer sectored plates are quite thin and transparent in this somewhat imperfect stellar dendrite.

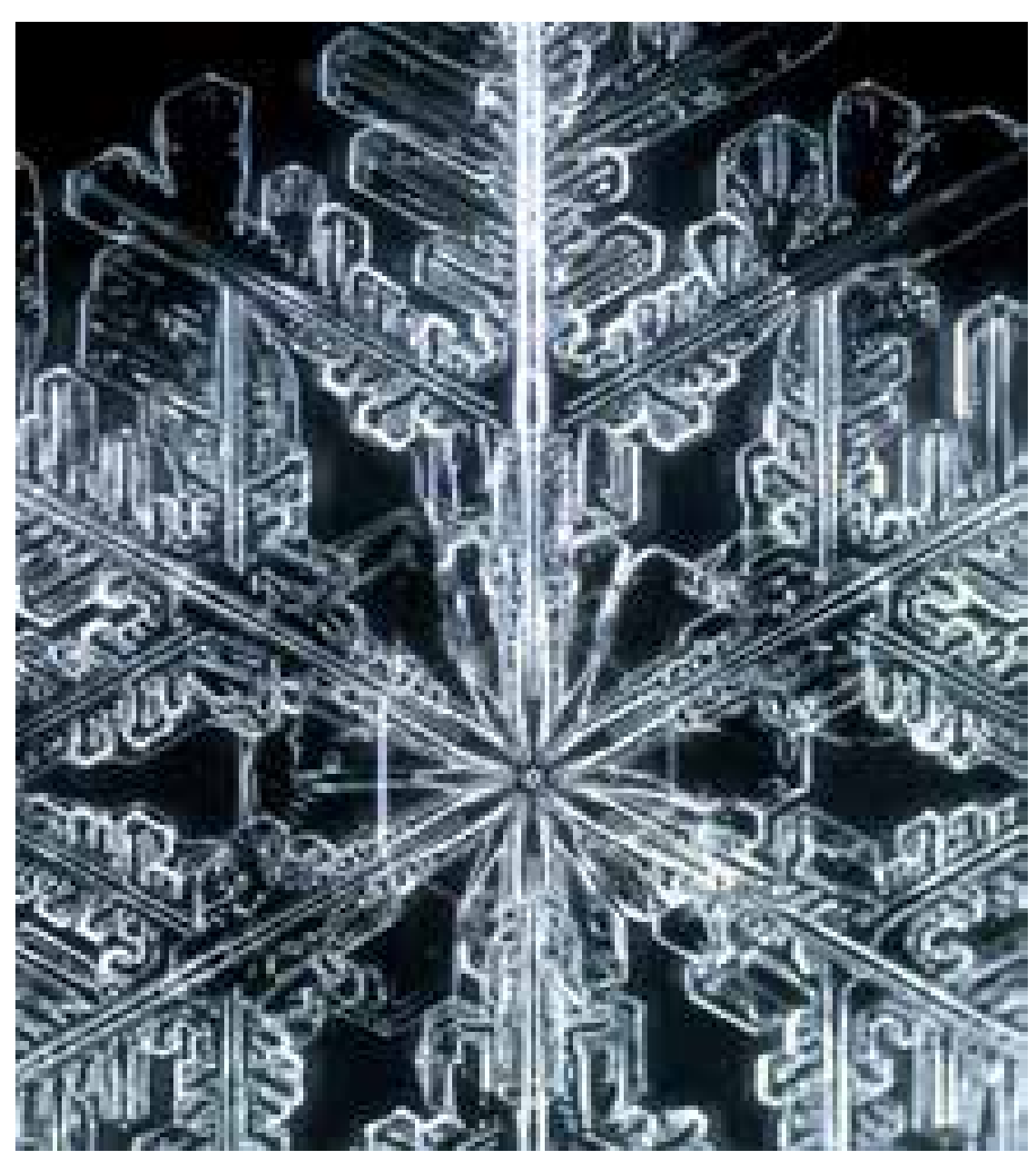

**Side Illumination#9, by the author.**
While side illumination with white light gives snow crystals a generally white appearance, the edges tend to dominate the photo, again giving a high-contrast look.

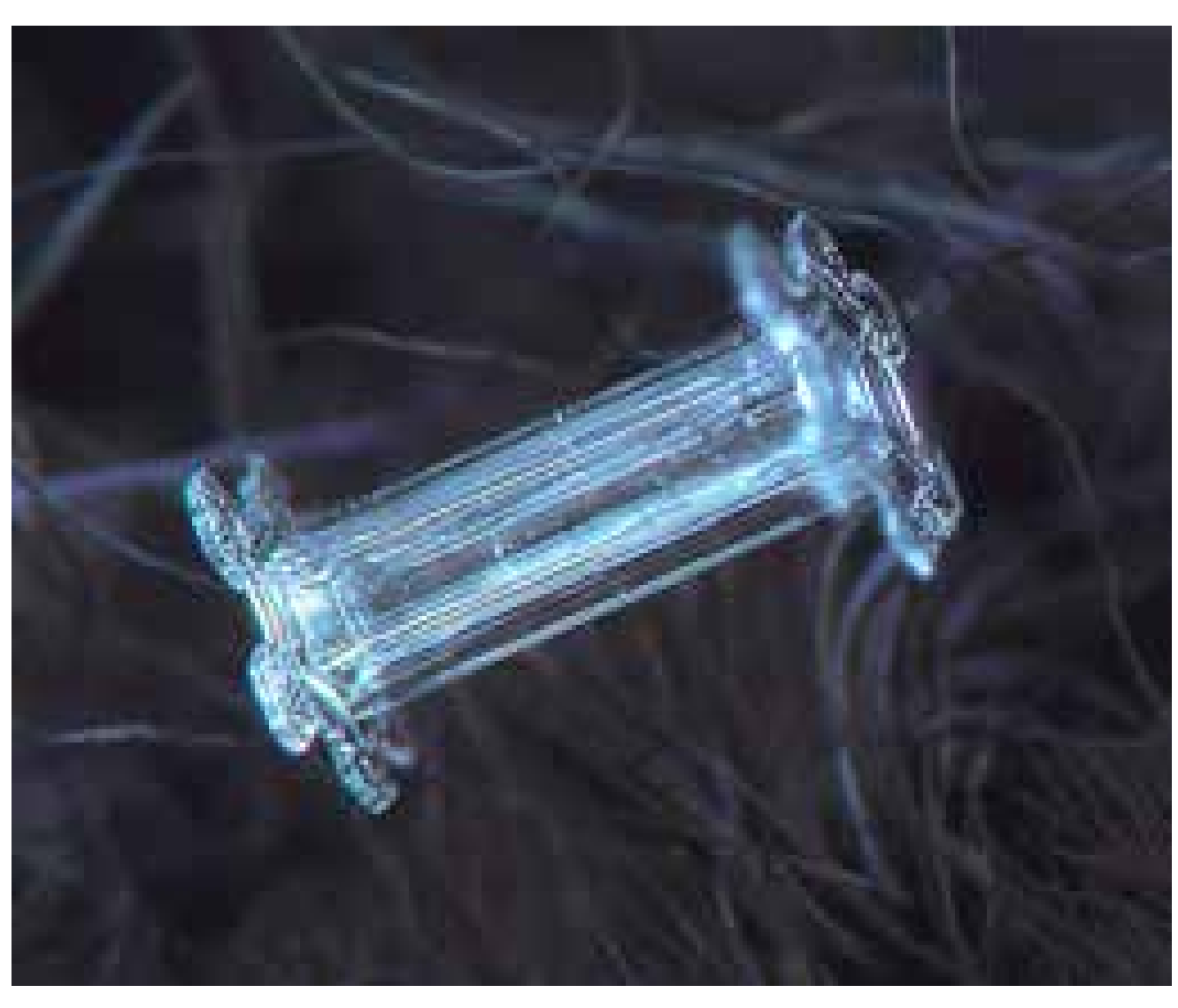

**Side Illumination#10, by Alexey Kljatov.**
Side illumination yields especially pleasing photos of thick crystals like this capped column. The body of the crystal scatters a good amount of light, and the out-of-focus regions give the picture an overall sense of depth.

**Side Illumination#11, Don Komarechka.**
(Below) With higher optical resolution and side illumination from a ring flash, this photo has an overall different appearance compared with SI#10, even though the crystals have fairly similar structures. Both these crystals were supported by fibers from woolen fabric, but Don often removes the background fibers in post-processing.

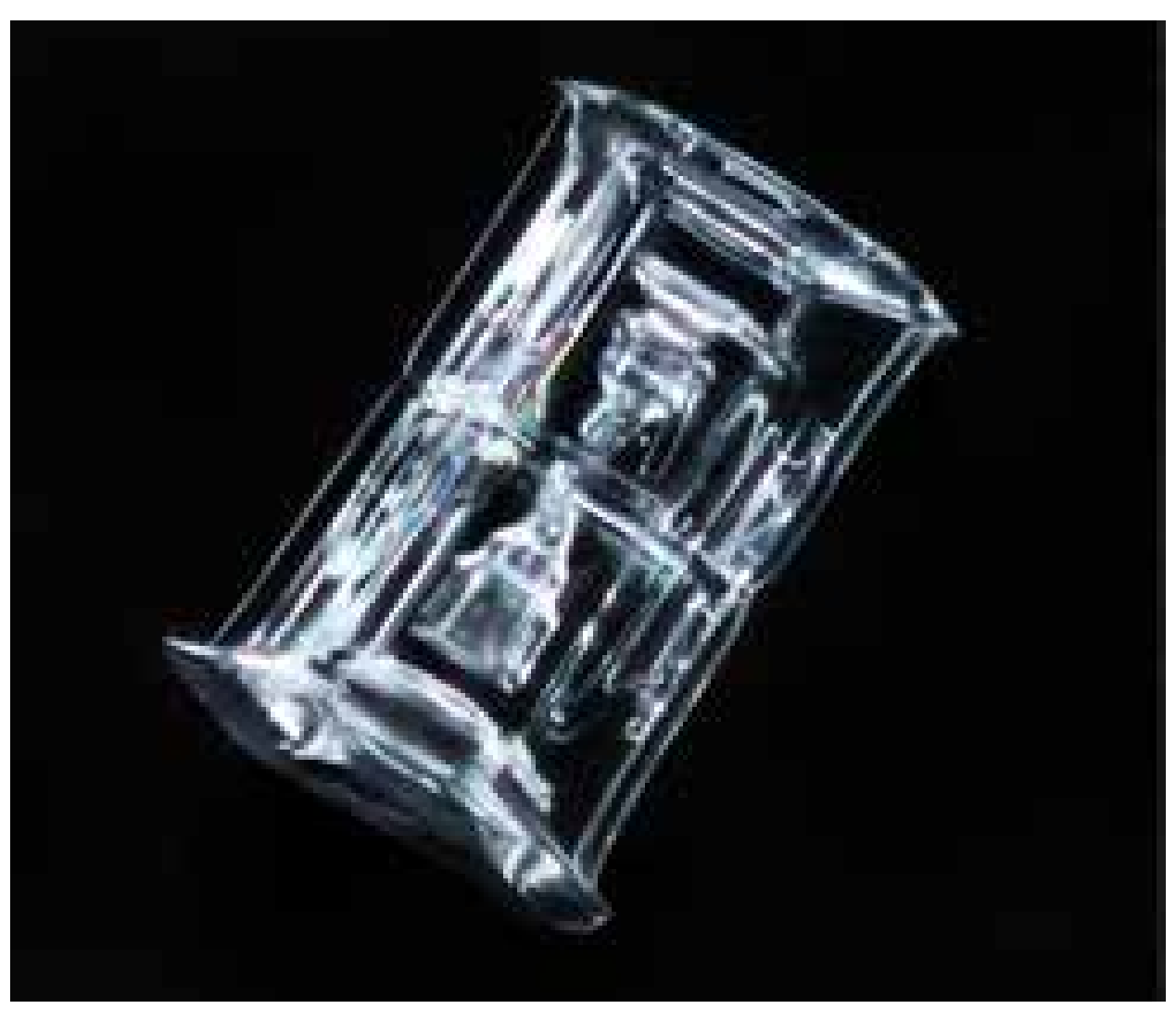

**Side Illumination#12, Don Komarechka.**
(Below) Side illumination gives thick crystals an appealing "watery" appearance. Even with a complex crystal like this one, it really looks like the highly structured block of clear ice that it is.

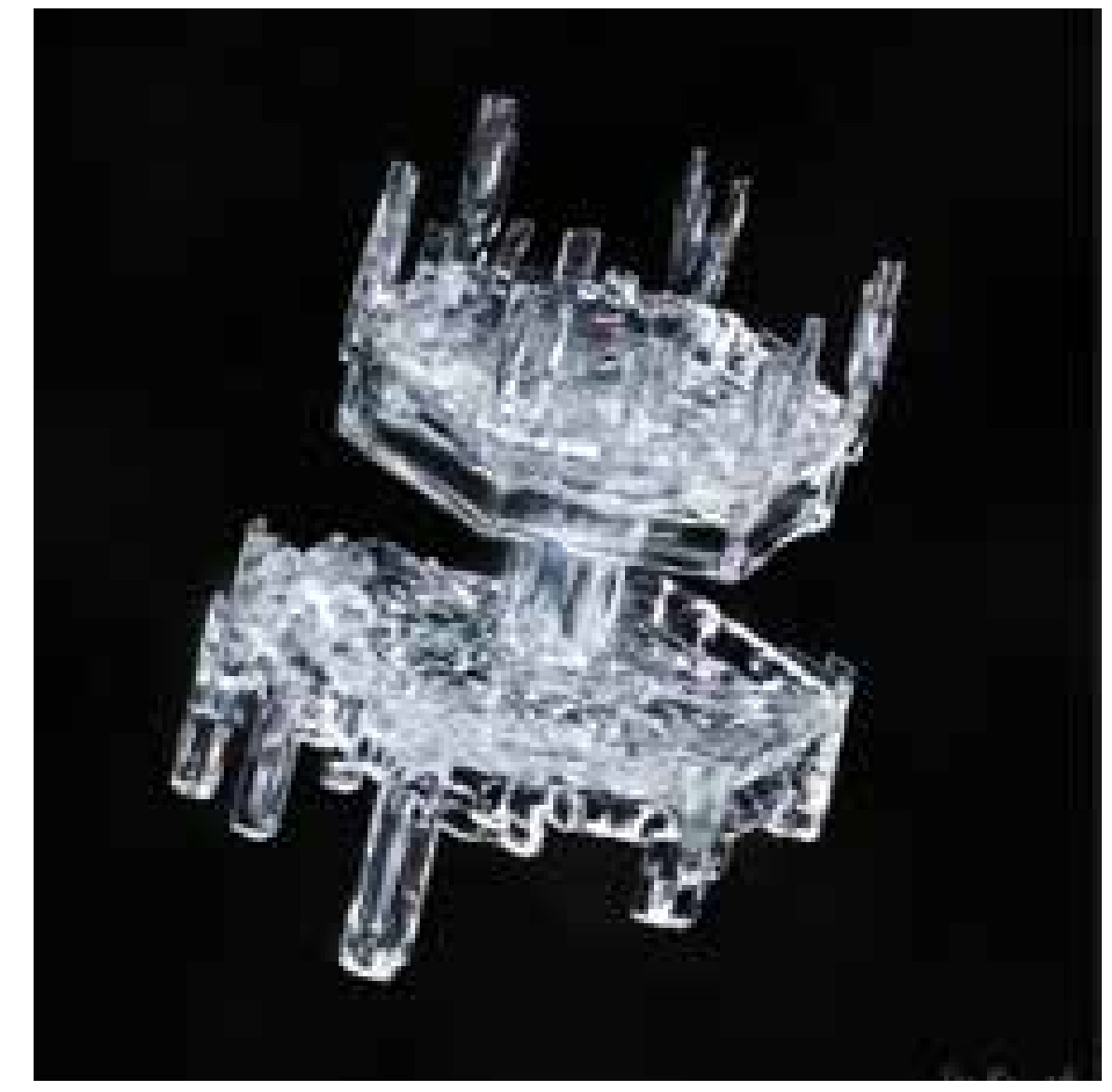

## Specular Reflection

Figure 11.17 illustrates another variation of side illumination, in which a plate-like snow crystal is tilted slightly compared with the face-on arrangement shown in Figure 11.16. Because of this tilt, flat basal surfaces will produce a mirror-like (specular) reflection of some of the side-illumination light into the camera lens. When this reflection is especially strong, the face of the crystal has a much brighter appearance compared to the side-illumination examples described above. Like a glass pane, a thin ice plate is effectively invisible under normal side illumination, but the specular reflection can be seen if the angle is right.

This added specular reflection solves, or greatly reduces, the main problem associated with using side-illumination for plate-like crystals, namely that bright edges dominate the photo while flat plates are nearly invisible. In Specular Reflection #1 (SR#1), for example, the main body of this stellar crystal is quite bright, so the image is not dominated by high-contrast edges. Moreover, this gives the crystal an overall white appearance, which satisfies many viewers' desire that snowflakes should look white. What is somewhat lost in the process, however, is the glassy look that gives one the (correct) impression that a snow crystal is not intrinsically white, but is rather made from a sliver of transparent ice. Reducing the intensity of the specular reflection can alter this effect, as illustrated in SR#2 and SR#3.

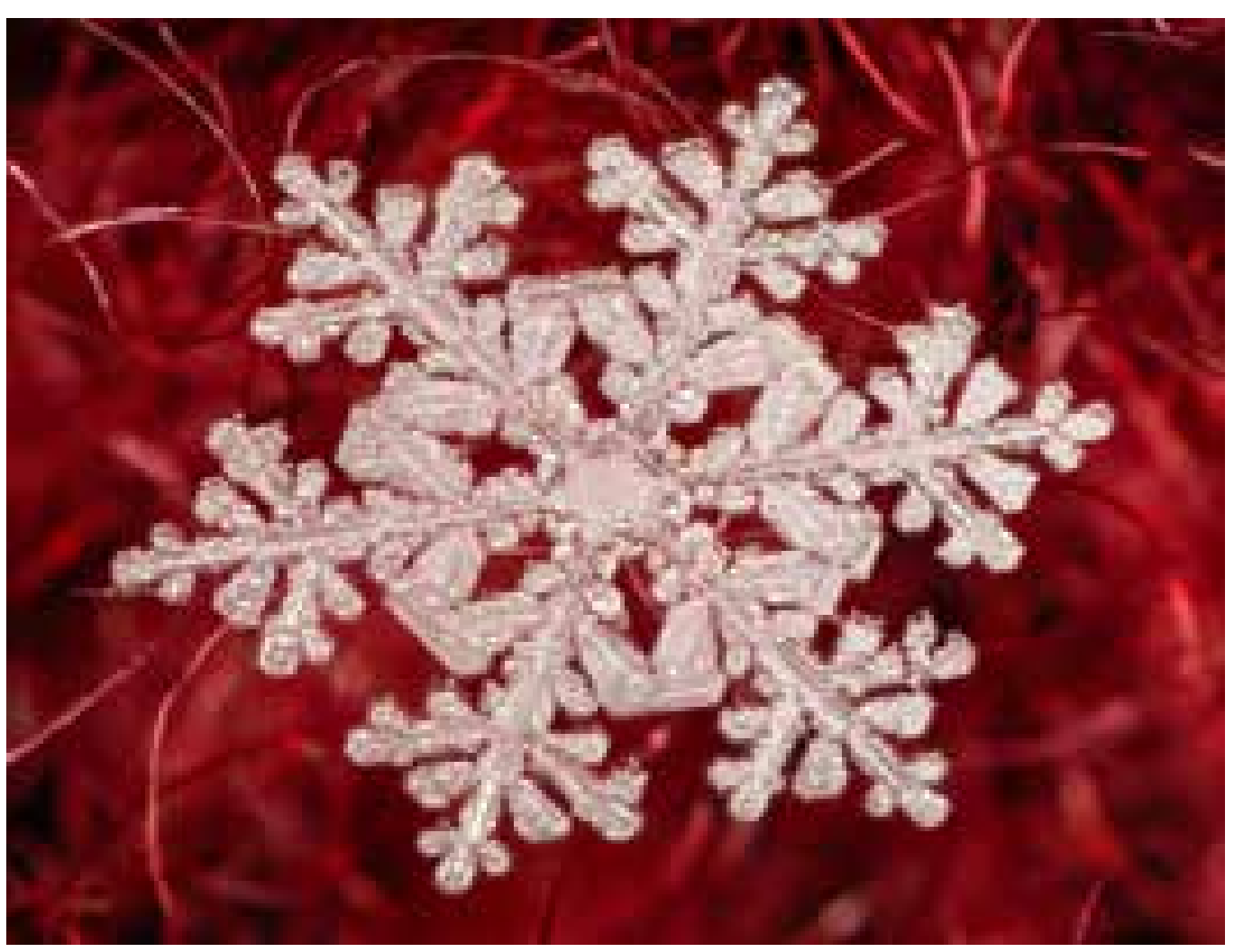

**Specular Reflection#1, by Pam Eveleigh.** In this image, the mostly-flat face of a stellar dendrite produces a bright reflection into the camera lens and onto the imaging sensor. Using white-light illumination, this mirror-like reflection gives the crystal an overall bright white appearance. Structural detail is somewhat obscured by the bright reflection, however, and the photo does not give the viewer the impression that the crystal is made of clear ice.

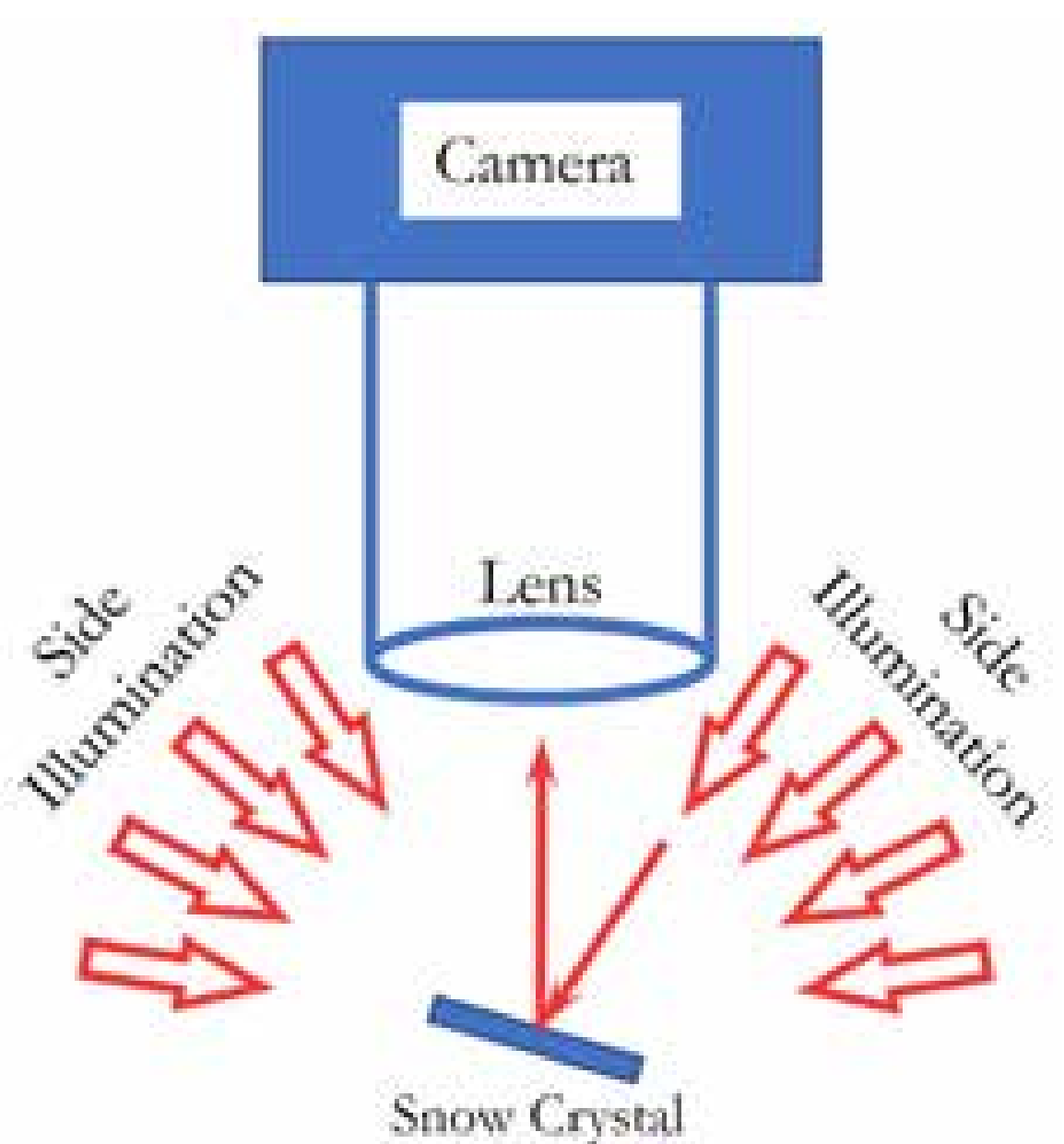

**Figure 11.17: Specular Reflection. In this variation of side illumination, a plate-like snow crystal is tilted so the flat face of the crystal reflects light into the camera lens. With this arrangement, thin plates appear much brighter than with the face-on side illumination method shown in Figure 11.16.**

Specular-reflection illumination is especially popular for point-and-shoot snowflake photographers working at moderate (10-20 micron) resolution, as illustrated in SR#4, SR#5, and SR#6. Finer structural details are diminished at the lower resolution, but there are endless opportunities for artistically placed crystals on colorful, textured backgrounds.

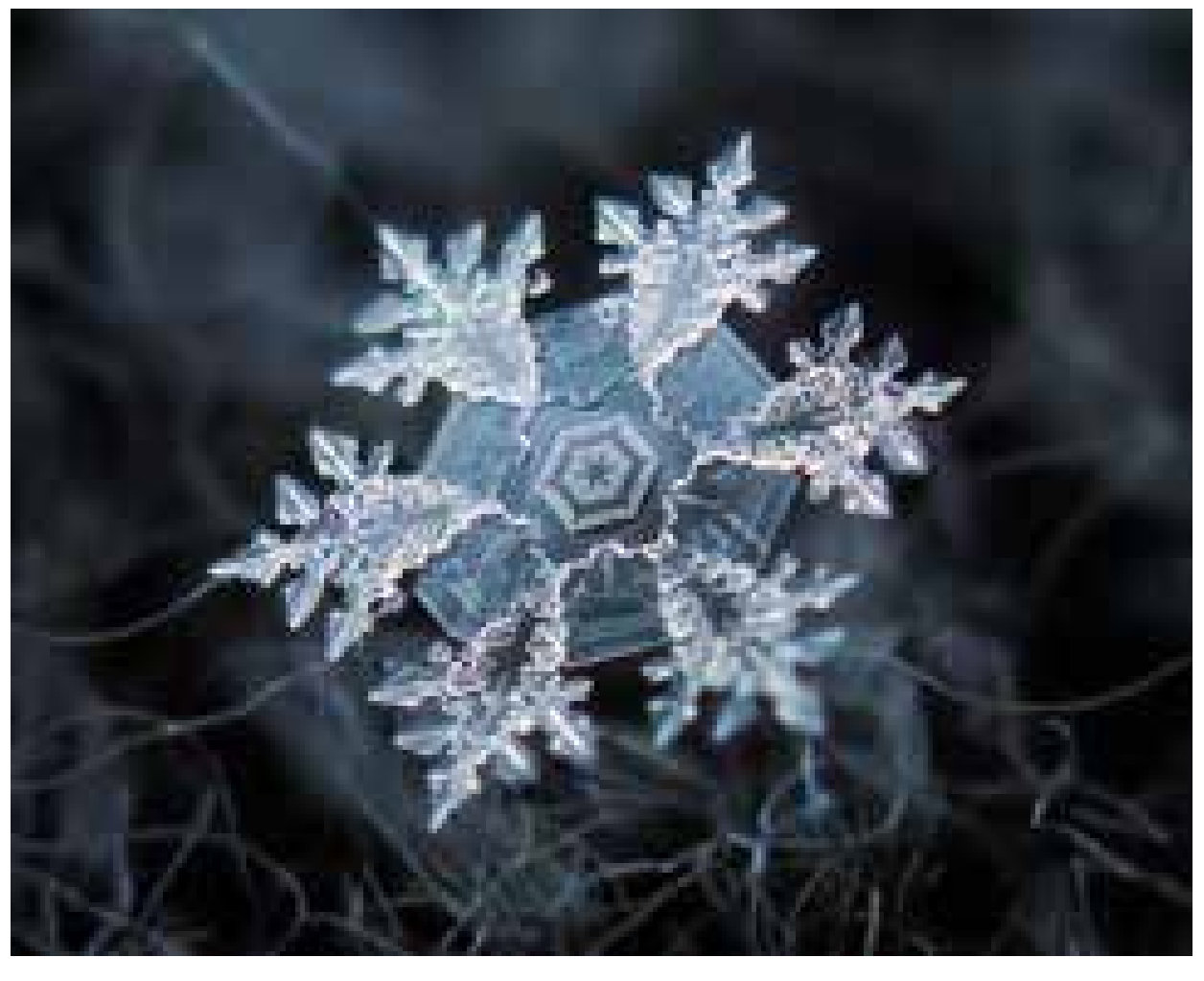

**Specular Reflection#2, by Alexey Kljatov.**
In this photo, Alexey found an exceptionally beautiful stellar crystal and used a pleasing balance of normal side illumination and specular reflection. The result is a somewhat bright-white crystal that still exhibits structural detail clearly and gives the impression of clear ice. Some of the branch tips are slightly out of focus, adding a sense of depth to the photo, while the cloth background provides some textural context.

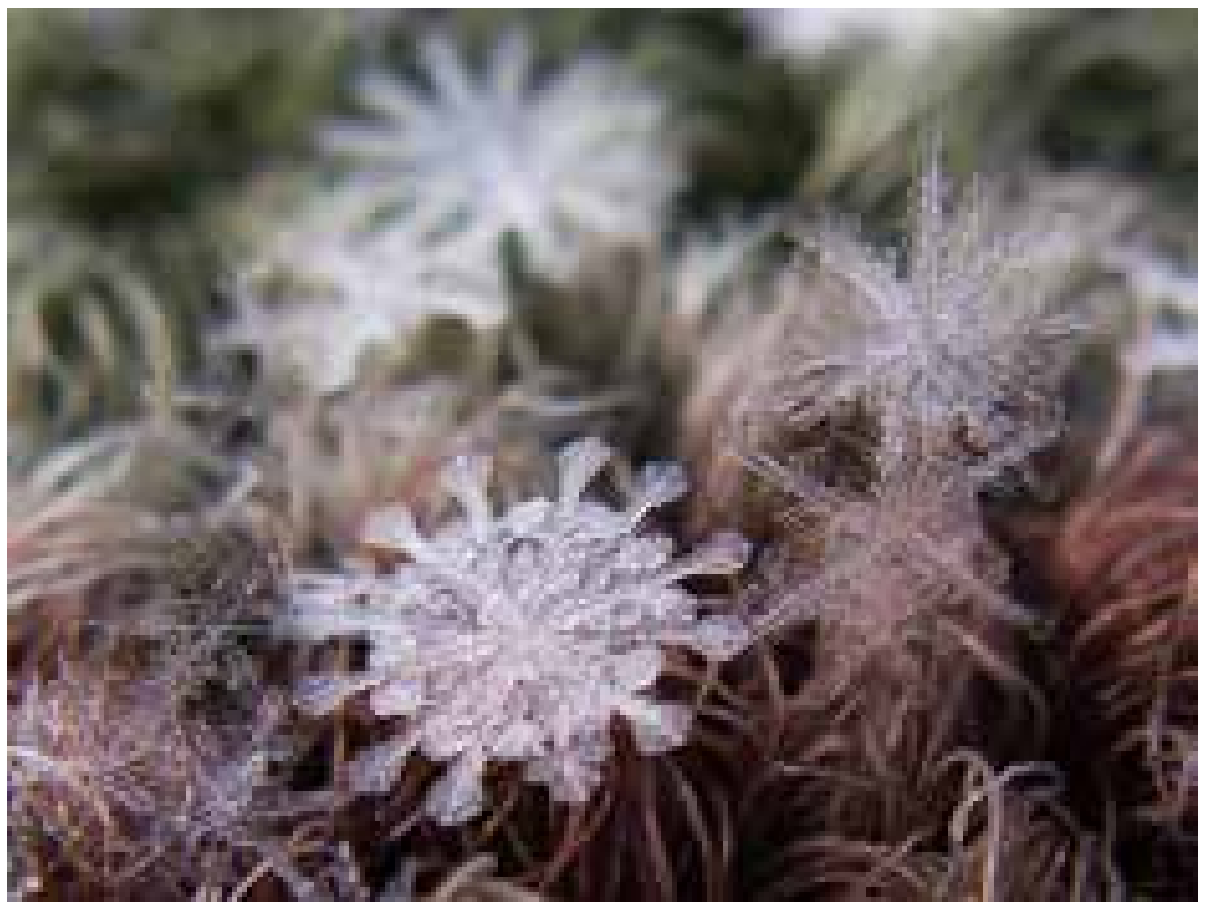

**Specular Reflection#3, by Olga Sytina.**
Here we see a strong specular reflection on the main 12-branched crystal, together with several crystals on the right under normal side illumination. (Olga is Alexey Kljatov's mother, so superb snow-crystal photography seems to run in this family.)

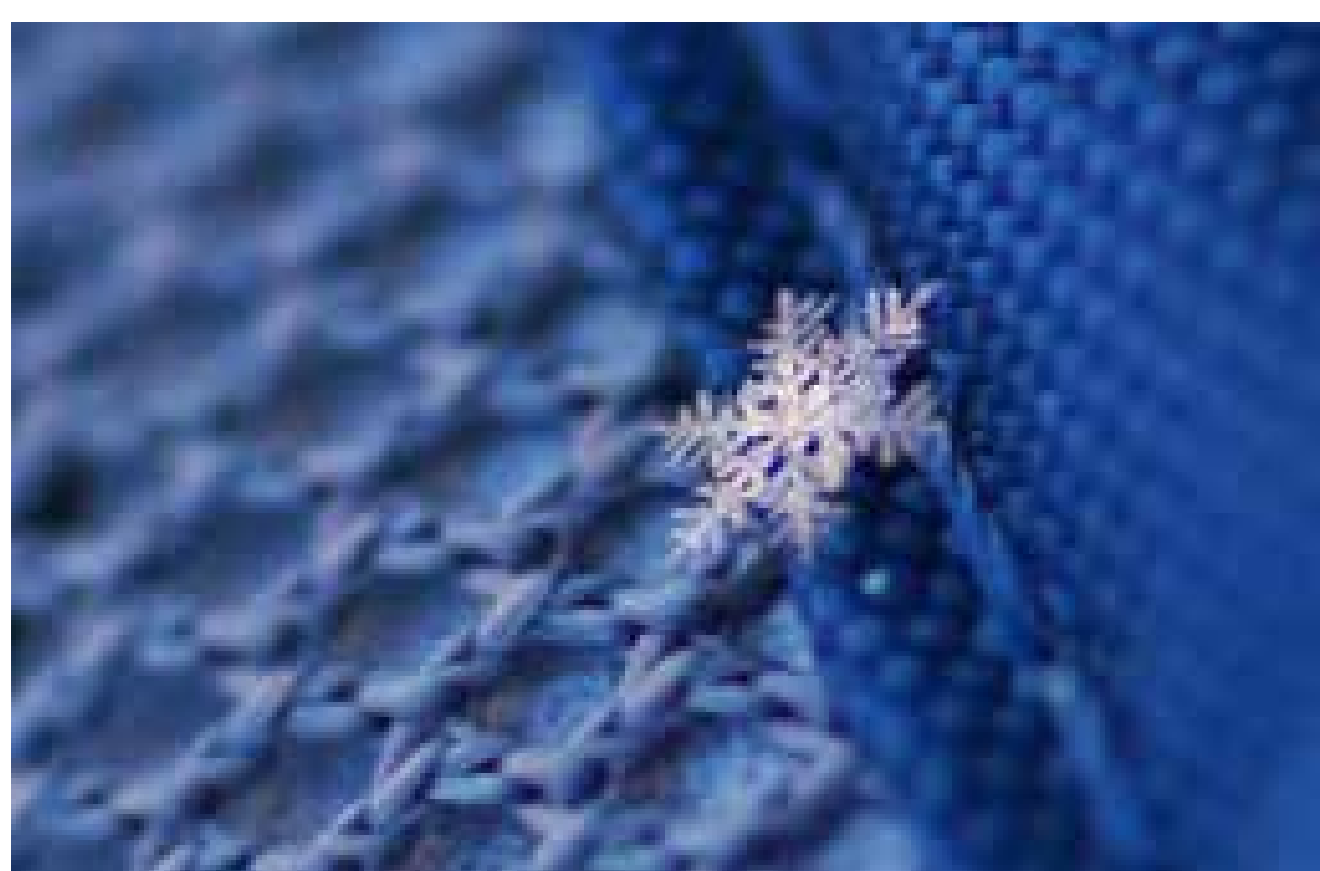

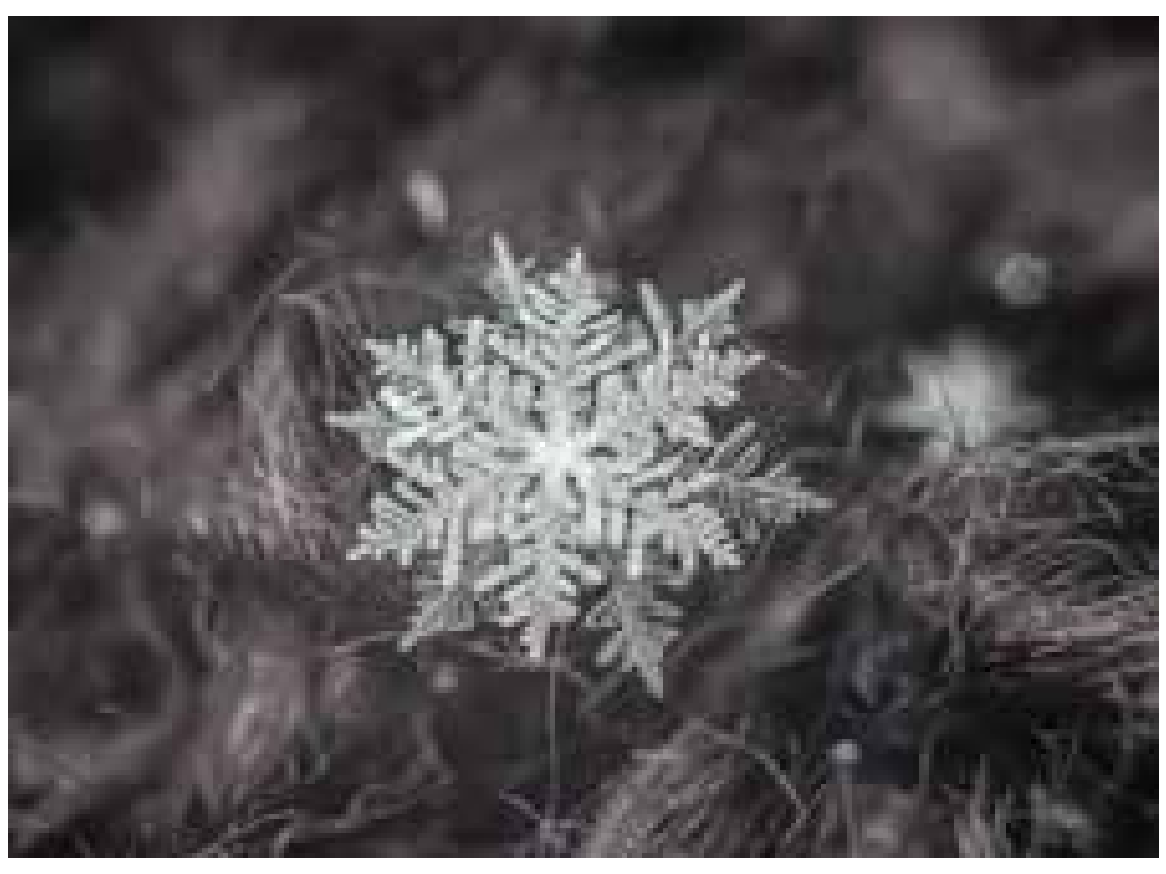

**(Above) Specular Reflection#4, by Jackie Novak.**

**(Left) Specular Reflection#5, by Delena-Jane Lane.**

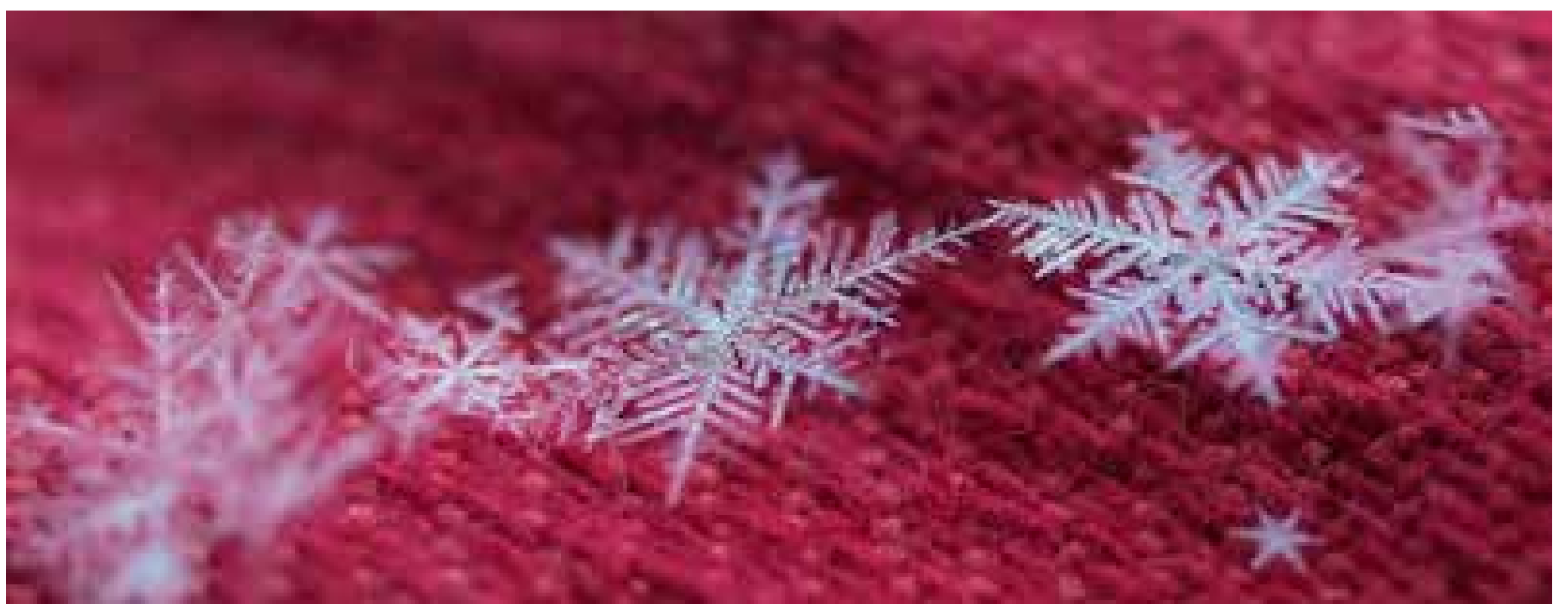

**Specular Reflection#6, by Elizabeth Akers.**

Perhaps the biggest drawback associated with specular-reflection illumination is that a tilted flat-plate crystal (see Figure 11.17) does not lie in the image plane of the camera+lens. When one is working at high resolution (5-10 microns), the smaller depth-of-focus from Equation 11.4 means that it is impossible for the entire crystal to be in focus at once. At exceptionally high resolution (<5 microns), only a small portion of a tilted, plate-like crystal can be in focus in a single photo, as illustrated in Figure 11.18.

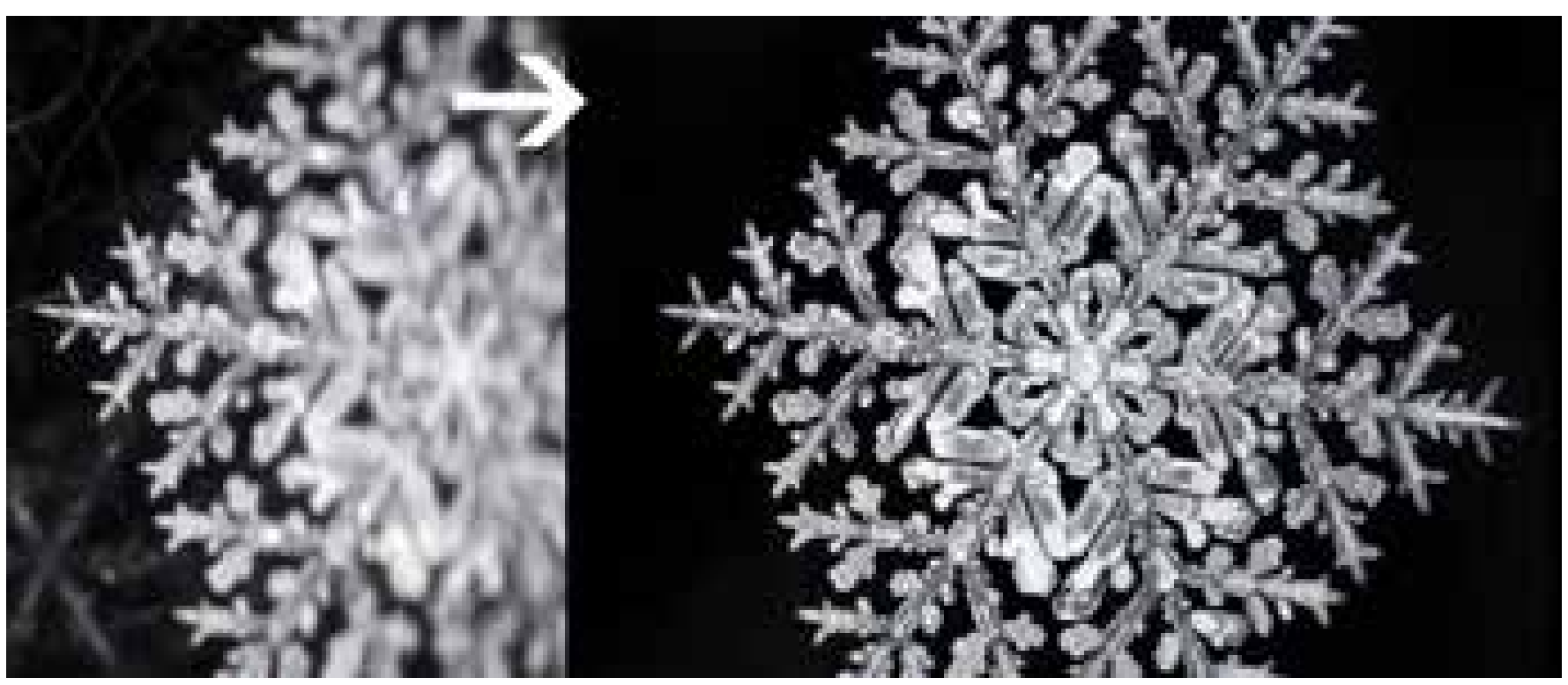

**Figure 11.18: Don Komarechka took these images using the Canon MP-E 65mm macro lens at 5X, giving an optical resolution of $R_{xy} \approx 4$ µm and a depth-of-focus of $R_z \approx 50$ µm (see Figures 11.11 and 11.12). With these parameters and specular-reflection illumination, only a small sliver of the tilted snow crystal is in focus in a single shot (left image). Don manages this problem by focus stacking 30-50 images into a single composite photo (right image) [2013Kom].**

Don Komarechka has been a pioneering proponent of high-resolution snow-crystal photography using specular-reflection illumination, managing the depth-of-focus problem by using an unprecedented amount of focus stacking. Don typically captures 30-50 images in quick succession using a hand-held camera with a high-speed ring flash, and he then combines all the images together into a single composite image in post-processing [2013Kom].

Don's technique is perhaps the most demanding in terms of hardware capabilities, software, computing power, and overall effort. Shooting 30-50 images in burst mode at six frames per second or faster [2013Kom] requires a high-end camera, a rapid-refresh ring flash, and a high-resolution lens, none of which is cheap. Moreover, a fairly powerful computer running first-rate software is needed for focus stacking so many images, and Don estimates that he spends up to four hours processing a single composite image [2013Kom]. This amount of expense and effort is not for everyone, but Don has captured some world-class snow-crystal photographs using this innovative technique, as illustrated in SR#7 and the following several images.

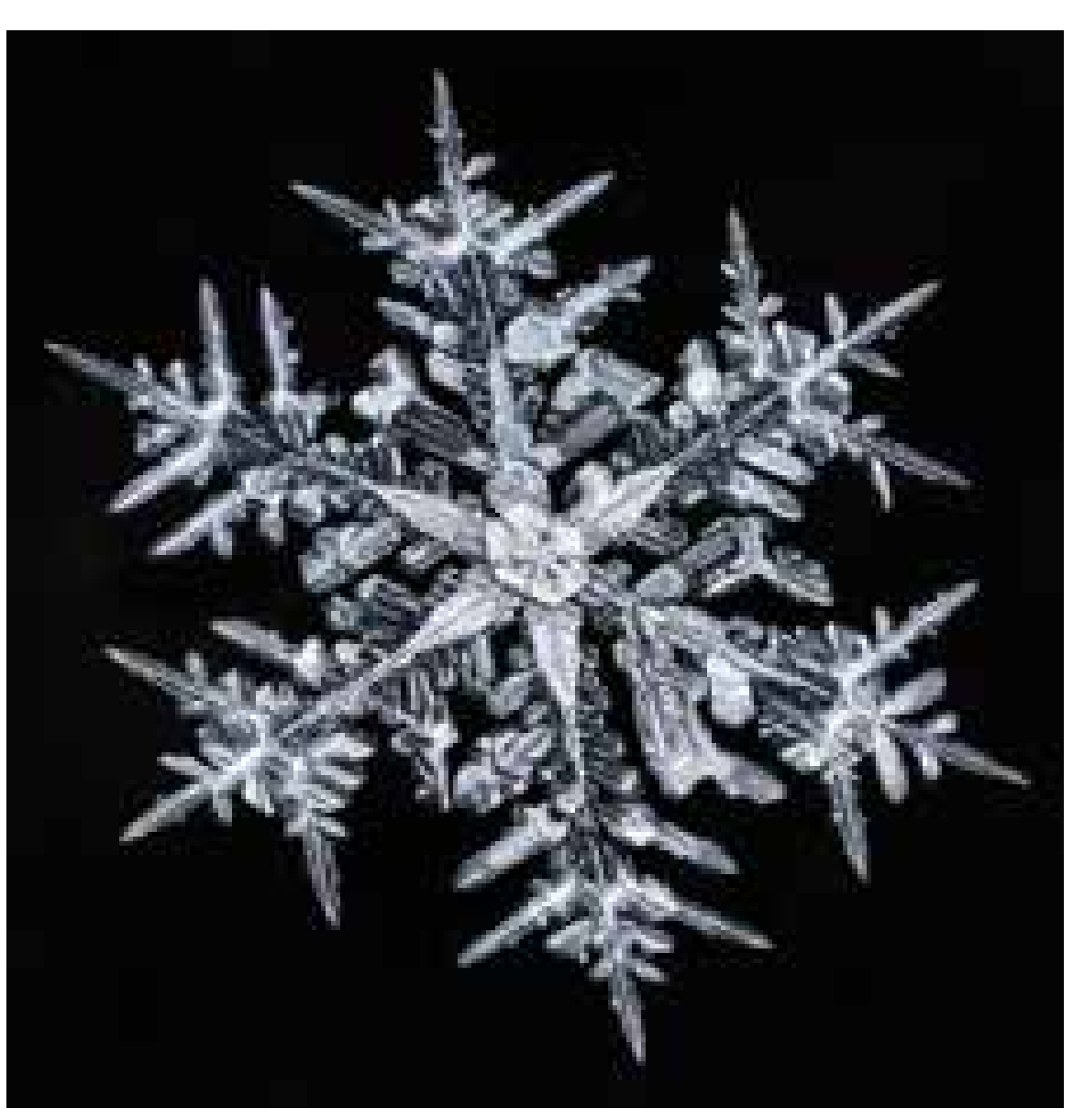

**Specular Reflection#7, by Don Komarechka.**
(Right) Don employs extensive focus stacking to capture high-resolution images using specular-reflection illumination of tilted snow crystals. This small-size reproduction does not do the photo justice, but many similar images can be viewed at high resolution on Don's Flickr page. Don clearly has a strong preference for the Bentley-esque style of a jet-black background, so he often digitally removes background cloth fibers that support the crystals, so they seem to float through the night [2013Kom].

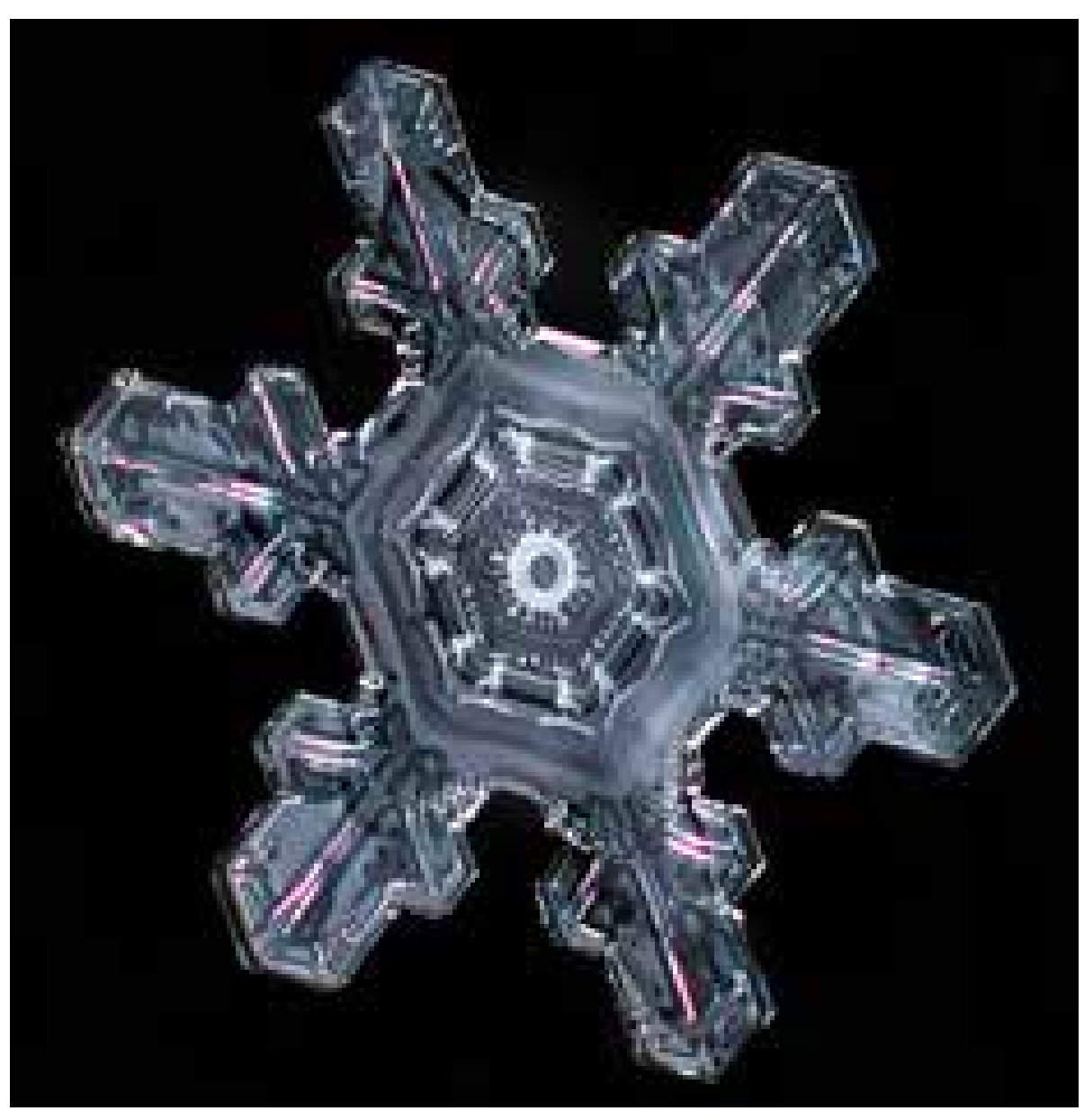

**Specular Reflect#8, by Don Komarechka.** The tilt angle of a crystal can be adjusted to vary the brightness of the specular reflection. In this example, the overall reflection intensity is lower than in SR#7, giving the crystal a somewhat glassier look.

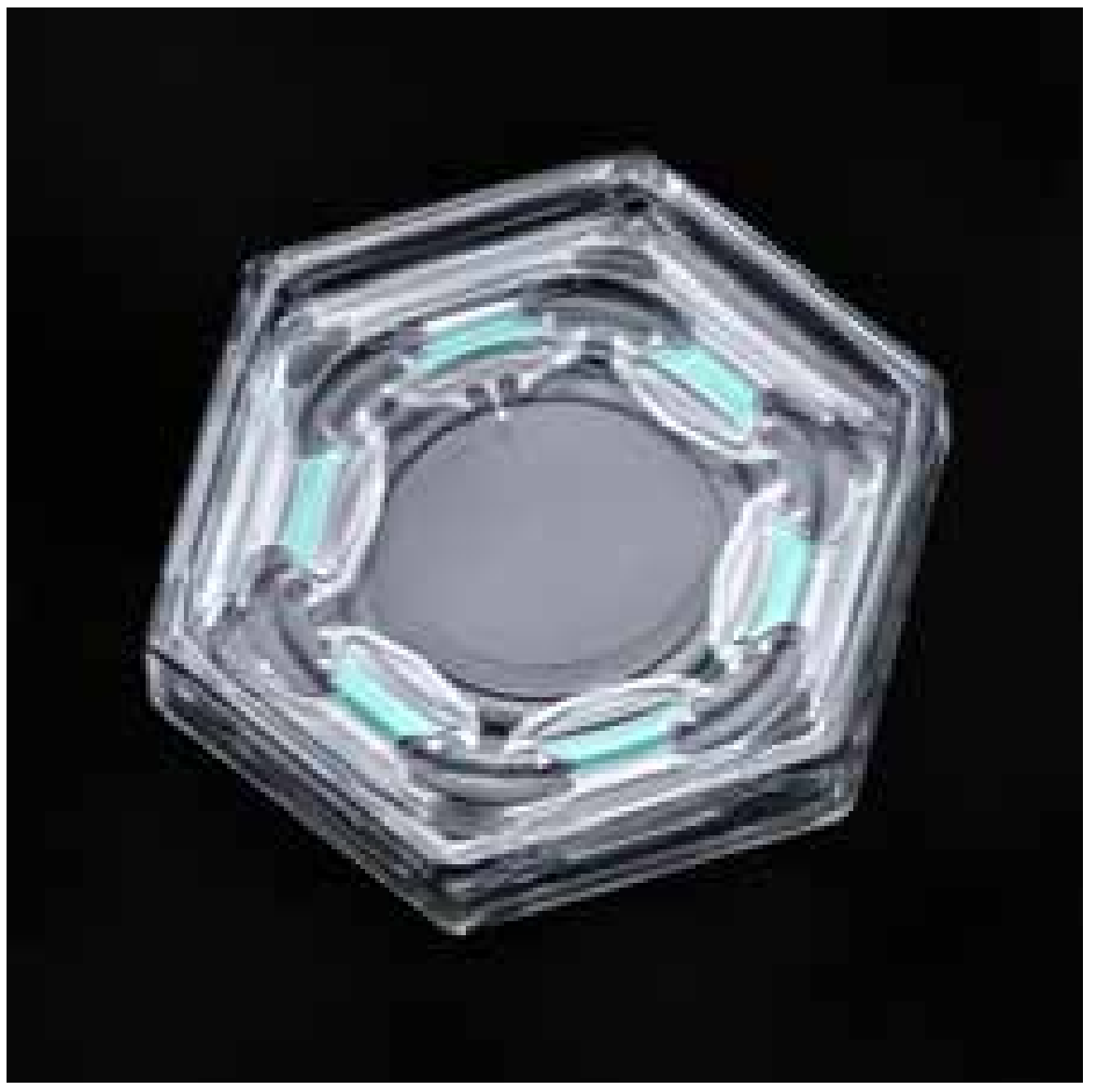

**Specular Reflect#9, by Don Komarechka.** Tilting a thick crystal introduces several additional reflections, as internal features reflect off the rear faceted ice surfaces. These reflections, plus the view of the thick sides of the plate, give the image a pleasing sense of depth and 3D structure.

While experimenting with specular-reflection illumination, Don discovered the colorful appearance of internal bubbles and thin hollows in plate-like snow crystals, as illustrated in SR#10 and SR#11. The colors arise from light interference effects when specular reflections from the top and bottom faceted surfaces of a hollow regions interfere with one another (see Chapter 6). Similar colorful interference effects can be seen in photographs of thin-film soap bubbles, and Figure 11.19 shows a calculation of the interference color as a function of the soap film thickness.

The details of the calculation depend on several factors relating to the physics of optical interference, such as the spectrum of the incident light source and the RGB sensitivity of the camera sensor. These details

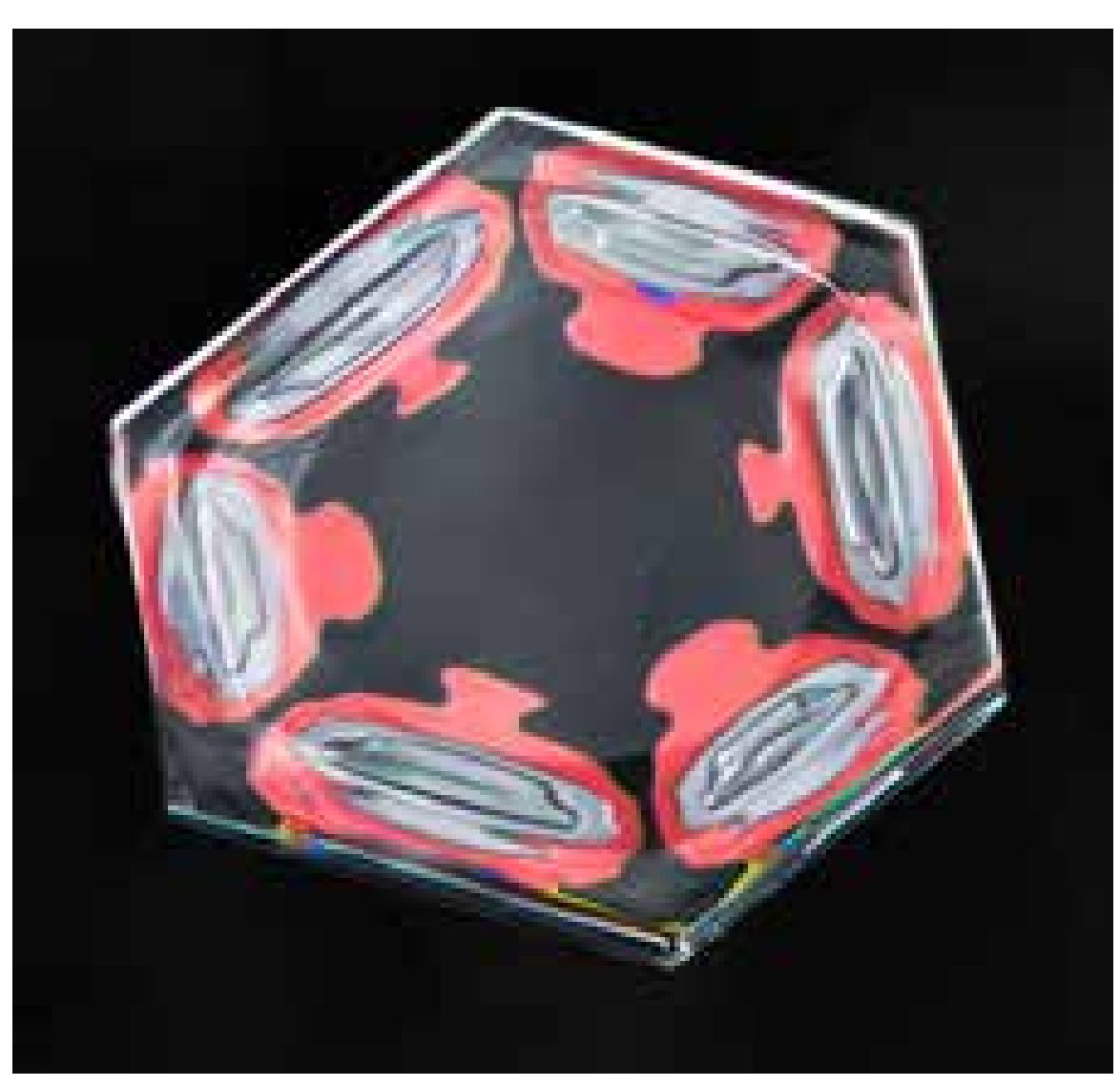

**Specular Reflection#10, by Don Komarechka.** (Left) The bright red color in this image results from the interference of two specular reflections from ice/air interfaces, at the top and bottom of broad, ~1-micron thick hollow regions within the body of the thick plate. The uniformity of the color indicates that these two internal ice/air surfaces are both flat basal facets.

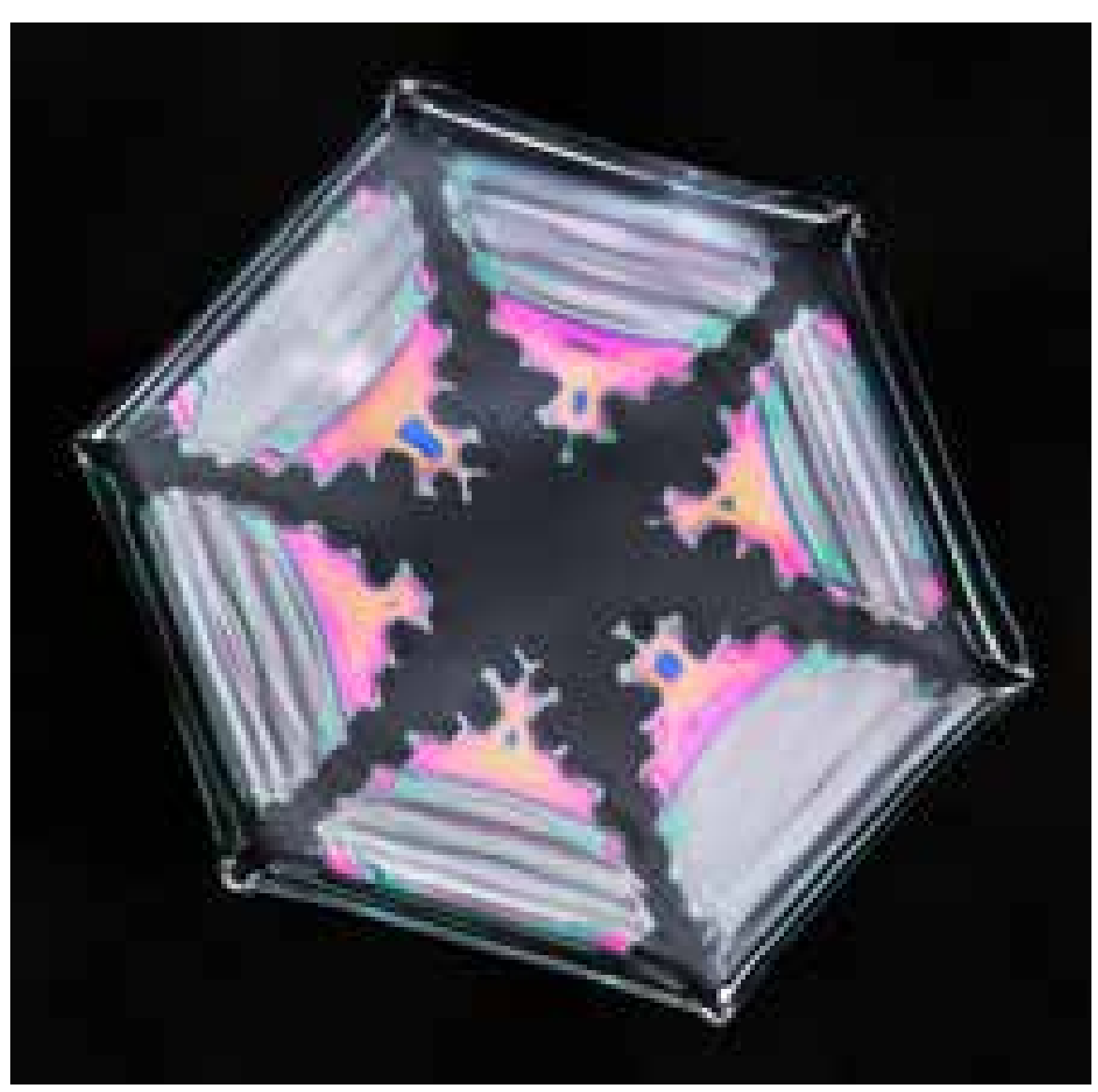

**Specular Reflect#11, by Don Komarechka.** In this crystal, the colors vary with the thickness of the hollow regions, and the colors disappear toward the edges of the crystal, where the thickness increases beyond several microns.

notwithstanding, Don's images suggest that the most colorful hollow regions and voids are roughly 0.5-1.5 microns in thickness. The structure and stability of such remarkably thin hollows and bubbles is described briefly in Chapter 3, but their formation is not well understood at present.

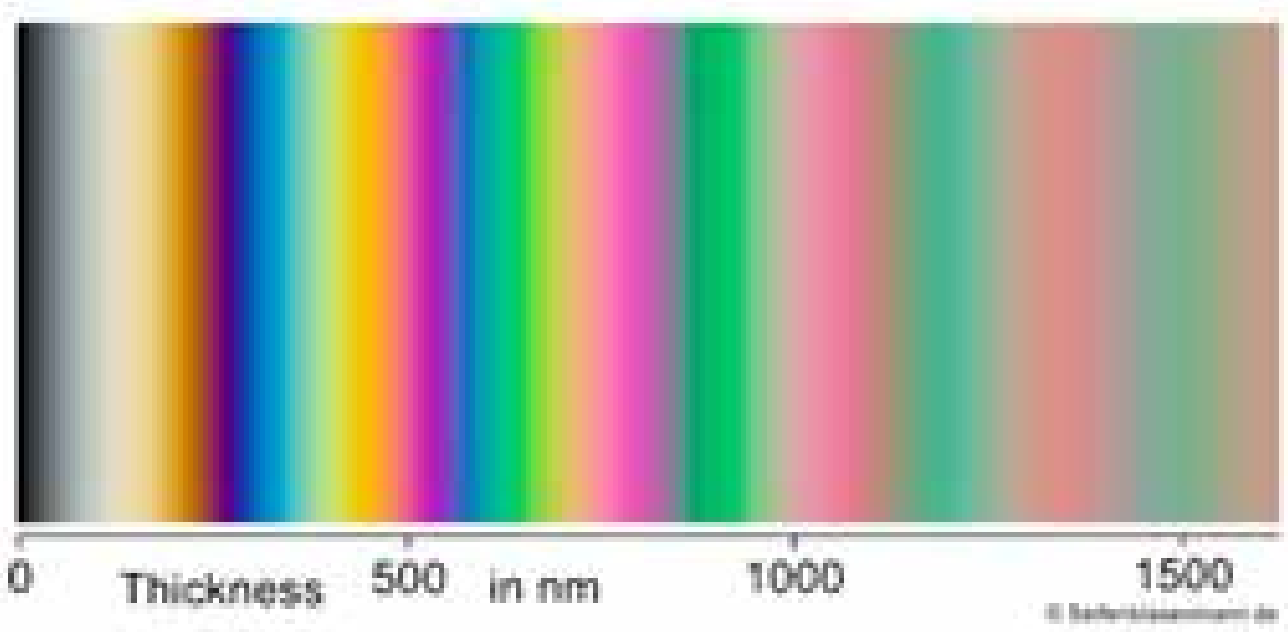

**Figure 11.19: The calculated interference colors for white light reflecting off a thin soap film, as a function of the thickness of the film in nanometers. Factoring in the index of refraction, the thickness of an internal air cavity in ice is roughly 1.3 times larger than the soap-film thickness with the same interference color. The calculation will change somewhat depending on the camera color response and the spectrum of the light source. Image from soapbubble.fandom.com/wiki/Color_and_Film_Thickness.**

## Front Illumination

Figure 11.20 shows a different type of specular-reflection illumination that avoids having to tilt plate-like snow crystals. With this geometry, the flake lies flat with respect to the image plane, so an entire thin-plate crystal can be brought into focus in a single photo. This avoids the need for extensive focus stacking, which is perhaps the biggest difficulty associated with specular-reflection illumination of stellar-plate snow crystals. From a scientific perspective, this technique has the additional advantage that a face-on view provides a more accurate depiction of the hexagonal geometry of plate-like crystals. With a tilted crystal, such as that shown in SR#11, the overall hexagonal shape in the image depends on viewing angle, so the measured angles between prism facets is no longer 120 degrees. A face-on view would make it possible to measure these and other angles accurately.

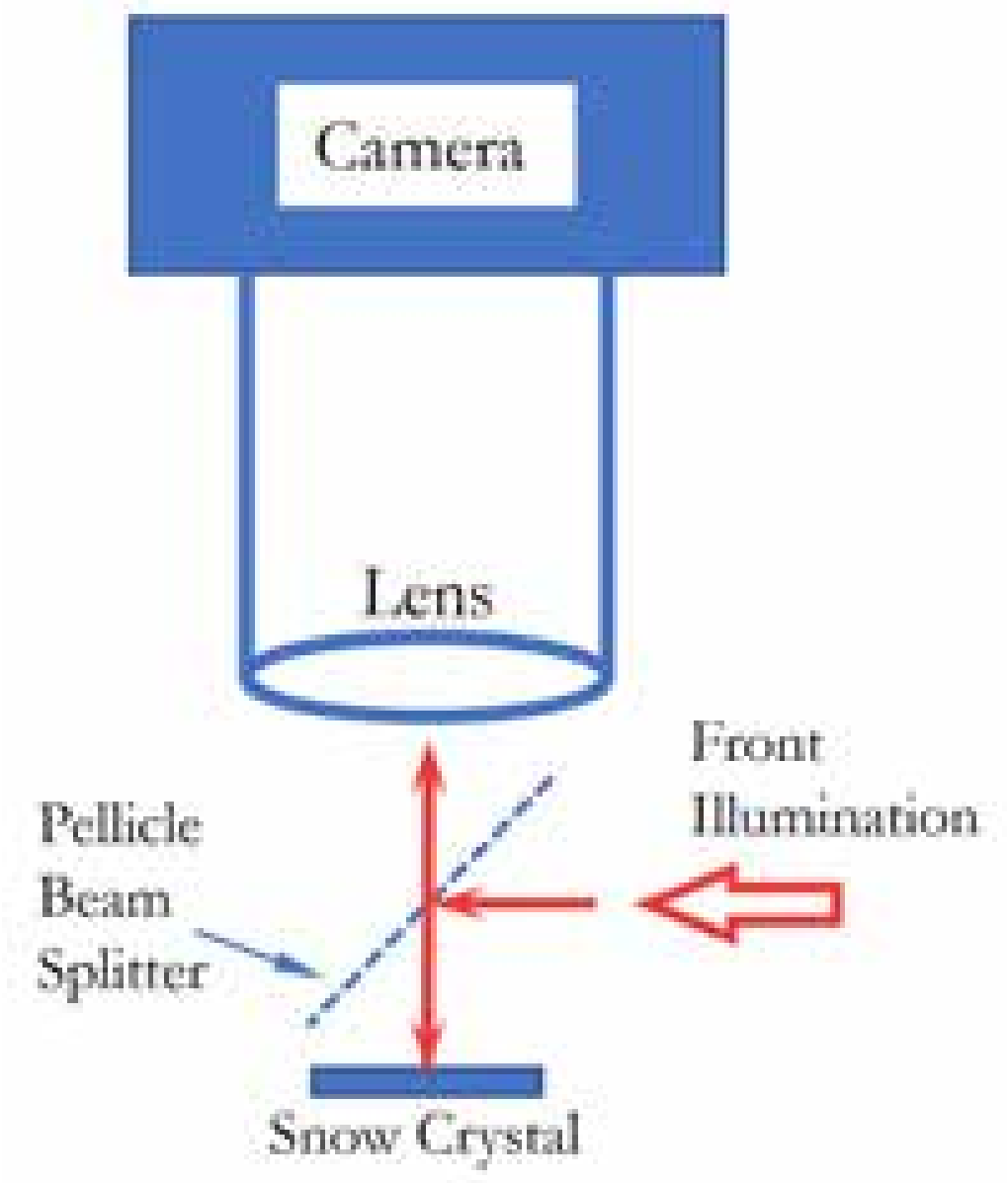

**Figure 11.20: (Left) Front Illumination. This variation of specular-reflection illumination uses a pellicle beamsplitter to direct light onto a plate-like crystal. In contrast to Figure 11.17, the face-on crystal lies nicely in the image plane, avoiding depth-of-focus issues.**

A pellicle beamsplitter is shown in Figure 11.20 because it is generally ill-advised to image through a glass-plate or cube beamsplitter. Because the pellicle is only a few microns in thickness, it introduces minimal image distortion compared to other beamsplitter options. Front illumination like this would be difficult to achieve using a hand-held setup, so it probably necessary to use a rigid mounting system for the various components.

This method of front illumination appears to have some significant advantages over tilted-crystal specular-reflection illumination, and it should be fairly straightforward to implement. However, I have no example images to show here because, to my knowledge, this technique has never been tried in snow-crystal photography.

## Back Illumination

Back illumination is perhaps the easiest-to-use and most adaptable method for photographing snow crystals. As illustrated in Figure 11.21, a specimen is placed on a clear substrate and photographed using light that is transmitted through the clear ice. Unlike all the previous illumination methods, refraction of the light by the ice now plays a major role in defining the overall appearance of the image, while specular reflection is unimportant.

The example images in BI#1, #2, #3, and #4 illustrate some of the many lighting effects that can be obtained using back illumination. In BI#1, I used plain, uniform white light impinging on the crystal from a broad range of angles. This is like covering a flashlight with waxed paper and placing it directly behind the crystal. As can be seen in the image, the uniform lighting tends to wash out the structural details in the crystal, producing a rather bland image.

Better results using the same crystal are shown in BI#2, where different colors of light were incident on the crystal from different angles. All places in the object plane received the same amount of light (because the illumination only varied by angle, not by position), so the image background shows a uniform color that is basically an average of all the incident light. Meanwhile the snow crystal acted like a complex lens that refracted the light through different angles. The refraction depended on the shape of the ice, so the resulting image shows a variety of subtle red and blue highlights that reveal more structural details and give the image a pleasing sense of depth in comparison to BI#1.

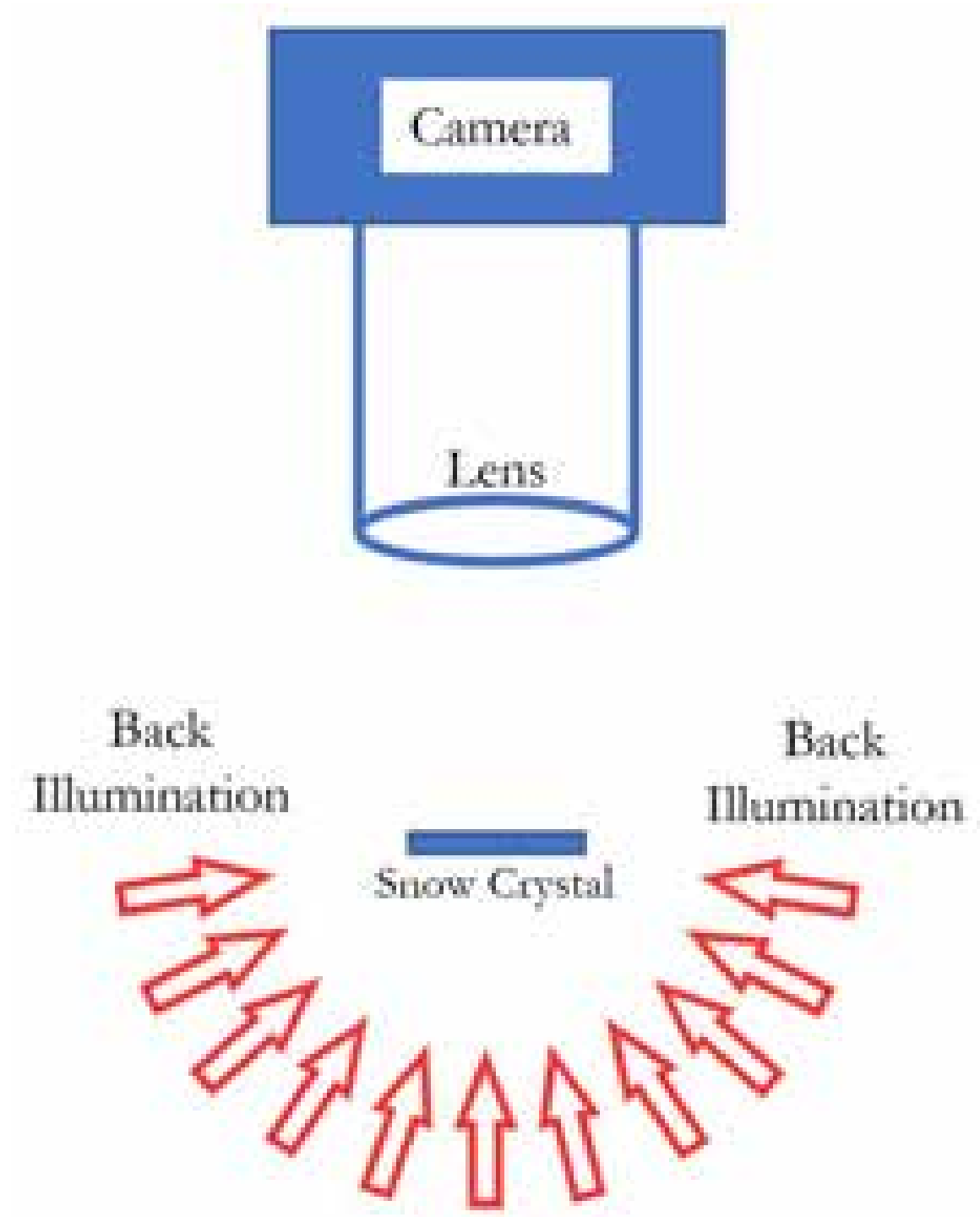

**Figure 11.21: Back Illumination. With this method, a snow crystal is photographed using light transmitted through the clear ice. It requires placing the crystal on a transparent substrate, typically a glass microscope slide, and it is best implemented using a rigid mounting system. Many illumination effects can be achieved by modifying the direction, intensity, and color of the illumination.**

BI#3 is similar to BI#2, but with greater variation in the illumination as a function of angle, resulting in stronger shading and higher contrast overall. Note how the inward-propagating rings (see Chapters 4 and 9) on the

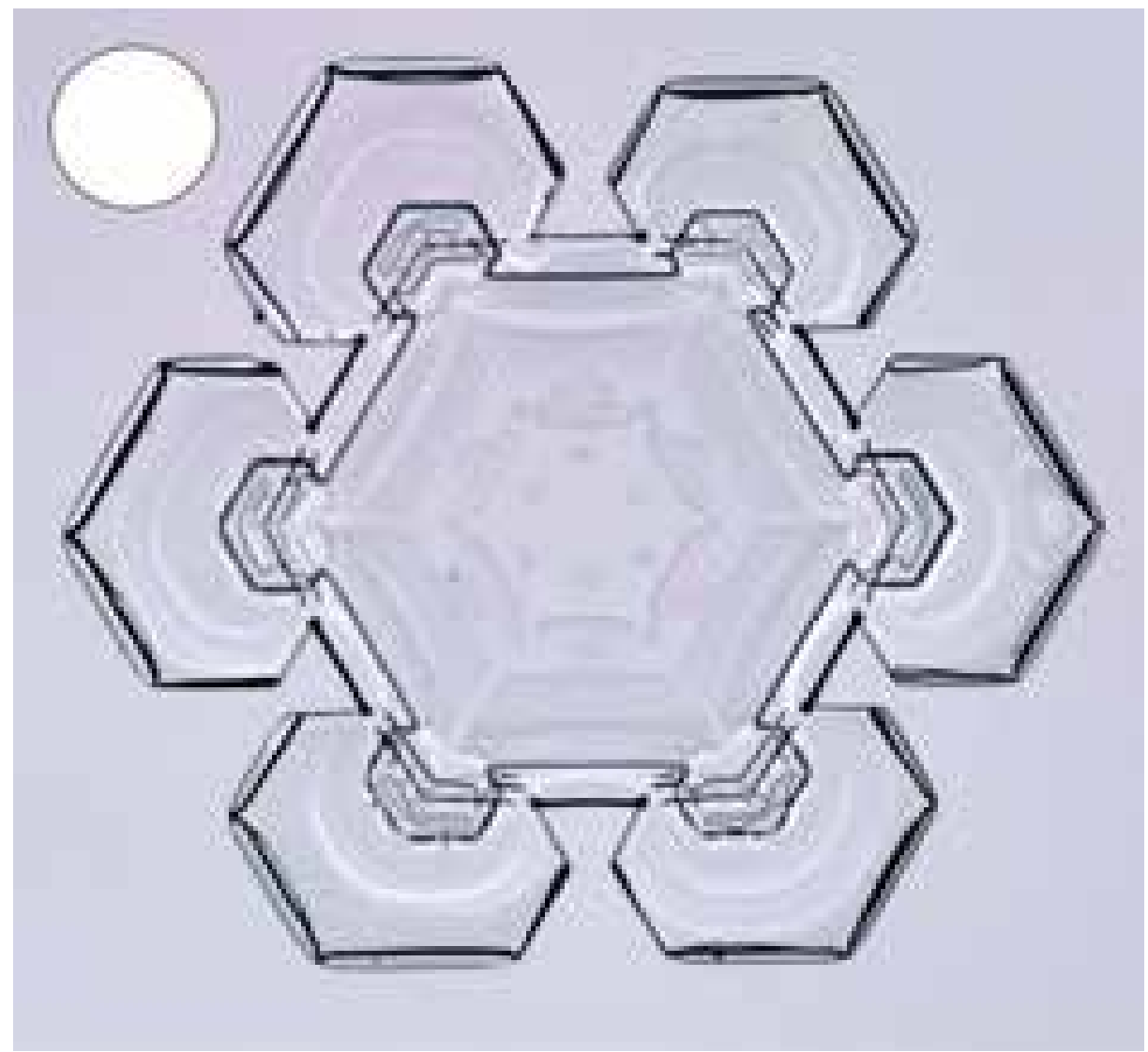

**Back Illumination#1, by the author.**
Plain white light was used to illuminate this crystal from behind, using the Rheinberg filter illustrated in the small inset image (Rheinberg illumination is described below). The uniformity of the illumination tends to wash out structural details, and it yields a drab, white-on-white image.

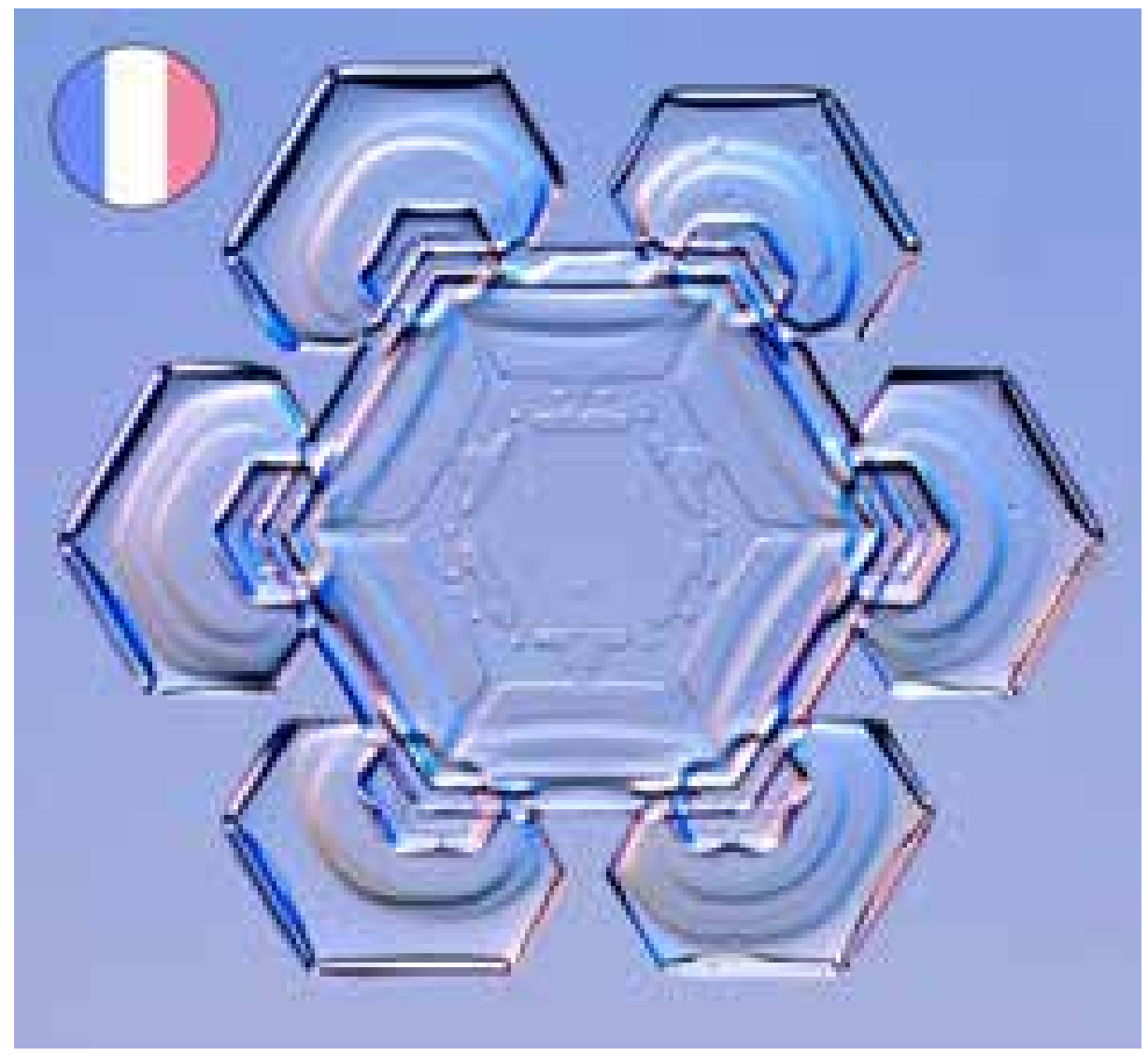

**Back Illumination#2, by the author.**
The same crystal as BI#1, but now using blue light shining in from one side and red light shining in from the other side. Here the angle-dependent illumination brings out structural details and adds a sense of depth to the photograph.

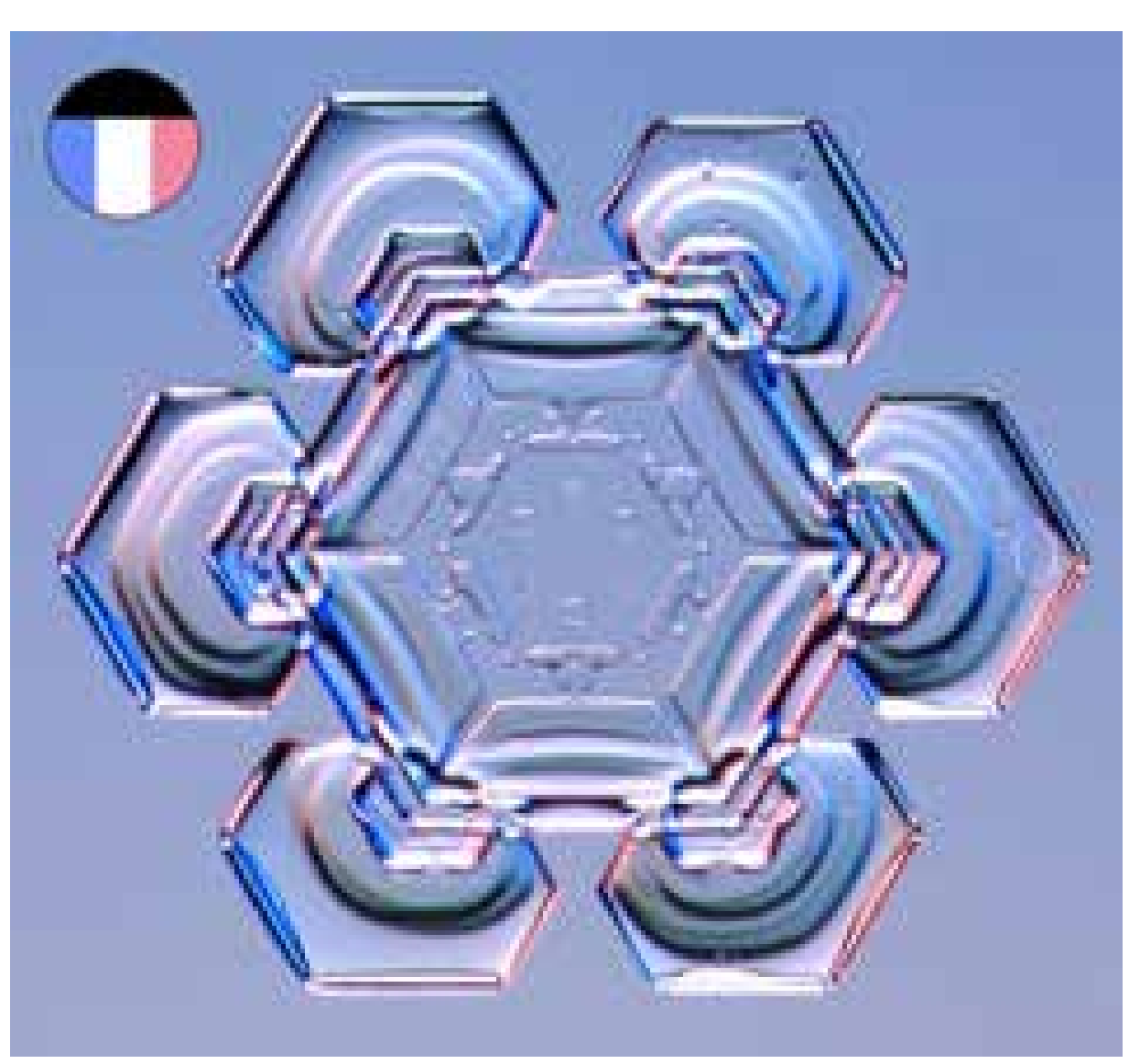

**Back Illumination#3, by the author.**
Similar to BE#2, but with no illumination (dark) from one side. The stronger variation in illumination with angle further accentuates structural details in this crystal.

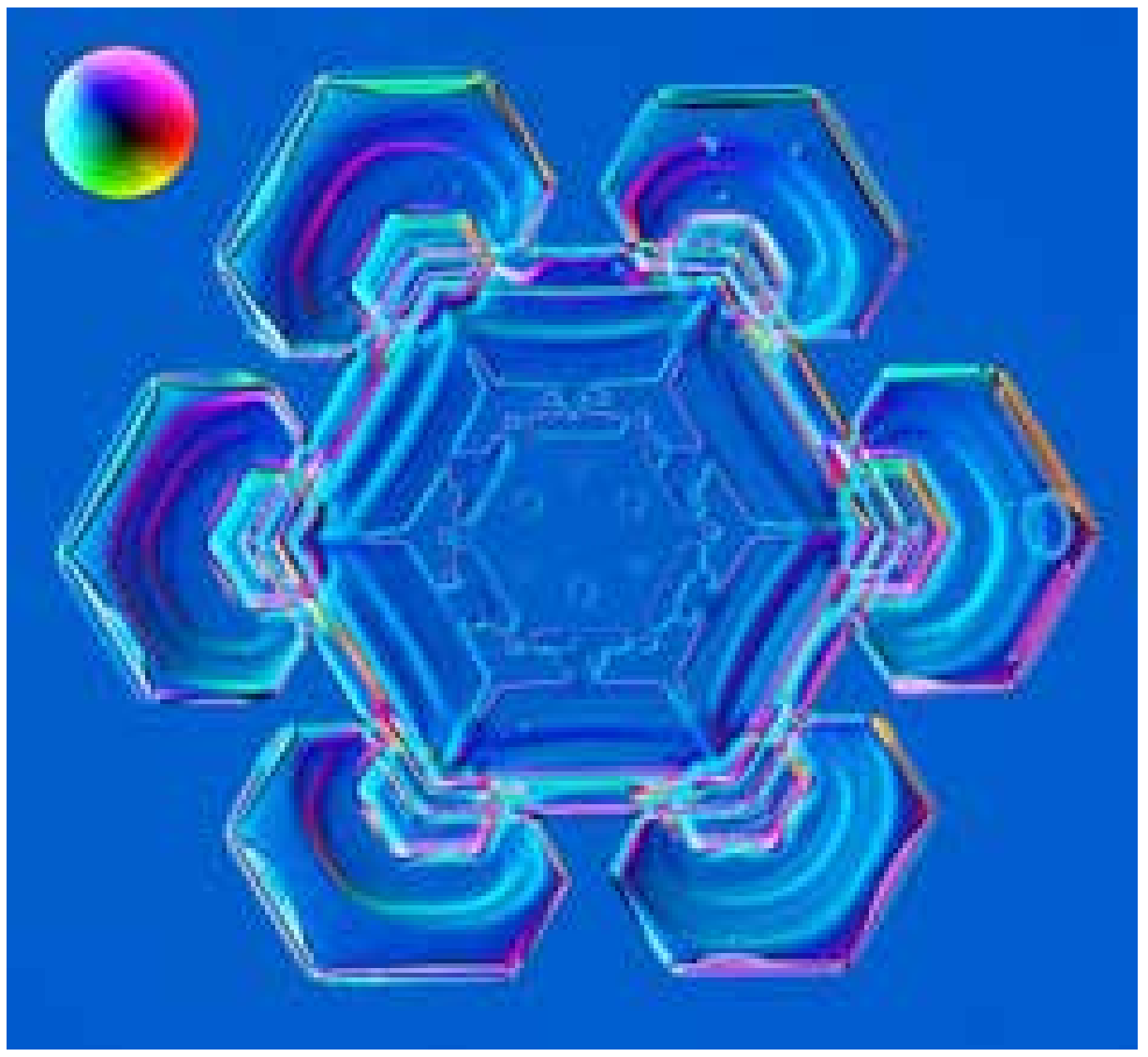

**Back Illumination#4, by the author.**
Here a (slightly offset) "rainbow" filter sent different colors of incident light in from different angles. Blue dominated, but with many colorful highlights on different parts of the crystal.

plate-like outer branches are barely visible in BI#1, quite clear in BI#2, and highly visible in BI#3. Generally, sharp edges and edge-like features stand out more with side illumination, while gentler structural features are more visible using back illumination.

BI#4 used a "rainbow" filter that shined many different colors in from different angles around the crystal, yielding a somewhat psychedelic effect. I would not use this as my go-to illumination method every day, but I personally like to photograph snowflakes using a full range of lighting techniques, just for variety.

There are many advantages and disadvantages associated with back illumination. For example, the crystals must be resting on a transparent background, and this means that a rigid mounting system is normally better than using point-and-shoot. Many colorful effects can be obtained with back illumination, but one loses the "natural" look that comes with side illumination. Colorful snowflakes get one out of the black-and-white habit, but colorful back-illuminated crystals tend to look like they are made of plastic. It all comes down to taste.

I usually photograph using a rigidly mounted back-illumination system (described below), in part because of its numerous technical advantages. Foremost is the ability to use the highest possible optical resolution, as rigid mounting largely eliminates image shake, and a face-on view requires little focus stacking when shooting stellar plates. Together, these make for easier shooting and much simpler post-processing in comparison to side illumination. This also allows me to push the resolution down to 2 µm using the Mitutoyo APO Plan 5X objective (see Figure 11.12).

## Dark-Field Illumination

Figure 11.22 shows a special variation of back illumination that can be used to good effect in snow crystal photography. Here the light rays impinge upon the subject from below, but only from the sides. There are no light rays arriving from directly below the lens. These side rays proceed undeflected if no snow crystal is present, and in that case no light enters the camera lens. Thus, the background is dark. But a snow crystal placed in the field will deflect some of these rays into the lens, thereby yielding a bright snow crystal on a dark background, as seen in DFI#1.

Side illumination from above the crystal (Figure 11.16 and SI#1) looks somewhat similar to side illumination from below the crystal (Figure 11.22 and DFI#1), but the underlying optics is quite different. In SI#1, reflected light produces the image, while in DFI#1 refracted light produces the image. Both effects depend on the surface structure of the crystal, and both produce bright edges and dark facets, so the final images look similar.

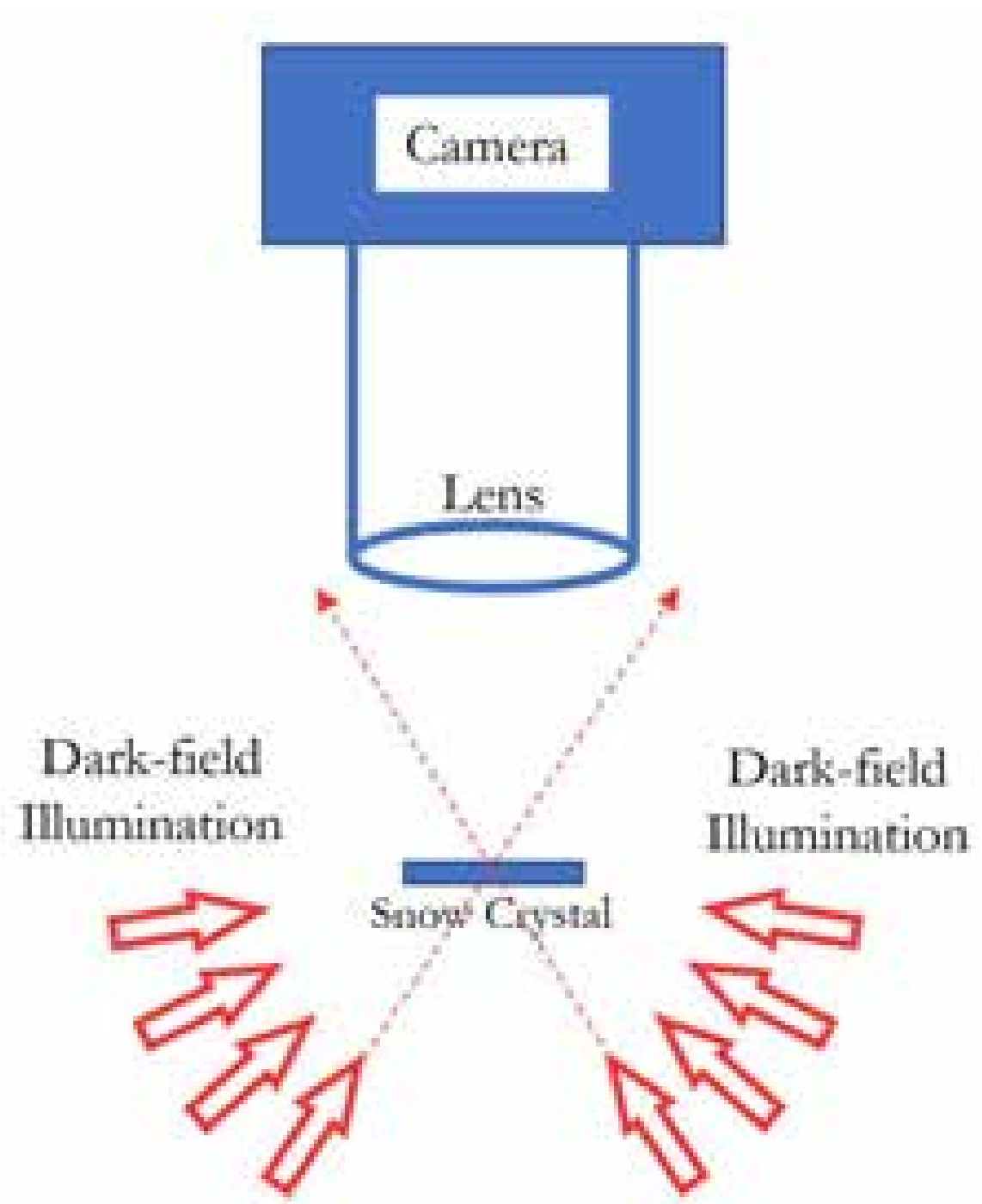

**Figure 11.22: Dark-field Illumination. This variation of back illumination uses only light that does not enter the lens in the absence of a snow crystal. If not deflected, the light rays shown in the sketch all pass outside the lens. A snow crystal will refract some of those rays, however, sending them into the lens. The result is a bright snow crystal on a dark background, as shown in DFI#1.**

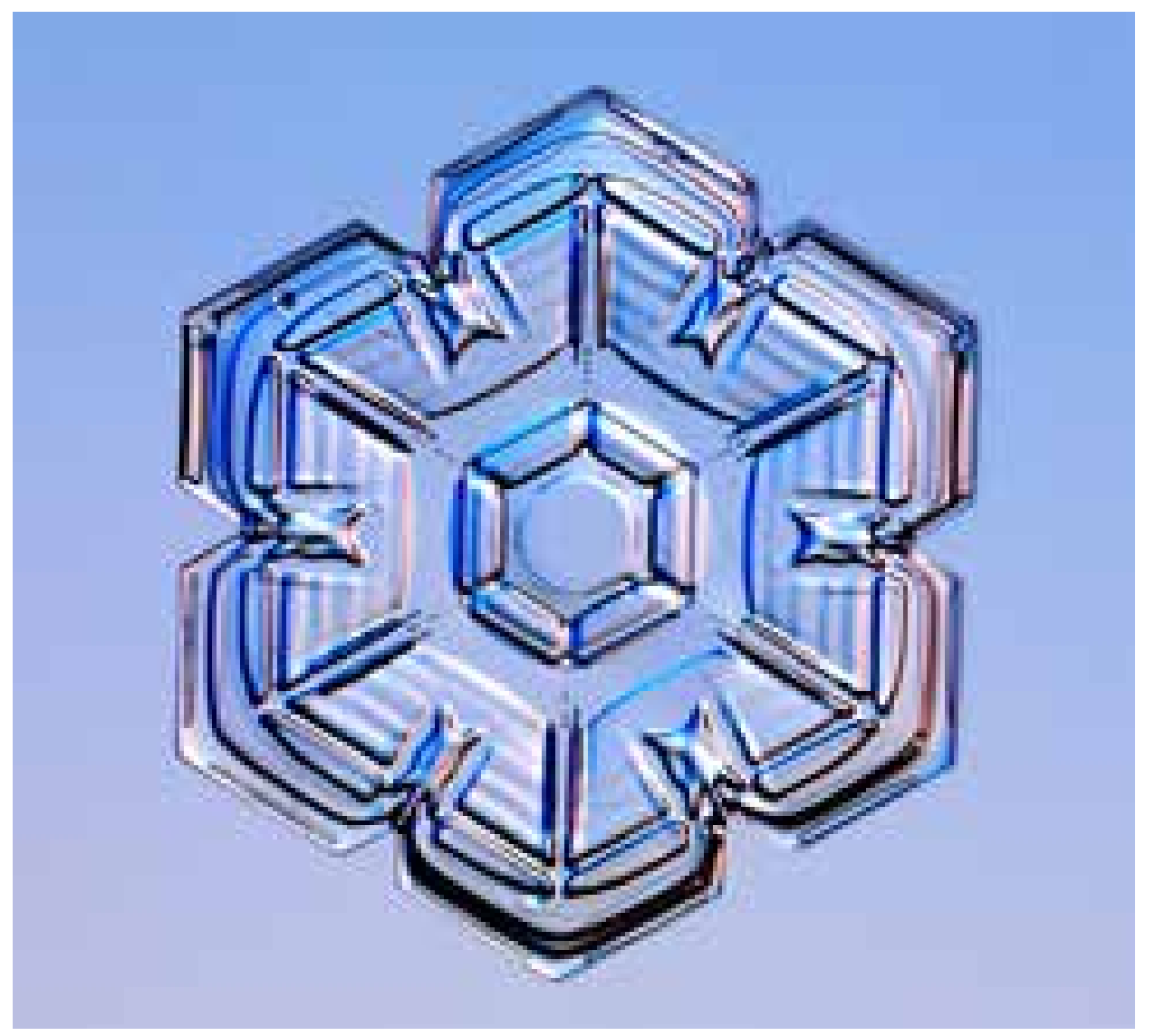

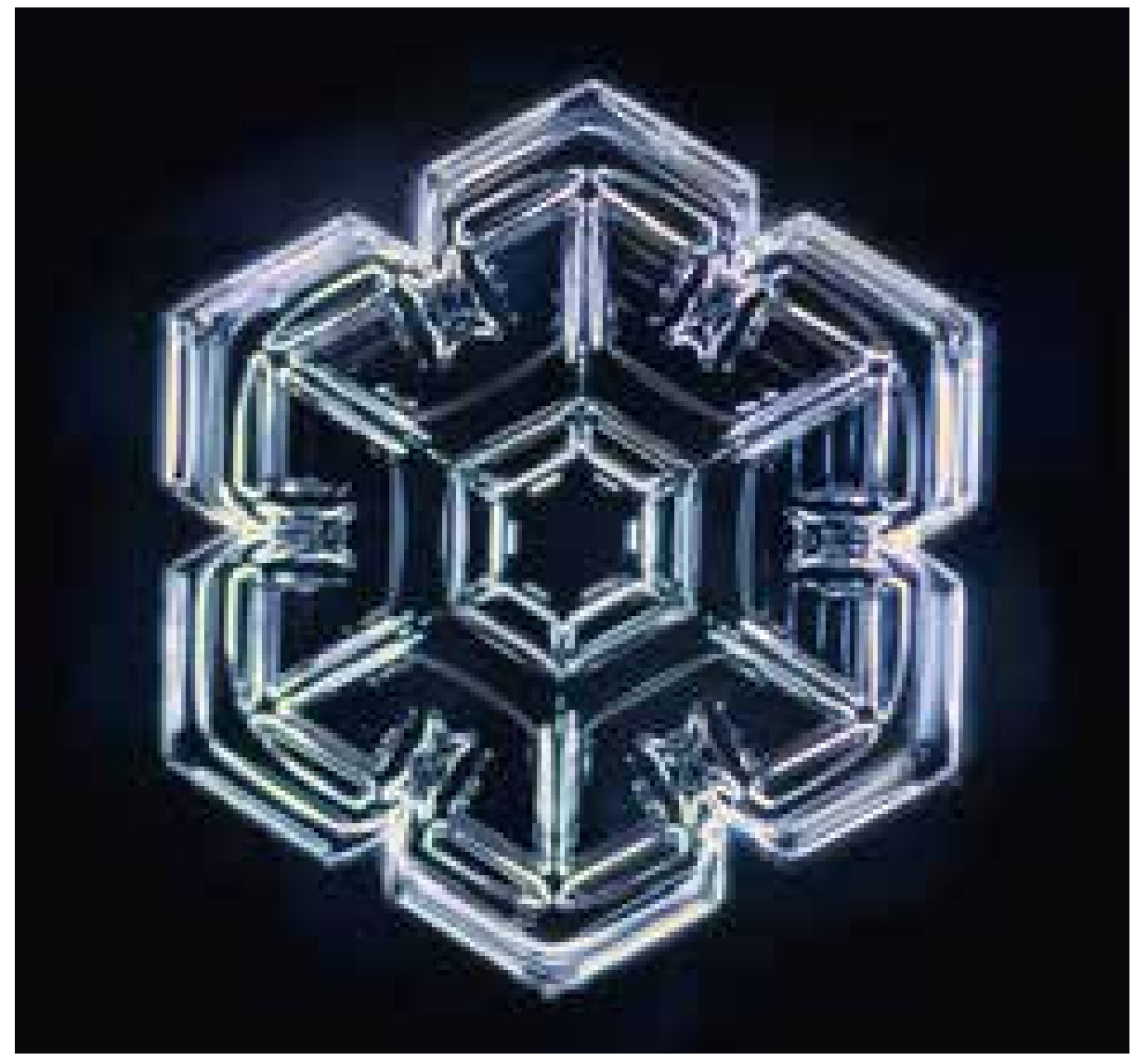

**Dark-field Illumination#1, by the author.** The left image above shows a photo taken using normal back illumination, while the right image shows the same crystal using dark-field illumination. Surface features and edges refract light through large angles, so these features appear especially bright in the dark-field image. Flat facets do not deflect the incoming light, so these regions appear dark. Note the central dark "hole" in the lower image, similar to that seen using side illumination.

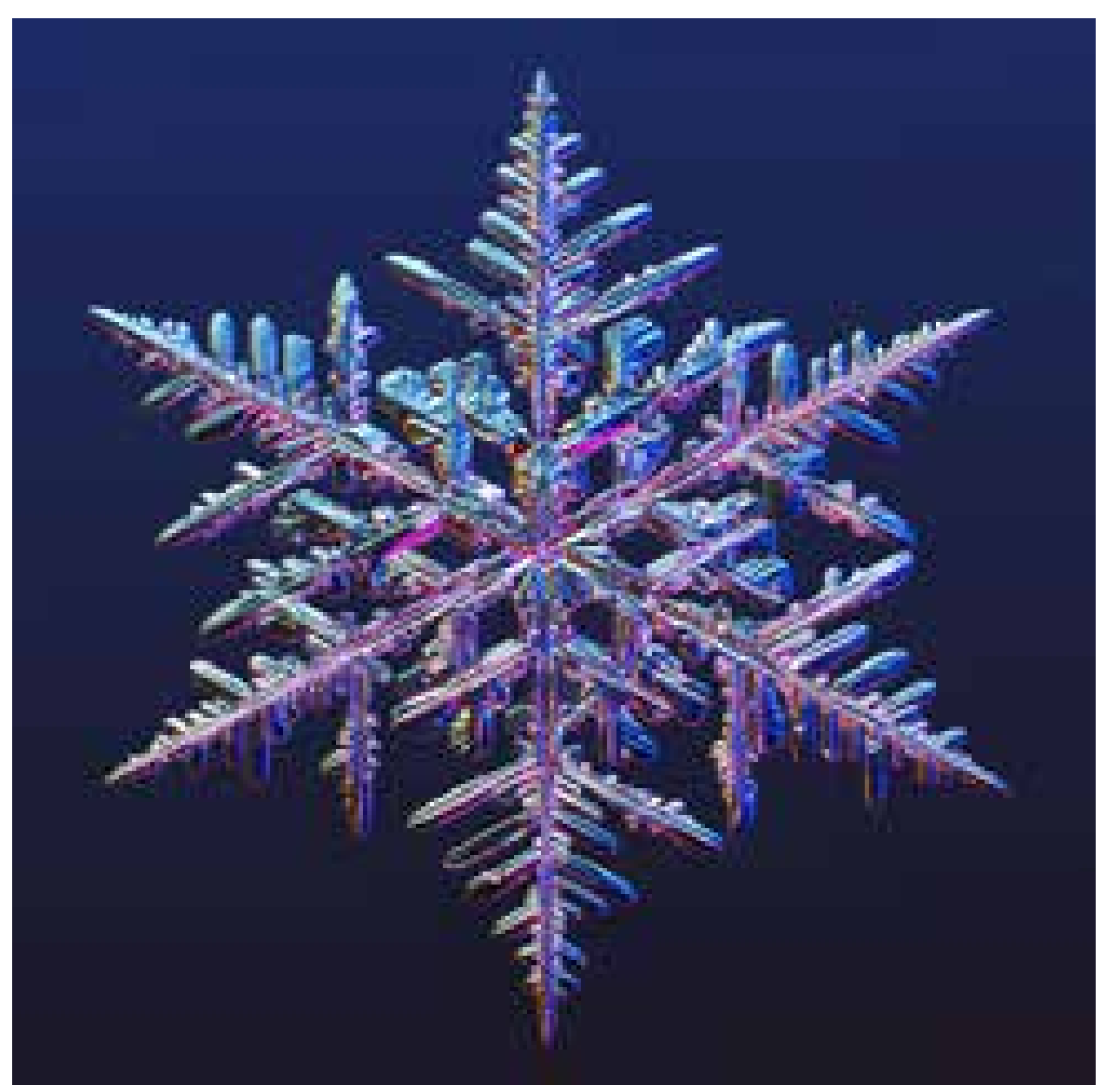

**Dark-field Illumination#2, by the author.** I took this photo using dark-field illumination with a twist of color by using different colored rays coming in from different directions. The added color makes the crystal look somewhat unnatural, but it adds variety to one's collection of snowflake photos. Setting the image saturation down to zero yields a black-and-white dark-field image.

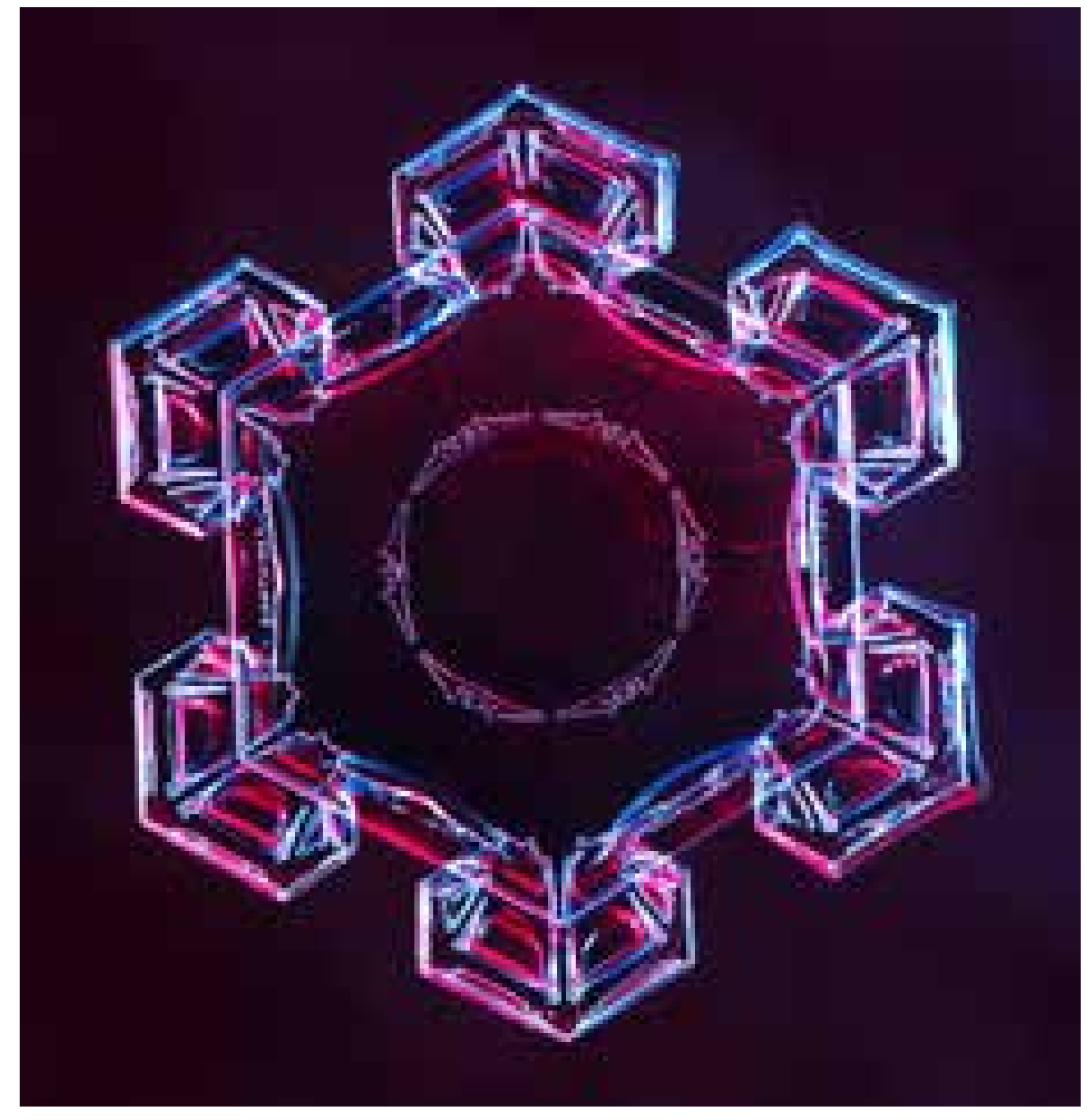

**Dark-field Illumination#3, by the author.** By using Rheinberg illumination (see below), it is straightforward to add a variety of pleasing colorful effects when using dark-field illumination, simply by sliding a color filter around to see what looks good with a given crystal.

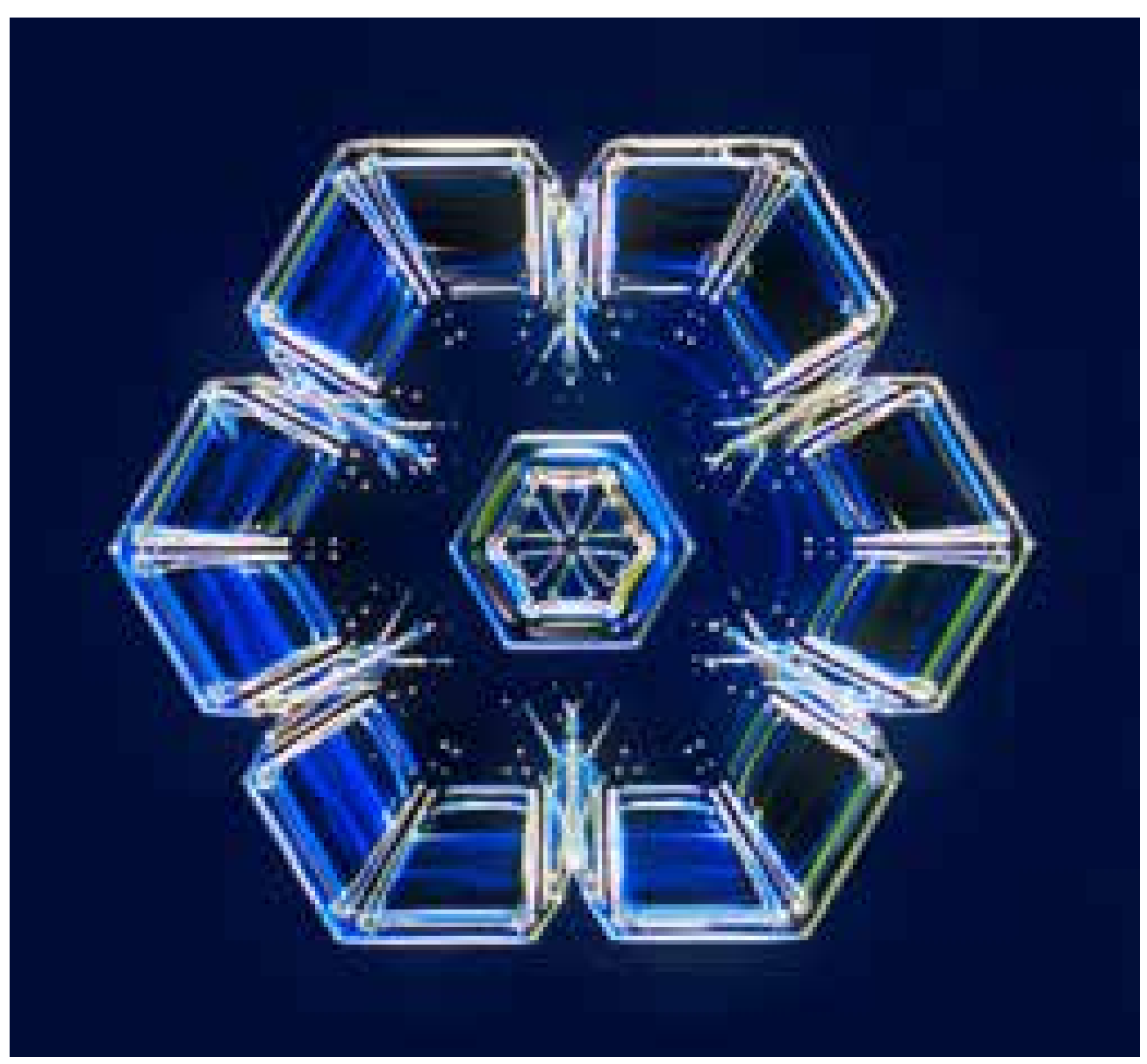

**Dark-field Illumination#4, by the author.** The color blue is generally associated with cold, so a splash of blue often adds an icy look to a snow-crystal photograph. SI#10 also incorporates a bluish hue.

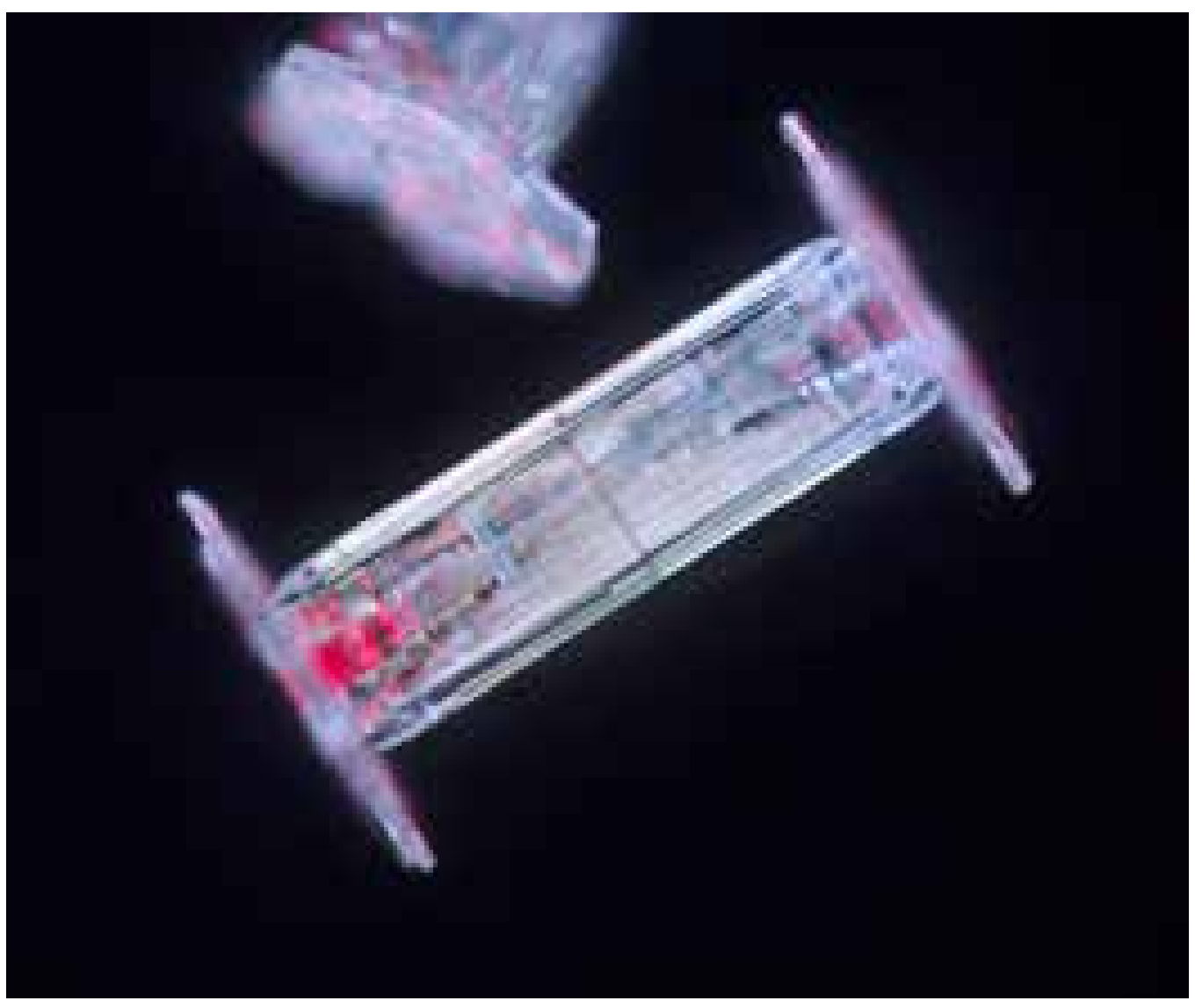

**Dark-field Illumination#5, by the author.** Capped columns and other thick crystals look especially nice with both dark-field illumination and side illumination. There are no flat-plate regions, so the images avoid that high-contrast, edge-heavy look seen with plate-like crystals under the same lighting. Here I used a splash of red light from one side to liven up the photograph.

## Rheinberg Illumination

Rheinberg illumination is not so much a separate illumination category, but rather a highly flexible method for achieving uniform-background back illumination. The overall optical layout is illustrated in Figure 11.23, and the key new addition is a *field lens* (near the image field) that images a color filter onto the optical pupil inside the microscope objective. (Information about what an optical pupil is, and what it's properties are, can be found online.) Adding this field lens complicates the optics, and it requires a rigidly mounted optical system. What one gains from this additional complexity is the ability to rapidly experiment with all different types of back illumination, simply by using different color filters. This capability is especially useful because different types of snow crystals tend to photograph better under different types of lighting. With Rheinberg illumination, switching the lighting around can be done simply by swapping in different color filters.

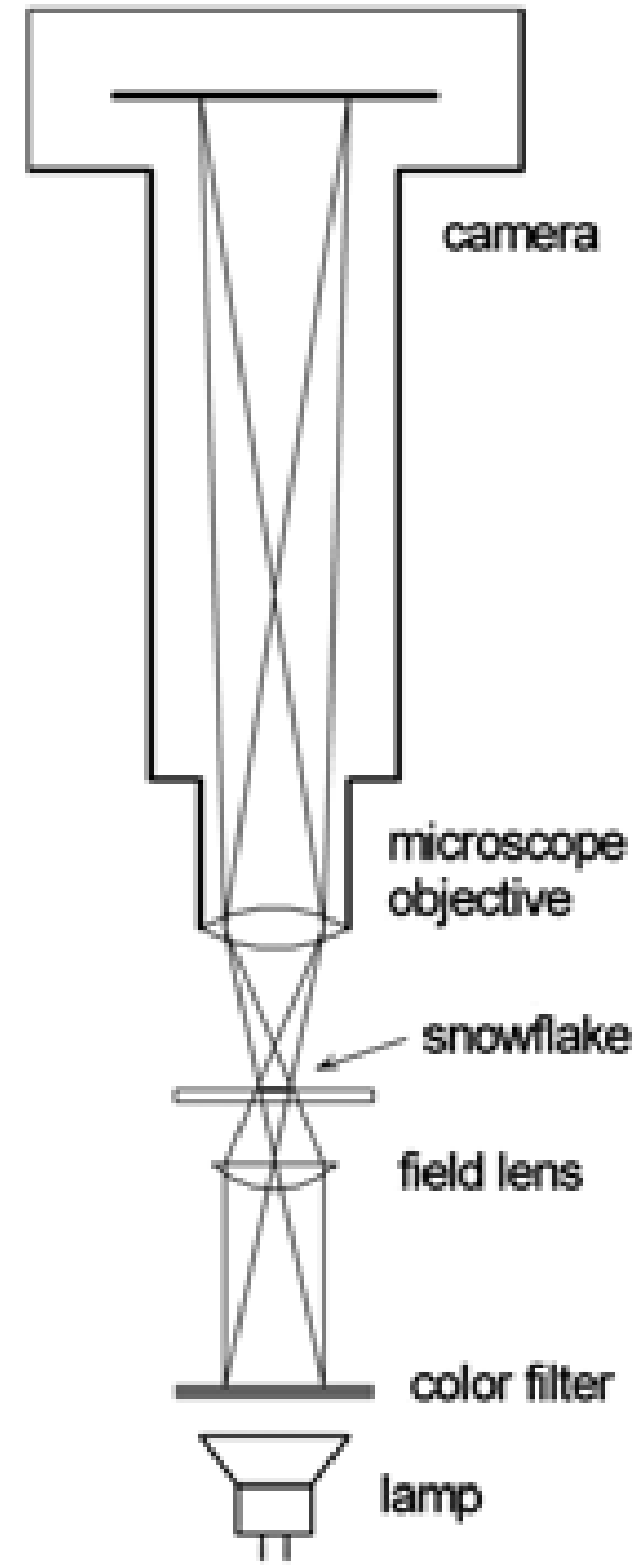

**Figure 11.23: This diagram shows the optical layout used for Rheinberg illumination. In addition to the usual camera+lens, the lower field lens images a color filter onto the optical pupil at the microscope objective. This layout yields a uniform background even with a highly pattered color filter.**

Figure 11.24 shows a variety of different color filters I have used with my Rheinberg illumination setup (see below). Looking at the fourth filter in this set, the rays in Figure 11.23 illustrate how the red side of the filter produces a set of nearly parallel red light rays impinging onto the snow crystal from the left, while the blue side of the filter produces rays impinging from the right. Both sets of rays fill the entire image plane uniformly, and these rays combine to yield an overall red+blue background color. The end result is angle-dependent back illumination, uniform across the image plane, with the colors of the different rays determined by the pattern on the filter. By imaging the filter onto the optical pupil, the overall background color is uniform regardless of the pattern in the filter.

Note that moving the red/blue filter left or right changes the red/blue balance, thus changing the background and the highlights in subtle ways. Moreover, any color filter can be inserted into the optical system and moved around with easy, while viewing the crystal. This makes it quick and easy to choose different optical effects, perhaps trying multiple filters on a single crystal. Dark-field illumination can be produced by using a filter with a central dark spot, for example seventh filter in Figure 11.24. Although Rheinberg illumination seems overly complicated, in practice it is easy to set up and easy to use, yielding a large diversity of unique lighting effects.

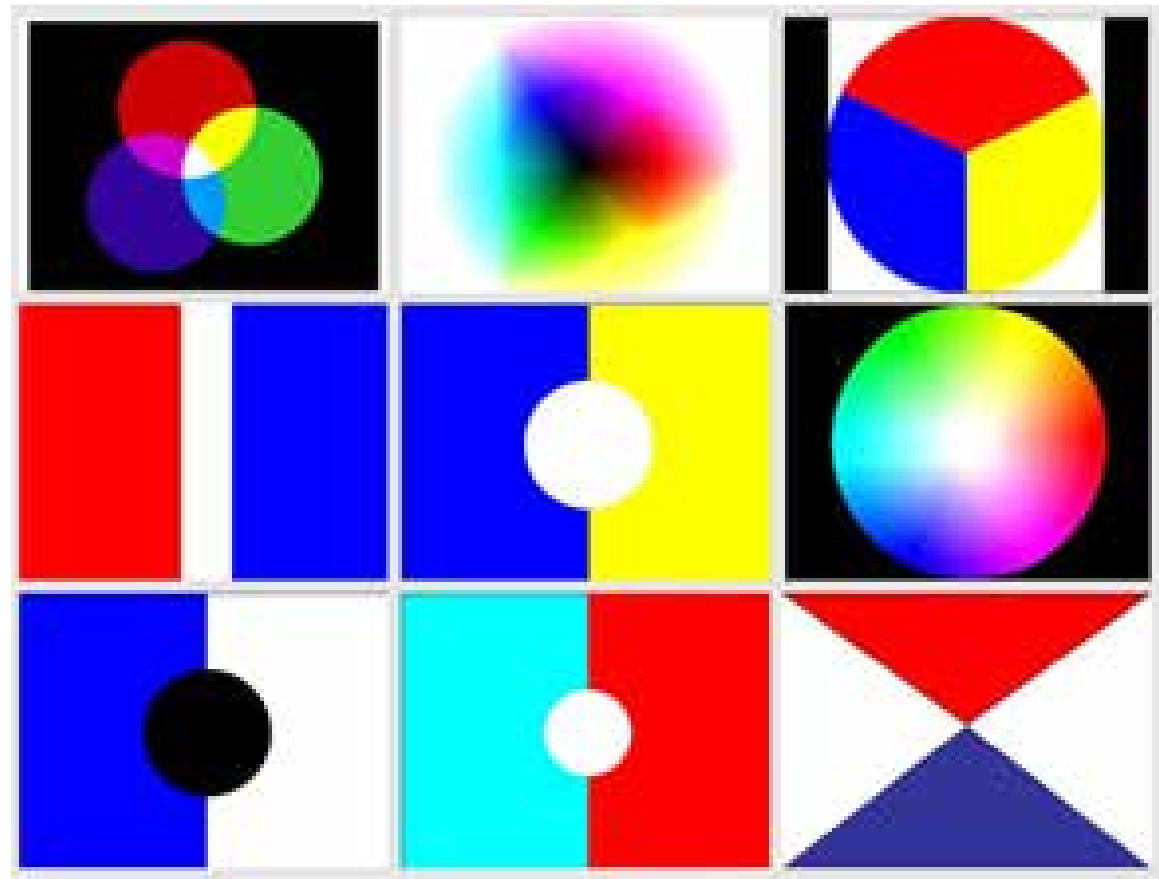

**Figure 11.24: Nine different color filters I have found useful with Rheinberg illumination. The second is a rainbow dark-field filter used for BI#4, and the fourth is a general-purpose red-blue filter I used for BI#2 and BI#3 (the difference being the placement of the filter relative to the optic axis.**

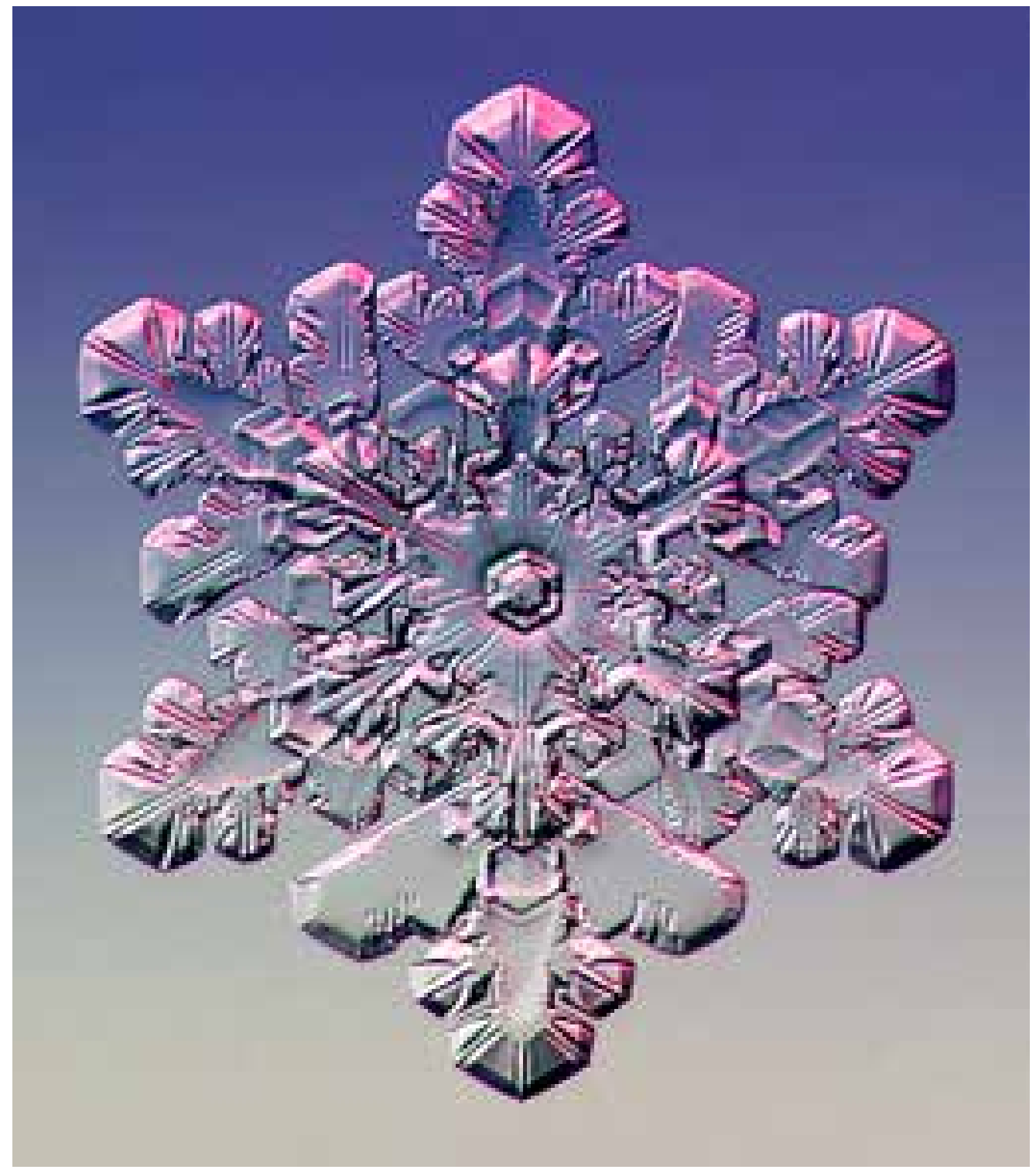

**Rheinberg Illumination#1, by the author.** Red highlights give this snow crystal a distinctive look. Many similar lighting effect can be obtained using Rheinberg illumination simply by swapping in different color filters and observing the results.

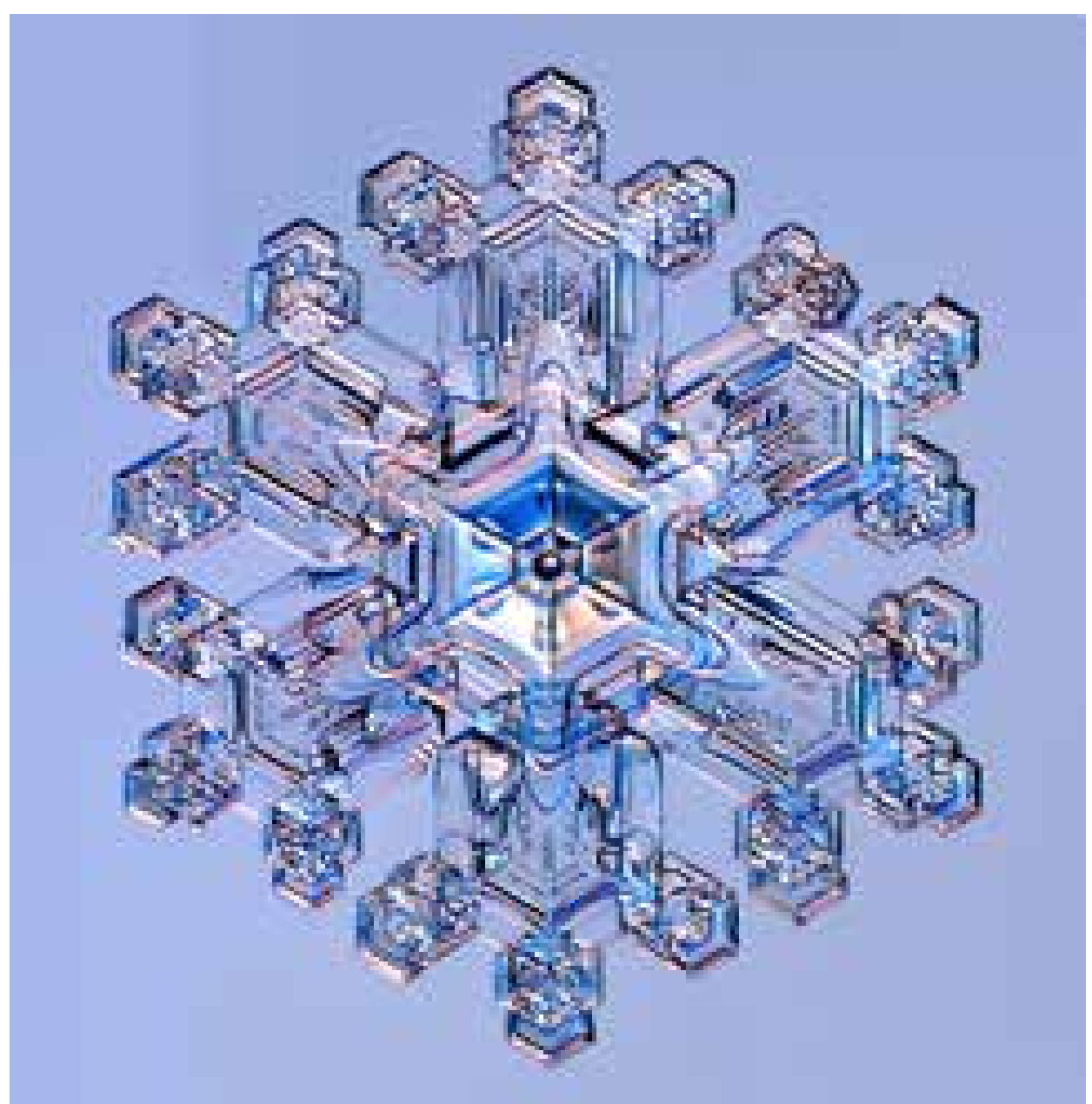
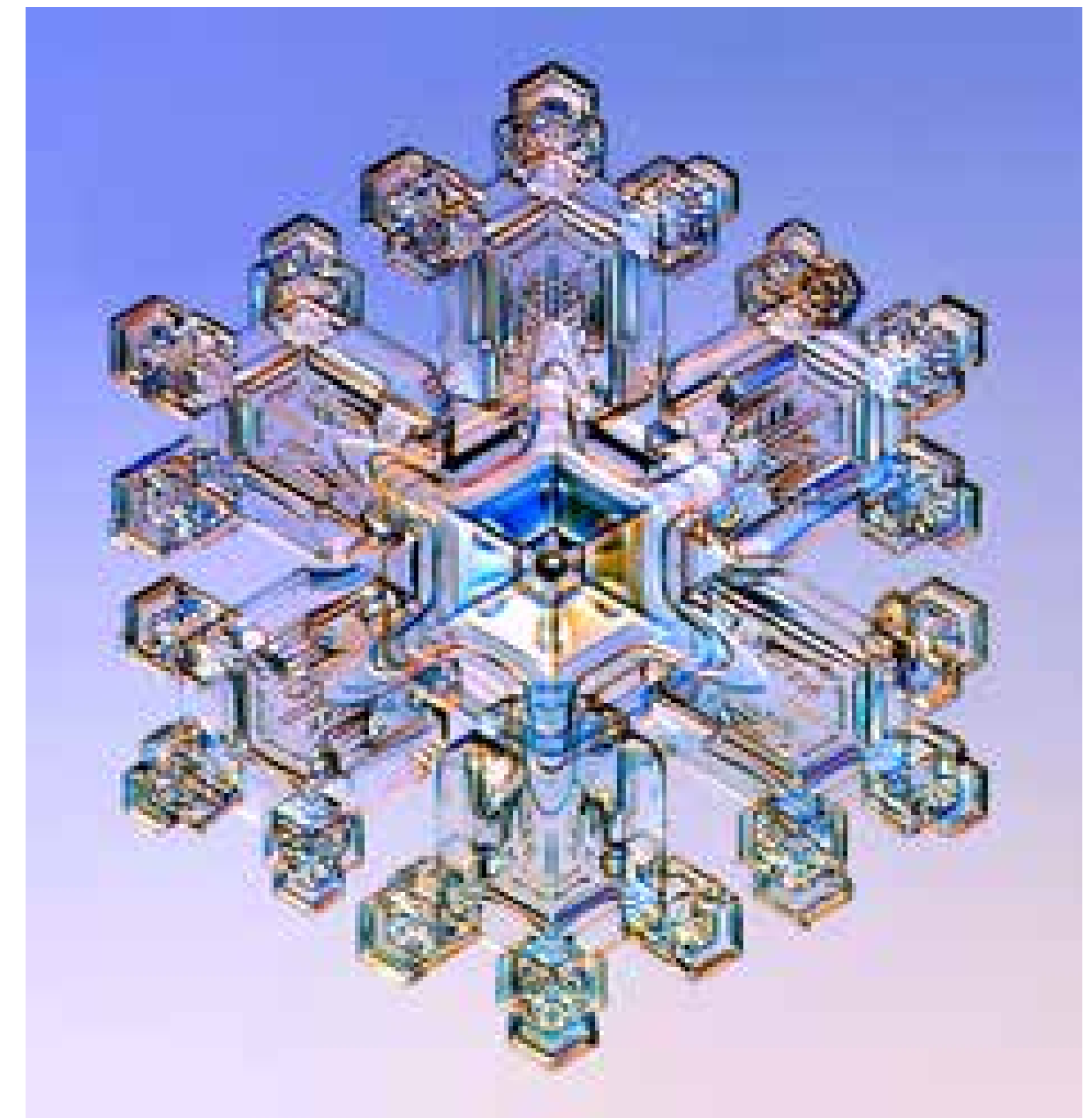

**Rheinberg Illumination#2, by the author.** The left image above shows the original, out-of-the camera image of a large stellar snow crystal (after minimal post-processing to remove dust specks from the background, level adjustments, etc.). The image shows some colored highlights and a uniform background color, which is typical for Rheinberg illumination. The image on the right was derived from the first mostly by adding a color gradient and goosing up the saturation a bit in post-processing. Adding gradients is a straightforward image-processing trick when the original background is uniform, but it can be quite difficult to remove an existing background gradient. The automatic uniformity of the background thus adds another layer of flexibility to the Rheinberg technique, as it allows easy experimentation with different types of superimposed color gradients in post-processing to create pleasing visual effects. Often just a bit of extra zing can turn a good photo into an eye-popping snow-crystal portrait.

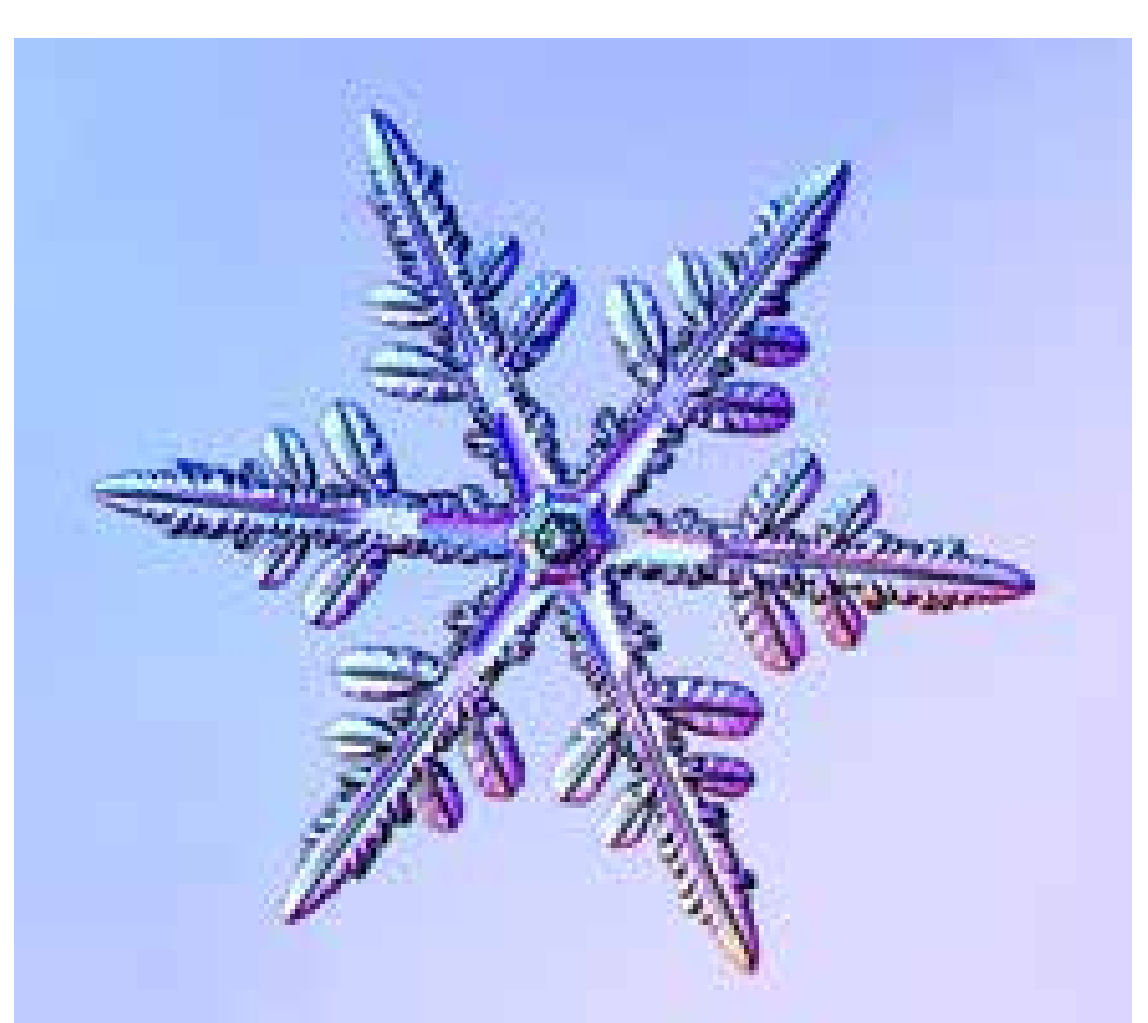
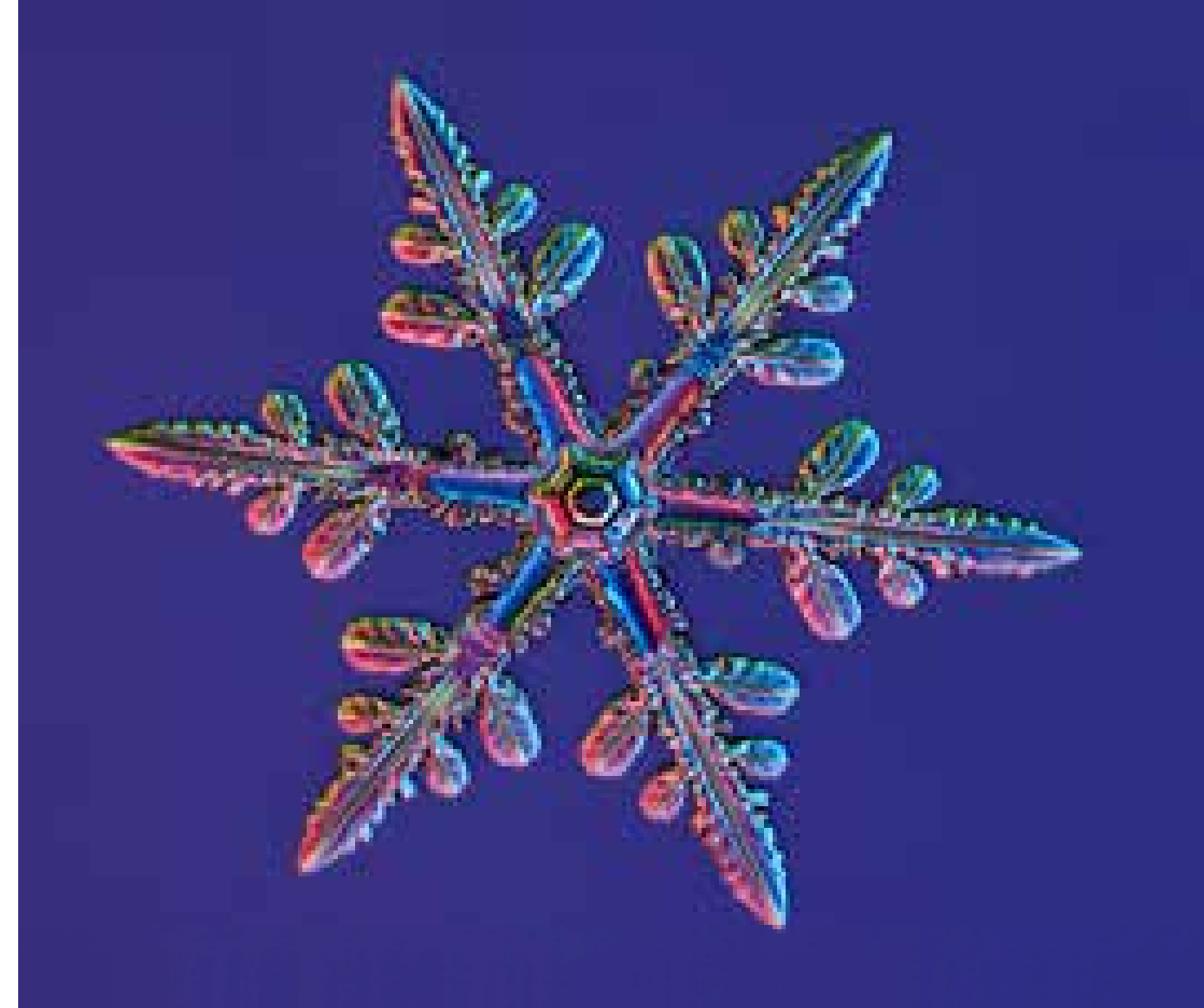

**Rheinberg Illumination#3, by the author.**
The left image above shows a relatively simple stellar crystal using normal back illumination. The colorful image on the right shows the same crystal using a rainbow color filter.

## THE SNOWMASTER 9000

For many years I have traveled to various cold locations to photograph snowflakes using the apparatus shown in Figure 11.25, constructed in 2003, which I call the *SnowMaster 9000*. The overall optical layout is basically that depicted in Figure 11.23, using Rheinberg illumination. This hardware has become a real workhorse for me, and I have already used it to take over 10,000 snowflake photos.

Starting at the top in Figure 11.25, the camera is an older Canon EOS single-lens reflex (SLR) model with a 20-Mpixel full-frame (36x24 mm) sensor. The image is viewed through a right-angle eyepiece that is better seen in Figure 11.14. It appears that SLR cameras are slowly being replaced by mirrorless systems with viewing screens, so this camera with its eyepiece is showing its age.

I put the camera in a styrofoam lined box and included a low-power heater to keep the camera temperature above freezing, as the camera specifications showed it rated only down to 0 C. I have since learned that this spec is highly conservative, and most digital cameras have no trouble working at temperatures down to -20 C, although the battery capacity tends to be reduced (temporarily) when the temperature is low.

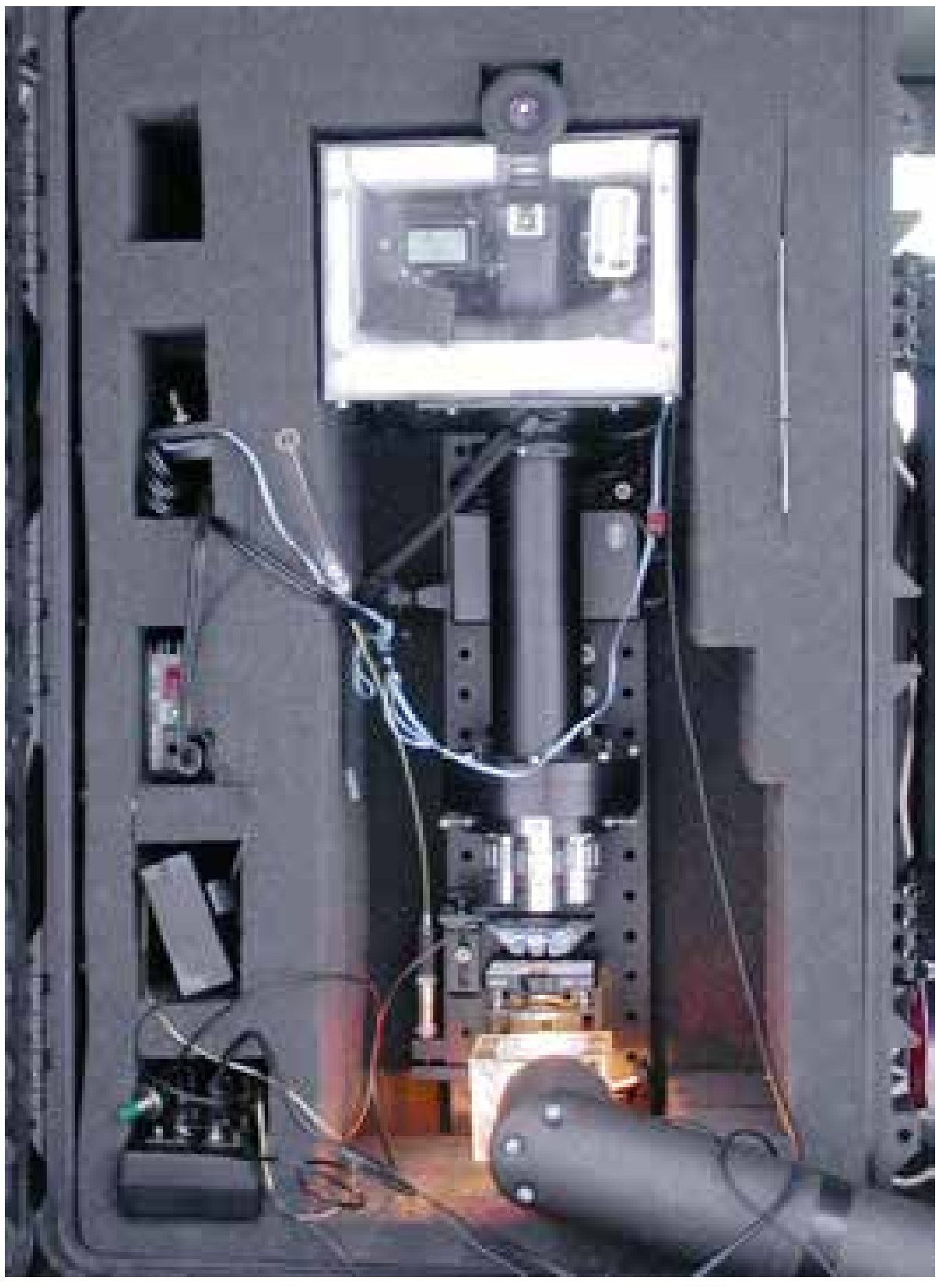

**Figure 11.25: The *SnowMaster 9000* apparatus, which has the basic optical layout shown in Figure 11.23. The hardware is mounted inside a hard-shell case for easy transport.**

Below the camera is a length of extension tube from Thorlabs, attached to the camera using an appropriate adaptor. Below this is a home-built turret complete with the Mitutoyo Plan APO 2X, 5X, and 10X objectives (see Figure 11.12). In hindsight, the turret was largely unnecessary, and I used the 5X objective about 90 percent of the time. On the other hand, the "monster" fernlike stellar dendrite described in Chapter 10 was so big that it required four separate photos at 2X, so I was happy to have the ability to switch objectives quickly that day. But I wouldn't recommend building your own turret, as I did, because that turned out to be something of a challenging mechanical-engineering task.

Directly below the microscope objectives there is an array of LEDs I used for experimenting with side illumination, although I was never very happy with the results I obtained using this light source. I mostly abandoned the LEDs in favor of Rheinberg illumination, which allowed for a much greater variety of illumination effects by changing color filters, as described above.

The LEDs point down toward a spot on the center of the image plane, where a snow crystal rests on a microscope slide. The slide rests on a 90-degree angle plate that is attached to a linear translation stage, both from Thorlabs. The micrometer on the stage is attached to a flexible cable that includes a large plastic handle at its end, used to adjust the focus. This thin plastic tube makes is fairly easy to change the focus using ungloved hands in the cold.

The field lens sits directly below the snow crystal, mounted to the bottom surface of the angle plate. The field lens is just a basic one-inch-diameter achromatic lens, as there is no need to focus the color filter onto the pupil with great precision. The incandescent light source is contained in the glowing box in the photo, although this was later replaced with an LED bulb to reduce the amount of heat generated.

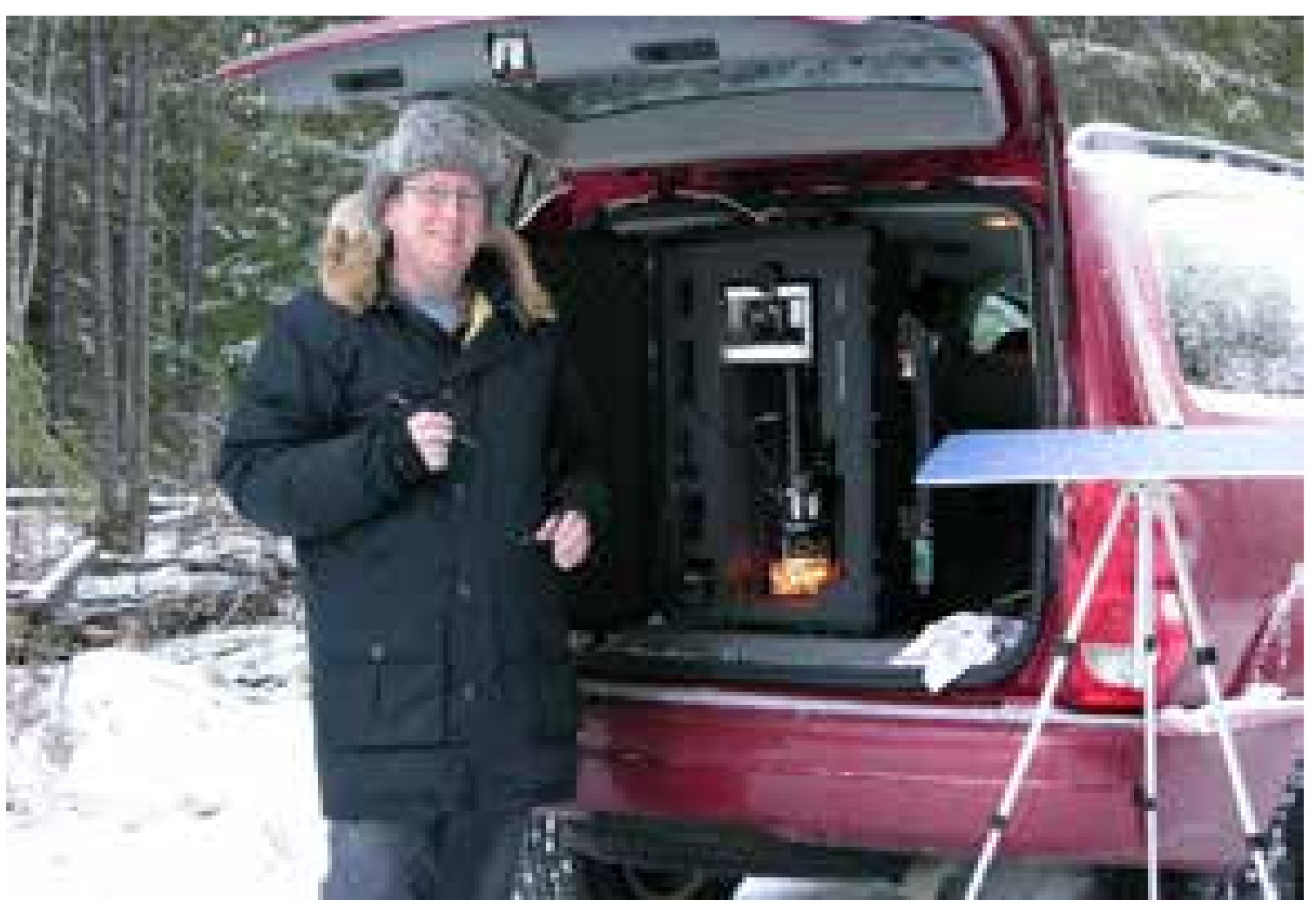

**Figure 11.26: The author photographing snowflakes using the SnowMaster 9000 in Cochran, Ontario. Note the blue foam-core collection board mounted at a convenient height using a tripod, the small paintbrush in one hand and a glass slide in the other. Alas, bare fingers are needed for dexterity, so only fingerless gloves can be used.**

Because this system provides plenty of light, I usually set my camera to ISO 100, as this reduced noise in the image relative to higher ISO settings. The shutter speed was quite slow, of order 1/100$^{th}$ second, which is plenty fast when using a rigid mounting system.

Although this hardware setup may look imposing, it is actually quite simple, as illustrated in Figure 11.23. If you remove the unnecessary heated camera box, get rid of the hard-shell case by putting the setup in an unheated garage or shed, and forego the turret, then what remains is not particularly difficult to build. One can assemble a similar system indoors (see Figure 11.14) for testing and debugging before moving it outside into the cold. You will not find microscope objectives or translation stages at your local camera store, but these tools are not difficult to work with, and they are no more expensive than traditional macro photography gear.

Figure 11.26 shows the author working with the SnowMaster 9000 under typical winter conditions. Once the system is set up, the workflow is straightforward: search for suitable crystals on the foam-core collection board by eye, pick up a promising specimen using a fine paintbrush, set it carefully onto a glass microscope slide, place the slide on the observing plate, adjust the focus, adjust the color filter, and take the shot. Take several shots for focus stacking as needed. If the crystal looks especially nice, try a few additional color filters for different visual effects. Then clean the slide and repeat. Fun!

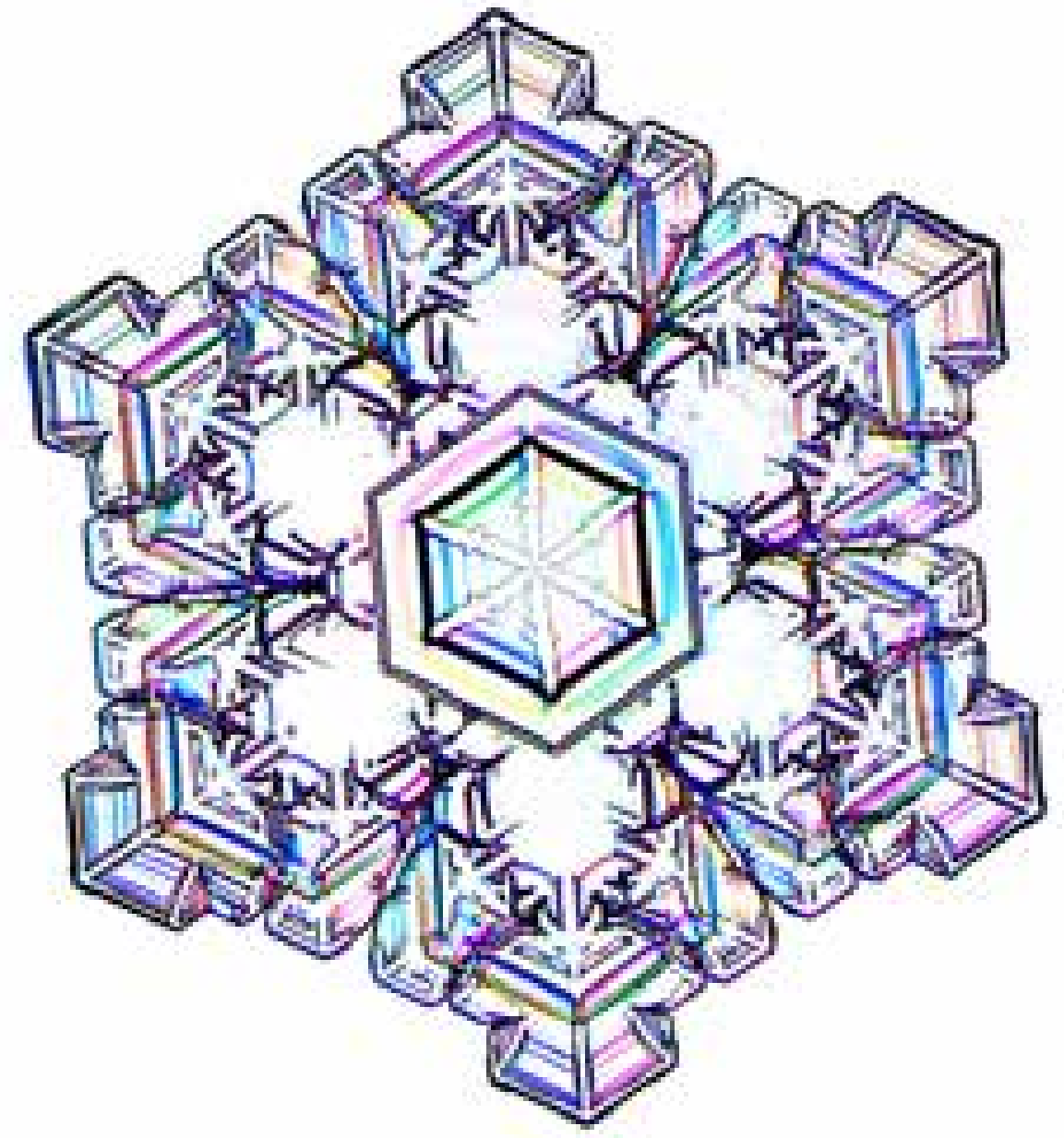

**Rheinberg Illumination#4, by the author.** Filter #6 in Figure 11.24 gives results similar to plain back illumination, but with color.

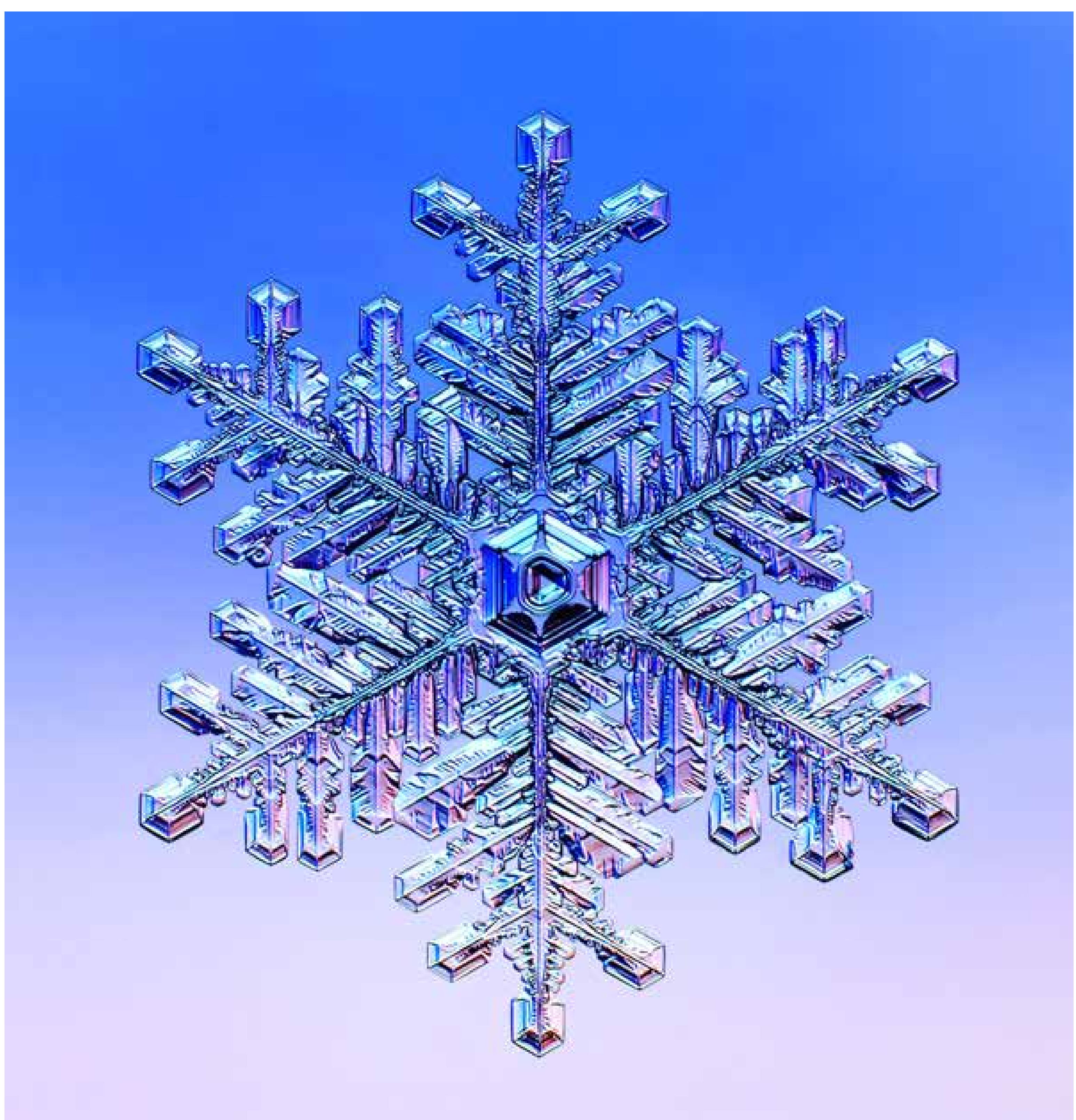

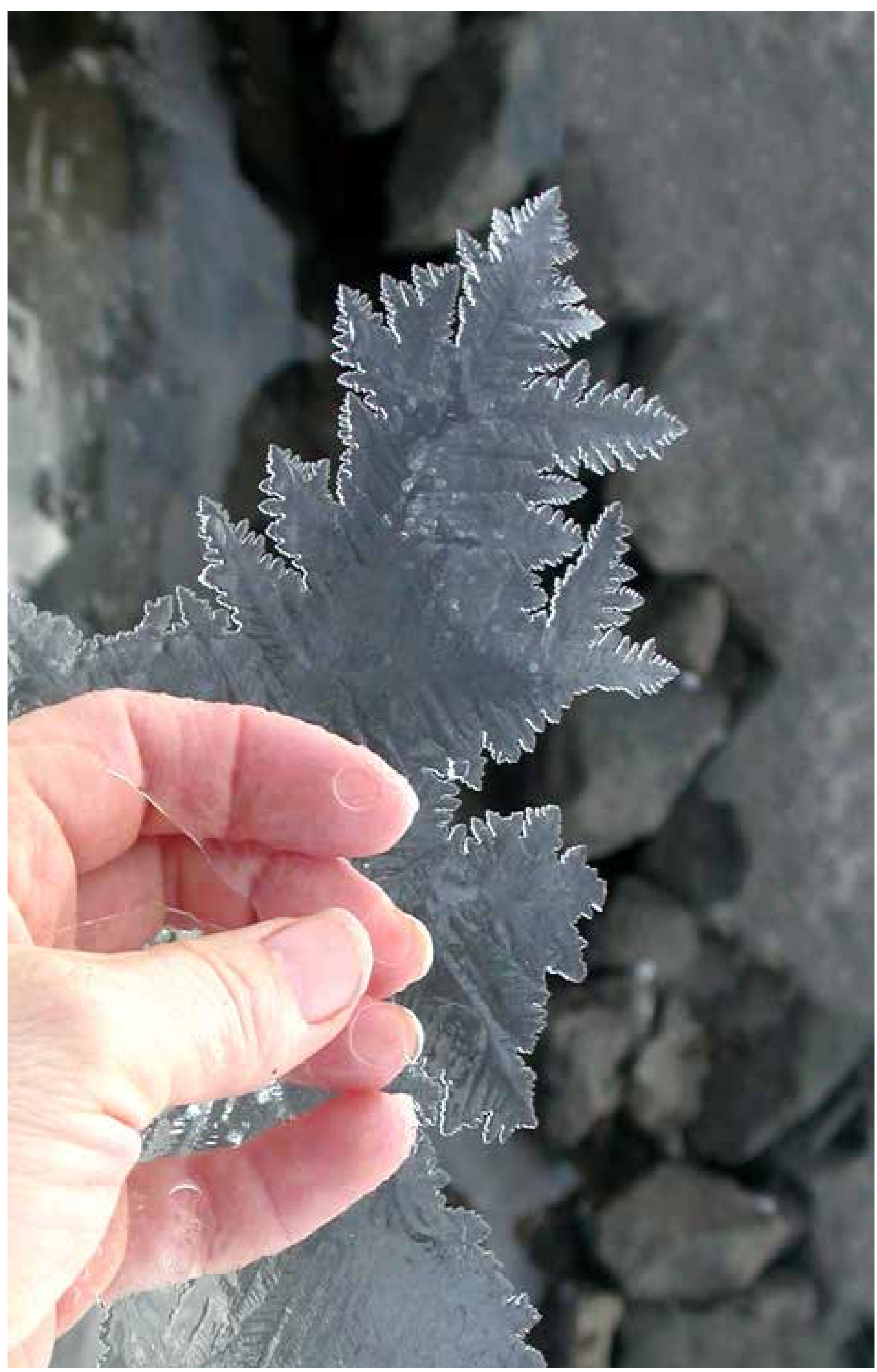

# Chapter 12

# Ice from Liquid Water

*We are all working for a common and well defined aim: to get more insight into the workings of nature. It is a constructive endeavor, where we build upon the achievements of the past; we improve but never destroy the ideas of our predecessors.*
– Victor Weisskopf
*The Privilege of Being a Physicist*, 1989

Ice growth from liquid water is related to ice growing from water vapor in some ways but can be quite unrelated in many other ways. The two subjects exhibit a clear correspondence in the attachment kinetics near 0 C, as the ice/water interface behaves much like the ice/vapor interface in the presence of substantial surface premelting, as I discussed in Chapter 4. Given the profound importance of the ice/water freezing transition in meteorology, the environment more generally, and even in our everyday lives, it is somewhat surprising that many fundamental aspects of the physical dynamics of ice freezing from liquid water remain relatively unexplored and unexplained, often even less so than the ice/vapor freezing transition.

My main focus in this chapter will be on the growth of ice from pure liquid water. In this simplest physical system, the growth dynamics are governed by a combination of attachment kinetics at the ice/water interface, the diffusion of latent heat generated by the freezing transition, and surface energy effects. The physics and mathematics are both similar to what we encountered with ice/vapor growth, so here the discussion will connect nicely to the corresponding areas in Chapters 3 and 4.

Chemical solutes add another dimension to the physics of ice/water solidification, along with considerable additional complexity. The heat-diffusion problem transforms to a combination of heat and particle diffusion, as solutes generally do not incorporate into the ice

**Facing page: Rachel Wing holds a pond crystal lifted from the surface of Lake Superior after a cold night. The thin-plate structure arises from the small ice/water basal kinetic coefficient, while the dendritic branching is the result of heat-diffusion-limited growth. Photo by the author.**

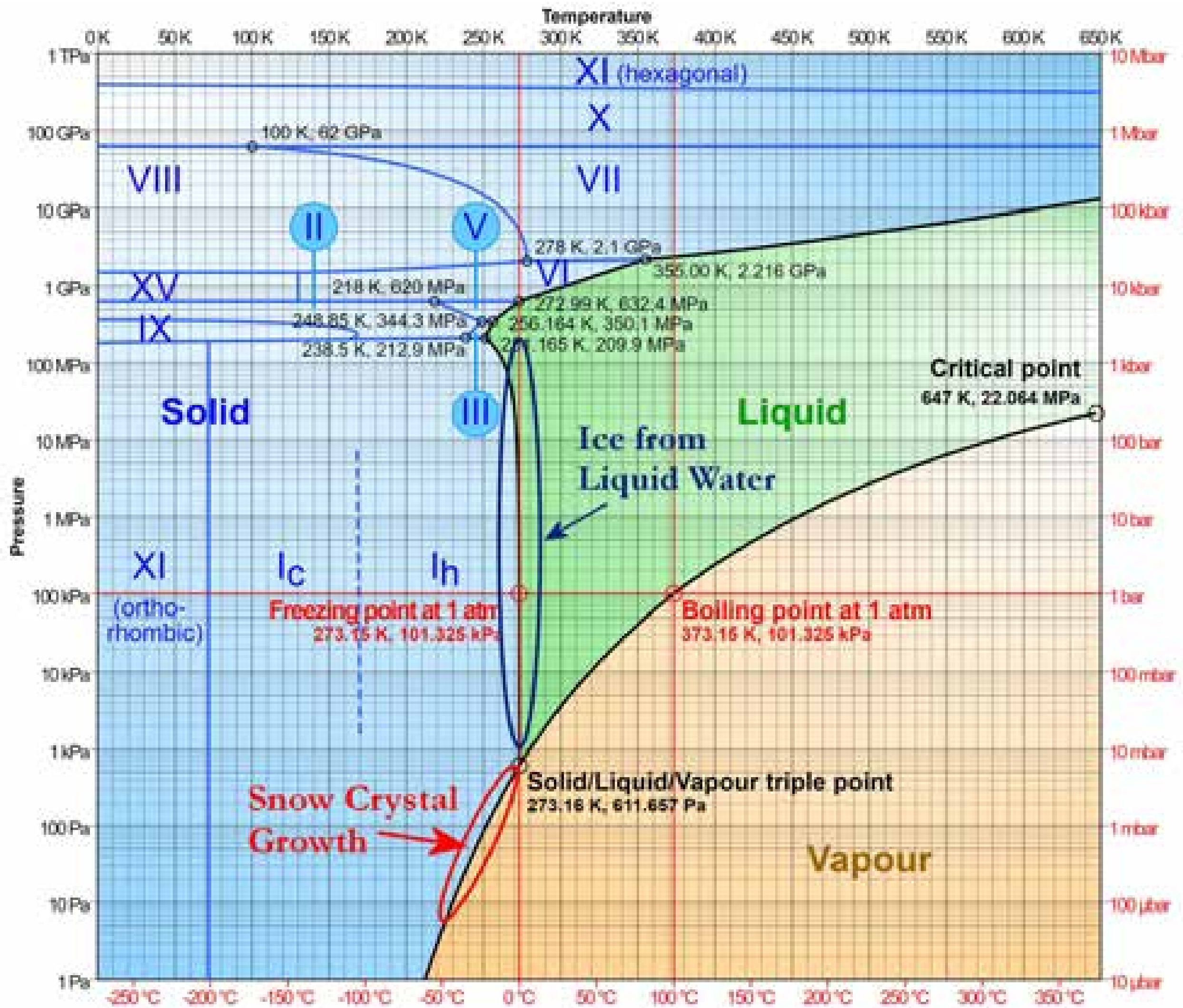

**Figure 12.1: The phase diagram of water as a function of temperature and pressure. Ice growth occurs near the liquid/solid phase boundary (blue oval) while snow crystal growth occurs near the vapor/solid boundary (red oval). The high-pressure portion of the liquid/solid boundary is difficult to access experimentally, so nearly all observations of ice growth from liquid water have taken place near one atmosphere. Image adapted from https://en.wikipedia.org/wiki/Ice.**

crystal lattice, and so will be pushed away by the advancing ice/water interface. In some especially interesting cases, however, solutes latch onto the ice surface and modify the attachment kinetics, dramatically changing the overall characteristics of the freezing process. Moving away from pure water quickly brings us into areas of cryobiology, lyophilization (freeze-drying), and food production that are both fascinating and rich with applications. I will only touch on these subjects here, however, as they are well covered in the scientific literature and clearly beyond the scope of this book.

## 12.1 Basic Phenomenology

The water phase diagram in Figure 12.1 shows that the ice/water phase transition occurs over a broad range of temperatures and pressures, as does the ice/vapor transition. However, while the ice/vapor coexistence line is easily accessed simply by observing snow-crystal growth as a function of temperature, probing the ice/water line in a significant way involves working in a high-pressure environment, which is both difficult and expensive in laboratory experiments. The ice/water freezing temperature changes only slowly with increasing pressure, and the overall crystal-growth behavior shows little pressure-dependence below 10 MPa. Things get more interesting above this pressure, but the high-pressure regime has been only slightly explored to date, as I describe briefly below.

For this reason, much of our discussion will be restricted to ordinary pressures, where there are a good number of observations and quantitative measurements to guide the discussion. It is an unfortunate restriction, however; while snow-crystal growth presents us with a fascinating 2D morphology diagram showing growth as a function of both temperature and supersaturation (see Figure 8.16), ice growth from liquid water gives us essentially a 1D morphology diagram describing growth only as a function of supercooling near 0 C. The upper reaches of the ice/water line in Figure 12.1 remain somewhat out of reach at present.

A key variable describing the solidification of ice from liquid water is the degree of *supercooling* at the ice/water surface, $\Delta T_{surf} = T_m - T_{surf}$, where $T_m \approx 0$ C is the temperature of the ice/water interface in equilibrium and $T_{surf}$ is the temperature of the interface during growth. Other important variables may include the surface orientation relative to the crystal axes, solute concentrations, overall crystal morphology, the presence of container walls or other foreign materials, or simply the initial conditions and/or boundary conditions describing a particular system. Except where noted, in what follows I focus mainly on free ice growth in a pure water bath, unconstrained by the presence of solutes or other materials.

Another commonly used measure of supercooling is $\Delta T_{bath} = T_m - T_\infty$, where $T_\infty$ is the temperature of the supercooled water far from the growing crystal. Latent heat created at the ice/water interface is usually removed via thermal diffusion, so generally $T_{surf} > T_\infty$. Remarkably, $\Delta T_{surf}$ and $\Delta T_{bath}$ are sometimes used interchangeably in the older scientific literature, which can lead to some confusion. Because $\Delta T_{surf}$ is often difficult to determine near a growing interface, $\Delta T_{bath}$ is commonly used when describing experimental results, especially when describing growth morphologies.

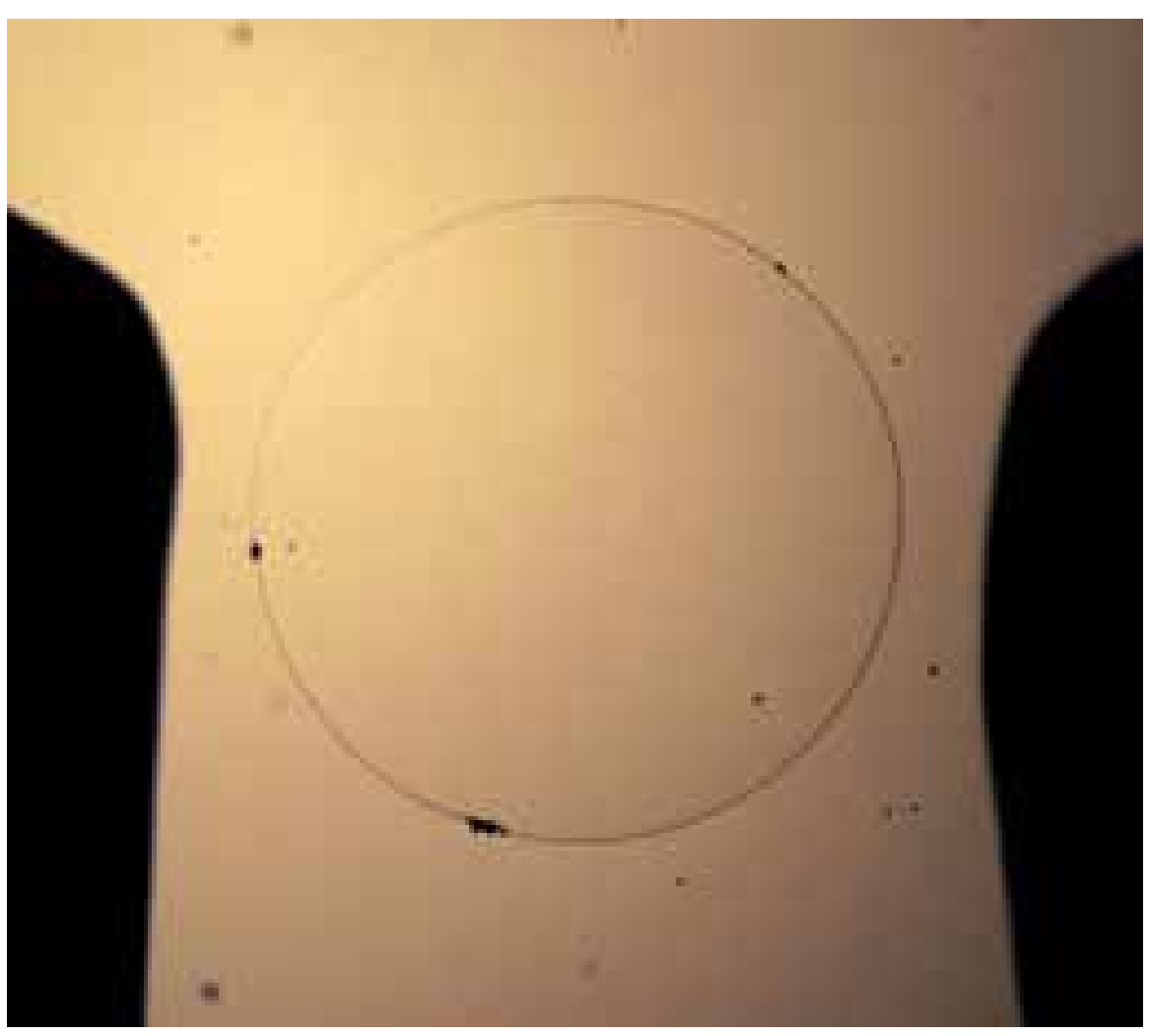

**Figure 12.2: A 2-mm-diameter disk of ice grows outward on the surface of a thin film of slightly supercooled water covering a glass plate. (The dark regions are copper support arms glued to the glass.) The c-axis of the oriented ice crystal is aligned perpendicular to the glass surface. It appears that the close proximity of the cold support plate stabilizes the circular shape against dendritic branching.**

For a small supercooling with $\Delta T_{bath} < 0.5$ C, a microscopic seed crystal tends to grow into the form of simple, circular disks. The slow-growing basal surfaces form the two faceted faces of the disk, indicating that the basal growth is limited primarily by attachment kinetics. Meanwhile the outward growth of the edge of the disk is limited primarily by the diffusion of latent heat in the system. Circular disks are especially likely when the plate forms close to a substrate surface, as illustrated in Figure 12.2.

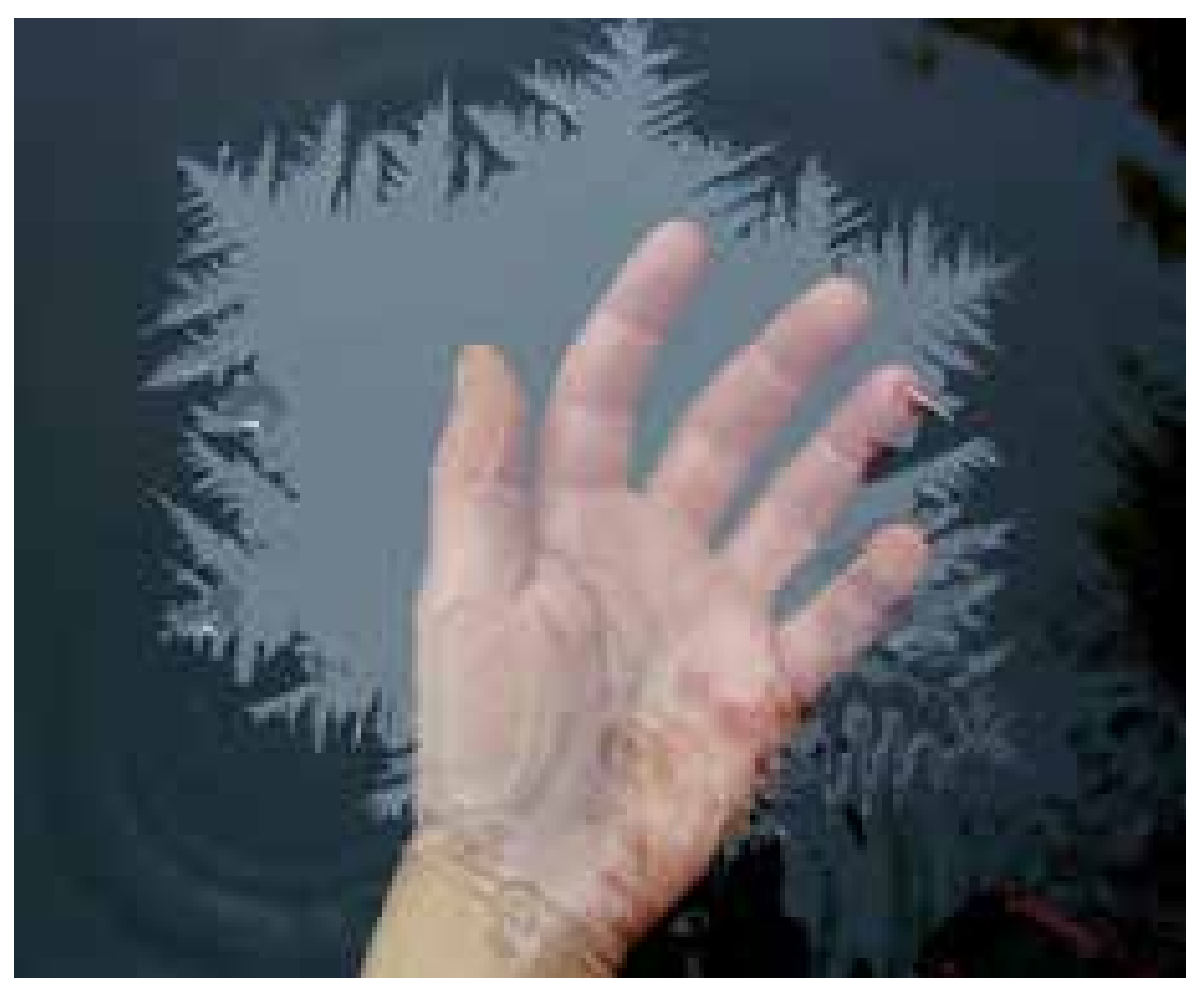

**Figure 12.3: This large pond crystal grew on the surface of an open body of water when the overnight temperature dipped slightly below 0 C. Attachment kinetics limited the basal growth, resulting in a thin plate of ice, while thermal diffusion brought about dendritic branching. The underlying six-fold symmetry of the ice crystal lattice guided the orientations of the branches and sidebranches, as it does with snow-crystal growth [Photo by Bathsheba Grossman.]**

As $\Delta T_{bath}$ is increased, the disk growth becomes unstable via the Mullins-Sekerka instability (see Chapter 3), resulting in dendritic branching, still mainly restricted to two dimensions by basal faceting. In an open body of supercooled water, such as a quiet pond or lake, buoyancy forces often push the ice disk to the water's surface, aligning the c-axis of the crystal along the vertical direction. The disk then grows outward along the water's surface to form large, often single-crystal, dendritic structures, such as the pond crystal shown in Figure 12.3. The process can easily be reproduced in the laboratory, yielding ice plates with thicknesses of 1 cm or more, with the c-axis conveniently oriented perpendicular to the faces of the plate. As described in Chapter 6, this provides an especially simple method for preparing large, single-crystal ice specimens for use in laboratory experiments [1996Kni].

Figure 12.4 shows several observations of different dendrite tip morphologies that formed when ice crystals grew inside a uniform bath of supercooled water [2004Shi]. Again the basal growth is inhibited to a large degree by slow attachment kinetics, while prism growth is limited mainly by the diffusion of latent heat released during solidification. The combination of attachment kinetics and diffusion-limited growth makes the ice/water growth problem somewhat more difficult to solve when compared to unfaceted solid/liquid systems, such as succinonitrile (see Chapter 3). To date, researchers have not been able to incorporate the necessary combination of heat diffusion, attachment kinetics, and surface energy terms into computational models of the solidification of liquid systems. Thus, much like snow-crystal growth, the structure formation that arises when ice grows in supercooled liquid water remains something of an unsolved problem.

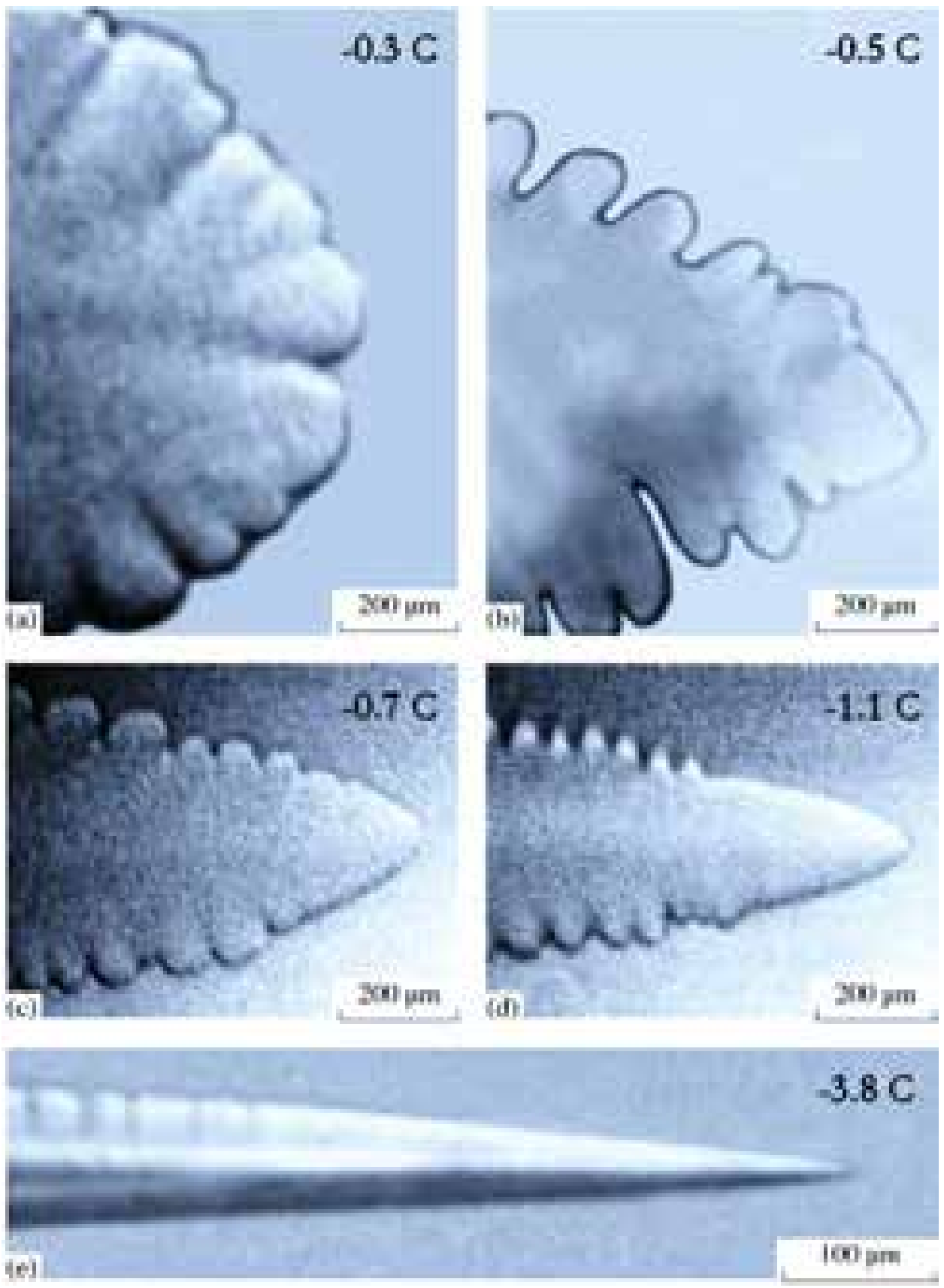

**Figure 12.4: Several photographs showing ice free-dendrite growth in a water bath at different bath temperatures [2004Shi]. Near the freezing point ($\Delta T_{bath} = -0.3$ C), the growth is relatively slow, the tip radius is large, and tip splitting can occur. As the bath temperature decreases, the growth rate increases, the tip radius decreases, and the overall morphology changes.**

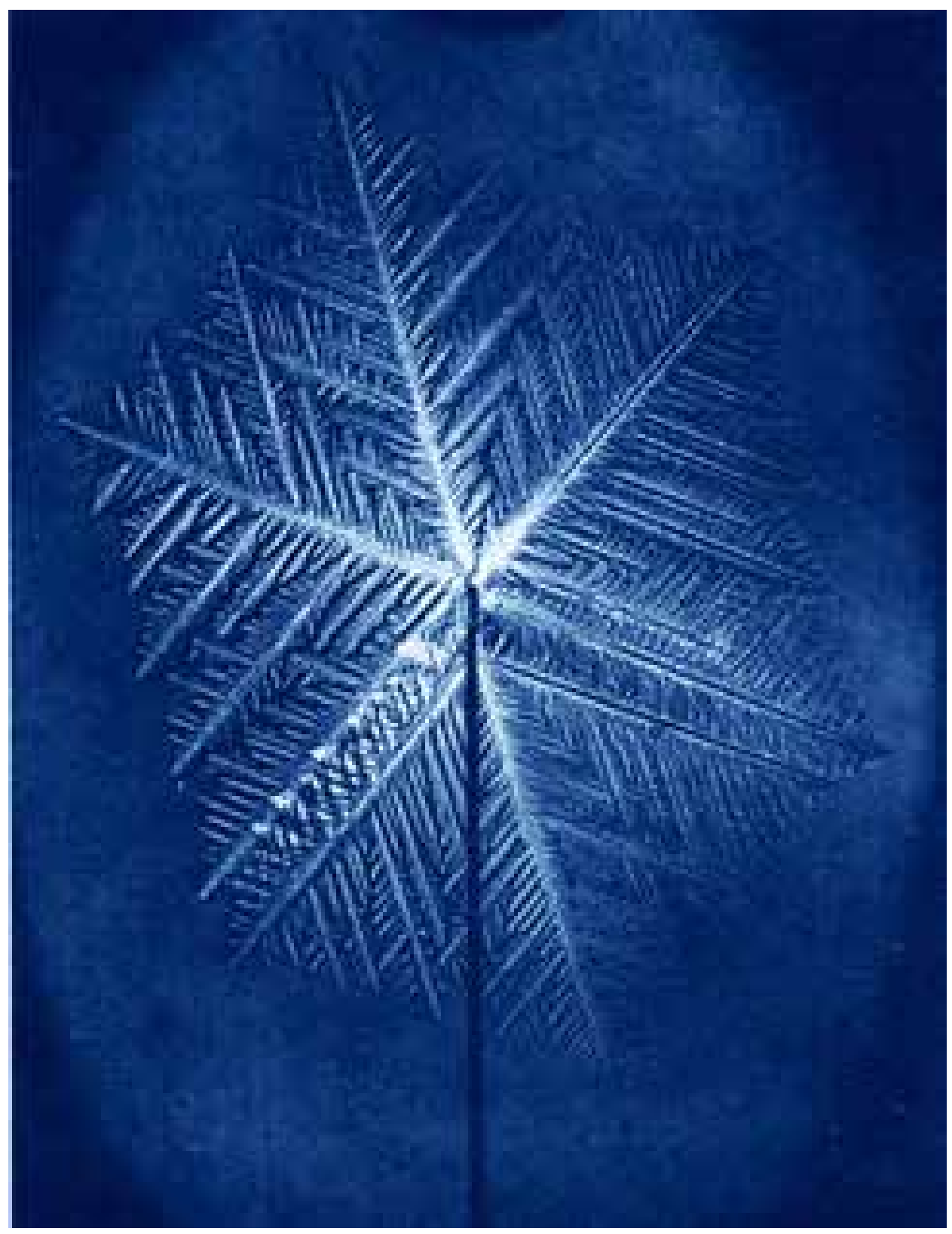

**Figure 12.5: Several ice dendrites growing in a bath of water at -2.3 C [1980Lan]. While the growth behavior at the dendrite tips is mainly determined by local conditions, the overall structure is likely influenced by the initial conditions near the capillary-tube support point.**

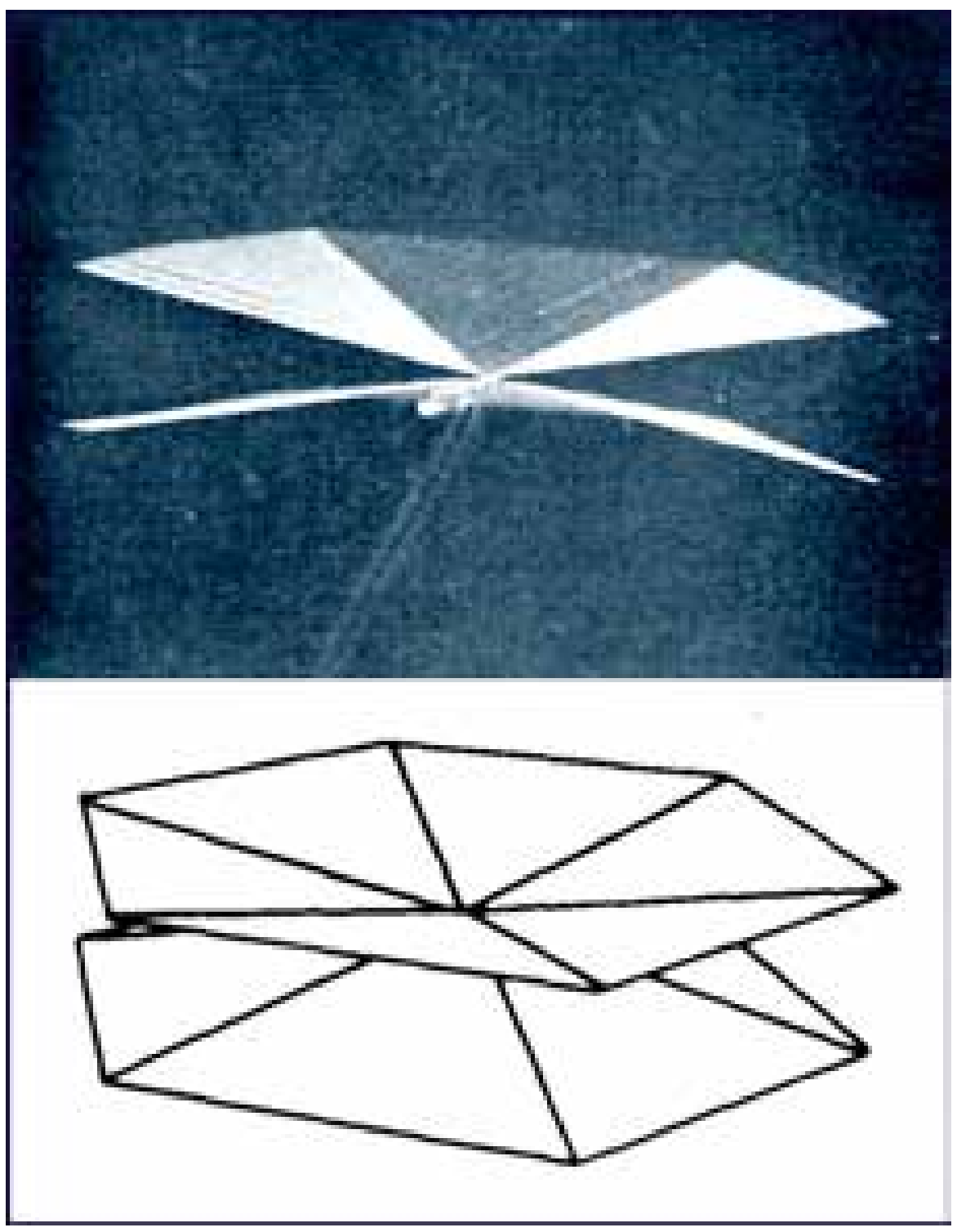

**Figure 12.6: This unusual ice structure, formed at -5.2 C supercooling, consists of plate-like dendrites where the sidebranches have fused together into nearly solid plates [1965Mac]. The dual-plate structure is somewhat analogous to the double-plate snow crystals described in Chapter 10. In both cases, the overall double-plate morphology reflects the full history of the growth and boundary conditions, particularly at early times.**

Note that ice-growth morphologies can depend on external boundary conditions to some extent. In Figure 12.2, the circular disk shape appears to be stabilized by the nearby cold plate, while the plate morphology in Figure 12.3 is affected by the fact that it floats on the surface of a large body of water. These both represent different boundary conditions compared to the less-constrained growth presented in Figure 12.4. Understanding the different morphologies and morphological transitions, and how they depend on supercooling, would require 3D analyses that included all the proper boundary conditions.

As the supercooling increases, the complexity of the resulting solidification structures increases as well. This is also true for snow-crystal growth (see Figure 8.16), and it is a general property of diffusion-limited growth. For example, Figure 12.5 shows a highly dendritic structure growing in a water bath cooled to -2.3 C. There is a clear resemblance with stellar-dendrite snow-crystal structures near -15 C, as basal faceting yields plate-like structures in both cases, while the underlying symmetry of the ice crystal lattice plays a role in defining the overall symmetry of the dendritic branches. From another experiment, Figure 12.6 shows a peculiar "bi-pyramidal" structure that consists of fused sheets of dendritic plate-like components. It appears that this structure is less fundamental than the planar dendrites shown in Figure 12.5, as the multiple sheets seem to arise from a specific history of growth and/or boundary conditions, much like double-plate snow crystals. An important take-away message from these experiments is that the growth and boundary conditions must be rather carefully specified if one wants to understand the detailed morphological structures that arise during crystal growth.

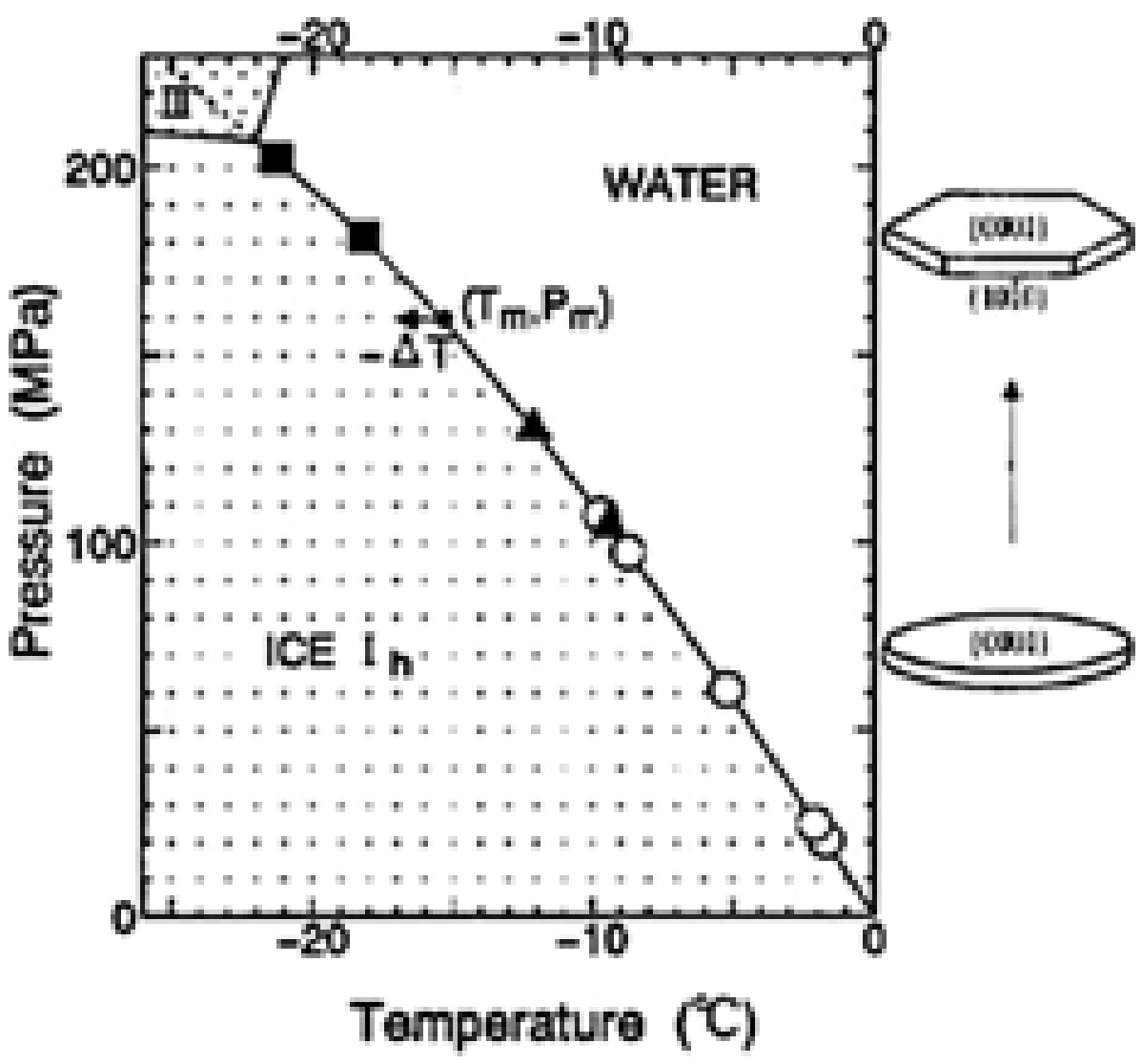

**Figure 12.7: The growth form of ice in water at low supercooling changes from thin disks (open circles) to faceted hexagonal plates (solid squares) with position on the ice/water coexistence line. The transition likely arises from a change in the prism attachment kinetics with temperature/pressure, specifically the terrace step energy on prism facets.**

As mentioned above, ice/water growth morphologies change substantially when experiments are done at high pressures, but there is are few laboratory observations describing this difficult-to-reach region of the water phase diagram. In a series of groundbreaking experiments, Minoru Maruyama and collaborators [1997Mar, 2005Mar] found that circular-disk morphologies evolve into faceted prism morphologies at high pressures and low temperatures, as illustrated in Figure 12.7. As with snow-crystal attachment kinetics (see Chapter 4), these experiments suggest the possibility of comparing measurements of terrace step energies on faceted ice surfaces with calculated values from molecular dynamics simulations over the entire ice/water coexistence line.

## 12.2 Attachment Kinetics

The attachment kinetics at the ice/water interface has received relatively little attention in the scientific literature, even though the appearance of basal faceting over a broad range of environmental conditions indicates that basal attachment kinetics clearly has a strong influence on growth morphologies. It is customary to write the growth velocity $v_n$ (normal to the surface) in the Wilson-Frenkel form,

$$v_n(T_{surf}) = K_T \Delta T_{surf} \tag{12.1}$$

where $K_T$ is the *kinetic coefficient*. This form automatically gives the equilibrium condition $v_n = 0$ when $\Delta T_{surf} = 0$, although in general $K_T$ will also be temperature dependent.

Measurements of basal growth rates [1958Hil, 1966Mic] indicate a functional form that agrees with a terrace-nucleation model, and one data set is reproduced in Figure 12.8. After correcting for small heating effects, the basal kinetic coefficient is given by $K_T = K_0 \exp\left(-\Delta T_0/\Delta T_{surf}\right)$, where $\Delta T_0 = 0.23$ C and $K_0 = 7.3 \times 10^{-4}$ m/sec [2014Lib]. As described in Chapter 4, these data give a basal step energy of $\beta_{sl,basal} \approx 5.6 \pm 0.7 \times 10^{-13}$ J/m, assuming classical nucleation theory [1958Hil, 2014Lib, 2017Lib]. (See also Chapter 4.) Once again, this measurement opens up the possibility of comparing measurements with step energies calculated directly using molecular dynamics simulations, which is a subject of current research [2012Fro, 2019Ben].

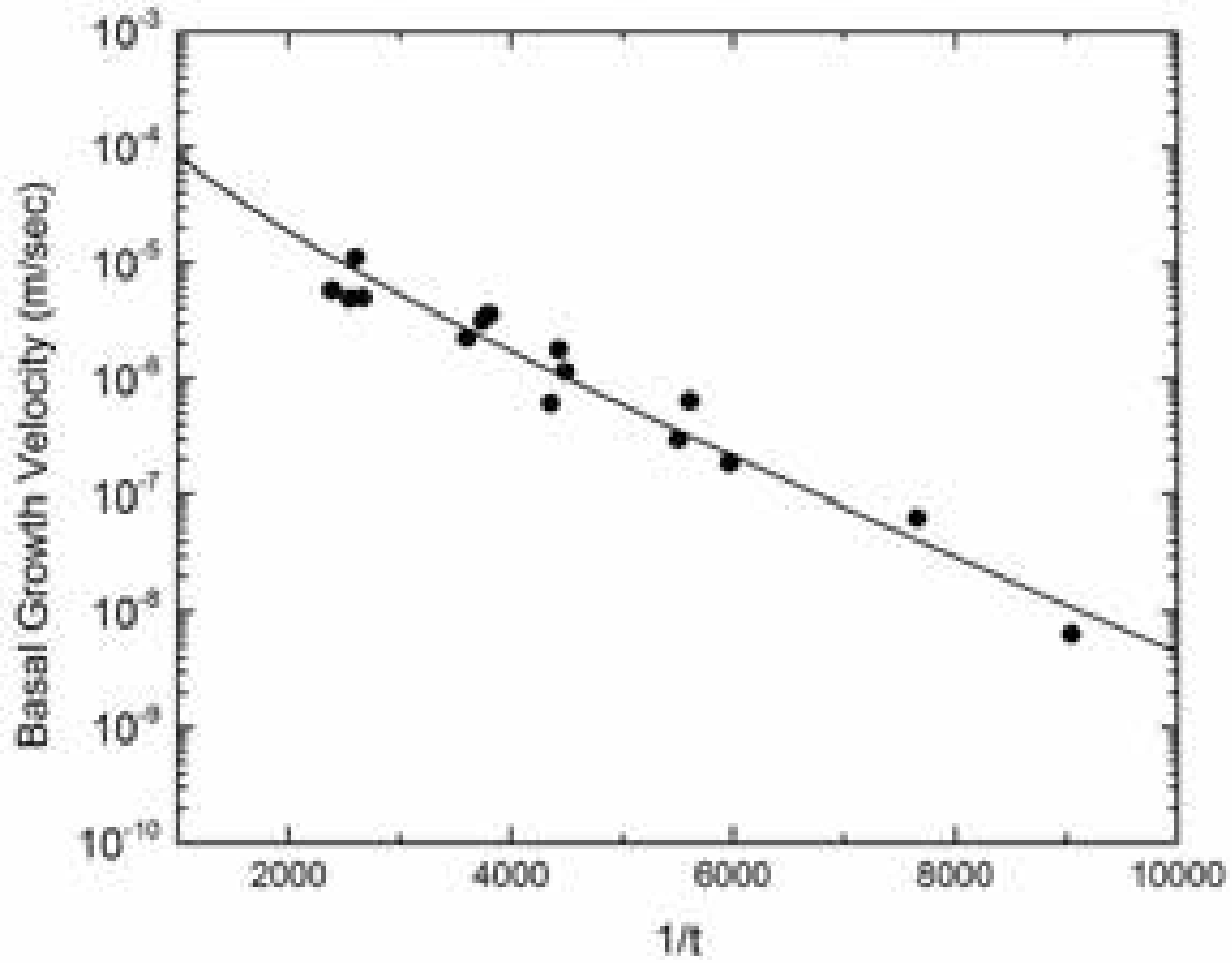

**Figure 12.8: Measurements of the basil growth velocity for ice in liquid water as a function of the dimensionless undercooling $t = (T_m - T_{surf})/T_{surf}$. The line through the data shows a terrace-nucleation model with an ice/water basal step energy of $\beta_{sl,basal} \approx 5.6 \times 10^{-13}$ J/m [1958Hil, 2014Lib, 2017Lib].**

The lack of any faceting on prism surfaces (at normal pressures) provides an indication that $\beta_{sl,prism} \ll \beta_{sl,basal}$, and $\beta_{sl,prism}$ is so small that it remains unmeasured. In fact, the kinetic coefficient $K_T$ for the prism facet is difficult to measure at all, because ice crystal growth is so strongly limited by thermal diffusion under most experimental conditions.

In the absence of a terrace nucleation barrier ($\Delta T_0 = 0$), one can produce a theoretical estimate of the kinetic coefficient from the Einstein-Stokes relation [1996Sai, 2017Lib] (see also Appendix B), which gives

$$K_T \approx \frac{L_{sl}\rho_{ice}a}{6\pi\eta_{eff}T_m} \qquad (12.2)$$

where $L_{sl}$ is the latent heat of melting per unit mass, $\rho_{ice}$ is the ice density, $a$ is the size of a water molecule, and $\eta_{eff}$ is the effective dynamical viscosity for liquid water near the ice surface. The kinetics of liquid water near an ice surface is nontrivial, and it is possible that $\eta_{eff}$ may differ substantially from the normal bulk viscosity. Nevertheless, using the bulk viscosity value gives the short-dotted line in Figure 12.9, which can be taken as a crude theoretical estimate for the kinetic coefficient of non-faceted ice surfaces.

The basal kinetic coefficient in Figure 12.9 is probably reasonably accurate for $\Delta T_{surf} < 0.1$ C, as accurate measurements are most feasible in this regime. The fast kinetics curve lies far above the basal kinetic curve at these low supercoolings, which is consistent with the observation of thin-plate crystals, both simple circular disks as well as plate-like dendritic structures. The curves in Figure 12.9 are progressively less accurate at higher supercoolings, where they should be considered only rough estimates. In this regime, the growth is strongly limited by heat diffusion, making it quite difficult to extract useful information regarding the attachment kinetics from experiments.

The ice/water attachment kinetics presents a relatively simple picture near 0 C, including 2D-nucleation-limited kinetics on basal facets and quite rapid kinetics for other surface orientations. It appears that a substantial nucleation barrier appears on prism facets only at higher pressures (and lower temperatures), but the data are insufficient to give a clear picture of how either $\beta_{sl,basal}$ or $\beta_{sl,prism}$ change with pressure+temperature along the ice/water coexistence line. This is in contrast to the ice/vapor case, where we have accurate data over a broad swath of the coexistence line, indicating a rich variety of physical processes underlying the attachment kinetics on both principal facets, as described in detail in Chapters 3 and 7.

## The Gibbs-Thomson Effect

If the ice/water surface is not flat, then Equation 12.1 is replaced by

$$v_n = K_T T_\Delta \left(u_{surf} - d_{sv}\kappa\right) \qquad (12.3)$$

where here $\Delta T_{surf}$ has been replaced with a dimensionless supercooling at the surface equal to

$$u_{surf} = \frac{T_m - T_{surf}}{T_\Delta} \qquad (12.4)$$

with

$$T_\Delta = \frac{L_{sl}}{c_{p,water}} \approx 78\,K \qquad (12.5)$$

and $c_{p,water}$ is the specific heat of water per unit mass (see Appendix A and Appendix B). The second term in Equation 12.3 incorporates the ice/water surface energy via the Gibbs-Thomson effect [1980Lan, 1989Lan, 1996Sai, 2017Lib], where $\kappa$ is the surface curvature (equal to $2/R$ for a spherical surface) and

$$d_{sv} = \frac{\gamma_{sl} T_m c_{p,water}}{\rho_{water} L_{sl}^2} \approx 0.4\ \text{nm} \qquad (12.6)$$

is called the *capillary length* and $\gamma_{sl}$ is the ice/water surface energy. The analogous Gibbs-Thomson term for ice/vapor growth is given in Chapter 2.

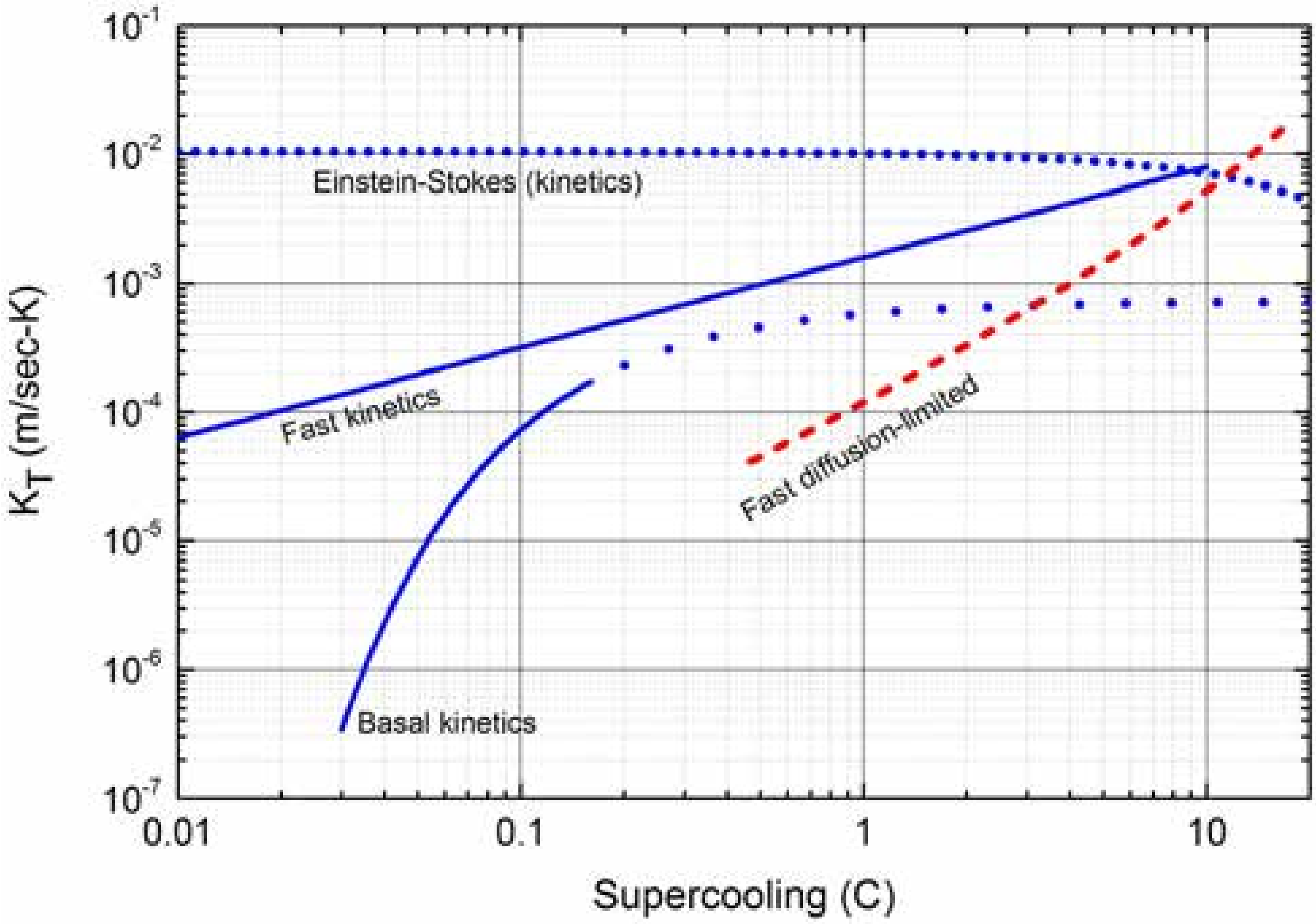

**Figure 12.9: The blue curves in this plot show the kinetic coefficient $K_T(\Delta T_{surf})$ for ice growth from liquid water for the basal facet (labeled basal kinetics) [2014Lib] and for fast-growing unfaceted surfaces (fast kinetics) [2017Lib]. The thick portion of the basal line derives from ice growth data, while the dotted portion is extrapolated to larger $\Delta T_{surf}$. The remaining blue short-dotted curve shows $K_T$ from Einstein-Stokes theory [1996Sai, 2017Lib]. For comparison, the red dashed curve shows $K_T(\Delta T_{bath})$ for the tips of growing dendrites [2005Shi]. Note that the abscissa in this graph does double duty, equal to either $\Delta T_{surf}$ (blue curves) or $\Delta T_{bath}$ (red curve).**

## 12.3 Diffusion-Limited Growth

As an ice crystal grows from supercooled liquid water, the release of latent heat raises the temperature at the ice/water interface, thus lowering $\Delta T_{surf}$ and slowing the growth rate via Equation 12.1. The growth is thus limited by the rate at which heat diffuses away from the growing crystal. The mathematics of this process is quite similar to that presented in Chapter 3, although with several differences arising from the validity of various physical approximations.

Heat diffusion is described by the thermal diffusion equation

$$\frac{\partial T}{\partial t} = D_{water} \nabla^2 T \qquad (12.7)$$

where $T(\vec{x})$ is the temperature field surrounding the crystal and

$$D_{water} = \frac{\kappa_{water}}{\rho_{water} c_{p,water}} \approx 1.4 \times 10^{-7} \text{ m}^2/\text{sec} \qquad (12.8)$$

where $\kappa_{water}$ is the thermal conductivity of water. In this form, I have ignored thermal diffusion through the ice compared to diffusion into the water bath, which is a reasonable assumption for a small ice crystal growing into a large bath. Using the dimensionless supercooling, the diffusion equation is

$$\frac{\partial u}{\partial t} = D_{water} \nabla^2 u \qquad (12.9)$$

### The Laplace Approximation

As was described in Chapter 3 for the ice/vapor case, it is useful to compare the diffusion time $\tau_{diff} \approx R^2/D_{water}$ with the crystal growth time $\tau_{grow} \approx R/v_n$, where $R$ is the approximate size of the growing crystal in question. The time $\tau_{diff}$ is roughly how long it takes for the temperature field to relax to a quasi-steady state, while $\tau_{grow}$ is the time it takes for the crystal size to change appreciably. The ratio of these two times is called the *Peclet number*

$$p \approx \frac{v_n R}{2 D_{water}} \qquad (12.10)$$

For the case of purely diffusion-limited growth of an ice sphere (see below), the diffusion equation gives $v_n R \approx (\Delta T_\infty / T_\Delta) D_{water}$, and in this case the Peclet number becomes

$$p \approx \frac{\Delta T_\infty}{2 T_\Delta} \qquad (12.11)$$

For all experiments done to date, $p$ has been less than unity, and $p \ll 1$ is a fairly accurate assumption overall. In this limit, the thermal diffusion equation simplifies to Laplace's equation

$$\nabla^2 u = 0 \qquad (12.12)$$

substantially simplifying the subsequent mathematics. The Laplace approximation is not as accurate for ice/water as it is for the ice/vapor system, where the Peclet number is exceedingly small. Nevertheless, the approximation is sufficiently accurate for drawing general conclusions regarding the importance of various physical effects that govern ice growth from liquid water.

### Surface Boundary Conditions

For solving the diffusion equation using the Laplace approximation (Equation 12.12), assume the fixed boundary condition $u = u_\infty$ far from the growing crystal. At a growing surface we have the Wilson-Frenkel relation (Equation 12.3) along with the fact that the heat generated at the surface is equal to that carried away from the surface via conduction, which gives

$$L_{sl} \rho_{water} v_n = \kappa_{water} (\nabla_n \mathrm{T})_{surf} \qquad (12.13)$$

where $(\nabla_n \mathrm{T})_{surf}$ is the temperature gradient normal to the surface, evaluated at the surface. Combining these two relations gives the surface boundary condition

$$K_T T_\Delta \left( u_{surf} - d_{sv}\kappa \right) = D_{water} (\nabla_n u)_{surf} \quad (12.14)$$

which relates $u_{surf}$ to $(\nabla_n u)_{surf}$.

## The Spherical Solution

As I described with the ice/vapor system in Chapter 3, examining the growth of a simple ice sphere provides many useful insights regarding more complex growth geometries, as the spherical case can be solved exactly and analytically (after applying several reasonable approximations). Fast-forwarding through the math (see Appendix B), the diffusion equation alone gives the general solution $u(r) = u_\infty - A_1/r$, where $u(r) = (T_m - T(r))/T_\Delta$ is the dimensionless supercooling field, $u_\infty$ is the supercooling at infinity, and $A_1$ is a constant. Applying the surface boundary condition to determine $A_1$ in the Laplace approximation gives the growth velocity

$$v_n = \frac{D_{water}}{R_{kin} + R} \left( u_\infty - \frac{2 d_{sv}}{R} \right) \quad (12.15)$$

where $R$ is the radius of the ice sphere and

$$R_{kin} = \frac{D_{water}}{K_T T_\Delta} \quad (12.16)$$

When the growth is largely diffusion limited ($R_{kin}, d_{sv} \ll R$), the surface supersaturation is given by

$$u_{surf} \approx \frac{D_{water}}{R K_T T_\Delta} u_\infty \quad (12.17)$$

which goes to zero for large $R$ or $K_T$, as one would expect.

Assuming $K_T \approx 4 \times 10^{-4}$ m/sec-K as a reasonable estimate for fast kinetics on prism surfaces (see Figure 12.9), gives $R_{kin} \approx 5$ µm, which is smaller than the tip radii for all but the fastest-growing tip structures (see Figure 12.4). This suggests that the tip growth will be largely diffusion-limited (and not kinetics limited), which is consistent with observations. However, $K_T$ is much smaller on basal surfaces, so there the growth will often be kinetics limited, which is also consistent with observations of basal facets. As with snow crystals, ice growth from liquid water is largely diffusion-limited, but with attachment kinetics factoring in substantially.

In most experiments examining ice growth from supercooled water, for example in Figure 12.4, the growth is mainly diffusion limited ($R_{kin}, d_{sv} \ll R$), so it makes sense to expand Equation 12.15 in the large-R limit to obtain

$$v_n \approx \frac{D_{water}}{R} \left( u_\infty - \frac{R_{kin} u_\infty}{R} - \frac{2 d_{sv}}{R} \right) \quad (12.18)$$

Because $d_{sv}$ is so small, the third term in the parentheses is much smaller than the second term in essentially all experimental circumstances. This reinforces the notion that thermal diffusion and attachment kinetics are both important in determining growth behavior, while surface-energy effects are small and possibly negligible in most experimental circumstances.

## Dendrite Growth

In considering the growth of dendritic structures like those in Figure 12.4, it is customary to assume that the dendrite tip can be approximately described by a paraboloid of revolution with a constant $R_{tip}$ and $v_{tip}$, as illustrated in Figure 12.10. Although this approximation is perhaps overly simplistic in the case of ice growth, the model is a useful tool for better understanding the underlying physical processes.

As described in Chapter 3 for the ice/vapor system, solving the diffusion equation for a growing paraboloid yields what is known as the Ivantsov solution [1947Iva], which is well described in the crystal-growth literature, including for arbitrary Peclet numbers [1980Lan, 1989Lan, 1996Sai]. Because the Peclet number is fairly small in the ice/water system, this simplifies the problem at hand, allowing us to write the Ivantsov solution in a form that looks quite similar to the spherical solution in Equation 12.18, giving

$$v_{tip} \approx \frac{D_{water}}{B_0 R_{tip}} \left( u_\infty - \frac{R_{kin} u_\infty}{R_{tip}} - \frac{2d_{sv}}{R_{tip}} \right) \quad (12.19)$$

where $B_0$ is a dimensionless parameter that depends only weakly on $u_\infty$ and other factors. To a reasonable approximation, we can take $B_0$ to be roughly constant for the present discussion.

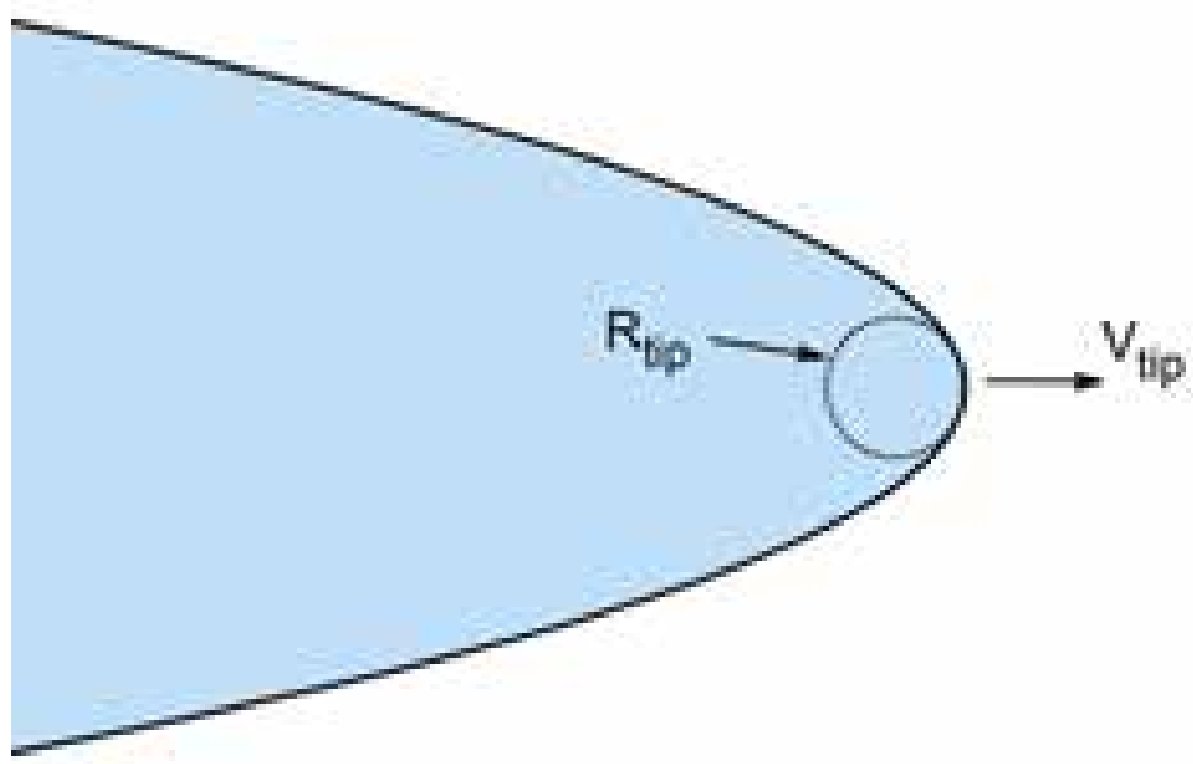

**Figure 12.10: The Ivantsov solution to the diffusion equation describes a crystalline paraboloid of revolution with an unchanging parabolic shape and a constant tip radius $R_{tip}$, the entire structure growing forward with a constant tip velocity $v_{tip}$. If viewed from a frame of reference that moves in the growth direction with velocity $v_{tip}$, the system would appear completely static.**

The Ivantsov solution does not specify either $R_{tip}$ or $v_{tip}$, but rather provides a relationship between these quantities. For mainly diffusion-limited growth like ice/water dendrites, $v_{tip} \sim 1/R_{tip}$, which is consistent with observations to the extent that the dendrite tip approximates the Ivantsov form. As $u_\infty$ increases, $R_{tip}$ generally decreases while $v_{tip}$ then increases as $v_{tip} \sim 1/R_{tip}$. However, the diffusion equation alone does not provide a physical solution for $R_{tip}(u_\infty)$.

The parabolic growth problem turned out to be remarkably subtle, both mathematically and physically, playing out in the 1980s with the development of what has become known as *solvability theory*. A full discussion of the theory is beyond the scope of this book (see [1996Sai, 1996Kas, 1999Pim]), but the essential physics can be seen in Equation 12.19.

Given that any value $R_{tip}$ will yield a suitable parabolic solution to the diffusion equation (the Ivantsov solution), the Mullins-Sekerka instability (see Chapter 3) tends to drive the parabolic form to smaller values of $R_{tip}$, as these grow faster. The drive to ever smaller $R_{tip}$ is eventually balanced, however, by the second and third terms in the parentheses in Equation 12.19. For both these factors, decreasing $R_{tip}$ reduces the effective $u_\infty$, and this negative feedback eventually stabilizes $R_{tip}$ and $v_{tip}$. The devil is very much in the details of this subtle problem, and that is the subject of full-blown solvability theory.

One important result from solvability theory is that some amount of anisotropy is needed to stabilize the parabolic needle solution shown in Figure 12.10. For purely isotropic surface tension and/or attachment kinetics, the parabolic solution is unstable to tip splitting. The example in Figure 12.4 with a supercooling of 0.3 C shows one instance of tip splitting, and additional examples can be found in Chapter 3.

Another result from solvability theory is that either an anisotropic surface energy or an anisotropic attachment kinetics can provide the necessary stabilization needed to produce stable, needle-like dendritic growth. Given that such structures are clearly seen in Figures 12.4 and 12.5, we know that one of these effects is in play. But it is not immediately obvious which is more important for ice/water dendrites.

It has become something of a standard practice in the scientific literature pertaining to dendritic solidification to ignore attachment kinetics and focus on surface energy, effectively setting $R_{kin}=0$ in Equation 12.19 [1978Lan, 2005Shi]. Making this assumption, solvability theory then yields the relation

$$v_{tip}R_{tip}^2 \approx \frac{2D_{water}d_{sv}}{s_1} \tag{12.20}$$

which can be used along with the Ivantsov relation to determine $R_{tip}(u_\infty)$ and $v_{tip}(u_\infty)$, giving

$$\begin{aligned} R_{tip}(u_\infty) &\approx \frac{B_0}{s_1 u_\infty} d_{sv} \\ v_{tip}(u_\infty) &\approx \frac{2s_1}{B_0^2}\frac{D_{water}}{d_{sv}} u_\infty^2 \end{aligned} \tag{12.21}$$

which can be compared with the experimental data shown in Figure 12.11. The unknown solvability parameter $s_1$ was adjusted to fit the data, and overall the agreement is quite good over a broad range of supercoolings.

While the fit to the data looks good, the underlying assumption of ignoring attachment kinetics seems terrible. The fact that $d_{sv} \ll R_{kin}$ suggests that attachment kinetics is far more important physically than surface energy, and this suggestion is reinforced by the appearance of strongly faceted basal surfaces at lower supercoolings. Moreover, prism faceted growth appears at lower temperatures (higher pressures), as shown in Figure 12.7, so again it seems like a poor assumption to ignore the anisotropy in even the prism attachment kinetics. Faceted prism growth is not observed at normal pressures, but that does not mean that the prism attachment kinetics is entirely negligible, especially given the small size of $d_{sv}$.

Given that $d_{sv} \ll R_{kin}$, as described above, I am inclined to ignore the $d_{sv}$ term entirely in Equation 12.19, instead keeping only the $R_{kin}$ term for dendrite stability. Solvability theory then yields the relation

$$v_{tip}R_{tip}^2 \approx \frac{D_{water}R_{kin}u_\infty}{s_2} \tag{12.22}$$

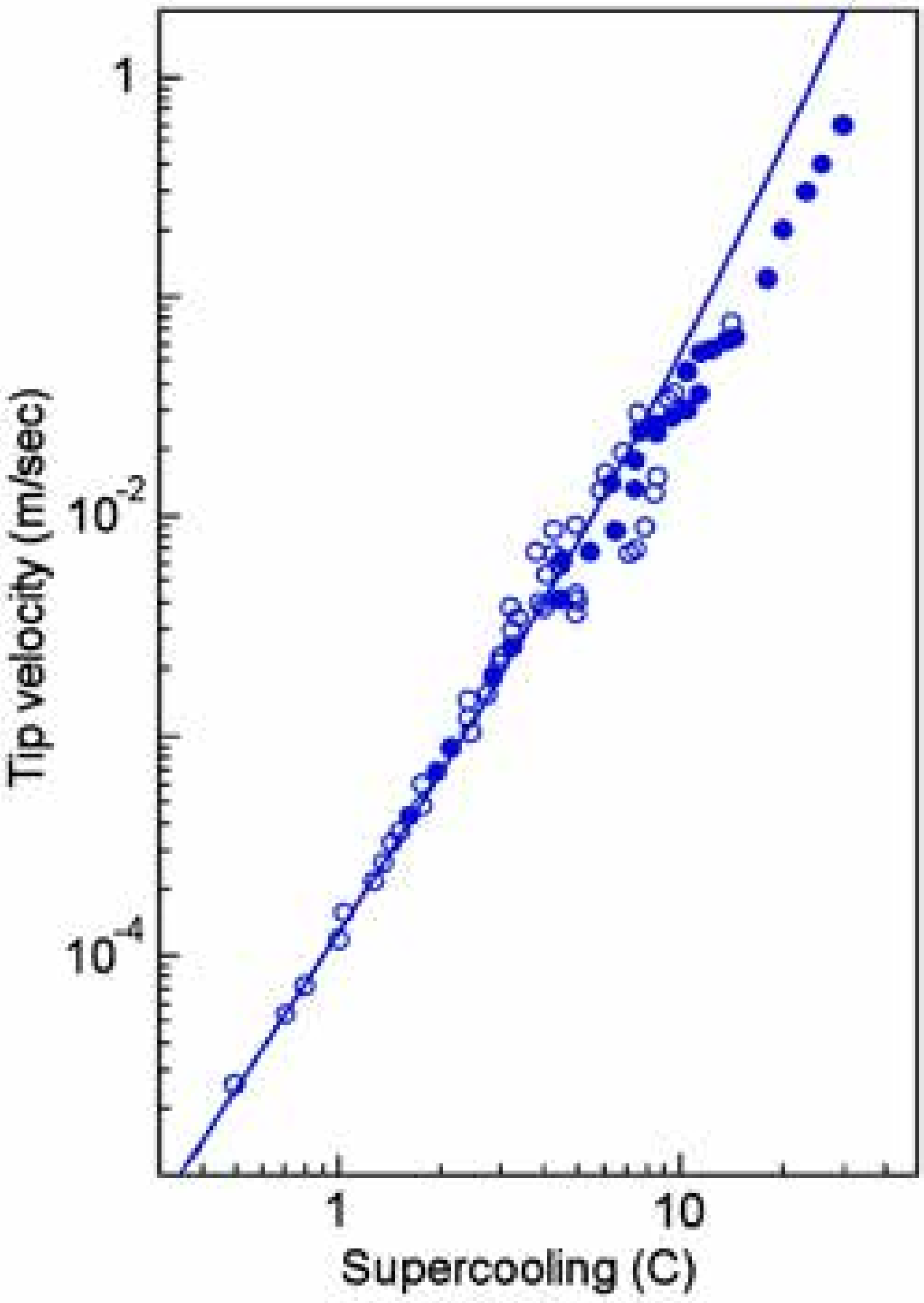

**Figure 12.11: Measurements of the free dendrite tip velocity $v_{tip}$ for ice growing from liquid water as a function of the bath supercooling. The line shows a calculation using a surface-energy-dominated version (neglecting attachment kinetics) of solvability theory. Figure adapted from [2005Shi].**

which, along with the Ivantsov relation, gives

$$\begin{aligned} R_{tip}(u_\infty) &\approx \frac{B_0}{2s_2} R_{kin} \\ v_{tip}(u_\infty) &\approx \frac{2s_2}{B_0^2}\frac{D_{water}}{R_{kin}} u_\infty \end{aligned} \tag{12.23}$$

Unfortunately, this attachment-kinetics-dominated version of solvability theory gives $R_{tip}(u_\infty) \sim$ constant with $v_{tip}(u_\infty) \sim u_\infty$, and a comparison with Figure 12.11 reveals that these functional forms provide a poor fit to the experimental data.

Thus we see a significant dilemma. On one hand, a surface-tension-dominated theory provides a nice fit to the data, but physical considerations suggest that surface-tension effects should be dwarfed by much larger attachment kinetics effects. On the other hand, an attachment-kinetics-dominated theory does not fit the data very well.

One way to resolve this dilemma is to note that the solvability parameters are known to depend strongly on the underlying anisotropies in their respective physical effects. If the anisotropy in the attachment kinetics is temperature dependent (in the right way), this could make the attachment-kinetics-dominated model fit the data. The ice/water attachment kinetics is quite poorly understood at present, so it is at least possible that a suitable parameterization of the attachment kinetics as a function of temperature could reproduce the observations.

Personally, my feeling is that the surface-energy effects are extremely small in the ice/water system, while attachment-kinetics effects are clearly large (as evidenced by strong basal faceting and weaker prism faceting away from the triple point). Ignoring attachment kinetics seems like it must be the wrong approach, so I prefer an attachment-kinetics-dominated model, even if the simplest possible form of that model does not immediately fit the observations.

Unfortunately, the solvability parameters are not only unknown at present, but are almost unknowable, and will remain so for the indefinite future. There is no straightforward way to measure anisotropies in either the attachment kinetics or the surface energy, and there is not a great deal of motivation to expend substantial resources into this topic. I predict that attachment kinetics will win out in the long run, but I also predict that it will be many decades before we know the full answer.

One possible path to progress is similar to the snow-crystal case – make precise measurements of growth rates and morphologies over a broad range of conditions, in well-controlled experiments, and compare the results with full 3D numerical models that include both surface-energy and attachment-kinetics effects. At present, the numerical models are having difficulties incorporating attachment kinetics in solidification from liquid systems (see Chapter 5), but this situation will not last indefinitely. So there is some promise for gaining an improved understanding of the basic physics underlying ice/water growth, much like there is for ice/vapor growth.

## 12.4 Chemically-Mediated Growth

Expanding our horizon briefly from the pure-water case, the freezing of ice from water solutions has received considerable attention in the scientific literature. For dilute solutions, the presence of an ordinary solute typically yields a freezing point depression proportional to concentration (Blagden's law), and the growth behavior is also affected by the fact that solvated molecules are not readily incorporated into the ice lattice. Freezing thus results in segregation effects as solute molecules are pushed away from the solution/ice interface. In these systems, both heat and solute diffusion need to be included in a solidification model, which significantly complicates the problem.

At higher solute concentrations, segregation can result in the formation of complex 3D patterns in the processes of freeze casting [2007Dev, 2009Mun] and lyophilization (also known as freeze drying) [2000Wan]. In many of these systems, there is little chemical interaction between solute molecules and the growing ice interface, so the overall pattern-forming behavior is dominated by segregation effects.

In some cases, solute molecules can also affect the ice/water attachment kinetics, although the changes are usually small, and the phenomenon of chemically mediated attachment kinetics in ice/water has not been well studied over a broad range of solutes [1968Rya, 1969Rya, 2002Sei, 2018Shi]. A notable, and quite dramatic, exception is the case of antifreeze proteins (AFPs), which bind strongly to certain ice surfaces and can halt growth altogether at low supercoolings [1991Kni, 2001Kni]. Over a broad spectrum of phenomena, many commercial applications have arisen from these chemically more complex systems.

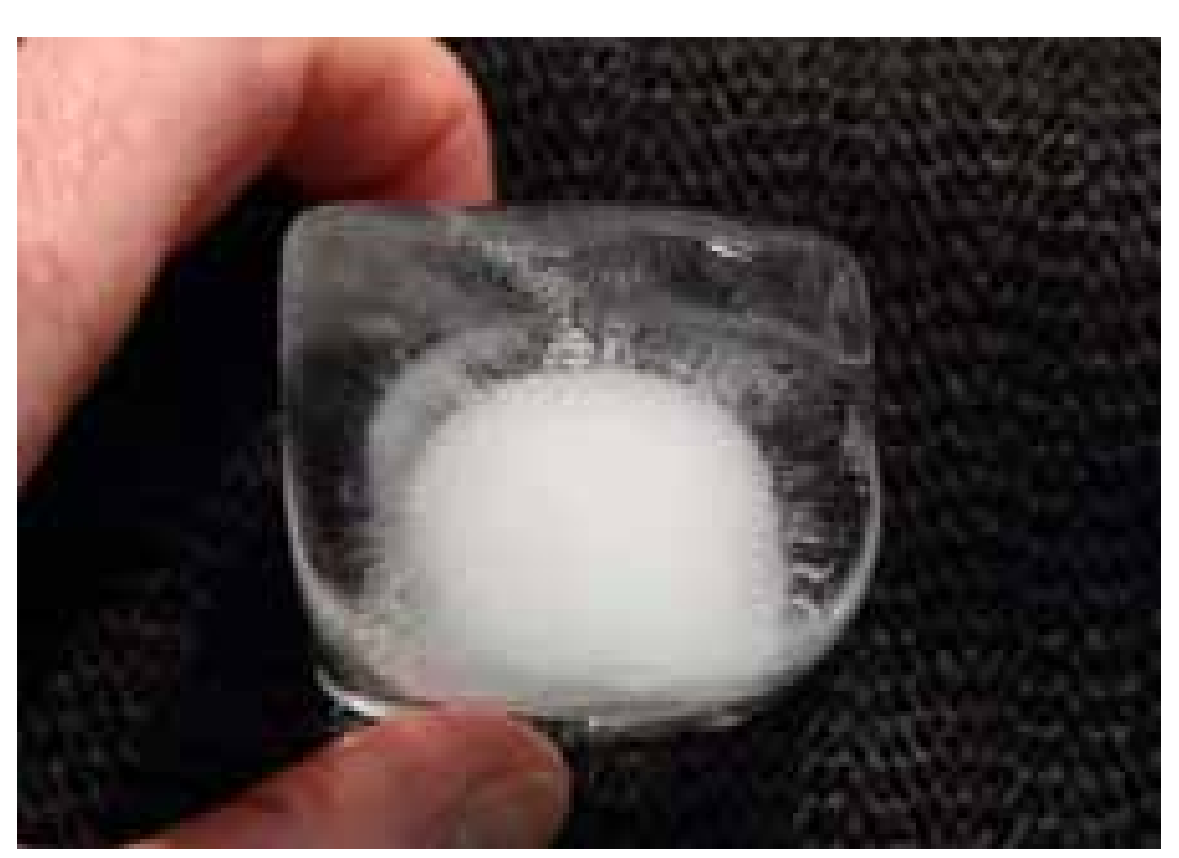

**Figure 12.12: When an ice cube freezes from the surface inward, dissolved gases and other chemical impurities are pushed toward the center by segregation at the moving ice/water interface. Thus, the outer/upper parts of this cube are made from relatively pure, clear ice, as these regions froze first. As the solute concentration increased, eventually the dissolved gases created small bubbles in the ice, giving the center a cloudy appearance. When ice like this melts in a beverage, chemical impurities from the center region release a harsh taste.**

## Segregation Phenomena

Perhaps the most familiar example of chemical segregation during the freezing of liquid water can be found in the production of ice cubes in one's freezer. As shown in Figure 12.12, dissolved gases and chemical impurities are pushed toward the center via chemical segregation, yielding a characteristic cloudy interior region. Ice cubes that are clear throughout can be made by freezing the water slowly from one side only, while providing an escape route for excluded solute molecules. Some expensive ice-making machines can make clear pure ice cubes this way, and simple silicone molds are available for making clear ice in household freezers. Segregation methods are also used for making especially pure ice samples for scientific investigations [1996Kni, 2016Bis].

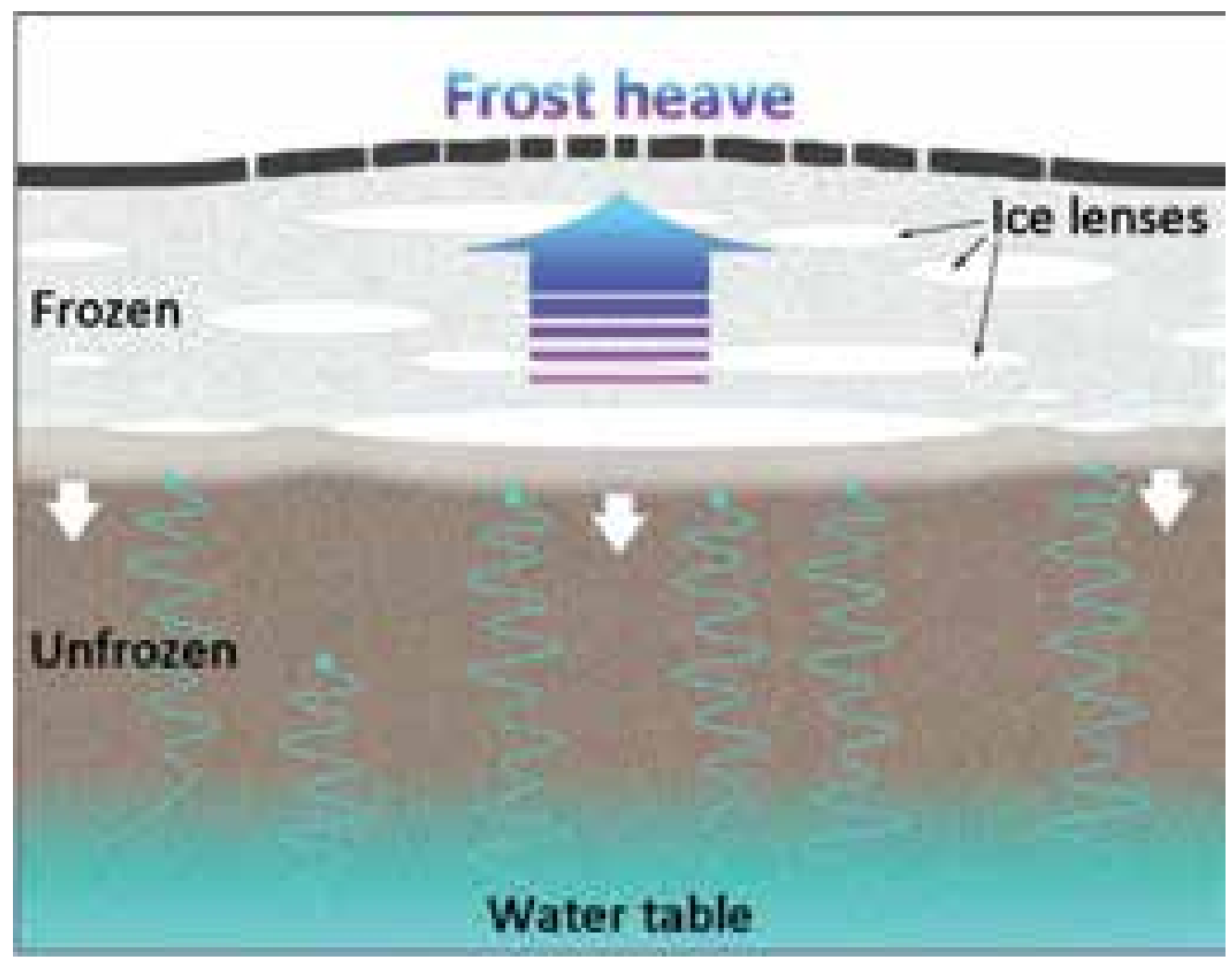

**Figure 12.13: The phenomenon of *frost heave* is often responsible for pushing up road surfaces in cold weather, producing potholes in the process. Contrary to popular belief, the well-known expansion of water upon freezing does not produce frost heave. Segregation provides the dominant mechanism, as ice "lenses" push out porous soil particles, increasing the overall volume of the frozen layer relative to its previously unfrozen state. Image adapted from https://www.dmr.nd.gov/ndgs/documents/newsletter/2011Summer/FrostHeave.pdf**

Frost heave is another phenomenon that results from segregation effects during freezing, as illustrated in Figure 12.13. Beginning with water-logged, porous soil, a growing "lens" of ice will exclude the surrounding soil as it grows via a macroscopic form of segregation. The underlying physics is similar to chemical segregation, as the small soil particles are not readily incorporated into the ice crystal lattice. As capillary action moves water through the porous soil to attach to the ice lenses, the ice grows larger while the soil loses water without losing volume substantially. As a result, the ice lenses along with the surrounding soil+air fills a higher total volume than the initial, unfrozen soil+water mixture. In other words, the soil expands as it freezes, producing frost heave. Note that this phenomenon is mainly a segregation effect, while the well-known expansion of water as it freezes is negligible in comparison. The maximum pressure generated during frost heave, as calculated from basic thermodynamic principles [2001Rem, 2008Lib5], can be high enough to damage heavy roads and bridges.

The production of *ice wine* is another amusing application of ice segregation that has been practiced for many centuries. In a nutshell, freezing a grape at relatively high temperatures can result in a collection of many small, nearly pure ice crystallites inside the grape, surrounded by unfrozen material with a lower water concentration, as a result of segregation. If one crushes the grapes while they are still frozen, discarding the ice, the resulting liquid will have a higher sugar content than would have been obtained crushing unfrozen grapes. Fermenting the concentrated mixture yields exceptionally sweet wines known as ice wines. The process is far less efficient compared to other concentration methods, notably distillation, but ice wines are still being produced for a niche market.

The field of lyophilization (freeze-drying) makes use of extensive segregation during solution freezing. Depending on the particular materials involved, freezing a dense solution often results in a fine matrix of nearly pure ice structures intertwined with the inverse matrix of what had been dissolved in the water. Pulling a vacuum on the frozen block then removes the ice via sublimation and leaves behind the solid solute matrix. In many instances, this process is less harmful to the chemical structure of the solute material than separation via evaporation or other means. Thus lyophilization has found many commercial applications in the food and pharmaceutical industries, and the overall process is much studied in the scientific literature.

As should be apparent from this discussion, the list of phenomena involving ice segregation is long and varied, and proceeding much further in this direction would take us far outside the scope of this book. As a final and quite important example, segregation effects influence many biological systems during freezing. The usual result is to cause damage at the cellular level, as the formation of ice crystallites can easily rupture cell membranes and organelles at many levels. The fact that some cells (for example, frozen embryos of various species) can survive freezing better than others, and whether freezing damage can be ameliorated via chemical or biological methods, brings us well into the field of cryobiology, where segregation effects in freezing solutions are of central importance.

## Ice Nucleation

Pure water must be cooled to quite low temperatures below it spontaneously turns to ice via the process of *homogeneous nucleation*. The nucleation rate turns on very rapidly as the temperature is reduced, and pure water droplets must be supercooled to nearly -40 C before the nucleation timescale drops to near zero. Foreign materials will facilitate *heterogeneous nucleation* that can greatly increase the typical nucleation temperature. For example, the dust particles included in cloud droplets (see Chapter 1) typically induce ice nucleation at temperatures between -6 C and -10 C, depending on the type of dust.

Crystalline materials that exhibit a good lattice match to ice can induce higher nucleation temperatures, and silver iodide is well known in this regard. The crystalline structure of silver iodide is a hexagonal lattice that provides a close physical match with ice, providing a template that encourages ice nucleation at its surface. Silver iodide thus induces a nucleation temperature of about -4 C, which is the highest of any known inorganic crystalline material.

The bacterium *Pseudomonas syringae* is another well-known ice nucleator, as it produces proteins that nucleate freezing at temperatures up to -2 C, which is higher than any other known material. Although the mechanism is not well understood, it appears that the protein's helical structure has been bioengineered (via natural evolution) to attract and hold water molecules in such a way as to promote high-temperature nucleation.

The biological motivation for developing this ice-nucleation ability appears to be related to an evolutionary desire to inflict frost damage in plants. *Pseudomonas syringae* is a plant pathogen, and damaging its hosts makes them more susceptible to bacterial infection. To improve its lot in life, *Pseudomonas syringae* therefore adapted and honed its ice-nucleation abilities.

Scientists were studying the bacterium in part to better understand and hopefully mitigate frost damage in commercial crops. During this process, however, it was soon realized that this microbe's unique talents could be harnessed to improve methods for making artificial snow at ski resorts. As a result of these studies, *Pseudomonas syringae* is sometimes mixed with water being using in snow-making equipment, so the liquid droplets can be coaxed into freezing at higher temperatures, thus allowing snow-making at higher temperatures and at lower costs.

## Growth Inhibition

While ice nucleators promote freezing, several naturally-occurring proteins have also been found that strongly inhibit the growth of ice crystals, stopping ice growth entirely in some cases down to temperatures as low as about -2 C. These are called anti-freeze proteins, (AFPs), and they are found in some fish and insects that have adapted to life in cold temperatures [2017Voe, 2014Dav, 2015Dum]. For example, if a fish ingests small ice crystallites in sub-freezing ocean waters, then there is a danger that the ice will grow within the fish and inflict damage to it internal organs. But a suitable density of AFPs will halt the ice growth, thereby allowing the fish to expel the still-small crystallites before they becomes harmful.

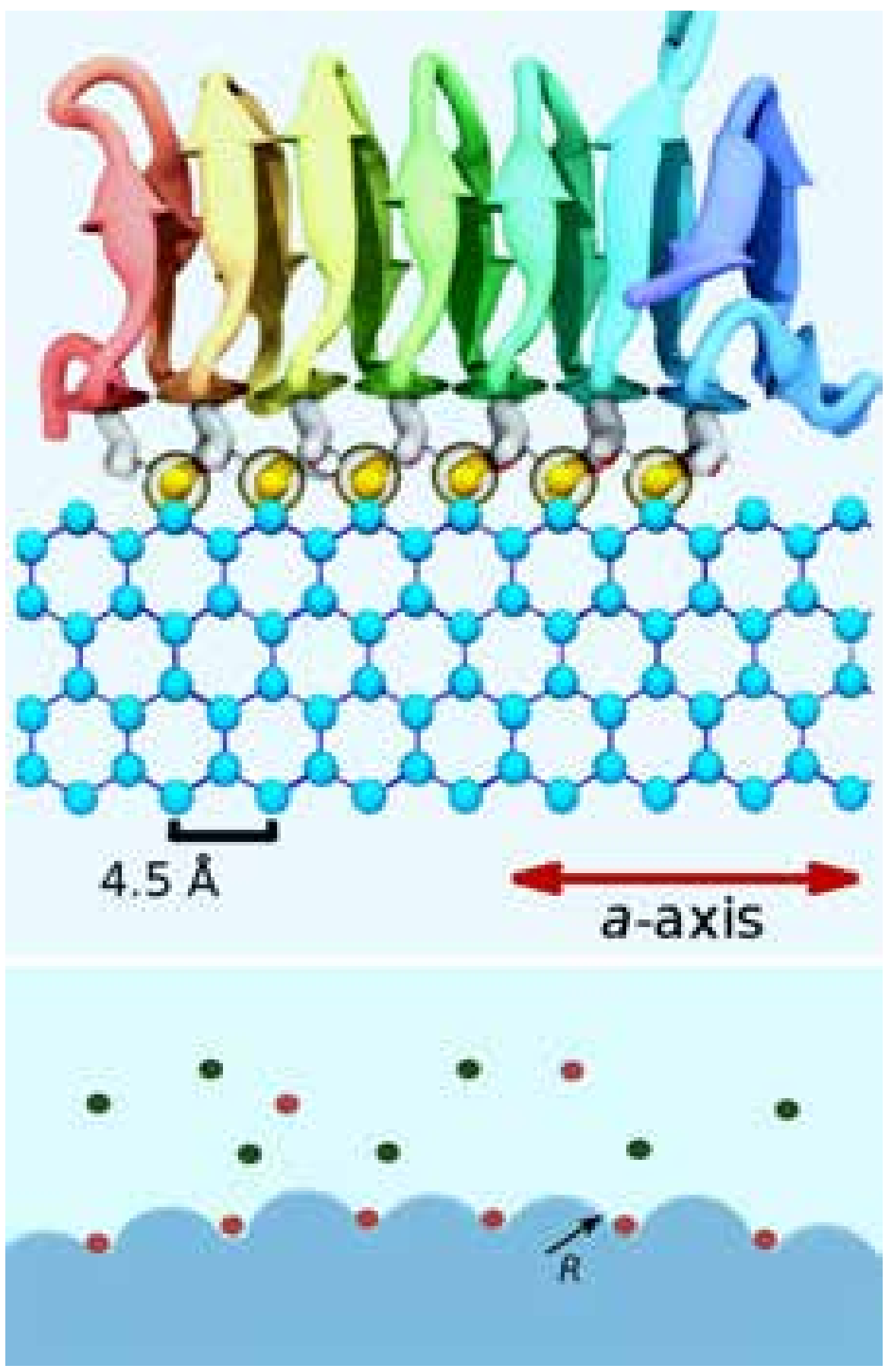

**Figure 12.14: Top: An illustration of how an antifreeze protein might bind to an ice crystal surface. Bottom: Even a sparse density of growth inhibitors on an ice surface (red dots) can halt growth over the whole surface via the Gibbs-Thomson effect, as the induced surface curvature lowers the temperature needed for additional growth. Image adapted from [2015Kui].**

The mechanism responsible for growth inhibition appears to be that the protein structure has again been bioengineered via natural evolution to latch onto the crystalline structure of the ice surface, as illustrated in the top part of Figure 12.14. Even an incomplete layer of anchor points like this on the ice surface can stop further growth via the Gibbs-Thomson effect, as illustrated in the lower part of Figure 12.14. For example, with a supercooling of $\Delta T_{bath}$, growth would be halted if the anchor spacing was roughly

$$R < \frac{T_\Delta}{2\Delta T_{bath}} d_{sv} \qquad (12.24)$$

which is of order 15 nm for $\Delta T_{bath} = 1$ C. Antifreeze proteins seem to prefer pyramidal facets for latching, and Figure 12.15 shows an example of a small ice crystallite that has had its growth halted in a supercooled solution of AFPs in water.

While the phenomenon of ice growth inhibition via AFPs clearly has potential applications in many areas of cryobiology, perhaps the most familiar commercial application at present is in the production of ice cream. When normal ice cream partially melts and then refreezes, segregation and recrystallization effects tend to produce fairly large ice crystals from what had been a homogeneous ice-cream suspension. This undesirable outcome can be reduced by using a small concentration of AFP additives in the recipe, as these suppress large ice crystals in favor of a much greater number of tiny crystallites during refreezing.

Bioengineered yeast has been developed to produce AFPs at an industrial scale, and these are now being added to many ice creams. Being naturally occurring biomolecules, the body digests AFPs as easily and harmlessly as other proteins, and the concentration is too low to significantly affect the taste. This allows manufactures to use a somewhat lower fat content in their recipes while yielding a product with an overall "creamier" texture – in other words, better ice cream at a lower cost.

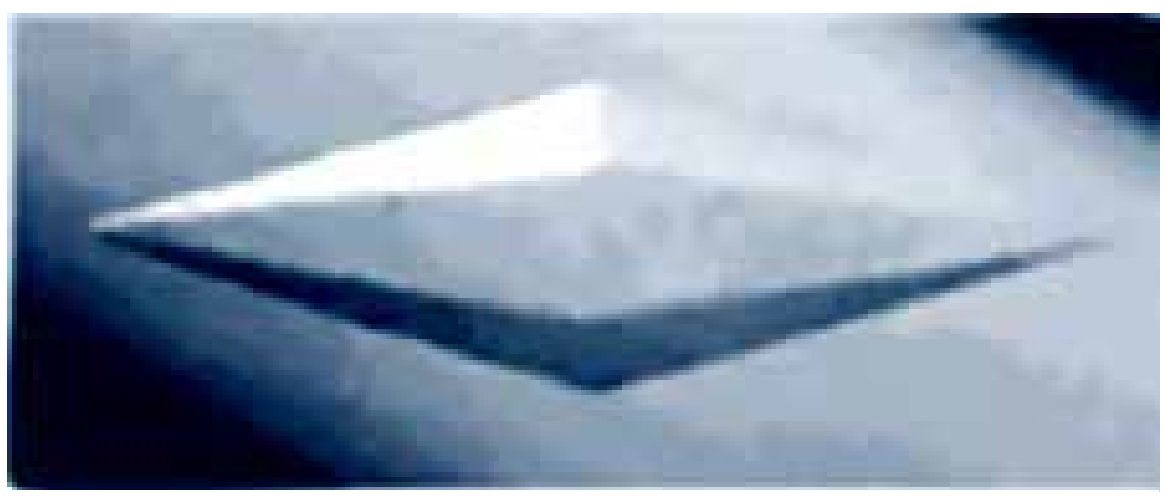

**Figure 12.15: A photograph of a small ice crystallite for which further growth has been halted by the attachment of antifreeze proteins. Image adapted from [2004Str].**

## 12.5 Constrained Growth

I will end this chapter with a brief discussion of ice growth within a water bath constrained by closely spaced vessel walls, as this may have application in future experimental investigations. I have not found much work on this topic in the scientific literature (perhaps simply because it is difficult to find every paper), but my interest was piqued by an investigation in which Hillig (see Figure 12.8) commented that he was able to promote oriented ice growth perpendicular to the basal plane in a small-diameter sapphire capillary tube. In other words, the ice somehow oriented itself so that the c-axis of the growing crystal was aligned along the central axis of the tube. He did not provide an explanation for this self-orientation, but it was quite useful in his measurement of the basal attachment kinetics.

To this experimental observation I add the additional images shown in Figure 12.16, illustrating ice growing between two closely spaced plastic plates during a directional solidification experiment in my lab (unpublished). The ice nucleation was randomly initiated, but the crystal orientation soon aligned itself so the ice c-axis was nearly perpendicular to the image plane in all cases.

I believe this phenomenon results from a form of dislocation-mediated crystal-growth reorientation, but I have not yet been able to develop a detailed physical model of such a process. I conjecture that random dislocations facilitate a continuous reorientation of the ice crystal axes, in such a way that the surface normal vector on the fastest-growing ice surface aligns itself to point along the highest available temperature gradient.

For example, in the first sketch in Figure 12.17, ice grows in a thin capillary tube made from a material with a thermal conductivity higher than that of either water or ice (such as sapphire). The tube is cooled from the outside, thus reproducing the overall physical conditions in Hillig's experiment. As the ice grows upward into the supercooled water, latent heat is generated at the interface, which is then conducted way to the walls of the tube. The relatively low heat conductivity of water and ice results in a higher concentration of heat along the tube axis, as shown in the sketch. The highest temperature gradients at the ice/water interface, therefore, are along the radial direction (indicated by arrows).

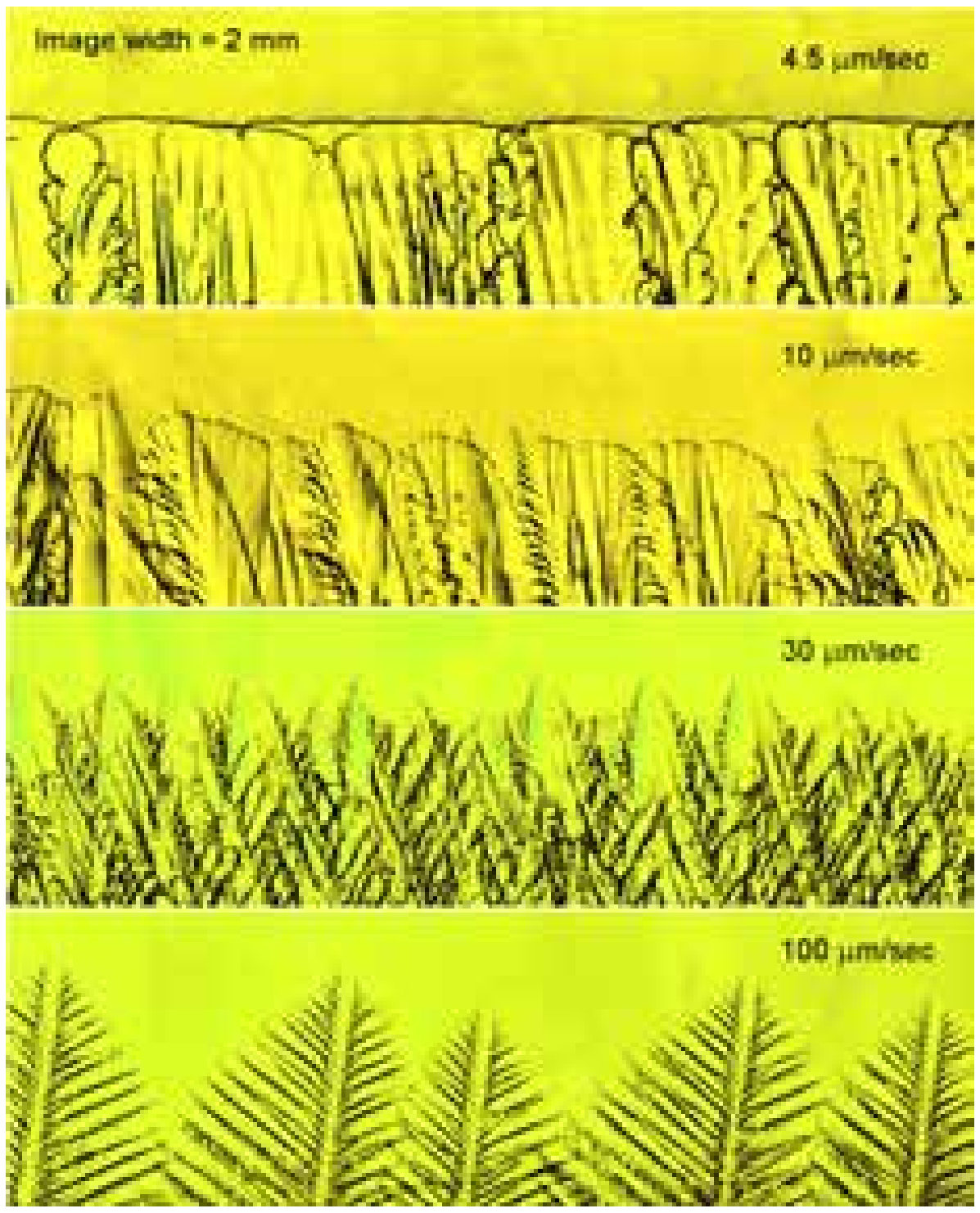

**Figure 12.16: Several photographs of ice growing in water held between two closely spaced (~200 μm) plastic plates. The plates were pulled at constant velocities (shown) between hot and cold regions, creating a temperature gradient from top (warm) to cold (bottom) within each panel.**

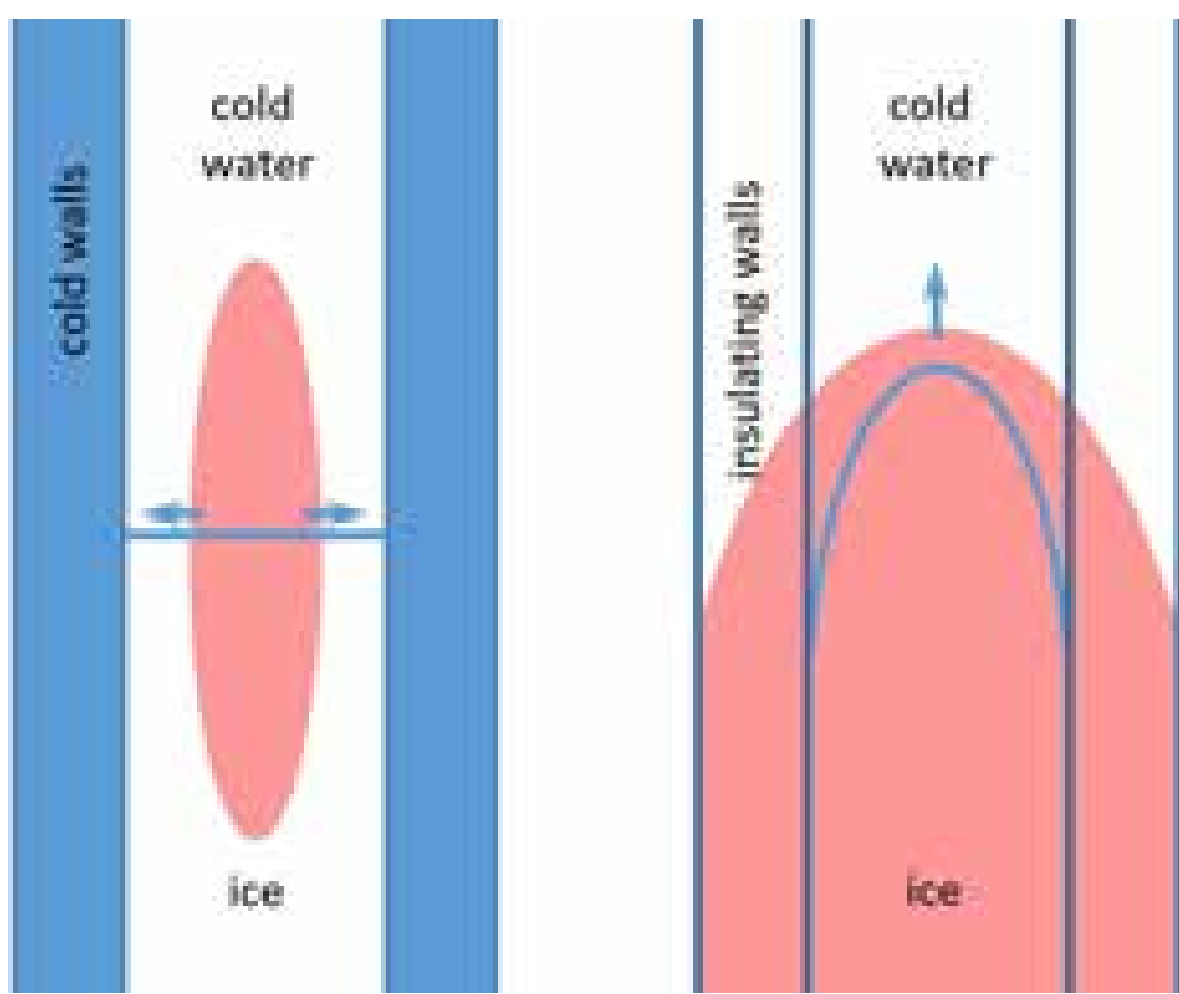

**Figure 12.17: A possible model to explain the orientation of ice growth in constrained channels, here a small-diameter capillary tube. In the left sketch, the tube conducts heat well and is cold. In the right sketch, the tube walls are insulating. Red shows areas of concentrated heating from latent heat deposition for the two cases. The arrows show the maximum temperature gradients at the growing ice surfaces.**

Assuming the stated conjecture, the random occurrence of dislocations during the growth process tends to align the fastest ice-growth vectors with the highest temperature-gradient vectors. Because the prism surfaces grow much faster than the basal surfaces, this process ends up aligning the basal growth direction with the central axis of the tube, in agreement with Hillig's observations.

In the second sketch in Figure 12.17, the thermal conductivity of the tube walls is now lower than that of either water or ice. The generated latent heat now conducts preferentially into the cold water, so the steepest temperature gradient is aligned along the tube axis. Once again, assuming the stated conjecture, this aligns the fastest growing crystalline a-axis with the central axis of the tube. The planar geometry in Figure 12.17 is not quite the same as the tubes in Figure 12.18, but the conjecture explains the observed growth behavior nevertheless.

A potential use of this phenomenon could be in future experiments investigating growth rates and growth morphologies of ice crystals in well controlled conditions, for comparison with detailed numerical models. Achieving high precision in such experiments will require creating well-defined initial conditions, and the easily implemented dislocation-mediated reorientation phenomenon described here may be a useful tool, as it clearly was in Hillig's measurements of the basal attachment kinetics.

# Appendix A
## List of Variables and Physical Constants

$a$ – Size of a water molecule:

$$a \approx c_{ice}^{-1/3} \approx 0.32 \text{ nm}$$

$a_0$ – Ice lattice parameter

$A$ – Nucleation parameter

$\alpha$ – Kinetics attachment coefficient

$\alpha_{basal}$ – Attachment coefficient on a basal faceted surface

$\alpha_{diff}$ – Attachment coefficient for diffusion-limited growth of a sphere

$$\alpha_{diff} = X_0/R$$

$\alpha_{heat}$ – Substrate heat-flow coefficient

$\alpha_{diff,heat}$ – Heat diffusion effective attachment coefficient

$\alpha_{diffcyl}$ – Attachment coefficient for diffusion-limited growth of a cylinder

$\alpha_{facet}$ – Attachment coefficient on a faceted surface

$\alpha_{prism}$ – Attachment coefficient on a prism faceted surface

$B$ – Cylinder/parabolic diffusion parameter:

$$B = \log(r_{far}/R)$$

$\beta$ – Step energy on an ice/vapor surface

$\beta_0$ – Fictitious "rigid terrace" step energy:

$$\beta_0 \equiv a\gamma_{sv}$$
$$\approx 3 \times 10^{-11} \text{ J/m}$$

$c$ – Water vapor number density

$c_0$ – Ice lattice parameter

$c_{ice}$ – Ice molecular number density:

$$c_{ice} = \rho_{ice}/m_{mol}$$
$$c_{ice} \approx 3.1 \times 10^{28} \text{ m}^{-3}$$

$c_{sat}$ – Saturated water vapor number density

$c_{sat,water}$ – Saturated water vapor density of supercooled water

$c_{surf}$ – Water vapor number density at the surface

$C_{diff}$ – 2nd derivative of $c_{sat}(T)$:

$$C_{diff} = \frac{1}{2}\frac{1}{c_{sat}}\frac{d^2 c_{sat}}{dT^2}$$

$c_{p,air}$ – Heat capacity of air:

$$c_{p,air} \approx 1.0 \text{ kJ/kg-K}$$

$c_{p,ice}$ – Heat capacity of ice:

$$c_{p,s} \approx 2.1 \text{ kJ/kg-K}$$

$c_{p,water}$ – Heat capacity of water:

$$c_{p,l} \approx 4.2 \text{ kJ/kg-K}$$

$c_{p,wv}$ – Heat capacity of water vapor:

$$c_{p,v} \approx 2.0 \text{ kJ/kg-K}$$

$\chi_0$ – Thermal parameter (dimensionless):

$$\chi_0 = \frac{\eta D L_{sv} \rho_{ice}}{\kappa_{air}} \frac{c_{sat}}{c_{ice}}$$

$D$ – Diffusion constant

$D_{air}$ – Particle diffusion constant in air:

$$D_{air} \approx 2 \times 10^{-5} \text{ m}^2\text{/sec}$$

$D_{therm}$ – Thermal diffusion constant in air:

$$D_{therm} \approx 2 \times 10^{-5} \text{ m}^2\text{/sec}$$

$D_{water}$ – Thermal diff constant in water:

$$D_{water} \approx 1.4 \times 10^{-7} \text{ m}^2\text{/sec}$$

$d_{sv}$ – Gibbs-Thomson length:

$$d_{sv} = \gamma_{sv}/c_{ice}kT$$
$$\approx 1 \text{ nm}$$

$\delta$ – Step density (number per unit length) on a faceted surface

$\eta$ – Logarithmic change in $c_{sat}$ with $T$:

$$\eta = (1/c_{sat})(dc_{sat}/dT) \quad \text{or} \ldots$$

$\eta$ – Part of the $(\eta, \xi, \varphi)$ parabolic coordinate system

$\eta_{eff}$ – Dynamical viscosity for liquid water

$F$ – Flux

$f_{\#}$ – Lens f-number

$\gamma$ – Surface energy

$\gamma_0$ – Ice solid/vapor surface energy of rough, unfaceted surface

$\gamma_{lv}$ – Ice liquid/vapor surface energy:

$$\gamma_{lv} \approx 76 \text{ mJ/m}^2$$

$\gamma_{sl}$ – Ice solid/liquid surface energy:

$$\gamma_{sl} \approx 30 \pm 5 \text{ J/m}^2$$

$\gamma_{sv}$ – Ice solid/vapor surface energy:

$$\gamma_{sv} \approx 106 \pm 15 \text{ mJ/m}^2$$

$k$ – Boltzmann constant:

$$k \approx 1.38 \times 10^{-23} \text{ m}^2 \text{ kg sec}^{-2} \text{ K}^{-1}$$
$$kT_{-15} \approx 3.6 \times 10^{-21} \text{ Joules}$$

$K_T$ – Kinetic coefficient for ice growth from liquid water

$\kappa$ – Surface curvature, or …

$\kappa$ – Thermal conductivity coefficient

$\kappa_{air}$ – Thermal conductivity of air:

$$\kappa_{air} \approx 0.025 \text{ W m}^{-1}\text{ K}^{1}$$

$\kappa_{ice}$ – Thermal conductivity of ice:

$$\kappa_{ice} \approx 2.3 \text{ W m}^{-1}\text{ K}^{1}$$

$\kappa_{water}$ – Thermal conductivity of water:

$$\kappa_{water} \approx 0.6 \text{ W m}^{-1}\text{ K}^{1}$$

$\kappa_{wv}$ – Thermal conductivity of water vapor:

$$\kappa_{wv} \approx 0.02 \text{ W m}^{-1}\text{ K}^{1}$$

$\ell$ – Latent heat per molecule, or …

$\ell$ – A generic length

$\ell_{vicinal}$ – Step spacing on a vicinal surface

$L$ – A generic length

$L_{sl}$ – Latent heat of melting (solid/liquid):

$$L_{sl} \approx 0.33 \times 10^{6} \text{ J/kg}$$

$L_{lv}$ – Lat. heat of evaporation (liquid/vapor):

$$L_{sl} \approx 2.5 \times 10^{6} \text{ J/kg}$$

$L_{sv}$ – Latent heat of sublimation (solid/vapor)

$$L_{sv} \approx 2.8 \times 10^{6} \text{ J/kg}$$

$\lambda$ – Wavelength of light

$m_{mol}$ – Mass of water molecule:

$$m_{mol} \approx 3.0 \times 10^{-26} \text{ kg}$$

$\mu$ – Dynamical viscosity of air:

$$\mu \approx 1.8 \times 10^{-5} \text{ kg/(m-sec)}$$

$N_A$ – Lens numerical aperture

$\nu_{kinematic}$ – Kinematic viscosity of air:

$$\nu_{kinematic} = \mu/\rho_{air} \approx 1.4 \times 10^{-5} \text{ m}^2\text{/sec}$$

at 1 atm pressure

$P_{ice}$ – Ice equilibrium vapor pressure

$P_{water}$ – Water equilibrium vapor pressure

$R$ – Generic radius

$R_{tip}$ – Tip radius of parabolic dendrite

$R_{xy}$ – Optical resolution (resolving power)

$R_z$ – Optical depth-of-focus

$\rho_{air}$ – Mass density of air:

$$\rho_{air} \approx 1.2 \text{ kg/m}^3$$

$\rho_{ice}$ – Mass density of ice:

$$\rho_{ice} \approx 917 \text{ kg/m}^3$$

$\rho_{water}$ – Mass density of water:

$$\rho_{water} \approx 1000 \text{ kg/m}^3$$

$s_0$ – Solvability parameter

$\sigma$ – Supersaturation

$\sigma_0$ – Nucleation parameter

$\sigma_\infty$ – Supersaturation at infinity

$\sigma_{surf}$ – Supersaturation at the surface

$\sigma_{water}$ – The supersaturation of supercooled liquid water relative to ice:

$$\sigma_{water} = \frac{c_{sat,water} - c_{sat,ice}}{c_{sat,ice}}$$

$\Delta t$ – A small time step

$T$ – Temperature

$T_K$ – Temperature in Kelvin

$T_0$ – Ice/water nucleation parameter

$T_{-15}$ – 258 K (-15 C)

$T_\infty$ – Temperature at infinity

$T_m$ – The ice/water melting point (0 C)

$T_{sl}$ – solid/liquid transition temperature

$T_{surf}$ – Surface temperature

$\Delta T_{surf}$ – Surface supercooling

$u_{term}$ – Terminal velocity

$\theta_{surf}$ – Surface normal relative to a faceted surface

$v$ – Crystal growth velocity

$v_n$ – Crystal growth velocity normal to the surface

$v_{kin}$ – Kinetic velocity:

$$v_{kin} = \frac{c_{sat}}{c_{ice}} \sqrt{\frac{kT}{2\pi m_{mol}}}$$

$v_{mol}$ – Water vapor molecular velocity

$v_{tip}$ – Growth velocity of dendrite tip

$w$ – Width of a top prism terrace

$X_0$ – Characteristic diffusion length:

$$X_0 = \frac{c_{sat}}{c_{ice}} \frac{D}{v_{kin}}$$

In air at -15 C, $X_0 \approx 0.145\ \mu$m

$x_{basal}$ – Spacing between basal layers

$x_{prism}$ – Spacing between prism layers

$x_{diff}$ – Diffusion length for a water molecule on an ice surface

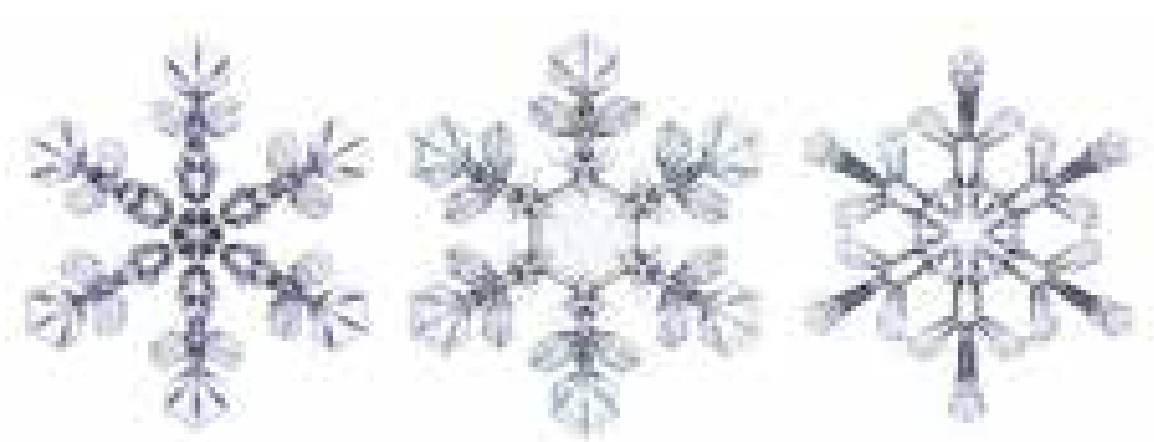